\documentclass{book}

\usepackage[greek, russian, german, french, italian, english]{babel}

\usepackage[T3, T1]{fontenc}
\usepackage[utf8]{inputenc}
\usepackage[italian]{varioref}

\usepackage{CJKutf8} 
\newcommand{\ZhSimplified}[1]{\begin{CJK}{UTF8}{gbsn}#1\end{CJK}}
\newcommand{\ZhTraditional}[1]{\protect\begin{CJK*}{UTF8}{bsmi}#1\end{CJK*}}
\newcommand{\Ja}[1]{\begin{CJK}{UTF8}{min}#1\end{CJK}}
    
\usepackage{todonotes}

\usepackage{titlesec}
\usepackage{titletoc}

\renewcommand{\footnotesize}{\scriptsize}
\renewcommand{\thefootnote}{\alph{footnote}}

\usepackage[perpage]{footmisc}

\newcommand{\symbolfootnote}[1]{%
\let\oldthefootnote=\thefootnote%
\stepcounter{mpfootnote}%
\addtocounter{footnote}{-1}%
\renewcommand{\thefootnote}{\fnsymbol{mpfootnote}}%
\footnote{#1}%
\let\thefootnote=\oldthefootnote%
}

\usepackage{emptypage}
\usepackage{microtype} 
\usepackage{fancyhdr}
\usepackage{indentfirst}
\usepackage{changepage}

\renewcommand{\chaptermark}[1]{\markboth{\thechapter. { }{#1}}{}}
\usepackage{empheq} 

\usepackage{tikz-cd}
\usetikzlibrary{decorations.fractals}
\usetikzlibrary{lindenmayersystems} 

\usepackage{booktabs} 
\usepackage{tabularx} 
\usepackage{longtable}
\usepackage{graphicx} 
\usepackage{pict2e, curve2e}
\usepackage{listings}
\usepackage{enumitem}

\usepackage[all]{xy}

\usepackage{caption}
\usepackage{xcolor}
\definecolor{eggplant}{HTML}{800080}
\definecolor{mallard}{HTML}{008080}
\definecolor{dusky-cerulean}{HTML}{004080} 
\definecolor{Chinese-loquat}{HTML}{F7C015}
\definecolor{pumpkin}{HTML}{FF7518}
\definecolor{cyan-blue}{HTML}{18A2FF}
\definecolor{vernal-green}{HTML}{03E364}
\definecolor{magenta}{HTML}{E30382}
\definecolor{cynara-violet}{HTML}{8803E3}

\usepackage{amsfonts}
\usepackage{amsmath}
\usepackage{amssymb}
\usepackage{amsthm}
\usepackage{mathrsfs}
\usepackage[scr]{rsfso} 
\usepackage{esstixfrak}
\usepackage{eufrak}
\usepackage{fixmath}

\usepackage{scalerel}
\newlength\bshft
\def\pseudobold#1{\ThisStyle{\ooalign{$\SavedStyle#1$\cr%
  \kern-\bshft$\SavedStyle#1$\cr%
  \kern\bshft$\SavedStyle#1$}}}
  
\usepackage[version=4]{mhchem}

\usepackage[bbgreekl]{mathbbol}
\DeclareSymbolFontAlphabet{\mathbb}{AMSb}
\DeclareSymbolFontAlphabet{\mathbbl}{bbold}
                        
\usepackage{mathtools}
\usepackage{latexsym}
\usepackage{siunitx}
\usepackage{icomma}
\usepackage{slashed}
\usepackage{braket}
\usepackage{xfrac} 

\usepackage{bbding}

\usepackage{textcomp}

\titleformat{\chapter}[display]
{\Large\filcenter}
{\bfseries\thechapter}
{0.3pc}
{\Large
\bfseries}
\assignpagestyle{\chapter}{empty} 

\titleformat{\section}[block]
{\large\itshape
\bfseries}
{\thesection.}{0.5em}{}

\titleformat{\subsection}[block]
{\normalsize\itshape
\bfseries}
{\thesubsection.}{0.5em}{}

\titleformat{\subsubsection}[block]
{\normalsize\itshape
\bfseries}
{\thesubsubsection.}{0.5em}{}

\contentsmargin{2.65em}

\titlecontents{chapter}
				[0pc]
				{\addvspace{.8pc}}
				{\bfseries\thecontentslabel{. }}
				{\hspace*{0em}}
				{\hfill\contentspage}
				[\addvspace{.3pc}]
				
\titlecontents{section}
				[0pc]
				{}
				{\thecontentslabel{. }}
				{\hspace*{0em}}
				{\titlerule*[0.4pc]{\thinspace.}\contentspage}

\titlecontents{subsection}
				[1.65pc]
				{}
				{\thecontentslabel{. }}
				{\hspace*{0em}}
				{\titlerule*[0.4pc]{\thinspace.}\contentspage}

\titlecontents{subsubsection}
				[3.95pc]
				{}
				{\thecontentslabel{. }}
				{\hspace*{0em}}
				{\titlerule*[0.4pc]{\thinspace.}\contentspage}

\newtheoremstyle{personal1}
  {\topsep}   
  {\topsep}   
  {\upshape}  
  {0pt}       
  {\itshape}  
  {.}         
  {5pt plus 1pt minus 1pt} 
  {\bfseries\thmname{#1}\thmnumber{ #2}\normalfont\thmnote{ \itshape(#3)}}
\theoremstyle{personal1}
\newtheorem{definitio}{Definition}[section]
\newtheorem{definitiones}{Definitions}[section]

\newtheoremstyle{personal2}
  {\topsep}   
  {\topsep}   
  {\itshape} 
  {0pt}      
  {\itshape} 
  {.}       
  {5pt plus 1pt minus 1pt} 
  {\bfseries\thmname{#1}\thmnumber{ #2}\normalfont\thmnote{ \itshape(#3)}} 
\theoremstyle{personal2}
\newtheorem{coniectura}{Conjecture}[section]
\newtheorem{corollarium}{Corollary}[section]
\newtheorem{lemma}{Lemma}[section]
\newtheorem{paradox}{Paradox}[section]
\newtheorem{problema}{Problem}[section]
\newtheorem{propositio}{Proposition}[section]

\newtheorem{subpropositio}{Subproposition}[section]
\newtheorem{theorema}{Theorem}[section]
\newtheorem{theoremata}{Theorems}[section]

\newtheoremstyle{personal3}
  {\topsep}   
  {\topsep}   
  {\upshape}  
  {0pt}       
  {\itshape} 
  {.}         
  {5pt plus 1pt minus 1pt} 
  {\thmname{#1}\thmnumber{ \itshape#2}\thmnote{ (#3)}} 
\theoremstyle{personal3}

\newtheorem{exemplum}{Example}[section]
\newtheorem{margo}{Margo}[section]
\newtheorem{quaestio}{Question}[section]
\newtheorem{scholium}{Scholium}[section]

\usepackage{float}

\graphicspath{figures}

\usepackage{url}

\usepackage{hyperref}
\hypersetup{colorlinks = true,
			citecolor = mallard,
			linkcolor = eggplant,
			urlcolor = dusky-cerulean}
               
\usepackage{endnotes}
\renewcommand*{\notesname}{\textbf{Endnotes}}
\makeatletter
\renewcommand*{\enoteheading}{%
\chapter{\notesname}%
\thispagestyle{empty}
\markboth{Endnotes}{Endnotes}%
\@afterindenttrue}
\makeatother
  
\makeatletter
\newif\ifenotelinks
\newcounter{Hendnote}
\let\savedhref\href
\let\savedurl\url
\def\endnotemark{%
\@ifnextchar[\@xendnotemark{%
\stepcounter{endnote}%
\protected@xdef\@theenmark{\theendnote}%
\protected@xdef\@theenvalue{\number\c@endnote}%
\@endnotemark
}%
}%
\def\@xendnotemark[#1]{%
\begingroup\c@endnote#1\relax
\unrestored@protected@xdef\@theenmark{\theendnote}%
\unrestored@protected@xdef\@theenvalue{\number\c@endnote}%
\endgroup
\@endnotemark
}%
\def\endnotetext{%
\@ifnextchar[\@xendnotenext{%
\protected@xdef\@theenmark{\theendnote}%
\protected@xdef\@theenvalue{\number\c@endnote}%
\@endnotetext
}%
}%
\def\@xendnotenext[#1]{%
\begingroup
\c@endnote=#1\relax
\unrestored@protected@xdef\@theenmark{\theendnote}%
\unrestored@protected@xdef\@theenvalue{\number\c@endnote}%
\endgroup
\@endnotetext
}%
\def\endnote{%
\@ifnextchar[\@xendnote{%
\stepcounter{endnote}%
\protected@xdef\@theenmark{\theendnote}%
\protected@xdef\@theenvalue{\number\c@endnote}%
\@endnotemark\@endnotetext
}%
}%
\def\@xendnote[#1]{%
\begingroup
\c@endnote=#1\relax
\unrestored@protected@xdef\@theenmark{\theendnote}%
\unrestored@protected@xdef\@theenvalue{\number\c@endnote}%
\show\@theenvalue
\endgroup
\@endnotemark\@endnotetext
}%
\def\@endnotemark{%
\leavevmode
\ifhmode
\edef\@x@sf{\the\spacefactor}\nobreak
\fi
\ifenotelinks
\expandafter\@firstofone
\else
\expandafter\@gobble
\fi
{%
\Hy@raisedlink{%
\hyper@@anchor{Hendnotepage.\@theenvalue}{\empty}%
}%
}%
\hyper@linkstart{link}{Hendnote.\@theenvalue}%
\makeenmark
\hyper@linkend
\ifhmode
\spacefactor\@x@sf
\fi
\relax
}%
\long\def\@endnotetext#1{%
\if@enotesopen
\else
\@openenotes
\fi
\immediate\write\@enotes{%
\@doanenote{\@theenmark}{\@theenvalue}%
}%
\begingroup
\def\next{#1}%
\newlinechar='40
\immediate\write\@enotes{\meaning\next}%
\endgroup
\immediate\write\@enotes{%
\@endanenote
}%
}%
\def\theendnotes{%
\immediate\closeout\@enotes
\global\@enotesopenfalse
\begingroup
\makeatletter
\edef\@tempa{`\string>}%
\ifnum\catcode\@tempa=12
\let\@ResetGT\relax
\else
\edef\@ResetGT{\noexpand\catcode\@tempa=\the\catcode\@tempa}%
\@makeother\>%
\fi
\def\@doanenote##1##2##3>{%
\def\@theenmark{##1}%
\def\@theenvalue{##2}%
\par
\smallskip
\begingroup
\def\href{\expandafter\savedhref}%
\def\url{\expandafter\savedurl}%
\@ResetGT
\edef\@currentlabel{\csname p@endnote\endcsname\@theenmark}%
\enoteformat
}%
\def\@endanenote{%
\par\endgroup
}%
\renewcommand*\@makeenmark{%
\hbox{\normalfont\@theenmark~}%
}%
\enoteheading
\enotesize
\input{\jobname.ent}%
\endgroup
}%
\def\enoteformat{%
\rightskip\z@
\leftskip1.8em
\parindent\z@
\leavevmode\llap{%
\setcounter{Hendnote}{\@theenvalue}%
\addtocounter{Hendnote}{-1}%
\refstepcounter{Hendnote}%
\ifenotelinks
\expandafter\@secondoftwo
\else
\expandafter\@firstoftwo
\fi
{\@firstofone}%
{\hyperlink{Hendnotepage.\@theenvalue}}%
{\makeenmark}%
}%
}%
\makeatother
\enotelinkstrue

\newcommand{\enumerationisinitium}{\begin{enumerate}[nolistsep, wide, label = \textnormal{($\mathnormal{\arabic{*}}$)}, ref = \textnormal{($\mathnormal{\arabic{*}}$)}]}
\newcommand{\enumerationisfinis}{\end{enumerate}}
\newcommand{\subenumerationisinitium}{\begin{enumerate}[nolistsep, wide, label = \textnormal{(\roman{*})}, ref = \textnormal{(\roman{*})}]}
\newcommand{\subenumerationisfinis}{\end{enumerate}}

\newenvironment{indent paragraph: 15pt}{%
  \par%
  \leftskip=15pt%
  \noindent\ignorespaces}{%
  \par}

\newenvironment{indent paragraph: 30pt}{%
  \par%
  \leftskip=30pt%
  \noindent\ignorespaces}{%
  \par}

\makeatletter
\newcommand{\xMapsto}[2][]{\ext@arrow 0599{\Mapstofill@}{#1}{#2}}
\def\Mapstofill@{\arrowfill@{\Mapstochar\Relbar}\Relbar\Rightarrow}
\makeatother

\DeclareSymbolFont{tipa}{T3}{cmr}{m}{n}
\DeclareMathAccent{\invertedbreve}{\mathalpha}{tipa}{16}

\DeclareMathOperator{\dAlembertian}{\square}
\DeclareMathOperator{\divergence}{div}
\DeclareMathOperator*{\esssup}{ess\,sup}
\DeclareMathOperator{\gradient}{grad}
\DeclareMathOperator{\Laplacian}{\bigtriangleup}
\DeclareMathOperator{\modulo}{mod}
\DeclareMathOperator{\subscript}{sub}
\DeclareMathOperator{\trace}{tr}

\DeclareRobustCommand{\background}{\textnormal{\hspace{0.6pt}\reflectbox{\textgreek{β}}}}
\DeclareRobustCommand{\reflectedepsilon}{\reflectbox{$\epsilon$}}
\DeclareRobustCommand{\textcyr}[1]{\foreignlanguage{russian}{#1}}
   
\newcommand{\adj}{\mathrm{adj}} 
\newcommand{\adjoint}{\overline}
\newcommand{\antimatter}{\overline}
\newcommand{\aprx}{\stackrel{\textnormal{\tiny{aprx}}}{=\joinrel=}}
\newcommand{\atmo}{\stackrel{\textnormal{\tiny{atmo}}}{=\joinrel=}}
\newcommand{\aut}{\mathfrak{aut}}

\newcommand{\Bessel}{J}
\newcommand{\binomcurly}{\genfrac{\{}{\}}{0pt}{}}
\newcommand{\BochnerRiesz}{B}
\newcommand{\cardinality}{\mathrm{card}}
\newcommand{\Chi}{\textgreek{\textit{Χ}}}
\newcommand{\Cl}{C\ell}
\newcommand{\corr}{\stackrel{\textnormal{\tiny{corr}}}{=\joinrel=}}
\newcommand{\couplingconstant}{{\textcyrillic{\textit{в}}_{(g)}}}
\newcommand{\Cotton}{C}
\newcommand{\cutofffunction}{%
  \text{\ooalign{\hidewidth\raisebox{0.0ex}{$\smallsetminus$}\hidewidth\cr$\jmath$\cr}}%
}
\newcommand{\Diracdelta}{\delta}
\newcommand{\diag}{\mathrm{diag}} 
\newcommand{\diameter}{\mathrm{diam}} 
\newcommand{\Diff}{\mathrm{Diff}}
\newcommand{\dist}{\mathrm{dist}}
\newcommand{\distance}{\bbrho}
\newcommand{\Einsteinconstant}{\textgreek{\text{κ}}}
\newcommand{\equival}{\stackrel{\tiny\textgreek{\textnormal{ιδ}}}{=}}
\newcommand{\EulerLagrange}{E}
\newcommand{\expansion}{\mathrm{exn}}
\newcommand{\F}{F}
\newcommand{\Galois}{\mathit{Gal}}
\newcommand{\geodesic}{\mathrm{geo}}
\newcommand{\gravitation}{G}
\newcommand{\Greenfunction}{\mathscr{G}}
\newcommand{\Hamiltonian}{\mathscr{H}}
\newcommand{\heatoperator}{%
  \text{\ooalign{\hidewidth\raisebox{0.4ex}{\tiny$\mathrm{h}$}\hidewidth\cr$\square$\cr}}%
}
\newcommand{\Hol}{\mathit{Ho\ell}}
\newcommand{\homologygroup}{\mathit{H}}
\newcommand{\hyperbolic}{\mathbb{H}}
\newcommand{\id}{\mathrm{id}}
\newcommand{\idem}{\mathbb{I}}
\newcommand{\idempotent}{\textcyrillic{\textit{и}}}
\newcommand{\indice}{\mathrm{index}}
\newcommand{\Inequality}{%
  \text{\ooalign{\hidewidth\raisebox{0.10ex}{$\smallfrown$}\hidewidth\cr$I$\cr}}%
}
\newcommand{\Jordancurve}{C_\textsc{j}}
\newcommand{\Kahlerpotential}{\rotatedvarpi}
\newcommand{\kernel}{\mathrm{ker}}
\newcommand{\Kochcurve}{C_\textsc{k}}
\newcommand{\Lagrangian}{\mathscr{L}}
\newcommand{\Langlands}{L}
\newcommand{\Lebesgue}{L}
\newcommand{\length}{\ell}
\newcommand{\lepton}{{\textgreek{\textit{λ}}}}
\newcommand{\Liederivative}{\pounds}
\newcommand{\Lorentz}{\textcyrillic{\textit{Л}}}
\newcommand{\Moebius}{\mathfrak{M\ddot{o}b}}
\newcommand{\momentum}{p}
\newcommand{\neutrino}{\nu}
\newcommand{\OrnsteinUhlenbeck}{\textgreek{\textit{Ϙ}}}
\newcommand{\overbar}[1]{\mkern 1.5mu\overline{\mkern-1.5mu#1\mkern-1.5mu}\mkern 1.5mu}
\newcommand{\pstroke}{\raisebox{-1.5ex}[\height][0pt]{$\mathchar'26$}\mkern-6mu p}
\newcommand{\prj}{\mathrm{prj}}
\newcommand{\quadrupole}{Q}
\newcommand{\quaternion}{\mathbbl{H}}
\newcommand{\rank}{\mathrm{rk}}
\newcommand{\refo}{\stackrel{\textnormal{\tiny{refo}}}{=\joinrel=}}
\newcommand{\Ricci}{R}
\newcommand{\Ric}{\mathrm{Ric}}
\newcommand{\Rie}{\mathrm{Rie}}
\newcommand{\Riemann}{R}
\newcommand{\Rorder}{%
  \text{\ooalign{\hidewidth\raisebox{0.19ex}{$-$}\hidewidth\cr$R$\cr}}%
	}
\newcommand{\rotatedell}{{\mathpalette\rotell\relax}}\newcommand{\rotell}[2]{\rotatebox[origin=c]{180}{$#1\ell$}}
\newcommand{\rotatedeta}{{\mathpalette\roteta\relax}}\newcommand{\roteta}[2]{\rotatebox[origin=c]{180}{$#1\eta$}}
\newcommand{\rotatedg}{{\mathpalette\rotg\relax}}\newcommand{\rotg}[2]{\rotatebox[origin=c]{180}{$#1g$}}
\newcommand{\rotatedgamma}{{\mathpalette\rotgamma\relax}}\newcommand{\rotgamma}[2]{\rotatebox[origin=c]{180}{$#1\gamma$}}
\newcommand{\rotatedheartsuit}{{\mathpalette\rotheartsuit\relax}}\newcommand{\rotheartsuit}[2]{\rotatebox[origin=c]{180}{$#1\heartsuit$}}
\newcommand{\rotatedm}{{\mathpalette\rotm\relax}}\newcommand{\rotm}[2]{\rotatebox[origin=c]{180}{$#1m$}}
\newcommand{\rotatedpi}{{\mathpalette\rotpi\relax}}\newcommand{\rotpi}[2]{\rotatebox[origin=c]{180}{$#1\pi$}}
\newcommand{\rotatedpsi}{{\mathpalette\rotpsi\relax}}\newcommand{\rotpsi}[2]{\rotatebox[origin=c]{180}{$#1\psi$}}
\newcommand{\rotatedPsi}{{\mathpalette\rotPsi\relax}}\newcommand{\rotPsi}[2]{\rotatebox[origin=c]{180}{$#1\Psi$}}

\newcommand{\rotatedtau}{{\mathpalette\rottau\relax}}\newcommand{\rottau}[2]{\rotatebox[origin=c]{180}{$#1\tau$}}
\newcommand{\rotatedtriangle}{{\mathpalette\rottriangle\relax}}\newcommand{\rottriangle}[2]{\rotatebox[origin=c]{180}{$#1\triangle$}}\newcommand{\scal}{\mathrm{scal}}
\newcommand{\rotatedupsilon}{{\mathpalette\rotupsilon\relax}}\newcommand{\rotupsilon}[2]{\rotatebox[origin=c]{180}{$#1\upsilon$}}
\newcommand{\rotatedvarpi}{{\mathpalette\rotvarpi\relax}}\newcommand{\rotvarpi}[2]{\rotatebox[origin=c]{180}{$#1\varpi$}}
\newcommand{\rotatedvarrho}{{\mathpalette\rotvarrho\relax}}\newcommand{\rotvarrho}[2]{\rotatebox[origin=c]{180}{$#1\varrho$}}
\newcommand{\rotatedw}{{\mathpalette\rotw\relax}}\newcommand{\rotw}[2]{\rotatebox[origin=c]{180}{$#1w$}}
\newcommand{\rotatedxi}{{\mathpalette\rotxi\relax}}\newcommand{\rotxi}[2]{\rotatebox[origin=c]{180}{$#1\xi$}}
\newcommand{\scalarcurvature}{\Ricci_\mathrm{s}}
\newcommand{\setsingularities}{\mathit{Sing}}
\newcommand{\sezione}{\textgreek{\textit{ς}}}
\newcommand{\sigmamodel}{\textgreek{\textit{σ}}}
\newcommand{\sigmaPauli}{\textgreek{\text{σ}}}
\newcommand{\simplex}{\mathbbl{s}}
\newcommand{\singularity}{{\mathpalette\rots\relax}}\newcommand{\rots}[2]{\rotatebox[origin=c]{180}{$#1s$}}
\newcommand{\Sobolev}{W}
\newcommand{\Spin}{\mathit{Spin}}
\newcommand{\splittingfieldK}{K^\mathrm{s}}
\newcommand{\splittingfieldL}{L^\mathrm{s}}
\newcommand{\surface}{S}
\newcommand{\symmetric}{\mathfrak{S}}
\newcommand{\tensorM}{M}
\newcommand{\tensorP}{P}
\newcommand{\topological}{\mathrm{top}}
\newcommand{\supp}{\mathrm{supp}}
\newcommand{\Tau}{\textgreek{Τ}}
\newcommand{\Taustroke}{\raisebox{-0.65ex}[\height][0pt]{$\mathchar'26$}\mkern-10mu\textgreek{Τ}}
\newcommand{\torus}{\mathbb{T}}
\newcommand{\Vitali}{V}
\newcommand{\viz}{\stackrel{\textnormal{\tiny{viz}}}{=}}
\newcommand{\volume}{\mathrm{vol}}
\newcommand{\Wey}{\mathrm{Wey}}
\newcommand{\Weyl}{W}

\makeatletter
\newcommand\incircbin
{%
  \mathpalette\@incircbin
}
\newcommand\@incircbin[2]
{%
  \mathbin%
  {%
    \ooalign{\hidewidth$#1#2$\hidewidth\crcr$#1\bigcirc$}%
  }%
}
\newcommand{\KulkarniNomizu}{\incircbin{\land}}
\makeatother

\makeatletter

\newcommand*\sNeg[2][0mu]{\Neginternal{#1}{\snegslash}{#2}}

\newcommand*\Neginternal[3]{\mathpalette\Neg@{{#1}{#2}{#3}}}
\newcommand*\Neg@[2]{\Neg@@{#1}#2}
\newcommand*\Neg@@[4]{%
  \mathrel{\ooalign{%
    $\m@th#1#4$\cr
    \hidewidth$\m@th#3{#1}\mkern\muexpr#2*2$\hidewidth\cr
  }}%
}

\newcommand*\negslash[1]{\m@th#1\not\mathrel{\phantom{=}}}
\newcommand*\snegslash[1]{\rotatebox[origin=c]{60}{$\m@th#1-$}}
\newcommand*\ssnegslash[1]{\rotatebox[origin=c]{60}{$\m@th#1{\dabar@}\mkern-7mu{\dabar@}$}}
\newcommand*\sssnegslash[1]{\rotatebox[origin=c]{60}{$\m@th#1\dabar@$}}
\makeatother 

\makeatletter
\newcommand\footnoteref[1]{\protected@xdef\@thefnmark{\ref{#1}}\@footnotemark}
\makeatother

\usepackage{accents}
\newcommand*\underdot[1]{%
\underaccent{\dot}{#1}}

\newcommand\definition[1]{
  \leavevmode\unskip\penalty9999 \hbox{}\nobreak\hfill
  \quad\hbox{#1}}
\newcommand{\definitiosymbol}{\definition{\tiny\JackStarBold}} 

\newcommand\example[1]{
  \leavevmode\unskip\penalty9999 \hbox{}\nobreak\hfill
  \quad\hbox{#1}}
\newcommand{\exemplumsymbol}{\example{\tiny\FourStar}} 

\newcommand\examples[1]{
  \leavevmode\unskip\penalty9999 \hbox{}\nobreak\hfill
  \quad\hbox{#1}}
\newcommand{\exemplasymbol}{\examples{\tiny\FourStar}} 

\newcommand\apostils[1]{
  \leavevmode\unskip\penalty9999 \hbox{}\nobreak\hfill
  \quad\hbox{#1}}
\newcommand{\margosymbol}{\apostils{\tiny\SixFlowerPetalRemoved}}  

\newcommand\question[1]{
  \leavevmode\unskip\penalty9999 \hbox{}\nobreak\hfill
  \quad\hbox{#1}}
\newcommand{\quaestiosymbol}{\question{\tiny\AsteriskThin}}  

\newcommand\scholion[1]{
  \leavevmode\unskip\penalty9999 \hbox{}\nobreak\hfill
  \quad\hbox{#1}}
\newcommand{\scholiumsymbol}{\scholion{$\diamond$}}

\addto{\captionsenglish}{\renewcommand{\bibname}{\textbf{Bibliography \\ 
	\normalsize List of Works Mentioned in the Text}
	}} 

\let\OLDthebibliography\thebibliography
\renewcommand\thebibliography[1]
{\OLDthebibliography{#1}
\setlength{\parskip}{0pt}
\setlength{\itemsep}{0pt plus 0.3ex}}

\usepackage[useregional]{datetime2}

\begin{document}
\frontmatter
\begin{titlepage}
    \centering
    {\LARGE{\textbf{Edoardo Niccolai}}\par}
    \vspace{1cm}
    {\Huge{\textbf{Notes in Pure Mathematics \& \\ Mathematical Structures in Physics}}\par}
\end{titlepage}

\begingroup
\small
\noindent \textsc{Current version}: \DTMnow \\
\textsc{The latest revision}: \emph{Download} page at \href{https://edoardoniccolai.com}{https://edoardoniccolai.com} \\

\endgroup

\thispagestyle{empty}
\vspace*{\fill}

\begingroup
\small
\noindent \textit{Notes in Pure Mathematics \& Mathematical Structures in Physics} \\
\noindent With 32 Figures

\endgroup

\vspace{1mm}

\begingroup
\footnotesize
\noindent © 2021-2023 · Edoardo Niccolai 

\endgroup

\vspace{6mm}

\begingroup
\footnotesize
\noindent The Greek typeface used on the next page in my \emph{motto}, as well as in the series of colored random letters, is called \texttt{GFS Gazise.otf}, by Greek Font Society (\textsc{gfs}), \textgreek{Σπ. Μερκούρη 33, 116 34, Αθήνα}. The Chinese typeface, for numbers \ZhTraditional{〇一} (01), and the words \ZhTraditional{內階一} (Ursæ Majoris, an egg hunt for my Muscìda, aka ‘Mumù’, viz. $\mu\mu$), \ZhTraditional{茶樹} (Tea tree/Camellia sinensis), \ZhTraditional{自然。} (Nature.), \ZhTraditional{黃龍} (Huanglong), \ZhTraditional{紫砂壺} (Yixing clay teapot), is known as \texttt{TieLan.ttf} font.

\endgroup

\newpage

\thispagestyle{empty}
\vspace*{\fill}

\begin{figure}
\centering
\includegraphics{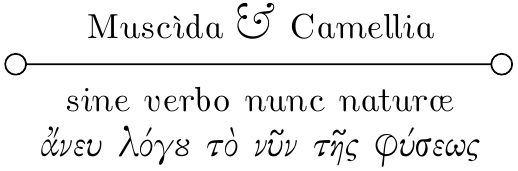}
\end{figure}

\begin{figure}[h]
\centering
\includegraphics{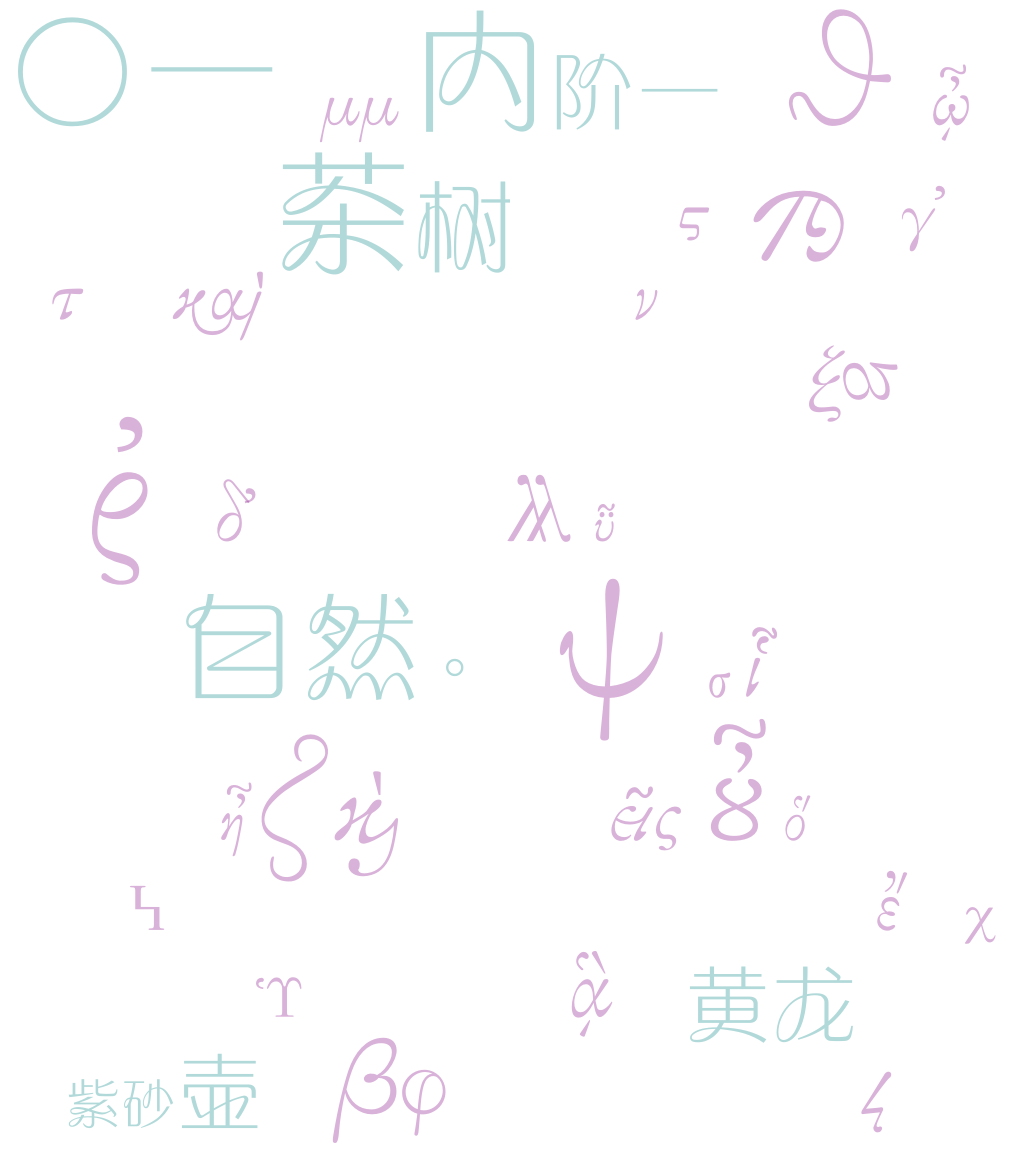}
\end{figure}


\begingroup
\hypersetup{linktocpage}
\chapter[\textbf{Dedication}]{Dedication}

\thispagestyle{empty}

\begingroup
\begin{center}
To all those who, scientists, literates, artists and musicians, \\
after going through many and obscure sorrows \& difficulties \\
because of shackles of an (un)civil society, \\
woefully ever more fragmented, \\
inhumanly technocratic, viz. blindly/dully bureaucratic, \\
to all those who, tired of the infamous living conditions,\footnote{
	P. Calderón de la Barca \cite[Jornada primera, Escena VIII, 910, p. 118]{Calderon de la Barca "Life Is a Dream / La vida es sueno}: «[\,\dots] que vida infame no es vida».
	} \\
resulting from the ineptitude of ignominious charlatans, of chatty politicians, \\
or just sick of being a gear deprived of a \emph{proper motion}, \\
of being a Kafkian rotating wheel—without its own sense \& a unique identity— \\
of a hideously larger mechanism, \\
have dared to follow Sophocles' words \cite[1225-1228,\footnote{
	Alternative numbering in this edition (\textsc{mdccclv}): 1220-1223.
	} 
pp. 149-150]{Sophocles "Oedipus at Colonus"}: \\
\textgreek{μὴ φῦναι τὸν ἅπαντα νι- \\
κᾷ λόγον· τὸ δ᾽, ἐπεὶ φανῇ, \\
βῆναι κεῖθεν, ὅθεν περ ἥ- \\
κει, πολὺ δεύτερον ὡς τάχιστα.}\footnote{
	«Not to be born \\
	wins every gain [\textgreek{λόγον}]; but once in existence, \\
	go back fleetly from where one came, \\ 
	it is certainly the best remedy», or, more literally, «the best next [step]».}\textsuperscript{,}\footnote{
	Even before Sophocles's \textgreek{μὴ φῦναι}, there is that of Theognis \cite[\textgreek{Ελεγείων Α´}, 425-428, pp. 46, 48]{Theognis "Poemes elegiaques"}: «Of all things the best for a earthly men is not to be born / nor to see the rays of the blazing Sun, / [but] once he is born [it is best] to pass the gates of Hades as quickly as possible / and lie under a sizable heap of earth (\textgreek{Πάντων μὲν μὴ φῦναι ἐπιχθονίοισιν ἄριστον / μηδ᾽ ἐσιδεῖν αὐγὰς ὀξέος ἠελίου, / φύντα δ᾽ ὅπως ὤκιστα πύλας Ἀίδαο περῆσαι / καὶ κεῖσθαι πολλὴν γῆν ἐπαμησάμενον})».

	In the Roman antiquity, the unfussy \& veracious Greek wisdom finds new resonance in M.T. Cicero \cite[Liber I, 48, 114, p. 71]{Cicero "Tusculanarum disputationum"}: «[N]on nasci homini longe optimum esse, proximum autem quam primum mori».
	} \\
\end{center}
\endgroup
\vspace*{\fill}

\newpage\null\thispagestyle{empty}\newpage

\chapter[\textbf{Tribute}]{Tribute}

\thispagestyle{empty}

\begingroup
\begin{center}
to Muscìda ‘Mumù’, \\
'cos, into the delight of her presence, \\
the intuitions of these pages \\
have taken the convenient consistency and form; \\
no words, no equations, \\
and yet a deeper delight, already within the laws of nature; \\
supreme incarnation of every epic of \textgreek{φύσις}.
\end{center}
\endgroup

\vspace*{\fill}

\newpage\null\thispagestyle{empty}\newpage
\tableofcontents
\thispagestyle{empty}
\endgroup

\chapter[\textbf{Précis}]{Précis}
\markboth{Précis}{Précis}

This book is divided into two but closely related parts. The first part (Chapters 1-19), mathematically rigorous, offers an insight into variegated topics of pure mathematics and mathematical physics, where the themes of \emph{dimensional continuum} and \emph{discreteness} intertwine, to disappear and reappear as in a karstic flow, passing through some crucial problems, such as the concept of \emph{point}, of \emph{infinity}, and that of mathematical and physico-mathematical \emph{primitiveness}. This part works a bit like a \textgreek{προπαιδεία}, a preparatory road: a comprehension of mathematics \& mathematical physics needfully entails an «active experience in mathematics itself», unescorted by imbroglios, hocus-pocus, and fuggy quibbles. Scilicet: the best way to learn mathematics is to do it, without being harnessed/cheated by fumesophy (cf. Chapter \ref{chapter "Outro—Parva Mathematica: Libera Divagazione 4/8"}).

The second part (Chapters 20-27, which are all linked to the Intro, plus a special and short closing Chapter, entitled \textit{Ulterius Elementum in Cauda}) is nevertheless tightly tied into the first, and it is more accessible to non-experts in equational language. It is a rich reflection on the jagged concept of mathematics (one among very many issues, is the \textit{vexata quæstio}: is mathematics invented or discovered? Or both? Or rather, is it a form of \emph{art}, id est \textgreek{τέχνη}, as I believe?), and on the relationships, more or less paradoxical, between mathematics and physics, \& natural sciences (e.g. what is the much-talked-about “connection” between mathematics and nature?).

It is only by means of this kind of musing, and by finding an answer to certain related questions—primarily, the preponderant role of the \emph{creative imagination/inspiration}, and the prominence of the personal \emph{intuition} rooted into the abyssality of the unconscious, or what is hidden behind words such as “analogy”, “dream”, “beauty”, “faith”,  “figment”, “metaphor”, and “myth”, in mathematics \& in the mathematical structures in physics—that the first part of the book makes sense, so as to have a \emph{more truthful understanding} of what one does when one does mathematics. References to literature, art, and music are not lacking, with the purpose to reaffirm that the “two cultures” (scientific and humanistic professions) are intimately knotted, and, at times, the same thing.

Otherwise, in physics and in the physico-mathematical area, there is the risk of easily falling into the sinkholes of, say, a conglomerate of fashionable theories, accompanied by \textit{balocchi meccanici},\footnote{
	Collodian-tasting expression, which stands for “mechanical toys”, or rather,  “(quantum) mechanical model”, “(quantum) mechanical formalism”, and so on.
	} 
and of getting caught in their, often hysterical, debates on the “images” (\textgreek{φαντάσματα}) or “representations” of nature, which may be phantasmagoric, false and distorted.

\setcounter{secnumdepth}{0}  
\chapter[\textbf{Intro}]{Intro}
\markboth{Intro}{Intro}

\begingroup
\footnotesize
We know too much for one man to know much, we live too variously to live as one. \\
\indent — \textsc{J.R. Oppenheimer} \cite[p. 44]{Oppenheimer "Prospects in the Arts and Sciences"}

\vspace{2mm}

[I]t is difficult, but interesting to master ten percent of the information and specific methods in any field of the [mathematical, physical \& natural] sciences, but this is essential in order to begin independent work, or at least to calmly get oriented. Further, the path from ten to ninety percent understanding is pure pleasure and genuine creativity. And to go through the next nine percent is infinitely difficult, and far from everyone's ability. The last percent is hopeless. It is more reasonable to switch to a new problem before it is too late, and have the joy of continuous creation. \\
\indent — \textsc{Ya.B. Zeldovich} based on the \textsc{R.A. Sunyaev}'s memories \cite[p. 238]{Sunyaev "When we were young"}

\endgroup

\subsection{A Kaleidoscope-Opus, and the Briar-patch of Mathematics}

\begingroup
\footnotesize
The name Kaleidoscope, which I have given to a new Optical Instrument, for creating and exhibiting beautiful forms, is derived from the Greek words \textgreek{καλός}, \emph{beautiful}; \textgreek{εἶδος}, \emph{a form}; and \textgreek{σκοπέω}, to see. \\
\indent — \textsc{D. Brewster} \cite[p. 1]{Brewster "A Treatise on the Kaleidoscope"}\footnote{
	The Brewster's \textit{Treatise on the kaleidoscope} bears the Latin motto \textit{Nihil tangit quod non ornat}.
	}

\vspace{2mm}
\textgreek{[Μ]ὴ εἶναι βασιλικὴν ἀτραπὸν ἐπὶ γεωμετρίαν} · There [is] no royal road to geometry. \\
\indent — \textsc{Euclid} in \textsc{Proclus} commentary on the first Book of Euclid's \textit{Elements} \cite[p. 154]{Various authors "Greek Mathematical Works I: from Thales to Euclid"}

\vspace{2mm}

In Mathematicis spinetis via verè Regia\,\dots \\
\indent — \textsc{E. Torricelli} \cite[\textit{Quadratura Parabolæ}, p. 56]{Torricelli "De Dimensione Parabolae"}

\endgroup

\vspace{2mm}
 
\enumerationisinitium
\item These \textit{Notes} are a prodigal act of individual freedom; they represent a personal pleasure in \emph{roaming} along the lands of thought.\footnote{
	\label{footnote "Ravings of bibliometrics"}
	And it could not be otherwise: I totally shun the rules for measuring the research “productivity” and “citation impact”, aka the dopey dictate of “publish or perish”, through the ravings of bibliometrics (under the auspices of agencies such as the \textsc{anvur}); see e.g. \cite{Adler Ewing Taylor "Citation Statistics"} \cite{Molinie and Bodenhausen "Bibliometrics as Weapons of Mass Citation La bibliometrie comme arme de citation massive"} \cite{Ernst "The Follies of Citation Indices and Academic Ranking Lists: A Brief Commentary to 'Bibliometrics as Weapons of Mass Citation'"} \cite{Arnold and Fowler "Nefarious Numbers"} \cite[appendice]{Israel "Chi sono i nemici della scienza?"}. 
	
	In this regard, it is fair to talk of—it is a provocation—«cretinoid idiocy» that strangles the circus of a large part of science. The latter is an expression of the language of medicine, see e.g. A. Mitchell \cite[note 20, p. 177]{Mitchell "Notes"} or C. Taruffi \cite{Taruffi "Intorno ad un idiota cretinoide"}.
	}
\item As a result of the above point, I am constantly looking (\textgreek{σκοπέω}) for \emph{immeasurable} shapes \& forms (\textgreek{εἶδος}) of beauty (\textgreek{καλός}); it is precisely a \emph{kal-eîdo-skopéō} view, under which mathematics is a piece of art (in the original Greek meaning: \textgreek{τέχνη}, tékhnē), and art is a piece of mathematics, and together they create a multiplicity of symmetrical patterns \emph{randomly} (see below).
\item Fortunately, \emph{there is no} a Royal road in the briar-patch\footnote{
	In Italian it sounds better: \textit{spinaio}, or \textit{ginepraio} (juniper thicket).
	} 
of mathematics (\textit{in Mathematicis spinetis via verè Regia}), to adopt, and negate, a sentence of Torricelli.\footnote{
	But Torricelli also knew it well: Euclid docet.
	} 
The image of the kaleidoscope leaps out from here.
\enumerationisfinis

\subsection{Math-art: the Truthfulness of a Lie}

\begingroup
\footnotesize
We all know that [mathematics] is not truth. [Mathematics] is a lie that makes us realize truth, at least the truth that is given us to understand. \\
\indent — Modified sentence of \textsc{P. Picasso} in \cite[p. 10]{Barr Jr. (Ed.) "Picasso Forty Years of his Art"}\footnote{
	The original Picasso's sentence is: «We all know that art is not truth. Art is a lie that makes us realize truth, at least the truth that is given us to understand. The artist must know the manner whereby to convince others of the truthfulness of his lies. If he only shows in his work that he has searched, and re-searched, for the way to put over his lies, he would never accomplish anything». We replaced the word \emph{art} with the word \emph{mathematics}, but the sentence is equally valid.
	}

\endgroup

\vspace{2mm}

Mathematics and art share, through intuition and imagination, the ability to \emph{generate concepts and forms} for the understanding of what surrounds us. They are our main organ, or instrument (\textgreek{ὄργανον}) of knowledge, capable of deciphering the world, and connecting several objects and events simultaneously. This aspect is deepened without qualms by L. Boi \cite[pp. 8, 12]{Boi "Pensare l'impossibile. Dialogo infinito tra arte e scienza"}; he writes: 

\vspace{2mm}

\begingroup
\footnotesize
A mathematician uses a conceptual instrument to create mathematical objects essentially in the same way that an artist, e.g. a painter, uses a brush or a textile fabric to create a piece [of art]. The transforming intuition of space and a certain introspective “vision” of objects, as events imbued with movement and meaning, constitute a common territory for mathematics and art, where they discreetly and intensely meet and discover some of their deep affinities. 

It can be said that in mathematics, in contrast to the experimental sciences, there is something profound that unites it to the free artistic creation. Both develop a certain sensitive singularity, which resides in the ability to generate and organize forms. The meaning of artistic concepts, as well as that of mathematical concepts, is done and undone, is modified and reconfigured continuously, in conjunction with the generation and organization of the forms [\,\dots]. 

[A]rt and mathematics have an authentic hermeneutic function: both are a form of knowledge of objects and events; both contribute to reveal a part of their intrinsic history, objectifying themselves in a human culture and symbolic practice [\,\dots]. [B]oth mathematics and art offer a plurality of points of view of reality, alternative ways of “looking” at the world. The mathematical and artistic language reveal to us something about the “interiority” of our space which is invisible to our eyes, and at the same time open us up to a possible horizon on spaces that go beyond our three-dimensional physical world.

\endgroup

\vspace{2mm}

The closeness between mathematics and art (cf. Section \ref{subsection "Math-art: Prigioni of the Mind: a Suggestion by Weil"}) springs from man's ability to dream, or from his predisposition to inventiveness, from the ability to give birth to concepts with an \textit{inventŭs} (cf. footnote \ref{footnote "invention/creation and discovery"}, p. \pageref{footnote "invention/creation and discovery"}), that is, with a \emph{stratagem}, that of mathematics/art, in fact.

\subsection{Fantasy, or the Ability to Dream}

\begingroup
\footnotesize
The spacers will create in the spaces and through the spaces the new fantasies of [mathematics]. \\
\indent — Modified sentence of \textsc{L. Fontana} \cite{Fontana "Invito Mostra 1949"}\footnote{
	The original words of Fontana are: «Gli spaziali creeranno negli spazi e attraverso gli spazi le nuove fantasie dell'arte», in \textit{Invitation} to the Exhibit Inauguration, \textit{L'ambiente spaziale di Lucio Fontana}, Galleria del Naviglio, sabato 5 febbraio 1949. Again: by exchanging \emph{mathematics} for \emph{art}, the meaning of the slogan is not altered.
	}

\endgroup

\vspace{2mm}

Mathematics, first and foremost, arises from a \emph{free creation}; it is \emph{fantasy}, a manifestation of the \emph{dream}, as we shall see more thoroughly in Chapters \ref{chapter "Outro—Parva Mathematica: Libera Divagazione 1/8"} \& \ref{chapter "Outro—Parva Mathematica: Libera Divagazione 2/8"}, and Section \ref{subsection "Inexplicability and Ideality of Forms: in the Homeland of the Dream"}. There is a passage from C. Casolo \cite{Casolo "Matematica e sogni nella letteratura"} that may be useful to evoke  into this matter:

\vspace{2mm}

\begingroup
\footnotesize
If mathematics has been used to ensure support for a fantasy or a dream, it has happened that, although perhaps more rarely, the reciprocal function has also been practiced. Instead of calling the certainty of the rational proposition to underpin the dream and the imaginary, assuring them a visibility and plausibility, it is the [dream and imaginary] territory that is called to give citizenship, more precisely asylum, to those rational forms or truths, which would otherwise be taken into account with greater difficulty.

\endgroup

\vspace{2mm}

This is because mathematics is something much more than a logical-formal operation: it is \textgreek{φαντασία},\footnote{
	Verbal noun of \textgreek{φαντάζω}, “make visible, or present to the mind”, and \textgreek{φαίνω}, “cause to appear”, “reveal”, “disclose”.
	} 
which is literally an “appearing”, a “(creative) imagination”, intended as a faculty of the human mind to create images, to represent appearances, things, or facts, with the creation e.g. of multidimensional spaces, of algebraic or topological structures, or even of new concepts and symbols to underset the genesis of a new idea. Moreover, semantic—often ambiguous—nuances occur, and \emph{overflow} as a sign of freedom of the creative act.

\subsection{The Bearable Lightness of (Mathematical) Creativity}

It is nice to apprehend that the creative temper also hovers over jubilantly rigorous intellects; to give one example, G. Peano's name comes out without delay. Here is his great granddaughter L. Romano collects her memories \cite[pp. 9-10, e.m.]{Kennedy "Peano. Storia di un matematico"}:

\vspace{2mm}

\begingroup
\footnotesize
[T]he unrepeatable originality of [G. Peano's] nature was still in this: that the rigor of the mind[,]\endnote{
	«Mathematical rigor is very simple», Peano writes \cite[p. 275]{Peano "Sui fondamenti dell'analisi"}: «It consists in affirming true statements and in not affirming what we know is not true. It does not consist in affirming every truth possible. Science, or truth, is infinite; we know only a finite, and [an] infinitesimal part of it compared to the whole». The problem then becomes this: what is a “true statement”, in the restriction of mathematical rigor? Have a look at Section \ref{subsection "Peano's, Enriques', and Pieri's Non-defined Source Geometry"}.
	} 
was accompanied by another equally rare peculiarity: \emph{fantasy}.

Poets are men who have not lost the \emph{ability to wonder \textnormal{[}which belongs\textnormal{]} to childhood}; well, even true scientists—\emph{creators}—enjoy this privilege. In fact, \emph{science starts in wonder like art}. Albert Einstein wrote: “The study and in general the pursuit of truth and beauty is a sphere of activity [\textit{Gebiet}] in which we are permitted to remain children all our lives”.\endnote{
	This aphorism by Einstein is written in the diary of Adriana Enriques, daughter of Federigo, in memory of their meeting. It was Adriana who welcomed Einstein at the Bologna railway station, on the occasion of the Bolognese lectures (22, 24 and 26 October 1921) by the scientist from Ulm, cf. point ($\mathnormal{5}$) in endnote \ref{endnote "Contribution by Ricci and Levi-Civita in Einstein's theory of gravitation"}. The original sentence appearing in her diary is this:

	\setlength\parindent{8pt}
	“Das Studium und allgemein das Streben nach Wahrheit und Schönheit ist ein Gebiet, auf dem wir das ganze Leben lang Kinder bleiben dürfen.

	\vspace{1mm}
	
	Adriana Enriques zum Andenken an die Bekanntschaft von Oktober 1921

	\vspace{1mm}
	
	\hspace{180pt} Albert Einstein”.
	}

These researchers called Platonically “of truth and beauty” have something that makes them similar to children: a readiness, actually a passion for playing. Because \emph{the creative spirit is light}.

Peano possessed this wonderful gift: an affinity with children that led him to understand them. This is generally denied to the various pedagogues, rhetoricians, sadistic-sentimentalists. He \emph{really} knew how to play with a child. He put on larky speed competitions rushing down the stairs of the building, with a child who lived in the garret above his apartment on the fourth floor.\endnote{
	Original It. version: «[L]'originalità irri­petibile della natura [di G. Peano] era ancora in questo: che il rigore della mente si accompagnava a un'altra peculiarità altrettanto rara: la fantasia.

	\setlength\parindent{8pt}
	I poeti sono uomini che non hanno perso la facoltà di mera­vigliarsi [che è] propria dell'infanzia; ebbene, anche gli scienziati veri — creatori — godono di questo privilegio. Infatti la scienza nasce dalla meraviglia come l'arte. Albert Einstein ha scritto: “Lo studio e la ricerca della verità e della bellezza rappresentano una sfera di attività in cui è permesso di rimanere bambini per tutta la vita”.

	Questi ricercatori detti platonicamente “della verità e della bellezza” hanno qualcosa che li apparenta ai bambini: una disponibilità, anzi una passione per il gioco. Perché \emph{lo spirito creativo è leggero}.

	Peano possedeva questa dote meravigliosa: un'affinità con i bambini che lo portava a capirli. Cosa negata in genere ai vari pedagoghi, retori, sadico-sentimentali. Lui sapeva giocare \emph{dav­vero} con un bambino. Faceva allegre gare di velocità correndo giù dallo scalone del palazzo, con un bambino che abitava nelle soffitte sopra il suo alloggio del quarto piano». 
	}

\endgroup

\subsection{Metaphorical Procedure: the Example of the Crystals}
\label{subsection "Metaphorical Procedure: the Example of the Crystals"}

\begingroup
\footnotesize
Their ideas [of people of Laputa] are perpetually conversant in lines and figures. If they would, for example, praise the beauty of a woman or any other animal, they describe it by rhombs, circles, parallelograms, ellipses, and other geometrical terms, or else by words of art drawn from music [\,\dots]. I observed in the King's kitchen all sorts of mathematical and musical instruments, after the figures of which they cut up the joints that were served to his Majesty's table [\,\dots]. They are very bad reasoners and vehemently given to opposition, unless when they happen to be of the right opinion, which is seldom their case. Imagination, fancy, and invention, they are wholly strangers to, nor have any words in their language by which those ideas can be expressed. \\
\indent — \textsc{J. Swift} \cite[III, p. 165]{Swift "Travels into Several Remote Nations of the World"} 

\endgroup

\vspace{2mm}

\enumerationisinitium
\item The aforesaid themes of inventiveness, fantasy, and dream, which are denotative of mathematics, will be looked at carefully in Chapters \ref{chapter "Outro—Parva Mathematica: Libera Divagazione 1/8"} \& \ref{chapter "Outro—Parva Mathematica: Libera Divagazione 2/8"}, \ref{chapter "Outro—Parva Mathematica: Libera Divagazione 5/8"}, \ref{chapter "Outro—Parva Mathematica: Libera Divagazione 6/8"}, \ref{chapter "Outro—Parva Mathematica: Libera Divagazione 8/8"}. For now, it is sufficient to condense my stance by asserting that mathematics is a \emph{metaphor} of the regularity of nature;\footnote{
	\label{footnote "Metaphor"}
	By \emph{metaphor} we mean what is recurrently written in Dictionaries: a veiled “similarity”, an “analogical relationship” between images, a, say, psycholinguistic process through which two different ideas are \emph{associated}, so there is a symbolic “transposition” of images, of ideas, from the noun \textgreek{μεταφορά}, which comes, in turn, from  the vb. \textgreek{μεταφέρω}, “transport”.
	} 
it is an \emph{artefact} emerging from the possibilities of human perception. 
\item Let us take the example of the crystals, which are considered to be the prototypes of symmetry, with their patterns of lattice symmetry; but no crystal is perfect, and each of them has some imperfection; there are crystal-patterns, or crystallographic regularities, together with their defects, which we read \emph{as} mathematically corresponding to this or that crystallographic group, or crystal class. There is no mathematics in the crystals: no crystalline solid was formed because it performs—we pick two sets drawn at random—the abstract group: $\mathbb{V} \cong \mathbb{Z}_2 \times \mathbb{Z}_2$, or $\mathbbl{D}_6 \cong \mathbb{Z}_3 \rtimes \mathbb{Z}_2$. 

The first one ($\mathbb{V}$) is the Klein four-group \cite[Kap. I, § 5, \textit{Die Vierergruppe}]{Klein "Vorlesungen uber das Ikosaeder und die Auflosung der Gleichungen vom funften Grade"} of order 4; it is a finite non-cyclic abelian group; in the crystallographic area, it fulfils the centrosymmetric, enantiomorphic, or polar point symmetry, with prismatic, rhombic-disphenoidal, or rhombic-pyramidal class, respectively; the family to which it belongs is the monoclinic or orthorhombic system. The other set ($\mathbbl{D}_6$) is the dihedral group of order 6; in the crystallographic area, it fulfils the enantiomorphic or polar point symmetry, with trigonal-trapezohedral or ditrigonal-pyramidal class, respectively; the family to which it belongs is the hexagonal/trigonal system.
 
 These groups are only expressions of a model, a pattern, or a \emph{reading category}, for which they are our \emph{construction}, nay, they are an illusion, an \emph{artifice} of our mind, adequate for representing the regularities of the crystals. 
\item The cornerstone is that the various symmetries of the universe (cf. Section \ref{subsection "Anthropoid Ways Ib. Symmetry and Invariance in Physics: the Impact of Group Theory"}), videlicet, the regularities of nature, from which mathematics takes its lifeblood, do not exist because underneath there is a mathematical \textgreek{σκελετός} that supports them; mathematics is solely the “sieve” with which we read some regularities, or uniformities, of the world around us. If we overturn this concept, and believe that the bone structure of the universe is mathematical, we fall into Swift's satire, when Gulliver meets the loony inhabitants who live on the flying island of Laputa. This debases the imaginative and fanciful part of mathematics, which is its vital nervation, and transforms it into a comically arid discipline, closed in on itself, in a \emph{cocoon-like} tenet.
\item The very concept of symmetry,\footnote{
	The Gr. \textgreek{συμμετρία} (from \textgreek{σύν} plus \textgreek{μέτρον}) is for “commensurability”, “due proportion”. 
	} 
when it fades into abstract dimensions \textit{via mathematica}, is openly a \emph{human invention}, because it is part of \emph{our sensitivity}, it is \emph{our manner of understanding} some natural rules, with the most repetitive, periodic, or harmonic motifs. As the crystallographer A.L. Mackay \cite[p. 22]{Mackay "Generalised Crystallography"} points out: 
\vspace{2mm}

\begingroup
\footnotesize
We have a Pythagorean strain in our culture which has continually made congenial the idea that somehow the symmetrical geometrical figures—the Five Platonic Solids in particular—are at the bottom of things. This attitude was caricatured in Swift's \textit{Gulliver's Travels}, where the philosophers of Laputa carried about actual solid models of the concepts which they wished to discuss. If we wish to discuss spatial structure then we have effectively to do the same [\,\dots]. Discourse about solid structures is impossible without effectively being able to call up pre-fabricated concepts, level upon level, the simplest being the Platonic solids, as we will. Literary labels, such as the words “rhombic triacontahedron” or “para di-chloro-benzene” have precise meanings. If we do not know enough of them, then we cannot even begin to use the hierarchically structured tree of concepts which is modern science.

\endgroup

\vspace{2mm}

Which is not an indication that we are all Laputans or Platonists (cf. Section \ref{subsection "Logomachy of Mathematicians, and Cock-and-Bull Stories"}). Laputanity, as we might call it, is a convenient attitude (as a close relative of Platonism), under our proto-mathematics, in keeping with a bio-evolutive and physiological origin (cf. Section \ref{subsubsection "A Task for Other Scientists: Proto-mathematics"}). So when we use «pre-fabricated concepts» or «literary labels» in mathematics, we must not forget this pre-eminent fact.
\enumerationisfinis

\subsection{Allegorical Figurations of Reality, Idealizations, and Techniques of Transcendence Playing: Physics Affairs}
\label{subsection "Allegorical Figurations of Reality, Idealizations, and Techniques of Transcendence Playing: Physics Affairs"}

\begingroup
\footnotesize
I am enough of an \emph{artist} to draw freely upon my imagination. Imagination is more important than knowledge. Knowledge is limited. Imagination encircles the world. \\
\indent — \textsc{A. Einstein} \cite[p. 117, e.a.]{Einstein "What Life Means to Einstein. An Interview by George Sylvester Viereck"}  

\vspace{2mm}

The formulation of a problem is often more essential than its solution, which may be merely a matter of mathematical or experimental skill. To raise new questions, new possibilities, to regard old problems from a new angle, requires \emph{creative imagination} and marks real advance in science. \\
\indent — \textsc{A. Einstein \& L. Infeld} \cite[p. 95, e.a.]{Einstein and Infeld "The Evolution of Physics"} 

\vspace{2mm}

This \emph{imaginative vision} and \emph{faith} in the ultimate success are indispensable. The pure rationalist has no place [in science] [cf. footnote \ref{footnote "Two examples of physico-mathematical faith or belief"} on p. \pageref{footnote "Two examples of physico-mathematical faith or belief"}]. \\
\indent — \textsc{M. Planck} \cite[p. 215, e.a.]{Planck "Where Is Science Going? The Universe In The Light of Modern Physics"}

\endgroup

\vspace{2mm}

Since physics, at least the one where theoretical knowledge stands out, i.e. mathematical physics and theoretical physics, relies on a mathematical language (doing fundamental physics without mathematics is like playing a game of billiards without cue sticks), what has been argued in the previous Sections, regarding mathematics, is comparably valid and repeatable also in physics research (cf. Sections \ref{section "Mathematics in the Physical Sciences, and Nature of Reality I"}, \ref{section "Mathematics in the Physical Sciences, and Nature of Reality II"} and \ref{section "Mathematics in the Physical Sciences, and Nature of Reality III"}). Along these lines, there are some blunt and forthright pages from G. Vignale \cite[pp. vii, ix, e.a.]{Vignale "The Beautiful Invisible: creativity imagination and theoretical physics"}:

\vspace{2mm}

\begingroup
\footnotesize
Physics [\,\dots] is the military academy of \emph{liberal arts} [\,\dots]. [I]t is too rich in ideas, too loaded with philosophical content, too intertwined with the history of culture, to be considered merely a technical subject. A physics question rarely involves the details of a specific phenomenon—rather it concerns general patterns, regularities, laws. And the creation of new concepts in physics requires a \emph{power of imagination} comparable with, if not superior to, that found in the most abstract arts, for example poetry [cf. Section \ref{subsection "Fluvial- and Æolian-like Processes"}] [\,\dots].
 
I think it is delirium to believe that our theories describe \emph{literally} the world as it is. The success of a theory at explaining or predicting the facts in no way proves the objective reality of that theory. It simply demonstrates the power of our brain to successfully adjust to a reality on which we wish to prevail [\,\dots]. 

A good scientific theory is like a \emph{symbolic tale}, an \emph{allegory of reality}.\footnote{
	The word \textit{allegory} comes from the Gr. \textgreek{ἀλληγορία}, “metaphorical (figurative) language”, a portmanteau of \textgreek{ἄλλος}, “another”, and \textgreek{ἀγορεύω}, “speak”, “say”. This \textgreek{ἄλλος} (\textit{alius}), if you like, is the pintle on which the whole “mechanism” of mathematics \& physics is inserted.
	} 
Its characters are \emph{abstractions} that may not exist in reality; yet they give us a way of thinking more deeply about reality. Like a fine work of art, the theory \emph{creates its own world} [an \emph{artificial and fictional world}, cf. pp. 7, 10]: it transforms reality into \emph{something else}—an illusion perhaps, but an illusion that has more value than the literal [description] [of a] fact [\,\dots]. 

The world of a physical theory [\,\dots] is a tangential world, which makes contact with the world of facts in a limited region, but eventually flies off on an infinite plane, further and further from any observable reality. On this infinite plane we meet invisible actors ruled by invisible principles.

\endgroup

\vspace{2mm}

In the Preface to the It. edition \cite[p. xiii]{Vignale "La bellezza dell'invisibile. Creativita e immaginazione nella fisica}, he adds that

\vspace{2mm}

\begingroup
\footnotesize
physics [can, indeed must, be exposed] not as an explanation, even if only provisionally, of real facts, but as an \emph{artistic reconstruction} of the latter. The idea of a \emph{second reality}, invisible but more real than the other, has always fascinated me.

\endgroup

\vspace{2mm}

This second, invisible, reality is «somehow more real than reality», because it is built by some physico-mathematical theory, which removes the chaotic skein of phenomena, and seeks the essential, a distillate (cf. Section \ref{subsection "Physico-mathematical Reality"}); the theory itself produces, ipso facto, a «fictional» but «metaphorically exact world»—although it has nothing to do with the «innermost reality» \cite[pp. 15, 24]{Vignale "The Beautiful Invisible: creativity imagination and theoretical physics"}. Here, theoretical physics is conceived as the «science of the invisible», a sort of «modern form of theology» (see below), where “point particles”, “light rays”, “minimum principles”, “conservation laws”, and “force fields” «strongly remind us of something real, yet are nowhere to be seen»; so much that each of which sprouts from an exercise of «abstraction» \cite[p. 3, e.a.]{Vignale "The Beautiful Invisible: creativity imagination and theoretical physics"}: 

\vspace{2mm}

\begingroup
\footnotesize
Francis Bacon, one of the founders of modern empirical science, wrote that “Nature cannot be commanded except by being obeyed”. I would add that nature cannot be understood except by being \emph{transcended} [\,\dots]. When physicists work on a theory, they are not dealing directly with nature, but with an abstract model in which they have already decided which aspects of reality must be absolutely \emph{retained}, and which ones can be \emph{dismissed}.\footnote{
	Cf. endnote \ref{endnote "Difalcare gli impedimenti della materia"} and Scholium \ref{scholium "Concept of model in ancient Greek science"}. This is why \cite[p. 27]{Vignale "The Beautiful Invisible: creativity imagination and theoretical physics"} «the laws of physics are never laws about the world as it is, but [\,\dots] under a certain idealization». Different idealizations are consonant with different laws, different theories, and different ways of understanding of the world.
	} 
Often, in creating this model, they make bold and quite implausible assumptions, which can only be validated by the consistency of the results. But, to take such bold steps one cannot rely on calculation alone: it takes \emph{passion}, imagination,\footnote{
	Cf. e.g. G. Holton \cite[p. 184]{Holton "On the Art of Scientific Imagination"}: «Of course, the primary tools of the trade, which a scientist can be taught to use, are indispensable: perseverance, the use of one's rational faculties while forming and testing hypotheses, mathematics and instrumentation, judicious modeling, looking skeptically for flaws or disconfirmations, etc. But in truth, all these are not sufficient to explain the daring and risky leaps of speculation that are often the crucial ingredient, or even the initial impetus, for a project. There must be a second, complementary set of forces at work—an \emph{art} of the imagination». See also his \cite[chap. 3. \textit{Dionysians, Apollonians, and the scientific imagination}, pp. 84-110]{Holton "The Scientific Imagination: Case Studies}.
	} 
a sense of beauty—all things that we grasp with our whole \emph{personality}, and definitely with our heart.

\endgroup

\vspace{2mm}

The task of theoretical physics, drawing on the mathematical models (abstractions of  geometry, algebra and analysis),\footnote{
	But let us go back a bit to the notion of metaphor (cf. footnote \ref{footnote "Metaphor"}, p. \pageref{footnote "Metaphor"}). Be wary: we do not want to confuse the concept of “mathematical model” with that of “metaphor”, e.g. by identifying the plurality of mathematical models, which are created for some physical phenomenon, with the process of metaphorical creation. What we are saying is that one and the same model—one and the same conceptual representation—can be generated by different metaphors, sometimes conflicting with each other, into the bargain. On the role of the metaphors in theoretical physics, see e.g. G. Parisi \cite{Parisi "Scambi di metafore tra fisica e biologia"}.
	} 
seems to be that of «transcending the world» \cite[p. 293]{Vignale "The Beautiful Invisible: creativity imagination and theoretical physics"}.\footnote{
	The mathematization of physical phenomena is part of the modus operandi of a theoretical physicist. Nonetheless, an \emph{abstractive ability} is also strikingly present in the experimental mentality. For example, M. Faraday \cite[p. 353]{Faraday "On the Conservation of Force"}, a great experimentalist, but devoid of mathematical culture, writes: «For instance, \emph{time} is growing up daily into importance as an element in the exercise of force. The earth moves in its orbit in time; the crust of the earth moves in time; light moves in time; an electro-magnet requires time for its charge by an electric current: to inquire, therefore, whether power, acting either at sensible or insensible distances, always acts in time, is not to be metaphysical [\,\dots]. To inquire after the possible time in which gravitating, magnetic, or electric force is exerted, is no more metaphysical than to mark the times of the hands of a clock in their progress».
	} 
The \textit{punctum saliens} is that the nature of reality cannot be defined; it has its own \emph{undefinability}, or \emph{equivocality}. «La voce di \emph{Natura} è voce Equivoca» (The entry \emph{Nature} is an Equivocal entry), as declared by G. Crivelli \cite[p. 19]{Crivelli "Elementi di fisica"}.\footnote{
	We can take as an exemplary definition of physics, the one given by Crivelli \cite[p. 19]{Crivelli "Elementi di fisica"} with his eighteenth-century Italian (which we advisedly leave untranslated): «Fisica si dice quella Scienza, c'ha per oggetto i Corpi Naturali, e cerca le loro proprietà, così detta dalla voce Greca \textgreek{φύσις}, che significa Natura. La voce di \emph{Natura} è voce Equivoca. Imperocchè talvolta si prende per la Essenza delle cose, che contempliamo [\,\dots]. Talvolta si adopera per significare la Causa Universale di tutte le cose [\,\dots]. Talvolta significa lo stesso Universo, ed i Corpi, de' quali egli è composto, ed in questo senso la prende il Fisico, quando dice di essere il Contemplatore della Natura».
	} 
Hence the need for an act of transcendence.

One \emph{transcends} the external world of objects/happenings—the ultimate nature of reality—to fall into the \emph{Magic Mathematics}, which is \emph{a world of Carrollian fantasy}, and discover \emph{unanticipated relations}. For that, mathematics is the true act of \textit{religio}, i.e. of binding (\textit{religāre}), apropos of the notion of \textgreek{φύσις}, which justifies the previous reference to the «theological» facet of mathematical modeling. The good G.-C. Rota \cite[p. xviii]{Rota "Introduction to The Mathematical Experience"} comes to our rescue:

\vspace{2mm}

\begingroup
\footnotesize

Mathematics [along with all its physico-mathematical articles], like theology and all free creations of the Mind, obeys the inexorable laws of the imaginary.

\endgroup

\vspace{2mm}

This mirrors what C.S. Peirce \cite[p. 3659, e.a.]{Peirce "Mathematics"} had already said in \textit{The Century Dictionary and Cyclopedia} (1904):

\vspace{2mm}

\begingroup
\footnotesize
Mathematics [is] the study of \emph{ideal constructions (often applicable to real problems)}, and the discovery thereby of relations between the parts of these constructions, \emph{before unknown}. The observations [are] upon \emph{objects of imagination merely}.

\endgroup

\vspace{2mm}

An annotation of clarification. Such a parallelism, between physics and theology, may sound quirky, or like an urticant comparison, as I myself am a deeply irreverent and caustic man—I highly recommend reading G. Papini's \emph{libello} \cite{Papini "Le memorie d'Iddio"}, a masterpiece of sarcasm, a real laceration (cf. the vb. \textgreek{σαρκάζω}), soaked with exhilarating atheism, against baboonish-Catholic idolatry. But ultimately, system of “reason” (science) and system of “faith” (religion) are both human activities, and it is normal that they share some similarities, with their diversity, for better or worse: cf. Section \ref{subsection "Axioms of Faith"}, point \ref{item "Physics is not mathematics (and vice versa), axioms of faith"} on p. \pageref{item "Physics is not mathematics (and vice versa), axioms of faith"} \& footnote \ref{footnote "Two examples of physico-mathematical faith or belief"} on p. \pageref{footnote "Two examples of physico-mathematical faith or belief"}, and Section \ref{subsection "Dissolution of the Objective World: the Clamorous Incident of the Wave Function. An Authentic Story of Aesopian Fables and Theater of the Absurd"}. The so-called \textit{attractiones electivæ}, a happy expression from T. Bergman \cite{Bergman "De Attractionibus Electivis"}, later assimilated by J.W. von Goethe in his \textit{Die Wahlverwandtschaften}, do not exist only in chemistry and in novels.

\subsection{Blind Specialism: Cocoon Syndrome}
\label{subsection "Blind Specialism: Cocoon Syndrome"}

\begingroup
\footnotesize
By and large mathematicians write for the exclusive benefit of other mathematicians in their own field even when they lapse into “expository” work [\,\dots]. Such “expositions” are more often than not brilliant displays of virtuosity, designed to show the rest of the community (a half-dozen individuals) how much more elegantly and simply the author would have proved somebody else's results were it not for his more important commitments. \\
\indent — \textsc{G.-C. Rota} \cite[p. 243]{Rota "Indiscrete Thoughts"}

\endgroup

\vspace{2mm}

We said that mathematics and, along with it, any theoretical knowledge of the physical universe, is fantasy, force of creative imagination; which would suggest that its horizon of action is very vast. Conversely, if mathematics, together with its physico-natural apparatus, ends up shutting itself in specialism, it might bring the risk of self-referentiality, and becomes blind to a broader vision oriented toward critical thinking.\footnote{
	Cf. C. Bartocci \cite[pp. 11-13]{Bartocci "Dimostrare l'impossibile. La scienza inventa il mondo"}.
	} 
The sharp-edged judgment of G.-C. Rota's \cite[p. 243, e.a.]{Rota "Indiscrete Thoughts"}—already anticipated in the above epigraph—is delectably enjoyable:

\vspace{2mm}

\begingroup
\footnotesize
A specialist in quantum groups will write only for the benefit and approval of other specialists in quantum groups. A leader in the theory of pseudo-parabolic partial differential equations in quasi-convex domains \emph{will not stoop to being understood} by specialists in quasi-parabolic partial differential equations in pseudo-convex domains.

\endgroup

\vspace{2mm}

H. Poincaré \cite[I, chap. II,\endnote{
	This chap., entitled \textit{L'avenir des Mathématiques}, is a redrafting of a lecture held in Roma, 10 aprile (Atti del IV Congresso Internazionale dei Matematici, Roma, 6-11 Aprile 1908, Vol. I, Parte I—Relazione sul congresso, per cura di G. Castelnuovo, pp. 167-182), and subsequently published in various journals.
	}
pp. 25-26, 34-35]{Poincare "Science et methode"} 
had already denounced it much earlier:

\vspace{2mm}

\begingroup
\footnotesize
Mathematicians attach great importance to the elegance of their methods and results [\,\dots]. Elegance can springs up from unforeseen feelings [\textit{sentiment de l'imprévu}] caused by the unexpected combination of objects that we usually do not see associated with each other [\,\dots]; it is fecund, since it thus reveals to us relationships hitherto unrecognized [\,\dots].

To the extent that the science develops, it becomes more difficult to embrace it in its entirety; so we try to cut it into pieces, [and] to settle for one of these pieces: in a word, to specialize [\textit{à se spécialiser}]. If we continued with this orientation, it would be a serious obstacle to the advancement of science [\textit{ce serait un obstacle fâcheux aux progrès de la Science}]. As I have said, it is by unexpected concurrences [\textit{rapprochements inattendus}] between its various parts which makes this progress possible. Too much specializing would be to refrain from these concurrences.

\endgroup

\vspace{2mm}

Confusing a series of mathematical reasonings with a string of isolated procedures, each closed in its own sector (\emph{cocooning syndrome}), is like confusing the art of architecture with the activity of superimposing one brick upon another without a \emph{superior design}. In a great theorem, and in its proof, there is an \emph{inventive content}, comparable to a burning magma that expands and creates new rock, corresponding to the unfailing formation of \emph{new imaginary worlds}.

The disappearance of multidisciplinarity, which links different areas of study together, is reaffirmed by N. Wiener \cite[pp. 2-3]{Wiener "Cybernetics: or the Control and Communication in the Animal and the Machine"}: 

\vspace{2mm}

\begingroup
\footnotesize
Today there are few scholars who can call themselves mathematicians or physicists or biologists without restriction. A man may be a topologist or an acoustician or a coleopterist. He will be filled with the jargon of his field, and will know all its literature and all its ramifications, but, more frequently than not, he will regard the next subject as something belonging to his colleague three doors down the corridor, and will consider any interest in it on his own part as an unwarrantable breach of privacy. 

These specialized fields are continually growing and invading new territory [\,\dots]. There are fields of scientific work [\,\dots] which have been explored from the different sides of pure mathematics, statistics, electrical engineering, and neurophysiology; in which every single notion receives a separate name from each group, and in which important work has been triplicated or quadruplicated, while still other important work is delayed by the unavailability in one field of results that may have already become classical in the next field. 

It is these boundary regions of science which offer the richest opportunities to the qualified investigator [\,\dots]. If a physiologist who knows no mathematics works together with a mathematician who knows no physiology, the one will be unable to state his problem in terms that the other can manipulate, and the second will be unable to put the answers in any form that the first can understand [\,\dots]. [A] proper exploration of these blank spaces on the map of science could only be made by a team of scientists, each a specialist in his own field but each possessing a thoroughly sound and trained acquaintance with the fields of his neighbors [\,\dots]. The mathematician need not have the skill to conduct a physiological experiment, but he must have the skill to understand one, to criticize one, and to suggest one. The physiologist need not be able to prove a certain mathematical theorem, but he must be able to grasp its physiological significance and to tell the mathematician for what he should look.

\endgroup

\subsection{Bibliophily, Philological Care, and Lateral Spurs}

\enumerationisinitium
\item Being that I am a bibliophile (scilicet: avid reader), just about all books, articles \& papers in the \texttt{thebibliography} are present in my library (and represent a small part of it). A consideration should be added, that every bibliophile has the flaw of being, at least in a stage in life, a bibliomane.\footnote{
	Regrettably I lack the nobiliary possibility of purchase of B. Boncompagni Ludovisi (1821-1894), in addition to the adventurous brazenness of G. Libri Carucci dalla Sommaja (1803-1869). However—having to make the comparison with their prestigious and large private mathematical libraries (with thousands and thousands of manuscripts, printed volumes, and scientific articles)—in today's times there are alternative ways, to recover these deficiencies.
	}
\item The epigraphs placed hither and thither, under the Chapter and Section headings, are not for a historical scenario, since these \textit{Notes} are not concerned with the history of mathematics and physical theories; but, characterized by a philological care, they act, so to speak, as \emph{lateral spurs} for reading what follows, and for recalling the hidden affinities that lie behind the growth of a particular idea. There is no interest here in the chronological line; rather, it is more inspiring to make \emph{deliberately anachronistic} use of certain thoughts from the distant or recent past.
\item It is a secret of Polichinelle: for a good research it is advisable to read good masters, as N. Abel already explicitly suggested.\footnote{
	The phrase “Read/study the masters, not the pupils”, or something similar, is generally attributed to Abel, but there are no reliable sources other than the anecdotal narrative.
	}
\enumerationisfinis

\subsection{Working Method}

\begingroup
\footnotesize
Catching sight of the things I wish to know with my own enlightenment, rather than relying on notions [\,\dots] coming from a more or less large clan of which I found myself a member [\,\dots]. It is in [an] act of “going beyond”, of being something in oneself in short, and not simply the expression of a law of consensus, of not remaining locked in an imperative pinwheel that others have set—it is in this \emph{solitary act} that “\emph{creativity}” is found.\endnote{
	Extended original Fr. version: «[A]border par mes propres lumières les choses que je veux connaître, plutôt que de me fier aux idées et aux consensus, exprimés ou tacites, qui me viendraient d'un groupe plus ou moins étendu dont je me sentirais un membre, ou qui pour toute autre raison serait investi pour moi d'autorité. Des consensus muets m'avaient dit, au lycée comme à l'université, qu'il n'y avait pas lieu de se poser de question sur la notion même de “volume”, présentée comme “bien connue”, “évidente”, “sans problème”. J'avais passé outre, comme chose allant de soi—tout comme Lebesgue, quelques décennies plus tôt, avait dû \emph{passer outre}. C'est dans cet acte de “passer outre”, d'être soi-même en somme et non pas simplement l'expression des consensus qui font loi, de ne pas rester enfermé à l'intérieur du cercle impératif qu'ils nous fixent—c'est avant tout dans cet acte solitaire que se trouve “\emph{la création}”».
	} \\
\indent	— \textsc{A. Grothendieck} \cite[2.2. \textit{L'importance d'être seul}, p. 6 otm, e.m.]{Grothendieck "Recoltes et Semailles. Reflexions et temoignage sur un passe de mathematicien"}

\vspace{2mm} 

\ZhSimplified{我们相信已经找到了沟通情感的媒介，那就是茶}.\footnote{
	«We believe we have found a \emph{medium} to communicate emotions, and that is the tea». Then the sentence goes on like this:\ZhSimplified{「寻找茶的源头，也就是找寻中国人精神的源地」}«Finding the source of tea means finding the source of the Chinese spirit».
	} \\
\indent	— \ZhSimplified{茶之路} (\textit{Chá Zhī Lù}) · \textit{The Road of Tea}\footnote{
	The sentence above comes from the back cover of the book \ZhSimplified{茶之路} · \textit{The Road of Tea} /\ZhSimplified{《生活月刊》著}, published by Guangxi Normal University Press, Guilin, 2019\textsuperscript{re}, edited by \ZhSimplified{王澍, 朱赢椿, 马可, 金宇澄, 徐冰, 李宗盛, 黄永松, 林怀民, 阮义忠}.
	}

\vspace{2mm}

(\textit{Mathematics: creativity}) I think at the origin of creativity [\,\dots] there is what I call the ability, or willingness [\textit{disponibilità}], to \emph{dream}, to imagine different worlds, different things, to try to combine them inside one's own imagination in various ways. \\
\indent — \textsc{E. De Giorgi}\footnote{
	Video interview to E. De Giorgi, Pisa, July 1996, by M. Emmer, on behalf of the \textsc{umi} (Unione Matematica Italiana).
	}

\vspace{2mm}

\dots, what / does anyone want / but to feel a little more \emph{free} [to dream]? \\
\indent — \textsc{Godspeed You! Black Emperor}, from one of the two inner \textsc{lp} record sleeves in \textit{Lift Your Skinny Fists Like Antennas to Heaven} (Constellation · \textsc{cst}012, 2000)
 
\vspace{2mm}

Con frementi tormente di petali di meli / e di ciliegi con rapide rapide nubi di petali [\,\dots] petali petali amatamente dissolti [\,\dots] dà che solo in mitezza per te mi pensi / e in reciproco scambio di sonni amori e sensi [\,\dots] io ti individui per sempre e in te mi assuma.\footnote{
	«With quivering blizzards of apple / and cherry petals with rapid rapid cumuli of petals [\,\dots] lovingly dissolved petals petals [\,\dots] in such a way that only in mildness for you I think of myself / and in reciprocal exchange of sleeps loves and senses [\,\dots] I identify you forever and I assume myself in you».
	} \\
\indent	— \textsc{A. Zanzotto} \cite[p. 840]{Zanzotto "Ligonas II"}
 
\endgroup

\vspace{2mm}

My working method follows four simple guidelines: 

· self-denial,\endnote{
	There is a little piece by Miłosz \cite[p. 39]{Milosz "Road-side Dog"}, entitled \textit{Pursuing a goal}, which sums up nicely the spirit that animated my intent throughout the meditation and writing of this work: «In order to accomplish something, one must dedicate oneself to it totally, so much that our fellow men cannot even imagine such an exclusivity. And that does not mean at all the amount of time consumed. There are also the innumerable emotional subterfuges practiced toward oneself, slow transformations of personality, as if one supreme goal, beyond one's will and knowledge, pulled in a single direction and organized destiny».
	} 
and claustral seclusion—contrary to Galileian belief—in a “world on paper”,\footnote{
	\cite[p. 106]{Galilei "Dialogo sopra i due Massimi Sistemi del Mondo Tolemaico e Copernicano"}: «[I] discorsi nostri hanno a essere intorno al mondo sensibile, e non sopra un mondo di carta» (Our discourses must relate to the sensible world, and not to on a world on paper).
	}
as a consequence of bibliophily;

· cups of Chinese tea, engine of odorous emotions easily convertible into intellectual stimuli;\footnote{
	There are many phrases about tea; I choose to reveal two lines from G. Ceronetti \cite[p. 10]{Ceronetti "Pensieri del Te"}: «Man drinks Tea because he is distressed by man. / Tea drinks man, the most bitter herb [\textit{L'uomo beve il Tè perché lo angoscia l'uomo. / Il Tè beve l'uomo, l'erba più amara}]».
	}\textsuperscript{,}\endnote{
	My favorite combined formulistic ritual, other than mathematics \& music, is mathematics \& tea (see Sections \ref{section "Bohr's Tea Principle—Uncertainty and Entropy"} and \ref{section "Non-perfect Fluid in the Teapot and Brownian Motion"}). And since mathematics is a human activity—a complex of techniques and cognitive experiences—in view of a specific understanding, I find Ceronetti's \cite[pp. 9-10]{Ceronetti "Pensieri del Te"} (daily) story close to my way of being. Tea and mathematics, as I see it, have this in common, they are a fight against the “darkness” of the unknown. And since his Italian is highly gratifying, I opted not to scar (disfigure) it by translating it into English (I am sorry): «Due volte al giorno [\,\dots] tazza ripetuta di Tè verde della Cina arriva con la sua infallibile virtù unitiva, confirmativa, risuscitativa, a disincagliarmi e a preservarmi da ogni specie d'inerzia, d'inebetimento, di abbattimento [\,\dots].

	\setlength\parindent{8pt}
	Non sono un Orientale. Ma di Oriente orientante mi resta la fiducia che nell'uscire in giusta misura da se stessi, e abitualmente, non c'è nulla di pericoloso, e che vedere, sentire e incontrare spiriti non è inquietante.

	Lo Spirito del Tè comincia appena disceso ad operare. Leggere pressioni interne, agopunture invisibili, scatti tempestivi del sensorio, sampàn di lumettini, coloriture improvvise di silenzi, un susseguirsi puntuale di eccitamenti che vanno dall'occhio interno (che forse è un orecchio o una mano) lungo le disirrigidite vertebre, al coccige resurrecturo. Allora nel buio molte finestrine tornano vive, e le parole faticano meno a ritrovare il loro principio negli spazi lontani. Pace del massaggio, radice del suono [\,\dots]. Guardare da una pausa di connessione quel che è sconnesso e lacerato, è un momento senza morte. Fare arretrare di appena un poco il margine del finito, per molte ore rischiara.

	Nel combattimento per contrastare mentalmente quel che nel tempo è verificabile come aggressione materialmente incontrastata della tenebra, da làmine liberatrici che il Tè aiuta a ritrovare e a decifrare, imparo a non aborrire in eccesso le tenebre, per non distruggere le poche possibilità di penetrarne il segreto [\,\dots].

	[Con il Tè] I pensieri non miei diventano miei con molta facilità; quelli miei chiunque se vuole può farli proprii, qualunque sia il suo eccitante, senza bisogno di nome: il pensiero non pronuncia né Tuo né Mio». 
	}

· experimental music;\footnote{
	From microtonal modalities, as in G. Scelsi, through the twelve-note technique (as in P. Boulez, L. Nono, and K. Stockhausen), to I. Xenakis' avant-garde compositions, representing the peak, to my ears. Besides that, there is the newer music with symphonic-styled instrumentals: Godspeed You! Black Emperor, which stand head and shoulders above the rest, and then bands such as Set Fire To Flames, Silent Whale Becomes A° Dream, The Evpatoria Report, or Sparrows Swarm and Sing, just so I am clear. 
	}\textsuperscript{,}\footnote{
	\label{footnote "Listening to (a piece of) music..."}
	Listening to (a piece of) music is like being projected into a \emph{dreamland} (cf. epigraph from J.J. Sylvester in Section \ref{section "Grainy Music and Chance"}) where the distinction between \textgreek{ἀκουσματικός} (acusmatikós) and \textgreek{μαθηματικός} (mathematikós), between listener (eager to hear) and fond of learning, is canceled, where the knowledge, be it felt, reasoned, or imagined, is immediately reached in the stirring transport of the \emph{structured succession of sounds}. Music is an \emph{aesthetic experience}, and not a series of punctual (disconnected) acoustic sensations, for we \emph{links}—here is the mathematical aspect—the different sounds in a melody; cf. Section \ref{subsection "Contextus I. Elements of Brachylogy—the Reverie of a Perfect Language, with a Margo on Music and Mathematics"}.
	}

· intimate hideaway amidst the scents, the colors, and the sounds of nature, and especially in amorous senses with flowering plants;\footnote{
	On the power of nature, supremely in the seduction of its inflorescences, I could not refrain from reminiscing about a few Zanzottian verses of my youth.
	} 
so, my interest for botany;\footnote{
	Here are seven books (by publication date) on botany, all embellished with ravishing images, that every bibliophile should have in the shelf:

· R. Kesseler \& M. Harley \cite{Kesseler and Harley "Pollen: The Hidden Sexuality of Flowers"}, 

· R. Kesseler \& W. Stuppy \cite{Kesseler and Stuppy "Seeds: Time Capsules of Life"},

· W. Stuppy \& R. Kesseler \cite{Stuppy and Kesseler "Fruit: Edible Inedible Incredible"},

· E. Koinberg \cite{Koinberg "Herbarium Amoris Floral Romance"},

· Various authors \cite{Various authors "Plant: Exploring the Botanical World"},

· Various authors \cite{Various authors "Flora: Inside the Secret World of Plants"},

· L. Biss \cite{Biss "The Hidden Beauty of Seeds and Fruits: The Botanical Photography of L. B."}.

Why this list? Does that make sense, in a volume of mathematics \& physical structures? Yes, it does: for me, the entire botanical world acts as an inexhaustible treasure of beauty, but also of usefulness (cf. Margo \ref{margo "Botanical nomenclature and demon of order: a uniting thread for mathematics and morality"}); it puts us in front of the limits of formal sciences (we are going to see this in the Outro Chapters): a lesson in \textit{humilitas}, or better, in \textit{humus}, as these are plants.
	}
 
· bent for (day)dreaming, which is, say, the ubiquitous corolla of the four previous points. «Life, what is it but a dream?» \cite[p. 317]{Carroll "Through the Looking-Glass"}.\footnote{
	In English literature, an almost twin verse is also found e.g. in J. Keats \cite[\textit{On Death}, p. 357]{Keats "The Poetical Works"}: «Life is but a dream». But—as everyone knows—there is an illustrious precedent, at the base of this question-type: Shakespeare \cite[Act IV, Scene I, p. 48]{Shakespeare "The Tempest"}, with his peerless verses declaimed by Prospero: «We are such stuff / As dreams are made on, and our little life / Is rounded with a sleep».
	}

\subsection{Through the Magnifier of Half-sleep}
\label{subsection "Through the Magnifier of Half-sleep"}

\begingroup
\footnotesize
“Mathematizing” may well be a \emph{creative activity} of man, like [poetic] language or music, of primary originality, whose historical decisions \emph{defy complete objective rationalization}. \\
\indent — \textsc{H. Weyl} \cite[p. 392, e.a.]{Weyl "Obituaries David Hilbert (1862-1943)"} 

\vspace{2mm}

[We need] to re-evaluate the most active, most \emph{creative} (but also, resultantly, the most \emph{adventurous}, \emph{imaginative}, \emph{subjective}) aspects of our way of thinking [in mathematics] [\,\dots]. Deplorably, a false modesty generally forbids [us] to mention the part of the discovery process that takes place more or less in the \emph{sphere of the unconscious}, or of the subconscious, to exhibit only the \emph{fossilized demonstration} in its skeletal form of stonily deductive and formalistic logic. \\
\indent — \textsc{B. de Finetti} \cite[p. 427, e.a.]{de Finetti "Convegno della C.I.I.M. Viareggio 1974"}

\endgroup

\vspace{2mm}

I should like to say that the most profitable moments (intuitions, ideas, and overall visions) in the elaboration of this book came in the state of \emph{half-sleep}, in the morning, preceding the noise, or the dirt, by accumulation of information, which inevitably increases—entropically—along the phases of full consciousness (wakefulness). Along this sleep-wakefulness transition, something notable happens. It is like \emph{seeing simultaneously}, in a \emph{single mental space}, everything I did the day before. With this, the power of the mathematical invention of Hadamardian memory \cite[chapp. I-III]{Hadamard "The Psychology of Invention in the Mathematical Field"} is experienced, on my own:\footnote{
	J. Hadamard gives the example of an illumination shooting out from the unconscious in mathematics, physics and chemistry; but he does not fail to report nearly identical experiences in other fields,  including classical music. He in \cite[p. 16, e.a.]{Hadamard "The Psychology of Invention in the Mathematical Field"} quotes a letter of W.A. Mozart: «[I]n the night when I cannot sleep, thoughts crowd into my mind as easily as you could wish. Whence and how do they come? I do not know and I have nothing to do with it. Those which please me, I keep in my head and hum them [\,\dots]. Once I have my theme, another melody comes, linking itself to the first one, in accordance with the needs of the composition \emph{as a whole}: the counterpoint, the part of each instrument, and all these melodic fragments at last produce the entire work. Then my soul is on fire with inspiration, if however nothing occurs to distract my attention. The work grows; I keep expanding it, conceiving it more and more clearly until I have the entire composition finished in my head though it may be long. Then my mind seizes it as a glance of my eye a beautiful picture or a handsome youth. It does \emph{not come to me successively}, with its various parts worked out in detail, as they will be later on, but it is \emph{in its entirety} that my imagination lets me hear it».
	}
the greatest creativity crops up mostly in the cleansing states of the mind \& brain combination, when such a combination has a renovation of virginity for the day that is about to begin.

A testimony, in this direction, is offered authoritatively by H. Poincaré \cite[I, chap. III,\endnote{
	This chap., entitled \textit{L'invention mathématique}, is a transcript, with slight variations, of a \emph{Conférence} held in Paris, at the Institut général psychologique, on 23 May 1908, and subsequently published in various journals.
	}
p. 53]{Poincare "Science et methode"},
who talks about \textit{apparences d'illumination subite}, clear signs of a \textit{long travail inconscient antérieur} that effloresces \textit{dans l'invention mathématique}. Morning thoughts, as well as other thoughts appertaining to the period of apparently unconscious work, collide with each other and end up hooking together—so as to resemble Epicurus' \textit{atomes crochus} \cite[p. 60]{Poincare "Science et methode"}—, giving life to unlooked-for aggregations, surprising combinations.\footnote{
	\cite[pp. 55-56, 62]{Poincare "Science et methode"}: «[T]he unconscious ego or, as they say, the subliminal ego, plays a capital role in mathematical invention [\,\dots]. The subliminal ego is in no way inferior to the conscious ego [\,\dots]; it is capable of discernment, it has tact, delicacy; it knows how to select, it knows how to divine [\,\dots]. The rules of calculations are strict and complicated; they demand discipline, attention, will, and hence consciousness. In the subliminal ego reigns, on the contrary, what I would call liberty [\textit{liberté}], if one could give this name to the sheer absence of discipline and to the disorder born of chance [\textit{désordre né du hasard}]. [But] only this very disorder permits [the inception of] unexpected couplings [\textit{accouplements inattendus}]».
	} 

Another account comes from L. Schwartz, when he describes \cite[p. 246]{Schwartz "Un Mathematicien aux prises avec le siecle"} = \cite[p. 232]{Schwartz "A Mathematician Grappling with His Century"} the night that inspired the invention of distributions: 

\vspace{2mm}

\begingroup
\footnotesize
I used to have insomnias lasting several hours, and never took sleeping pills. I remained in my bed, the light off, and without writing anything, I did mathematics. My inventive energy [\textit{énergie inventive}] was redoubled, and I advanced rapidly without tiring. I felt entirely free, without any of the brakes [\textit{totalement libre, sans aucun des freins}] imposed by daily realities and writing. 

\endgroup

\vspace{2mm}

A further attestation is left by C. Villani \cite[p. 155]{Villani "Theoreme vivant"}: he speaks of a matutinal \textit{illumination}, in mathematics, of a \textit{petite} but ineffable illumination, knocking on the door of the brain, before its drowning in the technique (\textit{l'illumination sera noyée dans la technique}).

The visionary propulsion of the unconscious is manifestly central also in theoretical physics, see e.g. G. Parisi \cite{Parisi "Come nascono le idee"}. 

It may be noticed in passing, that there is no lack of similar examples in literature.\footnote{
		\emph{Synoptic note}. I suggest reading the essay by T. McLeish \cite{McLeish "The Poetry and Music of Science: Comparing Creativity in Science and Art"}, which is an exhaustive study of various subject-matters outlined above. It heedfully investigates the close links between the scientific sphere (mathematics, physics, chemistry, biology) and the humanistic sphere (literature, art, music), as well as the weight of the unconscious and of the creative imagination in science.
		} 
One need only to remember the account of S.T. Coleridge \cite[pp. 96-97]{Coleridge "Ancient Mariner Kubla Khan and Christabel"}, who recounts how the poem \textit{Kubla Khan, a Vision in a Dream} (written in 1798, but published in 1816) came about.\footnote{
	«[T]he author [\,\dots] [i]n consequence of a slight indisposition, an anodyne had been prescribed, from the effects of which he fell asleep in his chair [and] continued for about three hours in a profound sleep, at least of the external senses, during which time he has the most vivid confidence that he could not have composed less than from two to three hundred lines; if that indeed can be called composition in which all the images rose up before him as \emph{things}, with a parallel production of the correspondent expressions, without any sensation or consciousness of effort. On awaking he appeared to himself to have a distinct recollection of the whole, and taking his pen, ink, and paper, instantly and eagerly wrote down the lines that are here preserved. At this moment he was unfortunately called out by a person on business [\,\dots], and detained by him above an hour, and on his return to his room, found, to his no small surprise and mortification, that though he still retained some vague and dim recollection of the general purport of the vision, yet, with the exception of some eight or ten scattered lines and images, all the rest had passed away like the images of the surface of a stream into which a stone has been cast». 
		}

\subsection{Behind the Scenes}

\enumerationisinitium
\item This is a \LaTeX-based book (input source files \texttt{.tex}): the document structure is a customized \texttt{book.cls}, with \texttt{pdfLaTeX} typeset engine; the \LaTeX{} editor in action is \texttt{Texifier} (formerly named \texttt{Texpad}), paired with \texttt{Smultron}, a plain text editor.
\item The font color, for hyperlinks, is codified by hexadecimal values via \texttt{xcolor} and \texttt{hyperref} packages:

· \textcolor{eggplant}{\texttt{sample linkcolor (eggplant \#800080)}} for links inside: Conjecture, Corollary, Definition, endnote (numerical order), Equation(s), Example, footnote (alphabetic order), Lemma, Margo, Proposition, Postulate, Question, Scholium, Subproposition, Theorem in the text,

· \textcolor{mallard}{\texttt{sample citecolor (mallard \#008080)}} for links inside: work mentioned (bibliographic reference) in the text,

· \textcolor{dusky-cerulean}{\texttt{sample urlcolor (dusky cerulean \#004080)}} for links outside: web resource available on the internet.

The colors for unlinked reference marks (with a fixed-pitch font via \texttt{\textbackslash{texttt}} command) are:

· \textcolor{Chinese-loquat}{\texttt{loquat (\ZhSimplified{枇杷}) yellow \#F7C015}}, 

· \textcolor{pumpkin}{\texttt{pumpkin \#FF7518}} and \textcolor{cyan-blue}{\texttt{cyan-blue \#18A2FF}}, 

· \textcolor{vernal-green}{\texttt{vernal green \#03E364}} and \textcolor{magenta}{\texttt{magenta \#E30382}},

· \textcolor{cynara-violet}{\texttt{artichoke violet \#8803E3}} (referring to the artichoke inflorescence: \textit{cynaræ scolymi calathus}).
\item Plotting and graphic elements:

· all diagrams: via \texttt{tikz-cd} package;

· geodesics on melon- and egg-shaped surfaces (Figg. \ref{figure "Geodesics on melon-shaped surface"} and \ref{figure "Geodesics on egg-shaped surface"}): \texttt{Sketch}, a vector graphics editor;

· tessellations of the disk model (Figg. \ref{figure "tessellation 1"} and \ref{figure "tessellation 2"}): graphing calculator plus \texttt{Sketch};

· tessellation of the upper half-plane (Section \ref{section "Tessellation of the Upper Half-Plane Plane by Modular Group"}): \texttt{tikzpicture} code;

· hyperboloid surfaces coexistence (Fig. \ref{figure "hyperboloid surfaces coexistence"}): code with \texttt{pst-solides3d} package processed with \texttt{XeLaTeX};

· Beltrami's pseudosphere (Fig. \ref{figure "Beltrami's pseudosphere"}): graphing calculator plus \texttt{Sketch};

· Klein bottle (Fig. \ref{figure "Klein bottle"}) and Möbius strip (Fig. \ref{figure "Möbius strip"}): \texttt{tikzpicture} codes plus \texttt{Sketch} for both, but the Klein bottle is processed/drawn with Lua\LaTeX; 

· 2-torus with triangulation (Fig. \ref{figure "torus with triangulation"}): \texttt{Sketch};

· granular spin network/nodal lump of space and its multi-colored hunks, viz. quanta of space (Figg. \ref{figure "Loop-like space graph"} and \ref{figure "Spin network with multi-colored hunks"}): \texttt{TikZ} code plus \texttt{Sketch}; 

· warping deformations under Ricci flow (Figg. \ref{figure "Warping deformation under Ricci flow: cross-section of a surface of revolution"} and \ref{figure "Warping deformation under Ricci flow: cross-section of a dumb-bell-shaped surface of revolution"}): are the work of J.H. Rubinstein and R. Sinclair \cite[Fig. 2, p. 290, and Fig. 5, p. 293]{Rubinstein and Sinclair "Visualizing Ricci Flow of Manifolds of Revolution"}; small modifications were made in their drawings with \texttt{Sketch};

· horocycles (Fig. \ref{figure "horocycles"}): \texttt{Sketch};

· Lorenz attractor (Fig. \ref{figure "Lorenz attractor"}): \texttt{luacode} environment plus \texttt{Sketch};

· Hilbert curve (Fig. \ref{figure "Hilbert curve"}): \texttt{PGF} and \texttt{TikZ} codes;

· snowflake curve of von Koch (Section \ref{subsubsection "Snowflake Curve of von Koch"}): \texttt{tikzpicture} code via \texttt{lindenmayer} \texttt{system} (L-system) and \texttt{decorations.fractals};

· hazy attractors (Figg. \ref{figure "Hazy attractor: quasi-random attractor of a stochastic-like Lorenz system I"} and \ref{figure "Hazy attractor: quasi-random attractor of a stochastic-like Lorenz system II"}): \texttt{luacode} environment plus \texttt{Sketch};

· icosa- and dodeca-hedron (Figg. \ref{figure "Icosahedron"} and \ref{figure "Dodecahedron"}): codes with \texttt{pst-solides3d} package elaborated with \texttt{XeLaTeX};

· rhombus tilings à la Penrose (Figg. \ref{figure "Penrose Rhombus tiling 1"}, \ref{figure "Penrose Rhombus tiling 2"}, and \ref{figure "Penrose Rhombus tiling 3"}): \texttt{TikZ} plus \texttt{Sketch};

· \textit{Helianthus}-like phyllotaxis (Figg. \ref{figure "phyllotaxis with 610 circles"} and \ref{figure "phyllotaxis with 4181 circles"}): \texttt{tikzpicture} code plus \texttt{Sketch}.
\item \emph{Margo} stands for \emph{annotation in the margins}, and is a catchy contraction importing the singular of \emph{marginalia}.
\item The final list of works mentioned and the citation style are manually compiled with the \texttt{thebibliography} environment via \texttt{\textbackslash{bibitem}} command. All works without \texttt{\textbackslash{bibitem}} form are present directly in the foot- and end-notes.
\item For those interested, here is the list of Chapters in the original writing order—also to explain the heterogeneity of the writings:
	Chapter \ref{chapter "Panoramic Miscellanea I. Theory of Connections, Differential Forms, Geodesics, and Holonomy Groups"}, 
	Chapter \ref{chapter "Panoramic Miscellanea II. Space Forms, Möbius (Projective) Transformations, and Fuchsian Group; Groupable Synopsis—On the Spin(or)"}, 
	Chapter \ref{chapter "Geometric and Topological Aspects of Complexity and Dynamics, Part I. Flows, Hyperbolicity, and Foliations"}, 
	Chapter \ref{chapter "Geometric and Topological Aspects of Complexity and Dynamics, Part II: Ergodicity and Entropy"}, 
	Chapter \ref{chapter "On the Chaos, Part I. Micro- and  Macro-scales"},
	Chapter \ref{chapter "On the Chaos, Part II. Non-linear Analysis"},
	Chapter \ref{chapter "Randomness and Stochastic Systems"},
	Chapter \ref{chapter "On Dimensional Continuum, Part I. Ricci Calculus (Calculus of Tensors and Curvature Tensors), Lorentz–Minkowski 4-Manifolds plus Spinor Representation, and Clifford Algebra"}, 
	Chapter \ref{chapter "On Dimensional Continuum, Part II. Action Principles, Variations and Radiation in Curved Space-Time—Mathematical Details of Field Theory of Gravitation (General Relativity)"}, 
	Chapter \ref{chapter "On Dimensional Continuum, Part III. Curvature of What?"},
	Chapter \ref{chapter "Foundational Issues"}, 
	Chapters (Outro) from \sfrac{1}{8} to \sfrac{8}{8} (\ref{chapter "Outro—Parva Mathematica: Libera Divagazione 1/8"}, \ref{chapter "Outro—Parva Mathematica: Libera Divagazione 2/8"}, \ref{chapter "Outro—Parva Mathematica: Libera Divagazione 3/8"}, \ref{chapter "Outro—Parva Mathematica: Libera Divagazione 4/8"}, \ref{chapter "Outro—Parva Mathematica: Libera Divagazione 5/8"}, \ref{chapter "Outro—Parva Mathematica: Libera Divagazione 6/8"}, \ref{chapter "Outro—Parva Mathematica: Libera Divagazione 7/8"}, \ref{chapter "Outro—Parva Mathematica: Libera Divagazione 8/8"}), even though they were written at various times, are to be considered as a single block, 
	Chapter \ref{chapter "The Ricci Flow, or the Hamilton–Perelman Metric Evolution Machinery"}, 
	Chapter \ref{chapter "Galois' Legacy—Rules over the Calculations: the Pursuit of Generality"}, 
	Chapter \ref{chapter "Variations on the Same Theme: Minima in the Calculus"},
	Chapter \ref{chapter "Ghé Drakónton, Part I. Quantum Field Space and Gravity"}, 
	Chapter \ref{chapter "Ghé Drakónton, Part IIa. Spatial Primitiveness"},
	Chapter \ref{chapter "Ghé Drakónton, Part IIb. Space-numeral Primitiveness"},
	Chapter \ref{chapter "Calabi–Yau Theorem: a Non-linear Complex Equation of Monge–Ampère Type on Compact Kähler Manifolds"},
	Chapter \ref{chapter "Toroidal Fourier Analysis"}. 
	
The first two Chapters (\ref{chapter "Panoramic Miscellanea I. Theory of Connections, Differential Forms, Geodesics, and Holonomy Groups"} and \ref{chapter "Panoramic Miscellanea II. Space Forms, Möbius (Projective) Transformations, and Fuchsian Group; Groupable Synopsis—On the Spin(or)"}) are the oldest, and deals with various topics; they look lectures-like, so they may appear, alas, a little too (pedantically) academic; they were indispensable to me—when I wrote them—to enter in medias res, though. It is quite the case to say, with the old adage: \textit{excusatio non petita, accusatio manifesta}.
\item Hereafter is the list of abbreviations.
\subenumerationisinitium
\item Emphasis abbreviations:

e.a. emphasis added,

e.m. emphasis modified.
\item Language abbreviations:

En. English, 

Fr. French, 

Ge. German, 

Gr. Ancient Greek,

It. Italian,

La. Latin,

Pt. Portuguese,

Ru. Russian,

Zh. Chinese.

Translations of cited passages in epigraph or elsewhere from Fr., Ge., Gr., It., La. and Ru. are under my management, unless otherwise indicated. Sometimes the original text is reproduced in an \texttt{\textbackslash{endnote}} (when, for one reason or another, the text deserves to be read even in the original language); other times, it appears directly in epigraph, and the En. translation is presented in a \texttt{\textbackslash{footnote}}.
\item Abbreviations concerning the books and publications: 

\textsuperscript{c} \hspace{0.3pt} corrected,

\textsuperscript{d} \hspace{0.3pt} digital,

\textsuperscript{ed} \hspace{0.3pt} edition,

\textsuperscript{pr} \hspace{0.3pt} printing,

\textsuperscript{r} \hspace{0.3pt} revised,

\textsuperscript{r.c} \hspace{0.3pt} revised and corrected, 

\textsuperscript{r.e} \hspace{0.3pt} revised and enlarged, 

\textsuperscript{r.u} \hspace{0.3pt} revised and updated, 

\textsuperscript{re} \hspace{0.3pt} reprint, reprinted. 
\subenumerationisfinis
\enumerationisfinis

\setcounter{secnumdepth}{3}

\chapter[\textbf{Acknowledgments}]{Acknowledgments}
\markboth{Intro}{Acknowledgments}

\vspace{-5.5mm}

\begingroup
\footnotesize
Truth\footnote{
	Unquestionably, as I see it, the word “truth” should be taken in \emph{latissimo} sensu; there is no truth, but there are truths (a myriad of truths), and they are changeable, erratic, into the bargain.
	} 
is the offspring of silence, unbroken meditations, and thoughts often revised and corrected \\
\indent — Motto of \textsc{Wollaston–Newton} \cite[sec. III, p. 60]{Wollaston "The Religion of Nature Delineated"} \cite[p. 22]{M. Keynes "The Personality of Isaac Newton"} \cite{Conduitt "Notes on Newton's character"}

\endgroup

\vspace{2mm}

\enumerationisinitium
\item Primarily, a sense of full gratitude to Maryon,\footnote{
	Death has now separated us forever, by bringing you back to an amniotic peace; but it cannot reduce your presence to ashes: you will always be, for me, a horizon on which life shines, my darling Maryon (added on 9 July 2022).
	}
the first and most important name, for the liberty and a fertilely solitary workspace granted to me (feral cats, not much unlike grizzlies, although capable of exhibiting a wide range of social relationships with one's fellows or individuals of other family, are generally not like pack animals, nor do they follow the herd); without such advantage, without a silent shelter, the present book would never have seen the living light of the day.
\item Then I wish to put on record my indebtedness, at a distance, to the \textcyrillic{борхесианская Библиотека} team, and to \textcyrillic{Алексáндра}, for their visionary and active courage (\textgreek{καλὸς γὰρ ὁ κίνδυνος}).
\item List of people (abc order) in the area of mathematics \& physics, and affiliated reflections.
\subenumerationisinitium
\item[·] I am thankful to Claudio Bartocci, say, for reasons of elective affinities and pure congeniality, through the alluring \textgreek{γαῖ} of mathematical and physico-mathematical conceptions, and beyond. 
\item[·] I am grateful to Jean-Michel Bismut for the keen interest shown in this work (he was the first one to admire it), but also and especially for a far-reaching conversation in a pleasant Parisian evening. Non-academic discussions are always the best.
\item[·] A heartfelt appreciation and a sense of \textit{adelphikós} sentiment go to Luciano Boi, for his genuine, immediate and great interest in my \textit{Notes}, without diaphragms and mental superstructures, which is amazingly rare. We travel on the same mental wave. I often find incredible that a single sentence, or a quote, can bring two men closer. That is being, at its source, men of culture. The essence of beauty and allurement of knowledge work these marvels.
\item[·] Marco Fabbrichesi did his utmost to suggest a publication on alternative channels, this being an extremely atypical written, which challenges the platitude (\textit{piattùme}) of standardization in the academic publishing industry,\footnote{
	The vast majority of the academic, or academic-like, publishing industry is not “politically” correct; it generates profits thanks to the aforementioned platitude. We are in the trite slipstream of the cocooned sectorization/minute specialization of which I have already complained on p. \pageref{subsection "Blind Specialism: Cocoon Syndrome"}.
	}  
geared to churn out clone-books (textbooks are a trite and sad example), thereby raising the banner of anonymity, the death of inventiveness.\footnote{
	Additional tips on this matter, in a more traditional context, also came from Luigi Ambrosio.
	}
\item[·] Des remerciements particuliers vont à Thierry Lehner, whom I met at the International Conference organized by the Accademia Vivarium Novum (October 2022). He was intrigued, without delay, by my \textit{Notes}, so that within a few hours we built a special mood that allowed us to walk with four legs conjointly. Through his friendship and brilliant irony, there also passes a perceptive \& robust thread of bijective suggestions on the meaning of mathematics and physics. Among the many curiosities: I discovered that I share his admiration for J.A. Wheeler.
\item[·] A grand recognition of my effort also came from Tom McLeish, who promptly replied to the report of my scientific-literary venture; what brings us together is our certainty of the creative aspect (in latissimo sensu) in scientific production.
\item[·] An email was enough, on the basis of a common vision on science \& other related arguments (what the Germans call \textit{Weltanschauung}), to resonate with Giovanni Vignale, thence with the deeper parts of us. His profound admiration for my «overwhelming» \textit{opera} has acted as a further incitement, for me.
\item[·] The lofty esteem for the present «disconcerting» literary-stone from Giuseppe Vitiello is an incentive to move forward, and to swallow many embitterments and unpalatable  morsels. Herein lies my gratidão: \textit{dignatio est modus inter viros, quos virtutem æstimant}.
\subenumerationisfinis
\item I should also like to thank Cristian Nicolás Mancilla Mardel for his enthusiasm towards the humanistic dimension of my work—it needs to be reiterated that the “sciences” (mathematics \& physics) and the “humanities” (literature, music, art) are two parts of one culture.\footnote{
	\label{footnote "C.P. Snow"}
	It is pretty hard not to hear the rumble of C.P. Snow's words \cite[pp. 3-4]{Snow "The Two Cultures: A Second Look}: «I believe the intellectual life of the whole of western society is increasingly being split into two polar groups. When I say the intellectual life, I mean to include also a large part of our practical life [\,\dots]. Two polar groups: at one pole we have the literary intellectuals, who incidentally while no one was looking took to referring to themselves as ‘intellectuals’ as though there were no others. I remember G.H. Hardy once remarking to me in mild puzzlement, some time in the 1930's: ‘Have you noticed how the word “intellectual” is used nowadays? There seems to be a new definition which certainly doesn't include Rutherford or Eddington or Dirac or Adrian or me. It does seem rather odd, don't y' know.' Literary intellectuals at one pole—at the other scientists, and as the most representative, the physical scientists. Between the two a gulf of mutual incomprehension—sometimes [\,\dots] hostility and dislike, but most of all lack of understanding. They have a curious distorted image of each other». 
	}
\item The family \& non-family members, near and far, who cheered, each in their own way, for this project should be mentioned (abc order): Alberto, Bob,\footnote{
	 \textit{Fortis est, qui liber est}: it is, concurrently, a one-to-one tuition and a wish. In a difficult moment of my life, he pulled my chestnuts out of the fire more than once. I am glad for his active support.
	} 
Brunello, Chiara,\footnote{
	Compared to me, she have the ability to be an action person. So, for me, her endorsement is flattering. Usually a “doer” does not fall prey to the mushy and gelatinous filth hidden in the depths of his/her own thoughts, and thrives better. On the other hand, those who, like me, live isolated in the \textgreek{βάθος} of thought, risk the inanity of the nightmares of the mind, something similar to \textit{śūnyatā} but with the strain of an indissoluble heaviness.
	} 
Daniela, Elena,\footnote{
	She was the \textgreek{Α}-\textgreek{Ω} behind the unorthodox fillip of this opus, with her starry-eyed affection.
	}
Filly, Francesco, Gabriele,\footnote{
	Deviser of the Tardigrade Superluminal Accelerator (\textsc{tsa}). A touch of color. One of his best virtues is the aptitude to get and keep things of one type (collecting), and fantasize in whole about them. This make my day, bro. And all this was combined with the aspect of reception. Every time I came down from Bologna, you welcomed me warmly. How can I disregard your frank fervour, your being “divinely inspired” (\textgreek{ἔν-θεος}) for my mathematical \& physical tales? 
	} 
Giulio,\footnote{
	Do you remember the indistinct and overflowing stream of emotions that resulted from that special day? I refer to 4 July 2006, in Piazza Castello, Ferrara, under the \textit{imaginifiche} notes of \textit{Glósóli} by Sigur Rós. How can we forget that night, from which many views of things—\textgreek{ὄψεις εἰδωλοποιαί}—were born, which became anchored to our delicacy of feeling? 
	
	It is amazing how a single day is enough to transform an entire life, just as a random event, say, an unannounced \& slanted event, is enough to completely change the trajectory of a mental age.
	
	And since life is an interleaving of (apparently) disconnected elements, I add another ingredient. Being united and exalted by a Papinian spirit is no small matter.
	} 
Laura, Lucia, Marco, Martine,\footnote{
	She had the credit of being bluntly and passionately interested in my writing, despite having a different background.
	} 
Miranda.\footnote{
	«Life is too short not to dedicate ourselves to what pleases us [\textit{ci aggrada}] naturally». Thank you for this sentence that you have addressed to me. I might add, especially when this pleasure coincides not in doing things, already done by others, in a better way (or to the best of one's ability), with the underlying encumbrance of replicating the obvious, but rather in doing those things in a—completely—unusual way, which is the source of originality.
	} 
\item Albeit out of context, the next name is that of Daniele, my best youthful friend. I have not seen him for many years (sadly we kind of lost touch), but I can confidently state that he is the smartest guy I have ever met. He was a genius-drenched person. Our Pisan and Livornian face-to-face discussions (2005-2006) were in two voices (\textit{dialogues à double}), he and I; and yet, when something was captured by the attractive force of our confabs, no matter the topic, there was like a single self-feeding brain projected onto a tubular mirror curved on itself (\textit{incurvatus in se}), which, suddenly, turned into a mirrored bridge towards the world; every time, both of us, we were witnessing a sort of peerless and unrepeatable meta-understanding. There is still a lot of him in me (and maybe vice versa too).
\item I have to pay homage to five other people, who has nothing to do with mathematics, at least in a direct and overt way, let us think about the old \textit{quadrivium}. To do this I need the time machine (memory), and to look back over the years, when this book was still a long way from conception. Which, by the way, proves that mental \& effective life is a chain whose links are often stronger than one expects.
\enumerationisinitium
\item[·] I am starting with Takaakira ‘Taka’ Goto (\Ja{後藤孝顕}). We met on 29 November 2007 at Madonna dell'Albero (Ravenna): he is the first man in whom I saw \emph{intensity} (at the highest level) and \emph{passion} (love) fused into a single process. The transition from backstage (with the tranquility of an ordinary person) to frontstage (the eruption of emotions \emph{out} of the ordinary) was impressive. It is been a long time, so he will not remember me, but that moment of \emph{fusion} lives here, then as now, inside me. 
\item[·] Then the \textgreek{Τύχη} played in doubles: I saw (Bologna, 26 May 2008) a fusion of this kind revisited in Munaf Rayani. He is a majestic union of flesh, sound, and rhythm.
\item[·] The third person is Efrim M. Menuck. We met on 23 October 2008 by chance at the Fortezza da Basso, in Florence, between the swarm of people and the beginning of a kermesse. He does not remember me, for sure: a fleeting glance, and a few words. But it is also because of his music that I am \emph{still here}. 
\item[·] The fourth person is Philip Jamieson, along with other comrades, in particular, Erin L. Burke-Moran. The pre-show goliardic spirit is quite fresh in my remembrance, which is opposed to the concert (Turin, 30 November 2010), in a constant crescendo for the climactic closing moment. 
\item[·] The fifth person is Jóhann G. Jóhannsson. I can say that I was lucky enough to see him in Turin (3 September 2011) before his death: a unique lesson of delicacy, made up of \emph{freedom} and \emph{hierarchical harmony} with his ensemble of instrumentalists. 
\enumerationisfinis
\item A separate story is my gratefulness to Davide Lo Iacono, from Eimog—we are in the musical field, again. There was a missed evening, on 13 March 2010, which should have been held in Borgo San Lorenzo, at Villa Pecori Giraldi. We later spoke via email message, and he sent me, as a gift, a copy of their new CD-Digipak, \textit{Scenario} (Subway Productions, 2010), to apologize for the incident. However, the skipped encounter spawned another, more powerful, event. Butterfly effect, as the saying goes.
\item Finally, a sweet smile of complacency is all for Marsili-Alissi Gálakta, who accompanied the completion of the book with her unexpected arrival (the most welcome gift), and her sly and awe-inspiring silence.
\enumerationisfinis

\chapter[\textbf{Glossary: Acronyms and Symbols}]{Glossary \\
	\normalsize Acronyms and Symbols}
	\thispagestyle{empty}
\markboth{Glossary}{Glossary}

\setcounter{footnote}{0}

\begingroup
\footnotesize
\emph{G.e.} is for \emph{Generic element: function, index, map(ping), number, value}. The meaning of each generic type notation arises by referring to specific aspects of the context in which it is used.

The abbreviation [\texttt{cb}] indicates that, for the Greek letters, the \texttt{cbgreek} fonts is used instead of \texttt{cm}-based Greek math fonts. A typographical difference, which corresponds to a different command, e.g. \texttt{\textbackslash{varsigma}} vs. \texttt{\textbackslash{newcommand}\{\dots\}}\texttt{\{cb}-based \texttt{varsigma\}}, is the easiest way to avoid possible confusion.

\endgroup

\begingroup
\renewcommand{\arraystretch}{1.1} 		
\begin{longtable}[l]{p{75pt} p{252pt}}
\tiny\JackStarBold
	& End of definition \\ 

\tiny\FourStar
	& End of example \\ 

\tiny\SixFlowerPetalRemoved 
	& End of margo \\

\tiny$\square$
	& End of proof (\textgreek{ὅπερ ἔδει δεῖξαι}, quod erat demonstrandum) \\ 

\tiny\AsteriskThin
	& End of question \\

$\diamond$
	& End of scholium \\

$\#$ 
	& Cardinality: e.g. $\#G$ is the cardinality of the set $G$ \\

$\backslash$ 
	& The symbols $\backslash$ and $\setminus$ are typeset via \texttt{\textbackslash{backslash}} and \texttt{\textbackslash{setminus}} commands, respectively, and appear typographically the same, but they are characterized by a different blank area of separation (e.g. $A \backslash B$ vs. $A \setminus B$). The first, in that it has a neutral meaning, is universally usable (and therefore, more often than not, it is preferable); the other is originally designed to indicate the set difference \\

$\nmid$ 
	& Not a divisor of, e.g. $a \nmid b$ means that $a$ does not divide $b$ \\

$\leqslant$
	& $\leq$ \\ 
	
$\geqslant$
	& $\geq$ \\

$\viz$
	&
Videlicet, that is (to say), namely, the same as \\

$\equival$	
	& Identity expressed by an equivalence relation; the letters \textgreek{ιδ} are derived from the noun \textgreek{ισοδυναμία} (equivalence) \\
	 
$\aprx$
	& Approximation with equality sign \\

$\atmo$
	& At most \\

$\corr$ 
	& Correspondent with/to \\

$\refo$
	& Reformulated in \\

$\xrightarrow{1:1}$ 
	& One-to-one correspondence (bijection) \\
	
$\sim$ 
	& Asymptotic equivalence. — Equivalence relation \\

$\approx$
	& Approximation \\

$\cong$
	& Congruence; isomorphism (homeomorphism, diffeomorphism) \\  
	
$\propto$ 
	& Proportionality \\

$\varnothing$
	& Empty set \\

$\cap$ 
	& Intersection \\ 

$\cup$ 
	& Union \\

$\bigcup$ 
	& Big union \\

$\sqcup$
	& Disjoint union \\

$\in$
	& Set membership \\
	
$\notin$
	& Negation of set membership \\

$\subset$
	& Subset. More specifically, we talk about an improper, or broad, subset (e.g. $H \subseteqq G$ means $H \subset G$ and $H = G$) and a proper, or strict, subset (e.g. $H \subsetneqq G$ means $H \subset G$ and $H \neq G$) \\ 
	
$\Subset$ or $\subset\subset$ 
	& Compactly contained, compact embedding \\

$\supset$
	& Supset. More specifically, we talk about an improper, or broad, supset (e.g. $G \supseteqq H$ means $G \supset H$ and $G = H$) and a proper, or strict, supset (e.g. $G \supsetneqq H$ means $G \supset H$ and $G \neq H$) \\

$\hookrightarrow$
	& Embedding. — Immersion. — Inclusion map \\

$\sNeg[1mu]{\rightarrow}$ 
	& Negation of $\to$ \\

$\perp$
	& Orthogonal(ity) \\

$\circ$
	& Composition of functions or relations \\ 

$\rtimes$
	& Semidirect product (right normal factor) \\

$\oplus$
	& Direct sum \\ 
	
$\otimes$ 
	& Tensor product \\ 

$\bigotimes$ 
	& Big product \\ 

$\KulkarniNomizu$
	 & Kulkarni–Nomizu product \\

$\lrcorner$ 
	& Interior product (or contraction) \\

$\vee$
	& \texttt{\textbackslash{vee}} command: supremum, also called “least upper bound”, or “join” in lattice and order theory \\

$\bigvee$
	& \texttt{\textbackslash{bigvee}} command \\

$\wedge$
	& \texttt{\textbackslash{wedge}} command: exterior or wedge product in exterior algebra, in differential geometry, and vector calculus. — Infimum,  also called “greatest lower bound”, or “meet” in lattice and order theory \\

$\bigwedge$ 
	& \texttt{\textbackslash{bigwedge}} command \\
	
$\bigwedge^2$, $\bigwedge^k$ 
	& \texttt{\textbackslash{bigwedge}} command: second exterior power, $k$th exterior power \\

$\bigwedge^k_{\mathrm{c}}(\cdot) \viz \Omega_\mathrm{c}^k(\cdot)$ 
	& Set of all $k$-forms having compact support \\
	
$[\cdot \:, \cdot]$ 
 	& Lie bracket (of vector fields) \\

$\langle\cdot \:, \cdot\rangle$ 
	& Inner product, Hermitian product. — Dot or scalar product, a case of the inner product when one considers the field of real numbers, i.e. Euclidean spaces. — Note: in physics (quantum mechanics) the Dirac bra-ket notation \cite[p. 417]{Dirac "A new notation for quantum mechanics"} $\langle\cdot|\cdot\rangle$ is adopted \\  

$\{\cdot \:, \cdot\}$ 
	& Anti-commutator. — Poisson bracket \\

$\|\cdot\|$
	& Norm, vd. $\|x\|$ and $\mathfrak{X} \viz (\mathfrak{X}, \|\cdot\|)$ \\

$\Lbrack:\cdots:\Rbrack$ 
	& Repeat sign, inspired by the musical symbols for the left and right repeat sign \\

$\Langle\cdots\Rangle$ 
	& Term of omission: what is inside these chevrons is omitted \\

$\textcolor{cynara-violet}{\lfloor} \cdots \textcolor{cynara-violet}{\rceil}$
	& \texttt{\textbackslash{Biggl}} $\textcolor{cynara-violet}{\lfloor}$ \emph{floor} and \texttt{\textbackslash{Biggr}} $\textcolor{cynara-violet}{\rceil}$ \emph{ceiling} notations are employed in Chapter \ref{chapter "Toroidal Fourier Analysis"} improperly, viz. as simple splitters of an equation, when the latter is too long, and does not fit on a single line \\

$\forall$
	& For all/any/each/every element(s) (universal quantification) \\

$\lnot$
	& Negation (logical complement, via \texttt{\textbackslash{lnot}} command; idem via \texttt{\textbackslash{neg}} command) \\

$\nexists$ 
	& Negation of existential quantification, and it means “there does not exist” \\

$\flat$ 
	& Bemolle (flat) map, $\flat$-map in musical isomorphism \\

$\sharp$ 
	& Diesis (sharp) map, $\sharp$-map in musical isomorphism \\

$\natural$ 
	& Bequadro (natural) dimension of $\mathbb{CM}^\natural$ and $\mathbb{CM}^\natural_{(\infty)}$ \\

$\angle$ 
	& Angle \\

$\measuredangle$ 
	& Vertex angle \\

$\angle_\mathrm{d}$ 
	& Dihedral angle \\

$\angle_\mathrm{i}$ 
	& Interior angle \\

$\rotatedtriangle$ 
	& Triangle \\

$\heatoperator$ 
	& Heat operator\footnote{
	In literature it is denoted by the square/quabla $\square$ symbol (but there is the risk of being mistaken with the d'Alembertian) or by the Latin letter $L$ supposedly because of Laplace \cite{Laplace "Memoire sur divers points d'analyse"}.
	}, or Laplace–Poisson operator \\
$\heatoperator^*$ 
	& Adjoint of the heat operator $\heatoperator$ \\

$\dAlembertian$ 
	& d'Alembert operator, aka wave or box (and even, occasionally, quabla) operator, or simply d'Alembertian \\

$\Laplacian$ 
	& Laplace operator, or Laplacian \\ 

$\Laplacian_\textsc{l}$ 
	& Lichnerowicz Laplacian \\

$\rotatedheartsuit$ 
	& $\delta$-variation of a value, adding the symbol $\rotatedheartsuit$ in superscript (e.g. $\tau^\rotatedheartsuit = \delta\tau$) \\ 

$\mathbbl{1}$ 
	& Indicator function: $\mathbbl{1} = \idem$ \\ 

$\{\mathbbl{1}, \mathbbl{i}, \mathbbl{j}, \mathbbl{k}\}$ 
	& Basis elements of quaternion algebra \\
	
$^{(1, 3)^+}_{(1, 3)^+}$ and $^{(1, 3)^-}_{(1, 3)^-}$ 
	& Metric signatures $(+, -, -, -)$ and $(-, +, +, +)$, respectively \\

$2^{\aleph_0}$ 
	& Cardinality of the continuum $\mathsf{C}_\mathbb{R}$, or of the set of real numbers $\mathbb{R}$ \\

$\aleph_0$ 
	& Aleph null: lowest/smallest transfinite cardinal number; Cantor's number of algebraic numbers; it is also the cardinality of the set of all natural numbers, or cardinality of $\omega_\mathbb{N}$ (for the ordinal aspect) \\

$\aleph_1$ 
	& Aleph one: first uncountable cardinal number, designating the cardinality of the set of all countable ordinal numbers, i.e. of $(\omega_\mathbb{N})_1$ \\

$\beth_0$ 
	& Beth null: $\beth_0 = \aleph_0$ \\ 

$\beth_1$ 
	& Beth one: $\beth_1 = \aleph_1$ \\

$\beth_{\alpha_k + 1}$ 
	& Beth number with an ordinal number ($\alpha_k$): $\beth_{\alpha_k + 1} = 2^{\beth_{\alpha_k}}$ \\

$\alpha$
	& G.e. — Hölder exponent. — Parameter in the Kerr metric: $\alpha = \frac{J}{m}$ \\ 
 
$\alpha$ 
	& Multi-index \\

$\alpha_\mathrm{em}$ 
	& Fine-structure constant for electromagnetic interactions \\

$\alpha_k$ 
	& Ordinal number \\

$\alpha_{H^0}$ 
	& Parameter for Higgs boson, via scattering processes \\

$\alpha_\mathsf{T}$ 
	& Thermal diffusivity \\

$\textgreek{\textit{α}}$ 
	& \textsuperscript{[\texttt{cb}]} Rationalized area of a black hole: $\textgreek{\textit{α}} = \frac{A_\mathrm{h}}{4\pi}$ (in black hole thermodynamics) \\

$\beta$ 	
	& G.e. \\

\reflectbox{\textgreek{β}} 
	& Background, e.g. $\Ricci^\background_{\mu\nu}$, $g^\background_{\mu\nu}$, $\scalarcurvature^\background$ \\
	
$\gamma$ 
	& G.e. — Photon\endnote{
	Historical and philological note. The word \emph{photon} (\textgreek{φῶς, φωτός}) was coined by G.N. Lewis \cite[p. 874]{Lewis "The Conservation of Photons"} in 1926, but by “photon” he understands something different from what Einstein had imagined in 1905 \cite[pp. 144-145]{Einstein "Uber einen die Erzeugung und Verwandlung des Lichtes betreffenden heuristischen Gesichtspunkt"}, with his \textit{Lichtquant} (literally “light quantum”, in the sense of a “corpuscle/particle of light”).
	} \\

$\gamma_{(\alpha)(\beta)(\lambda)}$ 
	& Ricci rotation coefficients in the tetrad basis \\

$\gamma_{\hat{\alpha}\hat{\beta}\hat{\lambda}}$ 
	& The same as $\gamma_{(\alpha)(\beta)(\lambda)}$ \\

${\gamma^\mu}_{\nu\xi}$ 
	& Ricci rotation coefficients \\

$\gamma^\mu \viz \gamma_\textsc{d}^\mu$ 
	& Gamma (Dirac) matrices: $\gamma^\mu = \{\gamma^0, \gamma^1, \gamma^2, \gamma^3\}$, plus $\gamma^5 = i\gamma^0, \gamma^1, \gamma^2, \gamma^3$ \\

$\gamma_\textsc{w}^\mu$ 
	& Gamma matrices in the Weyl representation \\

$\gamma_\textsc{m}^\mu$ 
	& Gamma matrices in the Majorana representation \\

$\gamma_\varepsilon(x)$ 
	& Sequence of functions \\

$\gamma_\mathrm{c}$
	& Curve (curved line), path or 
	\begin{equation*}
	\text{geodesic}
	\begin{cases}
	\text{path}, \\
	\text{line}, \\
	\text{curve/curvature}
	\end{cases}
\end{equation*} 
(the subscript $c$ stands for \emph{curve}) \\

$\dot{\gamma}_\mathrm{c}$ 
	& Velocity (vector field) of a curve, path or geodesic \\

$\ddot{\gamma}_\mathrm{c}$
	& Acceleration (vector field) of a curve, path or geodesic: $\ddot{\gamma}_\mathrm{c} \equival D_t\dot{\gamma}_\mathrm{c} \viz \nabla_{\dot{\gamma}_\mathrm{c}}\dot{\gamma}_\mathrm{c}$ \\

$\gamma_\mathrm{null}$ 
	& Null geodesic, ray of light \\

$\gamma_{\mathrm{s}+}$, $\gamma_{\mathrm{s}-}$ 
	& Variables in the vibrating string model: $1\mathrm{D}$ wave phenomenon, in accordance with the method of d'Alembert \\

$\gamma_{(\mathbb{PT}\mathbbl{w})}$ 
	& Line in $\Upsilon_{\mathbb{PT}\mathbbl{w}}$ \\

$\gamma^\textsc{k}_{(\mathbb{PT}\mathbbl{w})}$ 
	& Kodaira's holomorphic line \\

$\textgreek{\textit{γ}}$ 
	& \textsuperscript{[\texttt{cb}]} Lorentz factor \\

$\bbgamma$
	& Gaussian measure \\

$\rotatedgamma^{(3)}_{\mu\nu}$, $\rotatedgamma^{(4)}_{\mu\nu}$ 
	& 3- and 4-metric tensor (field), i.e. 3- and 4-dimensional metric, respectively \\

$\Gamma_{\mu\nu\xi}$
	& (or other indices) Christoffel symbols (or component of the $\Gamma$ connection) of the first kind \\ 

$[\mu\nu, \xi]$
	& (or other letters) The same as $\Gamma_{\mu\nu\xi}$ \\

${\Gamma^\xi}_{\mu\nu}$
	& (or other indices) Christoffel symbols (or component of the $\Gamma$ connection) of the second kind \\ 

$\binomcurly{\xi}{\mu\nu}$ 
	& (or other letters) The same as ${\Gamma^\xi}_{\mu\nu}$ \\
	
$\Gamma_{\sezione}$
	& Space of sections; e.g. $\Gamma_{\sezione}\bigl(\bigwedge^k\mathring{\mathcal{T}}^*\mathcal{M} \otimes \mathring{\mathcal{E}}\bigr)$ is the space of sections of the bundle $\bigwedge^k\mathring{\mathcal{T}}^*\mathcal{M} \otimes \mathring{\mathcal{E}}$ \\

$\Gamma(\mathring{\mathcal{P}}_\textit{ß})$ 
	& Connection on a spinor bundle \\

$\vec{\Gamma}_\textsc{fp}$ 
	& Faddeev–Popov (gauge) ghost field associated with the photon \\

$\Gamma_\mathbb{R}$ 
	& Group of real numbers (non-Archimedean analysis): $\Gamma_\mathbb{R} = \{\mathbbl{r}^1, \mathellipsis, \mathbbl{r}^n\}$ \\

$\textgreek{\text{Γ}}(\cdot)$ 
	& Gamma function \\

$\mathbbl{\Gamma}$ 
	& Discrete group of isometries of $\widetilde{\mathcal{M}}^n_\kappa$ (see), group of covering transformations on $\widetilde{\mathcal{M}}^n_\kappa$. — Discrete group of (Möbius) $\Moebius$-transformations (see). — Fuchsian group:  discrete subgroup of $PSL_2(\mathbb{R})$ or conjugate of this group in $PSL_2(\mathbb{C})$ \\

$\mathbbl{\Gamma} \backslash \mathbb{U}^2_\mathbb{C}$ 
	& Riemann (or modular) surface, orbit space, see $\mathcal{S}_\mathbbl{\Gamma}$ \\

$\mathbbl{\Gamma} \backslash PSL_2(\mathbb{R})$ 
	& Orbit projective space \\

$\delta$
	& G.e. — $\delta$-variation (calculus of variations), change(s) in the value of a variable \\

$\Diracdelta$ 
	&
	\footnote{
	Via \texttt{\textbackslash{Diracdelta}} command.
	}
	Dirac delta function \\

$\delta^\angle$ 
	& Angular deficit \\

$\delta^\angle(\textit{\ng}) \viz \delta^\angle_{\textit{\ng}}$ 
	& Angular deficit at $\textit{\ng}$ \\

$\delta^\angle(\mathrm{ver}) \viz \delta^\angle_{\mathrm{ver}}$ 
	& Angular deficit at $\mathrm{ver}$ \\

$\delta_{\mu\varrho}, {\delta^\mu}_\varrho$
	& (or other indices) Kronecker delta: $\delta_{\mu\varrho} \equival {\delta^\mu}_\varrho$ \\
	
$\delta_\textsc{k}$
	& The same as $\delta_{\mu\varrho}, {\delta^\mu}_\varrho$, Kronecker delta \\

$\delta_\mathsf{T}$ 
	& Thermal fluctuation \\

$\delta\mathscr{S}_\textsc{eh}$ 
	& Variation of $\mathscr{S}_\textsc{eh}$ \\

$\delta\mathscr{S}_\textsc{p}$ 
	& Palatini variation of the gravitational action \\

$\mathrm{\Delta}$ 
	& G.e. — Positive or negative amount; finite change, displacement, or length of path \\

$\mathrm{\Delta}_\psi(\cdot)$ 
	& Uncertainty in a measurement of some observable $(\cdot)$, deviation  correlated with a measurement of $(\cdot)$ \\

$\mathrm{\Delta}_\psi(\hat{\mathrm{x}})$, $\mathrm{\Delta}_\psi(\hat{\mathrm{p}})$ 
	& Uncertainties of $\hat{\mathrm{x}}$ and $\hat{\mathrm{p}}$ of a wave function $\psi$ \\

$\nabla$ 
	& Connection (generic notation) on the fiber bundle, e.g. vector bundle, tangent bundle, see $\omega_\mathfrak{h}$. — Covariant derivative. — Levi-Civita connection. — Nabla or del: vector differential operator, e.g. gradient (operator), $\gradient\varphi = \nabla\varphi$ \\

$\nabla^2$ 
	& The same as $\Laplacian$ (Laplacian), so $\Laplacian = \nabla^2 = \nabla \cdot \nabla$ \\

$\nabla^{\mathring{\mathcal{E}}}$ 
	& Connection on the vector bundle \\

$\nabla^{\mathring{\mathcal{T}}}$ 
	& Connection on the tangent bundle (linear connection); Levi-Civita connection in explicit notation \\

$\nabla_{\vec{X}}\sezione$ 
	& Covariant derivative of $\sezione$ in the direction of $\vec{X}$ \\

$\nabla \times (\cdot)$ 
	& Curl of a vector field, i.e. $\mathrm{curl}(\cdot)$ \\  

$\epsilon$
	& G.e. \\

$\epsilon_\mathbbl{r}$ 
	& Monosemio number in Levi-Civita's non-Archimedean field ($\epsilon, \mathbbl{r} \in \mathbb{R}$, with $\mathbbl{r} = 0$) \\

$\reflectedepsilon$ 
	& Constant value \\

$\varepsilon$
	& G.e. \\

$\varepsilon \viz \mathrm{Leb}(\mathcal{U})$ 
	& Lebesgue number for the open covering ($\mathcal{U}$) of a compact metric space \\

$\{\varepsilon_{(\alpha)}\}$
	& Tetrad formalism \\

$\{\varepsilon_{\hat{\alpha}}\}$
	& The same as $\{\varepsilon_{(\alpha)}\}$ \\

$\varepsilon_{\mu\nu\xi}$ 
	& (or other indices) Levi-Civita symbol in three dimensions \\

$\textgreek{\textit{\ddigamma}}$ 
	& Scalar field, scalar function \\

$\textgreek{\textit{\ddigamma}}_{H^0}^0$, $\textgreek{\textit{\ddigamma}}_{H^0}^+$, $\textgreek{\textit{\ddigamma}}_{H^0}^-$ 
	& Higgs scalar fields \\

$\textgreek{\text{\ddigamma}}$ 
	& \textsuperscript{[\texttt{cb}]} Vector-valued function, vector function (generic notation); map, vector function of position or vector field: $\textgreek{\text{\ddigamma}} =\textgreek{\text{\ddigamma}}_1\sigma_1 + \cdots +\textgreek{\text{\ddigamma}}_n\sigma_n$ \\

$\textgreek{\text{\ddigamma}}_{\vec{X}}$ 
	& \textsuperscript{[\texttt{cb}]} Fundamental vector field \\

$\zeta$
	& G.e. — Single 2-component spinor, see $\psi$ as Dirac 4-spinor \\ 
		
$\{{\zeta_\bbmu}_n\}_{n \in \mathbb{N}}$
	& Non-decreasing sequence of $\bbmu$-integrable functions \\

$\mathring{\zeta}$ 
	& Fiber/vector bundle: $\mathring{\zeta} = (\mathring{\mathcal{E}}, \pi, \mathcal{M}, \mathring{\mathcal{F}})$, $\mathring{\zeta} = (\mathring{\mathcal{E}}, \pi, \mathcal{M})$ \\

$\mathring{\zeta}_{\mathbb{C}^r}$ 
	& Complex fiber/vector bundle of rank $r$ \\ 

$\mathring{\zeta}_{\mathbb{R}^r}$ 
	& Real fiber/vector bundle of rank $r$ \\

$\textgreek{Ζ}^{\mu\nu}$, ${\textgreek{Ζ}_{\mu\nu}}^{\xi\varrho}$ 
	& \textsuperscript{[\texttt{cb}]} 2-tensor, 4-tensor (generic notation) \\

$\eta$
	& G.e. \\
		
$\eta$, $\eta^{\mu\nu}$, $\eta_{\mu\nu}$ 
	& Metric tensor of Minkowski space(-time); it is a pseudo-Euclidean metric \\

$\eta_{\mu\nu}^{(1, 3)^+}$, $\eta_{\mu\nu}^{(1, 3)^-}$ 
	& Minkowski metric tensor with  signatures $(+, -, -, -)$ and $(-, +, +, +)$, respectively \\

$\eta_c$ 
 	& Dimensionless constant of order unity (in black hole thermodynamics) \\

$\eta^+_t$ 
	& 1-parameter groups related to the stable horocycle flow \\

$\eta^-_t$
	& 1-parameter groups related to the unstable horocycle flow \\

$\{\eta^+_t\}_{t \in \mathbb{R}}$ 
	& Stable horocycle flow \\

$\{\eta^-_t\}_{t \in \mathbb{R}}$ 
	& Unstable horocycle flow \\

$\rotatedeta$ 
	& Function (notation for the Poisson bracket) \\

$\textgreek{Η}_\mathrm{o}$ 
	& \textsuperscript{[\texttt{cb}]} Operator (Calabi conjecture) \\

$\theta$ 
	& G.e. — Angle (of rotation). — Polar angles (colatitude) \\

$\theta^\measuredangle$ 
	& Vertex angle \\

$\theta_\mathrm{w}$ 
	& Mixing angle, aka Weinberg angle (parameter of the electroweak) \\

$\vartheta$ 
	& G.e. \\

$\vartheta^\nu$ 
	& $\binom{1}{0}$-tensor valued 1-form; vector-valued 1-form \\ 

$\vartheta_1, \mathellipsis, \vartheta_n$ 
	& 1-forms, dual coframe of $\vec{E}_1, \mathellipsis, \vec{E}_n$, forming an orthonormal basis of the cotangent space $\mathcal{T}^*_p\mathcal{M}$ \\

$\vartheta_\Upsilon$
	& $\mathfrak{g}$-valued 1-form on $\Upsilon$ \\

$\vartheta \viz \vartheta_\mathbb{T}$ 
	& Dynamical system on $\mathcal{Q}$ (that is a group of homeomorphisms under composition on $\mathcal{Q}$): $\vartheta \viz \vartheta_\mathbb{T} = \{\vartheta_t\}_{t \in \mathbb{T}}$ \\

$(\vartheta, \varphi) \viz (\vartheta_\mathbb{T}, \varphi_\mathbb{T})$ 
	& Non-autonomous dynamical system; random dynamical system; skew product flow  \\

$\vartheta_w$ 
	& Vector: $\vartheta_w \viz \vartheta(w)$ \\

$\vartheta_\textsc{h}$ 
	& Hölder homeomorphism \\

$\Theta(\cdot)$ 
	& Heaviside step function \\

$\Theta_\rho$ 
	& Radial function \\

$\Theta_\tau$ 
	& Torsion form (of a connection form); vector-valued 2-form \\

$\textgreek{\text{Θ}}$ 
	& \textsuperscript{[\texttt{cb}]} Noise intensity, from \textgreek{Θ[όρῠβος]}, “noise” \\

$\iota$ 
	& G.e. \\
	
$\iota_{\vec{X}}$ 
	& (or other subscript) The same as $\lrcorner$ \\

$\kappa$
	& Curvature. — (Constant) sectional curvature, or Gaussian curvature \\

$\kappa_\bbmu$ 
	& $\bbmu$-integrable function: $\kappa_\bbmu \in \Lebesgue^1(\mathring{\mathcal{T}}^1\mathcal{S}_\mathbbl{\Gamma}, d\bbmu)$ \\

$\kappa_\Omega$
	& The same as $\Omega^2_\mathfrak{g}$. — The $H$-module decomposition for the $\kappa$-curvature is $\kappa_\Omega = \kappa_\mathfrak{h} \oplus \kappa_\mathfrak{p}$ \\

$\kappa_\mathrm{(tr)}$ 
	& Trace of the extrinsic curvature \\

$\varkappa$ 
	& G.e. \\

$\Einsteinconstant$ 
	& Einstein (gravitational) constant: $\Einsteinconstant = \frac{8\pi\gravitation}{c^4}$ \\

$\lambda$ 
	& G.e. — Scalar \\ 

$\lambda_1$ 
	& Constant (Poincaré inequality) \\

$\lambda_\bbmu$ 
	& $\bbmu$-integrable function: $\lambda_\bbmu \in \Lebesgue^1(\mathring{\mathcal{T}}^1\mathcal{S}_\mathbbl{\Gamma}, d\bbmu)$  \\

$\lambda_\textsc{l}$, $\lambda_\textsc{l}^+$, $\lambda_\textsc{l}^-$ 
	& Lyapunov exponents \\

$\lambda_\volume$ 
	& (Exponential) growth rate of volume, volume growth rate \\

$\lepton^{\varsigma, \tau}$ 
	& \textsuperscript{[\texttt{cb}]} Lepton, where $\varsigma, \tau = 1, \mathellipsis, 3$ are the generation indices \\

$\textgreek{\text{λ}}_\textsc{g-m}^\alpha \viz \textgreek{\text{λ}}_{\chi_i\chi_j}^\alpha$ 
	& \textsuperscript{[\texttt{cb}]} Gell-Mann matrices, with $\alpha = 1, \mathellipsis, 8$ \\ 	

$\bblambda$ 
	& Lebesgue measure \\

$\Lambda$ 
	& G.e. \\
	
$\Lambda = [{\Lambda^\mu}_\nu]$ 
	& Matrix of an orthogonal transformation, i.e. general, proper, orthochronous Lorentz transformation \\

$\Lambda_q$ 
	& $q$-fiber of $\mathbbl{\Lambda}$, i.e. $\Lambda_q = \{x \in \mathcal{X}\mid (q, x) \in \mathbbl{\Lambda}\}$, \\

$\Lambda_t$ 
	& $t$-fiber of $\mathbbl{\Lambda}$, i.e. $\Lambda_t = \{x \in \mathcal{X}\mid (t, x) \in \mathbbl{\Lambda}\}$ \\

$\mathrm{\Lambda}$ 
	& Cosmological constant \\

$\mathbbl{\Lambda}$ 
	& Attractor. — Non-autonomous set of a process: $\mathbbl{\Lambda} \subset (\mathbb{T}\mathcal{X})_\bbmu = \mathbb{T} \times \mathcal{X}$. — Non-autonomous set of a skew product flow: $\mathbbl{\Lambda} \subset (\mathcal{X}\mathcal{Q})_\bbmu = \mathcal{Q} \times \mathcal{X}$ \\

$\mathbbl{\Lambda}_\singularity$ 
	& Singular hyperbolic attractor, or repeller; singular Lorenz-like hyperbolic attractor \\

$\mu$
	& G.e. \\ 
	
$\textgreek{\text{μ}}$ 
	& \textsuperscript{[\texttt{cb}]} Coefficient of friction \\

$\bbmu$ 
	& Measure on a space/set; probability measure. Some examples: Borel measure, Liouville-like measure, Riemannian measure (induced by the Riemannian structure); Sinai–Ruelle–Bowen (\textsc{srb}) measure \\

$\nu$
	& G.e. \\ 

$\neutrino^{(\wr)}_\varsigma$ 
	& Image of a neutrino $\neutrino_\varsigma$ under \textsc{cpt} symmetry, i.e. transformations of charge conjugation (\textsc{c}), parity transformation (\textsc{p}), and time reversal (\textsc{t}) \\

$\bbnu$
	& Correlation dimension (measure for the strangeness of attractors; it is analogous to the fractal dimension) according to Grassberger–Procaccia algorithm: $\bbnu = \lim_{\distance \to 0}\frac{\log{C}(\distance)}{\log\distance}$ \\

$\xi$
	& G.e. \\ 
	
$\rotatedxi$ 
	& Function (notation for the Poisson bracket) \\
	
$\textgreek{\textit{ο}}$
	& Holomorphic function, map \\

$\pi$ 
	& G.e. — Irrational and transcendental number: $3.14159\cdots$. — Projection (map) \\ 

$\pi \viz \pi_\mathbb{T}$ 
	& Skew product flow (autonomous semi-dynamical system); $\pi = (\vartheta, \varphi)$ \\

$\pi_1$ 
	& First factor projection, projection onto the first factor (alternative notation: $\prj_1$) \\ 

$\pi_1(\mathcal{X})$ 
	& Fundamental group, first homotopy group, or Poincaré group \cite[§ 12]{Poincare "Analysis Situs"} of some topological space $\mathcal{X}$ (it is a topological and homotopy invariant) \\

$\pi_2$ 
	& Second factor projection, projection onto the second factor (alternative notation: $\prj_2$) \\

$\rotatedpi$ 
	& Permutation \\

$\rotatedpi^{\varphi_{(F)}} \in \symmetric_n$ 
	& Permutation belonging to the symmetric group \\

$\varpi$
	& G.e. \\ 

$\underdot{\varpi}$ 
	& Continuous function $\underdot{\varpi} \colon \mathcal{X} \to \mathbb{R}$ (notation adopted for the topological pressure) \\

$\Kahlerpotential$ 
	& Kähler potential (smooth real function) \\

$\Pi$ 
	& Perimeter, e.g. $\Pi(\invertedbreve{C})$ is the perimeter of $\invertedbreve{C}$ \\

$\OrnsteinUhlenbeck$ 
	& \textsuperscript{[\texttt{cb}]} Ornstein–Uhlenbeck process \\

$\rho$ 	
	& G.e. — Radius \\ 

$\rho_Q$ 
	& Source density (for the quadrupole moment tensor) \\

$\varrho$ 	
	& G.e. \\

$\rotatedvarrho$ 
	& G.e. \\

$\distance$ 	
	& Distance: $\distance = \dist$ \\ 

$\distance(x, y)$ 
	& Distance between two points $x$ and $y$ of a metric space $(\mathcal{X}, \distance)$; we can even write $\distance(x, y) \viz |x - y|_\mathcal{X}$ \\

$\sigma$
	& G.e. \\

$\sigma$-, $_\sigma$
	& $\sigma$-algebra ($\sigma$-field): sigma-algebra (sigma-field) \\

$\sigma_{\mathring{\mathcal{E}}}$ 
	& Section of $\mathring{\mathcal{E}}$. — Local and global frame field for the tangent/vector bundle, $\sigma_{\mathring{\mathcal{E}}} \in \mathfrak{E}(\Upsilon)$ and $\sigma_{\mathring{\mathcal{E}}} \in \mathfrak{E}(\mathcal{M})$, respectively, $\sigma_{\mathring{\mathcal{E}}} = \{\sigma_1, \mathellipsis, \sigma_n\}$ \\

$\sigmamodel$
	& \textsuperscript{[\texttt{cb}]} $\sigmamodel$-model, sigma model \\

$\sigmaPauli_\mu$
	& \textsuperscript{[\texttt{cb}]} Pauli (spin) matrices: $\sigmaPauli_\mu = \{\sigmaPauli_0, \sigmaPauli_1, \sigmaPauli_2, \sigmaPauli_3\}$ \\

$\vec{\sigmaPauli}$ 
	& \textsuperscript{[\texttt{cb}]} Spin vector having the Pauli matrices as its components: $\vec{\sigmaPauli} = (\sigmaPauli_1, \sigmaPauli_2, \sigmaPauli_3)$\footnote{
	In these \textit{Notes}, vectors, especially when they are written in lower case letters, are usually written without an arrow above. However, in some cases, it is desirable to use the arrow command to avoid semantic overlaps, distinguishing between $\sigmaPauli_\mu$ (Pauli matrices) and $\vec{\sigmaPauli}$ (Pauli vectors), or e.g. between $p$ (point) and $\vec{p}$ (momentum). 
	} \\

$\sigmaPauli^\mu$, $\underline{\sigmaPauli}^\mu$
	& \textsuperscript{[\texttt{cb}]} Sigma (Pauli) matrices: $\sigmaPauli^\mu = (1, \vec{\sigmaPauli})$, $\underline{\sigmaPauli}^\mu = (1, -\vec{\sigmaPauli})$ \\

$\sigmaPauli \cdot \vec{p}$ 
	& \textsuperscript{[\texttt{cb}]} Longitudinal polarization of a particle \\

$\bbsigma$ 
	& Probability measure: $\bbsigma \in \mathbbl{M}$ \\

$\varsigma$ 
	& G.e. — Spinor map: $\varsigma \colon SU_2(\mathbb{C}) \cong \Spin_3(\mathbb{R}) \to SO_3(\mathbb{R})$ \\

$\varsigma(\mathbb{R})_\mathcal{M}$ 
	& Real spinor field on $\mathcal{M}$ with decomposition into elements $\bigl(\varsigma(\mathbb{R})^+_\mathcal{M}\bigr) + \bigl(\varsigma(\mathbb{R})^-_\mathcal{M}\bigr) = \varsigma(\mathbb{R})_\mathcal{M}$ \\

$\sezione$ 
	& \textsuperscript{[\texttt{cb}]} Section of a fiber/vector/tangent bundle \\

$\sezione\bigl(\mathring{\mathcal{P}}_\textit{ß}^\pm\bigr)$ 
	& \textsuperscript{[\texttt{cb}]} Spinor field, or section of $\mathring{\mathcal{P}}_\textit{ß}$ \\

$\sezione(\mathring{\mathcal{T}}^r_s\mathcal{M})$ 
	 & \textsuperscript{[\texttt{cb}]} Section of $\mathring{\mathcal{T}}^r_s\mathcal{M}$ \\

$\Sigma$ 
	& Set of points \\

$\tau$ 
	& G.e. — Time (generic notation). — Proper time (interval) in Minkowski space(-time) diagram. — Torsion tensor: $\binom{1}{2}$-tensor field or tensor of type $(1, 2)$ \\

$\tau \in \mathfrak{T}^1_2(\mathcal{M})$ 
	& Torsion tensor of the the connection $\nabla$ \\

$\tau^\rotatedheartsuit$ 
	& $\tau^\rotatedheartsuit = \delta\tau$ \\

$\tau^r_s$ 
	& Tensor fields (or simply tensor) of type $(r, s)$: $\tau^r_s = \Tau^{\mu_1 \cdots \mu_r}_{\nu_1 \cdots \nu_s}$ \\

$\tau_\bbmu$ 
	& Measurable function: $\tau_\bbmu \in \Lebesgue^1(\invertedbreve{\Omega}, \bbmu)$ \\

$\tau_{\bbmu(\mathscr{U})}$ 
	& $\mathscr{U}_\sigma$-measurable function: $\tau_{\bbmu(\mathscr{U})} \in \Lebesgue^1(\invertedbreve{\Omega}, \bbmu)$ \\

$\tau_\textsc{qg}$ 
	& Particle's flight time due to quantum-mechanical fluctuations of space-time \\

$\rotatedtau$ 
	& G.e. \\

\textgreek{\text{τ}} 
	& \textsuperscript{[\texttt{cb}]} Reciprocal of the golden ratio: $\textgreek{\text{τ}} = \frac{1}{\textgreek{\text{φ}}}$ \\
	
$\Tau$
	& \textsuperscript{[\texttt{cb}]}\footnote{
	From Greek (\textgreek{ταῦ}, by way of a \texttt{\textbackslash{Tau}} command) and not Latin alphabet (although the two T-fonts are indistinguishable).
	} 
	Tensor \\

$\Tau^{\mu\nu}$, $\Tau_{\mu\nu}$ 
	& Energy-momentum tensor, also called stress-energy tensor, or stress-energy-momentum tensor \\

$\Tau^{\mu\nu}$ 
	& Total energy-momentum tensor: $\Tau^{\mu\nu} = \Tau_\mathrm{em}^{\mu\nu} + \Tau_\mathrm{p}^{\mu\nu}$ \\

$\Tau_\varepsilon^{\mu\nu}$ 
	& Regularized energy-momentum tensor (for any $\varepsilon > 0$) \\

$^\mathrm{div}\Tau_\varepsilon^{\mu\nu}$ 
	& Divergent tensor of $\Tau_\varepsilon^{\mu\nu}$ \\

$\Tau_\mathrm{em}^{\mu\nu}$ 
	& Energy-momentum tensor of the electromagnetic ($\mathrm{em}$) field \\

$\Taustroke_\mathrm{em}^{\mu\nu}$ 
	& Renormalized electromagnetic energy-momentum tensor \\

$\Tau^\mathrm{gw}_{\mu\nu}$ 
	& Energy-momentum tensor of a gravitational wave ($\mathrm{gw}$) \\

$\Tau_\mathrm{p}^{\mu\nu}$ 
	& Energy-momentum tensor of a particle, where the letter \emph{p} is for \emph{particle(s)} \\

$\Tau^\mathrm{m}_{\mu\nu}$ 
	& Energy-momentum tensor of matter \\

$\Tau^{\mu_1 \cdots \mu_r}_{\nu_1 \cdots \nu_s}$ 
	& Components of $\tau^s_r$ in relation to a system of coordinates $x^1, \mathellipsis, x^n$ \\

$\Tau^{\nu_1 \cdots \nu_r}_{\mu_1 \cdots \mu_s}$ 
	& Components of $\Tau$ in relation to $\mathcal{B}_v$ \\

$\widetilde{\Tau}^{\nu_1 \cdots \nu_r}_{\mu_1 \cdots \mu_s}$ 
	& Components of $\Tau$ in relation to $\mathcal{B}_w$ \\

$\widetilde{\Tau}^{\xi_1 \cdots \xi_r}_{\varrho_1 \cdots \varrho_s}$ 
	& Components of $\tau^s_r$ in relation to a system of coordinates $y^1, \mathellipsis, y^n$ \\

$\Tau^r_s$ 
	& $\binom{r}{s}$-tensor, tensor of type $(r, s)$, with $r, s \in \mathbb{N}_0$, which is $r$-contravariant (contravariant of degree $r$), i.e. $r$ times contravariant, and $s$-covariant (covariant of degree $s$), i.e. $s$ times covariant \\

$\Tau^{-1}g_{\mu\nu}$ 
	& Metric coupling, i.e. a Riemannian metric, which determines the coupling constant in the Friedan's non-linear $\sigmamodel$-model \\

$\upsilon$
	& G.e. — Function (generic notation). — In the heat, or diffusion, equation it represents a function (of temperature); or even a heat kernel (fundamental solution of the heat equation). — In the wave equation, it is the unknown function, that is the dependent variable, indicating the vertical displacement of the string \\

$\breve{\upsilon}(x)$ 
	& Extremal (function, in this case $\upsilon$) of a functional (or rather, of an integral functional) \\

$\rotatedupsilon$ 
	& Potential function \\

$\rotatedupsilon^\rotatedheartsuit$ 
	& $\rotatedupsilon^\rotatedheartsuit = \delta\rotatedupsilon$ \\

$\textgreek{\text{υ}}$
	& 
	\textsuperscript{[\texttt{cb}]}
	Baiocchi–Duvaut transformation \\

$\Upsilon$ 
	& Open neighborhood; open (sub)set (same meaning as $\Omega$, hence $\Omega \equival \Upsilon$) \\

$\Upsilon_{\Lambda}$ 
	& Neighborhood of $\Lambda_q$ \\

$\Upsilon_\mathbb{CM}$ 
	& Open neighborhood in $\mathbb{CM}^\natural$ \\

$\Upsilon_{\mathbb{PT}\mathbbl{w}}$ 
	& Tubular neighborhood in $\mathbb{PT}\mathbbl{w}$ \\

$(\Upsilon, \vartheta_\Upsilon)$ 
	& Cartan gauge \\
 			
$(\Upsilon^\pm_\mu, \varphi^\pm_\mu)$
	& Charts ($n$-charts) arranged in pair \\

$\phi$ 
	& Angle (of rotation) \\

$\varphi$ 
	& Application, function, map (or mapping), operator, transformation \\ 

$\varphi \viz \varphi_\mathbb{T}$ 
	& Dynamical system, that is a continuous function or mapping with time set $\mathbb{T}$ \\

$\varphi_\mathbbl{\Gamma}$ 
	& Covering transformation, where $\mathbbl{\Gamma}$ is a discrete group of isometries \\

$\varphi_\bbmu$ 
	& Measure preserving transformation, e.g. automorphisms of the Lebesgue–Rohlin space, transformation (or flow) of a compact topological space \\

$\varphi_\textcyrillic{\textit{м}}$ 
	& Continuous function with compact support \\

$\varphi_{(F)}$ 
	& $F$-automorphism of $\splittingfieldK$, $\varphi_{(F)} \in \Galois_F(\splittingfieldK)$ \\

$\varphi_\mathrm{m}$ 
	& Matter field: space-time function representing matter \\

$\varphi_\mathrm{m}^{(3)}$, $\varphi_\mathrm{m}^{(4)}$ 
	& 3- and 4-dimensional matter fields, respectively \\

$\varphi_t$, $\{\varphi_t\}_{t \in \mathbb{R}}$ 
	& Geodesic flow \\

$\varphi^\omega_t$, $\{\varphi^\omega_t\}_{t \in \mathbb{R}}$ 
	& 1-parameter group of $\ast$-automorphisms on $\mathfrak{H}$ (Hilbert space), modular group, modular flow, thermal time flow \\
 
 $\varphi^\mathcal{F}_t$, $\varphi^\mathcal{W}_t$ 
	& Foliation for a geodesic flow, foliated geodesic flow; geodesic foliation \\

$\varphi^\Hamiltonian_t$ 
	& Hamiltonian flow \\

$\varphi[\mathscr{A}]$ 
	& Anosov diffeomorphism \\

$\varphi_t[\mathscr{A}]$ 
	& Anosov flow \\

$\varphi^\mathcal{W}_t[\mathscr{A}]$ 
	& The same as $\varphi^\mathcal{W}_t$ but in the Anosov (or hyperbolic) system \\

$\textgreek{\text{φ}}$ 
	& 
	\textsuperscript{[\texttt{cb}]} Golden ratio (\textit{proportio aurea}): $\textgreek{\text{φ}} = \frac{1}{2}(1 + \sqrt{5})$ \\
	
$\Phi$ 
	& Diffeomorphism. — $\mathscr{C}^r$ map \\

$\textgreek{\text{Φ}}$ 
	& 
	\textsuperscript{[\texttt{cb}]} Golden section \\

$\textgreek{\text{Φ}}_{\mathbb{R}}$ 
	& 
	\textsuperscript{[\texttt{cb}]} Elliptic function \\

$\chi$
	& G.e. — Single 2-component spinor, see $\psi$ as Dirac 4-spinor \\

$\chi_{\mathring{\mathcal{E}}}$ 
	& Dual local frame field for the cotangent/vector bundle; $\chi_{\mathring{\mathcal{E}}} = \{\chi_1, \mathellipsis, \chi_n\}$ \\

$\textgreek{\textit{χ}}$ 
	&
	\textsuperscript{[\texttt{cb}]} Characteristic function (probability theory) \\

$\Chi$ 
	& \textsuperscript{[\texttt{cb}]}\footnote{
	From Greek (\textgreek{χῖ}, via \texttt{\textbackslash{Chi}} command) and not Latin alphabet, although the two X-fonts are  almost identical.
	} 
	$\Chi \Subset \Omega$ \\

$\psi$
	& G.e. — State vector, or wave function of a quantum-mechanical system \\

$\psi = 
\left(\begin{smallmatrix}
\psi^\alpha \\ 
\psi^\beta
\end{smallmatrix}\right)$ 
	& Pauli spinor or 2-component spinor (column vector with two complex components), 2-component wave function, $\psi^\alpha_+$ and $\psi^\beta_-$ \\

$\psi = 
\left(\begin{smallmatrix}
\psi_1 \\ 
\psi_2
\end{smallmatrix}\right)$
	& Weyl spinor, 2-component object/wave function \\

$\psi = 
\left(\begin{smallmatrix}
\psi_1 \\ \psi_2 \\ \psi_3 \\ \psi_4
\end{smallmatrix}\right)$
	& Dirac 4-spinor, pair of 2-spinors: object of a 4-dimensional complex vector space, 4-component wave function. This a column-like spinor, and it is indicate by $\psi \in \mathbb{C}^4$ \\

$\psi = \left(\begin{smallmatrix}
\zeta^\alpha \\ 
\tilde{\chi}_{\dot{\alpha}}
\end{smallmatrix}\right)$ 
	& Dirac 4-spinor (alternative notation), where the left-handed spinor is denoted by $\zeta^\alpha$, whilst the right-handed spinor by $\tilde{\chi}_{\dot{\alpha}}$ \\

$\psi \in {^4\mathbb{C}}\idempotent$
	& Dirac 4-spinor in the form of $\mathbb{C}^{4 \times 4}$ matrix spinor \\

$\psi \in \mathbb{C} \otimes \Cl_{1, 3}(\mathbb{R})$ 
	& Clifford–Dirac algebraic spinor \\

$\psi = 
\left(\begin{smallmatrix}
\psi_\textsc{l} \\ 
i\sigmaPauli_2\bar{\psi}_\textsc{l}
\end{smallmatrix}\right)$	
	&  Majorana spinor, 2-component object/wave function \\
	
$\psi = \left(\begin{smallmatrix}
\zeta^\alpha \\ 
\tilde{\zeta}_{\dot{\alpha}}
\end{smallmatrix}\right)$
	& Majorana spinor (alternative notation) \\

$\adjoint{\psi}$ 
	& Adjoint spinor (field) or Dirac adjoint: $\adjoint{\psi} = \psi^\dag\gamma^0$ \\

$\left(\adjoint{\psi}\slashed{\partial}\right)$ 
	& $\left(\adjoint{\psi}\slashed{\partial}\right) = \partial_\mu\adjoint{\psi}\gamma^\mu$ \\
	
$\psi_\mathrm{C}$ 
	& Charge conjugate of $\psi$ (in the context of spin theory) \\

$\widehat{\psi}$ 
	& Fourier transform of a function $\psi$ (some authors use the tilde symbol $\tilde{\psi}$); do not confuse the wide hat symbol $\widehat{\mathtt{l}}$ with the hat symbol $\hat{\mathtt{l}}$, used e.g. for the operators ($\mathtt{l}$ is for $\mathtt{letter}$) \\
	
$\psi_\bbmu$ 
	& Complex-valued measurable function \\

$\psi_t$ 
	& Diffeomorphism, flow \\

$\rotatedpsi$ 
	& Orlicz function \\

$\Psi$ 
	& Wave functional, for example in Wheeler–DeWitt and Hartle–Hawking equations \\

$\rotatedPsi$ 
	& Vector field of unit length \\

$\omega$ 
	& G.e. (see the context). — Angular frequency per unit time, angular speed \\

$\omega$, $\omega_\mathbb{R}$
	& Differential form or $k$-form; function over a $k$-dimensional manifold \\

$\omega$, $\omega$-state 
	& 
	State of an operator system, positive linear functional of norm 1 (1-form) \\

$\omega_\nabla$
	& Connection matrix: $\omega_\nabla = \left[{\omega^\xi}_\nu\right]$ \\

${\omega^\nu}_\mu$ 
	& Differential 1-form; connection 1-form: ${\omega^\nu}_\mu = {\Gamma^\nu}_{\mu\xi}\omega^\xi$ \\

${\omega^\xi}_\nu$ 
	& The same as ${\omega^\nu}_\mu$ \\

$\omega_\mathbb{N}$ 
	& As a number, it is the first, lowest/smallest transfinite ordinal number; as a set, it is set of all natural numbers in the ordinal construction \\

$(\omega_\mathbb{N})_1$ 
	& As a number, it is the first, lowest/smallest uncountable ordinal number; as a set, it is the set of all countable ordinal numbers; \\

$\omega_\mathbb{R}$
	& Differential form or $k$-form (see above) \\
	
$\omega_{\mathring{\mathcal{P}}}$ 
	& Connection on a principal $G$-bundle \\

$\omega_\mathfrak{g}$ 
	& Cartan connection, differential form on $\mathring{\mathcal{P}}$ ($\omega_\mathfrak{g} = \omega_\mathfrak{h} \oplus \omega_\mathfrak{p}$) \\

$\omega_\mathfrak{h}$
	& Ehresmann connection. — Principal connection equivalent to the Levi-Civita connection: $\omega_\mathfrak{h} \in \Omega^1(\mathring{\mathcal{P}}, \mathfrak{h}) \equival \nabla$ \\

$\omega_\mathfrak{p}$ 
	& Solder form \\

$\omega_G$ 
	& Maurer–Cartan form; left invariant form: $\omega_G \in \Omega^1 (G, \mathfrak{g}) \viz \omega^\textsc{l}_G \in \Omega^1 (G, \mathfrak{g})$, or $\omega^\textsc{l}_G \in \Omega^1(G) \otimes \mathfrak{g}$; right invariant form: $\omega^\textsc{r}_G \in \Omega^1 (G, \mathfrak{g})$, or $\omega^\textsc{r}_G \in \Omega^1(G) \otimes \mathfrak{g}$ \\

$\omega_\Ricci$ 
	& Ricci form: 2-form of type $(1, 1)$ \\

$\{\omega_\Ricci\}$ 
	& Cohomology class of the Ricci form: $\{\omega_\Ricci\} = 2\pi\mathring{\mathscr{C}}_1(\mathcal{M})_\mathbb{R}$ \\

$\omega_\mathrm{s}$ 
	& Symplectic form (non-degenerate closed differential 2-form): $\omega_\mathrm{s} = dx^1_\mathsf{q} \wedge dy^1_\mathsf{p} + \cdots + dx^n_\mathsf{q} \wedge dy^n_\mathsf{p}$. — Kähler form (if $d\omega = 0$, the Hermitian form is symplectic, $\omega = \omega_\mathrm{s}$) \\ 

$\{\omega_\mathrm{s}\}$ 
	& (de Rham) cohomology class as a Kähler class: $\{\omega_\mathrm{s}\} \in H_\mathrm{c}^2(\mathcal{M})_\mathbb{R} \cap H_\mathrm{c}^{1, 1}(\mathcal{M})_\mathbb{C}$ \\

$\omega_\mathrm{v}$ 
	& Magnitude of the vorticity \\

$\textgreek{\textit{ω}}$ 
	& 
	\textsuperscript{[\texttt{cb}]} Value related to the oscillation (Bessel function) \\

$\Omega$ 
	& Open (sub)set \\

$\Omega \subset \mathbb{M}^4$ 
	& Domain of integration of dimension 4, region or portion of Minkowski space-time \\

$\Omega \subset \mathbb{R}^n$ 
	& Subset of a Euclidean space \\

$\Omega^\nabla$ 
	& Curvature 2-form of $\omega_{\mathring{\mathcal{P}}}$ \\

$\Omega_\nabla$ 
	& Curvature matrix: $\Omega_\nabla = \bigl[{\Omega^\nu}_\mu\bigr]$ \\

${\Omega^\nu}_\mu$ 
	& Curvature form \\

$\Omega_{\xi, \varpi, \mathellipsis}$ 
	& Subspaces of $\mathbb{R}^n$ \\

$\Omega^2_\mathfrak{g}$ 
	& Curvature form of the Cartan connection \\

$\Omega^k(\mathcal{M}; \mathring{\mathcal{E}})$ 
	& Vector space of all $k$-forms (on a manifold $\mathcal{M}$ with values in $\mathring{\mathcal{E}} \equival \mathring{\zeta}$); alternative notation: $\mathfrak{E}^k(\mathcal{M})$ \\

$\invertedbreve{\Omega}$
	& 
	\footnote{
	The breve inverted diacritic mark $\invertedbreve{\cdot}$ here and elsewhere has the sole purpose of indicating that we are looking at the \emph{measurable aspect} of quantities and objects, for which $\invertedbreve{\Omega} = \Omega$.
	}
	Set in measure space \\

$\invertedbreve{\Omega}_\mathbbl{E}(\invertedbreve{\omega})$ 
	& $\invertedbreve{\omega}$-limit set of $\mathbbl{E}$ \\

$(\invertedbreve{\Omega}, \bbmu)$ 
	& $\sigma$-finite measure space \\

$(\invertedbreve{\Omega}, \mathscr{B}_\sigma, \bbmu)$ 
	& Probability space \\

$\mathrm{\Omega_\Lambda}$ 
	& Dark energy density parameter \\

$\mathrm{\Omega_m}$ 
	& Matter density parameter (the subscript $m$ stands for \emph{matter}) \\

$\textit{\ng}$ 
	& \footnote{
	The letter \textit{\ng}, not to be confused with Greek small letter $\eta$ (eta), is the cursive of \ng, engma letter (or n-hook) and velar nasal speech sound of \emph{ng} (agma).
	} 
Hinge, bone, or subsimplix \\

$\textit{\NG}_\mu$, $\{\textit{\NG}_\mu\}$ 
	& Partition of unity \\

$\textcyrillic{\textit{в}}$ 
	& Warping function \\

$\couplingconstant$ 
	& (Gauge) coupling constant \\

$\couplingconstant_{\mathrm{d}}$ 
	& Dimensionful coupling constant (in the Nambu–Jona-Lasinio model) \\

$\couplingconstant_{\mathrm{s}}$ 
	& Strong/\textsc{qcd} (gauge) coupling constant \\

$\textcyrillic{\textit{Г}}$ 
	& Homomorphism \\	

$\textcyrillic{\textit{г}}$ 
	&
	\footnote{
	From the Ru. \textcyrillic{г[ладкая]}.
	}
$\mathscr{C}^\infty$ function, smooth function \\

$\slashed{\partial}$ 
	& Partial derivative with Feynman slash notation \\  

$\partial\hyperbolic^n$ 
	& Boundary $\{x^n = 0\}$ of $\hyperbolic^n$ \\

$\partial_\infty\mathbb{B}^n$ 
	& Boundary at infinity of $\mathbb{B}^n$, $\partial_\infty\mathbb{B}^n \equival \mathbb{S}^{n - 1}_\infty$ \\
$\partial_\infty\mathbb{U}^n$ 
	& Boundary at infinity of $\mathbb{U}^n$ \\

$\partial^-\invertedbreve{C}$ 
	& Reduced boundary (à la De Giorgi) of a Caccioppoli set $\invertedbreve{C}$ \\

$\textcyrillic{\textit{Д}}_t$ 
	& 1-parameter family of diffeomorphisms \\

$\idempotent$
	&\footnote{
	Cursive Russian letter \textcyrillic{и}.
	}
	Idempotent \\

$\textcyrillic{\textit{л}}$ 
	& Lebesgue (measurable) function \\

$\textcyrillic{\textit{к}}$ &
	\footnote{
	From the Ru. \textcyrillic{к[орень]}.
	}
	Root \\

$\textcyrillic{\textit{к}}_\alpha$ 
	& Primitive $n$-th root of unity \\

$\Lorentz$ 
	& Lorentz group: $\Lorentz = \Lorentz_+^\uparrow \cup \Lorentz_-^\uparrow \cup \Lorentz_+^\downarrow \cup \Lorentz_-^\downarrow$, where $\Lorentz_+$ (proper L. g.), $\Lorentz_-$ (improper L. g.), $\Lorentz^\uparrow$ (orthochronous L. g.), $\Lorentz^\downarrow$ (non-orthochronous, or heterochronous, L. g.), $\Lorentz_+^\uparrow$ (proper orthochronous, or restricted, L. g.), $\Lorentz_-^\uparrow$ (improper orthochronous L. g.), $\Lorentz_+^\downarrow$ (proper non-orthochronous, or heterochronous, L. g.), $\Lorentz_-^\downarrow$ (improper non-orthochronous, or heterochronous, L. g.) appear \\

$\textcyrillic{\textit{м}}$ 
	& Möbius transformation: $\textcyrillic{\textit{м}}(z) = \frac{\alpha z + \beta}{\gamma z + \delta}$, $\textcyrillic{\textit{м}} \in \mathbbl{\Gamma}$, element in $\mathbbl{\Gamma}$, that is a discrete group of (hyperbolic) Möbius transformations \\

$\textcyrillic{\textit{п}}$ 
	& Polynomial function (on Hilbert space) \\

\textcyrillic{(\textit{У})} 
	& Condition \textcyrillic{(\textit{У})} for \textcyrillic{(\textit{У})}-flow [\textcyrillic{(\textit{У})-потоком}], better known as \emph{Anosov flow} (geodesic flow on a manifold of negative curvature) \\

$\textcyrillic{\textit{ф}}_\textsc{r}$ 
	& Ricci flow \\
	
$\textcyrillic{\textit{ф}}_\textsc{r}^\mathrm{nor}$ 
	& Normalized Ricci flow \\

$\textcyrillic{\textit{ф}}_\textsc{r}^\mathrm{unn}$
	& Unnormalized Ricci flow \\ 

$\textcyrillic{\textit{ф}}_{\textsc{rd}\text{e}}$ 
	& Ricci–DeTurck flow \\

$\textcyrillic{\textit{ф}}_\textsc{kr}$ 
	& Kähler–Ricci flow \\

$\textcyrillic{\textit{ф}}_\textsc{kr}^\mathrm{nor}$ 
	& Normalized Kähler–Ricci flow \\

$\textcyrillic{\textit{ф}}_\textsc{kr}^\mathrm{unn}$ 
	& Unnormalized Kähler–Ricci flow \\

$\textcyrillic{\textit{ф}}_\textsc{r}^\mathrm{s}$ 
	& Ricci soliton \\

$\textcyrillic{\textit{ф}}_\textsc{r}^{\mathring{\mathcal{L}}_\mathbb{C}^{-k}}$ 
	& Expanding soliton on $\mathring{\mathcal{L}}_\mathbb{C}^{-k \in \mathbb{Z}}$  \\

$\textcyrillic{\textit{ф}}_\textsc{r}^{\mathrm{cig}|\mathbb{R}^2}$ 
	& Cigar soliton, or Euclidean Witten's black hole \\

$\textcyrillic{\textit{ч}}_\mathbb{C}$ 
	& Complex number corresponding to a probability amplitude for $\Bbbk \subset \mathbb{R}^4_{1, 3}$ \\

$\textcyrillic{\textit{ч}}^\volume_k$ 
	& Number corresponding to the volume spectrum (according to the law of Weyl) of some space (e.g. manifold, surface) \\

$\textcyrillic{\textit{Ч}}_\mathbb{C}$ 
	& Complex number corresponding to a probability amplitude functional of a path \\

$\textcyrillic{\textit{ш}}$ 
	& Determinant of the induced metric on the boundary surface $\partial\Omega$ \\

$\textcyrillic{\textit{ъ}}$ 
	& Transformation $\textcyrillic{\textit{ъ}} \colon \alpha, \dot{\alpha} \mapsto \textcyrillic{\textit{ъ}}(\alpha), \textcyrillic{\textit{ъ}}(\dot{\alpha})$ on $(\partial\mathbb{D}_\mathbb{C} \times \partial\mathbb{D}_\mathbb{C}) \equival (\partial\mathbb{B}^2_\mathbb{C} \times \partial\mathbb{B}^2_\mathbb{C})$ \\

$\textcyrillic{\textit{ь}}$ 
	& Smooth real function \\

$\textcyrillic{\textit{Ь}}_a$, $\textcyrillic{\textit{Ь}}_b$, $\mathellipsis$ 
	& $(\textcyrillic{\textit{Ь}}_a, \textcyrillic{\textit{Ь}}_b, \textcyrillic{\textit{Ь}}_c, \textcyrillic{\textit{Ь}}_d) \subset \Omega$, all subsets of $\Omega$ \\

$\textcyrillic{\textit{э}}(x)$ 
	& Function almost continuous in $\Omega$  relative to a De Giorgi class, more precisely: $\textcyrillic{\textit{э}}(x) \in \mathscr{C}^{(2)}_\textsc{dg}(\Omega)$, and $\textcyrillic{\textit{э}}(x) \in \mathscr{C}_\textsc{dg}(\Omega, \varepsilon)$; in a local Sobolev space: $\textcyrillic{\textit{э}}(x) \in \Sobolev^{1, p}_\mathrm{loc}(\Omega)$ \\

$\breve{\textcyrillic{\textit{э}}}(x)$ 
	& Extremal (function, in this case $\textcyrillic{\textit{э}}$) of a functional (or rather of an integral functional) \\

$\textcyrillic{\textit{ю}}$
	& Function (generic notation) \\

$\textcyrillic{\textit{я}}$
	& Function (generic notation) \\

$\textcyrillic{\textit{Я}}$ 
	& Subset of $[0, 1]$ \\

$\textit{\dh}$ 
	& Density (for generic definitions); solute density function \\

$\textit{\DH}$ 
	& $\textit{\DH} = \frac{\|\nabla\upsilon\|_p}{\|\upsilon\|_p}$ \\

$\textit{\th}$ 
	& Function, e.g. greater than or equal to zero ($\textit{\th} \geqslant 0$) \\

$\mathbbl{A}$ 
	& Subset $\mathbbl{A} \subset (\mathbb{T}\mathcal{X})_\bbmu = \mathbb{T} \times \mathcal{X}$ induced by $\check{\mathbbl{A}}$ \\

$\check{\mathbbl{A}}$ 
	& Family $\check{\mathbbl{A}} = \{A_t\}_{t \in \mathbb{T}}$ of subsets of $(\mathcal{X}, \distance)$ \\

$\mathcal{A}$ 
	& Atlas \\

$\mathfrak{aps}$ 
	& Algebra of physical space, see $\Cl_{3, 0}(\mathbb{R})$ and $\mathfrak{pau}_{(3, 0)}$ \\

$\aut(\mathbb{D}_\mathbb{C})$ 
	& Automorphism group of $\mathbb{D}_\mathbb{C}$ (complex-valued disk) \\ 

$\aut(\mathbb{U}^2_\mathbb{C})$ 
	& Automorphism group of $\mathbb{U}^2_\mathbb{C}$ (complex-valued upper half-plane) \\

$\mathfrak{A}_5$ 
	& Alternating group of degree $5$, the smallest non-solvable group, as well as the the smallest non-Abelian simple group of order 60; it is isomorphic to the isometry group of the icosahedral group \\

$\mathfrak{A}_n$ 
	& Alternating group of degree $n$ \\

$a^\mu$ 
	& 4-acceleration vector \\

$\adj$ 
	& Adjoint action or adjoint representation \\

$\ast$-algebra 
	& Star-algebra, subform of $C^\ast$-algebra \\ 

$\hat{A}$ 
	& Hirzebruch $\hat{A}$-genus, or just $\hat{A}$-genus \\ 

$\vec{A}$ 
	& Vector potential \\
	
$A^{(\gamma)}$ 
	& Photon field, electromagnetic field: $A^{(\gamma)}_\mu$, $A^{(\gamma)}_\nu$. Cf. vector potential; it is a vector $A$-field: $A^\gamma_\mu A^\gamma_\nu \viz \vec{A}^\gamma_\mu\vec{A}^\gamma_\nu$ \\

$A_\varphi$ 
	& Function (generic notation) \\ 

$A_\mathrm{h}$ 
	& (Surface) area of the event horizon of a black hole, or just horizon area \\

$A_\mathbbl{Tor}$ 
	& Surface area of $\mathbbl{Tor}$ \\

$AdS_n$ 
	& Anti-de Sitter space, or anti-de Sitter space-time, of dimension $n$ \\

AdS/\textsc{cft} 
	& Anti-de Sitter/conformal field theory correspondence, Maldacena duality, or gauge/gravity duality \\

$\mathbb{B}^n$ 
	& Beltrami–Poincaré (open) ball model \\

$\overbar{\mathbb{B}}^n$ 
	& Closed ball model \\

$\overbar{\mathbb{B}}^n_\mathfrak{X}$ 
	& Closed unit ball of a normed vector space $\mathfrak{X} \viz (\mathfrak{X}, \|\cdot\|)$ \\

$\mathbbl{b}(x)$ 
	& Bounded measurable function \\

$\overbar{\mathbbl{B}}$ 
	& Set of points (by a theorem of De Giorgi) \\

$\mathcal{B}(\mathbbl{\Lambda})$ 
	& Attractor's basin of attraction \\

$\mathcal{B}_n$ 
	& Countable basis for the topology of a space \\

$\mathcal{B}_v$, $\mathcal{B}_w$ 
	& Bases of $\mathfrak{V}$: $\mathcal{B}_v = \{v_1, \mathellipsis, v_n\}$, $\mathcal{B}_w = \{w_1, \mathellipsis, w_n\}$, with the basis vectors $v_1, \mathellipsis, v_n$ and $w_1, \mathellipsis, w_n$ \\

$\mathscr{B}_\sigma$ 
	& Borel $\sigma$-algebra, sometimes called \emph{Borel $\sigma$-field} \\

$b$ 
	& Angle of a boost \\
	
$B_\varphi$ 
	& Function (generic notation) \\ 

$\BochnerRiesz_\Rorder^\alpha$ 
	& Bochner–Riesz means of index (or degree) $\alpha$ \\

$\mathbb{C}$ 
	& Field of complex numbers \\ 

$\hat{\mathbb{C}}$ 
	& Extended complex numbers (or extended complex plane): $\hat{\mathbb{C}} \equival \mathbb{C} \cup \{\infty\} \cong \mathbb{CP}^1$ (it is identifiable with the Riemann sphere) \\

$\mathbb{C}^2 \cong \mathfrak{H}$ 
	& Spin space \\

${^2\mathbb{C}}$ 
	& Algebra of $\mathbb{C}^{2 \times 2}$ matrices: ${^2\mathbb{C}} \viz \mathbb{C}(2) \viz M_2(\mathbb{C})$ \\

${^4\mathbb{C}}$ 
	& Algebra of $\mathbb{C}^{4 \times 4}$ matrices:  ${^4\mathbb{C}} \viz \mathbb{C}(4) \viz M_4(\mathbb{C})$ \\

$\mathbb{CM}^\natural$ 
	& Complex Minkowski space(-time) \\

$\mathbb{CM}^\natural_{(\infty)}$ 
	& Complex compactified Minkowski space(-time) \\

$\mathbb{CP}^1$ 
	& Complex projective line: $\mathbb{CP}^1 \cong \hat{\mathbb{C}} = \mathbb{C} \cup \{\infty\}$ (it is identifiable with the Riemann sphere) \\

$\mathbb{CP}^n$
	& Complex projective $n$-space: $\mathbb{CP}^n \viz \mathbb{P}^n(\mathbb{C}) \viz \mathbb{P}^n_\mathbb{C}$ \\

$\mathbb{CP}^4[5]$
	& Quintic 3-fold: hypersurface of degree 5 and dimension 3, with a real 6-dimensionality; to wit, quintic Calabi–Yau 3-fold in a 4-dimensional projective space $\mathbb{P}^4$ \\

$\mathbb{C}^n/\mathbb{Z}_{k > n}$ 
	& Metric cone \\ 

$(\mathbb{C}^n\backslash\{0\})/\mathbb{Z}_k$ 
	& Quotient of $\mathbb{C}^n/\mathbb{Z}_k$ \\

$\mathbb{C}\mathbbl{o}^7$ 
	& $\mathbb{C}\mathbbl{o}^7 = \left\{x \in \mathbb{R}^8 \mid x^2_1 + x^2_2 + x^2_3 + x^2_4 < x^2_5 + x^2_6 + x^2_7 + x^2_8\right\}$: 7-dimensional Simons' cone in $\mathbb{R}^8$ \\
$\mathbb{C}\mathbbl{o}^{2m}$ 
	& $\mathbb{C}\mathbbl{o}^{2m} = \left\{x \in \mathbb{R}^{2m} \mid x^2_1 + \cdots + x^2_m = x^2_{m + 1} + \cdots + x^2_{2m}\right\}$: cone in $\mathbb{R}^{2m}$ \\

$\mathcal{C}_\eta$ 
	& Horocyclic region, open region bounded by a horocycle tangent to $\partial\mathbb{U}^2_\mathbb{C}$ \\

$\mathfrak{C}$ 
	& Cartan geometry: $\mathfrak{C} = (\mathring{\mathcal{P}}, \omega_\mathfrak{g})$ \\
	
$\mathscr{C}^\infty$
	& Class (set) of infinitely (or $\infty$-times continuously) differentiable functions, maps, etc.; smooth atlas, function, functor, map, structure \\
	
$\mathscr{C}^0$ 
	& Class (set) of all continuous function, maps, etc. \\
	
$\mathscr{C}^1$ 
	& Class (set) of differentiable functions, maps, etc. whose derivative is continuous \\
	
$\mathscr{C}^2$ 
	& Class (set) of differentiable functions, maps, etc. whose first and second derivatives are continuous \\
	
$\mathscr{C}^\omega$
	& Class (set) of (real) analytic functions, maps, etc. \\

$\mathscr{C}^r$ 
	& Class (set) of $r$-times continuously differentiable functions, maps, etc. \\
	
$\mathscr{C}_\mathrm{c}$ 
	& Collection, or rather, (linear) space of all continuous functions of compact support (the subscript $c$ is for \emph{compact}) \\

$\mathscr{C}^\infty_\mathrm{c}$ 
	& Class (set) of all functions with compact supported in $\mathscr{C}^\infty_\mathrm{c}$-form, creating a vector $\mathbb{R}$-space \\

$\mathscr{C}^\infty_0$ 
	& Equally to $\mathscr{C}^\infty_\mathrm{c}$, the class (set) $\mathscr{C}^\infty_0$ concerns all bump functions (with continuous derivatives of any order and compactly supported) on a Euclidean space, forming a vector $\mathbb{R}$-space \\

$\mathring{\mathscr{C}}_1(\mathcal{M)}$ 
	& First Chern class (in this case referred to $\mathcal{M}$); more detailed notation: $\mathring{\mathscr{C}}_1(\mathcal{M)}_\mathbb{R}$ is for first real Chern class \\

$\mathscr{C}^{(2)}_\textsc{dg}(\Omega)$ 
	& De Giorgi class (set) of functions $\textcyrillic{\textit{э}}(x)$ almost continuous in a subset of a Euclidean space \\

$\mathscr{C}_\textsc{dg}(\Omega, \varepsilon)$ 
	& De Giorgi class (set) of functions $\textcyrillic{\textit{э}}(x)$ almost continuous in a subset of a Euclidean space, with a positive number $\varepsilon$: $\mathscr{C}_\textsc{dg}(\Omega, \varepsilon) = \bigl\{\left[\mathscr{C}_\textsc{dg}\right]^\pm_p(\Omega, \varepsilon) = \left[\mathscr{C}_\textsc{dg}\right]^+_p(\Omega, \varepsilon) \cap \left[\mathscr{C}_\textsc{dg}\right]^-_p(\Omega, \varepsilon)\bigr\}$, i.e. $\mathscr{C}_\textsc{dg}^\pm = \mathscr{C}_\textsc{dg}^+ \cap \mathscr{C}_\textsc{dg}^-$ \\

$\mathscr{C}_\textsc{dg}(\textcyrillic{\textit{Ь}}_d, \varepsilon)$ 
	& De Giorgi class (note that $\textcyrillic{\textit{Ь}}_d \subset \Omega$) \\

$\mathscr{C}^L_\alpha$, $\mathscr{C}^L_\beta$ 
	& Classes of $L_{\overline{\alpha\beta}} \viz \overline{\alpha\beta}$ \\

$\mathsf{C}_\mathbb{R}$ 
	& Continuum, that is, the set of real numbers $\mathbb{R}$ \\ 

$\mathtt{c}$ 
	& Characteristic class \\

$c$ 
	& Constant (generic notation). — Wave speed, constant concerning the speed of propagation of the wave. — Speed of light in vacuum \\
	
${c^\mu}_{\nu\xi}$ 
	& Structure constants \\

$c_\textsc{h}$ 
	& Hölder constant \\

$c_\textsc{l}$ 
	& Lipschitz constant \\

$c_\textsc{s}$ 
	& Sobolev constant \\

$c_\mathrm{t}$ 
	& Tadpole 1-loop Feynman diagram (constant) \\ 

$\cardinality$ 
	& Cardinality \\

$\mathrm{curl}$ 
	& Curl of a vector field \\

$\invertedbreve{C}$ 
	& Caccioppoli set \\

${C}(\distance)$ 
	& Correlation integral \\

$C^\mathcal{X}$, $K^\mathcal{X}$, $W^\mathcal{X}$ 
	& Partitions of a metric space $(\mathcal{X}, \distance)$ \\

$\Jordancurve$ 
	& Jordan curve \\

$\Kochcurve$ 
	& Snowflake curve of von Koch \\

$C_\mathrm{u}$ 
	& Uniform constant \\

$\mathit{Cat}$
	& Category in Grothendieck topoi \\

$\Cl_{0, 1}$ 
	& $\Cl_{0, 1} \cong \mathbb{C}$ \\

$\Cl_{0, 3}$ 
	& $\Cl_{0, 3} \cong \quaternion \oplus \quaternion$ \\

$\Cl_{0, 2}$ 
	& $\Cl_{0, 2} \cong \quaternion$ \\

$\Cl_{1, 0}$ 
	& $\Cl_{1, 0} \cong \mathbb{R} \oplus \mathbb{R}$ \\

$\Cl_{1, 3}(\mathbb{R})$ 
	& Space-time algebra, see $\mathfrak{sta}$ \\

$\Cl_{1, 3}^{+[0]}(\mathbb{R})$ 
	& Even subalgebra of $\Cl_{1, 3}(\mathbb{R})$ \\

$\Cl_{3, 0}(\mathbb{R})$ 
	& $\Cl_{3, 0} = \Cl_{3, 0}^{+[0]} + \Cl_{3, 0}^{-[1]} \cong (\Cl_{2, 0} \cong {^2\mathbb{R}}) \otimes (\Cl_{0, 1} \cong \mathbb{C}) \cong {^2\mathbb{C}}$, i.e algebra of physical space, denoted by $\mathfrak{aps}$, which corresponds to the Pauli algebra, denoted by $\mathfrak{pau}_{(3, 0)}$ \\

$\Cl_{3, 0}^{+[0]}$ 
	& Even subalgebra of $\Cl_{3, 0}$ \\ 

$\Cl_{3, 0}^{-[1]}$ 
	& Odd subalgebra of $\Cl_{3, 0}$ \\

$\Cl_{3, 1}(\mathbb{R})$ 
	& Majorana algebra, denoted by $\mathfrak{maj}_{(3, 1)}$: $\Cl_{3, 1}(\mathbb{R}) \cong \Cl_{1, 1} \otimes \Cl_{2, 0}$ \\

$\Cl_{4\times}(\mathbb{C})$ 
	& $\Cl_{4\times}(\mathbb{C}) \viz \Cl_{4, 1, \mathbb{C} \otimes}(\mathbb{R})$ \\

$\Cl_{4, 0}$ 
	& $\Cl_{4, 0} \cong \Cl_{2, 0} \otimes \Cl_{0, 2} \cong \Cl_{1, 3}$ \\

$\Cl_{4, 1, \mathbb{C} \otimes}(\mathbb{R})$ 
	& $\Cl_{4, 1, \mathbb{C} \otimes}(\mathbb{R}) \viz \Cl_{4\times}(\mathbb{C})$, that is Dirac algebra, see $\mathfrak{dir}_{(4, 1)}$: $\Cl_{4, 1, \mathbb{C} \otimes}(\mathbb{R}) \cong \Cl_{1, 1} \otimes \Cl_{3, 0}$ \\

$C^\ast$-algebra 
	& $C$ star-algebra, Banach-type algebra \\ 

$\mathrm{C}$ 
	& Charge conjugation \\

$\mathbb{D}$ 
	& Beltrami–Poincaré (open) disk model: $\mathbb{D} \viz \mathbb{D}^2 \equival \mathbb{B}^2$, i.e. real-valued $\mathbb{D}_\mathbb{R} \equival \mathbb{B}^2_\mathbb{R}$ or complex-valued $\mathbb{D}_\mathbb{C} \equival \mathbb{B}^2_\mathbb{C}$ disk \\
	
$\mathbb{D}_\mathbb{R}$ 
	& Real disk model: $\mathbb{D}_\mathbb{R} \equival \mathbb{B}^2_\mathbb{R}$ \\

$\mathbb{D}_\mathbb{C}$ 
	& Complex disk model: $\mathbb{D}_\mathbb{C} \equival \mathbb{B}^2_\mathbb{C}$ \\

$\overbar{\mathbb{D}}$ 
	& Closed unit disk: $\overbar{\mathbb{D}} \equival \overbar{\mathbb{B}}^2$, real-valued $\overbar{\mathbb{D}}_\mathbb{R} \equival \overbar{\mathbb{B}}^2_\mathbb{R}$ or complex-valued $\overbar{\mathbb{D}}_\mathbb{C} \equival \overbar{\mathbb{B}}^2_\mathbb{C}$ closed disk \\
	
$\mathbb{D}_\pi$ 
	& Projective disk model, or Beltrami–Klein disk model \\
	
$\mathbbl{D}_6$ 
	& Dihedral group of order 6 ($\mathbbl{D}_6 \cong \mathbb{Z}_3 \rtimes \mathbb{Z}_2$) \\

$\mathcal{D}(\cdot)$ 
	& Space of test functions \\

$\mathcal{D}^*(\cdot)$ 
	& Space of distributions, dual space of $\mathcal{D}(\cdot)$ \\

$\mathfrak{dir}_{(4, 1)}$ 
	& Dirac algebra: $\mathfrak{dir}_{(4, 1)} \viz \mathbb{R}_{4, 1} \viz \Cl_{4, 1, \mathbb{C} \otimes}(\mathbb{R}) \viz \Cl_{4\times}(\mathbb{C})$, $\mathfrak{dir}_{(4, 1)} \cong {^4\mathbb{C}}$, and $\mathfrak{dir}_{(4, 1)} \cong \mathbb{C} \otimes \mathfrak{maj}_{(3, 1)}$, or $\cong \mathbb{C} \otimes \mathfrak{sta}$ \\

$\mathfrak{dir}_{(4, 1)}^{+[0]}$ 
	& Even algebra of $\mathfrak{dir}_{(4, 1)}$ \\

$\mathfrak{D}(\cdot)$ 
	& Linear subspace of $\mathfrak{H}$, that is the domain of $\cdot$ \\ 

$\mathscr{D}^{(j)}$ 
	& Irreducible representation (for spinor representation) \\

$d\bbmu$ 
	& Volume element \\

$d\omega$ & Dual basis: $d\omega = \{dx^1|_p, \mathellipsis,dx^n|_p\}$ (differential in local coordinates of a function $\omega$); $d\omega_\mathbb{R}$ \\
	
$dS_n$ 
	& de Sitter space of dimension $n$ \\

$dV_g$ 
	& Volume form of $g$ \\

$\diag$ 
	& Diagonal matrix \\

$\diameter$ 
	& Diameter \\

$\dim$ 
	& Dimension \\

$\det$ 
	& Determinant \\

$\det(\mathcal{J})$ 
	& Jacobian determinant \\ 

$\dist$
	& Distance: $\dist = \distance$ \\
	
$\divergence$ 
	& Divergence (vector calculus): e.g. $ 	\divergence{\textgreek{\text{\ddigamma}}} = \nabla \cdot\textgreek{\text{\ddigamma}}$ \\

$D$
	& Covariant derivative, see $\nabla$. — Differential operator \\

$D^2$ 
	& Second covariant derivative \\

$D^2\upsilon$ 
	& Hessian matrix (of second derivatives) of a function $\upsilon$ \\

$D^\alpha$ 
	& Weak $\alpha$-th partial derivative \\

$D_\textsc{das}$ 
	& Dirac–Atiyah–Singer operator \\

$D_g$ 
	& Diagonal group related to the geodesic flow \\

$D_m$ 
	& Coefficient of diffusion (mass diffusivity) \\

$D_\mathrm{q}$ 
	& G.e. (quantity) \\

$D_t\dot{\gamma}_\mathrm{c}$ 
	& The same as $\ddot{\gamma}_\mathrm{c}$, with $D_t\dot{\gamma}_\mathrm{c} \viz \nabla_{\dot{\gamma}_\mathrm{c}}\dot{\gamma}_\mathrm{c}$ \\	

$D_t\sezione$ 
	& Covariant derivative of $\sezione$ along $\gamma_\mathrm{c}$ \\

$\mathrm{D}$
	& Dimension \\

$\mathrm{D_f}$ 
	& Fractal dimension, also known as Hausdorff or Hausdorff–Besicovitch dimension \\

$\mathrm{D_L}$ 
	& Lyapunov dimension, or Kaplan–Yorke dimension \\

$\Diff^1(\mathcal{M})$ 
	& Set of $\mathscr{C}^1$ diffeomorphisms $\varphi$ on $\mathcal{M}$ in the $\mathscr{C}^1$ topology \\

$\Diff^r(\mathcal{M})$ 
	& Set of $\mathscr{C}^r$ diffeomorphisms $\varphi$ on $\mathcal{M}$ in the $\mathscr{C}^r$ topology \\
	
\textsc{dgnm} 
	& De Giorgi–Nash–Moser theorem \\

$\mathbb{E}^n$ 
	& Euclidean $n$-space; flat space in any number of dimensions, see $\mathbb{R}$ \\ 

$\mathbbl{E}$ 
	& Subset $\mathbbl{E} \subset \invertedbreve{\Omega} \times \mathcal{X}$, or family $\mathbbl{E} = \{E_{\invertedbreve{\omega}}\}_{\invertedbreve{\omega} \in \invertedbreve{\Omega}}$ of subsets of $\mathcal{X}$ \\

$\mathcal{E}_{\Omega_\xi, \varpi, \mathellipsis}$ 
	& Subspaces of $\mathbb{R}^n$ \\

$\mathcal{E}^+(v), \mathcal{E}^-(v)$ 
	& 1-dimensional linear subspaces of $\mathcal{T}_v\mathring{\mathcal{T}}^1\mathcal{M}$, with $v \in \mathring{\mathcal{T}}^1\mathcal{M}$ \\

$\mathcal{E}^0(x)$ 
	& 1-dimensional subspace of $\mathcal{T}_x\mathcal{M}$ \\

$\mathcal{E}^\mathrm{s}(x)$, $\mathcal{E}^\mathrm{u}(x)$ 
	& Stable ($^\mathrm{s}$) and unstable ($^\mathrm{u}$) subspaces of $\mathcal{T}_x\mathcal{M}$, respectively \\

$\mathring{\mathcal{E}}$ 
	& Vector bundle or, more generally, fiber bundle on an open neighborhood $\Upsilon$ \\

$\mathring{\mathcal{E}}^0$ 
	& 1-dimensional subbundles of $\mathring{\mathcal{T}}\mathcal{M}$ \\

$\mathring{\mathcal{E}}^\mathrm{s}$, $\mathring{\mathcal{E}}^\mathrm{u}$ 
	& Stable ($^\mathrm{s}$) and unstable ($^\mathrm{u}$) subbundles of $\mathring{\mathcal{T}}\mathcal{M}$, respectively \\
 
$\mathring{\mathcal{E}}^\mathrm{cu}_\mathbbl{\Lambda}$ 
	& Central unstable direction of $\mathring{\mathcal{T}}_\mathbbl{\Lambda}\mathcal{M}$ \\

$\mathring{\mathcal{E}}^\mathrm{s}_\mathbbl{\Lambda}$, $\mathring{\mathcal{E}}^\mathrm{u}_\mathbbl{\Lambda}$ 
	& Stable ($^\mathrm{s}$) and unstable ($^\mathrm{u}$) subbundles of $\mathring{\mathcal{T}}_\mathbbl{\Lambda}\mathcal{M}$, respectively \\

$\mathring{\mathcal{E}}_p$ 
	& Fiber over $p$ (or other letter): $\mathring{\mathcal{E}}_p \equival \pi^{-1}(p)$ \\

$\mathfrak{E}(\Upsilon)$ 
	& Vector space of sections of the tangent/vector bundle on $\Upsilon$ (local section) \\

$\mathfrak{E}(\mathcal{M})$ 
	& Vector space of sections of the tangent/vector bundle on $\mathcal{M}$ (global section) \\

$\mathfrak{E}^k(\mathcal{M})$ 
	& The same as $\Omega^k(\mathcal{M}; \mathring{\mathcal{E}})$ \\

$e$ 
	& Euler's (or Napier's) number, that is irrational and transcendental number: $2.71828\cdots$; base for exponential functions (see $e^x$) and logarithms \\

$e^{i\theta}$
	& Phase factor (complex exponential factor); unit complex number \\

$e^x$ 
	& Exponential function: $e^x \viz \exp{\{x\}}$; $\underbrace{e \times \cdots \times e}_{x \text{ terms}}$ \\

$\mathrm{e}$ 
	& Edge \\
	
$\esssup$ 
	& Essential supremum \\

$\expansion$ 
	& Expansion \\

$\exp{\{x\}}$ 
	& The same as $e^x$ \\

$\invertedbreve{E}$ 
	& Measurable set, e.g. subset of $\invertedbreve{\Omega}$. — Borel set. — Subset of a metric space $(\mathcal{X}, \distance)$ \\

$\vec{E}_1, \mathellipsis, \vec{E}_n$ 
	& Frame fields ($n$-tuple of vector fields), that is, orthonormal basis for the tangent space $\mathcal{T}_p\mathcal{M}$ \\ 

$E_\gamma$ 
	& Photon ($\gamma$) energy \\

$\EulerLagrange^{\mu\nu}$ 
	& Euler–Lagrange operator, that is, tensor density: $\EulerLagrange^{\mu\nu} \viz \EulerLagrange^{\mu\nu}(\Lagrangian)$ \\

$E_k$
	& Kinetic energy \\

$E_\mathrm{em}$ 
	& Electrostatic self-energy, or electromagnetic energy, of a charged point-particle \\

$E_\mathrm{tot}$ 
	& Total energy \\

$E_u$
	& Potential energy \\

$\mathbb{F}$
	& Set of real or complex numbers, $\mathbb{R}$-field or $\mathbb{C}$-field, etc. \\

$\mathcal{F}$ 
	& Foliation (generic notation): $\mathcal{F} \viz \mathcal{W}^\mathrm{s} \text{ and/or } \mathcal{W}^\mathrm{u}$ \\ 
	
$\mathcal{F}_\iota$ 
	& Nullity foliation of $\iota$ (immersion) \\
	
$\mathring{\mathcal{F}}$
	& Fiber of the fiber bundle \\

$\mathring{\mathcal{F}}_\mathfrak{h}$ 
	& Fiber as $\mathfrak{h}$-space \\ 

$\mathfrak{F}$ 
	& Filtration: collection of linear subspaces $E_\nu \subset \mathbb{R}^n$, i.e. $\mathfrak{F} = \{E_\nu \mid \nu = 0, \mathellipsis, k\}$ \\

$\mathscr{F}$, $\mathscr{F}$-functional 
	& Perelman's entropy $\mathscr{F}$-functional, or just $\mathscr{F}$-entropy, but, more correctly, entropy-energy $\mathscr{F}$-functional, aka steady soliton entropy functional \\

$f^{\alpha\beta\gamma\delta\epsilon}$ 
	& (Gauge group) structure constants, or structure coefficients, of the Lie algebra $\mathfrak{su}_3$ of $SU_3$ \\

$f(\invertedbreve{\omega})$ 
	& Random variable \\

$f_{\text{ev}}$
	& Evaluation isomorphism \\
 
$f_\mathrm{(inv)}$
	& Involution \\

$F \subset \splittingfieldK$ 
	& Field extension, as a pair of fields, where $\splittingfieldK$ is an extension field of $F$ and $F$ is a subfield of $\splittingfieldK$ \\

$\vec{F}$
	& Force \\

$\F^{\mu\nu}$, $\F_{\mu\nu}$ 
	& Electromagnetic (field) tensor, or Maxwell tensor, also called (field) strength tensor, which represents the electromagnetic field \\

$\F_\varepsilon^{\mu\nu}$	
	& Regular distribution of the electromagnetic tensor \\

$\F_\mathrm{ext}^{\mu\nu}$ 
	& External field \\

$\F_\textsc{lw}^{\mu\nu}$ 
	& Liénard–Wiechert field \\

$\F_{\textsc{lw}(\varepsilon)}^{\mu\nu}$ 
	& Regularized Liénard–Wiechert field \\

$\F_{\mu\nu}^\alpha$
	& Yang–Mills (field) strength tensor: $\F_{\mu\nu}^\alpha = \partial_\mu A^\alpha_\nu - \partial_\nu A^\alpha_\mu + f^{\alpha\beta\gamma}A_\mu^\beta A_\nu^\gamma$ \\

$F_{\textgreek{\text{μ}}}$ 
	& Resistance of a rotating surface \\

$F_\textsc{gr}$
	& Functor in Grothendieck topoi \\

$\mathit{Fib}_n$ 
	& Fibonacci number(s) \\ 

\textsc{fpu+t} 
	& Fermi–Pasta–Ulam \& Tsingou problem \\ 

$\mathbb{G}^{4(\mathbb{C})}$ 
	& Complex 4-space for the \textsc{lh} non-linear graviton (quantum of gravity) \\

$\mathfrak{g}$
	& Lie algebra of $G$ (the corresponding Lie group); vector space $\mathfrak{g}$ together with a map $[\cdot \:, \cdot] \colon \mathfrak{g} \times \mathfrak{g} \to \mathfrak{g}$ \\
 
$\mathfrak{g}_2$ 
	& Lie algebra of $G_2$ \\

$\mathfrak{gl}_n(\mathbb{K})$
	& Lie algebra of $GL_n(\mathbb{K})$ \\

$\Greenfunction$, $\Greenfunction(x, y)$ 
	& Green's function \\

$g$, $g_{\mu\nu}$ 
	& (or other indices) Metric tensor (field): e.g. Einstein, Gödel, Kerr, Lorentzian, Riemannian, Schwarzschild metric (tensor)  \\

$g \viz g_\textsc{g}$ 
	& Gödel metric (tensor) \\

$g \viz g_\textsc{h}$ 
	& Hermitian metric (tensor) \\

$g \viz g_{\textsc{k}\textnormal{ä}}$ 
	& Kähler metric (tensor) \\

$g \viz g_{\textsc{k}\mathrm{e}}$ 
	& Kerr metric (tensor) \\

$g \viz g_\textsc{l}$ 
	& Lorentzian metric tensor (tensor) \\

$g \viz g_\textsc{m}$
	& Minkowski metric (tensor) \\

$g \viz g_\textsc{s}$ 
	& Schwarzschild metric (tensor) \\

$\tilde{g}$ 
	& Generic transformation involving a metric $g$. — Metric of the universal covering (space) $\widetilde{\mathcal{M}}^n_\kappa$ (for completeness, $\tilde{g} \equival \pi^*g$) \\
	
$g^\ddagger$ 
	& $g^\ddagger = g^{\mu\nu}g^\rotatedheartsuit_{\mu\nu}$ \\

$g^\rotatedheartsuit_{\mu\nu}$ 
	& $g^\rotatedheartsuit_{\mu\nu} = \delta g_{\mu\nu}$ \\

$g_0$ 
	& Initial metric \\

$g_\textcyrillic{\textit{в}}$ 
	& Warped product metric \\

$g^\background_{\mu\nu}$ 
	& Flat or curved background metric \\

$g_{\mathbb{S}^{n - 1}}$ 
	& Metric related to the $(n - 1)$-sphere \\

$g^1_\mathbb{Y}$ 
	& Metric on the Minkowski–Lorentz (or hyperboloid) model \\

$g^2_\mathbb{B}$ 
	& Metric on the Beltrami–Poincaré ball model \\

$g^3_\mathbb{U}$ 
	& Metric on the Beltrami–Poincaré half-space model \\

$g_t \viz g(t)$ 
	& Solution of the Ricci flow \\
	
$g_t \equival
	\left(\begin{smallmatrix}
	e^{\frac{t}{2}} & 0 \\
	0 & e^{-\frac{t}{2}}
	\end{smallmatrix}\right)$ 
	& 1-parameter group related to the geodesic flow \\

$g_\mathrm{cig}$ 
	& Metric of $\textcyrillic{\textit{ф}}_\textsc{r}^{\mathrm{cig}|\mathbb{R}^2}$ (cigar soliton, or Euclidean Witten's black hole) \\

$\rotatedg$ 
	& $\rotatedg = \bigl(g^\dag \viz \bar{g}^\textsc{t}\bigr)^{-1}$ \\

$g\ell$ 
	& Gluon; $g\ell$ is accompanied by superscripts $(^\alpha)$, $(^\beta)$, $(^\gamma)$, $(^\delta)$, or $(^\epsilon)$, where $\alpha\beta\gamma\delta\epsilon = 1, \mathellipsis, 8$ (eight types of gluons) \\

$G$ 
	& Lie group, non-empty set ($G \neq \varnothing$), with a binary operation $\varphi \colon G \times G \to G$ \\

$G/H$ 
	& Left coset of $H$ in $G$ (space of $H$ in $G$), i.e. $gH = \{gh \mid h \in H\}$, where $G$ is a group and $H$ is a subgroup of $G$; left coset space(s) of $G$ modulo $H$ \\

$G_2$ 
	& Exceptional simple Lie group \\

$\gravitation_{\mu\nu}$ 
	& Einstein tensor \\

$\gravitation_{\mu\nu,\xi\varrho}$ 
	& Wheeler–DeWitt metric \\

$\gravitation_{\mu\nu,\xi\varrho}(\simplex^3)$ 
	& Discrete version (via tetrahedral-like shape in $3\mathrm{D}$) of the Wheeler–DeWitt metric \\

$G_\textsc{n}$ 
	& Gravitational constant, Newtonian constant of gravitation \\

$\Galois$ 
	& Galois group \\

$\Galois_\mathbb{Q}(\splittingfieldK)$ 
	& Galois group over $\mathbb{Q}$, referring to a polynomial with rational coefficients $\mathrm{p}(x) \in \mathbb{Q}[x]$ \\

$\Galois_\mathbb{Q}\bigl(\mathbb{Q}(\sqrt[3]2, \textcyrillic{\textit{к}})\bigr)$ 
	& Galois group for a polynomial with rational coefficients: $\Galois_\mathbb{Q}\bigl(\mathbb{Q}(\sqrt[3]2, \textcyrillic{\textit{к}})\bigr) \viz \Galois\bigl(\mathbb{Q}(\sqrt[3]2, \textcyrillic{\textit{к}})/\mathbb{Q}\bigr)$ \\

$\Galois_F(\splittingfieldK)$ 	
	& Galois group of $\splittingfieldK$ over $F$, or Galois group of the $\splittingfieldK$-extension, sometimes also written $\Galois(\splittingfieldK/F)$ \\

$\Galois_F(\splittingfieldL)$ 
	& Galois group of $\splittingfieldL$ over $F$, with an intermediate splitting field $\splittingfieldL$ of the extension $F \subset \splittingfieldK$ \\

$\vec{G}\ell_\textsc{fp}^\alpha$ 
	& Faddeev–Popov (gauge) ghost fields, with $\alpha = 1, \mathellipsis 8$, associated with the eight types of gluons \\

$GL_n(\mathbb{K})$
	& General linear group of degree $n$ over $\mathbb{K} = \mathbb{R}$ or $\mathbb{K} = \mathbb{C}$, set of all regular $n \times n$ real or complex matrices \\

$\hyperbolic^n$
	& Hyperbolic $n$-space \\
	
$\mathbbl{h}$ 
	& Hausdorff measure \\

$\quaternion$ 
	& Quaternion algebra; field of quaternions \\ 

$\quaternion^\times$ 
	& Multiplicative group of non-zero quaternions \\

${^2\quaternion}$ 
	& Algebra of $\quaternion^{2 \times 2}$ matrices: ${^2\quaternion} \viz \quaternion(2) \viz M_2(\quaternion)$ \\

$\quaternion_\mathbb{C}$ 
	& Complex quaternions \\

$\quaternion\mathbb{P}^n$
	& Quaternionic projective $n$-space: $\quaternion\mathbb{P}^n \viz \mathbb{P}^n(\quaternion) \viz \mathbb{P}^n_{\quaternion}$ \\

$\mathfrak{h}$ 
	& Lie (sub)algebra of $H$ (the corresponding Lie (sub)group)\\

$\mathfrak{hol}_x(\omega_{\mathring{\mathcal{P}}})$ 	
	& Holonomy Lie algebra of $\Hol_x(\omega_{\mathring{\mathcal{P}}})$ \\

$\mathfrak{H}$ 
	& Hilbert space \\

$\mathfrak{H}^{\rotatedell + 1}$ 
	& $\rotatedell$-dimensional representation space, complex Hilbert space of dimension $2j + 1$ \\

$\Hamiltonian$ 
	& Hamiltonian (function) \\
	
$\hat{\Hamiltonian}$ 
	& Hamiltonian (operator) in quantum mechanics  \\

$\hat{\Hamiltonian}_{\varphi_\mathrm{m}}$ 
	& Matter Hamiltonian operator \\

$\mathsf{h}_\bbmu$ 
	& Metric entropy or measure-theoretic entropy by Kolmogorov–Sinai \\

$\mathsf{h}_\bbmu(\varphi_\bbmu)$ 
	& Entropy of a measure preserving transformation $\mathcal{X} \xrightarrow{\varphi_\bbmu} \mathcal{X}$ (Kolmogorov–Sinai metric entropy) \\

$\mathsf{h}_\geodesic(g)$ 
	& Geodesic entropy of the Riemannian metric \\

$\mathsf{h}_\topological$
	& Topological entropy \\

$\mathsf{h}_\topological(\varphi_\bbmu)$ 
	& Topological entropy of some continuous transformation $\mathcal{X} \xrightarrow{\varphi_\bbmu} \mathcal{X}$ of a compact topological or metric space \\

$\mathsf{h}_\topological(\varphi_t)$ 
	& Topological entropy of some geodesic flow, e.g. on the unit tangent bundle $\varphi_t \colon \mathring{\mathcal{T}}^1\mathcal{M} \to \mathring{\mathcal{T}}^1\mathcal{M}$ \\

$\mathsf{H}$ 
	& Boltzmann's quantity; function of the condition of a system: entropy of thermodynamics; Boltzmann's $\mathsf{H}$-theorem \\

$\mathsf{H}_\bbmu$, $\mathsf{H}_\bbsigma$ 
	& Entropy with reference to probability measures $\bbmu$, $\bbsigma$ \\

$\mathsf{H}(C^\mathcal{X})$ 
	& Entropy of a partition $C^\mathcal{X}$. Other examples: $\mathsf{H}_\bbmu(K^\mathcal{X})$, $\mathsf{H}_\bbsigma(W^\mathcal{X})$ \\

$\mathsf{H}_\textsc{s}$ 
	& Shannon entropy \\

$h$ 
	& Planck constant \\ 

$\hbar$ 
	& Reduced Planck constant: $\hbar = \frac{h}{2\pi}$ \\ 

$h_{\mu\nu}$ 
	& Symmetric tensor field of rank 2 (it represents small perturbations or deviations from the flatness; gravitational wave field) \\
	
$\homologygroup$
	&
	\footnote{ 
	Via \texttt{\textbackslash{homologygroup}} command. 
	} Homology group \\

$H < G$ 
	& Proper subgroup of $G$ ($H \neq G$) \\

$H \leqslant G$ 
	& Improper subgroup of $G$ ($H$ may or may not equal $G$) \\

 $\mathit{Hei}^3$ 
 	& Heisenberg group of $3 \times 3$ upper triangular matrices of dimension 3 \\

$\mathit{Hes}$ 
	& Hessian (matrix) \\

$\Hol_p(\nabla)$ 
	& Holonomy group of a connection: $\Hol_p(\nabla) \viz \Hol_p(\mathcal{M}, \nabla)$ at a point $p \in \mathcal{M}$, e.g. holonomy group of a vector bundle connection, with $\nabla \equival \nabla^{\mathring{\mathcal{E}}}$, or holonomy group of a linear connection (or holonomy group of a connection on the tangent bundle), with $\nabla \equival \nabla^{\mathring{\mathcal{T}}}$ \\

$\Hol_p(g)$ 
	& Holonomy group of a Riemannian manifold or Riemannian holonomy group: $\Hol_p(g) \viz \Hol_p(\mathcal{M}, g) \equival \Hol_p(\nabla)$ at a point $p \in \mathcal{M}$, where $\nabla \equival \nabla^{\mathring{\mathcal{T}}}$ \\

$\Hol^0_p(g)$ 	
	& Restricted holonomy group of a Riemannian manifold: $\Hol^0_p(g) \viz \Hol^0_p(\mathcal{M}, g)$ \\

$\Hol_x(\omega_{\mathring{\mathcal{P}}})$ 
	& Holonomy group of a connection on a principal $G$-bundle \\

$\Hol^0_x(\omega_{\mathring{\mathcal{P}}})$ 	
	& Restricted holonomy group of a connection on a principal $G$-bundle \\

$H^0$ 
	& Higgs boson \\

$H_\mathrm{c}$ 
	& (de Rham) cohomology group \\

$H_\mathrm{c}^2(\mathcal{M})_\mathbb{R}$ 
	& Second (de Rham) cohomology group of $\mathcal{M}$ with coefficients in $\mathbb{R}$ \\

$\mathrm{H}_0$ 
	& Hubble constant \\

$\idem$
	& 1 (number). — Identity matrix, function, or operator; indicator function. — Identity element or neutral element of a group. — $\idem = \mathbbl{1}$ \\ 

$\idem_\textsc{b}$ 
	& Bianchi identities \\

$\idem_\textsc{p}$ 
	& Palatini identity \\

$\Im$
	& Imaginary part $\Im(z)$ of a complex number $z$ \\

$\Im(\mathbb{O})$ 
	& Imaginary part of $\mathbb{O}$, set of (purely) imaginary octonions \\

$\mathfrak{isom}(\mathcal{M}^n_\kappa)$ 
	& Group of isometries of $\mathcal{M}^n_\kappa$ \\

$\mathfrak{isom}^+(\mathbb{D}_\mathbb{C})$ 
	& Group of orientation preserving isometries of the Beltrami–Poincaré disk: $\mathfrak{isom}^+(\mathbb{D}_\mathbb{C}) \equival PSU_2(\mathbb{C})$ \\

$\mathfrak{isom}^+(\mathbb{U}^2_\mathbb{C})$ 
	& Group of orientation preserving isometries of the Beltrami–Poincaré upper half-plane: $\mathfrak{isom}^+(\mathbb{U}^2_\mathbb{C}) \equival PSL_2(\mathbb{R})$ \\

$\mathfrak{Ic}$ 
	& Icosahedral group (group of isometries of a regular icosahedron) \\

$\widetilde{\mathfrak{Ic}}$ 
	& Binary icosahedral group \\
	
$\mathscr{I}[\cdot]$ 
	& Functional, that is, integral functional (mathematical analysis, calculus of variations) \\

$\mathscr{I}_\textsc{d}[\cdot]$ 
	& Functional, known as \emph{energy functional}, or \emph{Dirichlet energy} \\

$\mathscr{I}_{\textsc{d}\mathrm{o}}[\cdot]$ 
	& Douglas functional \\

$\mathscr{I}_t[\cdot]$ 
	& Brachistochrone curve functional \\

$i$
	& G.e. — Imaginary unit (number) \\

$i_g$ 
	& Index of $g$ \\
	
$\id_\Upsilon$
	& Identity map of $\Upsilon$\\ 

$\id_\mathcal{M}$
	& Identity map of $\mathcal{M}$ \\ 

iff
	& If and only if \\ 

$\indice$ 
	& Index (Atiyah–Singer index theorem) \\

$I$
	& Interval \\

$\Inequality^\mathrm{Har}$ 
	& Harnack inequality \\

$\Inequality^\mathrm{Har}_\textsc{ly}$ 
	& Li–Yau's Harnack inequality \\

$\Inequality^\mathrm{Har}_\textsc{lyh}$ 
	& Li–Yau–Hamilton's Harnack Inequality, aka Hamilton's Matrix inequality \\

$\mathcal{J}$ 
	& Compact set of $\mathbb{U}^2_\mathbb{C}$ \\

$\mathcal{J}_{\mathbb{C}|}$ 
	& Almost complex structure \\

$j$ 
	& G.e. — Eigenvalue of the total angular momentum \\ 

$\cutofffunction$ 
	& Cutoff function \\

$J$ 
	& (Total) angular momentum \\

$\hat{J}$ 
	& (Total) angular momentum, represented by an operator \\

$\vec{J}$ 
	& Jacobi (vector) field. — Geodesic deviation vector field (it is a Jacobi field) \\

$\Bessel_\alpha$ 
	& Bessel function \\

\textsc{jsj} 
	& \textsc{jsj} (Jaco–Shalen–Johannson) or toral decomposition \\

$\Bbbk$ 
	& Region of space-time: $\Bbbk \subset \mathbb{R}^4_{1, 3}$ \\

$\mathbb{K}\mathbbl{l}$ 
	& Klein bottle: $\mathbb{K}\mathbbl{l} \cong \mathbb{RP}^2 \# \mathbb{RP}^2 = 2\mathbb{RP}^2$ \\

$\mathcal{K}_\mathbbl{\Gamma}$ 
	& Compact core of $\mathcal{S}_\mathbbl{\Gamma}$, that is $\mathcal{K}_\mathbbl{\Gamma} \equival \mathbbl{\Gamma} \backslash \widetilde{\mathcal{K}}_\mathbbl{\Gamma}$ \\

$\widetilde{\mathcal{K}}_\mathbbl{\Gamma}$ 
	& Reduced Nielsen region: $\widetilde{\mathcal{K}}_\mathbbl{\Gamma} \equival \widetilde{\mathcal{N}}_\mathbbl{\Gamma} - \{\mathcal{C}_\eta\}$ \\

$\mathfrak{K}_\pm$ 
	& Kre\u{\i}n space: $\mathfrak{K}_\pm = \mathfrak{K}_+ \dotplus \mathfrak{K}_-$, where $(\mathfrak{K}_+ = 0) \subset \mathfrak{K}_{++}$ and $(\mathfrak{K}_- = 0) \subset \mathfrak{K}_{--}$ \\

$k$ 
	& G.e. (e.g. $k \in \mathbb{N}$, or $k \in \mathbb{Z}$). — Wave number, that is, spatial angular frequency of the wave, or magnitude of the wave vector \\

$k_\mathnormal{2}$ 
	& Base-2 (binary number system) notation, $k_\mathnormal{2} \in \mathbb{N}_0$ and/or $k_\mathnormal{2} \in \mathbb{N}_+ = \mathbb{N} \backslash \{0\}$ \\  

$k_\textsc{b}$ 
	& Boltzmann constant \\

$\kernel$ 
	& Kernel \\

$\splittingfieldK$ 
	& Splitting field; minimal, or smallest, field extension of a field $F$ \\

\textsc{kam} 
	& Kolmogorov–Arnold–Moser theorem \\

\textsc{kms} 
	& Kubo–Martin–Schwinger boundary condition \\

$\mathbbl{L}$ 
	& Subset of $\mathcal{X}$ \\

$\mathbb{L}^4$ 
	& Lorentzian space-time, see $\mathfrak{L}^4$ \\ 

$\mathbb{L}^n$ 
	& Lorentzian space, see $\mathfrak{L}^n$ \\ 

$\mathring{\mathcal{L}}$ 
	& Line bundle \\
	
$\mathring{\mathcal{L}}_\mathbb{C}$ 
	& Complex line bundle \\

$\mathring{\mathcal{L}}_\mathbb{C}^{-k \in \mathbb{Z}}$ 
	& Negative line bundle \\ 

$\mathring{\mathcal{L}}_\mathbb{R}$ 
	& Real line bundle \\

$\mathfrak{L}^4$ 
	& Lorentzian space-time or Lorentz–Minkowski space-time: $\mathfrak{L}^4 \viz \mathbb{L}^4 = \mathbb{R}^4_{1, 3}$ or $\mathbb{R}^4_{3, 1}$, i.e., $\mathcal{M}^4$ or $\mathfrak{M}^4 \viz \mathbb{M}^4$ \\ 

$\mathfrak{L}^n$ 
	& Lorentzian space: $\mathfrak{L}^n \viz \mathbb{L}^n = \mathbb{R}^{1, n - 1}$ or $\mathbb{R}^{n - 1, 1}$ \\

$\Liederivative$ 
	& Lie derivative \\ 

$\Lagrangian$ 
	& Lagrangian (function) \\ 

$\Lagrangian(g_{\mu\nu})$ 
	& Lagrangian as a scalar density \\

$\Lagrangian_\mathrm{m}$ 
	& Lagrangian density of the matter-energy \\

$\Lagrangian_\textsc{njl}$ 
	& Nambu–Jona-Lasinio-type Lagrangian density \\

$\Lagrangian_\textsc{sm}$
	& Lagrangian density in the Standard Model (of particle physics) \\

$\Lagrangian_\textsc{ym}$	
	& Yang–Mills-type Lagrangian density \\
	
$\length$ 
	& Length \\

$\length(\gamma_\mathrm{c})$ 
	& Length of a curve, path or geodesic \\

$\length(\mathrm{e})$ 
	& Edge length \\

$\length(\mathrm{s})$ 
	& Length of a string \\

$\length$-length 
	& Perelman's $\length$-length functional \\

$\length$-geodesics 
	& Perelman's $\length$-geodesics, whose equation is the Euler–Lagrange equation for critical curves about the $\length$-length \\
	
$\rotatedell$ 
	& G.e. \\

$\dot{\mathtt{l}} = \dot{\mathtt{letter}}$
	& $\mathtt{l}$ is any letter with a dot above. It denotes the first derivative of a function. — It also denotes a related or derived value, indicating e.g. a change of a quantity, a rotation (movement of an object), a translation (geometric transformation), a reflection (mapping from a space to itself), or else representing different inertial frames of reference (systems of coordinates). The origin of the (over)dot notation is in Newton \cite{Newton "Tractatus de Quadratura Curvarum"}, $\ddot{\mathtt{l}}, \dddot{\mathtt{l}}, \ddddot{\mathtt{l}}$ (second, third, and fourth derivatives). Nota bene. This notation, for the aesthetic issues, is \emph{usually preferred here}, except special cases, to the Lagrange's \cite{Lagrange "Nouvelle methode pour resoudre les equations litterales par le moyen des series"} prime notation, $\mathtt{l}', \mathtt{l}'', \mathtt{l}'''$, $\mathtt{l}''''$ (and then double, triple, and quadruple prime symbols), or the Roman numeral superscript notation, $\mathtt{l}^{(\mathrm{i})}, \mathtt{l}^{(\mathrm{ii})}, \mathtt{l}^{(\mathrm{iii})}$, $\mathtt{l}^{(\mathrm{iv})}$. \\

$\antimatter{\texttt{l}} = \antimatter{\texttt{letter}}$
	&
	\footnote{ 
	Via \texttt{\textbackslash{antimatter}} command. 
	}
	Antimatter (antiparticle, antifield) \\

$\ln$ 
	& Natural logarithm \\

$\hat{L}$ 
	& Orbital angular momentum, represented by an operator \\

$L/$
	& Limit point, or cluster point \\

$\Lebesgue^1$ 
	&
	\footnote{
	\label{footnote "Lebesgue command"}
	$L$ via \texttt{\textbackslash{Lebesgue}} command.
	}
	or $\Lebesgue^1$-space: function space, forming part of $\Lebesgue^p$ Lebesgue space(s) \\

$\Lebesgue^2$
	& 
	\footnoteref{footnote "Lebesgue command"}
	or $\Lebesgue^2$-space: Hilbert space \\ 

$\Lebesgue^2(\mathbb{R}, \mathbb{C}, \bblambda)$ 
	& 
	\footnoteref{footnote "Lebesgue command"}
	Hilbert space on the real line of complex-valued and square-integrable functions, with respect to the Lebesgue measure \\

$L_{\overline{\alpha\beta}} \viz \overline{\alpha\beta}$ 	
	& Line segment between points $\alpha$ and $\beta$ (non-Archimedean continuum) \\

$\Lebesgue^\rotatedpsi$ 
	& Orlicz space, aka Birnbaum–Orlicz space \\

${^\Langlands}G$, or $\Langlands$-group 
	& Langlands dual group of a reductive algebraic group $G$, $\Langlands$-group of $G$ \\

$\Lebesgue^p$
	&
	\footnoteref{footnote "Lebesgue command"} 
	or $\Lebesgue^p$-space: $\Lebesgue^p$ Lebesgue space(s); function space(s) \\

$\underline{L}_p$ 
	& Loop group, group of loops at a point $p$ \\

$\underline{L}^0_p$ 
	& Null-homotopic group loop, group of null-homotopic loops at $p$ \\

$L^\textsc{pmns}_{\varsigma\tau, \tau\varsigma}$ 
	& Pontecorvo–Maki–Nakagawa–Sakata matrix: unitary lepton (neutrino) mixing matrix in the weak interactions \\

$\splittingfieldL$ 
	& Intermediate splitting field satisfying $F \subset \splittingfieldL \subset \splittingfieldK$ \\
	
$\mathrm{L}$ 
	& Side \\

$\textsc{l}$ 
	& Left action \\

\textsc{lh} 
	& Left-handed \\

\textsc{lot} 
	& Lower order terms \\

\textsc{lqg} 
	& Loop quantum gravity \\

\textsc{lyh} 
	& Li–Yau–Hamilton's Harnack inequality, aka Hamilton's Matrix inequality \\

$\mathbb{M}^4$ 
	& Minkowski space-time, see $\mathfrak{M}^4$ \\ 

$\mathbb{M}^{4(\mathbb{R})}_{(\infty)}$ 
	& Conformally compactified Minkowski space(-time): ordinary flat Minkowski 4-space with a closed light cone, or null cone, at infinity, embedded in a bent pseudo-Euclidean 6-space $\mathbb{E}^{2, 4}$ \\

$\mathbb{M}^n$ 
	& Minkowski $n$-space, see $\mathfrak{M}^n$ \\ 

$\mathbbl{M}$ 
	& Set or family of $\varphi_\bbmu$-invariant Borel probability measures \\

$\mathcal{M}$ 
	& Generic differentiable manifold, for example Riemannian or pseudo-Riemannian manifold: $(\mathcal{M}, g)$ \\

$\mathcal{M}_0$ 
	& Conull set \\

$\mathcal{M}^n_\kappa$ 
	& Complete simply connected Riemannian manifold $n$-dimensional of constant (sectional) curvature \\

$\widetilde{\mathcal{M}}^n_\kappa$ 
	& Universal covering (space) of $\mathcal{M}^n_\kappa$, for which $\widetilde{\mathcal{M}}^n_\kappa \equival \mathbb{E}^n$ ($\kappa = 0$), $\widetilde{\mathcal{M}}^n_\kappa \equival \mathbb{S}^n$ ($\kappa > 0$) or $\widetilde{\mathcal{M}}^n_\kappa \equival \hyperbolic^n$ ($\kappa < 0$) \\

$\mathcal{M}_\torus^{2n}$ 
	& Kähler toric manifold: $\mathcal{M}_\torus^{2n} \viz (\mathcal{M}, \omega_\mathrm{s}, \torus^n, \mathcal{J}_{\mathbb{C}|})$ \\

$(\mathcal{M}, \omega_\mathrm{s})$ 
	& Symplectic manifold \\
	
$(\mathcal{M}, \omega_\mathrm{s}, \mathcal{J}_{\mathbb{C}|}, g)$ 
	& Kähler manifold, where $g \viz g_{\textsc{k}\textnormal{ä}}$ \\

$(\mathcal{M}, g)$
	& See $\mathcal{M}$ \\

$\mathfrak{maj}_{(3, 1)}$ 
	& Majorana algebra: $\mathfrak{maj}_{(3, 1)} \viz \mathbb{R}_{3, 1} \cong \Cl_{3, 1}(\mathbb{R}) \cong {^4\mathbb{R}}$ \\

$\mathfrak{maj}_{(3, 1)}^{+[0]}$ 
	& Even subalgebra of $\mathfrak{maj}_{(3, 1)}$ \\

$\mathfrak{M}^4$ 
	& Minkowski space-time: $\mathfrak{M}^4 \viz \mathbb{M}^4 = \mathbb{R}^4_{1, 3}$ or $\mathbb{R}^4_{3, 1}$ \\ 

$\mathfrak{M}^n$ 
	& Minkowski $n$-space: $\mathfrak{M}^n \viz \mathbb{M}^n = \mathbb{R}^{1, n - 1}$ or $\mathbb{R}^{n - 1, 1}$ \\

$\Moebius(\hat{\mathbb{C}})$ 
	& Möbius group; group of Möbius transformations \\

$\Moebius(\mathbb{D}_\mathbb{C})$ 
	& Group of all Möbius transformations leaving $\mathbb{D}_\mathbb{C}$ invariant \\ 

$\Moebius(\mathbb{U}^2_\mathbb{C})$ 
	& Group of all Möbius transformations leaving $\mathbb{U}^2_\mathbb{C}$ invariant \\

$\Moebius^+_2(\mathbb{R})$
	& See $\Moebius(\mathbb{U}^2_\mathbb{C})$ \\

$\Moebius^+_{\mathbb{B}^2}(\mathbb{C})$ 
	& See $\Moebius(\mathbb{D}_\mathbb{C})$ \\ 

$m$
	& Mass \\

$\rotatedm^\nu$
	 & Basis 1-form \\

$\max$, $\min$ 
	& Minimum (smallest) value, maximum (largest) value \\

$\modulo$ 
	& Modulo operation \\
	 
$M$ 
	& Example of a $2 \times 2$ real or complex matrix: $M \viz [M]^{2 \times 2} = \bigl(\begin{smallmatrix}\alpha & \beta \\ \gamma & \delta\end{smallmatrix}\bigr)$ \\

$\tensorM_{\mu\nu}$ 
	& 
	\footnote{
	$M$ via \texttt{\textbackslash{tensorM}} command.
	}
	2-tensor \\

$[M]^{\mathcal{B}_v}_{\mathcal{B}_w}$ 
	& Matrix used for the change of basis with $\mathcal{B}_v$ and $\mathcal{B}_w$ \\

$M^\textsc{ckm}_{\varsigma\tau, \tau\varsigma}$ 
	& Cabibbo–Kobayashi–Maskawa matrix: $3 \times 3$ unitary matrix concerning the strength of flavour-changing weak decays, with the six types of quarks \\

$M_\mathbb{C}$ 
	& Mandelbrot set \\

\textsc{mpp} 
	& Morales–Pacífico–Pujals scheme \\

$\mathbb{N}$, $\mathbb{N}_0$, $\mathbb{N}_+$
	& Set of natural numbers: $\{0, 1, 2, 3, \mathellipsis\}$: $\mathbb{N}_0 = \mathbb{N} \cup \{0\}$ (naturals with zero), $\mathbb{N}_+ = \mathbb{N} \backslash \{0\}$ (naturals without zero). \\

$\mathcal{N}$ 
	& Subset of $\mathcal{M}$ (generic notation) \\

$\widetilde{\mathcal{N}}_\mathbbl{\Gamma}$ 
	& Nielsen region of a Fuchsian group $\mathbbl{\Gamma}$ \\

$\mathcal{N}_\mathbbl{\Gamma}$ 
	& Convex core of $\mathcal{S}_\mathbbl{\Gamma}$, that is $\mathcal{N}_\mathbbl{\Gamma} \equival \mathbbl{\Gamma} \backslash \widetilde{\mathcal{N}}_\mathbbl{\Gamma}$ \\

$\mathscr{N}_\mathrm{eu}$ 
	& von Neumann algebra \\

$\hat{N}$ 
	& Unit normal vector \\

$\hat{N}_{\invertedbreve{C}}$ 
	& Inner normal vector, or unit inner normal vector, to $\invertedbreve{C}$ \\

$\invertedbreve{N}$ 
	& Measurable subset of $\mathcal{M}$, i.e. $\invertedbreve{N} \viz \mathcal{N} \subset \mathcal{M}$ \\

$[N^\mu_\nu]$ 
	& Invertible matrix of $[M^\mu_\nu]$ ($MN = NM = \idem_n$) \\

$N(\distance_n, \varepsilon)$ 
	& Maximum number of points of an $(n, \varepsilon)$-separated set $E \subset \mathcal{X}$ \\

$N_\mathrm{e}$, $N_\mathrm{h}$ 
	& Non-Archimedean numbers of elliptic or hyperbolic type, respectively \\

$N_\tau(x, y)$ 
	& (Average) number of geodesics (or geodesic segments) between two points $x$ and $y$ with length $\leqslant \tau$ \\

$N_\mathrm{ic}$ 
	& Number of independent components \\

$N^\textsc{m}$ 
	& Matrix containing Majorana mass parameters for neutrinos \\

$\mathit{Nil}$ 
	& One of the eight homogeneous Thurston 3-geometries ($\mathit{Nil}$ geometry, $\mathit{Nil}$ space, Nilpotent Lie group) \\

\textsc{nas} 
	& Non-autonomous dynamical system \\ 

\textsc{nlc} 
	& No local collapsing (theorem) \\

$\mathbb{O}$ 
	& Octonion algebra (Cayley algebra); field of octonions (Cayley numbers) \\ 

$\ddot{\mathbb{O}}$ 
	& Möbius strip: $\ddot{\mathbb{O}} \cong \mathbb{S}^1 \times_{\mathbb{Z}/2} \mathbb{R}$ \\

$\mathscr{O}$ 
	& Bachmann–Landau notation or Landau symbol \\

$O_{1, 3}(\mathbb{R})$ 
	& Indefinite orthogonal group of linear transformations of $\mathfrak{M}^4 = \mathbb{R}^4_{1, 3}$, stating that $O_{1, 3}(\mathbb{R}) = \Lorentz$ \\

$O_{2, 4}(\mathbb{R})$ 
	& Pseudo-orthogonal group for $\mathbb{M}^{4(\mathbb{R})}_{(\infty)}$ of linear transformations in $\mathbb{E}^{2, 4}$ \\

$O(\mathcal{T}_p\mathcal{M})$ 
	& Orthogonal group of the tangent space \\

$O(n)$ or $O_n$ 
	& Orthogonal group: $O_n(\mathbb{R})$ (o.g. over the field of real numbers), $O_n(\mathbb{C})$ (o.g. over the field of complex numbers) \\

$\wp_{\gamma_\mathrm{c}}$
	& Parallel transport map: $\wp^{\gamma_\mathrm{c}(1)}_{\gamma_\mathrm{c}(0)} \viz \wp_{\gamma_\mathrm{c}(0) \to \gamma_\mathrm{c}(1)} \colon \mathring{\mathcal{E}}_p \to \mathring{\mathcal{E}}_p$ \\

$\mathring{\mathcal{P}}$ 
	& Principal bundle, or principal $G$-bundle; non-short notations of $\mathring{\mathcal{P}}$ are the map $\pi \colon \mathring{\mathcal{P}} \to \mathcal{M}$ and the quadruple $(\mathring{\mathcal{P}}, \pi, \mathcal{M}, G)$ \\

$\mathring{\mathcal{P}}_\Lorentz$ 
	& Component of the principal $\Lorentz_+^\uparrow$-bundle over $\mathbb{R}^4_{1, 3}$, aka Lorentz bundle \\

$\mathring{\mathcal{P}}_\textit{ß}$ 
	& Component of the principal $SL_2(\mathbb{C})$-bundle over $\mathbb{R}^4_{1, 3}$, aka spinor bundle \\

$\mathring{\mathcal{P}}_\Spin$
	& $\Spin(n)$-principal bundle \\

$\mathbb{P}^n(\mathbb{C})$, $\mathbb{P}^n_\mathbb{C}$
	& The same as $\mathbb{CP}^n$ \\

$\mathbb{P}^n(\quaternion)$, $\mathbb{P}^n_{\quaternion}$
	& The same as $\quaternion\mathbb{P}^n$ \\

$\mathbb{P}^n(\mathbb{R})$, $\mathbb{P}^n_\mathbb{R}$
	& The same as $\mathbb{RP}^n$ \\

$\mathbb{P}^4[5]$
	& Alternative notation to $\mathbb{CP}^4[5]$ \\

$\mathbb{PT}\mathbbl{w}$ 
	& Projective twistor space: real 6-space $\mathbb{PT}\mathbbl{w}^{6(\mathbb{R})}$, or complex 3-space $\mathbb{PT}\mathbbl{w}^{3(\mathbb{C})} \cong \mathbb{CP}^3$ \\

$\mathbb{PT}\mathbbl{w}^{3(\mathbb{C})}_\pm$ 
	& Complex 3-spaces of $\mathbb{PT}\mathbbl{w}^6$ for 0-mass particles of helicity$_\pm$ \\

$\mathbb{PT}\mathbbl{w}^{5(\mathbb{R})}_\mathrm{null}$
	& Projective null twistor space, real 5-space \\

$\mathbb{PT}\mathbbl{w}\mathring{\mathcal{Z}}$ 
	& Projective space in which the fibers of the projection of $\mathring{\mathcal{Z}}$ lie \\

$\mathfrak{p}$
	& Lie-like structure on $\mathfrak{p}$-space (vector space), in combination with $H$- or $\mathfrak{h}$-module decomposition: $\mathfrak{g} = \mathfrak{h} \oplus \mathfrak{p}$ \\ 

$\mathfrak{pau}_{(3, 0)}$ 
	& Pauli algebra: $ \mathfrak{pau}_{(3, 0)} \viz \mathbb{R}_{3, 0} \viz \Cl_{3, 0}(\mathbb{R}) \cong \mathbb{C} \otimes \quaternion \cong {^2\mathbb{C}}$ \\

$\{\mathfrak{P}\}_\mathrm{set}$ 
	& Poincareian set of recurrent points \\

$\mathscr{P}$ 
	& Probability density function: $\mathscr{P}(\mathrm{\Delta}, \tau)$. —
	Probability function in path integrals: $\mathscr{P}(x_1, \mathellipsis, x_j, x_{j + 1}, \mathellipsis, x_\rotatedell)$ \\

$\mathsf{P}$ 
	& Pressure \\
	
$\mathsf{P}_\bbmu$ 
	& Pressure of $\bbmu \in \mathbbl{M}(\mathcal{X})$ \\

$\mathsf{P}_\topological$ 
	& Topological pressure \\

$\mathsf{P}_\topological(\underdot{\varpi})$ 
	& Topological pressure of a continuous function $\underdot{\varpi} \colon \mathcal{X} \to \mathbb{R}$ with respect to $\varphi_\bbmu$, i.e. $\mathsf{P}_\topological(\underdot{\varpi}) \viz \mathsf{P}_\topological(\varphi_\bbmu, \underdot{\varpi})$ \\

$p$
	& (or another letter) Point\footnote{
	\label{footnote "Letter p"}
	Attention: “point” and “4-momentum” (\texttt{\textbackslash{momentum}}) are designated by the same cursive letter, i.e., $p$. It is therefore the context that  illuminates its meaning.
	} \\
	
$\momentum^\mu$ 
	& 4-momentum vector\textsuperscript{\ref{footnote "Letter p"}} \\

$\vec{p}$ 
	& Momentum (vector) \\
	
$\hat{\mathrm{p}}$ 
	& Momentum operator: $\hat{\mathrm{p}} \colon \mathfrak{D}(\hat{\mathrm{p}}) \to \Lebesgue^2(\mathbb{R})$ \\

$\mathrm{p}(x)$ 
	& Polynomial; polynomial equation \\

$\prj$ 
	& Projection, see $\pi$ \\

$\invertedbreve{P}$ 
	& Set: $\invertedbreve{P}^{k \in \mathbb{N}} \equival \{x \in \invertedbreve{E} \mid \varphi_\bbmu^{n \geqslant k}({x} \notin \invertedbreve{E})\}$ \\

$\tensorP_{\mu\nu\xi}$
	& \footnote{
	$P$ via \texttt{\textbackslash{tensorP}} command.
	}
	3-tensor \\

$P_\rotatedell$ 
	& Probability of state $\rotatedell$ \\

$P_\mathrm{em}^\mu$ 
	& Total 4-momentum vector of the electromagnetic field \\

$PSL_2(\mathbb{C})$ 
	& Projective special linear group of $2 \times 2$ matrices over the complex field; $PSL_2(\mathbb{C}) \cong \Moebius(\hat{\mathbb{C}})$ \\

$PSL_2(\mathbb{F}_4)$ 
	& Projective special linear group isomorphic to $\mathfrak{A}_5$ \\	

$PSL_2(\mathbb{F}_5)$ 
	& Projective special linear group isomorphic to $\mathfrak{A}_5$ \\	

$PSL_2(\mathbb{R})$ 
	& Projective special linear group of $2 \times 2$ matrices over the real field; group of all biholomorphic maps of $\varphi \colon \mathbb{U}^2_\mathbb{C} \to \mathbb{U}^2_\mathbb{C}$ \\

$PSL_2(\mathbb{Z})$ 
	& Modular group: $PSL_2(\mathbb{Z}) \equival \mathbbl{\Gamma}$ \\ 

$PSU_2(\mathbb{C})$ 
	& Projective special unitary group of degree 2 over the complex field \\

$\mathrm{Pr}$ 
	& Prandtl number: the ratio of kinematic viscosity to thermal diffusivity (dimensionless number) \\ 

\textsc{p+fm} 
	& Theorem of Pesin \& Freire–Mañé \\

$\mathbbl{q}$ 
	& Quaternion \\ 

$\mathbb{Q}$
	& Set of rational numbers: $q = \frac{z}{n}$, with $z \in \mathbb{Z}$ and $n \in \mathbb{N}_+$ \\

$\mathcal{Q}$
	& Base space \\
	 
$\mathfrak{Q}_8$ 
	& Quaternion group: $\mathfrak{Q}_8 = \{\mathbbl{\pm1, \pm{i}, \pm{j}, \pm{k}}\}$ \\ 

$\mathsf{Q}$ 
	& Heat \\

$qk^{\varsigma, \tau}_\chi$ 
	& Quark, where the superscripts $\varsigma, \tau = 1, \mathellipsis, 3$ designate the flavor property, or the three quark generations, while the subscript $\chi = 1, \mathellipsis, 3$ is the color charge \\

$\invertedbreve{Q}$ 
	& Measurable subset embedded in a Riemannian space \\

$\quadrupole_{\mu\nu}$ 
	& Quadrupole moment tensor \\

$Q_f$ 
	& Quadratic form (map): $Q_f \colon \mathfrak{M}^4 \viz \mathbb{R}^4_{1, 3} \to \mathbb{R}$ \\

\textsc{qcd} 
	& Quantum chromodynamics \\

\textsc{qed}
	& Quantum electrodynamics \\ 

\textsc{qft}
	& Quantum field theory \\ 

\textsc{qg} 
	& Quantum gravity \\

$\mathbb{R}$
	& Euclidean space: $\mathbb{R}$ (line), $\mathbb{R}^2$ (plane), $\mathbb{R}^3$ (3-dimensional space), etc. \\
	
$\mathbb{R}_-, \mathbb{R}_+, \mathbb{R}_*$ 
	& Set of real numbers: $\mathbb{R}_-$ (negative reals), $\mathbb{R}_+$ (positive reals), $\mathbb{R}_* = \{0\} \cup \mathbb{R}_+$ (non-negative reals: zero or positive numbers) \\

$\mathbb{R}^4 \cong \quaternion$ 
	& Quaternionic space \\ 

${^4\mathbb{R}}$ 
	& Algebra of $\mathbb{R}^{4 \times 4}$ matrices: ${^4\mathbb{R}} \viz \mathbb{R}(4) \viz M_4(\mathbb{R})$ \\
	
$\mathbb{RP}^n$
	& Real projective $n$-space: $\mathbb{RP}^n \viz \mathbb{P}^n(\mathbb{R}) \viz \mathbb{P}^n_\mathbb{R}$ \\

$\Re$ 
	& Real part $\Re(z)$ of a complex number $z$ \\

$\mathfrak{R}^4$ 
	& Set of $\mathbb{C}^{2 \times 2}$ matrices \\

$\rank$ 
	& Rank \\

$\Riemann$
	&
	\footnote{
	\label{footnote "Riemann command"}
	$R$ via \texttt{\textbackslash{Riemann}} command.
	} 
	Riemann curvature tensor, notation without indices \\

$R^\theta$ 
	& Rotation (matrix) through an angle $\theta$ \\

$\Ricci_{\mu\nu}$, $\Ricci^\mu_\nu$
	&
	\footnote{
	\label{footnote "Ricci command"}
	$R$ via \texttt{\textbackslash{Ricci}} command.
	}
	Ricci curvature tensor, notation with indices \\

${\Riemann_{\mu\nu\xi}}^\varrho$ 
	& 
	\footnoteref{footnote "Riemann command"}
	Riemann curvature tensor in the form of $\binom{1}{3}$-tensor \\ 

$\Riemann_{\mu\nu\xi\varrho}$ 
	& 
	\footnoteref{footnote "Riemann command"}
	Riemann curvature tensor in the (fully covariant) form of $\binom{0}{4}$-tensor \\

$\Ricci_{(\mathrm{a})}$ 
	& Average scalar curvature \\

$\scalarcurvature$ 
	&
	\footnoteref{footnote "Ricci command"}	
	Scalar curvature or Ricci scalar \\

$(\scalarcurvature)_\mathrm{min}$ 
	& Scalar curvature at its global minimum \\

$\Rorder$ 
	& $\Rorder$-th order in Bochner–Riesz means \\

$\mathrm{R}$ 
	& Distance. — Radius \\

$\mathrm{Ra}$ 
	& Rayleigh number: the ratio of buoyancy-driven flow to viscous and thermal dissipation (dimensionless number) \\

$\Ric$ 
	& Ricci curvature tensor, notation without indices \\
	 
$\Rie$ 
	& The same as $\Riemann$ \\

$\textsc{r}$ 
	& Right action \\

$\textsc{r}_h$ 
	& Right action of $h \in H$ on $\mathring{\mathcal{P}}$ \\

\textsc{rds} 
	& Random dynamical system \\

\textsc{rh} 
	& Right-handed \\

$\mathbb{S}^1$ 
	& (Unit) circle: subset of the real 2-space, $\mathbb{S}^1 = \{x \in \mathbb{R}^2 \mid \|x\| = 1\}$ or, with complex numbers, $\{z = (x + iy) \in \mathbb{C} \mid \|z\| = 1\}$ \\	

$\mathbb{S}^3_\homologygroup$
	& Poincaré homology 3-sphere; spherical dodecahedral space (Poincaré–Weber–Seifert space) \\

$\mathbb{S}^n$ 
	& Spherical $n$-space: $\mathbb{S}^1, \mathellipsis, \mathbb{S}^n$ in $\mathbb{R}^2, \mathellipsis, \mathbb{R}^{n + 1}$ \\  

$\mathbb{S}^n_\mathfrak{X}$ 
	& Unit ($n$-dimensional) sphere of a normed vector space $\mathfrak{X} \viz (\mathfrak{X}, \|\cdot\|)$ \\

$\mathbb{S}^{n - 1}_\infty$ 
	& Sphere at infinity; boundary at infinity of the ball (hyperbolic space) \\

$\{\mathbb{S}^1 \times \mathbb{B}^1\}$ 
	& Pseudo-cylindrical (non-Euclidean) neck in a dumb-bell-shaped $\mathbb{S}^2$ \\

$\{\mathbb{S}^2 \times \mathbb{B}^1\}$ 
	& Pseudo-cylindrical (non-Euclidean) neck in a dumb-bell-shaped $\mathbb{S}^3$ \\

$\mathcal{S}_\mathbbl{\Gamma}$ 
	& $\mathcal{S}_\mathbbl{\Gamma} \equival \mathbbl{\Gamma} \backslash \mathbb{U}^2_\mathbb{C}$ \\

$\mathcal{S}_\mathfrak{c}$, $\mathcal{S}_\mathfrak{c}(\mathbb{R})$ 
	& Schwartz space; space of complex-valued functions \\

$\mathcal{S}(\simplex)$
	& Skeleton space, i.e. simplicial complex (or collection) of $\simplex$-simplexes \\

$\mathcal{S}_\textsc{r}(\simplex)$ 
	& Regge skeleton (discrete) 4-space \\

$\mathfrak{sl}_2(\mathbb{C})$ 
	& Lie algebra of $SL_2(\mathbb{C})$ \\ 

$\mathfrak{sl}_n(\mathbb{K})$
	& Lie algebra of $SL_n(\mathbb{K})$ \\

$\mathfrak{so}^+(1, 3)$ 
	& Lie algebra of $SO_{1, 3}^+(\mathbb{R})$ \\

$\mathfrak{so}(n)$ or $\mathfrak{so}_n$
	& Special orthogonal Lie algebra \\

$\mathfrak{sp}_1(\quaternion)$ 
	& Symplectic Lie algebra for quaternions \\

$\mathfrak{sp}_2(\mathbb{C})$ 
	& Lie algebra of the symplectic group $Sp_2(\mathbb{C})$ \\

$\mathfrak{spin}(n)$ or $\mathfrak{spin}_n$ 
	& Spin (Lie) algebra \\
	
$\mathfrak{sta}$ 
	& Space-time algebra: $\mathbb{R}_{1, 3} \viz \Cl_{1, 3}(\mathbb{R}) \cong {^2\quaternion}$ \\

$\mathfrak{su}(n)$ or $\mathfrak{su}_n$ 
	& Lie algebra of $SU(n)$ or $SU_n$ \\

$\symmetric_n$ 
	& Symmetric group of $n$-elements \\

$\mathscr{S}$ 
	& Action, or action functional (Lagrangian function, Hamilton's function, etc.) \\ 

$\mathscr{S}_\mathbb{E}$ 
	& Euclidean action \\

$\mathscr{S}_\mathrm{gw}^{(2)}$ 
	& Action for the propagation of a gravitational wave in a flat space (Minkowski space-time) \\

$\mathscr{S}^\mathsf{S}_\mathrm{g}$ 
	& Entropy (action) functional for gravity \\

$\mathscr{S}_\mathrm{m}$ 
	& Matter action \\

$\mathscr{S}^\mathsf{S}_\mathrm{m}$ 
	& Entropy (action) functional for matter \\

$\mathscr{S}_\textsc{eh}$ 
	& Einstein–Hilbert action \\

$\mathscr{S}_\textsc{f}$ 
	& Friedan action \\ 

$\mathscr{S}_\textsc{r}$ 
	& Regge (simplicial) action \\

$\mathsf{S}$ 
	& Entropy \\

$\mathsf{S}_\textsc{bh}$ 
	& Bekenstein–Hawking entropy, black hole entropy \\

$\mathsf{S}_\textsc{n}$ 
	& Nash's entropy \\

$\simplex$ 
	& Simplex: $\simplex^0$ (0-simplex: point), $\simplex^1$ (1-simplex: straight line segment), $\simplex^2$ (2-simplex: triangle), $\simplex^3$ (3-simplex: tetrahedron), $\simplex^4$ (4-simplex: 5-choron, or 5-tope), and so forth \\

$\singularity$ 
	& Singular, singularity \\

$\singularity(\textcyrillic{\textit{ф}}_\textsc{r})$ 
	& Singularities in/of the Ricci flow \\

$\singularity\{\mathbb{S}^1 \times \mathbb{B}^1\}$ 
	& Neckpinch singularity in $\textcyrillic{\textit{ф}}_\textsc{r}^{(2)}$ \\

$\singularity\{\mathbb{S}^2 \times \mathbb{B}^1\}$ 
	& Neckpinch singularity in $\textcyrillic{\textit{ф}}_\textsc{r}^{(3)}$ \\

$\scal$ 
	& Scalar, mathematical element or physical quantity \\

$\sin\theta_\mathrm{w}$, $\cos\theta_\mathrm{w}$ 
	& Sine and cosine of the weak mixing angle, or Weinberg angle $\theta_\mathrm{w}$ \\

$\subscript$
	& Subscript \\

$\supp$ 
	& Support of a function \\
	
$\surface$ 
	& Surface (generic notation) \\

$\hat{S}$
	& Spin angular momentum, or spin, for short, represented by an operator \\

$Sh(\mathcal{X}_{\textnormal{\textgreek{τ}}})$
	&
Category of sheaves of sets on $\mathcal{X}_{\textnormal{\textgreek{τ}}}$ \\

$Sh(\mathcal{X}^\mathrm{loc}_{\textnormal{\textgreek{τ}}})$
	&
Category of sheaves of sets on a local space \\

$Sh(\mathit{Cat}, \mathcal{X}_\topological)$ 
	& Topos on a site $(\mathit{Cat}, \mathcal{X}_\topological)$, having a category of sheaves of sets, with a Grothendieck topology $\mathcal{X}_\topological$ \\

$\setsingularities_{\vec{X}}(\mathbbl{\Lambda})$ 
	& Set of singularities of $\vec{X}$ in $\mathbbl{\Lambda}$ \\
	 
$SL_2(\mathbb{C})$ 
	& Special linear group of $2 \times 2$ matrices over the complex field; 2-fold covering of $SO_{1, 3}^+(\mathbb{R})$ \\

$SL_2(\mathbb{R})$ 
	& Special linear group of $2 \times 2$ matrices over the real field; 2-fold covering of $PSL_2(\mathbb{R})$ \\

$\widetilde{SL}_2(\mathbb{R})$
	& Universal cover of $SL_2(\mathbb{R})$ \\

$SL_n(\mathbb{K})$ 
	& Special linear group of degree $n$ over $\mathbb{K} = \mathbb{R}$ or $\mathbb{K} = \mathbb{C}$, set of all $n \times n$ real or complex matrices having determinant 1 \\

$SO(n)$ or $SO_n$ 
	& Special orthogonal group \\

$SO_{1, 3}(\mathbb{R})$ 
	& Indefinite special orthogonal group of linear transformations of $\mathfrak{M}^4 = \mathbb{R}^4_{1, 3}$, stating that $SO_{1, 3}(\mathbb{R}) = \Lorentz_+$ \\

$SO_{1, 3}^+(\mathbb{R})$ 
	& Indefinite special orthogonal group of linear transformations of $\mathfrak{M}^4 = \mathbb{R}^4_{1, 3}$, known as restricted Lorentz group, stating that $SO_{1, 3}^+(\mathbb{R}) = \Lorentz_+^\uparrow$ \\

$SO_{2, 4}$ 
	& Twistor group: $SO_{2, 4} \cong_{(1:2)} SU_{2, 2}$ \\

$SO_3(\mathbb{R})$
	& Lie group of all rotations of 3-dimensional Euclidean space, special orthogonal group in $3\mathrm{D}$ \\ 

$SO_4(\mathbb{R})$
	& Lie group of all rotations of 4-dimensional Euclidean space, special orthogonal group in $4\mathrm{D}$ \\ 

$\mathit{Sol}$ 
	& One of the eight homogeneous Thurston 3-geometries ($\mathit{Sol}$ geometry, $\mathit{Sol}$ space, solvable Lie group) \\

$Sp_1(\quaternion)$ 
	& Group of quaternions of norm 1 (compact symplectic group) \\

$Sp(n)$ or $Sp_n$ 
	& Symplectic group \\

$\Spin_{1, 3}^+(\mathbb{R})$ 
	& Spin group of signature $(1, 3)^+$ over the real field \\

$\Spin_3(\mathbb{R})$ 
	& Set of all unit quaternions $\{\mathbbl{q} \in \quaternion \mid \mathbbl{q}\bar{\mathbbl{q}} = 1\}$ \\ 

$\Spin(n)$ or $Spin_n$ 
	& Spin group \\

$SU_{2, 2}$ 
	& Pseudo-unitary group (spin group) \\ 

$SU(n)$ or $SU_n$
	& Special unitary group of $n \times n$ unitary matrices having determinant 1 \\

$S_\mathcal{X}$ 
	& Set of points and a discrete category of a Grothendieck topos: $S_\mathcal{X} = (x_1, \mathellipsis, x_n)$ \\

\textsc{sde} 
	& Stochastic differential equation \\

\textsc{srb} 
	& Sinai–Ruelle–Bowen measure \\

$\torus$ 
	&\footnote{
	\label{footnote "Torus command"}
	$\torus$ via \texttt{\textbackslash{torus}} command, so it is distinct from the \texttt{\textbackslash{mathbb\{T\}}} command used for the time group or time set.
	}
	1-torus, i.e. circle group: $\torus^1 \equival \mathbb{S}^1$ \\

$\torus^2$ 
	& 
	\footnoteref{footnote "Torus command"}
	2-torus: $\torus^2 \cong \mathbb{S}^1 \times \mathbb{S}^1$ \\

$\torus^n$
	&
	\footnoteref{footnote "Torus command"}
	$n$-torus: $\torus^n \cong \underbrace{\mathbb{S}^1 \times \cdots \times \mathbb{S}^1}_{n \text{ times}} \equival \mathbb{R}^n/\mathbb{Z}^n = (\mathbb{R}/\mathbb{Z})^n$ \\

$\mathbb{T}$, $\mathbb{T}_+$, $\mathbb{T}_*$ 
	& Time group or time set (see below). Attention, it is the same symbol denoting the torus; the context helps to make a distinction and avoid confusion \\ 
	
$\mathbb{T} = \mathbb{R}$ 
	& Two-sided continuous time: $\mathbb{T} = \mathbb{R} = \mathbb{R}_- \cup \{0\} \cup \mathbb{R}_+$ \\ 

$\mathbb{T}_+ = \mathbb{R}_+$ 
	& One-sided (positive sided) continuous time \\ 

$\mathbb{T}_* = \mathbb{R}_*$ 
	& One-sided continuous time: $\mathbb{T}_* = \mathbb{R}_* = \{0\} \cup \mathbb{R}_+ = \{t \in \mathbb{R} \mid t \geqslant 0\}$, i.e. $\mathbb{T}_* = \{0\} \cup \mathbb{T}_+ = \{t \in \mathbb{T} \mid t \geqslant 0\}$ \\
	
$\mathbb{T} = \mathbb{Z}$ 
	& Two-sided discrete time: $\mathbb{T} = \mathbb{Z} = \{0, \pm1, \pm2, \pm3, \mathellipsis\}$ \\ 

$\mathbb{T}_* = \mathbb{Z}_*$ 
	& One-sided discrete time:  $\mathbb{T}_* = \mathbb{Z}_* = \{0\} \cup \mathbb{Z}_+$ \\ 
	
$\mathbb{T}\mathbbl{w}$ 
	& Twistor space: vector space with a pseudo-Hermitian metric; complex 4-space $\mathbb{PT}\mathbbl{w}^{4(\mathbb{C})} \cong \mathbb{C}^4$, or real 8-space $\mathbb{T}\mathbbl{w}^{8(\mathbb{R})} \cong \mathbb{R}^8$ \\

$\mathbb{T}\mathbbl{w}^{3(\mathbb{C})}_\pm$ 
	& Complex 3-space of positive, $(\bar{x}_{\mathbb{T}\mathbbl{w}})_\mu(x_{\mathbb{T}\mathbbl{w}})^\mu > 0$, and negative, $(\bar{x}_{\mathbb{T}\mathbbl{w}})_\mu(x_{\mathbb{T}\mathbbl{w}})^\mu < 0$, twistors, respectively \\

$\mathbb{T}\mathbbl{w}^{7(\mathbb{R})}_\mathrm{null}$ 
	& Null twistor space: real 7-space; space of null twistors $(\bar{x}_{\mathbb{T}\mathbbl{w}})_\mu(x_{\mathbb{T}\mathbbl{w}})^\mu = 0$ \\

$\mathbb{T}\mathbbl{w} \oplus \mathbb{T}\mathbbl{w}^*$ 
	& Non-reduced (full) spin space for $SO_{2, 4}$ \\

$(\mathbb{T}\mathcal{X})_\bbmu$ 
	& Extended phase space (for a process): $(\mathbb{T}\mathcal{X})_\bbmu = \mathbb{T} \times \mathcal{X}$ \\

$\mathbbl{Tor}$ 
	& Torricelli's trumpet; acute hyperbolic solid \\

$\mathcal{T}_p\mathbb{R}^{1, 3}$ 
	& Tangent space of $\mathbb{R}^{1, 3}$ \\

$\mathcal{T}_{\idem}G$
	& Tangent space\footnote{
	Tangent and cotangent spaces are vector spaces, but, to distinguish them from the vector space as such, the \texttt{\textbackslash{mathcal\{T\}}} command is used instead of the \texttt{\textbackslash{mathfrak\{T\}}}.
	}  
of the Lie group $G$ at the identity (it is also denoted by $\mathcal{T}_1G$ or $\mathcal{T}_eG$) \\

$\mathcal{T}_gG$ 
	& Tangent space of the Lie group $G$ \\

$\mathcal{T}_{\gamma_\mathrm{c}(t)}\mathcal{M}$ 
	& Tangent space of a curve, path or geodesic \\

$\mathcal{T}_x\mathcal{M}$
	& Tangent space: set of all tangent vectors at $x$ (or other letter) in $\mathcal{M}$ \\ 
	
$\mathcal{T}^*_x\mathcal{M}$ 
	& Cotangent space (dual space to the tangent space $\mathcal{T}_x\mathcal{M}$) \\

$(\mathcal{T}_p)^k_\rotatedell\mathcal{M}$ 
	& Tensor space of type $(k, \rotatedell)$ from the tangent (vector) space: $(\mathcal{T}_p)^k_\rotatedell\mathcal{M} = \mathcal{T}_p\mathcal{M}^{\otimes^k} \otimes \mathcal{T}_p\mathcal{M}^{*\otimes^\rotatedell}$ \\
	
$(\mathcal{T}_p)^r_s\mathcal{M}$ 
	& Tensor space of type $(r, s)$ from the tangent (vector) space: $(\mathcal{T}_p)^r_s\mathcal{M} = \mathcal{T}_p\mathcal{M}^{\otimes^r} \otimes \mathcal{T}_p\mathcal{M}^{*\otimes^s}$ \\

$\mathcal{T}_v\mathring{\mathcal{T}}^1\mathcal{M}$ 
	& Second (or double) tangent space:  $\mathcal{T}_v\mathring{\mathcal{T}}^1\mathcal{M} = \mathcal{E}^+(v) \oplus \mathcal{E}^-(v)$, with $v \in \mathring{\mathcal{T}}^1\mathcal{M}$ \\

$\mathring{\mathcal{T}}\mathcal{F}$ 
	& Tangent bundle of a foliation, that is, subbundle \\

$\mathring{\mathcal{T}}\mathcal{M}$ 
	& Tangent bundle (disjoint union of the tangent spaces $\mathcal{T}_x\mathcal{M}$) \\

$\mathring{\mathcal{T}}^*\mathcal{M}$
	& Cotangent bundle (disjoint union of the cotangent spaces $\mathcal{T}^*_x\mathcal{M}$) \\

$\mathring{\mathcal{T}}^k_\rotatedell\mathcal{M}$ 
	& $\binom{k}{\rotatedell}$-tensor bundle or bundle of tensors of type $(k, \rotatedell)$ over $\mathcal{M}$ \\

$\mathring{\mathcal{T}}^r_s\mathcal{M}$ 
	& $\binom{r}{s}$-tensor bundle or bundle of tensors of type $(r, s)$ over $\mathcal{M}$ \\

$\mathring{\mathcal{T}}^1\mathbb{D}_\mathbb{C}$ 
	& Unit tangent bundle of $\mathbb{D}_\mathbb{C}$ (alternative notation: $\mathring{\mathcal{S}}^1\mathbb{D}_\mathbb{C}$) \\

$\mathring{\mathcal{T}}^1\mathbb{U}^2_\mathbb{C}$ 
	& Unit tangent bundle of $\mathbb{U}^2_\mathbb{C}$ (alternative notation: $\mathring{\mathcal{S}}^1\mathbb{U}^2_\mathbb{C}$) \\

$\mathring{\mathcal{T}}^1\mathcal{M}$ 
	& Unit tangent bundle of $\mathcal{M}$ (alternative notation: $\mathring{\mathcal{S}}^1\mathcal{M}$) \\

$\mathring{\mathcal{T}}^1\mathcal{S}_\mathbbl{\Gamma}$ 
	& Unit tangent bundle of $\mathcal{S}_\mathbbl{\Gamma} \equival \mathbbl{\Gamma} \backslash \mathbb{U}^2_\mathbb{C}$; $\mathring{\mathcal{T}}^1\mathcal{S}_\mathbbl{\Gamma} \cong \mathbbl{\Gamma} \backslash \mathring{\mathcal{T}}^1\mathbb{U}^2_\mathbb{C}$. — $\mathring{\mathcal{T}}^1\mathcal{S}_\mathbbl{\Gamma} \cong \mathbbl{\Gamma} \backslash PSL_2(\mathbb{R})$ \\

$\mathring{\mathcal{T}}_\mathbbl{\Lambda}\mathcal{M}$ 
	& Tangent bundle concerning the attractor $\mathbbl{\Lambda}$ \\

$\mathring{\mathcal{T}}_\mathrm{v}\mathring{\mathcal{E}}$, $\mathring{\mathcal{T}}_\mathrm{h}\mathring{\mathcal{E}}$ 
	& Vertical ($_\mathrm{v}$) and horizontal ($_\mathrm{h}$) bundles, respectively, that is, two subbundles of $\mathring{\mathcal{T}}\mathring{\mathcal{P}}$ \\

$\mathring{\mathcal{T}}\mathring{\mathcal{P}}$
	& Tangent bundle of $\mathring{\mathcal{P}}$ (principal $G$-bundle): $\mathring{\mathcal{T}}\mathring{\mathcal{P}} = \mathring{\mathcal{T}}_\mathrm{v}\mathring{\mathcal{E}} \oplus \mathring{\mathcal{T}}_\mathrm{h}\mathring{\mathcal{E}}$ \\

$\mathfrak{T}(\mathcal{M})$ 
	& Vector space of sections of $\mathring{\mathcal{T}}\mathcal{M}$, i.e. vector space of vector fields on $\mathcal{M}$ \\

$\mathfrak{T}^r_s(\mathcal{M})$ 
	& Tensor space of type $(r, s)$ over $\mathcal{M}$ \\ 

$\mathfrak{T}(\mathfrak{V})$ 
	& Tensor space, that is, tensor algebra of $\mathfrak{V}$ \\

$\mathfrak{T}^\bullet(\mathfrak{V})$, $\mathfrak{T}_\bullet(\mathfrak{V})$
	& Contravariant ($^\bullet$) and covariant ($_\bullet$) tensor algebra of $\mathfrak{V}$, respectively \\ 

$\mathfrak{T}^r_s(\mathfrak{V})$ 
	& Tensor space of type $(r, s)$, $\mathfrak{T}^r_s(\mathfrak{V}) = \mathfrak{V}^{\otimes^r} \otimes \mathfrak{V}^{*\otimes^s}$, vector space tensor product (i.e. tensor product of two vector spaces) between $r$ copies of a vector space $\mathfrak{V}$ and $s$ copies of its dual vector space $\mathfrak{V}^*$, that is, space of all tensors $\Tau^r_s$ \\

$\mathscr{T}_2$ 
	& Hausdorff space \\

$\mathsf{T}$ 
	& Temperature \\

$\mathsf{T}_\mathrm{inv}$ 
	& Inverse temperature \\

$\mathsf{T}_\mathrm{abs}$ 
	& Absolute temperature \\

$\topological$ 
	& Topological (property). — Topology of/on \\

$\trace$ 
	& Trace \\

\textsc{tt} 
	& Transverse-traceless gauge \\

\textsc{ttc} 
	& Tomita–Takesaki–Connes flow \\

$\mathbb{U}^2$ 
	& Beltrami–Poincaré half-plane model (open upper half-plane), i.e. real-valued $\mathbb{U}^2_\mathbb{R}$ or complex-valued $\mathbb{U}^2_\mathbb{C}$ half-plane \\
 
$\mathbb{U}^2_\mathbb{C}$ 
	& Complex upper half-plane \\

$\mathbb{U}^2_\mathbb{F}$ 
	& Real or complex upper half-plane (hyperbolic 2-space): $\mathbb{F} \viz (\mathbb{C} \cong \mathbb{R}^2)$ \\

$\mathbb{U}^2_\mathbb{R}$ 
	& Real upper half-plane \\

$\mathbb{U}^n$ 
	& Beltrami–Poincaré half-space model (open upper half-space) \\
 
$\overbar{\mathbb{U}}^n$ 
	& Closed upper half-space \\

$\mathcal{U}$ 
	& (Open) cover of a topological space: e.g. $\mathcal{U} = \{U_\nu\}$ of $\mathcal{X}$ such that $\mathcal{X} = \bigcup_\nu{U_\nu}$ \\

$\mathfrak{u}(n)$ or $\mathfrak{u}_n$ 
	& Lie algebra of $U(n)$ or $U_n$ \\

$\mathfrak{U}^n_{(\mathbb{C})}$ 
	& Vector pace of $n$-spinors \\

$\mathscr{U}_\sigma$ 
	& $\sigma$-subalgebra of $\mathscr{B}_\sigma$ \\

$u^\mu$ 
	& 4-velocity vector \\

$U$ 
	& Unitary matrix \\

$\hat{U}$ 
	& Time evolution operator, or propagator \\

$U_\mathbbl{\Gamma}$ 
	& Set of limit points of all orbits in $\mathbb{U}^2_\mathbb{C}$ \\

$U_{\mathbbl{E}}$ 
	& Universe of a tempered random set $\mathbbl{E}$ \\

$U(1)_\mathrm{em}$ or $U_{1(\mathrm{em})}$ 
	& Gauge group or symmetry (Lie) group of electromagnetism and, more generally, of Abelian gauge theories: $U(1)_\mathrm{em} \viz U(1)$ \\

$U(\textgreek{\textit{\ddigamma}})$ 
	& Potential (field) \\		

$U(n)$ or $U_n$ 
	& Unitary group \\
	
\textsc{uv} 
	& Ultraviolet \\

$\mathbb{V}$
	& Klein four-group of order 4 ($\mathbb{V} \cong \mathbb{Z}_2 \times \mathbb{Z}_2$) \\

$\mathring{\mathcal{V}}(\theta)$ 
	& Vertical fiber at $\theta$ \\
	
$\mathfrak{V} \viz (\mathfrak{V}, \mathbb{F})$ 
	& Vector space over $\mathbb{F}$ \\

$\mathfrak{V}^* \viz (\mathfrak{V}^*, \mathbb{F})$ 
	& Dual space of covectors \\

$v_{(\partial\surface)}$ 
	& Tangent vector (Stokes' theorem) \\

$\mathrm{v}_\varepsilon$ 
	& $\mathrm{v}_\varepsilon = 0, 1/\sqrt{\volume(\cdot)}$ \\

$\mathrm{v}_c$ 
	& Constant depending on the speed of moving surface \\

$\mathrm{ver}$ 
	& Vertex \\

$\volume$ 
	& Volume \\

$\volume\bigl(\mathbb{B}_\rho(x)\bigr)$ 
	& Volume of a ball of radius $\rho$ and center $x$  \\
	
$\Vitali$
	&\footnote{
	$\Vitali$ via \texttt{\textbackslash{Vitali}} command.
	}
	Vitali set \\

\textsc{vccm} 
	& Veltman–Connes–Chamseddine–Marcolli version for Lagrangian density (Standard Model) \\

$\mathcal{W}^\mathrm{s}$, $\mathcal{W}^\mathrm{u}$ 
	& Stable ($^\mathrm{s}$) and unstable ($^\mathrm{u}$) foliations (manifolds), respectively \\

$\mathcal{W}^\mathrm{s}(x)$, $\mathcal{W}^\mathrm{u}(x)$ 
	& Leaves of stable ($^\mathrm{s}$) and unstable ($^\mathrm{u}$) foliations, respectively, leafy stratifications \\ 

$\mathcal{W}^\mathrm{ss}$, $\mathcal{W}^\mathrm{su}$ 
	& Strongly stable ($^\mathrm{ss}$) and strongly unstable ($^\mathrm{su}$) foliations (manifolds), respectively \\

$\mathcal{W}^\text{ws}$, $\mathcal{W}^\text{wu}$ 
	& Weakly stable ($^\mathrm{ws}$) and weakly unstable ($^\mathrm{wu}$) foliations (manifolds), respectively \\

$\mathfrak{W}$ 
	& Affinely convex subset of a vector space \\

$\mathscr{W}$, $\mathscr{W}$-functional 
	& Perelman's entropy $\mathscr{W}$-functional, or just $\mathscr{W}$-entropy, but, more correctly, entropy-energy $\mathscr{W}$-functional, aka shrinker entropy functional \\

$\mathsf{W}$ 
	& Wiener process, $\mathsf{W}_t$, with $t \geqslant 0$ \\

$\rotatedw$, $\rotatedw_{\mu\nu}$ 
	& 2-form \\

$W$ 
	& Number of microscopic configurations, thermodynamic probability in the concept of Boltzmann–Planck's entropy \\

$W^\pm$ 
	& Charged positive/negative weak bosons \\

$\Weyl_{\mu\nu\xi\varrho}$, $\dots$
	& Weyl (conformal) curvature tensor: $\Weyl_{\mu\nu\xi\varrho}$, ${\Weyl^\mu}_{\nu\xi\varrho}$, ${\Weyl^{\mu\nu}}_{\xi\varrho}$ \\

${^\prj\Weyl^\mu}_{\nu\xi\varrho}$ 
	& Weyl projective (curvature) tensor \\

$\Sobolev^{1, 2}$, $\Sobolev^{1, 2}(\mathcal{M})$ 
	& Sobolev space on Riemannian $n$-manifolds; Sobolev space of first derivatives in $\Lebesgue^2$-space \\

$\Sobolev^{1, p}_0$, $\Sobolev^{1, p}_0(\Omega)$ 
	&
	\footnoteref{footnote "Sobolev space"}
	Sobolev space with zero trace \\

$\Sobolev^{k, p}$, $\Sobolev^{k, p}(\Omega)$ 
	& \footnote{
	\label{footnote "Sobolev space"}
	$W$ via \texttt{\textbackslash{Sobolev}} command. As a vector space, it should be denoted here by $\mathfrak{W}$, but in this case we follow the current and universal notation in literature without Fraktur typeface.
	}
	Sobolev space: $\Sobolev^{k, p}(\Omega) \viz \Sobolev^{k, p}(\Omega, \mathbb{R}^k)$, for $k \in \mathbb{Z}_*$ or  $k \in \mathbb{N}$, $\Omega \subset \mathbb{R}^n$ \\

$\Sobolev^{k, p}_\mathrm{loc}(\Omega)$ 
	& Local Sobolev space \\
	
$\vec{W}_\textsc{fp}^\pm$, $\vec{Z}_\textsc{fp}^0$ 
	& Faddeev–Popov (gauge) ghost fields associated with the positive/negative, and neutral weak bosons \\

$\Wey$ 
	& Weyl curvature tensor, notation with indices \\

Weak-$\star$ 
	& Weak-star topology \\

$\mathcal{X}$
	& (or another letter) Topological space, topological manifolds or topological vector space. One example is the Lindelöf space \\ 

$\mathcal{X}_{\textnormal{\textgreek{τ}}}$  
	& Category of sheaves of sets — on a topological space/on a site — in Grothendieck topoi \\

$\mathcal{X}_\topological$
	& Grothendieck topology \\

$\mathcal{X}^1_{\varphi_t}(\mathcal{M})$ 
	& Space of $\mathscr{C}^1$ flows $\varphi_t$ of Anosov type on $\mathcal{M}$ \\

$\mathcal{X}^r_{\varphi_t}(\mathcal{M})$ 
	& Space of $\mathscr{C}^r$ flows $\varphi_t$ of Anosov type on $\mathcal{M}$ \\

$\mathcal{X}^r(\mathcal{M})$ 
	& Space of $\mathscr{C}^r$ vector fields on $\mathcal{M}$ \\

$\mathcal{X}^r_\mathcal{F}(\mathcal{M})$ 
	& Space of $\mathscr{C}^r$ smooth foliations on $\mathcal{M}$ \\

$(\mathcal{X}, \distance)$ 
	& Metric space; geodesic (metric) space. — Separable metric space  \\

$(\mathcal{X}, \mathscr{B}_\sigma, \bbmu)$
	& Probability space: Lebesgue–Rohlin space, or Lebesgue (probability) space \\

$(\mathcal{X}\mathcal{Q})_\bbmu$ 
	& Extended phase space (for a skew product flow): $(\mathcal{X}\mathcal{Q})_\bbmu = \mathcal{Q} \times \mathcal{X}$ \\

$\mathfrak{X}$
	& (or another letter) Finite- or infinite-dimensional vector space \\ 

$\mathfrak{X} \viz (\mathfrak{X}, \|\cdot\|)$
	& Normed vector space \\

$\bar{x}$
	&\footnote{
	In physics literature the notation with asterisk in superscript, i.e. $x^*$, is preferred.
	}
	(or another letter) Complex conjugate of a complex number $x$ \\

$\hat{\mathrm{x}}$ 
	& Position operator: $\hat{\mathrm{x}} \colon \mathfrak{D}(\hat{\mathrm{x}}) \to \Lebesgue^2(\mathbb{R})$ \\

$\|x\|$
	& (or another letter) Norm of a vector $x$ \\ 

$x_\mathbb{CM}$ 
	& Point in $\mathbb{CM}^\natural_{(\infty)}$ \\

$x_\mathbb{M}$ 
	& Point in $\mathbb{M}^{4(\mathbb{R})}_{(\infty)}$ \\

$x_{\mathbb{T}\mathbbl{w}}$ 
	& Point, or twistor, in a twistor space \\

$\dot{x}^\nu_\mathsf{q}$, $\dot{y}^\nu_\mathsf{p}$ 
	& Hamilton's differential equations \\

$\vec{X}$
	& (or another letter) Vector field\footnote{
	In some cases the arrow above the upper case letter may be omitted, but when a vector field is denoted by a lower case letter, the arrow is usually omitted. Moreover, as appropriate, a boldface capital letter without an over arrow is used, see footnote \ref{footnote "Mathbold instead of the vector arrow"}, p. \pageref{footnote "Mathbold instead of the vector arrow"}.
	} \\

$\vec{X}_\Hamiltonian$ 
	& Hamiltonian vector field associated with $\Hamiltonian$ or symplectic gradient of $\Hamiltonian$ \\

$X^\textsc{t}$ 
	& (or another letter) Transpose of a matrix $X$ \\
	 
$X^\dag$ or $\bar{X}^\textsc{t}$ 
	& (or another letter) Conjugate transpose (Hermitian transpose, or even adjoint matrix) of a matrix $X$, for which $X^\dag \viz \bar{X}^\textsc{t}$ (where $\bar{X}$ indicates the complex conjugated entries and $X^\textsc{t}$ is the transpose). Note. Its generalization is the Hermitian adjoint (or adjoint operator) \\ 

$X_\mathbb{R}$ 
	& Group of real numbers (non-Archimedean analysis) \\

$\mathbb{Y}^n_+$ 
	& Upper sheet of the elliptic hyperboloid (2-sheeted hyperboloid) in the hyperboloid/Minkowski–Lorentz model \\
	
$\mathbb{Y}^n_-$ 
	& Lower sheet of the elliptic hyperboloid (2-sheeted hyperboloid) in the hyperboloid/Minkowski–Lorentz model \\

$\mathcal{Y}^n$ 
	& Set of not infinitely recurrent points $x \in \mathcal{B}_n$ \\

$\vec{Y}_\xi$ 
	& Vector field normal to $\partial\Omega$ \\

$\mathbb{Z}, \mathbb{Z}_-, \mathbb{Z}_+, \mathbb{Z}_*$
	& Set of integer numbers: $\mathbb{Z} = \{0, \pm1, \pm2, \pm3, \mathellipsis\}$: $\mathbb{Z}_-$ (negative integers), $\mathbb{Z}_+$ (positive integers), $\mathbb{Z}_* = \{0\} \cup \mathbb{Z}_+$ (non-negative integers: zero or positive numbers) \\
	
$\mathbb{Z}_2$ or $\mathbb{Z}/2\mathbb{Z}$
	& Cyclic group of order 2 \\
	
$\mathring{\mathcal{Z}}$ 
	& Curved twistor space \\

$\mathfrak{Z}$ 
	& $C^\ast$-algebra, i.e. linear space of bounded operators on $\mathfrak{H}$ (Hilbert space), or—depending on the context—algebra of quantum operators \\

$z$ 
	& G.e. \\

$z_\mathrm{rs}$ 
	& Redshift \\

$Z^0$ 
	& Neutral weak boson \\
\end{longtable}
\endgroup

\mainmatter
\chapter{Panoramic Miscellanea I. Theory of Connections, Differential Forms, Geodesics, and Holonomy Groups} 
\chaptermark{Panoramic Miscellanea I}{}
\label{chapter "Panoramic Miscellanea I. Theory of Connections, Differential Forms, Geodesics, and Holonomy Groups"}

\begingroup
\footnotesize
[Q]ueste due scientie (cioè l'Ari[t]metica [la cui parte maggiore è detta Algebra], e Geometria) hanno intra di loro tanta convenientia, che l'una è la prova dell'altra, e l'altra è la dimostration dell'una, n[é] già puote il Matematico esser perfetto, il quale in ambedue non sia versato.\footnote{
	«These two sciences (that is, Arithmetic [the major part of which is called Algebra], and Geometry) have so much convenience with each other, that one is the proof of the other, and the other is the demonstration of one, nor can the Mathematician be perfect, if he is not versed in both».
	} \\
\indent — \textsc{R. Bombelli} \cite[Problema \textsc{cclxxii}, p. 648]{Bombelli "L'Algebra parte maggiore dell'Arimetica divisa in tre libri di Rafael Bombelli da Bologna"}\endnote{
	Unchanged text and same page number in the second edition from 1579.
	}
	
\endgroup

\section{A Little Bit of Bundles} 
\label{section "A Little Bit of Bundles"}

In this Section we will explore the notions of tangent and cotangent bundles, which are typical examples of fiber bundles, known as \emph{vector bundles}; the very notions of fibrate and vector bundles, and what it calls a trivial chart. Finally, we will define a connection on the tangent bundle.

\subsection{Fiber Bundles (Tangent, Cotangent and Vector Bundle)}
\label{subsection "Fiber Bundles (Tangent, Cotangent and Vector Bundle)"}

\begin{definitio}[Tangent and cotangent bundles]
\label{definitio "Tangent and cotangent bundles"}
Let $\mathcal{M}$ be a differentiable manifold; then the tangent bundle of $\mathcal{M}$ is the disjoint union of the tangent spaces,
\begin{equation}\mathring{\mathcal{T}}\mathcal{M} = \bigcup_{p \in \mathcal{M}}\mathcal{T}_p\mathcal{M};
\end{equation}
analogously, the cotangent bundle (or dual bundle to the tangent bundle) of $\mathcal{M}$ is the disjoint union of the cotangent spaces, $\mathring{\mathcal{T}}^*\mathcal{M} = \bigcup_{p \in \mathcal{M}}\mathcal{T}^*_p\mathcal{M}$ (with the projection $\pi \colon \mathring{\mathcal{T}}^*\mathcal{M} \to \mathcal{M}$), where $\mathcal{T}_p\mathcal{M}$ and $\mathcal{T}^*_p\mathcal{M}$ are, respectively, the tangent and the cotangent spaces of $\mathcal{M}$ at a point $p \in \mathcal{M}$. \definitiosymbol
\end{definitio}

\begin{definitio}[Fiber bundle]
\label{definitio "Fiber bundle"}
Let $(\mathring{\mathcal{E}}, \pi, \mathcal{M}, \mathring{\mathcal{F}})$ be a \emph{fiber bundle}, where $\mathring{\mathcal{E}}$, $\mathcal{M}$ and $\mathring{\mathcal{F}}$ are the \emph{total space}, the \emph{base space}, and the \emph{fiber} of the (fiber) bundle ($\mathring{\mathcal{E}}$, $\mathcal{M}$ and $\mathring{\mathcal{F}}$ shall be clearly all topological manifolds); $\pi$ is a surjective submersion, called \emph{(bundle) projection map}. The map $\pi \colon \mathring{\mathcal{E}} \to \mathcal{M}$ is a \emph{fiber bundle} iff, for each $p \in \mathcal{M}$, there exists an open set $\Omega$ such that $\mathring{\mathcal{E}}|_\Omega \equival \pi^{-1}(\Omega)$ is diffeomorphic to $\Omega \times \mathring{\mathcal{F}}$, and the  diagram
\[
\begin{tikzcd}[row sep=large, column sep=large]
	\left(\mathring{\mathcal{E}}|_\Omega \equival \pi^{-1}(\Omega)\right) \arrow{r}{\varphi} \arrow[swap]{d}{\pi} & \left(\Omega \times \mathring{\mathcal{F}}\right) \arrow{dl}{\pi_1} \\
	\Omega
\end{tikzcd}
\]				
is commutative; $\pi_1$ (or $\prj_1$) is the projection onto the first factor. The diffeomorphism $\varphi \colon \mathring{\mathcal{E}}|_\Omega \equival \pi^{-1}(\Omega) \to \Omega \times \mathring{\mathcal{F}}$ is said to be \emph{local trivialization} of $\mathring{\mathcal{E}}$ over $\Omega$.	\definitiosymbol
\end{definitio}

The notion of vector bundle is totally in keeping with the definition above.

\begin{definitio}[Vector bundle]
\label{definitio "Vector bundle"}
Given a triple $\mathring{\zeta} = (\mathring{\mathcal{E}}, \pi, \mathcal{M})$, the (smooth) map $\pi \colon \mathring{\mathcal{E}} \to \mathcal{M}$ is called (real) \emph{vector bundle} of rank $r$ iff,
\enumerationisinitium
\item for each $p \in \mathcal{M}$, the set $\mathring{\mathcal{E}}_p \equival \pi^{-1}(p)$, otherwise said \emph{fiber over $p$}, is consistent with the vector space structure (this means that on each fiber $\mathring{\mathcal{E}}_p$ there is an $r$-dimensional vector space over $\mathbb{R}$);
\item for each $p \in \mathcal{M}$, there exists a neighborhood $\Upsilon \subset \mathcal{M}$ of $p$ and a homeomorphism $\varphi \colon \mathring{\mathcal{E}}|_\Upsilon \equival \pi^{-1}(\Upsilon) \to \Upsilon \times \mathbb{R}^r$, which is a diffeomorphism (here too $\varphi$ is a local trivialization of $\mathring{\mathcal{E}}$ over $\Upsilon$), such that the diagram
\[
\begin{tikzcd}[row sep=large, column sep=large]
	\left(\mathring{\mathcal{E}}|_\Upsilon \equival \pi^{-1}(\Upsilon)\right) \arrow[swap]{d}{\pi} \arrow{r}{\varphi}
    & \left(\Upsilon \times \mathbb{R}^r \right) \arrow{d}{\pi_1 \viz \pi_\Upsilon} \\
	\Upsilon \arrow[equal]{r}
	& \Upsilon 
\end{tikzcd}
\]				
is commutative, for which $\pi_1 \viz \pi_\Upsilon \circ \varphi = \pi$, with the projection $\pi_1 \viz \pi_\Upsilon \colon \Upsilon \times \mathbb{R}^r \to \Upsilon$;
\item the restriction of $\varphi$ to each fiber is an isomorphism between two spaces $\mathring{\mathcal{E}}_p \equival \pi^{-1}(p)$ and $\{p\} \times \mathbb{R}^r$, hence it is a linear map $\varphi \colon \mathring{\mathcal{E}}_p \equival \pi^{-1}(p) \to \{p\} \times \mathbb{R}^r$ (cf. Definition \ref{definitio "Differential form"}).	\definitiosymbol
\enumerationisfinis
\end{definitio}

It is clear from the Definition \ref{definitio "Vector bundle"} that the same concept of trivialization can be described with the notion of chart.

\begin{definitio}[Trivial chart]
Let $(\Upsilon, \varphi)$ be a chart on $\mathcal{M}$, and assume that $\Upsilon$ is a coordinate domain, or a coordinate neighborhood, and $\varphi(p) = x^1(p), \mathellipsis, x^n(p)$ are the local coordinates on $\Upsilon$. One says that $(\Upsilon, \varphi)$ is \emph{trivial} if it is a local trivialization of $\mathring{\mathcal{E}}$ defined on $\mathring{\mathcal{E}}|_\Upsilon \equival \pi^{-1}(\Upsilon)$. \definitiosymbol
\end{definitio}

\begin{scholium}
The distinction between (open) (sub)set $\Omega$ and (open) neighborhood $\Upsilon$ in this background are often used alternately, in fact they have the same meaning. The two notions can be substituted for each other, $\Omega \equival \Upsilon$. \scholiumsymbol
\end{scholium}

\begin{margo}
\label{scholium "Real and complex vector bundle"}
A vector bundle $\mathring{\zeta} = (\mathring{\mathcal{E}}, \pi, \mathcal{M})$ of rank $k$ is a space bundle over the field of real or complex numbers; the real and complex vector bundles are denoted by $\mathring{\zeta}_{\mathbb{R}^r}$ and $\mathring{\zeta}_{\mathbb{C}^r}$, respectively. \margosymbol
\end{margo}

Before going on, we need to introduce the concept of \emph{section} related to this specific kind of fiber bundles.

\begin{definitio}[Sections of the vector bundle]
Let $\pi \colon \mathring{\mathcal{E}} \to \mathcal{M}$ be a vector bundle over $\mathcal{M}$. A \emph{(global) smooth section} of $\mathring{\mathcal{E}}$ is a map 
\[
	\sigma_{\mathring{\mathcal{E}}} \colon \mathcal{M} \to \mathring{\mathcal{E}}
\]
such that $\pi \circ \sigma_{\mathring{\mathcal{E}}} = \id_\mathcal{M}$ (identity map of $\mathcal{M}$), namely $\sigma(p) \in \mathring{\mathcal{E}}_p$ for all $p \in \mathcal{M}$. If $\Upsilon$ is an open neighborhood of $\mathcal{M}$, a \emph{local smooth section} of $\mathring{\mathcal{E}}$ is a map 
\[
	\sigma_{\mathring{\mathcal{E}}} \colon \Upsilon \to \mathring{\mathcal{E}}
\]
such that $\pi \circ \sigma_{\mathring{\mathcal{E}}} = \id_\Upsilon$ (identity map of $\Upsilon$). A basis $\sigma_{\mathring{\mathcal{E}}} = \{\sigma_1(p), \mathellipsis, \sigma_n(p)\}$ of sections of $\mathring{\mathcal{E}}$ over $\Upsilon$ is a local frame field for the fiber $\mathring{\mathcal{E}}_p$ at all points $p \in \Upsilon$. \definitiosymbol
\end{definitio}

\subsection{Connection on the Tangent Bundle}
\label{subsection "Connection on the Tangent Bundle"}

We now can focus on the tangent bundle, as it is a prototypical disjoint union of the vector spaces (which are, more properly, the set of all tangent vectors at each point in a manifold). The connection on the tangent bundle is usually referred to as a linear connection on a smooth manifold (Definition \ref{definitio "Linear connection"}).

\begin{propositio}[Connection on the tangent bundle]
\label{propositio "Connection on the tangent bundle"} 
Let $\sigma_{\mathring{\mathcal{E}}} \in \mathfrak{E}(\Upsilon)$ be an orthonormal local frame for the tangent bundle $\mathring{\mathcal{T}}\mathcal{M}$, on which $\sigma_{\mathring{\mathcal{E}}} = \{\sigma_1, \mathellipsis, \sigma_n\}$ represents the sections of $\mathring{\mathcal{T}}\mathcal{M}$ on (defined over) an open neighborhood $\Upsilon \subset \mathcal{M}$ and $\mathfrak{E}(\Upsilon)$ is the vector space of sections of the tangent bundle.\footnote{
	A section $\sigma_{\mathring{\mathcal{E}}}$ of $\mathring{\mathcal{T}}\mathcal{M}$ is a vector field on $\mathcal{M}$.
	}
Then the required form by the Theorem \ref{theorema "Levi-Civita"} can be rewritten as 
\begin{equation}
\left\langle\nabla_{\sigma_\mu}, \sigma_\nu, \sigma_\xi\right\rangle = \frac{1}{2}\bigl\{\langle[\sigma_\mu, \sigma_\nu], \sigma_\xi\rangle - \langle[\sigma_\nu, \sigma_\xi], \sigma_\mu\rangle + \langle[\sigma_\xi, \sigma_\mu], \sigma_\nu\rangle\bigr\},
\end{equation}
taking a local $n$-tuple $\sigma_{\mathring{\mathcal{E}}}$ in place of $\vec{X}, \vec{Y}, \vec{Z}$.
\end{propositio}

Suppose $\bigl\{\frac{\partial}{\partial x^\mu}\bigr\}$ is the chosen coordinate basis for the tangent bundle $\mathring{\mathcal{T}}\mathcal{M}$ of a (pseudo-)Riemannian manifold.\footnote{
	\label{footnote "Riemannian manifold and pseudo-Riemannian manifolds"}
	The pair $(\mathcal{M}, g)$ is said to be a \emph{Riemannian manifold} if the metric tensor $g$ is a $\binom{0}{2}$-tensor field on a smooth manifold $\mathcal{M}$, and if $g$ is symmetric $g(v, w) = g(w, v)$, non-degenerate $g(v, w) = 0$, for any $v, w \in \mathcal{T}_p\mathcal{M}$, $p \in \mathcal{M}$, and positive definite $g(v, v) > 0$. The pair $(\mathcal{M}, g)$ is said to be a \emph{pseudo-Riemannian manifold} if it has these properties but $g$ is not necessarily positive definite.
	} 
In this way, for a connection $\nabla$-like on $\mathring{\mathcal{T}}\mathcal{M}$, we have 
\begin{equation}
\label{equation "Connection on the tangent bundle and Christoffel symbols"}
	\nabla_{\frac{\partial}{\partial x^\mu}}\frac{\partial}{\partial x^\nu} = \sum_\xi{\Gamma^\xi}_{\mu\nu}\frac{\partial}{\partial x^\xi},
\end{equation}
where the $\Gamma$-functions are the so-called \emph{coefficients of connection} in a coordinate basis or \emph{Christoffel symbols} for the Levi-Civita connection (Theorem \ref{theorema "Levi-Civita"}). They may be defined as follows.	

\section{Christoffel Symbols}
\label{section "Christoffel Symbols"}

\begingroup 
\footnotesize
The algorithm [consisting of the covariant derivative] of absolute differential Calculus [tensor calculus], that is to say the material instrument of the methods [\,\dots] is fully included [albeit still in nuce] in a remark by Mr. Christoffel \cite{Christoffel "Ueber die Transformation der homogenen Differentialausdrucke zweiten Grades"}.\endnote{
	Original Fr. version: «L'algorithme [de dérivation covariante] du Calcul différentiel absolu, c'est à dire l'instrument matériel des méthodes [\,\dots] se trouve tout entier [mais encore in nuce] dans une remarque due a M. Christoffel».
	} \\
\indent — \textsc{G. Ricci [Curbastro] et T. Levi-Civita} \cite[p. 127]{Ricci et Levi-Civita "Methodes de calcul differentiel absolu et leurs applications"}

\endgroup

\vspace{2mm}

The Christoffel symbols \cite{Christoffel "Ueber die Transformation der homogenen Differentialausdrucke zweiten Grades"} in \eqref{equation "Connection on the tangent bundle and Christoffel symbols"} are coefficients completely determining a  metric \emph{connection} with respect to a given coordinate system; it is about the \emph{smooth functions} in a local frame and a local chart. 

\begin{definitio}[Christoffel symbols of the second kind]
The expression that gives the aforenamed functions is
\begin{subequations}	
\label{subequations "Christoffel symbols for the Levi-Civita connection"}
\begin{align}
{\Gamma^\xi}_{\mu\nu} \viz \binomcurly{\xi}{\mu\nu} & = g^{\xi\varrho}{\Gamma^\varsigma}_{\mu\nu}\left\langle\frac{\partial}{\partial x^\varsigma}, \frac{\partial}{\partial x^\varrho}\right\rangle
	= g^{\xi\varrho}\left\langle\nabla_{\frac{\partial}{\partial x^\mu}}\frac{\partial}{\partial x^\nu}, \frac{\partial}{\partial x^\varrho}\right\rangle \\
	& = \frac{1}{2}g^{\xi\varrho} 
	\left\{
	\frac{\partial g_{\nu\varrho}}{\partial x^\mu} + \frac{\partial g_{\mu\varrho}}{\partial x^\nu} - \frac{\partial g_{\mu\nu}}{\partial x^\varrho}\right\} \\
	& = \frac{1}{2}g^{\xi\varrho}\bigl(g_{\nu\varrho, \mu} + g_{\mu\varrho, \nu} - g_{\mu\nu, \varrho}\bigr).
\end{align}
\end{subequations}
\definitiosymbol
\end{definitio}

The Christoffel symbols are named after Elwin B. Christoffel; but their concomitant exposition is also available in R. Lipschitz \cite{Lipschitz "Untersuchungen in Betreff der ganzen homogenen Functionen von n Differentialen"}.

According to Proposition \ref{propositio "Connection on the tangent bundle"}, the formula \eqref{equation "Connection on the tangent bundle and Christoffel symbols"} (for the connection on the tangent bundle) is clearly of the form $\nabla_{\sigma_\mu}, \sigma_\nu = {\Gamma^\xi}_{\mu\nu}\sigma_\xi$ or, equivalently, $\nabla_{\partial_\mu}, \partial_\nu = \binomcurly{\xi}{\mu\nu}\sigma_\xi$. 

\begin{definitio}[Christoffel symbols of the first kind]
In Eq. \eqref{subequations "Christoffel symbols for the Levi-Civita connection"} the Christoffel symbols are of the second kind and they are expressed in terms of the metric tensor and its derivatives, but they are not tensors themselves. Starting with this, one can define the Christoffel symbols of the first kind: 
\begin{equation}
g\left(\nabla_{\sigma_\mu}, \sigma_\nu, \sigma_\xi\right) \equival \Gamma_{\mu\nu\xi} \viz [\mu\nu, \xi] = g_{\varrho\xi}{\Gamma^\varrho}_{\mu\nu} = \frac{1}{2}\left(g_{\mu\xi, \nu} + g_{\nu\xi, \mu} - g_{\mu\nu, \xi}\right).
\end{equation} 
\definitiosymbol
\end{definitio}

\begin{corollarium}
The Christoffel symbols of the first kind are symmetric in the first two indices: $\Gamma_{\mu\nu\xi} = \Gamma_{\nu\mu\xi}$.
\end{corollarium}

\begin{proof}
$\Gamma_{\nu\mu\xi} = \frac{1}{2}\left(g_{\nu\xi, \mu} + g_{\mu\xi, \nu} - g_{\mu\nu, \xi}\right) = \frac{1}{2}\left(g_{\mu\xi, \nu} + g_{\nu\xi, \mu} - g_{\mu\nu, \xi}\right) = \Gamma_{\mu\nu\xi}$.
\end{proof}

\begin{corollarium}
The Christoffel symbols of the second kind are symmetric in the lower two indices: $\Gamma_{\mu\nu\xi} = \Gamma_{\nu\mu\xi}$.
\begin{equation}
\label{equation "Symmetry of the Christoffel symbols"}
{\Gamma^\xi}_{\mu\nu} = {\Gamma^\xi}_{\nu\mu}, \text{ for all } \mu, \nu, \xi = 1, \mathellipsis, n - 1.
\end{equation}
\end{corollarium}

\begin{proof}
The symbols of the second kind are rewritable as ${\Gamma^\xi}_{\mu\nu} = g^{\xi\varrho}\Gamma_{\mu\nu\varrho}$, whence we get 
\begin{equation}
	{\Gamma^\xi}_{\mu\nu} = g^{\xi\varrho}\Gamma_{\mu\nu\varrho} = g^{\xi\varrho}\Gamma_{\nu\mu\varrho} = {\Gamma^\xi}_{\nu\mu}.
\end{equation}
\end{proof}	

The Levi-Civita connection $\nabla$ is symmetric (Definition \ref{definitio "Torsion free connection"}) because of the symmetry of the Christoffel symbols in an arbitrary coordinate frame. Or rather: the connection $\nabla$ is torsion free iff, in any local coordinate chart $(\Upsilon, x^1, \mathellipsis, x^n)$, the $\Gamma$-functions satisfy the Eq. \eqref{equation "Symmetry of the Christoffel symbols"}.

\subsection[Matrix Notation and Kronecker $\delta$-Function]{Matrix Notation and Kronecker $\mathbold{\delta}$-Function}
\label{subsection "Matrix Notation and Kronecker delta-Function"}

A way of calculating the Christoffel symbols is to consider the \emph{matrix notation}, by writing the relation between the matrix $[g_{\mu\nu}]$ and its inverse $[g^{\mu\nu}]$ as 
\begin{equation}
	g_{\varrho\nu}g^{\nu\mu} = g^{\mu\nu}g_{\nu\varrho} = {g^\mu}_\varrho = {\delta^\mu}_\varrho,	
\end{equation}
where ${\delta^\mu}_\varrho \equival \delta_{\mu\varrho}$, i.e. $\delta_\textsc{k}$, is the \emph{Kronecker delta}, which has the value 1 when the indices are equal ($\delta_\textsc{k} = 1$, if $\mu = \varrho$) and 0 otherwise ($\delta_\textsc{k} = 0$, if $\mu \ne \varrho$). Then 
\begin{equation}
{\Gamma^\xi}_{\mu\nu} = g^{\xi\varrho}\Gamma_{\mu\nu\varrho} = \frac{1}{2}g^{\xi\varrho}(\partial_\mu g_{\nu\varrho} + \partial_\nu g_{\mu\varrho} - \partial_\varrho g_{\mu\nu}).
\end{equation}

\section{Parallel Transport of the Levi-Civita Connection}
\label{section "Parallel Transport of the Levi-Civita Connection"}
 
\begingroup 
\footnotesize
The parallel transport, along any path, of two concurrent directions preserves their angle. It  clearly means that the angle formed by two generic directions through the same point is also the angle formed by their parallels through another point.\endnote{
	Original It. version: «Il trasporto per parallelismo, lungo un cammino qualsiasi, di due direzioni concorrenti ne conserva l'angolo. Con ciò si vuol dire evidentemente che l'angolo formato da due generiche direzioni uscenti da un medesimo punto è anche l'angolo formato dalle loro parallele in un altro punto qualunque».
	} \\
\indent — \textsc{T. Levi-Civita} \cite[p. 175]{Levi-Civita "Nozioni di parallelismo in una varieta qualunque e conseguente specificazione geometrica della curvatura riemanniana"}

\endgroup

\subsection{Relativistic Gravitation as a Genesis of the Parallel Transport}
\label{subsection "Relativistic Gravitation as a Genesis of the Parallel Transport"}

\begingroup
\footnotesize
Einstein's theory of relativity [\,\dots] considers the geometrical structure of space as very tenuously, but also intimately, dependent on the physical phenomena taking place in it; differently from classical theories, which assume the whole physical space as given a priori. The mathematical development of Einstein's grandiose conception (which finds in Ricci's absolute differential calculus its natural algorithmic tool) draws on the curvature of a certain 4-dimensional manifold as an essential element and the related Riemann symbols. Meeting these symbols, or continuously using them [\,\dots] led me to investigate whether it would be possible to somewhat reduce the formal apparatus, which serves commonly to introduce them and to establish their covariant behaviour. \\
\indent — \textsc{T. Levi-Civita} \cite[p. 173]{Levi-Civita "Nozioni di parallelismo in una varieta qualunque e conseguente specificazione geometrica della curvatura riemanniana"}

\endgroup

\vspace{2mm}

\enumerationisinitium
\item The notion of parallel transport is strongly related to that of connection. We indeed be able to construct the parallel transport rules starting from a connection, and vice versa; we are allowed to use a previous knowledge about parallel transport to get a connection such as that concerning the Definition \ref{definitio "Connection on a vector bundle"} and its general \emph{covariance}, depending on the L. Bianchi explanations \cite[§§ 33-34]{Bianchi "Lezioni di geometria differenziale I (seconda edizione)"} \cite[§§ 36-37]{Bianchi "Lezioni di geometria differenziale I (terza edizione)"}.
\item An essential element of the (pseudo-)Riemannian geomerty is a \emph{connection on the tangent bundle} (Definition \ref{definitio "Tangent and cotangent bundles"}) of a manifold, drawn up by T. Levi-Civita \cite{Levi-Civita "Nozioni di parallelismo in una varieta qualunque e conseguente specificazione geometrica della curvatura riemanniana"}. It starts with an infinitesimal field, in order to characterize the parallel transport of two directions through two very close points, i.e. from one point to an infinitely close point (see Definition \ref{definitio "Parallel transport map"}). The angle between two tangents to a manifold at a point is equal to the angle between two parallel tangents at another point very close to the first.
\item It is worth noting that the \emph{mathematical considerations} of Levi-Civita for the parallelism stem from a review on Einstein's theory of gravitation (see Section \ref{section "Gravitational Field as a Curvature of the Space"}), which is algorithmically rooted in the tensor calculus, created by G. Ricci Curbastro and re-elaborated by Levi-Civita \cite{Ricci et Levi-Civita "Methodes de calcul differentiel absolu et leurs applications"} (see Sections \ref{section "Excerpts from Memory: Ricci Methods"} and \ref{section "Rudiments of Tensor Calculus"}). As we know, in the general relativistic framework, Einstein utilizes the curvature of a 4-dimensional manifold (space-time), the geometrical structure of which is intimately dependent on the \emph{physical phenomena} taking place in it.
\enumerationisfinis

\subsection{Parallel Transport: a Way of Viewing Euclidean Type Small Spaces in a Curved Space}
\label{subsection "Parallel Transport: a Way of Viewing Euclidean Type Small Spaces in a Curved Space"}

\begingroup
\footnotesize
Levi-Civita \cite{Levi-Civita "Nozioni di parallelismo in una varieta qualunque e conseguente specificazione geometrica della curvatura riemanniana"}, with his definition of parallelism, was the first to succeed in making the \emph{false} metric \emph{spaces} of Riemann, if not true Euclidean spaces, which is impossible, at least \emph{spaces with a Euclidean connection}, considered as collections of small pieces of Euclidean space, \emph{oriented with respect to each other in going from point to point} \cite[p. 297]{Cartan "Les recentes generalisations de la notion d'espace"}.\endnote{
	Original Fr. version: «C'est M. Levi-Civita qui le premier, par sa définition du parallélisme, réussit à faire des \emph{faux espaces} métriques de Riemann, non pas de vrais espaces euclidiens, ce qui est impossible, mais du moins des \emph{espaces à connexion euclidienne}, considérés comme des collections de petits morceaux d'espaces euclidiens, \emph{orientés de proche en proche les uns par rapport aux autres}».
	} \\
\indent A Riemannian space is ultimately formed by an infinity of small pieces of Euclidean spaces \cite[p. 2]{Cartan "La Geometrie des espaces de Riemann"}.\endnote{
	Original Fr. version: «[U]n espace de Riemann est, au fond, formé d'une infinité de petits morceaux d'espaces euclidiens».
	} \\
\indent — \textsc{É. Cartan}

\endgroup

\vspace{2mm}

To put it roughly, by \emph{parallel transport}, or \emph{Levi-Civita transport}, is meant a way of comparing tangent spaces at different points on a manifold with any metric; and it proves to be an analytic method for considering a Riemannian space \cite{Riemann "Ueber die Hypothesen welche der Geometrie zu Grunde liegen"} \cite{Riemann "On the Hypotheses which lie at the Bases of Geometry"} as \emph{an infinity of small pieces of Euclidean space}—against this backdrop, \emph{a Riemannian space is a set that locally resembles Euclidean space}. In so doing, a manifold endowed with a metric displaying a certain type of curvature, for every neighborhood of each of its points, looks like (and can be treated as) \emph{a mosaic of arbitrarily small flat spaces}.

\subsection{Severi's Theorem (Non-ambient Parallelism of Levi-Civita), and Other Contour Jottings}
\label{subsection "Severi's Theorem (Non-ambient Parallelism of Levi-Civita), and Other Contour Jottings"}

\begingroup
\footnotesize
The concept of parallelism between directions, in regard to a manifold $V_n$ with some metric, introduced with good idea by [my] Colleague [Levi-Civita], could be formulated in a geometric form entirely independent of [ambient] Euclidean space $S_N$ in which $V_n$ is immersed, thus the intrinsic character of that concept is manifest a priori, with respect to the given manifold; this, for Levi-Civita \cite{Levi-Civita "Nozioni di parallelismo in una varieta qualunque e conseguente specificazione geometrica della curvatura riemanniana"}, emerges a posteriori from the differential equations, that express a way of varying a parallel direction on a pre-assigned path.\endnote{
	Original It. version: «[I]l concetto di parallelismo fra direzioni, entro una varietà $V_n$, a metrica qualsiasi, introdotto con felice idea dal Collega [Levi-Civita], poteva presentarsi sotto una forma geometrica affatto indipendente dallo spazio euclideo [ambiente] $S_N$ in cui la $V_n$ è immersa, restando così a priori manifesto il carattere intrinseco di quel concetto, rispetto alla data varietà; cosa che al Levi-Civita risulta a posteriori dalle equazioni differenziali, che esprimon il modo di variare di una direzione parallela ad una data, lungo un assegnato cammino».
	} \\
\indent — \textsc{F. Severi} \cite[p. 227]{Severi "Sulla curvatura delle superficie e varieta"} 

\endgroup

\vspace{2mm}

\enumerationisinitium
\item Into the context of Levi-Civita's theory, a manifold can be viewed as \emph{embedded into a Euclidean space}, or as a \emph{submanifold of an affine flat space}. However, the behavior that can be detected by the parallelism system has an \emph{intrinsic} character; because it is dependent only on the metric manifold, and not on the auxiliary ambient Euclidean space as well \cite[pp. 174, 177]{Levi-Civita "Nozioni di parallelismo in una varieta qualunque e conseguente specificazione geometrica della curvatura riemanniana"}. But such a character is (for Levi-Civita) \emph{a posteriori result of differential equations}.
\item F. Severi's \cite[pp. 227, 254-256]{Severi "Sulla curvatura delle superficie e varieta"} was the first to speculate on the possibility of describing the intrinsic form about the parallel transport through \emph{a priori defined geometric procedures}, and to give a pure model of transport on a geodesic surface (see Section \ref{section "Geodesics, Straight Paths, and Euler–Lagrange Equations"}), i.e. an example of \emph{non-ambient parallelism}, in the sense that there is no embedding space, such as the $n$-dimensional (ambient) Euclidean field. This is what goes by the name of \emph{Severi's theorem}, see \cite[pp. 125-126]{Levi-Civita "Questioni di meccanica classica e relativistica"} \cite[pp. 194-195]{Levi-Civita "Lezioni di calcolo differenziale assoluto"}. 
\item See also the work of Bompiani \cite{Bompiani "Studi sugli spazi curvi: Del parallelismo di una varieta qualunque"}, which takes up  the approach of Severi and finds a new \emph{scalar invariant} (in addition to the one from the Riemann curvature) in parallel transport; hence he introduces an \emph{invariant vector} for summarizing the scalar properties with geometric evidence.
\item 
\label{item "Historical jotting"}
The idea of parallel transport of a vector along a positive and negative curvature, in non-Euclidean spaces, is outlined by L.E.J. Brouwer \cite[p. 133]{Brouwer "The force field of the non-Euclidean spaces with negative curvature"} \cite{Brouwer "The force field of the non-Euclidean spaces with positive curvature"}. But only with the Levi-Civita's paper there is a more definite formulation. Alternative outcomes in accordance with these interpretations are due to J.A. Schouten \cite{Schouten "Die direkte Analysis zur neueren Relativiteitstheorie"} and H. Weyl \cite{Weyl "Gravitation und Elektrizitat"} \cite{Weyl "Reine Infinitesimalgeometrie"}.
\item Modern developments on the invariant theory aimed at preserving the parallelism, under the law of linear transport (i.e. on the geometry of manifolds with affine connection), are in En. Bortolotti \cite{Bortolotti En. "Sulla geometria delle varieta a connessione affine. Teoria invariantiva delle trasformazioni che conservano il parallelismo"}.
\enumerationisfinis

\subsection{Koszul Connection and Linear Connections, \& Covariant Derivative}
\label{subsection "Koszul Connection, Linear Connection, and Covariant Derivative"}

Let us start by considering the general meaning of a connection on the vector bundle; it can be understood as a technique of differentiation of vector fields on some manifold.
\begin{definitio}[Connection on a vector bundle]
\label{definitio "Connection on a vector bundle"}
Let $\pi \colon \mathring{\mathcal{E}} \to \mathcal{M}$ be a vector bundle (see Definition \ref{definitio "Vector bundle"}) over a manifold $\mathcal{M}$, and let $\mathfrak{E}(\mathcal{M})$ denote the vector space of sections of $\mathring{\mathcal{E}}$, and $\mathfrak{T}(\mathcal{M})$ the vector space of vector fields on $\mathcal{M}$. A connection $\nabla$ on $\mathring{\mathcal{E}}$ is a map $\nabla^{\mathring{\mathcal{E}}} \colon \mathfrak{T}(\mathcal{M}) \times \mathfrak{E}(\mathcal{M}) \to \mathfrak{E}(\mathcal{M})$ (cf. Proposition \ref{propositio "Connection 1-form"}), which may be written as $(\vec{X}, \sezione) \mapsto \nabla_{\vec{X}}\sezione$, if
\enumerationisinitium
\item $\nabla_{\vec{X}}\sezione$ is $\mathscr{C}^\infty(\mathcal{M})$ linear in $\vec{X}$, where $\nabla_{\vec{X}}\sezione$ is the \emph{covariant derivative} \cite{Ricci "Sui parametri e gli invarianti delle forme quadratiche differenziali"} \cite{Ricci "Sulla derivazione covariante ad una forma quadratica differenziale"} \cite{Ricci "Delle derivazioni covarianti e controvarianti e del loro uso nella Analise applicata"} \cite{Ricci et Levi-Civita "Methodes de calcul differentiel absolu et leurs applications"} of the section $\sezione$ (cf. Definitions \ref{definitio "Parallel vector field along a curve"} and \ref{definitio "Parallel transport map"}) in the direction of the vector field $\vec{X}$; by introducing two functions $f_1$ and $f_2$, thus one obtains 
\begin{equation}
	\nabla_{f_1\vec{X}_1 + f_2\vec{X}_2}\sezione = f_1\nabla_{\vec{X}_1}\sezione + f_2\nabla_{\vec{X}_2}\sezione, 
\end{equation} 
for $\vec{X}_1, \vec{X}_2 \in \mathfrak{T}(\mathcal{M})$, $\sezione \in \mathfrak{E}(\mathcal{M})$ and $f_1, f_2 \in \mathscr{C}^\infty(\mathcal{M})$;
\item $\nabla_{\vec{X}}\sezione$ is $\mathbb{R}$-linear in $\sezione$, thus 
\begin{equation}
	\nabla_{\vec{X}}(\alpha\sezione_1 + \beta\sezione_2) = \alpha\nabla_{\vec{X}}\sezione_1 + \beta\nabla_{\vec{X}}\sezione_2, 
\end{equation}
for $\vec{X} \in \mathfrak{T}(\mathcal{M})$, $\sezione_1, \sezione_2 \in \mathfrak{E}(\mathcal{M})$ and $\alpha, \beta \in \mathbb{R}$;
\item $\nabla$ satisfies the product rule 
\begin{equation}
	\nabla_{\vec{X}}(f\sezione) = f\nabla_{\vec{X}}\sezione + (\vec{X}f)\sezione, 
\end{equation}
for $\vec{X} \in \mathfrak{T}(\mathcal{M})$, $\sezione \in \mathfrak{E}(\mathcal{M})$ and $f \in \mathscr{C}^\infty(\mathcal{M})$. \definitiosymbol
\enumerationisfinis
\end{definitio}
This type of connection (on a vector bundle) is referred to as a \emph{Koszul connection} \cite{Koszul "Homologie et cohomologie des algebres de Lie"}.

\begin{definitio}[Linear connection]
\label{definitio "Linear connection"}
A connection $\nabla$ on the tangent bundle $\mathring{\mathcal{T}}\mathcal{M}$ (see Definition \ref{definitio "Tangent and cotangent bundles"}) or, simply, a connection on $\mathcal{M}$, exactly like the one in the Definition \ref{definitio "Connection on a vector bundle"}, is commonly known as \emph{linear connection}, and it is a map $\nabla^{\mathring{\mathcal{T}}} \colon \mathfrak{T}(\mathcal{M}) \times \mathfrak{T}(\mathcal{M}) \to \mathfrak{T}(\mathcal{M})$. This is also described as a \emph{affine connection}; but, in some cases, it still possible to  distinguish between the affine-like and the linear-like connection. \definitiosymbol
\end{definitio}

\begin{propositio}[Operator determined by the connection]
\label{propositio "Operator determined by the connection"}
Let $\nabla$ be a connection on the vector bundle $\pi \colon \mathring{\mathcal{E}} \to \mathcal{M}$, and let $\gamma_\mathrm{c} \colon I \subset \mathbb{R} \to \mathcal{M}$ be a $\mathscr{C}^\infty$ curve in $\mathcal{M}$, setting 
\begin{equation}
	I = [\alpha, \beta] = \{x \in \mathbb{R} \mid \alpha \leqslant x \leqslant \beta\}.
\end{equation}
Let us also denote by $\mathfrak{E}(\gamma_\mathrm{c})$, or by $\mathfrak{T}(\gamma_\mathrm{c})$ (if $\mathring{\mathcal{E}} = \mathring{\mathcal{T}}\mathcal{M}$), the vector space of sections of $\mathring{\mathcal{E}}$. Then there exists an operator $D \colon \mathfrak{E}(\gamma_\mathrm{c}) \to \mathfrak{E}(\gamma_\mathrm{c})$, or $D \colon \mathfrak{T}(\gamma_\mathrm{c}) \to \mathfrak{T}(\gamma_\mathrm{c})$, satisfying these items:
\enumerationisinitium
\item $D$ is $\mathbb{R}$-linear, for which $D\left(\alpha\sezione_1 + \beta\sezione_2\right) = \alpha D\sezione_1 + \beta D\sezione_2$;
\item $D(f\sezione) = \dot{f}\sezione + fD\sezione$, for $f \in \mathscr{C}^\infty(I)$;
\item if $\sezione \in \mathfrak{E}(\gamma_\mathrm{c})$, or $\sezione \in \mathfrak{T}(\gamma_\mathrm{c})$, is extendible, and if $\tilde{\sezione}$ represents an extension of $\sezione$, it follows that $D_t\sezione(t) = \nabla_{\dot{\gamma}_\mathrm{c}(t)}\tilde{\sezione}$.
\enumerationisfinis
$D_t\sezione$ is called the \emph{covariant derivative} of $\sezione$ along $\gamma_\mathrm{c}$.
\end{propositio}

Recall that a section $\sezione$ of the tangent bundle $\mathring{\mathcal{T}}\mathcal{M}$ of $\mathcal{M}$ is a vector field on $\mathcal{M}$, namely a vector field $\sezione$ on a manifold $\mathcal{M}$ is the assignment of a tangent vector $\sezione_p \in \mathcal{T}_p\mathcal{M}$ to every point $p \in \mathcal{M}$.

\begin{definitio}
\label{definitio "Parallel vector field along a curve"}
Given a manifold $\mathcal{M}$ and a connection $\nabla$ in $\mathring{\mathcal{E}} \xrightarrow{\pi} \mathcal{M}$, a section $\sezione$ along a curve $\gamma_\mathrm{c} \colon I \to \mathcal{M}$ is \emph{parallel} along $\gamma_\mathrm{c}$ with respect to $\nabla$ if $D_t\sezione(t) = 0$, i.e. $\nabla_{\partial_t}\sezione(t) = 0$. \definitiosymbol
\end{definitio}

\begin{definitio}[Parallel transport map]
\label{definitio "Parallel transport map"}
Let $\nabla$ be a connection on the vector bundle $\pi \colon \mathring{\mathcal{E}} \to \mathcal{M}$, and let $\gamma_\mathrm{c} \colon [0, 1] \to \mathcal{M}$ be a smooth curve in $\mathcal{M}$, with $\gamma_\mathrm{c}(0) = p_\alpha$ and $\gamma_\mathrm{c}(1) = p_\beta$, where $p_\alpha, p_\beta \in \mathcal{M}$. Then
\enumerationisinitium
\item for each vector $w_{\mathring{\mathcal{E}}} \viz w \in \mathring{\mathcal{E}}_{p_\alpha}$, there is a unique parallel section $\sezione \in \mathfrak{E}(\gamma_\mathrm{c})$ along $\gamma_\mathrm{c}$ such that $\sezione(0) = w_{\mathring{\mathcal{E}}}$, and $\sezione$ is called a \emph{parallel extension}, or \emph{parallel translate}, of $w_{\mathring{\mathcal{E}}}$ along $\gamma_\mathrm{c}$;
\item the \emph{parallel transport} along $\gamma_\mathrm{c}$ with respect to $\nabla_{\dot{\gamma}_\mathrm{c}(t)}\sezione(t) = 0$, for all $t \in [0, 1]$, is the linear map $\tilde{\gamma}_\mathrm{c} \colon \mathring{\mathcal{E}}_{p_\alpha} \to \mathring{\mathcal{E}}_{p_\beta}$ defined by $\tilde{\gamma}_\mathrm{c}(w_{\mathring{\mathcal{E}}}) = \sezione(1)$. \definitiosymbol
\enumerationisfinis
\end{definitio}

This last point helps show that the parallel transport along $\gamma_\mathrm{c}$ is an isomorphism between $\mathring{\mathcal{E}}_{p_\alpha}$ and $\mathring{\mathcal{E}}_{p_\beta}$.

What we have seen so far is a way for constructing the parallel transport from a connection. We will present now the reverse procedure: the parallel transport can be used to get a connection as limit of an incremental ratio, according to a Knebelman's procedure \cite{Knebelman "Spaces of Relative Parallelism"}. 

\begin{propositio}[Connection as limit of an incremental ratio]
\label{propositio "Connection as limit of an incremental ratio"}
Given a connection $\nabla$ on the vector bundle $\pi \colon \mathring{\mathcal{E}} \to \mathcal{M}$, let $\gamma_\mathrm{c} \colon I \to \mathcal{M}$ be a (smooth) curve in $\mathcal{M}$, and let $t_0 \in I$. Then
\begin{align}
	& D_{t_0}\sezione \in \mathfrak{E}(\gamma_\mathrm{c}) = \frac{d}{dt}(\tilde{\gamma}_\mathrm{c})^{- 1}_t\sezione(t)\bigg|_{t = t_0}, \\
	& \nabla_{w_{\mathring{\mathcal{E}}}}\sezione \in \mathfrak{E}(\mathcal{M}) = \frac{d}{dt}(\tilde{\gamma}_\mathrm{c})^{-1}_t\sezione\bigl(\gamma_\mathrm{c}(t)\bigr)\bigg|_{t = t_0}.
\end{align}
Here the parallel transport is $\mathring{\mathcal{E}}_{\gamma_\mathrm{c}(t_0)} \xrightarrow{(\tilde{\gamma}_\mathrm{c})_t} \mathring{\mathcal{E}}_{\gamma_\mathrm{c}(t)}$, and $D$ denotes again the covariant derivative operator along $\gamma_\mathrm{c}$. \end{propositio}

\subsection{Levi-Civita Connection Theorem on a (pseudo-)Riemannian Manifold}
\label{subsection "Levi-Civita Connection Theorem on a (pseudo-)Riemannian Manifold"}

\begin{definitio}[Torsion tensor]
\label{definitio "Torsion tensor"}
Let $\tau \colon \mathfrak{T}(\mathcal{M}) \times \mathfrak{T}(\mathcal{M}) \to \mathfrak{T}(\mathcal{M})$ be a $\binom{1}{2}$-tensor field or a tensor of type $(1, 2)$, called \emph{torsion tensor} of the connection $\nabla$ on a (pseudo-)Riemannian manifold. The torsion tensor can be written as $\tau \in \mathfrak{T}^1_2(\mathcal{M})$, and it is defined by $\tau(\vec{X}, \vec{Y}) = \nabla_{\vec{X}}\vec{Y} - \nabla_{\vec{Y}}\vec{X} - [\vec{X}, \vec{Y}]$. \definitiosymbol
\end{definitio}

\begin{definitio}[Torsion free connection]
\label{definitio "Torsion free connection"}
If $\tau \in \mathfrak{T}^\nabla(\mathcal{M}) = 0$, one says the connection $\nabla$ is \emph{torsion free} and hence \emph{symmetric}. Put another way, the symmetry of the connection depends on the condition in which $\tau$ vanishes identically (and $\nabla$ is torsion free). \definitiosymbol
\end{definitio}

\begin{definitio}[Metric-compatible connection]
\label{definitio "Metric-compatible connection"}
For a given (pseudo-)Riemannian manifold, the connection $\nabla$ is compatible with the metric if $\nabla_{\vec{X}}\langle\vec{Y}, \vec{Z}\rangle = \langle\nabla_{\vec{X}}\vec{Y}, \vec{Z}\rangle + \langle\vec{Y}, \nabla_{\vec{X}}\vec{Z}\rangle$. \definitiosymbol
\end{definitio}

The connection $\nabla$, or more accurately $\nabla^{\mathring{\mathcal{T}}}$, is called a \emph{Levi-Civita connection} if both of the properties above can be verified, therefore the Definitions \ref{definitio "Torsion free connection"} and \ref{definitio "Metric-compatible connection"} holds.
	
The Riemannian metric tensor $g$ is \emph{covariantly constant} with respect to the Levi-Civita connection; and then the covariant derivative of the metric affine connection is zero:
\begin{align}
	g_{\mu\nu;\xi} & = \frac{\partial g_{\mu\nu}}{\partial x^\xi} - {\Gamma^\varrho}_{\mu\xi}g_{\varrho\nu} - {\Gamma^\varrho}_{\nu\xi}g_{\mu\varrho} \notag \\
	& = \frac{\partial g_{\mu\nu}}{\partial x^\xi} - g_{\varrho\nu}g^{\varrho\varpi}\Gamma_{\varpi;\mu\xi} - g_{\mu\varrho}g^{\varrho\varpi}\Gamma_{\varpi;\nu\xi} \notag \\
	& = 0,
\end{align}
where ${\Gamma^\xi}_{\mu\nu}$ are the Christoffel symbols, see Eq. \eqref{subequations "Christoffel symbols for the Levi-Civita connection"} and Section \ref{subsection "Fermi (Locally Geodesic Cartesian-like) Coordinates"}; the semicolon, with an index after it, indicates \emph{covariant differentiation}.

\begin{theorema}[Levi-Civita]
\label{theorema "Levi-Civita"}
If $(\mathcal{M}, g)$ is a (pseudo-)Riemannian manifold, then there exists a unique affine connection $\nabla$ for $\mathcal{M}$ which is symmetric and compatible with the metric $g$ such that 
\enumerationisinitium
\item $\nabla g = 0$ (the torsion of $\nabla$ vanishes identically),
\item $\langle\nabla_{\vec{X}}\vec{Y}, \vec{Z}\rangle = \frac{1}{2}\bigl\{\vec{X}\langle\vec{Y}, \vec{Z}\rangle + \vec{Y}\langle\vec{Z}, \vec{X}\rangle - \vec{Z}\langle\vec{X}, \vec{Y}\rangle + \langle[\vec{X}, \vec{Y}], \vec{Z}\rangle \\ - \langle[\vec{Y}, \vec{Z}], \vec{X}\rangle + \langle[\vec{Z}, \vec{X}], \vec{Y}\rangle\bigr\}$, for any vector fields $\vec{X}, \vec{Y}, \vec{Z} \in \mathfrak{T}(\mathcal{M})$, where $[\cdot \:, \cdot]$ is the Lie bracket (of vector fields) and $\mathfrak{T}(\mathcal{M})$ is the vector space of vector fields on $\mathcal{M}$.
\enumerationisfinis
\end{theorema}

\begin{proof}	
Supposing that $\nabla$ exists, then one has
\begin{subequations}
\begin{align}
	& \vec{X}\langle\vec{Y}, \vec{Z}\rangle = \langle\nabla_{\vec{X}}\vec{Y}, \vec{Z}\rangle + \langle\vec{Y}, \nabla_{\vec{X}}\vec{Z}\rangle, \\
	& \vec{Y}\langle \vec{Z}, \vec{X}\rangle = \langle\nabla_{\vec{Y}}\vec{Z}, \vec{X}\rangle + \langle\vec{Z}, \nabla_{\vec{Y}}\vec{X}\rangle, \\
	& \vec{Z}\langle \vec{X}, \vec{Y}\rangle = \langle\nabla_{\vec{Z}}\vec{X}, \vec{Y}\rangle + \langle\vec{X}, \nabla_{\vec{Z}}\vec{Y}\rangle;
\end{align}
\end{subequations}
from the symmetry of $\nabla$ it follows that
\begin{subequations}
\begin{align}
	& \vec{X}\langle\vec{Y}, \vec{Z}\rangle + \vec{Y}\langle\vec{Z}, \vec{X}\rangle - \vec{Z}\langle\vec{X}, \vec{Y}\rangle \\
	& = \langle\nabla_{\vec{X}}\vec{Z} - \nabla_{\vec{Z}}\vec{X}, \vec{Y}\rangle + \langle\nabla_{\vec{Y}} \vec{Z} - \nabla_{\vec{Z}}\vec{Y}, \vec{X}\rangle + \langle\nabla_{\vec{X}}\vec{Y} + \nabla_{\vec{Y}} \vec{X}, \vec{Z}\rangle \\
	& = - \langle[\vec{Z}, \vec{X}], \vec{Y}\rangle + \langle[\vec{Y}, \vec{Z}], \vec{X}\rangle - \langle[\vec{X}, \vec{Y}], \vec{Z}\rangle + 2\langle\nabla_{\vec{X}}\vec{Y}, \vec{Z}\rangle,
\end{align}
\end{subequations}
and this proves that the Theorem \ref{theorema "Levi-Civita"} is true.
\end{proof}

\subsection{Fermi (Locally Geodesic Cartesian-like) Coordinates}
\label{subsection "Fermi (Locally Geodesic Cartesian-like) Coordinates"}

\begingroup
\footnotesize
Per fare lo studio dei fenomeni che avvengono in vicinanza di una linea oraria, cioè, in linguaggio non relativistico, in una porzione di spazio, variabile eventualmente col tempo, ma sempre molto piccola in confronto alle divergenze dall'euclideità, della varietà spazio-tempo, converrà anzitutto ricercare un opportuno riferimento tale che, in vicinanza della linea studiata, il $ds^2$ della varietà prenda una forma semplice.\footnote{
	«To study phenomena that occur near a  world line, that is, in non-relativistic language, in a portion of space, eventually variable over time, but still very small compared to the divergences from the Euclidity, of the space-time manifold, first of all it will be convenient to look for a suitable reference such that, near the line under consideration, the $ds^2$ [the square of the line element, or the interval between two points] of the manifold takes a simple form».
	} \\
\indent — \textsc{E. Fermi} \cite[p. 21]{Fermi "Sopra i fenomeni che avvengono in vicinanza di una linea oraria"}
 
\endgroup

\vspace{2mm}

The condition that all the Christoffel symbols vanish at \emph{every point} of a curve, 
\begin{equation}
	{\Gamma^\xi}_{(\mu_j, \mathellipsis, \mu_1)\lambda}\big|_{\gamma_\mathrm{c}} = 0, 
\end{equation}
with  $\xi = 1, \mathellipsis, n$, and $\lambda, \mu_1, \mathellipsis, \mu_{j + 1} = 2, \mathellipsis, n$, 
\begin{equation}
	\partial_{\mu_{j + 1}}{\Gamma^\xi}_{(\mu_j, \mathellipsis, \mu_1)\lambda}\big|_{\gamma_\mathrm{c}} = 0, 
\end{equation}
is guaranteed by a theorem of E. Fermi \cite{Fermi "Sopra i fenomeni che avvengono in vicinanza di una linea oraria"} for a symmetric connection based on the so-called \emph{Fermi coordinates}, a reference frame that Levi-Civita \cite{Levi-Civita "Sur l'ecart geodesique"} exploits to the full. As a result, we can conveniently choose some coordinates which are \emph{locally geodesic}, so they basically behave \emph{like Cartesians} in the \emph{immediate vicinity} of a given point (whatever it is), and this applies to all points of the curve. In this way, the derivatives of the coefficients of the distance, i.e. the $ds^2$, vanish for a limited region of space; cf. \cite[§ 11, footnote on pp.  190-191]{Levi-Civita "Lezioni di calcolo differenziale assoluto"}.

\section{Connection Forms}
\label{section "Connection Forms"}

\begingroup
\footnotesize
[S]paces with a Euclidean connection allow of a \emph{curvature} and \emph{torsion}: in the spaces where parallelism is defined in the Levi-Civita way, the torsion is zero; in the spaces where parallelism is absolute (\emph{Fernparallelismus}) \cite[chap. XIII]{Weitzenbock "Invariantentheorie"} \cite{Vitali "Una derivazione covariante formata coll'ausilio di $n$ sistemi covarianti del primo ordine"}\footnote{
	The notion of \emph{absolute parallelism}, which is a system preserving the metric but with non-zero torsion, is considered in an accomplished manner by G. Vitali \cite{Vitali "Una derivazione covariante formata coll'ausilio di $n$ sistemi covarianti del primo ordine"} already in 1924; but it was not until 1929 that he communicates it to Einstein (letter from  Vitali to Einstein, 11 February 1929). The first and independently use of this notion by the German physicist is in \cite{Einstein "Riemann-Geometrie mit Aufrechterhaltung des Begriffes des Fernparallelismus"} \cite{Einstein "Neue Moglichkeit fur eine einheitliche Feldtheorie von Gravitation und Elektrizitat"} during an attempt to unify gravity with electromagnetism.
	}
\cite{Cartan and Schouten "On the Geometry of the of simple and groups", Cartan and Schouten "On Riemannian Geometries admitting an absolute parallelism"} the curvature is zero [flat metric], thus these are spaces without curvature and with torsion. \\
\indent — \textsc{É. Cartan} \cite[p. 7, letter to A. Einstein, originally in Fr., 8 May 1929]{Cartan and Einstein "Letters on absolute parallelism"}
 
\endgroup

\subsection{Connection 1-Form}
\label{subsection "Connection 1-Form"}

Section \ref{section "Parallel Transport of the Levi-Civita Connection"} provides us with the opportunity to better define a connection $\nabla$ on the tangent and the cotangent bundle.

\begin{propositio}[On the 1-form]
Given a local chart $\varphi_\mathcal{M} = (x^1, \mathellipsis, x^n)$ on the manifold $\mathcal{M}$ for a point $p \in \mathcal{M}$, let us define $\bigl\{\frac{\partial}{\partial x^1}|_p, \mathellipsis, \frac{\partial}{\partial x^n}|_p\bigr\}$ as the basis for the tangent space $\mathcal{T}_p\mathcal{M}$ and $\{dx^1|_p, \mathellipsis,dx^n|_p\}$ as the dual basis for the cotangent space $\mathcal{T}^*_p\mathcal{M}$. If we assume that $d\omega_\mathbb{R}$ represents the differential of a (differentiable) function in local coordinates, with $d\omega_\mathbb{R} \equival \{dx^1|_p, \mathellipsis,dx^n|_p\}$, then $d\omega_\mathbb{R}$ gives a local frame for the cotangent bundle or, equivalently, $d\omega_\mathbb{R}$ is a 1-form i.e. a smooth section of $\mathring{\mathcal{T}}^*\mathcal{M}$, such that $d\omega_\mathbb{R} \colon \mathfrak{T}(\mathcal{M}) \to \mathscr{C}^\infty(\mathcal{M})$ is a $\mathscr{C}^\infty(\mathcal{M})$ linear function.
\end{propositio}

The choice of local coordinates supplies a basis of the tangent space, that is why we are looking at an early expression of local reference frame for a vector bundle.

The function $\omega_\mathbb{R}$ is a differential $k$-form (with $0 \leqslant k \leqslant n$); for more on the issues involved, see the seminal work of É. Cartan \cite{Cartan "Sur certaines expressions differentielles et le probleme de Pfaff"}. The locally coordinate frames for $\omega_\mathbb{R}$ are
\begin{equation}
	\omega_\mathbb{R} = \sum_{1 \leqslant i_1 < \mathellipsis < i_k \leqslant n}\varphi_{i_1,\ldots ,i_k}dx^{i_1} \wedge \mathellipsis \wedge dx^{i_k},
\end{equation}
where $\varphi_{i_1, \mathellipsis, i_k} \colon \varphi \to \mathbb{R}^n$ is a continuously differentiable function and the symbol $\wedge$ indicates the exterior (or wedge) product.

The following assertions shall be used to determine the understanding of \emph{differential form} of degree $k$, with $k = \mathbb{Z}$, also called \emph{differential $k$-form} or just \emph{$k$-form}.

\begin{definitio}[Differential form]
\label{definitio "Differential form"}
~\enumerationisinitium
\item A differential form $\omega$ is a section of an algebra over a field $\mathbb{R}$ via the exterior product of the cotangent bundle $\mathring{\mathcal{T}}^*\mathcal{M}$ of a manifold $\mathcal{M}$. Therefore $\omega_\mathbb{R} \colon \mathcal{M} \to \bigwedge^k\mathring{\mathcal{T}}^*\mathcal{M}$. 
\item Let $\mathring{\mathcal{E}} \xrightarrow{\pi} \mathcal{M}$ or $\mathring{\zeta} = (\mathring{\mathcal{E}}, \pi, \mathcal{M})$ be a vector bundle. A differential form $\omega$ with values in $\mathring{\mathcal{E}} \equival \mathring{\zeta}$ is a section of the bundle $\bigwedge^k\mathring{\mathcal{T}}^*\mathcal{M} \otimes \mathring{\mathcal{E}}$. The set of all differential forms on a manifold $\mathcal{M}$ with values in $\mathring{\mathcal{E}} \equival \mathring{\zeta}$ is a vector space, and it will be denoted by $\Omega^k(\mathcal{M}; \mathring{\mathcal{E}})$ or $\mathfrak{E}^k(\mathcal{M})$.
\item The space of differential forms $\omega$ with values in $\mathring{\mathcal{E}} \equival \mathring{\zeta}$ is accordingly computed by the isomorphism of $\Omega^k$ with
\begin{equation}
\Omega^k(\mathcal{M}; \mathring{\mathcal{E}}) \cong \Gamma_{\sezione}\left(\bigwedge^k\mathring{\mathcal{T}}^*\mathcal{M} \otimes \mathring{\mathcal{E}}\right),
\end{equation}
which is the space of sections. \definitiosymbol
\enumerationisfinis
\end{definitio}

We consider for example the case where $\omega \in \Omega^k(\mathcal{M})$ is a (global) differential form and $\sezione \in \mathfrak{E}(\mathcal{M})$ is a (global) section of $\mathring{\mathcal{E}}$, with a section $\omega \otimes \sezione$ of $\bigwedge^k\mathring{\mathcal{T}}^*\mathcal{M} \otimes \mathring{\mathcal{E}}$; or we can give the example of the restricted case in which $\{\sigma_1, \mathellipsis, \sigma_r\}$ is a local frame for $\mathring{\mathcal{E}}$ on a neighborhood $\Upsilon \subset \mathcal{M}$, and $\omega_1, \mathellipsis, \omega_r \in \Omega^r(\Upsilon; \mathring{\mathcal{E}}) \viz \mathfrak{E}^r(\Upsilon)$ are differential forms on $\Upsilon$,
\begin{equation}
	\omega|_\Upsilon = \sum^r_{\nu = 1}\omega^\nu \otimes \sigma_\nu.
\end{equation}

\begin{propositio}[Connection 1-form]
\label{propositio "Connection 1-form"}
Let $\nabla \colon \mathfrak{T}(\mathcal{M}) \times \mathfrak{E}(\mathcal{M}) \to \mathfrak{E}(\mathcal{M})$ be a connection on the vector bundle $\mathring{\zeta} = (\mathring{\mathcal{E}}, \pi, \mathcal{M})$ of rank $r$ (Definition \ref{definitio "Vector bundle"}), and let $\varphi_\nabla \colon \mathfrak{E}(\mathcal{M}) \to \Omega^1(\mathcal{M}; \mathring{\mathcal{E}}) \viz \mathfrak{E}^1(\mathcal{M})$ be a bundle map, which builds on $\varphi_\nabla\sezione(\vec{X}) = \nabla_{\vec{X}}\sezione$, where $\nabla_{\vec{X}}\sezione$ is linear over $\mathscr{C}^\infty(\mathcal{M})$ in $\vec{X}$ and it is also linear over $\mathbb{R}$ in $\vec{X}$, such that $\sezione \in \mathfrak{E}(\mathcal{M})$, with a map $\vec{X} \mapsto \omega(\vec{X})\sezione$. Choosing a local frame $\{\sigma_1, \mathellipsis, \sigma_r\}$ for $\mathring{\mathcal{E}}$ on an open neighborhood $\Upsilon \subset \mathcal{M}$, i.e. establishing a restriction $\mathring{\mathcal{E}}|_\Upsilon \xrightarrow{\pi_\Upsilon} \Upsilon$ so that $\mathring{\mathcal{E}}|_\Upsilon \cong \Upsilon \times \mathbb{R}$, and setting $k = \xi$ for reasons of consistency with the Eqq. \eqref{equation "Connection on the tangent bundle and Christoffel symbols"} and \eqref{subequations "Christoffel symbols for the Levi-Civita connection"}, we have
\begin{equation}
		\varphi_\nabla\sigma_\nu = \sum^r_{\xi = 1}{\omega^\xi}_\nu \otimes \sigma_\xi, \enspace \nabla_{\vec{X}}\sigma_\nu = \sum^r_{\xi = 1}{\omega^\xi}_\nu(\vec{X})\sigma_\xi,
\end{equation}
with a uniquely connection matrix 
\begin{equation}
	\omega_\nabla = \left[{\omega^\xi}_\nu\right]_{\substack{1 \leqslant \xi \\ \nu \leqslant r}}
\end{equation}
of differential 1-forms ${\omega^\xi}_\nu$ defined on $\Upsilon_\mathbb{R}$, known generically as \emph{connection 1-forms}. Putting the Christoffel symbols for the Levi-Civita connection, the result is
\begin{equation}
	{\omega^\xi}_\nu = \sum^r_{\mu = 1}{\Gamma^\xi}_{\mu\nu}dx^\mu.
\end{equation}
\end{propositio}
	
\begin{scholium}
The connection 1-forms of the Proposition \ref{propositio "Connection 1-form"} satisfy the conditions 
\begin{subequations}	
\begin{align}
	{\omega^\xi}_\nu & = -{\omega^\nu}_\xi, \\
	d\omega^\nu & = \omega^\xi \wedge {\omega^\nu}_\xi.
\end{align}
\end{subequations}
\scholiumsymbol
\end{scholium}

\subsection{Curvature 2-Form and Structural Equations}
\label{subsection "Curvature 2-Form and Structural Equations"}

We can pursue this line and give a definition related to the proposition above.
	
\begin{definitio}[Curvature 2-form in the Levi-Civita connection scenario]
\label{definitio "Curvature 2-form in the Levi-Civita connection scenario"}
Let $\sigma_{\mathring{\mathcal{E}}} = \{\sigma_1, \mathellipsis, \sigma_n\}$ be a local frame field (i.e. an orthonormal basis) for the tangent bundle $\mathring{\mathcal{T}}\mathcal{M}$ on the usual trivializing neighborhood $\Upsilon \subset \mathcal{M}$, and let $\chi_{\mathring{\mathcal{E}}} = \{\chi_1, \mathellipsis, \chi_n\}$ be a dual local frame field for the cotangent bundle $\mathring{\mathcal{T}}^*\mathcal{M}$. If $\nabla$ is the Levi-Civita connection on a (pseudo-)Riemannian manifold, and
\begin{equation}
	{\omega^\nu}_\mu = {\Gamma^\nu}_{\mu\xi}\omega^\xi
\end{equation}
are its connection 1-forms (with the Christoffel symbols), one calls ${\Omega^\nu}_\mu$ the \emph{curvature forms}, which are 2-forms on $\Upsilon$, and $\Omega_\nabla = \bigl[{\Omega^\nu}_\mu\bigr]$ the \emph{curvature matrix}. \definitiosymbol
\end{definitio}

The 2-forms ${\Omega^\nu}_\mu$ allow to provide a description of the Riemann curvature tensor, $\Riemann_{\vec{X}\vec{Y}}\sigma_\mu = {\Omega^\nu}_\mu(\vec{X}, \vec{Y})\sigma_\nu$, for all $\vec{X}, \vec{Y} \in \mathfrak{T}(\Upsilon)$ and all $\mu, \nu = 1, \mathellipsis, n$. The 2-forms ${\Omega^\nu}_\mu$ can be written as 
\begin{align}
	{\Omega^\nu}_\mu & = d{\omega^\nu}_\mu - {\omega^\xi}_\mu \wedge {\omega^\nu}_\xi \notag \\ 
	& = \frac{1}{2}{\Riemann^\nu}_{\mu\xi\varrho}\omega^\xi \wedge \omega^\varrho, 
\end{align}
which corresponds to the \emph{second structural equation of É. Cartan} \cite[p. 133]{Cartan "Riemannian Geometry in an Orthogonal Frame"}; therefore they verify the property $d{\omega^\nu}_\mu = {\omega^\xi}_\mu \wedge {\omega^\nu}_\xi + {\Omega^\nu}_\mu$.

\begin{scholium}
The 1-forms ${\omega^\nu}_\mu$ of the Definition \ref{definitio "Curvature 2-form in the Levi-Civita connection scenario"} take values in the set $\mathfrak{gl}_n(\mathbb{R})$, and they satisfy the conditions
\begin{align}
	& {\omega^\nu}_\mu = -{\omega^\mu}_\nu, \\
	& d\omega^\mu = \omega^\nu \wedge {\omega^\mu}_\nu, \\
	& d\chi^\mu = -\sum^n_{\nu = 1} {\omega^\mu}_\nu \wedge \chi^\nu.
\end{align}
\scholiumsymbol
\end{scholium}

It is now possible to make some additional remarks on what has been said in the previous definition(s). 
\enumerationisinitium
\item For a Levi-Civita connection $\nabla$, one says that $\nabla$ is compatible with $g$ iff the connection 1-forms ${\omega^\nu}_\mu$ relative to an orthonormal frame $\{\sigma_1, \mathellipsis, \sigma_n\}$ give $g_{\nu\xi}{\omega^\xi}_\mu + g_{\mu\xi}{\omega^\xi}_\nu = dg_{\mu\nu}$. The matrix $\omega_\nabla = \left[{\omega^\nu}_\mu\right]$ is skew symmetric.
\item Here the same applies as described in the Definition \ref{definitio "Curvature 2-form in the Levi-Civita connection scenario"}. If $\tau$ is the torsion (tensor) of $\nabla$ (cf. Definition \ref{definitio "Torsion tensor"}), we can determine a map $\tau^\nu \colon \mathfrak{T}(\mathcal{M}) \times \mathfrak{T}(\mathcal{M}) \to \mathscr{C}^\infty(\mathcal{M})$, with $\nu = 1, \mathellipsis, n$, by $\tau(\vec{X}, \vec{Y}) = \tau^\nu(\vec{X}, \vec{Y})\sigma_\nu$. The 2-forms $\{\tau^1, \mathellipsis, \tau^n\}$ are called \emph{torsion forms}, which could be used to prove the \emph{first structural equation of É. Cartan}:
\begin{equation}
	d\varphi^\nu = \varphi^\mu \wedge {\omega^\nu}_\mu + \tau^\nu.	
\end{equation}
\enumerationisfinis

\section{Cartan Structure: Generalized Space and Lie Algebra-valued Form}
\label{section "Cartan Structure: Generalized Space and Lie Algebra-valued Form"}

\begingroup
\footnotesize
\emph{È data una varietà e in questa un gruppo di trasformazioni; studiare le forme appartenenti alla varietà per quanto concerne quelle proprietà che non si alterano nelle trasformazioni del gruppo dato \textnormal{[\,\dots]}. Si sviluppi la teoria invariantiva relativa al gruppo medesimo}.\footnote{
	«\emph{Given a manifoldness and a group of transformations of the same; to investigate the configurations belonging to the manifoldness with regard to such properties as are not altered by the transformations of the group \textnormal{[\,\dots]}; to develop the theory of invariants relating to that group}» \cite[pp. 218-219]{Klein "A Comparative Review of Recent Researches in Geometry"}.
	} \\
\indent — \textsc{F. Klein} \cite[p. 311]{Klein "Considerazioni comparative intorno a ricerche geometriche recenti"}, transl. into It. by \textsc{G. Fano} from Klein's \emph{Erlangen program} \cite[p. 67]{Klein "Vergleichende Betrachtungen uber neuere geometrische Forschungen"}, originally presented in 1872

\vspace{2mm}

I realized that \emph{most of the ordinary differential equations, the integration of which can be found using traditional methods, remain invariant by certain classes of transformations \textnormal{[Schaaren von Transformationen]} that are easily determinable} [\,\dots]. In other words, I understood that the concept of differential invariant of finite continuous group, even though only \emph{implicitly} and \emph{in special form}, is present in every textbook about ordinary differential equations. \\
\indent — \textsc{M.S. Lie} \cite[pp. iv-v]{Lie "Theorie der transformationsgruppen Erste Abschnitt"}

\vspace{2mm}

[A] generalized space (espace généralisé) in the sense of Cartan is a space of tangent spaces such that two infinitely near tangent spaces are related by an infinitesimal transformation of a given Lie group. Such a structure is known as a connection. \\
\indent — \textsc{S.-S. Chern} and \textsc{C. Chevalley} \cite[p. 221]{Chern and C. Chevalley "Elie Cartan and his mathematical work"} 

\endgroup

\subsection{\emph{Espace Généralisé} of the Klein geometry: Cartan Geometry and Connection}
\label{subsection "Espace Généralisé of the Klein Geometry: Cartan Geometry and Connection"}

In view of the above (Section \ref{section "Connection Forms"}), the algebra of differential forms gives us a way to reflect on the Levi-Cita connection along with Cartan space, which is a generalization of the theory of Lie groups in line with the Felix Klein's Erlanger program \cite{Klein "Vergleichende Betrachtungen uber neuere geometrische Forschungen"}; for a synopsis, see \cite{Cartan "La theorie des groupes finis et continus et l'Analysis situs"}.

Firstly, we recall briefly a few properties associated with the Klein geometry. 
\enumerationisinitium
\item Let $G$ be a Lie group and $H$ a closed (sub)group of $G$. A Klein geometry consists of a pair $(G, H)$ in which   the (left) coset space $G/H$ is connected, and it is called \emph{homogeneous space}, or even \emph{Klein space}. 
\item The group $G$ acts transitively on $G/H$ by the left action. 
\item For every Klein pair $(G, H)$ there exists a pair of Lie algebras $(\mathfrak{g, h})$; $\mathfrak{h}$ is a (sub)algebra of $\mathfrak{g}$, and it is closed (see below).
\enumerationisfinis

This is where we can begin to define the Cartan geometry, which is properly called \emph{espace généralisé} of the Klein geometry.

\begin{definitio}[Cartan structure]
Suppose that $\mathring{\mathcal{P}}$ is a principal $H$-bundle over a smooth manifold $\mathcal{M}$, $\pi \colon \mathring{\mathcal{P}} \xrightarrow{H} \mathcal{M}$ (cf. Definition \ref{definitio "Principal G-bundle"}). Then 
\enumerationisinitium
\item $\mathfrak{C} = (\mathring{\mathcal{P}}, \omega_\mathfrak{g})$ is a \emph{Cartan geometry}, 
\item $\omega_\mathfrak{g}$ is a differential form on $\mathring{\mathcal{P}}$ representing the \emph{Cartan connection} \cite{Cartan "Sur les varietes a connexion affine et la theorie de la relativite generalisee (premiere partie)", Cartan "Sur les varietes a connexion affine et la theorie de la relativite generalisee (premiere partie) (Suite)", Cartan "Sur les varietes a connexion affine et la theorie de la relativite generalisee (deuxieme partie)"} = \cite[pp. 23-193]{Cartan "On Manifolds with an Affine Connection and the Theory of General Relativity"} \cite{Cartan "Sur les varietes a connexion projective"} \cite{Cartan "Les groupes d'holonomie des espaces generalises"}. \definitiosymbol
\enumerationisfinis
\end{definitio}

The principal bundle $\pi \colon \mathring{\mathcal{P}} \xrightarrow{H} \mathcal{M}$, with group $H$, is a generalization of the bundle $\pi \colon G \xrightarrow{H} G/H$, where the connected coset space $G/H = \mathcal{M}$ is the space of the Klein geometry; and the differential form $\omega_\mathfrak{g}$ is a generalization of the Maurer–Cartan form $\omega_G$ (see Definition \ref{definitio "alpha. Maurer–Cartan form"}).

\subsection[From $\mathfrak{g}$-Space Homogeneity to Blob-like Space]{From $\protect\pseudobold{\mathfrak{g}}$-Space Homogeneity to Blob-like Space}
\label{subsection "From g-Space Homogeneity to Blob-like Space"}

\enumerationisinitium
\item The Cartan space is defined as \emph{homogeneous} or \emph{non-homogeneous} geometry. In the first case, the Cartan space is modeled, at least locally, on the Klein geometry/on the coset space $G/H$. In the second case, it adds \emph{deformations} to the Klein structure: from this point of view, the Cartan space is equivalent to a Klein geometry deformed by some curvature; the criterion by which to visualize a $\mathfrak{C}$-deformation is \emph{analogous} to that used for introducing the Riemannian curvatures in Euclidean space.
\item The Cartan space is defined as \emph{flat} or \emph{non-flat} geometry. In the first case, the Cartan geometry is a generalization of the standard Euclidean one, and hence the curvature in the Cartan structure is zero; this means that the curvature vanishes at all points; in the second case, it is a generalization of the Riemannian one, and the fabric of Cartan space has blob-like or, to borrow from Clifford \cite[p. 158]{Clifford "On the Space-Theory of Matter"} = \cite[p. 21]{Clifford "Mathematical Papers"}, hill-like values in $\mathfrak{C}$-shape. \scholiumsymbol
\enumerationisfinis

\begin{exemplum}[Commutative-homogeneity of space]
\label{exemplum "Commutative-homogeneity of space}
Consider a homogeneous $\mathfrak{g}$-space. It is possible to represent this type of space with a commutative diagram. Let $\mathring{\zeta} = (\mathring{\mathcal{P}}_\mathfrak{g}, \pi, \mathcal{M})$ a triple, where $\mathring{\mathcal{P}}_\mathfrak{g}$ is a principal $\mathfrak{g}$-space, and $\pi \colon \mathring{\mathcal{P}}_\mathfrak{g} \to \mathcal{M}$ is a map; and let $\mathring{\mathcal{F}}_\mathfrak{h}$ be a principal $\mathfrak{h}$-space, i.e. a fiber over each point $p \in \mathcal{M}$, with Lie algebra $\mathfrak{h} \subset \mathfrak{g}$. The graphic form is
\[
\begin{tikzcd}[row sep=large, column sep=large]
	\mathfrak{h} \arrow{d} \arrow{rr}{\varphi}
    && \mathfrak{O}(\mathring{\mathcal{F}}_\mathfrak{h}) \arrow{d} \\
	\mathfrak{g} \arrow{rr}{\varphi|_{\mathring{\mathcal{F}}_\mathfrak{h}}}
	&& \mathfrak{O}(\mathring{\mathcal{P}}_\mathfrak{g})
\end{tikzcd}
\]
with a homomorphism $\mathfrak{h} \xrightarrow{\varphi} \mathfrak{O}(\mathring{\mathcal{F}}_\mathfrak{h})$; the symbol $\varphi|_{\mathring{\mathcal{F}}_\mathfrak{h}}$ denotes the restriction of $\varphi \colon \mathfrak{g} \to \mathfrak{O}(\mathring{\mathcal{P}}_\mathfrak{g})$ to $\mathring{\mathcal{F}}_\mathfrak{h}$ (cf. Definition \ref{definitio "Vector bundle"}). \exemplumsymbol
\end{exemplum}

\subsection{Maurer–Cartan Forms and Equations}
\label{subsection "Maurer–Cartan Forms and Equations"}

\begin{definitio}[\textgreek{α}. Maurer–Cartan form]
\label{definitio "alpha. Maurer–Cartan form"}
We denote by $\mathfrak{g}$ the Lie algebra of the corresponding Lie group $G$, and let $\mathfrak{g}$ be a vector space together with a map $[\cdot \:, \cdot] \colon \mathfrak{g} \times \mathfrak{g} \to \mathfrak{g}$. The \emph{Maurer–Cartan form} \cite{Cartan "Sur la structure des groupes infinis de transformation"} $\omega_G$ is a $\mathfrak{g}$-valued 1-form on the group manifold $G$. \definitiosymbol
\end{definitio}

It is understandable that we are in the presence of a form in combination with the left invariant and right invariant actions of $G$ on itself; so the Maurer–Cartan form can be defined as a \textsc{lr}-map (in the sense that it is invariant under left and right translations in $G$). The left invariant Maurer–Cartan form is 
\[
	\omega_G \in \Omega^1 (G, \mathfrak{g}) \viz \omega^\textsc{l}_G \in \Omega^1 (G, \mathfrak{g}), \text{ or } \omega^\textsc{l}_G \in \Omega^1(G) \otimes \mathfrak{g}; 
\]
it can also be expressed by the map $\omega_G \colon g \mapsto \omega_G(g)$, or $\omega^\textsc{l}_G \colon g \mapsto \omega^\textsc{l}_G(g)$. The right invariant Maurer–Cartan form is 
\[
	\omega^\textsc{r}_G \in \Omega^1 (G, \mathfrak{g}), \text{ or } \omega^\textsc{r}_G \in \Omega^1(G) \otimes \mathfrak{g}, 
\]
and its map is $\omega^\textsc{r}_G \colon g \mapsto \omega^\textsc{r}_G(g)$.

In what follows we give more details about the left invariant case, corresponding to a trivialization of the tangent bundle of $G$ by \textsc{l}-translation; because the \textsc{l}-translation is what leads to generate a global parallel transport on a general Lie group, or a general manifold modelled on locally space.
\enumerationisinitium
\item The left Maurer–Cartan form can be written as $\omega_G(g) \colon \mathcal{T}_gG \to \mathfrak{g}$, identifying $\mathcal{T}_gG$ with the tangent space of $G$. 
\item It means that $\omega_G$, for each $g \in G$, is a linear map of tangent space at the identity $\idem$ with a Lie algebra-valued form; in fact, the tangent (i.e. vector) space $\mathfrak{g} = \mathcal{T}_\idem G$ is called the \emph{Lie algebra} of the  group $G$.
\item If we take a vector $v \in \mathcal{T}_gG$, so that $(\textsc{l}_h)_*v \in \mathcal{T}_{hg}G$, the Maurer–Cartan map above turns out to be described by 
\begin{equation}
	\omega_G[(\textsc{l}_{h})_*v] = (\textsc{l}_{h*}\omega_G)v = (\textsc{l}_{(hg)^{-1}})_*[(\textsc{l}_h)_*v] = (\textsc{l}_{g^{-1}})_*v = \omega_G(v).
\end{equation}
\enumerationisfinis

\begin{definitio}[\textgreek{α}. Maurer–Cartan equation]
\label{definitio "alpha. Maurer–Cartan equation"}
Let $\omega_1, \mathellipsis, \omega_n$ be the 1-forms on $G$, $\sigma_\mathfrak{g} = \{\sigma_1, \mathellipsis, \sigma_n\}$ a basis of $\mathfrak{g}$, and ${c^\mu}_{\nu\xi}$ the \emph{structure constants} of $\mathfrak{g}$ with respect to $\sigma_\mathfrak{g}$. By requiring that
\begin{equation}
	[\sigma_\nu, \sigma_\xi] = \sum^n_{\mu = 1} {c^\mu}_{\nu\xi}\sigma_\mu, \text{ for } 1 \leqslant \nu, \xi \leqslant n,
\end{equation}
is a left invariant bracket of Lie algebra-valued form, we define
\begin{equation}
\label{equation "Maurer–Cartan equation"}	
	d\omega_\mu = -\sum^n_{\nu < \xi} {c^\mu}_{\nu\xi}\omega_\nu \wedge \omega_\xi = -\frac{1}{2}\sum^n_{\nu, \xi = 1} {c^\mu}_{\nu\xi}\omega_\nu \wedge \omega_\xi.
\end{equation}
The Eq. \eqref{equation "Maurer–Cartan equation"} is the so-called \emph{Maurer–Cartan equation}, or even \emph{structural equation for the Maurer–Cartan form}, and it can be simplified by eliminating the upper and lower indices, $d\omega_G = -\frac{1}{2}[\omega_G, \omega_G]^\wedge$ (the Eq. above rests on the assumption that $\omega_G \equival \omega$). \definitiosymbol
\end{definitio}

\subsubsection{Ricci Rotation Coefficients and Tetrad Formalism}
\label{subsubsection "Ricci Rotation Coefficients and Tetrad Formalism"}

It is pertinent to note here the natural use of connection coefficients ${\gamma^\mu}_{\nu\xi}$, known as the \emph{Ricci rotation coefficients} \cite[p. 303]{Ricci "Dei sistemi di congruenze ortogonali in una varieta qualunque"}. For 
\begin{equation}
	{c^\mu}_{\nu\xi} = {\gamma^\mu}_{\nu\xi} - {\gamma^\mu}_{\xi\nu}, 
\end{equation}
we have 
\begin{equation}
	d\omega_\mu = \frac{1}{2}[{\gamma^\mu}_{\nu\xi} - {\gamma^\mu}_{\xi\nu}]\omega_\xi \wedge \omega_\nu.
\end{equation}

The Ricci rotation coefficients are widely used in combination with a set of four linearly independent vector fields, called (local) \emph{tetrad components}, to distinguish them from the coordinate components. That kind of approach, in combination with torsion coefficients, is a helpful tool for describing the gravitational field (i.e. the space-time vectors).

Let $\{\varepsilon_{(\alpha)}\}$ or $\{\varepsilon_{\hat{\alpha}}\}$ be the tetrad formalism of the vector fields, with $\alpha = 1, 2, 3, 4$, for the tetrad bases; the Ricci rotation coefficients are indicated by 
\begin{equation}
\label{equation "Connection coefficients in the tetrad basis"}
	\gamma_{\alpha\beta\lambda}
	\begin{cases}
	\gamma_{(\alpha)(\beta)(\lambda)} = {\varepsilon_{(\alpha)}}^\mu{\varepsilon_{(\lambda)}}^\nu\nabla_\nu\varepsilon_{(\beta)\mu}, \\
	\gamma_{\hat{\alpha}\hat{\beta}\hat{\lambda}} = {\varepsilon_{\hat{\alpha}}}^\mu{\varepsilon_{\hat{\lambda}}}^\nu\nabla_\nu\varepsilon_{\hat{\beta}\mu}.
	\end{cases}
\end{equation}
It is possible to adopt one or the other of these two notations; for instance, we can choose the one with the round brackets. From Eq. \eqref{equation "Connection coefficients in the tetrad basis"} follows an anti-symmetric property in the first pair of indices, 
\begin{equation}
\label{equation "Anti-symmetric property of Ricci rotation coefficients"}
	\gamma_{(\alpha)(\beta)(\lambda)} = -\gamma_{(\beta)(\alpha)(\lambda)}. 
\end{equation}
Thus there are $6 \times 4 = 24$ algebraically independent Ricci rotation coefficients (six combinations in the first two indices and four in the third index). Using the anti-symmetry \eqref{equation "Anti-symmetric property of Ricci rotation coefficients"} of $\gamma$, we see that 
\begin{align}
	\left[\varepsilon_{(\alpha)}, \varepsilon_{(\beta)}\right] = {c^{(\lambda)}}_{(\alpha)(\beta)}\varepsilon_{(\lambda)}, \\
	{c^{(\lambda)}}_{(\alpha)(\beta)} = {\gamma^{(\lambda)}}_{(\alpha)(\beta)} - {\gamma^{(\lambda)}}_{(\beta)(\alpha)},
\end{align}
for which 
\begin{equation}
\gamma_{(\alpha)(\beta)(\lambda)} = \frac{1}{2}\bigl(c_{(\alpha)(\beta)(\lambda)} + c_{(\beta)(\lambda)(\alpha)} - c_{(\lambda)(\alpha)(\beta)}\bigr).
\end{equation}

\begin{definitio}[\textgreek{β}. Maurer–Cartan form]
By taking the same set of  $\sigma$-elements of the Definition \ref{definitio "alpha. Maurer–Cartan equation"}, it is also possible to give an alternative description of the $\mathfrak{g}$-valued 1-form,
\begin{equation}
	\omega_G = \sum^n_\mu\omega_\mu \otimes \sigma_\mu,
\end{equation}
coinciding with the Maurer–Cartan form. \definitiosymbol
\end{definitio}

\begin{definitio}[\textgreek{β}. Maurer–Cartan equation]
Another way to write the Maurer–Cartan equation is
\begin{equation}
\label{equation "Cartan equation vanishing of the curvature I"}
	d\omega_\mu + \frac{1}{2} {c^\mu}_{\nu\xi}\omega_\nu \wedge \omega_\xi = 0,
\end{equation}
\begin{equation}
\label{equation "Cartan equation vanishing of the curvature II"}
	d\omega_G + \frac{1}{2}[\omega_G, \omega_G]^\wedge = 0,\footnote{
	Or $d\omega^\textsc{l}_G + \frac{1}{2}[\omega_G, \omega^\textsc{l}_G]^\wedge = 0$. Cf. \cite[p. 308]{Meinrenken "Clifford Algebras and Lie Theory"} \cite[p. 108]{Preston "Non-commuting Variations in Mathematics and Physics. A Survey"}.
	} \enspace d\omega^\textsc{r}_G + \frac{1}{2}[\omega_G, \omega^\textsc{r}_G]^\wedge = 0, \text{ with } \omega_G \in \mathfrak{g}^1.
\end{equation}
These equations are useful to remind us that such a form determines the vanishing of the curvature, and the geometry is flat. \definitiosymbol
\end{definitio}

The \emph{flatness} of the Cartan connection is associated with the Eqq. \eqref{equation "Cartan equation vanishing of the curvature I"} \eqref{equation "Cartan equation vanishing of the curvature II"}; meaning to say: the Cartan connection is flat when the components of the curvature tensor become zero \cite[p. 437]{Cartan "Geometry of Riemannian Spaces"}, $\kappa_\Omega = 0$, cf. Eq. \eqref{align "Cartan curvature"}, and the space is Euclidean.

\subsection{Connection with Lie Algebra Decomposition and Gauge Model}
\label{subsection "Connection with Lie Algebra Decomposition and Gauge Model"}

\begingroup
\footnotesize
It is the notion of parallelism that gives a Euclidean connection to the surface, to quote the words of H. Weyl [\,\dots]. In fact, what is essential in the idea of Levi-Civita \cite{Levi-Civita "Nozioni di parallelismo in una varieta qualunque e conseguente specificazione geometrica della curvatura riemanniana"} is that it allows to connect two small pieces of a manifold, which are infinitely close to each other, and it is this idea of \emph{connection} that is fruitful. We can therefore imagine, by developing this idea, the possibility of arriving at a general theory of manifolds with an \emph{affine}, \emph{conformal}, or \emph{projective} connection.\endnote{
	Original Fr. version: «[C]'est la notion de parallélisme qui doue la surface d'une \emph{connexion} euclidienne, pour employer un terme dû à M. H. Weyl [\,\dots]. En fait, ce qu'il y a d'essentiel dans l'idée de M. Levi-Civita, c'est qu'elle donne un moyen pour raccorder entre eux deux petits morceaux infiniment voisins d'une variété, et c'est cette idée de \emph{raccord} qui est féconde. On conçoit dès lors, en développant cette idée, la possibilité d'arriver à une théorie générale des variétés à connexion \emph{affine}, \emph{conforme}, \emph{projective}, etc.».
	} \\
\indent — \textsc{É. Cartan} \cite[pp. 205-206]{Cartan "Sur les varietes a connexion projective"}

\endgroup

\vspace{2mm}

We can direct our attention to the concept of \emph{connection of Cartan}, but not before we state some preparatory results for this purpose. 

\begin{definitio}
We indicate by $\mathfrak{h}$ the Lie (sub)algebra of $H$, for which $\mathfrak{h}$ is a (sub)vector space of vector space $\mathfrak{g}$. Once again, $\mathfrak{C} = (\mathring{\mathcal{P}}, \omega_\mathfrak{g})$ on $\mathcal{M}$ is the Cartan geometry with model pair ($\mathfrak{g, h}$). \definitiosymbol
\end{definitio}

This is why $\mathfrak{C}$ acts like the Lie algebra, by means of a decomposition $\mathfrak{g} = \mathfrak{h} \oplus \mathfrak{p}$ for the group of $H$-module algebra.

Since the Cartan connection $\omega_\mathfrak{g}$ is usually described as a $\mathfrak{g}$-valued 1-form on the principal bundle $\mathring{\mathcal{P}}$, it can consequently be decomposed as 
\begin{equation}
	\omega_\mathfrak{g} \in \Omega^1(\mathring{\mathcal{P}}, \mathfrak{g}) = \omega_\mathfrak{h} \oplus \omega_\mathfrak{p}, 
\end{equation}
in which $\omega_\mathfrak{h}$ is a $\mathfrak{h}$-valued 1-form on $\mathring{\mathcal{P}}$, called \emph{Ehresmann connection} \cite{Ehresmann "Les connexions infinitesimales dans un espace fibre differentiable"}, while $\omega_\mathfrak{p}$ is designated a \emph{solder form}.

\begin{scholium}[Ehresmann connection]
\label{scholium "Ehresmann connection"}
Introduce the symbols $\textsc{r}_h$ and $\adj$ to exhibit the right action of $h \in H$ on $\mathring{\mathcal{P}}$ and the adjoint action, respectively. Let us say that $\tilde{\eta}$ is the vector field for an infinitesimal action $\eta \mapsto \tilde{\eta}$ of the Cartan geometry $\mathfrak{C} = (\mathring{\mathcal{P}}, \omega_\mathfrak{g})$. The Ehresmann connection $\omega_\mathfrak{h}$ satisfies

\begin{equation} 
\label{equation "Ehresmann connection 1"} 
	(\textsc{r}_h)^*\omega_\mathfrak{h} = \adj(h^{-1})\omega_\mathfrak{h},
\end{equation} 

and

\begin{equation}
\label{equation "Ehresmann connection 2"}
	\omega_\mathfrak{h}(\tilde{\eta}) = \eta, \text{ for every } \eta \in \mathfrak{h}.
\end{equation}

If the geometry $\mathfrak{C}$ is a Riemannian-like structure, i.e. if we are considering a torsionless topological space, one talks about a Levi-Civita connection, therefore 
\[
	\omega_\mathfrak{h} \equival \nabla
\] 
(cf. Definition \ref{definitio "Cartan curvature"}), where $\omega_\mathfrak{h} \in \Omega^1(\mathring{\mathcal{P}}, \mathfrak{h})$. The Eqq. \eqref{equation "Ehresmann connection 1"} and \eqref{equation "Ehresmann connection 2"} occur also in the Levi-Civita connection. \scholiumsymbol
\end{scholium}

We now show the meaning of the Cartan connection, to find out that it is a connection form, which generalizes the Maurer–Cartan form.

\begin{propositio}[Cartan connection with decomposition]
\label{propositio "Cartan connection with decomposition"}
Let $\omega_\mathfrak{g} \in \Omega^1(\mathring{\mathcal{P}}, \mathfrak{g})$ be a $\mathfrak{g/h}$-Cartan connection on $\mathring{\mathcal{P}}$. The connection $\omega_\mathfrak{g}$ manifests itself in a Lie algebra $(\mathfrak{g, h})$ decomposition, from which we get three properties.
\enumerationisinitium
\item The restriction 
\begin{equation}
	\omega_\mathfrak{g}(p) \colon \mathcal{T}_p\mathring{\mathcal{P}} \xrightarrow{\mathfrak{h} \oplus \mathfrak{p}} \mathfrak{g}, 
\end{equation} 
for each point $p \in \mathring{\mathcal{P}}$, is a linear isomorphism. 
\item The equality 
\begin{equation}
	(\textsc{r}_h)^*\omega_\mathfrak{g} = \adj(h^{-1}) \circ \omega_\mathfrak{g}, 
\end{equation}
holds for all $h \in H$.
\item Let 
\begin{equation}
	\textgreek{\text{\ddigamma}}_{\vec{X}}(p) = \frac{dt}{d}\Big|_{t = 0}p\Bigl(\exp(t\textgreek{\text{\ddigamma}})\Bigr)
\end{equation}
be a fundamental vector field on $\mathcal{M}$; then $\omega_\mathfrak{g}\bigl(\textgreek{\text{\ddigamma}}_{\vec{X}}(p)\bigr) = \vec{X}$, with 
\begin{equation}
	\textgreek{\text{\ddigamma}}_{\vec{X}} \in \mathfrak{X}(\mathcal{M}) = \omega^{-1}_\mathfrak{g}(\vec{X}), 
\end{equation}
for all vector fields $\vec{X} \in \mathfrak{h}$. To lighten notation, one can just write 
\begin{equation}
	\omega_\mathfrak{g}({_\mathfrak{g}\tilde{\eta}}) = {_\mathfrak{g}\eta}
\end{equation}
(see Scholium \ref{scholium "Ehresmann connection"}).
\enumerationisfinis
\end{propositio}

\begin{scholium}
The fundamental vector field mapping for $\textsc{r}$-action is a Lie algebra homomorphism $\textgreek{\text{\ddigamma}} \colon \mathfrak{g} \to \mathfrak{X}(\mathcal{M})$. \scholiumsymbol
\end{scholium}

\begin{definitio}[Cartan curvature]
\label{definitio "Cartan curvature"}
Let 
\begin{align}
\label{align "Cartan curvature"}
	\kappa_\Omega \viz \Omega^2_\mathfrak{g} & = d\omega_\mathfrak{g} + \frac{1}{2}[\omega_\mathfrak{g}, \omega_\mathfrak{g}]^\wedge \notag \\
	& = d\omega_\mathfrak{g} + \omega_\mathfrak{g} \wedge \omega_\mathfrak{g},
\end{align}
be the \emph{curvature form} of the connection $\omega_\mathfrak{g}$, knowing that $\Omega^2_\mathfrak{g}$ is a $\mathfrak{g}$-valued 2-form on $\mathring{\mathcal{P}}$ (cf. Definition \ref{definitio "Curvature 2-form in the Levi-Civita connection scenario"}). \definitiosymbol
\end{definitio}

If $\Omega^2_\mathfrak{g}$ take values in the Lie algebra $\mathfrak{h}$, then the Cartan geometry is torsion free, and the $\kappa$-value lying in $\mathfrak{h}$ is precisely zero. When $\Omega^2_\mathfrak{g}$ is decomposable, it reflects a reductive geometry with an $H$-module decomposition: 
\begin{equation}
	\kappa_\Omega = \kappa_\mathfrak{h} \oplus \kappa_\mathfrak{p}.
\end{equation}

\begin{definitio}[Cartan gauge model]
Let $\Upsilon$ denote an open set of a manifold $\mathcal{M}$, and $\vartheta_\Upsilon$ a $\mathfrak{g}$-valued 1-form on $\Upsilon$ being part of the model above $(\mathfrak{g, h})$. The 1-form  
\begin{equation}
	\tilde{\vartheta}_\Upsilon \colon \mathcal{T}_\upsilon \xrightarrow{(\vartheta_\Upsilon)} \mathfrak{g}, 
\end{equation}
with canonical projection, viz. $\mathfrak{g} \xrightarrow{\pi} \mathfrak{g}/\mathfrak{h}$, is a linear isomorphism, for all $\upsilon \in \Upsilon$. The pair $(\Upsilon, \vartheta_\Upsilon)$ is an example of a \emph{Cartan gauge}. \definitiosymbol
\end{definitio}

\section{Geodesics, Straight Paths, and Euler–Lagrange Equations}
\label{section "Geodesics, Straight Paths, and Euler–Lagrange Equations"}

\begingroup
\footnotesize
[I]f we do not grant that the angles of incidence and reflection are equal, nature would be labouring in vain\footnote{
	The phrase «Nature does nothing in vain nor labours in vain (\textgreek{οὐδὲν μάτην ἐργάζεται ἡ φύσις οὐδὲ ματαιοπονεῖ})», which immediately precedes this portion of the epigraph, is a bad interpolation by Olympiodorus, and it certainly does not have a physical spice.
	} 
by following unequal angles, and instead of the eye apprehending the visible object by the shortest route [\textgreek{\textit{διὰ βραχείς περιόδου}}] it would do so by a longer.\endnote{
	Full original Gr. text: «\textgreek{ἐὰν μὴ δώσωμεν πρὸς ἴσας γωνίας γίνεσθαι τὴν ἀνάκλασιν, πρὸς ἀνίσους ματαιοπονεῖ ἡ φύσις, καὶ ἀντὶ τοῦ διὰ βραχείας περιόδου φθάσαι τὸ ὁρώμενον τὴν ὄψιν, διὰ μακρᾶς περιόδου τοῦτο φανήσεται καταλαμβάνουσα}».
	} \\
\indent — Theorem in \textsc{Heron}'s catoptrics \cite[pp. 496-499]{Various authors "Greek Mathematical Works II: from Aristarchus to Pappus"} according to Olympiodorus' commentary

\vspace{2mm}

Along the same geodesic, directions of the tangents are parallel, which generalizes an obvious feature of the straight line in Euclidean spaces.\endnote{
	Original It. version: «Lungo una medesima geodetica, le direzioni delle tangenti sono parallele, ciò che generalizza un'ovvia caratteristica della retta negli spazi euclidei».
	} \\
\indent — \textsc{T. Levi-Civita} \cite[p. 175]{Levi-Civita "Nozioni di parallelismo in una varieta qualunque e conseguente specificazione geometrica della curvatura riemanniana"}

\endgroup

\subsection{Geodesic: Some Features}
\label{subsection "Geodesic: Some Features"}

Bear in mind that 
\enumerationisinitium
\item the \emph{locally shortest (and the straightest) path between two points on a surface is a geodesic}, in a Beltrami-like manner \cite{Beltrami "Risoluzione del problema: riportare i punti di una superficie sopra un piano in modo che le linee geodetiche vengano rappresentate da linee rette"} \cite{Beltrami "Sulla teoria delle linee geodetiche"} (but be warned, see Scholium \ref{scholium "Geodesic: on the distance minimizing issue"}); 
\item a geodesic is a curve with zero acceleration (Definition \ref{definitio "Geodesic"}); parallel directions along a geodesic are always equally inclined with respect to the geodesic itself;
\item if the geodesics of a (Riemann) manifold are the same as those for the ambient space, then the manifold is called \emph{totally geodesic}, or \emph{Bompiani's space}; see E. Bompiani \cite{Bompiani "Spazi Riemanniani luoghi di varieta totalmente geodetiche"}, where he resumes and develops the insights of J. Hadamard \cite{Hadamard "Sur les elements lineaires a plusieurs dimensions"} and G. Ricci Curbastro \cite{Ricci "Formole fondamentali nella teoria generale delle varieta e della loro curvatura"}.
\enumerationisfinis

Let us cut to the chase, and get now to a more technical discussion. We can begin by saying what a geodesic (path, line or curve/curvature) is.

\begin{definitio}[Geodesic]
\label{definitio "Geodesic"}
Let 
\enumerationisinitium
\item $\nabla$ be a linear connection on a Riemannian-like manifold $\mathcal{M}$ (cf. Definition \ref{definitio "Connection on a vector bundle"}), 
\item $\gamma_\mathrm{c} \colon I \to \mathcal{M}$ be a smooth curve, or $\mathscr{C}^\infty$ curve, in the manifold $\mathcal{M}$, with an interval 
\begin{equation}
	I = [\alpha, \beta] = \{x \in \mathbb{R} \mid \alpha \leqslant x \leqslant \beta\}
\end{equation}
(see Proposition \ref{propositio "Operator determined by the connection"}),
\item $\dot{\gamma}_\mathrm{c} \colon I \to \mathring{\mathcal{T}}\mathcal{M}$ be a $\mathscr{C}^\infty$ curve contained in the total space of the tangent bundle—more precisely, we say that 
\begin{equation}
	\dot{\gamma}_\mathrm{c}(t) \in \mathcal{T}_{\gamma_\mathrm{c}(t)}\mathcal{M}
\end{equation}
at time $t \in I \subset \mathbb{R}$ is a \emph{velocity vector field} along $\gamma_\mathrm{c}$ (in fact $\pi_\mathcal{M} \circ \dot{\gamma}_\mathrm{c} = \gamma_\mathrm{c}$), where $\mathcal{T}_{\gamma_\mathrm{c}(t)}\mathcal{M}$ is the tangent space of the curve, and it is defined by 
\begin{equation}
	\dot{\gamma}_\mathrm{c}(t) = \frac{d}{dt}\gamma_\mathrm{c}(t),
\end{equation}
\item $D$ be a covariant differentiation operator, i.e. a covariant derivative.
\enumerationisfinis
Then
\enumerationisinitium
\item a curve $\gamma_\mathrm{c} \colon I \to \mathcal{M}$ with respect to $\nabla$ is a \emph{geodesic} (that can be arbitrarily parametrized) if the vector field 
\begin{equation}
	\ddot{\gamma}_\mathrm{c} \equival D_t\dot{\gamma}_\mathrm{c} \viz \nabla_{\dot{\gamma}_\mathrm{c}}\dot{\gamma}_\mathrm{c}, 
\end{equation} 
describing the tangential \emph{acceleration} of $\gamma_\mathrm{c}$, is equal to zero:
\begin{equation}
	\begin{rcases}
	\ddot{\gamma}_\mathrm{c}, \\
	D_t\dot{\gamma}_\mathrm{c}, \\
	\nabla_{\dot{\gamma}_\mathrm{c}}\dot{\gamma}_\mathrm{c}
	\end{rcases}
	= 0.
\end{equation}
In other words, $\gamma_\mathrm{c}$ is a geodesic iff the tangent vector $\dot{\gamma}_\mathrm{c} \colon I \to \mathring{\mathcal{T}}\mathcal{M}$ (see Definition \ref{definitio "Tangent and cotangent bundles"}) is \emph{parallel}, which means it is \emph{constant}, whereby the covariant derivative $D_t$ on sections of $\mathring{\mathcal{E}}$ along $\gamma_\mathrm{c}$ is zero;
\item a $\mathscr{C}^\infty$ curve whose speed is constant corresponds to a (regular) geodesic on a Riemannian-like manifold, ed it is plethorically called \emph{minimizing} or \emph{constant speed geodesic}; the (minimizing) geodesic has constant length of its velocity vector field; 
\item a geodesic curve (in the sense of a minimizing curve) with vanishing acceleration, or whose acceleration is identically zero, can be also expressed in working coordinates by the \emph{geodesic equation}
\begin{equation}
	\ddot{\gamma}_\mathrm{c}{}^\xi(t) + \sum_{\mu\nu}{\Gamma^\xi}_{\mu\nu}\dot{\gamma}_\mathrm{c}{}^\mu(t)\dot{\gamma}_\mathrm{c}{}^\nu(t) = 0,
\end{equation}
or, simply, 
\begin{equation}
	\ddot{\gamma}_\mathrm{c}{}^\xi + {\Gamma^\xi}_{\mu\nu}\dot{\gamma}_\mathrm{c}{}^\mu\dot{\gamma}_\mathrm{c}{}^\nu = 0, 
\end{equation}
where 
\begin{equation}
	\ddot{\gamma}_\mathrm{c}{}^\xi = - {\Gamma^\xi}_{\mu\nu}\dot{\gamma}_\mathrm{c}{}^\mu\dot{\gamma}_\mathrm{c}{}^\nu 
\end{equation}
is the (Euclidean) acceleration, technically it is the derivative of $\dot{\gamma}_\mathrm{c}(t)$, and 
\begin{equation}
	{\Gamma^\xi}_{\mu\nu} \viz \binomcurly{\xi}{\mu\nu} = \frac{1}{2}g^{\xi\varrho}(\partial_\mu g_{\nu\varrho} + \partial_\nu{g}_{\mu\varrho} - \partial_\varrho{g}_{\mu\nu})
\end{equation}
(cf. Section \ref{subsection "Matrix Notation and Kronecker delta-Function"}). Hence the acceleration of the curve is always \emph{orthogonal} (or normal) to the manifold, 
\begin{equation}
	\ddot{\gamma}_\mathrm{c}{}^\perp \equival (D_t\dot{\gamma}_\mathrm{c})^\perp \in \mathcal{M}, 
\end{equation}
for $t \in I \subset \mathbb{R} = [\alpha, \beta]$. \definitiosymbol
\enumerationisfinis

\end{definitio}

\begin{scholium}[Geodesic: on the distance minimizing issue]
\label{scholium "Geodesic: on the distance minimizing issue"}
~\enumerationisinitium
\item Let
\begin{equation}
	\length(\gamma_\mathrm{c}) = \int^{\beta}_{\alpha}|\dot{\gamma}_\mathrm{c}(t)|dt = \int^{\beta}_{\alpha}\sqrt{g_{(\alpha, \beta)}(\dot{\gamma}_\mathrm{c}{}_{(t)}, \dot{\gamma}_\mathrm{c}{}_{(t)})}dt
\end{equation}
be the length of a geodesic $\gamma_\mathrm{c} \colon [\alpha, \beta] \to \mathcal{M}$. Then $\gamma_\mathrm{c}$ is demanded to be \emph{locally the shortest path} in the surface iff its length is equal to the distance between the points $\gamma_\mathrm{c}(\alpha) = p_\alpha$ and $\gamma_\mathrm{c}(\beta) = p_\beta$, and $\gamma_\mathrm{c}$ is called \emph{minimizing} if 
\begin{equation}
	\length(\gamma_\mathrm{c}) \leqslant \length(\tilde{\gamma}_\mathrm{c})
\end{equation}
(every geodesic of the Levi-Civita connection is locally length minimizing, in this respect). However, it is not necessarily true that a geodesic is the shortest path between \emph{any} two points; e.g. two points on the unit sphere $\mathbb{S}^2$ can be connected by a segment of a great circle, and such a segment (of the sphere of dimension 2) is a geodesic but it is \emph{not minimizing} between its end-points. So \emph{a distance minimizing path is (always) a geodesic, but not every geodesic is a distance minimizing}.
\item The geometry of the \emph{cones} is a further good example: it illustrates that there are several geodesics that connect two points, so \emph{not all geodesics are the shortest paths}. \scholiumsymbol
\enumerationisfinis
\end{scholium}

\subsection{Visual Stimuli: Geodesics on Melon- and Egg-shaped Surfaces}

\begin{figure}[h!]
\centering
	\begin{minipage}[b]{0.550\textwidth}
	\includegraphics[width = \textwidth]{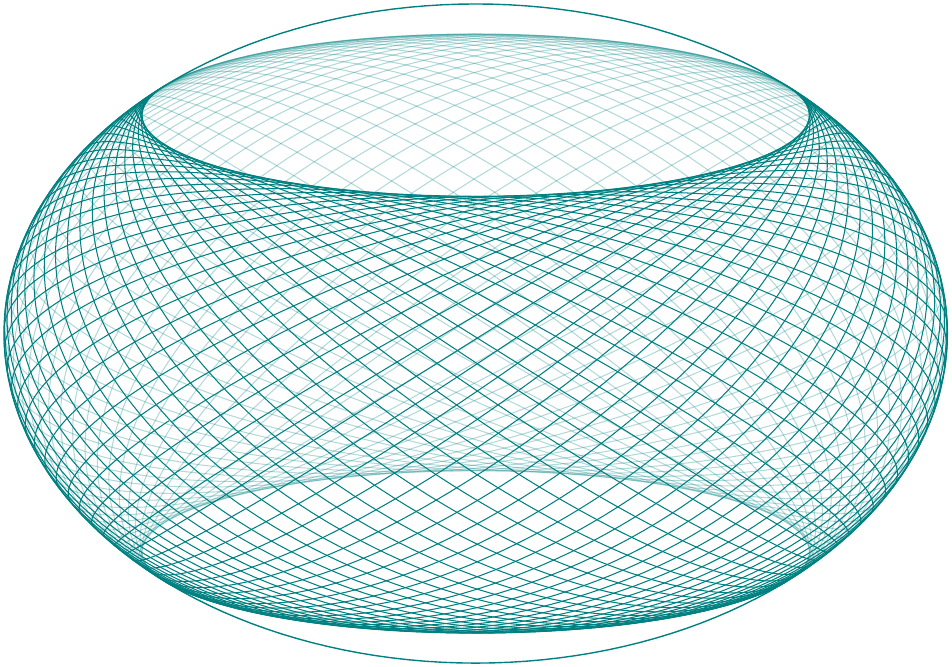}
	\caption{Geodesic in \textcolor{mallard}{\texttt{\#008080}}-tracks on an oblate spheroid, or a canary melon-like 2-space}
	\label{figure "Geodesics on melon-shaped surface"} 
	\end{minipage}
	\hspace{30pt}\par\vspace{3mm}
	\begin{minipage}[b]{0.425\textwidth}
	\includegraphics[width = \textwidth]{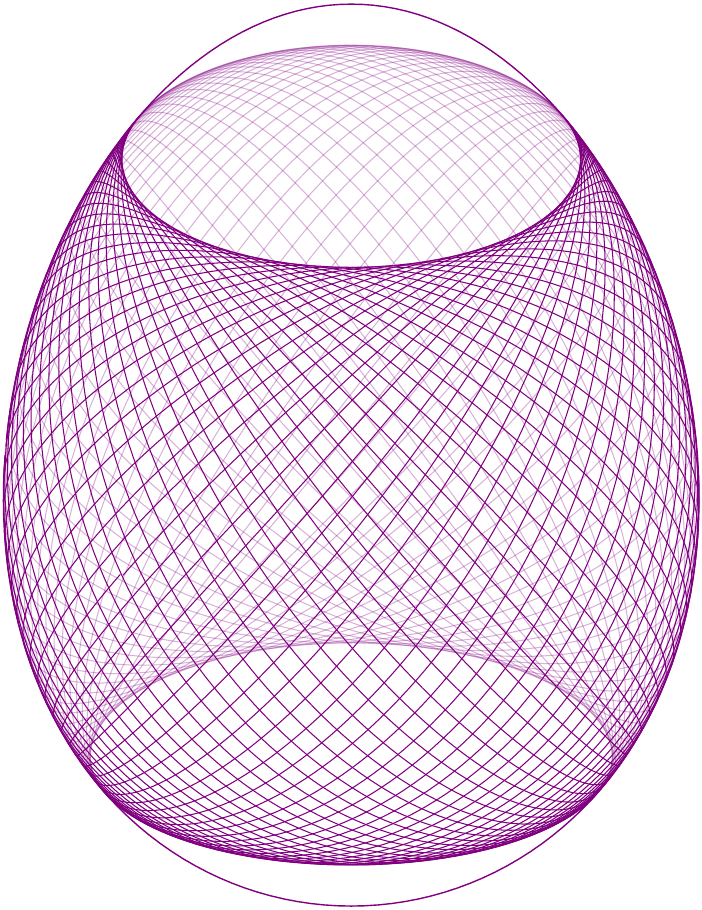}
	\caption{Geodesic in \textcolor{eggplant}{\texttt{\#800080}}-tracks on a quasi-prolate spheroid, or an egg-like 2-space, resembling an ovoidal surface}
	\label{figure "Geodesics on egg-shaped surface"}
	\end{minipage}
\end{figure}

We give some visual suggestions. In Figg. \ref{figure "Geodesics on melon-shaped surface"} and \ref{figure "Geodesics on egg-shaped surface"} geodesics on a canary melon-shaped surface and on an egg-shaped surface are drawn; to be exact: on an \emph{oblate spheroid}, flattened at the poles, and on a \emph{quasi-prolate spheroidal surface}, to wit, an \emph{ovate 2-space}, elongated in the direction of a polar diameter at the poles. There is a plain \emph{warping effect} (if compared to the sphere).

\subsection{Geodesics as Solutions of the Euler–Lagrange Equations}
\label{subsection "Geodesics as Solutions of the Euler–Lagrange Equations"}

\begingroup
\footnotesize
\emph{Methodus maximorum \& minimorum ad lineas curvas applicata}, est methodus inveniendi lineas curvas, quæ maximi minimive, proprietate quapiam proposita gaudeant.\footnote{
	«\emph{The method of maxima \& minima applied to curved lines}, is the method for finding curved lines on which maximum and minimum values for some proposed quantity are satisfied». 
	} \\
\indent — \textsc{L. Euler} \cite[p. 1]{Eulero "Methodus inveniendi lineas curvas Maximi Minimive proprietate gaudentes sive solutio problematis isoperimetrici latissimo sensu accepti"}

\vspace{2mm}

Now here is a method which requires only a very simple use of the principles of differential and integral Calculus.\endnote{
	Original Fr. version: «Maintenant voici une méthode qui ne demande qu'un usage fort simple des principes du Calcul différentiel et intégral».
	} \\
\indent — \textsc{J.L. Lagrange} \cite[p. 336]{Lagrange "Essai d'une nouvelle methode pour determiner les maxima et les minima des formules integrales indefinies"}

\endgroup

\vspace{2mm}

Because a geodesic is an extremal, and it is a possible solution when searching for a \emph{minimum} length, from the calculus of variations we know that a geodesic-like curve, say $\bigl(\gamma_\mathrm{c}(t)\bigr)_{0 \leqslant t \leqslant 1} \in \mathcal{M}$, obeys the Euler–Lagrange equations of motion \cite{Eulero "Curvarum maximi minimive proprietate gaudentium inventio nova et facilis"} \cite{Eulero "Methodus inveniendi lineas curvas Maximi Minimive proprietate gaudentes sive solutio problematis isoperimetrici latissimo sensu accepti"} \cite{Eulero "Elementa Calculi Variationum"} \cite{Eulero "Analytica explicatio methodi Maximorum et Minimorum"} \cite{Eulero "Methodus nova et facilis calculum variationum tractandi"} \cite{Lagrange "Essai d'une nouvelle methode pour determiner les maxima et les minima des formules integrales indefinies"} \cite{Lagrange "Sur la methode des variations"} \cite{Lagrange "Mecanique analytique. Tome premier", Lagrange "Mecanique analytique. Tome second"}, a suitable form of which is
\begin{subequations}	
\label{subequations "Suitable form of Euler–Lagrange equations"}
\begin{align}
	& \frac{\partial\Lagrangian}{\partial x^\mu} = \frac{d}{dt}\left(\frac{\partial\Lagrangian}{\partial\dot{x}^\mu}\right) \text{ or } -\frac{\partial E_u}{\partial x^\mu} = \frac{d}{dt}\left(\frac{\partial E_k}{\partial\dot{x}^\mu}\right), \text{ for } 1 \leqslant \mu \leqslant n, \\
	& \frac{\partial\Lagrangian}{\partial x^\mu}\Bigl(\gamma_\mathrm{c}(t), \dot{\gamma}_\mathrm{c}(t), t\Bigr) = \frac{d}{dt}\left(\frac{\partial\Lagrangian}{\partial v^\mu}\right)\Bigl(\gamma_\mathrm{c}(t), \dot{\gamma}_\mathrm{c}(t), t\Bigr),
\end{align}
\end{subequations}
with $\gamma_\mathrm{c}(t) = \{x^1(t), \mathellipsis, x^n(t)\}$, $\dot{\gamma}_\mathrm{c}(t) = \{\dot{x}^1(t), \mathellipsis, \dot{x}^n(t)\}$. Here, if we consider a particle of mass $m$ moving along $\gamma_\mathrm{c}(t)$ with a velocity $v$, the function $\Lagrangian(x, \dot{x}, t) \viz (x, v, t)$ is a Lagrangian on $\mathring{\mathcal{T}}\mathcal{M} \times [0, 1] \to \mathbb{R}$, and it corresponds to the difference between the kinetic energy and the potential energy. The kinetic energy is 
\begin{equation}
	E_k(x, \dot{x}, t) = \frac{1}{2}m_\mu\|\dot{x}_\mu\|^2
\end{equation}	 
or, for a system of $n$ particles, 
\begin{equation}
	E_k(x, v, t) = \frac{1}{2}\sum^n_{\mu = 1}m_\mu v^2_\mu,
\end{equation}	 
where $v_\mu = \frac{dx^\mu}{dt}$; the potential energy is $E_u(x, t) = - \frac{m(G_\textsc{n})}{\|x\|}$ ($G_\textsc{n}$ is a gravitational constant depending only on the massive nature of the particle) or, for a system of $n$ particles, it is equal to
\begin{equation}
\frac{1}{2}\sum^n_{1 \leqslant \mu < \nu \leqslant n}(E_u)_{\mu\nu}[|x_\mu - x_\nu|].
\end{equation}

It appears that the Lagrangian form $\Lagrangian(x, \dot{x}) = E_k(\dot{x}) - E_u(x)$ (this is a simplified formalism) is equivalent to Newton's second law \cite[p. 12]{Newton "Philosophiae Naturalis Principia Mathematica 1687"}\footnote{
	Axiomata sive Leges Motus, Lex II: «Mutationem motus proportionalem esse vi motrici impressæ, \& fieri secundum lineam rectam qua vis illa imprimitur».
	} 
\cite[p. 19]{Newton "The Mathematical Principles of Natural Philosophy I"}: since $-\frac{\partial E_u}{\partial x^\mu} = m_\mu\ddot{x}_\mu$, we conclude 
\begin{equation}
	\vec{F}_\mu = -\nabla(E_u)_{\mu\nu}, \text{ i.e. } \vec{F}_\mu = m_\mu\ddot{x}_\mu,
\end{equation} 
where the force is equal to $-\frac{\partial E_u}{\partial x^\mu}$.

It is noted that all this leads us to the \emph{Hamilton's principle}, also called \emph{law of least or stationary action} \cite[pp. 10-11]{Hamilton "On a general Method of expressing the Paths of Light and of the Planets by the Coefficients of a Characteristic Function"} \cite{Hamilton "On a General Method in Dynamics; by which the Study of the Motions of all free Systems of attracting or repelling Points is reduced to the Search and Differentiation of one central Relation or characteristic Function"} \cite{Hamilton "Second Essay on a General Method in Dynamics"}, which in turn may be used to obtain the Euler–Lagrange Eqq. \eqref{subequations "Suitable form of Euler–Lagrange equations"}. Letting $\mathscr{S}[\gamma_\mathrm{c}] \colon \mathcal{M} \to \mathbb{R}$ be the action functional of a Lagrangian system, 
\begin{equation}
	\mathscr{S}[\gamma_\mathrm{c}] = \int^{t_\beta}_{t_\alpha}\Lagrangian\Bigl(x(t), \dot{x}(t), t\Bigr) \equival \Lagrangian\Bigl(\tilde{\gamma}_\mathrm{c}(t), t\Bigr),
\end{equation}
and taking a time interval $I_t = [t_\alpha, t_\beta]$ for the motion of the system over $I_t$, we get
\begin{align}
\label{align "Equation with integration by parts"}
	& \frac{d}{d\epsilon}\bigg|_{\epsilon = 0}\mathscr{S}[(\gamma_\mathrm{c})_\epsilon] \notag \\
	& = \frac{d}{d\epsilon}\bigg|_{\epsilon = 0} \int^{t_\beta}_{t_\alpha} \Lagrangian\Bigl(x(t, \epsilon), \dot{x}(t, \epsilon), t\Bigr)dt \notag \\
	& = \sum^n_{\mu = 1}\int^{t_\beta}_{t_\alpha}\left(\frac{\partial\Lagrangian(x, \dot{x})}{\partial x^\mu}\delta x^\mu + \frac{\partial\Lagrangian(x, \dot{x})}{\partial\dot{x}^\mu}\delta\dot{x}^\mu\right)dt \notag \\
	& = \sum^n_{\mu = 1}\int^{t_\beta}_{t_\alpha}\left[\frac{\partial\Lagrangian(x, \dot{x})}{\partial x^\mu} - \frac{d}{dt}\left(\frac{\partial\Lagrangian(x, \dot{x})}{\partial\dot{x}^\mu}\right)\right]\delta x^\mu dt + \sum^n_{\mu = 1}\frac{\partial\Lagrangian(x, \dot{x})}{\partial\dot{x}^\mu}\delta x^\mu\bigg|^{t_\beta}_{t_\alpha}.
\end{align}
The Eq. \eqref{align "Equation with integration by parts"} shows the infinitesimal variation of the integral of the Lagrangian function. Hamilton's principle states that the infinitesimal variation of the Lagrangian function is \emph{stationary} (it vanishes): 
\begin{equation}
	\frac{d}{d\epsilon}\bigg|_{\epsilon = 0} \mathscr{S}[(\gamma_\mathrm{c})_\epsilon] = 0, \text{ or rather } \delta\mathscr{S}[(\gamma_\mathrm{c})_\epsilon] = 0. 
\end{equation}

Using the fundamental form of Euler–Lagrange equations,
\begin{subequations}
\label{subequations "Fundamental form of Euler–Lagrange equations"}
\begin{align}
	& \frac{\partial\Lagrangian(x, \dot{x})}{\partial x^\mu} - \frac{d}{dt}\left[\frac{\partial\Lagrangian(x, \dot{x})}{\partial\dot{x}^\mu}\right] = 0, \\
	& \frac{\partial\Lagrangian(x, v)}{\partial x^\mu}_{\bigg|\begin{subarray}{l}x = \gamma_\mathrm{c}(t) \\ v = \frac{d\gamma_\mathrm{c}(t)}{dt}\end{subarray}} - \frac{d}{dt}\left[\frac{\partial\Lagrangian(x, v)}{\partial x^\mu}_{\bigg|\begin{subarray}{l}x = \gamma_\mathrm{c}(t) \\ v = \frac{d\gamma_\mathrm{c}(t)}{dt}\end{subarray}}\right] = 0,
\end{align}
\end{subequations}
we have a \emph{geodesic equation} for the (geodesic) Lagrangian, according to which the covariant derivative of the tangent vector along the curve is vanishing,
\begin{align}
	& \frac{d^2x^\xi}{d\tau^2} + {\Gamma^\xi}_{\mu\nu}\frac{d^2x^\mu}{d\tau}\frac{d^2x^\nu}{d\tau} = 0, \\
	& \frac{d^2{\gamma_\mathrm{c}}^\xi(\tau)}{d\tau^2} + \binomcurly{\xi}{\mu\nu}_{\big|x = \gamma_\mathrm{c}(\tau)}\frac{d^2{\gamma_\mathrm{c}}^\mu(\tau)}{d\tau}\frac{d^2{\gamma_\mathrm{c}}^\nu(\tau)}{d\tau} = 0.
\end{align}
The letter $\tau$ indicates the \emph{proper time}, as represented in the setting of Minkowski space-time diagram \cite{Minkowski "Die Grundgleichungen fur die elektromagnetischen Vorgange in bewegten Korpern"}. Which means that, for calculating particle orbits in a curved space, the solution of the Euler–Lagrange equations for the Lagrangian $\Lagrangian = \frac{1}{2}m\dot{x}^2 - E_u(x)$ on a Riemannian-like $\mathscr{C}^\infty$ manifold is the geodesic (Definitions \ref{definitio "Geodesic"} and \ref{definitio "Parallel vector field along a curve"}) of a linear connection (Definition \ref{definitio "Connection on a vector bundle"} of which the Levi-Civita connection is the most frequently used example (Theorem \ref{theorema "Levi-Civita"}).

\section{On the Theory of Holonomy: Connections and Loops}
\label{section "On the Theory of Holonomy: Connections and Loops"}

\begingroup
\footnotesize
We see how much these illustrations enlighten and correct our ideas of even the most elementary portions of the theory of rotatory motion. Those who cultivate the geometrical properties of surfaces of the second order will draw from them without difficulty a great number of curious theorems relative to this kind of motion: for each proposition in Geometry gives a corresponding one in Dynamics. \\
\indent — \textsc{L. Poinsot} \cite[p. 72]{Poinsot "Outlines of a New Theory of Rotatory Motion"}, transl. from \cite[p. 56]{Poinsot "Theorie nouvelle de la rotation des corps"}

\vspace{2mm}

[§ 123] A material system between whose possible positions all conceivable continuous motions are also possible motions is called a holonomous system. The term means that such a system obeys integral (\textgreek{ὅλος}) laws (\textgreek{νόμος}), whereas material systems in general obey only differential conditions. \\
\indent [§ 190] In a holonomous system every geodesic path is a straightest path and, conversely, [every straightest path is a geodesic path]. \\
\indent — \textsc{H. Hertz} \cite[pp. 91, 116]{Hertz "Die Prinzipien der Mechanik in neuen Zusammenhange dargestellt"} = \cite[pp. 80, 103]{Hertz "The Principles of Mechanics Presented in a New Form"}

\endgroup

\subsection[Introductory Remarks with a Scholium on $n$-Torus]{Introductory Remarks with a Scholium on $\mathbold{n}$-Torus}
\label{subsection "Introductory Remarks with a Scholium on $n$-Torus"}

In this Section we will give a brief mathematical description of holonomy (group). The first to talk about it, at least in the modern meaning, was É. Cartan \cite{Cartan "Sur les varietes a connexion affine et la theorie de la relativite generalisee (premiere partie)", Cartan "Sur les varietes a connexion affine et la theorie de la relativite generalisee (premiere partie) (Suite)"} \cite{Cartan "Les groupes d'holonomie des espaces generalises"} \cite{Cartan "Sur une classe remarquable d'espaces de Riemann", Cartan "Sur une classe remarquable d'espaces de Riemann (suite et fin)"}.

The holonomy is a \emph{global invariant} of the connection, whereas the curvature is a \emph{local invariant}. Holonomy does not depend on base points, and it is called such, \textgreek{ὁλο(ς)νομία} (law of the entire [space]), because of that. The theory of holonomy group shows features which may nevertheless be analyzed locally in addition to global features; recall that the parallel transport is a way of transporting vectors with regard to the local geometry of a manifold along a certain curve. 

The holonomy of the connection gives informations on the failure of the parallel transport of a vector around a loop to return to its initial value. A \emph{loop} is a \emph{closed curve}, that is a \emph{path having the same starting and ending point}. In this sense, the concept of holonomy can of course be understood as a gap-element of the circle group and thus of 1-torus $\torus^1 \cong \mathbb{S}^1$, which measures the defects of an element of the fiber to be the identity.

\begin{scholium}
\label{scholium "Torus"}
A circle can be considered as a 1-torus, with the relation of being isomorphic. The direct product space of $n$ circles is topologically an $n$-torus: 
\begin{equation}
\label{equation "$n$-torus"}
(\mathbb{S}^1)^n \viz \underbrace{\mathbb{S}^1 \times \cdots \times \mathbb{S}^1}_{n \text{ times}} \: (n \text{ factors of } \mathbb{S}^1) \cong \torus^n.
\end{equation}
For example the product space of two circles is a 2-torus (a doughnut-shaped surface): $\mathbb{S}^1 \times \mathbb{S}^1 \cong \torus^2$, often described as $\mathbb{S}^1$-action on the 2-dimensional torus (cf. Example \ref{exemplum "Möbius strip vs. torus"}). \scholiumsymbol
\end{scholium}

\subsection{Bundles and Holonomy Groups; Ambrose–Singer Theorem}
\label{subsection "Bundles and Holonomy Groups; Ambrose–Singer Theorem"}

Let us start out being a bit more technical by taking the notion of holonomy together with the vector bundle.
 
\begin{definitio}[Holonomy group of a vector bundle connection]
Let $\wp_{\gamma_\mathrm{c}} \colon \mathring{\mathcal{E}} \to \mathcal{M}$, denoting by $\mathring{\mathcal{E}}$ a vector bundle over a manifold $\mathcal{M}$ and by $\nabla$ a connection on $\mathring{\mathcal{E}}$ (cf. Definition \ref{definitio "Connection on a vector bundle"}). Let $\gamma_\mathrm{c} \colon [0, 1] \to \mathcal{M}$ be a piecewise closed $\mathscr{C}^\infty$ curve. Setting $\gamma_\mathrm{c}(0) = \gamma_\mathrm{c}(1) = p \in \mathcal{M}$ (this means that $\gamma_\mathrm{c}$ starts and ends at $p$), one has a parallel transport 
\begin{equation}
	\wp^{\gamma_\mathrm{c}(1)}_{\gamma_\mathrm{c}(0)} \viz \wp_{\gamma_\mathrm{c}(0) \to \gamma_\mathrm{c}(1)} \colon \mathring{\mathcal{E}}_p \to \mathring{\mathcal{E}}_p,
\end{equation}
which is a linear and invertible map lying in $GL(\mathring{\mathcal{E}}_p)$. The \emph{holonomy group of a connection} $\nabla \equival \nabla^{\mathring{\mathcal{E}}}$ at a point $p \in \mathcal{M}$ is defined as  
\begin{align}
	\Hol_p(\nabla) & \viz \Hol_p(\mathcal{M}, \nabla) \notag \\
	& = \left\{\wp^{\gamma_\mathrm{c}(1)}_{\gamma_\mathrm{c}(0)} \viz \wp_{\gamma_\mathrm{c}(0) \to \gamma_\mathrm{c}(1)} \mathrel{\big|} \gamma_\mathrm{c} \text{ is a loop based at } p\right\} \in GL(\mathring{\mathcal{E}}_p).
\end{align}
\definitiosymbol
\end{definitio}

When dealing with a parallel transport $\wp_{\gamma_\mathrm{c}}$ around only loops at $p$ which are contractible or null-homotopic $\mathscr{C}^\infty$ curves (a space is said to be \emph{contractible} if it is homotopy equivalent to a point), one obtains the \emph{restricted holonomy group of} $\nabla \equival \nabla^{\mathring{\mathcal{E}}}$, denoted by
\begin{equation}
	\Hol^0_p(\nabla) \viz \Hol^0_p(\mathcal{M}, \nabla) \subset GL(\mathring{\mathcal{E}}_p) = \bigr\{\wp_{\gamma_\mathrm{c}} \mathrel{\big|} \gamma_\mathrm{c} \text{ is a null-homotopic loop at } p\bigl\}.
\end{equation}

If the smooth curve $\gamma_\mathrm{c}$ is a piecewise path from $\gamma_\mathrm{c}(0) = p_\alpha$ to $\gamma_\mathrm{c}(1) = p_\beta$ in $\mathcal{M}$, then the map is
\begin{equation}
	\wp^{\gamma_\mathrm{c}(1)}_{\gamma_\mathrm{c}(0)} \viz \wp_{\gamma_\mathrm{c}(0) \to \gamma_\mathrm{c}(1)} \colon \mathring{\mathcal{E}}_{p_\alpha} \to \mathring{\mathcal{E}}_{p_\beta},
\end{equation}
from which it is clear that the holonomy representations at $p_\alpha$ and $p_\beta$ are isomorphic, $\wp_{\gamma_\mathrm{c}}\Hol_{p_\alpha}(\nabla)\wp^{-1}_{\gamma_\mathrm{c}} = \Hol_{p_\beta}(\nabla)$, and $\wp_{\gamma_\mathrm{c}}\Hol^0_{p_\alpha}(\nabla)\wp^{-1}_{\gamma_\mathrm{c}} = \Hol^0_{p_\beta}(\nabla)$, in the restricted case.

Now we shall consider the holonomy concerning the tangent bundle of a differentiable manifold. 
\begin{definitio}[Holonomy group of a linear connection] 
Suppose that $\mathcal{M}$ is a manifold with fiber $\mathring{\mathcal{E}}_p \cong \mathbb{R}^k$. Let
\begin{equation}
	\wp^{\gamma_\mathrm{c}(1)}_{\gamma_\mathrm{c}(0)} \viz \wp_{\gamma_\mathrm{c}(0) \to \gamma_\mathrm{c}(1)} \colon \mathcal{T}_p\mathcal{M} \to \mathcal{T}_p\mathcal{M}
\end{equation}
be an \emph{isomorphism of vector spaces}, commonly referred to as \emph{parallel displacement} along $\gamma_\mathrm{c}(0) \to \gamma_\mathrm{c}(1)$, with $\gamma_\mathrm{c}(0) = \gamma_\mathrm{c}(1) = p \in \mathcal{M}$ fixed. The parallel displacement of a vector field $\sezione \mapsto \wp_{\gamma_\mathrm{c}(t)}(\sezione) \equival \sezione(t)$ along $\gamma_\mathrm{c}$ satisfies the equation $\nabla_{\dot{\gamma}_\mathrm{c}(t)}\sezione(t) = 0$ (see Definitions \ref{definitio "Parallel vector field along a curve"} and \ref{definitio "Parallel transport map"}), given that a connection is a tool to differentiate vector field covariantly.
Then the expressions
\begin{align}
	\Hol_p(\nabla) & \viz \Hol_p(\mathcal{M}, \nabla) \notag \\
	& = \left\{\wp^{\gamma_\mathrm{c}(1)}_{\gamma_\mathrm{c}(0)} \viz \wp_{\gamma_\mathrm{c}(0) \to \gamma_\mathrm{c}(1)} \colon \mathcal{T}_p\mathcal{M} \to \mathcal{T}_p\mathcal{M}\right\} \in GL_k(\mathbb{R})
\end{align}
specify the \emph{holonomy group of a connection} $\nabla \equival \nabla^{\mathring{\mathcal{T}}}$ at $p \in \mathcal{M}$. \definitiosymbol
\end{definitio}

Cf. \cite{Hano and Ozeki "On the holonomy groups of linear connections"}. Now let us examine the idea of holonomy by applying it to a connection on a principal bundle.

\begin{definitio}[Principal G-bundle]
\label{definitio "Principal G-bundle"}
A principal bundle is a case of a fiber bundle in which the fiber is given by $G$. Let $G$ be a Lie group. A principal bundle $\mathring{\mathcal{P}}$ over a base $\mathcal{M}$ is a surjective smooth map $\pi \colon \mathring{\mathcal{P}} \to \mathcal{M}$, i.e. a $\mathscr{C}^\infty$ projection $\pi$ of $\mathring{\mathcal{P}}$ onto $\mathcal{M}$, if there exists a smooth right $\textsc{r}$-action of $G$ on $\mathring{\mathcal{P}}$, which is 
\begin{subequations}
\begin{align}
	& \textsc{r}_G \colon \mathring{\mathcal{P}} \times G \to \mathring{\mathcal{P}}, \enspace \textsc{r}_G(x, g) = x \cdot g, \\
	& \textsc{r}_g \colon x \mapsto x \cdot g, 
\end{align}
\end{subequations}
for $x \in \mathring{\mathcal{P}}$ and $g \in G$. One has $\mathring{\mathcal{P}}/G = \mathcal{M}$, since the fiber bundle of a principal $G$-bundle is isomorphic to $G$-space. (This implies that a principal $G$-bundle can be subsumed in the class of smooth manifolds).

The basic assumption here is that $\mathring{\mathcal{P}}$ is a fiber bundle whose fiber is $G$. There is talk of \emph{principal $G$-bundle}, \emph{$\mathscr{C}^\infty$ (smooth) $G$-bundle} or, simply, \emph{principal bundle}, to be exact. More widely, such a principal bundle is the quadruple $(\mathring{\mathcal{P}}, \pi, \mathcal{M}, G)$. The conditions under which this fiber bundle takes place are the following.
\enumerationisinitium
\item The $\textsc{r}$-action of $G$ is free, transitive and fiber preserving, $\pi(x \cdot g) = \pi(x)$, for all $x \in \mathring{\mathcal{P}}$ and $g \in G$, so that for each $p \in \mathcal{M}$, any fiber $\pi^{-1}(p)$ is a vector space isomorphic to $G$. 
\item The base space $\mathcal{M}$ is the quotient of $\mathring{\mathcal{P}}$. The orbits of the right $G$-action coincide with the fibers of $\pi$, and that is $\pi^{-1}(p) = x \cdot G$, with $p \in \mathcal{M}$, $x \in \pi^{-1}(p)$.
\item $\mathring{\mathcal{P}}$ is locally trivial. For each $p \in \mathcal{M}$, there is an open neighborhood $\Upsilon_p$ in $\mathcal{M}$, namely a bundle chart $(\Upsilon_p, \varphi)$ with $p \in \Upsilon$, and a diffeomorphism $\Phi \colon \pi^{-1}(\Upsilon_p) \to \Upsilon_p \times G$, which may be written in the $G$-equivariant form $\Phi(x) = [\pi(x), \varphi(x)]$; in virtue of the diffeomorphism, the representation $\varphi(x \cdot g) = \varphi(x)g$ is satisfied, for all $x \in \pi^{-1}(\Upsilon_p)$ and $g \in G$. 

Note. A principal $G$-bundle can also be made explicit in another way, focusing on the action of $G$ on itself given by left translation: 
\begin{subequations}
\begin{align}
	& \textsc{l}_G \colon G \times G \to G, \\
	& \textsc{l}_g \colon G \to G \colon h \mapsto g \cdot h. 
\end{align}
\end{subequations}
The $G$-bundle is homeomorphic to the structural group $G$, if there is a left $G$-action on $G$ itself. \definitiosymbol
\enumerationisfinis
\end{definitio}

We can move on to the following statements.
	
\begin{definitio}
\label{definitio "Preparatory definition and horizontality"}
Let $\mathring{\mathcal{P}}$ a principal $G$-bundle over $\mathcal{M}$, and $\omega_{\mathring{\mathcal{P}}}$ a connection on $\mathring{\mathcal{P}}$. 
\enumerationisinitium
\item We may take $\gamma_\mathrm{c} \colon [0, 1] \to \mathring{\mathcal{P}}$ to be a $\mathscr{C}^\infty$ curve (a loop) in $\mathring{\mathcal{P}}$. Letting $\dot{\gamma}_\mathrm{c}(t) \in \mathcal{T}_{\gamma_\mathrm{c}(t)}\mathring{\mathcal{P}}$ be tangent to $\gamma_c(I)$, where $I = [0, 1] = \{n \in \mathbb{R} \mid 0 \leqslant n \leqslant 1\}$, at time $t \in [0, 1]$ (cf. Definition \ref{definitio "Geodesic"}), the tangent vector $\dot{\gamma}_\mathrm{c}(t) \in \mathcal{T}_{\gamma_\mathrm{c}(t)}\mathring{\mathcal{P}}$ is horizontal, then $\gamma_\mathrm{c}$ is also \emph{horizontal}, namely $\dot{\gamma}_\mathrm{c}(t) \in (\omega_{\mathring{\mathcal{P}}})_t\gamma_\mathrm{c}$.
\item Consider the piecewise $\mathscr{C}^\infty$ curve $\gamma_\mathrm{c} \colon [0, 1] \to \mathcal{M}$, with $\gamma_\mathrm{c}(0) = \gamma_\mathrm{c}(1) = p$ and $p \in \mathcal{M} = \pi(x)$. For each $x \in \mathring{\mathcal{P}} = \pi^{-1}(p)$, there is a unique \emph{horizontal lift} $\tilde{\gamma}_\mathrm{c} \colon [0, 1] \to \mathring{\mathcal{P}}$ of the curve in $\mathring{\mathcal{P}}$, with $x = \tilde{\gamma}_\mathrm{c}(0)$; the end-point of the horizontal lift, that is $\tilde{\gamma}_\mathrm{c}(1)$, is in the fiber over $p = \pi(x)$. \definitiosymbol
\enumerationisfinis
\end{definitio}

Finally, define the holonomy group of $\omega_{\mathring{\mathcal{P}}}$ on $\mathring{\mathcal{P}}$.

\begin{definitio}[Holonomy group of a principal G-bundle connection]
\label{definitio "Holonomy group of a principal G-bundle connection"}
The equivalence relation $x \sim y$, for $x, y \in \mathring{\mathcal{P}}$, is possible if 
there exists a unique horizontal, piecewise $\mathscr{C}^\infty$ curve $\gamma_\mathrm{c} \colon [0, 1] \to \mathring{\mathcal{P}}$ joining $x$ and $y$. 
Setting $g \in G$, the holonomy group of a connection on a principal $G$-bundle is
\begin{equation}
	\Hol_x(\omega_{\mathring{\mathcal{P}}}) = \{g \in G \mid x \sim x \cdot g\}.
\end{equation}
The corresponding restricted holonomy group at $x$ is the $\Hol^0_x(\omega_{\mathring{\mathcal{P}}})$ arising from the horizontal lift of null-homotopic curve in $\mathcal{M}$ which can be continuously contracted to a point. \definitiosymbol
\end{definitio}

From the binary relation $x \sim y$, it is evident that
\begin{align}
	&  \Hol_x(\omega_{\mathring{\mathcal{P}}}) = \Hol_y(\omega_{\mathring{\mathcal{P}}}), \\
	& \Hol_{x \cdot g}(\omega_{\mathring{\mathcal{P}}}) = g\Hol_x(\omega_{\mathring{\mathcal{P}}})g^{-1}.
\end{align}

In order to truly appreciate  the relation between the curvature and the holonomy group of a connection on a principal bundle, we recall the Ambrose–Singer theorem \cite[p. 438]{Ambrose and Singer "A theorem on holonomy"}, initially proposed, without proof, by É. Cartan \cite{Cartan "Les groupes d'holonomie des espaces generalises"}.

\begin{theorema}[Ambrose–Singer]
\label{theorema "Ambrose–Singer"}
Let us suppose that $(\mathring{\mathcal{P}}, \pi, \mathcal{M}, G)$ is a principal $G$-bundle with a connected and paracompact base space $\mathcal{M}$. Let $\Hol_x(\omega_{\mathring{\mathcal{P}}})$ be the holonomy group of a connection $\omega_{\mathring{\mathcal{P}}}$ on $\mathring{\mathcal{P}}$ and $\mathring{\mathcal{P}}(x)$ the holonomy bundle of $\omega_{\mathring{\mathcal{P}}}$ through $x \in \mathring{\mathcal{P}}$. Let also $\Omega^\nabla$ be a curvature 2-form of $\omega_{\mathring{\mathcal{P}}}$. Then the Lie algebra of $\Hol_x(\omega_{\mathring{\mathcal{P}}})$, denoted by $\mathfrak{hol}_x(\omega_{\mathring{\mathcal{P}}})$ and called \emph{holonomy Lie algebra}, or, quite simply, \emph{holonomy algebra}, is equal to the subalgebra (or subspace) of $\mathfrak{g}$, which is the Lie algebra of $G$, so that $\mathfrak{hol}_x(\omega_{\mathring{\mathcal{P}}})$ is spanned by the elements of the form $\Omega^\nabla_y(v, w)$, where the point $y \in \mathring{\mathcal{P}}(x)$ is joinable to $x$ by means of a horizontal curve (of the type described above), and $v, w \in \mathcal{T}_y\mathcal{M}$ are horizontal tangent vectors, that is
\begin{equation}
	\mathfrak{hol}_x(\omega_{\mathring{\mathcal{P}}}) = \mathrm{span}\left\{\Omega^\nabla_y(v, w) \in \mathfrak{g}\right\}.
\end{equation}
Said in other words, $\mathfrak{hol}_x(\omega_{\mathring{\mathcal{P}}})$ is the Lie algebra of the holonomy group of a connection on a principal $G$-bundle, and it is thereby a Lie subalgebra of the Lie algebra $\mathfrak{g}$ of $G$, i.e. $\mathfrak{hol}_x(\omega_{\mathring{\mathcal{P}}}) \subset \mathfrak{g}$.
\end{theorema}

\begin{proof}
We got to prove that $\mathfrak{hol} = \mathfrak{g}$. If we reduce the structure group $G$, and the connection to the holonomy group, we see that $G = \Hol(\omega_{\mathring{\mathcal{P}}})$. We claim that $\mathfrak{hol}$ is an ideal of $\mathfrak{g}$. Since $\Omega^\nabla$ is $G$-equivariant, the Lie algebra (subspace) $\mathfrak{hol}$ is invariant under the adjoint $G$-action. Consider the vector bundle $\pi \colon \mathring{\mathcal{E}} \to \mathcal{M}$. Let
\begin{equation}
	\mathring{\mathcal{T}}_\mathrm{v}\mathring{\mathcal{E}} \viz (\mathring{\mathcal{T}}_\mathrm{v}\mathring{\mathcal{E}}, \pi_{\mathrm{v}\mathring{\mathcal{E}}}, \mathring{\mathcal{E}})
\end{equation}
denote the vertical bundle of $\mathring{\mathcal{T}}\mathring{\mathcal{P}}$ or, to be accurate, the vertical subbundle of the tangent bundle $
	\pi_{\mathrm{v}\mathring{\mathcal{E}}} \colon \mathring{\mathcal{T}}\mathring{\mathcal{P}} \to \mathring{\mathcal{E}}$ of $\mathring{\mathcal{P}}$, defined by $\mathfrak{hol}$. We assert that $\mathring{\mathcal{T}}_\mathrm{v}\mathring{\mathcal{E}}$ is integrable. The tangent bundle $\mathring{\mathcal{T}}\mathring{\mathcal{P}}$ of $\mathring{\mathcal{P}}$ is the direct sum 
\begin{equation}
	\mathring{\mathcal{T}}\mathring{\mathcal{P}} = \mathring{\mathcal{T}}_\mathrm{v}\mathring{\mathcal{E}} \oplus \mathring{\mathcal{T}}_\mathrm{h}\mathring{\mathcal{E}}
\end{equation}
of vertical and horizontal subbundles. Taking two horizontal vector fields $\vec{V}$ and $\vec{W}$, the vertical projection of the brackets of $\vec{V}$ and $\vec{W}$ is provided by $\Omega^\nabla(\vec{V}, \vec{W})$, and it is a section of $\mathring{\mathcal{T}}_\mathrm{v}\mathring{\mathcal{E}}$. Now, the bracket of a vertical vector field and a horizontal vector field (both invariant vector fields, with values in the vertical and horizontal tangent spaces) is null, for which
\enumerationisinitium
\item $\mathring{\mathcal{T}}_\mathrm{v}\mathring{\mathcal{E}}$ is invariant under $G$-action,
\item all integral manifolds are invariant under $G$-action for such a subbundle, 
\item the horizontal subspace is in the tangent space. 
\enumerationisfinis
Finally, $\dim(\mathfrak{hol}) = \dim(\mathfrak{g})$, and then $\mathfrak{hol} = \mathfrak{g}$. Lie algebra of the the holonomy group of a connection of the frame bundle is the same as the Lie algebra spanned by the curvature 2-form.
\end{proof}

\subsection{Holonomy in Riemannian Spaces}
\label{subsection "Holonomy in Riemannian Spaces"}

In Riemannian geometry, the parallel transport, from one point to another, does not depend on the curve if the Riemann curvature tensor $g$ vanishes in this region (locally), condition in which a manifold turns out to be flat; but it depends on the curve if the curvature tensor does not vanish. The holonomy group (along with the non-vanishing of the Riemann curvature tensor) is a measure of the deviation of a Riemannian-like manifold from flatness, or rather from being locally isometric to Euclidean space.

\begin{definitio}[Holonomy group of a Riemannian manifold]
The parallel translation along $\gamma_\mathrm{c}$ from $\gamma_\mathrm{c}(0)$ to $\gamma_\mathrm{c}(1)$ coincides with an orthogonal transformation of the tangent space $\mathcal{T}_p\mathcal{M}$ to the manifold. The set of all translations forms a group, $\Hol_p(g) \viz \Hol_p(\mathcal{M}, g)$, which is a closed and compact subgroup \cite{A. Borel Lichnerowicz "Groupes d'holonomie des varietes riemanniennes"} of the orthogonal group of the tangent space $O(\mathcal{T}_p\mathcal{M}) \cong O_n(\mathbb{R})$ with respect to $g_p$. The group 
\[
	\Hol_p(g) \subset O(\mathcal{T}_p\mathcal{M})
\]
is called the \emph{holonomy group of $(\mathcal{M}, g)$} at $p \in \mathcal{M}$, and it is a Riemannian holonomy group, with $g \mapsto \wp_{\gamma_c} \circ g \circ \wp^{-1}_{\gamma_c}$. \definitiosymbol	
\end{definitio}

The \emph{restricted holonomy group} of $(\mathcal{M}, g)$ at $p$ is 
\[
	\Hol^0_p(g) \viz \Hol^0_p(\mathcal{M}, g). 
\]
If $(\mathcal{M}, g)$ is a simply connected space, then the holonomy group and the restricted holonomy group are the same algebraic structure: $\Hol_p(g) = \Hol^0_p(g)$.

Let $\nabla g = 0$ (the covariant derivative of the metric tensor vanishes). Since the holonomy of a Riemannian manifold is the holonomy group of the Levi-Civita connection on $\mathring{\mathcal{T}}\mathcal{M}$ (or vice versa, the holonomy of the Levi-Civita connection on the tangent bundle is the holonomy of a Riemannian manifold), we have 
\[
	\Hol_p(\nabla) \viz \Hol_p(\mathcal{M}, \nabla) \subset O_n(\mathbb{R}), \text{ with } \nabla \equival \nabla^{\mathring{\mathcal{T}}} = 0, 
\]
where $O_n(\mathbb{R})$ is the orthogonal group of transformations of $\mathcal{T}_p\mathcal{M}$ which preserves $g_p$.

\begin{scholium}[Holonomic automorphism]
\label{scholium "Holonomic automorphism"}
The parallel transport with loop $\gamma_\mathrm{c}(0) = \gamma_\mathrm{c}(1) = p$ is an automorphism of the tangent space $\mathcal{T}_p\mathcal{M}$ at a point $p$. Let us say that $\underline{L}_p$ and $\underline{L}^0_p$ are the sets of null-homotopic loops at $p \in \mathcal{M}$, respectively. So we can redefine the holonomy group of $(\mathcal{M}, g)$ at $p$ as the set of all automorphisms of $\mathcal{T}_p\mathcal{M}$ obtained by parallel translation in $\mathcal{M}$ associated with $\underline{L}_p$, or with $\underline{L}^0_p$ if the holonomy group is restricted. \scholiumsymbol
\end{scholium}

\begin{definitio}[Holonomy algebra of the metric tensor]
\label{definitio "Holonomy algebra of the metric tensor"}
Let $(\mathcal{M}, g)$ be a Riemannian manifold, with $p \in \mathcal{M}$ and Levi-Civita connection of the metric tensor $g$. The Lie algebra of $\Hol(g)$ of a connection on the tangent bundle of $\mathcal{M}$ is 
\[
	\mathfrak{hol}(g) \equival \mathfrak{hol}(\nabla), \text{ or } \Hol^\nabla_\mathfrak{hol}, 
\]
always putting $\nabla \equival \nabla^{\mathring{\mathcal{T}}}$, where $\Hol^\nabla_\mathfrak{hol}$ 

· is a Lie subalgebra of the special orthogonal Lie algebra $\mathfrak{so}(n)$—we would like to say that $\Hol^\nabla_\mathfrak{hol}$ is a subalgebra of the vector space of all $n \times n$ skew symmetric matrices $X^\textsc{t} = -X$, $X \in \mathfrak{so}(n)$;

· is a vector subspace of $\mathcal{T}_p\mathcal{M} \otimes \mathcal{T}^*_p\mathcal{M}$, or $\bigotimes^2\mathcal{T}^*_p\mathcal{M}$, such that $\Hol^\nabla_\mathfrak{hol}$ is lying in the subspace $\bigwedge^2\mathcal{T}^*_p\mathcal{M}$, 
\begin{equation}
	\bigwedge^2_\pm\mathring{\mathcal{T}}^*\mathcal{M} \cong \bigwedge_+\mathring{\mathcal{T}}^*\mathcal{M} \oplus \bigwedge_-\mathring{\mathcal{T}}^*\mathcal{M},
\end{equation}
taking into account that $\bigwedge^2\mathring{\mathcal{T}}^*\mathcal{M}$ is the second exterior power of the cotangent bundle of $\mathcal{M}$ splittable into \emph{self-dual} ($+$) and \emph{anti-self-dual} ($-$) parts. \definitiosymbol
\end{definitio}

\subsection{Berger's Classification; from (Hyper)kählerian to Spin Manifolds; Simons Theorem}
\label{subsection "Berger's Classification; from (Hyper)kählerian to Spin Manifolds; Simons Theorem"}

For purposes of exposition, we continue with the same topic as in the previous Section and mention the M. Berger's classification \cite[p. 318]{Berger "Sur les groupes d'holonomie homogenes de varietes a connexion affine et des varietes riemanniennes"}, a list of the possible restricted (Riemannian) holonomy groups of simply connected Riemannian manifolds which are \emph{irreducible}, or non-locally a product, and \emph{non-symmetric}, or non-locally a symmetric spaces.

\begin{theorema}[Berger]
\label{theorema "Berger"}
Let $(\mathcal{M}, g)$ be a connected Riemannian manifold and $n = \dim(\mathcal{M})$ the dimension of $\mathcal{M}$. Suppose that the restricted holonomy group $\Hol^0_p(g) \viz \Hol^0_p(\mathcal{M}, g)$ acts irreducibly on its tangent space $\mathcal{T}_p\mathcal{M}$ at $p \in \mathcal{M}$. Then $(\mathcal{M}, g)$ turns out to be locally isometric to a symmetric space, except if the holonomy group is conjugate (isomorphic) to one of the underlying subgroups of $SO(n)$: $U\left(\frac{n}{2}\right)$, $SU\left(\frac{n}{2}\right)$, $Sp\left(\frac{n}{4}\right)$, $Sp\left(\frac{n}{4}\right) \cdot Sp(1)$, $G_2$ and $\Spin(7)$.
\end{theorema} 

\begin{proof}
The methods of the demonstration are beyond the aim of this Chapter, so  we refer the reader to Berger \cite{Berger "Sur les groupes d'holonomie homogenes de varietes a connexion affine et des varietes riemanniennes"} \cite{Berger "Les espaces symetriques noncompacts"} and S. Salamon \cite[chap. 10]{Salamon S. "Riemannian geometry and holonomy groups"}.
\end{proof}

We can better describe the classification \cite{Berger "Sur les groupes d'holonomie homogenes de varietes a connexion affine et des varietes riemanniennes"} in the following manner.
\enumerationisinitium
\item $\Hol^0_p(g) \equival SO(n)$, for $\dim(\mathcal{M}) = n \geqslant 2$. 
\subenumerationisinitium
\item $(\mathcal{M}^{\geqslant 2}, g)$ is a \emph{generic Riemannian manifold}. If $\Hol^0_p(g)$ is a \emph{proper} subgroup of $SO(n)$, the invariant $\Hol$ is indicated as \emph{special holonomy}.
\item The special orthogonal group $SO(n) = O(n) \cap SL_n(\mathbb{R})$ is an automorphism group of the real field $\mathbb{R}^n$.
\item Lie algebra: $\mathfrak{so}(n)$; it is identified with the vector space of all $n \times n$ skew symmetric matrices.
\subenumerationisfinis
\item 
\label{item "Berger's classification unitary group"}
$\Hol^0_p(g) \equival U(m) \viz U\left(\frac{n}{2}\right)$ in $SO(2m)$, for $\dim(\mathcal{M}) = n = 2m \geqslant 4$, with $m \geqslant 2$.
\subenumerationisinitium
\item $(\mathcal{M}^{2m}, g)$ is a \emph{generic Kähler manifold} \cite{Kahler "Uber eine bemerkenswerte Hermitesche Metrik"} (see Scholium \ref{scholium "Kähler manifold, and almost complex structure"}); the existence of metrics having such a holonomy (with examples of compact Kähler-spaces) is ensured by S.-T. Yau's solution \cite{Yau "On the Ricci Curvature of a Compact Kahler Manifold and the Complex Monge-Ampere Equation I"} for the Calabi conjecture. 
\item The unitary group $U(m) = \{X \in GL_m(\mathbb{C}) \mid X^{\dag}X = \idem_m\}$ is an automorphism group of the complex field $\mathbb{C}^{m = \frac{n}{2}} \cong \mathbb{R}^{2m = n}$, where $X^\dag \viz \bar{X}^\textsc{t}$ is the conjugate transpose matrix of $X$. The inclusion chain for this set is $\Hol^0_p(g) \subset U(m) \in GL_m(\mathbb{C})$. Cf. Section \ref{subsection "Holonomy in Abelian Phase Factor: Gauge Group of Electromagnetic Interactions"} for a representations of the group with complex ($1 \times 1$)-matrices and its physical interpretation.
\item Lie algebra: $\mathfrak{u}(m) = \{X \in \mathfrak{gl}_m(\mathbb{C}) \mid X + X^\dag = 0\}$.
\subenumerationisfinis
\item $\Hol^0_p(g) \equival SU(m) \viz SU\left(\frac{n}{2}\right)$ in $SO(2m)$, for $\dim(\mathcal{M}) = n = 2m \geqslant 4$, with $m \geqslant 2$.
\subenumerationisinitium
\item $(\mathcal{M}^{2m}, g)$ is a \emph{Calabi–Yau manifold}, and thereby a \emph{Ricci-flat Kähler manifold} with vanishing first Chern class \cite{Chern "Characteristic classes of Hermitian Manifolds"}.
\item The special unitary group $SU(m) = U(m) \cap SL_m(\mathbb{C})$ is an automorphism group of the complex field $\mathbb{C}^{m = \frac{n}{2}} \cong \mathbb{R}^{2m = n}$.
\item Lie algebra: $\mathfrak{su}(m) = \mathfrak{u}(m) \cap \mathfrak{sl}_m(\mathbb{C})$.
\subenumerationisfinis
\item $\Hol^0_p(g) \equival Sp(m) \viz Sp\left(\frac{n}{4}\right)$ in $SO(4m)$, for $\dim(\mathcal{M}) = n = 4m \geqslant 8$, with $m \geqslant 2$.
\subenumerationisinitium
\item $(\mathcal{M}^{4m}, g)$ is a \emph{Ricci-flat hyperkähler manifold}, see E. Calabi \textnormal{\cite{Calabi "Metriques kahleriennes et fibres holomorphes"}}, i.e. a complex manifold (with Kähler structure) admitting a \emph{holomorphic symplectic form}.
\item The symplectic group $Sp(m) = \{X \in GL_m(\quaternion) \mid X^{\dag}X = \idem_m\}$ is an automorphism group of Hamilton's quaternions $\quaternion^{m = \frac{n}{4}} \cong \mathbb{R}^{4m = n}$.
\item Lie algebra: $\mathfrak{sp}(m) = \{X \in \mathfrak{gl}_m(\mathbb{\quaternion}) \mid X + X^\dag \viz \bar{X}^\textsc{t} = 0\}$.
\subenumerationisfinis	
\item 
\label{item "Holonomy in Berger's theorem, quaternionic objects, and more"}
$\Hol^0_p(g) \equival Sp(m) \viz Sp\left(\frac{n}{4}\right) \cdot Sp(1)$ in $SO(4m)$, for $\dim(\mathcal{M}) = n = 4m \geqslant 8$, with $m \geqslant 2$.
\subenumerationisinitium 
\item $(\mathcal{M}^{4m}, g)$ is both a \emph{quaternionic Kähler} (or \emph{quaternion-Kähler}) and an \emph{Einstein manifold with non-zero (positive/negative) Ricci curvature}, as defined by E. Bonan \textnormal{\cite{Bonan "Structure presque quaternale sur une variete differentiable"}}. Note that a (compact) quaternionic Kähler manifold with Ricci-flat Kähler metric (so we are talking about a metric of vanishing scalar curvature) is \emph{locally conformally hyperkählerian}.
\item The symplectic groups $Sp(m) \cdot Sp(1)$ are automorphism groups of the quaternion field $\quaternion^{m = \frac{n}{4}} \cong \mathbb{R}^{4m = n}$.
\item 
\label{item "Quaternionic projective space, quaternionic Hopf bundles, etc."}
The quaternionic projective space 
\begin{equation}
	\quaternion\mathbb{P}^r = \frac{Sp(r + 1)}{Sp(r) \cdot Sp(1)}
\end{equation}
is itself a quaternionic Kähler manifold (recall that $\quaternion\mathbb{P}^r$ is the quaternionic projective space; it is defined as quotient space of $\quaternion^{r + 1} - \{0\}$ by the action of the multiplicative group $\quaternion^\times$ of non-zero quaternions on the right). Placing an identification between $\quaternion\mathbb{P}^{1}$ and $\mathbb{S}^4$, the resulting \emph{quaternionic Hopf bundles} \cite{Hopf H. "Uber die Abbildungen der dreidimensionalen Sphare auf die Kugelflache"} are
\begin{equation}
Sp(1) \hookrightarrow \mathbb{S}^7 \xrightarrow{\pi_1 \text{ and } \pi_{-1}} \quaternion\mathbb{P}^{1} \cong \mathbb{S}^4 \cong \frac{Sp(2)}{Sp(1) \cdot Sp(1)}.
\end{equation}
One can see that the principal $Sp(1)$-bundle over $\quaternion\mathbb{P}^{r - 1}$ coincides with the sphere $\mathbb{S}^{4r - 1}$, i.e.
\begin{equation}
	Sp(1) \hookrightarrow \mathbb{S}^{4r - 1} \cong \frac{Sp(r)}{Sp(r - 1)} \xrightarrow{\pi} \quaternion\mathbb{P}^{r - 1}.
\end{equation}
\item Lie algebra: $\mathfrak{sp}(m) \viz \mathfrak{sp}\left(\frac{n}{4}\right) \cdot \mathfrak{sp}(1)$.
\subenumerationisfinis
\item $\Hol^0_p(g) \equival G_2$ in $SO(7)$, for $\dim(\mathcal{M}) = n = 7$, and this holonomy is called \emph{exceptional}; $G_2$ is a Lie subgroup of $GL_7(\mathbb{R})$.
\subenumerationisinitium
\item $(\mathcal{M}^7, g) \equival \mathbb{R}^4 \times \mathbb{S}^3$ \cite{Gibbons Page and Pope "Einstein Metrics on $S^3$ $R^3$ and $R^4$ Bundles"} is a \emph{Ricci-flat $G_2$-manifold}. Examples of application of $G_2$ to physics (string theory), can be found in E. Witten's papers in collaboration with B. Acharya and M. Atiyah \cite{Acharya Witten "Chiral Fermions from Manifolds of $G_2$ Holonomy"} \cite{Atiyah Witten "M-Theory Dynamics On A Manifold Of $G_2$ Holonomy"}.
\item The exceptional group $G_2$ is an automorphism group of the imaginary octonion field,\footnote{
	Historical margo: the first intuition of the octonions is ascribed to J.T. Graves \cite{Graves "On a Connection between the General Theory of Normal Couples and the Theory of Complete Quadratic Functions of Two Variables"} and it dates back to 1843 (26th of December), as W.R. Hamilton \cite{Hamilton "Note respecting the Researches of John T. Graves"} goes on to explain.
	} 
according to Cayley numbers \cite{Cayley "On Jacobi's Elliptic functions in reply to the Rev. Brice Bronwin; and on Quaternions"}, $\Im(\mathbb{O}) \cong \mathbb{R}^7 \rtimes SO(7)$ resulting from $\mathbb{O} \cong \mathbb{R} \oplus \Im(\mathbb{O}) \cong \mathbb{R}^7 \cong \mathbb{R}^3 \oplus \mathbb{C}^2$.
\item Lie algebra: $\mathfrak{g}_2$ in $\mathfrak{so}(7)$.
\subenumerationisfinis
\item $\Hol^0_p(g) \equival \Spin(7)$ in $SO(8)$, for $\dim(\mathcal{M}) = n = 8$ (this holonomy is called \emph{exceptional}).
\subenumerationisinitium
\item $(\mathcal{M}^8, g)$ is a \emph{Ricci-flat $\Spin(7)$-manifold}.
\item The spin group $\Spin(7)$ is an automorphism group of the octonions $\mathbb{O} \cong \mathbb{R}^8 \cong \mathbb{R} \oplus \mathbb{R}^7$. If $\varsigma(\mathbb{R})_\mathcal{M}$ is a real spinor field on $\mathcal{M}$, and if it is decomposable into parallel and non-trivial elements $\bigl(\varsigma(\mathbb{R})^+_\mathcal{M}\bigr) + \bigl(\varsigma(\mathbb{R})^-_\mathcal{M}\bigr) = \varsigma(\mathbb{R})_\mathcal{M}$, then the holonomy group is reducible to $\Spin^+(7) \cap \Spin^-(7) \cong G_2$. Furthermore, the isotropy structure $\Spin(7)/G_2$ and the 7-sphere (obviously equipped with a constant curvature metric) are holomorphic spaces:
\begin{equation}
	\frac{\Spin(7)}{G_2} \cong \mathbb{S}^7 \cong \frac{SO(8)}{SO(7)},
\end{equation}
with fibration $G_2 \to \Spin(7) \to \mathbb{S}^7$.

Which brings us to the real projective space: 
\begin{equation}
	\frac{SO(7)}{G_2} \cong \frac{SO(8)}{\Spin(7)} \cong \mathbb{RP}^7 \viz \mathbb{P}_7(\mathbb{R}) \cong \mathbb{S}^7/\mathbb{Z}_2 \viz \mathbb{Z}/2\mathbb{Z},
\end{equation}
where $\mathbb{Z}_2$ is the cyclic group of order 2.
\item Lie algebra: $\mathfrak{spin}(7)$ in $\mathfrak{so}(8)$.
\subenumerationisfinis
\enumerationisfinis

In each of the seven cases, the Riemannian manifolds are irreducible and non-symmetric.

\begin{margo}
The Berger's original list provided for the $\Spin(9)$-structure, for $\dim(\mathcal{M}) = n = 16$, but D.V. Alekseevsky \cite{Alekseevsky "Riemannian spaces with exceptional holonomy groups"} showed that a 16-dimensional Riemannian manifold with such holonomy is a locally symmetric space. On the incompleteness of Berger's classification, and the inclusion of an (infinite) series of candidates for the so-called \emph{exotic holonomies}, as well as on the adjustment of the classification of irreducible holonomy groups of a torsion free affine connection, and so of the possible holonomies of the Levi-Civita connection with respect to a (pseudo-)Riemannian metric, see \cite{Bryant "Two Exotic Holonomies in Dimension Four Path Geometries and Twistor Theory"} \cite{Bryant "Classical exceptional and exotic holonomies: a status report"} \cite{Chi Merkulov Schwachhofer "On the existence of infinite series of exotic holonomies"} \cite{Bryant "Recent advances in the theory of holonomy"} \cite{Merkulov and Schwachhofer "Classification of irreducible holonomies of torsion-free affine connections"}. \margosymbol
\end{margo}

Then there is a theorem by J. Simons \cite{Simons "On the Transitivity of Holonomy Systems"} helpful to give an algebraic generalization of the holonomy of a connection related to Berger's list. The theorem states that a space equipped with a non-transitive holonomy is necessarily locally symmetric.

\begin{theorema}[Simons]
\label{theorema "Simons"}
Given a simply connected and irreducible Riemannian manifold $(\mathcal{M}, g)$, if the holonomy group $\Hol_p(g) \viz \Hol_p(\mathcal{M}, g)$ is not transitive on the unit sphere in the tangent space $\mathcal{T}_p\mathcal{M}$, then the manifold $(\mathcal{M}, g)$ is (must be a) a locally symmetric space of rank $\geqslant 2$.
\end{theorema}

\begin{proof}
Suffice it to say that Simons \cite{Simons "On the Transitivity of Holonomy Systems"} supposes $\dim(\mathcal{M}) \geqslant 3$, seeing that the identity component is the only connected group of isometries which acts non-transitively on $\mathbb{S}^1$, but this conflicts with the irreducibility. The proof follows from previous theorems regarding the irreducible holonomy system (under which the non-transitivity of a connected holonomy group on $\mathbb{S}^{n-1}$ implies that $\mathbb{S}$ is symmetric) and the symmetry of a irreducible Riemannian manifold of dimension $\geqslant 3$ and the connected component of the holonomy group on the manifold. 
\end{proof}

The case where $\Hol_p(g)$ acts transitively on the unit sphere in $\mathcal{T}_p\mathcal{M}$, the holonomy group falls into the Berger's classification; but if $\Hol_p(g)$ does not act transitively on the unit sphere, a Riemannian manifold $(\mathcal{M}, g)$ is a locally symmetric space.

From the latter proposition, C.E. Olmos \cite{Olmos "A geometric proof of the Berger Holonomy Theorem"} has obtained a geometric proof of Theorem \ref{theorema "Berger"} thanks to the use of \emph{submanifold geometry of orbits in Euclidean space}, by this showing the existence of a link between the holonomy groups of the normal connection in submanifolds of Euclidean space and the holonomy groups of Riemannian manifolds.

\subsection{Holonomy in Abelian Phase Factor: Gauge Group of Electromagnetic Interactions}
\label{subsection "Holonomy in Abelian Phase Factor: Gauge Group of Electromagnetic Interactions"}

Berger's classification \ref{item "Berger's classification unitary group"} gets us motivated to take a closer look at the group $U(1)$, referred to as the \emph{group of 1-dimensional unitary transformations} or \emph{1-parameter (Abelian) unitary group}.

\enumerationisinitium
\item Let 
\[
	U(1) \hookrightarrow \mathring{\mathcal{P}} \xrightarrow{\pi} \mathcal{M}
\]
be a principal $U(1)$-bundle over $\mathcal{M}$. It is well-known that every element of $U(1)$ is a point on the circle group marked with a \emph{unit complex number} or \emph{phase factor} $e^{i\theta}$ in the complex plane $\mathbb{C}$ (where $i$ is the imaginary unit and $\theta$ is the angle of rotation), while the set of all phases $\exp{\{i\theta \mid \theta \in \mathbb{R}}\}$ generates a representation $\varphi_k$ of $U(1)$.

Given an inclusion map 
\begin{equation}
	\varphi_k \colon U(1) \cong \mathbb{S}^1 = \{e^{i\theta}\} \hookrightarrow GL_1(\mathbb{C}),
\end{equation}
with $k \in \mathbb{Z}$ ($k$ is an integer), we can then say that the \emph{phase} of a point electric (charged particle) is an element $e^{i\theta}$ of $U(1) \cong \mathbb{S}^1$ on $\mathbb{C}$. Specifically, this involves the group of \emph{phase transformations} for some complex-valued (wave)function.

The irreducible representation of the unit circle on $\mathbb{C}$ is a map $U(1) \xrightarrow{\varphi_k} GL_1(\mathbb{C})$, since there is $\varphi_k(g)z = g^k(z)$, for all $k \in \mathbb{Z}$, $g \in U(1)$ and $z \in \mathbb{C}$. So, in other words, the representation $\varphi_k$ of $U(1)$ acts on the circle by rotations, when 
\begin{equation}
	g = e^{i\theta}, \enspace 0 \leqslant \theta \in \mathbb{R} < 2\pi, 
\end{equation}
and its vector bundle has the form 
\[
	\pi_{\varphi_k} \colon \mathring{\mathcal{P}} \times \mathbb{C} \to \mathcal{M}.
\]
\item In physics, the group $U(1)_\mathrm{em} \viz U_{1(\mathrm{em})}$, or, simply, $U(1)$, corresponds to the \emph{gauge group} of \emph{electromagnetism} and, more generally, of \emph{Abelian gauge theories}, a representation of which in \textsc{qft} (when considering the interactions of relativistic point-like particles) is the quantum electrodynamics (\textsc{qed}).
\subenumerationisinitium
\item The group $U(1)_\mathrm{em}$ is a \emph{symmetry group}, as there is invariance under local unitary transformations. 

\item For instance, one could think of a (electrically) charged point-like particle moving in an electromagnetic field on space-time; and observe the line integral or, equivalently, the circulation of a connection $\vec{A}$, aka \emph{vector potential}, on a loop. The definition of $U(1)_\mathrm{em}$-holonomy descends from the exponential of the line integral along a loop $\gamma_\mathrm{c}$,
\begin{equation} 
	\Hol_{\vec{A}}(\gamma_\mathrm{c}) \equival \exp{\left\{i \oint_{\gamma_\mathrm{c}}\vec{A}_\mu dx^\mu\right\}} \in U(1)_\mathrm{em}.
\end{equation} 	
The Abelian phase factor $e^{i\oint_{\gamma_\mathrm{c}}\vec{A}}$ is, to put it briefly, the very $U(1)_\mathrm{em}$-holonomy.
\subenumerationisfinis
\item In this context, and nevertheless in a broader view of gauge transformations (Yang–Mills theory \& non-Abelian Lie groups, \textsc{qcd}, Chern–Simons theory, \textsc{lqg}), the development of the holonomy is related to the gauge connection using a gauge invariant observable, the so-called \emph{Wilson loop} \cite{Wilson "Confinement of quarks"}.
\enumerationisfinis

\vspace{10mm}

\setcounter{secnumdepth}{0}  
\section{References and Bibliographic Details}
\setcounter{secnumdepth}{3}
\markright{References and Bibliographic Details}

\begingroup
\footnotesize
\noindent Section \ref{subsection "Fiber Bundles (Tangent, Cotangent and Vector Bundle)"}

\begin{indent paragraph: 15pt}
Definition \ref{definitio "Fiber bundle"}: cf. \cite[p. 118]{Geoghegan "Topological Methods in Group Theory"} \cite[pp. 2-3]{Luke and Mishchenko "Vector Bundles and Their Applications"} \cite[p. 26]{Taubes "Differential Geometry. Bundles Connections Metrics and Curvature"}. — Definition \ref{definitio "Vector bundle"}: cf. \cite[sec. 3.1]{Abate Tovena "Geometria Differenziale"} \cite[chap. 3, secc. 1-2]{Husemoller "Fibre Bundles"} \cite[pp. 49-50, 76-77]{Kolar Michor Slovak "Natural Operations in Differential Geometry"} \cite[pp. 16-19]{Lee "Riemannian Manifolds: An Introduction to Curvature"} \cite[pp. 257-258, 270]{Lee "Manifolds and Differential Geometry"} \cite[pp. 249-250]{Lee "Introduction to Smooth Manifolds"} \cite[pp. 55-58]{Rudolph Schmidt "Differential Geometry and Mathematical Physics I"}. — Scholium \ref{scholium "Real and complex vector bundle"}: on the complex vector bundles, see e.g. \cite[chap. I]{Kobayashi "Differential Geometry of Complex Vector Bundles"}; on the vector bundles over $\mathbb{CP}$-like spaces, see \cite{Okonek Schneider Spindler "Vector Bundles on Complex Projective Spaces"}.
\end{indent paragraph: 15pt}

\noindent Section \ref{section "Christoffel Symbols"}

\begin{indent paragraph: 15pt}
On the Christoffel symbols, cf. \cite[p. 45]{Joshi "Global Aspects in Gravitation and Cosmology"} \cite[p. 34]{Joshi "Gravitational Collapse and Spacetime Singularities"} \cite[pp. 171-173, 195]{Jost "Riemannian Geometry and Geometric Analysis"} \cite[p. 141]{Kreyszig "Differential Geometry"} \cite[p. 142]{Manton Mee "The Physical World. An Inspirational Tour of Fundamental Physics"} \cite[p. 31]{Oliva "Geometric Mechanics"} \cite[pp. 78-80, 329]{Schafer Schmidt "Tensor Analysis and Elementary Differential Geometry for Physicists and Engineers"}. 
\end{indent paragraph: 15pt}

\noindent Section \ref{subsection "Matrix Notation and Kronecker delta-Function"}

\begin{indent paragraph: 15pt}
Cf. \cite[p. 240]{Banchoff Lovett "Differential Geometry of Curves and Surfaces"} \cite[p. 241]{Bishop Goldberg "Tensor Analysis on Manifolds"} \cite[p. 45]{Gasperini "Theory of Gravitational Interactions"} \cite[pp. 69-70]{Lee "Riemannian Manifolds: An Introduction to Curvature"} \cite[p. 74]{Vazquez Gilkey Nikcevic "Geometric Realizations of Curvature"}.
\end{indent paragraph: 15pt}

\noindent Section \ref{subsection "Relativistic Gravitation as a Genesis of the Parallel Transport"}

\begin{indent paragraph: 15pt}
On the Levi-Civita connection, see e.g. \cite[pp. 55-56]{do Carmo "Riemannian Geometry"} \cite{Davies and Yano "The influence of Levi-Civita's notion of parallelism on differential geometry"} \cite[§§ 18-19]{Gentili Podesta Vesentini "Lezioni di geometria differenziale"} \cite[chap. IV]{Kobayashi and Nomizu "Foundations of Differential Geometry I"} \cite[pp. 132-135]{Kriele "Spacetime. Foundations of General Relativity and Differential Geometry"}; about the connection in \textsc{qft}, see \cite[sec. 7.5]{Parker and Toms "Quantum Field Theory in Curved Spacetime"}. 
\end{indent paragraph: 15pt}

\noindent Section \ref{subsection "Parallel Transport: a Way of Viewing Euclidean Type Small Spaces in a Curved Space"}

\begin{indent paragraph: 15pt}
Cf. \cite[p. 1]{Besse "Einstein Manifolds"} \cite[p. 98]{Cartan "Riemannian Geometry in an Orthogonal Frame"} \cite[sec. 1.1]{Hebey "Nonlinear Analysis on Manifolds: Sobolev Spaces and Inequalities"} \cite[sec. 10.2]{Vargas "Differential Geometry For Physicists And Mathematicians. Moving Frames and Differential Forms: From Euclid Past Riemann"}; see also \cite[pp. 101-102]{Cordero Dodson de Leon "Differential Geometry of Frame Bundles"}.
\end{indent paragraph: 15pt}

\noindent Section \ref{subsection "Severi's Theorem (Non-ambient Parallelism of Levi-Civita), and Other Contour Jottings"}
 
\begin{indent paragraph: 15pt}
About the point \ref{item "Historical jotting"}, see \cite[p. 47]{Veblen "Invariants of Quadratic Differential Forms"} for a historical jotting.
\end{indent paragraph: 15pt}

\noindent Section \ref{subsection "Koszul Connection, Linear Connection, and Covariant Derivative"}

\begin{indent paragraph: 15pt}
Definition \ref{definitio "Connection on a vector bundle"}: cf. \cite[pp. 316-317]{Abate Tovena "Geometria Differenziale"} \cite[sec. 2.5]{Andrews Hopper "The Ricci Flow in Riemannian Geometry: A Complete Proof of the Differentiable 1/4-Pinching Sphere Theorem"} \cite[pp. 386-388]{Johannesen "Smooth Manifolds and Fibre Bundles with Applications to Theoretical Physics"} \cite[pp. 49-50]{Lee "Riemannian Manifolds: An Introduction to Curvature"} \cite[chap. 35, § 6]{Postnikov "Geometry VI. Riemannian Geometry"}. — On the Koszul connection, see e.g. \cite[sec. 3.2]{Kupeli "Singular Semi-Riemannian Geometry"}. — Definition \ref{definitio "Linear connection"}: see e.g. \cite[chap. III, sec. 3]{Kobayashi and Nomizu "Foundations of Differential Geometry I"} \cite[chap. I, sec. 3]{Nomizu Sasaki "Affine differential geometry: Geometry of Affine Immersions"}. — Definition \ref{definitio "Parallel vector field along a curve"}: cf. e.g. \cite[p. 10]{Chow Chu Glickenstein Guenther Isenberg Ivey Knopf Lu Luo Ni "The Ricci Flow: Techniques and Applications Part II: Analytic Aspects"}. — Definition \ref{definitio "Parallel transport map"}: cf. \cite[pp. 25-26]{Joyce "Compact Manifolds with Special Holonomy"} \cite[pp. 24-25]{Joyce "Riemannian Holonomy Groups and Calibrated Geometry"}. — Proposition \ref{propositio "Connection as limit of an incremental ratio"}: cf. \cite[pp. 309-310]{Arnold "Mathematical Methods of Classical Mechanics"} \cite[pp. 149-150]{Villani "Optimal Transport: Old and New"}.
\end{indent paragraph: 15pt}

\noindent Section \ref{subsection "Levi-Civita Connection Theorem on a (pseudo-)Riemannian Manifold"}

\begin{indent paragraph: 15pt}
Definition \ref{definitio "Metric-compatible connection"}: cf. \cite[11.2b]{Frankel "The Geometry of Physics: An Introduction"} \cite[pp. 90-91]{Langlois Deville "Slow Viscous Flow"} \cite[chap. III, sec. 1.1]{Thi and Fomenko "Minimal Surfaces Stratified Multivarifolds and the Plateau Problem"} \cite[pp. 232-233]{Zelikin "Control Theory and Optimization I: Homogeneous Spaces and the Riccati Equation in the Calculus of Variations"}.
\end{indent paragraph: 15pt}

\noindent Section \ref{subsection "Fermi (Locally Geodesic Cartesian-like) Coordinates"}

\begin{indent paragraph: 15pt}
See \cite[chap. II, sec. 25]{Eisenhart "Non-Riemannian Geometry"} \cite[chap. III, § 8]{Schouten "Ricci-Calculus. An Introduction to Tensor Analysis and Its Geometrical Applications"} \cite{Raifeartaigh "Fermi Coordinates"} \cite[chap. 2]{Gray "Tubes"}.	
\end{indent paragraph: 15pt} 

\noindent Section \ref{section "Connection Forms"}

\begin{indent paragraph: 15pt}
On the absolute parallelism (\emph{Fernparallelismus}), see e.g. \cite{Eisenhart "Spaces admitting complete absolute parallelism"} \cite{Wolf "On the geometry and classification of absolute parallelisms. I", Wolf "On the geometry and classification of absolute parallelisms. II"} \cite[pp. 423-426]{Figueroa-O'Farrill "Lorentzian symmetric spaces in supergravity"}. 
\end{indent paragraph: 15pt}
	
\noindent Section \ref{subsection "Connection 1-Form"}

\begin{indent paragraph: 15pt}
On the differential form, see \cite[sec. 2.6]{Cartan H. "Differential Forms"}; then cf. \cite[sec. 6.2]{Abate Tovena "Geometria Differenziale"} \cite[pp. 370-371]{Lee "Manifolds and Differential Geometry"} \cite[p. 254]{Rudin "Principles of Mathematical Analysis"} \cite[chap. 4, §§ 21.1, 21.7]{Tu "Differential Geometry: Connections Curvature and Characteristic Classes"}. — Proposition \ref{propositio "Connection 1-form"}: cf. \cite[pp. 201-202]{Darling "Differential Forms and Connections"}.
\end{indent paragraph: 15pt}

\noindent Section \ref{subsection "Curvature 2-Form and Structural Equations"}

\begin{indent paragraph: 15pt}
On the connection form and the Cartan structure equations, cf. \cite[§ 5.5]{Akivis Rosenfeld "Elie Cartan"} \cite[pp. 306, 310]{Choquet-Bruhat DeWitt-Morette with Dillard-Bleick "Analysis Manifolds and Physics I"} \cite[p. 315]{Dubrovin Fomenko Novikov "Modern Geometry I"} \cite[p. 19]{Li "Geometric Analysis"} \cite[pp. 201-202]{Morita "Geometry of Differential Forms"} \cite[pp. 111-112]{Petersen "Riemannian Geometry"} \cite[chap. 2, § 11.1]{Tu "Differential Geometry: Connections Curvature and Characteristic Classes"}.
\end{indent paragraph: 15pt}

\noindent Section \ref{section "Cartan Structure: Generalized Space and Lie Algebra-valued Form"}
 
\begin{indent paragraph: 15pt}
About the Lie's sentence in epigraph, a partial En. transl. of the Sophus Lie's work \cite{Lie "Theorie der transformationsgruppen Erste Abschnitt"} is available in \cite{Lie "Theory of Transformation Groups I"}. 
\end{indent paragraph: 15pt}

\noindent Section \ref{subsection "Espace Généralisé of the Klein Geometry: Cartan Geometry and Connection"}

\begin{indent paragraph: 15pt}
On the Cartan structure (geometry, connection, etc.), see \cite{Kobayashi "Theory of Connections"} \cite[chap. IV, secc. 3-4]{Kobayashi "Transformation Groups in Differential Geometry"} \cite{Griffiths "On Cartan's method of Lie groups and moving frames as applied to uniqueness and existence questions in differential geometry"} \cite{Alekseevsky Michor "Differential geometry of Cartan connections"} \cite[chap. 5, §§ 1-3, chap. 6, § 2]{Sharpe "Differential Geometry Cartan's Generalization of Klein's Erlangen Program"} \cite{Cap Schichl "Parabolic geometries and canonical Cartan connections"} \cite{Lam "Totally Asymmetric Torsion on Riemann-Cartan Manifold"} \cite[sec. 1.5]{Cap Slovak "Parabolic Geometries I: Background and General Theory"} \cite[chapp. 6-7, sec. 10.3]{Crampin Saunders "Cartan Geometries and their Symmetries. A Lie Algebroid Approach"}. 
\end{indent paragraph: 15pt}

\noindent Section \ref{subsection "From g-Space Homogeneity to Blob-like Space"}

\begin{indent paragraph: 15pt}
Example \ref{exemplum "Commutative-homogeneity of space}: \cite[p. 116]{Bernshtein and Rozenfel'd "Homogeneous spaces of infinite-dimensional Lie algebras and characharacteristic classes of foliations"}; for completeness, see \cite[pp. 32-33]{Brocker tom Dieck "Representations of Compact Lie Groups"} \cite[chap. 1]{Timashev "Homogeneous Spaces and Equivariant Embeddings"}.
\end{indent paragraph: 15pt}

\noindent Section \ref{subsection "Maurer–Cartan Forms and Equations"}

\begin{indent paragraph: 15pt}
On the Maurer–Cartan forms, cf. \cite[chap. V, § IV]{Chevalley "Theory of Lie groups I"} \cite[pp. 71-74]{Olver "Equivalence Invariants and Symmetry"} \cite[pp. 98-100]{Sharpe "Differential Geometry Cartan's Generalization of Klein's Erlangen Program"} \cite[p. 339]{Tuynman "Supermanifolds and Supergroups"} \cite[sec. 9.4.7]{Zeidler "Quantum Field Theory III: Gauge Theory"}. — On the Maurer–Cartan equations, cf. e.g. \cite[p. 308]{Garcia-Diaz "Exact Solutions in Three-Dimensional Gravity"} \cite[p. 137]{Helgason "Differential Geometry Lie Groups and Symmetric Spaces"} \cite[pp. 91-92]{Morita "Geometry of Differential Forms"} \cite[p. 181]{Pommaret "Partial Differential Equations and Group Theory: New Perspectives for Applications"} \cite[p. 107]{Tondeur "Geometry of Foliations"} \cite[pp. 236-238]{Schucking Surowitz "Einstein's Apple: Homogeneous Einstein Fields"}.
\end{indent paragraph: 15pt}

\noindent Section \ref{subsubsection "Ricci Rotation Coefficients and Tetrad Formalism"} 

\begin{indent paragraph: 15pt}
On the Ricci rotation coefficients, see \cite[sec. 4.5]{de Felice Clarke "Relativity on curved manifolds"} \cite{Levy "Ricci's coefficients of rotation"} \cite[sec. 9.7.1]{Sharan "Spacetime Geometry and Gravitation"}. — On the tetrad basis, see \cite[sec. 8.4]{Alcubierre "Introduction to 3 + 1 Numerical Relativity"} \cite[chap. 1, sec. 7]{Chandrasekhar "The Mathematical Theory of Black Holes"} \cite[sec. 3.4]{Ellis "Cosmological models"} \cite[secc. 3-4]{Dhurandhar "Tetrads the Newman-Penrose formalism and spinors"} \cite[p. 416]{Shibata "Numerical Relativity"} \cite[p. 73]{Torres del Castillo "Spinors in Four-Dimensional Spaces"}. 
\end{indent paragraph: 15pt}

\noindent Section \ref{subsection "Connection with Lie Algebra Decomposition and Gauge Model"}

\begin{indent paragraph: 15pt}
On the Cartan decomposition of Lie algebras, see e.g. \cite[sec. 1.2.3]{Vilenkin and Klimyk "Representation of Lie Groups and Special Functions I"}, and more fully \cite[chap. VII]{Bourbaki "Elements of Mathematics: Lie Groups and Lie Algebras 7-9"}. — Proposition \ref{propositio "Cartan connection with decomposition"}: see e.g. \cite[pp. 61-64]{Audin "Torus Actions on Symplectic Manifolds"} \cite{Cap Slovak and Soucek "Bernstein-Gelfand-Gelfand sequences"} \cite{Ruh "Cartan Connections"}. — Definition \ref{definitio "Cartan curvature"}: cf. e.g. \cite[pp. 58-60, 92]{Azcarraga and Izquierdo "Lie groups Lie algebras cohomology and some applications in physics"} \cite[p. 340]{Clelland "From Frenet to Cartan: The Method of Moving Frames"}; see also \cite[pp. 491-492]{Cap Schichl "Parabolic geometries and canonical Cartan connections"} \cite[p. 7]{Cap Slovak "Parabolic Geometries I: Background and General Theory"} \cite[p. 9]{Cap Slovak and Soucek "Bernstein-Gelfand-Gelfand sequences"}.
\end{indent paragraph: 15pt}

\noindent Section \ref{subsection "Geodesic: Some Features"} 

\begin{indent paragraph: 15pt}
On the shortest geodesic path, see e.g. \cite{Rossl and Theisel "Couple Points: A Local Approach to Global Surface Analysis"}. — Definition \ref{definitio "Geodesic"}: cf. \cite[p. 251 ff.]{Abate Tovena "Curves and Surfaces"} \cite[p. 155]{Landsman "Mathematical Topics Between Classical and Quantum Mechanics"} \cite[pp. 223-224]{Montiel Ros "Curves and Surfaces"} \cite[sec. 5.2]{Petersen "Riemannian Geometry"} \cite[sec. 8.2]{Taubes "Differential Geometry. Bundles Connections Metrics and Curvature"} \cite[p. 153]{Villani "Optimal Transport: Old and New"}. — Scholium \ref{scholium "Geodesic: on the distance minimizing issue"}: see \cite[pp. 174-177]{Agricola Friedrich "Global Analysis: Differential Forms in Analysis Geometry and Physics"} \cite[p. 48]{Ryan Jr. and Shepley "Homogeneous Relativistic Cosmologies"}.
\end{indent paragraph: 15pt}

\noindent Section \ref{subsection "Geodesics as Solutions of the Euler–Lagrange Equations"}
 
\begin{indent paragraph: 15pt}
On the Euler–Lagrange equations, see \cite[chap. 5, §§ 31.1]{Dubrovin Fomenko Novikov "Modern Geometry I"} \cite[pp. 211-212]{Montgomery "A Tour of Subriemannian Geometries Their Geodesies and Applications"} \cite[p. 7]{Takhtajan "Quantum Mechanics for Mathematicians"}; see also \cite{Lee Leok McClamroch "Global Formulations of Lagrangian and Hamiltonian Dynamics on Manifolds. A Geometric Approach to Modeling and Analysis"} referring to the curves $[0, 1] \to \mathbb{R}^3$, $\to \mathbb{S}^1$, $\to \mathbb{S}^2$ and $\to SO_3(\mathbb{R})$ on pp. 128-129, 204-205, 268 and 310 respectively. — On the kinetic and potential energy, cf. \cite[sec. 7.2]{Awrejcewicz "Classical Mechanics: Dynamics"} \cite[sec. 3.9]{Calin Chang "Geometric Mechanics on Riemannian Manifolds: Applications to Partial Differential Equations"} \cite[p. 139]{Fasano Marmi "Analytical Mechanics"} \cite[p. 562]{Garling "A Course in Mathematical Analysis II: Metric and Topological Spaces Functions of a Vector Variable"} \cite[sec. 3.1]{Kibble Berkshire "Classical Mechanics"} \cite[p. 10]{Takhtajan "Quantum Mechanics for Mathematicians"}. — On the Euler–Lagrange and geodesic equations for the Lagrangian, cf. \cite[pp. 40-41]{Synge and Schild "Tensor Calculus"} \cite[pp. 169-170]{Das An. "Tensors: The Mathematics of Relativity Theory and Continuum Mechanics"} \cite[pp. 151-152]{Das As. "Lectures on Gravitation"}. — About the space-like and time-like geodesics, see \cite{Calabi and Markus "Relativistic Space Forms"}. — On the proper time in Minkowskian geometry, see e.g. \cite[sec. 2.6]{Rowe "Geometrical Physics in Minkowski Spacetime"} \cite[pp. 227-230]{Naber "The Geometry of Minkowski Spacetime: An Introduction to the Mathematics of the Special Theory of Relativity"}.
\end{indent paragraph: 15pt}

\noindent Section \ref{section "On the Theory of Holonomy: Connections and Loops"} 

\begin{indent paragraph: 15pt}
About the Poinsot's mathematical physics, see \cite{DiBenedetto "Classical Mechanics: Theory and Mathematical Modeling"}.
\end{indent paragraph: 15pt}

\noindent Section \ref{subsection "Introductory Remarks with a Scholium on $n$-Torus"}

\begin{indent paragraph: 15pt}
About the holonomy and loop group, see \cite[chap. 1]{Gambini Pullin "Loops Knots Gauge Theories and Quantum Gravity"} \cite[p. 51]{Pressley and Segal "Loop Groups"}. — For a general introduction to holonomy, especially in the context of foliations, see \cite[chap. 2]{Candel Conlon "Foliations I"}. — Scholium \ref{scholium "Torus"}: cf. \cite[pp. 18, 185]{Audin "Torus Actions on Symplectic Manifolds"} \cite[p. 131]{Buchstaber Panov "Toric Topology"}.
\end{indent paragraph: 15pt}

\noindent Section \ref{subsection "Bundles and Holonomy Groups; Ambrose–Singer Theorem"}

\begin{indent paragraph: 15pt}
About the holonomy and linear connection, see \cite[sec. 2.7]{Agricola "Non-integrable geometries torsion and holonomy"} \cite[app. A.2]{Baumann and McAllister "Inflation and String Theory"} \cite[sec. 2.1]{Galaev and Leistner "Holonomy groups of Lorentzian manifolds: classification examples and applications"} \cite[sec. 2.1]{Galaev and Leistner "Recent developments in pseudo-Riemannian holonomy theory"} \cite[chap. I, secc. 1.1, 1.4]{Maurin "The Riemann Legacy: Riemannian Ideas in Mathematics and Physics"} \cite[sec. 10.3]{Petersen "Riemannian Geometry"}. — Definition \ref{definitio "Principal G-bundle"}: cf. \cite[pp. 84-85]{Davis Kirk "Lecture Notes in Algebraic Topology"} \cite[sec. 1.9]{Dodson Galanis and Vassiliou "Geometry in a Frechet Context: A Projective Limit Approach"} \cite[pp. 26-27, 80]{Fatibene and Francaviglia "Natural and Gauge Natural Formalism for Classical Field Theories: A Geometric Perspective including Spinors and Gauge Theories"} \cite[pp. 265-266, 270-271]{Gadea Masque Mykytyuk "Analysis and Algebra on Differentiable Manifolds"} \cite[sec. 4.6]{Ivancevic Ivancevic "Applied Differential Geometry: A Modern Introduction"} \cite[pp. 141-142]{Mukherjee "Differential Topology"} \cite[pp. 16, 244]{Naber "Topology Geometry and Gauge Fields: Foundations"}. — Definitions \ref{definitio "Preparatory definition and horizontality"} and \ref{definitio "Holonomy group of a principal G-bundle connection"}: cf. \cite[pp. 190-193]{Mangiarotti Sardanashvily "Connections in Classical and Quantum Field Theory"} \cite[sec. 3.5]{Morgan "An Introduction to Gauge Theory"} \cite[sec. 5.6]{Moroianu "Lectures on Kahler Geometry"} \cite[sec. 2.3]{Joyce "Riemannian Holonomy Groups and Calibrated Geometry"}. — Theorem \ref{theorema "Ambrose–Singer"}: see \cite[sec. 1.4]{Nomizu "Recent Development in the Theory of Connections and Holonomy Groups"} \cite[chap. II, sec. 8]{Kobayashi and Nomizu "Foundations of Differential Geometry I"} \cite[pp. 167-168]{Ramanan "Global Calculus"} \cite[p. 221]{Epstein Elzanowski "Material Inhomogeneities and their Evolution: A Geometric Approach"}.
\end{indent paragraph: 15pt}

\noindent Section \ref{subsection "Holonomy in Riemannian Spaces"}

\begin{indent paragraph: 15pt}
Cf. \cite[p. 638]{Berger "A Panoramic View of Riemannian Geometry"} \cite[p. 394 ff.]{Berndt Console Olmos "Submanifolds and Holonomy"} \cite[pp. 280-281]{Besse "Einstein Manifolds"} \cite[pp. 213-214]{Huybrechts "Complex Geometry: An Introduction"} \cite[sec. 3.2]{Leung "Geometric Structures on Riemannian Manifolds"} \cite[pp. 62, 137]{Rudolph Schmidt "Differential Geometry and Mathematical Physics II"}. — Scholium: \ref{scholium "Holonomic automorphism"}: cf. \cite[sec. 7.4, and pp. 373-374]{Eschrig "Topology and Geometry for Physics"} \cite[p. 360]{Husemoller "Elliptic Curves"}; see also \cite[chap. III, § 1.2, pp. 213-214]{Bourbaki "Elements of Mathematics: Lie Groups and Lie Algebras 1-3"}. — Definition \ref{definitio "Holonomy algebra of the metric tensor"}: cf. \cite[p. 121]{Getzler "The Thom Class of Mathai and Quillen and Probability Theory"}.
\end{indent paragraph: 15pt}

\noindent Section \ref{subsection "Berger's Classification; from (Hyper)kählerian to Spin Manifolds; Simons Theorem"}

\begin{indent paragraph: 15pt}
On the Berger's classification/theorem, see e.g. \cite[lecture 8]{Bryant "An Introduction to Lie Groups and Symplectic Geometry"} \cite[sec. 3.1]{Bryant "Geometry of Manifolds with Special Holonomy: 100 Years of Holonomy"} \cite[sec. 1.2.2]{Calvaruso Castrillon Lopez "Pseudo-Riemannian Homogeneous Structures"} \cite[secc. 3.3-3.5]{Gross Huybrechts Joyce "Calabi-Yau Manifolds and Related Geometries"} \cite[sec. 3.4]{Joyce "Riemannian Holonomy Groups and Calibrated Geometry"} \cite[app. B, pp. 849-850]{Ortin "Gravity and Strings"} \cite[secc. 3.2.-3.5]{Schwachhofer "Holonomy Groups and Algebras"}. — Theorem \ref{theorema "Berger"}: on the Calabi–Yau and Ricci-flat Kähler manifolds, see \cite[sec. 3.5]{Font and Theisen "Introduction to String Compactification"} \cite[sec. 15.4]{Green Schwarz Witten "Superstring theory Vol. 2: Loop Amplitudes Anomalies and Phenomenology"} \cite[secc. 1.3-1.5]{Joyce "Constructing compact manifolds with exceptional holonomy"}; Kähler manifolds with vanishing first Chern class are studied in \cite{Michelsohn "Kahler manifolds with vanishing first Chern class}; on the symplectic structure of Kähler manifolds, see e.g. \cite[sec. 8]{Beauville "Varietes Kahleriennes dont la premiere classe de Chern est nulle"}, and on the hyperkähler manifold, see \cite[p. 138]{Hitchin "Hyperkahler manifolds"} \cite{Verbitsky Kaledin "Hyperkahler manifolds"}; about the holonomy, $Sp(m)$ and $Sp(1)$ groups, see \cite{Gray "A note on manifolds whose holonomy group is a subgroup of $Sp(n) [cdot] Sp(1)$"}; on the quaternionic Kähler and Kähler–Einstein manifolds, see \cite{Marchiafava Romani "Sui fibrati con struttura quaternionale generalizzata"} \cite{Salamon S. "Quaternionic Kahler manifolds"} \cite{Alekseevsky and Marchiafava "Hypercomplex Structures on Quaterionic Manifolds"} \cite[pp. 136-137]{Dragomir Ornea "Locally Conformal Kahler Geometry"} \cite[sec. 3, pp. 383-388]{Semmelmann and Weingart "Vanishing Theorems for Quaternionic Kahler Manifolds"} \cite{Dotti "Quaternion Kahler flat manifolds"} \cite[sec. 13.5.2]{Berger "A Panoramic View of Riemannian Geometry"} \cite[§ 9.6]{Brendle "Ricci Flow and the Sphere Theorem"}; about \ref{item "Quaternionic projective space, quaternionic Hopf bundles, etc."} of \ref{item "Holonomy in Berger's theorem, quaternionic objects, and more"} concerning the quaternionic projective space and quaternionic Hopf bundles, cf. \cite[p. 217]{Naber "Topology Geometry and Gauge Fields: Foundations"} \cite[pp. 26-27]{Naber "Topology Geometry and Gauge Fields: Interactions"}; on the exceptional holonomy, see \cite{Salamon S. "Manifolds with Exceptional Holonomy"}, and for a synopsis on $G_2$, see \cite{Agricola "Old and New on the Exceptional Group $G_2$"}; on the Ricci-flat $G_2$-manifold, as well as on $G_2$ holonomy, see \cite{Bonan "Sur des varietes riemanniennes d'holonomie $G_2$ ou $Spin(7)$"} \cite{Bryant "Metrics with exceptional holonomy"} \cite[§ 2]{Bryant and Salamon "On the construction of some complete metrics with exceptional holonomy"} \cite{Cabrera Monar and Swann "Classification of $G_2$-Structures"} \cite{Joyce "Compact Riemannian 7-manifolds with holonomy $G_2$. I", Joyce "Compact Riemannian 7-manifolds with holonomy $G_2$. II"} \cite[secc. 9.4.1-9.4.2]{Grassi and Rossi "Large N dualities and transitions in geometry"} \cite{Apostolov Salamon "Kahler Reduction of Metrics with Holonomy $G_2$"}, and for a report, see \cite{Duff "M-theory on manifolds of $G_2$ holonomy: the first twenty years"}; on the relationship with the imaginary octonions, cf. \cite{Ohashi "A method of determining the $SO(7)$-invariants for curves in ImO by their $G_2$-invariants"}, and on the octonions, see \cite{Baez "The Octonions"} \cite[chap. III]{Conway Smith "On Quaternions and Octonions: Their Geometry Arithmetic and Symmetry"}; regarding the manifold in $8\mathrm{D}$ with holonomy $\Spin(7)$, see \cite{Joyce "A new construction of compact 8-manifolds with holonomy $Spin(7)$"}; on the spin group $\Spin(7)$, $G_2$, $SO(7)$, $SO(8)$, and other related arguments, cf. \cite[p. 348]{Lawson Jr. and Michelsohn "Spin Geometry"} \cite[sec. 23.6, pp. 310-311]{Lounesto "Clifford Algebras and Spinors"} \cite[p. 159]{Dixon "Division Algebras: Octonions Quaternions Complex Numbers and the Algebraic Design of Physics"} \cite{Boya Campoamor-Stursberg "Composition algebras and the two faces of $G_2$"} \cite{Hashimoto and Ohashi "Realizations of subgroups of $G_2$ $Spin(7)$ and their applications"} \cite[secc. 2.3-2.5, pp. 51-62]{Nadirashvili Tkachev Vladut "Nonlinear Elliptic Equations and Nonassociative Algebras"} \cite[p. 28]{Schafer "Nearly Pseudo-Kahler Manifolds and Related Special Holonomies"}. — Theorem \ref{theorema "Simons"}: see \cite[p. 300]{Besse "Einstein Manifolds"} \cite[chap. 8]{Berndt Console Olmos "Submanifolds and Holonomy"}.
\end{indent paragraph: 15pt}

\noindent Section \ref{subsection "Holonomy in Abelian Phase Factor: Gauge Group of Electromagnetic Interactions"}
 
\begin{indent paragraph: 15pt}
On the group $U(1)$, see e.g. \cite[pp. 388-389]{Naber "Topology Geometry and Gauge Fields: Foundations"} \cite[pp. 39, 43]{Naber "Topology Geometry and Gauge Fields: Interactions"} \cite[chap. 2, sec. 38.2.1, chap. 45]{Woit "Quantum Theory Groups and Representations"}. — For the role of $U(1)$ in physics, see e.g. \cite[secc. 8.3, 9.3]{Costa Fogli "Symmetries and Group Theory in Particle Physics. An Introduction to Space-time and Internal Symmetries"}. — On the phase factor and holonomy of $U(1)$ in a physical (electromagnetic) perspective, cf. \cite[pp. 609-610]{Bohm "Quantum Mechanics: Foundations and Applications"} \cite[pp. 171-172, 231-242]{Baez Muniain "Gauge Fields Knots and Gravity"} \cite[pp. 62-63]{Ashtekar "Classical and Quantum Physics of Isolated Horizons: A Brief Overview"} \cite[sec. 3.1, pp. 243-247]{Majid "Meaning of Noncommutative Geometry and the Planck-Scale Quantum Group"} \cite[pp. 27-28]{Bohm Mostafazadeh Koizumi Niu Zwanziger "The Geometric Phase in Quantum Systems"} \cite[p. 121]{Thiemann "Lectures on Loop Quantum Gravity"} = \cite[p. 391]{Thiemann "Modern Canonical Quantum General Relativity"} \cite[p. 15]{Rovelli "Quantum Gravity"}. — About the holonomy of the set of loops in various contexts of physics (including the holonomy of the gauge connection), see e.g. \cite{Witten "Quantum Field Theory and the Jones Polynomials"} \cite[app. C.2, pp. 245-246]{Carlip "Quantum Gravity in 2 + 1 Dimensions"} \cite[pp. 30, 63]{Gambini Pullin "Loops Knots Gauge Theories and Quantum Gravity"} \cite[pp. 285-317]{Rovelli and Gaul "Loop Quantum Gravity and the Meaning of Diffeomorphism Invariance"} \cite[pp. 35, 61-62]{Bandyopadhyay "Geometry Topology and Quantum Field Theory"} \cite[pp. 27, 32]{Marino "Chern-Simons theory matrix models and topological strings"}.
\end{indent paragraph: 15pt}

\endgroup

\chapter{Panoramic Miscellanea II. Space Forms, Möbius (Projective) Transformations, and Fuchsian Group; Groupable Synopsis—On the Spin(or)}
\chaptermark{Panoramic Miscellanea II}{}
\label{chapter "Panoramic Miscellanea II. Space Forms, Möbius (Projective) Transformations, and Fuchsian Group; Groupable Synopsis—On the Spin(or)"}

\begingroup
\footnotesize
Geometry has its origin in the direct observation of objects in the external world, which is the physical space, and from the intuition of them it derives its first undemonstrable truths, necessary for its theoretical evolution, which are the axioms [\,\dots]. However, to be exact, geometry must represent the objects arising from the observation by way of pure abstract forms and the axioms with well-determined hypotheses, i.e. independent of the intuition, so that geometry becomes part of pure mathematics [\,\dots]. It is necessary to distinguish physical space from intuitive space, and the latter from geometric space [\,\dots]. Geometric space is precisely that part of the pure extension in which the physical and intuitive space is represented, but in turn it does not have a representation in the real world for all its forms. And \emph{while physical and intuitive space cannot be defined, geometric space can instead be defined} [\,\dots]. The three geometries [of Euclidean, elliptic, and hyperbolic spaces] in a very small field give the same results with good approximation [\,\dots]. It may be that, extending the field of our external observations, or with new more precise means of measuring equal quantities, the physical space is found to correspond to one of non-Euclidean geometries [\,\dots]. If an observer with Euclidean intuition enters a pseudo-spheric or spherical space, he would gain the impression, by moving, that objects move in certain ways, and dilate and shrink in certain directions, in the same way that, according to our movement, we see that the size of the objects changes, and we would have no chance to decide whether such a fact is apparent or real, if we did not know the laws of perspective under other conditions.\endnote{
	Original It. version: «La geometria ha la sua origine nell'osservazione diretta degli oggetti del mondo esteriore, che è lo spazio fisico, e dall'intuizione di essi trae le sue prime verità indimostrabili e necessarie al suo svolgimento teoretico, che sono gli assiomi [\,\dots]. Eppure, per essere esatta, la geometria deve rappresentare gli oggetti forniti dall'osservazione per mezzo di forme pure astratte e gli assiomi con ipotesi bene determinate, rese cioè indipendenti dell'intuizione, cosicchè la geometria diventi parte della matematica pura [\,\dots]. [È] necessario distinguere lo spazio fisico dallo spazio intuitivo, e questo dallo spazio geometrico [\,\dots]. Lo spazio geometrico è appunto quella parte dell'estensione pura nella quale è rappresentato lo spazio fisico e intuitivo, ma che a sua volta non ha per tutte le sue forme una rappresentazione nel mondo reale. E mentre lo spazio fisico e quello intuitivo non possono essere definiti, può essere invece definito lo spazio geometrico [\,\dots]. Le tre geometrie [quella euclidea, quella ellittica e quella iperbolica] in un campo piccolissimo dànno con grande approssimazione gli stessi risultati [\,\dots]. Può darsi che, estendendo il campo delle nostre osservazioni esteriori, o con nuovi mezzi più precisi di misura delle grandezze eguali, si trovi che lo spazio fisico corrisponda ad una delle geometrie non euclidee [\,\dots]. Se un osservatore coll'intuizione euclidea entrasse in uno spazio pseudosferico o sferico, avrebbe l'impressione, movendosi, che gli oggetti si spostano in determinati modi, e in determinate direzioni si dilatano e si restringono, nello stesso modo che noi, secondo che ci moviamo, vediamo cambiare la grandezza degli oggetti, e non avremmo modo di decidere se tale fatto è apparente o reale, se non conoscessimo per altre vie le leggi della prospettiva». 
	} \\
\indent — \textsc{G. Veronese} \cite[pp. 10, 12, 14, 15-16, e.a.]{Veronese "Il vero nella matematica"}

\endgroup

\section{Intrinsic Surface Property}

\subsection{Gauss' Theorema Egregium}
\label{subsection "Gauss' Theorema Egregium"}

\begingroup
\footnotesize
Si superficies curva in quamcunque aliam superficiem explicatur, mensura curvaturae in singulis punctis invariata manet.\footnote{
	«[Theorema egregium.] If a curved surface is developed upon any other surface whatever, the measure of curvature in each point remains unchanged» \cite[p. 20]{Gauss "General Investigations of Curved Surfaces of 1827 and 1825"}.	
	} \\
\indent — \textsc{C.F. Gauss} \cite[p. 24]{Gauss "Disquisitiones generales circa superficies curvas"}

\endgroup

\vspace{2mm}

In this Section we will briefly retrace the notion of \emph{Gaussian curvature}. It is an \emph{intrinsic quantity of a surface}, which means that it depends on the metric, and not on the manner in which the surface is immersed in 3-dimensional space (the ambient space).

\begin{theorema}[Theorema egregium]
\label{theorema "Theorema egregium"} 
We shall indicate by $\kappa$ the curvature, by $\surface \subset \mathbb{R}^3$ a surface,\footnote{
	A surface $\surface$, or, equally, a 2-manifold $\mathcal{M}^2$.
	} 
and by $p \in \surface$ some point on the surface, letting $p$ at the origin of $\mathbb{R}^3$. There is a map $\{x^1, \mathellipsis, x^n\} \colon \Upsilon \to \surface$, where $\Upsilon \subset \surface$, under which
\begin{equation}
	\kappa = \frac{\partial^2g_{12}}{\partial v\partial w} - \frac{\partial^2g_{22}}{2\partial v^2} - \frac{\partial^2g_{11}}{2\partial w^2},
\end{equation}
where $v, w$ are tangent vectors at $p$. 
\end{theorema}

\begin{proof}
Setting 
\[
	(v, w) \mapsto \varphi(v, w), 
\]
for parametric equations, one gets
\begin{equation}
	\begin{pmatrix}	
	g_{11}(v, w) & g_{12}(v, w) \\ 
	g_{21}(v, w) & g_{22}(v, w)
	\end{pmatrix} =
	\begin{pmatrix}	
	\frac{(\partial\varphi)^2}{(\partial v)^2} + 1 & \frac{\partial\varphi}{\partial v}\frac{\partial\varphi}{\partial w} \\
	\frac{\partial\varphi}{\partial v}\frac{\partial\varphi}{\partial w} & \frac{(\partial\varphi)^2}{(\partial w)^2} + 1	
	\end{pmatrix},
\end{equation} 
and finally
\begin{equation}
	\frac{\partial^2g_{12}}{\partial v\partial w} - \frac{\partial^2g_{22}}{2\partial v^2} - \frac{\partial^2g_{11}}{2\partial w^2} = \frac{\partial^2\varphi}{\partial v^2} \frac{\partial^2\varphi}{\partial w^2} - \frac{\partial^2\varphi}{\partial v\partial w} = \det(D^2)\varphi = \det(\surface) = \kappa,
\end{equation}
where $D^2$ is the second covariant derivative.
\end{proof}

If we define a local parametrization of $\surface$ as $\varphi \colon \Upsilon \to \surface$, putting $\varphi(x^1, x^2)$, we may define the Theorema egregium by resorting to the Christoffel symbols (Section \ref{section "Christoffel Symbols"}):
\begin{equation}
	\kappa = \frac{\left\{\frac{\partial\Gamma^1_{22}}{\partial x_1} - \frac{\partial\Gamma^1_{12}}{\partial x_2} + \sum^2_{m = 1}\left(\Gamma^m_{22}\Gamma^1_{1m} - \Gamma^m_{12}\Gamma^1_{2m}\right)\right\}}{g_{22}}.
\end{equation}

\subsection{Relative vs. Absolute Surfaces: Casorati's Observation}

\begingroup
\footnotesize
If one considers surfaces as flexible but inextensible, almost like veils, and one imagines different shapes that each one can assume in such conditions, one is led to distinguish the properties into two classes. That is, to distinguish the properties that are no troubled by alteration due to any change in shape of the surface to which they relate and that can be called \emph{absolute}, from those that depend on the individual shapes under which the surface can be conceived and that can be called \emph{relative} [\,\dots]. Means for the study of surfaces regardless of the shapes in which, with the conditions of flexibility and inextensibility, can be imagined, are offered by the use of curvilinear coordinates, for which the surfaces are considered in themselves and not referred to extraneous entities (such as for ordinary coordinate planes) which do not necessarily change position and form with them.\endnote{
	Original It. version: «Se si considerano le superficie come flessibili ma inestendibili, a guisa direi quasi di veli, e s'immaginano le forme differenti che ognuna può assumere in siffatte condizioni, si è condotti a distinguere le proprietà in due classi. A distinguere cioè le proprietà che non subiscono alterazione per qualsiasi cambiamento di forma della superficie cui si riferiscono e che si ponno chiamare assolute, da quelle che dipendono invece dalle singole forme sotto le quali la superficie può essere concepita e che si ponno chiamare relative [\,\dots]. Mezzi per lo studio delle superficie indipendentemente dalle forme nelle quali, colle condizioni di flessibilità ed inestendibilità, ponno essere immaginate, sono offerti dall'uso di coordinate curvilinee, per le quali le superficie vengano considerate in se stesse e non riferite ad enti estranei (come per gli ordinari piani coordinati) che non cambiano necessariamente con esse di posizione e di forma».
	} \\
\indent — \textsc{F. Casorati} \cite[p. 363, e.a.]{Casorati "Ricerca fondamentale per lo studio di una certa classe di proprieta delle superficie curve"}

\endgroup

\vspace{2mm}

Gauss' Theorema \ref{theorema "Theorema egregium"} on the \emph{conservation of curvature}, i.e. on the reciprocal product of the main radii of curvature expresses an \emph{absolute property} of a surface. 

It should be noted that a surface can be thought of in two ways: 
\enumerationisinitium
\item as the \emph{boundary of solids}, 
\item or as a \emph{flexible but inextensible solid}, almost like veils («a guisa quasi di veli»), in the words of F. Casorati \cite{Casorati "Ricerca fondamentale per lo studio di una certa classe di proprieta delle superficie curve"}. 
\enumerationisfinis

When choosing the second point of view, surfaces are divided into two classes, 
\subenumerationisinitium
\item one with \emph{relative properties}, for which surfaces hinge on the distinct shapes of conception, so that the properties change along with the particular shapes of surfaces;
\item and the other with \emph{absolute properties}, for which surfaces are seen as \emph{objects in themselves}, that is, surface properties are treated independently of any particular determination of the shape itself. Here what remains unchanged is the length of each linear element, and from this inalterability all properties of absolute value derive as just consequences; see also Beltrami \cite[XIII, in particular pp. 355-359]{Beltrami "Ricerche di analisi applicata alla geometria"}.
\subenumerationisfinis

\section[Space Forms of Constant Curvature, and Discrete Crystal-like Group]{Space Forms of Constant Curvature (Parabolic, Elliptic, Hyperbolic Type), and Discrete Crystal-like Group}
\sectionmark{Space Forms of Constant Curvature, and Discrete Crystal-like Group}

\begingroup
\footnotesize
[A]dimandiamo che ci sia concesso due linee rette non chiudere alcuna superficie [o alcuno spazio].\footnote{
	«We enjoin that we may [think that] two straight lines do not close any surface» or space.
	} \\
\indent — \textsc{N. Tartaglia}'s It. transl. of \textsc{Euclid} \cite[Lib. Primo, Petitione vi, Fo. XIII]{Tartaglia "Euclide Megarense"}\endnote{
	The sixth postulate is nowadays expunged; it nevertheless survived for centuries, see e.g. \textit{Gli elementi di Euclide}, per cura di E. Betti e F. Brioschi \cite[p. 5]{Betti e Brioschi (per cura di) "Gli elementi"}.
	}

\vspace{2mm}

In omni triangulo Sphærico, producto uno latere, angulus exterior minor erit utrisque interioribus eidem oppositis simul sumptis: \& tres anguli trianguli simul sumpti majores erunt duobus rectis.\footnote{
	«In every spherical triangle, [if] one of the sides [is] produced, [then] the exterior angle is less than either of the interior and opposite [angles] taken together: \& the three angles of the triangle taken together are greater than two right [angles]», i.e. $180^\circ$.
	} \\
\indent — \textsc{Menelaus of Alexandria} \cite[Lib. I, Prop. XI. Theor., p. 11]{Menelaus of Alexandria "Menelai sphaericorum libri III"}, cf. the 1558 \textsc{F. Maurolico}'s version \cite[p. 19]{Maurolico "Theodosii Sphaericorum Elementorum"}.\footnote{
		Menelaus' work is the first, as far as we know, to dealing the idea of a \emph{spherical triangle} and of \emph{geometry of figures on spherical surfaces}. The Greek text, lost in the original, have been preserved to us in an Arabic translation. The La. translations here consulted are those by Maurolico \cite{Maurolico "Theodosii Sphaericorum Elementorum"} and Halley \cite{Menelaus of Alexandria "Menelai sphaericorum libri III"}, based on the Arabic version; the latter is in \cite{"Menelaus' Spherics Early Translation and Arabic Version"}.
	
	After Menelaus of Alexandria, it is usual to jump to 1766, with J.H. Lambert \cite[§ 81, p. 353]{Lambert "Theorie der Parallellinien"}: «It seems remarkable to me that the [\,\dots] hypothesis [of the obtuse angle of a quadrilateral] holds if instead of a plane triangle we take spherical one, for in this case the sum of [interior] angles of a triangle is greater than 180 degrees and the excess is also proportional to the area of the triangle». But in the middle there is G. Saccheri \cite{Saccheri "Euclides ab omni naevo vindicatus"} \cite{Saccheri "Euclid Vindicated from Every Blemish"}; Beltrami \cite[pp. 444, 446]{Beltrami "Un precursore italiano di Legendre e di Lobatschewsky"} does not hesitate to consider the Italian mathematician the real founder of non-Euclidean geometry, as a precursor of Legendre \cite[propp. XX (pp. 19-20), XXII (pp. 21-22) in Livre I, note III (pp. 286-287)]{Legendre "Elements de geometrie avec des notes"} and Lobačevskij \cite{Lobacevskij "Novyye nachala geometrii"} \cite{Lobacevskij "New Principles of Geometry"}.
	}

\vspace{2mm}

Semmiből egy újj, más világot teremtettem · Out of nothing I have created a new, different world.\footnote{
	\cite[§ 15, pp. 7-8]{Bolyai "Appendix scientiam spatii absolute veram exhibens"}: \textit{Systema Geometriae, hypothesi veritatis Axiomatis Euclidei XI insitens dicatur \textgreek{Σ}; et hypotesi contrariae superstructum sit $S$. Omnia, quae expresse non dicentur, in \textgreek{\text{Σ}} \emph{vel} in $S$ esse;  absolute enuntiari, i.e. illa, sive \textgreek{\text{Σ}} sive $S$ reipsa sit, vera asseri intelligatur} («The System of Geometry founded upon the hypothesis of the truth of Euclid's Axiom XI is called \textgreek{Σ}; and the system founded upon the contrary hypothesis is $S$. Any [result] that is not expressly said to be in \textgreek{Σ} or in $S$, it is understood to be enunciated absolutely, i.e. it is supposed to be true, whether it placed in \textgreek{Σ} or $S$»).
	} \\
\indent — \textsc{J. Bolyai} \cite[p. 188]{Bolyai "Letter to his father 3 November 1823"}\endnote{
	About the young János Bolyai, Gauss in a letter to C.L. Gerling, dated 14 February 1832, puts it like this \cite{Gauss' letter to Gerling}: «These days I have received a short work from Hungary on non-Euclidean geometry, in which I find all my own ideas and \emph{results}, developed with great elegance [\,\dots]. The author is a very young Austrian officer, the son of a childhood friend of mine, with whom I often discussed about the subject in 1798, although my ideas, at that time, were still very far from the elaborateness and maturity they have achieved through this boy's own thinking. I consider this young geometer v. Bolyai a genius of the first magnitude [\textit{ein Genie erster Grösse}]».
 
	\setlength\parindent{8pt}
	To János' father, Farkas (also known as “Wolfgang”), Gauss, in a letter of 6 March 1832, writes \cite[pp. 220-221]{Gauss' letter to W. [F.] von Bolyai}: «Now something about your son's work. If I start by saying “that I am unable to praise it”, you will probably be shocked for a moment: but I cannot do otherwise; to praise it, would be to praise myself: the entire content of the work, the path that your son has taken and the results to which he has been led, coincide almost always with my own meditations, some of which have occupied me in part for 30-35 years. In fact, I am extremely surprised by this». On non-Euclidean geometry in Gauss, see \cite{Halsted "Gauss and the Non-Euclidean Geometry"}.
	}

\vspace{2mm}

The methods of ordinary [Euclidean] geometry always lead to results true but less extended than those given by the general geometric method to which I have given the name of \emph{Imaginary Geometry} (\textcyrillic{\emph{Воображаемая Геометрия}}) [the current hyperbolic geometry]. The difference between the equations of the one and the other comes from a new constant, that it is necessary to determine experimentally, and which so obtained, is found such that, without sensible error, ordinary [Euclidean] geometry more than suffices for the usual cases, while yet being possibly not rigorously true. \\
\indent — \textsc{N.I. Lobačevskij} \cite[p. 65]{Lobacevskij "Novyye nachala geometrii"} = \cite[pp. 17-18]{Lobacevskij "New Principles of Geometry"}

\endgroup

\vspace{2mm}

In this Section and in the next one (Section \ref{section "Space Forms as Triad of Riemannian Manifolds"}) we will give a cursory glance—as anticipated by the three epigraphic quotations—at the 

· \emph{Euclidean} and 

· \emph{non-Euclidean (elliptical/spherical and hyperbolic) spaces}, each characterized by \emph{constant sectional curvature}, otherwise known as \emph{space forms}.

\subsection{Geodesics on Space Forms}
\label{subsection "Geodesics on Space Forms"}

\begingroup
\footnotesize
The only surfaces which can be represented upon a plane in such a manner that [\,\dots] to every geodesic line [corresponds] a straight line, are those for which the curvature is everywhere constant (positive, negative or zero). When this is a constant null curvature, the correspondence law is no different from the ordinary homography.\endnote{
	Original It. version: «Le sole superficie suscettibili di essere rappresentate sopra un piano, in modo che [\,\dots] ad ogni linea geodetica [corrisponda] una linea retta, sono quelle la cui curvatura è dovunque costante (positiva, negativa o nulla). Quando questa curvatura costante è nulla, la legge di corrispondenza non differisce dall'ordinaria omografia».
	} \\
\indent — \textsc{E. Beltrami} \cite[p. 203]{Beltrami "Risoluzione del problema: riportare i punti di una superficie sopra un piano in modo che le linee geodetiche vengano rappresentate da linee rette"}

\endgroup

\vspace{2mm}

The upcoming definition extends the concept of geodesic to the notion of space. Let us first introduce the notion of length space.

\begin{definitio}[Length space]
Let $(\mathcal{X}, \distance)$ denote a \emph{metric space}, based on Fréchet's axioms \cite{Frechet "Sur quelques points du calcul fonctionnel"}, and let $x$ and $y$ be two arbitrary points of $(\mathcal{X}, \distance)$. The metric space $(\mathcal{X}, \distance)$ is said to be a \emph{length space} if the distance between $x$ and $y$ is the infimum of the lengths of all curves joining them. \definitiosymbol
\end{definitio}

\begin{definitio}[Geodesic space]
\label{definitio "Geodesic space"}
Let $(\mathcal{X}, \distance)$ denote a \emph{metric space} and $\gamma_\mathrm{c} \colon [\alpha, \beta] \to \mathcal{X}$ a geodesic path, where $\distance \colon \mathcal{X} \times \mathcal{X} \to \mathbb{R}$ is a \emph{metric} or a \emph{distance function} on $\mathcal{X}$ and $[\alpha, \beta]$ is a geodesic segment in $\mathcal{X}$. Let $\distance(x, y)$ be the distance between two points $x$ and $y$ of $(\mathcal{X}, \distance) \viz |x - y|_\mathcal{X}$. Then the pair $(\mathcal{X}, \distance)$
\enumerationisinitium
\item is said to be a \emph{geodesic (metric) space}, if in $(\mathcal{X}, \distance)$ there is a (constant speed) geodesic $\gamma_\mathrm{c}$  joining $x$ to $y$, for all $x, y \in \mathcal{X}$, for which $\gamma_\mathrm{c}(\alpha) = x$ and $\gamma_\mathrm{c}(\beta) = y$,
\item is called \emph{uniquely geodesic space}, if there exists one and only one geodesic connecting $x$ and $y$. \definitiosymbol
\enumerationisfinis
\end{definitio}

In summary, a metric space $(\mathcal{X}, \distance)$ is a geodesic space if $x$ and $y$ can be connected by a geodesic segment $[\alpha, \beta] \equival [x, y]$ whose length is at most a times the distance between $x$ and $y$.
 
\emph{A geodesic space $(\mathcal{X}, \distance)$ is at all times a length space}, but the converse is not true. Recall that the path $\gamma_\mathrm{c}$ is a distance minimizing curve of constant speed, and thereby is linearly reparameterized, so $\distance(\alpha - \beta) = \distance(x - y) = \length(\gamma_\mathrm{c})$.

Now, we remember what geodesic spaces are. They can be divided into six types: 
\enumerationisinitium
\item parabolic (Euclidean) space,
\item elliptic (spherical) space, 
\item hyperbolic space,
\item convex subsets of a normed vector space, 
\item Teichmüller space of Teichmüller's metric, 
\item Teichmüller space of Thurston's metric.
\enumerationisfinis

We want to consider the first three types of geodesic space (Section \ref{section "Space Forms as Triad of Riemannian Manifolds"}), the so-called \emph{space forms}, and, consequently, we will touch the fourth case (Section \ref{section "Certain Geodesically Convex Conditions"}), but we exclude the treatment of geodesics in other cases, ($\mathnormal{5}$) and ($\mathnormal{6}$), with Teichmüller spaces. We are therefore interested in studying the (connected) Riemannian spaces, in the three forms just mentioned, as geodesic (metric) spaces, for which $(\mathcal{X}, \distance) \equival (\mathcal{M}, \distance)$.

\begin{margo}[Apollonian–Kleinian nomenclature]
Space forms descend from a Kleinian nomenclature \cite[p. 344]{Klein "Sur la geometrie dite non euclidienne"}; \emph{parabola} is an open curve touching the line at infinity in one point. Beware that all these terms are a legacy of Apollonian treatise on conic sections: \textgreek{παραβολή} (parabola) \cite[p. 38]{Apollonius of Perga "Apollonii Pergaei quae graece exstant cum commentariis antiquis I"}, \textgreek{ἔλλειψις} (ellipse) \cite[p. 48]{Apollonius of Perga "Apollonii Pergaei quae graece exstant cum commentariis antiquis I"}, \textgreek{ὑπερβολή} (hyperbola) \cite[p. 42]{Apollonius of Perga "Apollonii Pergaei quae graece exstant cum commentariis antiquis I"}. \margosymbol	
\end{margo}

\begin{propositio}[Geodesics in constant curvature metrics]
\label{propositio "Geodesics in constant curvature metrics"}
Let $\mathcal{M}^n_\kappa$ be a complete simply connected Riemannian $n$-dimensional manifold with \emph{constant sectional curvature} $\kappa$. Let $\kappa \in \mathbb{R}_-$ and $\kappa \in \mathbb{R}_* = \{0\} \cup \mathbb{R}_+$. Then
\begin{equation}
	\mathcal{M}^n_\kappa \equival
	\begin{cases}
	\mathbb{E}^{n \geqslant 0}_{\kappa} \text{ if } \kappa \in \mathbb{R}_{\{0\}} = 0, \text{ parabolic type of Euclidean $(\mathcal{M}, \distance)$-space}, \\
	\mathbb{S}^{n \geqslant 0}_{\sqrt{\kappa}} \text{ if } \kappa \in \mathbb{R}_+ > 0, \text{ elliptic type of non-Euclidean $(\mathcal{M}, \distance)$-space}, \\
	\hyperbolic^{n \geqslant 2}_{\sqrt{-\kappa}} \text{ if } \kappa \in \mathbb{R}_- < 0, \text{ hyperbolic type of non-Euclidean $(\mathcal{M}, \distance)$-space}.
	\end{cases}
\end{equation}
Spaces of constant vanishing, positive and negative (sectional) curvature (see Margo \ref{margo "Minding and Codazzi"}), are geodesic for $n \geqslant 0$ both in the parabolic type and in the elliptic type, and for $n \geqslant 2$ in the hyperbolic type.
\end{propositio}

\begin{margo}
\label{margo "Minding and Codazzi"}
Among the first studies of constant curvature, there are F. Minding \cite{Minding "Wie sich entscheiden lasst"} and D. Codazzi \cite[p. 355]{Codazzi "Intorno alle superficie le quali hanno costante il prodotto de' due raggi di curvatura"}. \margosymbol	
\end{margo}

\subsection[Discrete $\mathbbl{\Gamma}$-Crystallographic Group, Killing–Hopf Theorem, and Isometric Action]{Discrete $\protect\pseudobold{\mathbbl{\Gamma}}$-Crystallographic Group, Killing–Hopf Theorem, and Isometric Action}
\label{subsection "Discrete Gamma-Crystallographic Group, Killing–Hopf Theorem, and Isometric Action"}

\begingroup
\footnotesize
The ordered internal structure of minerals normally determines their characteristic external form of polyhedra, that is of crystals. \\
\indent The particles (atoms, molecules) that constitute the minerals are, in many cases, arranged in a fixed and determined position, and form a geometric [microscopic] structure called crystal lattice. \\
\indent Crystals possess symmetry properties that manifest themselves in the [atomic] face arrangements with respect to certain planes, axes and the center. \\
\indent Taking into account the quantity and quality of the symmetry elements (planes, axes and center; crystallographic axes, parameters), the crystals are brought together under [crystallographic or Fedorov space] groups.\endnote{
	Exhibition panels originally written in It..
	} \\
\indent — \textsc{Palazzo Pompei} Museo Civico di Storia Naturale (Lungadige Porta Vittoria 9, Verona) 

\endgroup

\vspace{2mm}

An additional way to describe the Proposition \ref{propositio "Geodesics in constant curvature metrics"}, with respect to spaces of constant curvature, can be stated through the use of Killing–Hopf's theorem \cite{Killing "Ueber die Clifford-Klein'schen Raumformen"} \cite{Hopf H. "Zum Clifford-Kleinschen Raumproblem"}, originally referred (by Killing) to as \emph{Clifford–Klein space form problem}.
 
\begin{theorema}[Killing–Hopf]
\label{theorema "Killing–Hopf"}
Let $\mathcal{M}^n \viz \mathcal{M}^n_\kappa$. Let $\pi \colon \widetilde{\mathcal{M}}^n_\kappa \to \mathcal{M}^n_\kappa$ be a \emph{universal covering (space)} of $\mathcal{M}^n_\kappa$ ($\pi$ is a smooth covering map and a local isometry). Let $\widetilde{\mathcal{M}}^n_\kappa$ be equals to $\mathbb{E}^n$, $\mathbb{S}^n$ or $\hyperbolic^n$, with $\kappa \in \mathbb{R}$. Let $\mathbbl{\Gamma}$ denote a \emph{discrete group of fixed point free isometries} of $\widetilde{\mathcal{M}}^n_\kappa$. Then $\mathcal{M}^n_\kappa$ is a complete connected Riemannian manifold of constant sectional curvature and dimension $n \geqslant 2$ iff it is isometric to $\widetilde{\mathcal{M}}^n_\kappa/\mathbbl{\Gamma}$, i.e. iff $\mathcal{M}^n_\kappa$ is isometrically diffeomorphic to a topological quotient $\mathbb{E}^n/\mathbbl{\Gamma}$ ($\kappa = 0$), $\mathbb{S}^n/\mathbbl{\Gamma}$ ($\kappa > 0$) or $\hyperbolic^n/\mathbbl{\Gamma}$ ($\kappa < 0$). The group $\mathbbl{\Gamma}$ acts freely and properly discontinuously on $\widetilde{\mathcal{M}}^n_\kappa$.
\end{theorema}

\begin{proof}
Let be $\tilde{g} \equival \pi^*g$ the metric of the universal covering (space) $\widetilde{\mathcal{M}}^n_\kappa$, while $g$ is the metric on $\mathcal{M}$. The group $\mathbbl{\Gamma}$ is isomorphic to the fundamental group $\pi_1(\mathcal{M}^n_\kappa)$, and it operates on $\widetilde{\mathcal{M}}^n_\kappa$ as the group of covering transformations. So $\tilde{g}$ being invariant under every \emph{covering transformation} 
\begin{equation}
	\varphi_\mathbbl{\Gamma} \colon \left(\widetilde{\mathcal{M}}^n_\kappa\right)_1 \to \left(\widetilde{\mathcal{M}}^n_\kappa\right)_2,
\end{equation}
such that $\pi \circ \varphi_\mathbbl{\Gamma} = \pi$ and $\varphi^*_\mathbbl{\Gamma}\tilde{g} = \varphi^*_\mathbbl{\Gamma}\pi^*g = \pi^*g \equival \tilde{g}$, the group $\mathbbl{\Gamma}$ acts by isometries and discontinuously. Now, letting $\mathfrak{isom}(\mathcal{M}^n_\kappa)$ be the group of isometries of $\mathcal{M}^n_\kappa$, we can say that $\mathbbl{\Gamma}$ is a subgroup of $\mathfrak{isom}(\widetilde{\mathcal{M}}^n_\kappa)$, and it is a discrete set in $\mathfrak{isom}(\widetilde{\mathcal{M}}^n_\kappa)$ because there is no an accumulation point in $\widetilde{\mathcal{M}}^n_\kappa$ about $\mathbbl{\Gamma}$. In fact, any point $\tilde{x} \in \widetilde{\mathcal{M}}^n_\kappa$ maps onto (projects to) the same point in $\mathcal{M}^n_\kappa$. 
\end{proof}

The isometry group $\mathbbl{\Gamma}$ is a Fuchsian group (see Section \ref{section "Fuchsian Group (Properly Discontinuous Action)"}). It is also called a \emph{crystallographic} or \emph{Fedorov group} \cite{Fedorov "Symmetry of Crystals"}, and more specifically a \emph{non-Euclidean crystallographic group} (because it is Fuchsian), a set describing a \emph{lattice structure} in the group of isometries of $\widetilde{\mathcal{M}}^n_\kappa$. The name of it, crystallographic, is linked to the fact that it can be thought of as a \emph{symmetry group} and we are working with the symmetry of a crystal lattice in a given configuration space (whereas the crystal lattice is the most stable solid form).

\section{Space Forms as Triad of Riemannian Manifolds}
\label{section "Space Forms as Triad of Riemannian Manifolds"}

Let $\mathcal{M}^n_\kappa$ be a Riemannian manifold of dimension $n$, and $\kappa \in \mathbb{R}$  the sectional curvature of $n$-dimensional spaces tangent to $\mathcal{M}^n_\kappa$. To deepen what we just saw (Theorem \ref{theorema "Killing–Hopf"}), we concentrate on the main result \cite{Cartan "Sur une classe remarquable d'espaces de Riemann", Cartan "Sur une classe remarquable d'espaces de Riemann (suite et fin)"} showing that, for $n \geqslant 2$, there is a \emph{unique} complete simply connected Riemannian manifold, which is isometric to either 
\enumerationisinitium
\item Euclidean space, for $\kappa \in \mathbb{R}_{\{0\}} = 0$, 
\item spherical space of radius $\rho = \frac{1}{\sqrt{\kappa}}$, for $\kappa \in \mathbb{R}_+ > 0$, 
\item or hyperbolic space of curvature $\sqrt{-\kappa}$, for $\kappa < 0$. 
\enumerationisfinis
Let us get into these three types of space.

\subsection{Type I. Parabolic Case}

The parabolic case means that $\mathcal{M}^n_\kappa$ ($\kappa = 0$) is the \emph{Euclidean $n$-space} $\mathbb{E}^n$ representing the set of all ordered $n$-tuples $(x^1, \mathellipsis, x^n)$ of real numbers (coordinates). We are looking at a Riemannian manifold, with the Euclidean metric; it is clear that the Euclidean metric is flat, or, equivalently, it has \emph{constant zero sectional curvature} (all sectional curvatures equal to zero, given that the curvature tensor in a Euclidean space is identically zero);\footnote{
	The Euclidean plane $\mathbb{E}^2$ (as well as other surfaces of the same type of geometry, for instance the cylinder and the cone without apex) has constant zero Gaussian curvature. Note that if the cylinder is locally Euclidean, the first postulate of Euclid \cite[\textgreek{αἰτήματα, α´, Στοιχείων α´}, Book I, p. 8]{Euclidis "Elementa I"}—«Let it be postulated to draw a straight line from any point to any point (\textgreek{Ἠιτήσθω ἀπὸ παντὸς σημείου ἐπὶ πᾶν σημεῖον εὐθεῖαν γραμμὴν ἀγαγεῖν})»—fails: between two points of a cylindrical surface there is not a single straight line.
	} 
in here, the fifth (parallel) postulate, \cite[\textgreek{αἰτήματα, ε´, Στοιχείων α´}, Book I]{Euclidis "Elementa I"} applies.\footnote{
	\cite[p. 8]{Euclidis "Elementa I"}: «\textgreek{Καὶ ἐὰν εἰς δύο εὐθείας εὐθεῖα ἐμπίπτουσα τὰς ἐντὸς καὶ ἐπὶ τὰ αὐτὰ μέρη γωνίας δύο ὀρθῶν ἐλάσσονας ποιῇ, ἐκβαλλομένας τὰς δύο εὐθείας ἐπ' ἄπειρον συμπίπτειν, ἐφ' ἃ μέρη εἰσὶν αἱ τῶν δύο ὀρθῶν ἐλάσσονες}».\endnote{
	«And if a straight line falling across two straight lines makes interior angles on the same side less than two right angles, [then] the two straight lines, being produced boundlessly [infinitely], meet on that side on which the [sum of the interior] angles [is] less than two right angles». Compare with \cite[p. 155]{Euclid "The Thirteen Books of the Elements I Introduction and Books I-II"}.
	} 
	In the renowned It. translation from 1575 by F. Commandino \cite[p. 6 verso]{Commandino "De gli Elementi d'Euclide libri quindici"} the Euclid's fifth postulate sounds like this: «Et se sopra due rette linee cadendo una retta farà gli angoli interiori \& da una medesima parte minori di due retti, quelle linee prolungate in infinito congiungersi insieme da quella parte, dove sono gli angoli minori di due retti».
	}

\subsection{Type II. Elliptic Geometry (and Parabolic View)}
\label{subsection "Type II. Elliptic Geometry (and Parabolic View)"}

The elliptic case means that $\mathcal{M}^n_\kappa$ ($\kappa \in \mathbb{R}_+$) is a \emph{spherical $n$-space} 
\begin{align}
\label{align "Spherical $n$-space"}
	\mathbb{S}^n_\rho & = \{x \in \mathbb{R}^{n + 1} \mid \|x\| = \rho\} \notag \\
	& = \{(x^1, \mathellipsis, x^{n + 1}) \in \mathbb{R}^{n + 1} \mid (x^1)^2 + \mathellipsis + (x^{n + 1})^2 = \rho\} \subset \mathbb{R}^{n + 1},
\end{align} 
with radius $\rho > 0$. The sphere \eqref{align "Spherical $n$-space"} has \emph{constant sectional curvature $\kappa = \frac{1}{\rho^2}$}.\footnote{
	The 2-sphere $\mathbb{S}^2_\rho$ has constant (positive) Gaussian curvature $\kappa = \frac{1}{\rho^2}$.
	} 
The unit sphere (a sphere of radius 1) is just
\begin{equation}
\label{equation "Unit sphere"}
	\mathbb{S}^n_{(\rho = 1)} = \left\{(x^1, \mathellipsis, x^{n + 1}) \in \mathbb{R}^{n + 1} \mathrel{\Bigg|} \sum^{n + 1}_{\mu = 1} (x^\mu)^2 = 1\right\}.
\end{equation}

The metric coefficients on $\mathbb{S}^n_\rho$ is $g^{\mathbb{S}}_{\mu\nu} = \rho^2(\sin{\theta}^{i + 1} \cdots \sin{\theta}^n)^2$ if $\mu = \nu$, $g^{\mathbb{S}}_{\mu\nu} = 0$ if $\mu \neq \nu$.

The Eqq. \eqref{align "Spherical $n$-space"} \eqref{equation "Unit sphere"} tell that $\mathbb{S}^n_\rho$ is a submanifold of $\mathbb{R}^{n + 1}$. This is evident by the following. Take the atlases 
\begin{align}
	& \Upsilon^\pm_\mu = \{x \in \mathbb{S}^n_\rho \mid +x^\mu > 0 \text{ and } -x^\mu < 0\}, \\
	& \varphi^\pm_\mu(x) = (x^1, \mathellipsis,x^{\mu - 1}, x^{\mu + 1}, \mathellipsis, x^{n + 1}), 
\end{align}
with $\mu = 1, \mathellipsis, n + 1$, from which we get the charts $(\Upsilon^\pm_\mu, \varphi^\pm_\mu)$ arranged in pair. Let 
\begin{equation}
	\iota \colon \mathbb{S}^n_\rho = \bigcup^{n + 1}_{\mu = 1}(\Upsilon^\pm_\mu) \to \mathbb{R}^{n + 1} 
\end{equation}
be an inclusion map, and let $\varphi^\pm_\mu \colon \Upsilon^\pm_\mu \to \mathbb{B}^n_\rho \subset \mathbb{R}^n$, where $\mathbb{B}^n_\rho = \{y = (y^1, \mathellipsis, y^n) \in \mathbb{R}^n \mid \|y\| < \rho\}$ is the $n$-ball, so that $\varphi^\pm_\mu$ is a bijection between $\Upsilon^\pm_\mu$ and $\mathbb{B}^n_\rho$. The injective immersion 
\begin{equation}
	\iota \circ (\varphi^\pm_\mu)^{-1}(y) \colon \mathbb{B}^n_\rho \hookrightarrow \mathbb{R}^{n + 1} 
\end{equation}
is also an \emph{embedding}. Consequently, the metric of $\mathbb{S}^n_\rho$ is induced by the Euclidean metric on $\mathbb{R}^{n +1}$, and that is precisely why the topological space of $\mathbb{S}^n_\rho$ is metrizable; the spherical topology induced by the topology on $\mathbb{R}^{n +1}$ is the same as the topology induced by the ambient space (ambient manifold), for which $\mathbb{S}^n_\rho$ is a subspace (submanifold) of $\mathbb{R}^{n +1}$.

Another result, within this framework, is obtained by seeing the real projective $n$-space $\mathbb{RP}^n$ as a quotient of $\mathbb{S}^n_\rho$. Assume first that $\mathcal{A} = \{(\Upsilon_\mu, \varphi_\mu \mid \mu \in \mathcal{A})\}$ is an atlas consisting of an $n$-chart $(\Upsilon_\mu, \varphi_\mu)$. Let $\Upsilon_\mu = \{[x] \in \mathbb{RP}^n\}$, where $[x] = [x^0 \colon \cdots \colon x^n] \in \mathbb{RP}^n$ denotes the projection of $x = (x^1, \mathellipsis, x^n) \in \mathbb{R}^{n + 1} \backslash \{0\}$; and then let 
\begin{equation}
	\varphi_\mu[x] = \left(\frac{x^0}{x^\mu}, \mathellipsis, \frac{x^{\mu - 1}}{x^\mu}, \frac{x^{\mu + 1}}{x^\mu}, \mathellipsis, \frac{x^n}{x^\mu}\right),
\end{equation}
hence $\varphi^{-1}_\mu(y) = [y^1 \colon \cdots \colon y^{\mu - 1} \colon \idem \colon y^\mu \colon \cdots \colon y^n]$. Knowing that the projection $\pi \colon \mathbb{S}^n_\rho \to \mathbb{RP}^n$ is a smooth covering map, and since it is composed with the inclusion map $\iota \colon \mathbb{S}^n_\rho \to \mathbb{R}^{n + 1} \backslash \{0\}$ and the projection $\pi \colon \mathbb{R}^{n + 1} \backslash \{0\} \to \mathbb{RP}^n$, one can prove that the manifold topology induced by the smooth atlas $\mathcal{A}$ is the quotient topology induced by the projection $\mathbb{R}^{n + 1} \backslash \{0\} \xrightarrow{\pi} \mathbb{RP}^n$.

In general, we can say that the Eqq. \eqref{align "Spherical $n$-space"} \eqref{equation "Unit sphere"} are a representation of Riemann's elliptic theory \cite{Klein "Sur la geometrie dite non euclidienne"} in Euclidean space about the Riemann spherical geometry \cite{Riemann "Ueber die Hypothesen welche der Geometrie zu Grunde liegen"}. However, there are differences between the elliptic and spherical systems, as already noted and commented on by R. Bonola \cite[§§ 75-76]{Bonola "Non-Euclidean Geometry"}.

\begin{exemplum}[Möbius strip vs. torus]
\label{exemplum "Möbius strip vs. torus"}
Let 
\begin{equation}
	\varphi(\theta, t) = \left\{\left(2 - t\sin{\frac{\theta}{2}}\right)\sin{\theta}, \left(2 - t\sin{\frac{\theta}{2}}\right)\cos{\theta}, t\cos{\frac{\theta}{2}}\right\}
\end{equation}
be a parametrization from the map $\varphi \colon [0, 2\pi] \times (-1, 1)\to \mathbb{R}^3$. The image of $\varphi$ leads to an algebraic curve $\{\Moebius\}\varphi$ of genus 1, with the same properties of the elliptic plane, called the \emph{Möbius strip} \cite[§ 11, p. 41]{Mobius "Bestimmung des Inhalts eines Polyeders"}\footnote{
	\label{footnote "Listing–Möbius strip"} 
	Perhaps the Möbius strip should be called \emph{Listing strip}, or at least \emph{Listing–Möbius strip/non-orientable surface}, for it appears in all its glory in J.B. Listing \cite[Fig. 3, Abhandlungen Bd. X, Tab. I, after p. 182]{Listing "der Census raumlicher Complexe oder Verallgemeinerung des Euler'schen Satzes von den Polyadern"}. 
	}
$\ddot{\mathbb{O}} \cong \mathbb{S}^1 \times_{\mathbb{Z}/2} \mathbb{R}$, characterized by a \emph{one-sided surface} (see Fig. \ref{figure "Möbius strip"}).

On the other hand, let us get the \emph{2-torus}, a doughnut-object of genus 1 (cf. Scholium \ref{scholium "Torus"}) falling within the spherical system; it can be parametrized as 
\begin{subequations}
\begin{align}
	& x = (\mathrm{R} + \rho \cos{\theta}) \cos{\phi}, \\
	& y = (\mathrm{R} + \rho \cos{\theta}) \sin{\phi}, \\
	& z = \rho \sin{\phi}, \mathrm{R} > \rho, \theta, \phi \in (0, 2\pi), 
\end{align}	
\end{subequations}
from the map $\varphi \colon \mathbb{R} \to \torus^2 \cong \mathbb{S}^1 \times \mathbb{S}^1$, and it is characterized by a \emph{two-sided surface}. \exemplumsymbol
\end{exemplum}

\begin{scholium}[Myers–Cheng \& Grove–Shiohama sphere theorem]
\label{scholium "Myers–Cheng and Grove–Shiohama sphere theorem"}
We are mentioning a stimulating theorem due to S.-Y. Cheng \cite{Cheng "Eigenvalue Comparison Theorems and Its Geometric Applications"} concerning the isometry of the sphere $\mathbb{S}^n_\rho$ of radius $\rho > 0$. If $\mathcal{M}^n_\kappa$ is a complete Riemannian manifold of dimension $n \geqslant 2$ having Ricci curvature tensor
\begin{equation}
	\Ric \geqslant (n - 1) \kappa > 0
\end{equation}
and a diameter
\begin{equation}
	\diameter(\mathcal{M}) = \frac{\pi}{\sqrt{\kappa}} = \pi\left(\frac{1}{\sqrt{\kappa}}\right) = \pi\rho,
\end{equation}
then $\mathcal{M}^n_\kappa$ is \emph{isometric} to $\mathbb{S}^n_\rho$.
This result should be read in combination with the prior \emph{Myers's theorem} \cite{Myers "Riemannian manifolds with positive mean curvature"} (or sometimes \emph{Bonnet–Myers theorem}), which, for $\Ric \geqslant (n - 1) \kappa > 0$, demonstrates that 
\enumerationisinitium
\item $\mathcal{M}^n_\kappa$ ($n \geqslant 2$) is compact, 
\item $\diameter(\mathcal{M}) \leqslant \frac{\pi}{\sqrt{\kappa}} = \pi\rho$, 
\item the fundamental group $\pi_1(\mathcal{M}^n_\kappa)$ of $\mathcal{M}^n_\kappa$ is finite.
\enumerationisfinis

Another topological direction of the sphere theorem is adopted by the Rauch–Berger–Klingenberg solution \cite{Rauch "A Contribution to Differential Geometry in the Large"} \cite{Berger "Les varietes riemanniennes (1/4)-pincees"} \cite{Klingenberg "Uber Riemannsche Mannigfaltigkeiten mit nach oben beschrankter Krummung"}, under which, starting from a complete connected $n$-dimensional Riemannian manifold, with $\frac{1}{4} < \kappa \leqslant 1$, it is shown that $\mathcal{M}^n_\kappa$ is a \emph{twisted sphere} and, in particular, $\mathcal{M}^n_\kappa$ is \emph{homeomorphic} to $\mathbb{S}^n_\rho$. K. Grove and K. Shiohama \cite{Grove and Shiohama "A Generalized Sphere Theorem"}, see also \cite{Shiohama "A Sphere Theorem for Manifolds of Positive Ricci Curvature"}, bring a generalization of this proof (homeomorphism between $\mathcal{M}^n_\kappa$ and $\mathbb{S}^n_\rho$), setting $\kappa \geqslant 1$ and $\diameter(\mathcal{M}) > (\pi\sqrt{\kappa})/2$.
\scholiumsymbol
\end{scholium}

\begin{scholium}[Matryoshka-space forms]
Beltrami's proof \cite[p. 255]{Beltrami "Teoria fondamentale degli spazii di curvatura costante"} reminds us that the geodesic sphere in an $n$-space of constant negative curvature $-\frac{1}{R^2}$ is an $(n - 1)$-space of constant positive curvature $\frac{1}{D_\mathrm{q}^2}$, where $D_\mathrm{q} = R\sinh\frac{\rho}{R}$, and by that, \emph{the spherical geometry can be regarded as contained in the hyperbolic one}. \scholiumsymbol
\end{scholium}

\subsection{Type III. Beltrami–Poincaré Hyperbolic Model (and Parabolic View)}
\label{subsection "Type III. Beltrami–Poincaré Hyperbolic Model (and Parabolic View)"}

The third case means that $\mathcal{M}^n_\kappa$ ($\kappa \in \mathbb{R}_-$) is a \emph{hyperbolic $n$-space} $\hyperbolic^n_\rho$ having $\rho > 0$ and \emph{constant sectional curvature $\kappa = -\frac{1}{\rho^2}$}.\footnote{
	The hyperbolic plane $\hyperbolic^2_\rho$ has constant (negative) Gaussian curvature $\kappa = -\frac{1}{\rho^2}$.
	} 

\subsubsection{Upper Half-Space, Ball, and Hyperboloid}
\label{subsubsection "Upper Half-Space, Ball, and Hyperboloid"}

Next up, a subdivision into three models.
\enumerationisinitium
\item The $\hyperbolic^n_\rho$-space corresponds to the (open) \emph{upper half-space} in $\mathbb{R}^n$, which in the \emph{Beltrami–Poincaré half-space model} \cite{Beltrami "Saggio di interpetrazione della Geometria non-euclidea"} \cite{Beltrami "Teoria fondamentale degli spazii di curvatura costante"} \cite{Beltrami "Sulla superficie di rotazione che serve di tipo alle superficie pseudosferiche"} \cite{Poincare "Theorie des groupes fuchsiens"} \cite{Poincare "Memoire sur les groupes kleineens"} is written as 
\begin{equation}
	\mathbb{U}^n_\rho = \{x = (x^1, \mathellipsis, x^n) \in \mathbb{R}^n \mid x^n > 0\};
\end{equation}
the metric on $\mathbb{U}^n_\rho$ is given by 
\begin{equation}	
	g^3_\mathbb{U} = \rho^2(x^n)^{- 2}(dx^1 \otimes dx^1 + \mathellipsis + dx^n \otimes dx^n).
\end{equation}
The boundary of $\mathbb{U}^n$ is provided by the \emph{boundary at infinity} $\partial_\infty\mathbb{U}^n = (\mathbb{R}^{n-1} \times \{0\}) \cup \{\infty\}$. The closed upper half-space will be 
\begin{equation}
	\overbar{\mathbb{U}}^n = \{(x^1, \mathellipsis, x^n) \in \mathbb{R}^n \mid x^n \geqslant 0\}.
\end{equation}
\item The complementary version is the \emph{Beltrami–Poincaré} (open) \emph{ball model}, 
\begin{equation}
\label{equation "Beltrami–Poincaré (open) ball model"}
	\mathbb{B}^n_\rho = \{x = (x^1, \mathellipsis, x^n) \in \mathbb{R}^n \mid \|x\| < \rho\},
\end{equation}	
with $\rho = \frac{1}{\sqrt{-\kappa}}$. 
The metric on $\mathbb{B}^n_\rho$ is 
\begin{equation}
	g^2_\mathbb{B} = 4\rho^4(\rho^2 - \|x\|^2)^{- 2}(dx^1 \otimes dx^1 + \mathellipsis + dx^n \otimes dx^n).
\end{equation} 
The boundary of $\mathbb{B}^n$ coincides with the \emph{(boundary) sphere at infinity}
\begin{align}	
	\partial_\infty\mathbb{B}^n \equival \mathbb{S}^{n - 1}_\infty & = \{x \in \mathbb{R}^n \mid \|x\| = 1\} \notag \\
	& = \{(x^1, \mathellipsis, x^n) \in \mathbb{R}^n \mid (x^1)^2 + \mathellipsis + (x^n)^2 = 1\};
\end{align}
and the $(n - 1)$-sphere is finally the one-point compactification (in the sense of Aleksandrov \cite{Aleksandrov "Uber die Metrisation der im Kleinen kompakten topologischen Raume"}) of the plane $\mathbb{R}^{n-1} \cup \{\infty\}$.
\item There is a third model of hyperbolic space; it is the \emph{hyperboloid} 
\begin{equation}
\label{equation "Minkowski–Lorentz model"}
	\mathbb{Y}^n_+ \viz \mathbb{Y}^n_\rho = \{x \in \mathbb{R}^{n + 1} \mid (x^{n + 1})^2 - \|(x^1, \mathellipsis, x^n) \in \mathbb{R}^n\|^2 = \rho^2, x^{n + 1} > 0\}, 
\end{equation}	
also known as \emph{Minkowski–Lorentz model}. The hyperboloid $\mathbb{Y}^n_+$ is the \emph{upper sheet of the elliptic hyperboloid}, or \emph{2-sheeted hyperboloid}, in $\mathbb{R}^{n + 1}$. The metric on it is 
\begin{equation}
\label{equation "Metric on the Minkowski–Lorentz model"}
	g^1_\mathbb{Y} = \iota^*\eta, \enspace \iota \colon \mathbb{Y}^n_\rho \to \mathbb{R}^{n + 1},
\end{equation}
where $\iota$ indicates the inclusion map, while $\eta$ is a pseudo-Euclidean metric, more specifically, it is about the \emph{Minkowski metric} (see Section \ref{section "Lorentz–Minkowski 4-Manifolds"}, especially Fig. \ref{figure "hyperboloid surfaces coexistence"}).

All these three models (upper half-space, ball and hyperboloid) are representations of \emph{Lobačevskij–Bolyai's} geometry \cite{Lobacevskij "Novyye nachala geometrii"} \cite{Lobacevskij "New Principles of Geometry"} \cite{Bolyai "Appendix scientiam spatii absolute veram exhibens"} \cite{Bolyai "Sulla scienza dello spazio assolutamente vera"} \cite{Bolyai "The Science of Absolute Space"} \cite{Bolyai "The Science of Absolute Space in Bonola"} in the \emph{Euclidean system}, in addition to being examples of Riemannian manifolds. 

Nevertheless, Hilbert's theorem \cite{Hilbert "Ueber Flachen von Constanter Gaussscher Krummung"} sets a limit on the immersive structure of hyperbolic constructions,\footnote{
	Cf. G. Darboux \cite[§ 773, pp. 379-381]{Darboux "Lecons sur la theorie generale des surfaces et les applications geometriques du calcul infinitesimal III}, L. Bianchi \cite[§ 67, pp. 126-128]{Bianchi "Lezioni di geometria differenziale (prima edizione)"}, and U. Dini \cite[in particular § 11, pp. 184-185]{Dini "Sopra alcune formole generali della teoria delle superficie e loro applicazioni"}.
	} 
by establishing the non-existence of a complete analytic—belonging to the class $\mathscr{C}^\omega$—regular surface of constant negative Gaussian curvature in $\mathbb{R}^3$. The full hyperbolic plane does not admit an \emph{isometric immersion} in Euclidean 3-space. Note. Compare Hilbert's theorem with the publication of A. Genocchi \cite[pp. 390-404]{Genocchi "Sur un memoire de Daviet de Foncenex et sur les geometries non euclidiennes"}.
\enumerationisfinis

\subsubsection{Isometric Spaces}

$\mathbb{U}^n_\rho$, $\mathbb{B}^n_\rho$ and $\mathbb{Y}^n_\rho$ are \emph{isometric (Riemannian) spaces with mutual action}. For instance, by means of a \emph{hyperbolic stereographic projection} $\pi \colon \mathbb{Y}^n_\rho \to \mathbb{B}^n_\rho$, it can be proved that $\pi^*g^2_\mathbb{B} = g^1_\mathbb{Y}$, or $\pi^{-1}{}^*g^1_\mathbb{Y} = g^2_\mathbb{B}$. So we take the example of the first case.

\begin{propositio}
The Beltrami–Poincaré ball $\mathbb{B}^n_\rho$ and the upper sheet of the elliptic hyperboloid $\mathbb{Y}^n_\rho$ are isometric in a mutual relationship, for which the formula $\pi^*g^2_\mathbb{B} = g^1_\mathbb{Y}$ holds. 	
\end{propositio}

\begin{proof}
Let us start to build the aforementioned diffeomorphism, i.e. $\pi \colon \mathbb{Y}^n_\rho \to \mathbb{B}^n_\rho$. We write the map and its inverse as 
\begin{equation}	
	\pi(x) = \frac{\rho}{\rho + x^{n + 1}}\dot{x} \in \mathbb{B}^n_\rho, \text { and } \pi^{-1}(u) = \left(\frac{2\rho^2u}{\rho^2 - \|u\|^2}, \rho\frac{\rho^2 + \|u\|^2}{\rho^2 - \|u\|^2}\right),
\end{equation}	
where $x \in \mathbb{Y}^n_\rho$ is an element on the upper sheet of a 2-sheeted hyperboloid, and $\dot{x} = x^1, \mathellipsis, x^n$. Then
\begin{equation}
d\pi_x(v) = \frac{\rho}{\rho + x^{n + 1}}\left(\dot{v} - \frac{v^{n + 1}}{\rho + x^{n + 1}}\dot{x}\right).
\end{equation}
Also, let $v \in \mathcal{T}_x\mathbb{Y}^n_\rho$ be subject to the condition that $x^{n + 1}v^{n + 1} = \langle\dot{x}, \dot{v}\rangle$. We finally get to
\begin{align}
	\pi^*g^2_\mathbb{B}(v, v) & = g^2_\mathbb{B}\bigl(d\pi_x(v), d\pi_x(v)\bigr) \notag \\
	& = \frac{4\rho^4}{\bigl(\rho^2 - \|\pi(x)\|^2\bigr)^2}\|d\pi_x(v)\|^2 \notag \\
	& = \frac{4}{\left(1 - \frac{\|\dot{x}\|^2}{(\rho + x^{n + 1})^2}\right)^2}\frac{\rho^2}{(\rho + x^{n + 1})^2}\left\|\dot{v} - \frac{v^{n + 1}}{\rho + x^{n + 1}}\dot{x}\right\|^2 \notag \\
	& = \|\dot{v}\|^2 - \frac{2v^{n + 1}}{\rho + x^{n + 1}}\langle\dot{x}, \dot{v}\rangle + \frac{|v^{n + 1}|^2}{(\rho + x^{n + 1})^2}\|\dot{x}\|^2 \notag \\
	& = \|\dot{v}\|^2 - |v^{n + 1}|^2 \notag \\
	& = g^1_\mathbb{Y}(v, v).
\end{align}
\end{proof}
Likewise, in combination with a diffeomorphism $\varphi \colon \mathbb{B}^n_\rho \to \mathbb{U}^n_\rho$ it is possible to prove that $\varphi^*g^3_\mathbb{U} = g^2_\mathbb{B}$. The maps
\begin{subequations}
\begin{align}
	& \varphi(u) = \left(\frac{2\rho^2\dot{u}}{\|\dot{u}\|^2 + (u^n - \rho)^2}, \rho\frac{\rho^2 - \|\dot{u}\|^2 - |u^n|^2}{\|\dot{u}\|^2 + (u^n - \rho)^2}\right), \\
	& \varphi^{-1}(w) = \left(\frac{2\rho^2\dot{w}}{\|\dot{w}\|^2 + (w^n + \rho)^2}, \rho\frac{\|\dot{w}\|^2 + |w^n|^2 - \rho^2}{\|\dot{w}\|^2 + (w^n + \rho)^2}\right),
\end{align}
\end{subequations}  
coincide with the complex \emph{Cayley transform} \cite{Cayley "Sur quelques proprietes des determinants gauches"} and its inverse, imposing $\dot{u} = (u^1, \mathellipsis, u^{n - 1}) \in \mathbb{R}^{1-n}$.

\subsubsection{Upper Half-Plane and Disk}
\label{subsubsection "Upper Half-Plane and Disk"}

For the case of the 2-dimensional hyperbolic geometry, the Beltrami–Poincaré construction is defined by the (open) \emph{upper half-plane} and \emph{disk models}, $\mathbb{U}^2$ and $\mathbb{D} \viz \mathbb{D}^2 \equival \mathbb{B}^2$, respectively. In short, the real-valued or complex-valued \emph{upper half-plane} (cf. Section \ref{section "Tessellation of the Upper Half-Plane Plane by Modular Group"}),
\begin{align}
	& \mathbb{U}^2_\mathbb{R} = \{(x, y) \in \mathbb{R}^2 \mid y > 0\}, \\ 
	& \mathbb{U}^2_\mathbb{C} = \{z = (x + iy) \in \mathbb{C} \mid y = \Im(z) > 0\},
	\label{align "Complex-valued upper half-plane"}
\end{align} 
and the \emph{unit disk} in the real or complex plane,
\begin{align}	
	& \mathbb{D}_\mathbb{R} \equival \mathbb{B}^2_\mathbb{R} = \{(x, y) \in \mathbb{R}^2 \mid x^2 + y^2 < 1\}, \\ 
	& \mathbb{D}_\mathbb{C} \equival \mathbb{B}^2_\mathbb{C} = \{z = (x + iy) \in \mathbb{C} \mid \|z\| < 1\}. 
	\label{align "Unit disk in the complex plane"}
\end{align}

The \emph{geodesics} 
\enumerationisinitium
\item in $\mathbb{U}^2_\mathbb{C}$ are composed of \emph{vertical straight half-lines and half-circles with center on the real axis} in the upper half-plane,
\item in $\mathbb{D}_\mathbb{C}$ are composed of \emph{generalized $\mathbb{D}$-circles orthogonal to the (boundary) circle at infinity} $\partial\mathbb{D}_\mathbb{C} \equival \mathbb{S}^1_\infty \cong \mathbb{RP}^1$ of the unit disk.
\enumerationisfinis

\begin{scholium}
\label{scholium "Right (complex-valued) half-plane and closed unit disk"}
~\enumerationisinitium
\item The right (complex-valued) half-plane will be 
\begin{equation}
	 \{z = (x + iy) \in \mathbb{C} \mid x = \Re(z) > 0\}.
\end{equation}
\item The closed unit disk is 
\begin{align}
	& \overbar{\mathbb{D}} \equival \overbar{\mathbb{B}}^2 = \{(x, y) \in \mathbb{R}^2 \mid x^2 + y^2 \leqslant 1\} \text{ or } \\
	& \{z \in \mathbb{C} \mid \|z\| \leqslant 1\}. 
\end{align}
\scholiumsymbol
\enumerationisfinis
\end{scholium}

\begin{figure}
\centering
	\includegraphics[width = 0.550\textwidth]{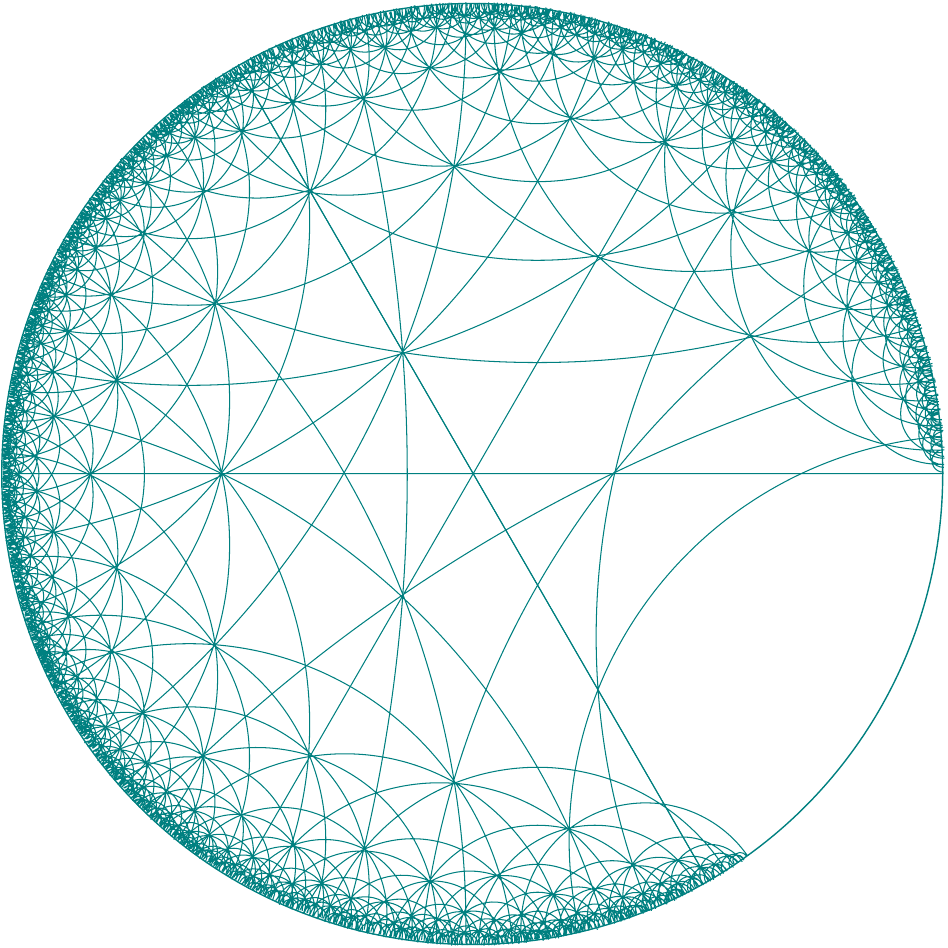}
	\caption{Process of tessellation, under construction, of the hyperbolic disk, i.e. of a $\mathbb{D}^2$- or $\mathbb{B}^2$-type space, via triangles, whose angles are $\frac{\pi}{2}$, $\frac{\pi}{3}$, and $\frac{\pi}{7}$}
	\label{figure "tessellation 1"} 
\end{figure}

\begin{figure}
\centering	
	\includegraphics[width = 0.550\textwidth]{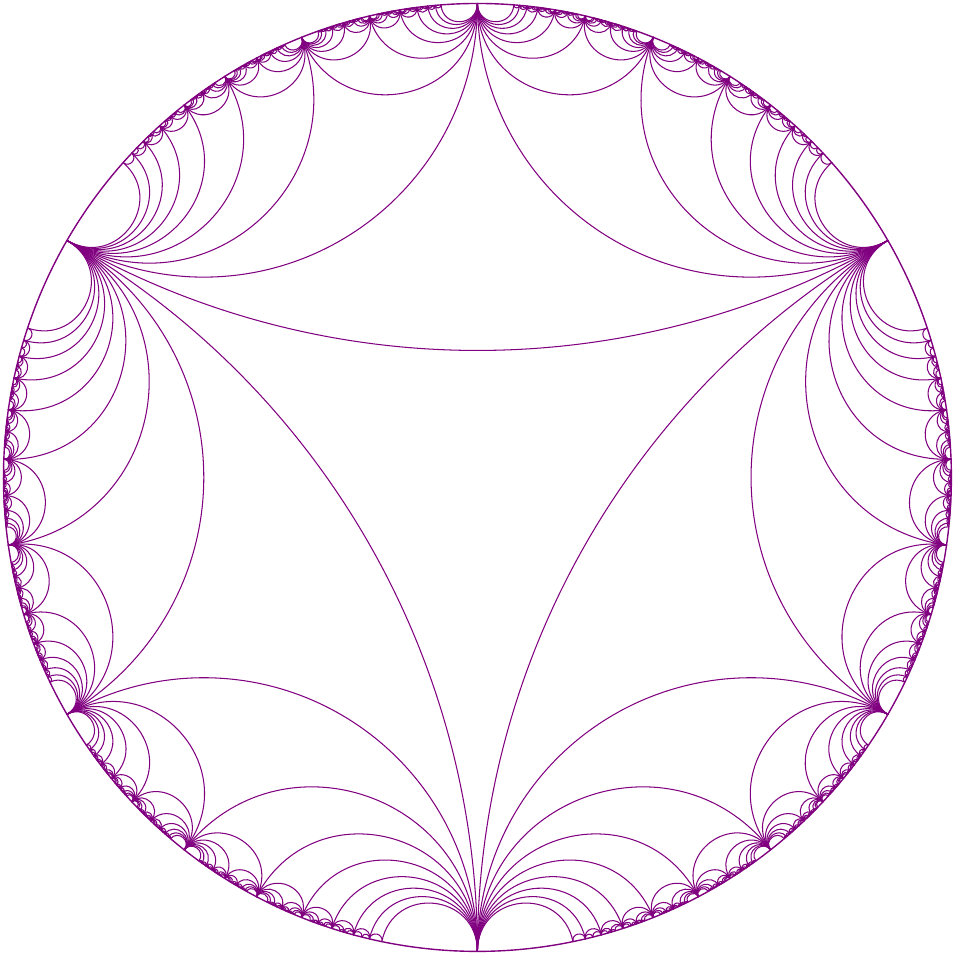}
	\caption{Process of tessellation of the hyperbolic disk, i.e. of a $\mathbb{D}^2$- or $\mathbb{B}^2$-type space, adopting an ideal triangular process, to wit, triply asymptotic triangles}
	\label{figure "tessellation 2"}
\end{figure}

In order to give also a visual indication of the 2-dimensional hyperbolic geometry, we present two tessellated models in Figg. \ref{figure "tessellation 1"} and \ref{figure "tessellation 1"}, since they give an  excellent epitomization of this geometry, along with its eye-appeal.
 
\subsubsection{Beltrami–Cayley–Klein Model}
\label{subsubsection "Beltrami–Cayley–Klein Model"}

A unifying \emph{projective} version of Beltrami's disk is previously prepared by F. Klein \cite{Klein "Sur la geometrie dite non euclidienne"} \cite{Klein "Ueber die sogenannte Nicht-Euklidische Geometrie"} \cite{Klein "Ueber die sogenannte Nicht-Euklidische Geometrie (Zweiter Aufsatz.)"} in the \emph{Beltrami–Klein model}, sometimes called \emph{Beltrami–Cayley–Klein model} \cite{Cayley "A Sixth Memoire upon Quantics"}:
\begin{equation}
	\mathbb{D}_\pi = \{x = (x^1, \mathellipsis, x^{n + 1}) \mid (x^1)^2 + \mathellipsis + (x^n)^2 < 1; x^{n + 1} = 1\}.
\end{equation}

\subsubsection{Equivalence via Cayley Transform}
\label{subsubsection "Equivalence via Cayley Transform"}

There is an equivalence between the \emph{Riemann surfaces} \eqref{align "Complex-valued upper half-plane"} \eqref{align "Unit disk in the complex plane"}. The \emph{Cayley transform} 
\begin{equation}
		\varphi_\mathbb{C} \colon z \mapsto \frac{z - i}{z + i}
\end{equation}
ensures a \emph{conformal mapping} of the upper half-plane onto the unit disk, to wit,
\[
	\{z \in \mathbb{C} \mid \Im(z) > 0\} \text{ onto } \{z \in \mathbb{C} \mid \|z\| < 1\},
\]
and $\varphi_\mathbb{C}$ is a \emph{linear fractional transformation}. 

The opposite Cayley's implication, 
\begin{equation}
		\varphi(z) = i\left(\frac{1 - z}{1 + z}\right),
\end{equation}
expects to map the disk $\mathbb{D}_\mathbb{C}$ onto the half-plane, with a one-to-one correspondence, such that 
\begin{equation}
	\varphi \colon \mathbb{D}_\mathbb{C} \to \{z \in \mathbb{C} \mid \Im(z) > 0\}. 
\end{equation}
So the spaces $\mathbb{U}^2_\mathbb{C}$ and $\mathbb{D}_\mathbb{C}$ are said to be \emph{conformally equivalent}.

\section{Certain Geodesically Convex Conditions}
\label{section "Certain Geodesically Convex Conditions"}

Let us also say something about the convexity, since is related to the above. First we will provide a number of definitions for the metric space, and then for the (normed) vector space.

\subsection{Geodesic and Strict Convexity; Convex Function on Metric and Vector Spaces}
\label{subsection "Geodesic and Strict Convexity; Convex Function on Metric and Vector Spaces"}

\begin{definitio}[Geodesic convexity]
\label{definitio "Geodesic convexity"}
Given a subset $\mathcal{W}$ of a metric space $(\mathcal{X}, \distance)$ and a (constant speed) geodesic $\gamma_\mathrm{c} \colon [x, y] \to \mathcal{X}$ (cf. Definition \ref{definitio "Geodesic space"}), the set $\mathcal{W}$ is said to be \emph{geodesically convex} if, for any two points $x$ and $y$ of $\mathcal{W}$, there is a geodesic segment $[x, y]$ in $\mathcal{X}$ joining them and the image of such a segment is entirely contained in $\mathcal{W}$. The set $\mathcal{W}$ is geodesically convex in the \emph{strongest sense} of the word, if the image of the segment is unique (there is one and only one geodesic between $x$ and $y$). \definitiosymbol
\end{definitio}

So let us look at when can one say that a function is convex.

\begin{definitio}[Convex function on a metric space]
Let $\mathcal{W}$ be once again a subset of $(\mathcal{X}, \distance)$. A function $\varphi \colon \mathcal{W} \to \mathbb{R} \cup \{\infty\}$ is \emph{convex} if the map $\varphi \circ \gamma_\mathrm{c} \colon [0, 1] \to \mathbb{R}$ is convex for every (constant speed) geodesic $(\gamma_\mathrm{c})_{0 \leqslant t \leqslant 1} \colon [0, 1] \to \mathcal{W}$, such that 
\begin{equation}
	\varphi\bigr(\gamma_\mathrm{c}(t)\bigl) \leqslant (1 - t)\varphi\bigl(\gamma_\mathrm{c}(0) = x\bigr) + t\varphi\bigl(\gamma_\mathrm{c}(1) = y\bigr), 
\end{equation}
where $\bigr(\gamma_\mathrm{c}(t)\bigl)$ has to do with an evaluation map at time $t \in [0, 1]$. \definitiosymbol	
\end{definitio}

Now, as for the vector space, the definition on convex function is the following.

\begin{definitio}[Convex function on a vector space]
Let $\mathfrak{X}$ be a vector space and $\mathfrak{W} \subset \mathfrak{X}$ an affinely convex subset of $\mathfrak{X}$. We choose two distinct points $x$ and $y$ in $\mathfrak{X}$. The function $\varphi \colon \mathfrak{W} \to \mathbb{R}$
\enumerationisinitium
\item is \emph{convex} if 
\begin{equation}
	\varphi\Bigl(t{y} + (1 - t)x\Bigr) \leqslant (1 - t)\varphi(x) + t\varphi(y),
\end{equation}
for all $x, y \in \mathfrak{W}$ and $t \in [0, 1]$, with $0 \leqslant t \leqslant 1$;
\item is \emph{strictly convex} if 
\begin{equation}
	\varphi\Bigl(t{y} + (1 - t)x\Bigr) < (1 - t)\varphi(x) + t\varphi(y),
\end{equation}
for all $x, y \in \mathfrak{W}$ and $t \in~]0, 1[$, with $0 < t < 1$. \definitiosymbol
\enumerationisfinis
\end{definitio}

Under this circumstance, it is sufficient to point out that the \emph{ball model}, whether open $\mathbb{B}^n$ \eqref{equation "Beltrami–Poincaré (open) ball model"} or closed $\overbar{\mathbb{B}}^n$, is a \emph{geodesically convex subset} (Definition \ref{definitio "Geodesic convexity"}) equipped with both hyperbolic and Euclidean metric, and \emph{one of many examples of convex subset of normed vector space(s)}. Remember that 
\enumerationisinitium
\item a closed unit ball $\overbar{\mathbb{B}}^n_{(\rho = 1)} = \{x = (x^1, \mathellipsis, x^n) \in \mathbb{R}^n \mid \|x\| \leqslant \rho = 1\}$ is a manifold with boundary $\partial\overbar{\mathbb{B}} = \mathbb{S}^{n - 1}$, and it is convex in a normed vector space;
\item a \emph{normed vector space} $\mathfrak{X} \viz (\mathfrak{X}, \|\cdot\|)$ is a vector space whose map $x \mapsto \|x\|$ goes from $\mathfrak{X}$ to $[0, \infty[$, and having the following properties, for all vectors $x, y \in \mathfrak{X}$:
\subenumerationisinitium
\item $\|x\| \geqslant 0$, and $\|x\| = 0$ iff $x = 0$,
\item $\|\lambda{x}\| = |\lambda|.\|x\|$, for any scalar $\lambda \in \mathbb{R}$ (homogeneity),
\item $\|x + y\| \leqslant \|x\| + \|y\|$ (triangle inequality).
\subenumerationisfinis 
\enumerationisfinis

When a normed vector space is finite-dimensional, we say it is a \emph{Minkowski space}.
The \emph{closed unit ball} and the \emph{sphere(s)} are \emph{strictly convex}, specifically. Let us see what this means.

\begin{definitio}[Strict convexity of a normed vector space]
Let a normed vector space $\mathfrak{X} \viz (\mathfrak{X}, \|\cdot\|)$, a unit sphere $\mathbb{S}^n_\mathfrak{X} = \{x \in \mathfrak{X} \mid \|x\| = 1\}$ \eqref{equation "Unit sphere"} in $\mathfrak{X}$, and $x, y \in \mathfrak{X}$ be given. A normed vector space $\mathfrak{X}$ is called \emph{strictly convex} on the basis of one of the following conditions:
\enumerationisinitium
\item if $\|x\| = \|y\| = 1$, with $x \neq y$, then $\|t{y} + (1 - t)x\| < 1$, for all $t \in~]0, 1[$,
\item if $\|x+y\| = \|x\| + \|y\|$, with $x \neq 0$, then $t \geqslant 0$, for which $y = tx$,
\item if there is a strictly convex function $x \xrightarrow{\varphi} \|x\|^2$,
\item if no affine segment (joining any two points) is contained in $\mathbb{S}^n_\mathfrak{X}$ of $\mathfrak{X}$. \definitiosymbol
\enumerationisfinis 
\end{definitio}

\emph{Every normed vector space—that is strictly convex—is uniquely geodesic} (see Definition \ref{definitio "Geodesic space"}). 

\section{Some Types of Möbius (Projective) Transformations}
\label{section "Some Types of Möbius (Projective) Transformations"}

\begingroup
\footnotesize
One of the fundamental concepts of modern geometry is by Möbius \cite{Mobius "Der barycentrische Calcul ein neues Hulfsmittel zur analytischen Behandlung der Geometrie"}, that is, the general concept of \emph{biunivocal correspondence} [bijection, one-to-one correspondence] or \emph{transformation}, on the plane and on the space. \\ 
\indent — \textsc{F. Enriques} \cite[p. 365]{Enriques "Lezioni di Geometria proiettiva"}\footnote{
	See also G. Castelnuovo \cite[§§ 38, 58, 136]{Castelnuovo "Lezioni di geometria analitica e proiettiva I"}.
	}

\endgroup

\subsection{Orientation Preserving Isometries of the Hyperbolic Plane and Disk}
\label{subsection "Orientation Preserving Isometries of the Hyperbolic Plane and Disk"}
 	
\subsubsection[$\mathbb{U}^2_\mathbb{C}$ Case]{$\protect\pseudobold{\mathbb{U}^2_\mathbb{C}}$ Case}
\label{subsubsection "I. Complex-Valued Upper Half-Plane Case"}

Let 
\begin{equation}
	\mathbb{U}^2_\mathbb{C} = \{z \in \mathbb{C} \cong \mathbb{R}^2 \mid \Im(z) > 0\}
\end{equation}
be the complex-valued upper half-plane, whose \emph{Poincaré metric} is
\begin{equation}	
	\frac{d\bar{z}dz}{\Im(z)^2} = \frac{dx^2 + dy^2}{y^2}, 
\end{equation}
with $z = (x + iy)$. The group acting smoothly and transitively on $\mathbb{U}^2_\mathbb{C}$ is the special linear group 
\begin{equation}
\label{equation "Special linear group of degree 2 matrices over the real field"} 
	SL_2(\mathbb{R}) = 
	\left\{
	\begin{pmatrix}	
	\alpha & \beta \\ 
	\gamma & \delta
	\end{pmatrix}
	\mathrel{\bigg|} \alpha, \beta, \gamma, \delta \in \mathbb{R}, \alpha\delta - \beta\gamma = 1
	\right\},
\end{equation}
composed of all $2 \times 2$ real matrices with unit determinant. As a set of points, the group $SL_2(\mathbb{R})$ is a subset of $\mathbb{R}^4$, i.e. can be regarded, topologically, as a submanifold of dimension 3 in the 4-space. The $SL_2$-action on the complex-valued upper half-plane is provided by a function of the form
\begin{align}	
\label{align "Möbius transformation of the complex upper half-plane"}
	& z \mapsto \varphi_{\alpha\beta\gamma\delta}[\mathbb{U}^2_\mathbb{C}](z) = \frac{\alpha z + \beta}{\gamma z + \delta} = 
	\begin{pmatrix}	
	\alpha & \beta \\ 
	\gamma & \delta
	\end{pmatrix}z, \notag \\
	& \text{for } \alpha, \beta, \gamma, \delta \in \mathbb{R}, \det(M) = 1, M \viz [M]^{2 \times 2} = \bigl(\begin{smallmatrix}\alpha & \beta \\ \gamma & \delta\end{smallmatrix}\bigr) \in SL_2(\mathbb{R}).
\end{align}
The function \eqref{align "Möbius transformation of the complex upper half-plane"} is called \emph{Möbius transformation}, or \emph{linear fractional transformation} of the $\mathbb{U}^2_\mathbb{C}$-hyperbolic space, mapping $\mathbb{U}^2_\mathbb{C}$ onto itself. Which means that $\varphi[\mathbb{U}^2_\mathbb{C}]$ is an \emph{orientation preserving isometry map}, and it leaves this model of (negative) geometry \emph{invariant}.

Here we are introducing another piece of the puzzle; it is the group $PSL_2(\mathbb{R})$, the projective special linear group of $2 \times 2$ matrices over $\mathbb{R}$. If the group $SL_2(\mathbb{R})$ acts by a Möbius transformation \eqref{align "Möbius transformation of the complex upper half-plane"}, then the group of \emph{orientation preserving isometries of the Beltrami–Poincaré upper half-plane} shall be represented by the quotient
\begin{equation}
\label{equation "Group $PSL_2(R)$"}
	PSL_2(\mathbb{R}) \cong \frac{SL_2(\mathbb{R})}
	{\left\{\pm
	\begin{pmatrix}
	1 & 0 \\
	0 & 1
	\end{pmatrix}
	\right\} \viz \{\pm\idem\}},
\end{equation}
where $\idem$ equals the $2 \times 2$ identity matrix. The expression \eqref{equation "Group $PSL_2(R)$"} is by \emph{lifting} to $SL_2(\mathbb{R})$ the group $PSL_2(\mathbb{R})$.

The set $PSL_2(\mathbb{R})$ is the group of Möbius transformations of the kind $\varphi\{\mathbb{U}^2_\mathbb{C}\}$, hence it is equal to the \emph{group of all biholomorphic (or conformal and bijective) maps} of $\mathbb{U}^2_\mathbb{C}$ to itself, 
\begin{equation}
	\varphi \colon \mathbb{U}^2_\mathbb{C} \to \mathbb{U}^2_\mathbb{C}.
\end{equation} 
We remark that the, in the topological sense, $PSL_2(\mathbb{R})$ is diffeomorphic to a solid torus $\mathbb{S}^1 \times \mathbb{D}$ (Cartesian product of the 1-dimensional circle with the disk). The group $SL_2(\mathbb{R})$ is topologically equivalent to a \emph{2-fold covering (double cover)} of $PSL_2(\mathbb{R})$. It can therefore be established a  correspondence between these two groups as follows.

\begin{propositio} 
\label{propositio "Universal cover of $PSL_2(R)$, and more"} 
The universal cover of $PSL_2(\mathbb{R})$ is $\widetilde{SL}_2(\mathbb{R})$. Using the $\Moebius$-group notation, we have
\begin{equation}	
	\pi \colon \left(\widetilde{\Moebius}^+_2(\mathbb{R}) \cong \widetilde{SL}_2(\mathbb{R})\right) \to \biggl(\left(\Moebius^+_2(\mathbb{R}) \equival \mathfrak{isom}^+(\mathbb{U}^2_\mathbb{C})\right) \cong PSL_2(\mathbb{R})\biggr).
\end{equation}
The notation $\mathfrak{isom}^+$ denotes the general group of orientation preserving isometries of the hyperbolic 2-space. Let $\mathbb{Z} = \pi_1\bigr(PSL_2(\mathbb{R})\bigr)$ be the center of $\widetilde{SL}_2(\mathbb{R})$ and the kernel of the projection from $\widetilde{SL}_2(\mathbb{R})$ into $PSL_2(\mathbb{R})$ (or, stated more correctly, the center of $\widetilde{SL}_2(\mathbb{R})$, the kernel projection $\kernel(\pi)$ and the fundamental group of $PSL_2(\mathbb{R})$ are all isomorphic to $\mathbb{Z}$). The group $\widetilde{SL}_2(\mathbb{R})$ is a central extension of $PSL_2(\mathbb{R})$, so the exact sequence is
\begin{equation}	
	0 \longrightarrow \mathbb{Z} \longrightarrow \left(\widetilde{\Moebius}^+_2(\mathbb{R}) \cong \widetilde{SL}_2(\mathbb{R})\right) \xrightarrow{\enspace \pi \enspace} \left(\Moebius^+_2(\mathbb{R}) \cong PSL_2(\mathbb{R})\right) \longrightarrow 0.
\end{equation} 
\end{propositio}

\begin{scholium}
Let $\varphi \colon \mathbb{U}^2_\mathbb{C} \to \mathbb{U}^2_\mathbb{C}$. If we are given a curve $\gamma_\mathrm{c}$ of the upper half-plane, the group isomorphism $\mathfrak{isom}^+(\mathbb{U}^2_\mathbb{C}) \cong PSL_2(\mathbb{R})$ is proven exhibiting the following integral,
\begin{equation}
	\int_{\varphi(\gamma_\mathrm{c})}\frac{|d\varphi_{\alpha\beta\gamma\delta}[\mathbb{U}^2_\mathbb{C}](z)|}{\Im\bigl(\varphi_{\alpha\beta\gamma\delta}[\mathbb{U}^2_\mathbb{C}](z)\bigr)} \leqslant \int_{\gamma_\mathrm{c}}\frac{|dz|}{\Im(z)}.
\end{equation}
\scholiumsymbol
\end{scholium}

\subsubsection[$\mathbb{D}_\mathbb{C}$ Case]{$\protect\pseudobold{\mathbb{D}_\mathbb{C}}$ Case}
\label{subsubsection "II. Unit Disk in the Complex Plane Case"}

Similar reasoning could be applied to the unit disk in the complex plane. The functions
\begin{subequations}
\begin{align}
	& z \mapsto \varphi_{\alpha\beta}[\mathbb{D}_\mathbb{C}](z) = \frac{\alpha z + \beta}{\bar{\beta} z + \bar{\alpha}} = 
	\begin{pmatrix}
	\alpha & \beta \\ 
	\bar{\beta} & \bar{\alpha}
	\end{pmatrix} z, \text{ for } \alpha, \beta \in \mathbb{C}, |\alpha|^2 - |\beta|^2 = 1, \\
	& z \mapsto \varphi_\alpha[\mathbb{D}_\mathbb{C}](z) = e^{i\theta}\frac{z - \alpha}{1 - \bar{\alpha}z}, \text{ for } \alpha \in \mathbb{C}, |\alpha| < 1, \theta \in \mathbb{R},
\end{align}
\end{subequations}
with complex numbers $\alpha$ and $\beta$, are \emph{Möbius invariant transformations}. In particular, they are a \emph{conformal self-mapping} of the unit disk (or 2-ball) $\mathbb{D}_\mathbb{C} \equival \mathbb{B}^2_\mathbb{C} = \{z = (x + iy) \in \mathbb{C} \mid \|z\| < 1\}$, i.e. a diffeomorphism from $\mathbb{D}_\mathbb{C}$ to itself, 
\begin{equation}
	\varphi \colon \mathbb{D}_\mathbb{C} \to \mathbb{D}_\mathbb{C}.
\end{equation}

Having regard to the distance preserving, it should be mentioned the \emph{group of orientation preserving isometries of the Beltrami–Poincaré disk}, with the \emph{projective special unitary group} 
\begin{equation}
	PSU_2(\mathbb{C}) \cong \frac{SU_2(\mathbb{C})}{\{\pm\idem\}} \cong \mathfrak{isom}^+(\mathbb{D}_\mathbb{C}).
\end{equation} 

Such a group is isomorphic to the group of rotations in ordinary Euclidean 3-space, $PSU_2(\mathbb{C}) \cong SO_3(\mathbb{R})$. So there is a link between the spatial rotation of $SO_3(\mathbb{R})$ and the 2-sphere $\mathbb{S}^2 \cong \hat{\mathbb{R}}^2 \equival \mathbb{R}^2 \cup \{\infty\}$, or the Riemann sphere, if we put $\mathbb{R}^2 \cong \mathbb{C}$ (see below).

\subsubsection{Abridgment of the Two Previous Cases}
\label{subsubsection "III. Abridgment of the Two Previous Cases"}

In the most schematic possible view of the above, we can rewrite in this way:
\begin{align}
	\Moebius^+_2(\mathbb{R}) & = \left\{\varphi_{\alpha\beta\gamma\delta}[\mathbb{U}^2_\mathbb{C}](z) = \frac{\alpha z + \beta}{\gamma z + \delta} \mathrel{\bigg|} \alpha, \beta, \gamma, \delta \in \mathbb{R}\right\} \notag \\
	& = \aut(\mathbb{U}^2_\mathbb{C}) \viz \Moebius(\mathbb{U}^2_\mathbb{C}) \equival \mathfrak{isom}^+(\mathbb{U}^2_\mathbb{C}) \cong PSL_2(\mathbb{R}) \cong \frac{SL_2(\mathbb{R})}{\{\pm\idem\}}, \\
	\Moebius^+_{\mathbb{B}^2}(\mathbb{C}) & = \left\{\varphi_{\alpha\beta}[\mathbb{D}_\mathbb{C}](z) = \frac{\alpha z + \beta}{\bar{\beta} z + \bar{\alpha}} \equival \varphi_\alpha[\mathbb{D}_\mathbb{C}](z) = e^{i\theta}\frac{z - \alpha}{1 - \bar{\alpha}z} \mathrel{\bigg|} \alpha, \beta \in \mathbb{C}\right\} \notag \\
	& = \aut(\mathbb{D}_\mathbb{C}) \viz \Moebius(\mathbb{D}_\mathbb{C}) \equival \mathfrak{isom}^+(\mathbb{D}_\mathbb{C}) \cong PSU_2(\mathbb{C}) \cong \frac{SU_2(\mathbb{C})}{\{\pm\idem\}}.
\end{align}
The \emph{automorphism groups} $\aut(\mathbb{U}^2_\mathbb{C})$ and $\aut(\mathbb{D}_\mathbb{C})$ are generated, under composition of mappings, by the \emph{set of automorphisms of $\mathbb{U}^2_\mathbb{C}$ and $\mathbb{D}_\mathbb{C}$}, as are the isomorphisms $\mathbb{U}^2_\mathbb{C} \to \mathbb{U}^2_\mathbb{C}$ and $\mathbb{D}_\mathbb{C} \to \mathbb{D}_\mathbb{C}$.

\subsection{Möbius Group and Stereographic Projection}
\label{subsection "Möbius Group and Stereographic Projection"}

\begingroup
\footnotesize
Cum sit possibile [\,\dots] \& plurimum necessarium, ut in plano repræsentetur circuli in sphæram corpoream incidentes, tanquam esset plana [\,\dots]. Cogit ergo huiusmodi ratio loco meridiani circuli rectis uti lineis.\footnote{
	«Because it is possible and very necessary to represent circles of a solid sphere on a plane [of the projection] as if it [the sphere] were flat [\,\dots]. Such a procedure therefore requires that straight lines are used in place of the meridian circle».
	} \\
\indent — \textsc{C. Ptolemy} \cite[Claudii Ptolemæi sphæræ a planetis proiectio in planum, p. 1]{Ptolemy "Ptolemaei planisphaerium"}

\vspace{2mm}

Bella, \& ingeniosa, \& utile inventione è stata quella degli antichi di gettare i punti, \& i circuli della sphera nei piani con proportione, \& rispondenza di ragione [\,\dots]. [Gli antichi] si sono fondati sopra la Perspettiva [\,\dots] si come ne insegna Tolomeo nel suo Trattato.\footnote{
	«Beautiful, \& ingenious, \& useful invention was that of the Ancients [of the way] of sending [projecting] the points, \& the circles of the sphere on the planes with proportion, \& respondence ratio [\,\dots]. [The Ancients] were based on the Perspective [\,\dots] as it is taught by Ptolemy in his Treatise».
	} \\
\indent — \textsc{D. Barbaro} \cite[Parte Sesta, Che si chiama Planispherio. Spiegatura, Descrittione, et Digradatione della Sphera, p. 163]{Barbaro "La pratica della Perspettiva"}

\endgroup

\vspace{2mm}

The groups $\Moebius(\mathbb{U}^2_\mathbb{C})$ and $\Moebius(\mathbb{D}_\mathbb{C})$ are subgroups of     
\begin{equation}
\label{equation "Möbius group"}
	\Moebius(\hat{\mathbb{C}}) \cong PSL_2(\mathbb{C}) \cong \frac{SL_2(\mathbb{C})}{\{\pm\idem\}},
\end{equation}
the set of all Möbius transformations, called \emph{Möbius group}. The hat on $\hat{\mathbb{C}} \equival \mathbb{C} \cup \{\infty\}$ represents the \emph{extended complex numbers} (more frequently, but inappropriately, referred to as \emph{extended complex plane}), the set of complex numbers augmented with a \emph{point at infinity in the projective space}. This brings us to identify $\hat{\mathbb{C}}$ with the \emph{complex projective line}, $\hat{\mathbb{C}} \cong \mathbb{CP}^1$. 

Since $\hat{\mathbb{C}}  \cong \mathbb{CP}^1$ can be characterized in terms of topological properties, we say that it is equivalent to the 2-sphere in the 3-dimensional real space, $\mathbb{S}^2 = \{(x^1, x^2, x^3) \in \mathbb{R}^3 \mid (x^1)^2 + (x^2)^2 + (x^3)^2 = 1\}$, through \emph{stereographic projection}. In such a case, the complex plane ($\mathbb{C}$) is describable with the plane $x^3 = 0$ in $\mathbb{R}^3$, and the number $z = (x + iy)$ can be identified with $(x, y, 0)$, for $z \in \mathbb{C} \cong \mathbb{R}^2$ and $x, y \in \mathbb{R}$. As a consequence, the system of the projective line extended by a point at infinity $\infty$ is also geometrically equivalent to the \emph{Riemann sphere}, 
\begin{equation}
	\hat{\mathbb{C}} \equival \mathbb{C} \cup \{\infty\} \cong \mathbb{CP}^1 \cong \mathbb{S}^2.
\end{equation}
The Möbius group \eqref{equation "Möbius group"} is thereby the set of all transformations 
\begin{equation}	
	z \mapsto \varphi_{\alpha\beta\gamma\delta}[\hat{\mathbb{C}}](z) = \frac{\alpha z + \beta}{\gamma z + \delta}, \text{for } \alpha, \beta, \gamma, \delta \in \mathbb{C}, \alpha\delta - \beta\gamma \neq 0, \bigl(\begin{smallmatrix}\alpha & \beta \\ \gamma & \delta\end{smallmatrix}\bigr) \in SL_2(\mathbb{C})
\end{equation}
of the Riemann sphere $\hat{\mathbb{C}} \cong \mathbb{CP}^1 \cong \mathbb{S}^2$, so
\begin{align}
\label{align "Möbius group and Riemann sphere"}
	\Moebius(\hat{\mathbb{C}}) & = \left\{\varphi \colon \hat{\mathbb{C}} \equival \mathbb{C} \cup \{\infty\} \to \hat{\mathbb{C}} \equival \mathbb{C} \cup \{\infty\} \mathrel{\bigg|} \varphi_{\alpha\beta\gamma\delta}[\hat{\mathbb{C}}](z) = \frac{\alpha z + \beta}{\gamma z + \delta}\right\} \notag \\
	& = \aut(\hat{\mathbb{C}}) \cong \aut(\mathbb{CP}^1) \cong PSL_2(\mathbb{C}), 
\end{align}
where the \emph{bijective meromorphic mapping} $\hat{\mathbb{C}} \equival \mathbb{C} \cup \{\infty\} \xrightarrow{\varphi} \hat{\mathbb{C}} \equival \mathbb{C} \cup \{\infty\}$ is an \emph{automorphism of the Riemann sphere} (which is merely a Möbius transformation).

Let us check out the value at $\infty$ of $\varphi_{\alpha\beta\gamma\delta}[\hat{\mathbb{C}}]$.

\begin{theorema}
\label{theorema "Theorem for the Möbius group and stereographic"}
Let $\varphi_{\alpha\beta\gamma\delta}[\hat{\mathbb{C}}](z) = \frac{\alpha z + \beta}{\gamma z + \delta}$ for $\alpha, \beta, \gamma, \delta \in \mathbb{C}, \alpha\delta - \beta\gamma \neq 0$, and let $z \in \mathbb{C}$.  
\enumerationisinitium
\item If $\gamma = 0$, one has $\varphi(\infty) = +\infty$, and $\varphi(z) = \left(\frac{\alpha}{\delta}\right)z + \frac{\beta}{\delta}$.
\item If $\gamma \neq 0$, one has 
\begin{equation}
\varphi(\infty) = \lim_{z \to \infty}\left|\frac{\alpha + \frac{\beta}{z}}{\gamma + \frac{\delta}{z}}\right| = \frac{\alpha}{\gamma}, \text{ and } \varphi(z) = \lim_{z \to -\frac{\delta}{\gamma}} \left|\frac{\alpha z + \beta}{\gamma z + \delta}\right| = +\infty.
\end{equation}  
\enumerationisfinis
\end{theorema}

\begin{margo}
Studies and insights on the projective geometry are disseminated in various works of A.F. Möbius \cite{Mobius "Der barycentrische Calcul ein neues Hulfsmittel zur analytischen Behandlung der Geometrie"} \cite{Mobius "Ueber eine neue Behandlungsweise der analytischen Spharik"} \cite{Mobius "Ueber die Grundformen der Linien der dritten Ordnung"} \cite{Mobius "Die Theorie der Kreisverwandtschaft in rein geometrischer Darstellung"} \cite{Mobius "Nachlass II. Theorie der symmetrischen Figuren"}. \margosymbol
\end{margo}

\section{Fuchsian Group (Properly Discontinuous Action)}
\label{section "Fuchsian Group (Properly Discontinuous Action)"}
  
Let us turn our attention to the group $\mathbbl{\Gamma} \leqslant PSL_2(\mathbb{R})$ called \emph{Fuchsian group} \cite{Fuchs "Ueber eine Klasse von Funktionen mehrerer Variablen"} \cite{Poincare "Theorie des groupes fuchsiens"}.
 It is interesting for us because $\mathbbl{\Gamma}$ is a subgroup of $PSL_2(\mathbb{R}) \cong \Moebius^+_2(\mathbb{R}) \equival \mathfrak{isom}^+(\mathbb{U}^2_\mathbb{C})$; an example of $\mathbbl{\Gamma}$ is the \emph{modular group} 
 \begin{equation}	
 	PSL_2(\mathbb{Z}) = 
 	\left\{
 	\begin{pmatrix}
 	\alpha & \beta \\ 
 	\gamma & \delta\end{pmatrix} 
 	\mathrel{\bigg|} \alpha, \beta, \gamma, \delta \in \mathbb{Z}, \alpha\delta - \beta\gamma = 1
 	\right\},
 \end{equation}
 with integer matrices. Here we list some of the main properties.
\enumerationisinitium
\item $\mathbbl{\Gamma}$ is a non-elementary \emph{discrete} (see below) subgroup of $PSL_2(\mathbb{R})$, or a conjugate of this group in $PSL_2(\mathbb{C})$. 
It is \emph{discrete} because
\enumerationisinitium
\item the $\mathbbl{\Gamma}$-identity element is not an accumulation point of $\mathbbl{\Gamma}$,
\item every convergent sequence of $\mathbbl{\Gamma}$ is eventually constant.
\enumerationisfinis
\item $\mathbbl{\Gamma}$ acts \emph{isometrically} and \emph{properly discontinuously} on $\mathbb{U}^2_\mathbb{C}$. In fact, it may be viewed as a group of isometries of the upper half-plane, or a group of isometries of a geodesically Riemannian surface locally isometric to $\mathbb{U}^2_\mathbb{C}$. The action of $\mathbbl{\Gamma}$ is said \emph{properly discontinuous}, if the set
\begin{equation}
	\left\{\epsilon \in \mathbbl{\Gamma} \mathrel{\Big|} \epsilon(\mathcal{J}) \cap \mathcal{J} \neq \varnothing\right\}
\end{equation}
is finite, where $\mathcal{J}$ is a compact set of $\mathbb{U}^2_\mathbb{C}$. We can express the same thing in another way. Take as (open) neighborhoods of $z_1, z_2 \in \mathbb{U}^2_\mathbb{C}$ the double set of points $\Sigma_{z_1}$ and $\Sigma_{z_2}$. According to the discontinuous action, $z_1$ and $z_2$ are in different orbits, so 
\begin{equation}	
	\epsilon(\Sigma_{z_1}) \cap \Sigma_{z_2} \neq \varnothing, 
\end{equation}	
for $\epsilon \in \mathbbl{\Gamma}$, with $\epsilon(z_1) \neq z_2$.
\item $\mathbbl{\Gamma}$ acts \emph{freely} on $\mathbb{U}^2_\mathbb{C}$, and that is because $\mathbbl{\Gamma}$ does not contain any non-trivial elements of finite order (it is torsion free).
\enumerationisfinis

\begin{propositio}
Given the properties stated above, the following equivalent conditions hold:
\enumerationisinitium
\item if $\mathbbl{\Gamma}$ is free and discrete, the quotient $\mathbb{U}^2_\mathbb{C}/\mathbbl{\Gamma}$ is a hyperbolic surface;
\item if $\mathbb{U}^2_\mathbb{C}/\mathbbl{\Gamma}$ is a compact Riemann surface for $\mathbbl{\Gamma}$, every element in $\mathbbl{\Gamma}$ is hyperbolic.	
\enumerationisfinis
\end{propositio}

But it means that $\mathbb{U}^2_\mathbb{C}$ is the universal cover of the hyperbolic surface, and $\mathbbl{\Gamma}$ is a torsion free group of covering transformations; the quotient map $\pi \colon \mathbb{U}^2 \to \mathbb{U}^2/\mathbbl{\Gamma}$ (canonical projection) is a holomorphic covering.

Since the other model for the hyperbolic plane is the (open) unit disk, the discrete action of $\mathbbl{\Gamma}$ is also applicable to $\mathbb{D} \equival \mathbb{B}^2$.

\section{Tessellation of the Upper Half-Plane by Modular Group}
\label{section "Tessellation of the Upper Half-Plane Plane by Modular Group"}

At this time, pointing out a tessellation of the Beltrami–Poincaré hyperbolic plane (Section \ref{subsubsection "Upper Half-Plane and Disk"}), is might be useful, because it can be achieved directly through the fundamental domain for the action on $\mathbb{U}^2$ of the modular group $PSL_2(\mathbb{Z})$:

\begin{center}
\pgfmathsetmacro{\myxlow}{-2}
\pgfmathsetmacro{\myxhigh}{2}
\pgfmathsetmacro{\myiterations}{6}

\begin{tikzpicture}[scale = 2]
	\draw (-1.5, 0) -- (1.5, 0);
	\filldraw[thin, fill = mallard, opacity = 0.30] (0, 0)
     arc[start angle = 0, end angle = 60, radius = 1] -- 
      ({-cos(60)}, 1.2) -- ({cos(60)}, 1.2) -- ({cos(60)}, {sin(60)}) 
     arc[start angle = 120, end angle = 180, radius = 1];
	
	\draw (\myxlow -0.1, 0) -- (\myxhigh +0.1, 0);
	\pgfmathsetmacro{\succofmyxlow}{\myxlow +0.5}
	\foreach \x in {\myxlow, \succofmyxlow,..., \myxhigh}
	{   \draw (\x, 0) -- (\x, -0.05) node[below, font = \tiny] {\x};
    }
    \foreach \y in {0.2, 0.4,..., 1}
    {   \draw (0, \y) -- (-0.05, \y) node[left, font = \tiny] {\pgfmathprintnumber{\y}};
    }
    \draw (0, -0.1) -- (0, 1.2);
    \clip (\myxlow, 0) rectangle (\myxhigh, 1.1);
    \foreach \i in {1,..., \myiterations}
    {   \pgfmathsetmacro{\mysecondelement}{\myxlow+1/pow(2,floor(\i/3))}
    \pgfmathsetmacro{\myradius}{pow(1/3, \i-1}
    \foreach \x in {-2, \mysecondelement,..., 2}
    {   \draw[thin, eggplant] (\x, 0) arc(0:180:\myradius);
    \draw[very thin, eggplant] (\x, 0) arc(180:0:\myradius);
    }   
    }
\end{tikzpicture}
\end{center}

\section{Groupable Synopsis via Commutative Diagram}
\label{section "Groupable Synopsis via Commutative Diagram"}
 
The diagram below provides a useful overall view on certain issues addressed in previous Sections.
Hence the inspiration for new arguments.	
\begin{equation}
\begin{tikzcd}
\label{tikzcd "Groupable Synopsis via Commutative Diagram"}
	\pi_1\bigl(SO_3(\mathbb{R})\bigr) \cong \mathbb{Z}_2 \arrow[dashed]{r}{f_{\text{ev}}} \arrow{d}[swap]{\text{seq.}} & \pi_1(\mathbb{RP}^3) \cong \mathbb{Z}_2 \arrow{r} \arrow[dotted]{d} & 1 \arrow[equal]{d} \\
	\mathbb{S}^1 \arrow[swap]{d}{\pi} \arrow[hookrightarrow]{r}{\iota} & \overbrace{\mathbb{S}^3 \cong SU_2(\mathbb{C})}^{\Spin_3(\mathbb{R})} \subset \mathbb{R}^4 \cong \quaternion \arrow{r}{\prj} \arrow{d}{\varsigma \text{ (2-sheeted covering)}} & \mathbb{S}^2 \cong \hat{\mathbb{C}} \cong \mathbb{CP}^1 \arrow{d}{\{\varphi\} = \Moebius(\hat{\mathbb{C}})} \\
	\mathbb{RP}^1 \cong \mathbb{S}^1 \arrow{r} & \mathbb{RP}^3 \cong SO_3(\mathbb{R}) \arrow{r}{\varpi} & \mathbb{S}^2 \cong \hat{\mathbb{C}} \cong \mathbb{CP}^1 
\end{tikzcd} 
\end{equation}

\subsection{Hopf Fibration}
\label{subsection "Hopf Fibration"}

\emph{Per row of \eqref{tikzcd "Groupable Synopsis via Commutative Diagram"}.}
\enumerationisinitium
\item $\mathbb{Z}_2 \viz \mathbb{Z}/2\mathbb{Z}$ is the cyclic group of order 2, and the fundamental group of $SO_3(\mathbb{R})$. For the action of $SO_3(\mathbb{R})$ on $\mathbb{RP}^3$, there is an evaluation isomorphism
\begin{equation}
	f_{\text{ev}} \colon \pi_1\bigl(SO_3(\mathbb{R})\bigr) \cong \mathbb{Z}_2 \dashrightarrow \pi_1(\mathbb{RP}^3) \cong \mathbb{Z}_2.
\end{equation}
\item The \emph{Hopf fibration} \cite{Hopf H. "Uber die Abbildungen der dreidimensionalen Sphare auf die Kugelflache"} states that 
\begin{equation}
	\iota \colon \mathbb{S}^1 \hookrightarrow SU_2(\mathbb{C}) \cong \mathbb{S}^3 \xrightarrow{\prj} \mathbb{S}^2. 
\end{equation} 
The 3-sphere $\mathbb{S}^3$ is intended as the total space of a fibration over a base space $\mathbb{S}^2$ with fiber made up of a (unit) circle 
\begin{equation}
	\mathbb{S}^1 = \{x \in \mathbb{R}^2 \mid \|x\| = 1\} \cong \{z = (x + iy) \in \mathbb{C} \mid \|z\| = 1\}.
\end{equation}
\item The bundle $\varpi \colon SO_3(\mathbb{R}) \to \mathbb{S}^2$ is a principal fiber bundle; looking closer, it is a $SO_2(\mathbb{R})$-bundle $SO_3(\mathbb{R}) \to SO_3(\mathbb{R})/SO_2(\mathbb{R}) \cong \mathbb{S}^2$.
\enumerationisfinis

\subsection{Spinorial Representation of the Orthogonal Group on a 3-Space}
\label{subsection "Spinorial Representation of the Orthogonal Group on a 3-Space"}

\begingroup
\footnotesize
Spinors were first used under that name, by physicists,\footnote{
	The first is P. Ehrenfest, as reported by S.-I. Tomonaga \cite[p. 129]{Tomonaga "The Story of Spin"}.
	}
in the field of Quantum Mechanics. In their most general mathematical form, spinors were discovered in 1913 \cite{Cartan "Les groupes projectifs qui ne laissent invariante aucune multiplicite plane"} by the author of this work, in his investigations on the linear representations of simple groups; they provide a linear representation of the group of rotations in a space with any number $n$ of dimensions, each spinor having $2^\nu$ components where $n = 2\nu + 1$ or $2\nu$. \\
\indent — \textsc{É. Cartan} \cite[intro]{Cartan "The Theory of Spinors"}

\endgroup

\vspace{2mm}

\emph{Per column of \eqref{tikzcd "Groupable Synopsis via Commutative Diagram"}.}
\enumerationisinitium
\item The projection $\pi \colon \mathbb{S}^1 \to \mathbb{RP}^1$ says that the real projective line $\mathbb{RP}^1$ is homeomorphic to the circle $\mathbb{S}^1$ (1-sphere), $\mathbb{RP}^1 \cong \mathbb{S}^1$. The fundamental group of $\mathbb{S}^1$ is isomorphic to $\mathbb{Z}$, for which $\mathbb{S}^1 \cong \mathbb{{R}/{Z}}$.
\item The covering map $\varsigma \colon SU_2(\mathbb{C}) \to SO_3(\mathbb{R})$ establishes a (smooth) surjective homomorphism between $SU_2(\mathbb{C})$ and $SO_3(\mathbb{R})$. It is known that $SU_2(\mathbb{C})$ is compact and simply connected, and $SO_3(\mathbb{R})$ is compact and connected but not simply connected; from that comes the significance of the map, which has a \emph{mathematical foundation} and a \emph{physical justification}. The set $SU_2(\mathbb{C})$ is the \emph{(simply connected) universal covering group} of $SO_3(\mathbb{R})$. Immediately below is a description of this result.
\item The automorphism of the Riemann sphere $\mathbb{S}^2 \cong \hat{\mathbb{C}} \cong \mathbb{CP}^1$ is inherent in the group $\Moebius(\hat{\mathbb{C}}) = \aut(\hat{\mathbb{C}})$, see Eq. \eqref{align "Möbius group and Riemann sphere"}.
\enumerationisfinis

\subsection{Pauli-like Spinors in the Complex Hilbert 2-Space; Angular Momentum in Quantum Mechanics and Topological Nature of the Electron Spin}
\label{subsection "Pauli-like Spinors in the Complex Hilbert 2-Space; Angular Momentum in Quantum Mechanics and Topological Nature of the Electron Spin"}

\begingroup
\footnotesize
One has often suggested that this formally possible representation, by means of two-valued eigenfunctions, is unfair to the true physical nature of things [\textit{wahren physikalischen Sachverhalt}]. [\,\dots]. On the other hand, a representation of the quantum-mechanical behavior of the magnetic electron using the method of eigenfunctions [\,\dots] is highly desirable. \\
\indent — \textsc{W. Pauli} \cite[p. 602]{Pauli "Zur Quantenmechanik des magnetischen Elektrons"}

\endgroup
 
\subsubsection[The Covering Morphisms $SU_2(\mathbb{C}) \to SO_3(\mathbb{R})$]{The Covering Morphisms $\protect\pseudobold{SU_2(\mathbb{C})} \protect\pseudobold{\to} \protect\pseudobold{SO_3(\mathbb{R})}$}
\label{subsubsection "The Covering Morphisms $SU_2(C)$ to $SO_3(R)$"}

Physically speaking, the covering map 
\begin{equation}
	\varsigma \colon SU_2(\mathbb{C}) \to SO_3(\mathbb{R}),
\end{equation}	
by making reference to an electron, or a spin-$\frac{1}{2}$ particle, is a \emph{spinor map} $\varsigma \colon SU_2(\mathbb{C}) \cong \Spin_3(\mathbb{R}) \to SO_3(\mathbb{R})$, where is possible to highlight the spin group $\Spin_3(\mathbb{R})$, the set of all unit quaternions 
\begin{equation}
	\{\mathbbl{q} \in \quaternion \mid \mathbbl{q}\bar{\mathbbl{q}} = 1\}, \text{ with } \mathbbl{q} = x^1\mathbbl{1} + x^2\mathbbl{i} + x^3\mathbbl{j} + x^4\mathbbl{k}
\end{equation}
(see below), and $\kernel(\pm\idem)$. The issue is that the universal covering of $SU_2(\mathbb{C}) \cong \Spin_3(\mathbb{R})$ is a 2-valued representation of $SO_3(\mathbb{R})$; and it is because \emph{two elements of $SU_2(\mathbb{C}) \cong \Spin_3(\mathbb{R})$ are exactly associated with each element of $SO_3(\mathbb{R})$}, so that the complex special unitary group of degree 2, or the real spin 3-group, is equivalent to a \emph{2-fold covering (double cover)} of the 3-dimensional Euclidean rotations group.

In a nutshell, $SU_2(\mathbb{C})$ and $SO_3(\mathbb{R})$ are \emph{not globally} but \emph{only locally isomorphic} (related to the composition of their infinitesimal transformations); they have the same universal covering group, with a corresponding Lie algebra $\mathfrak{su}_2(\mathbb{C}) \cong \mathfrak{so}_3(\mathbb{R})$ \cite[XI, pp. 352-355]{Cartan "Les groupes reels simples finis et continus"}.

Let us see what all the fuss is about. Consider a \emph{2-component spinor} or \emph{Pauli spinor} \cite{Pauli "Zur Quantenmechanik des magnetischen Elektrons"} 
\begin{equation}
	\psi = 
	\begin{pmatrix}
 	\psi^\alpha \\
 	\psi^\beta	
 	\end{pmatrix} \text{ with } 
 	\left\{
 	\psi^\alpha_+ =
 	\begin{pmatrix} 
 	\psi^\alpha \\ 
 	0	
 	\end{pmatrix}, \enspace
 	\psi^\beta_- = 
 	\begin{pmatrix}
 	0 \\
 	\psi^\beta
 	\end{pmatrix} \mathrel{\bigg|} \psi \in \mathbb{C}^2 \cong \mathfrak{H}
 	\right\}.
\end{equation}
Mathematically, it is a column vector with two complex components in the 2-dimensional complex Hilbert space, denoted by $\mathbb{C}^2 \cong \mathfrak{H}$ and properly called \emph{spin space}. In the language of quantum mechanics, the spinor is a \emph{2-component wave function} that serves to describe two states, $\psi\left[+\frac{1}{2}\right] = 
	\bigl(\begin{smallmatrix}
 	1 \\ 0
 	\end{smallmatrix}\bigr)$ and $\psi\left[-\frac{1}{2}\right] =
	\bigl(\begin{smallmatrix}
 	0 \\ 1
 	\end{smallmatrix}\bigr)$, of a \emph{non-relativistic electron} or any \emph{particle of one-half spin}, i.e. an object obeying Fermi–Dirac statistics, by the E. Schrödinger's method of eigenfunctions \cite{Schrodinger "Quantisierung als Eigenwertproblem (Erste Mitteilung)", Schrodinger "Quantisierung als Eigenwertproblem (Zweite Mitteilung)", Schrodinger "Quantisierung als Eigenwertproblem (Dritte Mitteilung)", Schrodinger "Quantisierung als Eigenwertproblem (Vierte Mitteilung)"}.
Applying a vector $\vec{x} = (x_1, x_2, x_3)$, we just write the spinor as
\begin{equation}
	\psi(x) =
	\begin{pmatrix}
 	\psi\Bigl(\vec{x}, +\frac{1}{2}\Bigr) \\
 	\psi	\Bigr(\vec{x}, -\frac{1}{2}\Bigr)
 	\end{pmatrix}.
\end{equation}

Let $R^\theta = \left\{R^\theta_1, R^\theta_2, R^\theta_3\right\}$ be the rotations about the axis $x_1$, $x_2$ and $x_3$ by an angle $\theta$,
\begin{align}
	R^\theta_1 = \left(\begin{smallmatrix}
	1 & 0 & 0 \\
	0 & \cos\theta^1 & -\sin\theta^1 \\
	0 & \sin\theta^1 & \cos\theta^1
	\end{smallmatrix}\right),
	R^\theta_2 =  \left(\begin{smallmatrix}
	\cos\theta^2 & 0 & \sin\theta^2 \\
	0 & 1 & 0 \\
	-\sin\theta^2 & 0 & \cos\theta^2
	\end{smallmatrix}\right),
	R^\theta_3 = \left(\begin{smallmatrix}
	\cos\theta^3 & -\sin\theta^3 & 0 \\
	\sin\theta^3 & \cos\theta^3 & 0 \\
	0 & 0 & 1
	\end{smallmatrix}\right).
\end{align}

Let $M \viz [M]^{2 \times 2} = 
	\bigl(
	\begin{smallmatrix}
	\alpha & \beta \\
	-\bar{\beta} & \bar{\alpha}
	\end{smallmatrix}
	\bigr)$ 	be a $2 \times 2$ matrix, with $\alpha, \beta \in \mathbb{C}$, relating to the special unitary 2-group.

The \emph{orthogonal transformations in a real 3-space} of $SO_3(\mathbb{R})$ are thus defined:
\begin{equation}
	\label{equation "Orthogonal transformations in real 3-space"}
	\dot{x}_\mu = \sum_{\nu}R^\theta_{\mu\nu}x_\nu,
\end{equation} 
from which
\begin{equation}
	\sum_\mu\dot{x}_\mu^2 = \sum_{\mu\nu\xi}R^\theta_{\mu\nu}R^\theta_{\mu\xi}x_{\nu}x_\xi = \sum_{\mu\nu\xi}\delta_{\nu\xi}x_{\nu}x_\xi = \sum_{\nu}x_\nu^2,
\end{equation}
to wit,
\begin{equation}
	\left\{\dot{\vec{x}} = (\dot{x}_1, \dot{x}_2, \dot{x}_3)\right\} = \left\{R^{\theta}\vec{x} = (x_1, x_2, x_3)\right\};
\end{equation}
whilst the \emph{unitary transformations in a complex 2-space} of $SU_2(\mathbb{C})$ is
\begin{equation}
	\label{align "Unitary transformations in complex 2-space"}
	\dot{\psi}^\mu = \sum_{\nu}M_{\mu\nu}\psi^\nu,
\end{equation}	
letting
\begin{equation}	
	\dot{\psi} =
	\begin{pmatrix}
	\alpha_\mathbb{C} & \beta_\mathbb{C} \\
	-\bar{\beta}_\mathbb{C} & \bar{\alpha}_\mathbb{C}
	\end{pmatrix}
	\psi.
\end{equation}
The complex conjugate of $\dot{\psi}^\mu$, together with the summation $\sum_\nu\bar{M}_{\mu\nu}\bar{\psi}^\nu$, gives $\bar{\psi}^\mu = \psi_\mu$, where $\bar{M} = 
	\bigl(\begin{smallmatrix}
	0 & 1 \\
	-1 & 0
	\end{smallmatrix}\bigr)
	M\bigl(\begin{smallmatrix}
	0 & 1 \\
	-1 & 0
	\end{smallmatrix}\bigr)^{-1}$.

We can set the correspondence between the two transformations \eqref{equation "Orthogonal transformations in real 3-space"} and \eqref{align "Unitary transformations in complex 2-space"}, by associating a $\mathbb{C}^{2 \times 2}$ Hermitian matrix $X$ to each 3-vector $\vec{x} = (x_1, x_2, x_3)$ in $\mathbb{R}^3$, 
\begin{align}
\label{align "Hermitian matrix and 3-vector of Pauli matrices"}
	X & = \left\{\vec{x} \cdot \vec{\sigmaPauli} = (\sigmaPauli_1, \sigmaPauli_2, \sigmaPauli_3)\right\} = x_1\sigmaPauli_1 + x_2\sigmaPauli_2 + x_3\sigmaPauli_3 \notag \\
	& = \Biggl\{
	\begin{pmatrix}
		x_3 & x_1 - ix_2 \\
		x_1 + ix_2 & x_3
	\end{pmatrix} \mathrel{\bigg|} x \in \mathbb{R}
	\Biggr\},
\end{align}
such that $\dot{X} = MXM^\dag$, where $\vec{\sigmaPauli} = (\sigmaPauli_1, \sigmaPauli_2, \sigmaPauli_3)$ is the \emph{3-spin vector} with the $\sigmaPauli$-matrices as its components. The matrices
\begin{equation}
\label{equation "Pauli matrices"}
	\sigmaPauli_1 = 
	\begin{pmatrix}
	0 & 1 \\
	1 & 0
	\end{pmatrix}, \enspace 
	\sigmaPauli_2 = 
	\begin{pmatrix*}[r]
	0 & -i \\
	i & 0
	\end{pmatrix*}, \enspace
	\sigmaPauli_3 = 
	\begin{pmatrix*}[r]
	1 & 0 \\
	0 & -1
	\end{pmatrix*}		
\end{equation}
are called \emph{Pauli (spin) matrices},\footnote{
	For a 4-vector $x^{\mu} = (x^0, x^1, x^2, x^3)$, in the Hermitian matrix 
	$X = \bigl\{x^\mu = (x^0, x^1, x^2, x^3) \cdot \sigmaPauli_\mu = (\sigmaPauli_0, \vec{\sigmaPauli})\bigr\} = x^0 + x^1\sigmaPauli_1 + x^2\sigmaPauli_2 + x^3\sigmaPauli_3
		= \left(\begin{smallmatrix}
		x^0 + x^3 & x^1 - ix^2 \\
		x^1 + ix^2 & x^0 - x^3
	\end{smallmatrix}\right)$, the matrix $\sigmaPauli_0 \viz \idem = 
		\left(\begin{smallmatrix}
		1 & 0 \\
		0 & 1
	\end{smallmatrix}\right)$ is the \emph{Pauli identity matrix}.
	}
	having trace equal to zero, $\trace(\sigmaPauli_\mu) = 0$, and satisfying the relation 
\begin{equation}
	\sigmaPauli_\mu \sigmaPauli_\nu = \delta_{\mu\nu}\idem + i\varepsilon_{\mu\nu\xi}\sigmaPauli_\xi,
\end{equation}
where $\delta_{\mu\nu}$ is the Kronecker delta and $\varepsilon_{\mu\nu\xi}$ is the \emph{Levi-Civita symbol} \cite[pp. 180-182]{Levi-Civita "Lezioni di calcolo differenziale assoluto"}, an anti-symmetric collection of indices,
\begin{equation}
	\varepsilon_{\mu\nu\xi}
	\begin{cases}
	1 & \text{for a cyclic permutation of } \{\mu, \nu, \xi\} = \{1, 2, 3\}, \\
	-1 & \text{for an anti-cyclic permutation of } \{\mu, \nu, \xi\} = \{1, 2, 3\}, \\
	0 & \text{otherwise}.	
	\end{cases}
\end{equation}
The Pauli matrices form a basis in the linear vector $\mathbb{R}$-space of the $2 \times 2$ Hermitian matrix \eqref{align "Hermitian matrix and 3-vector of Pauli matrices"}. 

Combining the rotation matrices and the matrix $M$ with its conjugate transpose $M^\dag$, we get 
\begin{equation}
	R^\theta_{\mu\nu} = \frac{1}{2}\trace(\sigmaPauli_\mu{M}\sigmaPauli_{\nu}M^\dag).
\end{equation}
Since it is possible to replace $M$ with $-M$ (without changing the equation), it became understandable that for a matrix $R^\theta$ of $SO_3(\mathbb{R})$ are set out two distinct matrices $M$ and $-M$ of $SU_2(\mathbb{C})$.

\subsubsection[Example. Irreducible Covering Spin-Space for $4\pi$]{Example. Irreducible Covering Spin-Space for 4$\mathbold{\pi}$}
\label{subsubsection "Example. Irreducible Covering Spin-Space for 4pi"}

Let us impose
\begin{equation}
\label{equation "Matrices for irreducible covering spin-space"}
	M = 
	\begin{pmatrix}
	e^{-\frac{1}{2}i\theta^3} & 0 \\
	0 & e^{\frac{1}{2}i\theta^3} 
	\end{pmatrix} \in SU_2(\mathbb{C})
	\longrightarrow R^\theta_3 = 
	\left(\begin{smallmatrix}
	\cos\theta^3 & -\sin\theta^3 & 0 \\
	\sin\theta^3 & \cos\theta^3 & 0 \\
	0 & 0 & 1
	\end{smallmatrix}\right) \in SO_3(\mathbb{R}),
\end{equation}
from which 
\begin{subequations}
\begin{align}	
	& \dot{x}_1 = \cos\theta^3x_1 - \sin\theta^3x_2, \\ 
	& \dot{x}_2 = \sin\theta^3x_1 + \cos\theta^3x_2, \\
	& \dot{x}^3 = x_3. 
\end{align}	
\end{subequations}
Let $\hat{S} = \frac{1}{2}\hbar\vec{\sigmaPauli}$, namely $S^1 = \frac{1}{2}\hbar\sigmaPauli_1, S^2 = \frac{1}{2}\hbar\sigmaPauli_2, S^3 = \frac{1}{2}\hbar\sigmaPauli_3$, be the spin (matrix) of an electron, where $\hat{S}$ is the spin angular momentum (or spin operator).

Let us recall the difference between the \emph{orbital angular momentum (operator)} $\hat{L}$ and the \emph{spin angular momentum (operator)} $\hat{S}$ of an electron, or a spin one-half particle: the first ($\hat{L}$) consists of an integer multiple of the reduced Planck constant, $\hbar = \frac{h}{2\pi}$, the second ($\hat{S}$) is a half-integer multiple of $\hbar$, which is $\frac{1}{2}\hbar$. The operator for the \emph{total angular momentum} ($\hat{J} = \hat{L} + \hat{S}$) will be a half-odd-integer multiple of $\hbar$.

There follows a remarkable occurrence. The rotation about, say, the $x_3$-axis through an angle $\theta^3 = 2\pi = \ang{360}$ on a coordinate system is different from that on a spinor. A coordinate system comes back to its original state after a rotation of $2\pi$, but a \emph{spinor returns to its original state after two full rotations}, i.e. after completing a rotation of $4\pi = \ang{720}$ about the same $x_3$-axis. A spinor representation, after a rotation of $2\pi$, is therefore only halfway, so to speak, to the \emph{identity (or neutral) symmetry element}, and it has negative value, $\dot{\psi} = -\psi$.

What the wording expresses is that the operations $R^\theta_3$ and $R^\theta_3 + 2\pi$ are the same (rotation) element of $SO_3(\mathbb{R})$, but they are also two distinct elements, or matrices, of $SU_2(\mathbb{C})$, one positive and one negative, $M$ \eqref{equation "Matrices for irreducible covering spin-space"} and $-M$.

Let $j = \bigl\{0, \frac{1}{2}, 1, \frac{3}{2}, 2, \frac{5}{2}, \mathellipsis\bigr\}$ be the \emph{eigenvalue of the
total angular momentum} ($\hat{J}$), be it integer or half-integer. The representation equipped with integer values of $j$, called \emph{even} (actually tensor) representation, can be written as
\begin{equation}
	\overbrace{\mathscr{D}^{(j)}(M) = +\mathscr{D}^{(j)}(-M)}^{\text{tensor/even rep. (integer $j$)}},
\end{equation}
and the representation equipped with half-integer values of $j$, called \emph{odd} (actually spinor) representation, as 
\begin{equation}
	\overbrace{\mathscr{D}^{(j)}(M) = -\mathscr{D}^{(j)}(-M)}^{\text{spinor/odd rep. (half-integer $j$)}}. 
\end{equation}
Representations of type $\mathscr{D}^{(j)}$ are irreducible. A representation is said to be \emph{irreducible} if it does not contain a non-trivial invariant subspace. The matrices $M_{\in SU_2(\mathbb{C})}$ and $-M_{\in SU_2(\mathbb{C})}$ are the same as a \emph{double-valued representation} of $SO_3(\mathbb{R})$. The space both for $\mathscr{D}^{(j)}$ of ${SO_3(\mathbb{R})}$ and for $\mathscr{D}^{(j)}$ of $SU_2(\mathbb{C})$ has the dimension $2j + 1$; it is a complex Hilbert space $\mathfrak{H}^{\rotatedell + 1}$, and it can be called \emph{$\rotatedell$-dimensional representation space}, where $\rotatedell = 2j$, with integer and half-integer $j$, respectively.

The \emph{existence}, in the \emph{mathematical sense}, of an electronic spin, or a Pauli-like spinor of fermionic particles, meaning objects with half-odd-integer spin, is guaranteed by the \emph{topological nature} of the rotation group in Euclidean 3-space. 
\enumerationisinitium
\item The representation $\mathscr{D}^{(j)}$ of $SO_3(\mathbb{R})$ amounts to \emph{integer} values of $j$.
\item The representation $\mathscr{D}^{(j)}$ of $SU_2(\mathbb{C})$ with \emph{integer} values of $j$ is a \emph{single-valued} representation of $SO_3(\mathbb{R})$.
\item The representation $\mathscr{D}^{(j)}$ of $SU_2(\mathbb{C})$ with \emph{half-odd-integer} values of $j$ is a \emph{double-valued} representation ($M$ and $-M$) of $SO_3(\mathbb{R}) \cong \Spin_3(\mathbb{R})$.
\item The $4\pi$-symmetry of a 2-component wave function has to do with the fact that the initial value, for a spin system, is reinstated with two successive rotations by $\ang{360}$, or a rotation of $\ang{720}$. 
\enumerationisfinis

\subsection{Unit Quaternions}

\begingroup
\footnotesize
[A]n \emph{under-current} of thought was going on in my mind [\,\dots]. An \emph{electric} circuit seemed to \emph{close}; and a spark flashed forth, the herald (as \emph{I foresaw, immediately}) of many long years to come of definitely directed thought and work. \\
\indent — \textsc{W.R. Hamilton}, in \cite[chap. XXVIII, pp. 434-435]{Graves "Life of Sir William Rowan Hamilton including selections from his poems correspondence and miscellaneous writings II"}

\endgroup
 
\vspace{2mm}

\emph{Center of \eqref{tikzcd "Groupable Synopsis via Commutative Diagram"}}. We will divide the investigation into two steps.

\subsubsection{Step I}
\label{subsubsection "Step I"}

The bijective mapping $\varphi^{-1} \colon \mathbb{S}^3 \subset \mathbb{R}^4 \to SU_2(\mathbb{C})$ is determined by
\begin{align}
	x \mapsto \varphi^{-1}(x) & = 
	\begin{pmatrix}
		x^1 + ix^4 & x^2 + ix^3 \\
		- x^2 + ix^3 & x^1 - ix^4
	\end{pmatrix}
	\equival \mathfrak{R}^4 = 
	\left\{
	\begin{pmatrix}
		\alpha & \beta \\
		-\bar{\beta} & \bar{\alpha}
	\end{pmatrix}	
	\mathrel{\bigg|} \alpha, \beta \in \mathbb{C}
	\right\} \notag \\
	& = x^1
	\bigl(\begin{smallmatrix}
	1 & 0 \\
	0 & 1
	\end{smallmatrix}\bigr) 
	+ x^2
	\bigl(\begin{smallmatrix*}[r]
	i & 0 \\
	0 & -i
	\end{smallmatrix*}\bigr)
	+ x^3
	\bigl(\begin{smallmatrix}
	0 & 1 \\
	-1 & 0
	\end{smallmatrix}\bigr)
	+ x^4
	\bigl(\begin{smallmatrix}
	0 & i \\
	i & 0
	\end{smallmatrix}\bigr),
\end{align} 
involving the 3-sphere $\mathbb{S}^3 = \{(x^1, x^2, x^3, x^4) \in \mathbb{R}^4 \mid (x^1)^2 + (x^2)^2 + (x^3)^2 + (x^4)^2 = 1\}$ in $\mathbb{R}^4$ and the special unitary group of degree 2 over the complex field. We will call the space $\mathbb{R}^4$ the \emph{quaternionic space $\mathbb{R}^4 \cong \quaternion$} (because the 3-sphere can be explained through the algebra of quaternions). The symbol $\mathfrak{R}^4$ denotes a set of $\mathbb{C}^{2 \times 2}$ matrices and represents a matrix model of $\mathbb{R}^4 \cong \quaternion$. Let us write the basis elements for $\mathfrak{R}^4$ in the form
\begin{equation}
\label{equation "Complex quaternions matrices"}
	\mathbbl{1} = 
	\begin{pmatrix}
	1 & 0 \\
	0 & 1
	\end{pmatrix}, \enspace 
	\mathbbl{i} = 
	\begin{pmatrix*}[r]
	i & 0 \\
	0 & -i
	\end{pmatrix*}, \enspace
	\mathbbl{j} = 
	\begin{pmatrix}
	0 & 1 \\
	-1 & 0
	\end{pmatrix}, \enspace
	\mathbbl{k} = 
	\begin{pmatrix}
	0 & i \\
	i & 0
	\end{pmatrix};
\end{equation}
the Eq. \eqref{equation "Complex quaternions matrices"} is a basis quaternions expressed with $\mathbb{C}^{2 \times 2}$ matrices, called \emph{complex quaternions} $\quaternion_{\mathbb{C}}$ (and therefore it falls under the complex quaternion algebra).
			
There exists a isomorphism $\varphi(\mathbbl{q} \mapsto x) \colon \mathfrak{R}^4 \to \mathbb{R}^4$ defined by $\{\mathbbl{1}, \mathbbl{i}, \mathbbl{j}, \mathbbl{k}\}$, where a quaternion expression of type 
\begin{equation}
	\mathbbl{q} = x^1\mathbbl{1} + x^2\mathbbl{i} + x^3\mathbbl{j} + x^4\mathbbl{k}
\end{equation}
and the set $x = (x^1, x^2, x^3, x^4)$ are at stake. It is noted that the quaternion $x^2\mathbbl{i} + x^3\mathbbl{j} + x^4\mathbbl{k}$ is the \emph{imaginary part} (or \emph{vector part}) of $\mathbbl{q}$, and $x^1\mathbbl{1} = x^1$ is the \emph{real part} (or \emph{scalar part}) of $\mathbbl{q}$. Thanks to $\varphi(\mathbbl{q} \mapsto x)$, the matrix multiplication is a quaternion multiplication on $\mathbb{R}^4$, and hence the 3-sphere $\mathbb{S}^3$ can be thought of as a group of \emph{unit quaternions} in $\mathbb{R}^4 \cong \quaternion$, for which $\mathbb{S}^3 \subset \mathbb{R}^4 \cong \quaternion$.
	
About the group $SU_2(\mathbb{C})$, all its elements are in $\mathfrak{R}^4$, so $SU_2(\mathbb{C}) \subset \mathfrak{R}^4$. Then $SU_2(\mathbb{C})$ is also a group of \emph{unit quaternions} in $\mathfrak{R}^4$ with a trivial bundle 
\begin{equation}
	SU_2(\mathbb{C}) \hookrightarrow \mathbb{R}^4 \times SU_2(\mathbb{C}) \to \mathbb{R}^4. 
\end{equation}
It is concluded that $\mathbb{S}^3 \cong SU_2(\mathbb{C})$.
	
\begin{scholium}[Fundamental formula of Hamilton, and non-Abelian quaternion group of order 8]
\label{scholium "Fundamental formula of Hamilton, and non-Abelian quaternion group of order 8"}
It is important to remember that the basis of the quaternions $\{\mathbbl{1}, \mathbbl{i}, \mathbbl{j}, \mathbbl{k}\}$, that is, the $\mathfrak{R}^4$-algebra of matrices
	$\bigl(\begin{smallmatrix}
	1 & 0 \\
	0 & 1
	\end{smallmatrix}\bigr)$, 
	$\bigl(\begin{smallmatrix*}[r]
	i & 0 \\
	0 & -i
	\end{smallmatrix*}\bigr)$,
	$\bigl(\begin{smallmatrix}
	0 & 1 \\
	-1 & 0
	\end{smallmatrix}\bigr)$,
	$\bigl(\begin{smallmatrix}
	0 & i \\
	i & 0
	\end{smallmatrix}\bigr)$,
satisfies the fundamental formula of Hamilton\endnote{
	It's the Brougham [Broome] Bridge formula for quaternion multiplication in a «flash of genius» discovered, dating back to October 16 1843, as inscribed on the plaque attached to the bridge; see Hamilton's account \cite[pp. 434-435]{Graves "Life of Sir William Rowan Hamilton including selections from his poems correspondence and miscellaneous writings II"} in the letter of August 5 1865 to his son Archibald H. (cf. epigraph in this Section).
	}
\begin{equation}	
	\mathbbl{i}^2 = \mathbbl{j}^2 = \mathbbl{k}^2 = \mathbbl{ijk} = -\mathbbl{1},
\end{equation}	
together with the following explicit relations,
\begin{subequations}
\begin{align}
	& \mathbbl{i}^2 = \mathbbl{j}^2 = \mathbbl{k}^2 = -\mathbbl{1}, \\
	& \mathbbl{ij} = -\mathbbl{ji} = \mathbbl{k}, \enspace \mathbbl{jk} = -\mathbbl{kj} = \mathbbl{i}, \enspace \mathbbl{ki} = -\mathbbl{ik} = \mathbbl{j},
\end{align}
\end{subequations}
in which the multiplication is associative and \emph{non-commutative}. Among the many works of W.R. Hamilton on his quaternions, see \cite{Hamilton "On Quaternions; or on a new System of Imaginaries in Algebra"} \cite{Hamilton "Elements of Quaternions"}. The \emph{quaternion group} is 
\begin{equation}
	\mathfrak{Q}_8 = \{\mathbbl{1, i, j, k, -1, -i, -j, -k}\}; 
\end{equation}
this is a \emph{non-Abelian group} of order 8, and its subgroups are: 
\begin{align*}
	& \{\mathbbl{1}\}, \text{ or the identity element (real quaternion)}, \\
	& \{\mathbbl{1, -1}\} \text{ of order 2, and} \\
	& \{\mathbbl{1, -1, i, -i}\}, \{\mathbbl{1, -1, j, -j}\}, \{\mathbbl{1, -1, k, -k}\} \text{ of order 4}.
\end{align*}
	 	
We should note that there is an isomorphism between the quaternion algebra and the Pauli matrices \eqref{equation "Pauli matrices"}. If $\mathbbl{i} = i\sigmaPauli_1, \mathbbl{j} = i\sigmaPauli_2, \mathbbl{k} = i\sigmaPauli_3$, then 
\begin{equation}
	\mathfrak{Q}_8 = \left\{\pm\mathbbl{[\mathbbl{1} \mapsto \idem]}^{2 \times 2}, \pm\sigmaPauli\right\} = \left\{\mathbbl{1}, \sigmaPauli_1, \sigmaPauli_2, \sigmaPauli_3, \mathbbl{-1}, -\sigmaPauli_1, -\sigmaPauli_2, -\sigmaPauli_3\right\}.
\end{equation}
\scholiumsymbol
\end{scholium}
	
\begin{scholium}
The complex special unitary group of degree 2 may also be realized as the \emph{group of quaternions with norm equal to 1}, which is denoted by $Sp_1(\quaternion)$ and it is a compact symplectic group; so 
\begin{align}
	& \mathbb{S}^3 \cong SU_2(\mathbb{C}) \cong Sp_1(\quaternion), \\ 
	& \mathfrak{su}_2(\mathbb{C}) \cong \mathfrak{so}_3(\mathbb{R}) \cong \mathfrak{sp}_1(\quaternion).
\end{align}
\scholiumsymbol
\end{scholium}
	
\subsubsection{Step II}
\label{subsubsection "Step II"}

The special orthogonal 3-group $SO_3(\mathbb{R})$ is homeomorphic to the real projective 3-space $\mathbb{RP}^3$, hence $\pi_1\bigl(SO_3(\mathbb{R})\bigr) \cong \mathbb{Z}_2$. The topological equivalence (continuous bijection) $\mathbb{RP}^3 \cong SO_3(\mathbb{R})$ is associated with what we saw before: the homomorphism from $SU_2(\mathbb{C})$ onto $SO_3(\mathbb{R})$, as well as the homeomorphism $\mathbb{S}^3 \cong SU_2(\mathbb{C})$. Note the following. 
\enumerationisinitium
\item Given a homomorphism 
\begin{equation}
	\xi \colon SU_2(\mathbb{C}) \to SU_2(\mathbb{C})/\mathbb{Z}_2 = \{\pm 1\}, 
\end{equation} 
we construct an isomorphism 
\begin{equation}
	\mu \colon SU_2(\mathbb{C})/\mathbb{Z}_2 = \{\pm 1\} \to SO_3(\mathbb{R});
\end{equation}
the quotient $SU_2(\mathbb{C})/\mathbb{Z}_2 = \{\pm 1\}$ is isomorphic to $SO_3(\mathbb{R})$. Here is a diagram,
	\[
	\begin{tikzcd}[row sep=large, column sep=large]
	\mathbb{S}^3 \cong SU_2(\mathbb{C}) \cong \Spin_3(\mathbb{R}) \arrow{r}{\varsigma} \arrow[swap]{d}{\xi} & SO_3(\mathbb{R}) \\
	\frac{SU_2(\mathbb{C})}{\mathbb{Z}_2 = \{\pm 1\}} \arrow[swap]{ru}{\mu} & 
	\end{tikzcd} 
	\] 
The morphism $\varsigma$ is a particular instance of the covering map $\varsigma$ previously stated. The group $SO_3(\mathbb{R})$ can be thought of as a quotient of $\mathbb{S}^3 \cong SU_2(\mathbb{C})$. But also the projective space $\mathbb{RP}^3$ is homeomorphic to the quotient of the 3-sphere with an antipodal map $\pi \colon \mathbb{S}^3 \to \mathbb{RP}^3$, from which the equivalence between $SO_3(\mathbb{R})$ and $\mathbb{RP}^3$.
\item The composition of 
\begin{equation}
	\mathbb{S}^3 \to \mathbb{RP}^3 \cong SO_3(\mathbb{R})
\end{equation} 
(action of $\mathbb{S}^3$ on $\mathbb{RP}^3$ through a map from the 3-sphere to the orthogonal group on a 3-space) and of
\begin{equation}
	\mathbb{RP}^3 \to \mathbb{S}^2 \cong SO_3(\mathbb{R})/SO_2(\mathbb{R})
\end{equation}
is but the above-mentioned Hopf bundle, or principal $\mathbb{S}^1$-bundle, $\mathbb{S}^1 \hookrightarrow \mathbb{S}^3 \to \mathbb{S}^2$.
\enumerationisfinis

\vspace{10mm}

\setcounter{secnumdepth}{0}  
\section{References and Bibliographic Details}
\setcounter{secnumdepth}{3}
\markright{References and Bibliographic Details}

\begingroup
\footnotesize
\noindent Section \ref{subsection "Geodesics on Space Forms"}

\begin{indent paragraph: 15pt}
Definition \ref{definitio "Geodesic space"}: cf. \cite[p. 37]{Ambrosio and Gigli "A User's Guide to Optimal Transport"} \cite[pp. 70-71]{Heinonen "Lectures on Analysis on Metric Spaces"} \cite[pp. 236-237]{Heinonen Koskela Shanmugalingam Tyson "Sobolev Spaces on Metric Measure Spaces: An Approach Based on Upper Gradients"} \cite[sec. 2.4]{Papadopoulos "Metric Spaces Convexity and Non-positive Curvature"}. — On the  hyperbolic geodesic space, see \cite[chap. 1]{Buyalo Schroeder "Elements of Asymptotic Geometry"}. — Proposition \ref{propositio "Geodesics in constant curvature metrics"}: cf. e.g. \cite[pp. 23-24]{Bridson Haefliger "Metric Spaces of Non-Positive Curvature"} \cite[pp. 148-150]{Ohta "Ricci curvature entropy and optimal transport"}. 
\end{indent paragraph: 15pt}

\noindent Section \ref{subsection "Discrete Gamma-Crystallographic Group, Killing–Hopf Theorem, and Isometric Action"}

\begin{indent paragraph: 15pt}
On the Killing–Hopf's Theorem \ref{theorema "Killing–Hopf"} (Clifford–Klein space form problem), cf. e.g. \cite[p. 69]{Wolf "Spaces of Constant Curvature"} \cite[p. 385]{Gehring Martin Palka "An Introduction to the Theory of Higher-Dimensional Quasiconformal Mappings"}. — On the crystallographic group, see \cite[pp. 145, 150-152, chap. 3]{Vinberg Shvartsman "Discrete Groups of Motions of Spaces of Constant Curvature"} and \cite[chap. 6]{Sunada "Topological Crystallography With a View Towards Discrete Geometric Analysis"}.
\end{indent paragraph: 15pt}

\noindent Section \ref{section "Space Forms as Triad of Riemannian Manifolds"}

\begin{indent paragraph: 15pt}
On the space forms, curvature and radius, cf. \cite[pp. 70-73]{Saucan "Metric Curvatures Revisited: A Brief Overview"}.
\end{indent paragraph: 15pt}

\noindent Section \ref{subsection "Type II. Elliptic Geometry (and Parabolic View)"} 

\begin{indent paragraph: 15pt}
Example \ref{exemplum "Möbius strip vs. torus"}: on the Möbius strip (and other Riemann surfaces) as a genus 1 algebraic curve, see \cite[pp. 5, 150-151, 180]{Seppala and Sorvali "Geometry of Riemann Surfaces and Teichmuller Spaces"}, whilst on the algebraic genus 1 and  other than 1, cf. O. Teichmüller \cite[pp. 438-439]{Teichmuller "Extremal quasiconformal mappings and quadratic differentials"}; on the Möbius strip in terms of field of numbers, cf. \cite[p. 20]{Audin "Torus Actions on Symplectic Manifolds"} \cite[p. 35]{Lee Raymond "Seifert Fiberings"}; about the 2-torus, cf. e.g. \cite[pp. 18-19, 83, 143]{Gadea Masque Mykytyuk "Analysis and Algebra on Differentiable Manifolds"}. — Scholium \ref{scholium "Myers–Cheng and Grove–Shiohama sphere theorem"}: on the sphere theorem, see \cite[chap. 1, sec. 9, and chap. 6]{Cheeger and Ebin "Comparison Theorems in Riemannian Geometry"}; a synopsis of Cheng and Grove–Shiohama theorem is available in \cite{Eschenburg "Diameter volume and topology for positive Ricci curvature"}; in \cite{Xia "A Generalization of the Classical Sphere Theorem"} there is a generalization of the Rauch–Berger–Klingenberg theorem for Riemannian spaces with partially positive curvature.
\end{indent paragraph: 15pt}

\noindent Section \ref{subsection "Type III. Beltrami–Poincaré Hyperbolic Model (and Parabolic View)"}

\begin{indent paragraph: 15pt}
A brief reconstruction is in \cite{Milnor "Hyperbolic geometry: The first 150 years"}.
\end{indent paragraph: 15pt}

\noindent Section \ref{subsubsection "Upper Half-Space, Ball, and Hyperboloid"}

\begin{indent paragraph: 15pt}
On the hyperbolic spaces, see \cite[pp. 9, 38-40]{Lee "Riemannian Manifolds: An Introduction to Curvature"}. — On the closed upper half-space, cf. \cite[p. 25]{Lee "Introduction to Smooth Manifolds"}. — About the boundary of the ball and the sphere at infinity, cf. \cite[p. 214]{Anderson "Hyperbolic Geometry"} \cite[p. xiv]{Daverman Venema "Embeddings in Manifolds"}; cf. also \cite[pp. 59, 67]{Thurston "Three-dimensional Geometry and Topology 1"} \cite[p. 40]{Kapovich "Hyperbolic Manifolds and Discrete Groups"}. — On the Hilbert's theorem, see \cite[sec. 5-11, pp. 451-457]{do Carmo "Differential Geometry of Curves and Surfaces"}.
\end{indent paragraph: 15pt}

\noindent Section \ref{subsubsection "Upper Half-Plane and Disk"} 

\begin{indent paragraph: 15pt}
On the upper half-plane, cf. \cite[chap. 3]{Terras "Harmonic Analysis on Symmetric Spaces-Euclidean Space the Sphere and the Poincare Upper Half-Plane"}. — About the geodesics in $\mathbb{U}^2_\mathbb{C}$ and $\mathbb{D}_\mathbb{C}$, cf. \cite[pp. 16, 20]{Aigon-Dupuy Buser Semmler "Hyperbolic Geometry"} \cite[pp. 4-5]{Dal'Bo "Geodesic and Horocyclic Trajectories"} \cite[pp. 66-68]{Bux "Finiteness properties of soluble $S$-arithmetic groups: a survey"}; for a wider summary table, see \cite[p. 130]{Kisil "Geometry of Mobius Transformations: Elliptic Parabolic and Hyperbolic Actions of $SL_2(R)$"}. — Scholium \ref{scholium "Right (complex-valued) half-plane and closed unit disk"}: on the right half-plane model, see \cite[sec. 1.4.4]{Goldman "Complex Hyperbolic Geometry"}.
\end{indent paragraph: 15pt}

\noindent Section \ref{subsubsection "Beltrami–Cayley–Klein Model"} 

\begin{indent paragraph: 15pt}
On the Beltrami–Cayley–Klein model, see \cite{Maracchia "Dalla geometria euclidea alla geometria iperbolica: il modello di Klein"} \cite{A'Campo and Papadopoulos "On Klein's So-called Non-Euclidean geometry"} \cite[p. 6]{Cano Navarrete Seade "Complex Kleinian Groups"} \cite{Onishchik Sulanke "Projective and Cayley-Klein Geometries"}.
\end{indent paragraph: 15pt}

\noindent Section \ref{subsubsection "Equivalence via Cayley Transform"}

\begin{indent paragraph: 15pt}
On the Cayley transform, cf. \cite[pp. 21, 187-189]{Greene Krantz "Function Theory of One Complex Variable"}.
\end{indent paragraph: 15pt}

\noindent Section \ref{subsection "Geodesic and Strict Convexity; Convex Function on Metric and Vector Spaces"}

\begin{indent paragraph: 15pt}
See \cite[p. 35]{Barbu Precupanu "Convexity and Optimization in Banach Spaces"} \cite[p. 4]{Borwein Lewis "Convex Analysis and Nonlinear Optimization: Theory and Examples"} \cite[chap. 5, sec. 13]{DiBenedetto "Real Analysis"} \cite[pp. 1-2]{Fabian Habala Hajek Montesinos Zizler "Banach Space Theory: The Basis for Linear and Nonlinear Analysis"} \cite[pp. 1-2, 107-108, 249]{Niculescu Persson "Convex Functions and Their Applications: A Contemporary Approach"} \cite[pp. 76-78, 135, 146, 176-178, 194-196]{Papadopoulos "Metric Spaces Convexity and Non-positive Curvature"} \cite[p. 435]{Villani "Optimal Transport: Old and New"}. — Definition \ref{definitio "Geodesic convexity"}: cf. \cite[p. 18]{Deza Deza "Encyclopedia of Distances"} \cite[p. 24]{Ratcliffe "Foundations of Hyperbolic Manifolds"}.
\end{indent paragraph: 15pt}

\noindent Section \ref{subsection "Orientation Preserving Isometries of the Hyperbolic Plane and Disk"}

\begin{indent paragraph: 15pt}
As a summary of the Möbius transformations and the hyperbolic space, see \cite[sec. 2]{Minsky "A Brief Introduction to Mapping Class Groups"} \cite[app. A]{Seppala and Sorvali "Geometry of Riemann Surfaces and Teichmuller Spaces"} \cite[chap. 1.2]{Toth "Finite Mobius Groups Minimal Immersions of Spheres and Moduli"}.
\end{indent paragraph: 15pt}

\noindent Section \ref{subsubsection "I. Complex-Valued Upper Half-Plane Case"} 

\begin{indent paragraph: 15pt}
On the special linear group of degree 2, see e.g. \cite[chap. I]{A. Borel "Automorphic Forms on $SL2(R)$"} \cite[p. 153]{Olver "Classical Invariant Theory"} \cite[pp. 1319-1320]{Palmer "Banach Algebras and the General Theory of *-Algebras Vol. II: *-Algebras"}. — On the projective linear group of degree 2, cf. \cite[pp. 25-27]{Jost "Compact Riemann Surfaces"} \cite[sec. 3, pp. 89-91]{Marklof "Selberg's Trace Formula: An Introduction"}. — Proposition \ref{propositio "Universal cover of $PSL_2(R)$, and more"}: cf. \cite[pp. 46-50]{Meeks III and Perez "Constant mean curvature surfaces in metric Lie groups"} \cite[pp. 56-57, 72-73]{Seade "On the Topology of Isolated Singularities in Analytic Spaces"} \cite[pp. 223-224]{Seppala and Sorvali "Liftings of Mobius Groups to Matrix Groups"}.
\end{indent paragraph: 15pt}

\noindent Section \ref{subsubsection "II. Unit Disk in the Complex Plane Case"}

\begin{indent paragraph: 15pt}
Cf. \cite[pp. 3-5]{Beardon "The geometry of Riemann surfaces"} \cite[chap. I, sec. 4, pp. 48-51]{A. Borel "Automorphic Forms on $SL2(R)$"} \cite[pp. 1-2]{Dal'Bo "Geodesic and Horocyclic Trajectories"} \cite[pp. 29-30]{Maclachlan "Introduction arithmetic of Fuchsian groups"} \cite[p. 30]{Schlag "A Course in Complex Analysis and Riemann Surfaces"}.
\end{indent paragraph: 15pt}

\noindent Section \ref{subsubsection "III. Abridgment of the Two Previous Cases"}

\begin{indent paragraph: 15pt}
Cf. \cite[pp. 68, 70]{Aramayona "Hyperbolic Structures on Surfaces"}.
\end{indent paragraph: 15pt}

\noindent Section \ref{subsection "Möbius Group and Stereographic Projection"}

\begin{indent paragraph: 15pt}
On the Möbius group and transformations, see e.g. \cite[pp. 2-4, 49-50]{Farkas Kra "Theta Constants Riemann Surfaces and the Modular Group"}. — On the Riemann sphere, Möbius transformation, and stereographic projection, see \cite{Arnold Rogness "Mobius Transformations Revealed"} \cite[pp. 18-19, 177]{Dubrovin Fomenko Novikov "Modern Geometry II"} \cite[sec. 2.1]{Jones and Singerman "Complex Functions: An Algebraic and Geometric Viewpoint"}. — Theorem \ref{theorema "Theorem for the Möbius group and stereographic"}: cf. \cite[pp. 28-29]{Anderson "Hyperbolic Geometry"} \cite[pp. 184-185]{Greene Krantz "Function Theory of One Complex Variable"}.
\end{indent paragraph: 15pt}

\noindent Section \ref{section "Fuchsian Group (Properly Discontinuous Action)"} 

\begin{indent paragraph: 15pt}
On the Fuchsian group, see \cite[p. 627]{Arbarello Cornalba Griffiths "Geometry of Algebraic Curves II"} \cite[chap. 8]{Beardon "The Geometry of Discrete Groups"} \cite{Fine Rosenberger "Classification of all generating pairs of two generator Fuchsian groups"} \cite[chap. 4]{Fine Rosenberger "Algebraic Generalizations of Discrete Groups: A Path to Combinatorial Group Theory Through One-Relator Products"} \cite[sec. 2.3]{Girondo Gonzalez-Diez "Introduction to Compact Riemann Surfaces and Dessins d'Enfants"} \cite[chapp. 2, 4-5]{Katok "Fuchsian Groups"} \cite[chap. 2]{Iwaniec "Spectral Methods of Automorphic Forms"}. — On the discrete groups of Möbius transformations and discontinuous groups acting on hyperbolic space, see \cite{Jorgensen "On Discrete Groups of Mobius Transformations"} \cite[chap. 2]{Elstrodt Grunewald Mennicke "Groups Acting on Hyperbolic Space: Harmonic Analysis and Number Theory"}. — About the Fuchsian group and the unit disk, see \cite[sec. 4.2]{Lehto "Univalent Functions and Teichmuller Spaces"}.	
\end{indent paragraph: 15pt}

\noindent Section \ref{section "Groupable Synopsis via Commutative Diagram"}

\begin{indent paragraph: 15pt}
Cf. \cite[pp. 48-49]{Seade "On the Topology of Isolated Singularities in Analytic Spaces"}; see also \cite[pp. 16, 19, 33, 57]{Arnold "The geometry of spherical curves and the algebra of quaternions"}.
\end{indent paragraph: 15pt}

\noindent Section \ref{subsection "Hopf Fibration"}

\begin{indent paragraph: 15pt}
On the evaluation map, cf. e.g. \cite[p. 35]{Lee Raymond "Seifert Fiberings"}.
\end{indent paragraph: 15pt}

\noindent Section \ref{subsection "Spinorial Representation of the Orthogonal Group on a 3-Space"} 

\begin{indent paragraph: 15pt}
See \cite[pp. 13-23]{Carmeli Malin "Theory of Spinors: An Introduction"} \cite[secc. 2.1-2.3]{Costa Fogli "Symmetries and Group Theory in Particle Physics. An Introduction to Space-time and Internal Symmetries"} \cite[chap. I, secc. 2-3, pp. 35-97]{Hladik "Spinors in Physics"} \cite[sec. 12.3.1, pp. 769-773]{Moretti "Spectral Theory and Quantum Mechanics"} \cite[pp. 50, 56, 187-189]{Schwarz "Topology for Physicists"}.
\end{indent paragraph: 15pt}

\noindent Section \ref{subsubsection "The Covering Morphisms $SU_2(C)$ to $SO_3(R)$"}

\begin{indent paragraph: 15pt}
On the spin group, $SU_2(\mathbb{C})$ and $SO_3(\mathbb{R})$, cf. \cite[p. 81]{Arnold "Lectures on Partial Differential Equations"} \cite[pp. 135-136]{Procesi "Lie Groups: An Approach through Invariants and Representations"} \cite[chap. XI, sec. XI.1]{Turaev "Torsions of 3-dimensional Manifolds"}. — See \cite{Plymen and Robinson "Spinors in Hilbert Space"} devoted to the Hilbert spin space. — On the Pauli spinor/electron spin, see e.g. \cite[chap. 4]{Lounesto "Clifford Algebras and Spinors"} \cite[sec. 19.1.7]{Zeidler "Quantum Field Theory III: Gauge Theory"}. — On the Pauli (spin) matrices, cf. e.g. \cite[app. to § 2.8, pp. 133-135]{Manoukian "Quantum Theory: A Wide Spectrum"} \cite[p. 238]{Schwarz "Quantum Field Theory and Topology"}.\end{indent paragraph: 15pt}

\noindent Section \ref{subsubsection "Example. Irreducible Covering Spin-Space for 4pi"} 

\begin{indent paragraph: 15pt}
On the spin angular momentum, cf. \cite[sec. 8.1]{Bongaarts "Quantum Theory: A Mathematical Approach"}; about the rotation in spin models, cf. e.g. \cite[sec. 4.5.3, pp. 136-139, and p. 255]{Joshi "Elements of Group Theory for Physicists"}. — On the linkage between the topological nature of the $3\mathrm{D}$ rotation group and the electronic spin, cf. \cite[pp. 146-147, 267-268]{Zeidler "Quantum Field Theory I: Basics in Mathematics and Physics"}.
\end{indent paragraph: 15pt}

\noindent Section \ref{subsubsection "Step I"}

\begin{indent paragraph: 15pt}
On quaternions in this context, cf. \cite[pp. 603-604, 660]{Ruzhansky Turunen "Pseudo-Differential Operators and Symmetries: Background Analysis and Advanced Topics II"}. — On the complex quaternions, cf. e.g. \cite[chap. 3]{Girard "Quaternions Clifford Algebras and Relativistic Physics"}. —	On the group $\mathbb{S}^3$, cf. e.g. \cite[sec. 1.2]{Burstall Ferus Leschke Pedit Pinkall "Conformal Geometry of Surfaces in $S^4$ and Quaternions"}. — Scholium \ref{scholium "Fundamental formula of Hamilton, and non-Abelian quaternion group of order 8"}: on the quaternion group, cf. \cite{Girard "The quaternion group and modern physics"}; on the quaternions and Pauli matrices, cf. \cite[pp. 162-168]{Koks "Explorations in Mathematical Physics: The Concepts Behind an Elegant Language"} \cite[p. 37]{Lenz "Topological Concepts in Gauge Theories"} \cite[p. 315]{Schucker "Forces from Connes' Geometry"} \cite[sec. 2.1.2]{Teodorescu and Nicorovici "Applications of the Theory of Groups in Mechanics and Physics"}.
\end{indent paragraph: 15pt}

\noindent Section \ref{subsubsection "Step II"}

\begin{indent paragraph: 15pt}
Diagram suggestions in \cite[pp. 398-401]{Naber "Topology Geometry and Gauge Fields: Foundations"}; for the rest, see e.g. \cite[pp. 110, 288]{Gallier "Geometric Methods and Applications: For Computer Science and Engineering"} \cite[pp. 269, 271]{Manetti "Topology"}.	
\end{indent paragraph: 15pt}

\endgroup

\chapter{On Dimensional Continuum, Part I. Ricci Calculus (Calculus of Tensors and Curvature Tensors), Lorentz–Minkowski 4-Manifolds plus Spinor Representation, and Clifford Algebra}
\chaptermark{On Dimensional Continuum, Part I}{}
\label{chapter "On Dimensional Continuum, Part I. Ricci Calculus (Calculus of Tensors and Curvature Tensors), Lorentz–Minkowski 4-Manifolds plus Spinor Representation, and Clifford Algebra"}

\begingroup
\footnotesize
[The] way of considering quantities in more than \emph{three dimensions} is just as exact as the other; in fact algebraic letters can always be regarded as representing numbers, whether they are rational or not. I already said that it is not possible to conceive more than three \emph{dimensions}. A clever person of my acquaintance believes however that duration [\textit{durée}] can be considered as a fourth \emph{dimension}.\endnote{
	Original Fr. version: «[\,\dots] J'ai dit plus haut qu'il n'étoit pas possible de concevoir plus de trois \emph{dimensions}. Un homme d'esprit de ma connoissance croit qu'on pourroit cependant regarder la durée comme une quatrieme \emph{dimension}».
	} \\
\indent — \textsc{J.-B. le R. D'Alembert} \cite[p. 1010]{d'Alembert "Dimension"}

\vspace{2mm} 

[F]unctions relate essentially to time, which we will always denote by $t$, and since the position of a point in space depends on [its distances from] three rectangular coordinates $x, y, z$, these coordinates, in mechanical problems, are supposed as being functions of $t$. So we may regard mechanics as a geometry of four dimensions [\textit{mécanique comme une géométrie à quatre dimensions}], and mechanical analysis as an extension of geometric analysis. \\
\indent — \textsc{J.L. Lagrange} \cite[№ 185, p. 223]{Lagrange "Theorie des fonctions analytiques"}

\endgroup

\section{Excerpts from Memory: Ricci Methods}
\label{section "Excerpts from Memory: Ricci Methods"}

\begingroup
\footnotesize
I will designate by the name of \emph{absolute differential calculus} [tensor calculus] the set of methods I have called another time of covariant and contravariant derivative, since they are applicable for each fundamental form regardless of the choice of independent variables and indeed require that these [variables] be fully general and arbitrary\endnote{
	Original It. version: «[D]esignerò col nome di \emph{calcolo differenziale assoluto} [calcolo tensoriale] l'insieme di metodi da me detti altra volta di derivazione covariante e controvariante, in quanto essi sono applicabili per ogni forma fondamentale indipendentemente dalla scelta delle variabili indipendenti ed esigono anzi che queste siano affatto generali ed arbitrarie». 
	} \cite[p. 1336]{Ricci "Di alcune applicazioni del calcolo differenziale assoluto alla teoria delle forme differenziali quadratiche binarie e dei sistemi a due variabili"}. \\
\indent On Analysis issues, which by their nature are not connected with the choice of independent variables, I have long availed myself of a tool, that I call \emph{absolute Differential Calculus}, leading to formulæ and equations, which always occur under the same form for any system of variables\endnote{
	Original It. version: «Nelle questioni di Analisi, che per loro natura non sono collegate colla scelta delle variabili indipendenti, io mi valgo da molto tempo di uno strumento, che chiamo \emph{Calcolo Differenziale assoluto}, il quale conduce a formule ed equazioni, che si presentano sempre sotto la identica forma per qualunque sistema di variabili».
	} \cite[p. 1]{Ricci "Lezioni sulla teoria delle superficie"}. \\
\indent — \textsc{G. Ricci Curbastro}

\vspace{2mm}

Nacque in questa casa il [12] 1 1853 Gregorio Ricci Curbastro, maestro insigne, matematico sommo. Diede alla scienza il calcolo differenziale assoluto, strumento indispensabile per la teoria della relatività generale, visione nuova dell'Universo.\footnote{
	«In this home Gregorio Ricci Curbastro was born on [12] 1 1853, a distinguished master, supreme mathematician. He gave to science the absolute differential calculus, an indispensable tool for the theory of general relativity, a new vision of the Universe».
	} \\
\indent — Memorial plaque of \textsc{G.R.C.} at his birthplace, Corso Garibaldi, 39-41, Lugo, Ravenna

\endgroup

\subsection{What is This Calculus?}

\subsubsection{A Multi-faceted Tool}

G. Ricci \cite{Ricci "Sui parametri e gli invarianti delle forme quadratiche differenziali"} \cite{Ricci "Sulla derivazione covariante ad una forma quadratica differenziale"} \cite{Ricci "Delle derivazioni covarianti e controvarianti e del loro uso nella Analise applicata"}\endnote{
	This \cite{Ricci "Delle derivazioni covarianti e controvarianti e del loro uso nella Analise applicata"} is the work where the notion of tensor, in the Ricci sense, emerges explicitly.
	} 
\cite{Ricci "Resume de quelques travaux sur les systemes variables des fonctions associes a une forme differentielle quadratique"} \cite{Ricci "Di alcune applicazioni del calcolo differenziale assoluto alla teoria delle forme differenziali quadratiche binarie e dei sistemi a due variabili"}, when he conceived the \emph{absolute differential calculus}—nowadays called \emph{Ricci calculus}, or \emph{tensor calculus} (tensor analysis)—as a wide extension of vector calculus (vector analysis), starting from the Riemannian metric and the symbolic equipment of Christoffel (Section \ref{section "Christoffel Symbols"}) (see Levi-Civita \cite[p. 5]{Levi-Civita "Lezioni di calcolo differenziale assoluto"} = \cite[p. vii]{Levi-Civita "The Absolute Differential Calculus (Calculus of Tensors)"}),\endnote{
	«Riemann's general metric and a formula of Christoffel constitute the premises of the absolute differential calculus. Its development as a systematic branch of mathematics was a later process, the credit for which is due to Ricci, who during the ten years 1887-1896 elaborated the theory and worked out the elegant and comprehensive notation which enables it to be easily adapted to a wide variety of questions of analysis, geometry, and physics».
	}
he was moved by the intention of finding \emph{invariant relations} and equations which do not lose their validity when there are arbitrary changes, according to the purposes of mathematical description (in the pre-relativistic era), from one coordinate system to another coordinate system.\footnote{
	A first yet strong stimulus for the conception of tensor calculus, in the Ricci–Levi-Civita form, is to be found in Beltrami's differential parameters (among the various uses of differential parameters, their extension to potential theory and theory of elasticity in Riemannian spaces is relevant in Beltrami's papers); for more in-depth information, see \cite{Tazzioli "The Role of Differential Parameters in Beltrami's Work"}.
	}
Only \emph{later} \cite{Dell'Aglio "Un case study nell'accettazione di teorie matematiche. Sviluppo e diffusione del calcolo differenziale assoluto in epoca pre-relativistica"} it will be understood that it also works well in the physical realm (general relativity).

The definition of the Ricci/tensor calculus can be divided into four ramifications, that are summarized here below. It is
\subenumerationisinitium
\item an algebraic theory of differential invariants (in which the differential quantities remain unchanged with respect to certain transformations of the coordinates);
\item a technique for solving partial differential equations, i.e. integrating these equations (with the operation of taking covariant and contravariant derivatives);
\item an apparatus of analytical procedures readily used in Riemannian geometry (reinterpretation of Christoffel's algorithms in an analytical perspective);
\item a tool for the expression of laws of nature that are invariant with respect to the displacement in space and time (general relativity under the Grossmann–Einstein principle).
\subenumerationisfinis

\begin{margo}
It is worth noting that many of these issues, which later converge under the category of tensor algebra (Section \ref{section "Rudiments of Tensor Calculus"}), are transversal; we find them mutatis mutandis e.g. in \emph{invariant theory} and \emph{combinatorics}. In this regard, G.-C. Rota writes \cite[p. 21]{Rota "Two turning points in invariant theory"}: 

\vspace{2mm}

\begingroup
\footnotesize
[I]f equations [in the tensor algebra] are to express geometric properties, then they must hold no matter what coordinate system is chosen; in other words, equations that describe geometric facts must be invariant under changes of coordinates. The program of invariant theory, from Boole to our day, is precisely the translation of geometric facts into invariant algebraic equations expressed in terms of tensors. 

\endgroup

\vspace{2mm}

This just to give an approximate idea of the wide range of applications. \margosymbol
\end{margo}

\subsubsection{Algorithmic Theory of Mechanics of Continuous Deformable Bodies}

To put the four definitions above more succinctly, and to merge them together, we can say that the Ricci/tensor calculus is an algorithm, or rather, an algorithmic theory, based on the notion of covariant and contravariant differentiation processes (applicable to covariant and contravariant tensors), for the \emph{mechanics of continuous deformable bodies} at the service of the intrinsic geometry of \emph{Riemannian spaces}, see A. Palatini \cite[§ 1]{Palatini "Sui fondamenti del calcolo differenziale assoluto"}. It allows to translate the geometro-physical structure of space into an analytic form independent of the particular choice of the system of variables. To this end, some elements (e.g. the distance between two infinitely close points, or the kinetic energy) are inserted and serve as absolute-term in the calculus.

\subsection{The Role of Geometry in Gravitational Physics}
\label{subsection "The Role of Geometry in Gravitational Physics"}

\begingroup
\footnotesize
\emph{\&} è forza confessare, che il voler trattar le quistioni naturali senza Geometria è un tentar di fare quello, che è impossibile ad esser fatto.\footnote{
	«\emph{\&} it is a forced confession, that the will to deal with natural issues without Geometry is an attempt to do what is impossible to do».
	} \\
\indent — \textsc{G. Galilei} \cite[Dialogo secondo, p. 198]{Galilei "Dialogo sopra i due Massimi Sistemi del Mondo Tolemaico e Copernicano"}

\vspace{2mm}

I attach special importance to the view of geometry [\,\dots], because without it I should have been unable to formulate the theory of relativity. Without it the following reflection would have been impossible: – In a system of reference rotating relatively to an inert system, the laws of disposition of rigid bodies do not correspond to the rules of Euclidean geometry on account of the Lorentz contraction; thus if we admit non-inert systems we must abandon Euclidean geometry. \\
\indent — \textsc{A. Einstein} \cite[pp. 6-7]{Einstein "Geometrie und Erfahrung"} \cite[p. 33]{Einstein "Geometry and Experience"}, cf. endnote \ref{endnote "Contribution by Ricci and Levi-Civita in Einstein's theory of gravitation"}

\vspace{2mm}

The space of old Euclidean geometry is comparable to a crystal [\textit{Kristall}], which is made up of uniform and immutable atoms [\textit{gleichen unveränderlichen Atomen}] in the regular and rigid, unchangeable arrangement of a lattice [\textit{starren, unveränderlichen Anordnung eines Gitters}]; the space of new Riemann–Einstein geometry is instead comparable to a liquid [\textit{Flüssigkeit}], which consists of the same uniform and immutable atoms, but in mobile positions and orientations, depending on the forces acting upon them. \\
\indent — \textsc{H. Weyl} \cite[p. 45]{Weyl "Mathematische Analyse des Raumproblems: Mathematische Analyse des Raumproblems Vorlesungen"} 
	
\endgroup

\vspace{2mm}

Einstein's subsequent understanding, on the advice of M. Grossman,\endnote{
	Cf. \cite[p. 278]{Kollros "Albert Einstein en Suisse: Souvenirs"}: «[Einstein] dit un jour: “Grossmann, Du mußt mir helfen, sonst werd' ich verrückt!”».
	} 
was to use the Ricci's covariance method to rethink the classes of inertial frame of reference outside the special relativity. Einstein (see epigraph) was aware that flat geometry had to be replaced, because of the Lorentz, or FitzGerald–Lorentz, contraction \cite{FitzGerald "The Ether and the Earth's Atmosphere"} \cite{Lorentz "De relatieve beweging van de aarde en den aether"}; but he lacked a study on Riemann's geometry and had no knowledge of Ricci's and Levi-Civita's investigations.\endnote{
	\label{endnote "Contribution by Ricci and Levi-Civita in Einstein's theory of gravitation"}
	To Ricci and his pupil Levi-Civita \cite{Ricci et Levi-Civita "Methodes de calcul differentiel absolu et leurs applications"} \cite{Levi-Civita "Nozioni di parallelismo in una varieta qualunque e conseguente specificazione geometrica della curvatura riemanniana"} \cite{Levi-Civita "Lezioni di calcolo differenziale assoluto"} \cite{Levi-Civita "Fondamenti di meccanica relativistica"} are also due applications of tensor calculus  to many problems of differential geometry and mathematical physics. Here are some testimonies that underlines the importance of their contribution in Einstein's theory of gravitation:
	
	\setlength\parindent{8pt}
	($\mathnormal{1}$) Einstein's letter \cite[5 March 1915]{Einstein "Letter in Ge. addressed to Levi-Civita 5 March 1915"} in Ge. addressed to Levi-Civita: «When I saw that you [Levi-Civita] had to object on the most important demonstration of the theory [of gravitation], which made me pour rivers of sweat, I was frightened not a little, since I know that you master these mathematical things much better than I [\textit{Als ich sah, dass Sie Ihren Angriff gegen den wichtigsten, mit Strömen von Schweiss erkauften Beweis der Theorie richten, erschrak ich nicht wenig, zumal ich weiss, dass Sie diese mathematischen Dinge weit besser beherrschen als ich}]».
	
	($\mathnormal{2}$) Theory of (general) relativity, Einstein writes \cite[p. 779]{Einstein "Zur allgemeinen Relativitatstheorie"} (4 November 1915), «represents an authentic triumph [\textit{wahren Triumph}] of the method of absolute differential calculus, founded by Gauss, Riemann, Christoffel, Ricci and Levi-Civita».
	
	($\mathnormal{3}$) Recollection of Einstein, from a lecture given in Kyoto on 14 December 1922 \cite[p. 47]{Einstein "How I created the theory of relativity"}: «[T]he idea [of E. Mach] that systems that have acceleration with respect to each other are equivalent [\,\dots] contradicts Euclidean geometry, since in the frame of reference with acceleration [or where there is the influence of the gravitational force] Euclidean geometry cannot be applied. Describing the physical laws without reference to geometry is similar to describing our thought without words. We need words in order to express ourselves. What should we look for to describe our problem? This problem was unsolved until 1912, when I hit upon the idea that the surface theory of Karl Friedrich Gauss might be the key to this mystery [\,\dots].  Until then I did not know that Bernhard Riemann [who was a student of Gauss'] had discussed the foundation of geometry deeply. I happened to remember the lecture on geometry in my student years [in Zürich] by Carl Friedrich Geiser who discussed the Gauss theory. I found that the foundations of geometry had deep physical meaning in this problem. When I came back to Zürich from Prague, my friend the mathematician Marcel Grossman was waiting for me. He had helped me before in supplying me with mathematical literature when I was working at the patent office in Bern and had some difficulties in obtaining mathematical articles. First he taught me the work of Curbastro Gregorio Ricci and later the work of Riemann». 
	
	($\mathnormal{4}$) Einstein's foreword (\textit{Vorwort des Autors zur Tschechischen Ausgabe}) to \textit{Theorie relativity speciální i obecná}. Lehce srozumitelný výklad. Se zvláštní předmluvou autorovou k českému vydání, Nakladatel Fr. Borový v Praze, 1923, p. 7: «[T]he decisive idea of the analogy between the mathematical formulation of the theory [of gravitation] and the Gaussian theory of surfaces came to me only in 1912 after my return to Zürich, without being aware at that time of the work of Riemann, Ricci, and Levi-Civita. This was first brought to my attention by my friend Grossmann when I posed to him the problem of looking for generally covariant tensors whose components depend only on derivatives of the coefficients of the quadratic fundamental invariant», i.e. components of the metric tensor. Text available in \cite[p. 42]{Bicak "Einstein in Prague: Relativity Then and Now"}. 
	
	($\mathnormal{5}$) Remarkable was the encounter between Einstein and Ricci, see e.g. this clipping from “Corriere della Sera”, 28 ottobre 1921: «Il prof. Einstein a Padova, 27 ottobre. Nell'Aula Magna della nostra Università Alberto Einstein ha tenuto oggi l'annunziata conferenza. L'aula è affollatissima. Il prof. Ricci-Curbastro, della Facoltà fisico-matematica, presenta il prof. Einstein con elevate parole ricordando anche come tre secoli or sono, in questa stessa aula, Galileo Galilei abbia insegnato la allora nuova dottrina della meccanica. Il prof. Einstein, che parla italiano, esprime anzitutto il suo compiacimento nel parlare nella città dove insegna il prof. Ricci al quale si deve il calcolo infinitesimale assoluto, ch'è l'arma matematica necessaria ad esprimere la teoria della relatività generale. Poscia espone in riassunto, nell'ordine tenuto nelle conferenze di Bologna, i tratti essenziali di questa teoria». The conferences at Bologna, upon the invitation of F. Enriques, are held on 22, 24 and 26 October in the Stabat Mater Hall of the Archiginnasio.
	
	($\mathnormal{6}$) Einstein's letter \cite[25 April 1949]{Einstein "Letter in En. addressed to L. Ricci Curbastro 25 April 1949"} in En. addressed to L. Ricci Curbastro (daughter of Gregorio): «The important investigations of your father together with Levi-Civita have helped me considerably in my work concerning the general theory of relativity».
	
	($\mathnormal{7}$) Einstein's letter \cite[15 December 1952]{Einstein "Letter in Ge. addressed to Poato 15 December 1952"} in Ge. addressed to J. Poato (Ricci's former student): «The theory of relativity is an amazing example of how mathematics provides a theoretical tool to a physical theory, without the problem of physics being  decisive for the corresponding creations in mathematics. The names of Gauss, Riemann, Ricci, Levi-Civita, together with their works, would belong to the fundamental contributions of Western thought even if they had not implied the overcoming of inertial systems».
	}

Tensor calculus is the foundation of Einstein's theory of general relativity, as it is required that the laws of gravitation reveal a tensorial behavior in space-time; here, it means that physical laws must be invariant, that is, \emph{absolute}, in the face of a generic change in the spatio-temporal frame of reference. This makes it possible to specify, in the heart of gravitational equations, the solidarity between phenomena and space-time, which is the theater where phenomena take place.

Incidentally, the Einstein's postulate for the gravitation is made of two passages.
\enumerationisinitium
\item Physical laws of gravity are expressed by means of geometric postulates of space(-time). Geometry thus guarantees to physics the request of \emph{absoluteness}, namely of independence of any reference frame (the geometric neutral-space is, by assumption, absolute). At this stage of postulation, the law of motion of a particle, with the equations of the geodesic (see Section \ref{section "Geodesics, Straight Paths, and Euler–Lagrange Equations"}), is used. In the absence of matter and energy, there is a pseudo-Euclidean space-time, aka Minkowski space-time, and the law of motion is the law of inertia (see Section \ref{subsection "Minkowski Space-Time (Flat Metric)"}); in the presence of matter and energy, with the law of motion it is expected that matter and energy induce a curvature in space-time.
\item 
\label{item "Energy-momentum tensor"}
Phenomena are described with a spatio-temporal tensor analysis, for which the character of physical laws is translated tensorially. The tensor representation then ensures the invariance of the laws in the change of reference frame. Here an \emph{energy-momentum tensor} \eqref{equation "Energy-momentum tensor as a variational derivative"}, often referred to as a \emph{stress-energy tensor}, or even \emph{stress-energy-momentum tensor}, is used, and represents the intrinsic \emph{kinetic energy} of matter (it works as \emph{matter-energy flow}), the linear invariant of which is the \emph{density} of matter-energy, multiplied by the square of the speed of light in vacuum. Since gravitational field Eqq. \eqref{subequations "Einstein field equations"} coincide with the definition of the energy-momentum tensor, the concept of energy corresponds to that of space-time curvature, or to an indicator of this curvature tensorially determined, and the density of matter-energy is but the Gaussian curvature (see Section \ref{subsection "Gauss' Theorema Egregium"}).
\enumerationisfinis

\section{Rudiments of Tensor Calculus}
\label{section "Rudiments of Tensor Calculus"}

\subsection{Tensor Multilinear Algebra and Tensor Analysis}
\label{subsection "Tensor Multilinear Algebra and Tensor Analysis"}
 
\begingroup
\footnotesize
The grouping of the physical properties of crystals [\,\dots] is made according to the states of matter [\,\dots] distinguished in scalar, vectorial and tensorial [\textit{tensorielle}], the last of which encompasses [states] that occur under stress and deformation [stress-strain relations] [\,\dots] of non-rigid bodies, [\,\dots] so the characteristic physical quantities may be called tensors [\textit{Tensoren}]. \\
\indent — \textsc{W. Voigt}	\cite[pp. v, 20]{Voigt "Die fundamentalen physikalischen Eigenschaften der Krystalle in elementarer Darstellung"}.

\endgroup

\vspace{2mm}

A tensor is an algebraic object generalizing the concept of vector and describing a linear transformation (between sets of algebraic objects). Among the various definitions, we will take a look at those of tensor 
\enumerationisinitium
\item as a multilinear map,
\item as an element of vector spaces (via tensor product),
\item as an element obtained under a change of basis.
\enumerationisfinis
Other definitions on tensor spaces and algebras, tensor bundle and tensor field, set out below.

\begin{margo}[From crystal structure to algebra]
The word \emph{tensor}, in the usual meaning, initially appears within the context of \emph{crystallography} studies, with W. Voigt, see J.F. Nye \cite[p. 5]{Nye "Physical Properties of Crystals: Their Representation by Tensors and Matrices"}. \margosymbol
\end{margo}

\subsubsection{Covariance and Contravariance in Ricci's Approach}

\begingroup
\footnotesize
If now all the linear equivalents of one of two associated forms are similarly related to corresponding linear equivalents of the other, so that each may be derived from each by the same law, the forms so associated will be said to be concomitant each to the other. This concomitance may be of two kinds [\,\dots]. The first species of concomitance is defined by the corresponding equivalents of the two associated forms being deduced by precisely similar, or, as we have expressed it, concurrent transformations or substitutions, each from its given primitive. The second species of concomitance is defined by the corresponding equivalents being deduced not by similar but by contrary, i.e. reciprocal or complementary substitutions. Concomitants of the first kind may be called co-variants; concomitants of the second kind may be called contra-variants. \\
\indent — \textsc{J.J. Sylvester} \cite[p. 290]{Sylvester "On the General Theory of Associated Algebraical Forms"}.

\endgroup

\vspace{2mm}

\emph{Covariance} and \emph{contravariance} are transformations that refer to the way in which the description of a geometric (or physico-geometric) tensor-like object changes when a change of basis or of coordinates is made. The origin of these two terms goes back to J.J. Sylvester.

\begin{definitio}[Via multilinear form and tensor product]
Let $\mathfrak{V} \viz (\mathfrak{V}, \mathbb{F})$ be a finite-dimensional vector space over a field $\mathbb{F}$, and $\mathfrak{V}^* \viz (\mathfrak{V}^*, \mathbb{F})$ the dual space of covectors, that is to say, of covariant vectors. An object $\Tau^r_s$ on $\mathfrak{V}$, which is $r$-contravariant (contravariant of degree $r$), i.e. $r$ times contravariant, and $s$-covariant (covariant of degree $s$), i.e. $s$ times covariant, is called \emph{$\binom{r}{s}$-tensor} or \emph{tensor of type $(r, s)$}, with $r, s \in \mathbb{N} \cup \{0\}$, under the following conditions. 
\enumerationisinitium
\item We say that $\Tau^r_s$ is a tensor of type $(r, s)$ since it is a \emph{multilinear map} (a function linear separately in each component), 
\begin{equation}
	\Tau \colon \underbrace{\mathfrak{V}^* \times \cdots \times \mathfrak{V}^*}_{r \text{ times/copies}} \times \underbrace{\mathfrak{V} \times \cdots \times \mathfrak{V}}_{s \text{ times/copies}} \to \mathbb{F},
\end{equation}
where there are $r$-contravariant copies (indices) of $\mathfrak{V}^*$ and $s$-covariant copies (indices) of $\mathfrak{V}$.
\item The second definition of tensor implies the fixing of concept of tensor space.\footnote{
		A tensor space is always a vector space, but the converse is not true (there are vector spaces that do not preserve the tensor nature along isomorphic transformations).
	} 
We call \emph{tensor space} the vector space $\mathfrak{T}(\mathfrak{V})$, and hence \emph{tensor space of type $(r, s)$} the vector space $\mathfrak{T}^r_s(\mathfrak{V})$, that is,
\begin{equation}
\label{equation "Tensor product space of type $(r, s)$"}
	\mathfrak{T}^r_s(\mathfrak{V}) \viz \mathfrak{T}^r_s(\mathfrak{V}, \mathbb{F}) = \left(\underbrace{\mathfrak{V} \otimes \cdots \otimes \mathfrak{V}}_{r \text{ times/copies}} \otimes \underbrace{\mathfrak{V}^* \otimes \cdots \otimes \mathfrak{V}^*}_{s \text{ times/copies}}\right) = \mathfrak{V}^{\otimes^r} \otimes \mathfrak{V}^{*\otimes^s}.
\end{equation}
\subenumerationisinitium
\item We say that $\Tau^r_s$ is a tensor of type $(r, s)$ since it is an element of $\mathfrak{T}^r_s(\mathfrak{V})$, as a result of a \emph{tensor product} of $r$-contravariant copies (indices) of $\mathfrak{V}$ and $s$-covariant copies (indices) of its dual space $\mathfrak{V}^*$, meaning that $\Tau^r_s$ is an element of the \emph{tensor product space} \eqref{equation "Tensor product space of type $(r, s)$"}. Comprehensibly, $\mathfrak{T}^r_s(\mathfrak{V})$ is the space of all tensors $\Tau^r_s$.
\item The rank of a tensor is the number of covariant and contravariant indices, and it is independent of the number of underlying space dimensions. Here one has $\rank(\mathfrak{V})^{r + s}$, and $\mathfrak{T}^r_s(\mathfrak{V}) = \dim(\mathfrak{V})^{r + s}$, whilst $\mathfrak{T}(\mathfrak{V})$ is finite-dimensional.
\item The $\binom{r}{s}$-tensor product of module homomorphism is
\begin{equation}
	\left(\bigotimes^r\mathfrak{V}\right) \otimes \left(\bigotimes^s\mathfrak{V}^*\right) \to \mathfrak{T}^r_s(\mathfrak{V}, \mathbb{F}).
\end{equation}
\definitiosymbol
\subenumerationisfinis
\enumerationisfinis
\end{definitio}

\begin{definitio}[Via change of basis matrix]
Let $\mathfrak{V}$ be a finite-dimensional vector space over a field $\mathbb{F}$. Let $\mathcal{B}_v = \{v_1, \mathellipsis, v_n\}$ and $\mathcal{B}_w = \{w_1, \mathellipsis, w_n\}$ be two bases of $\mathfrak{V}$, and $v_1, \mathellipsis, v_n$ and $w_1, \mathellipsis, w_n$ the basis vectors. Assume that $\mathcal{B}_v$ and $\mathcal{B}_w$ are related by a \emph{change of basis matrix} (also known as \emph{change of coordinates matrix}) $[M]^{\mathcal{B}_v}_{\mathcal{B}_w}$, given by $\mathcal{B}_v \mapsto \mathcal{B}_w$, putting $w_\nu = \sum^n_{\mu = 1}[M^\mu_\nu]{v}_\mu$ and $v_\nu = \sum^n_{\mu = 1}[N^\mu_\nu]{w}_\mu$, where $[N^\mu_\nu]$ is the invertible matrix of $[M^\mu_\nu]$ ($MN = NM = \idem_n$). Now, tensors (and tensor transformation) are by definition \emph{basis-independent}. Denoting by $\Tau = \Tau^{\nu_1 \cdots \nu_r}_{\mu_1 \cdots \mu_s}$ the components of $\Tau$ in relation to $\mathcal{B}_v$ and by $\widetilde{\Tau}^{\nu_1 \cdots \nu_r}_{\mu_1 \cdots \mu_s}$ the components of $\Tau$ in relation to $\mathcal{B}_w$, then
\begin{align}
	\Tau & = \Tau^{\nu_1 \cdots \nu_r}_{\mu_1 \cdots \mu_s}v_{\nu_1} \otimes \cdots \otimes v_{\nu_r} \otimes v^{\mu_1} \otimes \cdots \otimes v^{\mu_s} \notag \\
	& = \widetilde{\Tau}^{\nu_1 \cdots \nu_r}_{\mu_1 \cdots \mu_s}w_{\nu_1} \otimes \cdots \otimes w_{\nu_r} \otimes w^{\mu_1} \otimes \cdots \otimes w^{\mu_s}.
\end{align}
The coordinates of $\Tau$ with respect to $\mathcal{B}_v$ and $\mathcal{B}_w$ are
\begin{align}
	& \widetilde{\Tau}^{\nu_1 \cdots \nu_r}_{\mu_1 \cdots \mu_s} = \sum^n_{\substack{\xi_1, \mathellipsis, \xi_r, \\ \varrho_1, \mathellipsis, \varrho_s}}\left[M^{\nu_1}_{\xi_1}\right] \cdots \left[M^{\nu_r}_{\xi_r}\right]\left[N^{\varrho_1}_{\mu_1}\right] \cdots \left[N^{\varrho_s}_{\mu_s}\right]\Tau^{\xi_1 \cdots \xi_r}_{\varrho_1 \cdots \varrho_s}, \\ 	
	& \Tau^{\nu_1 \cdots \nu_r}_{\mu_1 \cdots \mu_s} = \sum^n_{\substack{\xi_1, \mathellipsis, \xi_r, \\ \varrho_1, \mathellipsis, \varrho_s}}\left[N^{\nu_1}_{\xi_1}\right] \cdots \left[N^{\nu_r}_{\xi_r}\right]\left[M^{\varrho_1}_{\mu_1}\right] \cdots \left[M^{\varrho_s}_{\mu_s}\right]\widetilde{\Tau}^{\xi_1 \cdots \xi_r}_{\varrho_1 \cdots \varrho_s}, 
\end{align}
where an arrangement of such relations turns up. \definitiosymbol
\end{definitio}

\begin{margo}[Upper and lower indices]
The upper (superscripts) $r$-indices are referred to as \emph{contravariant}, since the transformations of the tensor components are the \emph{inverse} of the change of basis. The lower (subscripts) $s$-indices are referred to as \emph{covariant}, since the transformations of the tensor components are the same as the transformations under change of basis.
\margosymbol
\end{margo}

\subsubsection{Tensor Spaces \& Tensor Algebras}

\begin{definitio}
Let us now give a list of some identical tensor spaces.
\begin{align}
	& \mathfrak{T}^0_0(\mathfrak{V}) = \mathfrak{T}_0(\mathfrak{V}) = \mathfrak{T}^0(\mathfrak{V}) = \mathbb{F}, \\
	& \mathfrak{T}^1(\mathfrak{V}) = \mathfrak{T}^1_0(\mathfrak{V}) = \mathfrak{V}, \enspace \mathfrak{T}_1(\mathfrak{V}) = \mathfrak{T}^0_1(\mathfrak{V}) = \mathfrak{V}^*, \\
	& \mathfrak{T}^r(\mathfrak{V}) = \mathfrak{T}^r_0(\mathfrak{V}) = \underbrace{\mathfrak{V}^{\otimes^r \cdots \otimes^r}}_{\text{ times/copies}}, \enspace \mathfrak{T}_s(\mathfrak{V}) = \mathfrak{T}^0_s(\mathfrak{V}) = \underbrace{\mathfrak{V}^{*\otimes^s \cdots \otimes^s}}_{\text{ times/copies}}, \\
	& \mathfrak{T}(\mathfrak{V}) = \bigoplus_{r, s \geqslant 0}\mathfrak{T}^r_s(\mathfrak{V}), 
	\enspace \mathfrak{T}^\bullet(\mathfrak{V}) = \bigoplus_{r \geqslant 0}\mathfrak{T}^r(\mathfrak{V}),  
	\enspace \mathfrak{T}_\bullet(\mathfrak{V}) = \bigoplus_{s \geqslant 0}\mathfrak{T}_s(\mathfrak{V}), \\
	& \mathfrak{T}^r_s(\mathfrak{V}) = \mathfrak{T}^r(\mathfrak{V}) \otimes \mathfrak{T}_s(\mathfrak{V}).
\end{align}
The tensor space $\mathfrak{T}(\mathfrak{V})$ is known as \emph{tensor algebra} of $\mathfrak{V}$, while $\mathfrak{T}^\bullet(\mathfrak{V})$ and $\mathfrak{T}_\bullet(\mathfrak{V})$ are, respectively, the \emph{contravariant} ($^\bullet$) and \emph{covariant} ($_\bullet$) \emph{tensor algebra} of $\mathfrak{V}$. \definitiosymbol 
\end{definitio}

\subsubsection{Tensor Bundle and Tensor Field}

Now, look at the notion of tensor bundle, which will be useful to define the tensor field.

\begin{definitio}[Tensor bundle]
\label{definitio "Tensor bundle"}
~\enumerationisinitium
\item Let $\mathcal{M}$ be a differentiable manifold, and $\mathring{\mathcal{T}}^r_s\mathcal{M}$\footnote{
	In a complex vector bundle (with a trivial 1-dimensional bundle), one has $\mathring{\mathcal{T}}^r_s\mathcal{M} = \mathbb{C} \otimes_\mathbb{R}(\mathring{\mathcal{T}}^r_s\mathcal{M})$.
	}
the vector bundle of tangent space $(\mathcal{T}_p)^r_s\mathcal{M}$ at a point $p \in \mathcal{M}$, with projector $(\mathcal{T}_p)^r_s\mathcal{M} \xrightarrow{\pi} \mathcal{M}$. The tangent space 
\begin{equation}
	(\mathcal{T}_p)^r_s\mathcal{M} = \mathcal{T}_p\mathcal{M}^{\otimes^r} \otimes \mathcal{T}_p\mathcal{M}^{*\otimes^s}
\end{equation}
is neatly consistent with the tensor space of type $(r, s)$. Then $\mathring{\mathcal{T}}^r_s\mathcal{M}$ is called \emph{$\binom{r}{s}$-tensor bundle} or \emph{bundle of tensor of type $(r, s)$} over $\mathcal{M}$, the standard fiber of which is a $\binom{r}{s}$-tensor space over the real numbers. Moreover, $\mathring{\mathcal{T}}^r_s\mathcal{M}$ is isomorphic to the tensor product of the tangent and cotangent bundles,
\begin{equation}
	\mathring{\mathcal{T}}^r_s\mathcal{M} \cong \bigotimes^r \mathring{\mathcal{T}}\mathcal{M} \otimes \bigotimes^s \mathring{\mathcal{T}}^*\mathcal{M} = \underbrace{\mathring{\mathcal{T}}\mathcal{M} \otimes \cdots \otimes \mathring{\mathcal{T}}\mathcal{M}}_{r \text{ times/copies}} \otimes \underbrace{\mathring{\mathcal{T}}^*\mathcal{M} \otimes \cdots \otimes \mathring{\mathcal{T}}^*\mathcal{M}}_{s \text{ times/copies}}.
\end{equation}
\item We can report particular cases of $\mathring{\mathcal{T}}\mathcal{M} = \mathring{\mathcal{T}}^1_0\mathcal{M}$ and $\mathring{\mathcal{T}}^*\mathcal{M} = \mathring{\mathcal{T}}^0_1\mathcal{M}$. \definitiosymbol
\enumerationisfinis
\end{definitio}

\begin{definitio}[Tensor field]
~\enumerationisinitium
\item A (smooth) section of $\mathring{\mathcal{T}}^r_s\mathcal{M}$, denoted by $\sezione(\mathring{\mathcal{T}}^r_s\mathcal{M})$, is said \emph{tensor field of type $(r, s)$}, denoted by $\tau^r_s$, so $\sezione(\mathring{\mathcal{T}}^r_s\mathcal{M}) = \tau^r_s(\mathcal{M})$; conversely, a $\binom{r}{s}$-tensor field over $\mathcal{M}$ is a (smooth) \emph{section} of $\mathring{\mathcal{T}}^r_s\mathcal{M}$, so $\tau^r_s(\mathcal{M}) = \sezione(\mathring{\mathcal{T}}^r_s\mathcal{M})$.
\item More properly, we speak of \emph{tensor field} (or simply \emph{tensor}) $\tau^s_r = \Tau^{\mu_1 \cdots \mu_r}_{\nu_1 \cdots \nu_s}$ when the components of a tensor are functions of each point of a differentiable manifold in a space, in relation to a system of coordinates $x^1, \mathellipsis, x^n$, in such a manner that $\tau^s_r$ varies continuously. For this purpose, it is required that the components of $\tau^s_r$ transform under coordinate changes $x^\mu = x^\mu(y^1, \mathellipsis, y^n)$ via partial derivatives; thusly, we obtain
\begin{equation}
	\tau^s_r = \Tau^{\mu_1 \cdots \mu_r}_{\nu_1 \cdots \nu_s} = \sum_{\xi, \varrho}\widetilde{\Tau}^{\xi_1 \cdots \xi_r}_{\varrho_1 \cdots \varrho_s}\frac{\partial{x^{\mu_1}}}{\partial{y^{\xi_1}}} \cdots \frac{\partial{x^{\mu_r}}\partial{y^{\varrho_1}}}{\partial{y^{\xi_r}}\partial{y^{\nu_1}}} \cdots \frac{\partial{y^{\varrho_s}}}{\partial{x^{\nu_s}}},
\end{equation}
if $\widetilde{\Tau}^{\xi_1 \cdots \xi_r}_{\varrho_1 \cdots \varrho_s}$ are the components of $\tau^s_r$ in relation to a system of coordinates $y^1, \mathellipsis, y^n$. \definitiosymbol
\enumerationisfinis
\end{definitio}

\subsection{Musical Isomorphism of Tensors}
\label{subsection "Musical Isomorphism of Tensors"}

Below we will analyze a tensor transformation dictated by a musical isomorphism. Let us first see what it is.

\subsubsection{Mapping with Key Signature (\emph{Armatura di Chiave}): Bemolle and Diesis Operators}

\begin{definitio}
~\enumerationisinitium
\item Let $g$ be a metric (tensor), which, as is generally known, is non-degenerate. An isomorphism is said \emph{musical} if between a vector space $\mathfrak{V}$ and its algebraic dual space $\mathfrak{V}^*$ there exists a \emph{bemolle} (\emph{flat}) map, i.e. $\flat$-map, or the reverse \emph{diesis} (\emph{sharp}) map, i.e. $\sharp$-map, 
\begin{subequations}
\begin{align}
	& \flat(g) \viz g_\flat \colon \mathfrak{V} \to \mathfrak{V}^*, \text { with } \flat = \sharp^{-1}, \\
	& \sharp(g) \viz g^\sharp \colon \mathfrak{V}^* \to \mathfrak{V}, \text { with } \sharp = \flat^{-1},
\end{align}
\end{subequations}
that is,
\begin{equation}
	\mathfrak{V} 
	\left\{\!\begin{aligned}
	& \xrightarrow[\flat = \sharp^{-1}]{\flat(g)} \\
	& \xleftarrow[\sharp = \flat^{-1}]{\sharp(g)}
	\end{aligned}\right\} 
	\mathfrak{V}^*.
\end{equation}
\item Same goes for a (pseudo-)Riemannian manifold $(\mathcal{M}, g)$. We talk about musical isomorphism, on this aspect, if there is a map between the tangent space $\mathcal{T}_p\mathcal{M}$ at $p \in \mathcal{M}$ and its dual space, the cotangent space $\mathcal{T}^*_p\mathcal{M}$, determined by a $\flat$-map or a $\sharp$-map; it is therefore possible to extend such an isomorphism to the tangent bundle $\mathring{\mathcal{T}}\mathcal{M}$ and the cotangent bundle $\mathring{\mathcal{T}}^*\mathcal{M}$,
\begin{subequations}
\begin{align}
	& \label{align "Musical isomorphism for tangent space/bundle: bemolle map"}
	\flat(g) \viz g_\flat \colon \mathcal{T}_p\mathcal{M} \to \mathcal{T}^*_p\mathcal{M} \mathrel{\big|}^\text{{et}} \mathring{\mathcal{T}}\mathcal{M} \to \mathring{\mathcal{T}}^*\mathcal{M}, \\
	& \label{align "Musical isomorphism for tangent space/bundle: diesis map"}
	\sharp(g) \viz g^\sharp \colon \mathcal{T}^*_p\mathcal{M} \to \mathcal{T}_p\mathcal{M} \mathrel{\big|}^\text{{et}} \mathring{\mathcal{T}}^*\mathcal{M} \to \mathring{\mathcal{T}}\mathcal{M}.
\end{align}
\end{subequations}
In \eqref{align "Musical isomorphism for tangent space/bundle: bemolle map"}, for all vectors $v = v^\mu\partial_\mu \in \mathcal{T}_p\mathcal{M}$ at $p \in \mathcal{M}$, it holds that
\begin{equation}
	\flat(v) \viz v_\flat =
	\begin{cases}
	g_p(w, v) \in \mathcal{T}^*_p\mathcal{M}, \\
	g_{\mu\nu}v^\mu{dx}^\nu = \omega_\nu dx^\nu, \text{with } \omega_\nu = g_{\mu\nu}v^\mu,
	\end{cases}
\end{equation} 
where $\omega_\nu$ is a covector field, that is, a 1-form, or a $\binom{0}{1}$-tensor, and the equivalence is in local coordinates. Note. If $v \in \mathcal{T}_p\mathcal{M}$, then there is a vector field $\vec{X}$ such that $\vec{X}_p = v$, for each $p \in \mathcal{M}$, from which it follows that we can rewrite the above in this way: 
\begin{subequations}
\begin{align}
	& \flat(\vec{Y}) \viz \vec{Y}_\flat(\vec{X}) = g(\vec{X}, \vec{Y}), \\
	& \flat(\vec{X}) \viz\vec{X}_\flat = g(\vec{X}^\mu\partial_\mu, \cdot) = g_{\mu\nu}\vec{X}^\mu{dx}^\nu, 
\end{align}
\end{subequations}
with a mapping $\flat \colon \vec{X} \mapsto \flat(\vec{X})$.

In \eqref{align "Musical isomorphism for tangent space/bundle: diesis map"}, for any 1-form $\omega = \omega_\mu dx^\mu$, one has
\begin{equation}
	\sharp(\omega) \viz \omega^\sharp = 
	\begin{cases}
\omega^\mu = g^{\mu\nu}\omega_\nu, \\
	 g^{\mu\nu}\omega_\mu\partial_\nu = v^\nu\partial_\nu, \text{with } v^\nu = g^{\mu\nu}\omega_\mu,
	\end{cases}
\end{equation}
with a mapping $\sharp \colon \omega \mapsto \sharp(\omega)$ fixed by $g\bigl(\sharp(\omega), \vec{Y}\bigr) = \omega(\vec{Y})$. If $\sharp(\omega) \in \mathring{\mathcal{T}}^*\mathcal{M}$, then $\omega(v) = \langle{v}, \sharp(\omega)\rangle$. \definitiosymbol
\enumerationisfinis
\end{definitio}

\begin{margo}[From key signature to mathematical language]
The isomorphism of $\flat$ is for lowering indices, whilst the isomorphism of $\sharp$ is for raising indices, just like, in music, the bemolle and diesis signs are, respectively, a lowering and raising of pitch of a note. The mathematical language is here inspired by the key signature (\emph{armatura di chiave}), the typical set, in musical notation, of flat-alterations and sharp-alterations, see e.g. \cite[A. IV.2, pp. 20-22]{Berger Gauduchon Mazet "Le Spectre d'une Variete Riemannienne"}. \margosymbol	
\end{margo} 

\subsubsection{Bemolle–Diesis Convention in Tensor Description}

Musical isomorphisms are applied by linearity to some cases of tensor algebra. For instance, take a tensor space $(\mathcal{T}_p)^k_\rotatedell\mathcal{M} = \mathcal{T}_p\mathcal{M}^{\otimes^k} \otimes \mathcal{T}_p\mathcal{M}^{*\otimes^\rotatedell}$ from the tangent space (cf. above). There exists an isomorphism $(\mathcal{T}_p)^k_\rotatedell\mathcal{M} \to (\mathcal{T}_p)^{k + 1}_{\rotatedell - 1}\mathcal{M}$, which is an analogue of the $(r, s)$ version, and it can also be expressed with the convention of raising–lowering indices. Then the $\sharp$–$\flat$ isomorphisms will be extended to bundles such as $\mathring{\mathcal{T}}^r_s\mathcal{M} \to \mathring{\mathcal{T}}^k_\rotatedell\mathcal{M}$, whenever $r + s = k + \rotatedell$.

\section{Curvature Tensors}
 
\subsection{Riemann Curvature Tensor}
\label{subsection "Riemann Curvature Tensor"}

\begin{definitio}
Take a (pseudo-)Riemannian manifold $(\mathcal{M}, g)$. The \emph{Riemann curvature tensor} \cite{Riemann "Ueber die Hypothesen welche der Geometrie zu Grunde liegen"} \cite{Riemann "Commentatio mathematica"}\footnote{
	In \cite{Riemann "Ueber die Hypothesen welche der Geometrie zu Grunde liegen"}  the concept of (Riemann) curvature is delineated, and in \cite{Riemann "Commentatio mathematica"} the (Riemann) tensor is presented. To be fair, it might better be called \emph{Riemann–Christoffel tensor}, in view of the importance of \cite{Christoffel "Ueber die Transformation der homogenen Differentialausdrucke zweiten Grades"} in its development and systematization.
	} 
is a tensor of rank 4 on $\mathcal{M}$, the expression of which, in terms of the Christoffel symbols (Section \ref{section "Christoffel Symbols"}), can be written in two main ways: 
\begin{align}
	& {\Riemann_{\mu\nu\xi}}^\varrho = \partial_\mu{\Gamma_{\nu\xi}}^\varrho - \partial_\nu{\Gamma_{\mu\xi}}^\varrho + {\Gamma_{\mu\varsigma}}^\varrho{\Gamma_{\nu\xi}}^\varsigma - {\Gamma_{\nu\varsigma}}^\varrho{\Gamma_{\mu\xi}}^\varsigma, \\
	& {\Riemann^\mu}_{\nu\xi\varrho} = {\Gamma^\mu}_{\nu\varrho, \xi} -  {\Gamma^\mu}_{\nu\xi, \varrho} + {\Gamma^\varsigma}_{\nu\varrho}{\Gamma^\mu}_{\varsigma\xi} - {\Gamma^\varsigma}_{\nu\xi}{\Gamma^\mu}_{\varsigma\varrho}.  
\end{align}
Let $\vec{X}, \vec{Y}, \vec{Z} \in \mathfrak{V}(\mathcal{M})$ be three smooth vector fields, where $\mathfrak{V}(\mathcal{M})$ is a real vector space. Under the Levi-Civita connection $\nabla$ (Section \ref{subsection "Levi-Civita Connection Theorem on a (pseudo-)Riemannian Manifold"}), the Riemann tensor is
\begin{align}
	\Riemann_{\vec{X}, \vec{Y}}\vec{Z} \viz \Riemann(\vec{X}, \vec{Y})\vec{Z} & = \nabla^2_{\vec{X}, \vec{Y}}\vec{Z} - \nabla^2_{\vec{Y}, \vec{X}}\vec{Z} \notag \\
	& = \nabla_{\vec{X}} \nabla_{\vec{Y}}\vec{Y} - \nabla_{\vec{Y}} \nabla_{\vec{X}} - \nabla_{[\vec{X}, \vec{Y}]}\vec{Z} \notag \\
	& = [\nabla_{\vec{X}}, \nabla_{\vec{Y}}]\vec{Z} - \nabla_{[\vec{X}, \vec{Y}]}\vec{Z}.
\end{align}
\definitiosymbol
\end{definitio}

\begin{scholium}
The Riemann curvature tensor describes the curvature of a (pseudo-)Riemannian manifold; so if it is \emph{zero}, the manifold is \emph{flat}. \scholiumsymbol
\end{scholium}

\subsubsection{Tensor Symmetries}

What follows are algebraic properties of the Riemann curvature tensor.
\enumerationisinitium
\item It is \emph{anti-symmetric} with respect to the interchange of the first two indices or of the last two indices:
\begin{subequations}
\begin{align}
	& {\Riemann_{\mu\nu\xi}}^\varrho = {\Riemann_{[\mu\nu]\xi}}^\varrho = -\Riemann_{\nu\mu\xi\varrho}, \\
	& {\Riemann_{\mu\nu\xi}}^\varrho = \Riemann_{\mu\nu[\xi\varrho]} = -\Riemann_{\mu\nu\varrho\xi},	\\
	& \Riemann_{\mu\nu\xi\varrho} = \Riemann_{[\mu\nu][\xi\varrho]} = -\Riemann_{\nu\mu\xi\varrho} = - \Riemann_{\mu\nu\varrho\xi}.
\end{align}
\end{subequations}
\item It is \emph{symmetric} with respect to the exchange of the first pair of indices with the second pair:
\begin{equation}
	\Riemann_{\mu\nu\xi\varrho} = {\Riemann_{\mu\nu\xi}}^\varsigma g_{\varsigma\varrho} = \Riemann_{\xi\varrho\mu\nu}.
\end{equation} 
\enumerationisfinis

\begin{scholium}
We are dealing here with two forms of the Riemann curvature tensor. The first is ${\Riemann_{\mu\nu\xi}}^\varrho$, under which it is a $\binom{1}{3}$-tensor. The second is $\Riemann_{\mu\nu\xi\varrho}$, under which it is a $\binom{0}{4}$-tensor (in fully covariant form). \scholiumsymbol
\end{scholium}

\subsubsection{Bianchi Identities}
\label{subsubsection "Bianchi Identities"}

The Riemann curvature tensor also satisfies the so-called Bianchi identities. The explicit form of these relations, accompanied by a purely analytical demonstration, is due to L. Bianchi \cite{Bianchi "Sui simboli a quattro indici e sulla curvatura di Riemann"} \cite[§ 161, p. 351]{Bianchi "Lezioni di geometria differenziale I (seconda edizione)"}.\endnote{
	The Bianchi identities were first discovered by Ricci \cite{Ricci "Delle derivazioni covarianti e controvarianti e del loro uso nella Analise applicata"}, and immediately commented, without demonstration, by E. Padova, who writes \cite[postilla, p. 176]{Padova "Sulle deformazioni infinitesime"}: «The prof. Ricci points out to me that in \cite[§ 5]{Ricci "Delle derivazioni covarianti e controvarianti e del loro uso nella Analise applicata"} [\,\dots] he has drawn up equations [\,\dots] from which [the equations] of this Note can be deduced, warning that the coefficients [\,\dots] are identically null». Neither of them, however, is inspired by a full consciousness of the importance of the physical meaning of such relations, even though they are both concerned with the applicative character of their formulations. 

	\setlength\parindent{8pt}
	Bianchi \cite[pp. 5, 7]{Bianchi "Sui simboli a quattro indici e sulla curvatura di Riemann"} mentions Ricci twice, but he does only to specify a «denomination», and misses the advantage of the equalities in a physical context. Bianchi's  consideration, about the identities, was focused solely on proving (with an analytic and geometric solution) the Schur's theorem \cite{Schur "Ueber den Zusammenhang der Raume constanten Riemann'schen Krummungsmaasses mit den projectiven Raumen"}.

	Ricci returns to the subject in \cite{Ricci "Sulle superficie geodetiche in una varieta qualunque e in particolare nelle varieta a tre dimensioni"}, recalling that the identities link the Riemann's formulæ, in the transformation of quadratic differential forms, to the Christoffel symbols.
	}

The \emph{first Bianchi identity}, or algebraic identity, is presented in two forms. Under the index notation, it is
\begin{subequations}
\label{subequations "First Bianchi identity"}
	\begin{empheq}[left = {\idem_\textsc{b} \equival \empheqlbrace}]{align}
	& {\Riemann_{\mu\nu\xi}}^\varrho + {\Riemann_{\nu\xi\mu}}^\varrho + {\Riemann_{\xi\mu\nu}}^\varrho = 0, \\
	& {\Riemann_{[\mu\nu\xi]}}^\varrho = 0, \\
	& \Riemann_{\mu\nu\xi\varrho} + \Riemann_{\mu\xi\varrho\nu} + \Riemann_{\mu\varrho\nu\xi} = 0, \\
	& \Riemann_{\mu[\nu\xi\varrho]} = 0. 
    \end{empheq}
\end{subequations}
Alternatively, it can be written as
\begin{equation}
\label{equation "First Bianchi Identity with vector fields"}
	\Riemann_{\vec{X}, \vec{Y}}\vec{Z} + \Riemann_{\vec{Y}, \vec{Z}}\vec{X} + \Riemann_{\vec{Z}, \vec{X}}\vec{Y} = 0.
\end{equation}

\begin{proof}[Proof of \eqref{equation "First Bianchi Identity with vector fields"}]
\begin{align}
	\Riemann_{\vec{X}, \vec{Y}}\vec{Z} + \Riemann_{\vec{Y}, \vec{Z}}\vec{X} + \Riemann_{\vec{Z}, \vec{X}}\vec{Y} & = (\nabla_{\vec{X}}\nabla_{\vec{Y}}\vec{Z} - \nabla_{\vec{Y}}\nabla_{\vec{X}}\vec{Z} - \nabla_{[\vec{X}, \vec{Y}]}\vec{Z}) \notag \\
	& \hskip 1.1em + (\nabla_{\vec{Y}}\nabla_{\vec{Z}}\vec{X} - \nabla_{\vec{Z}}\nabla_{\vec{Y}}\vec{X} - \nabla_{[\vec{Y}, \vec{Z}]}\vec{X}) \notag \\
	& \hskip 1.1em + (\nabla_{\vec{Z}}\nabla_{\vec{X}}\vec{Y} - \nabla_{\vec{X}}\nabla_{\vec{Z}}\vec{Y} - \nabla_{[\vec{Z}, \vec{X}]}\vec{Y}) \notag \\
	& = \nabla_{\vec{X}}(\nabla_{\vec{Y}}\vec{Z} - \nabla_{\vec{Z}}\vec{Y}) + \nabla_{\vec{Y}}(\nabla_{\vec{Z}}\vec{X} - \nabla_{\vec{X}}\vec{Z}) \notag \\
	& \hskip 1.1em + \nabla_{\vec{Z}}(\nabla_{\vec{X}}\vec{Y} - \nabla_{\vec{Y}}\vec{X}) - \nabla_{[\vec{X}, \vec{Y}]}\vec{Z} \notag \\
	& \hskip 1.1em - \nabla_{[\vec{Y}, \vec{Z}]}\vec{X} - \nabla_{[\vec{Z}, \vec{X}]}\vec{Y} \notag \\
	& = \nabla_{\vec{X}}[\vec{Y}, \vec{Z}] + \nabla_{\vec{Y}}[\vec{Z}, \vec{X}] + \nabla_{\vec{Z}}[\vec{X}, \vec{Y}] - \nabla_{[\vec{X}, \vec{Y}]}\vec{Z} \notag \\ 
	& \hskip 1.1em - \nabla_{[\vec{Y}, \vec{Z}]}\vec{X} - \nabla_{[\vec{Z}, \vec{X}]}\vec{Y} \notag \\
	& = [\vec{X}, [\vec{Y}, \vec{Z}]] + [\vec{Y}, [\vec{Z}, \vec{X}]] + [\vec{Z}, [\vec{X}, \vec{Y}]] = 0, 
\end{align}	
where 
\begin{equation}
	[\vec{X}, [\vec{Y}, \vec{Z}]] + [\vec{Y}, [\vec{Z}, \vec{X}]] + [\vec{Z}, [\vec{X}, \vec{Y}]] = 0 
\end{equation}
is the \emph{Jacobi identity} in the Lie bracket notation, so the Bianchi identity is the Jacobi identity for the covariant derivative.
\end{proof}

\begin{scholium}
~\enumerationisinitium
\item The algebraic Bianchi identity is true if the curvature of the connection is \emph{torsion free}, that is, if the torsion tensor of the connection is zero (Definition \ref{definitio "Torsion free connection"}).
\item The Riemann curvature tensor is invariant under isometries, or under all parallel translations (there is thus a bijective map preserving the distance). \scholiumsymbol 
\enumerationisfinis
\end{scholium}

The \emph{second Bianchi identity}, or differential identity, is consistent with
\begin{subequations}
\label{subequations "Second Bianchi identity"}
	\begin{empheq}[left = {\idem_\textsc{b} \equival \empheqlbrace}]{align}
	\label{subequations "Second Bianchi identity in explicit form"}
	& \nabla_\varkappa{\Riemann_{\mu\nu\xi}}^\varrho + \nabla_\mu{\Riemann_{\nu\varkappa\xi}}^\varrho + \nabla_\nu{\Riemann_{\varkappa\mu\xi}}^\varrho = 0, \\
	& \nabla_{[\varkappa}{\Riemann_{\mu\nu]\xi}}^\varrho = 0, \\
	& \nabla_\varsigma\Riemann_{\mu\nu\xi\varrho} + \nabla_\xi\Riemann_{\mu\nu\varrho\varsigma} + \nabla_\varrho\Riemann_{\mu\nu\varsigma\xi} = 0, \\
	& \nabla_{[\varsigma]}\Riemann_{\mu\nu[\xi\varrho]} = 0, \\
	& \label{subequations "Second Bianchi identity with covariant derivative"}
	\Riemann_{\mu\nu\xi\varrho;\varsigma} + \Riemann_{\mu\nu\varrho\varsigma;\xi} + \Riemann_{\mu\nu\varsigma\xi;\varrho} = 0, \\
	& \label{subequations "Second Bianchi identity with covariant derivative in a concise form"}
	\Riemann_{\mu\nu[\xi\varrho;\varsigma]} = 0.
	\end{empheq}
\end{subequations}
In \eqref{subequations "Second Bianchi identity with covariant derivative"} \eqref{subequations "Second Bianchi identity with covariant derivative in a concise form"} the semi-colon is for a covariant derivative. 

The Bianchi identities can also be made explicit with the symbolism of Cartan \cite[pp. 130, 133]{Cartan "Riemannian Geometry in an Orthogonal Frame"}, see Section \ref{section "Connection Forms"}. Given a covariant derivative $D$ of the torsion, the first Bianchi identity is equivalent to
\begin{subequations}
	\begin{empheq}[left = {\idem_\textsc{b} \equival \empheqlbrace}]{align}
	& \label{subequations "First Bianchi identity with Cartan's structure"}
	{\Riemann^\mu}_\nu \wedge \omega^\nu = 0, \\
	& {\Omega^\mu}_\nu \wedge \omega^\nu = 0, \\
	\label{subequations "First Bianchi identity with Cartan's structure + torsion form"}
	& D(\Theta_\tau)^\mu = {\Omega^\mu}_\nu \wedge \vartheta^\nu,
	\end{empheq}
\end{subequations}
where $\Theta_\tau \viz (\Theta_\tau)^\mu = d(\Theta_\tau)^\mu + {\omega^\mu}_\nu \wedge \vartheta^\nu$ is the \emph{torsion form} (i.e. the vector-valued 2-form) of the connection form $\omega^\nu$, and $\vartheta^\nu$ is the basis, that is, a $\binom{1}{0}$-tensor valued 1-form. 

\begin{proof}[Proof of \eqref{subequations "First Bianchi identity with Cartan's structure + torsion form"}] 
$D\Theta_\tau = d\Theta_\tau + \omega \wedge \Theta_\tau = d(d\vartheta + \omega \wedge \vartheta) + \omega \wedge (d\vartheta + \omega \wedge \vartheta) = d\omega \wedge \vartheta - \omega \wedge d\vartheta + \omega \wedge d\vartheta + \omega \wedge \omega \wedge \vartheta = \Omega \wedge \vartheta$.
\end{proof}

\begin{scholium}
If we are to treat e.g. a flat (Minkowskian) tangent space, we add a generic basis 1-form $\rotatedm^\nu$, and the first Bianchi identity \eqref{subequations "First Bianchi identity with Cartan's structure"} becomes 
\begin{align}
	\idem_\textsc{b} \equival D\Riemann^\mu & = d\Riemann^\mu + {\omega^\mu}_\nu \wedge \Riemann^\nu \notag \\
	& = d{\omega^\mu}_\nu \wedge \rotatedm^\nu - {\omega^\mu}_\nu \wedge d\rotatedm^\nu + {\omega^\mu}_\nu \wedge d\rotatedm^\nu + {\omega^\mu}_\xi \wedge {\omega^\xi}_\nu \wedge \rotatedm^\nu \notag \\
	& = {\Riemann^\mu}_\nu \wedge \rotatedm^\nu.	
\end{align}
\scholiumsymbol
\end{scholium}

The second Bianchi identity corresponds to 
\begin{subequations}
	\begin{empheq}[left = {\idem_\textsc{b} \equival \empheqlbrace}]{align}
	& d{\Riemann^\mu}_\nu + {\omega^\mu}_\xi \wedge {\Riemann^\xi}_\nu - {\Riemann^\mu}_\xi \wedge {\omega^\xi}_\nu = 0, \\
	& d\Omega + \omega \wedge \Omega - \Omega \wedge \omega = d\Omega + [\omega, \Omega] = 0, \\
	& d\Omega = \omega \wedge \Omega - \Omega \wedge \omega, \\
	\label{subequations "Second Bianchi identity with Cartan's structure + torsion form"}
	& D{\Omega^\mu}_\nu = 0.
	\end{empheq}
\end{subequations}

\begin{proof}[Proof of \eqref{subequations "Second Bianchi identity with Cartan's structure + torsion form"}] 
$D\Omega = d\Omega + \omega \wedge \Omega - \Omega \wedge \omega = d\Omega + \omega \wedge d\omega - d\omega \wedge \omega = d\Omega + [\omega, \Omega]$, and $d\Omega = d\omega \wedge \omega - \omega \wedge d\omega$.
\end{proof}

\begin{scholium}
\label{scholium "Independent components in the Riemann curvature tensor"}
The Riemann curvature tensor is, as we have said, a tensor of \emph{rank 4} with $(\mu\nu\xi\varrho)$-indices, and it has $4 \times 4 \times 4 \times 4 = 256$ independent components in $4\mathrm{D}$ (space-time). Nonetheless, the symmetries and identities outlined above consent us to reduce this number: the Riemann tensor can be firstly reconceived as a tensor product of two anti-symmetric tensors of \emph{rank 2}, with 36 components; then, thanks to the 1st Bianchi identity, there are 16 conditions on the components. The final number of components is hence $6 \times 6 - 4 \times 4 = 36 - 16 = 20$, or rather (with $4 = \dim$): $\frac{4(4 - 1)^2}{2^2} - \frac{1}{6}4^2(4 - 1)(4 - 2) = 4^2\frac{1}{12}(4^2 - 1) = 20$.
\scholiumsymbol
\end{scholium}

\subsection{Riccian Algebro-geometric Properties}
\label{subsection "Riccian Algebro-geometric Properties"}

For a start, let us define the $n$-tuple $\vec{E}_1, \mathellipsis, \vec{E}_n$ of smooth vector fields on (defined over) an open neighborhood $\Upsilon \subset \mathcal{M}$ as a frame field, that is, an orthonormal basis for the tangent space $\mathcal{T}_p\mathcal{M}$, for each point $p \in \Upsilon$. The collection $\vartheta_1, \mathellipsis, \vartheta_n$ will be the corresponding dual coframe field, i.e. 1-forms, providing an orthonormal basis of the cotangent space $\mathcal{T}^*_p\mathcal{M}$. 

Then we specify a tensor 
\begin{equation}
	\Tau = \Tau^{\mu_1 \cdots \mu_r}_{\nu_1 \cdots \nu_s}\vec{E}_{\mu_1} \otimes \cdots \otimes \vec{E}_{\mu_r} \otimes \vartheta^{\nu_1} \otimes \cdots \otimes \vartheta^{\nu_s},
\end{equation} 
fixing the vectors $v = \vartheta^\mu(v)\vec{E}_\mu = v^\mu\vec{E}_\mu$ in $\mathring{\mathcal{T}}\mathcal{M}$ and the covectors $\omega = \omega(\vec{E}_\nu)\omega^\nu = \omega_\nu\vartheta^\nu$ in $\mathring{\mathcal{T}}^*\mathcal{M}$. 

\subsubsection{Ricci Curvature Tensor}
\label{subsubsection "Ricci Curvature Tensor"}

The \emph{Ricci curvature tensor} \cite{Ricci "Resume de quelques travaux sur les systemes variables des fonctions associes a une forme differentielle quadratique"} is a tool that is used to measure the \emph{degree of flatness} or, which is the same, the \emph{value of non-flatness} of a certain surface, i.e. the difference between non-Euclidean space, or curved space, under the Riemannian geometry, and Euclidean space, whose curvature is zero. Note. E. Bompiani \cite{Bompiani "La geometrie des espaces courbes et le tenseur d'energie d'Einstein"} (see epigraph in Section \ref{subsection "Einstein Tensor"}) was the first to introduce officially the expression \emph{Ricci (curvature) tensor}.

\begin{definitio}
The Ricci curvature tensor is a symmetric tensor of rank 2, and it can be written in multiple ways. Let us see them one by one. 
\enumerationisinitium
\item Its three primary forms:
\begin{subequations}
	\begin{empheq}[left = {\Ric \equival \empheqlbrace}]{align}
	& {\Ricci_\nu}^\mu\vec{E}_\mu \otimes \vartheta^\nu \text{ as a $\tbinom{1}{1}$-tensor}, \\
	& \Ricci_{\nu\xi}\vartheta^\nu \otimes \vartheta^\xi = g_{\nu\mu}{\Ricci_\xi}^\mu\vartheta^\nu \otimes \vartheta^\xi \text{ as a $\tbinom{0}{2}$-tensor}, \\
	& \Ricci^{\mu\xi}\vec{E}_\mu \otimes \vec{E}_\xi = g^{\mu\nu}\vec{E}_\mu \otimes \vec{E}_\xi \text{ as a $\tbinom{2}{0}$-tensor},
	\end{empheq}
\end{subequations}
i.e. 
\begin{subequations}
	\begin{empheq}[left = {\Ric \in \empheqlbrace}]{align}
	& \Tau^1_1(\mathcal{M}), \\ 
	& \Tau^0_2(\mathcal{M}), \\
	& \Tau^2_0(\mathcal{M}).
	\end{empheq}
\end{subequations}
\item The Ricci tensor correlates to a \emph{contraction} of the Riemann curvature tensor:
\begin{subequations}
	\begin{empheq}[left = {\Ric \equival \empheqlbrace}]{align}
	& {\Ricci_\nu}^\mu\vec{E}_\mu \otimes \vartheta^\nu = {\Riemann_{\mu\xi}}^{\xi\nu}\vec{E}_\mu \otimes \vartheta^\nu = {\Riemann_{\mu\xi\varsigma}}^\nu{g}^{\varsigma\xi}\vec{E}_\mu \otimes \vartheta^\nu, \\
	& \Ricci_{\mu\nu}\vartheta^\mu \otimes \vartheta^\nu = g^{\xi\varrho}\Riemann_{\mu\xi\varrho\nu}\vartheta^\mu \otimes \vartheta^\nu.
	\end{empheq}
\end{subequations}
\item In an explicit solution, making use of the Christoffel symbols:
\begin{equation}
\label{equation "Ricci tensor in explicit form"}
		\Ric \equival \Ricci_{\mu\nu} = \partial_\xi{\Gamma_{\mu\nu}}^\xi - \partial_\nu{\Gamma_{\mu\xi}}^\xi + {\Gamma_{\mu\nu}}^\xi{\Gamma_{\xi\varrho}}^\varrho - {\Gamma_{\mu\xi}}^\varrho{\Gamma_{\nu\varrho}}^\xi.
\end{equation}
\item Ricci curvature, as a $\tbinom{0}{2}$-tensor, is also termed as the trace of a linear operator $\vec{Z} \mapsto \Riemann_{\vec{X}, \vec{Z}}\vec{Y}$, so 
\begin{equation}
	\Ric(\vec{X}, \vec{Y}) = \trace\left(\vec{Z} \mapsto \Riemann_{\vec{X}, \vec{Z}}\vec{Y}\right).
\end{equation}
It is with this in view that we say that $\Ric$ is a trace of the Riemann curvature tensor. \definitiosymbol
\enumerationisfinis
\end{definitio}

\subsubsection{Scalar Curvature} 
\label{subsubsection "Scalar Curvature"} 

The \emph{scalar curvature}, or \emph{Ricci scalar} \cite{Ricci "Lezioni sulla teoria delle superficie"} \cite{Ricci "Direzioni e invarianti principali in una varieta qualunque"}, is the trace of the Ricci curvature tensor, with respect to the Riemannian metric $g$, and one denotes it by $\scalarcurvature$. The scalar curvature represents the most elementary of \emph{local invariants} of $g$ on $\mathcal{M}$. The same definition there is by contraction with $g$. In formulæ:
\begin{subequations}
\label{subequations "Scalar curvature, i.e. Ricci scalar"}
	\begin{empheq}[left = {\scalarcurvature \in \mathscr{C}^\infty(\mathcal{M}) \equival \empheqlbrace}]{align}
	& \trace(\Ric), \\ 
	& {\Ricci^\mu}_\mu = g^{\mu\nu} \Ricci_{\mu\nu} = {\Ricci_\nu}^\nu, \\
	& g^{\mu\xi}g^{\nu\varrho}\Riemann_{\mu\nu\xi\varrho}.
	\end{empheq}
\end{subequations}

\subsection{Einstein Tensor}
\label{subsection "Einstein Tensor"}

\begingroup
\footnotesize
I will indicate a geometric construction of the Ricci tensor $\mathrm{R}_{ik} = \sum_h\{ih, hk\}$ from which it is easy to deduce the Einstein tensor $\mathrm{R}_{ik} -\frac{1}{2}g_{ik}\mathrm{R}$. \\
\indent — \textsc{E. Bompiani} \cite[p. 739]{Bompiani "La geometrie des espaces courbes et le tenseur d'energie d'Einstein"}

\endgroup

\vspace{2mm}

The \emph{Einstein tensor}\footnote{
	To tell the truth, it should be called \emph{Ricci–Einstein tensor}, as was the custom in the 1920s \cite[p. 157]{Ciliberto and Sallent Del Colombo "Enrico Bompiani: The Years in Bologna"}. 
	} 
is the combination of the Ricci curvature tensor ($\Ricci_{\mu\nu}$) and the scalar curvature ($\scalarcurvature$),
\begin{equation}
\label{equation "Einstein Tensor"}
	\gravitation_{[\mu\nu]} = \Ricci_{\mu\nu} - \frac{1}{2}g_{\mu\nu}\scalarcurvature, \enspace \gravitation_{[\mu\nu]} = \gravitation_{\mu\nu = \nu\mu}
\end{equation}
the physical purpose of which, in respect of the law of local energy-momentum conservation, is the description of the curvature of space-time in gravitational field Eqq. \eqref{subequations "Einstein field equations"}; $g_{\mu\nu}$ is the metric tensor. In consequence of the second Bianchi identity \eqref{subequations "Second Bianchi identity in explicit form"}, the covariant divergence of the Einstein tensor is null,\footnote{
	Exactly like the energy-momentum tensor \eqref{equation "Energy-momentum tensor as a variational derivative"}, see Eq. \eqref{equation "Energy-momentum conservation"}.
	}
\begin{equation}
\label{equation "Contracted Bianchi identity"}
	\nabla_\nu{\gravitation_\mu}^\nu = \nabla_\nu \gravitation^{\mu\nu} = 0.
\end{equation}
The latter expression is known as \emph{contracted Bianchi identity} dating back to A.  Voss \cite{Voss "Zur Theorie der Transformation quadratischer Differentialausdrucke und der Krummung hoherer Mannigfaltigkeiten"}. From \eqref{subequations "Second Bianchi identity in explicit form"}, if 
\begin{equation}
	{\Ricci_{\mu;\nu}}^\nu = \frac{1}{2}({\delta_\mu}^\nu\scalarcurvature)_{;\nu}
\end{equation}
is set as the covariant divergence of the Ricci curvature tensor, we get to
\begin{subequations}
	\begin{empheq}[left = {\text{equivalent Eqq.} \empheqlbrace}]{align}
	& \nabla_\nu{\Ricci_\mu}^\nu = \nabla_\nu{\Riemann_{\xi\mu}}^{\nu\xi} \\
	& - \nabla_\xi{\Riemann_{\mu\nu}}^{\nu\xi} - \nabla_\mu{\Riemann_{\nu\xi}}^{\nu\xi},
	\end{empheq}
\end{subequations}
and
\begin{equation}
	2\nabla_\nu{\Ricci_\mu}^\nu = \nabla_\mu\scalarcurvature,
\end{equation}
and finally
\begin{equation}
\label{equation "Einstein Tensor and contracted Bianchi identity"}
	\nabla_\nu\left({\Ricci_\mu}^\nu - \frac{1}{2}{\delta_\mu}^\nu\scalarcurvature\right) = 0, 
\end{equation}
ergo $\eqref{equation "Contracted Bianchi identity"} \equival \eqref{equation "Einstein Tensor and contracted Bianchi identity"}$, which also shows that it is possible to obtain the Einstein tensor from the contracted Bianchi identity; we immediately discover that 
\begin{equation}
	{\Ricci_\mu}^\nu - \frac{1}{2}{\delta_\mu}^\nu\scalarcurvature
\end{equation}
and
\begin{equation}
	\Ricci_{\mu\nu} - \frac{1}{2}g_{\mu\nu}\scalarcurvature 
\end{equation}
are equivalent. The use of the Bianchi identities in general relativity takes place with Levi-Civita \cite{Levi-Civita "Sulla espressione analitica spettante al tensore gravitazionale nella teoria di Einstein"}; for the role he played in all of this, see Section \ref{subsubsection "Invariantiveness and Tensorial Conservation: Levi-Civita's Analytical Expression"}.

\subsection{Weyl Curvature Tensors}
\label{subsection "Weyl Curvature Tensors"}

The Riemann curvature tensor has 20 independent components (see Scholium \ref{scholium "Independent components in the Riemann curvature tensor"}) in $4\mathrm{D}$, half of which is contained in the Ricci curvature tensor; the other half is captured by the so-called \emph{Weyl curvature tensor} (also known as \emph{conformal tensor})\footnote{
	This is why it is often denoted with the letter $C$.	
	} 
\cite[p. 404]{Weyl "Reine Infinitesimalgeometrie"} \cite[chap. IV]{Weyl "Space-Time-Matter"}. To him we owe the formulation of another tensor, said \emph{Weyl projective (curvature) tensor}.

\subsubsection{Conformal Curvature Tensor}

Let us begin to see the first of these two tensors.

\begin{definitio}
\label{definitio "Weyl curvature tensor"}
Given a pseudo-Riemannian manifold $(\mathcal{M}, g)$ of $\dim(\mathcal{M}) = n$, the \emph{Weyl (conformal) curvature tensor} is an algebraic object with the following forms:
\begin{subequations}
\label{subequations "Weyl curvature tensor of (0, 4)-type"}
\begin{align}
\label{align "Weyl curvature tensor I: D greater than or equal to 3"}
	& \Weyl_{\mu\nu\xi\varrho} = \Riemann_{\mu\nu\xi\varrho} - \frac{2}{n - 2}\left(g_{\mu[\xi}\Ricci_{\varrho]\nu} - g_{\nu[\xi}\Ricci_{\varrho]\mu}\right) + \frac{2}{(n - 1)(n - 2)}\scalarcurvature g_{\mu[\xi}g_{\varrho]\nu}, \\
\label{align "Weyl curvature tensor II: D greater than or equal to 3"}
	& \Weyl_{\mu\nu\xi\varrho} = \Riemann_{\mu\nu\xi\varrho} - \frac{1}{n - 2}(g_{\mu\xi}\Ricci_{\nu\varrho} - g_{\nu\xi}\Ricci_{\mu\varrho} - g_{\mu\varrho}\Ricci_{\nu\xi} + g_{\nu\varrho}\Ricci_{\mu\xi}) \notag \\
	& \hspace{38pt} + \frac{\scalarcurvature}{(n - 1)(n - 2)}(g_{\mu\xi}g_{\nu\varrho} - g_{\nu\xi}g_{\mu\varrho}), \\
	\label{align "Weyl curvature tensor in four dimensions I"}
	& \Weyl_{\mu\nu\xi\varrho} = \Riemann_{\mu\nu\xi\varrho} - \Ricci_{\xi[\mu}g_{\nu]\varrho} + \Ricci_{\varrho[\mu}g_{\nu]\xi} + \frac{1}{3}\scalarcurvature g_{\xi[\mu}g_{\nu]\varrho} \notag \\
	& \hspace{27pt} = \Riemann_{\mu\nu\xi\varrho} - \frac{1}{2}(\Ricci_{\mu\xi}g_{\nu\varrho} + \Ricci_{\mu\varrho}g_{\nu\xi} + \Ricci_{\nu\xi}g_{\mu\varrho} - \Ricci_{\nu\varrho}g_{\mu\xi}) \notag \\ 
	& \hspace{38pt} + \frac{1}{6}\scalarcurvature(g_{\mu\xi}g_{\nu\varrho} - g_{\mu\varrho}g_{\nu\xi}), \\
	\label{align "Weyl curvature tensor in four dimensions II"}
	& \Weyl_{\mu\nu\xi\varrho} = \Riemann_{\mu\nu\xi\varrho} - g_{\mu[\xi}\Ricci_{\varrho]\nu} + g_{\nu[\xi}\Ricci_{\varrho]\mu} + \frac{1}{3}\scalarcurvature g_{\mu[\xi}g_{\varrho]\nu}, 
\end{align}
\end{subequations}
\begin{equation}
\label{equation "Weyl curvature tensor with Kulkarni–Nomizu product"}
	\Wey = \Rie - \frac{\scalarcurvature}{2n(n - 1)}g \KulkarniNomizu g - \frac{1}{n - 2}\left(\Ric - \frac{\scalarcurvature}{n}g\right) \KulkarniNomizu g,
\end{equation}
where the Riemann and the Ricci tensors, plus the scalar curvature, appear. The expressions \eqref{align "Weyl curvature tensor I: D greater than or equal to 3"} \eqref{align "Weyl curvature tensor II: D greater than or equal to 3"} are used in $\mathrm{D} \geqslant 3$, whilst \eqref{align "Weyl curvature tensor in four dimensions I"} \eqref{align "Weyl curvature tensor in four dimensions II"} in $4\mathrm{D}$. We can also write the above expressions with indices in $\mu$- and $\mu\nu$-different positions: 
\begin{align}
\label{align "Weyl curvature tensor of (1, 3)-type"}
	{\Weyl^\mu}_{\nu\xi\varrho} = {} & {\Riemann^\mu}_{\nu\xi\varrho} - \frac{1}{n - 2}\left({\delta^\mu}_\xi\Ricci_{\nu\varrho} - g_{\nu\xi}{\Ricci^\mu}_\varrho - {\delta^\mu}_\varrho\Ricci_{\nu\xi} + g_{\nu\varrho}{\Ricci^\mu}_\xi\right) \notag \\ 
	& \hspace{26pt} + \frac{\scalarcurvature}{(n - 1)(n - 2)}({\delta^\mu}_\xi{g}_{\nu\varrho} - g_{\nu\xi}{\delta^\mu}_\varrho),
\end{align}
and similarly in \eqref{align "Weyl curvature tensor I: D greater than or equal to 3"}; the Eq. \eqref{align "Weyl curvature tensor in four dimensions I"} becomes
\begin{equation}
	{\Weyl^{\mu\nu}}_{\xi\varrho} = {\Riemann^{\mu\nu}}_{\xi\varrho} - 2{\delta^{[\mu}}_{\xi}{\Ricci^{\nu]}}_{\varrho]} + \frac{1}{3}\scalarcurvature{\delta^{[\mu}}_{\xi}{\delta^{\nu]}}_{\varrho]}.
\end{equation}
In \eqref{equation "Weyl curvature tensor with Kulkarni–Nomizu product"}, the   abbreviation $\Rie$ is for the Riemann tensor, i.e. 
\begin{equation}
	\Rie = \frac{\scalarcurvature}{2n(n - 1)}g \KulkarniNomizu g + \frac{1}{n - 2}\left(\Ric - \frac{\scalarcurvature}{n}g\right) \KulkarniNomizu g + \Wey,
\end{equation}
and the symbol $\KulkarniNomizu$ denotes the \emph{Kulkarni–Nomizu product} \cite{Kulkarni "Curvature structures and conformal transformations"} \cite{Kulkarni "On the Bianchi identities"} \cite{Nomizu "On the decomposition of generalized curvature tensor fields Codazzi Ricci Bianchi and Weyl revisited"} of a symmetric $\binom{0}{2}$-tensors building a curvature $\binom{0}{4}$-tensor. \definitiosymbol
\end{definitio}

\begin{margo}
Let $\Tau_a$ and $\Tau_b$ two symmetric covariant $\binom{0}{2}$-tensors. The Kulkarni–Nomizu product is consistent with the $\binom{0}{4}$-tensor $\Tau_a \KulkarniNomizu \Tau_b$ defined by 
\begin{align}
	\Tau_a \KulkarniNomizu \Tau_b(v_1, v_2, v_3, v_4) = {} & \Tau_a(v_1, v_3)\Tau_b(v_2, v_4) + \Tau_a(v_2, v_4)\Tau_b(v_1, v_3) \notag \\
	& - \Tau_a(v_1, v_4)\Tau_b(v_2, v_3) - \Tau_a(v_2, v_3)\Tau_b(v_1, v_4),
\end{align}
with four tangent $v$-vectors. \margosymbol
\end{margo}

\subsubsection{Projective Curvature Tensor}

Let us move on to the projective structure.

\begin{definitio}
The tensor the invariant form of which coincides with
\begin{align}
	{^\prj\Weyl^\mu}_{\nu\xi\varrho} & = {\Riemann^\mu}_{\nu\xi\varrho} - \frac{1}{n - 1}({\delta^\mu}_\varrho\Ricci_{\nu\xi} - {\delta^\mu}_\xi\Ricci_{\nu\varrho}) \notag \\ 
	& = {\Riemann^\mu}_{\nu\xi\varrho} + 1/(n - 1)({\delta^\mu}_\xi\Ricci_{\nu\varrho} - {\delta^\mu}_\varrho\Ricci_{\nu\xi}),
\end{align}
is called \emph{Weyl projective (curvature) tensor}. \definitiosymbol
\end{definitio}

\subsection[Conformal Flatness and $n$-Dimensionality]{Conformal Flatness and $\mathbold{n}$-Dimensionality}
\label{subsection "Conformal Flatness and $n$-Dimensionality"}

Pay attention, in Definition \ref{definitio "Weyl curvature tensor"}, to the \emph{dimensionality of the space}, which is crucial.
\enumerationisinitium
\item If $n = 1$ and $n = 2$, the Riemannian spaces in question (1- and 2- manifolds) are \emph{conformally flat}, in a sense that will be explained shortly (they are in fact mono- and bi-dimensional), and there is no Weyl curvature tensor.
\item If $n = 3$, the Weyl tensor vanishes identically, and it is the $\Ric$-tensor that completely determines the $\Rie$-tensor, for which the curvature in a 3-manifold is Ricci dependent. A 3-manifold is, instead, \emph{(locally) conformally flat} iff the Cotton tensor \eqref{equation "Cotton tensor"} vanishes. 
\item If $n \geqslant 4$, the Weyl tensor is typically not null, and the the reference space in called \emph{non-conformally flat}. In the event that the Weyl tensor is identically zero, the metric space is \emph{conformally flat}. So a Riemannian $n$-space, with $n \geqslant 4$, turns out to be \emph{conformally flat} iff the Weyl tensor associated is zero. This is the case of spaces of constant curvature (e.g. 4-sphere, Euclidean 4-space, and hyperbolic 4-space), all of which have an identically vanishing $\Wey$-solution.
\enumerationisfinis

\subsubsection{Details, and Cotton \& Cotton–York Tensors}

Some clarifications on the above. 
\enumerationisinitium
\item A smooth (pseudo-)Riemannian space $(\mathcal{M}, g)$ is called \emph{(locally) conformally flat} if
\subenumerationisinitium
\item every point of $\mathcal{M}$ has a neighborhood $\Upsilon \subset \mathcal{M}$ conformally equivalent to an open subset of (pseudo-)Euclidean space, i.e., equivalently,
\item if there is a map $\varphi(x) \colon \Upsilon \to \mathbb{R}$ such that $\Upsilon$ at every point is mapped conformally into a flat space. The  adjective \emph{conformal} means \emph{angle preserving}. 

Note. The flatness of a (pseudo-)Riemannian depends on the possibility of point-covering $\Upsilon$ by a metric coordinate system, putting $ds^2 = g_{\mu\nu}dx^\mu dx^\nu$, with $\mu, \nu = 1, \mathellipsis, n$. 

For further information, see  the basic writings of N.H. Kuiper \cite{Kuiper "On Conformally-Flat Spaces in the Large"} \cite{Kuiper "On Compact Conformally Euclidean Spaces of Dimension > 2"}; S.I. Goldberg \cite{Goldberg "On conformally flat spaces with definite Ricci curvature"}, and R. Schoen \& S.-T. Yau \cite{Schoen and Yau "Conformally flat manifolds Kleinian groups and scalar curvature"}.
\subenumerationisfinis
\item The vanishing of the Weyl tensor, we already know, causes the conformal flatness of the metric $g_{\mu\nu}$. More precisely, it is true that there is a conformal function $\varphi(x) \in \mathscr{C}^\infty(\mathcal{M})$, $0 < \varphi < \infty$, such that $g_{\mu\nu} = \varphi^2(x)\eta_{\mu\nu}$, where $\eta_{\mu\nu}$ is the flat space metric, i.e. the Minkowski $\binom{0}{2}$-tensor; and indeed, given a conformal transformation $g_{\mu\nu} \to \tilde{g}_{\mu\nu} = \varphi^2(x)g_{\mu\nu}$, the Weyl tensor acts invariantly,
\begin{equation}	
	\widetilde{\Weyl}^\mu{}_{\nu\mu\xi} = {\Weyl^\mu}_{\nu\mu\xi}.
\end{equation}
\item The \emph{Cotton tensor} \cite{Cotton "Sur les varietes a trois dimensions"} can be determined as a 3-tensor,
\begin{subequations}
\label{equation "Cotton tensor"}
	\begin{empheq}[left = {\Cotton_{\mu\nu\xi} = \empheqlbrace}]{align}
	& \nabla_\nu\left(\Ricci_{\mu\xi} - \frac{1}{4}\scalarcurvature g_{\mu\xi}\right) - \nabla_\xi\left(\Ricci_{\mu\nu} - \frac{1}{4}\scalarcurvature g_{\mu\nu}\right), \\
	& \left\{\nabla_{[\mu}\Ricci_{\nu]\xi} - \frac{1}{2(n - 1)}\nabla_{[\mu}\scalarcurvature g_{\nu]\xi}\right\}2,
	\end{empheq}
\end{subequations}
where the Ricci tensor and the scalar curvature appear, or as a 2-tensor, also called \emph{Cotton–York tensor} \cite{York Jr. "Gravitational Degrees of Freedom and the Initial-Value Problem"} \cite{York Jr. "Role of Conformal Three-Geometry in the Dynamics of Gravitation"} \cite{York Jr. "Conformally invariant orthogonal decomposition of symmetric tensors on Riemannian manifolds"},
\begin{equation}
\label{equation "Cotton–York tensor"} 
		{\Cotton_\mu}^\nu = \varepsilon^{\xi\varrho\nu}\nabla_\xi\left(\Ricci_{\varrho\mu} - \frac{1}{4}\scalarcurvature g_{\varrho\mu}\right).
\end{equation}
The tensor \eqref{equation "Cotton–York tensor"} is traceless, symmetric, and it has   zero divergence: $g_{\mu\nu}\Cotton^{\mu\nu} = 0$, $\Cotton^{[\mu\nu]}$, $\nabla_\nu\Cotton^{\mu\nu} = 0$. 
\item The Weyl tensor is equivalent to a $\binom{0}{4}$-tensor, as in \eqref{subequations "Weyl curvature tensor of (0, 4)-type"} \eqref{equation "Weyl curvature tensor with Kulkarni–Nomizu product"}, or to a $\binom{1}{3}$-tensor, as in \eqref{align "Weyl curvature tensor of (1, 3)-type"}. As a $\binom{0}{4}$-tensor, its properties are:
\begin{subequations}
\begin{align}
	& \Weyl_{\mu\nu\xi\varrho} = - \Weyl_{\nu\mu\xi\varrho} = - \Weyl_{\mu\nu\varrho\xi} = \Weyl_{\xi\varrho\mu\nu}, \\
	& \Weyl_{\mu\nu\xi\varrho} + \Weyl_{\mu\xi\varrho\nu} + \Weyl_{\mu\varrho\nu\xi} = 0, \\
	& {\Weyl^\mu}_{\nu\mu\xi} = 0.
\end{align}
\end{subequations}
Beware, however: the \emph{conformal invariance} property belongs only to the Weyl $\binom{1}{3}$-tensor and not to the Weyl $\binom{0}{4}$-tensor.
\enumerationisfinis

\begin{margo}[Independent components compared]
For 2- and 3-dimensional spaces the Riemann and Ricci curvature tensors have the same number $N_\mathrm{ic}$ of independent components at each point; for a 4-dimensional space the number no longer coincides.
\enumerationisinitium
\item In $1\mathrm{D}$ (1-manifold) the space is flat, and there is no intrinsic curvature.
\item In $2\mathrm{D}$ (2-manifold) one has $N_\mathrm{ic} = 1$ (the Gaussian curvature) for both Riemann and Ricci tensors; this one component is related to the scalar curvature.
\item In $3\mathrm{D}$ (3-manifold) one has $N_\mathrm{ic} = 6$ for both Riemann and Ricci tensors, according to the respective formulæ, $3^2\frac{1}{12}(3^2 - 1)$ and $3\frac{1}{2}(3 + 1)$.
\item In $4\mathrm{D}$ (4-manifold) the number $N_\mathrm{ic}$ for the Riemann tensor is $4^4 = 256$, reducible to $4^2\frac{1}{12}(4^2 - 1) = 20$; but for the $\Ric$-tensor, one has $N_\mathrm{ic} = 4\frac{1}{2}(4 + 1) = 10$, and the same for the Weyl tensor, based on the formula $4^2\frac{1}{12}(4^2 - 1) - 4\frac{1}{2}(4 + 1) = 10$ (note that in $\mathrm{D} \leqslant 3$, one has $N_\mathrm{ic} = 0$ for the Weyl tensor).
\enumerationisfinis

A synopsis of the number of independent components in the tensorial objects is in S. Weinberg \cite[pp. 142-146]{Weinberg "Gravitation and Cosmology: Principles and Applications of the General Theory of Relativity"}.
\margosymbol
\end{margo}

\section{Lorentz–Minkowski 4-Manifolds}
\label{section "Lorentz–Minkowski 4-Manifolds"}

We will introduce some notions of Minkowski and Lorentzian geometries in the following Sections. In general theory of relativity, 
\enumerationisinitium
\item space-time is a smooth \emph{Lorentzian 4-manifold manifold}, that is a special case of a pseudo-Riemannian manifold; 
\item if space-time is empty (space without matter-energy), the 4-manifold is equipped with a \emph{Ricci-flat Lorentzian} metric, and it coincides with a 4-dimensional \emph{Minkowski space-time}, which is a special case of a Lorentzian manifold. 
\enumerationisfinis

\subsection{Minkowski Space-Time (Flat Metric)}
\label{subsection "Minkowski Space-Time (Flat Metric)"}

\begingroup
\footnotesize
The concepts of time and space, which I want to develop, have arisen on experimental physical grounds [see Michelson–Morley experiment \cite{Michelson and Morley "On the Relative Motion of the Earth and the Luminiferous Ether"} against the existence of a luminiferous ether]. Herein is their strength. Their tendency is radical. Henceforth, space for itself and time for itself will completely reduce to shadows [\textit{Schatten}], and only a sort of union of the two will maintain an independence [\,\dots]. Three-dimensional geometry becomes a chapter of four-dimensional physics. \\
\indent — \textsc{H. Minkowski} \cite[pp. 75, 79]{Minkowski "Raum und Zeit"}.

\endgroup

\vspace{2mm}

Let us start by listing the fundamental properties of the Minkowski space(-time) formalism. 
\enumerationisinitium
\item \emph{Minkowski space} is a real vector space $\mathfrak{M}^n \viz \mathbb{M}^n = \mathbb{R}^{1, n - 1}$ or $\mathbb{R}^{n - 1, 1}$ of $n \geqslant 2$ dimension, characterized by a \emph{bilinear form} $g \viz g_\textsc{m} \colon \mathfrak{M}^n \times \mathfrak{M}^n \to \mathbb{R}$ on the tangent space at each point of $\mathfrak{M}$, stating that $g$ is symmetric $g(v, w) = g(w, v)$ and non-degenerate $g(v, w) = 0$, for any $v, w \in \mathfrak{M}^n$.\footnote{
	Taking a generic map $g \colon \mathbb{R}^n \times \mathbb{R}^n \to \mathbb{R}$, we determine $g(v, w) = v^0w^0 + v^1w^1 + v^2w^2 + \cdots + v^{n - 1}w^{n - 1} - v^nw^n$ or $g(v, w) = -v^0w^0 + v^1w^1 + v^2w^2 + \cdots +v^nw^n$.
	} 
The bilinear form $g(v, w)$ is called more properly a \emph{Minkowski (or Lorentzian) inner product}, or even \emph{Minkowski (or Lorentzian) metric tensor}.

Letting $\{{e_0, \mathellipsis, e_{n - k}, e_{n - k +1}, \mathellipsis, e_n}\}$ be a basis, $n = \dim(\mathfrak{M})$, with $v = v^0e_0 + \cdots + v^ne_n$ and $w = w^0e_0 + \cdots + w^ne_n$, then
\begin{equation}
	g(v, w) = v^0w^0 + v^1w^1 + v^2w^2 + \cdots + v^{n - k}w^{n - k} - v^{n - k + 1}w^{n - k + 1} - \cdots - v^nw^n, 
\end{equation} 
where $k$ is a non-negative integer. 
\item \emph{Minkowski space-time} \cite{Minkowski "Die Grundgleichungen fur die elektromagnetischen Vorgange in bewegten Korpern"} \cite{Minkowski "Raum und Zeit"} \cite{Minkowski "Das Relativitatsprinzip"} is simply a $4\mathrm{D}$ real vector space $\mathfrak{M}^4 \viz \mathbb{M}^4 = \mathbb{R}^4_{1, 3}$.\footnote{
	Therefore, there is a bilinear form $g$ on the tangent space $\mathcal{T}_p\mathbb{R}^{1, 3}$ at each $p$ of $\mathbb{R}^{1, 3}$.
	} 
Let $\{e_0, e_1, e_2, e_3\}$ be a basis, with $v = v^\mu e_\mu$ and $w = w^\nu e_\nu$, $\mu, \nu = 0, 1, 2, 3$. The signature of the Minkowski metric tensor is 2-fold. We denote by $^{(1, 3)^+}$ and by $^{(1, 3)^-}$ the Minkowski signature $(+, -, -, -)$ and $(-, +, +, +)$, respectively, whilst in the Euclidean signature the signs are all positive. The Minkowski inner product is thus
\begin{align}
	& g(v, w)^{(1, 3)^+} = v^0w^0 - v^1w^1 - v^2w^2 - v^3w^3 = \eta_{\mu\nu}v^\mu w^\nu, \\
	& g(v, w)^{(1, 3)^-} = -v^0w^0 + v^1w^1 + v^2w^2 + v^3w^3 = \eta_{\mu\nu}v^\mu w^\nu,
\end{align}
where $g$ has index 1, and
\begin{equation}
\label{equation "Minkowski metric values"}
	\eta_{\mu\nu} = g(e_\mu, e_\nu) =
	\begin{cases}
	1 \text{ if} 
		\begin{cases}
		\mu = \nu = 0, \text{ with } \eta^{(1, 3)^+}, \\
		\mu = \nu = 1, 2, 3, \text{ with } \eta^{(1, 3)^-},
		\end{cases} \\
	-1 \text{ if} 
		\begin{cases}
		\mu = \nu = 1, 2, 3, \text{ with } \eta^{(1, 3)^+}, \\
		\mu = \nu = 0, \text{ with } \eta^{(1, 3)^-}, \\
		\end{cases} \\
	0 \text{ if}
		\begin{cases}
		\mu \neq \nu, \text{ with } \eta^{(1, 3)^+}, \\
		\mu \neq \nu, \text{ with } \eta^{(1, 3)^-}.
		\end{cases}
	\end{cases}
\end{equation}
\item The Minkowski metric tensor is a pseudo-Riemannian or Lorentzian metric, and it is defined in the algebra of $4 \times 4$ matrices
\begin{equation}	
\label{equation "Matrices for Minkowski metric tensor"}	
	\eta_{\mu\nu}^{(1, 3)^+} =
	\begin{Bmatrix*}[r]
	1 & 0 & 0 & 0 \\
	0 & -1 & 0 & 0 \\
	0 & 0 & -1 & 0 \\
	0 & 0 & 0 & -1
	\end{Bmatrix*} \text{ and }
	\eta_{\mu\nu}^{(1, 3)^-} = 
	\begin{Bmatrix}
	-1 & 0 & 0 & 0 \\
	0 & 1 & 0 & 0 \\
	0 & 0 & 1 & 0 \\
	0 & 0 & 0 & 1
	\end{Bmatrix},
\end{equation}
according to the respective metric signatures.
\item We choose $x = x^0e_0 + x^1e_1 + x^2e_2 + x^3e_3$, with the time $(x^0)$ and the spatial $(x^1, x^2, x^3)$ coordinates. Putting $x^0 = ct$, $x^1 = x$, $x^2 = y$, $x^3 = z$, the Minkowski metric, or, better, the metric tensor of Minkowski space-stime, can be given in standard coordinates $(ct, x, y, z)$ by
\begin{align}
	& 
	\label{align "Line element of Minkowski metric with signature (1, 3)+"}
	ds^2 = c^2dt^2 - dx^2 - dy^2 - dz^2 = \eta_{\mu\nu}^{(1, 3)^+}dx^{\mu}dx^\nu, \\
	& 
	\label{align "Line element of Minkowski metric with signature (1, 3)-"}
	ds^2 = - c^2dt^2 + dx^2 + dy^2 + dz^2 = \eta_{\mu\nu}^{(1, 3)^-}dx^{\mu}dx^\nu,
\end{align}
where $c$ is the speed of light in vacuum, and $t$ is the time, with $-\infty < t, x, y, z < +\infty$, i.e. $(t, x, y, z) \in (-\infty, +\infty)$. The Eqq. \eqref{align "Line element of Minkowski metric with signature (1, 3)+"} \eqref{align "Line element of Minkowski metric with signature (1, 3)-"} are known as \emph{line elements}, which are line segments in the course of an infinitesimal displacement vector at each point in $\mathfrak{M}^4 \viz \mathbb{R}^4_{1, 3}$.

\begin{figure}[h!]
\centering
\includegraphics[width = 0.650\textwidth]{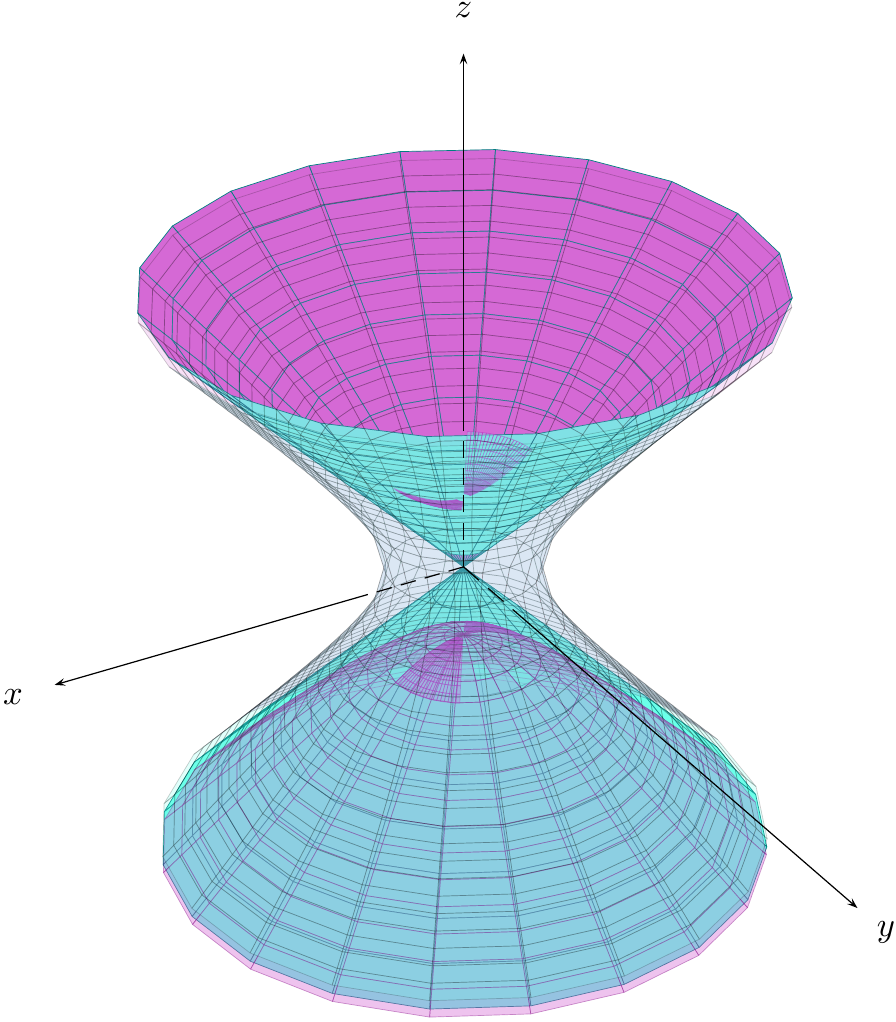}
\caption{In this image, along with a \emph{circular conical surface}, there is a coexistence of quadric surfaces, cf. Section \ref{subsubsection "Upper Half-Space, Ball, and Hyperboloid"}: \\
	(i) hyperbolic hyperboloid (gray), better known as \emph{1-sheeted hyperboloid}, along the $z$-axis, the Cartesian coordinates equation being: $\frac{x^2}{\alpha^2} + \frac{y^2}{\alpha^2} - \frac{z^2}{\gamma^2} = 1$; \\
	(ii) elliptic hyperboloid, better known as \emph{2-sheeted hyperboloid}, along the $z$-axis, with $\mathbb{Y}^n_+$ and $\mathbb{Y}^n_-$ (upper and lower sheets), the Cartesian coordinates equation being: $\frac{x^2}{\alpha^2} + \frac{y^2}{\alpha^2} - \frac{z^2}{\gamma^2} = -1$. \\
	Parametric equations in accordance with their commands: \\
	· circular conical surface: $\mathtt{(u, v){u*cos(v)}{u*sin(v)}{u}}$; \\
	· 1-sheeted hyperboloid (upper part): $\mathtt{(u, v){u*cos(v)}{u*sin(v)}{sqrt(u^2 - 1)}}$; \\
	· 1-sheeted hyperboloid (lower part): $\mathtt{(u, v){u*cos(v)}{u*sin(v)}{-sqrt(u^2 - 1)}}$; \\
	· 2-sheeted hyperboloid (upper sheet, only partial): $\mathtt{(u, v){u*cos(v)}{u*sin(v)}{sqrt(u^2 + 1)}}$; \\
	· 2-sheeted hyperboloid (lower sheet): $\mathtt{(u, v){u*cos(v)}{u*sin(v)}{-sqrt(u^2 + 1)}}$. \\
	Models for space-time in special relativity start from here}
\label{figure "hyperboloid surfaces coexistence"}
\end{figure}

It is also possible to use spherical coordinates $(ct, \rho, \theta, \phi)$, setting  $x^0 = t$, $x^1 = x = \rho\sin\theta\cos\phi$, $x^2 = y = \rho\sin\theta\sin\phi$, $x^3 = z = \rho\cos\theta$, where $\rho$ is the radius corresponding to the line segment moving from a point to the origin, $\theta$ is the colatitude, that is, the polar or zenith angle measured from the $z$-axis, and $\phi$ is the longitude, that is, the azimuthal angle measured from the $x$-axis within the $xy$ plane. Taking $0 \leqslant \rho < \infty$, $0 \leqslant \theta \leqslant \pi$, $0 \leqslant \phi < 2\pi$, i.e. $\rho \in [0, \infty), \theta \in [0, \pi], \phi \in [0, 2\pi)$, the metric Eqq. \eqref{align "Line element of Minkowski metric with signature (1, 3)+"} \eqref{align "Line element of Minkowski metric with signature (1, 3)-"} become
\begin{align}
	& ds^2_{(1, 3)^+} = c^2dt^2 - d\rho^2 - \rho^2(d\theta^2 - \sin^2{\theta}d\phi^2), \\
	& ds^2_{(1, 3)^-} = -c^2dt^2 + d\rho^2 + \rho^2(d\theta^2 + \sin^2{\theta}d\phi^2).
\end{align}
\item A vector in a Lorentz–Minkowski space-time is said \emph{4-vector}, also known as \emph{event}. A 4-vector 
\begin{equation}
	v \in \mathfrak{M}^4 \viz \mathbb{R}^4_{1, 3} \text{ is }
	\begin{cases}
    \text{space-like if}
    	\begin{cases}
		g(v, v)^{(1, 3)^+} < 0, \\
		g(v, v)^{(1, 3)^-} > 0,
    	\end{cases} \\ 
	\text{time-like if}
		\begin{cases}
    	g(v, v)^{(1, 3)^+} > 0, \\
		g(v, v)^{(1, 3)^-} < 0,	
		\end{cases} \\ 
	\text{light-like (null) if}
		\begin{cases}
		g(v, v)^{(1, 3)^+} = 0, \\
    	g(v, v)^{(1, 3)^-} = 0.	
		\end{cases}
	\end{cases}
\end{equation}
\item Minkowski space-time is a \emph{pseudo-Euclidean} vector 4-space: the topology of Minkowski space-time, unlike Euclidean 4-space, is not locally homogeneous; in $\mathfrak{M}^4 \viz \mathbb{R}^4_{1, 3}$ space vectors are separated from time vectors, see E.C. Zeeman \cite{Zeeman "The Topology of Minkowski Space"}. 
\item 
\label{item "Minkowski space-time and its characteristic"}
Minkowski space-time, in a sufficiently small neighborhood of a point, is a precise approximation of the 4-dimensional manifold in Einstein's theory of special relativity \cite{Einstein "Zur Elektrodynamik bewegter Korper"}, and it it represents a flat 4-space, or a 4-space without matter-energy, for which $\Ricci_{\mu\nu} = 0$, since if $\Tau_{\mu\nu} = 0$, then 
\begin{equation}
	g_{\mu\nu}\left(\Ricci_{\mu\nu} - g^{\mu\nu}\frac{\scalarcurvature}{2}\right) = 0. 
\end{equation}
In the presence of gravity, this space becomes \emph{non-Euclidean}, or, to say it better, \emph{non-pseudo-Euclidean} (cf. Section \ref{subsection "Concepts (?) of Space-Time"}).
\item Plainly, we are free to write the Minkowski space-time as $\mathfrak{M}^4 \viz \mathbb{M}^4 = \mathbb{R}^4_{3, 1}$, provided that the signs are changed, including \eqref{equation "Matrices for Minkowski metric tensor"}. We can also write like this: $x = x^1e_1 + x^2e_2 + x^3e_3 + x^4e_4$, where $(x^1, x^2, x^3)$ are the spatial coordinates and $(x^4)$ is the time coordinate, and e.g. $g(v, w)^{(3, 1)^+} = v^1w^1 + v^2w^2 + v^3w^3 - v^4w^4$, $\eta^{(3, 1)^+} = \diag(1, 1, 1, -1)$. It is a numbering more intuitive (cf. Section \ref{subsubsection "Step I"}) but, in this case, mathematically less efficient.
\item The Minkowski metric \eqref{equation "Metric on the Minkowski–Lorentz model"} contributes to form the hyperboloid (aka Minkowski–Lorentz) model \eqref{equation "Minkowski–Lorentz model"} $\mathbb{Y}^n_+$ in $\mathbb{R}^{n + 1}$ of hyperbolic space. Accordingly, the Minkowski space, although not Riemannian, may contain subspaces with a Riemannian metric directly related to hyperbolic geometry, see Fig. \ref{figure "hyperboloid surfaces coexistence"}.
\enumerationisfinis

\subsection{Lorentzian Generalization}
\label{subsection "Lorentzian Generalization"}

\begingroup
\footnotesize
Lorentz's idea \cite{Lorentz "Electromagnetic phenomena in a system moving with any velocity less than that of light"} can be summed up as follows: if we are able to bring a translation upon a whole system, without modification of any observable phenomena, it is because the equations of an electromagnetic medium are not altered by certain transformations, which we will call \emph{Lorentz transformations}; two systems, one of which is motionless, the other in translation, thus become exact images of each other. \\
\indent — \textsc{H. Poincaré} \cite[p. 130]{Poincare "Sur la dynamique de l'electron"}

\endgroup

\subsubsection[Vector Spaces $\mathbb{L}^4$]{Vector Spaces $\protect\pseudobold{\mathbb{L}^4}$}
\label{subsubsection "Vector Spaces L^4"}

Lorentzian geometry is a general case of Minkowskian geometry. It is sufficient to mention a few essential points.
\enumerationisinitium
\item \emph{Lorentzian space} is an $n$-dimensional real vector space $\mathfrak{L}^n \viz \mathbb{L}^n = \mathbb{R}^{1, n - 1}$ or $\mathbb{R}^{n - 1, 1}$, characterized by a \emph{Lorentzian inner product} $g \viz g_\textsc{l}$, that is, 
\begin{align}
	& 
	\label{align "Lorentzian inner product (+)"}
	g(v, w)^+ = v^0w^0 + v^1w^1 + v^2w^2 + \cdots + v^{n - 1}w^{n - 1} - v^nw^n, \\
	\label{align "Lorentzian inner product (-)"}
	& g(v, w)^- = -v^0w^0 + v^1w^1 + v^2w^2 + \cdots + v^nw^n,
\end{align}
on the tangent space at each point of $\mathfrak{L}$, for any $v, w \in \mathfrak{L}^n$, with the signatures $(+, -, -, \mathellipsis, -)$ and $(-, +, +, \mathellipsis, +)$.
\item The \emph{Lorentzian metric tensor} $g = g_{\mu\nu}dx^{\mu}dx^\nu$ is a pseudo-Riemannian metric, and the signature of the quadratic form is the same one that is present in \eqref{align "Lorentzian inner product (+)"} or in \eqref{align "Lorentzian inner product (-)"}.
\item Let $\mathcal{M}^n$ be an $n$-dimensional smooth differentiable \emph{pseudo-Riemannian manifold}, hence a connected separable metrizable $\mathscr{C}^\infty$ space; denote by $i_g$ the index of $g$ on $\mathcal{M}$, with the common value $i_g[0 \leqslant i_g \leqslant n = \dim(\mathcal{M})]$. If $i_g = 1$ and $n \geqslant 2$, then the pair $(\mathcal{M}^n, g)$ is a \emph{Lorentzian manifold},\footnote{
	If $i_g = 0$, $(\mathcal{M}, g)$ is a Riemannian manifold, and $g$ is symmetric, non-degenerate, and positive definite; if $i_g \neq 0$, the pair is pseudo-Riemannian, cf. footnote \ref{footnote "Riemannian manifold and pseudo-Riemannian manifolds"}, p. \pageref{footnote "Riemannian manifold and pseudo-Riemannian manifolds"}.
	} 
and $g \in \sezione(\mathring{\mathcal{T}}^0_2\mathcal{M})$ is the Lorentzian metric, which can be read as a section of the bundle of tangent spaces (i.e. a section of the tangent bundle) of $\mathcal{M}$ (see Definition \ref{definitio "Tensor bundle"}).
\item A \emph{Lorentzian space-time} $\mathfrak{L}^4 \viz \mathbb{L}^4 = \mathbb{R}^4_{1, 3} = \mathcal{M}^4$ is a connected $\mathscr{C}^\infty$ (smooth) manifold of dimension 4, as well as $\mathscr{T}_2$ Hausdorff 4-space (see Margo \ref{margo "Hausdorff space"}), with a Lorentzian metric $\binom{0}{2}$-tensor $g^{(1, 3)}$. Note. Alternately, the form is $\mathbb{R}^4_{3, 1}$ and $g^{(3, 1)}$. 
\item $\mathbb{L}^4 = \mathbb{R}^4_{1, 3}$ coincides with the Minkowski's description $\mathfrak{M}^4 \viz \mathbb{M}^4$ when the Riemann curvature tensor (Section \ref{subsection "Riemann Curvature Tensor"}) of the Levi-Civita connection is zero, and there is talk of \emph{flat Lorentzian space-time} or \emph{Lorentz–Minkowski space-time}; so the Minkowskian formalism in Section \ref{subsection "Minkowski Space-Time (Flat Metric)"} can be replicated here. Given a basis $\{e_0, e_1, e_2, e_3\}$, one has the same values of \eqref{equation "Minkowski metric values"}, $g(e_\mu, e_\nu) = \eta_{\mu\nu} = \diag(1, -1, -1, -1)$ and $g(e_\mu, e_\nu) = \eta_{\mu\nu} = \diag(-1, 1, 1, 1)$. For a system of coordinates $x^1, \mathellipsis, x^n$, we get
\begin{equation}
		g_{\mu\nu}(x) = \left(\frac{\partial}{\partial{x}^\mu}\Big|_x,\frac{\partial}{\partial{x}^\nu}\Big|_x\right) = \eta_{\mu\nu}.
\end{equation}	
\item 
\label{item "Minkowski space in general relativity"} 
$\mathbb{L}^4 = \mathbb{R}^4_{1, 3}$, intended as $\mathfrak{M}^4 \viz \mathbb{M}^4$, is incorporated also in general relativity. This is because the infinitesimal neighborhood around any point in a curved 4-space is \emph{still} a Minkowski-like flat 4-manifold, although there is a \emph{variation} of $g_\textsc{l}$ from point to point. See, about that, the local Fermi's coordinates in Section \ref{subsection "Fermi (Locally Geodesic Cartesian-like) Coordinates"}, and \cite{Manasse and Misner "Fermi Normal Coordinates and Some Basic Concepts in Differential Geometry"}.
\item A $\mathbb{L}^4$-type space provides a model for geometry of Einsteinian gravity (in which, to be more specific, the gravitational force is a manifestation of the curvature of space-time by the action of matter-energy) when it represents a hyperbolic 4-manifold, and there is talk of \emph{non-flat (curved) Lorentzian space-time}. Here Lorentzian space-time and Einstein's gravitationally curved space-time are the same mathematical object. 
\item In summary,
\[
	\mathbb{L}^4 = \mathbb{R}^4_{1, 3}
	\begin{cases}
	\text{flat pseudo-Euclidean Lorentzian space-time: } \mathfrak{M}^4 \viz \mathbb{M}^4, \\
	\text{non-flat (curved) non-pseudo-Euclidean Lorentzian space-time},
	\end{cases}
\]
\enumerationisfinis

\subsubsection{Lorentz Group plus Transformations}
\label{subsubsection "Lorentz Group plus Transformations"}

\enumerationisinitium
\item For an empty space-time, \emph{Lorentz transformations} \cite{Lorentz "Electromagnetic phenomena in a system moving with any velocity less than that of light"} leave the Maxwell's equations\footnote{
	Or, more correctly, Maxwell–Heaviside equations, since in the Maxwell's treatise there is no sign of the Maxwell's equations (at least as we know them today), but they are a later elaboration, mainly due to O. Heaviside \cite[XLIV (1889), LII (1891)]{Heaviside "Electromagnetic Theory I"} \cite{Heaviside "Electrical Papers II"}.
	} \cite{Maxwell "A Treatise on Electricity and Magnetism I"} \cite{Maxwell "A Treatise on Electricity and Magnetism II"} \emph{unchanged} between different inertial systems, i.e. if spatial and time coordinates are subjected to the Lorentz group, according to the intuition of Poincaré \cite[§§ 1, 4]{Poincare "Sur la dynamique de l'electron"}.
\item We say that $\{e_0, e_1, e_2, e_3\}$ and $\{e'_0, e'_1, e'_2, e'_3\}$ are two oriented and orthonormal time-oriented bases. If 
\begin{equation}	
	\Lambda = [{\Lambda^\mu}_\nu]_{\mu\nu = 0, 1, 2, 3} = 
	\begin{Bmatrix}
	{\Lambda^0}_0 & {\Lambda^0}_1 & {\Lambda^0}_2 & {\Lambda^0}_3 \\
	{\Lambda^1}_0 & {\Lambda^1}_1 & {\Lambda^1}_2 & {\Lambda^1}_3 \\
	{\Lambda^2}_0 & {\Lambda^2}_1 & {\Lambda^2}_2 & {\Lambda^2}_3 \\
	{\Lambda^3}_0 & {\Lambda^3}_1 & {\Lambda^3}_2 & {\Lambda^3}_3 \\  	
	\end{Bmatrix}
\end{equation}
designates a matrix of an orthogonal transformation, for which $e_\nu = {\Lambda^0}_{\nu}e'_0 + {\Lambda^1}_{\nu}e'_1 + {\Lambda^2}_{\nu}e'_2 + {\Lambda^3}_{\nu}e'_3 = {\Lambda^\mu}_{\nu}e'_\mu$, then $\Lambda = [{\Lambda^\mu}_\nu]$ imposes three conditions:
\subenumerationisinitium
\item \emph{orthogonality}, under a \emph{general} Lorentz transformation: 
\begin{equation}
\label{equation "General Lorentz transformation"}
	\Lambda^\textsc{t}\eta\Lambda = \eta,
\end{equation}
where $\Lambda^\textsc{t}$ is the transpose of $\Lambda$, and $\eta = \eta_{\mu\nu}^{(1, 3)^+}$ or $\eta_{\mu\nu}^{(1, 3)^-}$ in \eqref{equation "Matrices for Minkowski metric tensor"},
\item \emph{orientability}, under a \emph{proper} Lorentz transformation: $\det(\Lambda) = 1$,
\item \emph{time orientability}, under an \emph{orthochronous} Lorentz transformation: ${\Lambda^0}_0 \geqslant 1$; the adjective “orthochronous” (\textgreek{ὀρθός} [upright] + \textgreek{χρόνος}) means that the direction of time is preserved.
\subenumerationisfinis
\item $\Lambda = [{\Lambda^\mu}_\nu]$ is an element of the \emph{Lorentz group}, denoted by $\Lorentz$, that is the group of all Lorentz transformations of $\mathfrak{M}^4 \viz \mathbb{R}^4_{1, 3}$ (preserving the Minkowski metric). Here is a comprehensive outline of $\Lorentz$ with its four components,
\begin{equation}
	\Lorentz =
	\begin{cases}
	\begin{rcases}
	\Lorentz_+^\uparrow = \Lorentz_+ \cap \Lorentz^\uparrow \mid \det(\Lambda) = 1 \\
	\Lorentz_-^\uparrow = \Lorentz_- \cap \Lorentz^\uparrow \mid \det(\Lambda) = -1
	\end{rcases}
	{\Lambda^0}_0 \geqslant 1 \\
	\begin{rcases}
	\Lorentz_+^\downarrow = \Lorentz_+ \cap \Lorentz^\downarrow \mid \det(\Lambda) = 1 \\
	\Lorentz_-^\downarrow = \Lorentz_- \cap \Lorentz^\downarrow \mid \det(\Lambda) = -1
	\end{rcases}
	{\Lambda^0}_0 \leqslant -1,
	\end{cases}
\end{equation}
where $\uparrow$ and $\downarrow$ are the inequalities. The Lorentz group 
\begin{equation}
\label{equation "Lorentz group"}
	\Lorentz = \Lorentz_+^\uparrow \cup \Lorentz_-^\uparrow \cup \Lorentz_+^\downarrow \cup \Lorentz_-^\downarrow = O_{1, 3}(\mathbb{R})
\end{equation}	
corresponds to the indefinite orthogonal group of signature $(1, 3)$ of linear transformations of Minkowski space-time, and it can be written as a union of disjunct sets. Below are some details.
\begin{align}
	&
	\label{align "Lorentz_+"} 
	\Lorentz_+ = \{\Lambda \mid \det(\Lambda) = 1\} = SO_{1, 3}(\mathbb{R}), \\
	& 
	\label{align "Lorentz_-"}
	\Lorentz_- = \{\Lambda \mid \det(\Lambda) = -1\}, \\
	& 
	\label{align "Lorentz^uparrow"}
	\Lorentz^\uparrow = \{\Lambda \mid \textgreek{\textit{γ}} \geqslant 1\}, \\
	& 
	\label{align "Lorentz^downarrow"}
	\Lorentz^\downarrow = \{\Lambda \mid \textgreek{\textit{γ}} \leqslant -1\}, \\
	& 
	\label{align "Lorentz_+^uparrow"}
	\Lorentz_+^\uparrow = \{\Lambda \in \Lorentz_+ \mid {\Lambda^0}_0 \geqslant 1\} = SO_{1, 3}^+(\mathbb{R}), \\
	& 
	\label{align "Lorentz_-^uparrow"}
	\Lorentz_-^\uparrow = \{\Lambda \in \Lorentz_- \mid {\Lambda^0}_0 \geqslant 1\}, \\
	& 
	\label{align "Lorentz_+^downarrow"}
	\Lorentz_+^\downarrow = \{\Lambda \in \Lorentz_+ \mid {\Lambda^0}_0 \leqslant -1\}, \\
	& 
	\label{align "Lorentz_-^downarrow"}
	\Lorentz_-^\downarrow = \{\Lambda \in \Lorentz_- \mid {\Lambda^0}_0 \leqslant -1\}.
\end{align}

\eqref{align "Lorentz_+"} is the \emph{proper Lorentz group}, corresponding to the indefinite special orthogonal group of signature $(1, 3)$; 

\eqref{align "Lorentz_-"} is the \emph{improper Lorentz group};

\eqref{align "Lorentz^uparrow"} is the \emph{orthochronous Lorentz group}, where 
\begin{equation}
	\textgreek{\textit{γ}} = \frac{1}{\sqrt{1 - \frac{v^2}{c^2}}} = \frac{dt}{d\tau} 
\end{equation}
is the \emph{Lorentz factor};

\eqref{align "Lorentz^downarrow"} is the \emph{non-orthochronous} (or \emph{heterochronous}) \emph{Lorentz group};

\eqref{align "Lorentz_+^uparrow"} is the \emph{proper orthochronous Lorentz group}, corresponding to the indefinite special orthogonal group of signature $^+(1, 3)$, better known as \emph{restricted Lorentz group};

\eqref{align "Lorentz_-^uparrow"} is the \emph{improper orthochronous Lorentz group};

\eqref{align "Lorentz_+^downarrow"} is the \emph{proper non-orthochronous} (or \emph{heterochronous}) \emph{Lorentz group}; 

\eqref{align "Lorentz_-^downarrow"} is the \emph{improper non-orthochronous} (or \emph{heterochronous}) \emph{Lorentz group}.
\enumerationisfinis

\begin{scholium}
~\enumerationisinitium
\item Each element of $\Lorentz_+$ will be said \emph{proper Lorentz transformation}, each element of $\Lorentz_-$ will be said \emph{improper Lorentz transformation}, and so forth. 
\item Take two different frames of reference, $x^0, x^1, x^2, x^3$ and $y^0, y^1, y^2, y^3$, such that $y^\mu = {\Lambda^\mu}_{\nu}x^\nu$, $\mu = 0, 1, 2, 3$. When the group/transformation is proper, the orientation of the spatial coordinates $x^1, x^2, x^3$ is preserved, when the group/transformation is orthochronous, the direction  of time is preserved, since $x^0 \geqslant 0$, then $y^0 = {\Lambda^0}_\mu x^\mu \geqslant 0$, for time-like or light-like vectors. \scholiumsymbol
\enumerationisfinis
\end{scholium}

\section{Spinor Representation of the Lorentz Group}

Some historical background. 
\enumerationisinitium
\item The first analyses concerning the representation of the Lorentz group, and thus a classification of the possible representations in Lorentz frames, are due to
\subenumerationisinitium
\item Bargmann–Wigner programme \cite{Wigner "On Unitary Representations of the Inhomogeneous Lorentz Group"} \cite{Bargmann "Irreducible Unitary Representations of the Lorentz Group"} \cite{Bargmann and Wigner "Group Theoretical Discussion of Relativistic Wave Equations"}, which tackles these issues: (a) irreducible unitary representations of the Lorentz group, (b) proper Lorentz group, with homogeneous linear transformations in 4 variables, $x^0, x^1, x^2, x^3$, (c) inhomogeneous Lorentz group, (d) transformations of wave functions under the operations of the Lorentz group (invariance of functions), (e) Lorentz groups for a free particle in terms of a wave packet and relativistic wave equations;
\item I.M. Gel'fand and M.A. Naimark \cite{Gel'fand Naimark "Unitary representations of the Lorentz group"}, see also \cite{Gel'fand Minlos and Shapiro "Representations of the rotation and Lorentz groups and their applications"}: study of unitary representations of the Lorentz group;
\item H.-Chandra \cite{H.-Chandra "Infinite irreducible representations of the Lorentz group"}: infinite-dimensional irreducible representations of the Lorentz group bearing the spin properties of a particle in Dirac's exposition \cite{Dirac "Relativistic Wave Equations"} \cite{Dirac "Unitary representations of the Lorentz group"} of quantum mechanics.
\subenumerationisfinis
\item Results afferent to harmonic analysis, such as the application of the Plancherel theorem \cite{Plancherel "Contribution a l'etude de la representation d'une fonction arbitraire par les integrales definies"} to the complex special linear group (see below), are in \cite{Gel'fand Naimark "Unitary representations of the Lorentz group"}, and, more fully, in H.-Chandra \cite{H.-Chandra "Plancherel Formula for Complex Semisimple Lie Groups"} and Gel'fand \& Graev \cite{Gel'fand Graev "On a general method of decomposition of the regular representation of a Lie group into irreducible representations"}. 
\item Weyl's contribution \cite[III.8, IV.5-8]{Weyl "The Theory of Groups and Quantum Mechanics"} is also worthy of mention. 
\item A first systematic discussion of the spinor fields on a space-time in general relativity, is in R. Geroch \cite{Geroch "Spinor Structure of Space-Times in General Relativity. I"} \cite{Geroch "Spinor Structure of Space-Times in General Relativity. II"}.
\enumerationisfinis

\begin{margo}[Majorana's brainwave]
\label{margo "Majorana's brainwave"}
It should be added that the \emph{infinite-dimensional} unitary representations of the Lorentz group have their origin in the Majorana's paper \cite{Majorana "Teoria relativistica di particelle con momento intrinseco arbitrario"},\footnote{
	Cf. endnote \ref{endnote "Majorana's further studies on the spinor representation of the Lorentz group"}.
	} 
which anticipates Wigner \cite{Wigner "On Unitary Representations of the Inhomogeneous Lorentz Group"}\footnote{
	Wigner \cite[p. 87]{Pauli "Wissenschaftlicher Briefwechsel III: 1940-1949"}, unlike Pauli \cite[letters to M. Fierz № 598-600, 622, 628]{Pauli "Wissenschaftlicher Briefwechsel III: 1940-1949"}, does not fully understand Majorana's innovation.
	} 
and Dirac \cite{Dirac "Unitary representations of the Lorentz group"}, and sows seeds for successive ideas, see e.g. S.-J. Chang and L. O'Raifeartaigh \cite{Chang and O'Raifeartaigh "Spacelike Solutions of Infinite-Component Wave Functions"}, E.C.G. Sudarshan and N. Mukunda \cite{Sudarshan and N. Mukunda "Quantum Theory of the Infinite-Component Majorana Field and the Relation of Spin and Statistics"}, and A.O. Barut and I.H. Duru \cite{Barut and Duru "Introduction of internal coordinates into the infinite-component Majorana equation"}. \margosymbol
\end{margo}

\subsection[Spinor Map (6-Dimensional Homomorphism): the Covering $SL_2(\mathbb{C}) \to SO_{1, 3}^+(\mathbb{R})$]{Spinor Map (6-Dimensional Homomorphism): the Covering $\protect\pseudobold{SL_2(\mathbb{C})} \protect\pseudobold{\to} \protect\pseudobold{SO_{1, 3}^+(\mathbb{R})}$}
\label{subsection "Spinor Map (6-Dimensional Homomorphism): the Covering $SL_2(C)$ to $SO_{1, 3}^+(R)$"}

Two words about an important link between the restricted Lorentz group $SO_{1, 3}^+(\mathbb{R})$ and the special linear group $SL_2(\mathbb{C})$, which will be followed (Section \ref{subsection "Gamma Matrices and Type of Fermionic Spinor Fields"}) by some mathematical objects related to it.
\enumerationisinitium
\item The group $SL_2(\mathbb{C})$ is the set of all spin transformations in $\mathbb{C}^{2 \times 2}$, namely the set of $\mathbb{C}^{2 \times 2}$ matrices with determinant 1, represented by 
\begin{equation}	
\label{equation "Linear transformation on $C^2$ (complex matrices)"}	
	\begin{pmatrix}
	z^1 \\
	z^2
	\end{pmatrix} 
	\mapsto g =
	\begin{pmatrix}
	\alpha & \beta \\ 
	\gamma & \delta
	\end{pmatrix}
	\in SL_2(\mathbb{C})
	\begin{pmatrix}
	z^1 \\
	z^2	
	\end{pmatrix}
	=
	\begin{pmatrix}
	\alpha{z}^1 + \beta{z}^2 \\ 
	\gamma{z}^1 + \delta{z}^2
	\end{pmatrix}. 
\end{equation} 
Taking advantage of the \emph{irreducible representation} for spinors (see Section \ref{subsubsection "Example. Irreducible Covering Spin-Space for 4pi"}), we can write the linear transformation on $\mathbb{C}^2$ of \eqref{equation "Linear transformation on $C^2$ (complex matrices)"} as \emph{left- and right-handed spinor irreducible representations} of $SL_2(\mathbb{C})$, i.e. 
\begin{align}	
	& \mathscr{D}^{\left(\frac{1}{2}, 0\right)} \colon SL_2(\mathbb{C}) \to GL(\mathbb{C}^2), \enspace \mathscr{D}^{\left(\frac{1}{2}, 0\right)}(g) = g, \\
	& \mathscr{D}^{\left(0, \frac{1}{2}\right)} \colon SL_2(\mathbb{C}) \to GL(\mathbb{C}^2), \enspace \mathscr{D}^{\left(0, \frac{1}{2}, \right)}(g) = \rotatedg,
\end{align}
respectively, where 
\begin{equation}	
	\rotatedg = \left(\frac{1}{g^\dag \viz \bar{g}^\textsc{t}}\right),
\end{equation}	 
$\bar{g}^\textsc{t}$ is the conjugate transpose of $g$ (more accurately, $\bar{g}$ is the complex conjugated entries, and $g^\textsc{t}$ is the transpose).

With a direct sum, we get a representation on a complex 4-space:
\begin{align}
\label{align "Irreducible representation on a complex 4-space with a direct sum"}
	& \mathscr{D}^{\left(\frac{1}{2}, 0\right)} \oplus \mathscr{D}^{\left(0, \frac{1}{2}\right)} \colon SL_2(\mathbb{C}) \to GL(\mathbb{C}^4), \\
	& \mathscr{D}^{\left(\frac{1}{2}, 0\right)} \oplus \mathscr{D}^{\left(0, \frac{1}{2}\right)}(g) = 
	\begin{pmatrix}
	g & 0 \\
	0 & \rotatedg
	\end{pmatrix};
\end{align} 
and the tensor product has this form:
\begin{align}
	& \left(\mathscr{D}^{\left(\frac{1}{2}, 0\right)} \otimes \mathscr{D}^{\left(0, \frac{1}{2}\right)} = \mathscr{D}^{\left(\frac{1}{2}, \frac{1}{2}\right)}\right) \colon SL_2(\mathbb{C}) \to GL(\mathbb{C}^4), \\
	& \mathscr{D}^{\left(\frac{1}{2}, \frac{1}{2}\right)}(g) = 
	\begin{pmatrix}
	\alpha\rotatedg & \beta\rotatedg \\
	\gamma\rotatedg & \delta\rotatedg
	\end{pmatrix} 
	=
	\begin{Bmatrix}
	\alpha\bar{\delta} & -\alpha\bar{\gamma} & \beta\bar{\delta} & -\beta\bar{\gamma} \\	
	-\alpha\bar{\beta} & \alpha\bar{\alpha} & -\beta\bar{\beta} & \beta\bar{\alpha} \\
	\gamma\bar{\delta} & -\gamma\bar{\gamma} & \delta\bar{\delta} & -\delta\bar{\gamma} \\
	-\gamma\bar{\beta} & \gamma\bar{\alpha} & -\delta\bar{\beta} & \delta\bar{\alpha}
	\end{Bmatrix}.
\end{align} 
\item $SO_{1, 3}^+(\mathbb{R})$ and $SL_2(\mathbb{C})$ have the same dimension on the real field, that is 6. Let us see why. 
\subenumerationisinitium
\item 
\label{item "Indefinite orthogonal group, i.e. Lorentz group, and indefinite special orthogonal group, i.e. restricted Lorentz group: real 6-dimensional non-compact Lie spaces"} 
The groups $O_{1, 3}(\mathbb{R})$ and $SO_{1, 3}^+(\mathbb{R})$ are \emph{real 6-dimensional non-compact Lie spaces}. Eq. \eqref{equation "General Lorentz transformation"} has 10 independent components relating to a symmetric $4 \times 4$ matrix, so $16 - 10 = 6$.
\item The special linear group $SL_2(\mathbb{C})$ is a \emph{complex Lie group of dimension 3} and a \emph{real Lie group of dimension 6}. It contains 4 complex numbers, or 8 real numbers, inasmuch as $\mathbb{C}^4$ is equal to $\mathbb{R}^8$, but the unit determinant takes away 2 of its 8 degrees of freedom, so $8 - 2 = 6$. From a geometric point of view, $SL_2(\mathbb{C})$ is diffeomorphic to the 3-sphere\footnote{
	Under an inclusion map, one has $\iota \colon \mathbb{S}^3 \hookrightarrow \mathbb{C}^2 = \mathbb{R}^4$.
	}
multiplied by a real 3-space, $\mathbb{S}^3 \times \mathbb{R}^3$, and here too its real \emph{6-dimensionality} is evident. 
\enumerationisfinis
\item $SO_{1, 3}^+(\mathbb{R})$ and $SL_2(\mathbb{C})$ are topologically different, although their Lie algebras are isomorphic, 
\begin{equation}
	\mathfrak{so}^+(1, 3) \cong \Bigl\{\Bigl(\mathfrak{sl}_2(\mathbb{C}) = \mathfrak{su}(2) \oplus \mathfrak{su}(2)\Bigr)\cong \mathfrak{sp}_2(\mathbb{C})\Bigr\}.
\end{equation}

The  above-mentioned link consists in a continuous \emph{homomorphism} from  $SL_2(\mathbb{C})$ into $SO_{1, 3}^+(\mathbb{R})$, that becomes explicit by a \emph{spinor map} 
\begin{equation}	
\label{equation "Spinor map with restricted Lorentz group"}	
	\varsigma \colon \Bigl(SL_2(\mathbb{C}) \cong \Spin_{1, 3}^+(\mathbb{R})\Bigr) \longrightarrow \Bigl(SO_{1, 3}^+(\mathbb{R}) = \Lorentz_+^\uparrow\Bigr),
\end{equation}
where $SL_2(\mathbb{C}) \cong \Spin_{1, 3}^+(\mathbb{R})$ acts as a \emph{2-fold cover} of $SO_{1, 3}^+(\mathbb{R}) = \Lorentz_+^\uparrow$, for which the former doubly covers the latter. The special linear group of $2 \times 2$ complex matrices, which is simply connected, is the \emph{universal covering group} of the restricted Lorentz group. 

By adding a representation of the restricted Lorentz group on a real vector space $\mathfrak{M}$ as a homomorphism $f$ of $SO_{1, 3}^+(\mathbb{R})$ into a general linear group $GL(\mathfrak{M})$, it is possible to draw the following summary diagram,
\[
\begin{tikzcd}[row sep=large, column sep=large]
	SL_2(\mathbb{C}) \cong \Spin_{1, 3}^+(\mathbb{R}) \arrow{r}{\varsigma} \arrow[swap]{dr}{f \circ \varsigma} & SO_{1, 3}^+(\mathbb{R}) = \Lorentz_+^\uparrow \arrow{d}{f} \\
	& GL(\mathfrak{M})
\end{tikzcd}
\]
\item We define the Lorentz bundle and the spinor bundle. 
\subenumerationisinitium
\item The \emph{Lorentz bundle} is the principal $\Lorentz_+^\uparrow$-bundle over $\mathbb{R}^4_{1, 3}$ (space-time), with $\mathring{\mathcal{P}}_\Lorentz$,
\begin{equation}
	SO_{1, 3}^+(\mathbb{R}) = \Lorentz_+^\uparrow \hookrightarrow \Lorentz(\mathbb{R}^4_{1, 3}) \xrightarrow{\mathring{\mathcal{P}}_\Lorentz} \mathbb{R}^4_{1, 3}.
\end{equation}
\item The \emph{spinor bundle} is the principal $SL_2(\mathbb{C})$-bundle over $\mathbb{R}^4_{1, 3}$ (space-time), with $\mathring{\mathcal{P}}_\textit{ß}$, 
\begin{equation}
\label{equation "Spinor bundle"}
	SL_2(\mathbb{C}) \cong \Spin_{1, 3}^+(\mathbb{R}) \hookrightarrow \textit{ß}(\mathbb{R}^4_{1, 3}) \xrightarrow{\mathring{\mathcal{P}}_\textit{ß}} \mathbb{R}^4_{1, 3},
\end{equation}  
to which we can associate the irreducible representation \eqref{align "Irreducible representation on a complex 4-space with a direct sum"}. The \emph{spinor configuration} is the spinor bundle \eqref{equation "Spinor bundle"} plus a map $\varphi \colon \textit{ß}(\mathbb{R}^4_{1, 3}) \to \Lorentz(\mathbb{R}^4_{1, 3})$, satisfying three conditions, for any $p \in \textit{ß}(\mathbb{R}^4_{1, 3})$ and any $g \in SL_2(\mathbb{C})$: 

(a) $\mathring{\mathcal{P}}_\Lorentz\bigl(\varphi(p)\bigr) = \mathring{\mathcal{P}}_\textit{ß}(p)$, 

(b) $\mathring{\mathcal{P}}_\Lorentz \circ \varphi = \mathring{\mathcal{P}}_\textit{ß}(p)$, 

(c) $\varphi(p \cdot g) = \varphi(p) \cdot \Spin(g)$. 
\item We can finally summarize with a diagram:
\[
\begin{tikzcd}[row sep=large, column sep=large]
	SL_2(\mathbb{C}) \cong \Spin_{1, 3}^+(\mathbb{R}) \arrow[hookrightarrow]{r} & \textit{ß}(\mathbb{R}^4_{1, 3}) \arrow[bend left=50]{r}{\mathring{\mathcal{P}}_\textit{ß}} \arrow{d}{\varphi} & \mathbb{R}^4_{1, 3} \viz \mathfrak{M}^4 \\
	SO_{1, 3}^+(\mathbb{R}) = \Lorentz_+^\uparrow \arrow[hookrightarrow]{r} & \Lorentz(\mathbb{R}^4_{1, 3}) \arrow[swap]{ru}{\mathring{\mathcal{P}}_\Lorentz}
\end{tikzcd}
\]
from which $\varphi \times \Spin \colon \textit{ß}(\mathbb{R}^4_{1, 3}) \times SL_2(\mathbb{C}) \to \Lorentz(\mathbb{R}^4_{1, 3}) \times SO_{1, 3}^+(\mathbb{R}) = \Lorentz_+^\uparrow$. Note. The Lorentz and spinor bundles can also be obtained via product bundle, 
\begin{align}
	& SO_{1, 3}^+(\mathbb{R}) = \Lorentz_+^\uparrow \hookrightarrow \mathbb{R}^4_{1, 3} \times \left(SO_{1, 3}^+(\mathbb{R}) = \Lorentz_+^\uparrow\right) \to \mathbb{R}^4_{1, 3}, \\
	& 	SL_2(\mathbb{C}) \cong \Spin_{1, 3}^+(\mathbb{R}) \hookrightarrow \mathbb{R}^4_{1, 3} \times SL_2(\mathbb{C}) \to \mathbb{R}^4_{1, 3},
\end{align}
respectively.
\subenumerationisfinis
\item From Eq. \eqref{equation "Möbius group"}, we note that
\begin{equation}
	\Moebius(\hat{\mathbb{C}}) \cong PSL_2(\mathbb{C}) \cong \frac{SL_2(\mathbb{C})}{\{\pm\idem\}} \cong SO_{1, 3}^+(\mathbb{R}) = \Lorentz_+^\uparrow,
\end{equation}
in which the restricted group is associated as a further set in the isomorphism chain.
\item We also have
\begin{equation}
	SL_2(\mathbb{C})/\mathbb{Z}_2 \cong SO_{1, 3}^+(\mathbb{R}).
\end{equation}
We signal the implication of the \emph{Klein 4-group} \cite{Klein "Vorlesungen uber das Ikosaeder und die Auflosung der Gleichungen vom funften Grade"} = \cite{Klein "Lectures on the Ikosahedron and the Solution of Equations of the Fifth Degree"}, 
\begin{equation}
	\frac{O_{1, 3}(\mathbb{R})}{SO_{1, 3}^+(\mathbb{R})} \cong \mathbb{Z}_2 \times \mathbb{Z}_2.
\end{equation}
\item Let $Q_f \colon \mathfrak{M}^4 \viz \mathbb{R}^4_{1, 3} \to \mathbb{R}$ be a \emph{quadratic form} determined by $Q_f(v) = g(v, v)$. If $\Lorentz = O_{1, 3}(\mathbb{R})$ \eqref{equation "Lorentz group"} corresponds to $\mathbb{R}^{4 \times 4}$ matrices preserving $Q_f$, then the explicit form of the spinor map \eqref{equation "Spinor map with restricted Lorentz group"} is 
\begin{align}
	& \left\{M \viz [M]^{2 \times 2} \viz \left(\begin{smallmatrix}
	\alpha & \beta \\ 
	\gamma & \delta
	\end{smallmatrix}\right)\right\}
	\mapsto 
	\widetilde{M} \viz \widetilde{\left(\begin{smallmatrix}
	\alpha & \beta \\ 
	\gamma & \delta
	\end{smallmatrix}\right)} = \notag \\
	& \tfrac{1}{2}\left\{
	\begin{smallmatrix}
	\alpha\bar{\alpha} + \beta\bar{\beta} + \gamma\bar{\gamma} + \delta\bar{\delta} \\
	\alpha\bar{\gamma} + \gamma\bar{\alpha} + \beta\bar{\delta} + \delta\bar{\beta} \\
	i(\gamma\bar{\alpha} - \alpha\bar{\gamma} + \delta\bar{\beta} - \beta\bar{\delta}) \\
	\alpha\bar{\alpha} + \beta\bar{\beta} - \gamma\bar{\gamma} - \delta\bar{\delta}
	\end{smallmatrix}
	\begin{smallmatrix}
	\alpha\bar{\beta} + \beta\bar{\alpha} + \gamma\bar{\delta} + \delta\bar{\gamma} \\
	\alpha\bar{\delta} + \delta\bar{\alpha} + \beta\bar{\gamma} + \gamma\bar{\beta} \\
	i(\delta\bar{\alpha} - \alpha\bar{\delta} + \gamma\bar{\beta} - \beta\bar{\gamma}) \\
	\alpha\bar{\beta} + \beta\bar{\alpha} - \gamma\bar{\delta} - \delta\bar{\gamma}
	\end{smallmatrix}
	\begin{smallmatrix}
	i(\alpha\bar{\beta} - \beta\bar{\alpha} + \gamma\bar{\delta} -\delta\bar{\gamma}) \\
	i(\alpha\bar{\delta} - \delta\bar{\alpha} + \gamma\bar{\beta} - \beta\bar{\gamma}) \\
	\alpha\bar{\delta} + \delta\bar{\alpha} - \beta\bar{\gamma} - \gamma\bar{\beta} \\
	i(\alpha\bar{\beta} - \beta\bar{\alpha} + \delta\bar{\gamma} -  \gamma\bar{\delta})	
	\end{smallmatrix}
	\begin{smallmatrix}
	\alpha\bar{\alpha} - \beta\bar{\beta} + \gamma\bar{\gamma} - \delta\bar{\delta} \\
	\alpha\bar{\gamma} + \gamma\bar{\alpha} - \beta\bar{\delta} - \delta\bar{\beta} \\ 
	i(\gamma\bar{\alpha} - \alpha\bar{\gamma} + \beta\bar{\delta} - \delta\bar{\beta}) \\
	\alpha\bar{\alpha} - \beta\bar{\beta} - \gamma\bar{\gamma} + \delta\bar{\delta}
	\end{smallmatrix}
	\right\}M.
\end{align}
\item The connection $SL_2(\mathbb{C}) \to SO_{1, 3}^+(\mathbb{R})$ is analogous to the spinor map from $SU_2(\mathbb{C})$ to $SO_3(\mathbb{R})$, as explained in the previous Sections \ref{subsection "Spinorial Representation of the Orthogonal Group on a 3-Space"} and \ref{subsection "Pauli-like Spinors in the Complex Hilbert 2-Space; Angular Momentum in Quantum Mechanics and Topological Nature of the Electron Spin"}.
\enumerationisfinis

\subsection{Gamma Matrices and Type of Fermionic Spinor Fields}
\label{subsection "Gamma Matrices and Type of Fermionic Spinor Fields"}

\subsubsection{Dirac 4-Spinor Representation}
\label{subsubsection "Dirac 4-Spinor Representation"}

\begingroup
\footnotesize
[T]he Hamiltonian which describes the interaction of the atom and the electromagnetic waves can be made identical with the Hamiltonian for the problem of the interaction of the atom with an assembly of particles moving with the velocity of light and satisfying the Einstein–Bose statistics [\,\dots]. There is thus a complete harmony between the wave and light-quantum descriptions of the interaction. \\
\indent — \textsc{P.A.M. Dirac} \cite[p. 245]{Dirac "The Quantum Theory of the Emission and Absorption of Radiation"}

\vspace{2mm}

Until a few years ago it had been impossible to construct a theory of radiation which could account satisfactorily both for interference phenomena and the phenomena of emission and absorption of light by matter. The first set of phenomena was interpreted by the wave theory, and the second set by the theory of light quanta. It was not until in 1927 that Dirac \cite{Dirac "The Quantum Theory of the Emission and Absorption of Radiation"} \cite{Dirac "The Quantum Theory of Dispersion"} succeeded in constructing a quantum theory of radiation which could explain in an unified way both types of phenomena. \\
\indent — \textsc{E. Fermi} \cite[p. 87]{Fermi "Quantum Theory of Radiation"}

\endgroup

\vspace{2mm}

The Eq. \eqref{align "Irreducible representation on a complex 4-space with a direct sum"} is the representation for constructing the \emph{Dirac spinor}, or \emph{4-spinor}, as a solution of the free \emph{Dirac equation} \cite{Dirac "The Quantum Theory of the Electron"}, of which we show three possible ways of writing,
\begin{equation}
\label{equation "Dirac equation"}
	\begin{rcases}
	\left(i\hbar\gamma^\mu\frac{\partial}{\partial{x}^\mu} - mc\right)\psi \\
	\bigl(i\gamma^\mu\partial_\mu - m\bigr)\psi \\
	\bigl(i\slashed{\partial} - m\bigr)\psi
	\end{rcases}
	= 0,
\end{equation}
To follow some specifications.
\enumerationisinitium
\item $\gamma^\mu \viz \gamma_\textsc{d}^\mu = \{\gamma^0, \gamma^1, \gamma^2, \gamma^3\}$ are the $4 \times 4$ Dirac gamma matrices, 
\begin{equation}
\label{equation "Dirac gamma matrices"}
	\gamma^0 = 
	\begin{pmatrix*}[r]
	1 & 0 \\
	0 & -1
	\end{pmatrix*}, 
	\gamma^1 = 
	\begin{pmatrix}
	0 & \sigmaPauli_1 \\
	-\sigmaPauli_1 & 0
	\end{pmatrix},
	\gamma^2 = 
	\begin{pmatrix}
	0 & \sigmaPauli_2 \\
	-\sigmaPauli_2 & 0
	\end{pmatrix},
	\gamma^3 = 
	\begin{pmatrix}
	0 & \sigmaPauli_3 \\
	-\sigmaPauli_3 & 0
	\end{pmatrix},
	\end{equation}
plus 
$\gamma^5 = \bigl(\begin{smallmatrix}
	0 & 1 \\
	1 & 0
	\end{smallmatrix}\bigr)$, where $\gamma^0$ is the time-like matrix, $1 = \idem_2$, and $\sigmaPauli_{1, 2, 3}$ the Pauli matrices \eqref{equation "Pauli matrices"}; in extended formulation:
\begin{equation}
 	\gamma^0 = 
 	\Biggl\{\begin{smallmatrix} 
	1 & 0 & 0 & 0 \\
	0 & 1 & 0 & 0 \\ 
	0 & 0 & -1 & 0 \\
	0 & 0 & 0 & -1 
	\end{smallmatrix}\Biggr\},
	\gamma^1 =
	\Biggl\{\begin{smallmatrix}
	0 & 0 & 0 & 1 \\
	0 & 0 & 1 & 0 \\
	0 & -1 & 0 & 0 \\
	-1 & 0 & 0 & 0
	\end{smallmatrix}\Biggr\},
	\gamma^2 =
	\Biggl\{\begin{smallmatrix}
	0 & 0 & 0 & -i \\
	0 & 0 & i & 0 \\
	0 & i & 0 & 0 \\
	-i & 0 & 0 & 0
	\end{smallmatrix}\Biggr\}, 
	\gamma^3 =
	\Biggl\{\begin{smallmatrix}
	0 & 0 & 1 & 0 \\
	0 & 0 & 0 & -1 \\
	-1 & 0 & 0 & 0 \\
	0 & 1 & 0 & 0
	\end{smallmatrix}\Biggr\},
\end{equation} 
plus 
$\gamma^5 = i\gamma^0, \gamma^1, \gamma^2, \gamma^3 =
 	\Biggl\{\begin{smallmatrix} 
	0 & 0 & 1 & 0 \\
	0 & 0 & 0 & 1 \\ 
	1 & 0 & 0 & 0 \\
	0 & 1 & 0 & 0 
	\end{smallmatrix}\Biggr\}$, under the anti-commutation relation $\left\{\gamma^\mu, \gamma^\nu\right\} = \gamma^\mu\gamma^\nu + \gamma^\nu\gamma^\mu = 2\eta^{\mu\nu}\idem_{4 \times 4}$.
\item $\partial_\mu = \frac{\partial}{\partial{x}^\mu}$.
\item $\slashed{\partial} = \gamma^\mu\partial_\mu$ is the partial derivative in the Feynman slash notation, see e.g. \cite[13th lecture]{Feynman "Quantum Electrodynamics"}. 
\enumerationisfinis

Associated with various possible representations of gamma matrices, there are as many types of spinors (for instance, the ones in Dirac, Weyl, and Majorana representations). The Dirac spinor is a $(\sfrac{1}{2}, 0) \oplus (0, \sfrac{1}{2})$ representation of the Lorentz group; its form is 
\begin{equation}
\label{equation "Dirac spinor"}
	\psi\left[\mathscr{D}^{\left(\frac{1}{2}, 0\right)} \oplus \mathscr{D}^{\left(0, \frac{1}{2}\right)}\right] 
	= 
	\begin{pmatrix}
	\psi_\textsc{l} \\ 
	\psi_\textsc{r} 
	\end{pmatrix}
	= 
	\begin{pmatrix} 
	\psi_1 \\ \psi_2 \\ \psi_3 \\ \psi_4
	\end{pmatrix} =
	\begin{cases}
	\psi_\textsc{l} = 
	\begin{pmatrix} 
	\psi_1 \\ 
	\psi_2	
	\end{pmatrix} \\
	\psi_\textsc{r} =
	\begin{pmatrix} 
	\psi_3 \\
	\psi_4	
	\end{pmatrix},	
	\end{cases}
\end{equation}
having left- and right-handed states, which transform under $\mathscr{D}^{\left(\sfrac{1}{2}, 0\right)}$ and $\mathscr{D}^{\left(0, \sfrac{1}{2}\right)}$, respectively. Using a neat notation, one can write
\begin{equation}
	\begin{pmatrix}
	\zeta \\ \chi
	\end{pmatrix}e^{i(\vec{p}{x} - Et)}  
	\begin{cases}
	\zeta = \frac{\vec{\sigmaPauli} \cdot \vec{p}}{E - m}\chi, \\
	\chi = \frac{\vec{\sigmaPauli} \cdot \vec{p}}{E + m}\zeta,
	\end{cases}
\end{equation}
with energy $E \to i\hbar\frac{\partial}{\partial{t}}$, and momentum $\vec{p} \to -i\hbar\nabla$ (nabla here is the gradient operator), where $\vec{\sigmaPauli} \cdot \vec{p}$ is the longitudinal polarization of the particle, in which $\vec{\sigmaPauli} = (\sigmaPauli_1, \sigmaPauli_2, \sigmaPauli_3)$ is the spin vector having the Pauli matrices as its components; then 
\begin{equation}
	(E - m)\zeta = (\vec{\sigmaPauli} \cdot \vec{p}\,)\chi, \enspace (E + m)\chi = (\vec{\sigmaPauli} \cdot \vec{p}\,)\zeta.
\end{equation}

\begin{scholium}
~\enumerationisinitium
\item The Dirac spinor \eqref{equation "Dirac spinor"} is a 4-component object/wave function (splittable into two 2-component spinors), and is \emph{simultaneously} left- and right-handed; it treats \emph{massive} or \emph{massless} particles. 
\item Left- and right-handed components of a Dirac spinor correspond to Weyl spinors \eqref{align "Weyl spinor left"} \eqref{align "Weyl spinor right"}.
\item The Dirac spinor is a \emph{complex} (reducible) representation of the Lorentz group. The possibility of a real representation is delegated to the Majorana spinor \eqref{equation "Majorana spinor"} \eqref{equation "Majorana spinor with dotted/undotted notation"}.
\item We highlight an alternative notation to \eqref{equation "Dirac spinor"},
\begin{equation}
\label{equation "Dirac spinor with dotted/undotted notation"}
	\psi = \begin{pmatrix} 
	\zeta^\alpha \\ 
	\tilde{\chi}_{\dot{\alpha}}
	\end{pmatrix}
	=
	\begin{cases}
	\zeta^\alpha =
	\begin{pmatrix} 
	\zeta^1 \\ 
	\zeta^2	
	\end{pmatrix} = \psi_\textsc{l} \\
	\tilde{\chi}_{\dot{\alpha}} =
	\begin{pmatrix} 
	\tilde{\chi}_{\dot{1}} \\ 
	\tilde{\chi}_{\dot{2}}
	\end{pmatrix} = \psi_\textsc{r},
	\end{cases}
\end{equation}
by which the left-handed spinor ($\zeta^\alpha$), as well as the right-handed anti-spinor (anti-$\zeta^\alpha$), is indicated by a letter with upper indices without dot above, whilst the right-handed spinor ($\tilde{\chi}_{\dot{\alpha}}$), as well as the left-handed anti-spinor (anti-$\tilde{\chi}_{\dot{\alpha}}$), is indicated by a letter with lower indices and dot above.
\item The Lagrangian density for the Dirac field described by \eqref{equation "Dirac spinor"} is
\begin{equation}
	\Lagrangian = 
	\begin{cases}
	\adjoint{\psi}\bigl(i\gamma^\mu\partial_\mu - m\bigr)\psi, \\
	\frac{1}{2}\left(i\adjoint{\psi}\gamma^\mu\partial_\mu\psi - i\partial_\mu\adjoint{\psi}\gamma^\mu\psi\right) - m\adjoint{\psi}\psi, \\
	\frac{i}{2}\Bigl(\adjoint{\psi}\slashed{\partial}\psi - \left(\adjoint{\psi}\slashed{\partial}\right)\psi\Bigr) - m\adjoint{\psi}\psi, \enspace \left(\adjoint{\psi}\slashed{\partial}\right) = \partial_\mu\adjoint{\psi}\gamma^\mu,
	\end{cases}
\end{equation}
defining $\adjoint{\psi} = \psi^\dag\gamma^0$ as the \emph{adjoint spinor} (or \emph{Dirac adjoint}), where $\psi^\dag \viz \bar{\psi}^\textsc{t}$\footnote{
	Do not confuse $\bar{\psi}$, which is the complex conjugate (short overline, via \texttt{\textbackslash{bar}} command), with $\adjoint{\psi}$, the adjoint spinor (long overline, via \texttt{\textbackslash{adjoint}} overline-like command).
	} 
is the conjugate transpose of $\psi$. \scholiumsymbol
\enumerationisfinis
\end{scholium}

\subsubsection{Weyl (Chiral) Representation}

Gamma matrices \eqref{equation "Dirac gamma matrices"} in the \emph{Weyl} or \emph{chiral representation}\footnote{
		The term \emph{chiral}, from the Greek \textgreek{χείρ} (hand), was introduced by Lord Kelvin \cite[p. 619]{Kelvin "Baltimore Lectures on Molecular Dynamics and the Wave Theory of Light"}: «I call any geometrical figure, or group of points, \emph{chiral}, and say that it has chirality if its image in a plane mirror, ideally realized, cannot be brought to coincide with itself». A left hand and its mirror image, that is, a right hand, and vice versa, are not superimposable: «[R]ight and left hands are heterochirally similar».
	}
\cite{Weyl "Gravitation and the Electron"} have this $2 \times 2$ form:
\begin{equation}
\label{equation "Gamma matrices in the Weyl representation"}
	\gamma_\textsc{w}^0 = 
	\begin{pmatrix}
	0 & 1 \\
	1 & 0
	\end{pmatrix}, 
	\gamma_\textsc{w}^1 = 
	\begin{pmatrix}
	0 & \sigmaPauli_1 \\
	-\sigmaPauli_1 & 0
	\end{pmatrix},
	\gamma_\textsc{w}^2 = 
	\begin{pmatrix}
	0 & \sigmaPauli_2 \\
	-\sigmaPauli_2 & 0
	\end{pmatrix},
	\gamma_\textsc{w}^3 = 
	\begin{pmatrix}
	0 & \sigmaPauli_3 \\
	-\sigmaPauli_3 & 0
	\end{pmatrix},
\end{equation}
plus 
$\gamma_\textsc{w}^5 = \bigl(\begin{smallmatrix}
	-1 & 0 \\
	0 & 1
	\end{smallmatrix}\bigr)$, where $1 = \idem_2$, and $\sigmaPauli_{\mu = 1, 2, 3}$ are the Pauli matrices \eqref{equation "Pauli matrices"}. The correlated left- and right-handed Weyl spinors, i.e. $(\sfrac{1}{2}, 0)$ and $(0, \sfrac{1}{2})$, respectively, can be written as   
\begin{align}
	& 
	\label{align "Weyl spinor left"}
	\psi_\textsc{l} \to \exp{\left\{\frac{1}{2}\left(i\theta_k\sigmaPauli_k - b_k\sigmaPauli_k\right)\right\}}_{\left(\tfrac{1}{2}, 0\right)} =\left(1 + \frac{i\theta_k\sigmaPauli_k}{2} - \frac{1}{2}b_k\sigmaPauli_k\right)\psi_\textsc{l} = 
	\begin{pmatrix}
	\psi_{1} \\ 
	\psi_{2}
	\end{pmatrix}, \\
	&
	\label{align "Weyl spinor right"}
	\psi_\textsc{r} \to \exp{\left\{\frac{1}{2}\left(i\theta_k\sigmaPauli_k + b_k\sigmaPauli_k\right)\right\}}_{\left(0, \tfrac{1}{2}\right)} = \left(1 + \frac{i\theta_k\sigmaPauli_k}{2} + \frac{1}{2}b_k\sigmaPauli_k\right)\psi_\textsc{r} =
	\begin{pmatrix}
	\psi_{1} \\ 
	\psi_{2}
	\end{pmatrix},
\end{align}
where $\theta_k, b_k \in \mathbb{R}$ are the angles of rotation and of a boost, respectively (the \emph{boost} transformations are those between two inertial reference frames).

\begin{scholium}
~\enumerationisinitium
\item Weyl spinors \eqref{align "Weyl spinor left"} \eqref{align "Weyl spinor right"} are 2-component objects/wave functions, and treat \emph{massless} particles; they are \emph{alternatively} left- or right-handed. With a $\delta$-infinitesimal value, they can be written as $\delta\psi_\textsc{l} = \frac{1}{2}(i\theta_k - b_k)\sigmaPauli_k\psi_\textsc{l}$ and $\delta\psi_\textsc{r} = \frac{1}{2}(i\theta_k + b_k)\sigmaPauli_k\psi_\textsc{r}$. 
\item Let $\sigmaPauli^\mu = (1, \vec{\sigmaPauli})$ and $\underline{\sigmaPauli}^\mu = (1, -\vec{\sigmaPauli})$, by using them instead of \eqref{equation "Dirac gamma matrices"}, under the assumption that 
$\gamma^\mu = 
\left(\begin{smallmatrix}
0 & \sigmaPauli \\
\underline{\sigmaPauli}^\mu & 0	
\end{smallmatrix}\right)$. By dividing the spinor \eqref{equation "Dirac spinor"} into its two 2-components,
$\left(\begin{smallmatrix}
\psi_\textsc{l} \\ 
0
\end{smallmatrix}\right) = \frac{1}{2}(\idem - \gamma^5)\psi$ and 
$\left(\begin{smallmatrix}
0 \\ 
\psi_\textsc{r}
\end{smallmatrix}\right) = \frac{1}{2}(\idem + \gamma^5)\psi$, the Dirac equation \eqref{equation "Dirac equation"} in Weyl's reformulation is
\begin{align}
	& \begin{pmatrix}
	-m & i\sigmaPauli^\mu\partial_\mu \\
	i\underline{\sigmaPauli}^\mu\partial_\mu & -m
	\end{pmatrix}
	\begin{pmatrix}
	\psi_\textsc{l} \\
	\psi_\textsc{r}
	\end{pmatrix} = 0, \\
	& \begin{rcases}
	i\underline{\sigmaPauli}^\mu\partial_\mu\psi_\textsc{l} \\ i\sigmaPauli^\mu\partial_\mu\psi_\textsc{r}
	\end{rcases}
	= 0.
\end{align}
\item A couple of a left- and a right-handed Weyl spinor forms a Dirac spinor \eqref{equation "Dirac spinor"}, i.e., two Weyl spinors give a Dirac spinor. \scholiumsymbol
\enumerationisfinis 
\end{scholium}

\subsubsection{Majorana Symmetric Representation}
\label{subsubsection "Majorana Symmetric Representation"}

\begingroup
\footnotesize
The possibility of achieving a complete formal symmetrization of quantum theory of the electron and positron is shown by making use of a new quantization process. The meaning of Dirac equations is somewhat modified and there is no reason to speak of negative energy states \cite{Dirac "Discussion of the infinite distribution of electrons in the theory of the positron"} [see Section \ref{subsubsection "Dirac's Prediction of Anti-matter from Klein–Gordon Equation"}]; nor to assume for any other type of particles, particularly neutral ones, the existence of “antiparticles” corresponding to the “holes” of negative energy.\endnote{
	Original It. version: «Si dimostra la possibilità di pervenire a una piena simmetrizzazione formale della teoria quantistica dell'elettrone e del positrone facendo uso di un nuovo processo di quantizzazione. Il significato delle equazioni di Dirac ne risulta alquanto modificato e non vi è più luogo a parlare di stati di energia negativa; nè a presumere per ogni altro tipo di particelle, particolarmente neutre, l'esistenza di “antiparticelle” corrispondenti ai “vuoti” di energia negativa».
	} \\
\indent — \textsc{E. Majorana} \cite[p. 171]{Majorana "Teoria simmetrica dell'elettrone e del positrone"}

\endgroup

\vspace{2mm}

Gamma matrices \eqref{equation "Dirac gamma matrices"}, under the \emph{Majorana representation} \cite{Majorana "Teoria simmetrica dell'elettrone e del positrone"} = \cite{Majorana "A symmetric theory of electrons and positrons"},\endnote{
	\label{endnote "Majorana's further studies on the spinor representation of the Lorentz group"}
	Majorana's further studies on the spinor representation of the Lorentz group are in his unpublished \textit{Quaderno 1}, pp. 14-, 26-, 37-, and \textit{Quaderno 3}, p. 71-, as reported in \cite[pp. 180, 183]{Majorana "Lezioni all'Universita di Napoli"}, \textit{Appendice. Catalogo degli scritti di Ettore Majorana–Manoscritti scientifici inediti}, a cura di M. Baldo, R. Mignani, E. Recami.
	} 
are:
\begin{equation}
\label{equation "Gamma matrices in the Majorana representation"}
	\gamma_\textsc{m}^0 = 
	\begin{pmatrix}
	0 & \sigmaPauli^2 \\
	\sigmaPauli^2 & 0
	\end{pmatrix}, 
	\gamma_\textsc{m}^1 = 
	\begin{pmatrix}
	i\sigmaPauli^3 & 0 \\
	0 & i\sigmaPauli^3
	\end{pmatrix},
	\gamma_\textsc{m}^2 = 
	\begin{pmatrix}
	0 & -\sigmaPauli^2 \\
	\sigmaPauli^2 & 0
	\end{pmatrix},
	\gamma_\textsc{m}^3 = 
	\begin{pmatrix}
	-i\sigmaPauli^1 & 0 \\
	0 & -i\sigmaPauli^1
	\end{pmatrix},
\end{equation}
plus 
$\gamma_\textsc{m}^5 = \Bigl(\begin{smallmatrix}
	\sigmaPauli^2 & 0 \\
	0 & -\sigmaPauli^2
	\end{smallmatrix}\Bigr)$, and
$\mathrm{C}_\textsc{m} = -\gamma_\textsc{m}^0 = \Bigl(\begin{smallmatrix}
	0 & -\sigmaPauli^2 \\
	-\sigmaPauli^2 & 0
	\end{smallmatrix}\Bigr)$, with $\gamma_\textsc{m}^0 = \gamma^0\gamma^2$, $\gamma_\textsc{m}^1 = \gamma^2\gamma^1$, $\gamma_\textsc{m}^2 = -\gamma^2$, $\gamma_\textsc{m}^3 = \gamma^2\gamma^3$, $\gamma_\textsc{m}^5 = -i\gamma^0\gamma^1\gamma^3$, where $\sigmaPauli^{\mu = 1, 2, 3}$ are the Pauli matrices \eqref{equation "Pauli matrices"}, and $\mathrm{C}_\textsc{m}$ is the charge conjugation operator under Majorana. Here $\gamma_\textsc{m}^0$ is anti-symmetric, and $\gamma_\textsc{m}^\mu$, $\mu = 1, 2, 3$, are symmetric; then $\bar{\gamma}_\textsc{m}^\mu = -\gamma_\textsc{m}^\mu$ ($\bar{\gamma}_\textsc{m}^\mu$ is the complex conjugate), and $\gamma_\textsc{m} ^\mu = U\bar{\gamma}_\textsc{m}^\mu{U}^\dag$, in which $U = U^\dag = U^{-1} = \frac{1}{\sqrt{2}}\gamma^0(1 + \gamma^2)$ is a unitary matrix. Majorana spinor can be written in this way: 
\begin{equation}
\label{equation "Majorana spinor"}
	\psi \viz \psi_\textsc{m} = 
	\begin{pmatrix}
	\psi_\textsc{l} \\ 
	0
	\end{pmatrix} +
	\begin{pmatrix}
	0 \\ 
	i\sigmaPauli_2\bar{\psi}_\textsc{l}
	\end{pmatrix} =
	\begin{pmatrix}
	\psi_\textsc{l} \\ 
	i\sigmaPauli_2\bar{\psi}_\textsc{l}
	\end{pmatrix} =
	\begin{pmatrix}
	\psi_\textsc{l} \\ 
	\psi_{(\textsc{l}, \mathrm{C})}
	\end{pmatrix},
\end{equation}
where $\sigmaPauli_2\bar{\psi}_\textsc{l}$ acts as a right-handed spinor, and $\psi_{(\textsc{l}, \mathrm{C})} = i\sigmaPauli_2\bar{\psi}_\textsc{l}$. The same construction is for the right-handed spinor, with 
$\psi \viz \psi_\textsc{m} =
	\left(\begin{smallmatrix}
	0 \\
	\psi_\textsc{r}
	\end{smallmatrix}\right) \to 
	\left(\begin{smallmatrix}
	-i\sigmaPauli_2\bar{\psi}_\textsc{r} \\
	\psi_\textsc{r} \\ 
	\end{smallmatrix}\right)$.

\begin{scholium}
~\enumerationisinitium
\item Majorana gamma matrices \eqref{equation "Gamma matrices in the Majorana representation"} are all purely \emph{imaginary}, so the related Majorana–Dirac equation has no complex coefficients, and the Majorana spinor is a \emph{real} representation of the Lorentz group, for which $\bar{\psi} = \psi$. For a detailed overview, see D.Tz. Stoyanov and I.T. Todorov \cite{Stoyanov and Todorov "Majorana Representations of the Lorentz Group and Infinite-Component Fields"}.
\item The Majorana spinor \eqref{equation "Majorana spinor"} is a 2-component object/wave function; it is \emph{alternatively} left- or right-handed, and is still a $(\sfrac{1}{2}, 0) \oplus (0, \sfrac{1}{2})$ representation of the Lorentz group.
\item Rule of thumb: a spinor is called a \emph{Majorana spinor} when $\psi_\mathrm{C} = \psi$, i.e.
\begin{equation}
\label{equation "Majorana condition"}
	\psi_\mathrm{C} = i\gamma^2\bar{\psi} = 
	\begin{pmatrix}
	0 & -i\sigmaPauli^2 \\
	i\sigmaPauli^2 & 0
	\end{pmatrix} 
	\begin{pmatrix}
	\bar{\psi}_\textsc{l} \\ 
	i\sigmaPauli_2\bar{\psi}_\textsc{l}
	\end{pmatrix} =
	\begin{pmatrix}
	\psi_\textsc{l} \\ 
	i\sigmaPauli_2\bar{\psi}_\textsc{l}
	\end{pmatrix} = \psi.
\end{equation}
Eq. \eqref{equation "Majorana condition"} is illustrative of Majorana's condition under which a fermion particle is its own anti-particle. 
\item The Lagrangian helpful for obtaining the above Majorana formalism is
\begin{equation}
	\Lagrangian = \Biggl(i\psi_\textsc{l}^\dag\sigmaPauli_\mu\partial_\mu\psi_\textsc{l} + i\left(\frac{m}{2}\right) \cdot \psi_\textsc{l}^\dag\sigmaPauli_2\bar{\psi}_\textsc{l} - \psi_\textsc{l}^\textsc{t}\sigmaPauli_2\psi_\textsc{l}\Biggr). 	
\end{equation}
\item Two Majorana spinors give a Dirac spinor: as in \eqref{equation "Majorana spinor"} it is possible to construct a Majorana spinor from a Weyl spinor. Says otherwise, a Majorana spinor is a Dirac spinor \eqref{equation "Dirac spinor with dotted/undotted notation"} in chiral-Weyl form,
$\left(\begin{smallmatrix}
\zeta^\alpha \\ 
0
\end{smallmatrix}\right)$ and 
$\left(\begin{smallmatrix}
0 \\ 
\tilde{\chi}_{\dot{\alpha}}
\end{smallmatrix}\right)$, in which $\zeta\text{-component} = \chi\text{-component}$, so 
\begin{equation}	
\label{equation "Majorana spinor with dotted/undotted notation"}
	\psi_\textsc{m} = 
	\begin{pmatrix}
	\zeta^\alpha \\ 
	\tilde{\zeta}_{\dot{\alpha}}
	\end{pmatrix}.
\end{equation}
\scholiumsymbol
\enumerationisfinis
\end{scholium}

\begin{margo}[Electrically neutral spinors in nature and mathematical models in comparison]
\label{margo "Electrically neutral spinors in nature and mathematical models in comparison"}
The most appealing spinor in nature is, probably, the \emph{neutrino} (for a  historical reconstruction inherent in this particle, see E. Amaldi \cite[sec. 3, in particular footnote 277]{Amaldi "From the discovery of the neutron to the discovery of nuclear fission"}, through Pauli's theorization and Fermi's paper on \textgreek{β}-decay \cite{Fermi "Tentativo di una teoria dei raggi beta"} = \cite{Fermi "Versuch einer Theorie der beta-Strahlen. I} = \cite{Wilson "Fermi's Theory of Beta Decay"}); not only because it is useful for understanding the matter vs. anti-matter problem \cite{T2K Collaboration "Constraint on the matter-antimatter symmetry-violating phase in neutrino oscillations"}, thanks to \emph{Pontecorvo's neutrino oscillations} \cite{Pontecorvo "Mesonium and Antimesonium"} \cite{Pontecorvo "Neutrino Experiments and the Problem of Conservation of Leptonic Charge"}, by which the neutrino changes its lepton flavor (electron, muon, and tau(on) neutrino) during propagation; but also because it represents a crossroads between Dirac and Majorana solutions, in which forms of neutrino and anti-neutrino are and are not distinct particles, respectively. \margosymbol
\end{margo}

\section{Clifford (Geometric) Algebra and Spinoriality}
\label{section "Clifford (Geometric) Algebra and Spinoriality"}

One of the most exciting characteristics of mathematics is the creation/discovery of unexpected links between different fields of investigation and apparently distinct categories. Clifford algebra is an ingenious set of instruments, almost like a trick, that ties and holds several parts of algebra, and not only. In overall terms, \emph{Clifford algebra}, or \emph{geometric algebra}, is called an \emph{associative algebra} generated by an $n$-dimensional vector space equipped with a symmetric quadratic form. 

\begin{margo}
Historical reference writings for the Clifford algebra are \cite{Clifford "Preliminary Sketch of Biquaternions"} \cite{Clifford "Applications of Grassmann's Extensive Algebra"}. In I. Stringham \cite{Stringham "On the Geometry of Planes in a Parabolic Space of Four Dimensions"} quaternions and Clifford's view are mixed for a construction of hyper- or multi-dimensional geometry (initially within the $4\mathrm{D}$ field). The analytical roots of multi-dimensional geometry are in J. Plücker \cite{Plucker "Theoremes generaux concernant les equations d'un degre quelconque entre un nombre quelconque d'inconnues"} and, especially, in A. Cayley \cite{Cayley "Chapters in the Analytical Geometry of $(n)$ Dimensions"} and H. Grassmann \cite{Grassmann "Die lineale Ausdehnungslehre ein neuer Zweig der Mathematik"} \cite{Grassmann "Anhang III (1877) Kurze Uebersicht uber das Wesen der Ausdehnungslehre"}. A first and fundamental contribution for an algebraic (but not yet physical) theory of connection between spinors and Clifford algebra, it is due to C. Chevalley \cite[chap. III]{Chevalley "The Construction And Study Of Certain Important Algebras"} \cite[chap. II]{Chevalley "The Algebraic Theory of Spinors"}. \margosymbol
\end{margo}

\subsection{Abralgebras: Tricks of the Clifford's Associative Tool} 

Pauli matrices \eqref{equation "Pauli matrices"}, and Dirac \eqref{equation "Dirac gamma matrices"}, Weyl \eqref{equation "Gamma matrices in the Weyl representation"} \& Majorana \eqref{equation "Gamma matrices in the Majorana representation"} matrices, are all \emph{matrix representations of the geometric algebra of space}, i.e., they are included in the \emph{Clifford algebra}. Let us explore in more detail the various ramifications.

\enumerationisinitium
\item \emph{Pauli algebra}:
\begin{align}
	&
	\left(\mathfrak{pau}_{(3, 0)} \viz \mathbb{R}_{3, 0} \viz \Cl_{3, 0}(\mathbb{R})\right) \cong \mathbb{C} \otimes \quaternion \cong \left({^2\mathbb{C}} \viz \mathbb{C}(2)\right), \\
	& \text{ where } \Cl_{3, 0}(\mathbb{R}) \viz \mathfrak{aps}, \mathbb{C} \cong \Cl_{0, 1}, \text{ and } \quaternion \cong \Cl_{0, 2} \cong	\Cl^{+[0]}_{3, 0}. \notag
\end{align}
Pauli algebra (see Pauli spinor representation in Section \ref{subsection "Pauli-like Spinors in the Complex Hilbert 2-Space; Angular Momentum in Quantum Mechanics and Topological Nature of the Electron Spin"}), denoted by $\mathfrak{pau}_{(3, 0)} \viz \mathbb{R}_{3, 0}$, coincided with the \emph{algebra of physical space}, identified by $\mathfrak{aps}$, that is the real Clifford algebra $\Cl_{3, 0}(\mathbb{R})$ of the Euclidean (vector) 3-space $\mathbb{R}^3$.  

The $\mathfrak{aps}$-valued elements are isomorphic to the complexification of the quaternions (i.e., to the biquaternions), the algebra of which is given by the tensor product $\mathbb{C} \otimes \quaternion$, that is in turn isomorphic to the algebra ${^2\mathbb{C}} \viz \mathbb{C}(2)$ of $\mathbb{C}^{2 \times 2}$ (complex) matrices. Quaternions $\quaternion$ correspond to the even subalgebra $\Cl^{+[0]}_{3, 0}$ of $\Cl_{3, 0}$. Note that one has 
\begin{equation}
	\Cl_{3, 0} \cong (\Cl_{2, 0} \cong {^2\mathbb{R}}) \otimes (\Cl_{0, 1} \cong \mathbb{C}) \cong {^2\mathbb{C}}, 
\end{equation}
whereas $\Cl_{0, 3} \cong (\Cl_{0, 2} \cong \quaternion) \otimes (\Cl_{1, 0} \cong \mathbb{R} \oplus \mathbb{R} = \mathbb{R}^2) \cong \quaternion \oplus \quaternion$.
\item \emph{Dirac algebra}: 
\begin{equation}
\label{equation "Dirac algebra"}
	\left(\mathfrak{dir}_{(4, 1)} \viz \mathbb{R}_{4, 1} \viz \Cl_{4, 1, \mathbb{C} \otimes}(\mathbb{R}) \viz \Cl_{4\times}(\mathbb{C}) \right) \cong \left({^4\mathbb{C}} \viz \mathbb{C}(4)\right),
\end{equation}
where ${^4\mathbb{C}} \viz \mathbb{C}(4)$ is the algebra of $\mathbb{C}^{4 \times 4}$ (complex) matrices. Note that 
\subenumerationisinitium
\item $\Cl_{4, 1, \mathbb{C} \otimes}(\mathbb{R}) \cong \Cl_{1, 1} \otimes \Cl_{3, 0} \cong {^2\mathbb{R}} \otimes {^2\mathbb{C}} \cong {^4\mathbb{C}}$, 
\item $\mathbb{C} \otimes \quaternion \otimes \quaternion \cong \mathbb{C} \otimes {^4\mathbb{R}} \cong {^4\mathbb{C}}$.
\subenumerationisfinis
\item \emph{Space-time algebra}, introduced by D. Hestenes \cite{Hestenes "Real Spinor Fields"} \cite{Hestenes "Spacetime physics with geometric algebra"} \cite{Hestenes "Space-Time Algebra"}:
\begin{equation}
\label{equation "Space-time algebra"}
	\left(\mathfrak{sta} \viz \mathbb{R}_{1, 3} \viz \Cl_{1, 3}(\mathbb{R}) \cong \Cl_{1, 1} \otimes \Cl_{0, 2}\right) \cong \left({^2\quaternion} \viz \quaternion(2)\right),
\end{equation}
where ${^2\quaternion} \viz \quaternion(2)$ is the algebra of $\quaternion^{2 \times 2} = \quaternion^4 = \mathbb{R}^{16}$ (quaternionic) matrices. Note that 
\enumerationisinitium
\item $\Cl_{4, 0} \cong \Cl_{2, 0} \otimes \Cl_{0, 2} \cong \Cl_{1, 3} \cong {^2\mathbb{R}} \otimes \quaternion \cong {^2\quaternion}$,
\item $\Cl_{1, 3} + i\Cl_{1, 3} = {^4\mathbb{C}}$.
\enumerationisfinis

Space-time algebra is but a geometric algebra characterized by a coordinate-free reformulation, in a unified language, of relativistic theories (Dirac Eq. \eqref{equation "Dirac equation"}, Einstein field Eqq. \eqref{subequations "Einstein field equations"}, Einstein–Maxwell–Dirac equations). Among its features there is that of  providing a basis vectors in space, for the Pauli matrices \eqref{equation "Pauli matrices"}, and in space-time, for the Dirac matrices \eqref{equation "Dirac gamma matrices"}; plus there is that of giving a spatio-temporal origin, or a geometric dimension, to the imaginary unit in quantum mechanics. 
\item \emph{Majorana algebra}:
\begin{equation}	
	\left(\mathfrak{maj}_{(3, 1)} \viz \mathbb{R}_{3, 1} \viz \Cl_{3, 1}(\mathbb{R}) \cong \Cl_{1, 1} \otimes \Cl_{2, 0}\right) \cong \left({^4\mathbb{R}} \viz \mathbb{R}(4)\right),
\end{equation}
where ${^4\mathbb{R}} \viz \mathbb{R}(4) \cong \quaternion \otimes \quaternion \cong {^2{\mathbb{R}}} \otimes {^2{\mathbb{R}}}$ is the algebra of $\mathbb{R}^{4 \times 4}$ (real) matrices. The Majorana algebra is the Clifford algebra of $\mathbb{R}^{3, 1}$. Note that $\Cl_{1, 1} \otimes \Cl_{2, 0} \cong \Cl_{2, 2} \cong {^2\mathbb{R}} \otimes {^2\mathbb{R}}$. 
\enumerationisfinis

Summing up, these are the four Clifford algebras that are of interest to us: $\mathfrak{pau}_{(3, 0)} \viz \mathfrak{aps} = \Cl_{3, 0}$, $\mathfrak{dir}_{(4, 1)}$, $\mathfrak{sta} = \Cl_{1, 3}$, and $\mathfrak{maj}_{(3, 1)}$. Let us look at some intersections. 
\begin{align}
	& \left(\mathfrak{pau}_{(3, 0)} \viz \Cl_{3, 0}(\mathbb{R})\right) \cong \left(\mathfrak{sta}^{+[0]} \viz \Cl_{1, 3}^{+[0]}(\mathbb{R})\right) \cong \left(\mathfrak{maj}_{(3, 1)}^{+[0]} \viz \Cl_{3, 1}^{+[0]}(\mathbb{R})\right), \\
	& \left(\mathfrak{sta} \viz \Cl_{1, 3}(\mathbb{R})\right) \cong \left(\mathfrak{dir}_{(4, 1)}^{+[0]} \viz \Cl_{4, 1, \mathbb{C} \otimes}^{+[0]}(\mathbb{R}) \viz \Cl_{4\times}^{+[0]}(\mathbb{C})\right), \\
	& \left(\mathfrak{dir}_{(4, 1)} \viz \Cl_{4, 1, \mathbb{C} \otimes}(\mathbb{R})\right) \cong 
	\begin{cases}
	\mathbb{C} \otimes \left(\mathfrak{sta} \viz \Cl_{1, 3}(\mathbb{R})\right) = \Cl_{1, 3}(\mathbb{R})_\mathbb{C}, \\
	\mathbb{C} \otimes \left(\mathfrak{maj}_{(3, 1)} \viz \Cl_{3, 1}(\mathbb{R})\right) = \Cl_{3, 1}(\mathbb{R})_\mathbb{C},
	\end{cases} \\
	& \mathbb{C} \otimes \left(\mathfrak{sta} \viz \Cl_{1, 3}(\mathbb{R})\right) \cong \mathbb{C} \otimes \left(\mathfrak{maj}_{(3, 1)} \viz \Cl_{3, 1}(\mathbb{R})\right). 
\end{align}

Pauli algebra is the even subalgebra of the space-time and Majorana algebras, i.e.
\begin{align}
	& \mathfrak{aps} \cong (\mathfrak{sta} \cong \mathfrak{maj})^{+[0]}, \notag \\
	& \Cl\text{-}\mathfrak{pau}_{(3, 0)} \cong (\Cl_{1, 3} \cong \Cl_{3, 1})^{+[0]},
\end{align}
through the maps 
\begin{align}	
	& \varphi_{\Cl} \colon \mathfrak{sta}^{+[0]} \to \Cl_{3, 0}(\mathbb{R}), \\
	& \varphi_{\Cl} \colon \mathfrak{maj}_{(3, 1)}^{+[0]} \to \Cl_{3, 0}(\mathbb{R}). 
\end{align}	
The space-time algebra is the even subalgebra of the Dirac algebra; so the Dirac algebra is isomorphic to the complexification of the space-time algebra or of the Majorana algebra. The isomorphism 
\begin{equation}
	\varphi_{\Cl} \colon \Cl_{4, 1, \mathbb{C} \otimes}(\mathbb{R}) \to \mathbb{C} \otimes \Cl_{1, 3}(\mathbb{R})
\end{equation}
comes from the fact that $\Cl_{4, 1, \mathbb{C} \otimes}(\mathbb{R}) \cong {^4\mathbb{C}}$ (the Clifford algebra $\Cl_{4, 1, \mathbb{C} \otimes}(\mathbb{R})$ is isomorphic to the algebra of $\mathbb{C}^{4 \times 4}$ matrices), for which each element of the Dirac Eq. \eqref{equation "Dirac equation"} can be represented in $\Cl_{4, 1, \mathbb{C} \otimes}(\mathbb{R})$.

\begin{scholium}[Pauli algebra for the Minkowski space-time]
In W.E. Baylis and G. Jones \cite{Baylis and Jones "The Pauli-Algebra Approach to Special Relativity"} \cite{Baylis "Electrons photons and spinors in the Pauli algebra"} \cite[sec. 17.1]{Baylis "The Paravector Model of Spacetime"} there is a description of how the Pauli algebra, although related to multiplication of vectors in $3\mathrm{D}$ Euclidean space, can form a 4-dimensional space with a Minkowski-like metric. This is because the Minkowski space-time (Section \ref{subsection "Minkowski Space-Time (Flat Metric)"}), both in Dirac \eqref{equation "Dirac algebra"} or in space-time \eqref{equation "Space-time algebra"} algebra version, can be divided into two parts, and therefore it can be mapped onto the spinor $\mathfrak{pau}_{(3, 0)}$-valued algebra; in this way, the elements in the sum of scalars and vectors, called \emph{paravectors}, become elements of a real 4-space, moving from vector to 4-vector interpretation: the transformations of the paravectors are spin Lorentz-like transformations, and from here it is possible to (re)build the spinor map from $SL_2(\mathbb{C})$ into $SO_{1, 3}^+(\mathbb{R}) = \Lorentz_+^\uparrow$.\scholiumsymbol
\end{scholium}

\subsection[Spin(or) in $\Cl$-Algebra]{Spin(or) in $\protect\pseudobold{\Cl}$-Algebra}

\enumerationisinitium
\item The spin group in the spinor map \eqref{equation "Spinor map with restricted Lorentz group"}	has a precise identification in the Clifford algebra,
\begin{align}
	& \Spin_{1, 3}^+(\mathbb{R}) = \biggl\{\left\{s \in \Cl_{3, 0}(\mathbb{R})\right\} \cong
	\left\{s \in \Cl_{1, 3}^{+[0]}(\mathbb{R})\right\}\biggr\} \cong SL_2(\mathbb{C}), \\
	& \Spin_{1, 3}^+(\mathbb{R}) \subset
	\begin{cases}
	\Cl_{3, 0}(\mathbb{R}), \text{ with } \mathbb{R}^3 \oplus \bigwedge^2 \mathbb{R}^3, \\
	\Cl_{1, 3}^{+[0]}(\mathbb{R}), \text{ with } \bigwedge^2 \mathbb{R}^{1, 3}.  
	\end{cases}
\end{align}
Of course, $\mathfrak{maj}_{(3, 1)}$-side, one has
\begin{equation}	
	\Spin_{3, 1}^+(\mathbb{R}) = \left\{s \in \Cl_{3, 1}^{+[0]}(\mathbb{R})\right\} 
	\cong SL_2(\mathbb{C}). 
\end{equation}
\item The Dirac spinor \eqref{equation "Dirac spinor"} is a \emph{4-complex column} form; another form is in terms of \emph{$\mathbb{C}^{4 \times 4}$ matrix spinor},
\begin{equation}
\label{equation "Dirac spinor as column spinor and square matrix spinor"}
	\psi =
	\begin{cases}
	\left\{\begin{smallmatrix}
	\psi_1 \\ \psi_2 \\ \psi_3 \\ \psi_4
	\end{smallmatrix}\right\} \in \mathbb{C}^4, \\
	\left\{\begin{smallmatrix}
	\psi_1 & 0 & 0 & 0 \\ 
	\psi_2 & 0 & 0 & 0 \\ 
	\psi_3 & 0 & 0 & 0 \\ 
	\psi_4 & 0 & 0 & 0
	\end{smallmatrix}\right\} \in {^4\mathbb{C}}\idempotent, \text{ with } \idempotent = \frac{1}{2}(1 + \gamma^0)\frac{1}{2}(1 + i\gamma^1\gamma^2),
	\end{cases}
\end{equation}
where $\idempotent$ is the \emph{primitive idempotent}. 
There is a third possibility to write the Dirac spinor, called \emph{Clifford–Dirac algebraic spinor}, distinguishing between the real part and the complex conjugate in this way, 
\begin{equation}
	\psi \in \mathbb{C} \otimes \Cl_{1, 3}\idempotent 
	\left\{
	\Re(\psi)\tfrac{1}{2} = 
	\left\{\begin{smallmatrix*}[r]
	\psi_1 & -\bar{\psi}_2 & 0 & 0 \\ 
	\psi_2 & \bar{\psi}_1 & 0 & 0 \\ 
	\psi_3 & \bar{\psi}_4 & 0 & 0 \\ 
	\psi_4 & -\bar{\psi}_3 & 0 & 0
	\end{smallmatrix*}\right\}, \enspace
	\bar{\psi} = 
	\left\{\begin{smallmatrix*}[r]
	& 0 & -\bar{\psi}_2 & 0 & 0 \\ 
	& 0 & \bar{\psi}_1 & 0 & 0 \\ 
	& 0 & \bar{\psi}_4 & 0 & 0 \\ 
	& 0 & -\bar{\psi}_3 & 0 & 0
	\end{smallmatrix*}\right\}\right\}.
\end{equation}	

\begin{margo}
The real part and the complex conjugate of $\psi \in {^4\mathbb{C}}\idempotent$ in \eqref{equation "Dirac spinor as column spinor and square matrix spinor"} have instead the following form: 
\begin{equation}
	\Re(\psi) = 
	\left\{\begin{smallmatrix}
	\Re(\psi)_1 & 0 & 0 & 0 \\ 
	\Re(\psi)_2 & 0 & 0 & 0 \\ 
	\Re(\psi)_3 & 0 & 0 & 0 \\ 
	\Re(\psi)_4 & 0 & 0 & 0
	\end{smallmatrix}\right\}, \enspace
	\bar{\psi} = 
	\left\{\begin{smallmatrix}
	\bar{\psi}_1 & 0 & 0 & 0 \\ 
	\bar{\psi}_2 & 0 & 0 & 0 \\ 
	\bar{\psi}_3 & 0 & 0 & 0 \\ 
	\bar{\psi}_4 & 0 & 0 & 0
	\end{smallmatrix}\right\}.
\end{equation}
\margosymbol
\end{margo}
\enumerationisfinis

\vspace{10mm}

\setcounter{secnumdepth}{0}  
\section{References and Bibliographic Details}
\setcounter{secnumdepth}{3}
\markright{References and Bibliographic Details}

\begingroup
\footnotesize
\noindent Section \ref{section "Excerpts from Memory: Ricci Methods"}

\begin{indent paragraph: 15pt}
On the tensor calculus (Ricci calculus, absolute differential calculus) theory, see B. Finzi and M. Pastori \cite{Finzi "Applicazioni fisiche del calcolo tensoriale"} \cite{Finzi Pastori "Calcolo tensoriale e applicazioni"}. For a historical reconstruction, see \cite{Dell'Aglio "On the genesis of the concept of covariant differentiation"} \cite{Dell'Aglio "Un case study nell'accettazione di teorie matematiche. Sviluppo e diffusione del calcolo differenziale assoluto in epoca pre-relativistica"} and \cite{Bottazzini "Ricci and Levi-Civita: from differential invariants to general relativity"}.
\end{indent paragraph: 15pt}

\noindent Section \ref{subsection "Tensor Multilinear Algebra and Tensor Analysis"}

\begin{indent paragraph: 15pt}
For an overview of the Ricci (tensor) calculus, see \cite[secc. 1.1-1.4, 3.2]{Abate Tovena "Geometria Differenziale"} \cite[chapp. 3-4]{Dubrovin Fomenko Novikov "Modern Geometry I"} \cite[part I, chap. 3]{Hackbusch "Tensor Spaces and Numerical Tensor Calculus"} \cite[chap. I, sec. 2.5, chap. IV. sec. 2.4]{Iliev "Handbook of Normal Frames and Coordinates"} \cite[chapp. 2, 7]{Lee "Riemannian Manifolds: An Introduction to Curvature"} \cite[chap. 7]{Lee "Manifolds and Differential Geometry"} \cite[chap. 12]{Lee "Introduction to Smooth Manifolds"} \cite[chap. 2]{Renteln "Manifolds Tensors and Forms: An Introduction for Mathematicians and Physicists"} \cite[secc. 1.2-1.3, 2.5]{da Rocha and Vaz Jr. "An Introduction to Clifford Algebras and Spinors"} \cite[chap. 2]{Schafer Schmidt "Tensor Analysis and Elementary Differential Geometry for Physicists and Engineers"} \cite[chap. 3]{Sharafutdinov "Integral Geometry of Tensor Fields"} \cite[chapp. 1-2, 4-5]{de Souza Sanchez Filho "Tensor Calculus for Engineers and Physicists"} \cite[chap. II]{Yokonuma "Tensor Spaces and Exterior Algebra"}. A clean summary about the origin of the Ricci (tensor) calculus is in A. Tonolo \cite{Tonolo "Sulle origini del Calcolo di Ricci"}.

\end{indent paragraph: 15pt}

\noindent Section \ref{subsection "Musical Isomorphism of Tensors"}

\begin{indent paragraph: 15pt}
On the musical isomorphism in tensor frame, see \cite[sec. 5.2]{Garcia-Rio Vanhecke  Vazquez-Abal "Notes on Harmonic Tensor Fields"}.	
\end{indent paragraph: 15pt}

\noindent Sections \ref{subsection "Riemann Curvature Tensor"} and \ref{subsection "Riccian Algebro-geometric Properties"}

\begin{indent paragraph: 15pt}
On the Riemann and Ricci curvature tensors, see e.g. \cite[secc. 6.1-2]{Gasperini "Theory of Gravitational Interactions"} \cite[chap. 6]{Gron "Lecture Notes on the General Theory of Relativity: From Newton's Attractive Gravity to the Repulsive Gravity of Vacuum Energy"} \cite[sec. 2.2]{de Felice Bini "Classical Measurements in Curved Space-Times"} \cite[chap. 3]{de Felice Clarke "Relativity on curved manifolds"} \cite[sec. 7.3]{Gron Hervik "Einstein's General Theory of Relativity: With Modern Applications in Cosmology"} \cite[pp. 74-77, 87-89]{O'Neill "Semi-Riemannian Geometry With Applications to Relativity"} \cite[sec. 5.3.3]{Padmanabhan "Gravitation: Foundations and Frontiers"} \cite[secc. 1.5.1-2., 3.1.1, 3.1.4-5]{Petersen "Riemannian Geometry"} \cite[chap. 9]{Sharan "Spacetime Geometry and Gravitation"}.	
\end{indent paragraph: 15pt}

\noindent Section \ref{subsubsection "Bianchi Identities"}

\begin{indent paragraph: 15pt}
On the Bianchi identities, see e.g. \cite[secc. 15.5-8, app. D.18]{Straumann "General relativity"} \cite[sec. 14.1, and p. 217]{Taubes "Differential Geometry. Bundles Connections Metrics and Curvature"}; for a Bianchi identities version in tetrad components, see \cite[sec. 7.3]{Stephani Kramer MacCallum Hoenselaers Herlt "Exact Solutions of Einstein's Field Equations"}.
\end{indent paragraph: 15pt}

\noindent Section \ref{subsection "Weyl Curvature Tensors"}

\begin{indent paragraph: 15pt}
On the Weyl curvature tensor, see e.g. \cite[1G]{Besse "Einstein Manifolds"} \cite[VI.3]{Choquet-Bruhat "General Relativity and Einstein's Equations"} \cite[V.D. Problem 4]{Choquet-Bruhat DeWitt-Morette with Dillard-Bleick "Analysis Manifolds and Physics I"} \cite{Danehkar "On the Significance of the Weyl Curvature in a Relativistic Cosmological Model"} \cite{Frances "Essential conformal structures in Riemannian and Lorentzian geometry"} \cite{Friedrich "Conformal Einstein Evolution"} \cite[3.K.1]{Gallot Hulin Lafontaine "Riemannian Geometry"} \cite[pp. 160, 490]{Gron Hervik "Einstein's General Theory of Relativity: With Modern Applications in Cosmology"} \cite[sec. 4.16]{Hall "Symmetries and Curvature Structure in General Relativity"} \cite[§ 13.5]{Misner Thorne Wheeler "Gravitation"} \cite[pp. 110, 163]{Petersen "Riemannian Geometry"} \cite[sec. 9.5.2]{Sharan "Spacetime Geometry and Gravitation"} \cite[pp. 524. 610]{Thiemann "Modern Canonical Quantum General Relativity"}. — On the Weyl projective curvature tensor, see \cite[p. 135]{Eisenhart "Riemannian Geometry"} \cite{Ehlers Pirani Schild "The Geometry of Free Fall and Light Propagation"}, and \cite[2.2.4]{Borchers Sen "Mathematical Implications of Einstein-Weyl Causality"} \cite[1.5.3]{Sen "Causality Measurement Theory and the Differentiable Structure of Space-Time"}.
\end{indent paragraph: 15pt}

\noindent Section \ref{subsection "Conformal Flatness and $n$-Dimensionality"}

\begin{indent paragraph: 15pt}
On the Cotton and Cotton–York tensors \cite[I.14.8]{Choquet-Bruhat "Introduction to General Relativity Black Holes and Cosmology"}, see \cite{Garcia[-Diaz] Hehl Heinicke and Macias "The Cotton tensor in Riemannian spacetimes"} reworked in \cite[chap. 20]{Garcia-Diaz "Exact Solutions in Three-Dimensional Gravity"} \cite[sec. 7.1]{Gourgoulhon "3+1 Formalism in General Relativity: Bases of Numerical Relativity"}. — On the conformal transformation, see \cite{Dabrowski Garecki Blaschke Conformal transformations and conformal invariance in gravitation"} \cite[secc. 1.3.3, 3]{Fujii Maeda "The Scalar-Tensor Theory of Gravitation"} \cite[secc. 2.3.3, 4.4.1]{Griffiths Podolsky "Exact Space-Times in Einstein's General Relativity"}; see also \cite[sec. 17.1]{Frishman Sonnenschein "Non-Perturbative Field Theory: From Two-Dimensional Conformal Field Theory to QCD in Four Dimensions"} \cite[sec. 2.2]{Parker and Toms "Quantum Field Theory in Curved Spacetime"}.	
\end{indent paragraph: 15pt}

\noindent Section \ref{subsection "Minkowski Space-Time (Flat Metric)"}

\begin{indent paragraph: 15pt}
On the Minkowski space-time, see e.g. \cite[secc. 1.1-2]{Penrose and Rindler "Spinors and space-time I. Two-spinor calculus and relativistic fields"} \cite[summary of Vol. 1, secc. 9.1-2]{Penrose and Rindler "Spinors and space-time II. Spinor and twistor methods in space-time geometry"} \cite[secc. 2.2, 3.4]{Naber "Topology Geometry and Gauge Fields: Interactions"} \cite[sec. 1.2]{Naber "The Geometry of Minkowski Spacetime: An Introduction to the Mathematics of the Special Theory of Relativity"}.
\end{indent paragraph: 15pt}

\noindent Section \ref{subsubsection "Vector Spaces L^4"}

\begin{indent paragraph: 15pt}
About the Lorentzian geometry (manifolds and spaces), see \cite[sec. 3.9]{Adams "Dynamics on Lorentz Manifolds"} \cite[sec. 1.3]{Bar Ginoux Pfaffle "Wave Equations on Lorentzian Manifolds and Quantization"} \cite[sec. 3]{Beem Ehrlich Easley "Global Lorentzian Geometry"} \cite[secc. 2.4, 3]{Duggal and Bejancu "Lightlike Submanifolds of Semi-Riemannian Manifolds and Applications"} \cite[sec. 5]{O'Neill "Semi-Riemannian Geometry With Applications to Relativity"} \cite{Pfaffle "Lorentzian Manifolds"} \cite[secc. 8.1-2]{Sachs Wu "General Relativity for Mathematicians"}. 
\end{indent paragraph: 15pt}

\noindent Section \ref{subsubsection "Lorentz Group plus Transformations"}

\begin{indent paragraph: 15pt}
On the Lorentz groups, see e.g. \cite[sec. 3]{Costa Fogli "Symmetries and Group Theory in Particle Physics. An Introduction to Space-time and Internal Symmetries"} \cite[chap. 6]{Hall "Symmetries and Curvature Structure in General Relativity"} \cite[sec. 2.8.1]{Rodrigues Jr. and de Oliveira "The Many Faces of Maxwell Dirac and Einstein Equations. A Clifford Bundle Approach"} \cite[sec. 3.4.1]{Sundermeyer "Symmetries in Fundamental Physics"}.
\end{indent paragraph: 15pt}

\noindent Section \ref{subsection "Spinor Map (6-Dimensional Homomorphism): the Covering $SL_2(C)$ to $SO_{1, 3}^+(R)$"}

\begin{indent paragraph: 15pt}
On the representations of the Lorentz group, spinor map, and links between $SO_+(1, 3)$ and $SL_2(\mathbb{C})$, see e.g. \cite{Carmeli and Malin "Finite- and Infinite-Dimensional Representations of the Lorentz Group"} \cite[chapp. 3-4]{Carmeli Malin "Theory of Spinors: An Introduction"} \cite{Coddens "Spinors in the Lorentz group and their implications for quantum mechanics"} \cite{Crawford "Spinors: Lorentz Group"} \cite[sec. 2.4]{Naber "Topology Geometry and Gauge Fields: Interactions"}; see also the lecture course of S. Coleman on \textsc{qft} \cite[chap. 18]{Chen Derbes Griffiths Hill Sohn Ting (Eds.) "Lectures of Sidney Coleman on Quantum Field Theory"}.
\end{indent paragraph: 15pt}

\noindent Section \ref{subsection "Gamma Matrices and Type of Fermionic Spinor Fields"}

\begin{indent paragraph: 15pt}
· On the Dirac, Majorana and Weyl representations of gamma matrices plus spinors, see e.g. \cite[sec. 15.1]{Akhmedov "Majorana neutrinos and other Majorana particles: theory and experiment"} \cite[chap. 1]{Bandyopadhyay "Geometry Topology and Quantum Field Theory"} \cite[secc. 19.3-4, 20.4, 22.2]{Chen Derbes Griffiths Hill Sohn Ting (Eds.) "Lectures of Sidney Coleman on Quantum Field Theory"} \cite{Fradkin "Quantum Field Theory: An Integrated Approach"} \cite{Pal "Dirac Majorana and Weyl Fermions"} \cite[chap. 10, sec. 11.3]{Schwartz "Quantum Field Theory and the Standard Model"} \cite[chapp. 7-8]{Ticciati "Quantum Field Theory for Mathematicians"}. \\
· For a basic mathematical introduction to Dirac's and Weyl's formalism, you can see the book by Talagrand \cite[part II] {Talagrand "What Is a Quantum Field Theory? A First Introduction for Mathematicians"}, but it is a pity that Majorana's inventions are limited to a brief page. \\
· A broader study on the spinors is in P. Deligne \cite{Deligne "Notes on Spinors"}.
\end{indent paragraph: 15pt}

\noindent Section \ref{section "Clifford (Geometric) Algebra and Spinoriality"}

On the Clifford algebra, in relation to Pauli, Dirac, and Majorana theories, plus algebra of physical space, space-time algebra, and spinors, see \cite{Doran Lasenby Gull Somaroo and Challinor "Spacetime Algebra and Electron Physics"} \cite{Lounesto "Crumeyrolle's bivectors and spinors"} \cite{Lounesto "Clifford Algebras and Spinors"} \cite{Morgan "The massless Dirac equation Maxwell's equations and the application of Clifford algebras"} \cite{Piazzese "On the relationships between the Dirac spinors and Clifford subalgebra $Cl_{1, 3}^+$"} \cite{Porteous "Mathematical Structure of Clifford Algebras"} \cite{da Rocha and Vaz Jr. "Conformal structures and twistors in the paravector model of spacetime"}; see also \cite[sec. 4.6]{de Faria de Melo "Mathematical Aspects of Quantum Field Theory"} \cite[chapp. 1-2]{Lawson Jr. and Michelsohn "Spin Geometry"}.

\endgroup

\chapter{On Dimensional Continuum, Part II. Action Principles, Variations and Radiation in Curved Space-Time—Mathematical Details of Field Theory of Gravitation (General Relativity)}
\chaptermark{On Dimensional Continuum, Part II}{}
\label{chapter "On Dimensional Continuum, Part II. Action Principles, Variations and Radiation in Curved Space-Time—Mathematical Details of Field Theory of Gravitation (General Relativity)"}

\begingroup
\footnotesize
[According to Riemann] the axioms of plane geometry are true within the limits of experiment on the surface of a sheet of paper, and yet we know that the sheet is really covered with a number of small ridges and furrows, upon which (the total curvature not being zero) these axioms are not true [\,\dots]. I hold in fact \\
\indent ($\mathnormal{1}$) That small portions of space \emph{are} in fact of a nature analogous to little hills on a surface which is on the average flat; namely, [\,\dots] the ordinary laws of geometry are not valid in them. \\
\indent ($\mathnormal{2}$) That this property of being curved or distorted is continually being passed on from one portion of space to another after the manner of a wave. \\
\indent ($\mathnormal{3}$) That this variation of the curvature of space is what really happens in that phenomenon which we call the \emph{motion of matter}, whether ponderable or etherial. \\
\indent ($\mathnormal{4}$) That in the physical world nothing else takes place but this variation, subject (possibly) to the law of continuity \cite[pp. 157-158]{Clifford "On the Space-Theory of Matter"} = \cite[pp. 21-22]{Clifford "Mathematical Papers"}. \\
\indent Our space is perhaps really possessed of a curvature varying from point to point, which we fail to appreciate because we are acquainted with only a small portion of space, or because we disguise its small variations under [really geometrical] changes in our physical condition which we do not connect with our change of position [\,\dots]. We might even go so far as to assign to this variation of the curvature of space “what really happens in that phenomenon which we term the motion of matter” \cite[pp. 224-225]{Clifford "The Common Sense of the Exact Sciences"}. \\
\indent — \textsc{W.K. Clifford}

\vspace{2mm}

[I]f we \emph{assume} as an axiom that space resists curvature with a resistance proportional to the change, we find that waves of “space-displacement” are precisely similar to those of the elastic medium which we suppose to propagate light and heat. We also find that “space-twist” is a quantity exactly corresponding to magnetic induction, and satisfying relations similar to those which hold for the magnetic field. It is a question whether physicists might not find it simpler to assume that space is capable of a varying curvature, and of a resistance to that variation, than to suppose the existence of a subtle medium pervading an invariable homoloidal [flat or Euclidean] space. \\
\indent — \textsc{K. Pearson}, annotation on \textsc{W.K. Clifford} \cite[pp. 225-226]{Clifford "The Common Sense of the Exact Sciences"}
		
\vspace{2mm}

A further development of the theory of elastic media\footnote{
	Cf. Section \ref{subsubsection "Spatio-temporal/Gravitational Thermodynamics, and Entropic Gravity"}.
	} 
in curved spaces will perhaps allow to answer a question by Clifford: \emph{if it were not possible that we consider as physical variations certain effects actually due to changes in the curvature of our space; in other words, if some of the causes, that we call physical, and maybe all of them, were not by chance due to the geometric constitution of the space in which we live}.\endnote{
	Original It. version: «L'ulteriore svolgimento della teoria dei mezzi elastici negli spazii curvi permetterà forse di rispondere alla domanda di Clifford: \emph{se non potrebbe darsi che noi consideriamo come variazioni fisiche certi effetti realmente dovuti a cambiamenti della curvatura del nostro spazio; in altre parole, se alcune delle cause, che noi chiamiamo fisiche, e forse tutte, non fossero per avventura dovute alla costituzione geometrica dello spazio nel quale viviamo}».
	} \\
\indent — \textsc{E. Cesàro} \cite[p. 213]{Cesaro "Introduzione alla teoria matematica della elasticita"} 
		
\endgroup 

\section{Gravitational Field as a Curvature of the Space}
\label{section "Gravitational Field as a Curvature of the Space"}

We continue our discussion started in Sections \ref{section "Excerpts from Memory: Ricci Methods"}-\ref{section "Lorentz–Minkowski 4-Manifolds"}.

\subsection{Einstein Field Equations}

\begingroup
\footnotesize
The answer is that it does not seem possible to draw any distinction between the warping of physical space and the warping of physical objects which define space [\,\dots]. The law of gravitation is not a law in the sense that it restricts the possible behaviour of the substratum of the world; it is merely the definition of a vacuum. We need not regard matter as a foreign entity causing a disturbance in the gravitational field; the disturbance is matter. \\
\indent — \textsc{A.S. Eddington} \cite[pp. 126, 190]{Eddington "Space Time and Gravitation. An Outline of the General Relativity Theory"}

\endgroup

\vspace{2mm}

The theoretical nucleus of general relativity, and that is the point of union between space-time (geometry) and matter (physics), within the Ricci calculus of tensors, is composed of the \emph{Einstein field equations} \cite[pp. 844-845]{Einstein "Die Feldgleichungen der Gravitation"} \cite[§ 4]{Einstein "Kosmologische Betrachtungen zur allgemeinen Relativitatstheorie"}:
\begin{subequations}
\label{subequations "Einstein field equations"}
\begin{align}
	\label{align "Einstein–Levi-Civita field equations"}
	\gravitation_{\mu\nu} & = \Ricci_{\mu\nu} - \frac{1}{2}g_{\mu\nu}\scalarcurvature = \Einsteinconstant\Tau_{\mu\nu} \\ 
	& = \Ricci_{\mu\nu} - \frac{1}{2}g_{\mu\nu}\scalarcurvature = \frac{8\pi{G_\textsc{n}}}{c^4}\Tau_{\mu\nu}, \\
	& = \Ricci_{\mu\nu} = \Einsteinconstant\left(\Tau_{\mu\nu} - \frac{1}{2}g_{\mu\nu}\Tau\right), \\
	& = -\frac{\partial}{\partial_{x_\xi}}\binomcurly{\mu\nu}{\xi} + \binomcurly{\mu\xi}{\varrho}\binomcurly{\nu\varrho}{\xi} + \frac{\partial^2\log\sqrt{-g}}{\partial_{x_\mu}\partial_{x_\nu}} - \binomcurly{\mu\nu}{\xi}\frac{\partial\log\sqrt{-g}}{\partial_{x_\xi}},
\end{align}
\end{subequations}
where $\gravitation_{\mu\nu}$ is the Einstein tensor \eqref{equation "Einstein Tensor"}, $\Ricci_{\mu\nu}$ is the Ricci curvature tensor (Section \ref{subsubsection "Ricci Curvature Tensor"}), $g_{\mu\nu}$ is the metric tensor, $\scalarcurvature$ is the scalar curvature (Ricci scalar) \eqref{subequations "Scalar curvature, i.e. Ricci scalar"}, $\Einsteinconstant = \frac{8\pi\gravitation}{c^4}$ is the Einstein (gravitational) constant \cite{Einstein "Zur allgemeinen Relativitatstheorie"}, that is the strength of coupling between matter (physical dimension) and geometric space, $\Tau_{\mu\nu}$ is the energy-momentum tensor \eqref{equation "Energy-momentum tensor as a variational derivative"}, $G_\textsc{n}$ is the Newtonian constant of gravitation, and $\binomcurly{\mu\nu}{\xi}$ etc. are the Christoffel symbols of the second kind \eqref{subequations "Christoffel symbols for the Levi-Civita connection"}. With the addition of the cosmological constant, denoted by $\mathrm{\Lambda}$,\footnote{
	By $\textgreek{\textnormal{λ}}$ in \cite[p. 151]{Einstein "Kosmologische Betrachtungen zur allgemeinen Relativitatstheorie"}: $\gravitation_{\mu\nu} - \textgreek{\textnormal{λ}}g_{\mu\nu} = -\Einsteinconstant\left(\Tau_{\mu\nu} - \frac{1}{2}g_{\mu\nu}\Tau\right)$.
	} we get 
\begin{equation}
\label{equation "Einstein field equations with cosmological constant"}
	\Ricci_{\mu\nu} - \frac{1}{2}g_{\mu\nu}\scalarcurvature + \mathrm{\Lambda}g_{\mu\nu} = \Einsteinconstant\Tau_{\mu\nu}.
\end{equation}

\subsubsection{Invariantiveness and Tensorial Conservation: Levi-Civita's Analytical Expression}
\label{subsubsection "Invariantiveness and Tensorial Conservation: Levi-Civita's Analytical Expression"}

\begingroup
\footnotesize
The mechanical meaning of the [tensorial] system [in general relativity, so that it is admissible] implies an analytical structure with convenient invariant properties [\textit{convenienti proprietà invariantive}] in the face of any coordinate transformations. \\
\indent — \textsc{T. Levi-Civita} \cite[pp. 381-382]{Levi-Civita "Sulla espressione analitica spettante al tensore gravitazionale nella teoria di Einstein"}

\vspace{2mm}

$A_{ik} + \Tau_{ik} = 0$\footnote{
	Same formula, but the order of the addends is reversed, in \cite[p. 338]{Levi-Civita "Sulla espressione analitica spettante al tensore gravitazionale nella teoria di Einstein"}.
	}
i.e. inertial[-gravitational] tensor and energy tensor are balanced [\textit{si fanno equilibrio}]. Namely the curvature of the space is such that modifying itself it balances [\textit{equilibra}] any external physical action. \\
\indent — \textsc{R. Marcolongo} \cite[p. 181]{Marcolongo "Calcolo differenziale assoluto e Teoria della Relativita"}

\endgroup

\vspace{2mm}

Levi-Civita \cite[§ 7]{Levi-Civita "Sulla espressione analitica spettante al tensore gravitazionale nella teoria di Einstein"} shows that the Bianchi identities \eqref{subequations "First Bianchi identity"} \eqref{subequations "Second Bianchi identity"} contain a \textit{formal justification} of the gravitational field Eqq. \eqref{subequations "Einstein field equations"},\footnote{
	An intrinsic form of \eqref{subequations "Einstein field equations"} is summarised in U. Cisotti \cite{Cisotti "Forma intrinseca delle equazioni gravitazionali nella relativita generale"}.
	} 
correcting a misunderstanding in Einstein \cite[p. 696]{Einstein "Naherungsweise Integration der Feldgleichungen der Gravitation"}, which was unable to impress an invariant character to the whole, and asserting the idea that the gravitation field is, from the mathematical perspective, a tensor-like phenomenon. We focus on \eqref{align "Einstein–Levi-Civita field equations"}: $\Ricci_{\mu\nu} - \frac{1}{2}g_{\mu\nu}\scalarcurvature = \Einsteinconstant\Tau_{\mu\nu}$. 

The right-hand side of \eqref{align "Einstein–Levi-Civita field equations"}, i.e. $\Einsteinconstant\Tau_{\mu\nu}$, is a double system, that is, a system with covariant and contravariant indices, and \emph{zero divergence}, cf. e.g. Eddington \cite[IV, sec. 54]{Eddington "The Mathematical Theory of Relativity"}: 
\begin{equation}
\label{equation "Energy-momentum conservation"}
	\nabla_\nu{\Tau_\mu}^\nu = \nabla_\nu\Tau^{\mu\nu} = {\Tau^{\mu\nu}}_{;\nu} 	= 0.
\end{equation}
The vanishing covariant divergence of $\Tau^{\mu\nu}$ shall therefore guarantee the energy-momentum \emph{conservation}. The completeness of \eqref{align "Einstein–Levi-Civita field equations"} requires, nevertheless, that the divergence of the left-hand side, i.e. $\Ricci_{\mu\nu} - \frac{1}{2}g_{\mu\nu}\scalarcurvature$, is equally \emph{identically null}. And this is precisely what the Bianchi identities express, see Eqq. \eqref{equation "Contracted Bianchi identity"} \eqref{equation "Einstein Tensor and contracted Bianchi identity"}. It is therefore appropriate, and historically more correct, the proposal (of O. Onicescu) to call the system of gravitational equations under the name of \emph{Einstein–Levi-Civita field equations}.

\subsubsection{Energy-Momentum Tensor as a Variational Derivative}

The energy-momentum tensor is a quantity for the density of matter, which represents the presence of a \emph{disturbance} in space-time (see point \ref{item "Energy-momentum tensor"} in Section \ref{subsection "The Role of Geometry in Gravitational Physics"}):
\begin{equation}
\label{equation "Energy-momentum tensor as a variational derivative"}
	\Tau_{\mu\nu} = \frac{2}{\sqrt{-g}}\frac{\delta\Bigl[\sqrt{-g}\Lagrangian = \Bigl(\sqrt{-g}\left(\frac{1}{2}g^{\mu\nu}\partial_\mu\textgreek{\textit{\ddigamma}}\partial_\nu\textgreek{\textit{\ddigamma}} - U\right)\Bigr)\Bigr]}{\delta{g}^{\mu\nu}},
\end{equation}
where $\Lagrangian \viz \Lagrangian_\mathrm{m} = \frac{1}{2}\partial_\mu\textgreek{\textit{\ddigamma}}\partial^\mu\textgreek{\textit{\ddigamma}} - U(\textgreek{\textit{\ddigamma}})$ is the Lagrangian density of the matter-energy, in which $\partial^\mu\textgreek{\textit{\ddigamma}} = \frac{\partial\Lagrangian}{\partial(\partial_\mu\textgreek{\textit{\ddigamma}})}$, with a scalar field $\textgreek{\textit{\ddigamma}}$ and a potential $U(\textgreek{\textit{\ddigamma}})$. 

It can be described as a \emph{functional} in the calculus of variations. Indicating by $\mathscr{S}_\mathrm{m}$ the action functional, referred to as a \emph{matter action}, useful to describe the dynamics of gravitational fields within the doubly interactive scheme of \emph{matter-geometry}, by $\delta$ the metric $g$-variations, and by $\Omega \subset \mathbb{M}^4$ a domain of integration of dimension 4, i.e. a 4-volume corresponding to a region (or portion) of Minkowski space-time, we can write the variation of the action integral,
\begin{align}
\label{align "Variation of an action integral with energy-momentum tensor"}
	\delta\mathscr{S}_\mathrm{m} & = \int_{\Omega \subset \mathbb{M}^4}\left\{\frac{\partial \cdot \sqrt{-g}\Lagrangian}{\partial g^{\mu\nu}}\delta g^{\mu\nu} + \frac{\partial \cdot \sqrt{-g}\Lagrangian}{\partial(\partial_\xi g^{\mu\nu})}\partial_\xi\delta g^{\mu\nu} + \cdots \right\}d^4x \notag \\
	& = \int_{\Omega \subset \mathbb{M}^4}\left(\delta \cdot \sqrt{-g}\Lagrangian\right)d^4x = \left(\frac{1}{2}\int_{\Omega \subset \mathbb{M}^4}\sqrt{-g}\Lagrangian\Tau_{\mu\nu}\delta g^{\mu\nu}\right)d^4x,
\end{align}
where $d^4x = d(t, x, y, z)$, and again $\Lagrangian \viz \Lagrangian_\mathrm{m}$, in which the tensor \eqref{equation "Energy-momentum tensor as a variational derivative"} appears.

\subsection{Einstein–Hilbert (Gravitational) Action}

The variation of $\mathscr{S}_\mathrm{m}$ \eqref{align "Variation of an action integral with energy-momentum tensor"} brings us straight to the \emph{Einstein–Hilbert action} \cite{Hilbert "Die Grundlagen der Physik. (Erste Mitteilung)"} \cite{Hilbert "Die Grundlagen der Physik. (Zweite Mitteilung)"}, that is, the action from which it is possible to reconstruct the Eqq. \eqref{subequations "Einstein field equations"}:
\begin{equation}
\label{equation "Einstein–Hilbert (Gravitational) Action"}
	\mathscr{S}_\textsc{eh} = -\frac{c^4}{16\pi G_\textsc{n}} \int\scalarcurvature\sqrt{-g}d^4x = -\frac{1}{2\Einsteinconstant} \int_{\Omega \subset \mathbb{M}^4}\scalarcurvature\sqrt{-g}d^4x.
\end{equation}

The variation of $\mathscr{S}_\textsc{eh}$ will be:
\begin{align}
	\delta\mathscr{S}_\textsc{eh} & = -\frac{1}{2\Einsteinconstant} \int_{\Omega \subset \mathbb{M}^4}\delta\left(\scalarcurvature\sqrt{-g}\right)d^4x = -\frac{1}{2\Einsteinconstant} \int_{\Omega \subset \mathbb{M}^4}\delta\left(g^{\mu\nu}\Ricci_{\mu\nu}\sqrt{-g}\right)d^4x \notag \\
		& = -\frac{1}{2\Einsteinconstant}\int_{\Omega \subset \mathbb{M}^4} \left(\Ricci_{\mu\nu}\sqrt{-g}\delta g^{\mu\nu} + g^{\mu\nu}\Ricci_{\mu\nu}\delta\sqrt{-g} + g^{\mu\nu}\delta \Ricci_{\mu\nu}\sqrt{-g}\right)d^4x \notag \\
	& = -\frac{1}{2\Einsteinconstant}\int_{\Omega \subset \mathbb{M}^4}\sqrt{-g}\left\{\left(\Ricci_{\mu\nu} - \frac{1}{2}g_{\mu\nu}\scalarcurvature\right)\delta g^{\mu\nu} + g^{\mu\nu}\delta \Ricci_{\mu\nu}\right\}d^4x \notag \\
	& = -\frac{1}{2\Einsteinconstant}\int_{\Omega \subset \mathbb{M}^4}\sqrt{-g}\left(\gravitation_{\mu\nu}\delta g^{\mu\nu}\delta \Ricci_{\mu\nu}\right)d^4x.
\end{align}

\begin{margo}
On the priority (Einstein vs. Hilbert) in \cite{Hilbert "Die Grundlagen der Physik. (Erste Mitteilung)"} for the correct exposition of the gravitational field equations in an explicit form, see \cite{Corry Renn Stachel "Belated Decision in the Hilbert-Einstein Priority Dispute"}. \margosymbol
\end{margo}

\subsection{Palatini Identity, and the Variation of the Gravitational Action}

\begingroup
\footnotesize
After the gravitational equations were discovered, thanks to Einstein, an attempt was made to derive them from a variational principle, as, in ordinary mechanics, one derives the [Euler–]Lagrange equations from the Hamilton principle. The goal was achieved by Einstein himself, establishing a new principle of Hamilton, further articulated by Hilbert \cite{Hilbert "Die Grundlagen der Physik. (Erste Mitteilung)"} and Weyl \cite{Weyl "Zur Gravitationstheorie"} \cite[2ed]{Weyl "Raum-Zeit-Materie: Vorlesungen Uber Allgemeine Relativitatstheorie 1918"}. The procedures followed by these authors, however, do not conform to the spirit of the absolute differential Calculus, because invariant equations (in the face of changes of variables) arise from these procedures, passing through other formulæ that have no invariant character. I intend to achieve the same goal preserving the invariance in the subsequent formulæ [\textit{mi propongo di raggiungere il medesimo scopo conservando l'invarianza nelle successive formule}] being introduced. \\
\indent — \textsc{A. Palatini} \cite[pp. 203-204]{Palatini "Deduzione invariantiva delle equazioni gravitazionali dal principio di Hamilton"}

\endgroup

\vspace{2mm}

It is known that the gravitational action \eqref{equation "Einstein–Hilbert (Gravitational) Action"} can also be written in another way, with a method of variation concocted by A. Palatini \cite{Palatini "Deduzione invariantiva delle equazioni gravitazionali dal principio di Hamilton"} = \cite{Palatini "Invariant Deduction of the Gravitational Equations from the Principle of Hamilton"}, which, unlike the Einstein–Hilbert approach, maintains a \emph{total invariant} character throughout the process of deducing the gravitational Eqq. \eqref{subequations "Einstein field equations"} from the variational principle.

We first define the notion of identity according to Palatini. Setting the variation $\delta\Ricci_{\mu\nu}$ of the Ricci curvature tensor \eqref{equation "Ricci tensor in explicit form"}, plus the variation $\delta\Gamma$ of the Christoffel symbols (Section \ref{section "Christoffel Symbols"}), 
\begin{equation}
	\delta\Ricci_{\mu\nu} = \partial_\xi\delta{\Gamma_{\mu\nu}}^\xi - \partial_\nu\delta{\Gamma_{\mu\xi}}^\xi + \delta{\Gamma_{\mu\nu}}^\xi{\Gamma_{\xi\varrho}}^\varrho + {\Gamma_{\mu\nu}}^\xi\delta{\Gamma_{\xi\varrho}}^\varrho - \delta{\Gamma_{\mu\xi}}^\varrho{\Gamma_{\nu\varrho}}^\xi - {\Gamma_{\mu\xi}}^\varrho\delta{\Gamma_{\nu\varrho}}^\xi,
\end{equation}
and then adding the covariant derivative $\nabla$, finally we get to the \emph{Palatini identity}:
\begin{equation}
	\idem_\textsc{p} = \left\{\delta\Ricci_{\mu\nu} = \nabla_\xi \left(\delta{\Gamma_{\mu\nu}}^\xi\right) - \nabla_\nu \left(\delta{\Gamma_{\mu\xi}}^\xi\right)\right\}.
\end{equation}
The variation formula for the Einstein–Hilbert action turns into
\begin{subequations}
\begin{align}
	\delta\mathscr{S}_\textsc{p} & = -\frac{1}{2\Einsteinconstant}\int_{\Omega \subset \mathbb{M}^4} 
	\biggl\{ 
	\left(\Ricci_{\mu\nu} - \frac{1}{2}g_{\mu\nu}\scalarcurvature\right)\delta g^{\mu\nu}\sqrt{-g} + \sqrt{-g} \notag \\
	& \hspace{11pt} \times \left[\nabla_\xi\left(g^{\mu\nu}\delta{\Gamma_{\mu\nu}}^\xi\right) - \nabla^\mu \left(\delta{\Gamma_{\mu\xi}}^\xi\right)\right] 
	\biggr\}d^4x, \\
	& = -\frac{1}{2\Einsteinconstant}\int_{\Omega \subset \mathbb{M}^4}\left(\Ricci_{\mu\nu} - \frac{1}{2}g_{\mu\nu}\scalarcurvature\right)\delta g^{\mu\nu}\sqrt{-g}d^4x,
\end{align}	
\end{subequations}
which we may call the \emph{Palatini variation of the gravitational action}.

\begin{margo}
Palatini's flash of inspiration is in showing that the $\delta$-variations of the Christoffel symbols are the components of the curvature tensor, and this independently of any choice of a symmetric affine connection. Such a solution should not be confused, as frequently happens, with the so-called \emph{Palatini method of variation}, see e.g. \cite{El-Kholy Sexl and Urbantke "On the so-called "Palatini method" of variation in covariant gravitational theories"} \cite{Tsamparlis "On the Palatini method of variation"}, which was successively conceived by Einstein \cite{Einstein "Einheitliche Feldtheorie von Gravitation und Elektrizitat}, even if the ideational substratum of this proposal is already in Palatini. The Palatini method consists of assuming that the metric tensor field $g$ and the Levi-Civita connection (Section \ref{subsection "Levi-Civita Connection Theorem on a (pseudo-)Riemannian Manifold"}) are \emph{independent variables}, along with the intention of developing a \emph{non-symmetric} $\Gamma$-connection, but a procedure of this kind is not present in the original paper \cite{Palatini "Deduzione invariantiva delle equazioni gravitazionali dal principio di Hamilton"}. For a historical reconstruction, see \cite{Ferraris Francaviglia Reina "Variational formulation of general relativity from 1915 to 1925 "Palatini's method" discovered by Einstein in 1925"}. \margosymbol
\end{margo}

\subsection{\emph{Vetturale} of Energy Radiated in Gravitational Waves}
\label{subsection "Vetturale of Energy Radiated in Gravitational Waves"}

\begingroup
\footnotesize
Now what is there \emph{analogous} to magnetic force in the gravitational case? And if it have its analogue, what is there to correspond with electric current? [\,\dots] [R]results will be sensibly [\,\dots] expressed in terms of wave-propagation. \\
\indent — \textsc{O. Heaviside} \cite[pp. 457, 460, e.a.]{Heaviside "A Gravitational and Electromagnetic Analogy"}

\vspace{2mm}

[W]hen we talk about the position or velocity of the attracting body, it is its position or its velocity when the gravitational wave [\textit{l'onde gravifique}] leaves it; for the attracted body, on the contrary, it is its position or its velocity when the gravitational wave reaches it, assuming this wave propagates with the speed of light [\textit{vitesse de la lumière}]. \\
\indent — \textsc{H. Poincaré} \cite[§ 9, p. 174]{Poincare "Sur la dynamique de l'electron"}

\endgroup

\vspace{2mm}

When it is \emph{perturbed}, space-time—in spite of the Einstein's initial (and persevering) skepticism, or the Eddington's \cite{Eddington "The Propagation of Gravitational Waves"} refusal—is the \emph{vetturale}\footnote{
	Term used by Leonardo da Vinci about the water, in \textit{Manoscritto K} (1504-), Institut de France, Paris, foglio 2 recto: «lacqua el vetturale della natura» (\emph{water nature's cart}).
	}
(conveyor, carrier) of energy in the form of gravitational radiation. One thinks of the gravitational waves \cite{Einstein "Uber Gravitationswellen"} \cite{Einstein and Rosen "On gravitational waves"}, the ripples in the construction of space-time, which can be flat or curved.

\subsubsection{Quadrupolarity and Transverse-Traceless Gauge}
\label{subsubsection "Quadrupolarity and Transverse-Traceless Gauge"}

\begingroup 
\footnotesize
It turns out that the structure that Einstein was seeking was the gauge field: It is a geometrical structure [\,\dots]; the simplest Abelian gauge field is Maxwell's electromagnetic field [\,\dots]. That gauge fields are deeply related to the geometrical concept of connections on fiber bundles has been appreciated by physicists only in recent years.\footnote{
	But compare with point \ref{item "Yang on gauge fields and fiber bundles"} in Section \ref{subsection "Physics is (Not) Mathematics"}.
	} \\
\indent — \textsc{C.N. Yang} \cite[p. 44]{Yang "Einstein's impact on theoretical physics"}

\endgroup

\vspace{2mm}

We can think of the gravitational waves in \emph{analogy} to electrodynamics, in which electromagnetic waves, i.e. electric and magnetic fields, are produced by accelerated charges through empty space (void, vacuum) at the same constant speed of light.

Gravitational waves have a very similar behavior, but with the emergence of \emph{geometric fluctuations or disturbances} in respect of a curvature-undulation in vacuum, and with a \emph{quadrupole moment}, instead of a dipole moment, typical for electromagnetic waves. The quadrupolar radiation formula \cite{Einstein "Uber Gravitationswellen"} is 
\begin{equation}
	h_{\mu\nu} = \frac{2G_\textsc{n}}{c^4\distance}\ddot{\quadrupole}_{\mu\nu}, \text{ with } \quadrupole_{\mu\nu} = \int\rho_Q\Bigl(x_\mu x_\nu - \tfrac{1}{3}x^\xi x_\xi\delta_{\mu\nu}\Bigr)d^3x,
\end{equation}
where $h_{\mu\nu} \viz h^{\textsc{tt}}_{\mu\nu}$ represents \emph{small perturbations} or \emph{deviations} from the flatness, i.e. the \emph{gravitational wave-field}, corresponding to a symmetric tensor field of rank 2, $\distance = |x|$ is the length of a position vector, i.e. the distance to the source, and $\quadrupole_{\mu\nu} \viz \quadrupole^{\textsc{tt}}_{\mu\nu}$ is the quadrupole moment tensor, in which $\rho_Q$ is the source density; the double \textsc{tt} means \emph{transverse-traceless gauge}, and its conditions are the following:
\begin{equation}
\label{equation "Transverse-traceless gauge conditions"}
	h_{\mu 0} = 0, \enspace h_{\mu\mu} = 0, \enspace {h_\mu}^\mu = 0, \enspace h_{\mu\nu, \nu} = 0, \enspace \partial_\mu h^{\mu\nu} = 0, \text{ putting } h \viz h^{\textsc{tt}}.
\end{equation}

There are two types of propagation. Gravitational waves propagate 
\enumerationisinitium
\item in a \emph{flat space-time \textnormal{\background}-background (Minkowski vacuum)},
\item or in a \emph{curved space-time \textnormal{\background}-background}. 
\enumerationisfinis

\subsubsection{Flux of Gravitational Radiation in the Minkowski Vacuum}

If the propagation is in the Minkowski vacuum, the metric can be written as  
\begin{equation}
	g_{\mu\nu} = \left(\eta_{\mu\nu} \viz g^\background_{\mu\nu}\right) + h_{\mu\nu}, 
\end{equation}
where $\eta_{\mu\nu} \viz g^\background_{\mu\nu}$ and $h_{\mu\nu}$ are the background and perturbation parts, respectively.

Without a source energy-momentum tensor, namely $\Tau_{\mu\nu} = 0$, the metric tensor fluctuations in a flat background is expressed with the d'Alembert operator or, more commonly, d'Alembertian \cite{d'Alembert "Recherches sur la Courbe que forme une Corde tendue mise en vibration; Suite des Recherches sur la Courbe que forme une Corde tendue mise en vibration", d'Alembert "Addition au Memoire sur la Courbe que forme une Corde tendue mise en vibration}, cf. Eq. \eqref{equation "d'Alembertian"}
\begin{equation}
\label{equation "Fluctuation of the flat space-time metric"}
	\dAlembertian h_{\mu\nu} = 0, \enspace \dAlembertian = \eta^{\mu\nu}\partial_\mu\partial_\nu,
\end{equation}
under $\partial^\nu h_{\mu\nu} = \frac{1}{2}\partial_\mu h$, that is, a \emph{harmonic gauge condition}.

The action useful for determining the free dynamics propagation of a gravitational wave in the flat space-time is
\begin{align}
	\mathscr{S}_\mathrm{gw}^{(2)} & = -\frac{1}{2\Einsteinconstant}\int\left(\left(\sqrt{-g}g^{\nu\xi}\right)^{(2)}\Ricci^{(0)}_{\nu\xi} + \left(\sqrt{-g}g^{\nu\xi}\right)^{(0)}\Ricci^{(2)}_{\nu\xi} + \left(\sqrt{-g}g^{\nu\xi}\right)^{(1)}\Ricci^{(1)}_{\nu\xi}\right)dx^4 \notag \\
	 & = -\frac{1}{2\Einsteinconstant}\int\left(\left(\sqrt{-g}g^{\nu\xi}\right)^{(2)} + \eta^{\nu\xi}\Ricci^{(2)}_{\nu\xi} + \frac{1}{2}h^{\nu\xi}\dAlembertian h_{\nu\xi}\right)dx^4 \notag \\
	& = \frac{c^4}{32\pi G_\textsc{n}}\int\frac{1}{2}\left(\partial_\mu h^{\xi\varrho}\partial^\mu h_{\xi\varrho}\right)d^4x.
\end{align}

The variation $\delta\mathscr{S}_\mathrm{gw}^{(2)} = \frac{c^4}{32\pi G_\textsc{n}}\int\frac{1}{2}\left(\partial_\mu h^{\xi\varrho}\partial_\nu h_{\xi\varrho}\delta g^{\mu\nu}\cdots\right)\sqrt{-g}d^4x$ allows us to determine the energy-momentum tensor,
\begin{equation}
	\Tau^\mathrm{gw}_{\mu\nu} = \frac{c^4}{32\pi G_\textsc{n}}\partial_\mu h^{\xi\varrho}\partial_\nu h_{\xi\varrho}, 
\end{equation}
which guarantees the description of the flux of energy, relatively to the radiated wave-field, in the \textsc{tt} gauge \eqref{equation "Transverse-traceless gauge conditions"}, so ${\Tau_\nu}^\nu = 0$ and $\partial^\nu\Tau^\mathrm{gw}_{\mu\nu} = 0$, in accordance with the Eq. \eqref{equation "Fluctuation of the flat space-time metric"} of vacuum metric disturbances.

\begin{scholium}
~\enumerationisinitium
\item Note that the background metric $g^\background_{\mu\nu}$ is a \emph{neutral}, that is, it can be flat (Minkowski metric) or curved (non-pseudo-Euclidean metric).
\item The flat Minkowski background (vacuum) is like a rigid \emph{stage} of the propagative event (cf. Section \ref{subsection "Concepts (?) of Space-Time"}); it must also be remembered that the Minkowski space-time $\mathfrak{M}^4 \viz \mathbb{M}^4$ (Section \ref{subsection "Minkowski Space-Time (Flat Metric)"}) is a manifold equipped with a \emph{strongly asymptotic flatness}, having a \emph{global dynamic stability} \cite{Christodoulou and Klainerman "The Global Nonlinear Stability of the Minkowski Space"}. \scholiumsymbol
\enumerationisfinis
\end{scholium}

\subsubsection{Formalism for a Background Curvature}

The presence of perturbations of the metric tensor in a curved background is given by 
\begin{equation}
	g_{\mu\nu} \to g^\background_{\mu\nu} + (h_{\mu\nu} = \delta g_{\mu\nu}).
\end{equation}	
The \textsc{tt} gauge, in this case, is the same as \eqref{equation "Transverse-traceless gauge conditions"} but with covariant behavior, so 
\begin{equation}
	\begin{rcases}	
	\nabla^\nu h_{\mu\nu} \\ 
	g^{\mu\nu}h_{\mu\nu} \\
	g^{\mu\nu}\nabla_\mu\nabla_\nu h
	\end{rcases}
	= 0.
\end{equation}

As for the dynamics of the gravitational field, the reference equation is that of Einstein–Levi-Civita \eqref{align "Einstein–Levi-Civita field equations"}; with the energy-momentum tensor, it takes this form:
\begin{align}
	& \Ricci^\background_{\mu\nu} - \frac{1}{2}g^\background_{\mu\nu}\scalarcurvature^\background = 8\pi G_\textsc{n}\Tau^{\mathrm{gw}}_{\mu\nu}, \\
	& \Tau^\mathrm{gw}_{\mu\nu} = -\frac{1}{8\pi G_\textsc{n}}\left(\left\langle\Ricci^{(2)}_{\mu\nu}\right\rangle - \frac{1}{2}g^\background_{\mu\nu}\left\langle\scalarcurvature^{(2)}\right\rangle\right),
\end{align}
where $\scalarcurvature^{(2)} = \Ricci^{(2)}_{\mu\nu}g^{\background\hspace{0.5pt}\mu\nu}$.

\subsubsection{Contour Jottings}

\enumerationisinitium
\item We cite here three cardine studies.
\subenumerationisinitium
\item The Einstein–Rosen solution \cite{Einstein and Rosen "On gravitational waves"} deals with the theoretical problem of gravitational waves by reducing it to the cylindrical waves in a (flat) euclidean space. 
\item F.A.E. Pirani's research, with H. Bondi and I. Robinson, 

(a) examines and summarizes \cite{Pirani "Invariant Formulation of Gravitational Radiation Theory"} the gravitational wavefronts in terms of a \emph{discontinuity in the Riemann tensor} across a null 3-surface, establishing criteria for a general interpretation of the plane wave metric and the gravitational radiation; 

(b) introduces \cite{Bondi Pirani and Robinson "Gravitational waves in general relativity III. Exact plane waves"} plane gravitational waves as \emph{non-flat} solutions of Eqq. \eqref{subequations "Einstein field equations"} for an empty space-time, having the same symmetry as the plane electromagnetic waves.
\subenumerationisfinis
\item The first detection \cite{Abbott et al. LIGO Virgo "Observation of Gravitational Waves from a Binary Black Hole Merger"} \cite{Abbott et al. LIGO Virgo "Directly comparing GW150914 with numerical solutions of Einstein's equations for binary black hole coalescence"} of gravitational waves is related to the phenomenon of \emph{binary black hole coalescence}. A later detection \cite{Abbott et al. LIGO Virgo "GW170817: Observation of Gravitational Waves from a Binary Neutron Star Inspiral"} concerns a merging of two neutron stars. On a theoretical level, a first accurate simulation of a coalescence of binary black holes goes back to F. Pretorius \cite{Pretorius "Evolution of Binary Black-Hole Spacetimes"} \cite{Pretorius "Simulation of binary black hole spacetimes with a harmonic evolution scheme"}; equations for the \emph{Schwarzschild} \eqref{equation "Schwarzschild metric"} and \emph{Kerr} \eqref{equation "Kerr metric"} \emph{metrics} are in Section \ref{subsubsection "Schwarzschild, Gödel and Kerr Metrics"}, that are the basis for understanding the behavioral nature of black holes.
\enumerationisfinis

\section{Lovelock's Scalar Lagrangian Density of the Gravitational Field} 
\label{section "Lovelock's Scalar Lagrangian Density of the Gravitational Field"} 

D. Lovelock \cite{Lovelock "The Uniqueness of the Einstein Field Equations in a Four-Dimensional Space"} \cite{Lovelock "Degenerate Lagrange Densities Involving Geometric Objects"} \cite{Lovelock "Divergence-Free Tensorial Concomitants"} \cite{Lovelock "The Einstein Tensor and Its Generalizations"} \cite{Lovelock "The Four-Dimensionality of Space and the Einstein Tensor"} \cite{Lovelock "Divergence-Free Third Order Concomitants of the Metric Tensor in Three Dimensions"} \cite[sec. 8.3, co-written with H. Rund]{Lovelock and Rund "Tensors Differential Forms and Variational Principles"} showed that, in a \emph{4-dimensional $\mathscr{C}^\infty$ manifold}, i.e. for \emph{(2-, 3- and) 4-dimensional spaces} (the starting model is the flat Minkowski space-time with non-gravitational interactions), 
\enumerationisinitium
\item the Einstein field equations with cosmological constant $\mathrm{\Lambda}$, see Eq. \eqref{equation "Einstein field equations with cosmological constant"}, are the \emph{only} admissible second order Euler–Lagrange equations, see Eqq. \eqref{subequations "Suitable form of Euler–Lagrange equations"};
\item the Einstein tensor \eqref{equation "Einstein Tensor"} is the \emph{only} tensor of rank 2 having the properties of symmetry and zero divergence. 
\enumerationisfinis
The two statements are \emph{no longer valid in dimension higher than 4}. This has involved a generalization of the (geo)metric theory of gravitation.

\begin{propositio}
Let $\varphi^\alpha = \varphi(x^\mu)$, with $\alpha = 1, \mathellipsis, m$ and $\mu = 1, \mathellipsis, n$, denote $m$ quantities on an $n$-dimensional space. Setting the Lagrangian as $\Lagrangian \viz \Lagrangian\bigl(\varphi^\alpha, \varphi^\alpha_{\mu_1}, \varphi^\alpha_{\mu_1 \cdots \mu_r}\bigr)$, and putting $\varphi^\alpha_{\mu_1 \cdots \mu_j} = \frac{\partial^j\varphi^\alpha}{\partial{x}^{\mu_1} \cdots \partial{x}^{\mu_j}}$, with $1 \leqslant j \leqslant r$, the system of Euler–Lagrange equations will be 
\begin{equation}
	\frac{\partial\Lagrangian}{\partial\varphi^\alpha} + \sum^r_{j = 1} (-1)^j\frac{\partial^j}{\partial{x}^{\mu_1} \cdots \partial{x}^{\mu_j}}\left(\frac{\partial\Lagrangian}{\partial\varphi^\alpha_{\mu_1 \cdots \mu_j}}\right) = 0.
\end{equation}
Now, we consider the Lagrangian as a \emph{scalar density} (so that the Euler–Lagrange equations are tensorially determined), and we write it in these forms, 
\begin{subequations}
	\label{equation "Scalar Lagrangian density"}
	\begin{empheq}[left = {\Lagrangian \equival \empheqlbrace}]{align}
	& 
	\label{equation "Scalar Lagrangian density of type I"}
	\Lagrangian(g_{\mu\nu}, g_{\mu\nu, \xi_1}, g_{\mu\nu, \xi_1, \mathellipsis, \xi_r}), \\
	& \Lagrangian(g_{\mu\nu}, g_{\mu\nu, \xi}, g_{\mu\nu, \xi\varrho}),
	\end{empheq}
\end{subequations}
namely as a function of the metric tensor $g_{\mu\nu}$ and its first two derivatives. 

Let $\EulerLagrange^{\mu\nu} \viz \EulerLagrange^{\mu\nu}(\Lagrangian)$ be the \emph{Euler–Lagrange operator}, the components of which correspond to those of a symmetric \emph{$\binom{2}{0}$-tensor density}, i.e. $\EulerLagrange^{\mu\nu} = \EulerLagrange^{\nu\mu}$. More specifically, the Euler–Lagrange expression correlated to the scalar $\Lagrangian$-density is
\begin{equation}
\label{equation "Euler–Lagrange expression correlated to the scalar density"}
	\EulerLagrange^{\mu\nu} = \frac{\partial\Lagrangian}{\partial{g}_{\mu\nu}} + \sum^r_{j = 1} (-1)^j\frac{\partial^j}{\partial{x}^{\xi_1} \cdots \partial{x}^{\xi_j}}\left(\frac{\partial\Lagrangian}{\partial{g}_{\mu\nu, \xi_1 \cdots \xi_j}}\right).
\end{equation}
Then, 
\enumerationisinitium
\item for any scalar density, like that in \eqref{equation "Scalar Lagrangian density of type I"}, the divergence components of the Euler–Lagrange tensor density \eqref{equation "Euler–Lagrange expression correlated to the scalar density"} itself vanishes identically, is automatically null, 
\begin{equation}
	{\EulerLagrange^{\mu\nu}}(\Lagrangian)_{|\nu} = 0.
\end{equation}
\item If the space is a torsionless 4-space, the \emph{only} second order Euler–Lagrange equations arising from the scalar $\Lagrangian$-density \eqref{equation "Scalar Lagrangian density"} are
\begin{equation}  
	c_{(1)}\left(\Ricci^{\mu\nu} - \frac{1}{2}g^{\mu\nu}\scalarcurvature\right) + c_{(2)}g^{\mu\nu} = 0,	
\end{equation}
which coincides with the Einstein's Eq. \eqref{equation "Einstein field equations with cosmological constant"} with cosmological constant $\mathrm{\Lambda}$ (the values $c_{(1)}$ and $c_{(2)}$ are two simple constants). It also entails that the Einstein field equations with $\mathrm{\Lambda}$ are the \emph{only} obtainable second order Euler–Lagrange equations in $4\mathrm{D}$. (The same speech is worth for the sole Einstein tensor). This is false in $\mathrm{D}$-dimensional space if $\mathrm{D} > 4$. The Euler–Lagrange expression, rewritten with the parameters in general relativity, becomes
\begin{equation}
	\EulerLagrange^{\mu\nu} = c_{(1)}\sqrt{g}\left(\Ricci^{\mu\nu} - \frac{1}{2}g^{\mu\nu}\scalarcurvature\right) - \frac{1}{2}c_{(2)}\sqrt{g}g^{\mu\nu}. 
\end{equation}
\enumerationisfinis
\end{propositio}

\begin{scholium}
Lovelock's theory is one of the cases where the mathematical toolbox (in this context, we talk about the space dimensionality) \emph{constrains} the physics that it describes. To put it differently, it is a direct example in which physics \emph{comes straight} from mathematics. \scholiumsymbol
\end{scholium}

\section{Historico-Philological Remarks}
\label{section "Historico-Philological Remarks"}

\subsection{Concepts (?) of Space-Time}
\label{subsection "Concepts (?) of Space-Time"}

\enumerationisinitium
\item Minkowski space-time is, in the Einsteinian perspective, a rigid \emph{palcoscenico} (\emph{Bühne}) distinct or separate from matter and energy, because gravitational effects are supposed to be absent (see e.g. point \ref{item "Minkowski space-time and its characteristic"} in Section \ref{subsection "Minkowski Space-Time (Flat Metric)"}); about the Minkowski space-time, in proper parlance, it is said that it has an «existence independent» (\textit{selbständige Existenz}) of matter. However, in general relativity, this independence is fundamentally conceptual, mathematically useful, but it is denied on the physical level (but cf. Margo \ref{margo "Faradayian deformed space"}).

For Einstein, the condition of matter-energy free regions of space-time is an idealization: absence of gravitational forces means \emph{assuming} that, in certain cases, gravity has a negligible effect. Such a de facto condition is not consistent with experimental data, because there is no a candid way of testing whether a stand-alone space-time that is independent of its content of matter-energy really exist or not. On the subatomic scale, a separate existence or physical reality of space-time loses the sense of itself; besides, talking here about \emph{existence} has a finely \emph{math-operational significance}.

We quote some Einstein's passages: «Physical objects are not \emph{in space} [\textit{im Räume}], but these objects are \emph{spatially extended} [\textit{räumlich ausgedehnt}]» \cite{Einstein "Note to the 15th Edition"}. «[T]he geometric properties of space are not independent, but they are determined by matter [\,\dots]. There is no an empty space [\textit{leeren Raum}], i.e. a space without [gravitational] field» \cite[pp. 74, 107]{Einstein "Uber die Spezielle und Allgemeine Relativitatstheorie"}. Space-time has no physical existence (\textit{reale Existenz}), but it is only a metric set—that is to say, it is a \emph{mathematical structure}—of the gravitational field; or see \cite[p. 271]{Einstein "Kritisches zu einer von Hrn. De Sitter gegebenen Losung der Gravitationsgleichungen"}, which is a criticism against the de Sitter's solution \cite{de Sitter "On the curvature of space"} of gravity field equations: «[T]here cannot be a $g_{\mu\nu}$-field, i.e. a space-time continuum, without matter that generates it».

\item But beware: Einstein's view is fluctuating over the years. Contrary to what we have just read, he says in \cite[p. 610]{Einstein "Professor Einstein's Address at the University of Nottingham"}: «The strange conclusion to which we have come is this—that now it appears that space will have to be regarded as a primary thing and that matter is derived from it, so to speak, as a secondary result. Space is now turning around and eating up matter. We have always regarded matter as a primary thing and space as a secondary result. Space is now having its revenge»; and in \cite[p. 8 in the typescript with hand written corrections]{Einstein "Raum Ather und Feld in der Physik}: «Space [after swallowing ether, time, field and particles] remains the sole medium [substrate] of reality [\textit{alleiniger Träger der Realität}]».
\item We then have three types of space-time:
\subenumerationisinitium
\item \emph{Minkowski space-time} (in special relativity): 
	
	· stage space-time,
	
	· flat, empty, and ambient-like pseudo-Euclidean space, 

	· old-fashioned substratum (\textgreek{ὑποκειμένος χῶρος}), which underlies the matter and has an existence independent of matter;
	
\item \emph{Curved space-time}, or \emph{locally equivalent to (a set of flat frames identifying) a curved space-time} (in general relativity of the early period): 
	
	· non-pseudo-Euclidean space,
	
	· metric not capable of separate or independent existence of the matter fields;
\item 
\label{item "Space-entity"}
\emph{Space-entity} (in general relativity of the last period):

	· space with independent existence by itself, 
	
	· autonomous Reality, with a Newtonian reminiscence,\footnote{
	In Newton \cite[Definitiones, Scholium, p. 5]{Newton "Philosophiae Naturalis Principia Mathematica 1687"} \cite[p. 9]{Newton "The Mathematical Principles of Natural Philosophy I"}, both space and time are an \textit{absolutum}, \textit{in se \& natura sua absq\textnormal{[}ue\textnormal{]} relatione ad externum}: they are as something that is «absolute», «in its own nature, without regard to any thing external».
	} 
that destroys and devours the whole matter and time (neo-myth of a \textgreek{πανδαμάτωρ χῶρος}).
\subenumerationisfinis
\enumerationisfinis

\begin{margo}[Newtonianism and Euclidicity for gravity in Levi-Civita's view]
\label{margo "Newtonianism and Euclidicity for gravity in Levi-Civita's view"}
~\enumerationisinitium
\item The fact that space-time is, in an equivocal way, physically interpreted on several levels, is reflected, mathematically, in the possibility of guessing, as Levi-Civita shows in his Notes I-IX \cite{Levi-Civita "$ds^2$ einsteiniani in campi newtoniani. I-IX"} on Einsteinian $ds^2$ in Newtonian fields, a one-to-one correspondence between classical fields (starting from Newton) and relativistic fields, in a static model—and not just because they coincide locally (cf. point \ref{item "Minkowski space in general relativity"} in Section \ref{subsubsection "Vector Spaces L^4"}), but also because they \emph{globally} have in common some groups of symmetry. One of the Levi-Civita's exact solutions of field equations has the particularity that the relativistic space is \emph{strictly Euclidean}, and the \emph{gravitational field} is presented as a \emph{series of parallel planes}, under which \emph{relativistic-gravity phenomena} are associated with \emph{equipotential surfaces} the intensity of which varies from one plane to another \cite[Note V-VII]{Levi-Civita "$ds^2$ einsteiniani in campi newtoniani. I-IX"}. 
\item The classicality in the Einstein 4-space emerges e.g. by applying the \emph{d'Alembert–Levi-Civita version} of \eqref{align "Einstein–Levi-Civita field equations"}:
\begin{equation}
	\frac{1}{\Einsteinconstant} \cdot \Ricci_{\mu\nu} - \frac{1}{2}g_{\mu\nu}\scalarcurvature + \Tau_{\mu\nu} = 0.
\end{equation}
In this way of looking, cf. \cite[II, § 2, p. 96]{Levi-Civita "Fondamenti di meccanica relativistica"}, the general theory of relativity is but a correction of classical mechanics. 
\item We point out in passing that the mathematical predictions of gravitational waves, among the various methods at our disposal, may be made via \emph{post-Newtonian approximation} \cite{Blanchet Iyer Will and Wiseman "Gravitational waveforms from inspiralling compact binaries to second-post-Newtonian order"}, whose root is to be sought in the intuitions of Levi-Civita \cite{Levi-Civita "Le Probleme des n corps en relativite generale"} about the curved $n$-body problem. \margosymbol
\enumerationisfinis 
\end{margo}

\begin{margo}[Faradayian deformed space]
\label{margo "Faradayian deformed space"}
The idea of an empty space (void of matter) having physical properties, as opposed to Newton's absolute space, which is a “space-container” distinct from matter, is already, rudimentary, in M. Faraday \cite{Faraday "A speculation touching Electric Conduction and the Nature of Matter"}. It is not an idea introduced ex novo by Einstein, but comes, albeit in still nuanced forms, from the experimental scientific background sprouted around the notion of \emph{field}. 

Faraday \cite{Faraday "On The Physical Character of The Lines of Magnetic Force"} imagined (pure) space as something in which the magnetic and electric lines of force “acquire” a curvature: there are  thus curved lines of inductive action that run through the space, and connect all particles, or material masses, together. Consequently, Faraday's space is a \emph{space deformed} by the action of the lines of force, and these deformations produce influences on the physical state of the material masses: briefly, it is an \emph{active space, with a physical action}, since Faradayian space \cite[2787, p. 25]{Faraday "II. Experimental Researches in Electricity. Twenty-fifth Series par. 31. On the magnetic and diamagnetic condition of bodies"} «has a magnetic relation of its own». \emph{In Faraday}, however, \emph{there is still no identification between space and matter}. Insights in \cite[pp. 111, 127-128]{Bellone "I modelli e la concezione del mondo nella fisica moderna: da Laplace a Bohr"}. \margosymbol
\end{margo}

\subsection{Formulation with and without Coordinates}
\label{subsection "Formulation with and without Coordinates"}

\begingroup 
\footnotesize
The \emph{vector} (or, in general, \emph{geometric}) calculus should put the mathematician in a position to be able to resolve \emph{directly} any question of geometry, mechanics, physics, in \emph{absolute form}, that is, \emph{independent of any reference system} (with \emph{zero} coordinates).\endnote{
	Original It. version: «Il \emph{calcolo vettoriale} (o, in generale, \emph{geometrico}) deve porre il matematico in condizione di poter risolvere \emph{direttamente} una qualsiasi questione di geometria, di meccanica, di fisica, sotto \emph{forma assoluta}, cioè \emph{indipendente da qualsiasi sistema di riferimento} (con \emph{zero} coordinate)».
	} \\
\indent — \textsc{C. Burali-Forti and R. Marcolongo} \cite[p. 97]{Burali-Forti e Marcolongo "Elementi di Calcolo vettoriale con numerose applicazioni alla geometria alla meccanica e alla fisica-matematica"}

\endgroup

\vspace{2mm}

A crucial summary of the general relativity in tensor form is in R. Marcolongo's lectures \cite{Marcolongo "Calcolo differenziale assoluto e Teoria della Relativita"} of the academic year 1919-1920, and in V.A. Fock \cite[chapp. II-III]{Fock "The theory of Space Time and Gravitation}; see also \cite{Das An. "Tensors: The Mathematics of Relativity Theory and Continuum Mechanics"} \cite[chap. XII]{Dodson Poston "Tensor Geometry: The Geometric Viewpoint and its Uses"}. 

The figure of Marcolongo swings between two “worlds” of thought; although belonging to the school of Italian vectorialists, together with C. Burali-Forti, P. Burgatti, and T. Boggio, he is a mediator and a promoter of relativistic solutions \cite{Marcolongo "Relativita"} pertaining to the other mathematical school of his era (represented by Ricci Curbastro, Levi-Civita, and Palatini). 

The school of vectorialists treats vectors as \textit{absolute entities}, that is, as quantities perfectly free of the choice of arbitrary reference elements, so develops a method independent of coordinate systems, i.e. with all \textit{coordinates equal to zero}, or without fixed and moving coordinate axes \cite[p. viii]{Burali-Forti e Boggio "Meccanica razionale"}. Vectorialists' analysis is an \emph{absolute calculus without coordinates} or \emph{calculus of vector homographies} (vector invariants without coordinates), whilst the method of Ricci and prosecutors is \emph{absolute with coordinates} (with covariant derivatives of contravariant and covariant vectors) \cite[pp. v-vii]{Burgatti Boggio Burali-Forti "Analisi vettoriale generale e applicazioni II: Geometria differenziale"}.

Ricci's school, as is well-known, has prevailed, and it is the foundation of the exposition of relativistic mathematics; but the vectorialists group, beyond the controversies with the  opposing school of thought and against the theory of relativity, has bequeathed the idea of a formulation without coordinates for Einstein equations, which was expanded in several directions in the succeeding decades (see e.g. the skeleton calculus in Section \ref{subsection "Ex. 1. Regge Calculus: Simplicial Decompositions"}).

\vspace{10mm}

\setcounter{secnumdepth}{0}  
\section{References and Bibliographic Details}
\setcounter{secnumdepth}{3}
\markright{References and Bibliographic Details}

\begingroup
\footnotesize
\noindent Section \ref{section "Gravitational Field as a Curvature of the Space"}

\begin{indent paragraph: 15pt}
For a synopsis on geometry and relativity, see e.g. R. Penrose \cite{Penrose "Techniques of Differential Topology in Relativity"}, and K. Krasnov \cite{Krasnov "Formulations of General Relativity: Gravity Spinors and Differential Forms"}.
\end{indent paragraph: 15pt}

\noindent Section \ref{subsection "Vetturale of Energy Radiated in Gravitational Waves"} 

\begin{indent paragraph: 15pt}
For those who wish to delve further into the subject of gravitational waves, see e.g. \cite[chap. 9]{Gasperini "Theory of Gravitational Interactions"} \cite{Lichnerowicz "La relativite generale et les ondes"} \cite[chapp. 35-37]{Misner Thorne Wheeler "Gravitation"} \cite[chap. 9]{Padmanabhan "Gravitation: Foundations and Frontiers"} \cite[chap. 11]{Poisson and Will "Gravity: Newtonian Post-Newtonian Relativistic"}.	
\end{indent paragraph: 15pt}

\noindent Section \ref{section "Lovelock's Scalar Lagrangian Density of the Gravitational Field"} 

\begin{indent paragraph: 15pt}
For the Lovelock's theorem \cite{Lovelock "The Uniqueness of the Einstein Field Equations in a Four-Dimensional Space"} et seq., see \cite{Farhoudi "Lovelock tensor as generalized Einstein tensor"} \cite{Navarro Navarro "Lovelock's theorem revisited"} \cite[sec. 2.4.1]{Clifton Ferreira Padilla Skordis "Modified gravity and cosmology"} \cite[sec. 3.2.2]{Rakotomanana "Covariance and Gauge Invariance in Continuum Physics: Application to Mechanics Gravitation and Electromagnetism"}. — For an insight into the Lovelock gravity, see \cite{Bostani Dehghani Sheykhi "Lovelock Gravity and The Counterterm Method"} \cite[sec. 8.1]{Charmousis "Higher Order Gravity Theories and Their Black Hole Solutions"} \cite[sec. 2.2.1]{Charmousis "From Lovelock to Horndeski's Generalized Scalar Tensor Theory"} \cite[sec. 4.2]{Zanelli "Gravitation Theory and Chern-Simons Forms"} \cite[sec. 2]{Cognola Sebastiani Zerbini "Thermodynamics of Extended Gravity Black Holes"} \cite{Dadhich "A guiding criterion for gravitational equation in higher dimensions"} \cite[sec. 2]{Camanho Dadhich Molina "(Not so) pure Lovelock Kasner metrics"}.	
\end{indent paragraph: 15pt}

\noindent Section \ref{subsection "Concepts (?) of Space-Time"}

\begin{indent paragraph: 15pt}
Margo \ref{margo "Newtonianism and Euclidicity for gravity in Levi-Civita's view"}: on the classicality (as a classical mechanics) present in general theory of relativity under Levi-Civita's consideration, see \cite{Cattaneo "Leggi classiche e leggi relativistiche nel pensiero di Tullio Levi-Civita"}; for the $n$-body problem in general relativity according to Levi-Civita, see \cite{Lichnerowicz "Le probleme des $n$ corps en relativite generale et Tullio Levi-Civita"} \cite[I-V]{Hagihara "Tullio Levi-Civita's Works in Celestial Mechanics"}.
\end{indent paragraph: 15pt}

\noindent Section \ref{subsection "Formulation with and without Coordinates"}

\begin{indent paragraph: 15pt}
On the figures of Burali-Forti, Burgatti, and Boggio, see \cite{Freguglia "Burali-Forti e gli studi sul calcolo geometrico"} \cite[sec. 7, pp. 349-355]{Pizzocchero "Geometria differenziale"}	\cite[sec. 4. pp. 453-483]{Pastrone e Caparrini "Fisica Matematica e Meccanica razionale"}. 
\end{indent paragraph: 15pt}

\endgroup

\chapter{On Dimensional Continuum, Part III. Curvature of What?}
\label{chapter "On Dimensional Continuum, Part III. Curvature of What?"}

\section{Mathematics, Physics \& Reality of Space}

\begingroup
\footnotesize
[E.] Kretschmann \cite{Kretschmann "Uber den physikalischen Sinn der Relativitatspostulate A. Einsteins neue und seine ursprungliche Relativitatstheorie"} took the view that the postulate of general covariance does not make any assertions about the physical \emph{content} of the laws of nature [\textit{physikalischen \emph{Inhalt} der Naturgesetze}], but only about their mathematical \emph{formulation} [\textit{mathematische \emph{Formulierung}}]; and Einstein \cite{Einstein "Prinzipielles zur allgemeinen Relativitatstheorie"} entirely concurred with this view. The generally covariant formulation of the physical laws acquires a physical content only through the principle of equivalence, in consequence of which gravitation is described \emph{solely} by the $g_{ik}$ and these latter are not given independently from matter, but are themselves determined by field equations. Only for this reason can the $g_{ik}$ be described as \emph{physical quantities} [\textit{\emph{physikalische Zustandsgrößen}}], cf. H. Weyl \cite[§ 26, pp. 173-174, 182]{Weyl "Raum-Zeit-Materie: Vorlesungen Uber Allgemeine Relativitatstheorie 1918"}. \\
\indent — \textsc{W. Pauli} \cite[§ 52, p. 711]{Pauli "Relativitatstheorie"} = \cite[§ 52, p. 150]{Pauli "Theory of Relativity"}

\vspace{2mm}

[I]n regard to the nature of things, [the human] knowledge [of the physical laws brought on by the theory of relativity] is only an \emph{empty shell—a form of symbols}. It is \emph{knowledge of structural form}, and not \emph{knowledge of content}. All through the physical world runs that unknown content [\,\dots]. And, moreover, we have found that where science has progressed the farthest, \emph{the mind has but regained from nature that which the mind has put into nature}. We have found a strange foot-print on the shores of the unknown. We have devised profound theories, one after another, to account for its origin. At last, we have succeeded in reconstructing the creature that made the foot-print. And Lo! it is \emph{our own}. \\
\indent — \textsc{A.S. Eddington} \cite[pp. 200-201, e.a.]{Eddington "Space Time and Gravitation. An Outline of the General Relativity Theory"}

\endgroup

\vspace{2mm}

We talk about \emph{curved space-time} in general relativity; and often, erroneously, curvature is referred to as a property of \emph{real space}, since we are in the purview of physics. However, it is a \emph{representation}, a process of mathematics that realizes (in a mathematical sense) an abstract entity, which is the space-time, with its curvature. 

What did A.S. Eddington \cite{Dyson Eddington and Davidson "A Determination of the Deflection of Light by the Sun's Gravitational Field from Observations Made at the Total Eclipse of May 29 1919} discover, along the acclaimed observation of the solar eclipse of 29 May 1919? That space-time is curved? Actually, the precise relationship, expressed by Einstein's field equation, between (solar) mass and curvature, does not refer to real space, but to \emph{physical space}, which is a \emph{mathematical representation of the real one}. The conclusion of Eddington's book \cite{Eddington "Space Time and Gravitation. An Outline of the General Relativity Theory"} mentioned in epigraph, with an arcane relish, an almost mystical (\textgreek{μύω}) propensity, is it laudably resumptive.\footnote{
	Some tips are in V. Benci and P. Freguglia \cite[capp. 5-8]{Benci e Freguglia "Modelli e realta. Una riflessione sulle nozioni di spazio e tempo"}.
	}

\section{Beltrami–Nash's Teachings}

\begingroup
\footnotesize
[A] very notable consequence which can be deduced from expression $ds = R\frac{\sqrt{d\eta^2 + d\eta^2_1 + \cdots + d\eta^2_{n - 1}}}{\eta}$ is that an $(n - 1)$-dimensional space $\eta = \text{cost.}$ has its \emph{null} curvature at every point, since its linear element has the form $ds = \text{cost.}\sqrt{d\eta^2_1 + d\eta^2_2 + \cdots + d\eta^2_{n - 1}}$ [\,\dots] whence we conclude [\,\dots] that an $(n - 1)$-dimensional space $\eta = \text{cost.}$ is none other than one of the orthogonal trajectories of all geodesics converging towards the same point at infinity, that is, of a system of  geodesics [that are] \emph{parallel} to each other.\endnote{
	Original It. version: «[U]na conseguenza assai notabile che si deduce dalla espressione $ds = R\frac{\sqrt{d\eta^2 + d\eta^2_1 + \cdots + d\eta^2_{n - 1}}}{\eta}$ è che lo spazio ad $n - 1$ dimensioni $\eta = \text{cost.}$ ha la sua curvatura \emph{nulla} in ogni punto, poiché il suo elemento lineare ha la forma $ds = \text{cost.}\sqrt{d\eta^2_1 + d\eta^2_2 + \cdots + d\eta^2_{n - 1}}$ [\,\dots] donde si conclude [\,\dots] che lo spazio ad $n - 1$ dimensioni $\eta = \text{cost.}$ non è altro che una delle trajettorie ortogonali di tutte le geodetiche convergenti verso uno stesso punto all'infinito, cioè di un sistema di geodetiche \emph{parallele} fra loro».
	} \\
\indent — \textsc{E. Beltrami} \cite[pp. 419-420]{Beltrami "Teoria fondamentale degli spazii di curvatura costante"} 

\vspace{2mm}

To what extent are the abstract Riemannian manifolds a more general family than the sub-manifolds of euclidean spaces? \\
\indent — \textsc{J.F. Nash} \cite[p. 20]{Nash "The Imbedding Problem for Riemannian Manifolds"} 

\endgroup

\vspace{2mm}

We can draw a line over a surface, such that at each point the main normal (the normal vector to the curve) coincides with the normal to the surface at that point. We have built a geodesic. When (and if) a natural phenomenon follows this path, it expresses a differential paradigm, and is connectable to a non-Euclidean geometry. 

Nevertheless, as E. Beltrami \cite{Beltrami "Saggio di interpetrazione della Geometria non-euclidea"} \cite{Beltrami "Teoria fondamentale degli spazii di curvatura costante"} taught, it is possible to derive \emph{figures of non-Euclidean geometry within flat spaces of higher dimension}, so a \emph{non-Euclidean space has one less dimension} than the ambient (or surrounding) space—tersely, Beltrami constructs a mathematical 2-space for non-Euclidean geometry  in $\mathbb{R}^3$ (Euclidean 3-space), cf. Fig. \ref{figure "Beltrami's pseudosphere"}. And as J.F. Nash's embedding theorems \cite{Nash "$C^1$ Isometric Imbeddings"} \cite{Nash "The Imbedding Problem for Riemannian Manifolds"} (see Section \ref{section "On Nash's Embeddings: (Curved) Spaces in Euclidean Spaces"}) demonstrate, it is possible to \emph{isometrically embed any Riemannian manifold into a(n) (ambient) Euclidean space}. 

The concept of curvature, or the condition of a geometric object of departing from a plane, under the Euclidean axiomatic, from a \emph{mathematical} point of view—which \emph{coincides with that of physics}, because it makes use of the language of mathematics (see, more thoroughly, Sections \ref{section "Mathematics in the Physical Sciences, and Nature of Reality I"}, \ref{section "Mathematics in the Physical Sciences, and Nature of Reality II"} and \ref{section "Mathematics in the Physical Sciences, and Nature of Reality III"})—is still related to a flat structure. 

The catch of the issue is that, not infrequently, \emph{the physical space is intended as a real space (and confused with this), but instead it is just a space of representation of reality}. Space-time in relativity, which is a psuedo-Riemannian–Lorentzian manifold, as well as Nash's embedded spaces, are \emph{spaces of mathematics}, or spaces of mathematical physics, but are not directly identifiable as \emph{real spaces}.

\section{Nash Embeddings: (Curved) Spaces in Euclidean Spaces}
\label{section "On Nash's Embeddings: (Curved) Spaces in Euclidean Spaces"}

In this Section we will look into, although fleetingly, some theorems that are the cornerstone of the embedding theory in Riemannian geometry.

\subsection{Schläfli's Inverse Beltrami Problem}

\begingroup
\footnotesize
Mr. Beltrami \cite{Beltrami "Teoria fondamentale degli spazii di curvatura costante"} has shown that in the  linear element expression of an $n$-dimensional space  of constant curvature, the $n$ independent variables can be chosen so that each geodesic line within such a space is represented by $n - 1$ linear equations between the independent variables. And it was the reading of this interesting Memoir that led me to propose the following inverse problem: Find the definition of a space, whose geodetic lines are  represented each one by a system of $n - 1$ linear equations, $n$ being the number of variables independent in space.\endnote{
	Originally written in It. The end of Schläfli's Note \cite[pp. 192-193]{Schlaefli "Nota alla Memoria del sig. Beltrami "Sugli spazii di curvatura constante""} is intriguing: «Se si comincia una volta dal muover dubbio contro le ordinarie nozioni dello spazio, in quanto esso, insieme col tempo, è parte essenziale della serie dei fenomeni attuali, non capisco, perché si debba arrestarsi all'ipotesi che una porzione dello spazio, mediante un trasporto, sia suscettibile della congruenza con un'altra porzione dello stesso spazio. La forma di un corpo solido è il risultato istantaneo delle forze e delle velocità relative onde le sue molecole sono animate; e gli errori inerenti alla ipotesi che un tal corpo, dopo avvenuto un trasporto rispetto ad altri corpi che riputiamo in riposo, abbia serbato la sua forma, non sono in estremo grado minori di quelli inerenti alla presente astronomia pratica in connessione colle nozioni geometriche, anzi possono riguardarsi dello stess'ordine di piccolezza. E di quest'ordine, od almeno di un ordine comparabile con esso, sarebbe, parmi, anche la curvatura dello spazio, s'esso avesse, giusta la presunzione che il celeberrimo Riemann \cite{Riemann "Ueber die Hypothesen welche der Geometrie zu Grunde liegen"} sembra far tralucere, una tessitura di curvatura costante $\frac{1}{a^2}$ ovvero $-\frac{1}{a^2}$. Ma poiché una porzione infinitesima d'ogni tessuto a tre dimensioni intorno ad un punto preso ad arbitrio s'avvicina sotto tutti i rapporti allo spazio geometrico con un errore relativo anche infinitesimo, non vi sarebbe bisogno d'una curvatura costante talmente piccola da farne scendere una parte degli errori dell'astronomia moderna; anche un tessuto qualunque a grandissima unità lineare farebbe all'uopo, e nella sua formola definitrice i sei coefficienti [in the expression of an $n$-dimensional space] potrebbero essere funzioni sì del tempo che delle tre coordinate. Riflettendo poi che lo spazio della meccanica non è uno spazio assoluto, tale cioè che possa dirsi in riposo piuttosto che in istato di moto uniforme e rettilineo, si comprende che una nuova definizione dello spazio deve accommodarsi anzitutto a questa relatività fisica; ma allora chi ci dirà, che cosa dovremo sostituire alle espressioni della forma $\frac{\partial^2x}{\partial t^2}, x' - x, \frac{mm'}{r^2}$, quando avremo delle coordinate molto più accidentali di quelle dello spazio finora riputato il vero? Se, per es., la curvatura non fosse nulla, ma costante, e se per $r$ s'intende la distanza geodetica di due molecole, dovremo, nella correzione della formola $\frac{mm'}{r^2}$, sostituire alla $r$ primitiva $2a\sin\frac{r}{2a}$ o $2a\sinh\frac{r}{2a}$, a seconda della qualità positiva o negativa della curvatura, ovvero la distanza geodetica stessa?».
	} \\
\indent — \textsc{L. Schläfli} \cite[p. 178]{Schlaefli "Nota alla Memoria del sig. Beltrami "Sugli spazii di curvatura constante""} 

\endgroup

\vspace{2mm}

L. Schläfli \cite{Schlaefli "Nota alla Memoria del sig. Beltrami "Sugli spazii di curvatura constante""} was among the first to treat the question of finding a way of isometrically embedding a Riemannian manifold in a Euclidean space, albeit limited to a \emph{local} level, and it is a Note to Beltrami \cite{Beltrami "Teoria fondamentale degli spazii di curvatura costante"}. And there is a response from Beltrami \cite[p. 194]{Beltrami "Osservazione sulla precedente Memoria del sig.r prof. Schlafli"}: 

\vspace{2mm}

\begingroup
\footnotesize
The final result achieved by Mr. Schl[ä]fli in his previous Memoir is that the more general $n$-dimensional space, for which the property that each geodetic line is represented by the set of $n - 1$ linear equations holds, is obtained simply by making a homographic transformation on that space that I had already considered in the Memoir \cite{Beltrami "Teoria fondamentale degli spazii di curvatura costante"}, and for which I had demonstrated \emph{a posteriori} the existence of this property.

\endgroup

\vspace{2mm}

But it is only later, with the surveys of Whitney, that a greater framing is reached; and it will take the stream of Nash's visionarity, subsequently accompanied by the works of Kuiper, Gromov and Günther, to find a vast answer to the question posed.

\subsection{Whitney Embedding \& Immersion}

According to H. Whitney \cite{Whitney "Differentiable Manifolds"}, the starting question is: \emph{can any differentiable manifold be mapped, in an analytic manner, into a Euclidean space?} In fact, it should be remembered that a differentiable manifold can be defined either 

· as \emph{a set of points with neighborhoods homeomorphic with Euclidean space $\mathbb{R}^n$}, whose coordinates are related by differentiable transformations, 

· or as \emph{a subset of $\mathbb{R}^n$}, so that, near each point, the coordinates are expressed with differentiable functions.

\begin{theoremata}[Whitney]
A group of Whitney's propositions \textnormal{\cite[p. 654]{Whitney "Differentiable Manifolds"}} establishes the following.
\enumerationisinitium
\item Any $\mathscr{C}^r$ manifold $\mathcal{M}$ of dimension $m$, with $r \geqslant 1$ finite or infinite, is $\mathscr{C}^r$ homeomorphic with an analytic manifold in a Euclidean space $\mathbb{R}^{2m + 1}$.
\item Take two manifolds $\mathcal{M}$ and $\mathcal{N}$ of dimension $m$ and $n$, respectively, each of which is embedded in some Euclidean space. Let $\varphi$ be a $\mathscr{C}^r$ map of $\mathcal{M}$ into $\mathcal{N}$, with $r \geqslant 0$ finite, and $\textcyrillic{\textit{я}}^+$ a continuous function in $\mathcal{M}$. Then there exists a $\mathscr{C}^r$ map $\Phi$ of $\mathcal{M}$ into $\mathcal{N}$ under which
\subenumerationisinitium
\item $\Phi$ gives an approximation of $(\varphi, \mathcal{M}, r, \textcyrillic{\textit{я}}^+)$,
\item if $n \geqslant 2m$, $\Phi$ is completely regular,
\item $\Phi$ is analytic.
\subenumerationisfinis
In another group of statements he shows that
\item \textnormal{\cite[pp. 236-237]{Whitney "The Self-Intersections of a Smooth $n$-Manifold in $2n$-Space"}} any smooth $n$-manifold can be embedded in $2n$-space $\mathbb{R}^{2n}$, where for $n = 1$, the proof is trivial, while for $n = 2$, the embedding  process is about a sphere, a projective plane, and a Klein bottle \textnormal{\cite{Klein "Uber Riemann's Theorie der Algebraischen Functionen und ihrer Integrale"}} (a surface with self-intersection, see Fig. \ref{figure "Klein bottle"}), in $\mathbb{R}^4$,
\item \textnormal{\cite[p. 265]{Whitney "The Singularities of a Smooth $n$-Manifold in $(2n - 1)$-Space"}} any smooth $n$-manifold may be immersed in $(2n - 1)$-space $\mathbb{R}^{2n - 1}$, with $n \geqslant 2$.
\enumerationisfinis	
\end{theoremata}

\subsection[Nash $\mathscr{C}^1$ Isometric Embedding]{Nash $\protect\pseudobold{\mathscr{C}^1}$ Isometric Embedding}

The credit for the solution of the isometric embedding problem on a \emph{global} level goes to Nash in two papers.

\begin{theoremata}[Nash on $\mathscr{C}^1$ embeddings in Euclidean spaces]
Below we list some results in the first paper \textnormal{\cite{Nash "$C^1$ Isometric Imbeddings"}} by Nash.
\enumerationisinitium 
\item Any closed Riemannian $n$-manifold of class $\mathscr{C}^1$ may be always isometrically embedded in $2n$-space $\mathbb{R}^{2n}$. 
\item Any Riemannian $n$-manifold of class $\mathscr{C}^1$ may be always immersed in $2n$-space $\mathbb{R}^{2n}$, and isometrically embedded in $(2n + 1)$-space $\mathbb{R}^{2n + 1}$. 
\item Given an open Riemannian $n$-manifold having a short $\mathscr{C}^\infty$ immersion, or embedding, in $\mathbb{R}^k$, with $k \geqslant n + 2$, we claim that if the manifold does not meet its limit set, it possesses an isometric immersion, or embedding, in $\mathbb{R}^k$.
\item An $n$-manifold of class $\mathscr{C}^3$, for an embedding (of type $\mathscr{C}^3$), requires a space of $1\frac{1}{2}n^2 + 5\frac{1}{2}n$ dimensions.
\enumerationisfinis
\end{theoremata}

\subsection{Kuiper's Prosecution}

N.H. Kuiper completes the pathway of \textnormal{\cite{Nash "$C^1$ Isometric Imbeddings"}} in two works; the theorems concerned are called \emph{$\mathscr{C}^1$ embedding Nash–Kuiper theorems}. 

\begin{theoremata}[Kuiper on $\mathscr{C}^1$ embeddings in Euclidean spaces]
~\enumerationisinitium
\item In \textnormal{\cite{Kuiper "On C1-isometric imbeddings. I"}} Kuiper proves a Nash's conjecture: take a compact Riemannian $n$-manifold of class $\mathscr{C}^1$ with (void) boundary; if it has a $\mathscr{C}^1$ embedding in Euclidean $m$-space $\mathbb{R}^m$, with $m = 2n$, for $m \geqslant n + 1$, then our $n$-manifold may be isometrically embedded in $\mathbb{R}^m$ via $\mathscr{C}^1$ embedding.
\item In \textnormal{\cite{Kuiper "On C1-isometric imbeddings. II"}} he expresses the following propositions.
\subenumerationisinitium
\item Let us say that a $\mathscr{C}^\infty$ Riemannian $n$-manifold has a $\mathscr{C}^\infty$ embedding in $\mathbb{R}^m$, for $m = 2n$, and that it does not meet its limit set; then the $n$-manifold has a short $\mathscr{C}^\infty$ embedding in $\mathbb{R}^{m + 1}$ which in turn does not meet its limit set.
\item Let us assume that an open $n$-manifold, with $\mathscr{C}^\infty$ Riemannian metric, has a short $\mathscr{C}^\infty$ embedding in $\mathbb{R}^m$, $m > n$, which does not meet its limit set; then $n$-manifold has a $\mathscr{C}^1$ isometric embedding in $\mathbb{R}^m$ which in turn does not meet its limit set.
\item Let us say that an open $n$-manifold, with $\mathscr{C}^\infty$ Riemannian metric, has a $\mathscr{C}^\infty$ embedding in $\mathbb{R}^m$, $m \geqslant n$, which does not meet its limit set; then, putting $m = 2n$, the $n$-manifold has a $\mathscr{C}^1$ isometric embedding in $\mathbb{R}^{m + 1}$ which in turn does not meet its limit set.
\item The hyperbolic space $\hyperbolic^n$ admits a $\mathscr{C}^1$ isometric embedding in a Euclidean $(n + 1)$-space $\mathbb{R}^{n + 1}$. E.g. the hyperbolic plane $\hyperbolic^2$ can be isometrically embedded in $\mathbb{R}^{2 + 1}$. 
\subenumerationisfinis 
\enumerationisfinis
\end{theoremata}

\subsection[Nash $\mathscr{C}^k$ Isometric Embedding]{Nash $\protect\pseudobold{\mathscr{C}^k}$ Isometric Embedding}

Now we come to the second paper \cite{Nash "The Imbedding Problem for Riemannian Manifolds"} by Nash.

\begin{theoremata}[Nash on $\mathscr{C}^k$ embeddings in Euclidean spaces]
~\enumerationisinitium
\item Any compact Riemannian $n$-manifold, with $\mathscr{C}^k$ positive metric, as long as $3 \leqslant k \leqslant \infty$, has a $\mathscr{C}^k$ isometric embedding in a Euclidean space $\mathbb{R}^{\frac{n}{2}(3n + 11)}$, i.e. every compact Riemannian $n$-space is realizable as a submanifold of a Euclidean space of dimension $\frac{1}{2}n(3n + 11)$.
\item Any compact Riemannian $n$-manifold, with $\mathscr{C}^k$ positive metric, provided $3 \leqslant k \leqslant \infty$, admits a $\mathscr{C}^k$ isometric embedding in a Euclidean space $\mathbb{R}^{1\frac{1}{2}n^3 + 7n^2 + 5\frac{1}{2}n}$, i.e. every compact Riemannian $n$-space is realizable as a submanifold of a Euclidean space of dimension $1\frac{1}{2}n^3 + 7n^2 + 5\frac{1}{2}n$.
\enumerationisfinis
\end{theoremata}

The $\left\{\frac{1}{2}n(3n + 11)\right\}$- and $\left(1\frac{1}{2}n^3 + 7n^2 + 5\frac{1}{2}n\right)$-spaces are Nash's ambient (surrounding) spaces, coinciding with Euclidean spaces of those dimensionality, which can \emph{also contain non-Euclidean spaces}.

\subsection{Gromov and Günther Instances}

Let us move on to other subsequent contributions.

\begin{theoremata}[Gromov immersions]
~\enumerationisinitium
\item \textnormal{\cite{Gromov "Isometric immersions of Riemannian manifolds"}} Any $\mathscr{C}^\infty$ Riemannian manifold has an isometric $\mathscr{C}^\infty$ immersion in $\mathbb{R}^s$, for $s = m + 2n + 3$, namely into a Euclidean $(m + 2n + 3)$-space.
\item \textnormal{\cite[3.1.7]{Gromov "Partial Differential Relations"}} Given two $\mathscr{C}^\infty$ Riemannian manifolds $\mathcal{N}^n$ and $\mathcal{M}^s$, let be $\varphi^0 \colon \mathcal{N}^n \to \mathcal{M}^s$ a strictly short map between them, on the assumption that $s \geqslant n + 2\frac{n + 3}{2}$. Then $\varphi^0$ has a fine $\mathscr{C}^0$ approximation via free isometric $\mathscr{C}^\infty$ mapping 
\[ 
	\varphi \colon \mathcal{N}^n \to \mathcal{M}^{s \geqslant \frac{1}{2}(n + 2)(n + 3)}.
\]
\enumerationisfinis
\end{theoremata}

M. Günther \cite{Gunther "Isometric Embeddings of Riemannian Manifolds"} attains to this corollary, but that has the value of a statement: \emph{any  $\mathscr{C}^\infty$ Riemannian manifold admits a $\mathscr{C}^\infty$  isometric embeddding into $\mathbb{R}^s$, with $s = \max\left\{n\frac{n + 5}{2}, n\frac{n + 3}{2} + 5\right\}$}. We are therefore talking about a Euclidean space of dimension $\max\left\{\frac{1}{2}n(n + 5), \frac{1}{2}n(n + 3) + 5\right\}$.

\section{Appendix with Figures}

\subsection{Pseudosphere, or Tractricoid (an Example of No Embedding Hyperbolic 2-Space)}

\begin{figure}[h!]
\centering
\includegraphics[width = 0.50\textwidth]{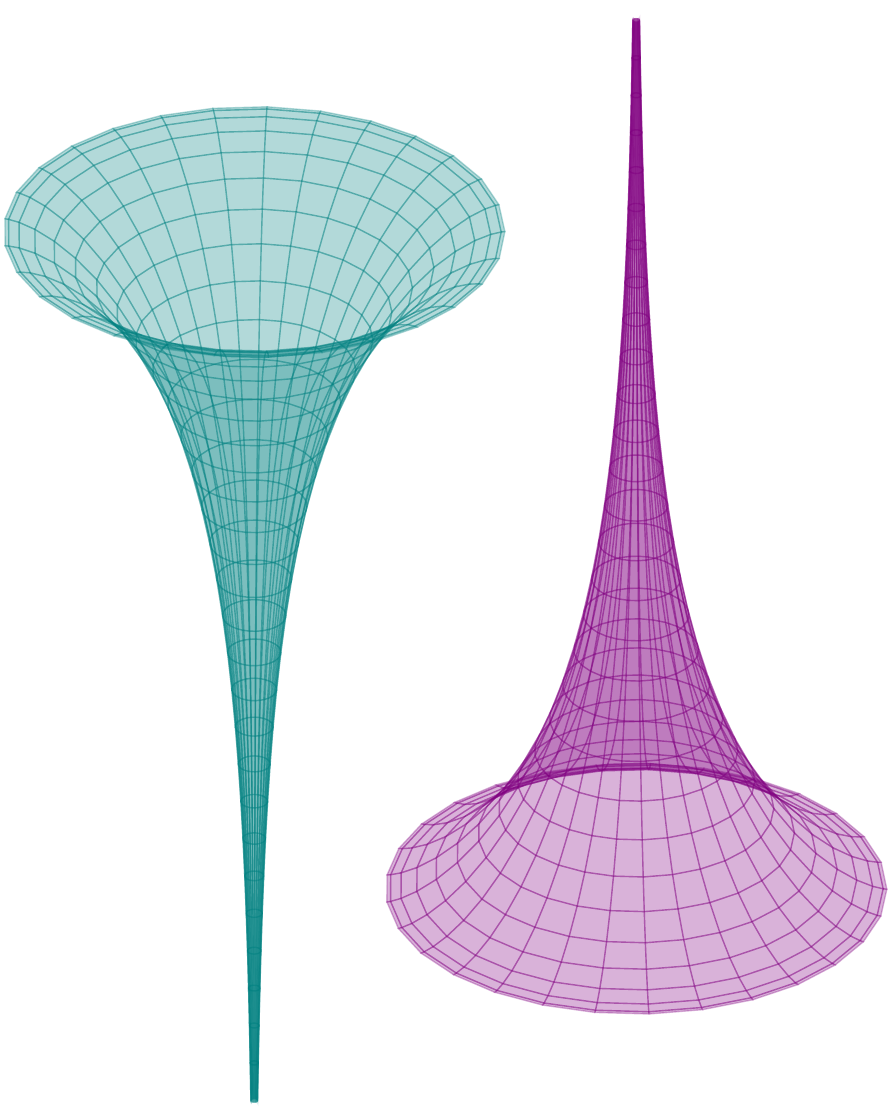}
\caption{Beltrami's pseudosphere, or tractricoid: same but inverted surface \& in different colors}
\label{figure "Beltrami's pseudosphere"}
\end{figure}

\emph{Beltrami's pseudosphere} \cite{Beltrami "Saggio di interpetrazione della Geometria non-euclidea"}, also called \emph{tractricoid}, is a surface (2-space) in $\mathbb{R}^3$ with constant negative curvature $\kappa = -\frac{1}{R^2}$. But beware: as we have already mentioned in Section \ref{subsubsection "Upper Half-Space, Ball, and Hyperboloid"}, D. Hilbert \cite{Hilbert "Ueber Flachen von Constanter Gaussscher Krummung"} proved that a complete,\footnote{
	We say that a topological space (then also a metric space) is \emph{complete} if any Cauchy sequence (see Definition \ref{definitio "Cauchy criterion"} and Theorem \ref{theorema "Cauchy criterion"}) in that space is convergent, or has a limit. 
	} 
regular (i.e. without singular points) surface of constant negative curvature cannot be isometrically embedded into Euclidean 3-space. 

The pseudosphere is no exception to this; moreover, it is not a complete space, because it ends with a rounded \emph{rim}. H. von Helmholtz \cite[pp. 13-14]{von Helmholtz "Ueber den Ursprung und die Bedeutung der geometrischen Axiome"} tells us of it, comparing Beltrami's pseudosphere to a \emph{Champagnerglas} with a tapering stem infinitely elongated, and he warns us that it is a surface bounded by a sharply edge beyond which a continuous extension of the calyx-shaped 2-space is not directly realizable. 

\subsection{Klein Bottle and Möbius Strip}

\emph{Klein bottle} \cite[Abschnitt II]{Klein "Uber Riemann's Theorie der Algebraischen Functionen und ihrer Integrale"} = \cite[part II]{Klein "On Riemann's Theory of Algebraic Functions and Their Integrals"} is a connected sum of two projective planes: 
\begin{equation}
	\mathbb{K}\mathbbl{l} \cong \mathbb{RP}^2 \# \mathbb{RP}^2 = 2\mathbb{RP}^2.
\end{equation}
It is a non-orientable surface, whose realization in $\mathbb{R}^3$ it possible exclusively as a space with self-intersection, so that its inside and its outside are the same; ergo it refers to a one-side 2-space. The \emph{Möbius strip} \cite[§ 11, p. 41]{Mobius "Bestimmung des Inhalts eines Polyeders"} 
\begin{equation}
	\ddot{\mathbb{O}} \cong \mathbb{S}^1 \times_{\mathbb{Z}/2} \mathbb{R}, 
\end{equation}
which we have already come across in Example \ref{exemplum "Möbius strip vs. torus"}, is equally a non-orientable surface. What is engaging is that, if we \emph{glue} two Möbius strips, we get a Klein bottle. 
 
\begin{figure}[H]
\centering
	\begin{minipage}[b]{0.410\textwidth}
	\includegraphics[width = \textwidth]{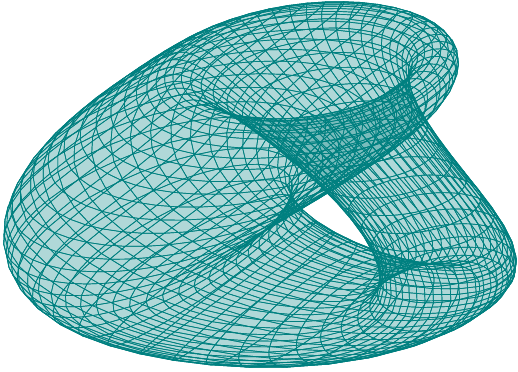}
	\caption{Klein bottle: one-side 2-space $\mathbb{K}\mathbbl{l} \cong \mathbb{RP}^2 \# \mathbb{RP}^2 = 2\mathbb{RP}^2$}
	\label{figure "Klein bottle"} 
	\end{minipage}
	\hspace{30pt}
	\begin{minipage}[b]{0.410\textwidth}
	\includegraphics[width = \textwidth]{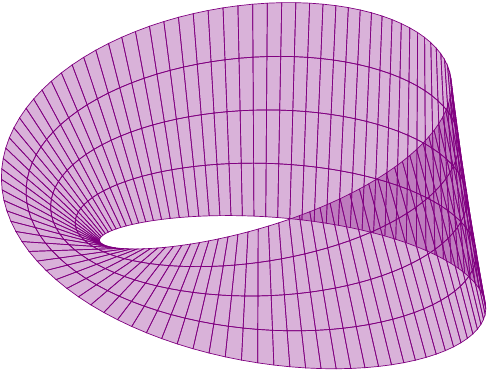}
	\caption{Möbius strip (but see footnote \ref{footnote "Listing–Möbius strip"}, p. \pageref{footnote "Listing–Möbius strip"}) $\ddot{\mathbb{O}} \cong \mathbb{S}^1 \times_{\mathbb{Z}/2} \mathbb{R}$}
	\label{figure "Möbius strip"}
	\end{minipage}
\end{figure}

A Klein bottle can whence be seen as a sphere with two disks removed and, in their place, two Möbius strips glued together; its coordinates are: $(x, y)$, $0 \leqslant x \leqslant 1$, $0 \leqslant y \leqslant 1$, in which, for any $x$, one has the points $(x, 0)$ and $(1 - x, 1)$, and, for any $y$, the points $(0, y)$ and $(1, y)$.

\vspace{10mm}

\setcounter{secnumdepth}{0}  
\section{References and Bibliographic Details}
\setcounter{secnumdepth}{3}
\markright{References and Bibliographic Details}

\begingroup
\footnotesize
\noindent Section \ref{section "On Nash's Embeddings: (Curved) Spaces in Euclidean Spaces"}

\begin{indent paragraph: 15pt}
On embeddings and immersions (as an overview), see the aforementioned Gromov book \cite{Gromov "Partial Differential Relations"}.	
\end{indent paragraph: 15pt}

\endgroup

\chapter[Foundational Issues: Discreteness \& Quantum Manifolds (Regge's Skeletonization, Topological Defects, Foam Substructure), Initial Boundary Problems in Cosmology, Wheeler's Pre-geometry and Pieri's \emph{Raw Materials}]{Foundational Issues: \\
Discreteness \& Quantum Manifolds \\ 
(Regge's Skeletonization, Topological Defects, Foam Substructure), \\
Initial Boundary Problems in Cosmology, \\ 
Wheeler's Pre-geometry and Pieri's \emph{Raw Materials}}
\chaptermark{Foundational Issues}{}
\label{chapter "Foundational Issues"}

\begingroup
\footnotesize
Diconsi stelle di \textsc{xvi} grandezza e tanto più lontane sono che la luce loro solo dopo \textsc{xxiv} secoli arriva a noi: visibili furono esse coi telescopi di Hærschel.\footnote{
	J.F.W. Herschel \cite{Herschel "A Treatise on Astronomy"} \cite{Herschel "Results of astronomical observations"} \cite{Herschel "Catalogue of one thousand new nebulae and clusters of stars"}.
	} 
Ma chi narrerà delle stelle anche più remote: atomi percettibili solo colle più meravigliose lenti che la scienza possegga o trovi? Quale cifra rappresenterà tal distanza che solo correndo per milioni d'anni la luce alata valicherebbe? Uomini udite: oltre quelle spaziano ancora i confini dell'Universo!\footnote{
	«They are called stars of \textsc{xvi} magnitude and are all the more distant that their light only after \textsc{xxiv} centuries reaches us: they were visible with Hærschel's telescopes. But who will tell about the stars even farther away: atoms, perceptible only with the most wonderful lenses that science possesses or finds? What numerical digit will represent such a distance that only by running for millions of years the winged light may cross? Hear ye, men: beyond those [stars] the edge of the Universe extends still further!».
	}\textsuperscript{,}\endnote{
	This is the inscription placed on the short side of the tomb, under the heading «Stelle di \textsc{xvi} grandezza». On the long side, under the heading «Stelle di \textsc{ix} grandezza», there is another inscription, unfortunately illegible: «[$\times$] grandezza e tanto sono lontane che solo nel [$\times$] loro raggio arriva a noi pur correndo la lu[ce] [$\times$] [chilom]tri al secondo. L'orbita annuale della Ter[ra] attorno il S[ole] [$\times$] orbita di \textsc{ccxvi} milioni di chilometri, vinta da quelle stel[le] [appa]rirebbe un punto e noi, uomini, atomi di questo punto [dell']universo, ci vantiamo di essere!».
	
	\setlength\parindent{8pt}
	The conclusion of the sentence brings to mind the popular—and dear to me—comment of C. Sagan \cite[pp. 6-7]{Sagan "Pale Blue Dot: A Vision of the Human Future in Space"} on the \emph{Pale Blue Dot} (photograph of Earth, February 14, 1990, by the \textit{Voyager 1}). Let us listen to his words. 
	
	«Look again at that dot. That's here. That's home. That's us. On it everyone you love, everyone you know, everyone you ever heard of, every human being who ever was, lived out their lives. The aggregate of our joy and suffering, thousands of confident religions, ideologies, and economic doctrines, every hunter and forager, every hero and coward, every creator and destroyer of civilization, every king and peasant, every young couple in love, every mother and father, hopeful child, inventor and explorer, every teacher of morals, every corrupt politician, every “superstar”, every “supreme leader”, every saint and sinner in the history of our species lived there—on a mote of dust suspended in a sunbeam.

	The Earth is a very small stage in a vast cosmic arena. Think of the rivers of blood spilled by all those generals and emperors so that, in glory and triumph, they could become the momentary masters of a fraction of a dot. Think of the endless cruelties visited by the inhabitants of one corner of this pixel on the scarcely distinguishable inhabitants of some other corner, how frequent their misunderstandings, how eager they are to kill one another, how fervent their hatreds.

	Our posturings, our imagined self-importance, the delusion that we have some privileged position in the Universe, are challenged by this point of pale light. Our planet is a lonely speck in the great enveloping cosmic dark. In our obscurity, in all this vastness, there is no hint that help will come from elsewhere to save us from ourselves [\,\dots]. 

	It has been said that astronomy is a humbling and character-building experience. There is perhaps no better demonstration of the folly of human conceits than this distant image of our tiny world».
	} \\
\indent — Epitaph on the anonymous sarcophagus of \textsc{C. Mattei} in his Rocchetta, on SP 62, in the vicinity of Riola, Bologna

\endgroup

\section[New \emph{Quantitates Sylvestres} I]{New \emph{Quantitates Sylvestres},\footnote{
	\label{footnote "Cardano's quantitas silvestris"}
	\emph{Quantitas silvestris}, «wild quantity», is an evocative expression that G. Cardano \cite[cap. X, p. 20]{Cardano "de Aliza regula Libellus"} has used for calling the imaginary number; in his time (1570) it was still an unknown or non-domestic mental entity, i.e. a «quantity that is not in any kind of roots, nor it is composed by those» (\textit{in quãtitate sylvestri, scilicet quæ non sit in aliquo genere radicum, nec composita ex illis}). Penetrating the wild and dark zones of thought, or constructing models applying to natural phenomena not yet understood, it is like an adventure that leads us to meet \emph{monstruosi} objects of mathematics, or unseizable theories (\emph{praedæ fugaces}) of physics. \\
	\indent The introduction of the term “imaginary” comes from R. Descartes \cite[Livre Troisième]{Descartes "La Geometrie"}, in a later period (1637); see his comment in the margin («Que les racines, tant vrayes que fausses peuvent être r[é]elles ou imaginaires») plus the paragraph on p. 380.
	}
	Part I
	}
	
We know many properties of (ordinary) geometric space ($3\mathrm{D}$) and space-time ($4\mathrm{D}$), including the space-relatedness of mathematical objects, by reason of number and symbol systems, and algebraic-topological structures; but, conceptually, a great deal of \emph{confusion} is prevailing about the notion of \emph{absolute quantity} (for instance like the Einsteinian–Newtonian entity, see \ref{item "Space-entity"} in Section \ref{subsection "Concepts (?) of Space-Time"}) or \emph{relational quantity} (on the model of \textit{ordo coexistendi}). But there is more. Once it has been accepted one of its natural dimensions, or physical realities, we do not know, \emph{within} this framework, if space(-time) is \emph{continuous} on all scales or if it is characterized by a \emph{Planckian discontinuity}, arranged in a \emph{discrete lattice shape}, with the formation of conical or curved \emph{singularities}, related in some way to the problem of the  extension of (cosmic) space, going back to the to the so-called \emph{primordial singularity}. Nor do we know where space(-time) emerges, if it emerges from something. 

What is certain is that such a range of possibilities is due to the \emph{transfer of our mathematical categories into nature}; and despite that, this is the charm of exploration, through mathematics, in the unknown regions of nature. Let us make seven examples in the Sections below.

\subsection{Ex. 1. Regge Calculus: Simplicial Decompositions}
\label{subsection "Ex. 1. Regge Calculus: Simplicial Decompositions"}

\begingroup
\footnotesize
Simplicial decompositions of Riemannian manifolds are introduced, which constitute higher-dimensional analogs of polyhedra [the so-called \emph{skeleton spaces}, and approximate smoothly curved spaces in general relativity]. It is hoped that this new formalism will make it possible to discuss solutions of Einstein's equations corresponding to highly complex topologies [like Wheeler's wormhole]. \\
\indent — \textsc{T. Regge} \cite[pp. 571, 558]{Regge "General Relativity without Coordinates"}\endnote{
	Originally written in It. (Riassunto).
	}

\vspace{2mm}

Regge's skeleton calculus puts within the reach of computation problems that in practical terms are beyond the power of normal analytical methods. It affords any desired level of accuracy by sufficiently fine subdivision of the space-time region under consideration. \\
\indent — \textsc{C.W. Misner, K.S. Thorne, J.A. Wheeler} \cite[p. 1179]{Misner Thorne Wheeler "Gravitation"}

\endgroup

\vspace{2mm}

The first example is dedicated to Regge calculus \cite{Regge "General Relativity without Coordinates"}. It is silent on the nature of space-time; it is a «computational» tool to facilitate the solution of Einstein field equations and calculus problems, thanks to the use of Riemannian manifolds \emph{without coordinates} (cf. Section \ref{subsection "Formulation with and without Coordinates"}). Nonetheless, it allows to build a \emph{quantum geometrization of space}, known as \emph{geometrodynamics}, see Wheeler \cite[chap. 8]{Wheeler "Geometrodynamics and the Issue of the Final State"} \cite{Wheeler "Superspace and the Nature of Quantum Geometrodynamics"}, and better understand quantum gravity. In essence, Regge calculus develops, according to Aleksandrov \cite[chapp. III-IV]{Aleksandrov "Combinatorial Topology I"}, a \emph{discrete} mathematical approach to Einstein curved spaces. 

\subsubsection{Skeletonization in a Piecewise Flat Space}

This approach \cite{Regge "General Relativity without Coordinates"}, in its basic version, consists in the \emph{partitioning-triangulation of a curvature of space(-time) into simplexes},\footnote{
	 From the La. \textit{simplex}, with \textit{sine-plica} as a probable derivation: “without” (\textit{sine}) a “fold” (\textit{plica}).
	} 
or rather, into a number $r$ of simplexes, denoted by $\simplex^n_1, \mathellipsis, \simplex^n_r$, i.e. in the \emph{decomposition} of a surface that has continuously varying curvature into \emph{flat pieces of Euclidean space}. Just so we are clear, 

· a point is a $\simplex^0$, i.e. 0-simplex in $0\mathrm{D}$, 

· a (straight) line segment is a $\simplex^1$, i.e. 1-simplex in $1\mathrm{D}$, 

· a triangle is a $\simplex^2$, i.e. 2-simplex in $2\mathrm{D}$, 

· a tetrahedron is a $\simplex^3$, i.e. 3-simplex in $3\mathrm{D}$, 

· a 5-hedroid, aka 5-choron, or 5-tope, is a $\simplex^4$, i.e. 4-simplex in $4\mathrm{D}$, and suchlike.

\begin{figure}[h!]
\centering
\includegraphics[width = 0.930\textwidth]{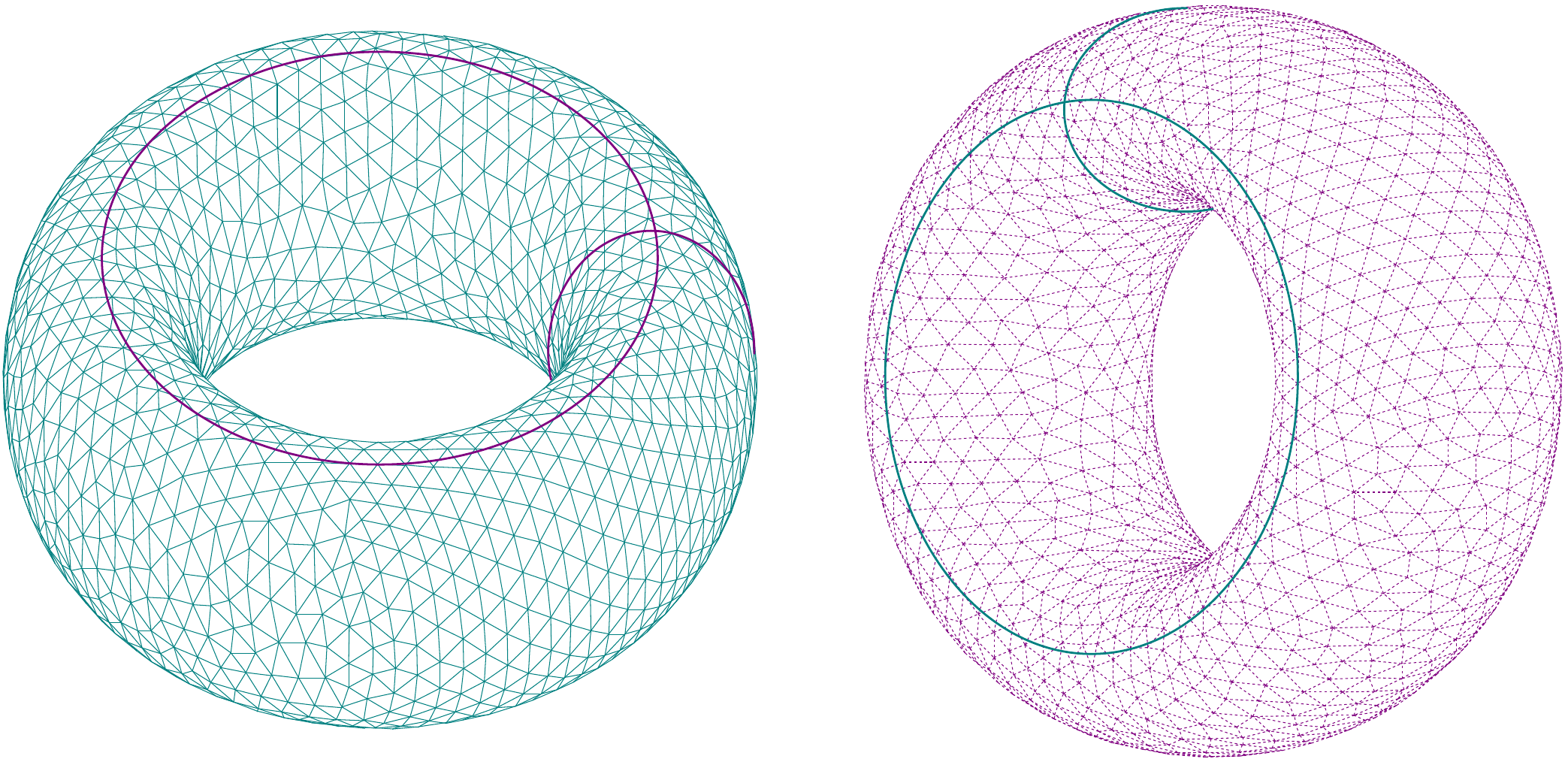}
\caption{2-torus $\mathbb{S}^1 \times \mathbb{S}^1 \cong \torus^2$ built with a net of triangular-like polygons (dashed version on the right). The more triangles, the better the approximation: a higher number of \emph{smaller} triangles allow a more accurate and smoother curved surface. Nothing new: it is an evolution of the great legacy of Archimedes \cite[\textgreek{κα´, κβ´, βδ´}]{Archimedes "Quadratura parabolae"}; it deals with a \emph{calculus process} via \emph{triangular approximants}, namely a parabolic segment (\textgreek{τμᾶμα περιεχόμενον ὑπὸ εὐθείας καὶ ὀρθογωνίου κώνου τομᾶς}) dissection into \emph{infinitely many triangular slices}, each of them is infinitesimally thin in an arbitrary manner}
\label{figure "torus with triangulation"}
\end{figure}

The result is that a continuous space, or the metric of space(-time) manifold, with the gravitational action and smoothness properties, can be approximated arbitrarily closely by a \emph{polyhedron} built with a \emph{net of triangular-like polygons} acting as a support structure (see Fig. \ref{figure "torus with triangulation"}), that is, a family of blocks gradually more and more small glued to each other, under a process of reduction called \emph{simplicial discretization} or, more effectively, \emph{skeletonization}.\footnote{
	This terminology is taken from the language of biology, on account of the similarity that a Regge geometry of this type has with the bone complex and the articular apparatus.
	}

If a collection of $n$-dimensional simplexes is usually known as \emph{simplicial complex}, reasonably, we can call \emph{skeleton space} the number $r$ of $n$-simplexes in the collection $\mathcal{S}(\simplex)$, i.e. 
\[
	\bigl\{\simplex^n_1, \mathellipsis, \simplex^n_r\bigr\} \in \mathcal{S}(\simplex), 
\]
forming all together a discrete approximation of a smooth variation concerning a certain geodetic. 

\subsubsection{Regge Action, and Discretum 4-Space: Skeleton for a Lattice Gravity Representation}

A great manner to approximate a smooth or $\mathscr{C}^\infty$ curved 4-space in general relativity with a skeleton space, consisting of pieces of $\simplex^4$ (4-dimensional simplexes), and therefore have a discrete representation of the geometrical part of Einstein field equations, is to use a variational principle into simplicial mode with which to calculate the continuum action functional \eqref{equation "Einstein–Hilbert (Gravitational) Action"}, and translate it into a skeleton-shaped manifold. The resulting action, 
\begin{equation}
\label{equation "Regge (simplicial) action"}
	\mathscr{S}_\textsc{r} = \sum_{\textit{\ng}_r}\volume(\textit{\ng})\delta^\angle(\textit{\ng}), \enspace \textit{\ng} = \simplex^2 = \rotatedtriangle,
\end{equation}
is in the form of a \emph{lattice} approximating the continuum limit in (classical) relativistic gravity, and reproducing the spatio-temporal curvature; in this action, $\textit{\ng} = \simplex^2$ is the so-called \emph{hinge} (or \emph{bone}) in $2\mathrm{D}$, which is a triangular subspace of codimension 2, i.e. a subsimplex $\textit{\ng} \subset \simplex^4$, while
\begin{equation}
	\delta^\angle(\textit{\ng}) \viz \delta^\angle_{\textit{\ng}} = 2\pi - \sum_{\simplex^4_r \text{ at } \textit{\ng}} \theta^{\angle_\mathrm{d}}_1, \mathellipsis, \theta^{\angle_\mathrm{d}}_j,
\end{equation}
is the \emph{angular deficit} on a hinge, which is equal to $2\pi$ minus the sum of the dihedral angles emerging from the gluing of various faces of the 4-simplexes (acting as rigid blocks) that meet at $\textit{\ng}$, so that $\delta^\angle(\textit{\ng})$ reflects the measure of the curvature of 4-space on $\textit{\ng}$.

\begin{scholium}
~\enumerationisinitium
\item A collection of $\simplex^n$ (simplexes of dimension $n$) meet on their flat faces of dimension $(n - 1)$, and the measure of the curvature (which must be reproduced-approximated) is given by $\textit{\ng}$ of dimension $(n - 2)$, according to the number of $\textit{\ng}$ necessary for the construction of triangles under the fixed skeletonization.
\item The angular deficit in $2\mathrm{D}$,
\begin{equation}
	\delta^\angle(\mathrm{ver}) \viz \delta^\angle_{\mathrm{ver}} = 2\pi - \sum_{\stackrel{\rotatedtriangle_r}{\text{ at } \mathrm{ver}}} \theta^\measuredangle_1, \mathellipsis, \theta^\measuredangle_j, \enspace \text{ where } \rotatedtriangle = \simplex^2,
\end{equation}
is a deficiency at a vertex, and it is equal to $2\pi$ minus the sum of the vertex angles $\theta^\measuredangle_1, \mathellipsis, \theta^\measuredangle_j$ of the triangles $\rotatedtriangle_1, \mathellipsis, \rotatedtriangle_r$ that meet on $\mathrm{ver}$.
\item In $2\mathrm{D}$ a hinge is (called) a \emph{side}, and in $3\mathrm{D}$ is (called) an \emph{edge}—recall that an edge, as a segment on the boundary, is an example of $\simplex^1$ (1-simplex). In $4\mathrm{D}$ a hinge is consistent with a \emph{triangle}, i.e. a $\simplex^2$ (2-dimensional simplex).

Alternatively and with greater fascination, a hinge can be thought of as the place where \emph{conical singularity} is formed, or as a limit-point in which, \emph{finitely}, the space is no longer smooth, see J.W. Barrett \cite{Barrett "The geometry of classical Regge calculus"}. \scholiumsymbol
\enumerationisfinis
\end{scholium}
 
Regge's Einsteinian type field expression is obtained, in a heuristic way, by varying the action \eqref{equation "Regge (simplicial) action"} in relation to the edge length; from here one has a variation of the angular deficit at the edges that vanishes identically:
\begin{equation}
\label{equation "Regge's Einsteinian type field expressions"}
	\mathcal{S}_\textsc{r}(\simplex)
	\begin{cases}
	\displaystyle
	\sum_{\textit{\ng}_r}\frac{\partial\volume(\textit{\ng})}{\partial\length(\mathrm{e})}\delta^\angle(\textit{\ng}) = 0.
	\end{cases}
\end{equation}
Eq. \eqref{equation "Regge's Einsteinian type field expressions"} is the simplicial equivalent of gravitational field Eqq. \eqref{subequations "Einstein field equations"} in keeping with Levi-Civita's thinking (Section \ref{subsubsection "Invariantiveness and Tensorial Conservation: Levi-Civita's Analytical Expression"}) for an \emph{empty discrete space} \cite[p. 570]{Regge "General Relativity without Coordinates"}, which can be named the \emph{Regge skeleton 4-space}. The skeleton space replaces, within the limits of the lattice representation, the smoothness of the 4-dimensional psuedo-Riemannian–Lorentzian–Einsteinian manifold (Section \ref{subsection "Lorentzian Generalization"}) with a rigid discreteness of flat block spaces; cf. A.P. Gentle \cite{Gentle "Regge Calculus: A Unique Tool for Numerical Relativity"} and H.W. Hamber \cite{Hamber "Simplicial Quantum Gravity"} \cite{Hamber "Simplicial Quantum Gravity from Two to Four Dimensions"}.

Finally, Eqq. \eqref{equation "Regge (simplicial) action"} and \eqref{equation "Regge's Einsteinian type field expressions"} allow the construction of \emph{piecewise linear Einstein 4-manifolds}. The papers of R. Friedberg and T.D. Lee \& collaborators \cite{Friedberg and Lee "Derivation of Regge's action from Einstein's theory of general relativity"} \cite{Feinberg Friedberg Lee and Ren "Lattice gravity near the continuum limit"} go in this direction; but especially see J.B. Hartle \cite{Hartle "Simplicial minisuperspace I. General discussion"} \cite{Hartle "Simplicial minisuperspace. II. Some classical solutions on simple triangulations"}, M. Roček, and R.M. Williams \cite{Rocek and Williams "Quantum Regge Calculus"} \cite{Rocek Williams "The Quantization of Regge Calculus"} \cite{Williams "Discrete quantum gravity: the Regge calculus approach"} \cite{Regge and Williams "Discrete structures in gravity"} \cite{Williams "Discrete quantum gravity"} \cite{Williams "Quantum Regge calculus"}, as well as R. Loll \cite{Loll "Discrete Approaches to Quantum Gravity in Four Dimensions"}.

\begin{margo}[Some indications and applications]
~\enumerationisinitium
\item In \cite{Hamber and Kagel "Exact Bianchi identity in Regge gravity"} there is a discrete (simplicial) analogue \cite[p. 566]{Regge "General Relativity without Coordinates"} of Bianchi identities under an order of rotation matrices in the product around null-homotopic loops.
\item The 4-geometry of Regge calculus holds two versions \cite[sec. 2.3]{Williams and Tuckey "Regge calculus: a brief review and bibliography"} of simplicial space-time: 
\subenumerationisinitium
\item in one version, time is discrete, with space-like hypersurfaces divided by a finite interval of time; 
\item in the other, time is continuous, or better, it presupposes at first a discrete time, and where the hypersurfaces are infinitesimally close the limit is taken; alternatively, the discretization process is direct.
\subenumerationisfinis
\item An investigation on the Regge calculus (skeleton manifolds) and functional/path integrals is e.g. in J. Fröhlich \cite{Frohlich "Regge calculus and discretized gravitational functional integrals"}. 
\item In D. Weingarten \cite{Weingarten "Euclidean quantum gravity on a lattice"} an integration over discrete space is built with 4-cubes embedded in a flat 5-cubic (5-dimensional hypercubic) lattice, having 80 edges + 80 faces.
\item A major application of simplicial geometry and Regge-like calculus is the \emph{simplicial quantum gravity} (\emph{dynamical triangulations approach}), see J. Ambjørn, M. Carfora, A. Marzuoli \cite{Ambjorn Carfora Marzuoli "The Geometry of Dynamical Triangulations"} \cite{Carfora Marzuoli "Quantum Triangulations: Moduli Space Quantum Computing Non-Linear Sigma Models and Ricci Flow"}.

\enumerationisfinis	
\end{margo}

\subsection{Ex. 2. Regge-like Discretization of Wheeler–DeWitt Equation}
\label{subsection "Ex. 2. Regge-like Discretization of Wheeler–DeWitt Equation"}

\begingroup
\footnotesize
I tend to assume that space-time and everything in it are in some sense emergent. By the way, you'll certainly find that that's what Wheeler expected in his essay \cite{Wheeler "Information Physics Quantum: The Search for Links"}. As you'll read, he thought the continuum was wrong in both physics and math. He did not think one's microscopic description of space-time should use a continuum of any kind—neither a continuum of space nor a continuum of time, nor even a continuum of real numbers.\footnote{
	\label{footnote "Thom on continuum"}
	Compare with R. Thom \cite[pp. 101-102, e.a.]{Connes Faltings Jones Smale Thom "Round-Table Discussion"}: Interlocutor: «Nothing in itself, not a single real world entity is continuous [\,\dots]. I cannot imagine the physical objects as being continuous». — R. Thom: «When you refer to a physical object, you mean an object which can be scientifically described. I would accept that, \emph{in any kind of description, we have a discrete element}, because \emph{a true continuum has no points}. We are unable to specify anything in the continuum. The continuum is something which cannot be described. It is a sort of \emph{unsayable}. It is a world in which one lives outside of symbolic description. But nevertheless, it exists [mathematically], despite the fact that we cannot describe it in any sense».
	} 
On the space and time, I'm sympathetic to that. On the real numbers, I've got to plead ignorance or agnosticism. It is something I wonder about, but I've tried to imagine what it could mean to not use the continuum of real numbers,\footnote{
	M. Picone and G. Fichera \cite[pp. 56-57]{Picone e Fichera "Lezioni di analisi matematica I"} write: «[In Mathematical Analysis] the ordered line on which a system of abscissae is introduced is identified [\,\dots] with the body $\mathbb{R}$ of the real numbers [\,\dots]. On a strictly logical level, the so-called geometric representation of the numbers of $\mathbb{R}$ by means of points of a straight line is completely inessential [\textit{del tutto inessenziale}] from the perspective of Mathematical Analysis, since it consists solely of replacing $\mathbb{R}$ with another model isomorphic to it (arithmetically and orderly) [thanks to a one-to-one correspondence]. However, this representation, by resorting to our geometric intuition, has such a force of suggestion that it is extremely useful under a purely practical aspect». 
	} 
and the one logician I tried discussing it with didn't help me. \\
\indent — \textsc{E. Witten}\endnote{
	Interview by N. Wolchover, “A Physicist's Physicist Ponders the Nature of Reality”, \textit{Quanta Magazine}, November 28, 2017.
	}

\endgroup

\vspace{2mm}

In H.W. Hamber and R.M. Williams \cite{Hamber and Williams "Discrete Wheeler-DeWitt equation"} the Wheeler–DeWitt equation \cite{Wheeler "Geometrodynamics and the Issue of the Final State"} \cite{DeWitt "Quantum Theory of Gravity. I. The Canonical Theory"} \cite {Wheeler "Superspace and the Nature of Quantum Geometrodynamics"} \cite{DeWitt "Dynamical Theory of Groups and Fields"} \cite{DeWitt "Quantum Theory of Gravity. II. The Manifestly Covariant Theory"} \cite{DeWitt "Quantum Theory of Gravity. III. Applications of the Covariant Theory"} is presented in the discretized formalism of Regge.

Let $|\Psi\rangle$ be a \emph{state vector}. Denoting by 
\begin{equation}
	|\Psi\rangle \to \Psi[\rotatedgamma_{\mu\nu}]
\end{equation}
a \emph{vacuum wave functional} of a 3-metric $\rotatedgamma_{\mu\nu}$, to wit, of 3-dimensional metric associated with a hypersurface, the Wheeler–DeWitt equation is written as
\begin{align}
\label{align "Wheeler–DeWitt equation"}
	\Biggl\{
	& -16\pi G_\textsc{n} \left(\gravitation_{\mu\nu,\xi\varrho} = \frac{1}{2\sqrt{\rotatedgamma}}\left(\rotatedgamma_{\mu\xi}\rotatedgamma_{\nu\varrho} + \rotatedgamma_{\mu\varrho}\rotatedgamma_{\nu\xi} - \rotatedgamma_{\mu\nu}\rotatedgamma_{\xi\varrho}\right)\right) \notag \\
	& \frac{\delta^2}{\delta \rotatedgamma_{\mu\nu}\delta \rotatedgamma_{\xi\varrho}} - \frac{1}{16\pi G_\textsc{n}} \sqrt{\rotatedgamma}\left(\scalarcurvature^{(3)} - 2\mathrm{\Lambda}\right) + \hat{\Hamiltonian}_{\varphi_\mathrm{m}}
	\Biggr\}\Psi[\rotatedgamma_{\mu\nu}] = 0, \text{ with } \rotatedgamma_{\mu\nu} \viz \rotatedgamma^{(3)}_{\mu\nu}, 
\end{align}	
where $\gravitation_{\mu\nu,\xi\varrho}$ is the Wheeler–DeWitt metric, $\scalarcurvature^{(3)}$ the Ricci curvature scalar relatively to $\rotatedgamma$-metric, $\mathrm{\Lambda}$ the cosmological constant (with a scaling behavior), and $\hat{\Hamiltonian}_{\varphi_\mathrm{m}}$ is the matter Hamiltonian operator, in which $\varphi_\mathrm{m}$ indicates a matter field, i.e. a space-time function representing matter. 

A \emph{discrete analogue} of \eqref{align "Wheeler–DeWitt equation"} is achieved through solutions of Regge's simplicial lattice model from which, in a piecewise linear space of dimension 3, an approximation of the continuum wave functional of the universe\footnote{
	\label{footnote "There is no wave function of the universe"}
	It is an \emph{exercise in style}. There is no wave function of the universe, if by \emph{universe} we mean the \emph{entire space}. This is because there are no observers/devices outside the all-space-universe for measuring the wave function of the universe. We are internal parts of the space-universe (\textgreek{πᾶς κόσμος}), which is our total space (\textgreek{πᾶσα χώρα}).	
	} 
descends:
\begin{align}
\label{align "Discrete analog of the Wheeler–DeWitt equation"} 
	\Biggl\{
	& -(16\pi G_\textsc{n})^2\gravitation_{\mu\nu}\bigl(\length^2(\mathrm{e})\bigr)\frac{\partial^2}{\partial\length^2_\mu(\mathrm{e})\partial\length^2_\nu(\mathrm{e})} \notag \\
	& \hspace{93pt} - \sqrt{\rotatedgamma\length^2(\mathrm{e})}\left(\scalarcurvature^{(3)}\length^2(\mathrm{e}) - 2\mathrm{\Lambda}\right)
	\Biggr\}\Psi\left[\length^2(\mathrm{e})\right] = 0,
\end{align}
where $\gravitation_{\mu\nu}$ is taken on the space of squared edge lengths $\length^2(\mathrm{e})$, with $\mu$- or $\nu$-edges of tetrahedra, that is, simplexes in $3\mathrm{D}$, under which the first term in \eqref{align "Discrete analog of the Wheeler–DeWitt equation"} can be obtained with a sum of edge contributions of a tetrahedron as a $(n - 1)$-simplex. If all the contributions are summed over all hinges $\textit{\ng}$ on the tetrahedral 3-simplex, for which all hinges are edges in 3 dimensions, one has
\begin{align} 
	\Biggl\{
	& -(16\pi G_\textsc{n})^2 \sum_{\mu, \nu \subset \simplex^3}
	\gravitation_{\mu\nu}(\simplex^3)\frac{\partial^2}{\partial\length^2_\mu(\mathrm{e})\partial\length^2_\nu(\mathrm{e})} \notag \\ 
	& - 2n_{\simplex^3(\textit{\ng})}\sum_{\textit{\ng} \subset \simplex^3}\length(\textit{\ng})\delta^\angle(\textit{\ng}) + 2\mathrm{\Lambda}\left(\volume(\simplex^3) = \sqrt{\rotatedgamma(\simplex^3)}\right)
	\Biggr\}\Psi\left[\length^2(\mathrm{e})\right] = 0,
\end{align}
where $\length(\textit{\ng})$ is the $\textit{\ng}$-edge length, $\delta^\angle(\textit{\ng})$ the angular deficit on $\textit{\ng}$, and $\volume(\simplex^3)$ is the volume of the tetrahedron. The discrete version, in tetrahedral-like shape, of the Wheeler–DeWitt metric is then
\begin{align}
	\gravitation_{\mu\nu,\xi\varrho}(\simplex^3) = \tfrac{1}{2\sqrt{\rotatedgamma}}(\simplex^3)\Bigl(\rotatedgamma_{\mu\xi}(\simplex^3)\rotatedgamma_{\nu\varrho}(\simplex^3) & + \rotatedgamma_{\mu\varrho}(\simplex^3)\rotatedgamma_{\nu\xi}(\simplex^3) \notag \\
	& - \rotatedgamma_{\mu\nu}(\simplex^3)\rotatedgamma_{\xi\varrho}(\simplex^3)\Bigr), \text{ with } \rotatedgamma_{\mu\nu} \viz \rotatedgamma^{(3)}_{\mu\nu}.
\end{align}

\subsection{Ex. 3. Hartle–Hawking Proposal: Euclidean Functional Integral for a No Initial Boundary of the Universe with Feynman Calculus}

\begingroup
\footnotesize
It is established that space [\textit{spacium}] [universe] be finite [\textit{finitum}]; [and suppose that] someone runs towards the fartherest verge [i.e., border of space] [\textit{ad oras \textnormal{/} Vltimus extremas}], and from thence he throws a winged dart [\textit{uolatile tellum}], with arbitrary vigorous force, should [this dart] reach the point designed, and fly away, or you think that something should stop or hinder [its flight?] [\textit{longeq\textnormal{[}ue\textnormal{]} uolare \textnormal{/} An prohibere aliquid censes obstareq\textnormal{[}ue\textnormal{]} posse}]. \\
\indent — \textsc{T. Lucretius Carus} \cite[Liber primus, 14v]{Lucretii Cari "De rerum natura"}

\vspace{2mm}

We put forward a proposal for the wave function of the “ground state” or state of minimum excitation: the ground-state amplitude for a three-geometry is given by a path integral over all compact positive-definite four-geometries which have the three-geometry as a boundary [\,\dots]. Our proposal is that [\,\dots] the Universe does not have any boundaries in space or time [\,\dots]. If this were the case, one would have solved the problem of the initial boundary conditions of the Universe: the boundary conditions are that it has no boundary. \\
\indent — \textsc{J.B. Hartle, S.W. Hawking} \cite[pp. 2960-2961, 2975]{Hartle Hawking "Wave function of the universe"}

\endgroup

\vspace{2mm}

Another example that combines the problem of space-time structure with the issue  of gravitational singularity, playing between mathematical categories and physical (experimental) knowledge, is the Hartle–Hawking proposal \cite{Hartle Hawking "Wave function of the universe"}, which, in accordance with the Wheeler–DeWitt Eq. \eqref{align "Wheeler–DeWitt equation"}, describes a quantum state of a \emph{spatially closed universe} through the use of a wave function (functional) on compact 3-manifolds and related matter fields; calculation goes deep into the \emph{Feynman path integral} \cite{Feynman "Space-Time Approach to Non-Relativistic Quantum Mechanics"} \cite{Feynman and Hibbs "Quantum Mechanics and Path Integrals"} technique (see Examples \ref{exemplum "Probability amplitude functional as sum over histories"} and \ref{exemplum "Hamiltonian path integral—phase space formulation"}). 

The wave function of Hartle–Hawking 

· is a function of the \emph{ground state}, this is, \emph{state of minimum excitation}, the amplitude of which is defined by a path integral over all compact positive definite \emph{Euclidean Riemannian} 4-geometries with a 3-geometry as a boundary; 

· is constructed as a \emph{Euclidean functional integral} of the form
\begin{equation}
	\Psi_0\left[\rotatedgamma^{(3)}_{\mu\nu}, \varphi_\mathrm{m}^{(3)}\right] = \sum_\mathcal{M}\int\delta\left[\rotatedgamma^{(4)}_{\mu\nu}, \varphi_\mathrm{m}^{(4)}\right]\exp\left\{-\mathscr{S}_\mathbb{E}\left[\rotatedgamma^{(4)}_{\mu\nu}, \varphi_\mathrm{m}^{(4)}\right]\right\},
\end{equation}
letting $\sum_\mathcal{M}$ be a summation over $\mathcal{M}$-topologies, where $\rotatedgamma^{(3)}_{\mu\nu}$ is the induced 3-metric of the 3-geometry, acting as a boundary space-like surface, and $\rotatedgamma^{(4)}_{\mu\nu}$ the 4-metric of the 4-geometry, whilst $\varphi_\mathrm{m}^{(3)}$ and $\varphi_\mathrm{m}^{(4)}$ are the 3- and 4-dimensional matter fields, respectively. Then 
\begin{align}
	\mathscr{S}_\mathbb{E}\left[\rotatedgamma^{(4)}_{\mu\nu}, \varphi_\mathrm{m}^{(4)}\right] = -\frac{1}{16\pi G_\textsc{n}}\int_\mathcal{M}\sqrt{\rotatedgamma}(\scalarcurvature - 2\mathrm{\Lambda})d^4x & - \frac{1}{8\pi G_\textsc{n}}\int_{\partial\mathcal{M}}\sqrt{\rotatedgamma^{(3)}_{\mu\nu}}\kappa_\mathrm{(tr)} d^3x \notag \\
	& - \int\sqrt{\rotatedgamma}\Lagrangian_\mathrm{m}d^4x,
\end{align}
is the Euclidean action for gravity with a cosmological constant $\mathrm{\Lambda}$, in which $\scalarcurvature$ is the Ricci curvature scalar, 
\begin{equation}
	\kappa_\mathrm{(tr)} = \rotatedgamma^{\mu\nu}\kappa_{\mu\nu}
\end{equation}
the trace in the $\rotatedgamma$-metric of the extrinsic curvature of (on) the bounding 3-surface, to wit, the boundary surface $\partial\mathcal{M}$ in 3-space,\footnote{
	The trace $\kappa_\mathrm{(tr)}$ of the extrinsic curvature is more often referred to as the \emph{mean curvature}. 
	}	
and $\Lagrangian_\mathrm{m}$ is the Lagrangian density of the matter-energy.  

\begin{margo}[No singular space-time ab ovo]
~\enumerationisinitium
\item The operation of Euclideanization of Lorentz–Minkowski geometry in the Hartle–Hawking's idea fully exploits the \emph{Wick rotation} \cite{Wick "Properties of Bethe-Salpeter Wave Functions"}, under which a line element 
\begin{equation}
	ds^2 = - dt^2 + dx^2 + dy^2 + dz^2
\end{equation}
transforms to 
\begin{equation}
	ds^2 = d\tau^2 + dx^2  + dy^2 + dz^2, 
\end{equation}
so a pseudo-Riemannian metric of Lorentz–Minkowski type becomes equivalent to a general Euclidean 4-metric, if the time coordinate is such that $t \to -i\tau$. 

This is due to the advantage of moving from a non-compact Lorentz model, with both the group $O_{1, 3}(\mathbb{R}) = \Lorentz$, and the restricted group $SO_{1, 3}^+(R) = \Lorentz_+^\uparrow$, having a possibly infinite-dimensional representation (see point \ref{item "Indefinite orthogonal group, i.e. Lorentz group, and indefinite special orthogonal group, i.e. restricted Lorentz group: real 6-dimensional non-compact Lie spaces"} in Section \ref{subsection "Spinor Map (6-Dimensional Homomorphism): the Covering $SL_2(C)$ to $SO_{1, 3}^+(R)$"}, and Margo \ref{margo "Majorana's brainwave"}), to a compact and finite model, with the group $SO_4(\mathbb{R})$ of all rotations of 4-dimensional Euclidean space.
 \item The Hartle–Hawking universe is \emph{devoid} of initial boundary conditions, \emph{with no initial singularity}, and the thorny problems arising therefrom—what we can call the \emph{Lucretius' dart dilemma}. It is a cosmogony, at the pre-Planck epoch, characterized by no-boundary Euclidean Riemannian smooth metrics. Topologically, this kind of universe is in the shape of a shuttlecock with a base of pure-space, where the earliest superposition of different (sub)space-times occurs, from which the quantum cosmic evolution in terms of path integrals follows; in the future, its expansion proceeds continuously in a \emph{de Sitter} type state \cite{de Sitter "On the curvature of space"}.
 \item Limitations and incompleteness of the Hartle–Hawking theory are discussed by J. Ambjørn et al. \cite{Ambjorn Loll Nielsen and Rolf "Euclidean and Lorentzian Quantum Gravity-Lessons from Two Dimensions"} and D.N. Page \cite{Page "Boundary Conditions and Predictions of Quantum Cosmology"}; more recently, see these opposing \cite{Feldbrugge Lehners Turok "Lorentzian quantum cosmology"} \cite{Feldbrugge Lehners Turok "No Smooth Beginning for Spacetime} and supporting \cite{Janssen Halliwell Hertog "No-boundary proposal in biaxial Bianchi IX minisuperspace"} arguments. \margosymbol
\enumerationisfinis	
\end{margo}

\subsection{Ex. 4. Topological and Cosmic Defects}

\begingroup
\footnotesize
The [topological] defects are formed, roughly speaking, because the directions of symmetry breaking are different in different regions of space. When these regions try to match at the boundaries, they sometimes run into topological problems, and as a result we get defects which trap the high-energy symmetric vacuum in their cores. \\
\indent — \textsc{A. Vilenkin} \cite[p. 1]{Vilenkin "Cosmic Defects"}

\endgroup

\vspace{2mm}

We will look at the fourth example. The image of a discreetness or fuzziness of space-time, together with the expectation of a speed variation inherent to the propagation in quantum space-time, can be read in parallel with a proposal that space(-time) is afflicted, on different scales, by some kind of flaws distinguishable from the classical continuum, the so-called \emph{topological defects} the classification of which is laid down in the homotopy groups; see A. Vilenkin and E.P.S. Shellard \cite{Vilenkin "Cosmic strings and domain walls"} \cite{Vilenkin Shellard "Cosmic Strings and Other Topological Defects"} \cite{Vilenkin "Cosmic Defects"}. Here is a list: 

· \emph{domain walls} in $2\mathrm{D}$ with $\pi_0(\mathcal{X})$, or defects in the form of surfaces; 

· \emph{strings} in $1\mathrm{D}$ with $\pi_1(\mathcal{X})$, or defects in the form of lines; 

· \emph{monopoles} in $0\mathrm{D}$ with $\pi_2(\mathcal{X})$, or defects in the form of points;

· \emph{textures} with $\pi_3(\mathcal{X})$, or defects without localized cores in a small and temporary region of high-energy vacuum;

· \emph{monopoles connected by strings} and  \emph{walls bounded by strings}, or  hybrid defects.

\subsection{Ex. 5. Riemann Busillis (Discretuum vs. Continuum), and the Quantum Manifold}

\begingroup
\footnotesize
We are [\,\dots] quite at liberty to suppose that the metric relations of space in the infinitely small [\textit{Unendlichkleinen}] do not conform to the hypotheses of geometry [\,\dots]; in a discrete manifoldness [\textit{discreten Mannigfaltigkeit}], the ground of its metric relations is given in the notion of it,\footnote{
	As it is contained in the concept of number.
	} 
while in a continuous [\textit{stetigen}] manifoldness, this ground must come from outside. Either therefore the reality which underlies space [\textit{Raume zu Grunde liegende Wirkliche}] must form a discrete manifoldness, or we must seek the ground of its metric relations outside it, in binding forces which act upon it [\,\dots]. This leads us into the domain of another science, of physic.\footnote{
	\label{footnote "Grothendieck: discrete and continuous"}
	See what A. Grothendieck \cite[2.20. \textit{Coup d'oeil chez les voisins d'en face}, note 71, p. 58 otm, e.a.]{Grothendieck "Recoltes et Semailles. Reflexions et temoignage sur un passe de mathematicien"} affirms about it: «[Riemann] observes that it may very well be that the ultimate structure of space is “discrete”, and that our “continuous” representations of it are perhaps a simplification [\,\dots] of a more complex reality; that for the human mind, the “continuous” [is] easier to grasp than the “discontinuous”, and that therefore we need it as an “approximation” for understanding the discontinuous [\,\dots]; in a strictly logical sense, it is rather the discontinuous which, traditionally, functioned as a method of technical approach to [understanding] of the continuous. 
	
	Developments in mathematics in recent decades showed a much more intimate symbiosis between continuous and discontinuous structures [\,\dots]. To finding a “satisfactory” model [\,\dots], which can be “continuous”, “discrete” or “mixed” nature—such work will undoubtedly involve a \emph{great conceptual imagination} [\,\dots]. I predict that the expected renewal (if it must yet come\,\dots) will come from someone who is a mathematician in the soul, well informed about the great problems of physics, rather than from a physicist. But above all, it will take a man with “philosophical openness” to grasp the crux of the problem. This is not a technical problem at all, but a fundamental problem of “natural philosophy”», the good one, thence, and not that in the hands of fumesophers (cf. Section \ref{section "Interludio Giocoso. Against the Fumesophers, or the Tragicomic Smoke-sellers"}).
	} \\
\indent — \textsc{B. Riemann} \cite[pp. 149-150]{Riemann "Ueber die Hypothesen welche der Geometrie zu Grunde liegen"} = \cite[p. 37]{Riemann "On the Hypotheses which lie at the Bases of Geometry"}

\vspace{2mm}

[M]acroscopic space-time may be the classical-geometrical limit of a causal quantum space.\footnote{
	«The problem is posed of giving finite quantum rules for the generation of quantum symbol sets such that the order of generation becomes, in the classical limit, the causal order of space-time». Finkelstein's full publication on the subject: \cite{Finkelstein "Space-Time Code", Finkelstein "Space-Time Code. II", Finkelstein "Space-Time Code. III", Finkelstein "Space-time code. IV"}, and \cite[with G. Frye and L. Susskind]{Finkelstein Frye and Susskind "Space-time code. V"}; cf. V. Kaplunovsky and M. Weinstein \cite{Kaplunovsky and Weinstein "Space-time: Arena or illusion?"}.
	} \\
\indent — \textsc{D. Finkelstein} \cite[p. 1261]{Finkelstein "Space-Time Code"} 
	
\endgroup

\vspace{2mm}

Finkelstein's words \cite[p. 1261]{Finkelstein "Space-Time Code"} are worth quoting at length, because they go to the crux of the problem without the need for further explanations.

\vspace{2mm}

\begingroup
\footnotesize
Until we find a satisfactory theory of space-time structure, we shall be beset by the dilemma of
the discrete versus the continuous, the dilemma already posed by Riemann \cite{Riemann "Ueber die Hypothesen welche der Geometrie zu Grunde liegen"}, in much the following terms:
\enumerationisinitium
\item[(a)] A discrete manifold [like a chessboard or honeycomb, a tesselation or graph] has finite properties, whereas a continuous manifold does not. Natural quantities are to be finite. The world must be discrete.
\item[(b)] A discrete manifold possesses natural internal metrical structure, whereas a continuous manifold must have its metrical structure imposed from without. Natural law is to be unified. The world must be discrete.
\item[(c)] A continuous manifold has continuous symmetries [see Lorentz symmetries], whereas a discrete manifold does not. Nature possesses continuous symmetries. The world must be continuous.
\enumerationisfinis

[\,\dots] The same question about matter, asked for two millen[n]ia—Is it continuous or is it discrete?—has at last been answered in this century: No. Matter is made neither of discrete objects nor [continuous] waves but of quanta [\,\dots]. A quantum is an object whose coordinates form a non-commutative algebra [\,\dots] [and] whose class calculus is neither a discrete nor a continuous Boolean algebra, but an algebra which is not even Boolean, being non-distributive. This non-distributive class calculus is the lattice of subspaces of a separable Hilbert space, and is naturally imbedded in (and defines) the algebra of operators on that Hilbert space. A quantum manifold is a third possibility for space-time too.\footnote{
	One of the first investigations into a Lorentz invariant discrete (quantum) space-time is owed to H.S. Snyder \cite{Snyder "Quantized Space-Time"}.
	} 
This possibility would pass us cleanly between the horns of Riemann's dilemma:
\enumerationisinitium
\item[(a)] A quantum manifold, like a discrete one, has better convergence than a continuous manifold—remember Planck and the black body.
\item[(b)] A quantum manifold, like a discrete one, is born with internal structure and is even more unified, being coherent.
\item[(c)] A quantum manifold, like a continuous one, possesses continuous symmetry groups.
\enumerationisfinis

\endgroup

\subsection{Ex. 6. Space-Time Foam Effect}

\begingroup
\footnotesize
 [I]t is possible [\,\dots] to construct a completely self-consistent quantum theory of gravity within the framework of special relativity (i.e. when the space-time continuum is “Euclidean”). However, within the domain of General Relativity theory, where deviations from “Euclideanness” can be arbitrary large, the situation is quite different.\endnote{
 	Original Ge. version: «[E]s möglich ist [\,\dots] Im Rahmen der speziellen Relativitätstheorie (d. h. wenn das raumzeitliche Kontinuum ein „Euklidisches“ ist) eine durchaus konsequente Quantentheorie der Gravitation aufzubauen. Im Gebiet der allgemeinen Relativitätstheorie, wo die Abweichungen von der „Euklidizität“ beliebig gross sein können, steht aber die Sache ganz anders».
 	} \\
\indent — \textsc{M.P. Bronštejn} \cite[p. 150]{Bronstein "Quantentheorie Schwacher Gravitationsfelder"} = \cite[p. 276]{Bronstein "Republication of: Quantum theory of weak gravitational fields"}
 
\vspace{2mm}

The events, in terms of which the world is to be described in general relativity theory, are thought of as intersection nodes of the coordinate “mollusc” \cite[§§ 28-29]{Einstein "Uber die Spezielle und Allgemeine Relativitatstheorie"}.\endnote{
	Cf. Einstein \cite[§ 28, pp. 65-66]{Einstein "Uber die Spezielle und Allgemeine Relativitatstheorie"}: «This non-rigid reference-body [\textit{nichtstarre Bezugskörper}], which might rightly be called a “reference-mollusc” [\textit{„Bezugsmolluske“}], is essentially equivalent to any Gaussian 4-dimensional coordinate system. That which gives the “mollusc” [\textit{„Molluske“}] a certain comprehensibility, as compared to the Gaussian coordinate system, is the (really unjustified) formal preservation of the separate existence of the space coordinates as opposed to the time coordinate [\textit{formale Wahrung der Sonderexistenz der räumlichen Koordinaten gegenüber der Zeitkoordinate}]. Every point on the mollusc is treated as a space-point [\textit{Raumpunkt}], and every material point which is at rest relatively to it is at rest, so long as the mollusc is considered as reference-body. The general principle of relativity demands that all these molluscs can be used as reference-bodies with equal rights and equal success in formulating the general laws of nature; the laws must be entirely independent of the choice of mollusc».
	
	\setlength\parindent{8pt}
	The abuse of Einstein's theory under the hands of improvisers is a \emph{cabaret show}. For example, you can see how moronic economists are by reading what G. Palomba (an economist, 1908-1986) writes on p. 179, in \textit{Le grandezze fondamentali dell'economica corporativa}, Giornale degli Economisti e Annali di Economia, Nuova Ser., Anno 2, № 3-4, 1940, pp. 168-181. I do not translate the text into En., because in It. his \textit{delirium} shines even better: «[\,\dots] Cotesto stato di cose mi aveva fatto, in un primo momento, intravedere la possibilità di ricondurre il nostro problema entro il quadro della relatività einsteiniana. E precisamente: il caso dell'equilibrio stazionario pensavo di inquadrarlo nei confini della relatività speciale, cercando qualcosa di analogo alla \emph{trasformazione del Lorentz} per poter passare da un punto all'altro di questo mondo economico analogo al mondo fisico del Minkowsky [\textit{sic!}]; il caso, invece, dell'equilibrio dinamico [economico] pensavo di inquadrarlo nei confini della relatività generale, cercando qualcosa di analogo alla suggestiva \emph{piovra di riferimento dell'Einstein} stesso [the “reference-mollusc”] conducente ad un continuo addirittura non[-]euclideo». Mamma mia, I am impressed\,\dots.
	} 
No matter what the [space-time] transformation of [the four] coordinates, the intersection nodes cannot be transformed away, but persist in all systems, and it is this invariant background of \emph{nodes of intersection} that corresponds to the physical “reality”. But there is no general relativity theory of what the nodes represent. The implication seems to be that they represent some sort of \emph{discreteness} or \emph{singularity} in the solution of the underlying equations, and that there is nothing more to be said about the situation than the mere fact of the existence of the \emph{discontinuities}. \\
\indent — \textsc{P.W. Bridgman} \cite[p. 199, e.a.]{Bridgman "The Way Things Are"}.

\vspace{2mm}

The dependence of quantum fluctuations in the geometry upon the scale of observation [in a domain of extension] $L$ suggests the following picture. Space is like an ocean which looks flat to an aviator who flies high above it (big $L$). On closer approach dynamic structure of the surface is seen (quantum fluctuations; smaller $L$). Under still closer examination wave crests are seen to be forming and breaking up foam with a scale of millimeters, governed by the surface tension. The topology the ocean surface is recognized to be non-Euclidean. It is natural to conclude the geometry of space is likewise impermeated every which way with \emph{worm-holes} at distances of the order of $10^{-33}$ cm. In other words, geometry in the small would seem to have to be considered as having a \emph{foam-like} character. \\
\indent — \textsc{J.A. Wheeler} \cite[p. 509, e.a.]{Wheeler "Geometrodynamics and the Issue of the Final State"}.

\endgroup

\vspace{2mm}

Regarding the sixth example, reference is made to the studies on the \emph{foam-like quantum gravity fluctuations}, that is, the \emph{fuzzy structure of space-time}; see G. Amelino-Camelia, J. Ellis, et al. \cite{Amelino-Camelia Ellis Mavromatos Nanopoulos and Sarkar "Tests of quantum gravity from observations of gamma-ray bursts"} \cite{Jacob and Piran "Lorentz-violation-induced arrival delays of cosmological particles"} \cite{Amelino-Camelia Fiore Guetta and Puccetti "Quantum-Spacetime Scenarios and Soft Spectral Lags of the Remarkable GRB130427A"} \cite{Vasileiou Granot Piran and Amelino-Camelia "A Planck-scale limit on spacetime fuzziness and stochastic Lorentz invariance violation"}. It is possible to determine the contribution to the light travel time due to quantum-mechanical fluctuations of space-time ($\tau_\textsc{qg}$), from a specific source (gamma-ray burst) to a given detector, searching for \emph{Lorentz invariance violation}:
\begin{equation}
	\tau_\textsc{qg} = -s_\pm\frac{E_\gamma}{m_\textsc{qg}}\frac{\mathrm{\Delta}z_\mathrm{rs} = \frac{c}{\mathrm{H}_0} \int^{z_\mathrm{rs}}_0 \left(\frac{1 + \zeta_\mathrm{rs}}{\sqrt{\mathrm{\Omega_m}(1 + \zeta_\mathrm{rs})^3 + \mathrm{\Omega_\Lambda}}}\right)d\zeta_\mathrm{rs}}{c},
\end{equation}
where $E_\gamma$ is the photon energy, $z_\mathrm{rs}$ the redshift, $c$ the speed of light, $\mathrm{H}_0$ the Hubble constant, $\mathrm{\Omega_m}$ the matter density parameter, and $\mathrm{\Omega_\Lambda}$ the dark energy density parameter, in which $\mathrm{\Lambda}$ is the cosmological constant \cite{Einstein "Kosmologische Betrachtungen zur allgemeinen Relativitatstheorie"}; the sign parameter $s_\pm$ and the scale $m_\textsc{qg}$ shall be determined experimentally. Clearly, $\zeta_\mathrm{rs} \viz z'_\mathrm{rs}$.

\section{New \emph{Quantitates Sylvestres}, Part II}

\subsection{Ex. 7.a. Wheeler's Pre-geometry, and its Fundamental Question: Where Does (Geometry of) Space-Time Come from?}

\begingroup
\footnotesize
[A]n elastic substance [as a piece of cloth] reveals at a crack that the concept of “ideal elastic medium” is a fiction. Cloth shows at a sel­vage that it is not a continuous medium but woven out of thread [\,\dots]. At big bang and at collapse [space-time] cannot be a continuum [\,\dots]. If the elastic medium [of space-time] is built out of electrons and nuclei and nothing more, if cloth is built out of thread and nothing more, we are led to ask out of what “pregeometry” [underlying structure] the geometry of space and space[-]time are built [\,\dots]. There is no such thing as “elasticity” in the space between the electron and the nucleus [\,\dots]. [T]he gates of time tell us that physics must be built from a foundation that has no physics [\,\dots]—to make up the grand structure that we call “reality”. \\
\indent — \textsc{J.A. Wheeler} \cite[pp. 1-2, 6]{Wheeler "Pregeometry: Motivations and Prospects"}

\endgroup

\vspace{2mm}

Going back to and seeing, by means of imagination, the condition in which the geometry of space-time is dynamically generated, in Wheeler's effort, is like thinking of what he calls \emph{pre-geometry}, a \emph{substratum} (underlying structure) that acts as a reservoir for the emergence of continuum space-time. There is, and remains, the \emph{difficult to conceive of something that predetermines a Riemann–Einstein geometry, without referring to geometric notions (such as the concept of distance)}. 

Into physics field-work, this entails that \emph{it is hard to conceive of something from which the physics of space-time might emerge, without referring to physical notions of (subsequent) limitations}. In this regard, Wheeler \cite[pp. 227, 243-244, e.m.]{Wheeler "From Relativity to Mutability"} notes that 

\vspace{2mm}

\begingroup
\footnotesize
[T]here is no such thing as space[-]time in the real world of quantum physics [\,\dots]. A “pregeometry” that is \emph{primordial chaos}, and law built upon this chaos [\,\dots]. Molecular chaos leads to concepts like temperature and entropy only when \emph{limitations} are imposed, such as fixity of volume and total energy. \emph{Otherwise chaos is chaos}. Does the chaos, the “pregeometry”, that we think of as underlying the universe, also fail to yield any law \emph{until} it is analogously \emph{limited}?

\endgroup

\subsection{Ex. 7.b. Backtracking to the Source of Pieri's \emph{Raw Materials} of Point and Motion}
\label{subsection "Ex. 7.b. Backtracking to the Source of Pieri's Raw Materials of Point and Motion"}

\begingroup
\footnotesize
Primitive ideas [in geometry] are perhaps comparable to the raw materials for the industry; as are the primitive propositions [axioms or postulates] to simple machines [\textit{Le idee primitive son forse paragonabili alle materie prime dell'industria; come le proposizioni primitive alle macchine semplici}]. \\
\indent — \textsc{M. Pieri} \cite[p. 171]{Pieri "Della Geometria elementare come sistema ipotetico deduttivo. Monografia del punto e del moto"} 

\endgroup

\vspace{2mm}

The most elementary system for a \emph{foundation of geometry} that I know is the M. Pieri's system \cite{Pieri "Della Geometria elementare come sistema ipotetico deduttivo. Monografia del punto e del moto"}, see also \cite{Pieri "I principii della Geometria di posizione composti in sistema logico deduttivo"} \cite{Pieri "La Geometria Elementare istituita sulle nozioni di 'punto' e 'sfera'"}. If in authors such as M. Pasch \cite{Pasch "Vorlesungen uber neuere Geometrie"}, G. Peano \cite{Peano "Sui fondamenti della geometria"}, and D. Hilbert \cite{Hilbert "Grundlagen der Geometrie"}, the axiomatic method of geometry is grounded on three or more fundamental entities,

· \emph{point}, 

· \emph{straight line}, 

· \emph{line segment}, 

· \emph{flat surface}, 

· \emph{motion}, 

· \emph{congruence}, or \emph{coincidence}, \\
in Pieri there are only two fundamental entities,

· \emph{point}, 

· \emph{motion}, \\
from which the whole spatio-geometric system arises. Pieri call his underlying binary-system (the notions of point and motion in its various formal declinations) in different ways: «mother, primitive, or undecomposed ideas» (\textit{idee madri, primitive, o indecomposte}) \cite[p. 170]{Pieri "Della Geometria elementare come sistema ipotetico deduttivo. Monografia del punto e del moto"}, «primary ideas» (\textit{idee prime}) \cite[p. 175]{Pieri "Della Geometria elementare come sistema ipotetico deduttivo. Monografia del punto e del moto"}, «raw materials» (\textit{materie prime}) \cite[pp. 171, 176]{Pieri "Della Geometria elementare come sistema ipotetico deduttivo. Monografia del punto e del moto"}, «fundamental», «simple» or «primitive» «concepts» or entities» (\textit{concetti} o \textit{enti fondamentali}, \textit{semplici} o \textit{primitivi}) \cite[§ 1]{Pieri "I principii della Geometria di posizione composti in sistema logico deduttivo"} \cite[pp. 171, 203]{Pieri "Della Geometria elementare come sistema ipotetico deduttivo. Monografia del punto e del moto"}. Pieri writes

\vspace{2mm}

\begingroup
\footnotesize
\cite[pp. 2, 6]{Pieri "I principii della Geometria di posizione composti in sistema logico deduttivo"} \emph{Primitive} or undecomposed concepts, around which all postulates are deposited [\,\dots], are like the raw material of every proposition [\,\dots]. The main character of  \emph{primitive entities} of any hypothetical-deductive system is that they are capable of arbitrary interpretations, within certain boundaries assigned by \emph{primitive propositions} (\emph{axioms or postulates}).\endnote{
	Original It. version: «I concetti \emph{primitivi} o indecomposti, intorno a cui versano tutti i postulati [\,\dots] sono come la materia prima d'ogni proposizione [\,\dots]. Carattere principalissimo degli \emph{enti primitivi} d'un qualsivoglia sistema ipotetico-deduttivo è l'esser questi capaci d'interpetrazioni arbitrarie, dentro certi confini assegnati dalle \emph{proposizioni primitive} (\emph{assiomi o postulati})».
	}

\cite[pp. 175, 180]{Pieri "Della Geometria elementare come sistema ipotetico deduttivo. Monografia del punto e del moto"} The system, which is now offered to the public's judgment, admits only two primary ideas [raw materials, i.e. general ideas, or classes]: the \emph{point} and the \emph{motion}; the latter being understood as a representation of points in points, and far from any mechanical meaning [\,\dots]. [E]ach motion is a representation of the “point” class [\,\dots], and acts on each point (it operates in all space). The “motion” is therefore an individual [entity] of the category that goes by the names of “function”, “representation”, “transformation”, etc.\endnote{
	Original It. version: «Il sistema, che or si offre al giudizio del pubblico, non ammette che due sole idee prime: il \emph{punto} ed il \emph{moto}; quest'ultimo inteso come rappresentazione dei punti in punti, e lungi da ogni qualunque significato meccanico [\,\dots]. [O]gni moto è rappresentazione della classe “punto” [\,\dots], ed agisce sopra ogni punto (opera in tutto lo spazio). Il “moto” è pertanto individuo della categoria che va sotto i nomi di “funzione”, “rappresentazione”, “trasformazione”, ecc.».
	}

\endgroup

\vspace{2mm}

The concept of point can be intended primitively as a \emph{position}, and does not need a spatial proto-definition, because it is an embryo-concept for the geometry space—it goes from zero to infinity (singularity); but this, however, poses a problem in terms of \emph{extension}. If, on the contrary, we use an \emph{analytic language}, i.e. a \emph{coordinate geometry}, the dimensionality of the point loses a part of its tearing enigmaticity, as well as of its charm: a point can nonetheless be regarded as a certain value, and its position-notion shall be represented e.g. by a \emph{pair of numbers}, or a couple of parameters.

\begin{quaestio}
Is it possible to \emph{build a pre-geometry} \cite{Wheeler "Pregeometry: Motivations and Prospects"}, starting from primary ideas (raw materials) of point and motion à la Pieri? Or from a primitive hodgepodge, with a kind of pre-space soup? But then the following interrogation is: \emph{how primitive does a Wheelerian pre-geometry have to be?} Or, \emph{how much primitiveness, or raw material, does a pre-geometry hold?} \quaestiosymbol
\end{quaestio}

\begin{quaestio}
Finally, is it possible to trace back over a primitive point-motion system in geometry, and hence \emph{generate}, from this \emph{undecomposed and atomic-like aggregate}, the subsequent \emph{formation and formulation of the Riemann–Einstein's space continuum?} 

Furthermore, it should be remembered, within a pre-geometry background, there is still no border (as we know it) between a metamathematics of Euclidean geometry and a metamathematics of non-Euclidean geometry. \quaestiosymbol
\end{quaestio}

\begin{quaestio}
How can we ensure that a foundation of geometry, which serves as the backbone to advance some considerations on the pre-geometry topic, is independent from other branches of mathematics? It may well be, for instance, that the foundation of geometry is preceded by that of \emph{arithmetic}.

This is the opinion of M. Pasch, when he abandons his original mindset presented in \cite{Pasch "Vorlesungen uber neuere Geometrie"}. In a letter (dated 11 February 1894) to F.L.G. Frege, Pasch writes: «Arithmetic must be placed on firm foundations before this can be done for geometry. In this respect I have not yet been able to make up my mind to regard arithmetic as merely part of logic».\footnote{
	From Correspondence XIII/1 Frege–Pasch, in Frege's correspondence \cite{Frege "Wissenschaftlicher Briefwechsel"} = \cite[p. 103]{Frege "Scientific Correspondence"}.
	} \quaestiosymbol
\end{quaestio}

\begin{quaestio}
\label{quaestio "Bridgman on the simplicity"}
Who guarantees that nature, at its core, is «simple», or «primitive»? Or that it needs a primitive conceptual toolkit to be understood? P.W. Bridgman icastically notes that «“simple” means simple to us, when stated in terms of our concepts», which shows that anthropomorphism pervades scientific dictates no less than religious ones; he writes \cite[pp. 198-203]{Bridgman "The Logic of Modern Physics"}: 

\vspace{2mm}

\begingroup
\footnotesize
The hypothesis of simplicity assumes several forms; some physicists are convinced that the laws which govern nature are simple, others that the ultimate stuff of which nature is composed is simple [\,\dots]. Practically all the history of physics is a history of the reduction of the complicated to the simpler [\,\dots]. One may find great justification here for the belief that all nature will ultimately be reduced to a similar simplicity, and, in particular, justification for the attempt to find the explanation of all nature in the action of mechanical laws. Now, of course, as a matter of physical and historical fact, this program could not be carried through, but obdurate physical phenomena were discovered [\,\dots]. In one respect it is obvious that nature is not simple, namely numerically—try counting the electrons or atoms or stars! [\,\dots] We have in the first place to notice that “simple” means simple to us, when stated in terms of our concepts [\,\dots]. A tempting question is whether there may not be some laws of nature that are \emph{really} simple, without relation to our mode of formulation, such as the law of the inverse square.\footnote{
	Cf. Section \ref{subsubsection "Inverse-square Laws"}.
	} \quaestiosymbol

\endgroup
\end{quaestio}

\vspace{10mm}

\setcounter{secnumdepth}{0}  
\section{References and Bibliographic Details}
\setcounter{secnumdepth}{3}
\markright{References and Bibliographic Details}

\begingroup
\footnotesize
\noindent Section \ref{subsection "Ex. 2. Regge-like Discretization of Wheeler–DeWitt Equation"}

\begin{indent paragraph: 15pt}
For a synopsis in discrete vs. continuum methods for quantum gravity in this context, see e.g. \cite{Gielen and Oriti "Discrete and Continuum Third Quantization of Gravity"}. — A compendium on the continuum in physics and the problems hidden therein is \cite{Baez "Struggles with the Continuum"}.
\end{indent paragraph: 15pt}

\noindent References for multiple Sections

\begin{indent paragraph: 15pt}
Useful books on various subjects (Regge calculus, discrete Wheeler–DeWitt equation, Hartle–Hawking wave function) are \cite{Carlip "Quantum Gravity in 2 + 1 Dimensions"} \cite{Hamber "Quantum Gravitation: The Feynman Path Integral Approach"}.
\end{indent paragraph: 15pt}

\endgroup

\chapter[\textgreek{Γῆ Δρακόντων}, Part I. Quantum Field Space and Gravity]{\textgreek{Γῆ Δρακόντων},\footnote{
	\textit{Terra draconum}, that is, unknown/dangerous «land of dragons», in \textgreek{Αιλιανού περί ζώων ιδιότητος Βιβλία ιζ'}—\textit{Aeliani De natura animalium Libri XVII} \cite[Lib. II, cap. XXI, p. 55]{Aeliani De natura animalium Libri XVII}.
	}
	Part I. Quantum Field Space and Gravity}
\label{chapter "Ghé Drakónton, Part I. Quantum Field Space and Gravity"}

\begingroup
\footnotesize
[L]a debolezza del nostro Intelletto intorno alle cose naturali, ed anco Geometriche, è tale che venendo noi interrogati di qualsivoglia Problema, se vogliamo rispondere per verità, ed aggiustatamente, non possiamo rispondere meglio che con un sincero e schietto \textsc{non lo so} [\,\dots]; ed insomma la nostra risposta non può essere assoluta, ma sibbene come si suole dire \emph{ex suppositione}.\footnote{
	«The weakness of our Intellect around natural and also Geometric things is such that when we are being interrogated on any Problem, if we want to answer truthfully, and properly, we cannot answer better than with a sincere and blunt \textsc{i do not know} [\,\dots]; in short, our answer cannot be absolute, but rather as is usually said \emph{ex suppositione}».
	} \\
\indent — \textsc{B. Castelli} \cite[p. 550]{Castelli "Discorso sopra la Calamita"}

\vspace{2mm}

[L]es opérations de la nature sont infiniment supérieures à celles que l'adresse humaine est capable de produire.\footnote{
	«[O]perations of nature are infinitely superior to those that human skill is capable of producing». 
	} \\
\indent — \textsc{L. Euler} \cite[lettre XIII, 24 May 1760, p. 49]{Eulero "Lettres a une princesse d'Allemagne I"}

\endgroup

\section{A Lot but Not Everything: the Lagrangian Farrago in Standard Quantum Fields}

\begingroup
\footnotesize
We're not building a machine that calculates answers; instead, we're discovering questions. Nature's shape-shifting laws seem to be the answer to an unknown mathematical question [\,\dots]. The ascension to the tenth level of intellectual heaven would be if we find the question to which the universe is the answer, and the nature of that question in and of itself explains why it was possible to describe it in so many different ways. \\
\indent — \textsc{N. Arkani-Hamed}\endnote{
	The text of the interview with N. Arkani-Hamed has been adapted for editorial purposes by N. Wolchover, “A Different Kind of Theory of Everything. Physicists used to search for the smallest components of the universe. What if that's not the point?”, \textit{The New Yorker}, February 19, 2019.
	}

\endgroup

\vspace{2mm}

In this Section, we will go towards the limits of knowledge in high energy physics, reporting the representation of a function of quantum fields  in Lagrangian formalism. Beyond this representation \emph{lands (still) unknown} open up.

\subsection{Lagrangian Density in 3-Interactions}

\begingroup
\footnotesize
Scientific ideas [of physics] are prisoners, and more than one thinks, of the experimental devices, just as musical ideas are of their musical instrumentation.\footnote{
	Cf. footnote \ref{footnote "We see the nature of the world on the strength of the technology that we are capable of building"} on p. \pageref{footnote "We see the nature of the world on the strength of the technology that we are capable of building"}.
	} \\
\indent — Inverted phrase from \textsc{P. Schaeffer}\footnote{
	The original version, in P. Schaeffer \cite[p. 17]{Schaeffer "Traite des objets musicaux: essai interdisciplines"}, reads as follows: «[L]es idées musicales sont prisonnières, et plus qu'on ne le croit, de l'appareillage musical, tout comme les idées scientifiques de leurs dispositifs expérimentaux». An En. transl. is in  \cite[p. 2]{Schaeffer "Treatise on Musical Objects: An Essay Across Disciplines"}.
	}

\endgroup

\vspace{2mm}

We will analyze below the Lagrangian density in the Standard Model of particle physics, which describes three of the four known fundamental interactions of nature: electromagnetic, weak, and strong forces, excluding gravity. We will organize the analysis into two parts, one preparatory, the other with the exposition of the Lagrangian.

\subsubsection{I. Preparatory Parts: Higgs Field and Adjustments for Neutrinos (Majorana Mass \& Pontecorvo's Oscillation)} 

The expression of the Lagrangian density is a \emph{Frankenstein's monster equation},\footnote{
	But not quite a modern Prometheus.
	} 
created by sewing pieces together from M. Veltman's book \cite[app. E]{Veltman "Diagrammatica. The Path to Feynman Rules"},\footnote{
	A Veltman's pre-\textit{Diagrammatica} is 't Hooft–Veltman's \textit{Diagrammar} \cite{'t Hooft and Veltman "Diagrammar"}.
	} 
according to Connes–Chamseddine–Marcolli revision \cite{Connes "Noncommutative geometry and the standard model with neutrino mixing"} \cite{Chamseddine Connes and Marcolli "Gravity and the standard model with neutrino mixing"} \cite[sec. 9.4]{Connes Marcolli "Noncommutative Geometry Quantum Fields and Motives"}, compare with \cite{Connes "Noncommutative Differential Geometry and the Structure of Space-Time"}.

\enumerationisinitium
\item Should we get a list of all mathematical objects to use for Frankenstein's equation in Section \ref{subsubsection "II. An Equation à la Frankenstein"}. 
\subenumerationisinitium
\item Let $\alpha_{H^0} = \frac{m_{H^0}^2}{4m_W^2}$ be a parameter for Higgs boson, via scattering processes, where $m_{H^0}$ and $m_W$ are the masses of the Higgs and weak ($W$-type) bosons, respectively (see below).  
\item Let $\theta_\mathrm{w}$ be the mixing angle, aka Weinberg angle, a parameter of the electroweak interaction under the \emph{Glashow–Weinberg–Salam theory} \cite{Glashow "The renormalizability of vector meson interactions"} \cite{Glashow "Partial-symmetries of weak interactions"} \cite{Salam Ward "Weak and electromagnetic interactions"} \cite{Weinberg "A Model of Leptons"}. 
\item Let $\textgreek{\text{λ}}_\textsc{g-m}^\alpha \viz \textgreek{\text{λ}}_{\chi_i\chi_j}^\alpha$, be the \emph{Gell-Mann matrices} \cite[p. 1074]{Gell-Mann "Symmetries of Baryons and Mesons"}, with $\alpha = 1, \mathellipsis, 8$,
\begin{subequations}
\begin{align}
 	\textgreek{\text{λ}}_\textsc{g-m}^1 = 
 	\left(\begin{smallmatrix} 
	0 & 1 & 0 \\
	1 & 0 & 0 \\ 
	0 & 0 & 0
	\end{smallmatrix}\right), \enspace
	\textgreek{\text{λ}}_\textsc{g-m}^2 =
	\left(\begin{smallmatrix}
	0 & -i & 0 \\
	i & 0 & 0 \\ 
	0 & 0 & 0
	\end{smallmatrix}\right), \enspace 
	\textgreek{\text{λ}}_\textsc{g-m}^3 =
	\left(\begin{smallmatrix}
	1 & 0 & 0 \\
	0 & -1 & 0 \\ 
	0 & 0 & 0
	\end{smallmatrix}\right), \enspace
	\textgreek{\text{λ}}_\textsc{g-m}^4 =
	\left(\begin{smallmatrix} 
	0 & 0 & 1 \\
	0 & 0 & 0 \\ 
	1 & 0 & 0
	\end{smallmatrix}\right), \\
	\textgreek{\text{λ}}_\textsc{g-m}^5 = 
 	\left(\begin{smallmatrix} 
	0 & 0 & -i \\
	0 & 0 & 0 \\ 
	i & 0 & 0
	\end{smallmatrix}\right), \enspace
	\textgreek{\text{λ}}_\textsc{g-m}^6 =
	\left(\begin{smallmatrix}
	0 & 0 & 0 \\
	0 & 0 & 1 \\ 
	0 & 1 & 0
	\end{smallmatrix}\right), \enspace 
	\textgreek{\text{λ}}_\textsc{g-m}^7 =
	\left(\begin{smallmatrix}
	0 & 0 & 0 \\
	0 & 0 & -i \\ 
	0 & i & 0
	\end{smallmatrix}\right), \enspace
	\textgreek{\text{λ}}_\textsc{g-m}^8 =
	\left(\begin{smallmatrix} 
	\frac{1}{\sqrt{3}} & 0 & 0 \\
	0 & \frac{1}{\sqrt{3}} & 0 \\ 
	0 & 0 & -\frac{2}{\sqrt{3}}
	\end{smallmatrix}\right).
\end{align}
\end{subequations}
\item Let $A^{(\gamma)}$ be the photon field, that is, the electromagnetic field—in particle terms, $A^{(\gamma)}$ corresponds to the photon, the gauge boson for electromagnetic interactions; $A^{(\gamma)}$ is accompanied by subscripts $(_\mu)$ or $(_\nu)$, i.e. $A^{(\gamma)}_\mu$, $A^{(\gamma)}_\nu$.
\item Let $c_\mathrm{t}$ be a constant in the so-called \emph{tadpole 1-loop} Feynman diagram. 
\item 
\label{item "Gauge group structure constants"}
Let $f^{\alpha\beta\gamma\delta\epsilon}$ be the (gauge group) structure constants, or structure coefficients, of the Lie algebra $\mathfrak{su}_3$ of $SU_3$, the special unitary group of degree 3, that find application in quantum chromodynamics (\textsc{qcd}).
\item Let $\couplingconstant$ be a (gauge) coupling constant, putting $\couplingconstant = \sqrt{4\pi\alpha_\mathrm{em}}$, $\alpha_\mathrm{em} = \frac{\sin_\mathrm{w}^2\couplingconstant^2}{4\pi}$ being the fine-structure constant, which determines the coupling strength for electromagnetism.
\item Let $\couplingconstant_{\mathrm{s}}$ be the strong (gauge) coupling constant, i.e. \textsc{qcd} coupling constant.
\item Let $g\ell$ be the gluon \cite[p. 1073]{Gell-Mann "Symmetries of Baryons and Mesons"}, the gauge boson acting in the strong interaction between quarks; $g\ell$ is accompanied by superscripts $(^\alpha)$, $(^\beta)$, $(^\gamma)$, $(^\delta)$, or $(^\epsilon)$, where $\alpha\beta\gamma\delta\epsilon = 1, \mathellipsis, 8$, since there are eight types of gluons, and subscripts $(_\mu)$ or $(_\nu)$.
\item Let $\vec{\Gamma}_\textsc{fp}$, $\vec{G}\ell_\textsc{fp}^\alpha$, $\vec{W}_\textsc{fp}^+$, $\vec{W}_\textsc{fp}^-$, and $\vec{Z}_\textsc{fp}^0$ be the \emph{Faddeev–Popov ghosts} \cite{Faddeev Popov "Feynman diagrams for the Yang-Mills field"}, i.e. ghost gauge fields, which are thought to make the quantum formalism consistent: $\Gamma_\textsc{fp}$ is the Faddeev–Popov ghost field associated with the photon; $\vec{G}\ell_\textsc{fp}^\alpha$, with $\alpha = 1, \mathellipsis 8$, are the Faddeev–Popov ghost fields associated with the eight types of gluons; $\vec{W}_\textsc{fp}^+$, $\vec{W}_\textsc{fp}^-$, and $\vec{Z}_\textsc{fp}^0$, are the Faddeev–Popov ghost fields associated with the positive/negative, and neutral weak bosons (see below). 
\item Let $H^0$, $\textgreek{\textit{\ddigamma}}_{H^0}^0$, $\textgreek{\textit{\ddigamma}}_{H^0}^+$, and $\textgreek{\textit{\ddigamma}}_{H^0}^-$ be the \emph{Higgs boson} and \emph{Higgs scalar fields}; refer to papers by F. Englert \& R. Brout \cite{Englert and R. Brout "Broken Symmetry and the Mass of Gauge Vector Mesons"}, P.W. Higgs \cite{Higgs "Broken Symmetries Massless Particles and Gauge Fields"} \cite{Higgs "Broken Symmetries and the Masses of Gauge Bosons"} \cite{Higgs "Spontaneous Symmetry Breakdown without Massless Bosons"},\footnote{
	About the Englert–Brout–Higgs mechanism, it is sufficient to recall here the theoretical core that animates it: 
	
	($\mathnormal{1}$) F. Englert \& R. Brout \cite[p. 321]{Englert and R. Brout "Broken Symmetry and the Mass of Gauge Vector Mesons"} examine a model based on a chiral invariant Lagrangian, with vector and pseudo-vector gauge fields, in such a way as to guarantee an invariance under local phase and local $\gamma^5$ phase transformations (cf. Section \ref{subsubsection "Dirac 4-Spinor Representation"}). «In this model the gauge fields themselves may break the $\gamma^5$ invariance leading to a mass for the original Fermi field», and it is possible to show that «the pseudovector field acquires mass».
	
	($\mathnormal{2}$) P.W. Higgs \cite[pp. 508-509]{Higgs "Spontaneous Symmetry Breakdown without Massless Bosons"} reports that «the spin-one quanta of some of the gauge fields acquire mass», imagining that «a spontaneous breakdown of symmetry under an internal Lie group occurs», and if «the conserved currents associated with the internal group are coupled to gauge fields». «[T]he longitudinal degrees of freedom of these particles (which would be absent if their mass were zero) go over into the [Nambu–]Goldstone bosons \cite{Nambu "Quasi-Particles and Gauge Invariance in the Theory of Superconductivity"} \cite{Goldstone "Field Theories with "Superconductor" Solutions"} when the coupling tends to zero». «It may be expected that when a further mechanism (presumably related to the weak interactions) is introduced in order to break [\,\dots] conservation, one of these gauge fields [in particle-form of weak bosons $W^\pm$ and $Z^0$] will acquire mass, leaving the photon as the only massless vector particle». The \emph{Higgs boson} makes its first appearance in \cite[p. 508, equation 2b]{Higgs "Spontaneous Symmetry Breakdown without Massless Bosons"}.
	}
and G.S. Guralnik, C.R. Hagen \& T.W.B. Kibble \cite{Guralnik Hagen Kibble "Global Conservation Laws and Massless Particles"}.
\item Let $\lepton^{\varsigma, \tau}$, $\neutrino^{\varsigma, \tau}$ be the leptons, where $\varsigma, \tau = 1, \mathellipsis, 3$ mean \emph{generation indices}; there are six types of leptons, with three lepton generations: \emph{electron} and \emph{electron neutrino} in the first generation, \emph{muon} and \emph{muon neutrino} in the second generation, \emph{tau(on)} and \emph{tau(on) neutrino} in the third generation.

Note. Let $\neutrino^{(\wr)}_\varsigma$ denotes the image of a neutrino $\neutrino_\varsigma$ under \textsc{cpt} symmetry, or under transformations of charge conjugation (\textsc{c}), parity transformation (\textsc{p}), and time reversal (\textsc{t}); in Majorana's case, $\neutrino = \neutrino^{(\wr)}_\varsigma$.
\item Let
\begin{equation}
	L^\textsc{pmns}_{\varsigma\tau, \tau\varsigma} = 
	\left(\begin{smallmatrix} 
	L_1 & L_2 & L_3 \\
	L_4 & L_5 & L_6 \\ 
	L_7 & L_8 & L_9
	\end{smallmatrix}\right)	
\end{equation} 
be the \emph{Pontecorvo–Maki–Nakagawa–Sakata matrix} \cite{Pontecorvo "Inverse beta processes and nonconservation of lepton charge"} \cite{Maki Nakagawa Sakata "Remarks on the Unified Model of Elementary Particles"}, a unitary lepton-neutrino mixing matrix in the weak interactions, for the phenomenon of \emph{neutrino mixing}, where 

$L_1 = \cos\theta_{12}\cos\theta_{13}$,
 
$L_2 = \sin\theta_{12}\cos\theta_{13}$,

$L_3 = \sin\theta_{13}e^{-i\delta_\textsc{cp}}$, 

$L_4 = -\sin\theta_{12}\cos\theta_{23} - \cos\theta_{12}\sin\theta_{23}\sin\theta_{13}e^{i\delta_\textsc{cp}}$,
 
$L_5 = \cos\theta_{12}\cos\theta_{23} - \sin\theta_{12}\sin\theta_{23}\sin\theta_{13}e^{i\delta_\textsc{cp}}$,
	
$L_6 = \sin\theta_{23}\cos\theta_{13}$,

$L_7 = \sin\theta_{12}\sin\theta_{23} - \cos\theta_{12}\cos\theta_{23}\sin\theta_{13}e^{i\delta_\textsc{cp}}$,

$L_8 = -\cos\theta_{12}\sin\theta_{23} - \sin\theta_{12}\cos\theta_{23}\sin\theta_{13}e^{i\delta_\textsc{cp}}$,

$L_9 = \cos\theta_{23}\cos\theta_{13}$, \\
with the mixing $\theta$-angles, $\delta_\textsc{cp}$ is the \emph{\textsc{cp} symmetry violating phase} \cite{Christenson Cronin Fitch and Turlay "Evidence for the 2pi Decay of the K02 Meson"} (\textsc{c} stands for \emph{charge conjugation symmetry}, and \textsc{p} stands for \emph{parity symmetry}). Here $L_\textsc{pmns} = L_\textsc{pmns}^\dag = L_\textsc{pmns}^{-1}$.

Note. This is related to \emph{Pontecorvo's neutrino oscillation} \cite{Pontecorvo "Mesonium and Antimesonium"} \cite{Pontecorvo "Neutrino Experiments and the Problem of Conservation of Leptonic Charge"}, under which the three types of neutrino (electron, muon, and tau(on) neutrino) change their lepton flavor as while they propagate (cf. Margo \ref{margo "Electrically neutral spinors in nature and mathematical models in comparison"}).
\item Let $m$ be the mass: e.g. $m_\neutrino$ (neutrino mass), $m_\lepton$ (lepton  mass), $m_W$ (mass of $W^\pm$);
\item Let
\begin{equation}
	M^\textsc{ckm}_{\varsigma\tau, \tau\varsigma} = 
	\left(\begin{smallmatrix} 
	M_{11} & M_{12} & M_{13} \\
	M_{21} & M_{22} & M_{23} \\ 
	M_{31} & M_{32} & M_{33}
	\end{smallmatrix}\right) =
	\left(\begin{smallmatrix} 
	M_{u(qk)d(qk)} & M_{u(qk)s(qk)} & M_{u(qk)b(qk)} \\
	M_{c(qk)d(qk)} & M_{c(qk)s(qk)} & M_{c(qk)b(qk)} \\ 
	M_{t(qk)d(qk)} & M_{t(qk)s(qk)} & M_{t(qk)b(qk)}
	\end{smallmatrix}\right)
\end{equation}
be the \emph{Cabibbo–Kobayashi–Maskawa matrix} \cite{Cabibbo "Unitary Symmetry and Leptonic Decays"} \cite{Kobayashi Maskawa "CP-Violation in the Renormalizable Theory of Weak Interaction"}, a $3 \times 3$ unitary matrix concerning the strength of flavour-changing weak decays, with the presence of the six types of quarks (see below). Here again $M_\textsc{ckm} = M_\textsc{ckm}^\dag = M_\textsc{ckm}^{-1}$.
\item Let $N^\textsc{m}$ be the matrix containing \emph{Majorana mass} parameters for \emph{neutrinos}, in pursuance of the \emph{Majorana representation} \cite{Majorana "Teoria simmetrica dell'elettrone e del positrone"} = \cite{Majorana "A symmetric theory of electrons and positrons"} (see Section \ref{subsubsection "Majorana Symmetric Representation"}).
\item Let $qk^{\varsigma, \tau}_\chi$ be the quark (see Gell-Mann's \cite[eightfold way]{Gell-Mann "The Eightfold Way. A Theory of Strong Interaction Symmetry"} \cite{Gell-Mann "A Schematic Model of Baryons and Mesons"} and G. Zweig's \cite{Zweig "An SU3 Model for Strong Interaction Symmetry and its Breaking"} \cite{Zweig "An SU3 Model for Strong Interaction Symmetry and its Breaking II"} theoretical papers), divided in ${_{(u)}qk}^{\varsigma, \tau}_\chi$ and ${_{(d)}qk}^{\varsigma, \tau}_\chi$, where the superscripts $\varsigma, \tau = 1, \mathellipsis, 3$ designate the \emph{flavor property}, or the three quark \emph{generations}: \emph{up} and \emph{down} in the first generation, \emph{strange} and \emph{charm} in the second generation, \emph{bottom} and \emph{top} in the third generation; while the subscript $\chi = 1, \mathellipsis, 3$ is the \emph{color charge} (blue, green, and red). 
\item Let 
\begin{equation}
	\sin\theta_\mathrm{w} = \frac{\couplingconstant_1}{\sqrt{\couplingconstant_1^2 + \couplingconstant_2^2}} \text{ and } \cos\theta_\mathrm{w} = \frac{\couplingconstant_2}{\sqrt{\couplingconstant_1^2 + \couplingconstant_2^2}}
\end{equation}
be the sine and cosine of the weak mixing angle (or Weinberg angle) $\theta_\mathrm{w}$.
\item Let $W^\pm$ and $Z^0$ be the positive/negative charged and neutral weak bosons, respectively, the vector bosons that mediate the weak interaction. Note. The $W^\pm$ and $Z^0$ particles have a definitive theorization within the electroweak model by S. Glashow \cite{Glashow "The renormalizability of vector meson interactions"} \cite{Glashow "Partial-symmetries of weak interactions"}, A. Salam \cite{Salam Ward "Weak and electromagnetic interactions"}, and S. Weinberg \cite{Weinberg "A Model of Leptons"}; nevertheless the prediction of $Z^0$ is already outlined in J. Leite Lopes \cite{Leite Lopes "A model of the universal Fermi interaction"}.
\subenumerationisfinis
\item So, concisely, the Standard Model, in its Lagrangian form, is modified with the addition of

· Higgs boson and Higgs scalar fields, 

· Majorana mass terms for neutrinos,

· Pontecorvo's neutrino oscillation, 

· neutrino mixing, with the Pontecorvo–Maki–Nakagawa–Sakata matrix.
\item Antimatter (antiparticle, antifield) is indicated with a horizontal line over the particle's and field's symbols.
\enumerationisfinis

\subsubsection{II. An Equation à la Frankenstein}
\label{subsubsection "II. An Equation à la Frankenstein"}

\begingroup
\footnotesize
But the Snark is at hand, let me tell you again! /
'Tis your glorious duty to seek it! [\,\dots] For the Snark's a peculiar creature, that won't / Be caught in a commonplace way. \\
\indent — \textsc{L. Carroll} \cite[pp. 39-40]{Carroll "The Hunting of the Snark: An Agony in Eight Fits"}

\endgroup

\vspace{2mm}

\textit{\textbf{A Revision of the Veltman–Connes–Chamseddine–Marcolli Version}}

\vspace{2mm}

We shall now proceed to the writing of the Lagrangian \textsc{sm} density  via \textsc{vccm} version (the acronym stands for Veltman, Connes, Chamseddine, and Marcolli), in which each part is marked with the  \textcolor{Chinese-loquat}{\texttt{yellow \#F7C015}} color in superscript.

· Eq. \textcolor{Chinese-loquat}{\texttt{[a]}} is the \emph{color Lagrangian}, with the gluonic interactions in \textsc{qcd};

· Eq. \textcolor{Chinese-loquat}{\texttt{[b]}} is the \emph{fermionic color Lagrangian}, with fermionic plus gluonic interactions, and Gell-Mann matrices;

· Eq. \textcolor{Chinese-loquat}{\texttt{[c]}} is the \emph{Faddeev–Popov ghost Lagrangian} in \textsc{qcd};

· Eq. \textcolor{Chinese-loquat}{\texttt{[d]}} is the \emph{weak Lagrangian}, with weak vector bosons \& weak nuclear force, and Englert–Brout–Higgs mechanism;

· Eq. \textcolor{Chinese-loquat}{\texttt{[e]}} is the \emph{fermion Lagrangian}, with interactions between fermions and weak vector bosons;

· Eq. \textcolor{Chinese-loquat}{\texttt{[f]}} is the \emph{fermion-Higgs Lagrangian}, with interactions of the fermions in the Englert–Brout–Higgs mechanism;

· Eq. \textcolor{Chinese-loquat}{\texttt{[g]}} is the \emph{Faddeev–Popov ghost Lagrangian} in the weak interaction.
\vspace{2mm}

\begin{math}
\Lagrangian_\textsc{sm} = \\
\textsuperscript{\textcolor{Chinese-loquat}{\texttt{[a]}}} 
- \frac{1}{2}\partial_\nu g\ell^\alpha_\mu \partial_\nu g\ell^\alpha_\mu
- \couplingconstant_\mathrm{s}f^{\alpha\beta\gamma}\partial_\mu g\ell^\alpha_\nu g\ell^\beta_\mu g\ell^\gamma_\nu 
- \frac{1}{4}\couplingconstant^2_\mathrm{s}f^{\alpha\beta\gamma} f^{\alpha\delta\epsilon}g\ell^\beta_\mu g\ell^\gamma_\nu g\ell^\delta_\mu g\ell^\epsilon_\nu \\
\textsuperscript{\textcolor{Chinese-loquat}{\texttt{[b]}}} 
+ \frac{1}{2}i\couplingconstant_\mathrm{s}\textgreek{\text{λ}}_{\chi_i\chi_j}^\alpha\left(\antimatter{qk}{^\varsigma_{\chi_i}} \gamma^\mu qk^\varsigma_{\chi_j}\right)g\ell^\alpha_\mu \\
\textsuperscript{\textcolor{Chinese-loquat}{\texttt{[c]}}} 
+ \antimatter{\vec{G}\ell}{_\textsc{fp}^\alpha}\partial^2 \vec{G}\ell_\textsc{fp}^\alpha + \couplingconstant_\mathrm{s}f^{\alpha\beta\gamma} \partial_\mu\antimatter{\vec{G}\ell}{_\textsc{fp}^\alpha}\vec{G}\ell_\textsc{fp}^\beta g\ell^\gamma_\mu \\
\textsuperscript{\textcolor{Chinese-loquat}{\texttt{[d]}}} 
- \partial_\nu W^+_\mu\partial_\nu W^-_\mu - m_W^2W^+_\mu W^-_\mu - \frac{1}{2}\partial_\nu Z^0_\mu \partial_\nu Z^0_\mu - \frac{1}{2\cos\theta_\mathrm{w}^2}m_W^2 Z^0_\mu Z^0_\mu \\ 
- \frac{1}{2}\partial_\mu A^{(\gamma)}_\nu\partial_\mu A^{(\gamma)}_\nu - \frac{1}{2}\partial_\mu H^0\partial_\mu H^0 - 2m_W^2\alpha_{H^0}(H^0)^2 
- \partial_\mu\textgreek{\textit{\ddigamma}}_{H^0}^+\partial_\mu \textgreek{\textit{\ddigamma}}_{H^0}^- \\
- \frac{1}{2}\partial_\mu\textgreek{\textit{\ddigamma}}_{H^0}^0\partial_\mu \textgreek{\textit{\ddigamma}}_{H^0}^0 - c_\mathrm{t}
\Bigl\{
\frac{2m_W^2}{\couplingconstant^2} + \frac{2m_W}{\couplingconstant}H^0 + \frac{1}{2}\Bigl((H^0)^2 + \textgreek{\textit{\ddigamma}}_{H^0}^0 \textgreek{\textit{\ddigamma}}_{H^0}^0 \\
+ 2\textgreek{\textit{\ddigamma}}_{H^0}^+\textgreek{\textit{\ddigamma}}_{H^0}^-\Bigr)
\Bigr\} + \frac{2m_W^4}{\couplingconstant^2}\alpha_{H^0} - i\couplingconstant\cos\theta_\mathrm{w}
\Bigl\{
\partial_\nu Z^0_\mu\left(W^+_\mu W^-_\nu - W^+_\nu W^-_\mu\right) \\
- Z^0_\nu\left(W^+_\mu\partial_\nu W^-_\mu - W^-_\mu\partial_\nu W^+_\mu\right) + Z^0_\mu\left(W^+_\nu\partial_\nu W^-_\mu - W^-_\nu\partial_\nu W^+_\mu\right)
\Bigr\} \\
- i\couplingconstant\sin\theta_\mathrm{w}
\Bigl\{
\partial_\nu A^{(\gamma)}_\mu\left(W^+_\mu W^-_\nu - W^+_\nu W^-_\mu\right) \\
- A^{(\gamma)}_\nu\left(W^+_\mu\partial_\nu W^-_\mu - W^-_\mu\partial_\nu W^+_\mu\right)  
+ A^{(\gamma)}_\mu\left(W^+_\nu\partial_\nu W^-_\mu - W^-_\nu\partial_\nu W^+_\mu\right)
\Bigr\} \\
- \frac{1}{2}\couplingconstant^2 W^+_\mu W^-_\mu W^+_\nu W^-_\nu
+ \frac{1}{2}\couplingconstant^2 W^+_\mu W^-_\nu W^+_\mu W^-_\nu \\
+ \couplingconstant^2\cos\theta_\mathrm{w}^2\left(Z^0_\mu W^+_\mu Z^0_\nu W^-_\nu - Z^0_\mu Z^0_\mu W^+_\nu W^-_\nu\right) \\
+ \couplingconstant^2\sin\theta_\mathrm{w}^2\left(A^{(\gamma)}_\mu W^+_\mu A^{(\gamma)}_\nu W^-_\nu - A^{(\gamma)}_\mu A^{(\gamma)}_\mu W^+_\nu W^-_\nu\right) \\
+ \couplingconstant^2\sin\theta_\mathrm{w} \cos\theta_\mathrm{w}
\left\{
A^{(\gamma)}_\mu Z^0_\nu\left(W^+_\mu W^-_\nu - W^+_\nu W^-_\mu\right) - 2A^{(\gamma)}_\mu Z^0_\mu W^+_\nu W^-_\nu
\right\} \\
- \couplingconstant\alpha_{H^0}m_W\Bigl((H^0)^3 + H^0\textgreek{\textit{\ddigamma}}_{H^0}^0\textgreek{\textit{\ddigamma}}_{H^0}^0 + 2H^0\textgreek{\textit{\ddigamma}}_{H^0}^+ \textgreek{\textit{\ddigamma}}_{H^0}^-\Bigr) \\ 
- \frac{1}{8}\couplingconstant^2\alpha_{H^0}
\Bigl\{
(H^0)^4 + (\textgreek{\textit{\ddigamma}}_{H^0}^0)^4 + 4(\textgreek{\textit{\ddigamma}}_{H^0}^+ \textgreek{\textit{\ddigamma}}_{H^0}^-)^2 + 4(\textgreek{\textit{\ddigamma}}_{H^0}^0)^2 \textgreek{\textit{\ddigamma}}_{H^0}^+ \textgreek{\textit{\ddigamma}}_{H^0}^- \\
+ 4(H^0)^2\textgreek{\textit{\ddigamma}}_{H^0}^+ \textgreek{\textit{\ddigamma}}_{H^0}^- + 2(\textgreek{\textit{\ddigamma}}_{H^0}^0)^2 (H^0)^2
\Bigr\} 
- \couplingconstant m_WW^+_\mu W^-_\mu H^0 \\
- \frac{1}{2}\couplingconstant\frac{m_W}{\cos\theta_\mathrm{w}^2}Z^0_\mu Z^0_\mu H^0 
- \frac{1}{2}i\couplingconstant
\Bigl\{
W^+_\mu\left(\textgreek{\textit{\ddigamma}}_{H^0}^0\partial_\mu \textgreek{\textit{\ddigamma}}_{H^0}^- - \textgreek{\textit{\ddigamma}}_{H^0}^-\partial_\mu \textgreek{\textit{\ddigamma}}_{H^0}^0\right) \\
- W^-_\mu\left(\textgreek{\textit{\ddigamma}}_{H^0}^0\partial_\mu \textgreek{\textit{\ddigamma}}_{H^0}^+ - \textgreek{\textit{\ddigamma}}_{H^0}^+ \partial_\mu \textgreek{\textit{\ddigamma}}_{H^0}^0\right)
\Bigr\}
+ \frac{1}{2}\couplingconstant
\Bigl\{
W^+_\mu\left(H^0\partial_\mu \textgreek{\textit{\ddigamma}}_{H^0}^- - \textgreek{\textit{\ddigamma}}_{H^0}^-\partial_\mu H^0\right) \\
+ W^-_\mu\left(H^0\partial_\mu\textgreek{\textit{\ddigamma}}_{H^0}^+ - \textgreek{\textit{\ddigamma}}_{H^0}^+\partial_\mu H^0\right)
\Bigr\} 
+ \frac{1}{2}\couplingconstant\frac{1}{\cos\theta_\mathrm{w}}Z^0_\mu\left(H^0\partial_\mu \textgreek{\textit{\ddigamma}}_{H^0}^0 - \textgreek{\textit{\ddigamma}}_{H^0}^0 \partial_\mu H^0\right) \\
+ m_W\left(\frac{1}{\cos\theta_\mathrm{w}}Z^0_\mu \partial_\mu \textgreek{\textit{\ddigamma}}_{H^0}^0 + W^+_\mu\partial_\mu \textgreek{\textit{\ddigamma}}_{H^0}^- + W^-_\mu\partial_\mu \textgreek{\textit{\ddigamma}}_{H^0}^+\right) \\
- i\couplingconstant\frac{\sin\theta_\mathrm{w}^2}{\cos\theta_\mathrm{w}}m_WZ^0_\mu\left(W^+_\mu \textgreek{\textit{\ddigamma}}_{H^0}^- - W^-_\mu \textgreek{\textit{\ddigamma}}_{H^0}^+\right) \\
+ i\couplingconstant\sin\theta_\mathrm{w}m_WA^{(\gamma)}_\mu\left(W^+_\mu \textgreek{\textit{\ddigamma}}_{H^0}^- - W^-_\mu \textgreek{\textit{\ddigamma}}_{H^0}^+\right) \\
- i\couplingconstant\frac{1 - 2\cos\theta_\mathrm{w}^2}{2\cos\theta_\mathrm{w}}Z^0_\mu\left(\textgreek{\textit{\ddigamma}}_{H^0}^+\partial_\mu\textgreek{\textit{\ddigamma}}_{H^0}^- - \textgreek{\textit{\ddigamma}}_{H^0}^-\partial_\mu \textgreek{\textit{\ddigamma}}_{H^0}^+\right) \\
+ i\couplingconstant\sin\theta_\mathrm{w} A^{(\gamma)}_\mu\left(\textgreek{\textit{\ddigamma}}_{H^0}^+\partial_\mu \textgreek{\textit{\ddigamma}}_{H^0}^- - \textgreek{\textit{\ddigamma}}_{H^0}^- \partial_\mu \textgreek{\textit{\ddigamma}}_{H^0}^+\right) \\ 
- \frac{1}{4}\couplingconstant^2 W^+_\mu W^-_\mu\Bigl((H^0)^2 + (\textgreek{\textit{\ddigamma}}_{H^0}^0)^2 + 2\textgreek{\textit{\ddigamma}}_{H^0}^+\textgreek{\textit{\ddigamma}}_{H^0}^-\Bigl) \\
- \frac{1}{8}\couplingconstant^2\frac{1}{\cos\theta_\mathrm{w}^2}Z^0_\mu Z^0_\mu\Bigl((H^0)^2 + (\textgreek{\textit{\ddigamma}}_{H^0}^0)^2 + 2(2\sin\theta_\mathrm{w}^2 - 1)^2 \textgreek{\textit{\ddigamma}}_{H^0}^+\textgreek{\textit{\ddigamma}}_{H^0}^-\Bigr) \\
- \frac{1}{2}\couplingconstant^2\frac{\sin\theta_\mathrm{w}^2}{\cos\theta_\mathrm{w}}Z^0_\mu \textgreek{\textit{\ddigamma}}_{H^0}^0\left(W^+_\mu \textgreek{\textit{\ddigamma}}_{H^0}^- + W^-_\mu \textgreek{\textit{\ddigamma}}_{H^0}^+\right) \\
- \frac{1}{2}i\couplingconstant^2\frac{\sin\theta_\mathrm{w}^2}{\cos\theta_\mathrm{w}}Z^0_\mu H^0\left(W^+_\mu \textgreek{\textit{\ddigamma}}_{H^0}^- - W^-_\mu \textgreek{\textit{\ddigamma}}_{H^0}^+\right) \\
+ \frac{1}{2}\couplingconstant^2 \sin\theta_\mathrm{w} A^{(\gamma)}_\mu \textgreek{\textit{\ddigamma}}_{H^0}^0\left(W^+_\mu \textgreek{\textit{\ddigamma}}_{H^0}^- + W^-_\mu \textgreek{\textit{\ddigamma}}_{H^0}^+\right) \\
+ \frac{1}{2}i\couplingconstant^2 \sin\theta_\mathrm{w} A^{(\gamma)}_\mu H^0\left(W^+_\mu\textgreek{\textit{\ddigamma}}_{H^0}^- - W^-_\mu\textgreek{\textit{\ddigamma}}_{H^0}^+\right) \\
- \couplingconstant^2\frac{\sin\theta_\mathrm{w}}{\cos\theta_\mathrm{w}}\left(2\cos\theta_\mathrm{w}^2 - 1\right)Z^0_\mu A^{(\gamma)}_\mu \textgreek{\textit{\ddigamma}}_{H^0}^+ \textgreek{\textit{\ddigamma}}_{H^0}^-
- \couplingconstant^2 \sin\theta_\mathrm{w}^2 A^{(\gamma)}_\mu A^{(\gamma)}_\mu \textgreek{\textit{\ddigamma}}_{H^0}^+ \textgreek{\textit{\ddigamma}}_{H^0}^- \\
\textsuperscript{\textcolor{Chinese-loquat}{\texttt{[e]}}}
- \antimatter{\lepton}{^\tau}\left(\gamma\partial + m_\lepton^\tau\right)\lepton^\tau - \antimatter{\neutrino}{^\tau}\left(\gamma\partial + m_\neutrino^\tau\right)\neutrino^\tau \\
- \antimatter{_{(u)}qk}{^\tau_{\chi_j}}\left(\gamma\partial + m_{{_{(u)}qk}}^\tau\right){_{(u)}qk}^\tau_{\chi_j} - \antimatter{_{(d)}qk}{^\tau_{\chi_j}}\left(\gamma\partial + m_{{_{(d)}qk}}^\tau\right){_{(d)}qk}^\tau_{\chi_j} \\
+ i\couplingconstant\sin\theta_\mathrm{w}A^{(\gamma)}_\mu
\Bigl\{
- \left(\antimatter{\lepton}{^\tau}\gamma^\mu\lepton^\tau\right) + \frac{2}{3}\left(\antimatter{_{(u)}qk}{^\tau_{\chi_j}}\gamma^\mu {_{(u)}qk}^\tau_{\chi_j}\right) \\
- \frac{1}{3}\left(\antimatter{_{(d)}qk}{^\tau_{\chi_j}}\gamma^\mu {_{(d)}qk}^\tau_{\chi_j}\right)
\Bigr\}
+ \frac{i\couplingconstant}{4\cos\theta_\mathrm{w}}Z^0_\mu
\Bigl\{
\Bigl(\antimatter{\neutrino}^{\tau}\gamma^\mu(1 + \gamma^5)\neutrino^\tau\Bigr) \\
+ \Bigl(\antimatter{\lepton}{^\tau}\gamma^\mu\left(4\sin\theta_\mathrm{w}^2 - 1 - \gamma^5\right)\lepton^\tau\Bigr) + \left(\antimatter{_{(d)}qk}{^\tau_{\chi_j}}\gamma^\mu\left(\frac{4}{3}\sin\theta_\mathrm{w}^2 - 1 - \gamma^5\right){_{(d)}qk}^\tau_{\chi_j}\right) \\ 
+ \left(\antimatter{_{(u)}qk}{^\tau_{\chi_j}}\gamma^\mu\left(1 - \frac{8}{3}\sin\theta_\mathrm{w}^2 + \gamma^5\right){_{(u)}qk}^\tau_{\chi_j}\right)
\Bigr\} \\
+ \frac{i\couplingconstant}{2\sqrt{2}}W^+_\mu
\left\{
\Bigl(\antimatter{\neutrino}{^\tau}\gamma^\mu(1 + \gamma^5)L_{\tau\varsigma}\lepton^\varsigma\Bigr) + \left(\antimatter{_{(u)}qk}{^\tau_{\chi_j}} \gamma^\mu(1 + \gamma^5)M_{\tau\varsigma}{_{(d)}qk}^\varsigma_{\chi_j}\right)
\right\} \\
+ \frac{i\couplingconstant}{2\sqrt{2}}W^-_\mu
\left\{
\Bigl(\antimatter{\lepton}{^\varsigma}L_{\varsigma\tau}^\dag\gamma^\mu(1 + \gamma^5)\neutrino^\tau\Bigr) + \left(\antimatter{_{(d)}qk}{^\varsigma_{\chi_j}}M_{\varsigma\tau}^\dag \gamma^\mu(1 + \gamma^5){_{(u)}qk}^\tau_{\chi_j}\right) \\
\right\} \\
\textsuperscript{\textcolor{Chinese-loquat}{\texttt{[f]}}}
+ \frac{i\couplingconstant}{2m_W\sqrt{2}}\textgreek{\textit{\ddigamma}}_{H^0}^+
\left\{
- m_\lepton^\varsigma\Bigl(\antimatter{\neutrino}{^\tau}L_{\tau\varsigma}(1 - \gamma^5)\lepton^\varsigma\Bigr) + m_\neutrino^\tau\Bigl(\antimatter{\neutrino}{^\tau}L_{\tau\varsigma}(1 + \gamma^5)\lepton^\varsigma\Bigr)
\right\} \\
+ \frac{i\couplingconstant}{2m_W\sqrt{2}}\textgreek{\textit{\ddigamma}}_{H^0}^-
\left\{
m_\lepton^\tau\Bigl(\antimatter{\lepton}{^\tau}L_{\tau\varsigma}^\dag(1 + \gamma^5)\neutrino^\varsigma\Bigr) - m_\neutrino^\varsigma\Bigr(\antimatter{\lepton}{^\tau}L_{\tau\varsigma}^\dag(1 - \gamma^5)\neutrino^\varsigma\Bigr)
\right\} \\
- \frac{\couplingconstant}{2}\frac{m_\neutrino^\tau}{m_W}H^0\left(\antimatter{\neutrino}{^\tau}\neutrino^\tau\right)
- \frac{\couplingconstant}{2}\frac{m_\lepton^\tau}{m_W}H^0\left(\antimatter{\lepton}{^\tau}\lepton^\tau\right)
+ \frac{i\couplingconstant}{2}\frac{m_\neutrino^\tau}{m_W}\textgreek{\textit{\ddigamma}}_{H^0}^0\left(\antimatter{\neutrino}{^\tau}\gamma^5\neutrino^\tau\right) \\
- \frac{i\couplingconstant}{2}\frac{m_\lepton^\tau}{m_W}\textgreek{\textit{\ddigamma}}_{H^0}^0\left(\antimatter{\lepton}{^\tau}\gamma^5\lepton^\tau\right)
- \frac{1}{4}\antimatter{\nu}{_\tau}N^\textsc{m}_{\tau\varsigma}(1 - \gamma_5)\neutrino^{(\wr)}_\varsigma
- \frac{1}{4}\overline{\antimatter{\nu}{_\tau}N^\textsc{m}_{\tau\varsigma}(1 - \gamma_5)\neutrino^{(\wr)}_\varsigma} \\
+ \frac{i\couplingconstant}{2m_W\sqrt{2}}\textgreek{\textit{\ddigamma}}_{H^0}^+ 
\Bigl\{
- m_{{_{(d)}qk}}^\varsigma\left(\antimatter{_{(u)}qk}{^\tau_{\chi_j}} M_{\tau\varsigma}(1 - \gamma^5){_{(d)}qk}^\varsigma_{\chi_j}\right) \\
+ m_{{_{(u)}qk}}^\tau\left(\antimatter{_{(u)}qk}{^\tau_{\chi_j}} M_{\tau\varsigma}(1 + \gamma^5){_{(d)}qk}^\varsigma_{\chi_j}\right)
\Bigr\} \\
+ \frac{i\couplingconstant}{2m_W\sqrt{2}}\textgreek{\textit{\ddigamma}}_{H^0}^-
\Bigl\{
m_{{_{(d)}qk}}^\tau\left(\antimatter{_{(d)}qk}{^\tau_{\chi_j}} M_{\tau\varsigma}^\dag(1 + \gamma^5){_{(u)}qk}^\varsigma_{\chi_j}\right) \\
- m_{{_{(u)}qk}}^\varsigma\left(\antimatter{_{(d)}qk}{^\tau_{\chi_j}} M_{\tau\varsigma}^\dag(1 - \gamma^5){_{(u)}qk}^\varsigma_{\chi_j}\right)
\Bigr\}
- \frac{\couplingconstant}{2}\frac{m_{{_{(u)}qk}}^\tau}{m_W}H^0\left(\antimatter{_{(u)}qk}{^\tau_{\chi_j}}{_{(u)}qk}^\tau_{\chi_j}\right) \\
- \frac{\couplingconstant}{2}\frac{m_{{_{(d)}qk}}^\tau}{m_W}H^0\left(\antimatter{_{(d)}qk}{^\tau_{\chi_j}}{_{(d)}qk}^\tau_{\chi_j}\right) + \frac{i\couplingconstant}{2}\frac{m_{{_{(u)}qk}}^\tau}{m_W}\textgreek{\textit{\ddigamma}}_{H^0}^0\left(\antimatter{_{(u)}qk}{^\tau_{\chi_j}}\gamma^5 {_{(u)}qk}^\tau_{\chi_j}\right) \\
- \frac{i\couplingconstant}{2}\frac{m_{{_{(d)}qk}}^\tau}{m_W}\textgreek{\textit{\ddigamma}}_{H^0}^0\left(\antimatter{_{(d)}qk}{^\tau_{\chi_j}}\gamma^5{_{(d)}qk}^\tau_{\chi_j}\right) \\
\textsuperscript{\textcolor{Chinese-loquat}{\texttt{[g]}}}
+ \antimatter{\vec{W}}{_\textsc{fp}^+}\left(\partial^2 - m_W^2\right)\vec{W}_\textsc{fp}^+ + \antimatter{\vec{W}}{_\textsc{fp}^-}\left(\partial^2 - m_W^2\right)\vec{W}_\textsc{fp}^- \\
+ \antimatter{\vec{Z}}{_\textsc{fp}^0}\left(\partial^2 - \frac{m_W^2}{\cos\theta_\mathrm{w}^2}\right)\vec{Z}_\textsc{fp}^0 + \antimatter{\vec{\Gamma}}{_\textsc{fp}}\partial^2 \vec{\Gamma}_\textsc{fp} \\
+ i\couplingconstant\cos\theta_\mathrm{w} W^+_\mu\left(\partial_\mu\antimatter{\vec{Z}}{_\textsc{fp}^0}\vec{W}_\textsc{fp}^- - \partial_\mu\antimatter{\vec{W}}{_\textsc{fp}^+}\vec{Z}_\textsc{fp}^0\right) \\
+ i\couplingconstant\sin\theta_\mathrm{w} W^+_\mu\left(\partial_\mu\antimatter{\vec{\Gamma}}{_\textsc{fp}}\vec{W}_\textsc{fp}^- - \partial_\mu\antimatter{\vec{W}}{_\textsc{fp}^+} \vec{\Gamma}_\textsc{fp}\right) \\
+ i\couplingconstant\cos\theta_\mathrm{w}W^-_\mu\left(\partial_\mu\antimatter{\vec{W}}{_\textsc{fp}^-}\vec{Z}_\textsc{fp}^0 - \partial_\mu\antimatter{\vec{Z}}{_\textsc{fp}^0} \vec{W}_\textsc{fp}^+\right) \\
+ i\couplingconstant\sin\theta_\mathrm{w}W^-_\mu\left(\partial_\mu\antimatter{\vec{W}}{_\textsc{fp}^-}\vec{\Gamma}_\textsc{fp} - \partial_\mu\antimatter{\vec{\Gamma}}{_\textsc{fp}}\vec{W}_\textsc{fp}^+\right) \\
+ i\couplingconstant\cos\theta_\mathrm{w} Z^0_\mu\left(\partial_\mu\antimatter{\vec{W}}{_\textsc{fp}^+}\vec{W}_\textsc{fp}^+ - \partial_\mu\antimatter{\vec{W}}{_\textsc{fp}^-}\vec{W}_\textsc{fp}^-\right) \\
+ i\couplingconstant\sin\theta_\mathrm{w}A^{(\gamma)}_\mu\left(\partial_\mu\antimatter{\vec{W}}{_\textsc{fp}^+}\vec{W}_\textsc{fp}^+ - \partial_\mu\antimatter{\vec{W}}{_\textsc{fp}^-}\vec{W}_\textsc{fp}^-\right) \\
- \frac{1}{2}\couplingconstant m_W\left(\antimatter{\vec{W}}{_\textsc{fp}^+} \vec{W}_\textsc{fp}^+ H^0 + \antimatter{\vec{W}}{_\textsc{fp}^-}\vec{W}_\textsc{fp}^- H^0 + \frac{1}{\cos\theta_\mathrm{w}^2}\antimatter{\vec{Z}}{_\textsc{fp}^0}\vec{Z}_\textsc{fp}^0 H^0\right) \\
+ \frac{1 - 2\cos\theta_\mathrm{w}^2}{2\cos\theta_\mathrm{w}}i\couplingconstant m_W\left(\antimatter{\vec{W}}{_\textsc{fp}^+}\vec{Z}_\textsc{fp}^0\textgreek{\textit{\ddigamma}}_{H^0}^+ - \antimatter{\vec{W}}{_\textsc{fp}^-}\vec{Z}_\textsc{fp}^0\textgreek{\textit{\ddigamma}}_{H^0}^-\right) \\
+ \frac{1}{2\cos\theta_\mathrm{w}}i\couplingconstant m_W\left(\antimatter{\vec{Z}}{_\textsc{fp}^0}\vec{W}_\textsc{fp}^- \textgreek{\textit{\ddigamma}}_{H^0}^+ - \antimatter{\vec{Z}}{_\textsc{fp}^0}\vec{W}_\textsc{fp}^+ \textgreek{\textit{\ddigamma}}_{H^0}^-\right) \\
+ i\couplingconstant m_W\sin\theta_\mathrm{w}\left(\antimatter{\vec{Z}}{_\textsc{fp}^0}\vec{W}_\textsc{fp}^- \textgreek{\textit{\ddigamma}}_{H^0}^+ - \antimatter{\vec{Z}}{_\textsc{fp}^0}\vec{W}_\textsc{fp}^+ \textgreek{\textit{\ddigamma}}_{H^0}^-\right) \\
+ \frac{1}{2}i\couplingconstant m_W\left(\antimatter{\vec{W}}{_\textsc{fp}^+}\vec{W}_\textsc{fp}^+ \textgreek{\textit{\ddigamma}}_{H^0}^0 - \antimatter{\vec{W}}{_\textsc{fp}^-}\vec{W}_\textsc{fp}^- \textgreek{\textit{\ddigamma}}_{H^0}^0\right).
\end{math}

\section{What is the Quantum Field Space?}
\label{section "What is the Quantum Field Space?"}

\subsection{Underlying Minkowskian Space}

We may think of space of the monster equation (Section \ref{subsubsection "II. An Equation à la Frankenstein"})—the reference system for the Lagrangian dynamics—as an \emph{underlying Minkowski manifold}, or \emph{space-time} (cf. Section \ref{section "Lorentz–Minkowski 4-Manifolds"}), for relativistic (without gravity) \textsc{qft}. 

\subsection{Function Multi-Space of Hilbertian Type}

\begingroup
\footnotesize
I expressed the concept that, for various researches of analysis, it is appropriate to consider the totality of analytic functions of one variable $x$ or—to fix the ideas better—the totality of series of positive integer powers of $x$, as a \emph{manifold} [\textit{varietà}] or \emph{space} of which every single series constitutes an element. To such a manifold, evidently with an infinite number of dimensions, we can give the name of \emph{function space}; any power series of $x$ will be a \emph{point} of this space and the coefficients of the series can be regarded as the coordinates of the point.\endnote{
	Original It. version: «[H]o espresso il concetto che, per varie ricerche d'analisi, è opportuno considerare la totalità delle funzioni analitiche di una variabile $x$ o — per meglio fissare le idee — la totalità delle serie di potenze intere positive di $x$, come una varietà o spazio di cui ogni singola serie costituisce un elemento. Ad una tale varietà, evidentemente ad un numero infinito di dimensioni, si può dare il nome di \emph{spazio funzionale}; ogni serie di potenze di $x$ sarà un \emph{punto} di questo spazio ed i coefficienti della serie si potranno riguardare come le coordinate del punto».
	} \\
\indent — \textsc{S. Pincherle} \cite[p. 85, e.a.]{Pincherle "Cenno sulla geometria dello spazio funzionale"}

\endgroup

\vspace{2mm}

We should also remember that the space structure on the quantum states corresponds to the \emph{Hilbert space},\footnote{
	J. von Neumann was the first to give a rigorous exposition of the Hilbert space in \cite{von Neumann "Allgemeine Eigenwerttheorie Hermitescher Funktionaloperatoren"} \cite{von Neumann "Zur Algebra der Funktionaloperationen und Theorie der normalen Operatoren"} \cite[chap. II]{Neumann "Mathematical Foundations of Quantum Mechanics"}, precisely named (\textit{abstrakten}) \textit{Hilbertschen Raum}; but his first studies on the Hilbert space are about a quantum context: \cite{von Neumann "Mathematische Begrundung der Quantenmechanik"} \cite{von Neumann "Wahrscheinlichkeitstheoretischer Aufbau der Quantenmechanik"} \cite{von Neumann "Thermodynamik quantenmechanischer Gesamtheiten"} \cite[written in collaboration with Hilbert and Nordheim]{Hilbert von Neumann und Nordheim "Uber die Grundlagen der Quantenmechanik"}.
	} 
one among the many \emph{function spaces}, see S. Pincherle in epigraph for a primal meaning. Hilbert space is thus a space by which quantum states are (geometrically) modeled. Alongside this set of functions there is a \emph{configuration space}—again with Hilbertian schemes, or even a \emph{Kre\u{\i}n space}\footnote{
	A \emph{Kre\u{\i}n space}, denoted by $\mathfrak{K}_\pm$, is a \emph{non-degenerate inner product space}, characterized by \emph{completeness}, and a \emph{decomposability} such as this one, $\mathfrak{K}_\pm = \mathfrak{K}_+ \dotplus \mathfrak{K}_-$, in which $(\mathfrak{K}_+ = 0) \subset \mathfrak{K}_{++}$ and $(\mathfrak{K}_- = 0) \subset \mathfrak{K}_{--}$ are \emph{(intrinsically complete) Hilbert spaces}. In a Kre\u{\i}n space there is, by definition, a rank of \emph{metric indefiniteness} \cite{Azizov Ginsburg and Langer "On M. G. Krein's papers in the theory of spaces with an indefinite metric"}.
	}—of the quantum fields. Hilbert space has properties of a \emph{generalized Euclidean geometry}, whilst a Kre\u{\i}n space is \emph{psuedo-Euclidean}.

\subsection{Spatiality in Particle Aspect vs. Wave-Field Aspect}

Here we would like to stress that a quantum space (intended as belonging to quantum mechanics and \textsc{qft}), either as a Minkowskian arena, or a Hilbert space from a functional analysis, is a system for \emph{discrete} particles, the punctual datum (see Section \ref{subsubsection "Particle(s) Kermesse"}), whose \emph{continuity} is expounded in the wave-field structure (of each particle): \emph{elementary particles} are \emph{vibrations}, or \emph{excitations}, under a \emph{discrete manner}, of the quantum fields, which in turn are fields of the  particles—it is a conceptual circularity: \emph{particles of the quantum fields} and \emph{quantum fields of the particles} are interchangeable notions. 

There is an infinite number of possible modes in which a field can vibrate.\footnote{
	The birth of \textsc{qft} can reasonably be traced back to E. Fermi, with his paper on \textgreek{β}-decay \cite{Fermi "Tentativo di una teoria dei raggi beta"} = \cite{Fermi "Versuch einer Theorie der beta-Strahlen. I} = \cite{Wilson "Fermi's Theory of Beta Decay"}. F. Wilczek \cite[pp. S86-S87, e.a.]{Wilczek "Quantum Field Theory"} writes: «The first conscious exploitation of the potential for quantum field theory to describe processes of transformation was Fermi's theory of beta decay. He turned the procedure around, inferring from the observed processes of particle transformation the nature of the \emph{underlying} local interaction of fields. Fermi's theory involved creation and annihilation not of photons, but of atomic nuclei and electrons (as well as neutrinos)—the ingredients of “matter”. It began the process whereby classic atomism, involving stable individual objects, was replaced by a more sophisticated and accurate picture. In this picture it is only the fields, and not the individual objects they create and destroy, that are \emph{permanent}». 
	} 
With clarity, it is a \emph{mathematical infinity}, as well as the  Hilbert space is an \emph{abstract (scilicet: mathematical) entity}, which has an infinite number of dimensions, because in \textsc{qft} it has an infinite number of degrees of freedom associated with the states of a system. In the real world, that is a different kettle of fish.

\subsection{Spectral Continuity, Discrete Set of Eigenvalues, and Punctual Relations in Function Spaces}

It is worth noting the following. In the definition of a function space, whatever the function space is, the double notion of \emph{discretum} and \emph{continuum} is explicitly co-present. If we take account of several eigenvalues of a certain system, and these values form a \emph{discrete set} of a linear operator, we may speak of a \emph{continuous set} of values, i.e. of a \emph{spectrum} of the operator (on a finite-dimensional vector space), based on \emph{punctual relations}—e.g. in the case of Pincherle, it is the homography.\footnote{
	He writes \cite[p. 87]{Pincherle "Cenno sulla geometria dello spazio funzionale"}: «Let us remember that an operation [\,\dots] which, applied to analytic functions, gives rise to analytic functions too [\,\dots], therefore gives a transformation of the function space which, for each linear manifold of finite order of this space, is reduced to a homography. Such an operation can be \emph{continuous} for the whole function space or for a part [\,\dots] of it».\endnote{
	Original It. version: «Ricordiamo che una operazione [\,\dots] la quale applicata alle funzioni analitiche, dà origine a funzioni pure analitiche [\,\dots], dà pertanto una trasformazione dello spazio funzionale la quale, per ogni varietà lineare d'ordine finito di questo spazio, si riduce ad una omografia. Una tale operazione può essere \emph{continua} per tutto lo spazio funzionale o per una parte [\,\dots] di esso».
	} 
	}

\subsection{Particle(s) Kermesse}
\label{subsubsection "Particle(s) Kermesse"}

\begingroup
\footnotesize
[T]he elementary particle is not an individual; it cannot be identified, it lacks “sameness” [\,\dots]. In technical language it is covered by saying that the particles “obey” new-fangled statistics, either Einstein–Bose or Fermi–Dirac statistics [\,\dots]. It is certainly useful to recall at times that all quantitative models or images conceived by the physicist are, epistemologically, only \emph{mathematical devices} [or rather, mathematical fictions] for computing observable events [\,\dots]. A [particle] lacks the most primitive property we associate with a piece of matter in ordinary life [\,\dots]: the modern atom[ism] consists of no stuff at all but is \emph{pure shape} [\,\dots]. I believe the situation is this. We have taken over from previous theory the idea of a particle and all the technical language concerning it. This idea is inadequate. It constantly drives our mind to ask information which has obviously no significance. Its \emph{imaginative structure} exhibits features which are alien to the real particle.\footnote{
	Cf. Section \ref{subsection "Scholium: Point-charge/Point-mass of Electricity: Singularities (or Quasi-singularities) of Fields"}.
	} \\
\indent — \textsc{E. Schrödinger} \cite[pp. 183, 185, 191, 188, e.a.]{Schrodinger "What is an Elementary Particle?"}

\endgroup

\vspace{2mm}

In physics, one does not have the faintest idea of what a \emph{particle} is (from experimental evidences), since if the definition of a particle starts from, or arrives at, the concept of \emph{point},\footnote{
	It is a mathematical concept, and as such it is not about nature.	
	} 
the leap of physics from the abstract world of mathematics to the phenomeno-observational one is itself determined by a theoretical burden (contradictory postulates, definitional inconsistencies, unclean logic, etc.), which is already present on a mathematical level, cf.  Sections \ref{subsection "Null Space No Infinite Time: the Point as a Capstone, and the Poincaré–Perelman Paradox"} and \ref{section "Continuity and Discreteness—Differential Equations and Numerical Computing"}.

Between the \emph{point} of mathematics and the \emph{particle} of physics there is, in the middle, a kermesse of convictions—all mathematical concepts, obviously—betraying a (con)fusion between mathematics and physics, see Chapters \ref{chapter "Outro—Parva Mathematica: Libera Divagazione 3/8"}, \ref{chapter "Outro—Parva Mathematica: Libera Divagazione 5/8"}, \ref{chapter "Outro—Parva Mathematica: Libera Divagazione 6/8"}, \ref{chapter "Outro—Parva Mathematica: Libera Divagazione 7/8"}. The main ones are those in which a particle is 

· a \emph{corpuscular ray},

· a \emph{wave function}, 

· an \emph{excitation of a quantum field},

· an \emph{algebraic structure} (set, or groups).

The four are one, from which the notorious \emph{wave-particle} duality \cite{Thomson and Reid "Diffraction of Cathode Rays by a Thin Film"} \cite{Davisson and Germer "Reflection of Electrons by a Crystal of Nickel"}, as reflected in the \emph{de Broglie's hypothesis of matter waves} \cite{De Broglie "Recherches sur la theorie des Quanta"}, from which \emph{Bohr's complementarity} \cite{Bohr "The Quantum Postulate and the Recent Development of Atomic Theory"} is affirmed (1927).
 
Schrödinger's comment \cite[pp. 472-473]{Schrodinger "Letter to Synge 9 November 1959"} is acrid, calling the complementarity character (of the quantum theory) a «thoughtless slogan»: «If I were not thoroughly convinced that the man [Bohr] is honest and really believes in the relevance of his—I do not say theory but—sounding word, I should call it intellectually wicked»; he concludes with these verses from Goethe's Faust I \cite[p. 72, vv. 1995-1996]{Goethe "Faust. Eine Tragodie"}:

\vspace{2mm}

\begingroup
\footnotesize
Denn eben wo Begriffe fehlen \\
\indent Da stellt ein Wort zur rechten Zeit sich ein,

\endgroup

\vspace{2mm}

which in En. means

\vspace{2mm}

\begingroup
\footnotesize 
For just where concepts are lacking, \\ 
\indent	a word appears at the right moment [to take its place].
	
\endgroup

\vspace{2mm}

Now stop and ponder, for a moment, over this fact.

\section{\emph{Intermezzo}. Limits of (Common) Language, and Verbal Ploys}

About the conclusion of the previous Section, there is a pronouncement from M. Born \cite[p. 97, e.a.]{Born "Atomic Physics"} that fits like a glove:

\vspace{2mm}

\begingroup
\footnotesize
The ultimate origin of the difficulty [inherent in the wave-particle duality] lies in the fact [\,\dots] that we are compelled to use the words of \emph{common language} when we wish to describe a phenomenon, not by logical or mathematical analysis, but by a picture appealing to the \emph{imagination}. Common language has grown by everyday experience and can never surpass these limits. Classical physics has restricted itself to the use of concepts of this kind; by analysing visible motions it has developed two ways of representing them by elementary processes: moving particles and waves. There is no other way of giving a pictorial description of motions—we have to apply it even in the region of atomic processes, where classical physics breaks down. Every process can be interpreted either in terms of corpuscles or in terms of waves, but on the other hand \emph{it is beyond our power to produce proof that it is actually corpuscles or waves with which we are dealing}, for we cannot simultaneously determine all the other properties which are distinctive of a corpuscle or of a wave, as the case may be. We can therefore say that the wave and corpuscular descriptions are only to be regarded as complementary ways of viewing one and the same objective process, a process which only in definite limiting cases admits of complete pictorial interpretation.

\endgroup

\vspace{2mm}

Let us dwell on the consideration of the limits of our (verbal/spoken) language. Similar thoughts are also in P.W. Bridgman \cite[p. 225, e.a.]{Bridgman "Remarks on the Present State of Operationalism"}: 

\vspace{2mm}

\begingroup
\footnotesize
[O]ne is impressed by the complexity of the verbal structure that mankind has erected through the ages. Here is an autonomous world in which a man can, and frequently does, live a more or less self-contained and independent existence. On the other hand, despite the complexity of the verbal world, the external world of objects and happenings is inconceivably more complex—\emph{so complex that all aspects of it can never be reproduced by any verbal structure}. Even in physics this is not sufficiently appreciated, as is shown, for example, by the \emph{reification of energy}. The totality of situations covered by various aspects of the energy concept is too complex to be reproduced by any simple verbal device. 

\endgroup

\vspace{2mm}
	
Here the energy concept is challenged. We run up against examples of inadequacy of language at every turn. The same goes for the notion of mass. This is what H.L. Jackson \cite[p. 278, e.a.]{Jackson "Presentation of the Concept of Mass to Beginning Physics Students"} declares, with a caustic tone: 

\vspace{2mm}

\begingroup
\footnotesize
Mass may be compared with an actor who appears on the stage in various disguises, but never as his true self. Actually mass—like the Deity—has a \emph{triune personality}. It may appear in the role of gravitational charge, or of inertia, or of energy; but nowhere does mass present itself to the senses as its unadorned self.

\endgroup

\vspace{2mm}

Regrettably, though, we cannot do without the language, despite the aforestated limits \cite[p. 31]{Bridgman "The Way Things Are"}: 

\vspace{2mm}

\begingroup
\footnotesize
The words in which the physicist defines the meaning of such concepts as “length” must be of the type that have nonverbal referents [\,\dots]. [But] even here we have to get back onto the verbal level if we wish to communicate the results of our nonverbal operations [\,\dots]. I as I write and you as you read cannot get away from words [\,\dots]. Any effects which I can here produce on you must be through the medium of words. It is a tautology to say that our verbal communication [\,\dots] cannot get away from words.

\endgroup

\vspace{2mm}

The search for, say, a pre-linguistic truth, which overcomes, or gets through, the language misunderstandings, whatever they may be, is just one of the many (Western) \textgreek{μῦθοι} anchored to the Platonic idea of Truth (cf. footnote \ref{footnote "Weyl Raum-Zeit-Materie, 4th edition"}, p. \pageref{footnote "Weyl Raum-Zeit-Materie, 4th edition"}). Mathematics, with its symbolism, and the physics that takes possession of it, do not escape the—entirely scientific—process of ambiguity of words and the multiplicity of their interpretations.\endnote{
	Because of that, in its auroral phase, in some thinkers, mathematics is, or pretends to be, \emph{with-no-words}; but, sooner or later, it must pass from no-words to words; see e.g. R. Penrose \cite[pp. 424-425, 427]{Penrose "The Emperor's New Mind: Concerning Computers Minds and The Laws of Physics"}: «Almost all my mathematical thinking is done visually [geometrically] and in terms of non-verbal concepts [\,\dots]. Often the reason is that there simply are not the words available to express the concepts that are required. In fact I often calculate using specially designed diagrams which constitute a shorthand for certain types of algebraic expression [\,\dots]. This is not to say that I do not sometimes think in words, it is just that I find words almost useless for \textit{mathematical} thinking [\,\dots]. 

	\setlength\parindent{8pt}
	I always had particular trouble with comprehending a verbal description of a formula, while many of my colleagues seemed to experience no such difficulty.

	A common experience, when some colleague would try to explain some piece of mathematics to me, would be that I should listen attentively, but almost totally uncomprehending of the logical connections between one set of words and the next. However, some guessed image would form in my mind as to the ideas that he was trying to convey—formed entirely on my own terms and seemingly with very little connection with the mental images that had been the basis of my colleague's own understanding [\,\dots]. It would be clear, at the end of it, that some genuine and positive communication had taken place. Yet the actual sentences that each one of us would utter seemed only very infrequently to be actually understood!».
	}

\section{Space and Time from No-space and No-time}
\label{section "Space and Time from No-space and No-time"}

All of this stuff, more correctly, must be examined in the limit of a combination of the monster equation with a doctrine in which the gravitational force coincides with the energy-momentum tensor, and the density of matter-energy is interpreted as a Gaussian-like curvature (see Chapters \ref{chapter "On Dimensional Continuum, Part I. Ricci Calculus (Calculus of Tensors and Curvature Tensors), Lorentz–Minkowski 4-Manifolds plus Spinor Representation, and Clifford Algebra"} and \ref{chapter "On Dimensional Continuum, Part II. Action Principles, Variations and Radiation in Curved Space-Time—Mathematical Details of Field Theory of Gravitation (General Relativity)"}). The most intriguing proposals, having the goal of imagining a quantum world, are 

· those in which space-time is a sort of  elastic entity decomposed into granules (we will mention the exemplary Sakharov case); 

· and those that seek to establish a connection between, on the one hand, curved space-time, and hence gravity (as it is ascribed to a curvature of a Lorentz-like manifold), or time (as an independent quantity, a concept distinct from space), and entropy, on the other.

The implication is that representations of space-time \& gravity, and time, as a one-directional flow, or \emph{time's arrow}, in the words of A.S. Eddington \cite[pp. 68-80]{Eddington "The Nature of the Physical World"}, cease to be fundamental entities and become \emph{emergent phenomena}.

\subsection{Emergent Gravity}

\subsubsection{Sakharov's Elasticity of Space, and Liquid Space-Time}	
\label{subsubsection "Sakharov's Elasticity of Space, and Liquid Space-Time"}

\begingroup
\footnotesize
Ingenious hypotheses [\,\dots] have been proposed to explain light, heat, magnetism, etc., considering the phenomena as produced by a reaction of space's opposition to the variability of its own curvature over time. And here it is important to observe that the additional term $\frac{B\Phi}{\nabla}$ in the effective part [\,\dots] of the elastic potential can be considered precisely as the expression of the energy of the reactions that the space, rigid in its geometric constitution, opposes to the elastic matter filling it, assuming [the matter] \emph{inert} in the sense that, forced to deform in the said space, it tends to do it as if the space itself were Euclidean.\endnote{
	Original It. version: «[I]ngegnose ipotesi [\,\dots] sono state proposte per spiegare luce, calore, magnetismo, ecc., considerandone i fenomeni come prodotti da una reazione che lo spazio opporrebbe alla variabilità della propria curvatura nel tempo. E qui importa osservare che il termine addizionale $\frac{B\Phi}{\nabla}$ nella parte efficace [\,\dots] del potenziale elastico si può considerare appunto come l'espressione dell'energia delle reazioni che lo spazio, rigido nella propria costituzione geometrica, oppone alla materia elastica che lo riempie, supponendo questa \emph{inerte} nel senso che, obbligata a deformarsi nel detto spazio, essa tende a farlo come se lo spazio stesso fosse euclideo».
	} \\
\indent — \textsc{E. Cesàro} \cite[p. 213]{Cesaro "Introduzione alla teoria matematica della elasticita"} 

\vspace{2mm}

The presence of the action $S(\Ricci) = -\frac{1}{16\pi\gravitation}\int(dx)\sqrt{-g}\Ricci$ [in Einstein's theory of gravitation] [where $\Ricci$ is the invariant of the Ricci tensor, with $(\Ricci) = \Lagrangian(0) + A\int kdk \cdot \Ricci + B\int\frac{dk}{k}\Ricci^2 + \cdots$, and $A \textnormal{ \& } B \sim 1$] leads to a “metrical elasticity” of space, i.e., to generalized forces which oppose the curving of space. Here we consider the hypothesis which identifies the [above] action with the change in the action of quantum fluctuations of the vacuum if space is curved. \\
\indent — \textsc{A.D. Sakharov} \cite[p. 394]{Sakharov "Vacuum quantum fluctuations in curved space and the theory of gravitation"}

\endgroup

\vspace{2mm}

The aforesaid issues (Section \ref{section "What is the Quantum Field Space?"}) push to look for a manner of combining the continuous part (quantum fields, including gravity) with the discrete part (particle units, or singularities). Among the many proposals, in this respect, an interesting one comes from Sakharov.

The above-mentioned piece of Cesàro's text precedes his sentence reported in epigraph under Chapter \ref{chapter "On Dimensional Continuum, Part II. Action Principles, Variations and Radiation in Curved Space-Time—Mathematical Details of Field Theory of Gravitation (General Relativity)"}. There, as suggested by the theory of elastic media, he wonders—on the heels of Clifford—whether the physical variations of certain phenomena can coincide with the effects due to changes in the curvature of space.

Curiously, in A.D. Sakharov \cite[p. 394]{Sakharov "Vacuum quantum fluctuations in curved space and the theory of gravitation"} the invocation of elasticity, for the physical significance of spatial curvature in \textsc{qft}, is pushed further. Space itself, for Sakharov, is identified with the concept of \emph{elastic continuum}, so the \emph{gravitational attraction} is considered as \emph{a sort of elasticity of space}; and the \emph{spatio-temporal continuum becomes a property emerging from the discrete cluster of the particles, analogously to the continuum mechanics of elastic bodies and/or of fluids emerging from the molecular aggregates}, see \cite{Pessa "On Sakharov's Theory of Gravitation"} \cite{Visser "Sakharov's induced gravity: a modern perspective"}. 

More recently, the postulation of a \emph{liquid space-time}, under the suggestion of a “smolecularization” into its alleged ground-constituents, is deepened by S. Liberati and L. Maccione \cite{Liberati and Maccione "Astrophysical Constraints on Planck Scale Dissipative Phenomena"}, through analysis of the \emph{space-time viscosity} (which appears to be very low, approaching zero), the possible dispersion relation, together with the phenomenology of the \emph{hydrodynamics-like dissipative effects}.\footnote{
	The analogy with water excites the imagination: water, macroscopically, is a continuous element; on a microscopic scale, it has a decomposition into \ce{H2O} molecules. Any molecule of water (one oxygen and two hydrogen atoms), taken by itself, is not water: the (\ce{H2O})-group is not a transparent substance in a liquid phase. So we say that water is a phenomenon of emergence from discrete constituents.
	}

\subsubsection{Spatio-temporal/Gravitational Thermodynamics, \& Entropic Gravity}
\label{subsubsection "Spatio-temporal/Gravitational Thermodynamics, and Entropic Gravity"}

\begingroup
\footnotesize
In thermodynamics, heat is energy that flows between degrees of freedom that are not macroscopically observable. In spacetime dynamics, we shall define heat as energy that flows across a causal horizon. It can be felt via the gravitational field it generates, but its particular form or nature is unobservable from outside the horizon. \\
\indent — \textsc{T. Jacobson} \cite[p. 1260]{Jacobson "Thermodynamics of Spacetime: The Einstein Equation of State"}

\vspace{2mm}

The key statement is simply that we need to have a temperature in order to have a force. Since we want to understand the origin of gravity, we need to know where the temperature comes from. \\
\indent — \textsc{E.P. Verlinde} \cite[p. 8]{Verlinde "On the origin of gravity and the laws of Newton"}

\endgroup

\vspace{2mm}

\enumerationisinitium
\item In the wake of (i) Bardeen–Carter–Hawking laws of black hole thermodynamics \cite{Bardeen Carter and Hawking "The Four Laws of Black Hole Mechanics"}, (ii) the black hole entropy, also called \emph{Bekenstein–Hawking entropy} \cite{Bekenstein "Black Holes and the Second Law"} \cite{Bekenstein "Black Holes and Entropy"}\footnote{
	For insights on the black hole entropy, see e.g. \cite{Carlip "Black hole thermodynamics"}, in particular sec. 2, entitled \textit{Prehistory: Black Hole Mechanics and Wheeler's Cup of Tea}, pp. 416-418. What happens if we throw a cup of (hot) tea into a black hole? As reported by Bekenstein \cite[pp. 2336-2337]{Bekenstein "Black Holes and Entropy"}, the black hole entropy increases, and since the Schwarzschild radius—the size of the event horizon—is proportional to the mass of the black hole, the cup of tea produces an expansion of its horizon: the black hole entropy is but the ratio of the black hole area to the square of the Planck length times a dimensionless constant of order unity $\eta_c$, 
\begin{equation}
	\mathsf{S}_\textsc{bh} = f(\textgreek{\textit{α}}) \viz \eta_c\hbar^{-1}\textgreek{\textit{α}}, 
\end{equation}
where $\textgreek{\textit{α}} = \frac{A_\mathrm{h}}{4\pi}$ symbolizes the rationalized area of a black hole.
	} 
\cite{Hawking "Gravitational Radiation from Colliding Black Holes"} \cite{Hawking "Black hole explosions?"},\footnote{
	Letting $c$ be the speed of light, $G_\textsc{n}$ the Newtonian constant of gravitation, and $A_\mathrm{h}$ the horizon area, or, to say it better, the (surface) area of the event horizon of a black hole.
	} 
\begin{equation}
	\mathsf{S}_\textsc{bh} = \frac{c^3}{4G_\textsc{n}\hbar}A_\mathrm{h},
\end{equation}
and (iii) Hawking's results on the quantum thermal radiation \cite{Hawking "Particle Creation by Black Holes"}, T. Jacobson \cite{Jacobson "Thermodynamics of Spacetime: The Einstein Equation of State"} establishes a proportionality of entropy and area. He finds that a \emph{horizon area, relative to the boundary of a null hypersurface in the Rindler frame \textnormal{\cite{Rindler "Essential Relativity: Special General and Cosmological"} \cite{Rindler "Relativity: Special General and Cosmological"}}, is proportional to the entropy of the space enclosed therein}. This allows him to derive the Einstein field Eqq. \eqref{subequations "Einstein field equations"} from the area-entropy relation 
\begin{equation}
	\delta\mathsf{Q} = \mathsf{T}d\mathsf{S},
\end{equation} 
intertwining heat $\mathsf{Q}$, temperature $\mathsf{T}$, and entropy $\mathsf{S}$. Consequently, \emph{expressions of mathematics of gravitation}, appear to be local equations of state, acquiring a \emph{thermodynamic interpretation}, for which \emph{the surface gravity is a(n indication of) temperature}, and \emph{the energy flux is a (kind of) heat flow}. See also Padmanabhan's articles \cite{Padmanabhan "Gravity from Spacetime Thermodynamics"} \cite{Padmanabhan "Thermodynamical Aspects of Gravity: New insights"}, and book \cite[chap. 16]{Padmanabhan "Gravitation: Foundations and Frontiers"}.
\item E.P. Verlinde \cite{Verlinde "On the origin of gravity and the laws of Newton"} sketches out an \emph{identification between gravity and entropy}, the latter seen as a force 
\begin{equation}
	\vec{F}\mathrm{\Delta}x = \mathsf{T}\mathrm{\Delta}\mathsf{S}, \text{ where } \mathrm{\Delta}x = \frac{\hbar}{mc}, \enspace \mathrm{\Delta}\mathsf{S} = 2\pi k_\textsc{b}\frac{mc}{\hbar}\mathrm{\Delta}x, 
\end{equation}
caused by changes in the \emph{information correlated to the position of collections of matter}, with their degrees of freedom, where $k_\textsc{b}$ is the Boltzmann constant. More precisely, the relationship between temperature $\mathsf{T}$ and entropy $\mathsf{S}$ is but the \emph{amount of information} resulting \emph{from the microscopic degrees of freedom}, as well as \emph{from the energy costs, or entropy changes in the amount of information}.\footnote{
	The entropy, as a thermodynamic system property, corresponds to $\mathsf{S}(E_\mathrm{tot}, x) = k_\textsc{b}\log\Omega(E_\mathrm{tot}, x)$, setting $\Omega(E_\mathrm{tot}, x)$ as the volume of the configuration space, in which a function of the total energy $E_\mathrm{tot}$ is placed.
	}

For this reason, we can speak of \emph{emergent gravity} as a general phenomenon arising from \emph{discrete distributions of matter}. Here, too, a spur of analogy produces its effects: the equations of gravity \emph{resemble} the laws of thermo- and hydro-dynamics. 

The pedestal on which Verlinde's thesis rests is the adoption of the tenets of the holographic model, already mentioned by Jacobson.\footnote{
	\cite[p. 1263]{Jacobson "Thermodynamics of Spacetime: The Einstein Equation of State"}: «Another argument that might be advanced in support of the proportionality of entropy and area comes from the holographic hypothesis, i.e., the idea that the state of the part of the universe inside a [3-dimensional] spatial region [spatio-temporal region] can be fully specified [stored] on the [2-dimensional] boundary of that [spatio-temporal] region».
	}
 The \emph{holographic principle}, or \emph{'t Hooft–Susskind principle} \cite{'t Hooft "Dimensional Reduction in Quantum Gravity"} \cite{Susskin "The world as a hologram"}, which has its first stimulus in the laws of black hole mechanics, and then in the AdS/\textsc{cft} correspondence, also called \emph{gauge/gravity duality}, which is a duality conceived by J. Maldacena \cite{Maldacena "The Large $N$ Limit of Superconformal field theories and supergravity"}, commonly believed to be an exemplary evidence of the holographic principle; see E. Witten \cite{Witten "Anti De Sitter Space And Holography"} and L. Smolin \cite{Smolin "The strong and weak holographic principles"}. 
 
 The AdS/\textsc{cft} is a duality conjecturing a connection between a negatively curved space, called \emph{anti-de Sitter space}, or \emph{anti-de Sitter space-time}, denoted by $AdS_n$,\footnote{
	de Sitter (and anti-de Sitter) spaces owe their name to W. de Sitter \cite[communicated in the meeting of March 31, 1917]{de Sitter "On the relativity of inertia. Remarks concerning Einstein's latest hypothesis"} \cite{de Sitter "On the curvature of space"} \cite{de Sitter "Further remarks on the solutions of the field-equations of Einstein's Theory of gravitation"}; but they were invented independently, and in the same year, by T. Levi-Civita \cite[seduta del 20 maggio 1917]{Levi-Civita "Realta fisica di alcuni spazi normali del Bianchi"}. Perhaps it would be more correct to call them \emph{de Sitter–Levi-Civita spaces}.
 	} 
and the \emph{conformal field theory} (\textsc{cft}),\footnote{
	A \textsc{qft} equipped with invariance under conformal transformations.
 	} 
assuming that the \textsc{cft} is a dataset image that may be defined on the boundary of $AdS_n$.
 Verlinde argues that \emph{a certain (not infinite) number of the microscopic degrees of freedom} associated with a collection of matter is \emph{representable holographically on the boundary of space-time, or on black hole horizons, under the $\mathsf{T}$-$\mathsf{S}$ relation}.
\enumerationisfinis

\subsubsection{Entropy Functionals for Matter and Gravity}
\label{subsubsection "Entropy Functionals for Matter and Gravity"}

The general form for the \emph{entropy (action) functional for matter}, as normal to a null hypersurface, is
\begin{equation}
	\mathscr{S}^\mathsf{S}_\mathrm{m} = \int_\Omega\left(\sqrt{-g} \Tau^\mathrm{m}_{\mu\nu}\vec{X}^\mu\vec{X}^\nu\right)d^nx,
\end{equation}
for $n \geqslant 4$, where $\Omega$ is a spatio-temporal region, $\Tau^\mathrm{m}_{\mu\nu}$ is the energy-momentum tensor of matter, and $\vec{X}^{\mu, \nu}$ is a null vector field; whilst the \emph{entropy (action) functional for gravity} is
\begin{equation}
	\mathscr{S}^\mathsf{S}_\mathrm{g} = -4\int_\Omega\sqrt{-g}\left({\textgreek{Ζ}_{\mu\nu}}^{\xi\varrho}\nabla_\xi\vec{X}^\mu\nabla_\varrho\vec{X}^\nu\right)d^nx,	
\end{equation}
where ${\textgreek{Ζ}_{\mu\nu}}^{\xi\varrho}$ is a Riemann-like 4-tensor; from which
\begin{equation}
	\mathscr{S}^\mathsf{S}_\mathrm{m|g} = - \int_\Omega\sqrt{-g}\left(4{\textgreek{Ζ}_{\mu\nu}}^{\xi\varrho}\nabla_\xi\vec{X}^\mu\nabla_\varrho\vec{X}^\nu - \Tau^\mathrm{m}_{\mu\nu}\vec{X}^\mu\vec{X}^\nu\right)d^nx,	
\end{equation}
where 
\begin{align}
	4{\textgreek{Ζ}_{\mu\nu}}^{\xi\varrho}\nabla_\xi\vec{X}^\mu\nabla_\varrho\vec{X}^\nu & = \Lbrack:4\nabla_\xi[{\textgreek{Ζ}_{\mu\nu}}^{\xi\varrho}\vec{X}^\mu\nabla_\varrho\vec{X}^\nu]:\Rbrack - 4\vec{X}^\mu{\textgreek{Ζ}_{\mu\nu}}^{\xi\varrho}\nabla_\xi\nabla_\varrho\vec{X}^\nu \notag \\
	& = \Lbrack:\cdots:\Rbrack - 2\vec{X}^\mu{\textgreek{Ζ}_{\mu\nu}}^{\xi\varrho}\nabla_{[\xi}\nabla_{\varrho]}\vec{X}^\nu \notag \\
	& = \Lbrack:\cdots:\Rbrack - 2\vec{X}^\mu{\textgreek{Ζ}_{\mu\nu}}^{\xi\varrho}{\Riemann^\nu}_{\varsigma\xi\varrho}\vec{X}^\varsigma \notag \\
	& = \Lbrack:\cdots:\Rbrack + 2\vec{X}^\mu\tensorP_{\mu\varsigma}\vec{X}^\varsigma.
\end{align}
where the symbols $\Lbrack:$ and $:\Rbrack$ signify that the expression within them must be repeated, inspired by the beginning and ending repeat signs in music notation, and $\Riemann$ is the Riemann tensor. Hence
\begin{align}
\label{align "Entropy functional for matter and gravity"}
	\mathscr{S}^\mathsf{S}_\mathrm{m|g} = & - \int_{\partial\Omega}\sqrt{\textcyrillic{\textit{ш}}}\left(4{\textgreek{Ζ}_{\mu\nu}}^{\xi\varrho}\vec{X}^\mu\nabla_\varrho\vec{X}^\nu\right)d^{n - 1}x\vec{Y}_\xi \notag \\
	& - \int_\Omega\sqrt{-g}\left\{\Bigl(2\tensorP_{\mu\nu} - \Tau^\mathrm{m}_{\mu\nu}\Bigr)\vec{X}^\mu\vec{X}^\nu\right\}d^nx,
\end{align}
putting $2\tensorP_{\mu\nu} = \Tau^\mathrm{m}_{\mu\nu}$, where $\textcyrillic{\textit{ш}}$ is the determinant of the induced metric on the \emph{boundary surface} $\partial\Omega$, and $\vec{Y}_\xi$ is the vector field normal to $\partial\Omega$. Inside the Eq. \eqref{align "Entropy functional for matter and gravity"} the gravitational entropy  functional is
\begin{align}
	\mathscr{S}^\mathsf{S}_\mathrm{g} = & - \int_\Omega\sqrt{-g}\left(4{\textgreek{Ζ}_{\mu\nu}}^{\xi\varrho}\nabla_\xi\vec{X}^\mu\nabla_\varrho\vec{X}^\nu\right)d^nx \notag \\
	& - \int_{\partial\Omega}\sqrt{\textcyrillic{\textit{ш}}}\left(4{\textgreek{Ζ}_{\mu\nu}}^{\xi\varrho}\vec{X}^\mu\nabla_\varrho\vec{X}^\nu\right)d^{n - 1}x\vec{Y}_\xi - \int_\Omega\sqrt{-g}\left(2\tensorP_{\mu\nu}\vec{X}^\mu\vec{X}^\nu\right)d^nx. 
\end{align}

In view of the foregoing, this defines the \emph{entropy flux vector}, of which we say it is proportional to the \emph{heat flux vector}—the entropy flux is related to the energy-momentum tensor of matter. The \emph{information}, \emph{measured by entropy}, about a \emph{region of space-time}, concerning \emph{matter} and its \emph{distribution}, combined with \emph{gravitational force}, is \emph{stored}, in terms of variable, \emph{on the boundary surface} of that region, i.e. \emph{on a holographic spatio-temporal screen},\footnote{
	The 2-dimensionality of the boundary-space justifies the term \emph{screen}.
	} 
relative to that part of the universe.

\subsubsection{Information Flow of What? An Entropy Flux Question}
\label{subsubsection "Information Flow of What? An Entropy Flux Question"}

There is a dark spot in the above topic (Sections \ref{subsubsection "Spatio-temporal/Gravitational Thermodynamics, and Entropic Gravity"} and \ref{subsubsection "Entropy Functionals for Matter and Gravity"}). 
\enumerationisinitium
\item We say that a (space-like) 2-dimensional surface is a set of screen-events. The area of a set of screen-events is a \emph{measure of the flow of quantum information}, i.e. a \emph{measure of the information flow capacity of a 2-surface}, to which the concept of quantum space-time (or discrete skein at the Planck scale) is connected; see e.g. F. Markopoulou and L. Smolin \cite{Markopoulou Smolin "Holography in a quantum spacetime"}. We keep the aforementioned analogies and identities:

· surface gravity = $\mathsf{T}$, 

· energy flux as a heat flow, 

· connection between temperature and entropy ($\mathsf{T}$-$\mathsf{S}$ relation).
\item By \emph{information} is to be understood, in the this context, not as the amount of information relating to a \emph{state of knowledge} (which is always a subjective state, within the limits of human knowledge), but a \emph{physical disposition}, in a system, to show \emph{change of state} (objective aspect), which are defined by entropy. We can take a Bateson's definition \cite[p. 231]{Bateson "A re-examination of "Bateson's rule""} as an example:

\vspace{2mm}

\begingroup
\footnotesize
 The technical term “information” may be succinctly defined as \emph{any difference which makes a difference in some later event}. This definition is fundamental for all analysis of cybernetic systems and organization. The definition links such analysis to the rest of science, where the causes of events are commonly not differences but forces, impacts and the like. The link is classically exemplified by the heat engine, where available energy (i.e. negative entropy)\footnote{
	See footnote \ref{footnote "Negative entropy"} on p. \pageref{footnote "Negative entropy"}. 
 	} 
 is a function of a \emph{difference} between two temperatures. In this classical instance “information” and “negative entropy” overlap.

\endgroup

\vspace{2mm}

\item Now, entropy is said to be a measure of the degree of disorder (chaos),\footnote{
	\label{footnote "order and chaos: separate concepts"}
	To be stickler, we should cautiously separate the concept of “disorder” from that of “chaos”, or “apparent randomness”, even if the two concepts end up merging into one, until we have given a satisfactory—for us—definition of “disorder”, see footnote \ref{footnote "Relativity of the order/disorder concept"}, p. \pageref{footnote "Relativity of the order/disorder concept"}. The identification between “chaos” and “state of disorder” is already in the Gr. word \textgreek{χάος},\endnote{
		Looking back through the history, the word \textgreek{χάος} peeks out into Hesiod's epic poetry \cite[p. 12, v. 116]{Hesiod "Theogony"}: «First of all \textgreek{Χάος} came to be (\textgreek{ἤτοι μὲν πρώτιστα Χάος γένετ'})», but, quite rightly, G.W. Most translates “\textgreek{Χάος}” as “Chasm” («gap», «opening») and not as “Chaos” («jumble of disordered matter»).
		} 
which means a “primordial space/state” (\textgreek{πρώτιστα χ.}), or a “limitless/infinite space” (\textgreek{ἄτρυτον χ.}).
	}
uncertainty, or \emph{mixed-up-ness} à la Gibbs \cite[p. 418]{Gibbs "Unpublished Fragments"} (Section \ref{subsection "The Entropy-Energy Roots"}). The crux of the issue is that information, even if it has to do with  a succession of real events, is a statistical succession of such events, in which the subjective aspect, or \emph{ignorance}, is mixed, at least partly, with the physical disposition:\footnote{
	Cf. P.W. Bridgman \cite[pp. 205-206, e.a.]{Bridgman "The Logic of Modern Physics"}: «[A] statistical method is used either to \emph{conceal} a vast amount of actual ignorance, or else to \emph{smooth out} the details of a vast amount of actual physical complication, most of which is unessential for our purposes».
	} 
entropy, in fact, is a lack of information on the microscopic state of the system. It is a question, therefore, of reconciling our ignorance with the emergence acting as a gravitational disposition. 
\item Where is the fringe between an amount of \emph{material information}, or a set of events in nature, and a \emph{representation/interpretation of information}, as an understanding of data? How can we discern \emph{information as (quantum) process} from \emph{information as knowledge}? To what extent are the two concepts separable? For instance, what is a sequences of binary (base-2) digits, 0s and 1s, without knowledge—of which physics and mathematics are part, of course? Information without knowledge is not physics (seen as rational description of natural phenomena), it is Nature: it is a process without observers, or conscious minds (physicists, or mathematicians), cf. point \ref{item "A physics without symbolic apparatus etc."} in Section \ref{subsection "Math-Language and its Reasonably Effectiveness"}.
\item In Shannon's theory \cite{Shannon "A Mathematical Theory of Communication"} (cf. Section \ref{definitio "Kolmogorov–Sinai metric entropy"}), \emph{information} is synonymous with \emph{(freedom of) choice}, \emph{uncertainty}, and \emph{entropy} (the four words are interchangeable).\footnote{
	Shannon asks \cite[p. 392]{Shannon "A Mathematical Theory of Communication"} = \cite[p. 49]{Shannon and Weaver "The Mathematical Theory of Communication"}, in a paradigmatic way: «Can we find a measure of how much “choice” [information] is involved in the selection of the event or of how uncertain we are of the outcome?». 
	} 
\emph{Shannon entropy}, in its simplest expression, can be written as 
\begin{equation}
\label{equation "Shannon entropy"}
	\mathsf{H}_\textsc{s} = - \sum^n_{\rotatedell = 1} P_\rotatedell\log_\mathrm{b}{P_\rotatedell} = - \sum_n P_\rotatedell\log_2{P_\rotatedell},
\end{equation}
letting $P_\rotatedell$ be the probability of state $\rotatedell$ (the base of the logarithm is usually $\mathrm{b} = 2$). From \eqref{equation "Shannon entropy"} it is almost immediate to return to the \emph{Gibbs entropy} \cite{Gibbs "Elementary principles in statistical mechanics"}, and then, originally, to the \emph{Boltzmann entropy} \cite{Boltzmann "Bemerkungen uber einige Probleme der mechanischen Warmetheorie"} \cite{Boltzmann "Uber die Beziehung zwischen dem zweiten Hauptsatze der mechanischen Warmetheorie und der Wahrscheinlichkeitsrechnung resp. den Satzen uber das Warmegleichgewicht"}, see footnote \ref{footnote "Boltzmann's grave formula"}, p. \pageref{footnote "Boltzmann's grave formula"}. W. Weaver, in the comments of Shannon's work, writes \cite[p. 15]{Shannon and Weaver "The Mathematical Theory of Communication"}:

\vspace{2mm}

\begingroup
\footnotesize
In the limiting case where one probability is unity (certainty) and all the others zero (impossibility), then $\mathsf{H}$ [entropy] is zero (no uncertainty at all—no freedom of choice—no information).

\endgroup

\vspace{2mm}

Without information, in the Shannon sense, there is no entropy; but this conception no longer fits for thermodynamic systems, since \emph{heat} is a form of energy, which is \emph{part of nature}, and \emph{exists even without information à la Gibbs–Shannon}.\footnote{
	Alternatively, we can consider information exclusively in its physical content. But what is information if not, first of all, the \emph{form} with which we know, or specify, the events in a statistical succession? Information has a natural dimension: it is physical; for us, however, information is mostly \emph{mathematical}, from the Gr. \textgreek{μάθ-ημα, -ματα}, “knowledge”, because by means of the \textgreek{μάθημα}-knowledge we become aware of the physical systems, in terms of laws and empirical data. No physicist sees the world outside his \textgreek{μάθημα}-knowledge. More details can be found in Chapters  \ref{chapter "Outro—Parva Mathematica: Libera Divagazione 3/8"}, \ref{chapter "Outro—Parva Mathematica: Libera Divagazione 5/8"}, \ref{chapter "Outro—Parva Mathematica: Libera Divagazione 6/8"}.
	}

The equivalence of \emph{information entropy}, in the Shannon, or Gibbs–Shannon, sense, and \emph{thermodynamic entropy}, in the Boltzmann sense (see Section \ref{subsection "Prior Knowledge: Maxwell–Boltzmann Probability Distribution, and Ergodic Hypothesis of Thermodynamics"}), is valid because \emph{our concept of information} is a \emph{mixture of material information and information as knowledge},\endnote{
	A further question then turns up: at what level the dividing border between the natural process (reality of the world) and the process of knowledge (rooted in the subjective perception) should be drawn. A text by von Neumann \cite[pp. 272-273]{Neumann "Mathematical Foundations of Quantum Mechanics"} acts as a reference: «We wish to measure the temperature. If we want, we can proceed numerically by looking to the mercury column in a thermometer, and then say: “This is the temperature as measured by the thermometer”. But we can carry the process further, and from the properties of mercury (which can be explained in kinetic and molecular terms) we can calculate its heating, expansion, and the resultant length of the mercury column, and then say: “This length is seen by the observer”. Going still further, and taking the light source into consideration, we could find out the reflection of the light quanta on the opaque mercury column, and the path taken by the reflected light quanta into the eye of the observer, their refraction in the eye lens, and the formation of an image on the retina, and then we would say: “This image is registered by the retina of the observer”. And were our physiological knowledge greater than it is today, we could go still further, tracing the chemical reactions which produce the impression of this image on the retina, and in the optic nerve and in the brain, and then in the end say: “These chemical changes of his brain cells are perceived by the observer”. But in any case, no matter how far we proceed—from the thermometer scale, to the mercury, to the retina, or into the brain—at some point we must say: “And this is perceived by the observer”. That is, we are obliged always to divide the world into two parts, the one being the observed system, the other the observer [\,\dots]. The boundary between the two is arbitrary to a very large extent. In particular, we saw [\,\dots] that the “observer” [\,\dots] need not be identified with the body of the actual observer: in one instance we included even the thermometer in it, while in another instance even the eyes and optic nerve were not included. That this boundary can be pushed arbitrarily far into the interior of the body of the actual observer is the content of the principle of psycho-physical parallelism».

	\setlength\parindent{8pt}
	A similar thought comes from J.S. Bell \cite[p. 687]{Bell "Subject and Object"}: «The subject-object distinction is indeed at the very root of the unease that many people still feel in connection with quantum mechanics. \emph{Some} such distinction is dictated by the postulates of the theory, but exactly \emph{where} or \emph{when} to make it is not prescribed [\,\dots]. [T]he theory is fundamentally about the results of ‘measurements’, and therefore presupposes in addition to the ‘system’ (or object) a ‘measurer’ (or subject). Now must this subject include a person? Or was there already some such subject-object distinction before the appearance of life in the universe? [\,\dots]. Whenever necessary a little more of the world can be incorporated into the object. In extremis the subject-object division can be put somewhere at the ‘macroscopic’ level».
	
	Undeniably, the subject-observer cannot be excluded from the measurement procedure. If this happened, there would be no information. E. Schrödinger's \cite[p. 162, e.a.]{Schrodinger "Mind and Matter"} comment is lapidary: «\emph{The observer is never entirely replaced by instruments; for if he were, he could obviously obtain no knowledge whatsoever} [\,\dots]. [A]ll information [of a certain measurement] goes back ultimately to the sense perceptions of some living person or persons [\,\dots]. The observer's senses have to step in eventually. The most careful record, when not inspected, tells us nothing». Attention: the same can be said for mathematics; a contribution by our mind is unavoidable (see Sections \ref{section "Mathematics in the Physical Sciences, and Nature of Reality I"}, \ref{section "Mathematics in the Physical Sciences, and Nature of Reality II"} and \ref{section "Mathematics in the Physical Sciences, and Nature of Reality III"}).
	}
 although the idea of entropy—including that of Boltzmann—is somehow related to a \emph{lack of form}, that is to say to a \emph{missing information}.
\enumerationisfinis

\subsubsection{Margo. In-depth Readings} 

\enumerationisinitium
\item On the Standard Model in curved space-time, and Lorentzian–Einsteinian quantum space-time in \textsc{qft} and beyond: see monographs \cite[sec. 8.2]{Dobado Gomez-Nicola Maroto Pelaez "Effective Lagrangians for the Standard Model"} \cite[secc. 6.1, 6.3]{Bandyopadhyay "Geometry Topology and Quantum Field Theory"}, and F. Finster's et al. papers \cite{Finster "From Discrete Space-Time to Minkowski Space: Basic Mechanisms Methods and Perspectives"} \cite{Finster Grotz Schiefeneder "Causal Fermion Systems: A Quantum Space-Time Emerging From an Action Principle"}.
\item To follow some indications on entropic gravity, in holographic and thermodynamic pictures, plus \emph{entanglement}.
\subenumerationisinitium
\item M. Van Raamsdonk \cite{Van Raamsdonk "Building up spacetime with quantum entanglement"} \cite{Van Raamsdonk "Building up spacetime with quantum entanglement II: It from BC-bit"}, and co-written papers \cite{Lashkari McDermott and Van Raamsdonk "Gravitational dynamics from entanglement "thermodynamics""} \cite{Faulkner Guica Hartman Myers and Van Raamsdonk "Gravitation from entanglement in holographic CFTs"} \cite{Swingle and Van Raamsdonk "Gravity from Entanglement"}: fulcrum of these studies is the claim that space-time arises from a \emph{set of nodes}, and \emph{discrete qubits} \cite{Schumacher "Quantum coding"}, in a network of tensor type, and that entanglement works as a ligand of the network—inevitably, \emph{entanglement} is thought to be the \emph{node-composition of space-time}.
\item Jacobson \cite{Jacobson "Entanglement Equilibrium and the Einstein Equation"}, and lecture book by M. Rangamani and T. Takayanagi \cite[part IV]{Rangamani Takayanagi "Holographic Entanglement Entropy"}.
\item For a construction of a spatio-temporal manifold, and its Lorentzian geometry, that is, for the \emph{emergence of space-time} and gravitational field equations, \emph{from the entanglement phenomenon} of a quantum state—the measure of which is the entanglement entropy—\emph{in an abstract Hilbert space}, see S.M. Carroll \& collaborators \cite{Cao Carroll and Michalakis "Space from Hilbert space: Recovering geometry from bulk entanglement"} \cite{Cao and Carroll "Bulk entanglement gravity without a boundary: Towards finding Einstein's equation in Hilbert space"}. 
\item For a succinct but exhaustive account on the entropic holographic/thermodynamic gravity, and entanglement, we refer to \cite{Carroll and Remmen "What is the entropy in entropic gravity?"}. 
\item Complementary, there are holographic (entropy) models with a spherical de Sitter space ($dS$), so the curvature is positive, obtained from anti-de Sitter spaces; which leads to the construction of a $dS/dS$ duality. The dual of $dS_{n + 1}$ contains two coupled \textsc{cft} sectors. See X. Dong et al. \cite{Dong Silverstein Torroba "De Sitter holography and entanglement entropy"}.
\subenumerationisfinis
\enumerationisfinis	

\subsection{Connes–Rovelli Time Flow in Heat Flow via von Neumann Algebra} 
\label{subsection "Connes–Rovelli Time Flow in Heat Flow via von Neumann Algebra"}

\begingroup
\footnotesize
We ascribe the temporal properties of the flow to thermodynamical causes, and therefore we tie the definition of time to thermodynamics [\,\dots]. [W]hat we intend to ascribe to thermodynamics is not the \emph{direction} of the time flow. Rather, it is the time flow itself. \\
\indent — \textsc{A. Connes and C. Rovelli} \cite[pp. 2901, 2908]{Connes and Rovelli "Von Neumann algebra automorphisms and time-thermodynamics relation in general covariant quantum theories"}

\endgroup

\vspace{2mm}

C. Rovelli and A. Connes \cite{Rovelli "Statistical mechanics of gravity and the thermodynamical origin of time"} \cite{Connes and Rovelli "Von Neumann algebra automorphisms and time-thermodynamics relation in general covariant quantum theories"} develop a conjecture according to which \emph{physical time flow}—as a description of the way in which events (sequence of facts, actions, or changes) have a regular occurrence, in a before-after succession—\emph{has a thermodynamic origin}.

Which merely means that \emph{the thermodynamic, or thermal, state of a system} is what \emph{determines, and defines, the time flow}. Thermal energy transfer, i.e. the flow of heat, causes the time evolution, and not the opposite (the passage of time does not “generate” the flow of heat). This proposal goes by the name of \emph{thermal time hypothesis}, under a \emph{time-thermodynamics relation}.

In a mechanical key, the characterization of a thermodynamic state is performed within the Maxwell–Boltzmann distribution \& Gibbs statistical ensemble (Section \ref{subsection "Prior Knowledge: Maxwell–Boltzmann Probability Distribution, and Ergodic Hypothesis of Thermodynamics"}), and it is, \emph{mathematically}, a \emph{statistical flow}. But, in \emph{our} interpretation of dimensionality in which we conceive the passing of things, a statistical/thermodynamic state is consistent with the \emph{physical} time flow. 

Let us see in three steps (divided into three Sections) how the thermal time hypothesis is technically articulated.

\subsubsection[I. Modular $\ast$-automorphisms, and Tomita–Takesaki Relation]{I. Modular $\protect\pseudobold{\ast}$-automorphisms, and Tomita–Takesaki Relation}

Connes–Rovelli's proposal is built on \emph{von Neumann algebra}, created by F.J. Murray \& J. von Neumann \cite{von Neumann "Zur Algebra der Funktionaloperationen und Theorie der normalen Operatoren"} \cite{Murray and von Neumann "On Rings of Operators"} \cite{Murray and von Neumann "On Rings of Operators. II*"} \cite{von Neumann "On infinite direct products"} \cite{von Neumann "On Rings of Operators. III"} \cite{Murray and von Neumann "On Rings of Operators. IV"} \cite{von Neumann "On Rings of Operators. Reduction Theory"}. 

\enumerationisinitium
\item For our purposes, we say that a von Neumann algebra $\mathscr{N}_\mathrm{eu}$ is a $\ast$-algebra (one of the subforms of $C^\ast$-algebra) of operators on a Hilbert space $\mathfrak{H}$. Let $\mathscr{N}_\mathrm{eu}$ be equipped with a 1-parameter group $\{\varphi^\omega_t\}_{t \in \mathbb{R}}$ of $\ast$-automorphisms on $\mathfrak{H}$, with $\omega$ denoting a \emph{state} of the operator system, that is, a positive linear functional of norm 1, over $\mathscr{N}_\mathrm{eu}$. 
\item Here the $\{\varphi^\omega_t\}$-group is a group of $\ast$-automorphisms of the weak-$\star$ operator topology closure of $\mathscr{N}_\mathrm{eu}$, called \emph{modular group}, or \emph{group of modular automorphisms}.
\item A state $\omega$, or $\omega$-state, over $\mathscr{N}_\mathrm{eu}$ determines a 1-parameter $\{\varphi^\omega_t\}$-group of $\ast$-automorphisms on $\mathfrak{H}$. Such a determination expresses a relation of the \emph{Tomita–Takesaki analysis} \cite{Tomita "Standard forms of von Neumann algebras"} \cite{Takesaki "Tomita's Theory of Modular Hilbert Algebras and its Applications"} \cite[chap. VI]{Takesaki "Theory of Operator Algebras II"}, see as a reference \cite{Inoue "Tomita-Takesaki Theory in Algebras of Unbounded Operators"}, and it is at the heart of  von Neumann algebra. 
\item From what we said above, we easily write the map 
\begin{equation}
\label{equation "Map with modular group"}
	\{\varphi^\omega_t\}_{t \in \mathbb{R}} \colon \{\mathscr{N}_\mathrm{eu}, \mathfrak{H}\} \to \{\mathscr{N}_\mathrm{eu}, \mathfrak{H}\}.
\end{equation}
Tomita–Takesaki relation requires that the map \eqref{equation "Map with modular group"} is specified by
\begin{equation}
\label{equation "Tomita–Takesaki relation"}
	\{\varphi^\omega_t\}_{t \in \mathbb{R}}Z = \Laplacian_\omega^{-it}Z\Laplacian_\omega^{it},
\end{equation}
with $Z \in \mathfrak{Z}$, where $\mathfrak{Z}$ is a $C^\ast$-algebra, which is a linear space of bounded operators on $\mathfrak{H}$, and $\Laplacian$ is a self-adjoint (Laplacian) positive operator. Eq. \eqref{equation "Tomita–Takesaki relation"} is what defines the modular group of a $\omega$-state.
\item At this stage, we are ready for a \emph{mental leap} from von Neumann abstract algebra to its \emph{observable} version, simply by applying it to phenomenal data (the normal flow of time), that is, by turing, with the mind, from a discussion of pure mathematics to one of mathematical physics (cf. Sections \ref{section "Mathematics in the Physical Sciences, and Nature of Reality I"}, \ref{section "Mathematics in the Physical Sciences, and Nature of Reality II"} and \ref{section "Mathematics in the Physical Sciences, and Nature of Reality III"}, for a better comprehension), with a view to combining the \emph{operator system} with the \emph{physical system}.
\subenumerationisinitium
\item We \emph{associate the $\omega$-state in $\ast$-algebra with a thermal state}, in keeping with a Gibbs (statistical) distribution \cite{Gibbs "Elementary principles in statistical mechanics"}, taking care to extend the system to a generally covariant context in quantum mechanics. E.g. the observables of a quantum system are a $C^\ast$-algebra plexus, and every $\omega$-state is a positive linear functional over the $C^\ast$-plexus. 
\item We associate the 1-parameter group $\{\varphi^\omega_t\}$ of $\ast$-automorphisms with the concept of physical time flow (the one is the representation of the other). 
\item We recall that the Connes–Rovelli hypothesis establishes a link between a thermal state, or heat flow, and the the time flow, asserting that time flow is \emph{determined by/made dependent on} the thermal state of the system. 
\item Now, the Tomita–Takesaki relation \eqref{equation "Tomita–Takesaki relation"}, which is basically an algebraic equality, can be interpreted as a \emph{modular flow} to \emph{derive} the time flow from a heat flow, i.e. from a thermal state of the physical system. 
\subenumerationisfinis
\enumerationisfinis

\subsubsection{II. Cocycle Radon–Nikodým Theorem, and \textsc{ttc} Flow}

Summarizing the above points in the previous Section, the modular flow \eqref{equation "Tomita–Takesaki relation"} of the $\omega$-state is (coincident with) the physical time flow of the thermal state, so that the modular group $\{\varphi^\omega_t\}$ of the $\omega$-state is (coincident with) a \emph{thermal time flow}.  

The \emph{cocycle Radon–Nikodým theorem} \cite{Radon "Theorie und Anwendugen der absolut additiven Mengenfunktionen"} \cite{Nikodym "Sur une generalisation des integrales de M. J. Radon"} assures that two modular $\ast$-automorphisms determined by two different $\omega$-states of a von Neumann algebra $\mathscr{N}_\mathrm{eu}$ are \emph{inner equivalent}, or that the difference between two modular flows under Tomita–Takesaki relation \eqref{equation "Tomita–Takesaki relation"} is an inner automorphism, see Connes \cite{Connes "Une classification des facteurs de type III"} \cite[chap. V.5]{Connes "Noncommutative Geometry"}. As a result, any modular-like flow describes an \emph{intrinsic property} of $\mathscr{N}_\mathrm{eu}$, since it is independent of all the $\omega$-states. In a quantum framework, a modular flow of this type can be called a \emph{Tomita–Takesaki–Connes flow} (\textsc{ttc}), and it provides greater completeness to the thermal time flow.

\subsubsection{III. Kubo–Martin–Schwinger Boundary Condition}

The relation between the modular $\ast$-automorphism group $\{\varphi^\omega_t\}$ and a $\omega$-state ensures that $\omega$ over $\mathfrak{Z}$ is $\{\varphi^\omega_t\}$-\textsc{kms} at the inverse temperature 
\begin{equation}
	\mathsf{T}_\mathrm{inv} = \frac{1}{k_\textsc{b}\mathsf{T}_\mathrm{abs}},
\end{equation}
with the specification that $\mathfrak{Z}$ plays here the role of an algebra of quantum operators ($k_\textsc{b}$ is the Boltzmann constant, and $\mathsf{T}_\mathrm{abs}$ the absolute temperature); i.e. a $\omega$-state respects a \emph{\textsc{kms} boundary condition} \cite{Kubo "Statistical-Mechanical Theory of Irreversible Processes. I. General Theory and Simple Applications to Magnetic and Conduction Problems"} \cite{Martin and Schwinger "Theory of Many-Particle Systems. I"}, an acronym of the initials of R. Kubo, and P.C. Martin \& J. Schwinger, so named by R. Haag, N.M. Hugenholtz and M. Winnink \cite{Haag Hugenholtz and Winnink "On The Equilibrium States In Quantum Statistical Mechanics"}—a \textsc{kms} condition is a representation of the $C^\ast$-algebra of observables equivalent to thermal equilibrium (or Gibbs state) of a system at a certain temperature $\mathsf{T}$. Lastly, one has a $\textsc{kms}_{\mathsf{T}_\mathrm{inv}}$ $\omega$-state for a Tomita–Takesaki–Connes $\{\varphi^\omega_t\}$-flow.

\subsubsection[Thermo-clocks and Entropic Flow: the Fundamental Limits of Measuring Time]{Thermo-clocks and Entropic Flow: the Fundamental Limits\footnote{
	With a better accuracy, these are thermodynamic limits and not limits \textit{tout court}. Since \emph{time is not an observable}, since there is no a time in and for itself, and resultantly its measurement must be \emph{indirect}, videlicet, it can only be measured in conjunction with other objects, the thermal time hypothesis does not exhaust the concept of time \textit{tout court}, but it is only a \emph{model} of time measurements \cite{Erker Mitchison Silva Woods Brunner and Huber "Autonomous Quantum Clocks: Does Thermodynamics Limit Our Ability to Measure Time?"}, via heat dissipation. Another problem is wondering if time, or rather, the word “time”, has any (physical) meaning outside the indirect measurements. This is something else entirely.
	}
	of Measuring Time}
	
The thermal time hypothesis, on the wake of the Connes–Rovelli's suggestion—under which it is not the time flow that “produces” a heat dissipation, but it is the heat dissipation that “produces” a time flow—can swimmingly be illustrated, and exemplified, saying that any sort of clock is a thermal device, and its flow is a measure of the \emph{entropic flow}, or of the heat flow. 

The most rudimentary case is that of a quantum 3-atomic thermo-clock. The stream of “ticks” of such a thermo-clock system is marked by its thermalization events, as representative of the one-way direction of time's arrow. The more accurate a clock is, the more energy it requires, and the greater is its entropy. A clock whose precision is, for argument's sake, absolute, needs, absurdly, an infinite quantity of energy, viz. an infinite entropy. See P. Erker, M. Huber, \& collaborators \cite{Erker Mitchison Silva Woods Brunner and Huber "Autonomous Quantum Clocks: Does Thermodynamics Limit Our Ability to Measure Time?"} \cite{Schwarzhans Lock Erker Friis and Huber "Autonomous Temporal Probability Concentration: Clockworks and the Second Law of Thermodynamics"}.

\subsection{Fragments of a Pattern}

The most complicated part of Section \ref{section "Space and Time from No-space and No-time"} is the task of putting together all the pieces of the mosaic, so as to devise a coherent pattern, at least in the mathematical intentions. Yes, because Jacobson's and E.P. Verlinde's  thermodynamic space-time (Section \ref{subsubsection "Spatio-temporal/Gravitational Thermodynamics, and Entropic Gravity"}) is not related to Connes–Rovelli's thermodynamic time (Section \ref{subsection "Connes–Rovelli Time Flow in Heat Flow via von Neumann Algebra"}) without forcing. These conjectures on a thermodynamic origin are, in their turn, separate from Sakharov's elastic space (Section \ref{subsubsection "Sakharov's Elasticity of Space, and Liquid Space-Time"}). A fortiori Minkowskian space-time for the monster equation (Section \ref{subsubsection "II. An Equation à la Frankenstein"}) does not bind with an entropic, or elastic, space-time.

There are fragments of hypotheses the meaning of which is precise within the fragments, while the conversion between the meanings of each theoretical proposal requires translation sacrifices. This happens because the notions of space and time differ, from one proposal to the next, on the basis of principles chosen, and preliminary conditions imposed, or, more generally, of mathematics adopted.

\chapter{\textgreek{Γῆ Δρακόντων}, Part IIa. Spatial Primitiveness}
\label{chapter "Ghé Drakónton, Part IIa. Spatial Primitiveness"}

\begingroup
\footnotesize
Little astonishment there should be [\,\dots] if the description of nature carries one in the end to logic, the ethereal eyrie at the center of mathematics. If, as one believes, all mathematics reduces to the mathematics of logic, and \emph{all physics reduces to mathematics}, what alternative is there but for all physics to reduce to the mathematics of logic? Logic is the only branch of mathematics that can “think about itself”.\footnote{
	Cf. G. Boole \cite[p. 13] {Boole "The Mathematical Analysis of Logic: Being an Essay Towards a Calculus of Deductive Reasoning"}: «Logic not only constructs a science, but also inquires into the origin and the nature of its own principles,—a distinction which is denied to Mathematics».
	} \\
\indent — \textsc{C.W. Misner, K.S. Thorne, J.A. Wheeler} \cite[p. 1212, e.a.]{Misner Thorne Wheeler "Gravitation"}, excerpt from a Box annotation about the Wheelerian concept of \emph{pregeometry} \cite{Wheeler "Pregeometry: Motivations and Prospects"}

\endgroup

\section{Spatial Primitiveness I. Primary vs. Secondary Spatio-temporal Archetypes}
\label{section "Spatial Primitiveness I. Primary vs. Secondary Spatio-temporal Archetypes"}

Another chance to arrive at an alternative conception of relativistic space-time, in the Minkowski sense, is the one undertaken by R. Penrose, with the elaboration of the so-called \emph{twistor space}. What we care for is that Penrose's space has, algebraically, a more \emph{primitive} nature in comparison to Minkowski space-time, for which the twistor space stands out as \emph{primary}, whilst Minkowski space becomes \emph{secondary}, or subsidiary.

\subsection{Pre-space(s) in Penrose's Twistor Algebra}

\begingroup
\footnotesize
Ordinary space-time concepts can then be translated into twistor terms. However, the geometrical expressions of the most immediate twistor concepts have a somewhat \emph{non-local} character. Thus, the \emph{primary} geometrical object will not be a point in Minkowski space-time, but rather a null straight line or, more generally, a twisting congruence of null lines. Points do, in fact, \emph{emerge}, but only at a \emph{secondary stage}. (It also turns out that a natural description of physical fields in twistor terms is given by quantities having a non-local space-time interpretation). However, any vector, tensor, or spinor operation \textit{can} be translated into twistor terms, if desired, and \textit{vice versa}. \\
\indent — \textsc{R. Penrose} \cite[p. 346, e.a.]{Penrose "Twistor Algebra"}

\vspace{2mm}

Space-time points \emph{arise as secondary concepts} corresponding to linear sets in twistor space. They, rather than the null [light] cones, should become “smeared out” on passage to a quantised gravitational theory [\,\dots]. Space-time points can then be reconstructed from the twistor space (being represented as certain linear subspaces), but they become secondary to the twistors themselves. \\
\indent — \textsc{R. Penrose and M.A.H. MacCallum} \cite[pp. 241, 244, e.a.] {Penrose and MacCallum "Twistor theory: An approach to the quantisation of fields and space-time"}

\endgroup

\vspace{2mm}

Reference works are: R. Penrose \cite{Penrose "Twistor Algebra"} \cite{Penrose "The Twistor Approach to Space-Time Structures"} \cite{Penrose "Twistor Theory as an Approach to Fundamental Physics"} \cite{Penrose "Twistor Theory: A Geometric Perspective for Describing the Physical World"}, R. Penrose and M.A.H. MacCallum \cite{Penrose and MacCallum "Twistor theory: An approach to the quantisation of fields and space-time"}, and R. Penrose \& W. Rindler \cite[chapp. 6, 9.3]{Penrose and Rindler "Spinors and space-time II. Spinor and twistor methods in space-time geometry"}. Following are some twistorial ingredients useful for our discussion.

\subsubsection{A \emph{Vice Versa} Structure: Projective Null Twistor Space \&  Compactified Minkowski Space(-Time)}

\enumerationisinitium
\item \emph{Twistor} is said to be any element $x_{\mathbb{T}\mathbbl{w}}$ in twistor spaces. We can think of the simplest type of twistor as a  \emph{representation of a null geodesic with a pair of 2-component spinors} (cf. Section \ref{subsubsection "The Covering Morphisms $SU_2(C)$ to $SO_3(R)$"})—one of which gives the direction of the geodesic, and the other its moment—with \emph{four complex components}, otherwise as a \emph{representation of the restricted conformal group} in dimension 4, or in dimension 8 if reflections are counted. 
\item We denote the \emph{twistor space} by $\mathbb{T}\mathbbl{w}$. It designates a \emph{complex vector space} with a pseudo-Hermitian metric and signature $(+, +, -, -)$. In the conventional assumption, $\mathbb{T}\mathbbl{w}$ is a complex 4-space $\mathbb{T}\mathbbl{w}^{4(\mathbb{C})} \cong \mathbb{C}^4$, or a real 8-space $\mathbb{T}\mathbbl{w}^{8(\mathbb{R})} \cong \mathbb{R}^8$.
\item By $\mathbb{PT}\mathbbl{w}$ is denoted the \emph{projective twistor space}. It is a real 6-space $\mathbb{PT}\mathbbl{w}^{6(\mathbb{R})}$, or a complex 3-space $\mathbb{PT}\mathbbl{w}^{3(\mathbb{C})} \cong \mathbb{CP}^3$. It derives from $\mathbb{T}\mathbbl{w}$.
\subenumerationisinitium
\item Letting $SO_{2, 4} \cong_{(1:2)} SU_{2, 2}$ be the twistor group for the previously mentioned spinorial object, under which the pseudo-unitary group $SU_{2, 2}$ is 2-fold covering (double cover) of $SO_{2, 4}$, the \emph{reduced spin space} for $SO_{2, 4}$ is $\mathbb{T}\mathbbl{w}$, and the \emph{non-reduced (full) spin space} for $SO_{2, 4}$ is $\mathbb{T}\mathbbl{w} \oplus \mathbb{T}\mathbbl{w}^*$, where the latter is the dual space of $\mathbb{T}\mathbbl{w}$.
\item Regarding the congruence with double covering, there is a composition preserving map onto the identity connected component of $SO_{2, 4}$ through a morphism 
\begin{equation}
	SU_{2, 2} \xrightarrow{\varsigma\text{-homomorphism}} SO_{2, 4}
\end{equation}
in a $(2:1)$ modality.
\subenumerationisfinis
\item $\mathbb{T}\mathbbl{w}^{7(\mathbb{R})}_\mathrm{null}$ indicates a \emph{null twistor space} (and a subspace of $\mathbb{T}\mathbbl{w}$). It is a non-complex and exclusively real 7-space,\footnote{
	A complex space requires an \emph{even} number of dimensions.
	} 
then we say that a space $\mathbb{T}\mathbbl{w}^{7(\mathbb{R})}_\mathrm{null}$ of null twistors $(\bar{x}_{\mathbb{T}\mathbbl{w}})_\mu(x_{\mathbb{T}\mathbbl{w}})^\mu = 0$, where $\bar{x}_{\mathbb{T}\mathbbl{w}}$ is a complex conjugate, divides $\mathbb{T}\mathbbl{w}$ into two parts, $\mathbb{T}\mathbbl{w}^{3(\mathbb{C})}_+$ and $\mathbb{T}\mathbbl{w}^{3(\mathbb{C})}_-$, i.e. a complex 3-space of positive (\textsc{rh}) twistors $(\bar{x}_{\mathbb{T}\mathbbl{w}})_\mu(x_{\mathbb{T}\mathbbl{w}})^\mu > 0$, and a complex 3-space of negative (\textsc{lh}) twistors $(\bar{x}_{\mathbb{T}\mathbbl{w}})_\mu(x_{\mathbb{T}\mathbbl{w}})^\mu < 0$.

Here is that the twistor space $\mathbb{T}\mathbbl{w}$, as a full space, is the \emph{disjoint union} of $\mathbb{T}\mathbbl{w}^{3(\mathbb{C})}_+$ and $\mathbb{T}\mathbbl{w}^{3(\mathbb{C})}_-$, plus $\mathbb{T}\mathbbl{w}^{7(\mathbb{R})}_\mathrm{null}$. It is an aggregate-space formed by these 3 parts-(sub)spaces.
\item $\mathbb{PT}\mathbbl{w}^{5(\mathbb{R})}_\mathrm{null}$ is the \emph{projective null twistor space}, which is a real 5-space, being a subspace of $\mathbb{PT}\mathbbl{w}^{6(\mathbb{R})}$.
\subenumerationisinitium
\item Each element of $\mathbb{PT}\mathbbl{w}^{5(\mathbb{R})}_\mathrm{null}$ is called \emph{projective null twistor}, and it is consistent with a \emph{point} of $\mathbb{PT}\mathbbl{w}^{5(\mathbb{R})}_\mathrm{null}$. 
\item A \emph{locus} in $\mathbb{PT}\mathbbl{w}^{5(\mathbb{R})}_\mathrm{null}$ is consistent with a \emph{Riemann sphere} $\mathbb{S}^2 \cong \mathbb{CP}^1$, since $\hat{\mathbb{R}}^2 \equival \mathbb{R}^2 \cup \{\infty\} \cong \mathbb{S}^2$, that is, $\hat{\mathbb{C}} \equival \mathbb{C} \cup \{\infty\} \cong \mathbb{CP}^1$.
\subenumerationisfinis
\item Let be $\mathbb{M}^{4(\mathbb{R})}_{(\infty)}$ a conformally \emph{compactified Minkowski space(-time)}, with topology $\mathbb{S}^1 \times \mathbb{S}^3$, endowed with a Lorentzian-like conformal metric, which is the ordinary \emph{flat} Minkowski 4-space with a closed light cone, or null cone, at infinity, assuming to embed the light cone in a \emph{bent} pseudo-Euclidean 6-space $\mathbb{E}^{2, 4}$.
\subenumerationisinitium
\item More precisely, $\mathbb{M}^{4(\mathbb{R})}_{(\infty)}$ is generated by the elements of a complex line in $\mathbb{PT}\mathbbl{w}^{3(\mathbb{C})} \cong \mathbb{CP}^3$, imagining such a complex line as a complex 2-subspace of $\mathbb{T}\mathbbl{w}^{4(\mathbb{C})}$ for the past and future null infinities of $\mathbb{M}^{4(\mathbb{R})}_{(\infty)}$.
\item The group acting on $\mathbb{E}^{2, 4}$ is the pseudo-orthogonal group $O_{2, 4}(\mathbb{R})$, which preserves the quadratic form, or distance between points in 6 dimensions. Hence a \emph{twistor} for $\mathbb{M}^{4(\mathbb{R})}_{(\infty)}$ is but a \emph{reduced spinor} whose group is $O_{2, 4}(\mathbb{R})$.
\item A \emph{point} in $\mathbb{M}^{4(\mathbb{R})}_{(\infty)}$ is consistent with an \emph{event}, which means that a point of $\mathbb{M}^{4(\mathbb{R})}_{(\infty)}$ is a generator of the light cone at infinity.
\item A \emph{locus} in $\mathbb{M}^{4(\mathbb{R})}_{(\infty)}$ is consistent with a \emph{ray of light}, or \emph{null geodesic}.
\subenumerationisfinis
\item Now, what is the relationship between $\mathbb{PT}\mathbbl{w}^{5(\mathbb{R})}_\mathrm{null}$ and $\mathbb{M}^{4(\mathbb{R})}_{(\infty)}$?
\subenumerationisinitium
\item A \emph{point}, that is, a projective null twistor, in $\mathbb{PT}\mathbbl{w}^{5(\mathbb{R})}_\mathrm{null}$ corresponds to a \emph{ray of light}, or \emph{null geodesic}, in $\mathbb{M}^{4(\mathbb{R})}_{(\infty)}$. Vice versa, a \emph{locus}, as a ray of light, in $\mathbb{M}^{4(\mathbb{R})}_{(\infty)}$ represents a \emph{point}, i.e. a twistor, in $\mathbb{PT}\mathbbl{w}^{5(\mathbb{R})}_\mathrm{null}$. 
\item A \emph{locus}, as a sphere $\mathbb{S}^2 \cong \mathbb{CP}^1$, in $\mathbb{PT}\mathbbl{w}^{5(\mathbb{R})}_\mathrm{null}$ corresponds to a \emph{point} (event) in $\mathbb{M}^{4(\mathbb{R})}_{(\infty)}$. Vice versa, a \emph{point} (event) in $\mathbb{M}^{4(\mathbb{R})}_{(\infty)}$ represents a \emph{sphere} $\mathbb{S}^2 \cong \mathbb{CP}^1$ at the origin of $\mathbb{E}^{2, 4}$ in $\mathbb{PT}\mathbbl{w}^{5(\mathbb{R})}_\mathrm{null}$. Note. there are two other ways to express the same thing: 
\item[(\textgreek{α})] a point in $\mathbb{M}^{4(\mathbb{R})}_{(\infty)}$ represents a \emph{projective line} in $\mathbb{PT}\mathbbl{w}^{5(\mathbb{R})}_\mathrm{null}$, since $\mathbb{S}^2 \cong \mathbb{CP}^1$ is a projective line in $\mathbb{PT}\mathbbl{w}$, but situated in $\mathbb{PT}\mathbbl{w}^{5(\mathbb{R})}_\mathrm{null}$, 
\item[(\textgreek{β})] a point in $\mathbb{M}^{4(\mathbb{R})}_{(\infty)}$ is a \emph{holomorphic entity} in $\mathbb{PT}\mathbbl{w}^{5(\mathbb{R})}_\mathrm{null}$, in the sense that it is related to a complex-valued function, since $\mathbb{S}^2 \cong \mathbb{CP}^1$ is a complex 1-space, recalling that $\hat{\mathbb{C}} \equival \mathbb{C} \cup \{\infty\} \cong \mathbb{CP}^1$.
\item Let $x_{\mathbb{T}\mathbbl{w}}$ be a point (twistor) in $\mathbb{PT}\mathbbl{w}^{5(\mathbb{R})}_\mathrm{null}$, and $x_\mathbb{M}$ a point in $\mathbb{M}^{4(\mathbb{R})}_{(\infty)}$, and let $\gamma_\mathrm{null}$ be a ray of light, or null geodesic. To summarize, 
\begin{align} 
	\text{ in } \mathbb{PT}\mathbbl{w}^{5(\mathbb{R})}_\mathrm{null}
	& \begin{cases}
	x_{\mathbb{T}\mathbbl{w}} \corr \gamma_\mathrm{null} \text{ in } \mathbb{M}^{4(\mathbb{R})}_{(\infty)}, \notag \\
	\mathbb{S}^2 \cong \mathbb{CP}^1 \corr x_\mathbb{M} \text{ in } \mathbb{M}^{4(\mathbb{R})}_{(\infty)}, \\
	\end{cases} \\
	\text{ in } \mathbb{M}^{4(\mathbb{R})}_{(\infty)}
	& \begin{cases}
	x_\mathbb{M} \corr \mathbb{S}^2 \cong \mathbb{CP}^1 \text{ (at the origin of $\mathbb{E}^{2, 4}$) in } \mathbb{PT}\mathbbl{w}^{5(\mathbb{R})}_\mathrm{null}, \notag \\
	\gamma_\mathrm{null} \corr x_{\mathbb{T}\mathbbl{w}} \text{ in } \mathbb{PT}\mathbbl{w}^{5(\mathbb{R})}_\mathrm{null}. \\
	\end{cases}
\end{align}
Ultimately, in a general perspective, the 5-space $\mathbb{PT}\mathbbl{w}^{5(\mathbb{R})}_\mathrm{null}$ is the \emph{space of all null geodesics} in the 4-space $\mathbb{M}^{4(\mathbb{R})}_{(\infty)}$.
\subenumerationisfinis
\item $\mathbb{PT}\mathbbl{w}^{5(\mathbb{R})}_\mathrm{null}$ divides $\mathbb{PT}\mathbbl{w}^{6(\mathbb{R})}$ into two parts, $\mathbb{PT}\mathbbl{w}^{3(\mathbb{C})}_+ \cong \mathbb{CP}^3_+$ and $\mathbb{PT}\mathbbl{w}^{3(\mathbb{C})}_- \cong \mathbb{CP}^3_-$, two complex 3-spaces, for 0-mass particles of positive (\textsc{rh}) and negative (\textsc{lh}) helicity$_\pm$. Accordingly, a point of $\mathbb{PT}\mathbbl{w}^{3(\mathbb{C})}_+$ is for a positive helicity$_+^\textsc{rh}$ 0-mass, and a point of $\mathbb{PT}\mathbbl{w}^{3(\mathbb{C})}_-$ is for a negative helicity$_-^\textsc{lh}$ 0-mass. On the whole, any point of $\bigl\{\mathbb{PT}\mathbbl{w}^{3(\mathbb{C})}_+, \mathbb{PT}\mathbbl{w}^{3(\mathbb{C})}_-\bigr\}$ is a vector 1-(sub)space of $\mathbb{T}\mathbbl{w}$.

Here is that the projective twistor space $\mathbb{PT}\mathbbl{w}^{6(\mathbb{R})}$, as a full space, is the \emph{disjoint union} of $\mathbb{PT}\mathbbl{w}^{3(\mathbb{C})}_+$ and $\mathbb{PT}\mathbbl{w}^{3(\mathbb{C})}_-$, plus $\mathbb{PT}\mathbbl{w}^{5(\mathbb{R})}_\mathrm{null}$. It is an aggregate-space formed by these 3 parts-(sub)spaces. 
\item Which brings us to a complexification of (real) $\mathbb{M}$-space, and its compactified version, symbolized by $\mathbb{CM}^\natural$ and  $\mathbb{CM}^\natural_{(\infty)}$, which are $\natural$-spaces (bequadro-dimensional), where the time and spatial coordinates, $x^0$ and $x^1, x^2, x^3$, are complex numbers. Note that:
\subenumerationisinitium
\item points of $\mathbb{CM}^\natural$ and $\mathbb{CM}^\natural_{(\infty)}$ correspond to complex projective lines in $\mathbb{PT}\mathbbl{w}$, 
\item in $\mathbb{CM}^\natural_{(\infty)}$ it creates a double \emph{complex 2-locus of points}, of which one, a \emph{self-dual locus} is incident with every non-zero twistor $x_{\mathbb{T}\mathbbl{w}}^\mu$, the other, an \emph{anti-self-dual locus}, is incident with every non-zero twistor $y_{\mathbb{T}\mathbbl{w}}^\mu$.  
\subenumerationisfinis
\enumerationisfinis

\subsection{Curved Twistor Space as a Projective Distortion: Quantum Complex 4-Space of Gravity from a Kodaira Analytic Family}

There is also an evolution of twistors in curved space. A \emph{curved twistor space} is a $\mathring{\mathcal{Z}}$-bundle, with the (fibers of the) projection of $\mathring{\mathcal{Z}}$ to a projective space $\mathbb{PT}\mathbbl{w}\mathring{\mathcal{Z}}$, for the \emph{anti-self-dual gravity} \cite{Penrose "Nonlinear Gravitons and Curved Twistor Theory"} \cite{Penrose "The Nonlinear Graviton"}, with the incorporation of a \textsc{lh} non-linear \emph{graviton}, the quantum particle which, as a \emph{complex 4-space} $\mathbb{G}^{4(\mathbb{C})}$, is entrusted with the \emph{task of carrying space-time curvature}. Here is how it happens, in layman's terms.
\subenumerationisinitium
\item Operating—not globally, but in portions—a \emph{distortion} (or deformation) of $\mathbb{PT}\mathbbl{w}$ is allowed. One starts from an open neighborhood $\Upsilon_\mathbb{CM}$ of a point $x_\mathbb{CM}$ in $\mathbb{CM}^\natural$, then a commensurate neighborhood $\Upsilon_{\mathbb{PT}\mathbbl{w}}$ of a line $\gamma_{(\mathbb{PT}\mathbbl{w})}$ in $\mathbb{PT}\mathbbl{w}$ is identified, known as \emph{tubular} region, with topology $\mathbb{S}^2 \times \mathbb{R}^4$. It should be noted that $\gamma_{(\mathbb{PT}\mathbbl{w})}$ corresponds to $x_\mathbb{CM}$ of $\mathbb{CM}^\natural$.
\item $\Upsilon_{\mathbb{PT}\mathbbl{w}}$ is the region subject to distortion. Distortion of $\Upsilon_{\mathbb{PT}\mathbbl{w}}$ results in the  breaking of $\gamma_{(\mathbb{PT}\mathbbl{w})}$.
\item A theorem of completeness of K. Kodaira \cite{Kodaira "A Theorem of Completeness of Characteristic Systems for Analytic Families of Compact Submanifolds of Complex Manifolds"}, see also \cite{Kodaira "On Stability of Compact Submanifolds of Complex Manifolds"}, ensures the existence of a family of 4-parameter curves $\gamma^\textsc{k}_{(\mathbb{PT}\mathbbl{w})}$ in $\mathbb{PT}\mathbbl{w}\mathring{\mathcal{Z}}$, from which the complex 4-space $\mathbb{G}^{4(\mathbb{C})}$, for the postulation of a \textsc{lh} non-linear graviton, is finally determined;\footnote{
	Difficulties and developments of the \textsc{rh} non-linear graviton are in \cite{Penrose "Towards a Twistor Description of General Space-Times; Introductory Comments"}.
	} 
$\mathbb{PT}\mathbbl{w}\mathring{\mathcal{Z}}$ is the space in which the fibers of the projection of $\mathring{\mathcal{Z}}$ lie.
\item Any point of $\mathbb{G}^{4(\mathbb{C})}$  is thus equal to the cross-sections by \emph{Kodaira's holomorphic lines}.
\item The \emph{Weyl curvature} (Section \ref{subsection "Weyl Curvature Tensors"}) of $\mathbb{G}^{4(\mathbb{C})}$ is \emph{anti-self-dual}, and it turns out to be \emph{Ricci-flat} ($\Ric = 0$). 
\subenumerationisfinis

\subsection{Twisotrial Hierarchy: Non-locality and Holomorphicity}

The fact that \emph{Minkowski space(-time) is treated as a space of complex lines in $\mathbb{PT}\mathbbl{w}^{5(\mathbb{R})}_\mathrm{null}$} decrees $\mathbb{PT}\mathbbl{w}^{5(\mathbb{R})}_\mathrm{null}$ as a \emph{primary space}, whilst Minkowski $\mathbb{M}$-space turns into a \emph{secondary} (\emph{subsidiary}) \emph{space}, namely a space that can be constructed/derived from $\mathbb{PT}\mathbbl{w}^{5(\mathbb{R})}_\mathrm{null}$ (cf. Penrose's epigraphs at the beginning of Section). This hierarchy, or classification, is also repeatable with other twistorial spaces, basically thanks to the use of complex numbers, and therefore to the \emph{holomorphic geometry} of twistors. In conclusion, twistors \& twistor spaces are, respectively, objects and structures relating to Minkowskian space-time in a \emph{global}, or \emph{non-local manner}, whereas classical vectors, tensors, and spinors act on a \emph{(single) point}.

\subsection{Summary Graph}

A summary graph getting together the main twistorial spaces:
\[
\begin{tikzcd}
	& \mathbb{PT}\mathbbl{w}^{5(\mathbb{R})}_\mathrm{null} \arrow[rd, no head, dotted, bend left] & \\
	 \left\{\mathbb{PT}\mathbbl{w}^{3(\mathbb{C})}_+, \mathbb{PT}\mathbbl{w}^{3(\mathbb{C})}_-\right\} \cong \mathbb{CP}^3_\pm \arrow[ru, no head, dotted]
	& \mathbb{CM}^\natural, \mathbb{CM}^\natural_{(\infty)}
	& \mathbb{M}^{4(\mathbb{R})}_{(\infty)} \text{ in } \mathbb{E}^{2, 4} \arrow[lld, no head, dotted] \arrow[l, no head, dotted] \\
	\mathbb{PT}\mathbbl{w}^{6(\mathbb{R})} \cong \mathbb{PT}\mathbbl{w}^{3(\mathbb{C})} \cong \mathbb{CP}^3 \arrow[u, no head, dotted]
	& \mathbb{T}\mathbbl{w}^{3(\mathbb{C})}_+, \mathbb{T}\mathbbl{w}^{3(\mathbb{C})}_-
	& \mathbb{T}\mathbbl{w}^{7(\mathbb{R})}_\mathrm{null} \arrow[l, no head, dotted] \\
	& \mathbb{T}\mathbbl{w}^{4(\mathbb{C})} \cong \mathbb{T}\mathbbl{w}^{8(\mathbb{R})} \arrow[ru, no head, dotted] \arrow[lu, no head, dotted] \arrow[u, no head, dotted] &                                                                                
\end{tikzcd}
\]

\section{Spatial Primitiveness II. The Loop Quantum Gravity Program: How Far Can We Go?}

\subsection{Spin Networks, Nodes, and Loops: a Relational Conception of Space}

\subsubsection{Topological Discrete Graph Structure}
\label{subsubsection "Topological Discrete Graph Structure"}

The Penrose hierarchy (Section \ref{section "Spatial Primitiveness I. Primary vs. Secondary Spatio-temporal Archetypes"}) does not address the \emph{problem of quantum space}, or the \emph{origin of space at the Planck scale}, with the related disagreements between continuous and discrete overtures. Moreover, twistorial spaces do not treat the Lagrangian with the same naturalness as field theories (\textsc{qft}), see the monster equation (Section \ref{subsubsection "II. An Equation à la Frankenstein"}); and indeed, the execution itself of a twistorialization of \textsc{qft} is in fieri.\footnote{
	Attempts to amalgamate space in twistor theory and holographic principle, by virtue of the point/line vice versa structure, are, for instance, in J.J. Heckman \& H. Verlinde \cite{Heckman and Verlinde Instantons Twistors and Emergent Gravity} and Y. Neiman \cite{Neiman "The holographic dual of the Penrose transform"}.
	}

Nevertheless, in Penrose's works can be found a theoretical arsenal, the so-called \emph{spin networks} \cite{Penrose "Angular momentum: an approach to combinatorial space-time"} \cite{Penrose "Applications of Negative Dimensional Tensors"} \cite{Penrose "On the Nature of Quantum Geometry"} \cite{Penrose "Combinatorial Quantum Theory and Quantized Directions"}, to tackle the \emph{quæstio} of space in its foundations. 

This arsenal, together with the \emph{(spinorial) Ashtekar variables} \cite{Ashtekar "New Variables for Classical and Quantum Gravity"}, have constituted  the driving force for the birth of the \emph{loop quantum gravity} (\textsc{lqg}), with main contributions from C. Rovelli \& L. Smolin \cite{Rovelli and Smolin "Knot Theory and Quantum Gravity"} \cite{Rovelli Smolin "Loop space representation of quantum general relativity"} \cite{Rovelli Smolin "Discreteness of area and volume in quantum gravity"}, and later from other researchers.

Spin network is a \emph{topological graph structure} of simple lines and vertices of a \emph{discrete} character but capable of representing a quantum state, under a combinatorial calculational rule. Penrose writes \cite[e.m., p. 151]{Penrose "Angular momentum: an approach to combinatorial space-time"}:

\vspace{2mm}

\begingroup
\footnotesize
The basic theme of these suggestions have been to try to \emph{get rid of the continuum} and build up physical theory from \emph{discreteness}. The most obvious place in which the continuum comes into physics is the structure of space-time. But, apparently independently of this, there is also another place in which the continuum is built into present physical theory. This is in quantum theory, where there is the superposition law [\,\dots]. One scarcely wants to take every concept in existing theory and try to make it combinatorial: there are too many things which look continuous in existing theory. And to try to eliminate the continuum by approximating it by some discrete structure would be to change the theory. The idea, instead, is to concentrate only on things which, in fact, are discrete in existing theory and try and use them as primary concepts—then to build up other things using these \emph{discrete primary concepts as the basic building blocks}. \emph{Continuous concepts could emerge in a limit}, when we take more and more complicated systems.

\endgroup

\vspace{2mm}

The inspiring plan is to understand space(-time) in a \emph{relational} way, \emph{à la Leibniz–Mach},\footnote{
	G.W. Leibniz \cite[\textit{Leibniz's Third Letter, Being an Answer
to Clarke's Second Reply}, 25 February 1716, p. 14, e.a.]{Leibniz and Clarke "Correspondence"}: «I hold space to be \emph{something purely relative}, as time is—[\,\dots] I hold it to be an order of coexistences, as time is an order of successions». The space(-time) itself (\textgreek{αὐτὸ καθ᾿ αὑτό})—on a range of Leibnizian inspirations—rests in its \emph{relational nature}.
	}
that is, not as an entity pre-existing in a background, but rather as a notion that derives, probabilistically, from the spin network. \emph{Space}, with its continuous nature, \emph{arises from the discretum-spin quantum combinatorics}. 

\subsubsection{Spatial Lumpiness via Node-Point (Spatiumculus)}
\label{subsubsection "Spatial Lumpiness via Node-Point (Spatiumculus)"}

Loop quantum gravity, in accordance with this (Section \ref{subsubsection "Topological Discrete Graph Structure"}), assumes that space is made up of \emph{(spin network) nodes} and \emph{links}, which are \emph{line segment} between one node to another. Nodes are \emph{points} where links are touching. 

The \textsc{lqg} space is necessarily an elementary set of \emph{two non-definite elements of elementary geometry} (Section \ref{section "Point and Line as Primitive Ideas"}): point and line segment, two primitive geometric ideas, one discrete and the another continuous (albeit within the segment); together they form a spin graph structure. 

Each spin graph structure is a \emph{granular} network, since its nodes form, by postulation, a loop. A set of loops forms, in turn, a complex of interwoven space-based lumps, for the generation of the quantum dimension of the gravitational field.

\emph{A (spin network) node, i.e. a point, is not located in space}, by definition; it is a \emph{quantum of space}, which is \emph{the space itself}—to be more exact, interactions between several nodes create space(s), without referring to an external space. Space, for its part, is an emergence of this \emph{relational network} of points and line segments (links); the (\textgreek{α-β-γ-δ-ε-ζ-η})-nodal space graph (Figg. \ref{figure "Loop-like space graph"} and \ref{figure "Spin network with multi-colored hunks"}) is just a small \emph{lump of emerging space}. 

\begin{figure}[h!]
\[
\begin{tikzcd}
	& \overset{\textgreek{\text{δ}}}{\text{·}} \arrow[rd, no head] \arrow[dd, no head] & \\
	\overset{\textgreek{\text{ε}}}{\text{·}} \arrow[ru, no head] \arrow[d, no head]
	\arrow[rd, no head] 
	&& \overset{\textgreek{\text{γ}}}{\text{·}} \arrow[d, no head] \arrow[ld, no head] \\
	  \overset{\textgreek{\text{ζ}}}{\text{·}} \arrow[rd, no head] \arrow[r, no head] 
	& \overset{\textgreek{\text{α}}}{\text{·}} 
	& \overset{\textgreek{\text{β}}}{\text{·}} \arrow[ld, no head] \arrow[l, no head] \\
	& \overset{\textgreek{\text{η}}}{\text{·}} \arrow[u, no head] &                                         
\end{tikzcd}
\]
\caption{Granular spin network: it is a loop-like graph. There are \emph{(spin network) nodes}, i.e. \emph{points acting as quanta of space}, and \emph{links}, for a (\textgreek{α-β-γ-δ-ε-ζ-η})-nodal lump of space. Note that \textgreek{αβγ} order is chosen with personal preference: there is no rotational privilege, because there is no spatial order \emph{ab initio}}
\label{figure "Loop-like space graph"}
\end{figure}

\begin{figure}[h!]
\centering
\includegraphics{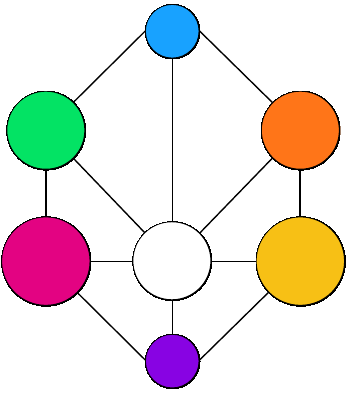}
\caption{Oversimplification of a lump of emerging space, highlighted with multi-colored hunks, which are but \emph{quanta of space}, sketched by the nodes of the granular spin network, the before-mentioned (\textgreek{α-β-γ-δ-ε-ζ-η})-nodal lump. Each hunk must be imagined \emph{adjacent} to each other, separated by surfaces, and \emph{not isolated} within a circular-shaped configuration; compare with Rovelli \cite[Fig. 5, p. 38]{Rovelli "Loop Quantum Gravity"}}
\label{figure "Spin network with multi-colored hunks"}
\end{figure}

One of the reasons why quantum gravity is elusive is that, in both relativity (continuum mechanics) and quantum theory (with continuous but above all discrete values), space, or space-time,\footnote{
	Here we neglect the distinction between space and space-time. Admittedly, see cosmology, it is possible to have a \emph{space} e.g. \emph{negatively curved} even in the presence of a \emph{space-time with zero curvature (flat space-time)}, if space is expanding (proportionally with the coordinate time). For the same reason, it is possible to think of a \emph{space with zero curvature for a curved space-time}, where space is in an expansive, or contractive, phase.
	} 
is an entity (more or less \emph{absolute}, cf. Section \ref{section "Historico-Philological Remarks"}), and not a \emph{relational} structure. \emph{Relationality} is one of the keys to open the casket of quantum space-time.

A relational conception of space is the best theory from which to think about space, or space-time. Certainly, a theory of relationality has many theoretical steps in need of convenient accommodations with the prevailing non-Leibniz–Mach view; but the bigger problem, anyhow, is that \textsc{lqg} is grounded on the concept of \emph{(spin network) node}, that is, of \emph{point}, a \emph{spatiumculus}, so to speak, which, in this context, takes on the guise of a gravitational quantum of space; but a node is, and remains, a \emph{geometric point}, exactly as it is pseudo-explained (\textgreek{Ὅρος, α´, Στοιχείων α´}) in Euclid, with all the captivating quandaries involved (see Section \ref{section "Point and Line as Primitive Ideas"}).

\begin{margo}
We report some others difficulties in \textsc{lqg}, of which we can only give a quick hint.
\enumerationisinitium
\item Ashtekar variables are a \emph{connection representation}; and they are usually defined starting from a 3-surface, so the spin bundle $\Gamma(\mathring{\mathcal{P}}_\textit{ß})$-connection, or rather, the \emph{connection on a spinor bundle} $\mathring{\mathcal{P}}_\textit{ß}$ (cf. Section \ref{subsection "Spinor Map (6-Dimensional Homomorphism): the Covering $SL_2(C)$ to $SO_{1, 3}^+(R)$"}) for the parallel transport (cf. Section \ref{section "Parallel Transport of the Levi-Civita Connection"}) of spinors, has a definitional approach on a 3-space. But, in strictly relativistic terms, a space+time connection is required, i.e. a $\Gamma(\mathring{\mathcal{P}}_\textit{ß})$-connection taken on a 4-space. This entails, in some instances, that a $\Gamma(\mathring{\mathcal{P}}_\textit{ß})$-connection is formulated with undetermined quantities. 
\item Directly related to what has just been said: Einstein Hamiltonian needs, in some cases, a more rigorous definition.
\item To manage loops variables for 4-spaces, it is possible to invoke the dynamic theory, with the so-called \emph{spin foam} models, see J.C. Baez \cite{Baez "Spin foam models"} \cite{Baez "An Introduction to Spin Foam Models of BF Theory and Quantum Gravity"}. The spin foam notion facilitates the transition from a \emph{kinematic} aspect (\emph{quantum geometry of space}) to a \emph{dynamic} one \emph{(quantum geometry of space-time}), for the quantum gravity. 

The difference is this: in a spin network, one has edges labeled by representations, and vertices labeled by intertwining operators; in a spin foam, one has a \emph{2-dimensional complex} with faces labeled by representations, and edges labeled by intertwining operators. \margosymbol
\enumerationisfinis
\end{margo}

\vspace{10mm}

\setcounter{secnumdepth}{0}  
\section{References and Bibliographic Details}
\setcounter{secnumdepth}{3}
\markright{References and Bibliographic Details}

\begingroup
\footnotesize
\noindent Section \ref{section "Spatial Primitiveness I. Primary vs. Secondary Spatio-temporal Archetypes"} (and subsequent Sections)

\begin{indent paragraph: 15pt}
· To learn more about mathematics of twistors, see: \cite{Baston and Eastwood "The Penrose Transform: Its Interaction with Representation Theory"} \cite{Bailey Baston (Eds.) "Twistors in Mathematics and Physics"} \cite{Ward and Wells Jr. "Twistor Geometry and Field Theories"} \cite[part II]{Mason and Woodhouse "Integrability Self-Duality and Twistor Theory"} \cite{Ward "Integrable Systems and Twistors"} \cite[chap. 9]{Krasnov "Formulations of General Relativity: Gravity Spinors and Differential Forms"}. \\
· An embryonic twistor string theory—that meshes, as far as possible, string theory and twistor theory (without exploiting the full hidden potential of twistor theory, because it is an amalgamation of theories)—is in Witten's paper \cite{Witten "Perturbative Gauge Theory as a String Theory in Twistor Space"}. \\
· New attempts to quantize gravity via twistor theory are being studied by P. Woit \cite{Woit "Euclidean Twistor Unification"} \cite{Woit "Notes on the Twistor P1"}.
\end{indent paragraph: 15pt}

\noindent Section \ref{subsubsection "Spatial Lumpiness via Node-Point (Spatiumculus)"}
\begin{indent paragraph: 15pt}
Painstaking information, and technical expositions, on the loop quantum gravity are in \cite[IV.2 (L. Bombelli, \textit{The weak field limit}), V.5-6 (L. Smolin, \textit{Exact solutions to the quantum Hamiltonian constraints}, and \textit{Knot theory, loop space and the diffeomorphism group}), V.8 (C. Rovelli, \textit{Loop space representation})]{Ashtekar "New Perspectives in Canonical Gravity"} \cite{Ashtekar "Lectures on Non-Perturbative Canonical Gravity"} \cite{Gambini Pullin "Loops Knots Gauge Theories and Quantum Gravity"} \cite{Rovelli "Quantum Gravity"} \cite{Thiemann "Modern Canonical Quantum General Relativity"}.	For a prompt recapitulation, see \cite{Rovelli "Quantum Spacetime"}. 
\end{indent paragraph: 15pt}

\endgroup

\chapter{\textgreek{Γῆ Δρακόντων}, Part IIb. Space-numeral Primitiveness}
\label{chapter "Ghé Drakónton, Part IIb. Space-numeral Primitiveness"}

\begingroup
\footnotesize
In re mathematica ars proponendi quaestionem pluris facienda est quam solvendi.\footnote{
	«In mathematics the art of proposing a problem must be [worth] more than solving it».
	} \\
\indent — \textsc{G. Cantor} \cite[p. 31]{Cantor "De aequationibus secundi gradus indeterminatis"}

\endgroup

\section{Point and Line as Primitive Ideas}
\label{section "Point and Line as Primitive Ideas"}

\begingroup
\footnotesize
\textgreek{Σημεῖόν}\footnote{
	\label{footnote "Semeión vs. stigmé"}
	The word \textgreek{σημεῖόν} (“sign”, hence “boundary”, “limit”, and mathematical “point”) is a Hellenistic term that stands for the previous—but later reintroduced—\textgreek{στιγμή} (“tittle”, “spot”, or mathematical “point”, from the vb. \textgreek{στίζω}, “mark”, “punctuate”), the La. calque of which is \emph{punctum}. 
	} 
\textgreek{ἐστιν, οὗ μέρος οὐθέν} · A point is that which has no part.\footnote{
	Be careful: M. Capella \cite[p. 150]{Capella "De Geometria"} offers a different translation: «Punctum verò est, cuius pars nihil est: quæ si duo fuerint, linea interiacente iunguntur» («A point is that, a part of which is nothing: if there be two [points], they are joined by an interconnecting line»). A similar interpretation is reflected in many other authors/translators in the following centuries, see e.g. O. Finé \cite[p. 1]{Finaei "In sex priores libros Geometricorum elementorum Euclidis Megaresis demonstrationes"}: «Punctum est, cuius pars nulla». Capella's transcription makes the Euclidean statement harder than it already is, as T.L. Heath \cite[p. 155]{Euclid "The Thirteen Books of the Elements I Introduction and Books I-II"} stresses: it is one thing to say that \emph{a part} of a point \emph{is nothing}, and it is another to say that a point has \emph{no part}.
	} \\	
\indent — \textsc{Euclid} \cite[\textgreek{Ὅροι, α´, Στοιχείων α´}, Book I, p. 2]{Euclidis "Elementa I"}\footnote{
	For a historical reconstruction of the first seven definitions of Book I of Euclid's \textit{Elements}, see L. Russo \cite{Russo "The Definitions of Fundamental Geometric Entities Contained in Book I of Euclid's Elements"} \cite[chap. 2.2]{Russo "The Forgotten Revolution: How Science Was Born in 300 BC and Why it Had to Be Reborn"}.
	}
or \cite[pp. 436-437]{Various authors "Greek Mathematical Works I: from Thales to Euclid"}

\vspace{2mm}

Il punto è quello, che non hà parte alcuna, \emph{cioè, che non occupa spatio alcuno}.\footnote{
	Compare with these previous translations: N. Tartaglia \cite[foglio iiii]{Tartaglia "Euclide Megarense"}: «Il Ponto è quello che non ha parte»; F. Commandino \cite[p. 1 verso]{Commandino "De gli Elementi d'Euclide libri quindici"}: «Il punto, è quello che non ha parte, ò vero che non ha grandezza alcuna»; or with this, which is subsequent: F. Gio. Ricci \cite[p. 1]{Ricci Gio. "De gli Elementi di Euclide"}: «Punto, è una cosa, che nella quantità continua hà positione, ma non ha parti».
		} \\		
\indent — \textsc{V. Giordano} \cite[p. 5]{Giordano "Euclide restituto"}

\endgroup

\vspace{2mm}

When a point is defined, its definition opens up to some ambiguities, if by \emph{definition} we mean a proposition that identifies and clearly describes, or explains, a concept from other terms having a supposedly known meaning, to wit, from concepts that are presumed to be primitive. The fact is that the point is a primitive concept, not derivable from a previous concept. 

Elementary geometry (definitions of point, straight line, surface, plane surface, etc.) has a \emph{foundational character}, anyway, and is tied in with spatial primitive ideas, as it deals with topics such as “nature”, “adequacy” and “validity” of geometric principles, cf. \cite[§ 2.1.3]{Acerbi "Il silenzio delle sirene. La matematica greca antica"}. We cannot go up the river of mathematical thought \emph{in infinitum}, in search of something that precedes the spatio-geometric primitiveness, if indeed it is a primitiveness; at most, we can replace one primitiveness with another. It is our source, the river-head, as it were.

This is to say that Euclid's definition \cite[p. 2]{Euclidis "Elementa I"}, \textgreek{\textit{Σημεῖόν ἐστιν, οὗ μέρος οὐθέν}},\footnote{
	See M. Simon's comment, \textit{Euclid und die sechs planimetrischen Bücher}, Teubner, Leipzig, 1901, p. 25, in \cite[pp. 157-158]{Euclid "The Thirteen Books of the Elements I Introduction and Books I-II"}: «The point is the limit of localisation; if this is more and more energetically continued, it leads to the limit-notion “point”, better “position” [\,\dots]. Content of \emph{space} vanishes, relative \emph{position} remains. “Point” then [\,\dots] is the extremest limit of that which we can still think of (not observe) as a \emph{spatial} presentation, and if we go further than that, not only does extension cease but even relative \emph{place}», and in this sense [\,\dots] a point has no part.
	}
is a non-definition; and the same is true for Heron's (or Diophantus') affirmation,\footnote{
	«A point is [\,\dots] an extremity without extension (\textgreek{ἢ πέρας ἀδιάστατον})», also quoted in epigraph of Section \ref{subsection "Hyperbolic Equation of a Vibrating String: d'Alembert's Formula for the 1-Dimensional Wave Phenomenon"}.
	} 
which is equally, and wonderfully, ambiguous.\footnote{
	Also noteworthy is the pre-Euclidean definitions of point in the Stagiriticus corpus: a point is a «geometric fiction» (\textgreek{δόγμα γεωμετρικόν}); \emph{dogma} (\textgreek{δόγμα}), from the vb. \textgreek{δοκέω}, “think”, “suppose”, “imagine”, is perceptively translated by Heath \cite[p. 156]{Euclid "The Thirteen Books of the Elements I Introduction and Books I-II"} with \emph{fiction}. Another valid translation is: «geometric supposition», see e.g. U. Cassina \cite[p. 113]{Cassina "Critica dei principi della matematica e questioni di logica}.
	} 
The repercussion is that a point—thus (not) defined—is a \emph{marvelous germination of problems and paradoxes} of various kinds,\footnote{
	Depending on how the continuum concept is understood, is quite simple to pull paradoxes out of the hat in which the point, even if it is intended as a nothing, \emph{equals the rest of space}. To take an example, Galileo, in controversy with B. Cavalieri, affirms \cite[p. 29]{Galilei "Discorsi e dimostrazioni matematiche"}: «thus, it appears that the circumference of an immense circle may be called equal to a single point (par dunque che la circonferenza di un cerchio immenso possa chiamarsi eguale à un sol punto)».
		}\textsuperscript{,}\footnote{
	Let us take a break, just trying to have a laugh. Worthy of being remembered is the amusing description given by E.A. Abbott in his novella \cite[p. 91]{Abbott "Flatland: A Romance of Many Dimensions"} of the Point, which lives in «the realm of Pointland, the Abyss of No Dimensions». Sphere (one of the characters) says: «[A] Point is a Being like ourselves, but confined to the non-dimensional Gulf. He is himself his own World, his own Universe; of any other than himself he can form no conception; he knows not Length, nor Breadth, nor Height, for he has had no experience of them; he has no cognizance even of the number Two; nor has he a thought of Plurality; for he is himself his One and All, being really Nothing». 
		} 
see Sections \ref{subsection "Ex. 7.b. Backtracking to the Source of Pieri's Raw Materials of Point and Motion"}, \ref{subsection "Null Space No Infinite Time: the Point as a Capstone, and the Poincaré–Perelman Paradox"}, and \ref{section "Continuity and Discreteness—Differential Equations and Numerical Computing"}. 

\subsection{Peano's, Enriques', and Pieri's Non-defined Source Geometry}
\label{subsection "Peano's, Enriques', and Pieri's Non-defined Source Geometry"}

\begingroup
\footnotesize
Which of the geometric entities can be defined, and which ones should be assumed without definition? And among the properties, experimentally true, of these entities, which ones must be assumed without proof, and which ones can be deduced as a consequence? [\,\dots] Starting from \emph{non-defined concepts of point and line segment}, we can define the infinite line, the plane and its parts, as well as the parts of space. It is also possible to recognize, among the propositions, those (axioms) that express the simplest properties of the entities considered, and those (theorems) that can be deduced from other simpler ones \cite[pp. 3-4, e.a.]{Peano "I principii di geometria logicamente esposti"}.\endnote{
	Original It. version: «Quali fra gli enti geometrici si possono definire, e quali occorre assumere senza definizione? E fra le proprietà, sperimentalmente vere, di questi enti, quali bisogna assumere senza dimostrazione, e quali si possono dedurre in conseguenza? [\,\dots] [P]artendo dai concetti non definiti di punto e retta limitata, si possono definire la retta illimitata, il piano e le sue parti, come pure le parti dello spazio. Riesce pure possibile riconoscere, fra le proposizioni, quelle (assiomi) che esprimono le più semplici proprietà degli enti considerati, e quelle (teoremi) che si possono dedurre da altre più semplici». 
	} \\
\indent Given an order to ideas of a science, not all of them can be defined. The first idea, which has no precedent, cannot be defined; we cannot define the = sign, which appears in any definition. An idea is said to be \emph{primitive}, with regard to a given order, if, in this order of ideas, one does not know how to define it. Therefore being a primitive idea, is not an absolute character, but only relative [to the group of ideas which are supposed to be known]. The question “is the idea of \emph{point} itself primitive, or can it be defined?”, makes no precise sense, if we do not fix in advance what ideas are supposed to be known [\,\dots]. Pieri \cite{Pieri "Della Geometria elementare come sistema ipotetico deduttivo. Monografia del punto e del moto"} [cf. Section \ref{subsection "Ex. 7.b. Backtracking to the Source of Pieri's Raw Materials of Point and Motion"}] came to express all the ideas of Geometry [\,\dots] via just two primitive ideas: “point, and distance between two points” \cite[p. 9]{Peano "Le definizioni in matematica"}.\endnote{
	Original It. version: «Dato un ordine alle idee d'una scienza, non tutte si possono definire. Non si può definire la prima idea, che non ha precedenti; non si può definire il segno =, che figura in ogni definizione. Si dice che una idea è \emph{primitiva}, relativamente a un dato ordine, se, in quest'ordine delle idee, essa non si sa definire. Perciò l'essere una idea primitiva, non è un carattere assoluto, ma solo relativo [al gruppo di idee che si suppongono note]. La questione “l'idea di \emph{punto} è essa stessa primitiva, o si può definire?”, non ha senso preciso, se noi non prefissiamo quali idee si suppongono note [\,\dots]. Il Pieri pervenne a[d] esprimere tutte le idee di Geometria [\,\dots] mediante due sole idee primitive: “punto, e distanza fra due punti” \cite{Pieri "Della Geometria elementare come sistema ipotetico deduttivo. Monografia del punto e del moto"}».
	} \\
\indent — \textsc{G. Peano}

\endgroup

\vspace{2mm}

Besides Peano, F. Enriques is also crystalline about the non-definiteness and primitiveness in mathematics. He writes \cite[pp. 2-3, e.m.]{Enriques "Conferenze di Geometria. Fondamenti di una Geometria iperspaziale"}:

\vspace{2mm}

\begingroup
\footnotesize
Geometry has as its object of study the relations inherent to the concept of space as it springs from our mind [\textit{quale esso scaturisce dalla nostra mente}] from the order of external sensitivity, that is, as it is presented to us by \emph{intuition}. The concept of space includes the notions of many geometric entities, such as \emph{points}, \emph{lines}, \emph{surfaces} [\,\dots]; some of these entities can be defined by establishing their relations with other [entities] already given [\,\dots], but some entities must be supposed to be given [\,\dots] as \emph{fundamental}, since they cannot be reduced to others without falling into a \emph{vicious circle}. There is \emph{arbitrariness} in the choice of the fundamental entities of space; relations intercede between them, some of which (theorems) are proved by logical deduction from other [relations] already known or supposedly known: but even here it is not possible to sail against the current [of thought] indefinitely [\textit{non si può risalire indefinitamente}], and some geometric relations between the fundamental entities must be given [\,\dots] as \emph{postulates}. Postulates are derived from intuition.

[\,\dots] In its principle and in its development, Geometry—in the above sense—is a subjective science [\textit{la Geometria\,\dots è scienza soggettiva}]. In its principle, because the fundamental entities and the postulates that refer to them reflect the concept of intuitive space as it is in our mind; in the development, because definitions and demonstrations with which other entities are added to the fundamental ones and theorems to postulates, are only logical explanations [\textit{definizioni e dimostrazioni\,\dots sono soltanto spiegazioni logiche}].\endnote{
	I have removed here the notion of a priori, as it can mislead Enriques' thought, in order to homogenise the two texts, \cite{Enriques "Conferenze di Geometria. Fondamenti di una Geometria iperspaziale"} and \cite{Enriques "Lezioni di Geometria proiettiva"}.
	}

\endgroup

\vspace{2mm}

Enriques reiterates his standpoint in a text \cite[pp. 1-3, e.m.]{Enriques "Lezioni di Geometria proiettiva"} a few years later:

\vspace{2mm}

\begingroup
\footnotesize
Some elements must be introduced \emph{as first or fundamental elements} of Geometry, without definition, since one could not give a (logical) definition of all without falling into a \emph{vicious circle}. The choice of fundamental elements of Geometry is not \emph{a priori}; one chooses the simplest elements with regard to \emph{psychological intuition}, that is, those [elements] of which the notion is formed in our mind as the content of the concept of space: such are e.g. the \emph{point}, the \emph{straight line} and the \emph{plane} [\,\dots]. Whatever the fundamental geometric elements may be, they are chosen in an arbitrary way and in a superabundant number, [for which] \emph{any other geometric entity at a later stage introduced will have to be logically defined through the fundamental elements} [\,\dots]. It is not possible to demonstrate all the properties that are assumed as postulates without falling into a vicious circle. It is thereby necessary to place some postulates in the beginning of geometry;\footnote{
	We could say: \textgreek{Ἐν ἀρχῇ ἐστίν ὁ ὅρος}, but the \textgreek{ὅρος} (definition), when it contains primitive concepts, is a pseudo-definition, a non-definition.
	} 
these are chosen from the properties that have more \emph{intuitive evidence}, but their choice is not \emph{a priori} determined.\endnote{
	Original It. version: «[A]lcuni elementi debbono essere introdotti \emph{come elementi primi o fondamentali} della Geometria, senza definizione, giacché non si potrebbe dare una definizione (logica) di tutti senza cadere in un circolo vizioso. La scelta degli elementi fondamentali della Geometria non è \emph{a priori} determinata; si scelgono come tali gli elementi più semplici rispetto alla \emph{intuizione psicologica}, cioè quelli di cui la nozione si trova formata nella nostra mente come contenuto del concetto di spazio: tali sono p. e. il punto, la retta e il piano [\,\dots]. Comunque però si sieno scelti gli elementi geometrici fondamentali in modo arbitrario ed in numero sovrabbondante, \emph{ogni altro ente geometrico successivamente introdotto dovrà essere definito logicamente mediante gli elementi fondamentali} [\,\dots]. [N]on è possibile dimostrare tutte le proprietà che si assumono come postulati senza cadere in un circolo vizioso. È dunque necessario porre in principio della Geometria alcuni postulati; questi si scelgono fra le proprietà che hanno maggiore \emph{evidenza intuitiva}, ma la loro scelta non è \emph{a priori} determinata».
	}

\endgroup

\vspace{2mm}

Something similar asserts M. Pieri \cite[pp. 170-171]{Pieri "Della Geometria elementare come sistema ipotetico deduttivo. Monografia del punto e del moto"}: 

\vspace{2mm}

\begingroup
\footnotesize
If by definition we mean a pure and simple imposition of names on things already known or acquired by the system, the primitive ideas will be the \emph{not defined} concepts [\,\dots]: thereupon we will say that the primitive concepts are not defined “otherwise than by postulates”. The latter indeed attribute to those certain properties sufficient to qualify them, for the deductive purposes we want to achieve. And to avoid any misunderstanding we will use the term “definition in the strict sense”, or “\emph{nominal} definition”, when we want to exclude the “real” definition, or [definition] “of what”.\endnote{
	Original It. version: «Se per definizione s'intenda una pura e semplice imposizione di nomi a cose già note od acquisite al sistema, le idee primitive saranno i concetti \emph{non definiti} [\,\dots]: perciò diremo che i concetti primitivi non sian definiti “altrimenti che per postulati”. Questi ultimi invero attribuiscono a quelli talune proprietà sufficienti a qualificarli, in ordine ai fini deduttivi che si voglion raggiungere. E per cansare ogni equivoco si userà il termine “definizione in senso stretto”, o “definizione \emph{nominale}”, quando si voglia escluder la definizione “reale”, o “di cosa”».
	}

\endgroup

\vspace{2mm}

Cf. Pieri in Section \ref{subsection "Ex. 7.b. Backtracking to the Source of Pieri's Raw Materials of Point and Motion"}. Peano also expresses himself \cite[p. 10]{Peano "Le definizioni in matematica"} in a similar manner: a primitive idea is determined, in its «fundamental properties», by «primitive propositions», wherefore by axioms or postulates, which are devoid of proof, but represent the first link of the deductive chain; all «primitive propositions serve in a certain way as definitions of primitive ideas».

As a conclusion, Peano's (but see Scholium \ref{scholium "More precise information on Peano"}), Enriques' and Pieri's observations leave no way of escape in this regard: something non-definite must be there (concepts as point and line), so that there is a beginning of geometric, or physico-geometric, reasoning. In C. Segre \cite{Segre "Studio sulle quadriche in uno spazio lineare ad un numero qualunque di dimensioni"} this mental make-up is already clear-cut; he leaves the nature of the point «indeterminate», when the point is not imagined as a «geometric entity», nor as an «analytic entity», but as an «entity in itself».\footnote{
	In his \cite[pp. 4 and 16]{Segre "Studio sulle quadriche in uno spazio lineare ad un numero qualunque di dimensioni"} words: «[T]he element or \textit{point} [of] space is not considered as a geometric entity of ordinary space (nor, which then is the same, as an analytic entity consisting of values of $n$ variable quantities), but rather as an entity in itself [\,\dots]. Let us consider any linear space with $n - 1$ dimensions. We will call each of its elements a \textit{point}, whatever its nature (which for us has absolutely no importance)».
	}

\begin{scholium}[More precise information on Peano]
\label{scholium "More precise information on Peano"}
Take heed: Peano is not always on the same line of thought. Not unlike Enriques and Pieri, when he enters into the issue of the primitive ideas, in geometry, he also speaks of arbitrariness in the choice and interpretation of such primitiveness, see \cite[§§ 1-2, esp. pp. 24-25]{Peano "I principii di geometria logicamente esposti"}; but elsewhere he feels the need to anchor the geometric practice into the physical reality, so that the arbitrary ingredient somehow disappears, or is candidly negated. In \cite[p. 141]{Peano "Sui fondamenti della geometria"} he observes that «[o]f course, anyone is allowed to preface those [primitive] hypotheses he wants, and develop the logical consequences contained in those hypotheses. But to deserve the name of Geometry in this work, those hypotheses or postulates must express the result of the simplest and most elementary observations of physical figures»; in \cite[p. 365]{Peano "Super theorema de Cantor-Bernstein"} he goes so far as to write: «si postulatos es hypothetico, et non respondentes ad factu reale», then «[p]roba de coexistentia de systema de postulatos pote es utile», and this is because we «[n]am nos non crea pustulatos ad arbitrio». \scholiumsymbol
\end{scholium}

\subsection{What is a Point, or a Line? Hilbert vs. Frege}
\label{subsection "What is a Point, or a Line? Hilbert vs. Frege"}

The correspondence between G. Frege and D. Hilbert is very instructive, in this regard. It is worth quoting some excerpts from their letters.  
 
Frege to Hilbert \cite[IV/3 · XV/3]{Frege "Wissenschaftlicher Briefwechsel"} = \cite[27 December 1899, pp. 34-35]{Frege "Scientific Correspondence"}: 

\vspace{2mm}

\begingroup
\footnotesize
I was interested to get to know your \emph{Festschrift} on the foundations of geometry \cite{Hilbert "Grundlagen der Geometrie"} [\,\dots] here the meanings of the words “point”, “line”, “between” are not given, but are assumed to be known in advance. At least it seems so. But it is also left unclear what you call a point. One first thinks of points in the sense of Euclidean geometry, a thought reinforced by the proposition that the axioms express fundamental facts of our intuition. But afterwards you think of a pair of numbers as a point. I have my doubts about the proposition that a precise and complete description of relations is given by the axioms of geometry and that the concept “between” is defined by axioms. Here the axioms are made to carry a burden that belongs to definitions. To me this seems to obliterate the dividing line between definitions and axioms in a dubious manner.

\endgroup

\vspace{2mm}

Hilbert to Frege \cite[IV/4 · XV/4]{Frege "Wissenschaftlicher Briefwechsel"} = \cite[29 December 1899, pp. 39-40, e.a.]{Frege "Scientific Correspondence"}: 

\vspace{2mm}

\begingroup
\footnotesize
If one is looking for other definitions of a “point”, e.g., through paraphrase in terms of extensionless, etc., then I must indeed oppose such attempts in the most decisive way; one is looking for something one can never find because there is nothing there; and everything gets lost and becomes vague and tangled and degenerates into a game of hide-and-seek [\,\dots]. You write: “I call axioms propositions \textellipsis From the truth of the axioms it follows that they do not contradict one another” [\,\dots]. I have been saying the exact reverse: if the arbitrarily given axioms do not contradict one another with all their consequences, then they are true and the things defined by the axioms exist. This is for me the criterion of truth and existence [\,\dots]. On the other hand, to try to give a definition of a point in three lines is to my mind an impossibility, for only the whole structure of axioms yields a complete definition. [Therefore: \emph{the definition of the concept of point is not complete till the structure of the system of axioms is complete}].\footnote{
	Draft or Excerpt by Hilbert \cite{Frege "Wissenschaftlicher Briefwechsel"} = \cite[p. 42]{Frege "Scientific Correspondence"}.
	}
That is right: \emph{every axiom contributes something to the definition}, and hence \emph{every new axiom changes the concept}. \emph{A “point” in Euclidean, non-Euclidean, Archimedean, and non-Archimedean geometry is something different in each case} [\,\dots]. [E]very theory is only a scaffolding or schema of concepts together with their necessary relations to one another, and that the basic elements can be thought of in any way one likes. If in speaking of my points I think of some system of things, e.g. the system: love, law, chimneysweep \textellipsis and then assume all my axioms as relations between these things, then my propositions, e.g. Pythagoras' theorem, are also valid for these things. 

\endgroup

\vspace{2mm}

Frege to Hilbert \cite[IV/5 · XV/5]{Frege "Wissenschaftlicher Briefwechsel"} = \cite[6 January 1900, p. 45]{Frege "Scientific Correspondence"}: 

\vspace{2mm}

\begingroup
\footnotesize
Given your definitions, I do not know how to decide the question whether my pocket watch is a point. The very first axiom deals with two points; thus if I wanted to know whether it held for my watch, I should first have to know of some other object that it was a point. But even if I knew this, e.g., of my penholder, I still could not decide whether my watch and my penholder determined a line, because I would not know what a line was. 

\endgroup

\vspace{2mm}

The mix-up between \emph{definition} and \emph{axiom}, or \emph{postulate}, which makes Frege doubtful—as a logician, he is unable to accept it—is a distinctive feature of primitive concepts. Hilbert, as a mathematician, seems instead to be fully in tune with Peano and Enriques.

\subsection{Margo. Grothendieckian topos-point}
\label{subsection "Margo. Grothendieckian topos-point"}

In Grothendieck topoi \cite{Artin Grothendieck Verdier "Theorie des Topos et Cohomologie Etale des Schemas I"}, which inhere in the categories of sheaves of sets on a topological space, the point 
\[
	x \in [\textnormal{\textgreek{τόπος}}]_{\mathcal{X} \to \mathcal{Y}} 
\]	
is but a (geometric) morphism 
\begin{equation}
	\varphi_x \colon \mathcal{X}_{\textnormal{\textgreek{τ}}} \to \mathcal{Y}_{\textnormal{\textgreek{τ}}},
\end{equation}
from a topos $\mathcal{X}_{\textnormal{\textgreek{τ}}}$ to topos $\mathcal{Y}_{\textnormal{\textgreek{τ}}}$, viz. a transformation of $\mathcal{X}_{\textnormal{\textgreek{τ}}}$ into $\mathcal{Y}_{\textnormal{\textgreek{τ}}}$ keeping unchanged the structurality of the two topoi; or, much more simply, a topos-point is a morphism from a topos of sets to $\mathcal{X}_{\textnormal{\textgreek{τ}}}$, viz. 
\begin{equation}
	\varphi_x \xrightarrow{\mathit{Sets}} \mathcal{X}_{\textnormal{\textgreek{τ}}}. 
\end{equation}
Another formalism to define a topos-point is this: let 
\[
	S_\mathcal{X} = (x_1, \mathellipsis, x_n)
\] 	
be a set of points and a \emph{discrete category} of a Grothendieck topos $\mathcal{X}_{\textnormal{\textgreek{τ}}}$ indexed by a set $\Sigma$, with a function 
\[
	\varphi_x \colon \Sigma \to S_\mathcal{X}. 
\]
The set $S_\mathcal{X}$ can be determined with a morphism 
\begin{equation}
	\tilde{\varphi}_x \colon [\Sigma, \mathit{Set}] \to \mathcal{X}_{\textnormal{\textgreek{τ}}}.
\end{equation}

The Grothendieckian point, having an initially geometric nature, thus becomes a functor 
\begin{equation}
	F_\textsc{gr}(\mathcal{X}_{\textnormal{\textgreek{τ}}}) \to F_\textsc{gr}\bigl(\{\mathcal{X}_{\langle*\rangle}\}\bigr).	
\end{equation}
Which allows us to make other sub-definitions:

· topos-point as a $\mathit{Set} \to [\mathit{Cat}, \mathit{Set}]$, given a topos $[\mathit{Cat}, \mathit{Set}]$;
  
· topos-point as a 
\[
	\mathcal{X}_\topological\text{-continuous functor } F_\textsc{gr}(\mathit{Cat}) \to \mathit{Set}, 
\]
for a topos $Sh(\mathit{Cat}, \mathcal{X}_\topological)$ on a site $(\mathit{Cat}, \mathcal{X}_\topological)$, having a category of sheaves of sets, with a Grothendieck topology $\mathcal{X}_\topological$.

· topos-point as a homomorphisms 
\[
	\mathcal{X}^\mathrm{loc}_{\textnormal{\textgreek{τ}}} \to \{0, 1\} 
\]
(we have to do here with a local space), given a topos $Sh(\mathcal{X}^\mathrm{loc}_{\textnormal{\textgreek{τ}}})$.

Conclusion: the concept of \emph{primitiveness} is \emph{not} in the least \emph{bypassed} by Grothendieck; his is just an \emph{operational proposal} to the concept of point.

\section{Space-numeral Primitiveness: On the Continuum of Real Numbers}
\label{section "Space-numeral Primitiveness: On the Continuum of Real Numbers"}

\subsection{Dedekind's Continuity Axiom}
\label{subsection "Dedekind's Continuity Axiom"}

\subsubsection{Euclidean Discretum, and Discrete Numerability}
\label{subsubsection "Euclidean Discretum, and Discrete Numerability"}

\begingroup
\footnotesize
If space has at all a real existence [\textit{Hat überhaupt der Raum eine reale Existenz}], it is not necessary to be continuous, for it; a huge amount of its properties would remain the same even if it were discontinuous [\textit{unstetig}]. And if we knew for sure that space was discontinuous, there would be nothing to block us [\,\dots] from filling its gaps in our thought, so as to make it continuous \cite[pp. 18-19]{Dedekind "Stetigkeit und irrationale Zahlen"}. \\
\indent I can imagine that the whole [Euclidean] space and every line in it [\,\dots] is entirely discontinuous [\,\dots]; the concept of space [\textit{Raumbegriff}] [including Euclidean space] is completely independent and separable from the idea of continuity [\textit{Vorstellung der Stetigkeit}] \cite[letter to R. Lipschitz, 27 July 1876, pp. 478-479]{Dedekind "Aus Briefen an R. Lipschitz"}. \\
\indent It is easy to see [that Euclidean space] is made up of points everywhere discontinuous [\textit{Punkten\,\dots überall unstetig}]; but in spite of the discontinuity [\,\dots], all constructions in Euclid's \textit{Elements} can be just as accurately carried out as in perfectly continuous space; the discontinuity [\textit{Unstetigkeit}] of this space would therefore not be noticed in Euclidean science, would not be felt, not a bit \cite[pp. xii-xiii]{Dedekind "Was sind und was sollen die Zahlen?"}. \\
\indent — \textsc{R. Dedekind}

\endgroup

\vspace{2mm}

\enumerationisinitium
\item Mathematics is an act of creation precisely for what we have just seen (Sections \ref{subsection "Peano's, Enriques', and Pieri's Non-defined Source Geometry"} and \ref{subsection "What is a Point, or a Line? Hilbert vs. Frege"}): it has a non-definite (scilicet: ambiguous) socle. Let us now take another example of a non-defined concept. But first a premise is needed.

\item Physicists often speak of the continuity of space, os space-time (macroscopic level), and try to imagine a granular scene (microscopic level), so as to theorize quantum gravity. See e.g. A. Schild \cite[p. 29, e.m.]{Schild "Discrete Space-time and Integral Lorentz Transformations"}, figuring the (macroscopic) \emph{space-time} as a (microscopic) \emph{discrete cubic lattice} in compliance with some Lorentz transformations (cf. Section \ref{subsubsection "Lorentz Group plus Transformations"}):
 
\vspace{2mm}

\begingroup
\footnotesize
The idea of introducing discreteness into space and time has occasionally been considered. \emph{It seems likely that a physical theory based on a discrete space-time background will be free of the infinities which trouble contemporary quantum mechanics}. The objection which is usually raised against such discrete schemes is that they are not invariant under the Lorentz group. The purpose of this investigation is to show that there is a simple model of discrete space-time which, although not invariant under all Lorentz transformations, does admit a surprisingly large number of Lorentz transformations.

\endgroup

\vspace{2mm}

Space, or space-time, that physicists dream up is not something other than what they can envisage by means of continuous/discrete mathematical tools (see Chapters \ref{chapter "Outro—Parva Mathematica: Libera Divagazione 3/8"}, \ref{chapter "Outro—Parva Mathematica: Libera Divagazione 5/8"}, \ref{chapter "Outro—Parva Mathematica: Libera Divagazione 6/8"}); however, for those who know mathematics, space—starting from Euclidean space, which is the basic space for every theory—can very well do without the concept of continuity.
\item 
\label{item "Continuity and discreteness"}
The concept of continuum in Euclid is no necessary, nor is any axiom of continuity explicitly defined, as R. Dedekind says (see epigraph), but it is only tacitly presupposed, even though the physical tradition believes, uncritically, that Euclidean space is the continuous structure pre-eminently; \emph{some results from Euclid}, by contrast, \emph{are achieved within discontinuous space-image}, and remain true even in \emph{non-continuous domains of algebraic numbers}.\footnote{
	See e.g. Euclid \cite[Proposition \textgreek{α´, Στοιχείων α´}, Book I, p. 10]{Euclidis "Elementa I"}: «Upon a given finite straight line to construct an equilateral triangle (\textgreek{Ἐπὶ τῆς δοθείσης εὐθείας πεπερασμένης τρίγωνον ἰσόπλευρον συστήσασθαι})». No continuous fabric of space is defined here; yet the proposition is implicitly ally with a continuous spatiality, at least intuitively.
	}
 
The concept of space is actually independent and separable from the notion of continuity. And no wonder why: continuity is part of the calculation; but calculation cannot be performed only with continuous values, since  we should be struggling with (a computability of) infinite quantities, or an infinite number of states. Which provides a perspicuous clue: \emph{real numbers are surreptitiously treated as a finite set of discrete values} (cf. footnote \ref{footnote "Thom on continuum"}, p. \pageref{footnote "Thom on continuum"}). There is more: \emph{every measurement}, to be such, \emph{relies upon on discrete numerical outcomes}. The moral is that our use of infinity is by its \emph{finite approximations}.

Not even the continuum à la Dedekind flees from this status, as we will see below. \emph{Dedekind's continuum}—in which «every real number is [considered to be] the representative of a certain \emph{partition} of rational numbers», to use the words of A. Capelli \cite[p. 209]{Capelli "Saggio sulla introduzione dei numeri irrazionali col metodo delle classi contigue"}—is also a \emph{false continuum}: it too has, say, holes, because not every distance, on the line, can be reworded via real numbers: the points of Dedekind's line-space are proof of this.
\enumerationisfinis

\begin{margo}
I find the same persuasion in P. Zellini's book \cite{Zellini "Discreto e continuo. Storia di un errore"}, published in the summer of 2022, according to which the continuum is an approximation of the discretum; rather than starting with an initially assigned continuum, which, at a later stage, is divided into a discrete number of parts, it is more correct to start with the idea that the continuum is defined and constructed via discrete operations. «What we actually know is only the discrete-datum, even if the behavior of discrete series of numbers must be analyzed, for the most part, by means of the continuum notion», writes Zellini, in perfect harmony with Thom, see the previously mentioned footnote \ref{footnote "Thom on continuum"} on p. \pageref{footnote "Thom on continuum"}. Ergo the discrete-datum precedes the continuous one, so the latter cannot be understood without the former, cf. footnote \ref{footnote "Grothendieck: discrete and continuous"} on p. \pageref{footnote "Grothendieck: discrete and continuous"} (Grothendieck's note), and Section \ref{subsubsection "The Ultimate Representation"}. \margosymbol
\end{margo}

\subsubsection{Schnitt and Bijection}
\label{subsubsection "Schnitt and Bijection"}

It is no accident that the \emph{Dedekindscher Schnitt} \cite[§ 4. \textit{Schöpfung der irrationalen Zahlen}]{Dedekind "Stetigkeit und irrationale Zahlen"}, as a construction of the \emph{real numbers} (indispensable for the scientific depth of arithmetic), thanks to the practice of \emph{cutting}, and the creation of irrational numbers, in the will to reach a compatibility between \emph{continuity of the numerical straight line} and \emph{punctual discreteness of each number}, is echoed from Euclid's \textit{Elements} of geometry, 5th Definition, Book V (see Margo \ref{margo "Euclidean theory of magnitudes: continuous quantities"}).

Here is the thing: given two sets, $A_1$ and $A_2$, of a discontinuous domain (\textit{unstetige Gebiet}) of rational numbers $\mathbb{Q}$, we can get a continuous $\mathbb{R}$-domain (\textit{stetige Gebiet}), as a completion of the $\mathbb{Q}$-domain. Dedekind \cite[pp. 19, 21]{Dedekind "Stetigkeit und irrationale Zahlen"} writes:\endnote{
	An unabridged En. transl. of \cite{Dedekind "Stetigkeit und irrationale Zahlen"} is in \cite[pp. 1-27]{Dedekind "Essays on the theory of numbers. I. Continuity and Irrational Numbers II. The Nature and Meaning of Numbers"}.
	} 

\vspace{2mm}

\begingroup
\footnotesize
Whenever there is a cut [\textit{Schnitt}] $(A_1, A_2)$ that is not produced by any rational number, we \emph{create} [\textit{erschaffen}] a new, an \emph{irrational number} $\alpha$, which we consider to be completely [defined] by this cut $(A_1, A_2)$; we shall say that the number $\alpha$ corresponds to this cut, or that it produces this cut. Hence, from now on, to every definite cut there corresponds one and only one definite rational or irrational number.

\endgroup

\vspace{2mm}

Dedekind's continuity is a completeness concept-axiom, flanked by the continuity of the (straight) line, under which \emph{all points of a line are placed into one-to-one correspondence (bijection) with all real numbers}, or elements of the set $\mathbb{R}$. Unlike the rational number line ($\mathbb{Q}$-line), the real number line ($\mathbb{R}$-line) is complete. In \cite[p. 18]{Dedekind "Stetigkeit und irrationale Zahlen"} Dedekind makes it clear that: 

\vspace{2mm}

\begingroup
\footnotesize
I find the essence of continuity [\,\dots] in the following principle: “If all points of the straight line fall into two classes in such a manner that every point of the first class lies to the left of every point of the second class, then there is one and only one point, which divides all points into two classes, this division of the straight line into two portions”. 

\endgroup

\vspace{2mm}

In T.L. Heath \cite[pp. 124-126]{Euclid "The Thirteen Books of the Elements I Introduction and Books III-IX"} \cite[pp. 326-327]{Heath "A History of Greek Mathematics I: From Thales to Euclid"}, but see also L. Russo \cite[sec. 2.5]{Russo "The Forgotten Revolution: How Science Was Born in 300 BC and Why it Had to Be Reborn"}, the Dedekindian theory of irrationals, for real numbers, is utterly coincident with the Euclidean theory of magnitudes (continuous quantities). It does not signify that the continuum of Dedekind is the same as the Euclidean continuum—when the congruence of segments is being examined, for instance: the two concepts of continuum are \emph{not identifiable}, although they have strong similarities; and this is all the more true in the non-Archimedean mathematics (we will make a brief mention of it in Section \ref{section "Non-Archimedean System"}), see e.g. V. Benci and P. Freguglia \cite[§ 3.2. \textit{Alla ricerca del continuo euclideo}]{Benci Freguglia "Alcune osservazioni sulla matematica non archimedea"}.

\begin{margo}[Euclidean theory of magnitudes: continuous quantities]
\label{margo "Euclidean theory of magnitudes: continuous quantities"}
In \cite[\textgreek{Ὅροι, ε´, Στοιχείων ε´}, Book V, p. 2]{Euclidis "Elementa II"} we read: «Magnitudes are said to be in the same ratio, the first to the second, and the third to the fourth, when equimultiples of the first and the third exceed, are  equal to, or are less than, equimultiples of the second and the fourth, being taken in corresponding order, respectively (\textgreek{Ἐν τῷ αὐτῷ λόγῳ μεγέθη λέγεται εἶναι πρῶτον πρὸς δεύτερον καὶ τρίτον πρὸς τέταρτον, ὅταν τὰ τοῦ πρώτου καί τρίτου ἰσάκις πολλαπλάσια τῶν τοῦ δευτέρου καὶ τετάρτου ἰσάκις πολλαπλασίων καθ᾿ ὁποιονοῦν πολλαπλασιασμὸν ἑκάτερον ἑκατέρου ἢ ἅμα ὑπερέχῃ ἢ ἅμα ἴσα ᾖ ἢ ἅμα ἐλλείπῇ ληφθέντα κατάλληλα})». Compare with \cite[p. 114]{Euclid "The Thirteen Books of the Elements I Introduction and Books III-IX"}. Namely: ratios between magnitudes—consider, for example, straight line segments—$\alpha, \beta$ and $\gamma, \delta$ are defined to be equal, $\alpha \colon \beta = \gamma \colon \delta$, if 
\vspace{-1.5mm}
\begin{subequations}
\begin{align}
	& k_2\alpha > k_1\beta \text{ and } k_2\gamma > k_1\delta, \\
	& k_2\alpha = k_1\beta \text{ and } k_2\gamma = k_1\delta, \\
	& k_2\alpha < k_1\beta \text{ and } k_2\gamma < k_1\delta, 
\end{align}
\end{subequations}
\vspace{-2mm}
for each pair of natural numbers $k_1$ and $k_2$. \margosymbol	
\end{margo}

\subsection{Cantorian Hierarchy: Transfinite Arithmetic, and Cardinality of the Continuum}
\label{subsection "Cantorian Hierarchy: Transfinite Arithmetic, and Cardinality of the Continuum"}

\begingroup
\footnotesize
I arrived at a well-ordered sequence of cardinalities [\textit{Mächtigkeiten}] or transfinite cardinal numbers [\textit{transfiniten Kardinalzahlen}], which I call “alephs”: $\aleph_0, \aleph_1, \aleph_2, \mathellipsis, \aleph_{\omega_0}, \mathellipsis$ [\,\dots]. The big question was whether there are other cardinalities, besides the alephs; [\,\dots] I have a evidence that there are no others, so that e.g. the arithmetic linear continuum [\textit{arithmetischen Linear-kontinuum}] (the totality of all real numbers) has a certain aleph [$\aleph_1$] as [its] cardinal number. \\
\indent — \textsc{G. Cantor} \cite[letter to R. Dedekind, 28 July 1899, p. 443]{Cantor "Letter from Cantor to Dedekind 28 July 1899"}

\endgroup

\vspace{2mm}

Closely linked to the aforementioned matters is the mathematical issue of the continuum pertaining to the infinite sets. The referent is G. Cantor, who was the first to establish an ambitious \emph{hierarchy for order types of infinite sets}. Let us schematically retrace the Cantorian thought, taking for granted the knowledge of all transfinite arithmetic; a summary, for a quick revision, is available in A. Kanamori \cite{Kanamori "The Mathematical Development of Set Theory from Cantor to Cohen"} \cite{Kanamori "The Higher Infinite: Large Cardinals in Set Theory from Their Beginnings"}.

\subsubsection{Transfinite Ordinals}

To follow the \emph{transfinite ordinal numbers}, see Cantor's papers \cite{Cantor "Ueber unendliche lineare Punktmannichfaltigkeiten"} \cite{Cantor "Beitrage zur Begrundung der transfiniten Mengenlehre (Erster Artikel)"} \cite{Cantor "Beitrage zur Begrundung der transfiniten Mengenlehre (Zweiter Arfikel)}. 
\enumerationisinitium
\item The transfinite ordinal number $\omega_\mathbb{N}$ 
\subenumerationisinitium
\item symbolizes a collection of natural numbers having some order, since they are labeled for their \emph{position},
\item more precisely, is the set of all natural numbers in the \emph{ordinal aspect}, for which is the order type of $\mathbb{N}_0 = \mathbb{N} \cup \{0\}$, cf. J. von Neumann \cite{Neumann "Zur Einfuhrung der transfiniten Zahlen"},
\begin{align}
	\omega_\mathbb{N}
	\begin{cases}
	0 = \varnothing, \\
	1 = 0 \cup \{0\} = \{0\} = \{\varnothing\}, \\
	2 = 1 \cup \{1\} = \{0, 1\} = \{\varnothing, \{\varnothing\}\}, \\
	3 = 2 \cup \{2\} = \{0, 1, 2\} = \{\varnothing, \{\varnothing\}, \{\varnothing, \{\varnothing\}\}\}, \\
	\mathellipsis \\
	k = k - 1 \cup \{k - 1\} = \{0, \mathellipsis, k - 1\} = \{\varnothing, \mathellipsis, \{\varnothing\}\cdots\}, \enspace k \in \mathbb{N},
	\end{cases}
\end{align}
\item is the first transfinite ordinal number, that is, the \emph{lowest/smallest countable infinity of $\mathbb{N}_0$}.
\subenumerationisfinis
\item The second transfinite ordinal number is $\omega_\mathbb{N} + 1$, the third one is $\omega_\mathbb{N} + 2$, the fourth one is $\omega_\mathbb{N} + 3$, and so on. 
\item With the multiplication and exponentiation, a pyramid of $\mathbb{N}$-infinities is constructible, generating new transfinite ordinal $\omega_\mathbb{N}$-numbers, for $k, n \in \mathbb{N}$: 
\begin{equation}
	\omega_\mathbb{N}, \omega_\mathbb{N} + 1, \omega_\mathbb{N} + 2, \omega_\mathbb{N} + 3, \mathellipsis, \omega_\mathbb{N} + k,
\end{equation}		

then
\begin{equation}
	\omega_\mathbb{N} \cdot 2, \omega_\mathbb{N} \cdot 3, \mathellipsis, \omega_\mathbb{N} \cdot k,
\end{equation}

then
\begin{subequations}
\begin{align}	
	& \omega_\mathbb{N} \cdot 2 + 1, \omega_\mathbb{N} \cdot 2 + 2, \mathellipsis, \omega_\mathbb{N} \cdot 2 + 3, \omega_\mathbb{N} \cdot 2 + k, \\ 
	& \omega_\mathbb{N} \cdot 3 + 1, \omega_\mathbb{N} \cdot 3 + 2, \omega_\mathbb{N} \cdot 3 + 3, \mathellipsis, \omega_\mathbb{N} \cdot 3 + k, \\
	& \mathellipsis \notag \\
	& \omega_\mathbb{N} \cdot k + 1, \omega_\mathbb{N} \cdot k + 2, \omega_\mathbb{N} \cdot k + 3, \mathellipsis, \omega_\mathbb{N} \cdot k + n, 
\end{align}	
\end{subequations}

then
\begin{equation}
	\omega_\mathbb{N}^2, \omega_\mathbb{N}^3, \mathellipsis, \omega_\mathbb{N}^k,
\end{equation}	

then
\begin{equation}
	\omega_\mathbb{N}^2 + 1, \omega_\mathbb{N}^2 + 2, \omega_\mathbb{N}^2 + 3, \mathellipsis, \omega_\mathbb{N}^2 + k, 	
\end{equation}

then
\begin{equation}
	\omega_\mathbb{N}^2 \cdot 2, \omega_\mathbb{N}^2 \cdot 3, \mathellipsis, \omega_\mathbb{N}^2 \cdot k,
\end{equation}

then
\begin{subequations}
\begin{align}	
	& \omega_\mathbb{N}^2 \cdot 2 + 1, \omega_\mathbb{N}^2 \cdot 2 + 2, \omega_\mathbb{N}^2 \cdot 2 + 3, \mathellipsis, \omega_\mathbb{N}^2 \cdot 2 + k, \\ 
	& \omega_\mathbb{N}^2 \cdot 3 + 1, \omega_\mathbb{N}^2 \cdot 3 + 2, \omega_\mathbb{N}^2 \cdot 3 + 3, \mathellipsis, \omega_\mathbb{N}^2 \cdot 3 + k, \\
	& \mathellipsis \notag \\
	& \omega_\mathbb{N}^2 \cdot k + 1, \omega_\mathbb{N}^2 \cdot k + 2, \omega_\mathbb{N}^2 \cdot k + 3, \mathellipsis, \omega_\mathbb{N}^2 \cdot k + n,
\end{align}
\end{subequations}

then
\begin{equation}
	\omega_\mathbb{N}^3 + 1, \omega_\mathbb{N}^3 + 2, \omega_\mathbb{N}^3 + 3, \mathellipsis, \omega_\mathbb{N}^3 + k,	
\end{equation}

then
\begin{equation}
	\omega_\mathbb{N}^3 \cdot 2, \omega_\mathbb{N}^3 \cdot 3, \mathellipsis, \omega_\mathbb{N}^3 \cdot k,
\end{equation}

then
\begin{subequations}
\begin{align}
	& \omega_\mathbb{N}^3 \cdot 2 + 1, \omega_\mathbb{N}^3 \cdot 2 + 2, \omega_\mathbb{N}^3 \cdot 2 + 3, \mathellipsis, \omega_\mathbb{N}^3 \cdot 2 + k, \\ 
	& \omega_\mathbb{N}^3 \cdot 3 + 1, \omega_\mathbb{N}^3 \cdot 3 + 2, \omega_\mathbb{N}^3 \cdot 3 + 3, \mathellipsis, \omega_\mathbb{N}^3 \cdot 3 + k, \\
	& \mathellipsis \notag \\
	& \omega_\mathbb{N}^3 \cdot k + 1, \omega_\mathbb{N}^3 \cdot k + 2, \omega_\mathbb{N}^3 \cdot k + 3, \mathellipsis, \omega_\mathbb{N}^3 \cdot k + n,
\end{align}
\end{subequations}

then
\begin{equation}
	\omega_\mathbb{N}^r + 1, \omega_\mathbb{N}^r + 2, \omega_\mathbb{N}^r + 3, \mathellipsis, \omega_\mathbb{N}^r + k,
\end{equation}

then
\begin{equation}
	\omega_\mathbb{N}^r \cdot 2, \omega_\mathbb{N}^r \cdot 3, \mathellipsis, \omega_\mathbb{N}^r \cdot k,
\end{equation}

then
\begin{subequations}
\begin{align}
	& \omega_\mathbb{N}^r \cdot 2 + 1, \omega_\mathbb{N}^r \cdot 2 + 2, \omega_\mathbb{N}^r \cdot 2 + 3, \mathellipsis, \omega_\mathbb{N}^r \cdot 2 + k, \\ 
	& \omega_\mathbb{N}^r \cdot 3 + 1, \omega_\mathbb{N}^r \cdot 3 + 2, \omega_\mathbb{N}^r \cdot 3 + 3, \mathellipsis, \omega_\mathbb{N}^r \cdot 3 + k, \\
	& \mathellipsis \notag
\end{align}
\end{subequations}
\enumerationisfinis

\subsubsection{Burali-Forti Paradox}

\begingroup
\footnotesize
The main purpose of this Note is to demonstrate that  there are \emph{transfinite numbers} \cite{Cantor "Beitrage zur Begrundung der transfiniten Mengenlehre (Erster Artikel)"} (or \emph{order types}) $a$, $b$ such that $a$ is not equal to, nor less and greater than $b$ [\,\dots]. If [in two different propositions] we put $\Omega$ instead of $a$ and [\,\dots] $\Omega + 1$  instead of $a$, we have [\,\dots] $\Omega + 1 > \Omega$; $\Omega + 1 < \Omega$ [\,\dots] [and thereby] we have arrived at an absurd [\,\dots]. It is therefore impossible to order the order types in general, and in particular also the ordinal numbers, in ascending sense, that is to say: the order types cannot provide a \emph{sample} class for the ordered classes. \\
\indent — \textsc{C. Burali-Forti} \cite[pp. 154, 164]{Burali-Forti "Una questione sui numeri transfiniti"} 

\endgroup

\vspace{2mm}

The most exciting thing, for us, is the origination of a typology of contradiction, known as \emph{Burali-Forti paradox} \cite{Burali-Forti "Una questione sui numeri transfiniti"}, \emph{with infinities of different magnitudo}, which have their gemmation from the fact that the set of all ordinal numbers $\omega_\mathbb{N}$ is, in itself, an ordinal number, for which e.g. $\omega_\mathbb{N} + 1$ is both greater than $\omega_\mathbb{N}$, because it has one more unit, and less than $\omega_\mathbb{N}$, because $\omega_\mathbb{N} + 1$ is an element contained in $\omega_\mathbb{N}$, being this latter the set of all ordinal numbers.\endnote{ 
	A primal intuition of the paradoxicality of infinite sets in comparison is in G. Galilei \cite[p. 32]{Galilei "Discorsi e dimostrazioni matematiche"}: «Quì nasce subito il dubbio, che mi pare insolubile; \emph{\&} è che sendo noi sicuri trovarsi linee una maggior dell'altra, tutta volta che amendue contenghino punti infiniti, bisogna confessare trovarsi nel medesimo genere una cosa maggior dell'infinito; perch[é] la infinità de i punti della linea maggiore eccederà l'infinità de i punti della minore. Ora questo darsi un infinito maggior dell'infinito mi par concetto da non poter' esser capito in verun modo». Transl. into En. reads like this: «Here the doubt arises immediately, which seems to me insoluble; since we are sure of finding lines one [of which] greater than another, both containing infinite points, we must confess that, within the same class, there is something greater than infinity; because the infinity of points in the major line will exceed the infinity of points in the minor [line]. Now this assigning to an infinity a [quantity] greater than infinity seems to me a concept that cannot be understood in any way».
	
	\setlength\parindent{8pt}
	Another name to make, on this issue, is B. Bolzano \cite[§ 20, pp. 28-29]{Bolzano "Paradoxien des Unendlichen"} = \cite[§ 20, p. 96]{B. Bolzano "Paradoxes of the Infinite"}: «Let us choose any two abstract quantities, say 5 and 12. Then the set of all quantities between zero and 5 [\,\dots] is clearly infinite, as also the set of all quantities less than 12», so the closed interval $[0, 5]$, as a set of real numbers, has “as many” points as does the closed interval $[0, 12]$. But «the latter set [is] greater than the former, seeing that the former constitutes a mere part of the latter». Now, let $x$ be an arbitrary quantity between 0 and 5, and write the equation $5y = 12x$, then $y$ is a quantity between 0 and 12; conversely, if $y$ lies between 0 and 12, then $x$ lies between 0 and 5. This implies that to any value of $x$ corresponds a single value of $y$, and conversely. Thus: to any quantity $x$, in the set between 0 and 5, there corresponds a quantity $y$, in the set between 0 and 12, such that «no constituent of either set remains uncoupled», and «none appears in two or more of the couples».
	}

Cantor copes with the paradox by distinguishing between \emph{absolutely infinite}, or \emph{inconsistent  multiplicity}, impregnated with the Burali-Forti's absurdity, and \emph{consistent multiplicity} (set, in the good acceptation).\footnote{
	In \cite[p. 443]{Cantor "Letter from Cantor to Dedekind 28 July 1899"} he writes: «A multiplicity can be such that the assumption that \emph{all} its elements are “together” leads to a contradiction, so that it is impossible to  conceive the multiplicity as a unity, as “a finished thing” [\textit{fertiges Ding}]. I call such multiplicities \emph{absolutely infinite} [\textit{absolut unendliche}] or \emph{inconsistent multiplicities} [\textit{inkonsistente Vielheiten}] [\,\dots]. If, on the other hand, the totality of the elements of a multiplicity can be thought of as “being together” without contradiction, so that it is possible to combine them [the elements] into “\emph{one} thing”, I call it a \emph{consistent multiplicity} [\textit{konsistente Vielheit}] or a “set” [\textit{Menge}]».
	} 
Cantor's solution is logically correct but mathematically dissatisfying, since it shifts what he calls «absolutely infinite» or «inconsistent multiplicity» onto a meta-mathematical plane; this multiplicity is a \emph{primitive type idea}, albeit necessary as an image on which to attack the rest of the transfinite theory.

\subsubsection{Transfinite Cardinals}

\enumerationisinitium
\item Cardinality is for the \emph{quantity}, or \emph{size}, of the elements of a set. 
\item The cardinality of the set of all natural numbers $\mathbb{N}_0$, or the cardinality of $\omega_\mathbb{N}$ (set of all natural numbers that can be well-ordered in a ordinal construction), is designated by $\aleph_0$ (aleph null). 
\subenumerationisinitium
\item $\aleph_0$ is the first infinite cardinal, i.e. the \emph{lowest/smallest transfinite cardinal number}. 
\item Then there are $\aleph_1, \aleph_2, \aleph_3, \mathellipsis$ (aleph one, two, three, etc.). 
\subenumerationisfinis
\item Since \cite{Cantor "Ein Beitrag zur Mannigfaltigkeitslehre"} two sets are said to have the same \emph{power} (\textit{Mächtigkeit}) if there is a \emph{bijection}, or a \emph{one-to-one correspondence}, so that two sets can be associated with each other in a complete way, element by element, we say that two natural sets, nay, two infinite sequences, like, e.g. 
\begin{gather*}
	\text{a } \textcolor{pumpkin}{\texttt{pumpkin}} \text{ set } \textcolor{pumpkin}{\texttt{P}} = \{0, 1, 2, 3, \mathellipsis\} \\
	\text{and} \\
	\text{a } \textcolor{cyan-blue}{\texttt{cyan-blue}} \text{ set } \textcolor{cyan-blue}{\texttt{C}} = \{1, 3, 5, 7, \mathellipsis\},
\end{gather*}
with different ordinalities, $\{\textcolor{pumpkin}{\texttt{p}}\}$-ordinality and $\{\textcolor{cyan-blue}{\texttt{c}}\}$-ordinality, respectively, have the same $\aleph$-cardinality, and they appear \emph{equally numerous}, when they respect this bijective correspondence:
\[
\begin{tikzcd}
	\textcolor{pumpkin}{\texttt{0}} \arrow[d] \arrow[r, no head, dotted] & \textcolor{pumpkin}{\texttt{1}} \arrow[d] \arrow[r, no head, dotted] & \textcolor{pumpkin}{\texttt{2}} \arrow[d] \arrow[r, no head, dotted] & \textcolor{pumpkin}{\texttt{3}} \arrow[d] \arrow[r, no head, dotted] & \textcolor{pumpkin}{\texttt{4}} \arrow[d] \arrow[r, no head, dotted] & \textcolor{pumpkin}{\texttt{5}} \arrow[d] \arrow[r, no head, dotted] & \textcolor{pumpkin}{\texttt{p}} \\
	\textcolor{cyan-blue}{\texttt{1}} \arrow[r, no head, dotted] & \textcolor{cyan-blue}{\texttt{3}} \arrow[r, no head, dotted] & \textcolor{cyan-blue}{\texttt{5}} \arrow[r, no head, dotted] & \textcolor{cyan-blue}{\texttt{7}} \arrow[r, no head, dotted] & \textcolor{cyan-blue}{\texttt{9}} \arrow[r, no head, dotted] & \textcolor{cyan-blue}{\texttt{11}} \arrow[r, no head, dotted] & \textcolor{cyan-blue}{\texttt{c}}
\end{tikzcd}   
\]
where 
\[
	\textcolor{pumpkin}{\texttt{P}} = \{\textcolor{pumpkin}{\texttt{p}}\}, \textcolor{pumpkin}{\texttt{p}} \in \mathbb{N}_0, 
\]
and 
\[
	\textcolor{cyan-blue}{\texttt{C}} = \{\textcolor{cyan-blue}{\texttt{c}}\}, \textcolor{cyan-blue}{\texttt{c}} \in \mathbb{N}_+ = \mathbb{N} \backslash \{0\}.
\]

But it does not end here. We can have fun with different base numeral systems, for setting up another countably infinite; an example:\footnote{
	Cantorian infinities, whether they are ordinal or cardinal, are nice kayos against the Anselm-like arguments: \textit{ens quo nihil majus cogitari potest}. It is always possible to think of a greater numerical entity. (To the contrary, the “absolute infinity”, which constitutes Cantor's religious part, is definitely Anselmian).
	}
\begin{align*}
\textcolor{pumpkin}{\texttt{0}} & \xrightarrow{\qquad 1:1 \qquad} \textcolor{cyan-blue}{\texttt{1}} \\
\textcolor{pumpkin}{\texttt{1}} & \xrightarrow{\qquad 1:1 \qquad} \textcolor{cyan-blue}{\texttt{11}}_\mathnormal{2} \\
\textcolor{pumpkin}{\texttt{10}}_\mathnormal{2} & \xrightarrow{\qquad 1:1 \qquad} \textcolor{cyan-blue}{\texttt{101}}_\mathnormal{2} \\
\textcolor{pumpkin}{\texttt{11}}_\mathnormal{2} & \xrightarrow{\qquad 1:1 \qquad} \textcolor{cyan-blue}{\texttt{111}}_\mathnormal{2} \\ 
\textcolor{pumpkin}{\texttt{100}}_\mathnormal{2} & \xrightarrow{\qquad 1:1 \qquad} \textcolor{cyan-blue}{\texttt{1001}}_\mathnormal{2} \\\
\textcolor{pumpkin}{\texttt{101}}_\mathnormal{2} & \xrightarrow{\qquad 1:1 \qquad} \textcolor{cyan-blue}{\texttt{1011}}_\mathnormal{2} \\
\textcolor{pumpkin}{\texttt{p}}_\mathnormal{2} & \xrightarrow{\qquad 1:1 \qquad} \textcolor{cyan-blue}{\texttt{c}}_\mathnormal{2}
\end{align*}
\enumerationisfinis

\subsubsection{Properties of Real Numbers: Archimedeanity, Completeness, Cauchy Criterion, and Cantor–Dedekind Axiom}
\label{subsubsection "Properties of Real Numbers: Archimedeanity, Completeness, Cauchy Criterion, and Cantor–Dedekind Axiom"}

One-to-one correspondence becomes focal with the treatment of real numbers. We have seen (Section \ref{subsubsection "Schnitt and Bijection"}) that, in Dedekind, the \emph{arithmetization of the continuum} can be considered \emph{independently of geometric intuition}, but it does not go against its \emph{geometric explanation}; it presupposes such a \emph{spatial construal}, asserting that every real number corresponds to a point on the line. So Dedekind constructs the continuity of real numbers on the \emph{analogy} of the continuity of the line.\footnote{
	For example H. Poincaré \cite[pp. 26-27]{Poincare "Le continu mathematique"} notes: «The continuum thus conceived [by mathematicians] is no more than a collection of individuals arranged in a certain order, in infinite number [\textit{Le continu ainsi conçu n'est plus qu'une collection d'individus rangés dans un certain ordre, en nombre infini}], it is true, but external to each other. This is not the ordinary conception, where one supposes that there is, among the elements of the continuum, a kind of intimate link which makes it a whole, in which the point does not exist before the line, but the line [before] the point. [In an analytic sense], the continuum is unity in multiplicity, only multiplicity subsists, unity has disappeared. Analysts are nonetheless right to define their continuous as they do [\,\dots]. But this is enough to warn us that the genuine mathematical continuum is something quite different from that of physicists».
	}

Conversely, Cantor \cite{Cantor "Ueber die Ausdehnung eines Satzes aus der Theorie der trigonometrischen Reihen"} \cite{Cantor "Ueber eine Eigenschaft des Inbegriffs aller reellen algebraischen Zahlen"}—since he believes that a \emph{Cauchy sequence} \cite[chap. VI]{Cauchy "Cours d'Analyse I.er Partie: Analyse Algebrique"} (see below) is, by definition, a representation of real numbers—considers the continuity of the line as a property reflecting the property of continuity of real numbers. Dedekind's and Cantor's positions are then \emph{equivalent but independent}. Let us examine  them in a little more detail, but not before having enunciated the Cauchy criterion.

\begin{definitio}[Cauchy criterion]
\label{definitio "Cauchy criterion"}
A sequence of real numbers 
\[
	\{x_n\} = x_1, \mathellipsis, x_n, 
\]
or of points $\{x_n\} \in \mathbb{R}$, is a \emph{Cauchy sequence} iff, for an arbitrary $\varepsilon > 0$, there is a fixed number $N_\varepsilon \in \mathbb{N}$ on $\varepsilon$ such that the inequality
\begin{equation}
	|x_n - x_j| < \varepsilon
\end{equation}
holds for all $n \geqslant N_\varepsilon$ and $j \geqslant N_\varepsilon$. \definitiosymbol
\end{definitio}

Cauchy's theorem on this criterion for sequences is as follows.
\begin{theorema}[Cauchy criterion]
\label{theorema "Cauchy criterion"}
A sequence of $\{x_n\} \in \mathbb{R}$ is convergent iff it is a Cauchy sequence. 
\end{theorema}

\begin{proof}
Denote by $L/ \in \mathbb{R}$ a limit point, or cluster point (in a rough way, a sequence of partial sums is said to be convergent if the partial sums are clustered together). We say that $\{x_n\}$ converges to $L/$ as $n \to \infty$. Since $\varepsilon > 0$, there is a number $N_\varepsilon$ such that 
\begin{equation}
	|x_n - L/| < \frac{\varepsilon}{2}, 
\end{equation}
for any $n \geqslant N_\varepsilon$. So if $n \geqslant N_\varepsilon$ and $j \geqslant N_\varepsilon$, then 
\begin{equation}
	|x_n - x_j| \leqslant |x_n - L/| + |x_j - L/| < \left(\frac{\varepsilon}{2} + \frac{\varepsilon}{2}\right) = \frac{\varepsilon + \varepsilon}{2} = \frac{2\varepsilon}{2} = \frac{2}{2}\varepsilon = \varepsilon.
\end{equation}
\end{proof}

Let us return to Dedekind's and Cantor's assiomatic positions.

\emph{Dedekind's axiom} \cite{Dedekind "Stetigkeit und irrationale Zahlen"}, aka \emph{completeness axiom of the real numbers}, states that \emph{any non-empty set of $\mathbb{R}$ with an upper bound has a least upper bound}. Dedekind's definition of the continuum \emph{implies} the Archimedean property, and the Dedekind reals are an ordered Archimedean field. 

The \emph{Archimedean property}, or \emph{axiom of Archimedes} \cite[\textgreek{Λαμβάνω δὲ ταῦτα, ε´}, p. 10, 18-28]{Archimedes "De sphaera et cylindro"} = \cite[p. 4]{Archimedes "On The Sphere and Cylinder I-II"} = \cite[p. 36]{Archimedes "The Works of Archimedes Vol. I: On the Sphere and the Cylinder"},\footnote{
	G. Veronese \cite[p. 83]{Veronese "Fondamenti di geometria a piu dimensioni e a piu specie di unita rettilinee} \cite[p. 198]{Veronese "La Geometria non-Archimedea"} reminds us that O. Stolz \cite{Stolz "Zur Geometrie der Alten insbesondere uber ein Axiom des Archimedes"} gave to this axiom the name of Archimedes because it is assumption № 5 in \textgreek{\textit{Περὶ σφαίρας καὶ κυλίνδρου}} (\textit{On the Sphere and the Cylinder}) \cite[\textgreek{Λαμβάνω δὲ ταῦτα, ε´}, p. 10]{Archimedes "De sphaera et cylindro"} of Archimedes. The axiom in question is also mentioned in \textgreek{\textit{Τετραγωνισμός παραβολής}} (\textit{Quadrature of the parabola}) \cite[proem, p. 296]{Archimedes "Quadratura parabolae"} and in \textgreek{\textit{Περί ἑλίκων}} (\textit{On Spirals}) \cite[proem, p. 14]{Archimedes "De lineis spiralibus"}.
	} 
is often read (and interpreted) in parallel with the \emph{axiom of Eudoxus}, which is present in Euclid's \textit{Elements} \cite[\textgreek{Ὅροι, δ´, Στοιχείων ε´}, Book V, p. 2]{Euclidis "Elementa II"}.\footnote{
	For a debate on the distinction, or similarity, between axiom of Eudoxus (Euclid) and axiom of Archimedes, see E.J. Dijksterhuis \cite[pp. 145-149, and especially 431-433]{Dijksterhuis "Archimedes"}, and F. Acerbi \cite[pp. 347-348]{Acerbi "Introduzione" a Euclide Tutte le opere"}.
	}
Let us see what we got.

\begin{theorema}[Eudoxus–Archimedes']
\label{theorema "axiom of Eudoxus–Archimedes"}
~\enumerationisinitium
\item Axiom of Eudoxus: \textnormal{«Magnitudes are said to have a ratio to one another which, being multiplied, are capable of exceeding one another (\textgreek{Λόγον ἔχειν πρὸς ἄλληλα μεγέθη λέγεται, ἃ δύναται πολλαπλασιαζόμενα ἀλλήλων ὑπερέχειν})»}.
\item Axiom of Archimedes: \textnormal{«Of unequal lines, unequal surfaces, and unequal solids, the greater exceeds the less by such a magnitude that is capable, when added to itself, of exceeding any  magnitude among those which are in a ratio [i.e. comparable] with [it and with] one another (\textgreek{Ἔτι δὲ τῶν ἀνίσων γραμμῶν καὶ τῶν ἀνίσων ἐπιφανειῶν καὶ τῶν ἀνίσων στερεῶν τὸ μεῖζον τοῦ ἐλάσσονος ὑπερέχειν τοιούτῳ, ὃ συντιθέμενον αὐτὸ ἑαυτῷ δυνατόν ἐστιν ὑπερέχειν παντὸς τοῦ προτεθέντος τῶν πρὸς ἄλληλα λεγομένων})»}. 
\item Translating everything into modern notation, we have:
\subenumerationisinitium
\item letting $x, y \in \mathbb{R}$ be two positive real numbers, there exists a natural number $n \in \mathbb{N}$ such that $nx > y$, or $y < nx$,
\item for any $\kappa > 0$, $\kappa \in \mathbb{R}$, there exists some $n \in \mathbb{N}$, with $n > \kappa$, and $\frac{1}{n} < \kappa$, 
\item the natural set $\mathbb{N}$ has no upper bound in $\mathbb{R}$, and no real number is an upper bound for $\mathbb{N}$, that is, the set $\mathbb{N}$ is bounded below but not above in $\mathbb{R}$. 
\subenumerationisfinis
\enumerationisfinis
\end{theorema}

\begin{proof}
Assume that $\mathbb{N}$ has an upper bound, and $\kappa$ is the least upper bound; seeing that $\kappa - 1$ is not an upper bound, there has to be some $n \in \mathbb{N}$, with $n > \kappa - 1$. Since $n + 1 > \kappa$, then $\kappa$ cannot be the least upper bound, and $\frac{y}{x}$ is not an upper bound, for which there is a number $n \in \mathbb{N}$, with $n > \frac{y}{x}$, and $nx > y$.
\end{proof}

\emph{Cantor's axiom} \cite{Cantor "Ueber die Ausdehnung eines Satzes aus der Theorie der trigonometrischen Reihen"} states that, \emph{given a sequence of nested intervals—i.e. a collection of sets of $\mathbb{R}$-numbers under which any interval is nested in each other—there exists at least one point common belonging to all interval of the sequence}. Unlike what happens for the Dedekind reals, Cantor's axiom does \emph{not} ensue from the Archimedean Property \ref{theorema "axiom of Eudoxus–Archimedes"}, so there is no implication between the Cantorian axiom and the axiom of Archimedes; and in fact Cantor's condition may be satisfied by a \emph{non-Archimedean field} (see Section \ref{section "Non-Archimedean System"}).

Which marks the independence between Dedekind's condition and Cantor's axiom. Their equivalence—on this subject, the \emph{Cantor–Dedekind axiom} is being talked about—is guaranteed by the fact that, in both formulations, \emph{a unique real number can be assigned to each and every point on a line, and each real number can be identified uniquely by a point on the line}.

\vspace{2mm}

\textit{\textbf{Compendium}}

\vspace{2mm}

We are now able to list some main properties of the real numbers $\mathbb{R}$:
\enumerationisinitium
\item Archimedean property,
\item Cauchy's criterion (test of convergent sequences of the reals),
\item Dedekind's axiom, or completeness axiom,
\item Archimedes–Dedekind axiom, to wit, \emph{completeness and archimedeanity},
\item Cantor–Dedekind axiom: the set of the reals can be put in \emph{one-to-one correspondence} with the points of a straight line—this correspondence is an \emph{order-isomorphism}, i.e. an isomorphism preserving the structures of ordered sets \emph{both} on $\mathbb{R}$ and the line too;
 \item general postulate in algebro-geometric analysis: the real number line, or $\mathbb{R}$-line, is the \emph{substratum of the continuum}.
\enumerationisfinis

\subsubsection{Cantor's Continuum Problem}

\begingroup
\footnotesize
Cantor's continuum problem is simply the question: \emph{How many points are there on a straight line in Euclidean space?} In other terms, the question is: How many different sets of integers do there exist? \\
\indent — \textsc{K. Gödel} \cite[p. 515, e.a.]{Godel "What is Cantor's Continuum Problem?"}

\endgroup

\vspace{2mm}

Alongside the axiom of correspondence the investigation, in Cantor, is imposed on the nature of the real set, which debouches into a fork: each infinite set of real numbers is countable or has the power of the continuum; and, as it is known, all this bring about the following Cantorian statements.
\enumerationisinitium
\item The \emph{continuum} $\mathsf{C}_\mathbb{R}$, or the \emph{set of real numbers} $\mathbb{R}$, \emph{cannot be numbered}, it is \emph{greater than any countable set}. \emph{Real numbers are not countable}, inasmuch as a set is said countable if there is a one-to-one correspondence between the set under consideration and the set of natural numbers. 
\item The \emph{continuum} $\mathsf{C}_\mathbb{R}$, which also contains a countably infinite subset, \emph{has a higher power than countable}.
\item The \emph{continuum} $\mathsf{C}_\mathbb{R}$ has the \emph{second power},
\begin{equation}
	\cardinality(\mathsf{C}_\mathbb{R}) = 2^{\aleph_0} > \cardinality(\mathbb{N}_0) = \aleph_0,
\end{equation}
where $\aleph_0$ is the cardinality of the set of all natural numbers, or of $\omega_\mathbb{N}$ (for the ordinal aspect); or even the cardinality of the (natural) integers $\cardinality(\mathbb{Z}_*)$. This is the Cantorian hypothesis \cite[p. 257]{Cantor "Ein Beitrag zur Mannigfaltigkeitslehre"}, but see also \cite[p. 574]{Cantor "Ueber unendliche lineare Punktmannichfaltigkeiten"}.
\item Between the cardinality of the continuum/the set of real numbers $\mathbb{R}$ and the cardinality of the set of natural numbers $\mathbb{N}_0$ there are \emph{no sets of intermediate cardinality},
\begin{equation}
	\nexists(S) \colon \cardinality(\mathbb{N}_0) = \aleph_0 < \cardinality(S) < \cardinality(\mathsf{C}_\mathbb{R}) = 2^{\aleph_0},
\end{equation}
where $S$ is some set.\footnote{
	\label{footnote "First Hilbert's mathematical problem: Cantor problem"}
	But here is how D. Hilbert \cite[p. 263, e.m.]{Hilbert "Mathematische Probleme"} = \cite[p. 70]{Hilbert "Sur les problemes futurs des Mathematiques"} describes the Cantorian hypothesis in the celebrated Parisian list of mathematical problems in 1900, see \cite[pp. 1-10]{Bartocci Betti Guerraggio Lucchetti (Eds.) "Mathematical Lives. Protagonists of the Twentieth Century From Hilbert to Wiles"}. It is the problem № 1 (Cantors Problem von der Mächtigkeit des Continuums): «Any system of infinitely many real numbers, i.e. any infinite set of numbers (or points), is either [equivalent to] the set of  natural integers $1, 2, 3, \mathellipsis$ or [to] the set of all real numbers and hence [to] the continuum, that is, to the points of a line; in the equivalence sense, there are [therefore] only two sets of numbers, the countable set and the continuum. From this assertion it would follow at once that the continuum has the next power [cardinal number] beyond the power [cardinal number] of the countable sets; the proof of this theorem would thus build a new \emph{bridge} [\textit{Brücke}] between the countable set and the continuum».
	}
\item Which translates into the \emph{continuum hypothesis},
\begin{align}
	& \cardinality(\mathsf{C}_\mathbb{R}) = 2^{\aleph_0} = \aleph_1, \\
	&
	\label{equation "Generalised continuum hypothesis"} 
	\cardinality(\mathsf{C}_\mathbb{R}) = 2^{\aleph_{\alpha_k}} = \aleph_{\alpha_k + 1},
\end{align}
where $\aleph_1$ is the the \emph{first uncountable cardinal number}, and also the cardinality of the set of all countable ordinal numbers, indicated by $(\omega_\mathbb{N})_1$. Intended as a number, $(\omega_\mathbb{N})_1$ is the \emph{lowest/smallest uncountable ordinal number}, so $(\omega_\mathbb{N})_1$ is the first ordinal of $\aleph_1$. The Eq. \eqref{equation "Generalised continuum hypothesis"}, for every ordinal number $\alpha_k$, is a generalised continuum hypothesis, see K. Gödel \cite{Godel "The Consistency of the Axiom of Choice and of the Generalized Continuum Hypothesis with the Axioms of Set Theory"}.
\item Some of the above rules may be imposed by beth numerical notation: 
\[ 
	\beth_0, \beth_1, \beth_2, \mathellipsis 
\]
(beth null, beth one, beth two, etc.), outlining a sequence of infinite cardinals. 
\subenumerationisinitium
\item Beth null is equal to aleph null: $\beth_0 = \aleph_0$. 
\item Beth one is equal to aleph one: 
\begin{equation}
	\Bigl\{\beth_1 = \aleph_1\Bigr\} \viz \Bigl\{\beth_1 = \cardinality(\mathsf{C}_\mathbb{R})\Bigr\},
\end{equation}
and this is the continuum hypothesis. 
\item Beth $\alpha_k$-number, where $\alpha_k$ is any ordinal number, is an auxiliary form of the generalised continuum hypothesis:  
\begin{equation}
	\beth_{\alpha_k + 1} = 2^{\beth_{\alpha_k}}.
\end{equation}	
\subenumerationisfinis 
\enumerationisfinis

\subsubsection{No Bridge between the Countable Set \& the Continuum}
 
\begingroup
\footnotesize
[T]he statement $2^{\aleph_0} = \aleph_\tau$, for $\tau$ in [a certain model] $\mathfrak{M}$, may not be capable of being expressed as a statement in Z[ermelo]–F[raenkel] or may have \emph{different interpretations in different countable models $\mathfrak{M}$ or $\mathfrak{N}$}. If $\tau$ is a particular natural number or $\omega_\mathbb{N}^2 + 1$, etc., then it can readily been expressed in Z[ermelo]–F[raenkel] and the proof sketched goes through. \\
\indent — \textsc{P.J. Cohen} \cite[p. 110, e.a.]{Cohen "The independence of the continuum hypothesis II"}

\endgroup

\vspace{2mm}
 
P.J. Cohen \cite{Cohen "The independence of the continuum hypothesis", Cohen "The independence of the continuum hypothesis II"}, availing himself of a technique called \emph{forcing}, showed that the continuum hypothesis is independent of the Zermelo–Fraenkel set theory \cite{Zermelo "Untersuchungen uber die Grundlagen der Mengenlehre. I"}\cite{Fraenkel "The notion "definite" and the independence of the axiom of choice"}, including the axiom of choice, for which the continuum hypothesis is not a theorem: the conundrum \emph{Is $\cardinality(\mathsf{C}_\mathbb{R}) = 2^{\aleph_0} = \aleph_1$?} appears to be \emph{undecidable}. There is no \emph{bridge} (as as wished by Hilbert, see footnote \ref{footnote "First Hilbert's mathematical problem: Cantor problem"} above, p. \pageref{footnote "First Hilbert's mathematical problem: Cantor problem"}) between 
\[
	\cardinality(\mathbb{N}_0) = \aleph_0, \text{ or } \cardinality(\mathbb{Z_*}) = \aleph_0, 
\]
on one side, and 
\[
	\cardinality(\mathsf{C}_\mathbb{R}) = 2^{\aleph_0}, 
\]
on the other. In fact, there is a \emph{set-gap}, a form of \emph{numeric discreteness}, that separates the countable set from the continuum. 

\subsubsection{The Ultimate Representation}
\label{subsubsection "The Ultimate Representation"}

It is very gripping to wonder which \emph{quantity} is better when the Cantorian system is adopted to give voice to the physical world; that is: which order types of infinite sets are more effective in the \emph{ultimate representation of natural phenomena}. The princely infinity, in mathematical physics, is the \emph{continuum of real numbers}; but one wonders whether, in more or less specific cases, an order of a \emph{discrete infinity} is better suited to the purpose, at least to enclose, in a more likely description, the core of certain laws of nature. 

\subsubsection{Back on Point(s), Line and Correspondence}
\label{subsubsection "Back on Point, Line and Correspondence"}

\begingroup
\footnotesize
A line is not composed of points as the forest is composed of trees, nor may a line be produced by putting together “all” the points in it. \\
\indent — \textsc{P.W. Bridgman} \cite[p. 229]{Bridgman "A Physicist's Second Reaction To Mengenlehre"}

\endgroup

\vspace{2mm}

Hilbert designates \cite[p. 170]{Hilbert "Uber das Unendliche"} the (Cantorian) set theory as a paradise: «Aus dem Paradies, das Cantor uns geschaffen, soll uns niemand vertreiben können». And in sooth it has its own enchantment; but it also has a dark side, ineluctably. B. de Finetti used to call the diabolical waste of that theory as \emph{insiemistificazione}, a playful crasis between “insiemi” (the It. word for “sets”) and “mystification”. 

But still and all, what interests us, is the concept of point, the definition of which, once again, remains eluded. There is, in this vein, an article by P.W. Bridgman \cite[pp. 227-229]{Bridgman "A Physicist's Second Reaction To Mengenlehre"}, which outlines the problem without frills, and gets down to the nitty-gritty. Let us read some of the highlights:

\vspace{2mm}

\begingroup
\footnotesize
It is said [in the spirit of the set theory] that there is a one-to-one correspondence between all the points of a line of unit length and all non-terminating decimals less than unity [\,\dots]. Given any terminating decimal, of no matter how many digits, then we can find by a perfectly definite geometrical procedure a unique corresponding point on the line. If one wants to know how I am sure the point corresponding to this construction exists, I believe the only answer is, “It exists by definition”. It would be silly to say that I know the point exists because I can reach it by actual construction with ruler and compass [\,\dots]; how shall we show that given any point on the line we can approximate as closely as we please to it by a variable point, itself determined by a terminating decimal of a continually increasing number of digits? The crux of the whole situation is contained in the expression “any point” [\,\dots]. If point is to mean anything, it must be identifiable, and this involves some operation or procedure for describing it. The simplest method perhaps is to specify its distance from one end of the line. This may be given in terms of terminating decimals or in terms of other things defined as numbers, such as the rationals. But there are other purely geometrical procedures possible. First there are Euclidean procedures expressible in terms of compass and rule; we can add to these other procedures involving the intersections of algebraic curves of any orders. All such procedures will obviously give us algebraic points. We can add other procedures, corresponding to integrations, and involving lengths of arcs or area [\,\dots]. But in any event, “point” has no meaning unless it is defined, and this involves the specification of some sort of procedure.  

\endgroup

\vspace{2mm}

And by that, Bridgman find cause for rejection of the Cantorian \textit{diagonal Verfahren} (the diagonal method) \cite{Cantor "Ueber eine elementare Frage der Mannigfaltigketislehre"} for proving the non-denumerability (by a one-to-one correspondence) of the points on a line, viz. the non-denumerability of the non-terminating decimals. He continues thusly:

\vspace{2mm}

\begingroup
\footnotesize
“All the points of a line” as a purely intuitional concept apart from the rules by which the points are determined, can have no operational meaning, and accordingly is to be held for mathematics an entirely meaningless concept.

A line is not composed of points as the forest is composed of trees, nor may a line be produced by putting together “all” the points in it. Points may be “determined” on a line, and the determination involves an operation according to a rule of some sort. “All the points of a line” means no more than “All the rules for determining points on a line” [\,\dots]. In other words, we have no more reason to describe the points on a line as non-denumerable than the non-terminating decimals.  

\endgroup

\vspace{2mm}

In this other passage \cite[pp. 98-99]{Bridgman "The Way Things Are"}, his stance is even more stringent: 

\vspace{2mm}

\begingroup
\footnotesize
[W]e talk about \emph{all} the points on a line between the origin and 1, for example [\,\dots] — a point is a curious thing and I do not believe that its nature is appreciated, even by many mathematicians. A line is not composed of points in any real sense. The above statement about all the points of the line between 0 and 1 is a paraphrase for “the entire line between 0 and 1”. We do not construct the line out of points, but, given the line, we may construct points on it. “All the points on the line” has the same sort of meaning that the “entire line” has [\,\dots]. We \emph{create} the points on a line just as we create the numbers, and we identify the points by the numerical values of the coordinates. The point \emph{is} the number, or, more generally, a point is an aggregate of three numbers [\,\dots]. And the point was not “there” before the numbers were given or determined. 

\endgroup

\section[Umbratile Elements: Non-measurable Spaces]{Umbratile\footnote{
	Being in the shade, shadowy.
	} Elements: Non-measurable Spaces}

\begingroup
\footnotesize
\emph{The problem of measure of \textnormal{[}sets\textnormal{]} of points on a straight line is impossible} [\,\dots]: the possibility of the problem of measure of [sets] of points of a straight line and that of well-ordering the continuum cannot coexist.\endnote{
	Original It. version: «\emph{\textnormal{[}I\textnormal{]}l problema della misura dei gruppi di punti di una retta è impossibile} [\,\dots]: la possibilità del problema della misura dei gruppi di punti di una retta e quella di bene ordinare il continuo non possono coesistere» (emphasis and bold modified).
	} \\
\indent — \textsc{G. Vitali} \cite[p. 5]{Vitali "Sul problema della misura dei gruppi di punti di una retta"}

\endgroup

\vspace{2mm}

The continuum of real numbers opens the doors to delightful oddnesses, which nestle in the shadowy zones of mathematical logic. One of the most debated theorems is that of S. Banach and A. Tarski \cite{Banach "Sur le probleme de la mesure"} \cite{Banach Tarski "Sur la decomposition des ensembles de points en parties respectivement congruentes"} directly related to Vitali sets.  

\subsection{Double Measure of a Volume: Banach–Tarski Paradox}

\begin{paradox}[Banach–Tarski]
\label{paradox "Banach–Tarski"}
Let $\overbar{\mathbb{B}}^3$ be a closed unit 3-ball of $\mathbb{R}^3$, that is to say, a volume $\volume(\overbar{\mathbb{B}}^3) > 0$ bounded by a 2-sphere $\mathbb{S}^2$ in Euclidean 3-space. It is stated that $\overbar{\mathbb{B}}^3$ can be decomposed, or splitted, into a finite number of disjoint sets (pieces) $S_{\mathbb{B}}$, 
\begin{equation}
	\overbar{\mathbb{B}}^3 = (S_\mathbb{B})_1 \sqcup \cdots \sqcup (S_\mathbb{B})_m,
\end{equation}
which appear to be mutually congruent, hence they can be reassembled through isometric motions of rotations and translations (under the group of isometries of Euclidean space), so as to form two 3-balls of the same radius, whose measure is therefore equal to that of the original 3-ball,
\begin{equation}
	\sum^{m\text{-sets}}_{\rotatedell = 1} \volume(\overbar{\mathbb{B}}^3)_\rotatedell = 2\volume(\overbar{\mathbb{B}}^3).
\end{equation}

This theorem (of dissection into finitely many closed bounded sets, reassembling and duplication of the measure of a volume) also works in higher dimensions, taking a hyperball $\overbar{\mathbb{B}}^n$ bounded by an $(n - 1)$-sphere in $\mathbb{R}^n$, for $n \geqslant 3$, but it stops working in the plane $\mathbb{R}^2$, and in the line $\mathbb{R}$.
\end{paradox}

What is the catch? It lies here, in the assemblage of these two previous results: 
\enumerationisinitium
\item each disjoint piece is non-measurable, thanks to G. Vitali \cite{Vitali "Sul problema della misura dei gruppi di punti di una retta"}, due to the fact that there are subsets of the real line $\mathbb{R}$ which are not measurable (Section \ref{subsection "Non-measurability of Vitali Sets"}), by virtue of the \emph{axiom of choice} \cite{Zermelo "Beweis dass jede Menge wohlgeordnet werden kann"};
\item in a (ordinary) 3-dimensional space, under a F. Hausdorff's \cite[pp. 401-402, 469-472]{Hausdorff "Grundzuge der Mengenlehre"} outcome, a 2-sphere can be decomposed, or splitted, into four disjoint sets (pieces), three of which are equal to each other and, at the same time, equal to their rearrangement. 
\enumerationisfinis

In view of the Vitali's non-measurability of certain real (sub)sets, which have \emph{no length}, the Theorem-Paradox \ref{paradox "Banach–Tarski"} is not a paradox at all, but a curious consequence of this fact brought to a further level: there is \emph{no volume} (since it is \emph{not well-defined}), so a volume may be doubled. That is how $\overbar{\mathbb{B}}^3$ is equally decomposable, or scissors-congruent, to the disjoint union of two (rotated and translated) copies of $\overbar{\mathbb{B}}^3$.

This discussion is only a glancing exploration of the Banach–Tarski paradox; for a deep study, see \cite{Tomkowicz Wagon "The Banach-Tarski Paradox"} \cite[chap. 17]{Drutu Kapovich "Geometric Group Theory"}.

\subsection{Non-measurability of Vitali Sets}
\label{subsection "Non-measurability of Vitali Sets"}

\begin{propositio}[Vitali]
A Vitali set \textnormal{\cite{Vitali "Sul problema della misura dei gruppi di punti di una retta"}} is a subset $\Vitali \subset [0, 1]$ of the real line $\mathbb{R}$ such that, for each real number $\epsilon$, there is properly one number $o \in \Vitali$ for which $o - \epsilon$ is a rational number $\frac{k}{z}$, where $k$ and $z \neq 0$ are integers. The claim is that $\Vitali \subset [0, 1]$ appears not measurable.	
\end{propositio}

\begin{proof}
Let 
\begin{equation}
	q_k = \mathbb{Q} \cap [-1, 1], \enspace k = 1, \mathellipsis, n, 
\end{equation}
be an enumeration of rational numbers in an interval $[-1, 1]$; we know that the set of $\mathbb{Q}$ is countable; let 
\begin{equation}
	\Vitali_k = \{o + k\}_{o \in \Vitali}, 
\end{equation}
under which 
\begin{equation}
	[0, 1] \subset \bigcup_{k \in q_k}\Vitali_k \subset [-1, 2],
\end{equation}
where 
\[
	\bigcup_{k \in q_k}\Vitali_k 
\]
is the countable system of sequences of pairwise disjoint sets. Suppose that $\Vitali$ and $\Vitali_k$ are measurable, it follows that $\bigcup_{k \in q_k}\Vitali_k$ will also be measurable. The measure of $\bigcup_{k \in q_k}\Vitali_k$ is 0 if $\Vitali$ is 0, and $\infty$ if the measure of $\Vitali$ is $\infty$. The first case is impossible, given that
\begin{equation}
	[0, 1] \subset \bigcup_{k \in q_k}\Vitali_k,
\end{equation} 
and the second one is impossible, considering that
\begin{equation}
	\bigcup_{k \in q_k}\Vitali_k \subset [-1, 2].
\end{equation}
Then $\Vitali$ is a \emph{non-Lebesgue-measurable} and \emph{non-countable} set.
\end{proof}

\section{Non-Archimedean System}
\label{section "Non-Archimedean System"}

\subsection{Veronese's Non-Archimedean Geometry: a Fully-holed Linear Continuum}

\begingroup
\footnotesize
According to Dedekind [and Cantor] to clarify [\,\dots] the representation of space continuum we need the numerical continuum [\,\dots]. In my opinion, however, it is the intuitive rectilinear continuum [as a simple and primitive representation] through the idea of a point without parts with respect to the continuum itself that serves to give us the abstract definitions of the continuum, of which the numerical one is only a particular case [\,\dots]. The rectilinear continuum is never made up of its points but of \emph{segments} connecting them two by two and which are also continuous. In this way the \emph{mystery} of continuity is pushed back from a given and constant part of the line to an indeterminate part, which is as small as we wish, and still continuous, and into which we are not allowed to enter further with our representation.\endnote{
	Extended original It. version: «G. Cantor, Dedekind [\,\dots] dicono che è arbitraria la corrispondenza univoca a partire da un punto della retta fra i punti della retta stessa e i numeri reali che costituiscono il continuo numerico ottenuto mediante una serie di definizioni astratte di segni, per quanto possibili arbitrarie sempre [\,\dots]. Secondo Dedekind per chiarire [\,\dots] la rappresentazione del continuo dello spazio occorre il continuo numerico [\,\dots]. Secondo me invece è il continuo intuitivo rettilineo mediante l'idea di punto senza parti rispetto al continuo stesso che serve a darci le definizioni astratte del continuo, di cui quello numerico non è che un caso particolare. In questo modo le definizioni non appariscono come uno sforzo della mente nostra, ma trovano la loro piena giustificazione nella rappresentazione sensibile dei continuo [\,\dots]. E d'altronde sarebbe veramente meraviglioso che una forma astratta cosi complessa qual'è [\textit{sic}] il continuo numerico ottenuto non solo senza la guida di quello intuitivo, ma come si fa oggi da alcuni autori, da pure definizioni di segni si trovasse poi d'accordo con una rappresentazione così semplice e primitiva qual'è quella del continuo rettilineo [\,\dots]. Il continuo rettilineo non è mai composto dai suoi punti ma dai \emph{tratti} che li congiungono due a due e che sono pur essi continui. In questo modo il \emph{mistero} della continuità viene ricacciato da una parte data e costante della retta ad una parte indeterminata quanto piccola si vuole, che è pur sempre continua, e dentro alla quale non ci è permesso di entrare più oltre colla nostra rappresentazione».
	} \\
\indent — \textsc{G. Veronese} \cite[p. 48, footnote]{Veronese "Fondamenti di geometria a piu dimensioni e a piu specie di unita rettilinee}	

\endgroup

\vspace{2mm}

Veronese's name is associated with the \emph{non-Archimedean} continuum;\footnote{
	But before him, P. Du Bois-Reymond \cite{Du Bois-Reymond "Die Allgemeine Functionentheorie I} and O. Stolz \cite{Stolz "Vorlesungen uber allgemeine Arithmetik I"} built a core of non-Archimedean systems.
	}
in addition to \cite[e.g. pp. xxix-xxx, 564]{Veronese "Fondamenti di geometria a piu dimensioni e a piu specie di unita rettilinee}, for Veronese's non-Archimedean geometry, see also his works \cite{Veronese "Intorno ad alcune osservazioni sui segmenti infiniti e infinitesimi attuali"} \cite{Veronese "Veronese "La geometria non Archimedea. Una questione di priorita"} \cite{Veronese "La Geometria non-Archimedea"}.

That which is important to consider, in this short Section, is that Veronese, in controversy with the Cantor–Dedekind axiom (Section \ref{subsubsection "Properties of Real Numbers: Archimedeanity, Completeness, Cauchy Criterion, and Cantor–Dedekind Axiom"}), defines the linear continuum within a non-Archimedean geometry, obtained through \emph{actually infinite (line) segments}, or \emph{actual infinitesimal (line) segments} \cite[cf. pp. 424-429]{Veronese "Intorno ad alcune osservazioni sui segmenti infiniti e infinitesimi attuali"}. In the last part of the above-mentioned epigraph, it can be read a hoary, and bewitching, diatribe between a continuum atomized in indivisible/elementary points and a continuum generated by segments having a continuous (not punctiform) nature.\footnote{
	The point, for Veronese \cite[p. 210]{Veronese "Fondamenti di geometria a piu dimensioni e a piu specie di unita rettilinee}, is consistent with a primitive definition (Section \ref{section "Point and Line as Primitive Ideas"}), and does not touch the question of whether or not it has parts in itself, although he makes no bones about his view.
	} 

Let us take a look at the non-Archimedean axiom in Verone's approach under the effective synthesis of F. Enriques \cite[§ 7, p. 38 (see also §§ 39-44)]{Enriques "Prinzipien der Geometrie"}.

\begin{propositio}[Veronese's non-Archimedean continuum]
Let $L_{\overline{\alpha\beta}} \viz \overline{\alpha\beta}$ be a line segment between points $\alpha$ and $\beta$. We will divide all the points $x_0, \mathellipsis, x_\omega \in L_{\overline{\alpha\beta}}$ into two classes, $\mathscr{C}^L_\alpha$ and $\mathscr{C}^L_\beta$, so that

· $\alpha \in \mathscr{C}^L_\alpha$,

· $\beta \in \mathscr{C}^L_\beta$,

· each point $x \in L_{\overline{\alpha\beta}}$ belongs to one or the other of these two classes,

· each point $x \in \mathscr{C}^L_\alpha$ is inside the segment that joins $\alpha$ with any point $x \in \mathscr{C}^L_\beta$. Then
\enumerationisinitium
\item $\mathscr{C}^L_\alpha$ has an end-point $x_\omega$, and $\mathscr{C}^L_\beta$ has a initial-point $x_0$, so there is a leap,
\item $\mathscr{C}^L_\alpha$ has an end-point $x_\omega$, and $\mathscr{C}^L_\beta$ has no initial-point $x_0$,
\item $\mathscr{C}^L_\alpha$ has no end-point $x_\omega$, and $\mathscr{C}^L_\beta$ has an initial-point $x_0$,
\item $\mathscr{C}^L_\alpha$ has no end-point $x_\omega$, and $\mathscr{C}^L_\beta$ has no initial-point $x_0$, so there is a gap, to wit, a hole. 
\enumerationisfinis

In Dedekind's theory (Section \ref{subsection "Dedekind's Continuity Axiom"}) there are no leaps nor gaps, i.e. holes; in Veronese's non-Archimedeanity there are no leaps but there are \emph{gaps}, or \emph{holes}.
\end{propositio}

\subsection{Non-Archimedean Analysis with the Levi-Civita's Monosemii}

Relying on the intuitions of Veronese \cite{Veronese "Fondamenti di geometria a piu dimensioni e a piu specie di unita rettilinee}, T. Levi-Civita \cite{Levi-Civita "Sugli infiniti ed infinitesimi attuali quali elementi analitici"} is the first to systematize the notion of \emph{non-Archimedean field}, wherefore with an \emph{analytical treatment} (and not exclusively via geometry, as it happens instead in Veronese), thanks to the introduction of the so-called \emph{monosemii} numbers. We start with the following distinction:
	
· numbers infinitely close to 0 are said to be \emph{infinitesimals},

· numbers infinitely close to any real number are said to be \emph{finite},

· reciprocal infinitesimal numbers are said to be \emph{infinite},

· ordinary real numbers having index 0 are said to be \emph{monosemii numbers}, e.g. $\epsilon_\mathbbl{r}$, where $\epsilon, \mathbbl{r} \in \mathbb{R}$, is a \emph{monosemio}, with $\mathbbl{r} = 0$ ($\epsilon$ is the characteristic, and $\mathbbl{r}$ the index).
	
The monosemii of index $\mathbbl{r} > 0$ are greater than all the (real) finite numbers (monosemii of index 0), and the monosemii of $\mathbbl{r} < 0$ are less than any finite number. So it is easy to sight in the monosemii the \emph{marks} of infinite and infinitesimal quantities \cite[footnote, p. 1768]{Levi-Civita "Sugli infiniti ed infinitesimi attuali quali elementi analitici"}.

A sum of monosemii has this form:
\begin{equation}
	\sum^n_{j = 1}(\epsilon_\mathbbl{r})^j = (\epsilon_\mathbbl{r})^1 + \mathellipsis + (\epsilon_\mathbbl{r})^n, 
\end{equation}
whilst, if we fix a group $\Gamma_\mathbb{R} = \{\mathbbl{r}^1, \mathellipsis, \mathbbl{r}^n\}$ of ordinary real numbers such that 
\begin{equation}
	(\Gamma_\mathbb{R})^\alpha = \left\{\mathbbl{r}^j \colon \mathbbl{r}^j > \alpha\right\},	
\end{equation}
for each $\alpha \in \mathbb{R}$, we can define, for a formal power series, the sum and product, 
\begin{equation}
	N_\mathrm{e} = \sum^\infty_{j = 1}(\epsilon_\mathbbl{r})^j, 
\end{equation}
with $(\epsilon_\mathbbl{r})^j \in \Gamma_\mathbb{R}$, assuming that $\Gamma_\mathbb{R}$ consists of a succession, the terms of which are arranged in \emph{descending} order. The number $N_\mathrm{e}$ is called a \emph{non-Archimedean number of elliptic type}. If instead we fix a group $X_\mathbb{R}$ such that 
\begin{equation}
	(E_\mathbb{R})^\alpha = \left\{\mathbbl{r}^j \colon \mathbbl{r}^j < \alpha\right\},
\end{equation}
assuming that $X_\mathbb{R}$ consists of a succession, the terms of which are arranged in \emph{ascending} order, then one has a \emph{non-Archimedean number of hyperbolic type}, indicated by $N_\mathrm{h}$.

\vspace{10mm}

\setcounter{secnumdepth}{0}  
\section{References and Bibliographic Details}
\setcounter{secnumdepth}{3}
\markright{References and Bibliographic Details}

\begingroup
\footnotesize
\noindent Section \ref{subsection "Peano's, Enriques', and Pieri's Non-defined Source Geometry"}
\begin{indent paragraph: 15pt}
For a historical examination on the foundations of geometry in Italian debate that raged in the late 19th century, with the germination of conflicts among the various authors (Veronese, Peano, Amodeo, Pieri, Segre, Enriques, Fano), refer to U. Bottazzini \cite{Bottazzini "I geometri italiani e il problema dei fondamenti (1889-1899)"}.
\end{indent paragraph: 15pt}

\noindent Section \ref{subsection "What is a Point, or a Line? Hilbert vs. Frege"}
\begin{indent paragraph: 15pt}
On Hilbert's foundations of geometry, along with the fervent theoretical context, see \cite[cap. 9]{Bartocci "Una piramide di problemi. Storie di geometria da Gauss a Hilbert"}; on the Hilbertian axiomatics, see also \cite{Troyanov "On the origin of Hilbert geometry"}.
\end{indent paragraph: 15pt}

\noindent Sections \ref{subsection "Margo. Grothendieckian topos-point"}
\begin{indent paragraph: 15pt}
Grothendieckian inventions on the notion of point in the topos theory are in \cite[pp. 378-384]{"Sheaves in Geometry and Logic: A First Introduction to Topos Theory"}.	
\end{indent paragraph: 15pt}

\noindent Sections \ref{subsection "Dedekind's Continuity Axiom"} and \ref{subsection "Cantorian Hierarchy: Transfinite Arithmetic, and Cardinality of the Continuum"}
\begin{indent paragraph: 15pt}
For a close examination of Dedekindian and Cantorian thought, see \cite[capp. II-III]{Lolli "Nascita di un'idea matematica"}.
\end{indent paragraph: 15pt}

\noindent Section \ref{subsubsection "Schnitt and Bijection"}
\begin{indent paragraph: 15pt}
The Euclidean definition in Margo \ref{margo "Euclidean theory of magnitudes: continuous quantities"} comes under the theory of proportions. For a critico-historical analysis of Euclid's exposition of such a theory in the fifth book of the \textit{Elements}, along many interpretations of it, and its legacy in the Galilean era, see \cite{Giusti "Euclides reformatus. La teoria delle proporzioni nella scuola galileiana"}.	
\end{indent paragraph: 15pt}

\noindent Section \ref{section "Non-Archimedean System"}
\begin{indent paragraph: 15pt}
A summing-up of the non-Archimedean mathematics is in \cite[§§ 12, 33]{Hilbert "Grundlagen der Geometrie"} \cite{Conrad "Several approaches to non-archimedean geometry"} \cite{Benci Freguglia "Alcune osservazioni sulla matematica non archimedea"}.
\end{indent paragraph: 15pt}

\endgroup

\chapter{The Ricci Flow, or the Hamilton–Perelman Metric Evolution Machinery}
\label{chapter "The Ricci Flow, or the Hamilton–Perelman Metric Evolution Machinery"}

\begingroup
\footnotesize
As far as Mechanics is concerned, I do not have to say how great a part Geometry plays and necessarily must play in it [\,\dots]. Indicating the \emph{vis viva} with $ds^2/dt^2$, the motion problem is equivalent to the geometric one of geodesics in a space of $n$-dimensions [under a representation of a mobile system with $n$ degrees of freedom by means of a point of $n$- or $2n$-dimensional space]. About the links between Geometry and Analysis, it can be said that they derive mainly from the fact that the \emph{objects} studied \emph{are largely the same} [\,\dots]. So what analyst calls a \emph{function} $y = f(x)$, geometer considers it as a \emph{curve}, or as a \emph{correspondence} between the points $x$ and $y$. What analyst calls \emph{differential equation} will be  for a geometer a \emph{variety of elements} in the sense of Sophus Lie. And the groups of linear transformations used in the study of automorphic functions by Poincaré and Klein [\,\dots] can be considered as particular groups of non-Euclidean \emph{movements}. \\
\indent — \textsc{C. Segre} \cite[pp. 109-110, slight textual amendments for purposes of improvement]{Segre "La Geometria d'oggidi e i suoi legami coll'Analisi"}

\endgroup

\section{Propaedeutics to Ricci Flow}
\label{section "Propaedeutics to Ricci Flow"}

\begingroup
\footnotesize
[S]ome geometrical object can be improved by evolving it with a parabolic partial differential equation. In the Ricci Flow we try to improve a Riemannian metric $g(x, y)$ by evolving it by its Ricci curvature $Rc(x, y)$ under the equation $\frac{\partial}{\partial t}g(X, Y) = -2Rc(X, Y)$. In local geodesic coordinates $\{x^i\}$ at a point $P$[,] where the metric is $ds^2 = g_{ij}dx^idx^j$[,] we find that the ordinary Laplacian of the metric is $``\Laplacian"g_{ij} = g^{pq}\frac{\partial^2}{\partial x^p\partial x^q}g_{ij} = -2Rc(X, Y)$[,] so the Ricci flow is [or rather, resembles] really the heat equation for a Riemannian metric $\frac{\partial}{\partial t}g = ``\Laplacian"g$. \\
\indent — \textsc{R.S. Hamilton} \cite[p. 7]{Hamilton "The Formation of Singularities in the Ricci Flow"}

\endgroup
	
\vspace{2mm}

Ricci flow is the object of our attention in this Chapter. Some starting information follows immediately.

\subsection{Quasi-linear Weakly Parabolic Equation}

\begin{definitio}[Ricci flow]
\label{definitio "Ricci flow"}
~\enumerationisinitium	
\item \emph{Ricci flow} was introduced by R.S. Hamilton \cite{Hamilton "Three-manifolds with positive Ricci curvature"}, and we can write as
\begin{align} 
	\textcyrillic{\textit{ф}}_\textsc{r}^\mathrm{unn}
	& \begin{cases}
	\label{equation "unnormalized Ricci flow"}
	\frac{\partial g_t}{\partial t} = -2\Ric(g_t), \enspace g_t \viz g(t), \\
		\partial_t g_{\mu\nu}(t) = -2\Ricci_{\mu\nu}(t), \\
		\text{with the initial condition } g(0) = g_0,
		\end{cases} \\
	\textcyrillic{\textit{ф}}_\textsc{r}^\mathrm{nor}
	& \begin{cases}
	\label{equation "normalized Ricci flow"}
	\frac{\partial g_t}{\partial t} = -2\Ric(g_t) + \frac{2}{n}\Ricci_{(\mathrm{a})}g_t, \enspace g_t \viz g(t), \\ 
		\partial_t g_{\mu\nu} = -2\Ricci_{\mu\nu} + \frac{2}{n}\Ricci_{(\mathrm{a})}g_{\mu\nu}, \\
	\text{with the initial condition } g(0) = g_0, \text{ and } \displaystyle \Ricci_{(\mathrm{a})} = \frac{\int_{\mathcal{M}^n}\scalarcurvature d\bbmu}{\int_{\mathcal{M}^n}d\bbmu},
	\end{cases}
\end{align}
where $g_t \viz g(t)$ and $g_{\mu\nu}$ denote the Riemannian metric, \emph{smoothly} structured in some interval $t \in [0, T) \subset \mathbb{R}$, with $T \in (0, \infty]$, $\Ric$ (notation without indices) and $\Ricci_{\mu\nu}$ (notation with indices) are the two notations for the the Ricci curvature tensor (Section \ref{subsubsection "Ricci Curvature Tensor"}), which gives its name \cite{Hamilton "The Ricci Flow on Surfaces"} to the flow, whereas $\mathcal{M}$ is a smooth $n$-manifold. Plainly, $g_t$ is a solution to $\textcyrillic{\textit{ф}}_\textsc{r}$ (Ricci flow)—on the basis of Eqq. \eqref{equation "unnormalized Ricci flow"} or \eqref{equation "normalized Ricci flow"}—for an arbitrary $\mathscr{C}^\infty$ initial metric $g_0$ (see Theorem \ref{theorema "Hamilton"}).
\item Eqq. \eqref{equation "unnormalized Ricci flow"} and \eqref{equation "normalized Ricci flow"} are the unnormalized and normalized versions of $\textcyrillic{\textit{ф}}_\textsc{r}$, respectively. The normalized flow is a \emph{volume preserving} equation with respect to $g_t$, in which, for a smooth measure $\bbmu$, the differential element $d\bbmu$ is the \emph{volume element}, or density ($\textit{\dh}$), i.e. 
\begin{subequations}
\label{subequations "volume element"}
	\begin{empheq}[left = {d\bbmu = \empheqlbrace}]{align}
	& \textit{\dh}(x)dx = \textit{\dh}(x)dx^1, \mathellipsis, dx^n, \text{ in local coordinates}, \\
	& \sqrt{\det(g_{\mu\nu})}dx,
	\end{empheq}
\end{subequations}
where $dx$ is the volume element in $\mathbb{R}^n$, and $\Ricci_{(\mathrm{a})} = \int_{\mathcal{M}^n}\scalarcurvature d\bbmu/\int_{\mathcal{M}^n}d\bbmu$ is the average of the scalar curvature, with the Ricci scalar $\scalarcurvature$ (Section \ref{subsubsection "Scalar Curvature"}).
\item Eq. \eqref{equation "unnormalized Ricci flow"}, to be more accurate, is a \emph{quasi-linear weakly parabolic equation}. \definitiosymbol
\enumerationisfinis
\end{definitio}

\subsection{Evolution of Curvature: Stretching-Shrinking Processes}
\label{subsection "Evolution of Curvature: Stretching-Shrinking Processes"}

What is it? Ricci flow is a \emph{geometric flow} with \emph{algebro-geometric descriptions}, or even a \emph{gradient flow} for a given functional, such as the one of Einstein–Hilbert \eqref{equation "Einstein–Hilbert (Gravitational) Action"}, on an $n$-space of Riemannian metric. 

\begin{figure}[h!]
\centering
	\begin{minipage}[b]{0.840\textwidth}
	\includegraphics[width = \textwidth]{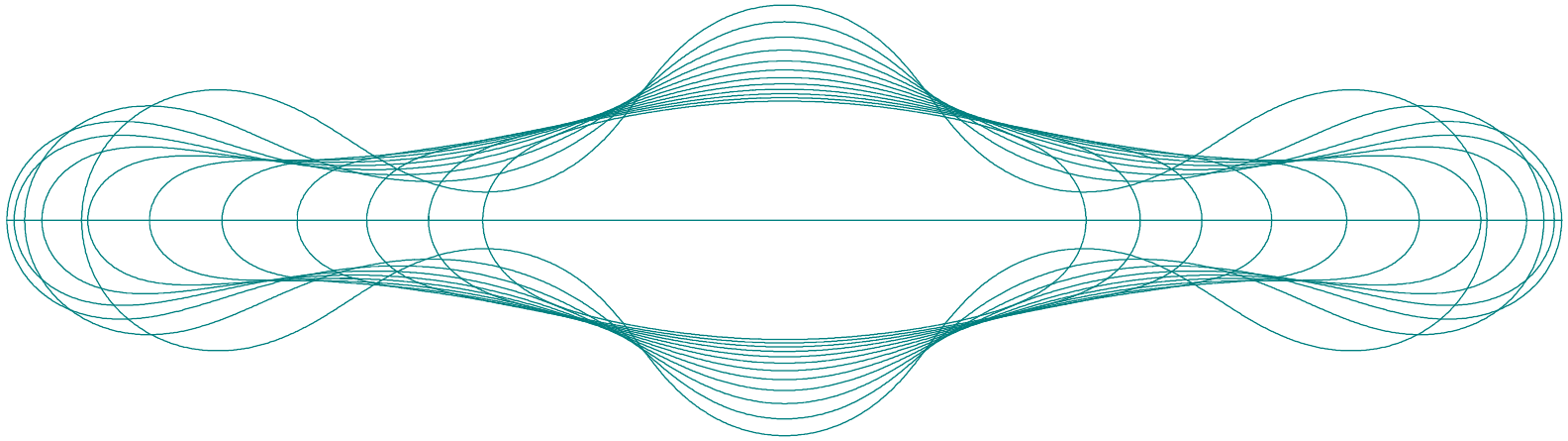}
	\caption{Warping deformation under Ricci flow: cross-section of a surface of revolution}
	\label{figure "Warping deformation under Ricci flow: cross-section of a surface of revolution"} 
	\end{minipage}
	\hspace{30pt}
	\begin{minipage}[b]{0.840\textwidth}
	\includegraphics[width = \textwidth]{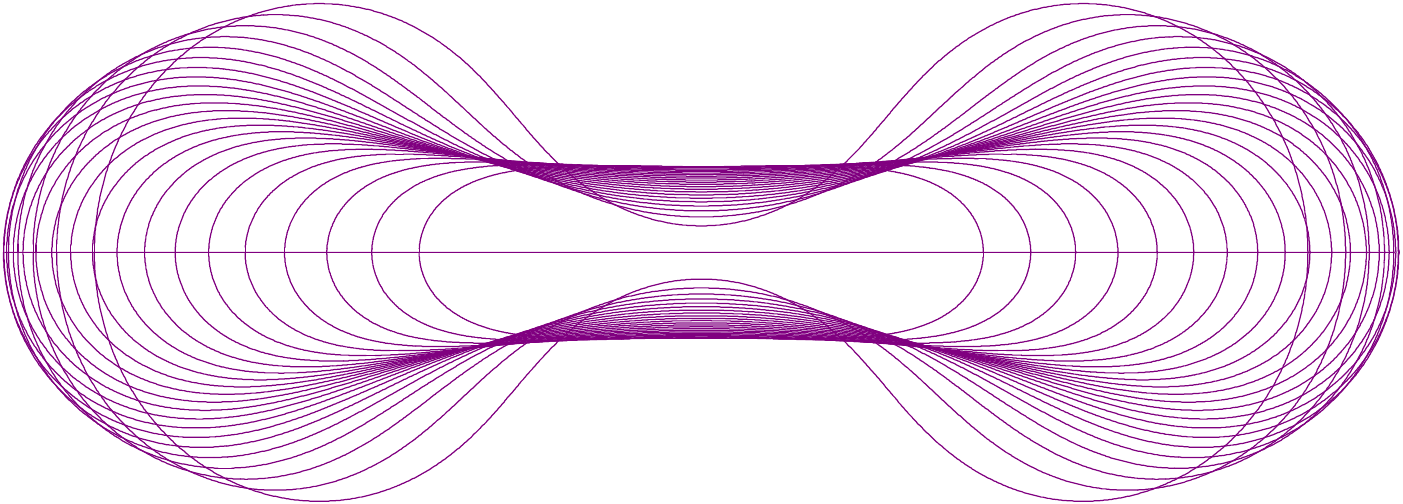}
	\caption{Warping deformation under Ricci flow: cross-section of a dumb-bell-shaped surface of revolution}
	\label{figure "Warping deformation under Ricci flow: cross-section of a dumb-bell-shaped surface of revolution"}
	\end{minipage}
\end{figure}

More specifically, it is a \emph{evolutionary process} that \emph{stretches}, or \emph{expands}, the metric $g$ of a manifold in directions corresponding to \emph{negative coefficients of the Ricci tensor}, and \emph{contracts}, or \emph{shrinks}, $g$ in directions corresponding to \emph{positive coefficients of the Ricci tensor}.\footnote{
	The metric is \emph{stationary} under $\textcyrillic{\textit{ф}}_\textsc{r}$ if there is a Ricci-flat metric ($\Ric = 0$).
	}
For a graphic supports, see—as illustrative examples—Figg. \ref{figure "Warping deformation under Ricci flow: cross-section of a surface of revolution"} and \ref{figure "Warping deformation under Ricci flow: cross-section of a dumb-bell-shaped surface of revolution"}.\footnote{
	These drawings are taken from J.H. Rubinstein and R. Sinclair \cite[Fig. 2, p. 290, and Fig. 5, p. 293]{Rubinstein and Sinclair "Visualizing Ricci Flow of Manifolds of Revolution"}. They deserve credit for the design. See also \cite{Ni Lin Luo and Gao "Community Detection on Networks with Ricci Flow"}.
	}

This is why Hamilton, \emph{by analogy}, associates \eqref{equation "unnormalized Ricci flow"} with a \emph{heat-like equation} in a tensor guise, although the geometric flow is \emph{quasi-linear}, here; and then he compares $\textcyrillic{\textit{ф}}_\textsc{r}$ to a \emph{Laplacian operator} $\Laplacian$ of the Riemannian metric (see epigraph). 

\subsection{Heat-like Diffusion from Harmonic Mappings}

\begingroup
\footnotesize
[Q]uales esse debeant proprietates corporis motum caloris determinantes et distributio caloris, ut detur systema linearum quae semper isothermae maneant [\,\dots]. Disquisitiones haece interpretatione quadam geometrica illustrari possunt, quae quamquam conceptibus inusitatis nitatur.\footnote{
	«[P]roperties of a body determining the conduction of heat and distribution of heat, such that there is a system of lines which remain isothermal [\,\dots]. These investigations may be illustrated with some geometrical interpretation, though this is founded on unusual concepts». 
	} \\
\indent — \textsc{B. Riemann} \cite[pp. 370, 382]{Riemann "Commentatio mathematica"}

\endgroup
	
\vspace{2mm}

Hamilton's idea \cite[p. 256]{Hamilton "Three-manifolds with positive Ricci curvature"} is openly inspired by the Eels–Sampson \cite{Eels Jr. and Sampson "Harmonic Mappings of Riemannian Manifolds"} \emph{harmonic mapping} of Riemannian spaces, $\varphi_\upsilon \colon \mathcal{M} \to \mathcal{N}$, leading to deformations of maps by the classical \emph{heat} (or \emph{diffusion}) \emph{equation}, 
\begin{equation}
\label{equation "Heat equation with spatial and temporal variables"}
	\frac{\partial\upsilon}{\partial t} = \alpha_\mathsf{T}\left(\frac{\partial^2\upsilon}{\partial x^2_1} + \cdots + \frac{\partial^2\upsilon}{\partial x^2_n}\right),
\end{equation}
\begin{subequations}
\label{subequations "Heat equation in Laplacian form"}
\begin{align}
	& \upsilon_t = \alpha_\mathsf{T}\Laplacian\upsilon, \enspace \upsilon_t = \Laplacian\upsilon, \\
	& \upsilon_t - \Laplacian\upsilon = 0, \\
	& \Laplacian = \nabla^2\upsilon - \tfrac{\partial\upsilon}{\partial t} = 0,
\end{align}
\end{subequations}
which can then either be written with spatial and temporal variables \eqref{equation "Heat equation with spatial and temporal variables"}, or in Laplacian form \eqref{subequations "Heat equation in Laplacian form"}, where 
\begin{equation}
	\upsilon \colon \Omega \to \mathbb{R}, \enspace \upsilon = \upsilon(x_1, \mathellipsis, x_n, t) = \upsilon(x, t), 
\end{equation}
is a function reliably identifiable with the temperature $\mathsf{T}$, defined on an open (sub)set $\Omega \subset \mathbb{R}^n$, $x \in \Omega \subset \mathbb{R}^n$, $t \in \mathbb{R}^n$, $\alpha_\mathsf{T}$ is the thermal diffusivity, namely the rate of diffusion for the medium through which heat flows (when we say that the heat \emph{flows} in a medium, we clearly use a liquid analogy), and $\Laplacian$ is the Laplacian. 

\begin{scholium}
The Eels–Sampson \emph{harmonic map heat flow}, for the associated energy functional, is the same as a \emph{gradient flow}. \scholiumsymbol
\end{scholium}

\begin{margo}[Heat \& geometry]
Riemann \cite{Riemann "Commentatio mathematica"} presents (see epigraph) an explicit forerunner of such an association, establishing a relationship between \emph{physical results (heat conduction problem)} and \emph{differential geometry issues}. \margosymbol
\end{margo}

\subsection{Modifying the Shapes of Space}
\label{subsection "Modifying the Shapes of Space"}

What is it for? Ricci flow is chiefly used to \emph{deform a (Riemannian) space} for \emph{non-linear/quasi-linear evolutions}, in order to better handle this space, in purely geometric terms, or analytically.

Otherwise stated: being some sort of \emph{quantity of evolution}, it is a mathematical tool for \emph{creating}, \emph{controlling}, or \emph{transforming}, within certain limits, the \emph{shapes of space}, and its \emph{volume growth}, also following geodesic paths; in the end, a tool for \emph{flatting-smoothing}, at least locally, possible \emph{irregularities} inherent in the topological space under examination. 

Hamilton investigates the Ricci flow properties on 2- \cite{Hamilton "The Ricci Flow on Surfaces"}, 3- \cite{Hamilton "Three-manifolds with positive Ricci curvature"} and 4-manifolds \cite{Hamilton "Four-manifolds with positive curvature operator"}, focusing on metrics of \emph{strictly positive}/\emph{non-negative} Ricci curvature.

\subsection{A Small \emph{Formulario} for Evolving Curvatures}

We will give three examples of curvature evolution for the Riemann and Ricci curvature tensors (Sections \ref{subsection "Riemann Curvature Tensor"} and \ref{subsubsection "Ricci Curvature Tensor"}), and scalar curvature (Section \ref{subsubsection "Scalar Curvature"}), in the presence of a solution to the Ricci flow $\frac{\partial g_t}{\partial t} = -2\Ric(g_t)$. It is about seeing what happens when the flow equation intervenes in this type of geometry.

\subsubsection[Riemann Curvature Tensor under $\normalfont{\textcyr{\textit{ф}}}_\textsc{r}$]{Riemann Curvature Tensor under $\protect\pseudobold{\normalfont{\textcyr{\textit{ф}}}_\textsc{r}}$}

\begin{propositio}
The evolution of the Riemann curvature in the $\binom{1}{3}$-form under the Ricci flow is
\begin{align}
	\frac{\partial}{\partial t}{\Riemann_{\mu\nu\xi}}^\varrho & = g^{\varrho\varsigma} \bigl(- \nabla_\mu\nabla_\nu\Ricci_{\xi\varsigma} - \nabla_\mu\nabla_\xi\Ricci_{\nu\varsigma} + \nabla_\mu\nabla_\varsigma\Ricci_{\nu\xi} \notag \\
	& \hspace{30pt} + \nabla_\nu\nabla_\mu\Ricci_{\xi\varsigma} + \nabla_\nu\nabla_\xi\Ricci_{\mu\varsigma} - \nabla_\nu\nabla_\varsigma\Ricci_{\mu\xi}\bigr) \\
	& 
	\label{align "Evolution of the Riemann curvature in the (1, 3)-form under the Ricci flow satisfying a heat-like equation"} 
	= \Laplacian{\Riemann_{\mu\nu\xi}}^\varrho + g^{\varsigma\tau}\bigl({\Riemann_{\mu\nu\varsigma}}^\rotatedtau{\Riemann_{\rotatedtau\tau\xi}}^\varrho - 2{\Riemann_{\varsigma\mu\xi}}^\rotatedtau{\Riemann_{\nu\tau\rotatedtau}}^\varrho + 2{\Riemann_{\varsigma\mu\rotatedtau}}^\varrho{\Riemann_{\nu\tau\xi}}^\rotatedtau\bigr) \notag \\
	& \hspace{50pt} - {\Ricci_\mu}^\varsigma{\Riemann_{\varsigma\nu\xi}}^\varrho - {\Ricci_\nu}^\varsigma{\Riemann_{\mu\varsigma\xi}}^\varrho - {\Ricci_\xi}^\varsigma{\Riemann_{\mu\nu\varsigma}}^\varrho + {\Ricci_\varsigma}^\varrho{\Riemann_{\mu\nu\xi}}^\varsigma.
\end{align}
\end{propositio}

\begin{proof}[Proof of \eqref{align "Evolution of the Riemann curvature in the (1, 3)-form under the Ricci flow satisfying a heat-like equation"} satisfying a heat-like Eq. \eqref{subequations "Heat equation in Laplacian form"}]
By the second Bianchi identity \eqref{subequations "Second Bianchi identity"}, we calculate
\begin{align}
	\Laplacian{\Riemann_{\mu\nu\xi}}^\varrho & = g^{\varsigma\tau}\nabla_\varsigma\nabla_\tau{\Riemann_{\mu\nu\xi}}^\varrho = g^{\varsigma\tau}\nabla_\varsigma\bigl(-\nabla_\mu{\Riemann_{\nu\tau\xi}}^\varrho -\nabla_\nu{\Riemann_{\tau\mu\xi}}^\varrho\bigr) \notag \\
	& = g^{\varsigma\tau}\bigl(-\nabla_\mu\nabla_\varsigma{\Riemann_{\nu\tau\xi}}^\varrho + {\Riemann_{\varsigma\mu\nu}}^\rotatedtau{\Riemann_{\rotatedtau\tau\xi}}^\varrho \notag \\
	& \hspace{30pt} + {\Riemann_{\varsigma\mu\tau}}^\rotatedtau{\Riemann_{\nu\rotatedtau\xi}}^\varrho + {\Riemann_{\varsigma\mu\xi}}^\rotatedtau{\Riemann_{\nu\tau\rotatedtau}}^\varrho - {\Riemann_{\varsigma\mu\rotatedtau}}^\varrho{\Riemann_{\nu\tau\xi}}^\rotatedtau \notag \\
	& \hspace{30pt} + \nabla_\nu\nabla_\varsigma{\Riemann_{\tau\mu\xi}}^\varrho - {\Riemann_{\varsigma\nu\mu}}^\rotatedtau{\Riemann_{\rotatedtau\tau\xi}}^\varrho \notag \\
	& \hspace{30pt} - {\Riemann_{\varsigma\nu\tau}}^\rotatedtau{\Riemann_{\mu\rotatedtau\xi}}^\varrho - {\Riemann_{\varsigma\nu\xi}}^\rotatedtau{\Riemann_{\mu\tau\rotatedtau}}^\varrho + {\Riemann_{\varsigma\nu\rotatedtau}}^\varrho{\Riemann_{\mu\tau\xi}}^\rotatedtau\bigr),
\end{align}
from which $g^{\varsigma\tau}\nabla_\varsigma{\Riemann_{\nu\tau\xi}}^\varrho = g^{\varsigma\tau}g^{\varrho\rotatedvarrho}\bigl(-\nabla_\xi\Riemann_{\nu\tau\rotatedvarrho\varsigma} - \nabla_\rotatedvarrho\Riemann_{\nu\tau\varsigma\xi}\bigr) = \nabla_\xi{\Ricci_\nu}^\varrho - \nabla^\varrho\Ricci_{\nu\xi}$. Hence
\begin{align}
	\Laplacian{\Riemann_{\mu\nu\xi}}^\varrho = & -\nabla_\mu\nabla_\xi{\Ricci_\nu}^\varrho + \nabla_\mu\nabla^\varrho\Ricci_{\nu\xi} + \nabla_\nu\nabla_\xi{\Ricci_\mu}^\varrho - \nabla_\nu\nabla^\varrho\Ricci_{\mu\xi} \notag \\
	& + g^{\varsigma\tau}\bigl({\Riemann_{\varsigma\mu\nu}}^\rotatedtau{\Riemann_{\rotatedtau\tau\xi}}^\varrho + {\Riemann_{\varsigma\mu\tau}}^\rotatedtau{\Riemann_{\nu\rotatedtau\xi}}^\varrho + {\Riemann_{\varsigma\mu\xi}}^\rotatedtau{\Riemann_{\nu\tau\rotatedtau}}^\varrho - {\Riemann_{\varsigma\mu\rotatedtau}}^\varrho{\Riemann_{\nu\tau\xi}}^\rotatedtau \notag \\
	& \hspace{29pt} - {\Riemann_{\varsigma\nu\mu}}^\rotatedtau{\Riemann_{\rotatedtau\tau\xi}}^\varrho - {\Riemann_{\varsigma\nu\tau}}^\rotatedtau{\Riemann_{\mu\rotatedtau\xi}}^\varrho - {\Riemann_{\varsigma\nu\xi}}^\rotatedtau{\Riemann_{\mu\tau\rotatedtau}}^\varrho + {\Riemann_{\varsigma\nu\rotatedtau}}^\varrho{\Riemann_{\mu\tau\xi}}^\rotatedtau\bigr). 
\end{align}
By the first Bianchi identity \eqref{subequations "First Bianchi identity"}, ${\Riemann_{\varsigma\mu\nu}}^\rotatedtau{\Riemann_{\rotatedtau\tau\xi}}^\varrho - {\Riemann_{\varsigma\nu\mu}}^\rotatedtau{\Riemann_{\rotatedtau\tau\xi}}^\varrho = - {\Riemann_{\mu\nu\varsigma}}^\rotatedtau{\Riemann_{\rotatedtau\tau\xi}}^\varrho$, for which
\begin{align}
	\Laplacian{\Riemann_{\mu\nu\xi}}^\varrho = & -\nabla_\mu\nabla_\xi{\Ricci_\nu}^\varrho + \nabla_\mu\nabla^\varrho\Ricci_{\nu\xi} + \nabla_\nu\nabla_\xi{\Ricci_\mu}^\varrho - \nabla_\nu\nabla^\varrho\Ricci_{\mu\xi} \notag \\
	& - {\Ricci_\mu}^\rotatedtau{\Riemann_{\nu\rotatedtau\xi}}^\varrho + {\Ricci_\nu}^\rotatedtau{\Riemann_{\mu\rotatedtau\xi}}^\varrho \notag \\
	& + g^{\varsigma\tau} \bigl(- {\Riemann_{\mu\nu\varsigma}}^\rotatedtau{\Riemann_{\rotatedtau\tau\xi}}^\varrho + {\Riemann_{\varsigma\mu\nu}}^\rotatedtau{\Riemann_{\nu\tau\rotatedtau}}^\varrho +  {\Riemann_{\varsigma\nu\rotatedtau}}^\varrho{\Riemann_{\mu\tau\xi}}^\rotatedtau \notag \\
	& \hspace{28pt} - {\Riemann_{\varsigma\mu\rotatedtau}}^\varrho{\Riemann_{\nu\tau\xi}}^\rotatedtau -   {\Riemann_{\varsigma\nu\xi}}^\rotatedtau{\Riemann_{\mu\tau\rotatedtau}}^\varrho\bigr).
\end{align}
And finally
\begin{align}
	\frac{\partial}{\partial t}{\Riemann_{\mu\nu\xi}}^\varrho = & -\nabla_\mu\nabla_\xi{\Ricci_\nu}^\varrho + \nabla_\mu\nabla^\varrho\Ricci_{\nu\xi} + \nabla_\nu\nabla_\xi{\Ricci_\mu}^\varrho - \nabla_\nu\nabla^\varrho\Ricci_{\mu\xi} \notag \\
		& + g^{\varrho\varsigma}\left({\Riemann_{\mu\nu\xi}}^\tau\Ricci_{\tau\varsigma} + {\Riemann_{\mu\nu\varsigma}}^\tau\Ricci_{\xi\tau}\right).
\end{align}
\end{proof}

\begin{propositio}
The evolution of the Riemann curvature in the $\binom{4}{0}$-form under the Ricci flow is
\begin{align}
	\frac{\partial}{\partial t}\Riemann_{\mu\nu\xi\varrho} & = \Lbrack:\Laplacian\Riemann_{\mu\nu\xi\varrho}:\Rbrack + g^{\varsigma\tau}\left({\Riemann_{\mu\nu\varsigma}}^\rotatedtau\Riemann_{\rotatedtau\tau\xi\varrho} - 2{\Riemann_{\varsigma\mu\xi}}^\rotatedtau\Riemann_{\nu\tau\rotatedtau\varrho} + 2\Riemann_{\varsigma\mu\rotatedtau\varrho}{\Riemann_{\nu\tau\xi}}^\rotatedtau\right) \notag \\
	& = \Lbrack:\cdots:\Rbrack + 2(\Tau_{\mu\nu\xi\varrho} - \Tau_{\mu\nu\varrho\xi} + \Tau_{\mu\xi\nu\varrho} - \Tau_{\mu\varrho\nu\xi}) \notag \\
	& - {\Ricci_\mu}^\varsigma\Riemann_{\varsigma\nu\xi\varrho} + {\Ricci_\nu}^\varsigma\Riemann_{\mu\varsigma\xi\varrho} + {\Ricci_\xi}^\varsigma\Riemann_{\mu\nu\varsigma\varrho} + {\Ricci_\varrho}^\varsigma\Riemann_{\mu\nu\xi\varsigma},
\end{align}
where $\Tau_{\mu\nu\xi\varrho} = \Tau_{\nu\mu\varrho\xi} = \Tau_{\xi\varrho\mu\nu}$ is a $\binom{4}{0}$-tensor.
\end{propositio}

\begin{proof}
By the first Bianchi identity \eqref{subequations "First Bianchi identity"},
\begin{align}
	g^{\varsigma\tau}{\Riemann_{\mu\nu\varsigma}}^\rotatedtau\Riemann_{\rotatedtau\tau\xi\varrho} & = g^{\varsigma\tau}g^{\rotatedtau\upsilon}\Riemann_{\rotatedtau\varsigma\nu\mu}\Riemann_{\upsilon\tau\xi\varrho} \notag \\
	& = g^{\varsigma\tau}g^{\rotatedtau\upsilon} \left(- \Riemann_{\rotatedtau\nu\mu\varsigma} - \Riemann_{\rotatedtau\mu\varsigma\nu}\right)\left(-\Riemann_{\upsilon\xi\varrho\tau} -\Riemann_{\upsilon\varrho\tau\xi}\right) \notag \\
	& = - \Tau_{\nu\mu\xi\varrho} + \Tau_{\nu\mu\varrho\xi} + \Tau_{\mu\nu\xi\varrho} - \Tau_{\mu\nu\varrho\xi},
\end{align}
where the symbols $\Lbrack:$ and $:\Rbrack$ indicate that the expression within them must be repeated, on inspiration of the musical beginning and ending repeat signs; so $g^{\varsigma\tau}{\Riemann_{\mu\nu\varsigma}}^\rotatedtau\Riemann_{\rotatedtau\tau\xi\varrho} = 2(\Tau_{\mu\nu\xi\varrho} - \Tau_{\mu\nu\varrho\xi})$. Here too a  heat-like Eq. \eqref{subequations "Heat equation in Laplacian form"} is satisfied.
\end{proof}

\subsubsection[Ricci Curvature Tensor under $\normalfont{\textcyr{\textit{ф}}}_\textsc{r}$]{Ricci Curvature Tensor under $\protect\pseudobold{\normalfont{\textcyr{\textit{ф}}}_\textsc{r}}$}

\begin{propositio}
The evolution of the Ricci curvature under the Ricci flow is
\begin{subequations}
\begin{align}
	\frac{\partial}{\partial t}\Ricci_{\nu\xi} & = \nabla\Ricci_{\nu\xi} + \nabla_\nu\nabla_\xi\scalarcurvature - g^{\varsigma\tau}(\nabla_\tau\nabla_\nu\Ricci_{\xi\varsigma} + \nabla_\tau\nabla_\xi\Ricci_{\nu\varsigma}) \\
	& 
	\label{align "Evolution of the Ricci curvature tensor with Lichnerowicz Laplacian"}
	= \Laplacian_\textsc{l}\Ricci_{\nu\xi} = \Laplacian\Ricci_{\nu\xi} + 2g^{\varsigma\tau}g^{\rotatedtau\upsilon}\Riemann_{\varsigma\nu\xi\rotatedtau}\Ricci_{\tau\upsilon} - 2g^{\varsigma\tau}\Ricci_{\nu\varsigma}\Ricci_{\tau\xi}, 
\end{align}
\end{subequations}
where $\Laplacian_\textsc{l}$ is the Lichnerowicz Laplacian \textnormal{\cite{Lichnerowicz "Propagateurs et commutateurs en relativite generale"}}. 
\end{propositio}

Note. Eq. \eqref{align "Evolution of the Ricci curvature tensor with Lichnerowicz Laplacian"} is a consequence of the recourse to $\idem_\textsc{b}$ \eqref{subequations "Second Bianchi identity"}.

\begin{proof}[Proof of \eqref{align "Evolution of the Ricci curvature tensor with Lichnerowicz Laplacian"}]
\begin{subequations}
\begin{align}
	\frac{\partial}{\partial t}\Ricci_{\nu\xi} & = \Lbrack:\Laplacian_\textsc{l}\Ricci_{\nu\xi} + \nabla_\nu\nabla_\xi\scalarcurvature:\Rbrack - g^{\varsigma\tau}(\nabla_\nu\nabla_\varsigma\Ricci_{\tau\xi} + \nabla_\xi\nabla_\varsigma\Ricci_{\nu\tau}) \\
	& = \Lbrack:\cdots:\Rbrack - \tfrac{1}{2}(\nabla_\nu\nabla_\xi\scalarcurvature + \nabla_\xi\nabla_\nu\scalarcurvature).
\end{align}
\end{subequations}
\end{proof}

\subsubsection[Scalar Curvature under $\normalfont{\textcyr{\textit{ф}}}_\textsc{r}$]{Scalar Curvature under $\protect\pseudobold{\normalfont{\textcyr{\textit{ф}}}_\textsc{r}}$}

\begin{propositio}
The evolution of the scalar curvature, or Ricci scalar, under the Ricci flow is
\begin{subequations}
\begin{align}
	\frac{\partial}{\partial t}\scalarcurvature & = 2\Laplacian\scalarcurvature - 2g^{\nu\xi}g^{\varsigma\tau}\nabla_\tau\nabla_\nu\Ricci_{\xi\varsigma} + 2|\Ric|^2 \\
	& 
	\label{align "Evolution of the scalar curvature after second Bianchi identity"}
	= \Laplacian\scalarcurvature + 2|\Ric|^2.
\end{align}	
\end{subequations}
\end{propositio}

Note. Eq. \eqref	{align "Evolution of the scalar curvature after second Bianchi identity"} comes from applying $\idem_\textsc{b}$ \eqref{subequations "Second Bianchi identity"}.

\subsection{Occurrence of Singularities of the Ricci Flow}
\label{subsection "Occurrence of Singularities of the Ricci Flow"}

The aforementioned control (Section \ref{subsection "Modifying the Shapes of Space"}) of the Ricci evolution of curvature encounters limitations: \emph{singularities} in finite time $T$ can be formed \cite{Hamilton "The Formation of Singularities in the Ricci Flow"} in some parts of the manifold and in a number of ways.

\subsubsection{Local Finite Time Neckpinch Singularity in 2- and 3D}

We choose two very simple examples, related to the so-called \emph{pinching singularity}, see M. Simon \cite{Simon "A class of Riemannian manifolds that pinch when evolved by Ricci flow"}, and S. Angenent \& D. Knopf \cite{Angenent and Knopf "An Example of Neckpinching for Ricci Flow on Sn+1"} \cite{Angenent and Knopf "Precise asymptotics of the Ricci flow neckpinch"}. 

We consider a 2-sphere $\mathbb{S}^2$, that is, a 2-dimensional surface of a 3-ball, or a positively curved space embedded in $3\mathrm{D}$ Euclidean space. Imagine that it is a rubbery sphere. The 2-sphere is now pinched, that is, squeezed, along its equatorial line, so it takes gradually the form of a dumb-bell. 

The process of pinching, or of squeezing-shrinking, that proceeds over time as $t \nearrow T < \infty$, gives rise to a \emph{pseudo-cylindrical (non-Euclidean) neck}, generating in it a series of cross-sectional 1-spheres $\mathbb{S}^1$, or 1-dimensional circumferences of a disk, with variable length (depending on where the circumference is taken).

Be careful: the neck looks like $\{\mathbb{S}^1 \times \mathbb{B}^1\}$, since  appear bounded $\mathbb{B}^1$-intervals, but it stretches into an \emph{increasingly (infinitely)} thin central part, leading in last steps to singularity $\singularity\{\mathbb{S}^1 \times \mathbb{B}^1\}$,
\begin{equation}
	\text{rubbery 2-surface } \mathbb{S}^2 \xrightarrow[\text{pinching of } \textcyrillic{\textit{ф}}_\textsc{r}^{(2)}]{t \nearrow T < \infty} {}\xrightarrow[\text{infinitely long \& thin}]{\{\mathbb{S}^1 \times \mathbb{B}^1\}\text{-neck}} \overset{\text{singularity of } \textcyrillic{\textit{ф}}_\textsc{r}^{(2)}}{\singularity\{\mathbb{S}^1 \times \mathbb{B}^1\}}. 	
\end{equation}

What we are witnessing, under the action of the Ricci flow, is the \emph{(de)formation}  of a $\mathbb{S}^2$-space in a \emph{constantly evolving} new space which gets longer and thinner all the time in the middle, and it is distinguished by a \emph{dynamic mix} of positive curvature (as the two lateral spheriform bulbs) and negative curvature (the collar zone, the almost tubular center portion of $\mathbb{S}^1 \times \mathbb{B}^1$); for that reason, in this case, it is referred to as \emph{local finite time neckpinch singularity} in $2\mathrm{D}$ Ricci flow.

In three dimensions the procedure is similar:
\begin{equation}
	\text{rubbery 3-boundary } \mathbb{S}^3 \xrightarrow[\text{pinching of } \textcyrillic{\textit{ф}}_\textsc{r}^{(3)}]{t \nearrow T < \infty} {}\xrightarrow[\text{infinitely long \& thin}]{\{\mathbb{S}^2 \times \mathbb{B}^1\}\text{-neck}} \overset{\text{singularity of } \textcyrillic{\textit{ф}}_\textsc{r}^{(3)}}{\singularity\{\mathbb{S}^2 \times \mathbb{B}^1\}}, 	
\end{equation}
where we have a 3-sphere $\mathbb{S}^3$ in $4\mathrm{D}$ Euclidean space, a $\{\mathbb{S}^2 \times \mathbb{B}^1\}$-neck in which the cross-sectional collars are 2-spheres  multiplied by bounded $\mathbb{B}^1$-intervals, and a local finite time neckpinch singularity in $3\mathrm{D}$ Ricci flow.

\subsubsection{Four Types of Singularities}

Let us draw up a classification of four types of Hamiltonian singularities \cite{Hamilton "The Formation of Singularities in the Ricci Flow"} that may arise in a generic smooth topological $n$-space of dimension, and an equally generic metric $g_t$ determining the $\textcyrillic{\textit{ф}}_\textsc{r}$-flow:
\begin{subequations}
	\begin{empheq}[left = {\text{type} \empheqlbrace}]{align}
	& \text{I $\singularity(\textcyrillic{\textit{ф}}_\textsc{r})$: } \displaystyle T < \infty \text{ and } \sup_{\mathcal{M}^n \times [0, T)}|\Rie(x, t)|(T - t) < \infty, \\	
	& \text{II\textgreek{α} $\singularity(\textcyrillic{\textit{ф}}_\textsc{r})$: } \displaystyle T < \infty \text{ and } \sup_{\mathcal{M}^n \times [0, T)}|\Rie(x, t)|(T - t) = \infty, \\
	& \text{II\textgreek{β} $\singularity(\textcyrillic{\textit{ф}}_\textsc{r})$: } \displaystyle T = \infty \text{ and } \sup_{\mathcal{M}^n \times [0, \infty)}|\Rie(x, t)|(t) = \infty, \\
	& \text{II\textgreek{γ} $\singularity(\textcyrillic{\textit{ф}}_\textsc{r})$: } \displaystyle T < \infty \text{ and } \sup_{\mathcal{M}^n \times [0, T)}|\Rie(x, t)|(t \to 0) = \infty, \\
	& \text{III $\singularity(\textcyrillic{\textit{ф}}_\textsc{r})$: } \displaystyle T = \infty \text{ and } \sup_{\mathcal{M}^n \times [0, \infty)}|\Rie(x, t)|(t) < \infty, \\
	& \text{IV $\singularity(\textcyrillic{\textit{ф}}_\textsc{r})$: } \displaystyle T < \infty \text{ and } \sup_{\mathcal{M}^n \times [0, T)}|\Rie(x, t)|(t \to 0) < \infty,
	\end{empheq}
\end{subequations}
where $\Rie$ is the Riemann curvature tensor.

\subsection{Geometro-topological Surgery of Cutting off and Gluing Back}
\label{subsection "Geometro-topological Surgery of Cutting off and Gluing Back"}

 \begingroup
\footnotesize
A “surgery” on a differentiable manifold $W$ of dimension $n = p + q + 1$ has the effect of removing an imbedded sphere of dimension $p$ from $W$, and replacing it by an imbedded sphere of dimension $q$. \\
\indent — \textsc{J. Milnor} \cite[p. 39]{Milnor "A Procedure for Killing Homotopy Groups of Differentiable Manifolds"}

\endgroup

\vspace{2mm}

The epigraph mentions the surgery technique, as it was historically introduced, out of the blue, by J. Milnor. \emph{Surgery}, in topological geometry, means cutting one or more parts of the manifold, and replacing it with others, and then, when necessary, gluing in the cut parts ad hoc spaces with artificial-prosthesis function. 

\subsubsection{Discrete Mechanism for Discontinuous Metrics}

It is this technique that provides the pedestal on which Hamilton \cite{Hamilton "The Formation of Singularities in the Ricci Flow"} \cite{Hamilton "Four-Manifolds with Positive Isotropic Curvature"} stands to address the problem of singularities.

In the specific case of the Ricci flow, it is a question of removing any singularity (one at a time), with the final purpose of recovering the geometro-analytical functionality compromised by points in the evolution field of the flow in which infinite values, or non-monovalued attributes, are assumed. That is, a \emph{surgically modified Ricci flow} is a \emph{stopped} movement-diffusion (as defined above) at \emph{discrete time intervals}, or, in technical words, a \emph{discontinuous piecewise smooth} operational mechanism. Consequently, the \emph{topology} of the manifold is being \emph{changed} little by little, following both \emph{geometro-topological and metric discontinuities}.

\subsection{Li–Yau's \& Hamilton's Harnack Inequalities, and Space-Time Gradient Estimate}
\label{subsection "Li–Yau's & Hamilton's Harnack Inequalities, and Space-Time Gradient Estimate"}

Singularities can be \emph{analyzed by comparing curvatures} for solutions of weakly parabolic equations \eqref{equation "unnormalized Ricci flow"} \eqref{equation "normalized Ricci flow"}, at different points ($x_1, x_2$) and times ($t_1, t_2$). For this, Hamilton \cite{Hamilton "The Harnack estimate for the Ricci flow"} implements an estimation method for calculating spatio-temporal differences in the Ricci topological description (Section \ref{subsubsection "Harnack–Hamilton Estimate for the Ricci Flow"}). This method is a \emph{type of Harnack inequality} \cite{Harnack "Die Grundlagen der Theorie des logarithmischen Potentiales und der eindeutigen Potentialfunktion in der Ebene"} (Section \ref{subsubsection "Classical Harnack Inequality"}) elaborated by P. Li and S.-T. Yau \cite{Li and Yau "On the parabolic kernel of the Schrodinger operator"} (Section \ref{subsubsection "Li–Yau's Harnack (Differential) Inequality"}). The \emph{Harnack inequality of Li–Yau–Hamilton} \cite{Hamilton "Matrix Harnack estimate for the heat equation"} (Section \ref{subsubsection "Li–Yau–Hamilton's Harnack Inequality, aka Hamilton's Matrix Inequality"}) is applied to the matrix of second derivatives in the scalar heat flow, to the Ricci flow on a surface, as well as to mean curvatures. Here is a synopsis.

\subsubsection{Classical Harnack Inequality}
\label{subsubsection "Classical Harnack Inequality"}

\begin{propositio}[Harnack inequality]
Take a smooth domain, which is a connected complete Riemannian manifold $\mathcal{M}$ of dimension $n$, and let $\upsilon \in \mathscr{C}^\infty\{\mathcal{M}^n \times (0, T)\}$ be a non-negative harmonic function, for $x_1, x_2 \in \mathcal{M}^n$ and $t_1 < t_2 \in (0, T)$. We say that $\upsilon$ solves the linear heat equation 
\begin{subequations}
\label{subequations "Linear heat equation"}
	\begin{empheq}[left = {\empheqlbrace}]{align}
	& \Laplacian\upsilon = \tfrac{\partial\upsilon}{\partial t}, \\	
	& \Laplacian\upsilon - \partial_t\upsilon = 0,
	\end{empheq}
\end{subequations}
on $\mathcal{M}$. Then, under the Harnack inequality $\Inequality^\mathrm{Har}$
\begin{equation}
\label{equation "Classical Harnack inequality"}
	\begin{rcases}
	\upsilon(x_1, t_1) \leqslant \upsilon(x_2, t_2) \\
	\text{or} \\
	\frac{\upsilon(x_2, t_2)}{\upsilon(x_1, t_1)} \geqslant \left(\frac{t_2}{t_1}\right)^{-\frac{n}{2}}
	\end{rcases}
	\exp{\left\{-\frac{d(x_1, x_2)^2}{4(t_2 - t_1)}\right\}},
\end{equation}
one has 
\begin{equation}
	\sup_{\mathcal{M}^n}\upsilon(x_1, t_1) \leqslant c\inf_{\mathcal{M}^n}\upsilon(x_2, t_2),
\end{equation}
where $c$ is a time- and $\mathcal{M}$-dependent constant.
\end{propositio}

\subsubsection{Li–Yau's Harnack (Differential) Inequality}
\label{subsubsection "Li–Yau's Harnack (Differential) Inequality"}

\begin{propositio}[Harnack inequality of Li–Yau]
Take a smooth domain with bounded curvature, $\partial\mathcal{M} \neq \varnothing$, and $\Ric \geqslant 0$. Let $\upsilon \colon \mathcal{M}^n \times [0, \infty) \to \mathbb{R}_+$ be a smooth non-negative solution for the heat Eq. \eqref{subequations "Linear heat equation"}. The Li–Yau's Harnack inequality gives a \emph{gradient estimate} for solutions of this kind, that is, if $\Laplacian\varphi_\upsilon - \frac{n}{2t} = \frac{\partial \varphi_\upsilon}{\partial t} + |\nabla\varphi_\upsilon|^2 \leqslant 0$, derives a \emph{heat kernel} estimate in a differential manner: 
\begin{subequations}
\label{subequations "Li–Yau's Harnack inequality"}
	\begin{empheq}[left = {\Inequality^\mathrm{Har}_\textsc{ly} \empheqlbrace}]{align}
	& \Laplacian\log\upsilon + \tfrac{n}{2t} = \tfrac{\partial}{\partial t}\log\upsilon - |\nabla\log\upsilon|^2 + \tfrac{n}{2t} \geqslant 0, \\
	& \tfrac{\partial_t\upsilon}{\upsilon} - \tfrac{|\nabla\upsilon|^2}{\upsilon^2} + \tfrac{n}{2t} \geqslant 0.
	\end{empheq}
\end{subequations}
\end{propositio}

\begin{subpropositio}
We can get identity \eqref{equation "Classical Harnack inequality"} from identity \eqref{subequations "Li–Yau's Harnack inequality"}. 
\end{subpropositio}

\begin{proof}
For a geodesic curve $\gamma_\mathrm{c}$, we put $\bigl|\frac{d\gamma_\mathrm{c}}{dt}(t)\bigr| \equival \frac{d(x_1, x_2)}{t_2 - t_1}$, if 
\begin{align}
	\log\frac{\upsilon(x_2, t_2)}{\upsilon(x_1, t_1)} & = \int^{t_2}_{t_1}\frac{d}{dt}\Bigl(\log\upsilon(\gamma_\mathrm{c}(t), t)\Bigr)dt \notag \\
	& = \int^{t_2}_{t_1}\biggl(\frac{\partial}{\partial t}\log\upsilon + \nabla\log\upsilon\left(\tfrac{d\gamma_\mathrm{c}}{dt}\right)\biggr)dt  \notag \\
	& \geqslant \int^{t_2}_{t_1}\biggl(|\nabla\log\upsilon|^2 - \tfrac{n}{2t} + \nabla\log\upsilon\left(\tfrac{d\gamma_\mathrm{c}}{dt}\right)\biggl)dt \notag \\
	& \geqslant -\tfrac{n}{2t}\log\left(\frac{t_2}{t_1}\right) - \frac{1}{4}\int^{t_2}_{t_1}\left|\frac{d\gamma_\mathrm{c}}{dt}(t)\right|^2dt.
\end{align}	
\end{proof}

\subsubsection{Li–Yau–Hamilton's Harnack Inequality, aka Hamilton's Matrix Inequality}
\label{subsubsection "Li–Yau–Hamilton's Harnack Inequality, aka Hamilton's Matrix Inequality"}

\begin{propositio}[\textsc{lyh}'s Harnack inequality]
Given a compact Riemannian manifold of dimension $n$, let $\upsilon \in \mathscr{C}^\infty(\mathcal{M}^n)$ be a non-negative solution for the heat equation $\Laplacian\upsilon = \frac{\partial\upsilon}{\partial t}$, see \eqref{subequations "Linear heat equation"}, for $t > 0$. If $\mathcal{M}$ is Ricci parallel and has weakly positive sectional curvatures, one gets
\begin{subequations}
\label{subequations "Li–Yau–Hamilton's Harnack inequality"}
	\begin{empheq}[left = {\Inequality^\mathrm{Har}_\textsc{lyh} \empheqlbrace}]{align}
	& \nabla_\mu\nabla_\nu\upsilon + \tfrac{1}{2t}\upsilon g_{\mu\nu} + \nabla_\mu\upsilon(\vec{X}_\nu) + \nabla_\nu\upsilon(\vec{X}_\mu) + \upsilon\vec{X}_\mu\vec{X}_\nu \geqslant 0, \\
	& \nabla_\mu\nabla_\nu\upsilon - \tfrac{\nabla_\mu\upsilon\nabla_\nu\upsilon}{\upsilon} + \tfrac{\upsilon}{2t}g_{\mu\nu} \geqslant 0,
	\end{empheq}
\end{subequations}
on $\mathcal{M}^n \times [0, T)$, for any vector field $\vec{X}$. 
 \end{propositio}
 The Li–Yau–Hamilton's Harnack inequality, or Hamilton's matrix inequality, via Eq. \eqref{subequations "Li–Yau–Hamilton's Harnack inequality"}, is the non-linear equivalent of \eqref{subequations "Li–Yau's Harnack inequality"}, and it is intended as a trace of a full matrix inequality. A summary is in L. Ni \cite{Ni "Monotonicity and Li-Yau-Hamilton Inequalities"}.

\subsubsection{Harnack–Hamilton Estimate for the Ricci Flow}
\label{subsubsection "Harnack–Hamilton Estimate for the Ricci Flow"}

\begin{theorema}[Hamilton's Harnack estimate of the Ricci flow]
Let $g_{\mu\nu}$ be a complete solution with bounded non-negative curvature to the Ricci flow $\frac{\partial}{\partial t}g_{\mu\nu} = -2\Ricci_{\mu\nu}$ on a manifold of dimension $n$, and assume that $g_{\mu\nu}$ has a weakly positive curvature operator, for some time interval, $0 < t < T$, i.e. $t \in (0, T)$, so that
\begin{equation}
	\Riemann_{\mu\nu\xi\varrho}\rotatedw_{\mu\nu}\rotatedw_{\xi\varrho} \geqslant 0,
\end{equation}
for each 2-form $\rotatedw \in \mathscr{C}^\infty$, where $\Riemann_{\mu\nu\xi\varrho}$ is the Riemann curvature tensor in the form of $\binom{0}{4}$-tensor. Let
\begin{equation}
	\tensorP_{\mu\nu\xi} = \nabla_\mu\Ricci_{\nu\xi} - \nabla_{\nu}\Ricci_{\mu\xi}
\end{equation}
be a 3-tensor, and let
\begin{equation}
	\tensorM_{\mu\nu} = \Laplacian\Ricci_{\mu\nu} - \frac{1}{2}\nabla_\mu\nabla_\nu\scalarcurvature + 2\Riemann_{\mu\xi\nu\varrho}\Ricci_{\xi\varrho} - \Ricci_{\mu\xi}\Ricci_{\nu\xi} + \frac{1}{2t}\Ricci_{\mu\nu}
\end{equation}
be a symmetric 2-tensor. Then, for all 1- and 2-forms $\omega, \rotatedw, \in \mathscr{C}^\infty$, respectively, we have
\begin{equation}
\label{equation "Hamilton's Harnack estimate of the Ricci flow"}
	\tensorM_{\mu\nu}\omega_\mu\omega_\nu + 2\tensorP_{\mu\nu\xi}\rotatedw_{\mu\nu}\omega_\xi + \Riemann_{\mu\nu\xi\varrho}\rotatedw_{\mu\nu}\rotatedw_{\xi\varrho} \geqslant 0,
\end{equation}
and, for any $\omega$-type 1-form,
\begin{equation}
	\frac{\partial\scalarcurvature}{\partial t} + \frac{\scalarcurvature}{t} + 2\nabla_\mu\scalarcurvature(\omega_\mu) + 2\Ricci_{\mu\nu}\omega_\mu\omega_\nu \geqslant 0.
\end{equation}
\end{theorema}
 
The proof of the previous propositions, with the central focus of Eqq. \eqref{equation "Classical Harnack inequality"} \eqref{subequations "Li–Yau's Harnack inequality"} \eqref{subequations "Li–Yau–Hamilton's Harnack inequality"} \eqref{equation "Hamilton's Harnack estimate of the Ricci flow"}, exceeds the scope of this Chapter, so we refer to the respective articles mentioned above.

\subsection{On the 3-Manifold with Positive Ricci Curvature Tensor}

Later on we will see a distinguished theorem of Hamilton and its proof.

\subsubsection{Hamilton's Main Theorem, plus Corollary}

\begin{theorema}[Hamilton]
\label{theorema "Hamilton"}
Let $(\mathcal{M}^3, g_0)$ be a compact Riemannian 3-manifold with an initial metric $g_0$ of strictly positive Ricci curvature tensor. Then there exists a unique maximal smooth solution $g_t$, $t \in [0, T)$, to the Ricci flow $\textcyrillic{\textit{ф}}_\textsc{r}$ (Definition \ref{definitio "Ricci flow"}) with $g(0) = g_0$, for all time $t \geqslant 0$ and a $T \leqslant T_\mathrm{max} \in (0, \infty]$. If in addition $t \to \infty$, subsequently $g_t$ converges exponentially fast to a $\mathscr{C}^\infty$ metric $g_\infty(t)$ of constant positive sectional curvature.\footnote{
	Hamilton's theorem in its original form is in \cite[№ 1.1]{Hamilton "Three-manifolds with positive Ricci curvature"}.
	}
\end{theorema}

\begin{corollarium}[Hamilton's special case of the Poincaré conjecture]
Let $\mathcal{M}^3$ be a closed Riemannian 3-manifold with strictly positive Ricci curvature. One gets that (by the main Theorem \ref{theorema "Hamilton"}) it admits a metric of constant positive sectional curvature. If $\mathcal{M}^3$ is simply connected, then it is diffeomorphic to the 3-sphere $\mathbb{S}^3$.
\end{corollarium}

We will pick this up in Section \ref{subsection "Poincaré Conjecture"}.

\subsubsection{Short Time Existence and Uniqueness in Ricci–DeTurck's Strictly Parabolic System}
\label{subsubsection "Short Time Existence and Uniqueness in Ricci–DeTurck's Strictly Parabolic System"}

Hamilton, for the purpose of verifying the validity of the statement \ref{theorema "Hamilton"}, relies \cite{Hamilton "The inverse function theorem of Nash and Moser"} on the \emph{Nash–Moser inverse function theorem} \cite{Nash "The Imbedding Problem for Riemannian Manifolds"} \cite{Moser "A rapidly convergent iteration method and non-linear partial differential equations - I"} \cite{Moser "A rapidly convergent iteration method and non-linear partial differential equations - II"}; in fact, the \emph{weakly parabolic} nature of the Ricci flow does not allow to have recourse to the standard parabolic theory. But we adopt a weakness-free modus, originally devised by D.M. DeTurck \cite{DeTurck "Deforming metrics in the direction of their Ricci tensors"}, to demonstrate a short time existence of $\textcyrillic{\textit{ф}}_\textsc{r}$, which is a \emph{strictly parabolic} version of the Ricci flow thanks to the pullback by diffeomorphisms.

\begin{proof}[Proof of the Theorem \ref{theorema "Hamilton"} (via DeTurck's modus)] 
Let
\begin{equation} 
\label{equation "Initial value problem of the Ricci flow"}
	\textcyrillic{\textit{ф}}_\textsc{r}
	\begin{cases}
	\partial_tg_{\mu\nu}(x, t) \viz \frac{\partial g_{\mu\nu}}{\partial t}(x, t) = -2\Ricci_{\mu\nu}(x, t), \\
	g_{\mu\nu}(0, x) = g_{\mu\nu}(x),
	\end{cases} 
\end{equation}	
be the initial value problem of $\textcyrillic{\textit{ф}}_\textsc{r}$ in the indices form, taking for granted that $g = g_t = g_{\mu\nu}(x, t)$. We have to prove that \eqref{equation "Initial value problem of the Ricci flow"} has a unique solution on $\mathcal{M}^3 \times [0, T)$, with $T > 0$, for $x \in \mathcal{M}^3$.
\enumerationisinitium
\item Let us begin to see why \eqref{equation "Initial value problem of the Ricci flow"}, combined with the Lie derivative $\Liederivative$ of $g$, is a parabolic system of equations, setting a \emph{DeTurck's parabolic system} of equations
\begin{equation}
\label{equation "DeTurck parabolic system"}
	\begin{cases}
	\partial_tg_{\mu\nu}(x, t) \viz \frac{\partial g_{\mu\nu}}{\partial t}(x, t) = -2\Ricci_{\mu\nu}(x, t) + \left[\Liederivative_{\vec{W}}g\right]_{\mu\nu}(x, t), \\
	g_{\mu\nu}(0, x) = g_{\mu\nu}(x),	
	\end{cases}
\end{equation}
with the vector field 
\begin{equation}
	\vec{W} = \vec{W}^\xi(x, t)\frac{\partial}{\partial x^\xi}, \enspace \vec{W}^\xi = g^{\varsigma\tau}\left({\Gamma_{\varsigma\tau}}^\xi - {\Gamma_{\varsigma\tau}}^\xi(0)\right), 
\end{equation}
where ${\Gamma_{\varsigma\tau}}^\xi$ and ${\Gamma_{\varsigma\tau}}^\xi(0)$ are the Christoffel symbols (Section \ref{section "Christoffel Symbols"}) of $g$ in \eqref{equation "Initial value problem of the Ricci flow"} and $g_0$, respectively. For a local coordinate system, we write 
\begin{equation}
\label{equation "Ricci–DeTurck flow"}
	\textcyrillic{\textit{ф}}_{\textsc{rd}\text{e}}
	\begin{cases}
	\partial_t g_{\mu\nu} \viz \frac{\partial g_{\mu\nu}}{\partial t} = -2\Ricci_{\mu\nu} + \nabla_\mu\vec{W}_\nu + \nabla_\nu\vec{W}_\mu, \\
	g(0) = g_0.
	\end{cases}
\end{equation}
The double formula in \eqref{equation "Ricci–DeTurck flow"} is what is called \emph{Ricci–DeTurck flow}, where $\nabla\vec{W}$ is the element of the covariant derivative of the dual 1-form of $\vec{W}$, that is, $\nabla\frac{\partial}{\partial x^\mu}\vec{W}_\mu dx^\mu$, with $\vec{W}_\mu = g_{\mu\xi}\vec{W}^\xi$. This is possible since, for two vector fields $\vec{X}$ and $\vec{Y}$, it is true that $\partial_t g_t(\vec{X}, \vec{Y}) = -2\Ric(g_t)(\vec{X}, \vec{Y}) + g_t\{\nabla_{\vec{X}}\vec{W}(t), \vec{Y}\} + g_t\{\vec{X}, \nabla_{\vec{Y}}, \vec{W}(t)\}$.

Using the acronym \textsc{lot} (which means \textit{lower order terms}) to denote all terms including the first derivatives relating to the metric $g$, we have 
\begin{equation}
	\begin{rcases}
	\nabla_\nu\vec{W}_\mu & = g_{\mu\xi}g^{\varsigma\tau}\partial_\nu{\Gamma_{\varsigma\tau}}^\xi \\
	& = \frac{1}{2}g_{\mu\xi}g^{\varsigma\tau}\partial_\nu\Bigl(g^{\xi\varrho}\left(\partial_\tau g_{\varsigma\varrho} + \partial_\varsigma g_{\tau\varrho} - \partial_\varrho g_{\varsigma\tau}\right)\Bigr) \\
	& = \frac{1}{2}g^{\varsigma\tau}\left(\partial_\nu\partial_\tau g_{\varsigma\mu} + \partial_\nu\partial_\varsigma g_{\tau\mu} - \partial_\nu\partial_\mu g_{\varsigma\tau}\right) 
	\end{rcases}
	+ \textsc{lot}.
\end{equation}
And then $-2\Ricci_{\mu\nu} + \nabla_\mu\vec{W}_\nu + \nabla_\nu\vec{W}_\mu = g^{\xi\varrho}\partial_\xi\partial_\varrho g_{\mu\nu} + \textsc{lot}$, once the Ricci curvature tensor is set as $\Ricci_{\mu\nu} = -\frac{1}{2}g^{\xi\varrho}(\partial_\mu\partial_\nu g_{\xi\varrho} - \partial_\xi\partial_\mu g_{\nu\varrho} - \partial_\xi\partial_\nu g_{\mu\varrho} + \partial_\xi\partial_\varrho g_{\mu\nu}) + \textsc{lot}$. The flow \eqref{equation "Ricci–DeTurck flow"} becomes thus 
\begin{equation}
	\partial_t g_{\mu\nu} \viz \frac{\partial g_{\mu\nu}}{\partial t} = g^{\xi\varrho}\partial_\xi\partial_\varrho g_{\mu\nu} + \textsc{lot},	
\end{equation}
and it is a quasi-linear \emph{strictly parabolic} equation; such is consequently also the the system of Eqq. \eqref{equation "DeTurck parabolic system"} which has a $\mathscr{C}^\infty$ solution on some (short) time interval $0 \leqslant t < T$, under the standard parabolic theory.
\item Let $\textcyrillic{\textit{Д}}_t$ be a 1-parameter family of diffeomorphisms; put\footnote{
	\label{footnote "Mathbold instead of the vector arrow"}
	The symbol $\mathbold{W}$ (\texttt{\textbackslash{mathbold}} command) is the same as the symbol $\vec{W}$ (\texttt{\textbackslash{vec}} command). The choice of a bold type is to avoid the double presence of the arrow and the tilde over the letter. 
	} 
\begin{equation}
\label{equation "1-parameter family of diffeomorphisms"}
	\frac{d\textcyrillic{\textit{Д}}_t}{dt} = \Bigl(\textcyrillic{\textit{Д}}_{t*}\vec{W}(\textcyrillic{\textit{Д}}_t, t) = \tilde{\mathbold{W}}(\textcyrillic{\textit{Д}}_t, t)\Bigr).
\end{equation}
We have to prove that there is a solution of the Ricci flow, as the theorem demands, that is, a solution 
\begin{equation}	
	\tilde{g}_t \viz \tilde{g}(t) = (\textcyrillic{\textit{Д}}_t^*)^{-1}g_t,
\end{equation}
to \eqref{equation "Initial value problem of the Ricci flow"}. Knowing that $g_t = \textcyrillic{\textit{Д}}_t^*\tilde{g}_t$, we calculate
\begin{subequations}
\label{subequations "Series of equations (computation)"}
\begin{align}
	\partial_tg_t & = \Lbrack:\partial_t\textcyrillic{\textit{Д}}_t^*\tilde{g}_t = \textcyrillic{\textit{Д}}_t^*\partial_t\tilde{g}_t:\Rbrack + \partial_t(\textcyrillic{\textit{Д}}_t^*\tilde{g}_t) \\
	& = \Lbrack:\cdots:\Rbrack + \textcyrillic{\textit{Д}}_t^*(\Liederivative_{\tilde{\mathbold{W}}(t)}\tilde{g}_t) \\
	& = \Lbrack:\cdots:\Rbrack + \Liederivative_{(\textcyrillic{\textit{Д}}_{t*})^{-1}\tilde{\mathbold{W}}(t)}\textcyrillic{\textit{Д}}_t^*\tilde{g}_t \\
	& = \Lbrack:\cdots:\Rbrack + \Liederivative_{\vec{W}(t)}g_t.
\end{align}
\end{subequations}
Thank to \eqref{equation "DeTurck parabolic system"}, it is clear that $\partial_tg_t = -2\Ric(g_t) + \Liederivative_{\vec{W}(t)}g_t = -2\textcyrillic{\textit{Д}}_t^*\bigl(\Ric(\tilde{g}_t)\bigr) + \Liederivative_{\vec{W}(t)}g_t$, from which 
\begin{equation}
	\partial_t\tilde{g}_t = -2\Ric(\tilde{g}_t).
\end{equation}	 
This last equation ensures the short time \emph{existence} of $\textcyrillic{\textit{ф}}_\textsc{r}$, for which  there is a solution $\tilde{g}_t$ to \eqref{equation "Initial value problem of the Ricci flow"} on a (short) time interval $[0, T)$, $T > 0$. 
\item It is the turn of the demonstration of \emph{uniqueness}. Firstly, we  define the Christoffel symbols of $g_t = \textcyrillic{\textit{Д}}_t^*\tilde{g}_t$\footnote{
	The metric $g_t = \textcyrillic{\textit{Д}}_t^*\tilde{g}_t$ appears to be a solution to \eqref{equation "DeTurck parabolic system"} if we do the inversion of Eqq. \eqref{subequations "Series of equations (computation)"}.
	} 
as
\begin{equation}
	{\Gamma_{\nu\varrho}}^\xi = \frac{\partial y^\alpha}{\partial x^\nu}\frac{\partial y^\beta}{\partial x^\varrho}\frac{\partial x^\xi}{\partial y^\gamma}\tilde{\Gamma}_{\alpha\beta}{}^\gamma + \frac{\partial x^\xi}{\partial y^\alpha}\frac{\partial^2y^\alpha}{\partial x^\nu\partial x^\varrho},
\end{equation}
where $\tilde{\Gamma}_{\alpha\beta}{}^\gamma$ are the Christoffel symbols of $\tilde{g}_t$.\footnote{
	Do not forget that the connection coefficients of the Levi-Civita connection, in a system of local coordinate, are the Christoffel symbols; this means that the metric connection described by $\tilde{\Gamma}$ acquires true meaning as a Levi-Civita connection of $\tilde{g}_t$.
	}
	The vector field $\vec{W}$ is now
\begin{equation}
	\vec{W}^\xi\frac{\partial}{\partial x^\xi} = g^{\nu\varrho}\left(\frac{\partial y^\alpha}{\partial x^\nu}\frac{\partial y^\beta}{\partial x^\varrho}\frac{\partial x^\xi}{\partial y^\gamma}\tilde{\Gamma}_{\alpha\beta}{}^\gamma + \frac{\partial x^\xi}{\partial y^\alpha}\frac{\partial^2y^\alpha}{\partial x^\nu\partial x^\varrho} - {\Gamma_{\nu\varrho}}^\xi(0)\right)\frac{\partial}{\partial x^\xi}. 
\end{equation}
Let $(\textcyrillic{\textit{Д}}_{t*}\vec{W})\textcyrillic{\textit{г}} = \vec{W}(\textcyrillic{\textit{г}} \circ \textcyrillic{\textit{Д}}_t)$, with $\textcyrillic{\textit{Д}}_t(x) = \left[y^1 (x, t), \mathellipsis, y^n(x, t)\right]$, where $\textcyrillic{\textit{г}}$ is a $\mathscr{C}^\infty$ function; we get
\begin{align}
	(\textcyrillic{\textit{Д}}_{t*}\vec{W})\textcyrillic{\textit{г}} & = \vec{W}^\xi\frac{\partial(\textcyrillic{\textit{г}} \circ \textcyrillic{\textit{Д}}_t)}{\partial x^\xi} = \vec{W}^\xi\frac{\partial\textcyrillic{\textit{г}}}{\partial y^\delta}\frac{\partial y^\delta}{\partial x^\xi} \notag \\
	& = g^{\nu\varrho}\left(\frac{\partial y^\alpha}{\partial x^\nu}\frac{\partial y^\beta}{\partial x^\varrho}\frac{\partial x^\xi}{\partial y^\gamma}\tilde{\Gamma}_{\alpha\beta}{}^\gamma + \frac{\partial x^\xi}{\partial y^\alpha}\frac{\partial^2y^\alpha}{\partial x^\nu\partial x^\varrho} - {\Gamma_{\nu\varrho}}^\xi(0)\right)\frac{\partial\textcyrillic{\textit{г}}}{\partial y^\delta}\frac{\partial y^\delta}{\partial x^\xi} \notag \\
	& = g^{\nu\varrho}\left(\frac{\partial y^\alpha}{\partial x^\nu} \frac{\partial y^\beta}{\partial x^\varrho}\tilde{\Gamma}_{\alpha\beta}{}^\delta + \frac{\partial^2y^\delta}{\partial x^\nu\partial x^\varrho} - {\Gamma_{\nu\varrho}}^\xi(0)\frac{\partial y^\delta}{\partial x^\xi}\right)\frac{\partial\textcyrillic{\textit{г}}}{\partial y^\delta}.  
\end{align}
In consequence, Eq. \eqref{equation "1-parameter family of diffeomorphisms"} becomes
\begin{equation}
	\begin{cases}
	\partial_t y^\delta = g^{\nu\varrho}\left(\frac{\partial^2y^\delta}{\partial x^\nu\partial x^\varrho} + \tilde{\Gamma}_{\gamma\beta}{}^{\delta}\frac{\partial y^\beta}{\partial x^\nu}\frac{\partial y^\gamma}{\partial x^\varrho} -  {\Gamma_{\nu\varrho}}^\xi(0)\frac{\partial y^\delta}{\partial x^\xi}\right) \\
	y^\delta(x, 0) = x^\delta,
	\end{cases}
\end{equation}
which states that Eq. \eqref{equation "1-parameter family of diffeomorphisms"} is a quasi-linear \emph{strictly parabolic} system with a \emph{unique} solution in $\mathscr{C}^\infty$ sense and for a short time, in the awareness that 
\begin{align}
	& g_{\nu\varrho} = \tilde{g}_{\alpha\beta}\frac{\partial y^\alpha}{\partial x^\nu}\frac{\partial y^\beta}{\partial x^\varrho}, \\
	& g^{\nu\varrho} = (g_{\nu\varrho})^{-1}.
\end{align}
\enumerationisfinis
\end{proof}

\section{Ricci Solitons: a Synoptic Classification}
\label{section "Ricci Solitons: a Synoptic Classification"}

There is a classification, based on the shape and other flow-evolutionary characteristics, which allows the Ricci flow to be classified into \emph{solitonic} categories.

\begin{margo}[Why solitons?]
The reference, in this geometro-topological context, is inspired by a physical phenomenon: a soliton is a solitary wave which \emph{does not change its shape} during propagation, according to the \emph{Korteweg–de Vries equation} \cite{Korteweg and de Vries "On the Change of Form of Long Waves advancing in a Rectangular Canal and on a New Type of Long Stationary Waves"} in the model of wave motions on the shallow water surface. The geometric soliton, as is the Ricci soliton, does something similar: it evolves, but maintains its original shape, establishing diffeomorphism symmetries for any specific flow. \margosymbol
\end{margo}

\begin{definitiones}[Ricci solitons]
\label{definitiones "Ricci solitons"}
~\enumerationisinitium
\item \emph{Via ordinary differential equation}. Let $\mathcal{M}$ be a Riemannian manifold, and $g_0$ the initial metric. The smooth space $(\mathcal{M}, g_0)$ is called \emph{Ricci soliton} $\textcyrillic{\textit{ф}}_\textsc{r}^\mathrm{s}$ if, for a constant $\lambda \in \mathbb{R}$ and a vector field $\vec{X}$, there is an equation like this 
\begin{equation}
	\Ric(g_0) + \tfrac{1}{2}\Liederivative_{\vec{X}}g_0 = \lambda g_0,
\end{equation}
under which if
\begin{align*}
	& \lambda > 0, \textcyrillic{\textit{ф}}_\textsc{r}^\mathrm{s} \text{ is said \emph{shrinking}}, \\
	& \lambda = 0, \textcyrillic{\textit{ф}}_\textsc{r}^\mathrm{s} \text{ is said \emph{steady}}, \\
	& \lambda < 0, \textcyrillic{\textit{ф}}_\textsc{r}^\mathrm{s} \text{ is said \emph{expanding}},
\end{align*}
at a time $t_0$, where $\Liederivative$ is the Lie derivative. So, for the sake of completeness, a Ricci soliton should be written as $\textcyrillic{\textit{ф}}_\textsc{r}^\mathrm{s} = (\mathcal{M}, g_0, \vec{X}, \lambda)$.
\item \emph{Via pullback of the metric}. 
\label{item "Via pullback of the metric"}
Choose a 1-parameter family of diffeomorphisms of the form $\textcyrillic{\textit{Д}}_t \colon \mathcal{M} \to \mathcal{M}$, i.e. of invertible functions between smooth manifolds, for some time evolution; and put 
\begin{equation}
	\frac{\partial}{\partial t}\textcyrillic{\textit{Д}}_t(x) = \frac{1}{1 - 2\lambda t}\vec{X}|_{\textcyrillic{\textit{Д}}_t(x)},
\end{equation}
$x \in \mathcal{M}$. Then a solution $g_t$ of $\partial_t g = -2 \Ric$ is called \emph{Ricci soliton}, such that $\textcyrillic{\textit{ф}}_\textsc{r}^\mathrm{s} = (\mathcal{M}, g_0)$, if there is a pullback of $g_0$,
\begin{equation}
\label{equation "Metric and its pullback"}
	g_t = (1 - 2\lambda t)\textcyrillic{\textit{Д}}_t^*(g_0). 
\end{equation}
\item 
\label{item "Potential function, and gradient soliton"}
\emph{Potential function, and gradient soliton}. Let 
\begin{equation}
	\rotatedupsilon \colon \mathcal{M} \to \mathbb{R}, \enspace \rotatedupsilon \in \mathscr{C}^\infty(\mathcal{M}), 
\end{equation}
be a \emph{potential function} of $\textcyrillic{\textit{ф}}_\textsc{r}$ on $\mathcal{M}$. Setting 
\begin{equation}
	-2\Ric(g_0) = \Liederivative_{\vec{X}}g_0 - 2\lambda g_0, 
\end{equation}
then
\begin{subequations}
\label{subequations "Ricci tensor and potential function"}
\begin{align}
	\Ric_\rotatedupsilon & = \lambda g_0, \text{ where } \Ric_\rotatedupsilon = \Ric(g_0) + \mathit{Hes}(\rotatedupsilon), \\
	\Ric(g_0) + \mathit{Hes}(\rotatedupsilon) & = \lambda g_0, \\ 
	\label{align "Ricci tensor with gradient of the function"}
	\Ric(g_0) + \nabla^2(\rotatedupsilon) & = \lambda g_0,
\end{align}
\end{subequations}
which are three equivalent forms, in which $\mathit{Hes}$ is the Hessian \cite{Hesse "Uber die Elimination der Variabeln aus drei algebraischen Gleichungen vom zweiten Grade mit zwei Variabein"} of $\rotatedupsilon$. A Ricci soliton $\textcyrillic{\textit{ф}}_\textsc{r}^\mathrm{s} = (\mathcal{M}, g_0, \vec{X}, \lambda)$ is known as \emph{gradient Ricci soliton} $\textcyrillic{\textit{ф}}_\textsc{r}^\mathrm{s} = (\mathcal{M}, g_0, \rotatedupsilon, \lambda)$, if $\vec{X} = \nabla\rotatedupsilon$ is both a vector potential field and the gradient of $\rotatedupsilon$. Note. When the $\mu\nu$-indices are expressed, Eq. \eqref{align "Ricci tensor with gradient of the function"} becomes 
\begin{equation}
	\Ricci_{\mu\nu} + \nabla_\mu\nabla_\nu(\rotatedupsilon) = \lambda g_{\mu\nu}.
\end{equation}
\definitiosymbol
\enumerationisfinis
\end{definitiones}

The three Sections that follow provide some examples of Ricci solitons.

\subsection[Shrinkers $(\normalfont{\textcyr{\textit{ф}}}_\textsc{r}^\mathrm{s})_{\lambda > 0}$]{Shrinkers $\protect\pseudobold{(\normalfont{\textcyr{\textit{ф}}}_\textsc{r}^\mathrm{s})_{\lambda > 0}}$}
\label{subsection "Shrinkers"}

\subsubsection{Gradient Shrinker with Potential Function} 

If $g_t = (T - t)\textcyrillic{\textit{Д}}_t^*(g_0)$, compare with Eq. \eqref{equation "Metric and its pullback"}, is a gradient shrinking soliton with potential $\rotatedupsilon$, see point \ref{item "Potential function, and gradient soliton"} in Definitions \ref{definitiones "Ricci solitons"}, the function $\rotatedupsilon$ meets these statements:
\begin{align}
	& \Bigl(\Ric(g_0) + \mathit{Hes}(\rotatedupsilon) = \Ric(g_0) + \nabla^2(\rotatedupsilon)\Bigr) - \frac{g_0}{2\tau} = 0, \\
	& \scalarcurvature + \Laplacian\rotatedupsilon - \frac{n}{2\tau} = 0, \\
	& \partial_t\rotatedupsilon = |\nabla\rotatedupsilon|^2 = -\partial_\tau\rotatedupsilon, \text{ where } \tau = -t.
\end{align}

\subsubsection{Shrinker from an Einstein Manifold} 

A shrinking Ricci soliton is an extension of an Einstein manifold (Scholium \ref{scholium "Einstein manifold"}), since the Ricci curvature tensor is \emph{invariant} under uniform dilation, or scaling, of the metric. We affirm that $g_0$ is an \emph{Einstein metric} if 
\begin{equation}
	\Ric(g_0) = \lambda g_0,
\end{equation}
for $\lambda \in \mathbb{R}$, and
\begin{equation}
\label{equation "$g$-metric in the context of an Einstein manifold"}
	g_t = (1 - 2\lambda t)g_0
\end{equation}
is a solution to $\textcyrillic{\textit{ф}}_\textsc{r}$ \eqref{equation "unnormalized Ricci flow"} \eqref{equation "normalized Ricci flow"} with $g_0$. 

\begin{scholium}[Einstein manifold]
\label{scholium "Einstein manifold"}
A (pseudo-)Riemannian manifold $(\mathcal{M}, g)$ is called an \emph{Einstein manifold}, and the metric $g$ is an \emph{Einstein metric}, see e.g. \cite{Anderson "A Survey of Einstein Metrics on 4-manifolds"}, if 
\begin{subequations}
\label{subequations "Einstein equations for Riemannian metric"}
	\begin{empheq}[left = {\empheqlbrace}]{align}
	& \Ric(v, w) = \lambda\langle v, w\rangle, \\
	& \Ric(g) = \lambda g,
	\end{empheq}
\end{subequations}
for all tangent vectors $v, w \in \mathring{\mathcal{T}}\mathcal{M}$, and for some constant $\lambda \in \mathscr{C}^\infty(\mathcal{M})$, with $\lambda = \frac{1}{n}\scalarcurvature$, where $n$ is the dimension of $\mathcal{M}$, ergo $\Ric = \frac{1}{n}\scalarcurvature(g)$. \scholiumsymbol
\end{scholium}

\subsubsection[Gaussian $(\lambda > 0)$-Soliton]{Gaussian $\protect\pseudobold{(\lambda > 0)}$-Soliton} 
\label{subsubsection "Gaussian Soliton"} 

The form of it is 
\begin{equation}
	\textcyrillic{\textit{ф}}_\textsc{r}^\mathrm{s} = \left(\mathbb{R}^n, g_\mathbb{E}, \rotatedupsilon(x) = \frac{\lambda > 0}{2}|x|^2\right),
\end{equation}
cf. G. Perelman \cite[sec. 2.1]{Perelman "The entropy formula for the Ricci flow and its geometric applications"} for the first appearance of this soliton.

\subsubsection{Shrinking Sphere} 

We are talking about an object in $2\mathrm{D}$ or in a higher dimension, $\textcyrillic{\textit{ф}}_\textsc{r}^\mathrm{s} = (\mathbb{S}^{n \geqslant 2}, g_{\mathbb{S}^n})$.

\subsubsection{Shrinking Cylinder} 

Its form is
\begin{equation}
	\textcyrillic{\textit{ф}}_\textsc{r}^\mathrm{s} = \left(\mathbb{S}^{n - 1} \times \mathbb{R}, g_t = (n - 2)\sqrt{t^2}g_{\mathbb{S}^{n - 1}} + d\rho^2\right),
\end{equation}
with $t \in (-\infty, 0)$, $n \geqslant 3$, or its $\mathbb{Z}_2 \viz \mathbb{Z}/2\mathbb{Z}$ quotient; the Ricci tensor here is 
\begin{equation}
	\Ric(g_t) = (n - 2)g_{\mathbb{S}^{n - 1}} = \frac{g_t}{2\sqrt{t^2}} - \frac{d\rho^2}{2\sqrt{t^2}}.
\end{equation}
The cylindrical-like shrinking Ricci soliton is, more generally, a Riemannian $\mathscr{C}^\infty$ manifold satisfying
\begin{equation}
	\Bigl(\Ric(g_0) + \mathit{Hes}(\rotatedupsilon) = \Ric(g_0) + \nabla^2(\rotatedupsilon)\Bigl) = \frac{1}{2}g_{\mathbb{S}^{n - 1}}.
\end{equation}

\subsubsection{Kähler–Ricci Shrinking Gradient Soliton} 

Identified and studied by N. Koiso \cite{Koiso "On Rotationally Symmetric Hamilton's Equation for Kahler-Einstein Metrics"} and H.-D. Cao \cite{Cao "Existence of Gradient Kahler-Ricci Solitons"} \cite{Cao "Limits of solutions to the Kahler-Ricci flow"}, they are constructed directly on the \emph{Kähler–Ricci flow},
\begin{subequations}
\label{align "Kähler–Ricci flow"}
\begin{align}
	& \textcyrillic{\textit{ф}}_\textsc{kr}^\mathrm{unn} = \frac{\partial g_{\mu\bar{\nu}}}{\partial t} = -\Ricci_{\mu\bar{\nu}}, \\
	& \textcyrillic{\textit{ф}}_\textsc{kr}^\mathrm{nor} = \frac{\partial g_{\mu\bar{\nu}}}{\partial t} = -\Ricci_{\mu\bar{\nu}} + g_{\mu\bar{\nu}},
\end{align}
\end{subequations}
this is, the Ricci flow on a Kähler manifold \cite{Kahler "Uber eine bemerkenswerte Hermitesche Metrik"} (see Scholium \ref{scholium "Kähler manifold, and almost complex structure"}). The metric 
\begin{equation}	
	g_{\mu\bar{\nu}} \viz g_{\mu\bar{\nu}}(t), \enspace t \geqslant 0,
\end{equation}	
represents a solution to $\textcyrillic{\textit{ф}}_\textsc{kr}$ in \eqref{align "Kähler–Ricci flow"} under a 1-parameter family of biholomorphisms. For a holomorphic vector field $\vec{X} = \vec{X}^\mu$, we say that a \emph{Kähler–Ricci soliton} and a \emph{gradient Kähler–Ricci soliton} are formed when, respectively, $\Ricci_{\mu\bar{\nu}} = \vec{X}_{\mu,\bar{\nu}} + \vec{X}_{\bar{\nu},\mu}$ and $\vec{X}$ is the gradient of a potential function $\rotatedupsilon$ (see above) such that 
\begin{align}
	\text{for } \textcyrillic{\textit{ф}}_\textsc{kr}^\mathrm{unn}
	& \begin{cases}
	\Ricci_{\mu\bar{\nu}} = \rotatedupsilon_{,\mu\bar{\nu}}, \\
	\rotatedupsilon_{,\mu\nu} = 0,
	\end{cases} \\
	\text{for } \textcyrillic{\textit{ф}}_\textsc{kr}^\mathrm{nor}
	& \begin{cases}
	\label{equation "Equation in the Kähler–Ricci soliton context for the normalized case"}
	\Ricci_{\mu\bar{\nu}} - g_{\mu\bar{\nu}} = \rotatedupsilon_{,\mu\bar{\nu}}, \\
	\rotatedupsilon_{,\mu\nu} = 0.
	\end{cases}
\end{align}
Eq. \eqref{equation "Equation in the Kähler–Ricci soliton context for the normalized case"} is for the condition on a compact Kähler space with positive first Chern class $\mathring{\mathscr{C}}_1(\mathcal{M})$ \cite{Chern "Characteristic classes of Hermitian Manifolds"}. For more details on the local and global generality of gradient Kähler–Ricci solitons, see Bryant \cite{Bryant "Gradient Kahler Ricci solitons"}.

\begin{scholium}[Kähler manifold, and almost complex structure]
\label{scholium "Kähler manifold, and almost complex structure"}
~\enumerationisinitium
\item A Kähler manifold is a complex structure but it can be defined as a Riemannian manifold of a special type. More accurately, a Kähler structure on a Riemannian manifold is a \emph{symplectic manifold} $(\mathcal{M}, \omega_\mathrm{s}) = (\mathbb{R}^{2n}, \omega_\mathrm{s})$ endowed with an integrable almost complex $\mathcal{J}_{\mathbb{C}|}$-structure compatible with a symplectic form $\omega_\mathrm{s}$, i.e. a non-degenerate closed real-valued differential 2-form (cf. Definition \ref{definitio "Hamiltonian vector field or symplectic gradient"}), the metric of which is Kählerian. From a Riemannian viewpoint, $\mathcal{J}_{\mathbb{C}|}$ is on a real manifold $\mathcal{M}$ of dimension $2n$.
\item The symbol $\mathcal{J}_{\mathbb{C}|}$ denotes an \emph{almost complex structure},
\begin{align}
	& 
	\label{align "Almost complex structure on the tangent spaces"}
	x \mapsto \mathcal{J}_{\mathbb{C}|} \colon \mathcal{T}_x\mathcal{M} \xrightarrow{\text{linear map}} \mathcal{T}_x\mathcal{M}, \enspace \mathcal{J}_{\mathbb{C}|}^2 = -\id_{\mathcal{T}_x\mathcal{M}},\\
	& 
	\label{align "Almost complex structure on the tangent bundle"}
	\mathcal{J}_{\mathbb{C}|} \colon \mathring{\mathcal{T}}\mathcal{M} \xrightarrow{\text{linear map}} \mathring{\mathcal{T}}\mathcal{M}, \enspace \mathcal{J}_{\mathbb{C}|}^2 = -\id_{\mathring{\mathcal{T}}\mathcal{M}},
\end{align}
coinciding with a smooth field of complex structures on the tangent spaces \eqref{align "Almost complex structure on the tangent spaces"}, or even with an automorphism on the tangent bundle \eqref{align "Almost complex structure on the tangent bundle"}. \scholiumsymbol
\enumerationisfinis
\end{scholium}

\subsubsection{Kähler–Ricci Soliton on Toric Fano Manifold} 

X.-J. Wang \& X. Zhu \cite{Wang and Zhu "Kahler-Ricci solitons on toric manifolds with positive first Chern class"} show that a Kähler–Ricci soliton exists on any \emph{Kähler geometry of toric manifolds} with $\mathring{\mathscr{C}}_1 > 0$, and that, iff Futaki invariant \cite{Futaki "An obstruction to the existence of Einstein Kahler metrics"} vanishes, the Kähler–Ricci soliton exists on \emph{toric Fano manifold} \cite{Fano "Sulle varieta algebriche a tre dimensioni aventi tutti i generi nulli} \cite{Fano "Su alcune varieta algebriche a tre dimensioni razionali e aventi curve-sezioni canoniche"}\footnote{
	Brutally said, an \emph{algebraic variety} (set of solutions to polynomial equations over $\mathbb{R}$ or $\mathbb{C}$ numbers) and a \emph{manifold} (topological space) have terms of coincidence but they are distinguished by the absence of singular points in the topological space.
	} 
with a \emph{Kähler–Einstein metric}. 

\begin{scholium}[Kähler toric manifold]
We remember that a Kähler toric manifold is a closed connected Kähler $2n$-space 
\begin{equation}
	\mathcal{M}_\torus^{2n} \viz (\mathcal{M}, \omega_\mathrm{s}, \torus^n, \mathcal{J}_{\mathbb{C}|})
\end{equation}
having a Hamiltonian holomorphic map 
\begin{equation}
	\textgreek{\textit{ο} } \colon \torus^n \to \Diff(\mathcal{M}_\torus^{2n})
\end{equation}	
of the real $n$-torus. We remind that $\mathcal{M}_\torus^{2n}$, because of its structure, is locally symplectomorphic to $\bigl(\mathbb{R}^{2n}, (\omega_\mathrm{s})_0\bigr)$. \scholiumsymbol
\end{scholium}

\begin{margo}[Kähler–Einstein on $\mathbb{CP}^n\#k\overbar{\mathbb{CP}}^n$, and on toric Fano 3- and 4-folds]
Among the first investigations in this direction, are to be mentioned the articles of G. Tian \& S.-T. Yau \cite{Tian and Yau "Kahler-Einstein metrics on complex surfaces with $C_1 > 0$"} and Y.-T. Siu \cite{Siu "The existence of Kahler-Einstein metrics on manifolds with positive anticanonical line bundle and a suitable finite symmetry group"}, in which there is a proof of the existence of Kähler–Einstein structures, with $\lambda > 0$ and $\mathring{\mathscr{C}}_1 > 0$, on any differential manifold of type $\mathbb{CP}^n$, that is, specifically,
\begin{equation*}
	\mathbb{CP}^1 \times \mathbb{CP}^1, \text{ or } \mathbb{CP}^2\#k\overbar{\mathbb{CP}}^2, \text{ for } 3 \leqslant k \leqslant 8,
\end{equation*}
i.e. over complex projective spaces, by blowing up $\mathbb{CP}^2$ at generic $k$-points, and then, by C. Real \cite{Real "Metriques d'Einstein-Kahler sur des varietes a premiere classe de Chern positive"}, for $n > 2$, $k = n + 1$. 

Kähler–Einstein metrics on toric Fano 3- and 4-folds (with vanishing of the Futaki invariant) are analyzed and demonstrated in T. Mabuchi \cite{Mabuchi "Einstein-Kahler forms Futaki invariants and convex geometry on toric Fano varieties"}, for Fano 3-folds \cite{Batyrev "Toroidal Fano 3-folds"}, and in Y. Nakagawa \cite{Nakagawa "Einstein-Kahler toric Fano fourfolds"} \cite{Nakagawa "Classification of Einstein-Kahler toric Fano fourfolds"} and Batyrev \& Selivanova \cite{Batyrev and Selivanova "Einstein-Kahler Metrics on Symmetric Toric Fano Manifolds"}, for Fano 4-folds \cite{Batyrev "On the classification of toric Fano 4-folds"} and symmetric toric Fano manifolds. \margosymbol
\end{margo}

\subsubsection{Feldman–Ilmanen–Knopf's Kähler–Ricci Shrinking (Gradient) Solitons} 
\label{subsubsection "Feldman–Ilmanen–Knopf's Kähler–Ricci Shrinking (Gradient) Solitons"} 

The \textsc{fik'kr} \cite{Feldman Ilmanen and Knopf "Rotationally Symmetric Shrinking and Expanding Gradient Kahler-Ricci Solitons"} solitons lie on complex line bundles over a $(n - 1)$-dimensional complex projective space $\mathbb{CP}^{n - 1}$, with $n \geqslant 2$. Some of these have an initial or final state equivalent to a metric cone $\mathbb{C}^n/\mathbb{Z}_k$, $k > n$, and its quotient flat metric. The part of interest of the \textsc{fik'kr} solitons is that they evolve for $t < 0$, as a \emph{non-compact Ricci flow} with shrinking and $\mathscr{C}^\infty$-flowing behavior, then become a cone at $t = 0$, and, finally, expand self-similarly for $t > 0$ (smooth evolution in space-time), until the formation of a spatio-temporal singular point causing the \emph{flow-down} (or \emph{blow-down}) of a $\mathbb{CP}^{n - 1}$ space.
 
\subsection[Steadies $(\normalfont{\textcyr{\textit{ф}}}_\textsc{r}^\mathrm{s})_{\lambda = 0}$]{Steadies $\protect\pseudobold{(\normalfont{\textcyr{\textit{ф}}}_\textsc{r}^\mathrm{s})_{\lambda = 0}}$}
\label{subsection "Steadies"}

\subsubsection{Gradient Steady Soliton} 

If $g_t = \textcyrillic{\textit{Д}}_t^*(g_0)$ represents a gradient steady solitonic wave with potential $\rotatedupsilon$, see point \ref{item "Potential function, and gradient soliton"} in Definitions \ref{definitiones "Ricci solitons"}, the function $\rotatedupsilon$ meets these statements: 
\begin{align}
	& \Ric(g_0) + \mathit{Hes}(\rotatedupsilon) = \Ric(g_0) + \nabla^2(\rotatedupsilon) = 0, \\
	& \scalarcurvature + \Laplacian\rotatedupsilon = 0, \\
	& \partial_t\rotatedupsilon = |\nabla\rotatedupsilon|^2.
\end{align}

\subsubsection{Steady Soliton from an Einstein Manifold} 

A steady Ricci soliton is a natural extension of an Einstein manifold, see Eqq. \eqref{subequations "Einstein equations for Riemannian metric"} \eqref{equation "$g$-metric in the context of an Einstein manifold"}.

\subsubsection{Gaussian Steady Soliton} 

It corresponds to Euclidean $n$-space $(\mathbb{R}^n, g_\mathbb{E})$, with flat metric (curvature zero) and stationary Ricci flow, with $\lambda = 0$.

\subsubsection{Cigar Soliton, or Euclidean Witten's Black Hole} 

This space, in mathematics literature is called \emph{cigar soliton}, see L.-F. Wu \cite{Wu "The Ricci Flow on Complete R^2"}, but in physics literature it is known as \emph{(Euclidean) Witten's black hole} \cite{Witten "String theory and black holes"} (Margo \ref{margo "Witten's semi-infinite cigar as a 2D black hole"}). The cigar soliton is given bay the $\mathbb{R}^2$-space, 
\begin{subequations}
\label{subequations "Cigar soliton"}
\begin{align}
	\textcyrillic{\textit{ф}}_\textsc{r}^\mathrm{s} \viz \textcyrillic{\textit{ф}}_\textsc{r}^{\mathrm{cig}|\mathbb{R}^2} & = \left\{\Lbrack:\mathbb{R}^2, \left(g_\mathrm{cig} \viz g_0\right):\Rbrack = \rho^2(dx^2 + dy^2)\right\}, \\
	& = \left\{\Lbrack:\cdots:\Rbrack = \frac{dx^2 + dy^2}{1 + x^2 + y^2}\right\},
\end{align}
\end{subequations}
with $\rho^2 = \frac{1}{1 + x^2 + y^2}$, $dx^2 = dx \otimes dx$, which is \emph{asymptotically equivalent to a cylinder} at spatial infinity. When there is a time dependence, the metric in \eqref{subequations "Cigar soliton"} becomes $g_\mathrm{cig} \viz g_t = \frac{dx^2 + dy^2}{e^{4t} + x^2 + y^2}$, $t \in (-\infty, +\infty)$.

The Ricci curvature tensor, with the initial metric $g_0$, can therefore take the Gaussian form, 
\begin{equation}
	\Ric(g_0) = \kappa(g_0), \enspace \kappa = \frac{2}{1 + x^2 + y^2},
\end{equation}
where $\kappa$ is the Gaussian curvature. Then 
\begin{equation}
	\Liederivative_{\left(-2 \cdot x\frac{\partial}{\partial x} + y\frac{\partial}{\partial y}\right)}g_0 = -\left(\frac{4}{1 + x^2 + y^2}\right)g_0.
\end{equation}

\begin{scholium}
If we use the metric $g_t = \textcyrillic{\textit{Д}}_t^*(g_0)$, cf. point \ref{item "Via pullback of the metric"} in Definitions \ref{definitiones "Ricci solitons"}, for the cigar-shaped space we have the map $\textcyrillic{\textit{Д}}_t \colon \mathbb{R}^2 \to \mathbb{R}^2$ in accordance with the determination 
\begin{equation}
	\textcyrillic{\textit{Д}}_t(x, y) = \left(\frac{x}{\sqrt{e^t}}, \frac{y}{\sqrt{e^t}}\right), \enspace x, y \in \mathbb{R}^2,
\end{equation}
wherefore $g^\mathrm{cig}_t = \textcyrillic{\textit{Д}}_t^*g^\mathrm{cig}_0$. \scholiumsymbol
\end{scholium}

\begin{margo}[Witten's semi-infinite cigar as a 2D black hole]
\label{margo "Witten's semi-infinite cigar as a 2D black hole"}
Witten imagines \cite[pp. 315-316]{Witten "String theory and black holes"} a semi-infinite cigar, which, in a polar coordinate system, is written as
\begin{equation}
	ds^2 = \tfrac{1}{1 + \rho^2} \cdot d\rho^2 + \rho^2d\theta^2 = \frac{d\rho^2 + \rho^2d\theta^2}{1 + \rho^2}, 
\end{equation}
$\rho$ and $\theta$ are the radial and angular coordinates; and he interprets it as a (Euclidean) black hole in a space-time of dimension 2 within a framework of correspondence between an exact conformal field theory and an $SL_2(\mathbb{R})/U(1)$ gauged Wess–Zumino–Witten model \cite{Wess and Zumino "Consequences of anomalous ward identities"} \cite{Witten "Global aspects of current algebra"} \cite{Witten "Non-abelian bosonization in two dimensions"}; see \cite{Lambert and Suneeta "Stability analysis of the Witten black hole (cigar soliton) under world-sheet renormalization group flow"}. \margosymbol
\end{margo}

\subsubsection{Bryant Soliton, and the Warped Product} 

R.L. Bryant \cite{Bryant "Ricci flow solitons in dimension three with $SO(3)$-symmetries"} finds a steady Ricci soliton on $\mathbb{R}^3$ rotationally symmetric with $SO_3(\mathbb{R})$-symmetry and $\kappa > 0$ (positive sectional curvature), which is \emph{asymptotically equivalent to a paraboloid}, see S. Brendle \cite{Brendle "Rotational symmetry of self-similar solutions to the Ricci flow"} \cite{Brendle "Rotational symmetry of Ricci solitons in higher dimensions"}. 

\begin{exemplum}[Bryant-like soliton]
The Bryant's radially symmetric Ricci soliton in $3\mathrm{D}$ can be conveniently generalized to $\mathbb{R}^n$-space. Let $g_{\mathbb{S}^{n - 1}}$ be the standard metric on the unit sphere $\mathbb{S}^{n - 1}$ in $\mathbb{R}^n$. A Bryant soliton is constructed as a \emph{steady Ricci soliton warped product} $(0, \infty) \times_\textcyrillic{\textit{в}} \mathbb{S}^{n - 1}$, $n > 1$, for a radial warping function $\textcyrillic{\textit{в}}$. By adopting the warped product metric 
\begin{equation}
\label{equation "Warped product metric"}
	g_\textcyrillic{\textit{в}} = d\rho^2 + \textcyrillic{\textit{в}}(\rho)^2g_{\mathbb{S}^{n - 1}},
\end{equation} 
with the radial coordinate $\rho$, it happens that 
\enumerationisinitium
\item the Ricci tensor of $g_\textcyrillic{\textit{в}}$ \eqref{equation "Warped product metric"} is
\begin{equation}
	\Ric(g_\textcyrillic{\textit{в}}) = -(n - 1) \frac{\ddot{\textcyrillic{\textit{в}}}}{\textcyrillic{\textit{в}}}d\rho^2 + \left\{(n - 2)(1 - \dot{\textcyrillic{\textit{в}}}^2) - \textcyrillic{\textit{в}}\ddot{\textcyrillic{\textit{в}}}\right\}g_{\mathbb{S}^{n - 1}},
\end{equation}
\item and, given some function $\rotatedupsilon$ of $\rho$, the Hessian of $\rotatedupsilon$ in relation to $g_\textcyrillic{\textit{в}}$ \eqref{equation "Warped product metric"} is
\begin{equation}
	\nabla\nabla\rotatedupsilon = \ddot{\rotatedupsilon}(\rho)d\rho^2 + \textcyrillic{\textit{в}}\dot{\textcyrillic{\textit{в}}}\dot{\rotatedupsilon}g_{\mathbb{S}^{n - 1}}.
\end{equation}
\enumerationisfinis
Then
\begin{equation}
	\Ric(g_\textcyrillic{\textit{в}}) + \nabla\nabla\rotatedupsilon = 0 \xrightarrow{\text{turns into}}
	\begin{cases}
	\ddot{\rotatedupsilon} = (n - 1)\frac{\ddot{\textcyrillic{\textit{в}}}}{\textcyrillic{\textit{в}}}, \\
	\textcyrillic{\textit{в}}\dot{\textcyrillic{\textit{в}}}\dot{\rotatedupsilon} = -(n - 2)(1 - \dot{\textcyrillic{\textit{в}}}^2) + \textcyrillic{\textit{в}}\ddot{\textcyrillic{\textit{в}}}.
	\end{cases}
\end{equation}
From here it is possible to sketch out a steady gradient Ricci soliton, whose potential function $\rotatedupsilon$ is exclusively on $\rho$. \exemplumsymbol
\end{exemplum}

Generalizations of the Bryant model are in 
\enumerationisinitium
\item T. Ivey \cite{Ivey "New examples of complete Ricci solitons"}, where 1-parameter family of solutions with complete non-compact Ricci solitons is shown; 
\item A.S. Dancer \& M.Y. Wang \cite{Dancer and Wang "Some New Examples of Non-Kahler Ricci Solitons"}, with non-Kähler complete steady gradient Ricci solitons; and \cite{Dancer Wang "On Ricci solitons of cohomogeneity one"}, with the production of complete shrinking, steady and expanding Kähler–Ricci solitons foliated by hypersurfaces, i.e. by circle bundles over an arbitrary product of Fano Kähler–Einstein spaces, or over a coadjoint orbit of a compact connected semi-simple Lie group;
\item Betancourt de la Parra, Dancer \& Wang \cite{Betancourt de la Parra Dancer and Wang "A Hamiltonian approach to the cohomogeneity one Ricci soliton equations"}, with the exhibition of non-Kähler Ricci solitons in $5\mathrm{D}$.
\enumerationisfinis

\subsubsection[Non-collapsed Non-Kähler \& Non-Einstein Steady Solitons in Dimensions $\geqslant$ 4D]{Non-collapsed Non-Kähler \& Non-Einstein Steady Solitons in Dimensions $\protect\pseudobold{\geqslant}$ 4D} 

In A. Appleton \cite{Appleton "A family of non-collapsed steady Ricci solitons in even dimensions greater or equal to four"} is constructed a family of non-collapsed Ricci steady solitons of dimension $\geqslant 4$, which are \emph{non-Kähler} and \emph{non-Einstein} but they exist on complex line bundles $\mathring{\mathcal{L}}_\mathbb{C}(k)$ (i.e., complex vector bundles of rank 1) of $\mathbb{CP}^1$ over Kähler–Einstein spaces with $\scalarcurvature > 0$. The Appleton solitons in dimension 4

· lie on $\mathbb{R}_{> 0} \times \mathbb{S}^3/\mathbb{Z}_k$ when a 2-sphere $\mathbb{S}^2$ is placed at the origin, where $\mathbb{Z}_k$ is a cyclic group of order $k \geqslant 3$, 

· are asymptotic to the quotient of the 4-dimensional Bryant soliton by $\mathbb{Z}_{k \geqslant 3}$.

\begin{scholium}[$\kappa$-non-collapsing condition]
\label{scholium "k-non-collapsing condition"}
Let $\textcyrillic{\textit{ф}}_\textsc{r}^\mathrm{s} = (\mathcal{M}, g)$. Taking a ball $\mathbb{B}(x, \rho) \subset \mathcal{M}$ of dimension $n$, with $\rho > 0$, and putting $|\Rie(y)| \leqslant \rho^{-2}$ on $\mathbb{B}(x, \rho)$ such that 
\begin{equation}
	\frac{\volume\bigl(\mathbb{B}(x, \rho)\bigr)}{\rho^n} \geqslant \kappa, 
\end{equation}
for any $y \in \mathbb{B}(x, \rho)$, where $\Rie$ is the Riemann curvature tensor, the $\kappa$-non-collapsing condition occurs if exists a constant curvature $\kappa > 0$. \scholiumsymbol
\end{scholium}

\subsection[Expanders $(\normalfont{\textcyr{\textit{ф}}}_\textsc{r}^\mathrm{s})_{\lambda < 0}$]{Expanders $\protect\pseudobold{(\normalfont{\textcyr{\textit{ф}}}_\textsc{r}^\mathrm{s})_{\lambda < 0}}$}

\subsubsection{Expander Soliton} 

For an expanding solitonic wave with potential $\rotatedupsilon$, see point \ref{item "Potential function, and gradient soliton"} in Definitions \ref{definitiones "Ricci solitons"}, the function $\rotatedupsilon$ meets these statements: 
\begin{align}
	& \Bigl(\Ric(g_0) + \mathit{Hes}(\rotatedupsilon) = \Ric(g_0) + \nabla^2(\rotatedupsilon)\Bigr) + \frac{g_0}{2(t - T )} = 0, \\
	& \scalarcurvature + \Laplacian\rotatedupsilon + \frac{n}{2(t - T )} = 0.
\end{align}

\subsubsection{Expanding Soliton from an Einstein Manifold} 

A steady Ricci soliton is a natural extension of an Einstein manifold, see Eqq. \eqref{subequations "Einstein equations for Riemannian metric"} \eqref{equation "$g$-metric in the context of an Einstein manifold"}.

\subsubsection[Gaussian $(\lambda < 0)$-Soliton]{Gaussian $\protect\pseudobold{(\lambda < 0)}$-Soliton} 

Its form is 
\begin{equation}
	\textcyrillic{\textit{ф}}_\textsc{r}^\mathrm{s} = \left(\mathbb{R}^n, g_\mathbb{E}, \rotatedupsilon(x) = \frac{\lambda < 0}{2}|x|^2\right).
\end{equation}  

\subsubsection{Feldman–Ilmanen–Knopf's Kähler–Ricci Expanding (Gradient) Solitons} 

See above, \textsc{fik'kr} \cite{Feldman Ilmanen and Knopf "Rotationally Symmetric Shrinking and Expanding Gradient Kahler-Ricci Solitons"} solitons in Section \ref{subsubsection "Feldman–Ilmanen–Knopf's Kähler–Ricci Shrinking (Gradient) Solitons"}. 

\begin{exemplum}[Basic expander on a negative complex line bundle]
Considering the complex coordinate space of dimension $n$, $\{\mathbb{C}^n \times \cdots \times \mathbb{C}^n\}$, a gradient Kähler–Ricci soliton of homothetically expanding type, cf. \cite{Cao "Limits of solutions to the Kahler-Ricci flow"}, emerges if $n \geqslant 2$, and it corresponds to
\begin{equation}
	\textcyrillic{\textit{ф}}_\textsc{r}^{\mathring{\mathcal{L}}_\mathbb{C}^{-k}} \viz \bigl\{\textcyrillic{\textit{ф}}_\textsc{r}^\mathrm{s}\bigr\}^{n, k, r}_t = \left(\mathring{\mathcal{L}}_\mathbb{C}^{-k}, g_t\right), \enspace 0 < t < \infty,
\end{equation} 
for a value $k > n$, $k \in \mathbb{Z}$, and a number $r > 0$, $r \in \mathbb{R}$, with a cone-like end on the quotient $(\mathbb{C}^n\backslash\{0\})/\mathbb{Z}_k$ of the metric cone $\mathbb{C}^n/\mathbb{Z}_{k > n}$ by $\mathbb{Z}_k$, and a negative line bundle $\mathring{\mathcal{L}}_\mathbb{C}^{-k}$. This is the so-called \emph{expanding soliton}. \exemplasymbol
\end{exemplum}

\section{Geometrization of Topology (or of Process of Creating the Space as a Geometry)}
\label{section "Geometrization of Topology (or of Process of Creating the Space as a Geometry)"}

\begingroup
\footnotesize
It is geometers' dream (first articulated by Heinz Hopf \cite{Hopf "Selecta Heinz Hopf: Herausgegeben zu seinem 70. Geburtstag von der Eidgenossischen Technischen Hochschule Zurich"}, I believe) to find a canonical metric $g_\mathrm{best}$ on a given smooth manifold $V$ so that all topology of $V$ will be captured by geometry. \\
\indent — \textsc{M.L. Gromov} \cite[p. 138]{Gromov "Spaces and questions"}

\endgroup

\subsection{Poincaré Conjecture}
\label{subsection "Poincaré Conjecture"}

\begingroup
\footnotesize
Is it possible that the fundamental group of [a 3-manifold] $V$ is reduced to the identical substitution, but $V$ is not simply connected [that is, homeomorphic to the 3-sphere]?\endnote{
	Original Fr. version: p. 110: «Est-il possible que le groupe fondamental de $V$ se réduise à la substitution identique, et que pourtant $V$ ne soit pas simplement connexe?», p. 46: «simplement connexe au sens propre du mot, c'est-à-dire homéomorphe à l'hypersphère».
	} \\
\indent — \textsc{H. Poincaré} \cite[pp. 110, 46]{Poincare "Cinquieme complement a l'analysis situs"}

\endgroup

\vspace{2mm}

As it is known, Hamilton's program (Section \ref{section "Propaedeutics to Ricci Flow"}) constitutes the basement on which G. Perelman (Section \ref{section "Perelman Tapestry"}) has erected the proof of the Poincaré conjecture. Let us see in brief (without dwelling on the backstories) what it is. For a historical reconstruction of the conjecture, which has survived for nearly 100 years, see \cite{Morgan "100 Years of Topology: Work Stimulated by Poincare's Approach to Classifying Manifolds"}.

\begin{coniectura}[Poincaré conjecture]
\label{coniectura "Poincaré conjecture"}
If a closed (compact) smooth 3-manifold, where any closed loop is shrinkable continuously to a point, is simply connected, then it is homeomorphic (diffeomorphic) to the 3-sphere $\mathbb{S}^3$.\footnote{
	In the formulation of Thurston \cite[p. 358]{Thurston "Three dimensional manifolds Kleinian groups and hyperbolic geometry"}: «Is every 3-manifold with trivial fundamental group homeomorphic to the 3-sphere?»
	}
\end{coniectura}

The \emph{simple connectivity}, as Poincaré writes (see epigraph), here means \emph{homeomorphic to the 3-sphere}, whilst, more broadly, a connected topological space $\mathcal{M}$ is \emph{simply} iff its fundamental group (first homotopy group, or Poincaré group) $\pi_1(\mathcal{M})$ turns out to be \emph{trivial}. Call to mind that a loop is a closed curve, or, even better, a path whose two initial- and end-points are equal to a fixed point. The property of being simple in the spatio-topological connectedness results in the fact that, visually, a \emph{loop} on the surface e.g. of an ordinary sphere could be \emph{deformed}, or \emph{shrunk} (as long as it remains on the surface), to a \emph{single point}.

\begin{margo}[Proofs of the Poincaré conjecture from dimensions 1 up to $\geqslant$ 7]
~\enumerationisinitium
\item[·] With manifolds of dimension 1 and 2, the Poincaré conjecture is trivial and classical, respectively. 
\item[·] J.R. Stallings \cite{Stallings "Polyhedral homotopy-spheres"} (1960) manages to demonstrate that in dimension $n \geqslant 7$ the Poincaré conjecture is true for piecewise linear $n$-manifolds which has the homotopy type of the $n$-sphere $\mathbb{S}^n$.
\item[·] E.C. Zeeman \cite{Zeeman "The generalised Poincare conjecture"} (1961) \cite{Zeeman "The Poincare Conjecture for n equal to or greater than 5"} (1962) gets a similar result with homotopy $n$-spheres in dimension $n \geqslant 5$ (5 and 6, to be precise).
\item[·] S. Smale's paper \cite{Smale "Generalized Poincare's Conjecture in Dimensions Greater Than Four"} (1961) \cite{Smale "On the Structure of Manifolds"} (1962) deals with differentiable homotopy $n$-sphere in dimensions $n \geqslant 7$, and $n \geqslant 5$, drawing on a differentiable method, known as \emph{handlebody procedure}, useful for building an $n$-ball to which certain handles are attached and then removed, piece by piece.\footnote{
	But note: the cases $n \geqslant 5$ are a discovery belonging to Smale, as he explains in \cite[p. 47]{Smale "The Story Of The Higher Dimensional Poincare Conjecture (What Actually Happened On The Beaches Of Rio)"}. The Stallings' and Zeeman's proofs come after.
	}
\item[·] In dimensions 4 the proof is the work of M.H. Freedman \cite{Freedman "The topology of four-dimensional manifolds"} (1982), and it involves the topological case. 
\item[·] M.H.A. Newman \cite{Newman "The Engulfing Theorem for Topological Manifolds"} (1966) gives a demonstration in dimension $\geqslant 5$ for topological manifolds within the Stallings' homotopic structures.
\item[·] The last and most difficult proof of the Poincaré conjecture  touches its primal enunciation (1904), which concerns spaces of dimension 3. The Perelman's demonstration \cite{Perelman "The entropy formula for the Ricci flow and its geometric applications"} \cite{Perelman "Ricci flow with surgery on three-manifolds"} \cite{Perelman "Finite extinction time for the solutions to the Ricci flow on certain three-manifolds"} (2002-2003) embraces three distinct categories: topological, piecewise linear, and differentiable manifolds. \margosymbol
\enumerationisfinis
\end{margo}

\subsection{Thurston's Conjecture: Decomposition into Pieces having Geometric Structures}
\label{subsection "Thurston's Conjecture: Decomposition into Pieces having Geometric Structures"}

\begingroup
\footnotesize
[A] geometric structure [is] a space $(X, G)$, where $X$ is a manifold and $G$ is a group of diffeomorphisms of $X$ such that the stabilizer of any point $x \in X$ is a compact subgroup of $G$ [\,\dots]. There are precisely eight homogeneous spaces\footnote{
	See Margo \ref{margo "The eight 3-geometries of Thurston"}.
	} 
$(X, G)$ which are needed for geometric structures on 3-manifolds [\,\dots]. To find a geometric structure for a particular manifold is a great help in understanding that manifold. \\
\indent — \textsc{W.P. Thurston} \cite[p. 358]{Thurston "Three dimensional manifolds Kleinian groups and hyperbolic geometry"}

\endgroup

\vspace{2mm}

A generalization of the Poincaré conjecture is the \emph{Thurston's geometrization conjecture} \cite[p. 357]{Thurston "Three dimensional manifolds Kleinian groups and hyperbolic geometry"}. Perelman, when he showed the resolution of the Poincaré conjecture, gave \cite[sec. 13]{Perelman "The entropy formula for the Ricci flow and its geometric applications"} the first proof, at the same time, of the Thurston's conjecture. 

\begin{coniectura}[Thurston's conjecture]
\label{coniectura "Thurston's conjecture"}
~\enumerationisinitium
\item[(Short exposition).] The interior of any compact and orientable 3-manifold can be cut along, or split into, pieces which have geometric structures. 
\item[(Detailed exposition).] Every prime 3-manifold of this type has a canonical decomposition into a finite collection (after a finite number of steps) of embedded 2-spheres $\mathbb{S}^2$ (Kneser–Milnor decomposition) and 2-tori $\torus^2$ (toral decomposition by Jaco–Shalen–Johannson). Next comes the gluing action: 3-balls are glued to the resulting boundary components, and then other pieces are glued along the boundary tori. The result is a collection of simpler 3-manifolds compared to the initial one. 
\enumerationisfinis
\end{coniectura}

\begin{margo}[The eight 3-geometries of Thurston]
\label{margo "The eight 3-geometries of Thurston"}
~\enumerationisinitium
\item Three \emph{geometries with constant curvatures} (Section \ref{section "Space Forms as Triad of Riemannian Manifolds"}), $\kappa = 0$, $\kappa > 0$, and $\kappa < 0$: (i) Euclidean $\mathbb{E}^3$, (ii) spherical $\mathbb{S}^3$, and (iii) hyperbolic $\hyperbolic^3$ simply connected spaces, with groups $\mathbb{R}^3 \times SO_3(\mathbb{R})$, $SO_4(\mathbb{R})$, and $PSL_2(\mathbb{C})$. 
\item Two \emph{product geometries}: (iv) $\mathbb{S}^2 \times \mathbb{E}^1$, and (v) $\hyperbolic^2 \times \mathbb{E}^1$, whose groups are the orientation preserving subgroup of $SO_3(\mathbb{R}) \times \mathfrak{isom}(\mathbb{E}^1)$ and $\mathfrak{isom}(\hyperbolic^2) \times \mathfrak{isom}(\mathbb{E}^1)$.
\item Three \emph{twisted product geometries}: (vi) geometry of $\widetilde{SL}_2(\mathbb{R})$, where the manifold is the universal cover of the unit sphere bundle of a hyperbolic surface $\hyperbolic^2$, and the group is $\mathbb{R} \times \widetilde{SL}_2(\mathbb{R})$; (vii) $\mathit{Nil}$ geometry, where the manifold is a nilpotent Heisenberg group $\mathit{Hei}^3$ of $3 \times 3$ upper triangular matrices of dimension 3, and the group is $\mathit{Hei}^3 \rtimes \mathbb{S}^1$, the semidirect product of $\mathit{Hei}^3$ acting on a 1-sphere; (viii) $\mathit{Sol}$ geometry, where the manifold is a  solvable Lie group of dimension 3, and the group is an extension of it. See W.P. Thurston \cite{Thurston "The Geometry and Topology of Three-Manifolds"} \cite{Thurston "Hyperbolic geometry and 3-manifolds"} \cite{Thurston "Three dimensional manifolds Kleinian groups and hyperbolic geometry"} \cite{Thurston "Hyperbolic structures on 3-manifolds I: Deformation of acylindrical manifolds"} \cite{Thurston "Three-dimensional Geometry and Topology 1"} \cite{Thurston "Hyperbolic Structures on 3-manifolds II: Surface groups and 3-manifolds which fiber over the circle"} \cite{Thurston "Hyperbolic Structures on 3-manifolds III: Deformations of 3-manifolds with incompressible boundary"}. \margosymbol
\enumerationisfinis
\end{margo}

\subsubsection{Kneser–Milnor Prime Decomposition and \textsc{jsj} Decomposition (with Atoroidal or Seifert Fiber Spaces)}

About the decomposition into spheres, the demonstration of the existence was made by H. Kneser \cite{Kneser "Geschlossene Flachen in dreidimensionalen Mannigfaltigkeiten"}, whereas that of the uniqueness by J. Milnor \cite{Milnor "A Unique Decomposition Theorem for 3-Manifolds"}: 

\begin{theorema}[Kneser–Milnor decomposition]
Any compact and orientable 3-manifold can be built up, with a connected sum, by prime 3-manifolds, or by 3-manifolds homeomorphic to $\mathbb{S}^1 \times \mathbb{S}^2$.	
\end{theorema}

The splitting techniques along the tori is due to W.H. Jaco \& P.B. Shalen \cite{Jaco and Shalen "A new decomposition theorem for irreducible sufficiently-large 3-manifolds"} \cite[chap. IV]{Jaco and Shalen "Seifert fibered spaces in 3-manifolds"}, and K. Johannson \cite[§§ 14-21]{Johannson "Homotopy Equivalences of 3-Manifolds with Boundaries"}. 

\begin{theorema}[Jaco–Shalen–Johannson decomposition]
A compact, irreducible, sufficiently large 3-manifold can be cut (split) along tori into 3-submanifolds, thus forming a finite and unique up to isotopy collection of disjoint and canonical embedded incompressible tori or annuli. All 3-submanifolds that are generated by the cutting (splitting) are atoroidal or Seifert fiber spaces \textnormal{\cite{Seifert "Topologie dreidimensionaler gefaserter Raume"}}.
\end{theorema}

\subsubsection{Geometrization Subconjectures}

Conjecture \ref{coniectura "Thurston's conjecture"} consists of four parts. Let us see them one by one.
\enumerationisinitium
\item \emph{Poincaré conjecture \ref{coniectura "Poincaré conjecture"}}, i.e., $\pi_1(\mathcal{M}^3) = 0 \Rightarrow \mathcal{M}^3 \cong \mathbb{S}^3$ (the 3-manifold is homeomorphic and homotopy equivalent to the 3-sphere).
\item \emph{Clifford–Klein spherical space form conjecture, or Hopf conjecture} \textnormal{\cite{Hopf H. "Zum Clifford-Kleinschen Raumproblem"}}, 
\subenumerationisinitium
\item A finite group of diffeomorphisms that acts freely on the 3-sphere $\mathbb{S}^3$ is conjugate to a group of linear isometries which replicates the symmetries of \emph{crystals}, see Section \ref{subsection "Discrete Gamma-Crystallographic Group, Killing–Hopf Theorem, and Isometric Action"}. 
\item It is evident that the Poincaré conjecture is a \emph{special case} of the spherical space form conjecture: every closed 3-manifold having a finite fundamental group is diffeomorphic to a spherical space form of dimension 3, that is, to a quotient of $\mathbb{S}^3$ under the free and linear action of a finite subgroup of the orthogonal group $O_4(\mathbb{R})$.
\subenumerationisfinis
\item \emph{Elliptization conjecture}. The previous conjecture takes the following formulation. Let $\mathcal{M}^3$ be an irreducible, orientable, and closed 3-manifold with \emph{finite fundamental group} $\pi_1(\mathcal{M}^3)$. As a result $\mathcal{M}^3$

· is diffeomorphic to a quotient $\mathbb{S}^3/\mathbbl{\Gamma}$ of $\mathbb{S}^3$ by a finite subgroup $\mathbbl{\Gamma}$ of the orthogonal group $O_4(\mathbb{R})$, 

· admits a metric with constant positive sectional curvature ($\kappa > 0$). 
\item \emph{Hyperbolization conjecture}, expressed in two ways. 
\subenumerationisinitium
\item (Thurston's version). Let $\mathcal{M}^3$ be a compact 3-manifold with non-empty boundary; its interior has a hyperbolic metric iff $\mathcal{M}^3$ is a prime and homotopically atoroidal, and if it is not homeomorphic to $\torus^2 \times I = [0, 1]/\mathbb{Z}_2 \viz \mathbb{Z}/2\mathbb{Z}$, which appears to be true for \emph{Haken 3-manifolds} (see Margo \ref{margo "Haken manifold"}). A sketch of proof is in Thurston \cite{Thurston "Three dimensional manifolds Kleinian groups and hyperbolic geometry"}, see C.T.C. Wall \cite{Wall "On the work of W. Thurston"} and J.W. Morgan \cite{Morgan "On Thurston's Uniformization Theorem for Three-Dimensional Manifolds"}.
\item We say that $\mathcal{M}^3$ is irreducible, orientable, and closed 3-manifold with \emph{infinite fundamental group} $\pi_1(\mathcal{M}^3)$, and $\pi_1(\mathcal{M}^3)$ does not have a subgroup isomorphic to $\mathbb{Z} \oplus \mathbb{Z} = \mathbb{Z}^2$. As a result $\mathcal{M}^3$ presents a hyperbolic structure of finite volume.
\subenumerationisfinis
\enumerationisfinis

Note. The proof of all the above conjectures are by Perelman, with the exception of the hyperbolization proposition for some special cases, see e.g. M. Kapovich \cite{Kapovich "Hyperbolic Manifolds and Discrete Groups"} and C.T. McMullen \cite{McMullen "The Evolution of Geometric Structures on 3-Manifolds"}.

\begin{margo}[Haken manifold]
\label{margo "Haken manifold"}
A \emph{Haken manifold} \cite{Haken "Theorie der Normalflachen: Ein Isotopiekriterium fur den Kreisknoten"} is a compact, irreducible, and orientable (sufficiently large) 3-manifold that is prime and it contains a properly embedded two-sided incompressible surface which is not a 2-dimensional sphere $\mathbb{S}^2$. \margosymbol
\end{margo}

\section{Perelman Tapestry} 
\label{section "Perelman Tapestry"}

\begingroup
\footnotesize
The Ricci flow has also been discussed in quantum field theory, as an approximation to the renormalization group (\textsc{rg}) flow for the two-dimensional nonlinear $\sigmamodel$-model\footnote{
	\label{footnote "Sigma model"}
	So named by M. Gell-Mann and M. Lévy \cite[p. 717]{Gell-Mann Levy "The Axial Vector Current in Beta Decay"} because of the scalar meson $\sigmamodel$, as already established by J. Schwinger \cite{Schwinger "A Theory of the Fundamental Interactions"}. See Section \ref{subsection "Margo. Non-linear sigma-Model and Ricci Flow (Renormalization Group Flow in Quantum Field Theory for Geometrical Couplings)"}.
	}
[\,\dots]. I would like to speculate on the Wilsonian picture \cite{Wilson "Renormalization Group and Critical Phenomena. I. Renormalization Group and the Kadanoff Scaling Picture"} \cite{Wilson "Renormalization Group and Critical Phenomena. II. Phase-Space Cell Analysis of Critical Behavior"} of the \textsc{rg} flow. In this picture, $t$ corresponds to the scale parameter; the larger is $t$, the larger is the distance scale and the smaller is the energy scale [\,\dots].\endnote{
	«[T]o compute something on a lower energy scale one has to average the contributions of the degrees of freedom, corresponding to the higher energy scale» (ibid.). Perelman \cite[p. 12]{Perelman "The entropy formula for the Ricci flow and its geometric applications"} then goes on to identify a \emph{statistical analogy}, which is «related to the description of the renormalization group flow»; in one case «one obtains various quantities by averaging over higher energy states», whereas in the other «those states are suppressed by the exponential factor [\,\dots]. The interplay of statistical physics and (pseudo)-[R]iemannian geometry occurs in the subject of Black Hole Thermodynamics, developed by Hawking et al.» (cf. Section \ref{subsubsection "Spatio-temporal/Gravitational Thermodynamics, and Entropic Gravity"}).
	} 
In other words, decreasing of $t$ should correspond to looking at our Space through a microscope with higher resolution, where Space is now described not by some ([R]iemannian or any other) metric, but by an hierarchy of [R]iemannian metrics, connected by the Ricci flow equation. Note that we have a paradox here: the regions that appear to be far from each other at larger distance scale may become close at smaller distance scale; moreover, if we allow Ricci flow through singularities, the regions that are in different connected components at larger distance scale may become neighboring when viewed through microscope. Anyway, this connection between the Ricci flow and the \textsc{rg} flow suggests that Ricci flow must be gradient-like. \\
\indent — \textsc{G. Perelman} \cite[p. 3]{Perelman "The entropy formula for the Ricci flow and its geometric applications"}

\endgroup

\vspace{2mm}

To follow a résumé of the key findings presented in the papers of G. Perelman \cite{Perelman "The entropy formula for the Ricci flow and its geometric applications"} \cite{Perelman "Ricci flow with surgery on three-manifolds"} \cite{Perelman "Finite extinction time for the solutions to the Ricci flow on certain three-manifolds"}, who draws and interweaves, on a single conceptual tapestry, the Ricci flow, in full possession of the \emph{Hamilton's techniques} (Section \ref{section "Propaedeutics to Ricci Flow"}), and the Poincaré \ref{coniectura "Poincaré conjecture"} and Thurston's \ref{coniectura "Thurston's conjecture"} Conjectures. For an exhaustive and close examination on the Perelman innovations, we refer to H.-D. Cao and X.-P. Zhu \cite{Cao and Zhu "A Complete Proof of the Poincare and Geometrization Conjectures - application of the Hamilton-Perelman theory of the Ricci flow"}, J.W. Morgan and G. Tian \cite{Morgan and Tian "Ricci Flow and the Poincare Conjecture"}, B. Kleiner and J. Lott \cite{Kleiner and Lott "Notes on Perelman's papers"}.

\subsection{Entropy-Energy Functionals: Variations, Monotonicity, and Gradient Flow}

Perelman \cite{Perelman "The entropy formula for the Ricci flow and its geometric applications"} develops a concept of \emph{entropy} for the Ricci flow, revisiting an idea of B. Chow \cite{Chow "On the entropy estimate for the Ricci flow on compact 2-orbifolds"}, already partly analyzed by Hamilton \cite[secc. 7-8]{Hamilton "The Ricci Flow on Surfaces"}, but whose roots lie in J.F. Nash \cite{Nash "Continuity of Solutions of Parabolic and Elliptic Equations"}.\footnote{
	That is why it is possible to connect the Nash's entropy, see below Eq.   \eqref{equation "Nash's entropy"}, with the Li–Yau's Harnack inequality (Section \ref{subsubsection "Li–Yau–Hamilton's Harnack Inequality, aka Hamilton's Matrix Inequality"}).
	} 
This type of entropy is associated with certain \emph{functionals} (see below, Sections \ref{subsubsection "The F-Functional (and Link with the Nash's Entropy)"} and \ref{subsubsection "The W-Functional"}), and it is not about the measure of the degree of atomic disorder (chaos) (see point \ref{item "Entropy in Prior Knowledge: Maxwell–Boltzmann Probability Distribution, and Ergodic Hypothesis of Thermodynamics"}, p. \pageref{item "Entropy in Prior Knowledge: Maxwell–Boltzmann Probability Distribution, and Ergodic Hypothesis of Thermodynamics"} in Section \ref{subsection "Prior Knowledge: Maxwell–Boltzmann Probability Distribution, and Ergodic Hypothesis of Thermodynamics"}, and Section \ref{subsection "The Entropy-Energy Roots"}), as a consequence of heat conduction; but it \emph{measures the geometric disorder (chaos) on space-time}. 

Not unlike the entropy of thermodynamics, the Perelman's entropy increases unidirectionally, and there is no reversibility process, except that it describes a \emph{quantity of energy combined with the Sobolev inequality} (Sections \ref{subsubsection "Carron–Akutagawa's Sobolev Embedding"} and \ref{subsubsection "Logarithmic Sobolev Inequality, and Lower Bound for the W-Entropy-Energy"}). Note. In L. Ni \cite{Ni "The Entropy Formula for Linear Heat Equation"} \cite{Ni "Addenda to "The Entropy Formula for Linear Heat Equation""} there is a simplified version of the Perelman's entropy without Ricci flow for a heat-like equation on a static space. 

\subsubsection[The $\mathscr{F}$-Functional (and Link with the Nash's Entropy)]{The $\protect\pseudobold{\mathscr{F}}$-Functional (and Link with the Nash's Entropy)}
\label{subsubsection "The F-Functional (and Link with the Nash's Entropy)"}

The first definition of entropy \cite[sec. 1]{Perelman "The entropy formula for the Ricci flow and its geometric applications"} is given in conjunction with an \emph{energy functional} $\mathscr{F} \colon \mathcal{X}_g \times \mathscr{C}^\infty(\mathcal{M}) \to \mathbb{R}$,\footnote{
	$\mathcal{X}_g$ is the space of the $g$-metric on $\mathcal{M}$.
	} 
so the formulation is 
\begin{equation}
\label{equation "Perelman's entropy-energy F-functional"}
	\mathscr{F}(g_{\mu\nu}, \rotatedupsilon) = \int_\mathcal{M}\left(\scalarcurvature + |\nabla\rotatedupsilon|^2\right)e^{-\rotatedupsilon}d\bbmu,
\end{equation}
for a Riemannian metric $g_{\mu\nu}$ and a function $\rotatedupsilon \in \mathscr{C}^\infty(\mathcal{M})$ on a closed manifold $\mathcal{M}$, where $d\bbmu$ is the volume element \eqref{subequations "volume element"}. Eq. \eqref{equation "Perelman's entropy-energy F-functional"} is what we can call \emph{entropy-energy $\mathscr{F}$-functional}.

We define the gradient flow of \eqref{equation "Perelman's entropy-energy F-functional"} as 
\begin{equation}
\label{equation "Gradient flow of entropy-energy F-functional"}
	\begin{cases}
		\partial_t g_{\mu\nu} = - 2(\Ricci_{\mu\nu} + \nabla_\mu\nabla_\nu\rotatedupsilon), \\
		\partial_t\rotatedupsilon = -\Laplacian\rotatedupsilon - \scalarcurvature.
	\end{cases}
\end{equation}
The gradient flow \eqref{equation "Gradient flow of entropy-energy F-functional"} is none other than the Ricci flow modified by a 1-parameter family of diffeomorphisms $\textcyrillic{\textit{Д}}_t$ (cf. point \ref{item "Via pullback of the metric"} in Definitions \ref{definitiones "Ricci solitons"}), an example of which is the DeTurck's modus (Section \ref{subsubsection "Short Time Existence and Uniqueness in Ricci–DeTurck's Strictly Parabolic System"}). In fact, Eq. \eqref{equation "Gradient flow of entropy-energy F-functional"} becomes $\partial_t g_{\mu\nu} = - 2\Ricci_{\mu\nu}$, that is \eqref{equation "unnormalized Ricci flow"}, plus the \emph{adjoint (or conjugate) heat equation}
\begin{equation}
\label{equation "Adjoint heat equation"}
	\heatoperator^*e^{-\rotatedupsilon} = \left[\left(-\partial_t \viz -\tfrac{\partial}{\partial t}\right) - \Laplacian + \scalarcurvature\right]e^{-\rotatedupsilon} = 0, \enspace e^{-\rotatedupsilon} = \upsilon,
\end{equation}
where 
\begin{equation}
\label{equation "Adjoint of the heat operator"}
	\heatoperator^* = -\partial_t \viz \frac{\partial}{\partial t} - \Laplacian + \scalarcurvature
\end{equation}	 
is the adjoint of the heat operator $\heatoperator = \partial_t \viz \frac{\partial}{\partial t} - \Laplacian$ on $\mathscr{C}^\infty \in \mathcal{M}_T$. The \emph{heat operator}, or \emph{Laplace–Poisson operator}, is the solution of the heat equation \eqref{subequations "Heat equation in Laplacian form"}, and goes back to P.-S. Laplace \cite{Laplace "Memoire sur divers points d'analyse"} (for the 1-dimensional space), and S.D. Poisson \cite{Poisson "Sur l'integrale de l'equation relative aux vibrations des plaques elastiques"}. 

\begin{margo}
Letting $e^{-\rotatedupsilon} = \upsilon$ be a solution of the adjoint heat Eq. \eqref{equation "Adjoint heat equation"}, we have $\int_\mathcal{M}e^{-\rotatedupsilon}d\bbmu = 1$ in \eqref{equation "Perelman's entropy-energy F-functional"}, which is therefore preserved. \margosymbol
\end{margo}

\enumerationisinitium
\item We can set down the first variation of \eqref{equation "Perelman's entropy-energy F-functional"} at $(g_{\mu\nu}, \rotatedupsilon)$. Putting $g^\rotatedheartsuit_{\mu\nu} = \delta g_{\mu\nu}$, $\rotatedupsilon^\rotatedheartsuit = \delta\rotatedupsilon$, and $g^\ddagger = g^{\mu\nu}g^\rotatedheartsuit_{\mu\nu}$, one has
\begin{align}
	\delta\mathscr{F}_{\left(g^\rotatedheartsuit_{\mu\nu}, \rotatedupsilon^\rotatedheartsuit\right)}(g_{\mu\nu}, \rotatedupsilon) & = \int_\mathcal{M}e^{-\rotatedupsilon}\biggl\{-\Laplacian g^\ddagger + \nabla_\mu\nabla_\nu g^\rotatedheartsuit_{\mu\nu} - \Ricci_{\mu\nu}g^\rotatedheartsuit_{\mu\nu} - g^\rotatedheartsuit_{\mu\nu}  \nabla_\mu\rotatedupsilon\nabla_\nu\rotatedupsilon \notag \\ 
	& \hspace{52pt} + 2\langle\nabla\rotatedupsilon, \nabla\rotatedupsilon^\rotatedheartsuit\rangle + \left(\scalarcurvature + |\nabla\rotatedupsilon|^2\right)\left(\frac{g^\ddagger}{2} - \rotatedupsilon^\rotatedheartsuit\right)\biggr\} \notag \\
	& = \int_\mathcal{M}e^{-\rotatedupsilon}\biggl\{-g^\rotatedheartsuit_{\mu\nu}(\Ricci_{\mu\nu} + \nabla_\mu\nabla_\nu\rotatedupsilon) + \left(\frac{g^\ddagger}{2} - \rotatedupsilon^\rotatedheartsuit\right) \notag \\
	& \hspace{52pt} \left(2\Laplacian\rotatedupsilon - |\nabla\rotatedupsilon|^2 + \scalarcurvature\right)\biggr\}.
\end{align}
Note the value $\tfrac{g^\ddagger}{2} - \rotatedupsilon^\rotatedheartsuit$ vanishes identically in a state where $e^{-\rotatedupsilon}d\bbmu$ is pointwise invariant.
\item The entropy-energy $\mathscr{F}$-functional \eqref{equation "Perelman's entropy-energy F-functional"} is a monotone quantity. Its \emph{monotonicity formula} is (without and without indices)
\begin{subequations}
\label{subequations "Entropy-energy monotonicity formula"}
\begin{align}
	\frac{d}{dt}\mathscr{F}\bigl(g_{\mu\nu}(t), \rotatedupsilon_t\bigr) & = 2\int_\mathcal{M}|\Ricci_{\mu\nu} + \nabla_\mu\nabla_\nu\rotatedupsilon|^2e^{-\rotatedupsilon}d\bbmu \geqslant 0, \\
	\partial_t\mathscr{F}(g_t, \rotatedupsilon_t) & = 2\int_\mathcal{M}|\Ric + \mathit{Hes}(\rotatedupsilon)|^2e^{-\rotatedupsilon}d\bbmu \geqslant 0,
\end{align}
\end{subequations}
under which $\mathscr{F}$ is a \emph{monotonically increasing} (or \emph{non-decreasing}) quantity along the Ricci flow, but also a \emph{constant value on steady Ricci solitons} (cf. Section \ref{subsection "Steadies"}) with potential $\rotatedupsilon$ with $\Ricci_{\mu\nu} + \nabla_\mu\nabla_\nu\rotatedupsilon = \Ric + \mathit{Hes}(\rotatedupsilon) = 0$. 
\item From \eqref{subequations "Entropy-energy monotonicity formula"} we derive the \emph{gradient flow} of the $\mathscr{F}$-functional: 
\begin{align}
	\begin{cases}
	\partial_t g_{\mu\nu} \viz \frac{\partial}{\partial t}g_{\mu\nu} = -2(\Ricci_{\mu\nu} + \nabla_\mu\nabla_\nu\rotatedupsilon), \\
	\partial_t\rotatedupsilon \viz \frac{\partial}{\partial t}\rotatedupsilon = -\scalarcurvature - \Laplacian\rotatedupsilon.
	\end{cases}
\end{align}
\enumerationisfinis

\begin{scholium}
The entropy-energy $\mathscr{F}$-functional \eqref{equation "Perelman's entropy-energy F-functional"}

· originally appears in the string theory literature as a \emph{low energy effective action}, and the function $\rotatedupsilon$ acts as a \emph{dilaton field}, see e.g. \cite[chap. 13]{Green Schwarz Witten "Superstring theory Vol. 2: Loop Amplitudes Anomalies and Phenomenology"};

· is an enhanced version of the Einstein–Hilbert action \eqref{equation "Einstein–Hilbert (Gravitational) Action"}. 

· is correlated with the Li–Yau's Harnack inequality $\Inequality^\mathrm{Har}_\textsc{ly}$ \eqref{subequations "Li–Yau's Harnack inequality"} by means of the Nash's entropy, which is also a monotone quantity.

We define the \emph{Nash's entropy} \cite[p. 936]{Nash "Continuity of Solutions of Parabolic and Elliptic Equations"} as
\begin{subequations}
\label{equation "Nash's entropy"}
\begin{align}
	\mathsf{S}_\textsc{n}(\rotatedupsilon, \rotatedtau) & = -\int_\mathcal{M}\upsilon\log{\upsilon}d\bbmu - \frac{n}{2}\log(4\pi\rotatedtau) - \frac{n}{2} = -\int_\mathcal{M}\rotatedupsilon\upsilon d\bbmu - \frac{n}{2}, \\
	\label{equation "Original Nash's entropy"}
	\mathsf{S}_\textsc{n}(\upsilon, t) & = -\int_\mathcal{M}\upsilon\log{\upsilon}d\bbmu,
\end{align}
\end{subequations}
assuming a quantity $\rotatedtau(\tau)> 0$ satisfying $\frac{d\rotatedtau}{d\tau} = 1$, and that the function $\rotatedupsilon$ is dictated by $\upsilon = (4\pi\rotatedtau)^{-\frac{n}{2}}e^{-\rotatedupsilon}$ with $\int_\mathcal{M}\upsilon d\bbmu = 1$, and $\upsilon$ coincides with the \emph{heat kernel}, or the fundamental (positive) solution, of the heat equation 
\begin{subequations}
	\begin{empheq}[left = {\empheqlbrace}]{align}
	& \heatoperator\upsilon = 0, \\ 
	& \Laplacian\upsilon = \tfrac{\partial\upsilon}{\partial\tau},
	\end{empheq}
\end{subequations}
on a closed manifold $\mathcal{M}$, cf. Eq. \eqref{subequations "Linear heat equation"}. Now, Eq. \eqref{equation "Perelman's entropy-energy F-functional"} gets to be
\begin{subequations}
	\begin{empheq}[left = {\empheqlbrace}]{align}
	& \mathscr{F}(\rotatedupsilon, \rotatedtau) = \tfrac{d}{d\tau}\mathsf{S}_\textsc{n}(\rotatedupsilon, \rotatedtau) \\
	& \mathscr{F}(g_{\mu\nu}, \rotatedupsilon) = \partial_t\mathsf{S}_\textsc{n}(\upsilon, t) \viz \tfrac{\partial}{\partial t}\mathsf{S}_\textsc{n}(\upsilon, t),
	\end{empheq}
\end{subequations}
that is, it can be regarded as the first derivative of \eqref{equation "Original Nash's entropy"}. \scholiumsymbol
\end{scholium}

\subsubsection[The $\mathscr{W}$-Functional]{The $\protect\pseudobold{\mathscr{W}}$-Functional}
\label{subsubsection "The W-Functional"}

Perelman's second notion of entropy \cite[sec. 3]{Perelman "The entropy formula for the Ricci flow and its geometric applications"} is a generalization of the previous one, with the insertions of a constant scale parameter $\tau = (T - t) > 0$, for the explicit purpose of handling a shrinker $(\textcyrillic{\textit{ф}}_\textsc{r}^\mathrm{s})_{\lambda > 0}$ (cf. Section \ref{subsubsection "Gaussian Soliton"}). Taking a closed Riemannian manifold $(\mathcal{M}^n, g)$ of dimension $n$, a function $\rotatedupsilon$ on $\mathcal{M}^n$, and a functional $\mathscr{W} \colon \mathcal{X}_g \times \mathscr{C}^\infty(\mathcal{M}) \times \mathbb{R}_+ \to \mathbb{R}$, the \emph{Perelman's entropy-energy functional}, serving as a shrinker entropy functional, is determined by
\begin{equation}
\label{equation "Perelman's entropy-energy W-functional"}
	\mathscr{W}(g_{\mu\nu}, \rotatedupsilon, \tau) = \int_{\mathcal{M}^n}\Bigl(\tau\left(\scalarcurvature + |\nabla\rotatedupsilon|^2\right) + (\rotatedupsilon - n)\Bigr)(4\pi\tau)^{-\frac{n}{2}}e^{-\rotatedupsilon}d\bbmu,
\end{equation}
provided that $\rotatedupsilon$ satisfies $\int_{\mathcal{M}^n}(4\pi\tau)^{-\frac{n}{2}}e^{-\rotatedupsilon}d\bbmu = 1$. The entropy-energy $\mathscr{W}$ is 

· \emph{monotonically increasing} (or \emph{non-decreasing}) quantity,

· \emph{invariant} under simultaneous change of the scale of both  $g_{\mu\nu}$ and $\tau$, under parabolic conditions, and also under diffeomorphism.
\enumerationisinitium
\item As in the previous case, we put $g^\rotatedheartsuit_{\mu\nu} = \delta g_{\mu\nu}$, $\rotatedupsilon^\rotatedheartsuit = \delta\rotatedupsilon$, $g^\ddagger = g^{\mu\nu}g^\rotatedheartsuit_{\mu\nu}$, and $\tau^\rotatedheartsuit = \delta\tau$, for writing the first variation of \eqref{equation "Perelman's entropy-energy W-functional"} at $(g_{\mu\nu}, \rotatedupsilon, \tau)$,
\begin{align}
\label{align "First variation of Perelman's shrinking soliton entropy functional"}
	\delta\mathscr{W}_{\left(g^\rotatedheartsuit_{\mu\nu}, \rotatedupsilon^\rotatedheartsuit, \tau^\rotatedheartsuit\right)}(g_{\mu\nu}, \rotatedupsilon, \tau) = & \int_{\mathcal{M}^n} -\tau g^\rotatedheartsuit_{\mu\nu}\left(\Ricci_{\mu\nu} + \nabla_\mu\nabla_\nu\rotatedupsilon - \frac{g_{\mu\nu}}{2\tau}\right)(4\pi\tau)^{-\frac{n}{2}}e^{-\rotatedupsilon}d\bbmu \notag \\
	 & + \int_{\mathcal{M}^n}\left(\frac{g^\ddagger}{2} - \rotatedupsilon^\rotatedheartsuit - \frac{n}{2\tau}\tau^\rotatedheartsuit\right)\Bigl(\tau\left(\scalarcurvature + 2\Laplacian\rotatedupsilon - |\nabla\rotatedupsilon|^2\right) \notag \\
	 & \hspace{10pt} + \rotatedupsilon - n - 1\Bigr)(4\pi\tau)^{-\frac{n}{2}}e^{-\rotatedupsilon}d\bbmu \notag \\
	 & + \int_{\mathcal{M}^n}\tau^\rotatedheartsuit\left(\scalarcurvature + |\nabla\rotatedupsilon|^2 - \frac{n}{2\tau}\right)(4\pi\tau)^{-\frac{n}{2}}e^{-\rotatedupsilon}d\bbmu.
\end{align}

We report a demonstration, in view of its single step conciseness.

\begin{proof}[Proof of \eqref{align "First variation of Perelman's shrinking soliton entropy functional"}]
\begin{align}
	\delta\mathscr{W}\left(g^\rotatedheartsuit_{\mu\nu}, \rotatedupsilon^\rotatedheartsuit, \tau^\rotatedheartsuit\right) & = \int_{\mathcal{M}^n}\Bigl\{\tau^\rotatedheartsuit\left(\scalarcurvature + |\nabla\rotatedupsilon|^2\right) + \tau\bigl(-\Laplacian g^\ddagger + \nabla_\mu\nabla_\nu g^\rotatedheartsuit_{\mu\nu} - \Ricci_{\mu\nu}g^\rotatedheartsuit_{\mu\nu} \notag \\
	& \hspace{10pt} - g^\rotatedheartsuit_{\mu\nu}\nabla_\mu\rotatedupsilon\nabla_\nu\rotatedupsilon + 2\langle\nabla\rotatedupsilon, \nabla\rotatedupsilon^\rotatedheartsuit\rangle\bigr) + \rotatedupsilon^\rotatedheartsuit\Bigr\}(4\pi\tau)^{-\frac{n}{2}}e^{-\rotatedupsilon}d\bbmu \notag \\
	& \hspace{10pt} + \Lbrack:\int_{\mathcal{M}^n}\biggl\{\Bigl(\tau\left(\scalarcurvature + |\nabla\rotatedupsilon|^2\right) + \rotatedupsilon - n\Bigr)\left(-\frac{n}{2}\frac{\tau^\rotatedheartsuit}{\tau} + \frac{g^\ddagger}{2} - \rotatedupsilon^\rotatedheartsuit\right)\biggr\} \notag \\
	& \hspace{22pt} (4\pi\tau)^{-\frac{n}{2}}e^{-\rotatedupsilon}d\bbmu:\Rbrack \notag \\
	& = \int_{\mathcal{M}^n}\Bigl\{\tau^\rotatedheartsuit\left(\scalarcurvature + |\nabla\rotatedupsilon|^2\right) + \rotatedupsilon^\rotatedheartsuit\Bigr\}(4\pi\tau)^{-\frac{n}{2}}e^{-\rotatedupsilon}d\bbmu \notag \\ 
	& \hspace{10pt} + \int_{\mathcal{M}^n}\Bigl\{-\tau g^\rotatedheartsuit_{\mu\nu}(\Ricci_{\mu\nu} + \nabla_\mu\nabla_\nu\rotatedupsilon) \notag \\
	& \hspace{22pt} + \tau\left(g^\ddagger - 2\rotatedupsilon^\rotatedheartsuit\right)\left(\Laplacian\rotatedupsilon - |\nabla\rotatedupsilon|^2\right)\Bigr\}(4\pi\tau)^{-\frac{n}{2}}e^{-\rotatedupsilon}d\bbmu \notag \\
	& \hspace{10pt} + \Lbrack:\cdots:\Rbrack \notag \\
	& = -\int_{\mathcal{M}^n}\tau g^\rotatedheartsuit_{\mu\nu}\left(\Ricci_{\mu\nu} + \nabla_\mu\nabla_\nu\rotatedupsilon - \frac{g_{\mu\nu}}{2\tau}\right)(4\pi\tau)^{-\frac{n}{2}}e^{-\rotatedupsilon}d\bbmu \notag \\
	& \hspace{10pt} + \int_{\mathcal{M}^n}\left(\frac{g^\ddagger}{2} - \rotatedupsilon^\rotatedheartsuit - \frac{n}{2\tau}\tau^\rotatedheartsuit\right)\Bigl\{\tau\left(\scalarcurvature + |\nabla\rotatedupsilon|^2\right) + \rotatedupsilon - n \notag \\ 
	& \hspace{22pt} + 2\tau\left(\Laplacian\rotatedupsilon - |\nabla\rotatedupsilon|^2\right)\Bigr\}(4\pi\tau)^{-\frac{n}{2}}e^{-\rotatedupsilon}d\bbmu \notag \\
	& \hspace{10pt} + \int_{\mathcal{M}^n}\biggl\{\tau^\rotatedheartsuit\left(\scalarcurvature + |\nabla\rotatedupsilon|^2 - \frac{n}{2\tau}\right) + \left(\rotatedupsilon^\rotatedheartsuit - \frac{g^\ddagger}{2}\frac{n}{2\tau}\tau^\rotatedheartsuit\right)\biggr\}\notag \\ 
	& \hspace{22pt}(4\pi\tau)^{-\tfrac{n}{2}}e^{-\rotatedupsilon}d\bbmu \notag \\
	& = \int_{\mathcal{M}^n} -\tau g^\rotatedheartsuit_{\mu\nu}\left(\Ricci_{\mu\nu} + \nabla_\mu\nabla_\nu\rotatedupsilon - \frac{g_{\mu\nu}}{2\tau}\right)(4\pi\tau)^{-\frac{n}{2}}e^{-\rotatedupsilon}d\bbmu \notag \\
	& \hspace{10pt} + \int_{\mathcal{M}^n}\left(\frac{g^\ddagger}{2} - \rotatedupsilon^\rotatedheartsuit - \frac{n}{2\tau}\tau^\rotatedheartsuit\right)\Bigl\{\tau\left(\scalarcurvature + 2\Laplacian\rotatedupsilon - |\nabla\rotatedupsilon|^2\right) \notag \\
	& \hspace{22pt} + \rotatedupsilon - n - 1\Bigr\}(4\pi\tau)^{-\tfrac{n}{2}}e^{-\rotatedupsilon}d\bbmu \notag \\ 
	& \hspace{10pt} + \int_{\mathcal{M}^n}\tau^\rotatedheartsuit\left(\scalarcurvature + |\nabla\rotatedupsilon|^2 - \frac{n}{2\tau}\right)(4\pi\tau)^{-\tfrac{n}{2}}e^{-\rotatedupsilon}d\bbmu.
\end{align}	
Patently $\delta\mathscr{W}\left(g^\rotatedheartsuit_{\mu\nu}, \rotatedupsilon^\rotatedheartsuit, \tau^\rotatedheartsuit\right) \viz \delta\mathscr{W}_{\left(g^\rotatedheartsuit_{\mu\nu}, \rotatedupsilon^\rotatedheartsuit, \tau^\rotatedheartsuit\right)}(g_{\mu\nu}, \rotatedupsilon, \tau)$.
\end{proof}

\item Consider
\begin{equation}
\label{equation "Ricci flow together with a backward heat-like equation"}
	\begin{cases}
	\partial_t g_{\mu\nu} \viz \frac{\partial}{\partial t}g_{\mu\nu} = -2\Ricci_{\mu\nu}, \\
	\frac{\partial\rotatedupsilon}{\partial t} = -\Laplacian\rotatedupsilon - \scalarcurvature + |\nabla\rotatedupsilon|^2 + \frac{n}{2\tau}, \\
	\frac{d\tau}{dt} = -1,
	\end{cases}
\end{equation}
that is, a Ricci flow together with a backward heat-like equation, and let $\upsilon = (4\pi\tau)^{-\frac{n}{2}}e^{-\rotatedupsilon}$ be a process that satisfies the adjoint heat equation 
\begin{equation}
	\heatoperator^*\upsilon = \left(-\partial_t \viz -\frac{\partial}{\partial t} - \Laplacian + \scalarcurvature\right)\upsilon = 0, 
\end{equation}	
cf. \eqref{equation "Adjoint heat equation"}, containing the adjoint of the heat operator \eqref{equation "Adjoint of the heat operator"}, i.e. $\heatoperator^* = -\partial_t \viz \frac{\partial}{\partial t} - \Laplacian + \scalarcurvature$. From the system \eqref{equation "Ricci flow together with a backward heat-like equation"} one arrives to the \emph{monotonicity formula} for the $\mathscr{W}$-functional (without and without indices),
\begin{subequations}
\label{subequations "Monotonicity formula for the W-functional"}
\begin{align}
	\frac{d}{dt}\mathscr{W}\bigl(g_{\mu\nu}(t), \rotatedupsilon_t, \tau_t\bigr) & = 2\tau\int_{\mathcal{M}^n}\left|\Ricci_{\mu\nu} + \nabla_\mu\nabla_\nu\rotatedupsilon - \frac{g_{\mu\nu}}{2\tau}\right|^2(4\pi\tau)^{-\frac{n}{2}}e^{-\rotatedupsilon}d\bbmu \geqslant 0, \\
	\partial_t\mathscr{W}(g_t, \rotatedupsilon_t, \tau_t\bigr) & = 2\tau\int_{\mathcal{M}^n}\left|\Ric + \mathit{Hes}(\rotatedupsilon) - \frac{g}{2\tau}\right|^2\upsilon d\bbmu \geqslant 0. 
\end{align}
\end{subequations}

The result $\frac{d}{dt}\mathscr{W}\bigl(g_{\mu\nu}(t), \rotatedupsilon_t, \tau_t\bigr) = 0$, or $\partial_t\mathscr{W}(g_t, \rotatedupsilon_t, \tau_t\bigr) = 0$, shall apply to gradient shrinker in which the potential $\rotatedupsilon$ fulfills the condition $\Ricci_{\mu\nu} + \nabla_\mu\nabla_\nu\rotatedupsilon - \frac{g_{\mu\nu}}{2\tau} = 0$, or $\Ric + \mathit{Hes}(\rotatedupsilon) - \frac{g}{2\tau} = 0$, respectively.
\enumerationisfinis

\subsection[The $\length$-Length Functional]{The $\protect\pseudobold{\length}$-Length Functional}

Alongside the geometric entropy formulæ (Sections \ref{subsubsection "The F-Functional (and Link with the Nash's Entropy)"} and \ref{subsubsection "The W-Functional"}), Perelman \cite[sec. 7]{Perelman "The entropy formula for the Ricci flow and its geometric applications"} outlines the notion of \emph{$\length$-length} \eqref{subequations "Perelman's length functional"}, an energy-like functional, and builds an overall picture of Riccian flow singularities. Incidentally, the paper \cite{Perelman "The entropy formula for the Ricci flow and its geometric applications"} offers solutions with non-negative curvature that may show up as \emph{blow-up limits of finite time singularities}, which satisfy a given non-collapsing condition (see Theorem \ref{theorema "Perelman's no local collapsing}) and correspond to a \emph{bounded entropy}. 

Let us now look at exactly what the $\length$-length is. Let $\partial_\tau g_\tau \viz \frac{\partial g_\tau}{\partial \tau} = 2\Ric(g_\tau)$, $g_\tau = g(\tau)$, be a \emph{backward Ricci flow}, where $\tau =  t_0 - t$, for a fixed time $t_0$. Let $\gamma_\mathrm{c} \colon [\tau_1, \tau_2] \to \mathcal{M}$ be a  $\mathscr{C}^1$ curve (parameterized by backward time), with $[\tau_1, \tau_2] \subset (0, \infty)$ and $\tau_1 \geqslant 0$, i.e. $0 \leqslant \tau_1 < \tau_2$\footnote{
	Perelman's original assumption is $0 < \tau_1 \leqslant \tau \leqslant \tau_2$.	}
(supposing that $\mathcal{M}$ is compact, or that $g_\tau$ is complete with uniformly bounded curvature). The $\length$-length is a \emph{local length functional} defined \emph{on the space of all space-time curves, or paths}; its form is
\begin{subequations}
\label{subequations "Perelman's length functional"}
\begin{align}
	\length(\gamma_\mathrm{c}) & = \int^{\tau_2}_{\tau_1}\sqrt{\tau}\left\{{\scalarcurvature}_{g_\tau}\bigr(\gamma_\mathrm{c}(\tau)\bigl) + \left|\frac{d\gamma_\mathrm{c}}{d\tau}\tau\right|^2_{g_\tau}\right\}d\tau, \\
	& = \int^{\tau_2}_{\tau_1}\sqrt{\tau}\left\{{\scalarcurvature}_{g_\tau}\bigr(\gamma_\mathrm{c}(\tau)\bigl) + |\dot{\gamma}_\mathrm{c}(\tau)|^2_{g_\tau}\right\}d\tau, \enspace \dot{\gamma}_\mathrm{c}(\tau) = \partial_\tau\gamma_\mathrm{c}(\tau) \viz \tfrac{\partial\gamma_\mathrm{c}}{\partial\tau},
\end{align}
\end{subequations}
in which the metric $g_\tau$ is used at time $t_0 - \tau$ for both the Ricci scalar ${\scalarcurvature}_{g_\tau}$ and the norm $\left|\dot{\gamma}_\mathrm{c}(\tau)\right|$. Setting the vectors fields $\vec{X}(\tau) = \dot{\gamma}_\mathrm{c}(\tau)$ and $\vec{Y}(\tau)$ along a curve $\gamma_\mathrm{c}(\tau)$ in $\mathcal{M}$, the variation formula of \eqref{subequations "Perelman's length functional"} is
\begin{equation}
	\delta_{\vec{Y}}\bigl(\length(\gamma_\mathrm{c})\bigr) = 2\sqrt{\tau}\langle\vec{X}, \vec{Y}\rangle\Big|^{\tau_2}_{\tau_1} + \int^{\tau_2}_{\tau_1}\sqrt{\tau}\left\langle\vec{Y}, \nabla\scalarcurvature - 2\nabla_{\vec{X}}\vec{X} - 4\Ric(\vec{X}, \cdot) - \frac{1}{\tau}\vec{X}\right\rangle d\tau,
\end{equation}
where $\langle\cdot \:, \cdot\rangle$ is the inner product by reference to $g_\tau$, whilst $\Ric(\vec{X}, \cdot)$ is a horizontal 1-form along $\gamma_\mathrm{c}(\tau)$ here equivalent to its dual $\Ric(\vec{X}, \cdot)^*$, which in turn is a tangent vector field. 

\begin{proof}
\begin{align}
	\delta_{\vec{Y}}\bigl(\length(\gamma_\mathrm{c})\bigr) & = \int^{\tau_2}_{\tau_1}\sqrt{\tau}\left(\langle\vec{Y}, \nabla\scalarcurvature\rangle + 2\langle\nabla_{\vec{Y}}\vec{X}, \vec{X}\rangle\right)d\tau \notag \\
	& = \int^{\tau_2}_{\tau_1}\sqrt{\tau}\left(\langle\vec{Y}, \nabla\scalarcurvature\rangle + 2\langle\nabla_{\vec{X}}\vec{Y}, \vec{X}\rangle\right)d\tau \notag \\
	& = \int^{\tau_2}_{\tau_1}\sqrt{\tau}\left(\langle\vec{Y}, \nabla\scalarcurvature\rangle + 2\frac{d}{d\tau}\langle\vec{Y}, \vec{X}\rangle - 2\langle\vec{Y}, \nabla_{\vec{X}}\vec{X}\rangle - 4\Ric(\vec{Y}, \vec{X})\right)d\tau \notag \\
	& = \int^{\tau_2}_{\tau_1}\biggl\{2\frac{d}{d\tau}\left(\sqrt{\tau}\langle\vec{Y}, \vec{X}\rangle\right) - \frac{1}{\sqrt{\tau}}\langle\vec{Y}, \vec{X}\rangle \notag \\ 
	& \hspace{11pt} + \sqrt{\tau}\left(\langle\nabla\scalarcurvature, \vec{Y}\rangle - 2\langle\vec{Y}, \nabla_{\vec{X}}\vec{X}\rangle - 4\Ric(\vec{X}, \vec{Y})\right)\biggr\}d\tau \notag \\
	& = 2\sqrt{\tau}\langle\vec{X}, \vec{Y}\rangle\Big|^{\tau_2}_{\tau_1} + \int^{\tau_2}_{\tau_1}\sqrt{\tau}\left\langle\vec{Y}, \nabla\scalarcurvature - 2\nabla_{\vec{X}}\vec{X} - 4\Ric(\vec{X}, \cdot) - \frac{1}{\tau}\vec{X}\right\rangle d\tau,
\end{align}
\end{proof}

\subsubsection[The $\length$-Geodesics, or the Euler–Lagrange Equation for a Critical Curve]{The $\protect\pseudobold{\length}$-Geodesics, or the Euler–Lagrange Equation for a Critical Curve}

We can then write the \emph{$\length$-geodesics equation}, that is, the \emph{Euler–Lagrange equation for critical curves} about the $\length$-length:
\begin{equation}
	\nabla_{\vec{X}}\vec{X} - \frac{1}{2}\nabla\scalarcurvature + \frac{1}{2\tau}\vec{X} + 2\Ric(\vec{X}, \cdot) = 0,
\end{equation}
where $\nabla\scalarcurvature$ designates a horizontal gradient.

\subsection{Perelman's No Local Collapsing Theorem}

 Hamilton in \cite{Hamilton "Four-Manifolds with Positive Isotropic Curvature"} had hoped for the possibility that, in treating surgically (Section \ref{subsection "Geometro-topological Surgery of Cutting off and Gluing Back"}) a space of dimension 3 with no prior conditions and uniformly bounded curvature, after a finite number of cutting/gluing operations, the normalized Ricci flow exists for all time $t \to \infty$, and that $\textcyrillic{\textit{ф}}_\textsc{r}^\mathrm{nor}$ is non-singular \cite{Hamilton "Non-singular solutions of the Ricci flow on three-manifolds"}. 

Perelman does not confirm the Hamilton's expectation; but he finds a way to \emph{control} the onset of unwanted singularities (Section \ref{subsection "Occurrence of Singularities of the Ricci Flow"}), through the so-called \emph{no local collapsing} (\textsc{nlc}) theorem \cite[secc. 4, 8]{Perelman "The entropy formula for the Ricci flow and its geometric applications"} \cite[sec. 7]{Perelman "Ricci flow with surgery on three-manifolds"} related to the action of the Ricci flow with or without surgery, but in any case with a classification of the asymptotic behavior of blow-ups of singularities in dimension 3, as Hamilton \cite{Hamilton "The Formation of Singularities in the Ricci Flow"} had already started to do, thanks to his reinterpretations of the Harnack inequality (Section \ref{subsection "Li–Yau's & Hamilton's Harnack Inequalities, and Space-Time Gradient Estimate"}). Let us find out what it is.

\subsubsection[$\kappa$-(Non-)collapsing, and Ball Volume Ratio]{$\mathbold{\kappa}$-(Non-)collapsing, and Ball Volume Ratio}

The \textsc{nlc} theorem is a consequence of the monotonicity \eqref{subequations "Monotonicity formula for the W-functional"} of the $\mathscr{W}$-functional. Before stating the theorem, there is need for definitions.

\begin{definitiones}[$\kappa$-non-collapsing and $\kappa$-collapsing]
\label{definitiones "k-non-collapsing and k-collapsing conditions"}
~\enumerationisinitium
\item \emph{The $\kappa$-non-collapsing and $\kappa$-collapsing on the topological space}.
\subenumerationisinitium
\item Let $(\mathcal{M}^n, g)$ be an $n$-dimensional Riemannian manifold. For for some positive constants $\reflectedepsilon \in (0, \infty]$ and $\kappa > 0$, a metric $g_{\mu\nu}$ is said to be \emph{$\kappa$-non-collapsed below the scale $\reflectedepsilon$} if, for a metric $n$-ball $\mathbb{B}(x, \rho) \subset \mathcal{M}^n$ of radius $\rho < \reflectedepsilon$, for any $x \in \mathcal{M}^n$, it happens that, equivalently, 
\begin{align}
\begin{cases}
	\displaystyle \frac{\volume\bigl(\mathbb{B}(x, \rho)\bigr)}{\rho^n} \geqslant \kappa, \\
	\volume\bigl(\mathbb{B}_{\rho^{-2}g}(x, 1)\bigr) \geqslant \kappa, \\
	\volume\bigl(\mathbb{B}(x, \rho)\bigr) \geqslant \kappa\rho^n,	
\end{cases}	
\end{align}
with $\rho > 0$, and the Riemann curvature tensor is $|\Rie(y)| \leqslant \rho^{-2}$, for any $y \in \mathbb{B}(x, \rho)$, cf. Scholium \ref{scholium "k-non-collapsing condition"}. The metric $g_{\mu\nu}$ is \emph{$\kappa$-non-collapsed at all the scales} if $g_{\mu\nu}$ is $\kappa$-non-collapsed below the scale $\reflectedepsilon < \infty$.
\item A metric $g_{\mu\nu}$, by contrast, is \emph{$\kappa$-collapsing at the scale $\rho$ and the point $x$} if 
\begin{equation}
	\frac{\volume\bigl(\mathbb{B}(x, \rho)\bigr)}{\rho^n} \leqslant \kappa.
\end{equation}
\subenumerationisfinis
\item \emph{The $\kappa$-non-collapsing and $\kappa$-collapsing in the Ricci flow}.
\subenumerationisinitium
\item Let $g_t \viz g(t)$, $t \in [0, T)$, be a solution to the Ricci flow $\textcyrillic{\textit{ф}}_\textsc{r}$ (Definition \ref{definitio "Ricci flow"}), where $T \in (0, \infty]$. Then $g_t$ is said to be \emph{$\kappa$-non-collapsed below the scale $\reflectedepsilon$} if $g_t$ keeps this $\kappa$-condition for all $t \in [0, T)$.
\item Given an $n$-dimensional manifold $\mathcal{M}^n$, a solution $g_t$, $t \in [0, T)$, is called \emph{locally collapsed at $T$} if there is a sequence 

· of points $x_k \in \mathcal{M}^n$, 

· of times $t_k \to T $,
 
· of metric balls $\mathbb{B}_{g(t_k)} = \mathbb{B}(x_k, \rho_k)$ at times $t_k$, with radius $\rho_k \in (0, \infty)$, such that $\rho^2_k/t_k$ is  uniformly bounded, $|\Rie|(g_{\mu\nu})_{t_k} \leqslant \rho^{-2}_k$ in $\mathbb{B}_{g(t_k)}$, and $\rho^{-n}_k\volume(\mathbb{B}_{g(t_k)}) \to 0$, or $\lim_{k \to \infty}\rho^{-n}_k\volume(\mathbb{B}_{g(t_k)}) = 0$. \definitiosymbol  
\subenumerationisfinis
\enumerationisfinis
\end{definitiones}

\begin{theorema}[Perelman's no local collapsing]
\label{theorema "Perelman's no local collapsing}
Let $g_t \viz g(t)$, $t \in [0, T)$, be a smooth solution to the Ricci flow $\textcyrillic{\textit{ф}}_\textsc{r}$, i.e. $\frac{\partial (g_{\mu\nu})_t}{\partial t} = -2\Ric(g_t)$ on a closed $n$-dimensional Riemannian manifold $\mathcal{M}^n$. Given a constant $\reflectedepsilon \in (0, \infty)$, i.e. a finite scale $\reflectedepsilon > 0$, if $T < \infty$, then $g_t$ is $\kappa$-non-collapsed below the scale $\reflectedepsilon$ (as described in Definitions \ref{definitiones "k-non-collapsing and k-collapsing conditions"}), for some constant $\kappa = \kappa(n, g_0, T, \reflectedepsilon) > 0$, $g(0) = g_0$, and for every $t \in [0, T)$. 
\end{theorema}

\begin{corollarium}
\label{corollarium "Corollarium of the Perelman's no local collapsing"}
If the manifold $\mathcal{M}^n$ is a closed space, and $g_t$ is a solution to $\textcyrillic{\textit{ф}}_\textsc{r}$ on $[0, T)$, $T < \infty$, ergo $g_t$ is not locally collapsing at $T$.  
\end{corollarium}

\begin{proof}[Proof of the Theorem \ref{theorema "Perelman's no local collapsing}]
~\enumerationisinitium
\item[(\textgreek{α}) — \textbf{step I}.] A first part of the demonstration is to show that, for a spatial dimensionality $n \geqslant 2$, there exists a form 
\begin{equation}	
\label{equation "bbmu-form"}	
	\bbmu(g, \rho^2) \leqslant \log{\frac{\volume\bigl(\mathbb{B}(x, \rho)\bigr)}{\rho^n}} + c(n, \reflectedepsilon), 
\end{equation}	
$g \viz g_{\mu\nu}$, for a constant $c\bigl(n, \reflectedepsilon, c(n)\bigr)$, and the previous quantities, that is, a finite scale $\reflectedepsilon \in (0, \infty)$, a point $x \in \mathcal{M}^n$, and the radius $\rho \in (0, \reflectedepsilon]$, such that the Ricci curvature tensor is $\Ric \geqslant - c(n)\rho^{-2}$, the Ricci scalar is $\scalarcurvature \leqslant c(n)\rho^{-2}$, and both are in the ball $\mathbb{B}(x, \rho)$. It follows that, for $\kappa > 0$ and $\rho > 0$, $g$ is $\kappa$-collapsed on the scale $\rho \leqslant \reflectedepsilon$, and that $\bbmu(g, \rho^2) \leqslant \log{\kappa} + c(n, \reflectedepsilon)$. 
\item[(\textgreek{β}) — \textbf{step II}.] Putting $t \in [\frac{T}{2}, T)$, one finds that 
\subenumerationisinitium
\item $\Ric_{g_t} \geqslant -c(n)\rho^{-2}$ and ${\scalarcurvature}_{g_t} \leqslant c(n)\rho^{-2}$ are effectively in $\mathbb{B}_{g_t}(x, \rho)$,
\item $t + \rho^2 \in [\frac{T}{2}, T + \reflectedepsilon^2)$, for $0 < \rho \leqslant \reflectedepsilon$, so (owing to the monotonicity structure) $-c(g_0, T, \reflectedepsilon) \leqslant \bbmu(g_0, t +\rho^2) \leqslant \bbmu(g_t, \rho^2)$. 
\subenumerationisfinis

This last expression, combined with \eqref{equation "bbmu-form"}, leads to 
\begin{equation}
	-c(g_0, T, \reflectedepsilon) \leqslant \bbmu\bigl(g_t,\rho^2\bigr) \leqslant \log{\frac{\volume_{g_t}\bigl(\mathbb{B}(x, \rho)\bigr)}{\rho^n}} + c(n, \reflectedepsilon), 
\end{equation}
and hence 
\begin{equation}
	\frac{\volume_{g_t}\bigl(\mathbb{B}(x, \rho)\bigr)}{\rho^n} \geqslant \exp{\bigl\{-c(g_0, T, \reflectedepsilon) - c(n, \reflectedepsilon)\bigr\}} > 0,
\end{equation}  
in which we can enforce the equality $\exp{\bigl\{-c(g_0, T, \reflectedepsilon) - c(n, \reflectedepsilon)\bigr\}} = \kappa_1(g_0, T, \reflectedepsilon)$. 

Remember that, as imposed by the theorem, $T < \infty$, thereby $-c(g_0, T, \reflectedepsilon) > -\infty = \inf_{\tau \in [\frac{T}{2}, T + \reflectedepsilon^2]}\bbmu(g_0, \tau)$, by which (inasmuch as $\frac{T}{2} < T$) we can write $\kappa_0 = \kappa_0 (g_0, T, \reflectedepsilon) > 0$, so that $g_t$ is $\kappa_0$-non-collapsed on the scale $\reflectedepsilon$, for any $t \in [0, \frac{T}{2}]$. The only conclusion is that $\kappa(g_0, T, \reflectedepsilon) = \min\{\kappa_0, \kappa_1\}$.
\item[(\textgreek{γ}) — \textbf{step III}.] Hints on step I. Its demonstration arises from the infimum statement
\begin{equation}
	\bbmu(g, \rho^2) \leqslant \mathscr{W}(g, \rotatedupsilon, \rho^2) = \int_{\mathcal{M}^n}\Bigl\{\rho^2\left(\scalarcurvature + |\nabla\rotatedupsilon|^2\right) + \rotatedupsilon - n\Bigr\}(4\pi\rho^2)^{-\tfrac{n}{2}}e^{-\rotatedupsilon}d\bbmu.
\end{equation} 
By placing $\textit{\th}^2 = (4\pi\rho^2)^{-\tfrac{n}{2}}e^{-\rotatedupsilon}$, with a function $\textit{\th} > 0$, one proceeds to
\begin{equation}
\label{equation "Equation with positive function for the NLC"}
	\bbmu(g, \rho^2) \leqslant \mathscr{W}_1(g, \textit{\th}, \rho^2) = \int_{\mathcal{M}^n}\Bigl\{\rho^2\left(4|\nabla\textit{\th}|^2 + \scalarcurvature\textit{\th}^2\right) + \rotatedupsilon\textit{\th}^2\Bigr\}d\bbmu,
\end{equation}
for $\int_{\mathcal{M}^n}\textit{\th}^2d\bbmu = 1$. By way of a cutoff function $\cutofffunction \colon [0, \infty) \to [0, 1]$, having $\cutofffunction = 1$ on $[0, \frac{1}{2}]$, $\cutofffunction = 0$ on $[1, \infty)$, and $|\cutofffunction^{(\mathrm{i})}| \leqslant 3$, and then choosing a constant $c_\cutofffunction = c_\cutofffunction(g, x, \rho)$, it is determined that
\begin{equation}
\label{equation "Equation with cutoff function for the NLC"}
	\textit{\th}^2_1(y) = (4\pi\rho^2)^{-\tfrac{n}{2}}\cutofffunction\Bigl(d(y, x)/\rho\Bigr)^2e^{-c_\cutofffunction}.
\end{equation}

To connect the quantity $\frac{\text{volume of } \mathbb{B}(x, \rho)}{\rho^n}$ and the constant $c_\cutofffunction$, it is necessary to impose an expression of this type,
\begin{equation}
\label{equation "Equation to connect the ball volume ratio and the cutoff function"}
	\log{\frac{\volume\bigl(\mathbb{B}(x, \rho)\bigr)}{\rho^n}} - c_1(n, \reflectedepsilon) \leqslant c_\cutofffunction \leqslant \log{\frac{\volume\bigl(\mathbb{B}(x, \rho)\bigr)}{\rho^n}},
\end{equation}
for a constant's value $c_1(n, \reflectedepsilon) < \infty$, and a radius $\rho \in (0, \reflectedepsilon)$. By means of Eq. \eqref{equation "Equation with cutoff function for the NLC"}, via $|\cutofffunction^{(\mathrm{i})}| \leqslant 3$ and Eq. \eqref{equation "Equation to connect the ball volume ratio and the cutoff function"}, we get to define that
\begin{equation}
	\rho^2|\nabla\textit{\th}_1|^2 \leqslant (4\pi\rho^2)^{-\tfrac{n}{2}}e^{-c_\cutofffunction}|\cutofffunction^{(\mathrm{i})}|^2 \leqslant 9(4\pi)^{-\tfrac{n}{2}}e^{-c_\cutofffunction}/\rho^n \leqslant \frac{9(4\pi)^{-\tfrac{n}{2}}e^{c_1(n, \reflectedepsilon)}}{\volume\bigl(\mathbb{B}(x, \rho)\bigr)}.
\end{equation} 
We would point out that the energetic-entropic values of the $\mathscr{W}$-functional are regulated by 
\begin{equation}
	\int_{\mathcal{M}^n}\rho^2\Bigl(4|\nabla\textit{\th}_1|^2 + \scalarcurvature\textit{\th}^2_1\Bigr)d\bbmu \leqslant c(n, \reflectedepsilon),
\end{equation}
and 
\begin{align}
	\int_{\mathcal{M}^n}\rotatedupsilon_1\textit{\th}^2_1d\bbmu & \leqslant c_\cutofffunction + \tfrac{1}{e}(4\pi)^{-\tfrac{n}{2}} \cdot \frac{e^{-c_\cutofffunction}\volume\bigl(\mathbb{B}(x, \rho)\bigr)}{\rho^n} \leqslant c_\cutofffunction + \tfrac{1}{e}(4\pi)^{-\tfrac{n}{2}}e^{c_1(n, \reflectedepsilon)} \notag \\
	& \leqslant \log{\frac{\volume\bigl(\mathbb{B}(x, \rho)\bigr)}{\rho^n}} + c(n, \reflectedepsilon).
\end{align}
So this means $\mathscr{W}_1(g, \textit{\th}_1, \rho^2) \leqslant \log{\frac{\text{volume of }\mathbb{B}(x, \rho)}{\rho^n}} + c(n, \reflectedepsilon)$. Since $\textit{\th}$ must be positive, cf. Eq. \eqref{equation "Equation with positive function for the NLC"}, it is appropriate to introduce a small value $\varepsilon \in \mathrm{v}_\varepsilon$, $\mathrm{v}_\varepsilon = 0, 1/\sqrt{\volume(\mathcal{M}^n)}$, by which $\textit{\th}_\varepsilon(y) c_\varepsilon(\textit{\th}_1 + \varepsilon) > 0$, and $\frac{1}{2} \leqslant c_\varepsilon \leqslant 1$, which accordingly gives 
\begin{equation}
	\bbmu(g, \rho^2) \leqslant \mathscr{W}_1(g, \textit{\th}_\varepsilon, \rho^2),
\end{equation}	 
for any $\varepsilon$. From the value $\varepsilon \in \mathrm{v}_\varepsilon$ it also follows that $\lim_{\varepsilon \to 0_+}c_\varepsilon = 1$, and $\lim_{\varepsilon \to 0_+}\textit{\th}_\varepsilon = \textit{\th}_1$. If $\lim_{\varepsilon \to 0_+}\varepsilon\log{\varepsilon} = 0$, then 
\begin{equation}
	\lim_{\varepsilon \to 0_+}\mathscr{W}_1(g, \textit{\th}_\varepsilon, \rho^2) = \mathscr{W}_1(g, \textit{\th}_1, \rho^2), 
\end{equation}	
and therefore $\bbmu(g, \rho^2) \leqslant \log{\frac{\text{volume of }\mathbb{B}(x, \rho)}{\rho^n}} + c(n, \reflectedepsilon)$. 
\item[(\textgreek{δ}) — \textbf{step IV}.] What happens next? For a constant $\kappa > 0$ and a radius $\rho > 0$, the metric tensor field $g$ is $\kappa$-collapsed on the scale $\rho$, and thus there is a volume ratio $\frac{\text{volume of }\mathbb{B}(x, \rho)}{\rho^n} \leqslant \kappa$ of a ball centered at $x$, in complete \emph{contradiction} to the monotonic request assumed under the Theorem \ref{theorema "Perelman's no local collapsing}.
\enumerationisfinis
\end{proof}

Corollary \ref{corollarium "Corollarium of the Perelman's no local collapsing"} is equivalent to the Theorem \ref{theorema "Perelman's no local collapsing}, so there is no need to prove it.

\subsection{Non-collapsing via Sobolev Inequalities}

The \textsc{nlc} Theorem \ref{theorema "Perelman's no local collapsing} can also be developed as a result of Sobolev embedding instances (cf. Theorems \ref{theorema "Carron–Akutagawa's"} and \ref{theorema "Sobolev inequality and the Ricci flow"}), so it appears related to certain Sobolev spaces; this is because the Perelman's entropy-energy $\mathscr{W}$-functional \eqref{equation "Perelman's entropy-energy W-functional"}, and its monotonicity \eqref{subequations "Monotonicity formula for the W-functional"}, entail specific Sobolev inequalities along the Ricci flow, known by the name \emph{logarithmic  Sobolev inequalities} (see Proposition \ref{propositio "Logarithmic Sobolev inequality on a Riemannian space"}).

\subsubsection[Sobolev Space $\Sobolev^{k, p}(\Omega)$]{Sobolev Space $\protect\pseudobold{\Sobolev^{k, p}(\Omega)}$}
\label{subsubsection "Sobolev Space $W^{k, p}(Omega)$"}

Let us provide some basic indications, in the most general terms, before going forward in the hot core of the matter.

\begin{definitio}[Sobolev space]
A \emph{Sobolev space} \cite{Sobolev "Sur un theoreme d'analyse fonctionnelle"} \cite{Sobolev "Some Applications of Functional Analysis in Mathematical Physics"} of type $\Sobolev^{k, p}(\Omega) \viz \Sobolev^{k, p}(\Omega, \mathbb{R}^k)$, for $k \in \mathbb{Z}_*$ (non-negative integer), or even $k \in \mathbb{N}$ (natural number), $1 \leqslant p \leqslant \infty$, with a domain (open subset) $\Omega \subset \mathbb{R}^n$ of real numbers, is a \emph{normed vector space of functions}, or the space of equivalence classes of functions, having \emph{weak derivatives}. Let us put it more technically.

· The space
\begin{equation}
	\Sobolev^{k, p}(\Omega) = \bigl\{\upsilon \in \mathscr{C}^k \mid D^\alpha\upsilon, \forall \alpha \in \mathbb{Z}_*, |\alpha| \leqslant k \in \{0, 1, 2, \mathellipsis\}\bigr\}
\end{equation}
is the set of all functions $\upsilon \colon \Omega \to \mathbb{R}^n$ locally summable, under which, fixed a multi-index $\alpha = (\alpha_1, \mathellipsis, \alpha_n) \in \mathbb{Z}_*$ of order $|\alpha| \leqslant k \in \mathbb{Z}_*$, where $|\alpha| = \alpha_1 + \cdots + \alpha_n$, there is a weak $\alpha$-th partial derivative 
\begin{equation}
\label{equation "Weak alpha-th partial derivative"}
	D^\alpha\upsilon = \frac{\partial^{|\alpha|}}{\partial x^\alpha} = \frac{\partial^{\alpha_1}}{\partial x^{\alpha_1}_1} \cdots \frac{\partial^{\alpha_n}}{\partial x^{\alpha_n}_n}\upsilon
\end{equation} 
of $\upsilon$ lying in the (Lebesgue) function space $\Lebesgue^p$, so $\Sobolev^{k, p}(\Omega) = \Lebesgue^p(\Omega)$; the norm of $\Sobolev^{k, p}(\Omega)$ is defined as
\begin{align}
	\|\upsilon\|_{\Sobolev^{k, p}(\Omega)} = 
	\begin{cases}
	\displaystyle
	\|\upsilon\|_{k, p} = \left\{\sum_{|\alpha| \leqslant k}\int_\Omega|D^\alpha\upsilon|^p_{\Lebesgue^p(\Omega)}\right\}^\frac{1}{p}, \enspace 1 \leqslant p < \infty \\
	\displaystyle
	\|\upsilon\|_{k, \infty} = \sum_{|\alpha| \leqslant k} \esssup_{\Omega}\|D^\alpha\upsilon\|^\infty_{\Lebesgue^\infty(\Omega)}, \enspace p = \infty,
	\end{cases}
\end{align}
where $\esssup$ indicates the essential supremum operator.

· The space 
\begin{equation}
	\Sobolev^{k, p}(\Omega) = \bigl\{\upsilon \in \Lebesgue^p(\Omega) \mid D^\alpha\upsilon, \forall \alpha \in \mathbb{N}^n_0, |\alpha| \leqslant k\bigr\}
\end{equation}
is the set of all functions $\upsilon \in \Lebesgue^p(\Omega) \viz \Lebesgue^p(\Omega, \mathbb{R}^k)$, $\Omega \subset \mathbb{R}^n$, with weak derivatives $D^\alpha\upsilon$ in $\Lebesgue^p(\Omega)$, for any $\alpha = (\alpha_1, \mathellipsis, \alpha_n) \in \mathbb{N}^n_0$, $|\alpha| \leqslant k \in \mathbb{N}$, the norm of which is 
\begin{equation}
	\|\upsilon\|_{\Sobolev^{k, p}(\Omega)} = \|\upsilon\|_{\Lebesgue^p(\Omega)} + \sum_{|\alpha| \leqslant k}\|D^\alpha\upsilon\|_{\Lebesgue^p(\Omega)}.
\end{equation}
\definitiosymbol
\end{definitio}

\begin{margo}[Tempered distribution]
The weak $\alpha$-th partial derivative \eqref{equation "Weak alpha-th partial derivative"} corresponds to what is called a \emph{tempered distribution}, which is just a \emph{slowly increasing} distribution, for any multi-index $\alpha$, whilst $\upsilon$ is a tempered function (by making explicit a tempered distribution). \margosymbol	
\end{margo}

\begin{scholium}
~\enumerationisinitium
\item For $p = 2$, the Sobolev space $\Sobolev^{k, p}(\Omega)$ is (coincided with) a \emph{Hilbert space}, i.e. $\Sobolev^{k, 2}(\Omega) = \mathfrak{H}^k(\Omega)$, and $\Lebesgue^2(\Omega) = \mathfrak{H}^0(\Omega)$.
\item As a function space, for $1 \leqslant p \leqslant \infty$, and $k_\mathbb{N} \in  \{1, 2, 3, \mathellipsis\}$, the Sobolev space $\Sobolev^{k, p}(\Omega)$ is (coincided with) a \emph{Banach space} \cite{Banach "Sur les operations dans les ensembles abstraits et leur application aux equations integrales"}.
\item The local Sobolev space is simply denoted by $\Sobolev^{k, p}_\mathrm{loc}(\Omega)$. We say that $\upsilon$ is a function of $\Sobolev^{k, p}_\mathrm{loc}(\Omega)$ if $\upsilon \in \Sobolev^{k, p}(\Chi)$, for any $\Chi \Subset \Omega$, and the embedding is $\Sobolev^{k, p}_\mathrm{loc}(\Omega) \hookrightarrow \Sobolev^{k, p}(\textgreek{\textit{Χ}})$, for any $\Chi \Subset \Omega$. \scholiumsymbol
\enumerationisfinis
\end{scholium}

\subsubsection{Sobolev Embedding for a Null Trace Space, Orlicz Space, and General Sobolev Inequality}
\label{subsubsection "Sobolev Embedding for a Null Trace Space, Orlicz Space, and General Sobolev Inequality"}

Let us endeavour to define the concept of Sobolev embedding, accompanied by that of inequality, for a generic space $\Sobolev^{k, p}_0(\Omega)$, with zero trace, and a Riemannian space $\Sobolev^{k, p}(\mathcal{M})$, respectively, but in both cases having $k = 1$, since they will come in handy for the purposes of this Section.

\begin{definitio}[Sobolev space with zero trace]
\label{definitio "Sobolev space with zero trace"}
Letting $\Omega \subset \mathbb{R}^n$, $k  \in \mathbb{N}$, and $1 \leqslant p < \infty$, the Sobolev space $\Sobolev^{k, p}_0(\Omega)$, is the closed subspace of functions $\upsilon \in \Sobolev^{k, p}(\Omega)$, that is, the closure in $\Sobolev^{k, p}(\Omega)$ of the vector (sub)space $\mathscr{C}^\infty_\mathrm{c}(\Omega)$ of forms compactly supported, with $\trace(\upsilon) = 0$. The case $k = 1$,
\begin{equation}
	\Sobolev^{1, p}_0(\Omega) = \bigl\{\upsilon \in \Sobolev^{k, p}(\Omega) \mid \mathscr{C}^\infty_\mathrm{c}(\Omega)\text{-form}, \trace(\upsilon) = 0 \bigr\},
\end{equation}
repeats the same definition. \definitiosymbol
\end{definitio}

\begin{propositio}[Sobolev embedding for a null trace space]
In light of Definition \ref{definitio "Sobolev space with zero trace"}, the Sobolev embedding for the space $\Sobolev^{1, p}_0(\Omega)$, given a domain $\Omega \subset \mathbb{R}^n$, is
\begin{subequations}
	\begin{empheq}[left = {\Sobolev^{1, p}_0(\Omega) \hookrightarrow \empheqlbrace}]{align}
	& 
	\label{empheq "Sobolev embedding for a null trace space 1"}
	\Lebesgue^{\frac{np}{n - p}}(\Omega), \\
	& 
	\label{empheq "Sobolev embedding for a null trace space 2"}
	\mathscr{C}^{1 - \frac{n}{p}}(\Omega),
	\end{empheq}
\begin{align}
	& 
	\label{align "Sobolev embedding for a null trace space 3"}
	\Sobolev^{1, p}_0(\Omega) \hookrightarrow \Lebesgue^\rotatedpsi(\invertedbreve{\Omega}) = \left\{\textcyrillic{\textit{л} } \colon \invertedbreve{\Omega} \to \mathbb{R}^n \mathrel{\Bigg|}\int_{\invertedbreve{\Omega}}\rotatedpsi(\lambda\textcyrillic{\textit{л}})d\bbmu(x) < \infty\right\},
\end{align}
\end{subequations}
for $1 \leqslant p < n$ in \eqref{empheq "Sobolev embedding for a null trace space 1"}, $p > n$ in \eqref{empheq "Sobolev embedding for a null trace space 2"}, and $p = n$ in \eqref{align "Sobolev embedding for a null trace space 3"}, which is an example of null trace Sobolev space embedded in Orlicz space.
\end{propositio}
 
\begin{scholium}[Orlicz space]
Given a $\sigma$-finite measure space $(\invertedbreve{\Omega}, \bbmu)$ (see Chapters \ref{chapter "Geometric and Topological Aspects of Complexity and Dynamics, Part I. Flows, Hyperbolicity, and Foliations"} and \ref{chapter "Geometric and Topological Aspects of Complexity and Dynamics, Part II: Ergodicity and Entropy"}), the \emph{Orlicz function space}, aka \emph{Birnbaum–Orlicz space} \cite{Birnbaum und Orlicz "Uber die Verallgemeinerung des Begriffes der zueinander konjugierten Potenzen"}, $\Lebesgue^\rotatedpsi(\invertedbreve{\Omega}) \viz \Lebesgue^\rotatedpsi(\invertedbreve{\Omega}, \bbmu)$, formed by the Orlicz function $\rotatedpsi \colon [0, \infty) \to [0, \infty]$, is the space of every \emph{Lebesgue (measurable) function} $\textcyrillic{\textit{л} } \colon \invertedbreve{\Omega} \to \mathbb{R}^n$ under which $\int_{\invertedbreve{\Omega} \subset \mathbb{R}^n} \rotatedpsi(\lambda\textcyrillic{\textit{л}})d\bbmu(x) < \infty$, for some $\lambda > 0$, where $\lambda\textcyrillic{\textit{л}} = \lambda|\textcyrillic{\textit{л}}(x)|$. \scholiumsymbol
\end{scholium}

We now come to the Sobolev inequalities. They are familiar when we have to provide the proof of embedding theorems, in particular for compact objects, and to verify connections with heat kernels. Here is a glimpse. 

\begin{exemplum}[General Sobolev inequality]
Given a Riemannian $n$-space $(\mathcal{M}^n, g)$, a function $\upsilon \in \Sobolev^{1, p}(\mathcal{M})$, and a constant $c$, we say that there is an inequality
\begin{equation}
	\left(\int_{\mathcal{M}^n}\upsilon^{\frac{np}{(n - p)}}d\bbmu\right)^{\frac{(n - p)}{np}} \leqslant c\left(\int_{\mathcal{M}^n}|\nabla\upsilon|^p d\bbmu\right)^\frac{1}{p} + c\left(\int_{\mathcal{M}^n}|\upsilon|^p d\bbmu\right)^\frac{1}{p},
\end{equation}
for all $p \in [1, n)$, $\Sobolev^{1, p}(\mathcal{M}) \subset \Lebesgue^{\frac{np}{(n - p)}}(\mathcal{M})$, by which $c = c(\mathcal{M}) > 0$. This is a \emph{Sobolev inequality in a generalized form}. \exemplumsymbol
\end{exemplum}

\subsubsection{Carron–Akutagawa's Sobolev Embedding}
\label{subsubsection "Carron–Akutagawa's Sobolev Embedding"}

We shall move on to the next embedding \& inequality statement by G. Carron \cite{Carron "Inegalites isoperimetriques de Faber-Krahn et consequences"} and K. Akutagawa \cite{Akutagawa "Yamabe metrics of positive scalar curvature and conformally flat manifolds"}. Its importance is due to the fact that it can be used as a \emph{first evidence for the non-collapsing request} within the Sobolev structures.

\begin{theorema}[Carron–Akutagawa's]
\label{theorema "Carron–Akutagawa's"}
Let $\mathcal{M}^n$ be a complete $n$-dimensional Riemannian manifold endowed with a metric tensor $g$. We are adopting the embedding $\Sobolev^{1, p}(\mathcal{M}) \hookrightarrow \Lebesgue^{\frac{np}{n - p}}(\mathcal{M})$, under which, for all $p \in [1, n)$, and for every function $\upsilon \in \Sobolev^{1, p}(\mathcal{M})$, one has the following inequalities
\begin{subequations}
	\begin{empheq}[left = {\empheqlbrace}]{align}
	& \|\upsilon\|_{\frac{np}{n - p}} \leqslant c\Bigl(\|\nabla\upsilon\|_p + \|\upsilon\|_p\Bigr), \\
	& \volume_g\bigl(\mathbb{B}(x, \rho)\bigr) \geqslant 
	\begin{cases}
	\min\left(\frac{1}{2c}\right)^n, \\
	\min\left(\frac{\rho}{2^{\frac{n + 2p}{p}}c}\right)^n,
	\end{cases}
	\end{empheq}
\end{subequations}
letting $\mathbb{B}(x, \rho) \subset \mathcal{M}^n$ be an $n$-ball, with $\rho > 0$.
\end{theorema}

\begin{proof}
From \emph{Hölder inequality} \cite{Holder "Ueber einen Mittelwertsatz"}, one obtains
\begin{subequations}
\begin{align}
	& \|\upsilon\|_p \leqslant \left[\volume_g\bigl(\mathbb{B}(x, \rho)\bigr)\right]^\frac{1}{n}\|\upsilon\|_{\frac{np}{n - p}} \leqslant \left[\volume_g\bigl(\mathbb{B}(x, \rho)\bigr)\right]^\frac{1}{n} c\Bigl(\|\nabla\upsilon\|_p + \|\upsilon\|_p\Bigr), \\
	& \|\upsilon\|_p - \left[\volume_g\bigl(\mathbb{B}(x, \rho)\bigr)\right]^\frac{1}{n} c\|\upsilon\|_p \leqslant \left[\volume_g\bigl(\mathbb{B}(x, \rho)\bigr)\right]^\frac{1}{n} c\|\nabla\upsilon\|_p, \\
	& 1 - \left[\volume_g\bigl(\mathbb{B}(x, \rho)\bigr)\right]^\frac{1}{n} c \leqslant \left[\volume_g\bigl(\mathbb{B}(x, \rho)\bigr)\right]^\frac{1}{n} c\textit{\DH}, \\
	& \frac{1}{\left[\volume_g\bigl(\mathbb{B}(x, \rho)\bigr)\right]^\frac{1}{n}} - c \leqslant c\textit{\DH}, \text{ putting } \textit{\DH} = \frac{\|\nabla\upsilon\|_p}{\|\upsilon\|_p}.
\end{align}
\end{subequations}
Here two possibilities branch out: either $\left[\volume_g\bigl(\mathbb{B}(x, \rho)\bigr)\right]^\frac{1}{n} \geqslant \frac{1}{2c}$ or $\leqslant \frac{1}{2c}$, so, in the latter case, $c \leqslant \frac{1}{2[\text{volume of }\mathbb{B}(x, \rho)]^{1/n}}$ and $\frac{1}{2[\text{volume of }\mathbb{B}(x, \rho)]^{1/n}} \leqslant c\textit{\DH}$. 

Considering that $\upsilon$ is a \emph{Lipschitz-type function}, we write
\begin{subequations}
\begin{align}
	& \|\upsilon\|_p = \left(\int_{\mathbb{B}(x, \rho)}|\upsilon|^p d\bbmu\right)^\frac{1}{p} \geqslant \left(\int_{\mathbb{B}(x, \frac{\rho}{2})}|\upsilon|^p d\bbmu\right)^\frac{1}{p}, \\
	& \|\upsilon\|_p \geqslant \left(\int_{\mathbb{B}(x, \frac{\rho}{2})}\left(\frac{\rho}{2}\right)^p d\bbmu\right)^\frac{1}{p} = \frac{\rho}{2}\left[\volume_g\Bigl(\mathbb{B}\left(x, \tfrac{\rho}{2}\right)\Bigr)\right]^\frac{1}{p}.
\end{align}
\end{subequations}
Since $\frac{1}{2[\text{volume of }\mathbb{B}(x, \rho)]^{1/n}} \leqslant c\textit{\DH}$,
\begin{subequations}
\begin{align}
	& \frac{1}{2\left[\volume\bigl(\mathbb{B}(x, \rho)\bigr)\right]^{\frac{1}{n}}} \leqslant c\frac{\left[\volume_g\bigl(\mathbb{B}(x, \rho)\bigr)\right]^{\frac{1}{p}}}{\frac{\rho}{2}\left[\volume_g\bigl(\mathbb{B}(x, \frac{\rho}{2})\bigr)\right]^\frac{1}{p}}, \\
	& \volume_g\bigl(\mathbb{B}(x, \rho)\bigr) \geqslant \left(\frac{\rho}{4c}\right)^{\frac{np}{n + p}}\left[\volume_g\Bigl(\mathbb{B}\left(x, \tfrac{\rho}{2}\right)\Bigr)\right]^{\frac{n}{n + p}}.
\end{align}
\end{subequations}
Let a distance $\mathrm{R} > 0$ be given. By induction,
\begin{align}
	& \volume_g\bigl(\mathbb{B}(x, \mathrm{R})\bigr) \geqslant \left(\frac{\mathrm{R}}{4c}\right)^{\frac{np}{n + p}}\left[\volume_g\Bigl(\mathbb{B}\left(x, \tfrac{\mathrm{R}}{2}\right)\Bigr)\right]^{\frac{n}{n + p}} \\
	& \volume_g\Bigl(\mathbb{B}\left(x, \tfrac{\mathrm{R}}{2}\right)\Bigr) \geqslant \left(\frac{\frac{\rho}{2}}{4c}\right)^{\frac{np}{n + p}}\left[\volume_g\Bigl(\mathbb{B}\left(x, \tfrac{\mathrm{R}}{4}\right)\Bigr)\right]^{\frac{n}{n + p}}, 
\end{align}	
and, for every $r \in \mathbb{N}\backslash\{0\}$, 
\begin{subequations}
\label{subequations "Group of items with addition of elements for the Carron–Akutagawa's theorem"}
\begin{empheq}[left = {\empheqlbrace}]{align}
	& 
	\textstyle 
	\volume_g\bigl(\mathbb{B}(x, \mathrm{R})\bigr) \geqslant \bigl(\frac{\mathrm{R}}{2c}\bigr)^{pZ_1(r)}, \text{ with } Z_1(r) = \sum^r_{j = 1}\bigl(\frac{n}{n + p}\bigr)^j, \\
	&
	\textstyle
	\left(\frac{1}{2}\right)^{pZ_2(r)}, \text{ with } Z_2(r) = \sum^r_{j = 1}j\bigl(\frac{n}{n + p}\bigr)^j, \\
	&
	\textstyle
	\left[\volume_g\bigl(\mathbb{B}(x, \frac{\mathrm{R}}{2^r})\bigr)\right]^{pZ_3(r)}, \text{ with } Z_3(r) = \bigl(\frac{n}{n + p}\bigr)^r, 
	\end{empheq}
\end{subequations}
by multiplying the above three items. 

We denote by $\volume_\mathbb{R}$ the volume of the Euclidean ball of radius $\rho^n$. Hence
\begin{subequations}
\begin{align}
	& \left[\volume_g\Bigl(\mathbb{B}\left(x, \tfrac{\mathrm{R}}{2^r}\right)\Bigr)\right] \geqslant \volume_{\mathbb{R}^n}/2\left(\tfrac{\mathrm{R}}{2^r}\right)^n, \\
	& \left[\volume_g\Bigl(\mathbb{B}\left(x, \tfrac{\mathrm{R}}{2^r}\right)\Bigr)\right]^{Z_3(r)} \geqslant \Bigl((\volume_{\mathbb{R}^n}/2)2^{-rn} \cdot \mathrm{R}^n\Bigr)^{\left(\frac{n}{n + p}\right)^r} \notag \\
	& = (\volume_{\mathbb{R}^n}/2)^{\left(\frac{n}{n + p}\right)^r} 2^{-rn\left(\frac{n}{n + p}\right)^r}\mathrm{R}^{n\left(\frac{n}{n + p}\right)^r},
\end{align}
\end{subequations}
and thus
\begin{equation}
	\lim_{r \to \infty}\left[\volume_g\Bigl(\mathbb{B}\left(x, \tfrac{\mathrm{R}}{2^r}\right)\Bigr)\right]^{Z_3(r)} = 1.
\end{equation}
From the group \eqref{subequations "Group of items with addition of elements for the Carron–Akutagawa's theorem"}, i.e. 
\begin{equation}
	\volume_g\bigl(\mathbb{B}(x, \mathrm{R})\bigr) \geqslant \left(\frac{\mathrm{R}}{2c}\right)^{pZ_1(r)} \left(\frac{1}{2}\right)^{pZ_2(r)} \left[\volume_g\Bigl(\mathbb{B}(x, \tfrac{\mathrm{R}}{2^r})\Bigr)\right]^{pZ_3(r)},
\end{equation}
being that $pZ_1(r) \to n$, and $pZ_2(r) \to \frac{n^2 + np}{p}$, letting $r \to \infty$, we observe that
\begin{equation}
	\volume_g\bigl(\mathbb{B}(x, \mathrm{R})\bigr) \geqslant \left(\frac{\mathrm{R}}{2c}\right)^n \left(\frac{1}{2}\right)^{\frac{n^2 + np}{p}} \left(\frac{\mathrm{R}}{2^{\frac{n + 2p}{p}}c}\right)^n,
\end{equation}
and $\volume_g\bigl(\mathbb{B}(x, \mathrm{R})\bigr) \geqslant \min\left(\frac{1}{2c}, \frac{\mathrm{R}}{2^{\frac{n + 2p}{p}}c}\right)^n$.
\end{proof}

\subsubsection[Logarithmic Sobolev Inequality, and Lower Bound for the $\mathscr{W}$-Entropy-Energy]{Logarithmic Sobolev Inequality, and Lower Bound for the $\protect\pseudobold{\mathscr{W}}$-Entropy-Energy}
\label{subsubsection "Logarithmic Sobolev Inequality, and Lower Bound for the W-Entropy-Energy"}

Also the following theorem contributes to reach a demonstration of non-collapse on topological spaces, but this time centered on the Ricci flow. In combination with Theorem \ref{theorema "Carron–Akutagawa's"}, it provides, within the Sobolevian context, an alternative proof to the Perelman's result \ref{theorema "Perelman's no local collapsing} for the no local collapsing issue.
 
\begin{theorema}[Sobolev inequality and the Ricci flow]
\label{theorema "Sobolev inequality and the Ricci flow"}
Take a metric $g = g_t$ evolving under the Ricci flow $\frac{\partial g_t}{\partial t} = -2\Ric(g_t)$ \eqref{equation "unnormalized Ricci flow"}
 on a compact $(n \geqslant 3)$-dimensional Riemannian manifold $\mathcal{M}^n$. Then, for each function $\upsilon \in \Sobolev^{1, 2}(\mathcal{M})$, the Sobolev inequality
\begin{equation}
	\left(\int_{\mathcal{M}^n}\upsilon^{\frac{2n}{n - 2}}d\bbmu(g_0)\right)^{\frac{n - 2}{n}} \leqslant c_{(1)} \int_{\mathcal{M}^n}|\nabla\upsilon|^2 d\bbmu(g_0) + c_{(2)} \int_{\mathcal{M}^n}\upsilon^2 d\bbmu(g_0)
\end{equation}
in respect of $(\mathcal{M}^n, g_0)$ is true, with two positive constants $c_{(1)}$ and $c_{(2)}$. In addition, the inequality
\begin{equation}
	\left(\int\upsilon^{\frac{2n}{n - 2}}d\bbmu(g_t)\right)^{\frac{n - 2}{n}} \leqslant c_{(1)}(t) \left(\int|\nabla\upsilon|^2 + \frac{1}{4}\scalarcurvature\upsilon^2\right) d\bbmu(g_t) + c_{(2)}(t) \int\upsilon^2 d\bbmu(g_t)
\end{equation}
in respect of $(\mathcal{M}^n, g_t)$ holds, with $t \in [0, T_0)$, $T_0 \leqslant \infty$, where $c_{(1)}(t)$ and $c_{(2)}(t)$ are treated as positive functions on $g_0$, and $\scalarcurvature$ is the scalar curvature relative to $g_t$.  
\end{theorema}

\begin{proof}[Proof (Frog's eye\footnotemark perspective)]\footnotetext{
	Frog's eye, with and against bird's eye, is a mental-math categorization of F. Dyson \cite{Dyson "Birds and Frogs"}.\endnote{
	F. Dyson \cite[p. 212, e.a.]{Dyson "Birds and Frogs"} writes: «Some mathematicians are birds, others are frogs. Birds fly high in the air and survey broad vistas of mathematics out to the far horizon. They delight in concepts that unify our thinking and bring together diverse problems from different parts of the landscape. Frogs live in the mud below and see only the flowers that grow nearby. They delight in the details of particular objects, and they solve problems one at a time. I happen to be a frog, but many of my best friends are birds [\,\dots]. Mathematics needs both birds and frogs. Mathematics is rich and beautiful because birds give it broad visions and frogs give it intricate details. \emph{Mathematics is both great art and important science}, because it combines generality of concepts with depth of structures. It is stupid to claim that birds are better than frogs because they see farther, or that frogs are better than birds because they see deeper. The world of mathematics is both broad and deep, and we need birds and frogs working together to explore it».
	
	\setlength\parindent{8pt}
	Such a distinction is somehow re-echoed, in physics, by L. Smolin \cite[chap. 18]{Smolin "The Trouble with Physics"}, who tells the difference between «seers» (great visionaries-philosophers of thought) and «craftspeople» (highly skilled people in math-techniques but suffering from myopia in wide-ranging visions); he dwells on the first category. The seers are \emph{dreamers} (à la H. Poincaré, A. Einstein, N. Bohr, E. Schrödinger, W. Heisenberg, J.A. Wheeler, J.S. Bell, D. Finkelstein, R. Penrose, and, more recently, H.B. Nielsen, G. 't Hooft and T. Jacobson, just to mention a few), some of which on the margins of the academic community, but active in a work of redeeming \emph{isolation}; they are people who are asking themselves questions on what they are doing when it comes to dealing with formulæ, rather than moving up with their heads down in the furrow of the “shut up and calculate” dictate. The seers forge a handrail for the rise of both the \emph{humanistic-cultural understanding} of the ongoing scientific operation and the more specialized understanding of the object/phenomenon/event to which this operation is aimed.
	}
	}
The entire demonstration is beyond the scope of this Chapter; see the works of Q.S. Zhang \cite{Zhang "A Uniform Sobolev Inequality Under Ricci Flow"} \cite[chap. 6.2]{Zhang "Sobolev Inequalities Heat Kernels under Ricci Flow and the Poincare Conjecture"} for a painstaking exposition. We are interested, rather, to draw attention, in the rest of the text, on few but chief points that, emerging from this Theorem, connect the Perelman's work with the Sobolevian toolkit.
\end{proof}

The foremost aspect to underline it is the opportunity to bring out the close implication between, on the one hand, the \emph{Perelman's $\mathscr{W}$-entropy-energy formula} \eqref{equation "Perelman's entropy-energy W-functional"}, together with the \emph{$\mathscr{W}$-monotonicity} \eqref{subequations "Monotonicity formula for the W-functional"}, and, on the other, the \emph{Sobolev inequality in logarithmic forms}.

To start with, the logarithmic Sobolev inequality is a dimensionless Sobolev-type inequality, and it was devised by L. Gross \cite{Gross "Logarithmic Sobolev Inequalities"}, according to the following meaning. 

\begin{definitio}
Given a \emph{Gaussian measure} $\bbgamma$, and a \emph{Lebesgue measure} $\bblambda$, both on $\mathbb{R}^n$, let 
\begin{equation}
	d\bbgamma(x) = (2\pi)^{-\frac{n}{2}}\exp\left\{-|x|^2/2\right\}d\bblambda. 
\end{equation}
The \emph{logarithmic Sobolev inequality} is originally defined as 
\begin{equation}
	\int_{\mathbb{R}^n}|\upsilon(x)|^2\ln|\upsilon(x)|d\bbgamma(x) \leqslant \int_{\mathbb{R}^n}|\nabla\upsilon(x)^2|d\bbgamma(x) + \|\upsilon\|^2_2\ln\|\upsilon\|_2.
\end{equation}
\definitiosymbol
\end{definitio}

We are looking for the logarithmic Sobolev inequality on a topological space  of Riemannian-type (because that is what counts in Theorem \ref{theorema "Sobolev inequality and the Ricci flow"}). And here is the statement for the closed case.

\begin{propositio}[Logarithmic Sobolev inequality on a Riemannian space]
\label{propositio "Logarithmic Sobolev inequality on a Riemannian space"}
The logarithmic Sobolev inequality on a closed Riemannian manifold $\mathcal{M}^n$, with metric $g$ and dimension $n > 2$, dictates that $\int_{\mathcal{M}^n}\upsilon^2 d\bbmu = 1$, for a function $\upsilon > 0$, and a constant $c(\lambda, g)$, $\lambda >0$, such that
\begin{equation}
\label{equation "Logarithmic Sobolev inequality on a closed smooth space"}
	\int_{\mathcal{M}^n}\upsilon^2\log\upsilon d\bbmu \leqslant \lambda\int_{\mathcal{M}^n}|\nabla\upsilon|^2 d\bbmu + c(\lambda, g).
\end{equation}   
\end{propositio}

\begin{proof}
Let $c(n)\log\upsilon \leqslant \upsilon^\frac{2}{n}$, so, for every $\varepsilon > 0$,
\begin{equation}
	c(n)\int_{\mathcal{M}^n}\upsilon^2\log\upsilon d\bbmu \leqslant \int_{\mathcal{M}^n}\upsilon^{2 + \frac{2}{n}}d\bbmu \leqslant \varepsilon \int_{\mathcal{M}^n}\upsilon^{2 + \frac{4}{n}}d\bbmu + \frac{1}{\varepsilon}\int_{\mathcal{M}^n}\upsilon^2 d\bbmu,
\end{equation}
because $\upsilon^{1 + \frac{2}{n}}\upsilon \leqslant \varepsilon\upsilon^{2 \left(1 + \frac{2}{n}\right)} + \frac{1}{\varepsilon}\upsilon^2$. Via \emph{Hölder inequality}, 
\begin{equation}
	\int_{\mathcal{M}^n}\upsilon^2\upsilon^\frac{4}{n}d\bbmu \leqslant \left(\int_{\mathcal{M}^n}\upsilon^\frac{2n}{n - 2}d\bbmu\right)^\frac{n - 2}{n}\left(\int_{\mathcal{M}^n}\upsilon^2 d\bbmu\right)^\frac{2}{n}.  
\end{equation}
By introducing the \emph{Sobolev constant} 
\begin{equation}
	c_\textsc{s}(\mathcal{M}^n, g) = \inf_{\rotatedupsilon \in \mathscr{C}^1(\mathcal{M})}\frac{\int_{\mathcal{M}^n}(|\nabla\rotatedupsilon| + |\rotatedupsilon|)d\bbmu}{\left(\int_{\mathcal{M}^n}\rotatedupsilon^\frac{n}{n - 1}d\bbmu\right)^\frac{n - 1}{n}},
\end{equation}
and imposing $c_\textsc{s}(\mathcal{M}^n, g) > 0$, we write an inequality known as \emph{$\Lebesgue^2$-type Sobolev inequality},
\begin{equation}
\label{equation "$L^2$-type Sobolev inequality"}
	c_\textsc{s}(\mathcal{M}^n, g)\left(\int_{\mathcal{M}^n}\upsilon^\frac{2n}{n - 2}d\bbmu\right)^\frac{n - 2}{n} \leqslant \int_{\mathcal{M}^n}|\nabla\upsilon|^2 d\bbmu + \volume_g(\mathcal{M}^n)^{-\frac{2}{n}}.	
\end{equation}
Putting $\varepsilon = \lambda c_\textsc{s}(\mathcal{M}^n, g)$, we conclude that
\begin{align}
	c(n)\int_{\mathcal{M}^n}\upsilon^2\log\upsilon d\bbmu & \leqslant \varepsilon \left(\int_{\mathcal{M}^n}\upsilon^\frac{2n}{n - 2}d\bbmu\right)^\frac{n - 2}{n} + \frac{1}{\varepsilon} \notag \\
	& \leqslant \frac{\varepsilon}{c_\textsc{s}(\mathcal{M}^n, g)}\left(\int_{\mathcal{M}^n}|\nabla\upsilon|^2 d\bbmu + \volume_g(\mathcal{M}^n)^{-\frac{2}{n}}\right) + \frac{1}{\varepsilon},
\end{align}
under which $c(\lambda, g) = \lambda\volume_g(\mathcal{M}^n)^{-\frac{2}{n}} + \frac{1}{\lambda c_\textsc{s}(\mathcal{M}^n, g)}$.
\end{proof}

We are now able to give a new form to the entropy-energy $\mathscr{W}$-functional \eqref{equation "Perelman's entropy-energy W-functional"}: 
\begin{align}
	\mathscr{W}(g_{\mu\nu}, \rotatedupsilon, \tau) & + \left(\frac{n}{2}\log(4\pi\tau) + n\right) \notag \\
	& = \int_{\mathcal{M}^n}\Bigl\{\tau\left[4\left|\nabla(4\pi\tau)^{-\frac{n}{4}}e^{-\frac{\rotatedupsilon}{2}}\right|^2 + \scalarcurvature\left((4\pi\tau)^{-\frac{n}{4}}e^{-\frac{\rotatedupsilon}{2}}\right)^2\right] \notag \\
	& - \left((4\pi\tau)^{-\frac{n}{4}}e^{-\frac{\rotatedupsilon}{2}}\right)^2\log\left((4\pi\tau)^{-\frac{n}{4}}e^{-\frac{\rotatedupsilon}{2}}\right)^2\Bigr\}d\bbmu.
\end{align}

The \emph{logarithmic Sobolev inequality} \eqref{equation "Logarithmic Sobolev inequality on a closed smooth space"} \emph{ensures a lower bound for the $\mathscr{W}$-entropy-energy} by exploiting the $\Lebesgue^2$-structure—see Eq. \eqref{equation "$L^2$-type Sobolev inequality"}—of the first derivative of the entropic functional. If $\lambda = 2\tau$, then
\begin{align}
	\mathscr{W}(g_{\mu\nu}, \rotatedupsilon, \tau) & \geqslant \int_{\mathcal{M}^n}\Bigl\{4\tau\left|\nabla(4\pi\tau)^{-\frac{n}{4}}e^{-\frac{\rotatedupsilon}{2}}\right|^2 \notag \\
	& \hspace{4mm} - \left((4\pi\tau)^{-\frac{n}{4}}e^{-\frac{\rotatedupsilon}{2}}\right)^2\log\left((4\pi\tau)^{-\frac{n}{4}}e^{-\frac{\rotatedupsilon}{2}}\right)^2\Bigr\}d\bbmu \notag \\
	& \hspace{4mm} + \tau(\scalarcurvature)_\mathrm{min} - \left(\frac{n}{2}\log(4\pi\tau) + n\right) \notag \\
	& \geqslant -2c(2\tau, g) + \tau(\scalarcurvature)_\mathrm{min} - \left(\frac{n}{2}\log(4\pi\tau) + n\right) > - \infty, 
\end{align}
where the scalar curvature has its global minimum; this inequality shows that 
\begin{equation}
	c(2\tau, g) = 2\tau\volume_g(\mathcal{M}^n)^{-\frac{2}{n}} + \frac{1}{2\tau c_\textsc{s}(\mathcal{M}^n, g)}, 
\end{equation}
in line with the above proof.

\subsection{Null Space No Infinite Time: the Point as a Capstone, and the Poincaré–Perelman Paradox}
\label{subsection "Null Space No Infinite Time: the Point as a Capstone, and the Poincaré–Perelman Paradox"}

\begingroup
\footnotesize
\textgreek{[Τ]ὸ μὴ ἐνδέχεσθαι τὰ ἄπειρα διελθεῖν [\,\dots] ἐν πεπερασμένῳ χρόνῳ} · No possibility is given to traverse an infinity [of singularities] [\,\dots] in a finite time. \\
\indent — \textsc{Zeno of Elea} in the testimony of Stagirite's Physics \cite[VI.2.10, p. 124]{Aristotle "Naturalis Auscultationis libri VIII"}

\endgroup

\vspace{2mm}

The non-collapsing assumption is but a way to have a bounded entropy, and this ties in with being able to have a flow-solution, for non-negative curvatures, in which appear blow-up limits of finite time singularities of the Ricci flow. 

The crux of the matter, as illustrated by Perelman, is that the rule of a bounded entropy must be conceived in conjunction with the \emph{surgery operation} \cite{Perelman "Ricci flow with surgery on three-manifolds"}, already tested by Hamilton, as we have seen previously (Section \ref{subsection "Geometro-topological Surgery of Cutting off and Gluing Back"}). We take e.g. a quasi-circular cylindrical neck; we pinch it, and we cut it, or we open the neck; and then we glue small caps to each of the boundaries, after which the Ricci flow normally (re)starts to run, \emph{until the solution meets the next singularity (in the next time)}. The whole operation is repeated from the beginning: pinch/cut/glue, and the solution \emph{goes singular again as $t \to T$}. The process of forming new singularities is \emph{(apparently) continuous}. We are faced with a \emph{Zeno-like paradox}, with the generation of \emph{infinite singularities}. The no local collapsing Theorem \ref{theorema "Perelman's no local collapsing} should therefore be mixed with a self-contradictory assertion.

The possibility found by Perelman \cite{Perelman "Finite extinction time for the solutions to the Ricci flow on certain three-manifolds"} of avoiding the paradox, that is, of finding a solution to the Ricci flow with surgery that \emph{becomes extinct in finite time}, is directly related to the proof of the Poincaré conjecture (Section \ref{subsection "Poincaré Conjecture"}), by virtue of estimate values on the time of formation of the singularity on the 3-space under consideration. This singularity-time is \emph{small enough} to allow an approximate evaluation on the existence of a minimal disk (or a family of disks) that can be continuously deformed or shrank to a \emph{point}—i.e. to a \emph{0-dimensional spatial entity}—\emph{in finite time}. In that point, which is equal to a \emph{null space coordinate}, the solution to the Ricci flow with surgery \emph{finally is extinguished}, and the Perelman tapestry is completed.

\begin{scholium}
We end up with the occurrence of an additional \emph{paradox}—a kind of Zeno's revenge. The \emph{point is the identifier of the singularity}, the position in which the topological space \emph{explodes} to infinity, with an unpredictable (pathological) behavior; here the Ricci flow produces a singularity-point (in the sense that it becomes singular), namely it runs into a singularity-point. Per contra, the \emph{point is also what furnishes a solution of the singularity problem}, with the reduction of a closed loop to a null space, or single point (verification of the Poincaré conjecture). \scholiumsymbol	
\end{scholium}

\subsection{Margo. Non-linear $\sigmamodel$-Model and Ricci Flow (Renormalization Group Flow in Quantum Field Theory for Geometrical Couplings)}
\label{subsection "Margo. Non-linear sigma-Model and Ricci Flow (Renormalization Group Flow in Quantum Field Theory for Geometrical Couplings)"}

\begingroup
\footnotesize
The infinitesimal form of the renormalization group [flow] is the renormalization group equation $\frac{d}{dt}g_{ij} = - \beta_{ij}(g)$, where $\beta$ is a vector field on [the space] $\tilde{\textbf{R}}$ [of Riemannian metrics], called the $\beta$-function. The tangent vector $\beta(g)$ to the space of metrics at $g$ is the symmetric tensor field $\beta_{ij}(g)$ on [the manifold] $M$ [\,\dots]. When $M$ is a homogeneous space $G/H$ [the quotient $G/H$ of a Lie group $G$ by a compact subgroup $H$], the $\beta$-function is shown to be a gradient on the finite[-]dimensional space of $G$-invariant metric couplings on $M$. And, when $M$ is a two[-]dimensional compact manifold, the $\beta$-function is shown to be a gradient on the infinite[-]dimensional space of metrics on $M$.\footnote{
	Compare with the Perelman's epigraph at the beginning of Section \ref{section "Perelman Tapestry"}.
	} \\
\indent — \textsc{D.H. Friedan} \cite[pp. 390-391, 318]{Friedan "Nonlinear Models in 2 + varepsilon Dimensions"}

\endgroup

\vspace{2mm}

A first definition, although still not explicit and collateral, of the Ricci flow, under the approach later known as \emph{Ricci–DeTurck} in the Hamilton's system, as well as a first proto-description of the Ricci soliton, are not in differential geometry, but both are descended from quantum field theory, for work of D.H. Friedan \cite{Friedan "Nonlinear Models in 2 + epsilon Dimensions"} \cite{Friedan "Nonlinear Models in 2 + varepsilon Dimensions"}. 

Friedan's attention, being a physicist (and not a mathematician), is mainly paid to a generalization of the non-linear $\sigmamodel$-model (part of particle physics, see footnote \ref{footnote "Sigma model"}, p. \pageref{footnote "Sigma model"}, aimed at theorising \emph{quantum strings} flowing in a dynamic arena of space-time), pushing forward on the ideas of J. Honerkamp \cite{Honerkamp "Chiral multi-loops"} and A.M. Polyakov \cite{Polyakov "Interaction of Goldstone particles in two dimensions. Applications to ferromagnets and massive Yang-Mills fields"}. Following  are some pieces of the puzzle that he considers. 
\enumerationisinitium
\item A (high energy) model renormalizable in $(2 + \epsilon$)-dimensions; specifically, it is a scalar field $\textgreek{\textit{\ddigamma}}(x)$ on Euclidean ($2 + \epsilon$)-space the values of which lie in a finite-dimensional smooth manifold $\mathcal{M}$. 
\item In the wake of the intuition of K. Meetz \cite{Meetz "Realization of Chiral Symmetry in a Curved Isospin Space"}, a \emph{geometric character} of the \emph{coupling constant}, for which the dimensionless coupling is but a Riemannian metric $\Tau^{-1}g_{\mu\nu}$ on $\mathcal{M}$, called a \emph{metric coupling}.
\item An action having an energy integral
\begin{equation}
	\mathscr{S}_\textsc{f}(\textgreek{\textit{\ddigamma}}_x) = \int\frac{1}{2}\left(\Tau^{-1}g_{\mu\nu}\right)\partial_\rotatedell\textgreek{\textit{\ddigamma}}^\mu(x)\partial_\rotatedell \textgreek{\textit{\ddigamma}}^\nu(x)dx.
\end{equation}
\item Topological properties of the \emph{renormalization group flow}, that is, the group acting on the infinite-dimensional space of Riemannian metrics, necessary for the purpose of the above generalization; renormalization respects the action of the reparametrizations of $\mathcal{M}$, i.e. the action of the diffeomorphism group as a group of equivalence transformations.
\enumerationisfinis

Note. For an in-depth study and advances on the connection between the non-linear $\sigmamodel$-model and the Ricci flow, see the writings of M. Carfora \cite{Carfora "The conjugate linearized Ricci flow on closed 3-manifolds"} \cite{Carfora "Renormalization Group and the Ricci Flow"} \cite{Carfora "The Wasserstein geometry of nonlinear sigma models and the Hamilton-Perelman Ricci flow"} \cite[chap. 4]{Carfora Marzuoli "Quantum Triangulations: Moduli Space Quantum Computing Non-Linear Sigma Models and Ricci Flow"}.

\vspace{10mm}

\setcounter{secnumdepth}{0}  
\section{References and Bibliographic Details}
\setcounter{secnumdepth}{3}
\markright{References and Bibliographic Details}

\begingroup
\footnotesize
\noindent Sections \ref{section "Propaedeutics to Ricci Flow"} and \ref{section "Ricci Solitons: a Synoptic Classification"}

\begin{indent paragraph: 15pt}
On the Ricci flow, see \cite{Chow "Ricci Flow and Einstein Metrics in Low Dimensions"} \cite[pre-Perelman collection]{Cao Chow Chu and Yau (Eds.) "Collected Papers on Ricci Flow"} \cite{Chow Knopf "The Ricci Flow: An Introduction"} \cite{Chow Lu and Ni "Hamilton's Ricci Flow"} \cite{Muller "Differential Harnack Inequalities and the Ricci Flow"}  \cite{Topping "Lectures on the Ricci Flow"} \cite{Chow Chu Glickenstein Guenther Isenberg Ivey Knopf Lu Luo Ni "The Ricci Flow: Techniques and Applications Part I: Geometric Aspects"} \cite{Cao Chen Zhu "Recent Developments on Hamilton's Ricci flow"} \cite{Chow Chu Glickenstein Guenther Isenberg Ivey Knopf Lu Luo Ni "The Ricci Flow: Techniques and Applications Part II: Analytic Aspects"} \cite{Brendle "Ricci Flow and the Sphere Theorem"} \cite{Morgan Tsz-Ho Fong "Ricci Flow and Geometrization of 3-Manifolds"} \cite{Zhu "Analytic Aspect of Hamilton's Ricci Flow"} \cite{Andrews Hopper "The Ricci Flow in Riemannian Geometry: A Complete Proof of the Differentiable 1/4-Pinching Sphere Theorem"} \cite{Chow Chu Glickenstein Guenther Isenberg Ivey Knopf Lu Luo Ni "The Ricci Flow: Techniques and Applications Part IV: Long-Time Solutions and Related Topics"} \cite{Sinestrari "Singularities of Three-Dimensional Ricci Flows"} \cite{Chen "A survey on Ricci solitons on Riemannian submanifolds"} \cite{Topping "Ricci Flow and Ricci Limit Spaces"}. — On the Kähler–Ricci flow, see \cite{Boucksom Eyssidieux Guedj (Eds.) "An Introduction to the Kahler-Ricci Flow"} \cite{Tian "Notes on Kahler-Ricci Flow"}. — For the role played by the Ricci/gradient flow in optimal transport problems, compatible with entropic and heat evolution contexts, see e.g. \cite{Carfora "Fokker-Planck dynamics and entropies for the normalized Ricci flow"} \cite{McCann and Topping "Ricci flow entropy and optimal transportation"} \cite{Gigli Mantegazza "A flow tangent to the Ricci flow via heat kernels and mass transport"} \cite{Daneri and Savare "Lecture notes on gradient flows and optimal transport"} \cite{Topping "Ricci flow: the foundations via optimal transportation"}. — It is beneficial to also keep an eye on the Ricci tensor system in metric measure spaces, since notions such as \emph{heat flow}, \emph{gradient flow}, and \emph{entropy} are automatically crossed: see e.g. \cite{Lott Villani "Ricci curvature for metric-measure spaces via optimal transport"} \cite{Ambrosio Gigli and Savare "Metric measure spaces with Riemannian Ricci curvature bounded from below"}.
\end{indent paragraph: 15pt}

\noindent Section \ref{section "Geometrization of Topology (or of Process of Creating the Space as a Geometry)"}

\begin{indent paragraph: 15pt}
For the Poincaré conjecture, and its correlations with the Thurston's geometrization and Ricci flow, see \cite{Anderson "Geometrization of 3-Manifolds via the Ricci Flow"} \cite{Morgan and Tian "Completion of the Proof of the Geometrization Conjecture"} \cite{Morgan notes by Lipyanskiy "Ricci Flow and Thurston's Geometrization Conjecture"} \cite{Bessieres Besson Maillot Boileau Porti "Geometrisation of 3-Manifolds"} \cite{Zhang "Sobolev Inequalities Heat Kernels under Ricci Flow and the Poincare Conjecture"} \cite{Boileau "Thick/Thin Decomposition of Three-Manifolds and the Geometrisation Conjecture"}.
\end{indent paragraph: 15pt}

\noindent Section \ref{subsection "Thurston's Conjecture: Decomposition into Pieces having Geometric Structures"}

\begin{indent paragraph: 15pt}
A digest of Thurston's theory of 3-manifolds is in \cite{Kojima Thurston's "Theory of 3-Manifolds"}.
\end{indent paragraph: 15pt}

\noindent Section \ref{subsubsection "Sobolev Space $W^{k, p}(Omega)$"}

\begin{indent paragraph: 15pt}
About the Sobolev space(s) and Sobolev inequalities, see e.g. \cite[chapp. 2-3]{Hebey "Nonlinear Analysis on Manifolds: Sobolev Spaces and Inequalities"} \cite{Saloff-Coste "Aspects of Sobolev-Type Inequalities"} \cite[chapp. 3-4]{Adams and Fournier "Sobolev Spaces"} \cite[chapp. 5, 8-9]{Tartar "An Introduction to Sobolev Spaces and Interpolation Spaces"} \cite[chap. 3]{Haroske Triebel "Distributions Sobolev Spaces Elliptic Equations"} \cite[chapp. 1, § 3, 10]{Krylov "Lectures on Elliptic and Parabolic Equations in Sobolev Spaces"} \cite[chap. 5]{Evans "Partial Differential Equations"} \cite[chap. 8]{Diening Harjulehto Hasto Ruzicka "Lebesgue and Sobolev Spaces with Variable Exponents"} \cite[chap. 7]{Heinonen Koskela Shanmugalingam Tyson "Sobolev Spaces on Metric Measure Spaces: An Approach Based on Upper Gradients"}.
\end{indent paragraph: 15pt}

\noindent Section \ref{subsubsection "Sobolev Embedding for a Null Trace Space, Orlicz Space, and General Sobolev Inequality"}

\begin{indent paragraph: 15pt}
On the Orlicz spaces, cf. \cite[chap. III]{Rao Ren "Theory of Orlicz spaces"} \cite[chap. 3]{Harjulehto Hasto "Orlicz Spaces and Generalized Orlicz Spaces"}.
\end{indent paragraph: 15pt}

\noindent Section \ref{subsection "Margo. Non-linear sigma-Model and Ricci Flow (Renormalization Group Flow in Quantum Field Theory for Geometrical Couplings)"}

\begin{indent paragraph: 15pt}
Further applications of Ricci/gradient flow for non-linear $\sigmamodel$-models are e.g. in \cite{Oliynyk Suneeta Woolgar "A gradient flow for worldsheet nonlinear sigma models"} \cite{Tseytlin "Sigma model renormalization group flow "central charge" action and Perelman's entropy"} \cite{Bakas and Sourdis "Dirichlet sigma models and mean curvature flow"}. — For an interpretation of the Ricci flow in relativistic cosmology (initial-value problem, evolution of curvatures of space-time), see \cite{Carfora Buchert "Ricci Flow Deformation of Cosmological Initial Data Sets"} \cite{Carfora "Ricci-flow-conjugated initial data sets for Einstein equations"}.
\end{indent paragraph: 15pt}

\endgroup

\chapter{Calabi–Yau Theorem: a Non-linear Complex Equation of Monge–Ampère Type on Compact Kähler Manifolds}
\chaptermark{Calabi–Yau Theorem: Complex Monge–Ampère Eq. on Compact Kähler Manifolds}
\label{chapter "Calabi–Yau Theorem: a Non-linear Complex Equation of Monge–Ampère Type on Compact Kähler Manifolds"}

\begingroup
\footnotesize
Let $M^n$ be a closed, $n$-dimensional complex manifold. We assume that $M^n$ admits at least one Kähler metric $g_{\alpha\beta^*}$; its associated closed exterior form $\omega = \sqrt{-1}g_{\alpha\beta^*}dz^\alpha \wedge dz^{\beta^*}$ determines a real cohomology class, called the principal class of the metric. Consider the space $\Omega$ of all infinitely differentiable Kähler metrics in $M^n$ with the same principal class; the topology of $\Omega$ is defined by the $\Lebesgue^2$ topology of the tensorial components of metrics in $\Omega$ [\,\dots]. If $\Ricci_{\alpha\beta^*}$ is the Ricci tensor of any metric in $\Omega$, then the Ricci form $\sqrt{-1}\Ricci_{\alpha\beta^*}dz^\alpha \wedge dz^{\beta^*}$ is closed and its cohomology class is $2\pi C^{(1)}$ ($C^{(r)}$ = $r$th Chern class). \emph{Theorem} 1. Given in $M^n$ any real, closed, infinitely differentiable exterior form $\Sigma$ of type $(1, 1)$ and cohomologous to $2\pi C^{(1)}$, there exists exactly one Kähler metric in $\Omega$ whose Ricci form equals $\Sigma$. \\
\indent — \textsc{E. Calabi} \cite[pp. 206-207]{Calabi "The space of Kahler metrics"}

\endgroup

\section{Ricci Form on the Space of a Kählerian Metric}
\label{section "Ricci Form on the Space of a Kählerian Metric"}

We propose a \emph{pocket} reconstruction of E. Calabi's conjecture, and its proof under the line dictated by T. Aubin and S.-T. Yau.

\subsection{Calabi Conjecture}

\enumerationisinitium
\item Let $(\mathcal{M}, \mathcal{J}_{\mathbb{C}|}, g)$ be a complex manifold, where $\mathcal{J}_{\mathbb{C}|}$ is an almost complex structure, and $g$ is a Riemannian metric. We can define $g \viz g_\textsc{h}$ a \emph{Hermitian metric} if $g(\vec{X}, \vec{Y}) = g(\mathcal{J}_{\mathbb{C}|}\vec{X}, \mathcal{J}_{\mathbb{C}|}\vec{Y})$, for any vector field $\vec{X}$ and $\vec{Y}$ on $\mathcal{M}$.
 
Once we have established that $g$ is Hermitian, we can determine on $\mathcal{M}$ a 2-form $\omega$ as a \emph{Hermitian form} of $g$ such that $\omega(\vec{X}, \vec{Y}) = g(\mathcal{J}_{\mathbb{C}|}\vec{X}, \vec{Y})$,

Thus one can see that $\omega$ is, in typological terms, a $(1, 1)$-form, i.e. a differential 2-form of type $(1, 1)$, whose Hermitianity is given only and exclusively by $\omega > 0$.
\item Take a complex manifold $(\mathcal{M}, \mathcal{J}_{\mathbb{C}|}, g)$, with a Hermitian metric $g$, and a Hermitian form $\omega$ of $g$.  We characterize $g \viz g_\textsc{h}$ as a \emph{Kähler metric} $g \viz g_{\textsc{k}\textnormal{ä}}$ on $\mathcal{M}$ if the equality $d\omega = 0$ holds, which implies that $\omega$ is a symplectic form $\omega_\mathrm{s}$ (cf. Definition \ref{definitio "Hamiltonian vector field or symplectic gradient"}), and therefore $\omega = \omega_\mathrm{s}$ is called \emph{Kähler form}. We can then define $(\mathcal{M}, \omega_\mathrm{s}, \mathcal{J}_{\mathbb{C}|}, g)$ a \emph{Kähler manifold} \cite{Kahler "Uber eine bemerkenswerte Hermitesche Metrik"}, with $g \viz g_{\textsc{k}\textnormal{ä}}$ (cf. Scholium \ref{scholium "Kähler manifold, and almost complex structure"}).\footnote{
	Letting $\nabla$ be a Levi-Civita connection (see Sections \ref{section "Christoffel Symbols"} and \ref{section "Parallel Transport of the Levi-Civita Connection"}), in a manifold with $g \viz g_\textsc{h}$ and a Hermitian form $\omega$ it is true that $\nabla\mathcal{J}_{\mathbb{C}|} = 0$ and $\nabla\omega = 0$.
	}
\enumerationisfinis

\begin{coniectura}[Calabi conjecture]
\label{coniectura "Calabi conjecture"}
Original references are \textnormal{\cite{Calabi "The space of Kahler metrics"} \cite{Calabi "On Kahler Manifolds with Vanishing Canonical Class"}}. Let $(\mathcal{M}, \omega_\mathrm{s}, \mathcal{J}_{\mathbb{C}|}, g)$ be a compact Kähler manifold, where $\omega_\mathrm{s}$ is a Kähler/symplectic form, with $g \viz g_{\textsc{k}\textnormal{ä}}$ on $\mathcal{M}$. Let $\omega_\Ricci$ be a Ricci form, and $\mathring{\mathscr{C}}_1(\mathcal{M})_\mathbb{R}$ the first Chern class \textnormal{\cite{Chern "Characteristic classes of Hermitian Manifolds"}}. The conjecture says that 
\enumerationisinitium
\item there exists a unique Kähler metric $\tilde{g}$ on $\mathcal{M}$ with a Kähler form $\tilde{\omega}_\mathrm{s}$, under which the cohomology class $\{\tilde{\omega}_\mathrm{s}\} = \{\omega_\mathrm{s}\} \in H_\mathrm{c}^2(\mathcal{M})_\mathbb{R} \cap H_\mathrm{c}^{1, 1}(\mathcal{M})_\mathbb{C}$ is a Kähler class—where $\{\omega_\mathrm{s}\}$ is a (de Rham) cohomology class as a Kähler class of $g$, and $H_\mathrm{c}^2(\mathcal{M})_\mathbb{R}$ the second (de Rham) cohomology group of $\mathcal{M}$ with coefficients in $\mathbb{R}$,
\item $\tilde{\omega}_\Ricci$, in the cohomology class $\{\tilde{\omega}_\Ricci\} = 2\pi\mathring{\mathscr{C}}_1(\mathcal{M})_\mathbb{R}$, is the Ricci form of $\tilde{g}$.
\enumerationisfinis
Besides, if $\mathring{\mathscr{C}}_1(\mathcal{M})_\mathbb{R} = 0$, that is, if the first Chern class is a vanishing class, then the metric of $(\mathcal{M}, \omega_\mathrm{s}, \mathcal{J}_{\mathbb{C}|}, g)$ is Ricci-flat Kähler.
\end{coniectura}

Conjecture \ref{coniectura "Calabi conjecture"} is tantamount to asking, directly, and more simply, the following.

\begin{quaestio}
\label{quaestio "Calabi conjecture"}
On a compact Kähler manifold, is \emph{any} closed real 2-form of type $(1, 1)$, having a cohomology class $\{\omega_\Ricci\} = 2\pi\mathring{\mathscr{C}}_1(\mathcal{M})_\mathbb{R}$, the Ricci form of a Kähler metric? Note. If $\mathring{\mathscr{C}}_1(\mathcal{M})_\mathbb{R} = 0$, then there is a flatness of the Ricci Kähler metric. \quaestiosymbol
\end{quaestio}

\subsection{Aubin–Calabi–Yau Theorem}

Demonstration of the Calabi conjecture is the work of T. Aubin \cite{Aubin "Equations du type Monge-Ampere sur les varietes kahleriennes compactes"}—see also \cite[chap. 7]{Aubin "Nonlinear Analysis on Manifolds. Monge-Ampere Equations}—and, independently, of S.-T. Yau \cite{Yau "Calabi's conjecture and some new results in algebraic geometry"} \cite{Yau "On the Ricci Curvature of a Compact Kahler Manifold and the Complex Monge-Ampere Equation I"}. Which turns the conjecture into a theorem. This theorem is usually called \emph{Calabi–Yau theorem}, but maybe it would be more correct to refer to it as \emph{Calabi–Aubin–Yau theorem}. 

The keystone to resolve the Calabi conjecture is to transform the conjecture's equational ensemble into a \emph{Monge–Ampère equation} \cite{Monge "Memoire sur le calcul integral des equations aux differences partielles"} \cite{Ampere "Memoire contenant l'application de la theorie exposee dans le XVII"}, i.e. a \emph{non-linear (second order) partial differential}. See, as preparatory studies, Calabi \cite{Calabi "Improper affine hyperspheres of convex type and a generalization of a theorem by K. Jorgens"}, A.V. Pogorelov \cite{Pogorelov "Monge-Ampere Equations of Elliptic Type"} \cite{Pogorelov "On the regularity of generalized solutions of the equation etc."} \cite{Pogorelov "The Dirichlet problem for the $n$-dimensional analogue of the Monge-Ampere equation"}, and S.‐Y. Cheng \& S.‐T. Yau \cite{Cheng and Yau "On the Regularity of the Monge-Ampere Equation etc."}. Guiding hints are in G. Tian \cite[chap. 5]{Tian "Canonical Metrics in Kahler Geometry"}.

\begin{proof}[Proof of the Conjecture \ref{coniectura "Calabi conjecture"}] 
~\enumerationisinitium
\item[(\textgreek{α}) — \textbf{step I. Reductio to Monge–Ampère equation}.]
~\subenumerationisinitium
\item Given a Kählerian space, i.e. a compact complex manifold $(\mathcal{M}, \omega_\mathrm{s}, \mathcal{J}_{\mathbb{C}|}, g)$, with a Kähler metric $g \viz g_{\textsc{k}\textnormal{ä}}$ on $\mathcal{M}$, and a Kähler form $\omega_\mathrm{s} = g_{\mu\bar{\nu}}dz_\mu \wedge d\bar{z}_{\bar{\nu}}$.\footnote{
	\label{footnote "Omission of the negative square root"}
	But actually $\sqrt{-1}g_{\mu\bar{\nu}}dz_\mu \wedge d\bar{z}_{\bar{\nu}}$, and $\omega_\Ricci = -\sqrt{-1}\partial\bar{\partial}\log\det(g_{\mu\bar{\nu}})$. The omission of the negative square root is dictated by agility requirements.
	} 
Let $\omega_\Ricci = -\partial\bar{\partial}\log\det(g_{\mu\bar{\nu}})$ be a Ricci form,\footnoteref{footnote "Omission of the negative square root"} $\Kahlerpotential$ a smooth real function representing the \emph{Kähler potential}, and $\textcyrillic{\textit{ь}}$ some smooth real function on $\mathcal{M}$, such that 
\begin{equation}
\label{equation "Effect with volume form"}
	\int_\mathcal{M}\Kahlerpotential dV_g = 0,
\end{equation}
where $dV_g$ is the volume form on $\mathcal{M}$ induced by the metric $g$. 

Note. Bar notation is for the complex conjugate. Choosing local holomorphic—complex—coordinates $z_1, \mathellipsis, z_n$, we put 
\begin{align}
	& -\partial\bar{\partial}\log\det\left(g_{\mu\bar{\nu}} + \frac{\partial^2\Kahlerpotential}{\partial z_\mu\partial\bar{z}_{\bar{\nu}}}\right) = -\partial\bar{\partial}\log\det(g_{\mu\bar{\nu}}) - \partial\bar{\partial}\textcyrillic{\textit{ь}}, \\
	& \partial\bar{\partial}\log\left\{\frac{\det\left(g_{\mu\bar{\nu}} + \frac{\partial^2\Kahlerpotential}{\partial z_\mu\partial\bar{z}_{\bar{\nu}}}\right)}{\det(g_{\mu\bar{\nu}})}\right\} = \partial\bar{\partial}\textcyrillic{\textit{ь}},
\end{align} 
in local and global coordinates, respectively, with indices $\mu\bar{\nu} = 1, \mathellipsis, n$; hence
\begin{equation}
	\frac{\det\left(g_{\mu\bar{\nu}} + \frac{\partial^2\Kahlerpotential}{\partial z_\mu\partial\bar{z}_{\bar{\nu}}}\right)}{\det(g_{\mu\bar{\nu}})} = e^{\textcyrillic{\textit{ь}} + c},
\end{equation}
that is, $(\omega_\mathrm{s} + \partial\bar{\partial}\Kahlerpotential)^n = e^{\textcyrillic{\textit{ь}} + c}\omega_\mathrm{s}^n$, with a constant $c$.
\item Inasmuch as $(\omega_\mathrm{s} + \partial\bar{\partial}\Kahlerpotential)^n - \omega_\mathrm{s}^n$, thanks to the \emph{Stokes–Cartan's theorem} \cite[question № 8, p. 320]{Stokes "Smith's Prize Examination Papers N. 8 February 1854"} \cite[§§ 29-32, pp. 38-43]{Cartan "Les systemes differentiels exterieurs et leurs applications geometriques"} (see Section \ref{subsection "Stokes–Cartan's Theorem (a Foundation of Exterior Calculus)"}), we fix
\begin{equation}
\label{equation "Via Stokes' theorem"}
	\int_\mathcal{M}e^{\textcyrillic{\textit{ь}} + c}\omega_\mathrm{s}^n = \int_\mathcal{M}(\omega_\mathrm{s} + \partial\bar{\partial}\Kahlerpotential)^n = \volume_g(\mathcal{M}) = \int_\mathcal{M}\omega_\mathrm{s}^n,
\end{equation}
where $\volume_g(\mathcal{M})$ is the volume of $\mathcal{M}$ with the volume form. So 
\begin{subequations}
\label{subequations "Monge–Ampère equations"}
\begin{align}
	& (\omega_\mathrm{s} + \partial\bar{\partial}\Kahlerpotential)^n = e^{\textcyrillic{\textit{ь}}}\omega_\mathrm{s}^n, \\
	& \det\left(g_{\mu\bar{\nu}} + \frac{\partial^2\Kahlerpotential}{\partial z_\mu\partial\bar{z}_{\bar{\nu}}}\right) = c \cdot e^{\textcyrillic{\textit{ь}}}\det(g_{\mu\bar{\nu}}),
\end{align}
\end{subequations}
which are two non-linear partial differential \emph{equations of Monge–Ampère type} in $\Kahlerpotential$ (cf. Section \ref{subsection "Postilla on the Monge–Ampère Equation"}), expressing the same inhomogeneous  content in different forms.
\subenumerationisfinis
\item[(\textgreek{β}) — \textbf{step II. Uniqueness}.]
~\subenumerationisinitium
\item Take two positive 2-form of type $(1, 1)$, $(\omega_\mathrm{s})_1 = \omega_\mathrm{s} + \partial\bar{\partial}\Kahlerpotential_1$ and $(\omega_\mathrm{s})_2 = \omega_\mathrm{s}\partial\bar{\partial}\Kahlerpotential_2$. Then there is no more than a function $\Kahlerpotential \in \mathscr{C}^3(\mathcal{M})_\mathbb{R}$ under which \eqref{equation "Effect with volume form"} is true.
\item If $\Kahlerpotential_1 = \Kahlerpotential$, and $\Kahlerpotential_2 = 0$, $\Kahlerpotential_1, \Kahlerpotential_2, \in \mathscr{C}^3(\mathcal{M})_\mathbb{R}$, since $(\omega_\mathrm{s})_1^n = c \cdot e^\textcyrillic{\textit{ь}}\omega_\mathrm{s}^n = (\omega_\mathrm{s})_2^n$, with $\textcyrillic{\textit{ь}} \in \mathscr{C}^1(\mathcal{M})_\mathbb{R}$, one has
\begin{align}
	0 = (\omega_\mathrm{s})_2^n - (\omega_\mathrm{s})_1^n & = \omega_\mathrm{s}^n - (\omega_\mathrm{s} + \partial\bar{\partial}\Kahlerpotential)^n \notag \\
	& = - \partial\bar{\partial}\Kahlerpotential \wedge \Bigl(\omega_\mathrm{s}^{n - 1} + \omega_\mathrm{s}^{n - 2} \wedge (\omega_\mathrm{s})_1 + \cdots + (\omega_\mathrm{s})_1^{n - 1}\Bigr).
\end{align} 
Under Stokes' theorem again, we get
\begin{equation}
	\int_\mathcal{M}\partial\Kahlerpotential \wedge \bar{\partial}\Kahlerpotential \wedge \Bigl(\omega_\mathrm{s}^{n - 1} + \cdots + (\omega_\mathrm{s})_1^{n - 1}\Bigr) = 0.
\end{equation}
From the last equation, and setting $\int_\mathcal{M}\omega_\mathrm{s}^n$ as the volume, we see that
\begin{equation}
	\frac{1}{\int_\mathcal{M}\omega_\mathrm{s}^n}\partial\Kahlerpotential \wedge \bar{\partial}\Kahlerpotential \wedge \Bigl(\omega_\mathrm{s}^{n - 1} + \cdots + (\omega_\mathrm{s})_1^{n - 1}\Bigr) \geqslant \frac{1}{\int_\mathcal{M}\omega_\mathrm{s}^n} \partial\Kahlerpotential \wedge \bar{\partial}\Kahlerpotential \wedge \omega_\mathrm{s}^{n - 1},
\end{equation} 
and that
\begin{equation}
	0 \geq \frac{1}{\int_\mathcal{M}\omega_\mathrm{s}^n}\partial\Kahlerpotential \wedge \bar{\partial}\Kahlerpotential \wedge \omega_\mathrm{s}^{n - 1} = \frac{1}{2n\left(\int_\mathcal{M}\omega_\mathrm{s}^n\right)}\int_\mathcal{M}|\nabla\Kahlerpotential|^2\omega_\mathrm{s}^n.
\end{equation}
This says that 

· $\nabla\Kahlerpotential = 0$, and

· $\Kahlerpotential_1 - \Kahlerpotential_2$ is constant, being the Kählerian manifold connected. 

Since 
\begin{equation}
	\int_\mathcal{M}\Kahlerpotential_1dV_g = \int_\mathcal{M}\Kahlerpotential_2dV_g = 0,
\end{equation}
it is clear that $\Kahlerpotential_1 - \Kahlerpotential_2 = 0$, and $\Kahlerpotential_1 = \Kahlerpotential_2$.
\subenumerationisfinis
\item[(\textgreek{γ}) — \textbf{step III. Existence}.]
Let 
\begin{equation}
\label{equation "Reference equation in existence step"}	
	(\omega_\mathrm{s} + \partial\bar{\partial}\Kahlerpotential)^n = e^{\textcyrillic{\textit{ь}}_\varsigma}\omega_\mathrm{s}^n 
\end{equation}	
be the reference equation, with 

· $\textcyrillic{\textit{ь}}_\varsigma = \varsigma\textcyrillic{\textit{ь}} + c_\varsigma$, for $0 \leqslant \varsigma \leqslant 1$,

· $\textcyrillic{\textit{ь}}_0 = 0$, and

· $\textcyrillic{\textit{ь}}_1 = \textcyrillic{\textit{ь}}$.

We introduce the set $\textcyrillic{\textit{Я}}$ as a subset of $[0, 1]$. It is self-evident that $0 \in \textcyrillic{\textit{Я}}$, and if $1 \in \textcyrillic{\textit{Я}}$, then there is, unique up to a constant, a $\mathscr{C}$-solution of \eqref{equation "Reference equation in existence step"}. The solvability of such an equation lies in showing that $\textcyrillic{\textit{Я}}$ is an open and closed subset of $[0, 1]$.
~\subenumerationisinitium
\item We prove that $\textcyrillic{\textit{Я}}$ is open. If we solve Eq. \eqref{equation "Reference equation in existence step"} via $\Kahlerpotential_\varsigma$, the corresponding form is $(\omega_\mathrm{s})_\varsigma = \omega_\mathrm{s} + \partial\bar{\partial}\Kahlerpotential_\varsigma$, and $(\omega_\mathrm{s})^n_\varsigma = e^{\textcyrillic{\textit{ь}}_\varsigma}\omega_\mathrm{s}^n$. But assume $(\omega_\mathrm{s} + \partial\bar{\partial}\Kahlerpotential_\tau)^n = e^{\textcyrillic{\textit{ь}}_\tau}\omega_\mathrm{s}^n$, for an imaginable solution $\Kahlerpotential_\tau$, so that 
\begin{equation}
	\Bigl((\omega_\mathrm{s})_\varsigma + \partial\bar{\partial}(\Kahlerpotential_\tau - \Kahlerpotential_\varsigma)\Bigr)^n = e^{\textcyrillic{\textit{ь}}_\tau - \textcyrillic{\textit{ь}}_\varsigma}(\omega_\mathrm{s})_\varsigma^n,
\end{equation}
and, for a certain $\eta = \Kahlerpotential_\tau - \Kahlerpotential_\varsigma$, 
\begin{equation}
	\log\frac{\left[(\omega_\mathrm{s})_\varsigma + \partial\bar{\partial}\eta\right]^n}{(\omega_\mathrm{s})_\varsigma^n} = \textcyrillic{\textit{ь}}_\tau - \textcyrillic{\textit{ь}}_\varsigma.
\end{equation}
Let us define an operator 
\begin{equation}
	\textgreek{Η}_\mathrm{o}(\eta) = \log\frac{\left[(\omega_\mathrm{s})_\varsigma + \partial\bar{\partial}\eta\right]^n}{(\omega_\mathrm{s})_\varsigma^n} \colon \mathscr{C}^{2, \frac{1}{2}}_0(\mathcal{M})_\mathbb{R} \to \mathscr{C}^{0, \frac{1}{2}}(\mathcal{M})_{\mathbb{R}(\cdot)},
\end{equation}
where $\mathscr{C}^{2, \frac{1}{2}}_0$ and $\mathscr{C}^{0, \frac{1}{2}}_0$ are subspaces of any continuous function $\upsilon \colon \mathcal{M} \to \mathbb{R}$ in the $\mathscr{C}^{k, \frac{1}{2}}(\mathcal{M})_\mathbb{R}$-topology. 

Whenever $\|\textcyrillic{\textit{ь}}_\tau - \textcyrillic{\textit{ь}}_\varsigma\|_{0, \frac{1}{2}}$ is sufficiently small, there exist a function $\eta$ such that
\begin{equation}
\label{equation "Eta operator and function"}
	\textgreek{Η}_\mathrm{o}(\eta) = \textcyrillic{\textit{ь}}_\tau - \textcyrillic{\textit{ь}}_\varsigma,
\end{equation}
and 
\begin{equation}
	D\textgreek{Η}_\mathrm{o}|_{\eta = 0} \colon \Lbrack:\mathscr{C}^{2, \frac{1}{2}}_0(\mathcal{M})_\mathbb{R} \to \mathscr{C}^{0, \frac{1}{2}}_0(\mathcal{M})_\mathbb{R}:\Rbrack.
\end{equation}
Imposing 
\begin{equation}
	D\textgreek{Η}_\mathrm{o}|_{\eta = 0}(\upsilon) = \Laplacian_{(\omega_\mathrm{s})_\varsigma}\upsilon,
\end{equation}
as the map $\Laplacian_{(\omega_\mathrm{s})_\varsigma}\upsilon \colon \Lbrack:\cdots:\Rbrack$ implies an inverse function, where $\Lbrack:$ and $:\Rbrack$ are for a repeat sign (see Glossary), we find the solvability of \eqref{equation "Eta operator and function"} for $|\tau - \varsigma|$. The set $\textcyrillic{\textit{Я}}$ is therefore open, being that $\tau \leqslant \varsigma$.
\item We must now show that $\textcyrillic{\textit{Я}}$ is closed. We can start from an equation like this, 
\begin{equation}
	\Bigl(\omega_\mathrm{s} + \partial\bar{\partial}(\Kahlerpotential_\mu - c_\mu)\Bigr)^n = e^{\textcyrillic{\textit{ь}}_{\varsigma_\mu}}\omega_\mathrm{s}^n, 
\end{equation}
supposing a convergence to $\Kahlerpotential_\infty$ by $\Kahlerpotential_\mu - c_\mu$, for which $(\omega_\mathrm{s} + \partial\bar{\partial}\Kahlerpotential_\infty)^n = e^{\textcyrillic{\textit{ь}}_{\varsigma_\infty}}\omega_\mathrm{s}^n$, such that $\varsigma_\infty \in \textcyrillic{\textit{Я}}$, given a sequence $\varsigma_\mu \in \textcyrillic{\textit{Я}}$, $\lim_{\mu \to \infty}\varsigma_\mu = \varsigma_\infty$, which serves to finally detect the closed nature of $\textcyrillic{\textit{Я}}$ in $\mathscr{C}^{2, \frac{1}{2}}$. To do this, we have to use an \emph{a priori estimate}.

By virtue of the \emph{Ascoli–Arzelà theorem} \cite{Ascoli "Le curve limite di una varieta data di curve"} \cite{Arzela "Un'osservazione intorno alle Serie di funzioni"} \cite{Arzela "Sulle funzioni di linee"}, we adopt the priori estimate $\|\Kahlerpotential_\mu - c_\mu\|_3 \leqslant C_\mathrm{u}$, for a uniform constant $C_\mathrm{u}$, by selecting two work tools.

The first one is a \emph{Green's function} \cite[§ 3]{Green "An Essay on the Application of Mathematical Analysis to the Theories of Electricity and Magnetism"} $\Greenfunction(x, y)$ so that
\begin{equation}
	\Kahlerpotential(x) \in \mathscr{C}^\infty(\mathcal{M}) = \frac{1}{\int_\mathcal{M}\omega_\mathrm{s}^n}\int_\mathcal{M}\Kahlerpotential(y)\omega_\mathrm{s}^n(y) - \frac{1}{\int_\mathcal{M}\omega_\mathrm{s}^n}\int_\mathcal{M}\Laplacian_g\Kahlerpotential(y)\Greenfunction(x, y)\omega_\mathrm{s}^n(y). 
\end{equation} 

The second tool is the \emph{Sobolev inequality} (cf. Sections \ref{subsubsection "Sobolev Embedding for a Null Trace Space, Orlicz Space, and General Sobolev Inequality"} and \ref{subsubsection "Carron–Akutagawa's Sobolev Embedding"}); for some constant $c_{(1)}$ and $c_{(2)}$, we write
\begin{equation}
	c_{(1)}\left(\frac{1}{\int_\mathcal{M}\omega_\mathrm{s}^n}\int_\mathcal{M}|\textcyrillic{\textit{ь}}|^\frac{2n}{n - 1}\omega_\mathrm{s}^n\right)^{\frac{n - 1}{n}} - \frac{c_{(2)}}{\int_\mathcal{M}\omega_\mathrm{s}^n}\int_\mathcal{M}|\textcyrillic{\textit{ь}}|^2\omega_\mathrm{s}^n \leqslant \frac{1}{\int_\mathcal{M}\omega_\mathrm{s}^n}\int_\mathcal{M}|\nabla\textcyrillic{\textit{ь}}|^2\omega_\mathrm{s}^n.
\end{equation}
At this point, estimates are outlined.
\item[(a)] We say that $x_0$ is the supremum for the Kähler potential $\Kahlerpotential$, by selecting $\sup_\mathcal{M}\Kahlerpotential = -1$. Consequently
\begin{subequations}
\begin{align}
	& \Kahlerpotential(x_0) = -\frac{1}{\int_\mathcal{M}\omega_\mathrm{s}^n}\int_\mathcal{M}|\Kahlerpotential(y)|\omega_\mathrm{s}^n - \Lbrack:\frac{1}{\int_\mathcal{M}\omega_\mathrm{s}^n}\int_\mathcal{M}\Laplacian\Kahlerpotential(y)\Greenfunction(x_0, y)\omega_\mathrm{s}^n:\Rbrack = -1, \\
	& - \Lbrack:\cdots:\Rbrack \leqslant \frac{n}{\int_\mathcal{M}\omega_\mathrm{s}^n}\int_\mathcal{M}\Greenfunction(x_0, y)\omega_\mathrm{s}^n = -1 + \frac{1}{\int_\mathcal{M}\omega_\mathrm{s}^n}\int_\mathcal{M}|\Kahlerpotential(y)|\omega_\mathrm{s}^n,
\end{align}
\end{subequations}
after a $0 < n + \Laplacian\Kahlerpotential$ value is entered in the second equation. As a result, the inequality $\frac{1}{\int_\mathcal{M}\omega_\mathrm{s}^n}\int_\mathcal{M}|\Kahlerpotential(y)|\omega_\mathrm{s}^n \leqslant C_\mathrm{u}$ is clarified.

If $\Kahlerpotential_- = -\Kahlerpotential \geqslant 1$, and $\tilde{\omega}_\mathrm{s} = \omega_\mathrm{s} - \partial\bar{\partial}\Kahlerpotential_-$ are specified, then
\begin{align}
	\left(e^{\textcyrillic{\textit{ь}}_\varsigma} - 1\right)\omega_\mathrm{s}^n & = \tilde{\omega}_\mathrm{s}^n - \omega_\mathrm{s}^n \notag \\
	& = -\partial\bar{\partial}\Kahlerpotential_- \wedge \left(\tilde{\omega}_\mathrm{s}^{n - 1} + \tilde{\omega}_\mathrm{s}^{n - 2} \wedge \omega_\mathrm{s} + \cdots + \tilde{\omega}_\mathrm{s} \wedge \omega_\mathrm{s}^{n - 2} + \omega_\mathrm{s}^{n - 1}\right).
\end{align}
For some real number $\epsilon \geqslant 1$, we register
\begin{align}
	& -\frac{1}{\int_\mathcal{M}\omega_\mathrm{s}^n}\int_\mathcal{M}\Kahlerpotential^\epsilon_-\partial\bar{\partial}\Kahlerpotential_- \wedge \left(\tilde{\omega}_\mathrm{s}^{n - 1} + \cdots + \omega_\mathrm{s}^{n - 1}\right) \notag \\
	& \geqslant \frac{1}{\int_\mathcal{M}\omega_\mathrm{s}^n}\int_\mathcal{M}\partial\Kahlerpotential^\epsilon_- \wedge \bar{\partial}\Kahlerpotential_- \wedge \omega_\mathrm{s}^n = \frac{\epsilon}{\int_\mathcal{M}\omega_\mathrm{s}^n}\int_\mathcal{M}\Kahlerpotential^{\epsilon - 1}_-\partial\Kahlerpotential_- \wedge \bar{\partial}\Kahlerpotential_- \wedge \omega_\mathrm{s}^{n - 1} \notag \\
	& = \frac{\epsilon}{\int_\mathcal{M}\omega_\mathrm{s}^n}\int_\mathcal{M}\Kahlerpotential^\frac{\epsilon - 1}{2}_-\partial\Kahlerpotential_- \wedge \Kahlerpotential^\frac{\epsilon - 1}{2}_-\bar{\partial}\Kahlerpotential_- \wedge \omega_\mathrm{s}^{n - 1} \notag \\
	& = \frac{4\epsilon}{\int_\mathcal{M}\omega_\mathrm{s}^n(\epsilon + 1)^2}\int_\mathcal{M}\partial\Kahlerpotential^\frac{\epsilon + 1}{2}_- \wedge \bar{\partial}\Kahlerpotential^\frac{\epsilon + 1}{2}_- \omega_\mathrm{s}^{n - 1} \notag \\
	& \geqslant \mathrm{F}\left\{c_{(1)}\left(\frac{1}{\int_\mathcal{M}\omega_\mathrm{s}^n}\int_\mathcal{M}\left|\Kahlerpotential^\frac{\epsilon  + 1}{2}_-\right|^\frac{2n}{n - 1}\omega_\mathrm{s}^n\right)^\frac{n - 1}{n} - \frac{c_{(2)}}{\int_\mathcal{M}\omega_\mathrm{s}^n}\int_\mathcal{M}\left|\Kahlerpotential^\frac{\epsilon + 1}{2}_-\right|^2\omega_\mathrm{s}^n\right\},
\end{align}
with $\mathrm{F} = \frac{4\epsilon}{n(\epsilon + 1)^2}$. As 
\begin{align}
	& -\frac{1}{\int_\mathcal{M}\omega_\mathrm{s}^n}\int_\mathcal{M}\Kahlerpotential^\epsilon_-\partial\bar{\partial}\Kahlerpotential_- \wedge \left(\tilde{\omega}_\mathrm{s}^{n - 1} + \cdots + \omega_\mathrm{s}^{n - 1}\right) = \frac{1}{\int_\mathcal{M}\omega_\mathrm{s}^n}\int_\mathcal{M}\Kahlerpotential^\epsilon_-	\left(e^{\textcyrillic{\textit{ь}}_\varsigma} - 1\right)\omega_\mathrm{s}^n \notag \\
	& \leqslant \frac{c}{\int_\mathcal{M}\omega_\mathrm{s}^n}\int_\mathcal{M}\Kahlerpotential^{\epsilon + 1}_-\omega_\mathrm{s}^n,
\end{align} 
we obtain
\begin{subequations}
\begin{align}
 	& \mathrm{F}\left\{c_{(1)}\left(\frac{1}{\int_\mathcal{M}\omega_\mathrm{s}^n}\int_\mathcal{M}\left|\Kahlerpotential^\frac{\epsilon  + 1}{2}_-\right|^\frac{2n}{n - 1}\omega_\mathrm{s}^n\right)^\frac{n - 1}{n} - \frac{c_{(2)}}{\int_\mathcal{M}\omega_\mathrm{s}^n}\int_\mathcal{M}\left|\Kahlerpotential^\frac{\epsilon + 1}{2}_-\right|^2\omega_\mathrm{s}^n\right\} \notag \\
 	& \leqslant \frac{c}{\int_\mathcal{M}\omega_\mathrm{s}^n}\int_\mathcal{M}\left|\Kahlerpotential^{\epsilon + 1}_-\right|\omega_\mathrm{s}^n, \\
 	& \left(\frac{1}{\int_\mathcal{M}\omega_\mathrm{s}^n}\int_\mathcal{M}\left|\Kahlerpotential^{\epsilon + 1}_-\right|^\frac{n}{n - 1}\omega_\mathrm{s}^n\right)^\frac{n - 1}{n} \leqslant \frac{C_\mathrm{u}(\epsilon + 1)}{\int_\mathcal{M}\omega_\mathrm{s}^n}\int_\mathcal{M}\left|\Kahlerpotential^{\epsilon + 1}_-\right|\omega_\mathrm{s}^n.
\end{align}
\end{subequations}

By look to the \emph{Nash–Moser iteration technique} \cite{Nash "The Imbedding Problem for Riemannian Manifolds"} \cite{Moser "A rapidly convergent iteration method and non-linear partial differential equations - I"} \cite{Moser "A rapidly convergent iteration method and non-linear partial differential equations - II"}, for $\epsilon_0 = 1$ and $\epsilon_\mu + 1 = \frac{n}{n - 1}(\epsilon_{\mu - 1} + 1)$, we expound the formulæ
\begin{subequations}
\begin{align}
	& \|\Kahlerpotential_-\|_{\Lebesgue^{\epsilon_\mu + 1}} \leqslant \prod^{\mu - 1}_{\nu = 0}C_\mathrm{u}(\epsilon_\nu + 1)^\frac{1}{\epsilon_\nu + 1}\|\Kahlerpotential_-\|_{\Lebesgue^2}, \\
	& \sup_\mathcal{M}|\Kahlerpotential_-| = \lim_{\mu \to \infty}\|\Kahlerpotential_-\|_{\Lebesgue^{\epsilon_\mu + 1}} \leqslant \prod^\infty_{\nu = 0}C_\mathrm{u}(\epsilon_\nu + 1)^\frac{1}{\epsilon_\nu + 1}\|\Kahlerpotential_-\|_{\Lebesgue^2} < \infty.
\end{align}
\end{subequations}
The result is that $\sup_\mathcal{M}|\Kahlerpotential| = \sup_\mathcal{M}|\Kahlerpotential_-| \leqslant C_\mathrm{u}\|\Kahlerpotential_-\|_{\Lebesgue^2}$. By means of the \emph{Poincaré inequality} \cite{Poincare "Sur les Equations aux Derivees Partielles de la Physique Mathematique},
\begin{align}
	\frac{c}{\int_\mathcal{M}\omega_\mathrm{s}^n}\int_\mathcal{M}|\Kahlerpotential|\omega_\mathrm{s}^n & \geqslant \frac{1}{\int_\mathcal{M}\omega_\mathrm{s}^n}\int_\mathcal{M}\Kahlerpotential\left(1 - e^{\textcyrillic{\textit{ь}}_\varsigma}\right)\omega_\mathrm{s}^n \notag \\
	& = \frac{1}{\int_\mathcal{M}\omega_\mathrm{s}^n}\int_\mathcal{M}\Kahlerpotential\Bigl(\omega_\mathrm{s}^n - (\omega_\mathrm{s} + \partial\bar{\partial}\Kahlerpotential)^n\Bigr) \notag \\
	& \geqslant \frac{1}{n\left(\int_\mathcal{M}\omega_\mathrm{s}^n\right)}\int_\mathcal{M}|\nabla\Kahlerpotential|^2\omega_\mathrm{s}^n \notag \\
	& \geqslant \frac{\lambda_1\omega_\mathrm{s}}{n\left(\int_\mathcal{M}\omega_\mathrm{s}^n\right)}\left\{\int_\mathcal{M}|\Kahlerpotential|^2\omega_\mathrm{s}^n - \left(\int_\mathcal{M}\Kahlerpotential\left(\omega_\mathrm{s}^n\right)\right)^2\right\}, \\ 
	& \Kahlerpotential \in \Sobolev^{1, 2}(\mathcal{M})_\mathbb{R}, \notag
\end{align}
where $\lambda_1$ is a constant, and $\Kahlerpotential$ belongs to Sobolev space $\Sobolev^{1, 2}(\mathcal{M})$ (see Section \ref{subsubsection "Sobolev Space $W^{k, p}(Omega)$"}), we conclude that $\|\Kahlerpotential\|_{\Lebesgue^2} \leqslant C_\mathrm{u} \cdot \|\Kahlerpotential\|_{\Lebesgue^1} + 1$, and $\sup_\mathcal{M}|\Kahlerpotential| \leqslant C_\mathrm{u}$. 
\item[(b)] We shall now proceed to the second estimate step. What we need is to bound $\trace_{{\omega}_\mathrm{s}}(\tilde{\omega}_\mathrm{s}) = n + \Laplacian\Kahlerpotential = \|g_{\mu\bar{\nu}} + \frac{\partial^2\Kahlerpotential}{\partial z_\mu\partial\bar{z}_{\bar{\nu}}}\| \leqslant \trace_g\bigl(g_{\mu\bar{\nu}} + \frac{\partial^2\Kahlerpotential}{\partial z_\mu\partial\bar{z}_{\bar{\nu}}}\bigr)$. Recalling Eq. \eqref{equation "Reference equation in existence step"}, let $ \textcyrillic{\textit{ь}}_\varsigma + \log\det(g_{\mu\bar{\nu}}) = \log\det\bigl(g_{\mu\bar{\nu}} + \frac{\partial^2\Kahlerpotential}{\partial z_\mu\partial\bar{z}_{\bar{\nu}}}\bigr)$. If we want to find the derivative, we will write
\begin{subequations}
\begin{align}
	\label{align "Differentiation A"}
	& \frac{\partial\textcyrillic{\textit{ь}}_\varsigma}{\partial z_\xi} = \tilde{g}^{\mu\bar{\nu}}\left(\frac{\partial g_{\mu\bar{\nu}}}{\partial z_\xi} + \frac{\partial^3\Kahlerpotential}{\partial z_\mu\partial\bar{z}_{\bar{\nu}}\partial z_\xi}\right) - g^{\mu\bar{\nu}}\frac{\partial g_{\mu\bar{\nu}}}{\partial z_\xi}, \\
	&
	\label{align "Differentiation B"}
	 \frac{\partial^2\textcyrillic{\textit{ь}}_\varsigma}{\partial z_\xi\partial\bar{z}_{\bar{\varrho}}} = \tilde{g}^{\mu\bar{\nu}}\left(\frac{\partial^2g_{\mu\bar{\nu}}}{\partial z_\xi\partial\bar{z}_{\bar{\varrho}}} + \frac{\partial^4\Kahlerpotential}{\partial z_\mu\partial\bar{z}_{\bar{\nu}}\partial z_\xi\partial\bar{z}_{\bar{\varrho}}}\right) + g^{\tau\bar{\nu}}g^{\mu\bar{\varsigma}}\left(\frac{\partial g_{\tau\bar{\varsigma}}}{\partial\bar{z}_{\bar{\varrho}}}\frac{\partial g_{\mu\bar{\nu}}}{\partial z_\xi}\right) - g^{\mu\bar{\nu}}\frac{\partial^2g_{\mu\bar{\nu}}}{\partial z_\xi\partial\bar{z}_{\bar{\varrho}}} \notag \\
	& -\tilde{g}^{\tau\bar{\nu}}\tilde{g}^{\mu\bar{\varsigma}}\left(\frac{\partial g_{\tau\bar{\varsigma}}}{\partial\bar{z}_{\bar{\varrho}}} + \frac{\partial^3\Kahlerpotential}{\partial z_\tau\partial\bar{z}_{\bar{\varsigma}}\partial\bar{z}_{\bar{\varrho}}}\right)\left(\frac{\partial g_{\mu\bar{\nu}}}{\partial z_\xi} + \frac{\partial^3\Kahlerpotential}{\partial z_\mu\partial\bar{z}_{\bar{\nu}}\partial z_\xi}\right), 
\end{align}
\end{subequations}
with respect to $\frac{\partial}{\partial z_\xi}$, in Eq. \eqref{align "Differentiation A"}, and to $\frac{\partial}{\partial \bar{z}_{\bar{\varrho}}}$, in Eq. \eqref{align "Differentiation B"}.

We thus consider
\begin{align}
	& \Laplacian\textcyrillic{\textit{ь}}_\varsigma = g^{\xi\bar{\varrho}}\tilde{g}^{\mu\bar{\nu}}\left(\frac{\partial^2g_{\mu\bar{\nu}}}{\partial z_\xi\partial\bar{z}_{\bar{\varrho}}} + \frac{\partial^4\Kahlerpotential}{\partial z_\mu\partial\bar{z}_{\bar{\nu}}\partial z_\xi\partial\bar{z}_{\bar{\varrho}}}\right) \notag \\
	& - g^{\xi\bar{\varrho}}\tilde{g}^{\tau\bar{\nu}}\tilde{g}^{\mu\bar{\varsigma}}\left(\frac{\partial^3\Kahlerpotential}{\partial z_\tau\partial\bar{z}_{\bar{\varsigma}}\partial\bar{z}_{\bar{\varrho}}}\frac{\partial^3\Kahlerpotential}{\partial z_\mu\partial\bar{z}_{\bar{\nu}}\partial z_\xi}\right) - g^{\xi\bar{\varrho}}g^{\mu\bar{\nu}}\frac{\partial^2g_{\mu\bar{\nu}}}{\partial z_\xi\partial\bar{z}_{\bar{\varrho}}}, 
\end{align}
and
\begin{subequations}
\begin{align}
	& \widetilde{\Laplacian}(\Laplacian\Kahlerpotential) = -g^{\xi\bar{\varrho}}\tilde{g}^{\mu\bar{\nu}}\frac{\partial^2g_{\mu\bar{\nu}}}{\partial z_\xi\partial\bar{z}_{\bar{\varrho}}} + g^{\xi\bar{\varrho}}\tilde{g}^{\tau\bar{\nu}}\tilde{g}^{\mu\bar{\varsigma}}\left(\frac{\partial^3\Kahlerpotential}{\partial z_\tau\partial\bar{z}_{\bar{\varsigma}}\partial\bar{z}_{\bar{\varrho}}}\frac{\partial^3\Kahlerpotential}{\partial z_\mu\partial\bar{z}_{\bar{\nu}}\partial z_\xi}\right) \notag \\
	& \hspace{36pt} + g^{\xi\bar{\varrho}}g^{\mu\bar{\nu}}\frac{\partial^2g_{\mu\bar{\nu}}}{\partial z_\xi\partial\bar{z}_{\bar{\varrho}}} + \Laplacian\textcyrillic{\textit{ь}}_\varsigma + \tilde{g}^{\xi\bar{\varrho}}\left(\frac{\partial^2g^{\mu\bar{\nu}}}{\partial z_\xi\partial\bar{z}_{\bar{\varrho}}}\frac{\partial^2\Kahlerpotential}{\partial z_\mu\partial\bar{z}_{\bar{\nu}}}\right), \\
	& \widetilde{\Laplacian}(\Laplacian\Kahlerpotential) = \Laplacian\textcyrillic{\textit{ь}}_\varsigma + g^{\xi\bar{\varrho}}\tilde{g}^{\tau\bar{\nu}}\tilde{g}^{\mu\bar{\varsigma}}\Kahlerpotential_{\tau\bar{\varsigma}\varrho}\Kahlerpotential_{\mu\bar{\nu}\xi} + \tilde{g}^{\mu\bar{\nu}}\Riemann_{\mu\bar{\nu}\xi\bar{\varrho}} - g^{\mu\bar{\nu}}\Riemann_{\mu\bar{\nu}\xi\bar{\varrho}} + \tilde{g}^{\xi\bar{\varrho}}\Riemann_{\mu\bar{\nu}\xi\bar{\varrho}}\Kahlerpotential_{\mu\bar{\nu}}, 
\end{align}
\end{subequations}
by replacing in the last equation $\frac{\partial^2g_{\mu\bar{\nu}}}{\partial z_\xi\partial\bar{z}_{\bar{\varrho}}}$ with $-\Riemann_{\mu\bar{\nu}\xi\bar{\varrho}}$, and $\frac{\partial^2g^{\mu\bar{\nu}}}{\partial z_\xi\partial\bar{z}_{\bar{\varrho}}}$ with $\Riemann_{\mu\bar{\nu}\xi\bar{\varrho}}$, where $\Riemann$ is the Riemann curvature tensor.

Return to coordinate geometry, and formulate the above equations a second time:
\begin{align}
	\widetilde{\Laplacian}(\Laplacian\Kahlerpotential) & = \left(\frac{1}{1 + \Kahlerpotential_{\mu\bar{\mu}}}\right)\left(\frac{1}{1 + \Kahlerpotential_{\nu\bar{\nu}}}\right)\Kahlerpotential_{\mu\bar{\nu}\xi}\Kahlerpotential_{\bar{\mu}\nu\bar{\xi}} + \Laplacian\textcyrillic{\textit{ь}}_\varsigma \notag \\
	& + \Lbrack:\Riemann_{\mu\bar{\mu}\xi\bar{\xi}}\left(-1 + \frac{1}{1 + \Kahlerpotential_{\mu\bar{\mu}}} + \frac{\Kahlerpotential_{\mu\bar{\mu}}}{1 + \Kahlerpotential_{\xi\bar{\xi}}}\right):\Rbrack, \\
	& \text{ where } \Lbrack:\cdots:\Rbrack = \frac{1}{2}\Riemann_{\mu\bar{\mu}\xi\bar{\xi}}\left\{\frac{(\Kahlerpotential_{\xi\bar{\xi}} - \Kahlerpotential_{\mu\bar{\mu}})^2}{(1 + \Kahlerpotential_{\mu\bar{\mu}})(1 + \Kahlerpotential_{\xi\bar{\xi}})}\right\} \notag \\
	& \geqslant \frac{C_\mathrm{u}}{2}\left\{\frac{(1 + \Kahlerpotential_{\xi\bar{\xi}} - 1 - \Kahlerpotential_{\mu\bar{\mu}})^2}{(1 + \Kahlerpotential_{\mu\bar{\mu}})(1 + \Kahlerpotential_{\xi\bar{\xi}})}\right\} \notag \\
	& = C_\mathrm{u}\left(\frac{1 + \Kahlerpotential_{\mu\bar{\mu}}}{1 + \Kahlerpotential_{\xi\bar{\xi}}} - 1\right), \text{ with } C_\mathrm{u} = \inf_{\mu \neq \xi}\Riemann_{\mu\bar{\mu}\xi\bar{\xi}}.
\end{align}
Via \emph{Schwarz lemma} \cite{Schwarz "Zur Theorie der Abbildung"} \cite{Caratheodory "Untersuchungen uber die konformen Abbildungen von festen und veranderlichen Gebieten"} it is possible to infer that
\begin{align}
	\widetilde{\Laplacian}\Bigl(e^{-\zeta\Kahlerpotential}(n + \Laplacian\Kahlerpotential)\Bigr) & = e^{-\zeta\Kahlerpotential}\widetilde{\Laplacian}(\Laplacian\Kahlerpotential) - \zeta e^{-\zeta\Kahlerpotential}\tilde{g}^{\mu\bar{\mu}}\Kahlerpotential_\mu(\Laplacian\Kahlerpotential)_{\bar{\mu}} \notag \\
	& \hspace{10pt} - \zeta e^{-\zeta\Kahlerpotential}\tilde{g}^{\mu\bar{\mu}}\Kahlerpotential_{\bar{\mu}}(\Laplacian\Kahlerpotential)_\mu - \zeta e^{-\zeta\Kahlerpotential}\widetilde{\Laplacian}\Kahlerpotential(n + \Laplacian\Kahlerpotential) \notag \\
	& \hspace{10pt} + \zeta^2e^{-\zeta\Kahlerpotential}\tilde{g}^{\mu\bar{\mu}}\Kahlerpotential_\mu\Kahlerpotential_{\bar{\mu}}(n + \Laplacian\Kahlerpotential) \notag \\
	& \geqslant e^{-\zeta\Kahlerpotential}\widetilde{\Laplacian}(\Laplacian\Kahlerpotential) - e^{-\zeta\Kahlerpotential}\tilde{g}^{\mu\bar{\mu}}(n + \Laplacian\Kahlerpotential)^{-1}(\Laplacian\Kahlerpotential)_\mu(\Laplacian\Kahlerpotential)_{\bar{\mu}} \notag \\
	& \hspace{10pt} - \zeta e^{-\zeta\Kahlerpotential}\widetilde{\Laplacian}\Kahlerpotential(n + \Laplacian\Kahlerpotential).
\end{align}
And with the aid of the \emph{Cauchy–Schwarz inequality} \cite{Cauchy "Sur les Formules qui resultent de l'emploi du signe > ou < et sur les Moyennes entre plusieurs quantites"} \cite{Schwarz "Uber ein Flachen kleinsten Flacheninhalts betreffendes Problem der Variationsrechnung"}, we can draw up a relation of this kind,
\begin{align}
	& (n + \Laplacian\Kahlerpotential)^{-1} \cdot \frac{1}{1 + \Kahlerpotential_{\mu\bar{\mu}}}\left|\frac{\Kahlerpotential_{\xi\bar{\xi}\mu}}{(1 + \Kahlerpotential_{\xi\bar{\xi}})^\frac{1}{2}}(1 + \Kahlerpotential_{\xi\bar{\xi}})^\frac{1}{2}\right|^2 \notag \\
	& \leqslant (n + \Laplacian\Kahlerpotential)^{-1}\left(\frac{1}{1 + \Kahlerpotential_{\mu\bar{\mu}}}\right)\left(\frac{1}{1 + \Kahlerpotential_{\xi\bar{\xi}}}\right)\Kahlerpotential_{\xi\bar{\xi}\mu}\Kahlerpotential_{\xi\xi\bar{\mu}} \cdot 1 + \Kahlerpotential_{\varrho\bar{\varrho}},
\end{align}
and $-(n + \Laplacian\Kahlerpotential)^{-1}\frac{1}{1 + \Kahlerpotential_{\mu\bar{\mu}}}(\Laplacian\Kahlerpotential)_\mu(\Laplacian\Kahlerpotential)_{\bar{\mu}} + \widetilde{\Laplacian}\Laplacian\Kahlerpotential \geqslant \Laplacian\textcyrillic{\textit{ь}}_\varsigma + C_\mathrm{u}(n + \Laplacian\Kahlerpotential)\frac{1}{1 + \Kahlerpotential_{\mu\bar{\mu}}}$. It follows that $\widetilde{\Laplacian}\bigl(e^{-\zeta\Kahlerpotential}(n + \Laplacian\Kahlerpotential)\bigr) \geqslant e^{-\zeta\Kahlerpotential}\bigl(\Laplacian\textcyrillic{\textit{ь}}_\varsigma + C_\mathrm{u}(n + \Laplacian\Kahlerpotential)\frac{1}{1 + \Kahlerpotential_{\mu\bar{\mu}}}\bigr) - \zeta e^{-\zeta\Kahlerpotential}\widetilde{\Laplacian}(n + \Laplacian\Kahlerpotential)$.

There is a further next inequality to require:
\begin{equation}
	\sum_\mu\frac{1}{1 + \Kahlerpotential_{\mu\bar{\mu}}} \geqslant \left\{\frac{\sum_\mu(1 + \Kahlerpotential_{\mu\bar{\mu}})}{\prod_\mu(1 + \Kahlerpotential_{\mu\bar{\mu}})}\right\}^\frac{1}{n - 1} = \exp{\left\{-\tfrac{\textcyrillic{\textit{ь}}_\varsigma}{n - 1}\right\}}(n + \Laplacian\Kahlerpotential)^\frac{1}{n - 1},
\end{equation}
for $e^{-\zeta\Kahlerpotential} \sum_\mu\frac{1}{1 + \Kahlerpotential_{\mu\bar{\mu}}}(n + \Laplacian\Kahlerpotential) \geqslant \exp{\left\{-\frac{\textcyrillic{\textit{ь}}_\varsigma}{n - 1}\right\}}\exp{\left\{-\frac{\zeta}{n - 1}\right\}}\bigl[e^{-\zeta\Kahlerpotential}(n + \Laplacian\Kahlerpotential)\bigr]^\frac{n}{n - 1}$.

We can bring to an end this sub-step. Conceding that $\upsilon = e^{-\zeta\Kahlerpotential}(n + \Laplacian\Kahlerpotential)$, we have  $\widetilde{\Laplacian}\upsilon \geqslant - c_{(1)} - c_{(2)}\upsilon + c_{(0)}\upsilon^\frac{n}{n - 1}$, under the condition that $\Kahlerpotential \leqslant -1$ and $e^{-\zeta\Kahlerpotential} \geqslant 1$. If $x_0$ is the maximum of $\upsilon$, do we conclude that 
\begin{equation}
	0 \leqslant (n + \Laplacian\Kahlerpotential)(x) \leqslant e^{\zeta\Kahlerpotential(x)}\upsilon(x_0) \leqslant C_\mathrm{u},
\end{equation}
which is the second estimate of the Kähler potential.
\item[(c)] The last step is to find an estimate for third-order derivatives of the Kähler potential, and an a priori bound for $\|\nabla^3\Kahlerpotential\|_{\mathscr{C}^0}$, accompanied by another estimate $\|\Kahlerpotential_\mu - c_\mu\|_3 \leqslant C_\mathrm{u}$, with which, finally, we can assert that $\textcyrillic{\textit{Я}}$ is closed. The chain of sub-steps is in Aubin \cite[pp. 410-411]{Aubin "Metriques riemanniennes et courbure} and Yau \cite[app. A, pp. 403-406]{Yau "On the Ricci Curvature of a Compact Kahler Manifold and the Complex Monge-Ampere Equation I"}.
\subenumerationisfinis
\item[(\textgreek{δ}) — \textbf{step IV. The end}.] The Conjecture \ref{coniectura "Calabi conjecture"} becomes a theorem, which also answers positively the Question \ref{quaestio "Calabi conjecture"}.
\enumerationisfinis	
\end{proof}

\section{Addendum}

In this Addendum there are a few clarifications we should like to make on the Monge–Ampère equation. After this, we shall enunciate the Stokes' theorem \cite[p. 320]{Stokes "Smith's Prize Examination Papers N. 8 February 1854"} mentioned above, cf. Eq. \eqref{equation "Via Stokes' theorem"}, reformulate in the Cartan's language \cite[§§ 29-32, pp. 38-43]{Cartan "Les systemes differentiels exterieurs et leurs applications geometriques"}, complete with a proof.

\subsection{\emph{Postilla} on the Monge–Ampère Equation}
\label{subsection "Postilla on the Monge–Ampère Equation"} 

We saw above the Monge–Ampère equation \cite{Monge "Memoire sur le calcul integral des equations aux differences partielles"} in dual form \eqref{subequations "Monge–Ampère equations"} serviceably adapted, in the field of complex numbers, for the Calabi conjecture. Here we want to mention, very briefly, some of its general forms. 
\enumerationisinitium
\item The first one is
\begin{equation}
	\det\bigl(D^2\upsilon(x)\bigr) = \varphi\bigl(x, \upsilon, \nabla\upsilon(x)\bigr) \text{ on } \Omega,
\end{equation}
the archetypal Monge–Ampère statement \cite{Monge "Memoire sur le calcul integral des equations aux differences partielles"} \cite{Ampere "Memoire contenant l'application de la theorie exposee dans le XVII"}, which is a non-linear degenerate elliptic partial differential equation, letting 
\begin{align}
	D^2\upsilon(x) = [D_{\mu\nu}\upsilon(x)] = \frac{\partial^2\upsilon(x)}{\partial x_\mu\partial x_\nu}, \enspace 1 \leqslant \mu, \nu \leqslant n,
\end{align}
be the Hessian matrix \cite{Hesse "Uber die Elimination der Variabeln aus drei algebraischen Gleichungen vom zweiten Grade mit zwei Variabein"} (of second derivatives) of a convex function $\upsilon \colon \Omega \to \mathbb{R}$ in a domain (open set) $\Omega \subset \mathbb{R}^n$, and $\varphi \colon \Omega \times \mathbb{R} \times \mathbb{R}^n \to \mathbb{R}_+$ some positive function, where $\nabla\upsilon(x)$ is the gradient of $\upsilon$ at $x$. 
\item If $\upsilon \in \mathscr{C}^2(\Omega)$, by imposing an elliptic function $\textgreek{\text{Φ}}_{\mathbb{R}}$ with respect to $\upsilon$ on 
\begin{equation}
	x \mapsto \bigl\{x, \upsilon(x), \nabla\upsilon(x), D^2\upsilon(x)\bigr\},
\end{equation}
then the Monge–Ampère equation takes the form
\begin{equation}
	\textgreek{\text{Φ}}_{\mathbb{R}}\bigl(\upsilon(x)\bigr) = \det\bigl(D^2\upsilon(x)\bigr) - \varphi\bigl(x, \upsilon, \nabla\upsilon(x)\bigr) = 0.
\end{equation}
\enumerationisfinis

\subsection{Stokes–Cartan's Theorem (a Foundation of Exterior Calculus)}
\label{subsection "Stokes–Cartan's Theorem (a Foundation of Exterior Calculus)"}

\begingroup
\footnotesize
If $X, Y, Z$ be functions of the rectangular co-ordinates $x, y, z$, $dS$ an element of any limited surface, $l, m, n$ the cosines of the inclinations of the normal at $dS$ to the axes, $ds$ an element of the bounding line, shew that $\iint\bigl\{l\bigl(\frac{dZ}{dy} - \frac{dY}{dz}\bigr) + m\bigl(\frac{dX}{dz} - \frac{dZ}{dx}\bigr) + n\bigl(\frac{dY}{dx} - \frac{dX}{dy}\bigr)\bigr\}dS = \int\bigl(X\frac{dx}{ds} + Y\frac{dy}{ds} + Z\frac{dz}{ds}\bigr)ds$, the differential coefficients of $X, Y, Z$ being partial, and the single integral being taken all round the perimeter of the surface. \\
\indent — \textsc{G.G. Stokes} \cite[p. 320]{Stokes "Smith's Prize Examination Papers N. 8 February 1854"}\footnote{
	Stokes theorem looks like this; it appears, historically, as a question № 8 in \textit{Smith's Prize Examination Papers, February 1854}. The examination was completed by J.C. Maxwell, who writes \cite[p. 27]{Maxwell "A Treatise on Electricity and Magnetism I"}: «This theorem was given by Professor Stokes, \textit{Smith's Prize Examination}, 1854, question 8». 
	}

\vspace{2mm}

It is demonstrated that with analogous orientation conventions, we have a completely general Stokes formula $\int\omega = \int d\omega$; the first integral is extended to the $p$-dimensional boundary of a $(p + 1)$-dimensional domain, and it is to this last domain that the second integral extends. \\
\indent — \textsc{É. Cartan} \cite[§ 29, p. 40]{Cartan "Les systemes differentiels exterieurs et leurs applications geometriques"}

\endgroup

\vspace{2mm}

\begin{theorema}[Stokes–Cartan]
Let $\mathcal{M}$ be a smooth $n$-dimensional manifold, or an open set oriented by $\mathbb{R}^n$, $\partial\mathcal{M}$ the boundary of $\mathcal{M}$ with the induced orientation of $\mathbb{R}^n$ (that is, $\partial\mathcal{M}$ is oriented by an exterior normal vector to $\mathcal{M}$), and $\iota \colon \partial\mathcal{M} \hookrightarrow \mathcal{M}$ the inclusion map, i.e. the injection of the boundary $\partial$ into $\mathcal{M}$. Given a smooth $(n - 1)$-form of class $\mathscr{C}^1$ (cf. Section \ref{section "Connection Forms"}), indicated by $\omega \in \bigwedge^{n - 1}_{\mathrm{c}}(\mathcal{M})$, with compact support on $\mathcal{M}$—and in fact 
\[
	\bigwedge^k_{\mathrm{c}}(\cdot) \viz \Omega_\mathrm{c}^k(\cdot)
\]
denotes the set of all $k$-forms having compact support—,\footnote{
	$\omega \in \Omega_\mathrm{c}^{n - 1}(\mathcal{M})$.
	}	
then
\begin{equation}
\label{equation "Stokes–Cartan theorem"}
	\int_\mathcal{M}d\omega = \int_{\partial\mathcal{M}}\iota^*\omega,
\end{equation}
by setting $\iota^*\omega$ as a pullback under the inclusion map, where if $\partial\mathcal{M} = \varnothing$, $\int_{\partial\mathcal{M}}\iota^*\omega = 0$. 
\end{theorema}

\begin{proof}
We will show three demonstrations, for three cases different cases.
\enumerationisinitium
\item The first one analyzes, \emph{locally}, a smooth $(n - 1)$-form
\begin{equation}
	\omega = \textcyrillic{\textit{я}} dx^1 \wedge \cdots \wedge dx^{n - 1}, \\
\end{equation}	
on $\mathcal{M} = \mathbb{R}^n$, and the exterior differential
\begin{equation}
	d\omega = (-1)^{n - 1}\frac{\partial\textcyrillic{\textit{я}}}{\partial x^n}dx^1 \wedge \cdots \wedge dx^n.
\end{equation}	
Via \emph{Fubini's theorem} \cite{Fubini "Sugli integrali multipli"}, one sees that
\begin{align}
	& \int_{\mathbb{R}^n} d\omega = (-1)^{n - 1}\int_{\mathbb{R}^{n - 1}}\left\{\int^{+\infty}_{-\infty}\frac{\partial\textcyrillic{\textit{я}}}{\partial x^n}dx^n\right\}dx^1 \cdots dx^{n - 1}, \\
	& \int^{+\infty}_{-\infty}\frac{\partial\textcyrillic{\textit{я}}}{\partial x^n}(x^1, \mathellipsis, x^{n - 1}, x^n)dx^n \notag \\
	& = \lim_{s \to \infty}\Bigl\{\textcyrillic{\textit{я}}(x^1, \mathellipsis, x^{n - 1}, s) -\textcyrillic{\textit{я}}(x^1, \mathellipsis, x^{n - 1}, -s)\Bigr\} = 0,
\end{align}
considering that $\textcyrillic{\textit{я}}$ has compact support; ergo 
\begin{equation}
	\int_{\mathbb{R}^n} d\omega = 0,
\end{equation}
which gives a proof of \eqref{equation "Stokes–Cartan theorem"} for $\mathbb{R}^n$ without a boundary.
\item The second case is \emph{local} again, but in the geometric context of the upper half-space (cf. Section \ref{subsubsection "Upper Half-Space, Ball, and Hyperboloid"}); it provides that $\mathcal{M} = \hyperbolic^n = \{(x^1, \mathellipsis, x^n) \in \mathbb{R}^n \mid x^n \geqslant 0\}$, with the boundary $\partial\hyperbolic^n = \{x^n = 0\}$. We designate 
\subenumerationisinitium
\item a smooth $(n - 1)$-form as
\begin{equation}
	\omega = \sum^n_{\nu = 1}(-1)^{\nu - 1}\textcyrillic{\textit{я}}_\nu dx^1 \wedge \cdots \wedge \Langle dx^\nu\Rangle \wedge \cdots \wedge dx^n,
\end{equation}
where $\textcyrillic{\textit{я}}_\nu$ is equipped with a compact support, and the chevrons $\Langle\cdots\Rangle$ indicates that this $\nu$-element is omitted (it is not part of the exterior product), 
\item its exterior differential as
\begin{align}
	d\omega & = \left\{\sum^n_{\nu - 1}(-1)^{\nu = 1}\frac{\partial\textcyrillic{\textit{я}}_\nu}{\partial x^\nu}(x^1, \mathellipsis, x^{n - 1}, x^n)\right\}dx^1 \wedge \cdots \wedge dx^n \notag \\ 
	& = \left\{\sum^n_{\nu = 1}\frac{\partial\textcyrillic{\textit{я}}_\nu}{\partial x^\nu}\right\}dx^1 \wedge \cdots \wedge dx^n,
\end{align}
\subenumerationisfinis
After that we fix
\begin{equation}
	\int_{\partial\hyperbolic^n}\omega = (-1)^n\int_{\mathbb{R}^{n - 1}}\textcyrillic{\textit{я}}_n(x^1, \mathellipsis, x^{n - 1}, 0)dx^1 \cdots dx^{n - 1},\footnotemark
\end{equation}
\footnotetext{
	Note that $\omega|_{\partial\hyperbolic^n} = \textcyrillic{\textit{я}}_n(x^1, \mathellipsis, x^{n - 1}, 0)dx^1 \wedge \cdots \wedge dx^{n - 1}$.
	}
and, for $1 \leqslant \nu \leqslant n - 1$,
\begin{subequations}
\begin{align}
	& \Lbrack:\int^{+\infty}_{-\infty}\frac{\partial\textcyrillic{\textit{я}}_\nu}{\partial x^\nu}(x^1, \mathellipsis, x^{n - 1}, x^n)dx^\nu:\Rbrack = \lim_{x^\nu \to \infty}\Bigl\{\textcyrillic{\textit{я}}_\nu(x^1, \mathellipsis, x^\nu, \mathellipsis, x^n) \notag \\
	& \hspace{170pt} - \textcyrillic{\textit{я}}_\nu(x^1, \mathellipsis, -x^\nu, \mathellipsis, x^n)\Bigr\} = 0, \\
	&
	\label{align "Second equation inherent to the particular hyperbolic case"}
	\int_{\hyperbolic^n}\frac{\partial\textcyrillic{\textit{я}}_\nu}{\partial x^\nu}(x^1, \mathellipsis, x^{n - 1}, x^n)dx^1 \cdots dx^n \notag \\
	& = \int^\infty_0\left\{\int_{\mathbb{R}^{n - 2}}\Lbrack:\cdots:\Rbrack dx^1 \cdots \Langle dx^\nu\Rangle \cdots dx^{n - 1}\right\}dx^n = 0,
\end{align}
\end{subequations}
for $\nu = 1, \mathellipsis, n - 1$ in \eqref{align "Second equation inherent to the particular hyperbolic case"}, through the repeat sign $\Lbrack:$ and $:\Rbrack$. Hence
\begin{align}
	& \int^\infty_0\frac{\partial\textcyrillic{\textit{я}}_n}{\partial x^n}(x^1, \mathellipsis, x^{n - 1}, x^n)dx^n \notag \\
	& = \lim_{s \to \infty}\textcyrillic{\textit{я}}_n(x^1, \mathellipsis, x^{n - 1}, s) - \textcyrillic{\textit{я}}_n(x^1, \mathellipsis, x^{n - 1}, 0).
\end{align}
By determining
\begin{equation}
	\iota^*\omega = (-1)^{n - 1}(\textcyrillic{\textit{я}}_n \circ \iota)\iota^*(dx^1 \wedge \cdots \wedge dx^{n - 1}),
\end{equation}
finally, 
\begin{align}
	\int_{\hyperbolic^n}d\omega & = (-1)^{n - 1}\int_{\hyperbolic^n}\frac{\partial\textcyrillic{\textit{я}}_n}{\partial x^n}(x^1, \mathellipsis, x^{n - 1}, x^n)dx^1 \cdots dx^n \notag \\
	& = (-1)^{n - 1}\int_{\mathbb{R}^{n - 1}}\left\{\int^\infty_0\frac{\partial\textcyrillic{\textit{я}}_n}{\partial x^n}(x^1, \mathellipsis, x^{n - 1}, x^n)dx^n\right\}dx^1 \cdots dx^{n - 1} \notag \\
	& = (-1)^{n - 1}\int_{\mathbb{R}^{n - 1}}\underbrace{\textcyrillic{\textit{я}}_n(x^1, \mathellipsis, x^{n - 1}, 0)}_{\textcyrillic{\textit{я}}_n \circ \iota}dx^1 \cdots dx^{n - 1} = \int_{\partial\hyperbolic^n}\iota^*\omega. 
\end{align}
\item The third case is \emph{global}. Let $\mathcal{A} = \{(\Upsilon_\mu, \varphi_\mu \mid \mu \in \mathcal{A})\}$ be an atlas with an $n$-chart $(\Upsilon_\mu, \varphi_\mu)$, for each index $\mu$, and $\{\textit{\NG}_\mu\}_{\mu \in \mathcal{A}} \subset \mathscr{C}^\infty(\mathcal{M})$ be a partition of unity subordinate to $\mathcal{A}$ on $\mathcal{M}$, i.e. a set such that 
 
· $0 \leqslant \textit{\NG}_\mu \leqslant 1$ on $\mathcal{M}$, $\forall\mu \in \mathcal{A}$, 

· $\supp(\textit{\NG}_\mu)$, that is, $\textit{\NG}_\mu$ has compact support.

If $d\omega = \sum_{\mu \in \mathcal{A}}d(\textit{\NG}_\mu\omega)$ and $\iota^*_\mathcal{M}\omega = \sum_{\mu \in \mathcal{A}^{(\mathrm{i})}}\iota^*_\mathcal{M}(\textit{\NG}_\mu\omega)$, it would appear that
\begin{align}
	\int_\mathcal{M}d(\textit{\NG}_\mu\omega) & = \sum_{\mu \in \mathcal{A}}\int_{\Upsilon_\mu}d(\textit{\NG}_\mu\omega) \notag \\
	& = \sum_{\mu \in \mathcal{A}}\int_{\varphi_\mu(\Upsilon_\mu)}(\varphi_\mu^{-1})^*d(\textit{\NG}_\mu\omega) = \sum_{\mu \in \mathcal{A}}\int_{\varphi_\mu(\Upsilon_\mu)}d(\varphi_\mu^{-1})^*(\textit{\NG}_\mu\omega) \notag \\
	& = \int_{\partial\varphi_\mu(\Upsilon_\mu)}(\varphi_\mu^{-1})^*(\textit{\NG}_\mu\omega) = \int_{\varphi_\mu(\Upsilon_\mu \cap \partial\mathcal{M})}(\varphi_\mu^{-1})^*(\textit{\NG}_\mu\omega) \notag \\
	& = \sum_{\mu \in \mathcal{A}^{(\mathrm{i})}}\int_{\varphi_\mu(\Upsilon_\mu \cap \partial\mathcal{M})}(\varphi_\mu^{(\partial)-1})^*\iota^*_\mathcal{M}(\textit{\NG}_\mu\omega) \notag \\
	& = \int_{\partial\mathcal{M}}\textit{\NG}_\mu\omega = \int_{\partial\mathcal{M}}\iota^*_\mathcal{M}\omega.
\end{align}
\enumerationisfinis	
\end{proof}
	
\begin{margo}[Stokes' theorem in terms of flux and circulation]
Outside the differential forms language, Stokes' theorem is as follows. {\itshape Given 

· a compact surface $\surface \subset \mathbb{R}^3$, i.e. a piecewise smooth surface in 3-space, whose orientation is established by the unit normal vector $\hat{N}$,

· a boundary $\partial\surface$ endowed with a positive orientation, via tangent vector $v_{(\partial\surface)}$, which is consistent with the orientation of $\surface$, and

· a smooth vector field  $\vec{X}$ of class $\mathscr{C}^1$ on an open set containing $\surface$, \\
then 
\begin{equation}
	\int_{\surface \subset \mathbb{R}^3} \mathrm{curl}(\vec{X}) \cdot \hat{N} = \oint_{\partial\surface} \vec{X} \cdot v_{(\partial\surface)}, \enspace \mathrm{curl}(\vec{X}) = \nabla \times \vec{X},
\end{equation}
where $\mathrm{curl}(\vec{X})$ is the curl (vector operator for the infinitesimal circulation) of $\vec{X}$}. 

Its meaning: the flux of $\mathrm{curl}(\vec{X})$ passing through $\surface$ is equal to the line integral of $\vec{X}$ around the surface's boundary $\partial\surface$, that is, the flux of the curl of $\vec{X}$ equals the circulation of the tangential component of $\vec{X}$ about the (closed) line bounding $\surface$ in 3-space.\footnote{
	The concept of \emph{curl}, already sketched out by J. MacCullagh \cite[lemma II]{MacCullagh "An Essay towards a Dynamical Theory of Reflexion and Refraction"}, is clarified, as well as terminologically conceived, by Maxwell \cite[pp. 231-232]{Maxwell "Remarks on the Mathematical Classification of Physical Quantities"}.
	} \margosymbol
\end{margo}	

\vspace{10mm}

\setcounter{secnumdepth}{0}  
\section{References and Bibliographic Details}
\setcounter{secnumdepth}{3}
\markright{References and Bibliographic Details}

\begingroup
\footnotesize
\noindent Section \ref{section "Ricci Form on the Space of a Kählerian Metric"}

\begin{indent paragraph: 15pt}
· For a look at the contributions of E. Calabi in geometry, see \cite{de Bartolomeis Tricerri Vesentini (Eds.) "Manifolds and Geometry"}. \\
· Insights on the Calabi conjecture, space of Kähler metrics, and Calabi–Yau structures, are in \cite{Mabuchi "Einstein Metrics in Complex Geometry: An Introduction"} \cite[chap. 7]{Ballmann "Lectures on Kahler Manifolds"} \cite[chap. 6]{Joyce "Riemannian Holonomy Groups and Calibrated Geometry"} \cite[chap. 18]{Moroianu "Lectures on Kahler Geometry"} \cite{Blocki "The Calabi-Yau Theorem"}. \\
· On the Calabi–Yau spaces, see \cite{Gross Huybrechts Joyce "Calabi-Yau Manifolds and Related Geometries"} \cite{Hitchin "Generalized Calabi-Yau manifolds"}.
\end{indent paragraph: 15pt}

\noindent Section \ref{subsection "Postilla on the Monge–Ampère Equation"} 

\begin{indent paragraph: 15pt}
· For a thorough examination on the Monge–Ampère equation (together with its many applications), see \cite{Pogorelov "Monge-Ampere Equations of Elliptic Type"} \cite{Caffarelli "A Priori Estimates and the Geometry of the Monge Ampere Equation"} \cite{Caffarelli Milman (Eds.) "Monge Ampere Equation: Applications to Geometry and Optimization} \cite{Gutierrez "The Monge-Ampre Equation"} \cite{Figalli "The Monge-Ampere Equation and its Applications"} \cite{Figalli "On the Monge-Ampere equation"}.
\end{indent paragraph: 15pt}

\endgroup

\chapter{Geometric and Topological Aspects of Complexity and Dynamics, Part I. Flows, Hyperbolicity, and Foliations}
\chaptermark{Geometric and Topological Aspects of Complexity and Dynamics, Part I}{}
\label{chapter "Geometric and Topological Aspects of Complexity and Dynamics, Part I. Flows, Hyperbolicity, and Foliations"}

\begingroup
\footnotesize
Plasmare dunque concetti in modo da potere introdurre la misura; misurare quindi; dedurre poi delle leggi; risalire da esse ad ipotesi; dedurre da queste, mercè l'analisi, una scienza di enti ideali sì, ma rigorosamente logica; confrontare poscia colla realtà; rigettare o trasformare, man mano che nascono contraddizioni fra i resultati del calcolo ed il mondo reale, le ipotesi fondamentali che han già servito; e giungere così a divinare fatti ed analogie nuove, o dallo stato presente arrivare ad argomentare quale fu il passato e che cosa sarà l'avvenire; ecco, nei più brevi termini possibili, riassunto il nascere e l'evolversi di una scienza avente carattere matematico.\footnote{
	«Shape concepts so as to be able to introduce the measure(ment); hence measure; then deduce laws; from them trace back to hypotheses; deduce from these, by means of analysis, a science of ideal entities, yes, but [a] rigorously logical [science]; later, compare with reality; reject or transform, as contradictions arise between the results of the calculation and the real world, the fundamental hypotheses already used; and thus get to predict new facts and analogies, or from the present state argue about what the past was and what the future will be; here, in the shortest possible terms, a summary of the birth and evolution of a science with a mathematical character».
	} \\
\indent — \textsc{V. Volterra} \cite[pp. 442-443]{Volterra "Sui tentativi di applicazione delle matematiche alle scienze biologiche e sociali"}

\endgroup

\section[Geodesic Flow on the Unit Tangent Bundle of a Negatively Curved Surface by the $\mathbbl{\Gamma}$-Action on the Hyperbolic Half-Plane]{Geodesic Flow on the Unit Tangent Bundle of a Negatively Curved Surface by the $\protect\pseudobold{\mathbbl{\Gamma}}$-Action on the Hyperbolic Half-Plane}
\sectionmark{Geodesic Flow on the Unit Tangent Bundle of a Negatively Curved Surface}
\label{section "Geodesic Flow on the Unit Tangent Bundle of a Negatively Curved Surface by the Gamma-action on the Hyperbolic Half-Plane"}

\subsection{Geodesic Flow from Lobačevskijan Geometry}
\label{subsection "Geodesic Flow from Lobačevskijan Geometry"}

\begin{definitio}
Let $\gamma_{\mathrm{c}(z, v)}(t)$ be a geodesic for a fixed initial conditions
\begin{equation}
	\begin{cases}
	\gamma_{\mathrm{c}(z, v)}(0) = z, \\
	\dot{\gamma}_{\mathrm{c}(z, v)}(0) = v.
	\end{cases}
\end{equation}
Let $\mathring{\mathcal{T}}^1\mathbb{U}^2_\mathbb{C}$\footnote{
	It is sometimes denoted by $\mathring{\mathcal{S}}^1[\mathrm{manifold}]$, so $\mathring{\mathcal{S}}^1\mathbb{U}^2_\mathbb{C}$.
	}  
be the \emph{unit tangent bundle} of $\mathbb{U}^2_\mathbb{C} = \{z \in \mathbb{C} \mid \Im(z) > 0\}$, that is the Beltrami–Poincaré upper half-plane \eqref{align "Complex-valued upper half-plane"}. The vector $v \in \mathcal{T}^1_z\mathbb{U}^2_\mathbb{C}$ is the unit tangent vector at $z \in \mathbb{U}^2_\mathbb{C}$ to the geodesic at unit speed; this means that $\gamma_{\mathrm{c}(z, v)}(t)$ passes through the point $z$ in the direction of the unit tangent vector, i.e. with $v$ as its tangent. The flow on $\mathbb{U}^2_\mathbb{C}$ that moves each $v$ along its geodesic at unit speed is called \emph{geodesic flow}, and it is more aptly defined as the \emph{flow on the unit tangent bundle} of a surface of constant negative curvature, 
\begin{equation}
\label{equation "Geodesic flow"}
	\{\varphi_t\}_{t \in \mathbb{R}} \colon \mathring{\mathcal{T}}^1\mathbb{U}^2_\mathbb{C} \to \mathring{\mathcal{T}}^1\mathbb{U}^2_\mathbb{C}.
\end{equation}
(We remind that $\mathbb{U}^2_\mathbb{C}$ is one of the best-known examples of Riemann surfaces, along with the unit disk). The geodesic flow on $\mathring{\mathcal{T}}^1\mathbb{U}^2_\mathbb{C}$ is the diffeomorphism defined as 
\begin{equation}
\label{equation "Diffeomorphism geodesic flow"} 
	\varphi_t(z, v) \equival \bigl(\gamma_{\mathrm{c}(z, v)}(t), \dot{\gamma}_{\mathrm{c}(z, v)}(t)\bigr),
\end{equation}
for 
\[
	(z, v) \in \mathring{\mathcal{T}}^1\mathbb{U}^2_\mathbb{C} = \{(z, v) \in \mathbb{U}^2_\mathbb{C} \times \mathbb{C} \mid \|v\|_z = 1\},
\] 
where $\dot{\gamma}_{\mathrm{c}(z, v)}(t)$ is the velocity vector of a point-mass (particle) at an instant of time $t$. The tangent bundle $\mathring{\mathcal{T}}^1\mathbb{U}^2_\mathbb{C}$ is invariant, because the speed of a geodesic is constant; one can think of $\mathring{\mathcal{T}}^1\mathbb{U}^2_\mathbb{C}$ as a \emph{constant-energy hypersurface} of a point-mass flowing (moving smoothly) along $\mathbb{U}^2_\mathbb{C}$. \definitiosymbol
\end{definitio}

\subsection{Projective Linear Transforms: Dynamics on the Modular Surface, and Horocycle Flow}
\label{subsection "Projective Linear Transforms: Dynamics on the Modular Surface, and Horocycle Flow"} 

Carrying on the above speech, we are alert to the fact that there exists an identification between the unit tangent bundle and the projective special linear group of $2 \times 2$ matrices over the real field. This last one, denoted by 
\begin{equation}
	PSL_2(\mathbb{R}) \cong \frac{SL_2(\mathbb{R})}{\{\pm\idem\}} \cong \Moebius^+_2(\mathbb{R}) \equival \mathfrak{isom}^+(\mathbb{U}^2_\mathbb{C}), 
\end{equation}
where $\pm\idem$ is equal to $\bigl(\begin{smallmatrix}
	1 & 0 \\
	0 & 1
	\end{smallmatrix}\bigr)$, is the Möbius group of all biholomorphic maps of $\varphi \colon \mathbb{U}^2_\mathbb{C} \to \mathbb{U}^2_\mathbb{C}$, i.e. it is the orientation preserving isometry group of $\mathbb{U}^2_\mathbb{C}$, see Eq. \eqref{equation "Group $PSL_2(R)$"}. So if $PSL_2(\mathbb{R})$ acts isometrically on $\mathbb{U}^2_\mathbb{C}$ is isometric, it also acts freely and transitively on the unit tangent bundle $\mathring{\mathcal{T}}^1\mathbb{U}^2_\mathbb{C}$ of $\mathbb{U}^2_\mathbb{C}$, thereby we can identify $\mathring{\mathcal{T}}^1\mathbb{U}^2_\mathbb{C}$ with $PSL_2(\mathbb{R})$, i.e. $\mathring{\mathcal{T}}^1\mathbb{U}^2_\mathbb{C} \cong PSL_2(\mathbb{R})$.

\begin{figure}[h!]
\centering
\includegraphics[width = 0.550\textwidth]{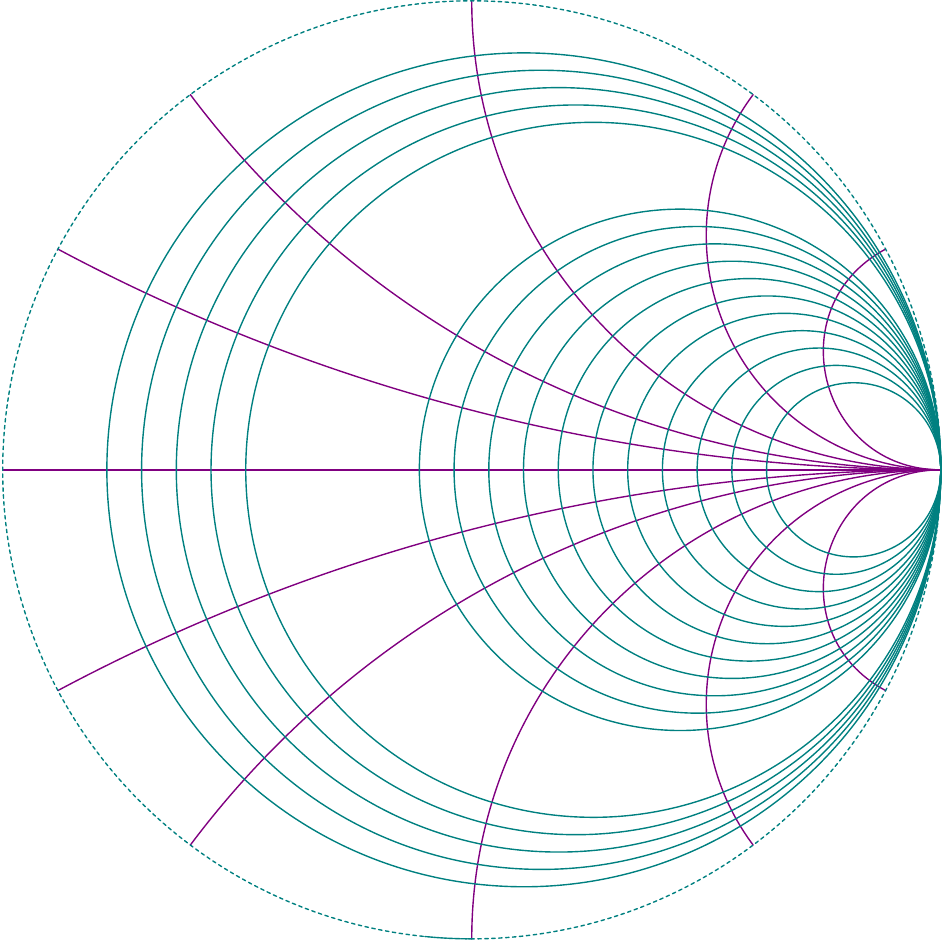}
\caption{Some horocycles (in \textcolor{mallard}{\texttt{mallard}} color) in the Beltrami–Poincaré disk model intersecting some asymptotic parallel lines (in \textcolor{eggplant}{\texttt{eggplant}} color)}
\label{figure "horocycles"}
\end{figure}

We care about the \emph{right multiplication} on $\mathring{\mathcal{T}}^1\mathbb{U}^2_\mathbb{C} \cong PSL_2(\mathbb{R})$; it provides us with two flows. 
\enumerationisinitium
\item The first is the \emph{geodesic flow} $\varphi_t$ as illustrated in \eqref{equation "Geodesic flow"}; however, under the identification between $\mathring{\mathcal{T}}^1\mathbb{U}^2_\mathbb{C}$ and $PSL_2(\mathbb{R})$, the action of $\varphi_t$ on $\mathring{\mathcal{T}}^1\mathbb{U}^2_\mathbb{C}$ corresponds to the action by right multiplication of the diagonal (1-parameter) group 
\begin{equation}
	D_g = \left\{\varphi_t(g) = g_t \equival
	\begin{pmatrix}
	e^{\frac{t}{2}} & 0 \\
	0 & e^{-\frac{t}{2}}
	\end{pmatrix} \mathrel{\bigg|} t \in \mathbb{R}\right\},
\end{equation}
on $PSL_2(\mathbb{R})$, so the flow is 
\begin{equation}
	\{\varphi_t\}_{t \in \mathbb{R}} \colon PSL_2(\mathbb{R}) \to PSL_2(\mathbb{R}),
\end{equation}
and the Eq. \eqref{equation "Diffeomorphism geodesic flow"} is of course $\varphi_t(z, v) = (z, v)g_t$. We can talk about geodesic flow arising from the the 1-parameter group $g_t$ in this instance. Therefore, the geodesic flow on $\mathring{\mathcal{T}}^1\mathbb{U}^2_\mathbb{C}$ can be described as the geodesic flow on $PSL_2(\mathbb{R})$ by the right translations $g \mapsto g\varphi_t(g) = g_t$.
\item The other one is the \emph{horocycle flow}.\footnote{
	Horocycle flows on a compact negatively curved manifold are uniquely ergodic (see below) iff $\mathbbl{\Gamma}$-invariant measures are the constant multiples of Lebesgue measure, as has been demonstrated by H. Furstenberg \cite{Furstenberg "The unique ergodicity of the horocycle flow"}.
	}
Keep in mind that a \emph{horocycle} (Fig. \ref{figure "horocycles"}) lying in $\mathbb{U}^2_\mathbb{C}$ is 
\subenumerationisinitium
\item a horosphere in dimension 2, that is to say a (Euclidean) circle tangent to the boundary\footnote{
	The name betrays the definition: \textgreek{ὅρι[ον]-κύκλος} (border-circle). 
	}
$\partial_\infty\mathbb{U}^2$ passing through a certain point $x \in \mathbb{RP}^1$, or a curve tangent to the real axis at $x$, so that the horocyclic curve is an orbit of $PSL_2(\mathbb{R})$,
\item a horizontal line at $\infty \in \mathbb{RP}^1$, form which we can distinguish between a horocycle at $\gamma_{\mathrm{c}(v)}(+\infty)$, concerning $\varphi_t$ of a \emph{positive-stable} manifold with $t \to +\infty$, and a horocycle at $\gamma_{\mathrm{c}(v)}(-\infty)$, concerning $\varphi_t$ of a \emph{negative-unstable} manifold with $t \to -\infty$.
\subenumerationisfinis

A horocycle can be \emph{stable} or \emph{unstable}; it is a set contains vectors orthogonal to the horocycle, in both cases; but in the stable set there are vectors pointing \emph{inward}, in the unstable set the normal vectors point \emph{outward}. Consequently, the horocycle flow can be \emph{stable} or \emph{unstable}, denoted by $\{\eta^+_t\}_t$ and $\{\eta^-_t\}_t$, respectively,
\begin{equation}
	\{\eta_t\}_{t \in \mathbb{R}} \colon \mathring{\mathcal{T}}^1\mathbb{U}^2_\mathbb{C} \to \mathring{\mathcal{T}}^1\mathbb{U}^2_\mathbb{C}, \text{ with } \eta_t = \eta^+_t \text{ or } \eta_t = \eta^-_t, 
\end{equation}
given by right multiplication by the groups
\begin{equation}
\label{equation "1-parameter groups for horocycle flow"}
	H^+_\eta = \left\{\eta^+_t \equival 
	\begin{pmatrix}
	1 & t \\
	0 & 1
	\end{pmatrix}\right\}_{t \in \mathbb{R}} \text{ or }
	H^-_\eta = \left\{\eta^-_t \equival 
	\begin{pmatrix}
	1 & 0 \\
	t & 1
	\end{pmatrix}\right\}_{t \in \mathbb{R}},
\end{equation}
respectively. Put another way, the horocycle flow is a \emph{positive or negative horocycle field} on $\mathring{\mathcal{T}}^1\mathbb{U}^2_\mathbb{C}$ that moves each $v \in \mathring{\mathcal{T}}^1\mathbb{U}^2_\mathbb{C}$ along $\{\eta_t\}_t$ at unit speed. The 1-parameter group $\eta^+_t$ generates the stable horocycle flow; the 1-parameter group $\eta^-_t$ generates the unstable horocycle flow. But even here, the action of $\eta^+_t \viz \{\eta^+_t\}_t$ and $\eta^-_t \viz \{\eta^-_t\}_t$ on $\mathring{\mathcal{T}}^1\mathbb{U}^2_\mathbb{C}$ corresponds to the action by right multiplication of $H^-_\eta$ and $H^+_\eta$ on $PSL_2(\mathbb{R})$; the flow is $\{\eta_t\}_{t \in \mathbb{R}} \colon PSL_2(\mathbb{R}) \to PSL_2(\mathbb{R})$, and it is represented by the right translations $g \mapsto g\eta^+_t$ or $g \mapsto g\eta^-_t$.
\enumerationisfinis

We shall now proceed with the modular surface still on the subject of the projective special linear group. Let 
\begin{equation}
	\mathbbl{\Gamma} \subset PSL_2(\mathbb{R}) \cong \Moebius^+_2(\mathbb{R}) \equival \mathfrak{isom}^+(\mathbb{U}^2_\mathbb{C}) 
\end{equation}
be a discrete cocompact subgroup of $PSL_2(\mathbb{R})$ (see Section \ref{subsection "Discrete Gamma-Crystallographic Group, Killing–Hopf Theorem, and Isometric Action"}), namely a \emph{Fuchsian group} (see Section \ref{section "Fuchsian Group (Properly Discontinuous Action)"}). Let 
\begin{equation}
	\mathcal{S}_\mathbbl{\Gamma} \equival \mathbbl{\Gamma} \backslash \mathbb{U}^2_\mathbb{C}
\end{equation}
be a \emph{Riemann surface}; more technically, 
\begin{equation}
	\mathcal{S}_\mathbbl{\Gamma} = \{\mathbbl{\Gamma}(z) \mid z \in \mathbb{U}^2_\mathbb{C}\}, 
\end{equation}
also called \emph{modular surface}, is an \emph{orbit space} of the modular group $\mathbbl{\Gamma}$.\footnote{
	We also write the space of $\mathbbl{\Gamma}$-orbit(s) as a \emph{quotient surface} $\mathbb{U}^2_\mathbb{C}/\mathbbl{\Gamma}$, so $\pi_z \equival \mathbbl{\Gamma}(z) \colon \mathbb{U}^2_\mathbb{C} \to \mathbb{U}^2_\mathbb{C}/\mathbbl{\Gamma}$.
	} 

We must take account of the action of $PSL_2(\mathbb{R})$ by \emph{left multiplication} on $\mathring{\mathcal{T}}^1\mathbb{U}^2_\mathbb{C} \cong PSL_2(\mathbb{R})$, whereas this is a derivative action by isometries through $PSL_2(\mathbb{R})$ on the unit tangent bundle. Since the group $\mathbbl{\Gamma}$ acts on $PSL_2(\mathbb{R})$ by multiplication to the left (by the projective special linear group), one understands that $\mathring{\mathcal{T}}^1\mathbb{U}^2_\mathbb{C} \cong PSL_2(\mathbb{R})$ leads to another identification, 
\begin{equation}
	\mathring{\mathcal{T}}^1\mathbbl{\Gamma} \backslash \mathbb{U}^2_\mathbb{C} \cong \mathbbl{\Gamma} \backslash PSL_2(\mathbb{R}), 
\end{equation}
and, that is, 
\begin{equation}
	\mathring{\mathcal{T}}^1\mathcal{S}_\mathbbl{\Gamma} \cong \mathbbl{\Gamma} \backslash PSL_2(\mathbb{R}), 
\end{equation}
whereby the unit tangent bundle of $\mathcal{S}_\mathbbl{\Gamma} \equival \mathbbl{\Gamma} \backslash \mathbb{U}^2_\mathbb{C}$ equates with the compact quotient $\mathbbl{\Gamma} \backslash PSL_2(\mathbb{R})$, with the specification that $\mathbbl{\Gamma}$ is a \emph{lattice} in $PSL_2(\mathbb{R})$\footnote{
	As long as $\mathbbl{\Gamma} \backslash PSL_2(\mathbb{R})$ has finite measure.
	} 
and $\mathbbl{\Gamma} \backslash PSL_2(\mathbb{R})$ is the $\mathbbl{\Gamma}$-\emph{orbit projective space}. Considering that $\mathbbl{\Gamma} \backslash PSL_2(\mathbb{R})$ is determined by the action of $\mathbbl{\Gamma}$ on the left, the flows $\varphi_t$, $\eta^+_t$ and $\eta^-_t$ are consequently also defined on $\mathring{\mathcal{T}}^1\mathcal{S}_\mathbbl{\Gamma} \cong \mathbbl{\Gamma} \backslash PSL_2(\mathbb{R})$. So there we have it: 
\enumerationisinitium
\item the flow $\varphi_t \colon \mathring{\mathcal{T}}^1\mathcal{S}_\mathbbl{\Gamma} \to \mathring{\mathcal{T}}^1\mathcal{S}_\mathbbl{\Gamma}$, or else, in the long form,
\begin{equation}
	\left(\mathring{\mathcal{T}}^1\bigl(\mathcal{S}_\mathbbl{\Gamma} \equival \mathbbl{\Gamma} \backslash \mathbb{U}^2_\mathbb{C}\bigr)\right) \cong \mathbbl{\Gamma} \backslash PSL_2(\mathbb{R}) \to \left(\mathring{\mathcal{T}}^1\bigl(\mathcal{S}_\mathbbl{\Gamma} \equival \mathbbl{\Gamma} \backslash \mathbb{U}^2_\mathbb{C}\bigr)\right) \cong \mathbbl{\Gamma} \backslash PSL_2(\mathbb{R}),
\end{equation}
is equivalent to the flow $\mathbbl{\Gamma} \backslash PSL_2(\mathbb{R}) \to \mathbbl{\Gamma} \backslash PSL_2(\mathbb{R})$ and the right translations $\mathbbl{\Gamma}g \mapsto \mathbbl{\Gamma}g\varphi_t(g) = g_t$ or
\begin{equation}
	\varphi_{\mathcal{S}_\mathbbl{\Gamma}}(\mathbbl{\Gamma}g) \mapsto \mathbbl{\Gamma}g
	\begin{pmatrix}
	e^{\frac{t}{2}} & 0 \\
	0 & e^{-\frac{t}{2}}
	\end{pmatrix};
\end{equation}
\item the flow linked to orbits of the stable or unstable horocycle,
\begin{equation}
	\eta_t \colon \mathring{\mathcal{T}}^1\mathcal{S}_\mathbbl{\Gamma} \to \mathring{\mathcal{T}}^1\mathcal{S}_\mathbbl{\Gamma}, \text{ with } \eta_t = \eta^+_t \text{ or } \eta_t = \eta^-_t, 
\end{equation}
is equivalent to the flow $\mathbbl{\Gamma} \backslash PSL_2(\mathbb{R}) \to \mathbbl{\Gamma} \backslash PSL_2(\mathbb{R})$ and the right translations $\mathbbl{\Gamma}g \mapsto \mathbbl{\Gamma}g\eta^+_t$ or $\mathbbl{\Gamma}g \mapsto \mathbbl{\Gamma}g\eta^-_t$.
\enumerationisfinis

\subsection{Continuous Function on the Unit Tangent Bundle with Compact Support}
\label{subsection "Continuous Function on the Unit Tangent Bundle with Compact Support"}

\enumerationisinitium
\item Let 
\begin{equation}
	\textcyrillic{\textit{м}} = 
	\begin{pmatrix}
	\alpha & \beta \\ 
	\gamma & \delta	
	\end{pmatrix} 
	\in \mathbbl{\Gamma}
\end{equation} 
be a matrix; the corresponding Möbius transformation is 
\begin{equation}
	\textcyrillic{\textit{м}}(z) = \frac{\alpha z + \beta}{\gamma z + \delta},
\end{equation}
with $\textcyrillic{\textit{м}} \in \mathbbl{\Gamma}$ acting on $z$ in $\mathbb{U}^2_\mathbb{C}$ by $\Moebius$-transformation, and $\mathbbl{\Gamma}$ here represents a discrete group of a hyperbolic $\Moebius$-transformation (see Section \ref{subsection "Orientation Preserving Isometries of the Hyperbolic Plane and Disk"}). It is noted that if $\textcyrillic{\textit{м}} \in \mathbbl{\Gamma}$ is a hyperbolic $\Moebius$-transformation preserving length minimizing paths, we obviously have a \emph{unique geodesic}, called the \emph{axis} of $\textcyrillic{\textit{м}}$. Let $d\bbmu = y^{-2}dxdy$ be an \emph{invariant measure} on $\mathbb{U}^2_\mathbb{C}$ under the action of $SL_2(\mathbb{R})$, see. Eq. \eqref{equation "Special linear group of degree 2 matrices over the real field"}. From the measure $d\zeta = d\bbmu(z)d\theta$ on $\mathring{\mathcal{T}}^1\mathbb{U}^2_\mathbb{C} \cong PSL_2(\mathbb{R})$, where $\theta \in [0, 2\pi[$, with $0 \leqslant \theta < 2\pi$, one obtains a measure $\tilde{\zeta}$ on $\mathring{\mathcal{T}}^1\mathcal{S}_\mathbbl{\Gamma} \cong \mathbbl{\Gamma} \backslash \mathring{\mathcal{T}}^1\mathbb{U}^2_\mathbb{C}$ fixed by
\begin{equation}
	\int_{\mathring{\mathcal{T}}^1\mathbb{U}^2_\mathbb{C}}\left(\varphi_\textcyrillic{\textit{м}} \in \mathscr{C}_\mathrm{c}(\mathring{\mathcal{T}}^1\mathbb{U}^2_\mathbb{C})\right)d\zeta = \int_{\mathbbl{\Gamma} \backslash \mathring{\mathcal{T}}^1\mathbb{U}^2_\mathbb{C}}\sum_{\textcyrillic{\textit{м}} \in \mathbbl{\Gamma}}\varphi_\textcyrillic{\textit{м}}\bigl(\textcyrillic{\textit{м}}(x)\bigr)d\tilde{\zeta}(x),
\end{equation} 
where $\varphi_\textcyrillic{\textit{м}}$ is a continuous function on $\mathring{\mathcal{T}}^1\mathbb{U}^2_\mathbb{C}$ of \emph{compact support} by $\mathscr{C}_\mathrm{c} \equival \{\varphi_\textcyrillic{\textit{м}} \in \mathscr{C}_\mathrm{c}(\mathring{\mathcal{T}}^1\mathbb{U}^2_\mathbb{C}) \mid \supp(\varphi_\textcyrillic{\textit{м}}) \text{ is compact}\}$, and $\mathscr{C}_\mathrm{c}$ is the space of $\varphi_\textcyrillic{\textit{м}}$, in the sense that the class $\mathscr{C}_\mathrm{c}$ of compactly supported continuous $\varphi_\textcyrillic{\textit{м}}$-functions is made into (or forms) a vector space over $\mathring{\mathcal{T}}^1\mathbb{U}^2_\mathbb{C}$. The support of $\varphi_\textcyrillic{\textit{м}}$, indicated with $\supp(\varphi_\textcyrillic{\textit{м}})$, is the closure of the set (or interval) on which the function is non-zero, $\varphi_\textcyrillic{\textit{м}} \neq 0$ or, equivalently, is the smallest closed set (or interval) outside which the function vanishes, $\varphi_\textcyrillic{\textit{м}} = 0$. The bundle $\mathring{\mathcal{T}}^1\mathcal{S}_\mathbbl{\Gamma} \cong \mathbbl{\Gamma} \backslash PSL_2(\mathbb{R})$ is found to be measurable with a \emph{finite measure}, and the associated geodesic flow establishes the structure of \emph{measure preserving (group) actions} on spaces of this type.
\item Let $z = (x + iy) \in \mathbb{U}^2_\mathbb{C}$. Let 
\begin{equation}
	\textcyrillic{\textit{м}} =
	\begin{pmatrix}
	\alpha & \beta \\ 
	\gamma & \delta	
	\end{pmatrix}
 	\in SL_2(\mathbb{R}), \enspace \text{for } \alpha, \beta, \gamma, \delta \in \mathbb{R}, \det(\textcyrillic{\textit{м}}) = 1, 
\end{equation}
be an element acting via Möbius (or linear fractional) transformation, $z \mapsto \textcyrillic{\textit{м}}(z) = (\alpha{z} + \beta)(\gamma{z} + \delta)^{-1}$ on $\mathbb{U}^2_\mathbb{C}$, such that $\textcyrillic{\textit{м}}(z) \in \mathbb{U}^2_\mathbb{C}$, for $\textcyrillic{\textit{м}} \in SL_2(\mathbb{R})$ and $z \in \mathbb{U}^2_\mathbb{C}$. It follows that
\begin{equation}
	\Im\bigl({\textcyrillic{\textit{м}}(z)}\bigr) = \frac{\Im(z)}{|\gamma{z} + \delta|^2},
\end{equation}
for which $\textcyrillic{\textit{м}}(\mathbb{U}^2_\mathbb{C})$ is a subset of $\mathbb{U}^2_\mathbb{C}$. Since the Jacobian \cite{Jacobi "De Determinantibus functionalibus"} of $\textcyrillic{\textit{м}}(z)$ reads $\det(\mathcal{J}_\textcyrillic{\textit{м}}) = \frac{1}{|\gamma{z} + \delta|^4}$, using $\Im\bigl({\textcyrillic{\textit{м}}(z)}\bigr)$ and making a change of variables (integration by substitution), we see that
\begin{equation}
	\int_{\mathbb{U}^2_\mathbb{C}}\varphi_\textcyrillic{\textit{м}}\bigl(\textcyrillic{\textit{м}}^{-1}(z)\bigr)\frac{dxdy}{y^2} = \int_{\mathbb{U}^2_\mathbb{C}}\varphi_\textcyrillic{\textit{м}}(z)\frac{1}{|\gamma{z} + d|^4}\frac{|\gamma{z} + \delta|^4}{y^2}dxdy = \int_{\mathbb{U}^2_\mathbb{C}}\varphi_\textcyrillic{\textit{м}}(z)\frac{dxdy}{y^2},
\end{equation}
for a compactly supported continuous function $\varphi_\textcyrillic{\textit{м}}$.
\enumerationisfinis

\section{Stable and Unstable Foliations with Leaves and Totally Geodesic Foliations}

\subsection{Leafy Stratifications for the Geodesic Flow by Tangent Vectors}
\label{subsection "Leafy Stratifications for the Geodesic Flow by Tangent Vectors"}

The geodesic and horocycle flows satisfy $\varphi_{t + s} = \varphi_t \circ \varphi_s$ and $\eta_{t + s} = \eta_t \circ \eta_s$, for any $t, s \in \mathbb{R}$, and they have commutation relations of the form
\begin{equation}
	\varphi_{-t}\eta^+_s\varphi_t = \eta^+_{se^{-t}}, \enspace \varphi_{-t}\eta^-_s\varphi_t = \eta^-_{se^t},
\end{equation}
and this is reflected in the fact that the orbits of the stable and unstable horocycle flows \eqref{equation "1-parameter groups for horocycle flow"} are (coincide with) the stable and unstable manifolds of the geodesic flow \eqref{equation "Diffeomorphism geodesic flow"}. 

\begin{exemplum}
\label{exemplum "Stable and unstable foliations, geodesic and horocycle flows"}
Denote by $\mathcal{W}^\mathrm{s}$ the stable ($^\mathrm{s}$) manifold and by $\mathcal{W}^\mathrm{u}$ the unstable ($^\mathrm{u}$) manifold. We shall say that $\mathcal{W}^\mathrm{s}$ and $\mathcal{W}^\mathrm{u}$ are the \emph{stable} and \emph{unstable foliations} for a geodesic flow $\varphi^\mathcal{W}_t$. Let $v \in \mathring{\mathcal{T}}^1\mathcal{S}_\mathbbl{\Gamma} \cong \mathbbl{\Gamma} \backslash PSL_2(\mathbb{R})$ be a vector, where $\mathcal{S}_\mathbbl{\Gamma} \equival \mathbbl{\Gamma} \backslash \mathbb{U}^2_\mathbb{C}$ (see above). Indicating by $\mathcal{W}^\mathrm{s}(v)$ the \emph{leaf} of $\mathcal{W}^\mathrm{s}$ and by $\mathcal{W}^\mathrm{u}(v)$ the \emph{leaf} of $\mathcal{W}^\mathrm{u}$, the stable and unstable leaves of $v$ for $\varphi^\mathcal{W}_t$ are defined by 
\begin{align}
\label{align "Stable and unstable manifolds for the geodesic foliation"}
	& \mathcal{W}^\mathrm{s}(v) = \left\{w \in \mathring{\mathcal{T}}^1\mathcal{S}_\mathbbl{\Gamma} \cong \mathbbl{\Gamma} \backslash PSL_2(\mathbb{R}) \mathrel{\Big|} \distance\left(\varphi^\mathcal{W}_t(v), \varphi^\mathcal{W}_t(w)\right) \xrightarrow[t \to +\infty]{}0\right\}, \\
	& \mathcal{W}^\mathrm{u}(v) = \left\{w \in \mathring{\mathcal{T}}^1\mathcal{S}_\mathbbl{\Gamma} \cong \mathbbl{\Gamma} \backslash PSL_2(\mathbb{R}) \mathrel{\Big|} \distance\left(\varphi^\mathcal{W}_t(v), \varphi^\mathcal{W}_t(w)\right) \xrightarrow[t \to -\infty]{}0\right\},
\end{align}
respectively, where $w = \dot{v}$ and $\distance = \dist$. The manifolds $\mathcal{W}^\mathrm{s}$ and $\mathcal{W}^\mathrm{u}$ of $v \in \mathring{\mathcal{T}}^1\mathcal{S}_\mathbbl{\Gamma}$ are
\enumerationisinitium
\item \emph{foliated with (smooth) leaves} or \emph{leafy stratifications} (of $\mathcal{W}$-stable and $\mathcal{W}$-unstable foliations) in $\mathring{\mathcal{T}}^1\mathcal{S}_\mathbbl{\Gamma}$, 
\item determined by the positive and negative horocycle flows: the positive $\eta$-flow is for the stable case,
\begin{equation} 
	\lim_{t \to +\infty}\Bigl\{\distance\bigl(\varphi_t(v), \varphi_t(w)\bigr)_\mathcal{W}\Bigr\} = 0,
\end{equation} 
the negative $\eta$-flow is for the unstable case,
\begin{equation} 
	\lim_{t \to -\infty}\Bigl\{\distance\bigl(\varphi_t(v), \varphi_t(w)\bigr)_\mathcal{W}\Bigr\} = 0,
\end{equation} 
\item are (coincide with) framed horocycles by inwardly and outwardly directed tangent vectors to geodesics centered at the points $\gamma_{\mathrm{c}(v)}(+\infty)$ and $\gamma_{\mathrm{c}(v)}(-\infty)$, respectively. In short, the $\mathcal{W}$-foliations consist of orbits of the horocycle flows.
\enumerationisfinis
Consider that each leaf is a $\mathscr{C}^1$ immersed surface, so the averages take constant values, and the foliations in $\mathring{\mathcal{T}}^1\mathcal{S}_\mathbbl{\Gamma}$ are \emph{invariant} under $\varphi^\mathcal{W}_t$, whether they are stable and unstable. \exemplumsymbol
\end{exemplum}

\subsection{Invariant and Geodesible Foliations}
\label{subsection "Invariant and Geodesible Foliations"}

We summarize some more or less recent results through a series of Scholia. First of all we say that 

· a \emph{folition} is a \emph{decomposition} of an $n$-manifold in terms of submanifolds of lower dimension, called \emph{leaves}; 

· a \emph{folition} is a concept made of several parts, for which it is comparable to a \emph{higher dimensional dynamical system}. 

\begin{scholium}[Totally geodesic foliation]
Let $\mathcal{F}$ denote a foliation on a Riemannian manifold $(\mathcal{M}, g)$, $g = g_\mathcal{M}$. Let $\mathring{\mathcal{T}}\mathcal{F}$ be a  tangent bundles to the leaves of $\mathcal{F}$, under which $\mathring{\mathcal{T}}\mathcal{F}$ could be thought of as a subbundle of $\mathring{\mathcal{T}}\mathcal{M}$. A noteworthy remark formalized by P. Tondeur \cite[Corollary 6.6, Theorem 10.6]{Tondeur "Foliations on Riemannian Manifolds"} is that 

· \emph{iff the induced metric $g_{(\mathring{\mathcal{T}}\mathcal{F})}$ along the leaves of $\mathcal{F}$ is invariant under the flow of vector fields orthogonal to $\mathcal{F}$, then $\mathcal{F}$ is a totally $\varphi^\mathcal{F}_t$-geodesic foliation}, 

· \emph{there is a $g$-metric such that $\mathcal{F}$ is a geodesible foliation, and the vector flow is a leaf preserving flow}. 

Note. Cf. \emph{Johnson–Whitt theorem} \cite[Theorem 1.6]{Johnson Whitt "Totally geodesic foliations"}, about a geometrization of a totally geodesic foliation in which $g_\mathcal{M}$ is a fiber-like metric for $\mathcal{F}$. In Ghys \cite{Ghys "Classification des feuilletages totalement geodesiques de codimension un"} can be found a classification of totally geodesic foliations of codimension 1 on a complete, not necessary compact, Riemannian manifold. \scholiumsymbol
\end{scholium}

\begin{scholium}[No totally geodesic foliations]
~\enumerationisinitium
\item F.G.B. Brito \cite[Théorèmes 4.2-3]{Brito "Une obstruction geometrique a l'existence de feuilletages de codimension 1 totalement geodesiques"}: \emph{there exist no totally geodesic $\mathscr{C}^\infty$ foliations of codimension 1 on a closed Riemannian manifold with non-zero (positive or negative) sectional curvature}. In this case, the obstruction to this type of existence also includes positively curved spaces.
\item P.G. Walczak \cite[Theorem 4]{Walczak "Dynamics of the geodesic flow of a foliation"}: \emph{given an oriented $\mathscr{C}^3$ foliation $\mathcal{F}$ on a compact Riemannian manifold of negative sectional curvature, then there is a positive number $\varepsilon > 0$ such that there exist no non-trivial foliations for which the second fundamental form of $\mathcal{F}$ and the norm of its covariant derivative are less than $\varepsilon$}.
\item E. Ghys \cite{Ghys "Dynamique des flots unipotents sur les espaces homogenes"}: \emph{there exist no totally geodesic $\mathscr{C}^0$ foliations of dimension $\geqslant 2$ on a compact hyperbolic space}. Generalizing, there are no totally geodesic foliations on the hyperbolic structure of a compact Riemann manifold.
\item A. Zeghib \cite[Théorème A]{Zeghib "Sur les feuilletages geodesiques continus des varietes hyperboliques"}: \emph{there exist no geodesic $\mathscr{C}^0$ foliations of dimension 1 on a closed hyperbolic 3-manifold with geodesic leaves}; see \cite{Zeghib "Feuilletages geodesiques appliques"} for $\varphi$-geodesic foliations of class $\mathscr{C}^1$. In \cite[Théorème B]{Zeghib "Feuilletages geodesiques des varietes localement symetriques"} it is proven that \emph{a locally Lipschitz foliation having totally geodesic leaves is not admitted on a complete negatively curved manifold of finite volume which carries a locally symmetric metric}.
\item P. Tondeur and L. Vanhecke \cite[Proposition 5.10]{Tondeur and Vanhecke "Jacobi fields Riccati equation and Riemannian foliations"}, see also \cite[Proposition 5.91]{Tondeur "Geometry of Foliations"}: \emph{given a Riemannian foliation $\mathcal{F}$ with bundle-like metric $g_\mathcal{M}$ on $(\mathcal{M}, g)$, and a Ricci curvature tensor $\Ric_{(\mathring{\mathcal{T}}\mathcal{F})} < 0$ at least one point of $\mathcal{M}$, for a tangent bundles $\mathring{\mathcal{T}}\mathcal{F} \subset \mathring{\mathcal{T}}\mathcal{M}$ to the leaves of $\mathcal{F}$, then there are no totally $\varphi^\mathcal{F}_t$-geodesic foliations on $(\mathcal{M}, g)$}. For the demonstration it is advantageous to make use of the notion of \emph{partial Ricci curvature} \cite{Kim and Tondeur "Riemannian foliations on manifolds with non-negative curvature"} \cite{Rovenski "On solutions to equations with partial Ricci curvature"}. \scholiumsymbol
\enumerationisfinis
\end{scholium}

\begin{scholium}[Totally geodesic foliations of hyperbolic space, and totally geodesic non-smooth foliations of codimension 1]
~\enumerationisinitium
\item Proofs of existence of totally geodesic foliations of hyperbolic space $\hyperbolic^n$ relate to hypersurfaces that are orthogonal to a geodesic curve. Proceeds in this way e.g. the following theorem by D. Ferus \cite[cf. Theorems 1, 3-4]{Ferus "On Isometrie Immersions between Hyperbolic Spaces"}. \emph{Let $\iota \colon \hyperbolic^n \to \hyperbolic^{n + 1}$ denote an isometric immersion free of umbilics (points on the surface where the normal curvature is equal in all directions) between hyperbolic spaces, and suppose $\iota$ is a $\mathscr{C}^\infty$ (smooth) map; let $\gamma_\mathrm{c} \colon I \subset \mathbb{R} \to \hyperbolic^n$ be a (constant speed) geodesic of curvature $\leqslant 1$. Then every (complete) totally geodesic hypersurface of $\hyperbolic^n$ orthogonal to $\gamma_\mathrm{c}$ forms a totally geodesic foliation $\mathcal{F}_\iota$ of $\hyperbolic^n$. This foliation is known as \emph{nullity foliation} of $\iota$}.
\item A classification of totally geodesic non-smooth foliations of codimension 1 of $\hyperbolic^n$ is drafted by H. Browne \cite[Theorem 5.5]{Browne "Codimension one totally geodesic foliations of $H^n$"}. \emph{Let $\gamma_\mathrm{c} \colon I \subset \mathbb{R} \to \hyperbolic^n$ be a geodesic curve, where $I$ is an interval, possibly infinite; let $\rotatedPsi$ be a unit-length vector field along $\gamma_\mathrm{c}$ such that $\langle\rotatedPsi, \dot{\gamma}_\mathrm{c}\rangle > 0$ and 
\begin{equation}
	\limsup_{\alpha \to \beta}\frac{\distance_{\gamma_\mathrm{c}}\Bigl(\rotatedPsi(\alpha), \rotatedPsi(\beta)\Bigr)}{|\beta - \alpha|} \leqslant \langle\rotatedPsi(\beta), \dot{\gamma}_\mathrm{c}(\beta)\rangle,
\end{equation}
for any $\beta \in I \subset \mathbb{R}$, where $\distance$ is the distance for which $\rotatedPsi$ is parallel by maintaining a constant angle with $\dot{\gamma}_\mathrm{c}$. Then the hypersurfaces orthogonal to $\rotatedPsi$ extend to a unique totally geodesic foliation of $\hyperbolic^n$}. \scholiumsymbol
\enumerationisfinis
\end{scholium}

\section{Anosov Diffeomorphism and Flow}

\begingroup
\footnotesize
[A] geodesic flow on a closed Riemannian manifold of negative curvature satisfies a certain “condition \textcyrillic{(\textit{У})}” [\,\dots] a dynamical system satisfies the condition \textcyrillic{(\textit{У})} if near an arbitrary fixed trajectory the behavior of the neighboring trajectories with respect to the fixed one is similar to the behavior of the trajectories close to a saddle. \\
\indent — \textsc{D.V. Anosov} \cite[p. 1]{Anosov "Geodesic flows on closed Riemannian manifolds of negative curvature"} (p. 3. in the Ru. version)

\endgroup

\subsection{Diffeomorphism and Flow on Negatively Curved Surfaces}
\label{subsection "Diffeomorphism and Flow on Negatively Curved Surfaces"}

In general, the geodesic flow on (the tangent bundle of) a compact Riemannian manifold of \emph{negative curvature} is called an \emph{Anosov flow} \cite{Anosov "Roughness of geodesic flows on compact Riemannian manifolds of negative curvature"} \cite{Anosov "Ergodic properties of geodesic flows on closed Riemannian manifolds of negative curvature"} \cite{Anosov "Geodesic flows on closed Riemannian manifolds of negative curvature"} of which \eqref{equation "Geodesic flow"} is part; see P.B. Eberlein \cite{Eberlein "When is a geodesic flow of Anosov type? I"} and S. Smale \cite[II.3]{Smale "Differentiable Dynamical Systems"}. Nevertheless, there are also other examples of Anosov flows. One may start with a propaedeutic description of an \emph{Anosov system}, forming \emph{part of non-trivial and complex dynamical systems}; see the works of D.V. Anosov in collaboration with Ya.G. Sinai \cite{Anosov and Sinai "Some smooth ergodic system"} and V.V. Solodov \cite[pp. 70-84]{Anosov Solodov "Hyperbolic Sets"}.

\begin{definitio}[Anosov diffeomorphism]
\label{definitio "Anosov diffeomorphism"}
Let $\varphi \colon \mathcal{M} \to \mathcal{M}$ be a diffeomorphism of a compact Riemannian manifold. The map $\varphi \viz \varphi[\mathscr{A}]$ is an \emph{Anosov diffeomorphism}, imagining e.g. a $\mathscr{C}^1$ Anosov system, so $\varphi \in \Diff^1(\mathcal{M})$,
\enumerationisinitium
\item if there is a direct sum decomposition $\mathcal{T}_x\mathcal{M} \equival \mathcal{E}^\mathrm{s}(x) \oplus \mathcal{E}^\mathrm{u}(x)$, i.e. a splitting of the tangent space into two $d\varphi$-invariant \emph{distributions}, or rather, \emph{subspaces}, the stable ($^\mathrm{s}$) subspace $\mathcal{E}^\mathrm{s}(x)$ and the unstable ($^\mathrm{u}$) subspace $\mathcal{E}^\mathrm{u}(x)$, at each point $x \in \mathcal{M}$,
\item if there exist some constants $c > 0$ and $0 < \lambda < 1 < \epsilon$  such that 
\begin{subequations}
\begin{align}
	& d\varphi\mathcal{E}^\mathrm{s}(x) = \mathcal{E}^\mathrm{s}\bigl(\varphi(x)\bigr) \text{ and } d\varphi\mathcal{E}^\mathrm{u}(x) = \mathcal{E}^\mathrm{u}\bigl(\varphi(x)\bigr), \\
	& \|d\varphi^n(v)\| \leqslant c\lambda^n\|v\|, \text{ for all } v \in \mathcal{E}^\mathrm{s}(x) \text{ and } n \in \mathbb{N} \geqslant 0, \\ 
	& \|d\varphi^{-n}(v)\| \leqslant c\epsilon^{-n}\|v\|, \text{ for all } v \in \mathcal{E}^\mathrm{u}(x) \text{ and } n \in \mathbb{N} \geqslant 0,
\end{align}
\end{subequations}
for which $\varphi$ is \emph{hyperbolic} and the intersection $\mathcal{E}^\mathrm{s}(x) \cap \mathcal{E}^\mathrm{u}(x) = 0$ is \emph{transversal}, where $\mathcal{E}^\mathrm{s}$ is \emph{uniformly contracted} and $\mathcal{E}^\mathrm{u}$ \emph{uniformly expanded} by $d\varphi$, and $\|\cdot\|$ is the norm induced on $\mathcal{T}_x\mathcal{M}$ by the Riemannian metric. \definitiosymbol
\enumerationisfinis
\end{definitio}

In conclusion, $\mathcal{M} \xrightarrow{\varphi[\mathscr{A}]} \mathcal{M}$ is of Anosov type if $\varphi[\mathscr{A}]$ is hyperbolic on the whole of $\mathcal{M}$.

\begin{exemplum}[Hyperbolic toral automorphism]
\label{exemplum "Hyperbolic toral automorphism"}
The hyperbolic automorphism of the $n$-dimensional torus is a basic model of Anosov diffeomorphism \cite[pp. 7-12]{Anosov "Geodesic flows on closed Riemannian manifolds of negative curvature"} \cite[pp. 140-141]{Anosov and Sinai "Some smooth ergodic system"} in which the map $\varphi[\mathscr{A}] \colon \torus^n \to \torus^n$ is a space preserving (Anosov) diffeomorphism of $\torus^n$, such that $|\lambda_1|\leqslant \cdots \leqslant|\lambda_\rotatedell|< 1 <|\lambda_{\rotatedell + 1}|\leqslant \cdots \leqslant|\lambda_n|$, with $1 \leqslant \rotatedell < n$. A Franks–Newhouse theorem \cite{Franks "Anosov Diffeomorphisms on Tori"} \cite{Newhouse "On Codimension One Anosov Diffeomorphisms"} states that \emph{an Anosov diffeomorphism $\varphi[\mathscr{A}]$ of codimension 1, for which $\dim\bigl(\mathcal{E}^\mathrm{s}_x\bigr) = 1$ or $\dim\bigl(\mathcal{E}^\mathrm{u}_x\bigr) = 1$, is topologically conjugate to a hyperbolic toral automorphism, so the manifold equipped with this diffeomorphism is homeomorphic to $\torus^n \cong \mathbb{S}^1 \times \cdots \times \mathbb{S}^1$}. \exemplumsymbol
\end{exemplum}

\begin{scholium}[$\mathscr{C}^r$ Anosov diffeomorphism]
It is understood that, by choosing a smooth compact connected Riemannian $\mathscr{C}^\infty$ manifold, the Anosov diffeomorphism $\mathcal{M} \xrightarrow{\varphi} \mathcal{M}$ is a $\mathscr{C}^r$ diffeomorphism, with $1 \leqslant r \leqslant \infty$, so that $\varphi[\mathscr{A}] \in \Diff^r(\mathcal{M})$ is $\mathscr{C}^{r \geqslant 1}$ differentiable. \scholiumsymbol
\end{scholium}

Similarly to the continuous ($\mathcal{T}\varphi$)-invariant splitting of $\mathcal{T}_x\mathcal{M}$ for the Anosov map, we are able to express the Anosov flow, using three subspaces rather than two.

\begin{definitio}[Anosov flow]
\label{definitio "Anosov flow"}
Let us exemplify the case of a hyperbolic $\mathscr{C}^r$ flow, e.g. with $r = 2$. More specifically, we shall use the class $\mathscr{C}^2$ to prove the Hölder continuity of subspaces (Theorem \ref{theorema "Hölder continuity of subspaces in Anosov map"}) and the ergodicity (Theorem \ref{theorema "Anosov's Ergodic Theorem"}) in maps with the Anosov property . Given a smooth compact Riemannian manifold $\mathcal{M}$ and a vector field $\vec{X}$, consider the \emph{non-singular smooth $\mathscr{C}^r$ flow} $\varphi_t \colon \mathcal{M} \to \mathcal{M}$ such that
\begin{equation}
	\vec{X}(t) = \frac{d}{dt}\bigl(\varphi_t(x)\bigr)\big|_{t = 0},
\end{equation}  
having continuous time $t \in \mathbb{R}$. The flow $\varphi_t \viz \varphi_t[\mathscr{A}]$ is said to be an \emph{Anosov flow}
\enumerationisinitium
\item if there is a decomposition of the tangent space into three $d\varphi_t$-invariant subspaces,
\begin{equation}
	\mathcal{T}_x\mathcal{M} \equival \mathcal{E}^\mathrm{s}(x) \oplus \mathcal{E}^0(x) \oplus \mathcal{E}^\mathrm{u}(x),
\end{equation}
i.e. if $\mathcal{T}_x\mathcal{M}$ splits into a direct sum of the stable subspace $\mathcal{E}^\mathrm{s}(x)$, the 1-dimensional subspace $\mathcal{E}^0(x)$ spanned by $\vec{X}(t)$, and the unstable subspace $\mathcal{E}^\mathrm{u}(x)$, at each point $x \in \mathcal{M}$,
\item if there exist some constants $c > 0$ and $0 < \lambda < 1 < \epsilon$ such that 
\begin{subequations}
\begin{align}
	& d_x\varphi_t\mathcal{E}^\mathrm{s}(x) = \mathcal{E}^\mathrm{s}\bigl(\varphi_t(x)\bigr) \text{ and } d_x\varphi_t\mathcal{E}^\mathrm{u}(x) = \mathcal{E}^\mathrm{u}\bigl(\varphi_t(x)\bigr), \\
	& \|d_x\varphi_t(v)\| \leqslant c\lambda^t\|v\|, \text{ for all } v \in \mathcal{E}^\mathrm{s}(x) \text{ and } t \in \mathbb{R} \geqslant 0, \\
	& \|d_x\varphi_{-t}(v)\| \leqslant c\epsilon^{-t}\|v\|, \text{ for all } v \in \mathcal{E}^\mathrm{u}(x) \text{ and } t \in \mathbb{R} \geqslant 0,
\end{align}
\end{subequations}
where $\|\cdot\|$ is again a Riemannian metric on $\mathcal{T}_x\mathcal{M}$. \definitiosymbol 
\enumerationisfinis
\end{definitio}

An Anosov flow  $\varphi_t \viz \varphi_t[\mathscr{A}]$ is topologically \emph{transitive} if it contains a dense orbit in $\mathcal{M}$, i.e. if there exists the condition that $\Omega_\mathcal{N} \cap \varphi_t \Omega_\mathcal{O} \neq \varnothing$ of $\mathcal{M}$, for all non-empty open subsets $\Omega_\mathcal{N}, \Omega_\mathcal{O} \subset \mathcal{M}$.

\begin{scholium}
\label{scholium "Tangent bundle in the Anosov system"}
When speaking of tangent bundle $\mathring{\mathcal{T}}\mathcal{M}$ instead of tangent space, we refer to the Anosov system through a ($\mathring{\mathcal{T}}\varphi_t$)-invariant splitting of $\mathring{\mathcal{T}}\mathcal{M}$ into a direct sum of flow invariant \emph{subbundles}, of course, i.e. 
\begin{equation}
	\mathring{\mathcal{T}}\mathcal{M} \equival \mathring{\mathcal{E}}^\mathrm{s} \oplus \mathring{\mathcal{E}}^0 \oplus \mathring{\mathcal{E}}^\mathrm{u}.
\end{equation}
The subbundles $\mathring{\mathcal{E}}^\mathrm{s}$, $\mathring{\mathcal{E}}^\mathrm{u}$, $\mathring{\mathcal{E}}^\mathrm{s} \oplus \mathring{\mathcal{E}}^0$ and $\mathring{\mathcal{E}}^\mathrm{u} \oplus \mathring{\mathcal{E}}^0$ are uniquely integrable. The Anosov foliations (manifolds) that occurred in these integral spaces are \emph{strongly stable} ($^\mathrm{ss}$), \emph{strongly unstable} ($^\mathrm{su}$), \emph{weakly stable} ($^\mathrm{ws}$) and \emph{weakly unstable} ($^\mathrm{wu}$), denoted by $\mathcal{W}^\mathrm{ss}$, $\mathcal{W}^\mathrm{su}$, $\mathcal{W}^\mathrm{ws}$ and $\mathcal{W}^\text{wu}$, respectively. The stable and unstable leaves of an Anosov-like foliation can be written as
\begin{equation}
	\mathcal{W}^\mathrm{s}(x) = \bigcup_{t \in \mathbb{R}}\mathcal{W}^\mathrm{s}\Bigl(\varphi^\mathcal{W}_t \viz \varphi^\mathcal{W}_t[\mathscr{A}](x)\Bigr), \enspace \mathcal{W}^\mathrm{u}(x) = \bigcup_{t \in \mathbb{R}}\mathcal{W}^\mathrm{u}\Bigl(\varphi^\mathcal{W}_t \viz \varphi^\mathcal{W}_t[\mathscr{A}](x)\Bigr),
\end{equation}
cf. Eqq. \eqref{align "Stable and unstable manifolds for the geodesic foliation"}. \scholiumsymbol
\end{scholium}

\subsection{Hölder Continuity of Subspaces in a Map with the Anosov Property}
\label{subsection "Hölder Continuity of Subspaces in a Map with the Anosov Property"}

The stable and unstable subspaces $\mathcal{E}^\mathrm{s}(x)$ and $\mathcal{E}^\mathrm{u}(x)$ satisfy the \emph{Hölder condition}, for which they are \emph{not always smooth} in the analytic category. This statement is a theorem, and it is proven by Anosov \cite{Anosov "Tangent fields of transversal foliations in U-systems"}, showing that tangential fields in term of transversal contractible and extensible foliations in $\mathscr{C}^2$ Anosov systems are \emph{Hölder continuous}.
	
We will follow the presentation of M. Brin's \cite{Brin "[Appendix A.] Holder Continuity of Invariant Distributions"} proof. (The core of Brin's argument is already present, in broad terms, in one of his paper in collaboration with Yu. Kifer \cite[§ 5. App.]{Brin and Kifer "Dynamics of Markov chains and stable manifolds for random diffeomorphisms"} on the dynamics of Markov chains). Firstly, we remind the meaning of Hölder continuity in this context; after which it will be necessary to introduce two Lemmas \ref{lemma "First lemma"} and \ref{lemma "Second lemma"}, and then move on to the theorem.

\begin{definitio}[Hölder condition]
\label{definitio "Hölder condition}
~\enumerationisinitium
\item Let $(\mathcal{M}, g)$ be a (smooth) Riemannian manifold. We assume that $\mathcal{M} \hookrightarrow \mathbb{R}^n$, an embedding of $\mathcal{M}$ in the Euclidean space $\mathbb{R}^n$, for all sufficiently large $n$. Let $\mathcal{E}$ be a $k$-dimensional \emph{distribution} on a subset $\mathcal{N}$ of $\mathcal{M}$. The distribution $\mathcal{E}$ forms a set of $k$-dimensional subspaces $\mathcal{E}(x)$ in $\mathcal{T}_x\mathcal{M}$, at each point $x \in \mathcal{N}$. Distances in the tangent space $\mathcal{T}_x\mathcal{M}$, and those in the space of $k$-dimensional subspaces in $\mathcal{T}_x\mathcal{M}$, are the induced distances from the Riemannian metric $g$ on $\mathcal{M}$. If we suppose that $\mathcal{M}$ is compact, then $g$ is equal to the Euclidean distance $\|x - y\|$ in the embedded space.
\item Taking a subspace $\Omega_\xi \subset \mathbb{R}^n$ and a vector $v \in \mathbb{R}^n$, we have a distance $\distance(v, \Omega_\xi)$, that is the length of the difference between $v$ and its orthogonal projection $w = \dot{v}$ onto $\Omega_\xi$,
\begin{equation}
	\distance(v, \Omega_\xi) = \min_{w \in \Omega_\xi} \| v - w\|.
\end{equation}
Given two subspaces $\Omega_\xi, \Omega_\varpi \subset \mathbb{R}^n$, we write
\begin{equation}
	\distance(\Omega_\xi, \Omega_\varpi) = \max\left\{\max_{\substack{v \in \Omega_\xi \\ \|v\| = 1}} \distance(v, \Omega_\varpi), \max_{\substack{w \in \Omega_\varpi \\ \|w\| = 1}} \distance(w, \Omega_\xi)\right\}.
\end{equation}
\enumerationisfinis
Let $\alpha \in (0, 1]$ be the Hölder exponent. A distribution $\mathcal{E}$ of dimension $k$ on $\mathcal{N}_{\mathbb{R}^n}$ is said to be \emph{Hölder continuous} if there exists a value $\varepsilon_0 > 0$ such that
\begin{equation}
	\distance\Bigl(\mathcal{E}(x), \mathcal{E}(y)\Bigr) \leqslant c_\textsc{h}\|x - y\|^\alpha,
\end{equation}
for a positive Hölder constant $c_\textsc{h} > 0$ and any $x, y \in \mathcal{N}$, with $\|x - y\| \leqslant \varepsilon_0$. \definitiosymbol
\end{definitio}

\begin{lemma}
\label{lemma "First lemma"}
Consider two sequences $\Omega^k_\xi$ and $\Omega^k_\varpi$ of $n \times n$ real matrices, with $k = \mathbb{Z}_*$, namely $k = 0, 1, 2, \mathellipsis$, such that $\|\Omega^k_\xi - \Omega^k_\varpi\| \leqslant \mathrm{\Delta}\mu^k$, for fixed $\mathrm{\Delta} \in (0, 1)$ and $\mu > 1$. If there exist subspaces $\mathcal{E}_{\Omega_\xi}, \mathcal{E}_{\Omega_\varpi} \subset \mathbb{R}^n$, and values $0 < \lambda < \epsilon$ and $\zeta > 1$, with $\lambda < \mu$, such that
\begin{subequations}
\begin{align}
	& \|\Omega^k_\xi(v)\| \leqslant \zeta\lambda^k\|v\|, \text{ if } v \in \mathcal{E}_{\Omega_\xi}, \\ 
	& \|\Omega^k_\xi(w)\| \geqslant \zeta^{-1}\epsilon^k\|w\|, \text{ if } w \in (\mathcal{E}_{\Omega_\xi})^\perp \viz w \perp \mathcal{E}_{\Omega_\xi}, \\
	& \|\Omega^k_\varpi(v)\| \leqslant \zeta\lambda^k\|v\|, \text{ if } v \in \mathcal{E}_{\Omega_\varpi}, \\ 
	& \|\Omega^k_\varpi(w)\| \geqslant \zeta^{-1}\epsilon^k\|w\|, \text{ if } w \in (\mathcal{E}_{\Omega_\varpi})^\perp \viz w \perp \mathcal{E}_{\Omega_\varpi},
\end{align}
\end{subequations}
then
\begin{equation}
	\distance\left(\mathcal{E}_{\Omega_\xi}, \mathcal{E}_{\Omega_\varpi}\right) \leqslant 3\zeta^2\frac{\epsilon}{\lambda}\mathrm{\Delta}^{\frac{\log{\epsilon} - \log{\lambda}}{\log{\mu} - \log{\lambda}}}.
\end{equation}
\end{lemma}

\begin{proof}
Let us set
\begin{align}
	& \Lambda^k_{\Omega_\xi} = \Bigl\{v \in \mathbb{R}^n \mathrel{\big|} \|\Omega^k_\xi(v)\| \leqslant 2\zeta\lambda^k\|v\|\Bigr\}, \\
	& \Lambda^k_{\Omega_\varpi} = \Bigl\{v \in \mathbb{R}^n \mathrel{\big|} \|\Omega^k_\varpi(v)\| \leqslant 2\zeta\lambda^k\|v\|\Bigr\},
\end{align}
and write $v = v_i + v_j$, where $v_i = v^\lambda \in \mathcal{E}_{\Omega_\xi}$ and $v_j = v^\perp \in (\mathcal{E}_{\Omega_\xi})^\perp \viz v^\perp \perp \mathcal{E}_{\Omega_\xi}$, for every $v \in \mathbb{R}^n$. If $v \in \Lambda^k_{\Omega_\xi}$, one sees that
\enumerationisinitium
\item $\|\Omega^k_\xi(v)\| = \|\Omega^k_\xi(v_i + v_j)\| \geqslant \|\Omega^k_\xi(v_j)\| - \|\Omega^k_\xi(v_i)\| \geqslant \zeta^{-1}\epsilon^k\|v_j\| - \zeta\lambda^k\|v_i\|$,
\item $\|v_j\| \leqslant \zeta\epsilon^{-k}(\|\Omega^k_\xi(v)\| + \zeta\lambda^k\|v_i\|) \leqslant 3\zeta^2\left(\frac{\lambda}{\epsilon}\right)^k\|v\|$,
\enumerationisfinis
and hence $\distance(v, \mathcal{E}_{\Omega_\xi}) \leqslant 3\zeta^2\left(\frac{\lambda}{\epsilon}\right)^k\|v\|$. A fixed $\delta = \frac{\lambda}{\mu} < 1$ determines a unique non-negative integer $k$ so that $\delta^{k + 1} < \mathrm{\Delta} \leqslant \delta^k$. If $w \in \mathcal{E}_{\Omega_\varpi}$, then 
\begin{align}
	\|\Omega^k_\xi(w)\| & \leqslant \|\Omega^k_\varpi(w)\| + \|\Omega^k_\xi - \Omega^k_\varpi\| \cdot \|w\| \notag \\
	& \leqslant \zeta\lambda^k\|w\| + \mathrm{\Delta}\mu^k\|w\| \leqslant \Bigl(\zeta\lambda^k + (\delta\mu)^k\Bigr)\|w\| \leqslant 2\zeta\lambda^k\|w\|.
\end{align} 
Therefore $w \in \Lambda^k_{\Omega_\xi}$, plus $\Lambda^k_{\Omega_\xi}$ includes $\mathcal{E}_{\Omega_\varpi}$ and, inversely, $\Lambda^k_{\Omega_\varpi}$ includes $\mathcal{E}_{\Omega_\xi}$. We finally get to 
\begin{equation}
	\distance\left(\mathcal{E}_{\Omega_\xi}, \mathcal{E}_{\Omega_\varpi}\right) \leqslant 3\zeta^2\left(\frac{\lambda}{\epsilon}\right)^k \leqslant 3\zeta^2\frac{\epsilon}{\lambda}\mathrm{\Delta}^{\frac{\log{\epsilon} - \log{\lambda}}{\log{\mu} - \log{\lambda}}}.
\end{equation}
\end{proof}

\begin{scholium}
Note that if $\|v_i - v_j\| \geqslant \theta$, then for a value $\theta > 0$ there exist subspaces $\mathcal{E}_i$ and $\mathcal{E}_j$ which we call \emph{$\theta$-transverse}, for all unit vectors $v_i \in \mathcal{E}_i$ and $v_j \in \mathcal{E}_j$. \scholiumsymbol
\end{scholium}

\begin{lemma}
\label{lemma "Second lemma"}
Let $\varphi \colon \mathcal{M} \to \mathcal{M}$ denote a $\mathscr{C}^{1 + \beta}$ map of a compact $m$-dimensional $\mathscr{C}^2$ manifold $\mathcal{M} \subset \mathbb{R}^n$. For each
\begin{equation}
	\mu > \left(\max_{z \in \mathcal{M}} \|d_z\varphi\|\right)^{1 + \beta}
\end{equation}
there is $\tau_i > 1$ such that $\|d_x\varphi^n - d_y\varphi^n\| \leqslant \tau_i\mu^n\|x - y\|^\beta$, for any $n \in \mathbb{N}$ and $x, y \in \mathcal{M}$.
\end{lemma}
\begin{proof}
We should impose $\tau_j$ so that $\|d_x\varphi - d_y\varphi\| \leqslant \tau_j\|x - y\|^\beta$. Let
\enumerationisinitium
\item $\nu = \max_{z \in \mathcal{M}}\|d_z\varphi\| \geqslant 1$,
\item $\|\varphi^n(x) - \varphi^n(y)\| \leqslant \nu^n\|x - y\|$, for any $x, y \in \mathcal{M}$,
\item $\mu > \nu$. 
\enumerationisfinis
It ensues that the lemma is true for $n = 1$ and $\tau_i \geqslant \tau_j$. Then
\begin{align}
	\|d_x\varphi^{n + 1} - d_y\varphi^{n + 1}\| & \leqslant \|d_{\varphi^n(x)}\varphi\| \cdot \|d_x\varphi^n - d_y\varphi^n\| \notag \\
	& + \|d_{\varphi^n(x)}\varphi - d_{\varphi^n(y)}\varphi\| \cdot \|d_y\varphi^n\| \notag \\
	& \leqslant \nu{\tau_i}\mu^n\|x - y\|^\beta + \tau_j\bigl(\nu^n\|x - y\|\bigr)^\beta\nu^n \notag \\
	& \leqslant \tau_i\mu^{n + 1}\|x - y\|^\beta\left(\frac{\nu}{\mu} + \frac{\tau_j}{\tau_i}\frac{(\nu^{1 + \beta})^n}{\mu^{n + 1}}\right).
\end{align}
If $\mu > \nu^{1 + \beta}$, there exists $\tau_i \geqslant \tau_j$ and the factor in parentheses is always smaller than 1. 
\end{proof}
 
\begin{theorema}
\label{theorema "Hölder continuity of subspaces in Anosov map"}
Let $\mathcal{M}$ be a compact $m$-dimensional $\mathscr{C}^2$ manifold in $\mathbb{R}^n$, with $m < n$, and $\varphi \colon \mathcal{M} \to \mathcal{M}$ a $\mathscr{C}^{1 + \beta}$ map, with $\beta \in (0,1)$. Suppose that there exist a subset $\mathcal{N} of \mathcal{M}$, real numbers $0 < \lambda < \epsilon$, $c > 0$, and $\theta > 0$, so that there are $\theta$-transverse subspaces $\mathcal{E}^\mathrm{s}(x), \mathcal{E}^\mathrm{u}(x)$ in $\mathcal{T}_x\mathcal{M}$, at each $x \in \mathcal{N}$, and the following properties hold:
\enumerationisinitium
\item $\mathcal{T}_x\mathcal{M} \equival \mathcal{E}^\mathrm{s}(x) \oplus \mathcal{E}^\mathrm{u}(x)$, 
\item $\|d_x\varphi^k(v^\mathrm{s})\| \leqslant c\lambda^k\|v^\mathrm{s}\|$ and $\|d_x\varphi^k(v^\mathrm{u})\| \geqslant c^{-1}\epsilon^k\|v^\mathrm{u}\|$, for all $v^\mathrm{s} \in \mathcal{E}^\mathrm{s}(x)$ and $v^\mathrm{u} \in \mathcal{E}^\mathrm{u}(x)$, and every $\mathbb{Z}_*$.
\enumerationisfinis
Then the distribution $\mathcal{E}$ is $\alpha$-Hölder continuous, with 
\begin{equation}
	\alpha = \frac{\log{\epsilon} - \log{\lambda}}{\log{\mu} - \log{\lambda}}\beta, 
\end{equation}
for each $\mu > (\max_{z \in \mathcal{M}} \|d_z\varphi\|)^{1 + \beta}$.
\end{theorema}

\begin{proof}
Let $\mathcal{E}(x)^\perp$ be the orthogonal complement to the tangent plane $\mathcal{T}_x\mathcal{M}$ in $\mathbb{R}^n$, for $x \in \mathcal{M}$. Given that $\mathcal{E}^\mathrm{s}(x)$ and $\mathcal{E}^\mathrm{u}(x)$ are $\theta$-transverse and moreover of complementary dimensions in $\mathcal{T}_x\mathcal{M}$, there is $d > 1$ such that $\|d_x\varphi^k(w)\| \geqslant d^{-1}\epsilon^k\|w\|$ at each $x \in \mathcal{N}$ and for every $w \perp \mathcal{E}(x)$. Let $\Omega^k_\xi$ and $\Omega^k_\varpi$ be two sequences of $n \times n$ real matrices, with $k = \mathbb{Z}_*$, for any $x, y \in \mathcal{N}$, such that 
\enumerationisinitium
\item $\Omega^k_\xi(v) = d_x\varphi^k(v)$, if $v \in \mathcal{T}_x\mathcal{M}$, and $\Omega^k_\xi(w) = 0$, if $w \in (\mathcal{T}_x\mathcal{M})^\perp \viz w \perp \mathcal{T}_x\mathcal{M}$, 
\item $\Omega^k_\varpi(v) = d_y\varphi^k(v)$, if $v \in \mathcal{T}_y\mathcal{M}$, and $\Omega^k_\varpi(w) = 0$, if $w \in (\mathcal{T}_y\mathcal{M})^\perp \viz w \perp \mathcal{T}_y\mathcal{M}$. The Lemma \ref{lemma "Second lemma"} tell us that $\|\Omega^k_\xi - \Omega^k_\varpi|\ \leqslant \tau_i\mu^k\|x - y\|^\beta$ (by replacing the natural dimensionality with a non-negative integer).
\enumerationisfinis

Observe the case with the stable distribution. The theorem is true since it is related to Lemma \ref{lemma "First lemma"} by putting $\widetilde{\mathcal{E}}^\mathrm{s} = \mathcal{E}^\mathrm{s} \oplus \mathcal{E}^\perp$, with $\mathrm{\Delta}(x) = \tau_i\|x - y\|^\beta$, $\mathcal{E}_{\Omega_\xi} = \widetilde{\mathcal{E}}^\mathrm{s}(x)$, $\mathcal{E}_{\Omega_\varpi} = \widetilde{\mathcal{E}}^\mathrm{s}(y)$, and $\zeta = \max(c, d)$. By reversing the time we get the result inherent in the continuous Hölderianity of the unstable distribution.
\end{proof}

Essentially, through the Theorem \ref{theorema "Hölder continuity of subspaces in Anosov map"} it can be stated that the \emph{stable and unstable subspaces} $\mathcal{E}^\mathrm{s}(x)$ and $\mathcal{E}^\mathrm{u}(x)$ of an Anosov diffeomorphism $\varphi[\mathscr{A}]$ \emph{depend $\alpha$-Hölder continuously} on the point $x$ in the manifold $\mathcal{M}$. These subspace distributions may also be non-smooth. See B. Hasselblatt \cite{Hasselblatt "Regularity of the Anosov splitting and of horospheric foliations"}. 

\subsection{Structural Stability: Andronov–Pontrjagin Criterion} 
\label{subsection "Structural Stability: Andronov–Pontrjagin Criterion"}

\begingroup
\footnotesize
[The] definition of the coarseness [\textcyrillic{\textit{грубости}}]\footnote{
	Systems that are structurally stable are denominated as \emph{coarse systems}.
	} 
of a system can be considered as the determination of the stability [\textcyrillic{\textit{устойчивости}}] of a set of orbits of a dynamic system with respect to sufficiently small perturbations in the right-hand sides of equations [such as] $\frac{dx}{dt} = P(x, y); \frac{dy}{dt} = Q(x, y)$. \\
\indent — \textsc{A. Andronov and L. Pontrjagin} \cite[p. 248]{Andronov and Pontrjagin "Coarse systems"}

\vspace{2mm}

The idea of [structural stability] is that qualitative properties of structurally stable diffeomorphisms [or flows] are unchanged by small $\mathscr{C}^1$ perturbations. \\
\indent — \textsc{J.W. Robbin} \cite[p. 447]{Robbin "A Structural Stability Theorem"}

\endgroup

\vspace{2mm}

An Anosov system is called \emph{structurally stable}, or a \emph{coarse system}, whether it be a diffeomorphism or a flow, by applying the \emph{Andronov–Pontrjagin criterion} \cite{Andronov and Pontrjagin "Coarse systems"} in the $\mathscr{C}^1$ topology. Simply stated, the criterion establishes a condition of \emph{equivalence}, claiming that \emph{small perturbations} of a hyperbolic (Anosov) dynamical system from the initial state \emph{do not alter} the future behavior of its orbits. Therefore, a dynamical system is structurally stable if the qualitative properties that characterize the system remain the same after some form of perturbation.

\begin{exemplum}[Harmonic and van der Pol oscillator]
\label{exemplum "Harmonic and van der Pol oscillator"}
Some primary examples of structurally stable dynamic systems are:
\enumerationisinitium
\item the \emph{equation(s) of a harmonic oscillator}
\begin{subequations}
\begin{align}
	& \ddot{x} + (\textgreek{\textnormal{μ}})\dot{x} + x = 0, \\
	& \dot{x} = y, \enspace \dot{y} = - x - (\textgreek{\textnormal{μ}})y,
\end{align}	
\end{subequations}
under the influence of a \emph{dissipative force}, e.g. by \emph{friction} or \emph{damping}, that is, mechanically, the motion of a pendulum in the phase plane, for a coefficient of friction $\textgreek{\textnormal{μ}}$, whereas a pendulum without friction ($\textgreek{\textnormal{μ}} = 0$) is proving to be structurally unstable. The behavior of a damped pendulum (with the dissipation of energy) remains the same, figuring that the friction, combined with the mass and length, is changed by a sufficiently small amount.
\item the \emph{van der Pol's equation} of the second order \cite{van der Pol "On Oscillation Hysteresis in a Triode Generator with Two Degrees of Freedom"} \cite{van der Pol "On Relaxation-Oscillations"}
\begin{equation}
\label{equation "van der Pol non-linear differential equations of the second order"}
	\ddot{x} - \varepsilon(1 - x^2)\dot{x} + x = 0,
\end{equation}
where $\varepsilon$ is a parameter, with a \emph{linear restoring force} for \emph{damped non-linear electronic oscillator} in a triode vacuum tube circuit, despite the fact that it is a model of \emph{deterministic chaos}. For a sufficiently small value of $\varepsilon$, the Eq. \eqref{equation "van der Pol non-linear differential equations of the second order"} contains a hyperbolic chaotic invariant set, and it has stability: except for 
\[
	x = \dot{x} = 0, 
\]
there exists a solution lying on a closed curve, which is a periodic orbit describing a harmonic-like oscillation; see the pioneering works of \emph{Cartwright-Littlewood} \cite{Cartwright and Littlewood "On Non-Linear Differential Equations of the Second Order: I"} \cite{Cartwright and Littlewood "On Non-Linear Differential Equations of the Second Order: II"} \cite{Cartwright "On non-linear differential equations of the second order: III"} \cite{Littlewood "On non-linear differential equations of the second order: III"} \cite{Littlewood "On non-linear differential equations of the second order: IV"}, the contributions e.g. by P.J. Holmes and D.A. Rand \cite{Holmes and Rand "Bifurcations of the forced van der Pol oscillator"} \cite{Holmes "A nonlinear oscillator with a strange attractor"}, and the most recent developments by R. Haiduc \cite{Haiduc "Horseshoes in the forced van der Pol system"}. \exemplumsymbol
\enumerationisfinis
\end{exemplum}

Now on to the formal definition of the concept of structural stability.
	
\begin{definitio}[Structurally stable Anosov system]
\label{definitio "Structurally stable Anosov system"}
~\enumerationisinitium
\item Let $\vec{X}$ be a vector field in the $\mathscr{C}^1$ topology. We say that a diffeomorphism $\varphi \viz \varphi[\mathscr{A}]$ in $\Diff^1(\mathcal{M})$ is \emph{structurally stable}
\subenumerationisinitium
\item if there is a $\mathscr{C}^1$ neighborhood $\vec{Y}$ of $\vec{X}$ and a $\mathscr{C}^1$ diffeomorphism $\psi \in \vec{Y}$ such that every $\psi \colon \mathcal{M} \to \mathcal{M}$ is \emph{orbitally topologically equivalent} to $\varphi$, i.e. topological properties of $\varphi$ are preserved under sufficiently $\mathscr{C}^1$ small perturbations of $\varphi$,
\item if there is a Hölder (continuous) homeomorphism $\vartheta_\textsc{h} \colon \mathcal{M} \to \mathcal{M}$ conjugating $\psi$ and $\varphi$, such that $\psi \circ \vartheta_\textsc{h} = \vartheta_\textsc{h} \circ \varphi$.
\subenumerationisfinis
\item Let $\mathcal{X}^1_{\varphi_t}(\mathcal{M})$ be the space of all $\mathscr{C}^1$ flows $\varphi_t$ of Anosov type on $\mathcal{M}$. Let $\vec{Y}$ be a vector field determining a flow $\psi_t$, and let us say that $\vec{Y}$ is sufficiently $\mathscr{C}^1$ close to $\vec{X}$, and hence $\psi_t$ is sufficiently $\mathscr{C}^1$ close to $\varphi_t$ in $\mathcal{X}^1_{\varphi_t}(\mathcal{M})$. A flow $\varphi_t \viz \varphi_t[\mathscr{A}]$ is said to be \emph{structurally stable} if there is a Hölder (continuous) homeomorphism $\vartheta_\textsc{h} \colon \mathcal{M} \to \mathcal{M}$ for which all $(\varphi_t)$-orbits and $(\psi_t)$-orbits are \emph{order preserving}. This means that $\vartheta_\textsc{h}$ maps the $(\varphi_t)$-orbits to $(\psi_t)$-orbits, and preserves the count of points of each orbit by guaranteeing a \emph{one-to-one orbital correspondence} between $\varphi_t$ and $\psi_t$ (in short, these flows shall be topologically equivalent). \definitiosymbol
\enumerationisfinis
\end{definitio}

In the case of $\mathscr{C}^r$ diffeomorphisms and $\mathscr{C}^r$ flows, for $r \geqslant 1$, it is clear that the structural stability is developed with open subsets $\Diff^r(\mathcal{M})$ of $\mathscr{C}^r\bigl(\mathcal{M} \xrightarrow{\varphi} \mathcal{M}\bigr)$, $\mathcal{X}^r_{\varphi_t}(\mathcal{M})$ spaces and any sufficiently $\mathscr{C}^r$ close vector field. 

Let us take the example of the foliations, whose definition of structural stability is similar to the one we have just saw, but it can be given in a more intuitive manner. 

\begin{definitio}[Structurally stable foliation]
\label{definitio "structurally stable foliation"}
Let 
\begin{equation}
	\mathcal{F} = \bigl(\varphi_t, f_\mathrm{(inv)}, \mathcal{M}\bigr)
\end{equation}
be a triple expressing a $\mathscr{C}^r$ foliation, where $\varphi_t \colon \mathcal{M} \times \mathbb{R} \to \mathcal{M}$ is a $\mathscr{C}^r$ flow and $f_\mathrm{(inv)} \colon \mathcal{M} \to \mathcal{M}$ is an involution, with the identity map $(f_\mathrm{(inv)})^2 = \id_\mathcal{M}$, assuming $\varphi_t$ and $f_\mathrm{(inv)}$ are both smooth on the manifold $\mathcal{M}$. Let $\mathcal{X}^r_\mathcal{F}(\mathcal{M})$ be the space of all $\mathscr{C}^r$ smooth foliations on $\mathcal{M}$. We say that a foliation $\mathcal{F}_{\mu_i} \in \mathcal{X}^r_\mathcal{F}(\mathcal{M})$ is \emph{structurally stable} if there is a neighborhood $\Upsilon_\mathcal{F}$ of $\mathcal{F}_{\mu_i}$ in $\mathcal{X}^r_\mathcal{F}(\mathcal{M})$ so that any foliation $\mathcal{F}_{\mu_j} \in \Upsilon_\mathcal{F}$ is topologically conjugate to $\mathcal{F}_{\mu_i}$. \definitiosymbol
\end{definitio}

A rich classification of structurally stable systems came into being in the Smale's \cite{Smale "Differentiable Dynamical Systems"} and Palis–Smale's \cite{Palis and Smale "Structural stability theorems"} studies, the (geometric) sufficient conditions of which are shown especially in J.W. Robbin \cite{Robbin "A Structural Stability Theorem"}, with emphasis on $\mathscr{C}^r$ diffeomorphisms, for $r \geqslant 2$; W. de Melo \cite{de Melo "Structural Stability of Diffeomorphisms on Two-Manifolds"}, focusing on $\mathscr{C}^1$ diffeomorphisms on 2-dimensional manifolds, referring to Palis–Smale tubular families; and C. Robinson \cite{Robinson "Structural stability of $C^1$ diffeomorphisms"}, with attention to $\mathscr{C}^r$ diffeomorphisms, for $r = 1$. The \emph{hyperbolicity} of structural stability of $\mathscr{C}^1$ diffeomorphisms is proved by R. Mañé \cite{Mane "A proof of the $C^1$ stability conjecture"},\footnote{
	A deficiency in the Mañé's proof is filled by Y. Zhang and S.B. Gan \cite{Zhang and Gan "On Mane's Proof of the $C^1$ Stability Conjecture"}.
	} 
while that of $\mathscr{C}^1$ flows by S. Hayashi \cite{Hayashi "Connecting Invariant Manifolds and the Solution of the $C^1$ Stability and Omega-Stability Conjectures for Flows"}.

\begin{margo}
The structural stability of Anosovian $\mathscr{C}^1$ flows, or of flows with hyperbolic-like dynamics, can be proved in several ways; the \emph{geometrical} demonstration by Arnold–Avez \cite[§ 16]{Arnold and Avez "Ergodic Problems of Classical Mechanics"} and the \emph{analytical} resolutive stages in J. Moser \cite{Moser "On a theorem of Anosov"} are classic examples of proofs. \margosymbol
\end{margo}

\begin{scholium}[Morse–Smale system, or hyperbolic strange attractors]
\label{scholium "Morse–Smale system, or hyperbolic strange attractors"}
In reference to the dimensionality, the structural stability is \emph{typical} (\emph{generic}) in low dimensions, as we know from \emph{Peixoto's theorem} \cite{Peixoto "On Structural Stability"} \cite{Peixoto "Structural stability on two-dimensional manifolds"}:
it is valid for a certain class of 2-dimensional differential systems on the plane; but \emph{in higher dimensions ($n \geqslant 3$) no such stability exists (it is theoretically incomplete)}, and this leads to \emph{bifurcation sets} and non-structurally stable dynamical systems.

Take the case of the \emph{Morse–Smale} system \cite{Smale "On Gradient Dynamical Systems"} \cite{Smale "Differentiable Dynamical Systems"}, with which it is possible to carry  over the definition of structural stability into the multi-dimensional space. The Morse-Smale approach specifies the criteria for identifying structurally stable dynamical systems with a finite number of equilibrium points and orbits of hyperbolic type; the related stable and unstable manifolds have mutual transverse intersections. But in dimension $n \geqslant 3$ appear \emph{strange attractors} within the structurally stable systems that, giving different examples of structurally unstable vector fields, prevent an extension of the typicality of Morse–Smale criteria. \scholiumsymbol	
\end{scholium}
 
\section{Integrability and Recurrence of Flow: Models in Comparison}
\label{section "Integrability and Recurrence of Flow: Models in Comparison"}

\enumerationisinitium
\item We shall now refer back to Section \ref{section "Geodesic Flow on the Unit Tangent Bundle of a Negatively Curved Surface by the Gamma-action on the Hyperbolic Half-Plane"}. Let $\mathbb{U}^2_\mathbb{F}$ be the hyperbolic 2-space, with $\mathbb{F} \viz (\mathbb{C} \cong \mathbb{R}^2)$, in which a unit tangent vector $v$ of $\mathbb{U}^2_\mathbb{F}$ identifies a geodesic 
\begin{equation}
	\gamma_{\mathrm{c}(x, v)}(t) \colon \mathbb{R} \to \mathbb{U}^2_\mathbb{F}, \text{ with } \gamma_{\mathrm{c}(x, v)}(0) = x \text{ and } \dot{\gamma}_{\mathrm{c}(x, v)}(0) = v, 
\end{equation} 
passing through two points, $\gamma_\mathrm{c}(+\infty), \gamma_\mathrm{c}(-\infty) \in \mathbb{R} \cup \{\infty\}$, on the boundary $\partial_\infty\mathbb{U}^2$, where $\dot{\gamma}_{\mathrm{c}(x, v)}$ is the derivative with respect to time $t$. In this model, $\gamma_{\mathrm{c}(x, v)}(t)$ is interchangeably a semicircle based on $\gamma_\mathrm{c}(+\infty)$ and $\gamma_\mathrm{c}(-\infty)$, or a vertical line based on the real $x$-axis. So, we can identify two pairs of points: 
\subenumerationisinitium
\item the first pair is $x$ and $\gamma_\mathrm{c}(+\infty)$, and it determines the positive horocycle flow $\eta^+_t\bigl(x, \gamma_\mathrm{c}(+\infty)\bigr)$; 
\item the second pair of points is $x$ and $\gamma_\mathrm{c}(-\infty)$, and it determines the negative horocycle flow $\eta^-_t\bigl(x, \gamma_\mathrm{c}(-\infty)\bigr)$. 
\subenumerationisfinis

The horocycle $\eta^+_t$ is the circle tangent to the real line at the end-point of the geodesic; the framing of $\eta^+_t$, with vectors (as the vector $v$) perpendicular to $\eta^+_t$ and pointing inwardly to $\gamma_\mathrm{c}(+\infty)$, is the stable foliation $\mathcal{W}^\mathrm{s}(x, v)$ through $(x, v)$. The horocycle $\eta^-_t$ is the circle tangent to the real line at the origin of the geodesic; the framing of $\eta^-_t$, with vectors (as the vector $v$) perpendicular to $\eta^-_t$ and pointing outwardly to $\gamma_\mathrm{c}(-\infty)$, is the unstable foliation $\mathcal{W}^\mathrm{u}(x, v)$ through $(x, v)$.

The geodesic flow
\begin{equation} 
	\varphi_t \colon (x, v) \mapsto \bigl(\gamma_{\mathrm{c}(x, v)}(t), \dot{\gamma}_{\mathrm{c}(x, v)}(t)\bigr)
\end{equation}
on $\mathbb{U}^2_\mathbb{F}$ is of Anosov type (since it is an $\mathbb{R}$-action on a negatively curved surface).
\item When the \emph{geodesic flow} is considered \emph{on some surface}, it is stated that the flow acts \emph{on the unit tangent bundle} which moves, with unit speed, a tangent vector along its geodesic $\gamma_{\mathrm{c}(x, v)}(t) \colon \mathbb{R} \to \mathcal{M}$. Letting $\mathcal{M}$ be a Riemannian manifold, we can visualize the unit tangent bundle $\mathring{\mathcal{T}}^1\mathcal{M}$ as coinciding with the \emph{unit sphere bundle} for the tangent bundle $\mathring{\mathcal{T}}\mathcal{M}$, and that can be done by defining $\mathring{\mathcal{T}}^1\mathcal{M}$ as the collection of all unit spheres $\mathbb{S}^{n - 1}(x)$ in all of the tangent spaces $\mathcal{T}_x\mathcal{M}$ to $\mathcal{M}$. For a natural projection $\pi \colon \mathring{\mathcal{T}}^1\mathcal{M} \to \mathcal{M}$, it is easy to see that each fiber $\pi^{-1}(x)$ of  $\mathring{\mathcal{T}}\mathcal{M}$ at any point $x \in \mathcal{M}$ is a unit sphere $\mathbb{S}^{n - 1}(x)$ included in $ \pi^{-1}(x)$.
\item In general, the \emph{geodesic flow} could be described, with a similar reasoning, as a \emph{dynamical $\mathbb{R}$-system on the tangent, or cotangent, bundle of a Riemannian manifold $\mathcal{M}$}. There is talk of tangent and cotangent bundles (Definition \ref{definitio "Tangent and cotangent bundles"}) because we can identify the tangent bundle of $\mathcal{M}$ at each point $x \in \mathcal{M}$ with the corresponding cotangent bundle through the isomorphism $\varphi \colon \mathring{\mathcal{T}}\mathcal{M} \to \mathring{\mathcal{T}}^*\mathcal{M}$, and therefore by identifying any tangent vector $v \in \mathcal{T}_x\mathcal{M}$ with the cotangent vector 
\begin{equation}
	\varphi_{\mathring{\mathcal{T}}\mathcal{M} \to \mathring{\mathcal{T}}^*\mathcal{M}}(v) \in \mathcal{T}^*_x\mathcal{M}. 
\end{equation} 
The tangent bundle is associated with the velocity vector field $\dot{\gamma}_\mathrm{c}(t) \in \mathcal{T}_{\gamma_\mathrm{c}(t)}\mathcal{M}$ at time $t \in I \subset \mathbb{R}$ along $\gamma_\mathrm{c}(t) \in \mathcal{M}$ (see Definition \ref{definitio "Geodesic"}), the cotangent bundle with the \emph{momentum} 
\begin{equation}
\label{equation "Momentum: cotangent bundle"}
	\varphi_{\mathring{\mathcal{T}}\mathcal{M} \to \mathring{\mathcal{T}}^*\mathcal{M}}\bigl(\dot{\gamma}_\mathrm{c}(t)\bigr) \in \mathcal{T}^*_{\gamma_\mathrm{c}(t)}\mathcal{M}.
\end{equation}
\item Flows of this kind, such as those discussed above, are characterized by
\subenumerationisinitium
\item a \emph{(complete) integrability in the sense of Liouville}, by viewing these flows as motions of a \emph{dynamical system of Hamiltonian type},
\item a \emph{chaoticity}, implying that the behavior of its foliations becomes \emph{unpredictable} (see Margo \ref{margo "Hamiltonian chaos through a geometrization of dynamics"}), whenever we are dealing with a discrete set, like imposing the Fuchsian group $\mathbbl{\Gamma} \subset PSL_2(\mathbb{R})$ (see Section \ref{subsection "Projective Linear Transforms: Dynamics on the Modular Surface, and Horocycle Flow"}). One other case of emerging chaos (for which the behavior of geodesics ceases to be regular) is that of a closed spherical surface with \emph{deformations}, e.g. a ping-pong ball with a bump or a dip. Later, we will analyze the chaotic (Chapters \ref{chapter "On the Chaos, Part I. Micro- and Macro-scales"} and \ref{chapter "On the Chaos, Part II. Non-linear Analysis"}) and even random roots (Chapter \ref{chapter "Randomness and Stochastic Systems"}) of some attractorial flows. 

Note that the property of Liouville integrability naturally results in the concept of \emph{complete integrability} primarily when doing a behavioral survey into the Hamiltonian systems. Two examples of classes of Riemannian manifolds whose geodesic flows are integrable are the \emph{Liouville manifolds} and \emph{Kähler–Liouville manifolds} in the Kiyohara nomenclature \cite{Kiyohara "Two Classes of Riemannian Manifolds Whose Geodesic Flows Are Integrable"}.
\subenumerationisfinis
\enumerationisfinis 

\begin{margo}[Eulerian equations of motion for rigid bodies and incompressible fluids as geodesic flows]
\label{margo "Eulerian equations of motion for rigid bodies and incompressible fluids as geodesic flows"}
We mainly used in this Chapter the geometry of Lobačevskij with the \emph{Beltrami–Poincaré 2-space}. The issue may nonetheless be extended to higher-dimensional manifolds; e.g. V.I. Arnold \cite{Arnold "Sur la geometrie differentielle des groupes de Lie de dimension infinie et ses applications a l'hydrodynamique des fluides parfaits"} identifies some \emph{examples of geodesic flow}
living on a Riemannian manifold, i.e. a Lie group, provided with a left invariant, or right invariant, riemannian metric:
\enumerationisinitium
\item \emph{Euler equations \textnormal{\cite{Euler "Decouverte d'un nouveau principe de mecanique"} \cite{Eulero "Theoria motus corporum solidorum seu rigidorum ex primis nostrae cognitionis principiis stabilita et ad omnes motus qui in huiusmodi corpora cadere possunt accomodata"}} of motion of a 3-dimensional rigid body}, which are equations of motion along geodesics in the group of rotations of Euclidean $\mathbb{R}^3$-space;
\item \emph{Euler equations \textnormal{\cite{Euler "Principia motus fluidorum"}\endnote{
	This Euler's study continues and develops in two other papers: \textit{Sectio secunda de principiis motus fluidorum} (1770) and \textit{Sectio tertia de motu fluidorum lineari potissimum aquae} (1771) divided into five Chapters (chap. I. \textit{De principiis motus linearis fluidorum}, chap. II. \textit{De motu aquae in tubis aequaliter ubique amplis}, chap. III. \textit{De motu aquae in tubis inaequaliter amplis}, chap. IV. \textit{De elevatione aquae antliarum ope}, chap. V. \textit{De motu aquae per tubos diverso caloris gradu infectos}). They were presented to the St. Petersburg Academy on March 17, 1766.
	} 
\cite{Euler "Principes generaux du mouvement des fluides"}}, originally in 2 and 3 dimensions, for the flow of an inviscid} (zero or very low viscosity) \emph{incompressible fluid}.
\enumerationisfinis
These two families of Eulerian equations are actually geodesic flows. \margosymbol
\end{margo}

\subsection{Liouville Measure: Integral Invariant of Hamiltonian Dynamics}
\label{subsection "Liouville Measure: Integral Invariant of Hamiltonian Dynamics"}

Let us focus on the integrability mentioned in the previous Section. It has to do with the fact that the geodesic flow $\varphi_t$ preserves 
\enumerationisinitium
\item the \emph{(Riemannian) volume form} induced by the metric $g$, 
\item the related smooth measure, known as \emph{Liouville measure} \textnormal{\cite{Liouville "Note sur la Theorie de la Variation des constantes arbitraires"}}, on the manifold.
\enumerationisfinis

In case the manifold is compact, the Liouville measure is finte (so the flow also exhibits a non-trivial recurrence) and normalizable. The invariance of the Liouville measure under the flow correlates with Hamiltonian conservative systems in the phase space. Indeed, \emph{the geodesic flow can also be thought of (and defined) as a special case of a Hamiltonian flow}, assuming $\mathcal{M}$ is endowed with a symplectic form. Before proceeding we need a quick remark.

\begin{scholium}[On the correspondence between the geodesic and Hamiltonian flows]
\label{scholium "On the correspondence between the geodesic and Hamiltonian flows"}
~\enumerationisinitium
\item It should be recalled that the \emph{phase space} is the space in which the points correspond to the possible states of the system, and each point of the phase space represents one state of the system, and one only.
\item The Hamiltonian flow, to be more precise, is said to be a \emph{cogeodesic flow}, because it is created when the system acts \emph{on the (unit) cotangent bundle} with the momentum \eqref{equation "Momentum: cotangent bundle"}. The expression “geodesic flow”, for a Hamiltonian system, is thus comprehensive.
\item \emph{A flow, when it is geodesic, is also Hamiltonian; but a Hamiltonian flow may not be geodesic}. Some manifolds may have singularities in the geodesic structure, and a distinction exists between locally and globally geodesic spaces; e.g. C. McCord, K.R. Meyer and D. Offin \cite{McCord Meyer Offin "Are Hamiltonian Flows Geodesic Flows?"} show that the \emph{flows of the planar $n$-body and spatial 3-body problems, both on the reduced spaces, are not geodesic flows}, excluding the case in which the components of the orbital angular momentum are zero (with no rotation at all) and the energy levels of a system are positive. \scholiumsymbol	
\enumerationisfinis
\end{scholium}

\subsection{Symplectic Geometry; Liouville's, Noether's and Poisson's Theorems}
\label{subsection "Symplectic Geometry; Liouville's, Noether's and Poisson's Theorems"}

\begingroup
\footnotesize
The name “complex group” formerly advocated by me in allusion to line complexes, as these are defined by the vanishing of antisymmetric bilinear forms, has become more and more embarrassing through collision with the word “complex” in the connotation of complex number. I therefore propose to replace it by the corresponding Greek adjective “symplectic” [\textgreek{συμπλεκτικός}]. Dickson calls the group the “Abelian linear group” in homage to Abel who first studied it. \\
\indent — \textsc{H. Weyl} \cite[p. 165]{Weyl "The Classical Groups: Their Invariants and Representations"}

\endgroup

\vspace{2mm}

Now let us examine in detail the aforementioned issues, with remembering some important concepts.

\begin{definitio}[Hamiltonian vector field or symplectic gradient]
\label{definitio "Hamiltonian vector field or symplectic gradient"}
~\enumerationisinitium
\item 
\label{item "Symplectic manifold"}
Let $(\mathcal{M}, \omega_\mathrm{s}) = (\mathbb{R}^{2n}, \omega_\mathrm{s})$ be a \emph{symplectic manifold}, with a pair of a smooth manifold $\mathcal{M}$ of dimension $2n$ and a symplectic form
\begin{align}
\label{align "Non-degenerate closed differential 2-form"}
	\omega_\mathrm{s} \in \Lambda^2(\mathcal{M}) & = dx^1_\mathsf{q} \wedge dy^1_\mathsf{p} + \cdots + dx^n_\mathsf{q} \wedge dy^n_\mathsf{p} \\ 
	& \viz \sum^n_{\nu = 1}dx^\nu_\mathsf{q} \wedge dy^\nu_\mathsf{p},
\end{align}
that is a \emph{non-degenerate closed differential 2-form}, for which $d\omega_\mathrm{s} = 0$ and ${\omega_\mathrm{s}}_{(x)}$ is a symplectic tensor, namely an alternating covariant 2-tensor, at each point $x \in \mathcal{M}$, where $(x^1_\mathsf{q}, \mathellipsis, x^n_\mathsf{q}, y^1_\mathsf{p}, \mathellipsis, y^n_\mathsf{p})$ are the standard coordinates on $\mathcal{M} = \mathbb{R}^{2n}_{x, y}$.
\item Let $\Hamiltonian \colon \mathcal{M} \to \mathbb{R}$ be a smooth function. Then there is a vector field
\begin{align}
	\vec{X}_\Hamiltonian(x^1_\mathsf{q}, \mathellipsis, x^n_\mathsf{q}, y^1_\mathsf{p}, \mathellipsis, y^n_\mathsf{p}) & = \left(\frac{\partial\Hamiltonian}{\partial{y^1_\mathsf{p}}}, \mathellipsis, \frac{\partial\Hamiltonian}{\partial{y^n_\mathsf{p}}}, -\frac{\partial\Hamiltonian}{\partial{x^1_\mathsf{q}}}, \mathellipsis, -\frac{\partial\Hamiltonian}{\partial{x^n_\mathsf{q}}}\right), \\
	& \vec{X}_\Hamiltonian|_{\Upsilon_x \subset \mathcal{M}} 
	\label{align "Hamiltonian vector field in Darboux coordinates"}
	= \sum^n_{\nu = 1}\left(\frac{\partial\Hamiltonian}{\partial{y^\nu_\mathsf{p}}}\frac{\partial}{\partial{x^\nu_\mathsf{q}}} - \frac{\partial\Hamiltonian}{\partial{x^\nu_\mathsf{q}}}\frac{\partial}{\partial{y^\nu_\mathsf{p}}}\right),
\end{align} 
determined by
\begin{equation}
\label{equation "Hamiltonian vector field"}
	\vec{X}_\Hamiltonian\lrcorner \thinspace \omega_\mathrm{s} = d\Hamiltonian \text{ and }\iota_{\vec{X}_\Hamiltonian}\omega_\mathrm{s} = -d\Hamiltonian,
\end{equation}
where the symbol $\lrcorner$ and the letter $\iota$ indicate, indistinctly, an operator called \emph{interior product} or \emph{contraction}, and finally $\Hamiltonian$ is a \emph{Hamiltonian} for $\vec{X}$.
\subenumerationisinitium
\item There are Hamilton's differential equations in use, corresponding to the the well-known formulæ
\begin{equation}
\label{equation "Hamilton's differential equations"}
	\begin{cases}
	\dot{x}^\nu_\mathsf{q} = \frac{dx^\nu_\mathsf{q}}{dt} = \frac{\partial \Hamiltonian}{\partial{y^\nu_\mathsf{p}}}, \\
	\dot{y}^\nu_\mathsf{p} = \frac{dy^\nu_\mathsf{p}}{dt} = -\frac{\partial \Hamiltonian}{\partial{x^\nu_\mathsf{q}}}.
	\end{cases}
\end{equation}
\item The Eq. \eqref{align "Hamiltonian vector field in Darboux coordinates"} is the mathematical relation for $\vec{X}$ of $\Hamiltonian$ in \emph{Darboux coordinates} on an open neighborhood $\Upsilon_x \subset \mathcal{M}$, such that we might find a local chart $(\Upsilon_x, \varphi)$ in $x$ with $\varphi_\mathcal{M} = (x^1_\mathsf{q}, \mathellipsis, x^n_\mathsf{q}, y^1_\mathsf{p}, \mathellipsis, y^n_\mathsf{p})$, for any $x \in \mathcal{M}$. 
\item The time dependent condition of $\vec{X}_\Hamiltonian$ can be written $\iota_{\vec{X}_{\Hamiltonian_t}}\omega_\mathrm{s} = -d\Hamiltonian_t$, with $\Hamiltonian_t(x) = \Hamiltonian(t, x)$, by putting $\Hamiltonian = \Hamiltonian(t, x) \colon [0, 1] \times \mathbb{R}^{2n} \to \mathbb{R}$.
\subenumerationisfinis
\item Let $\hat{\omega}_\mathrm{s} \colon \mathring{\mathcal{T}}\mathcal{M} \to \mathring{\mathcal{T}}^*\mathcal{M}$ be a bundle (fiberwise) isomorphism induced by $\omega_\mathrm{s}$, which is an identification of the tangent bundle and the cotangent bundle through the 2-form.

Under these three conditions, the following properties and results hold. 
\enumerationisfinis
\enumerationisinitium
\item One calls
\begin{equation}
	\vec{X}_\Hamiltonian = {\hat{\omega}_\mathrm{s}}^{-1}(d\Hamiltonian)
\end{equation}
on $\mathcal{M}$ the \emph{(globally) Hamiltonian vector field} associated with $\Hamiltonian$ or the \emph{symplectic gradient} of $\Hamiltonian$, for any $\Hamiltonian \in \mathscr{C}^\infty(\mathcal{M})$. The vector field $\vec{X}_\Hamiltonian$ is said to be \emph{locally Hamiltonian} if for each $x \in \mathcal{M}$ there exists a neighborhood $\Upsilon_x \subset \mathcal{M}$ on which $\vec{X}$ is Hamiltonian. It is obvious that every (global) Hamiltonian vector field is also a locally Hamiltonian vector field.
\item $\Hamiltonian$ is constant on the integral curves of $\vec{X}_\Hamiltonian$, and $\vec{X}_\Hamiltonian$ is tangent to the smooth level sets of $\Hamiltonian$, being that
\begin{equation}
	\bigl(\vec{X}_\Hamiltonian\bigr)\Hamiltonian = d\Hamiltonian\bigl(\vec{X}_\Hamiltonian\bigr) = \omega_\mathrm{s}\bigl(\vec{X}_\Hamiltonian, \vec{X}_\Hamiltonian\bigr) = 0.
\end{equation}
\definitiosymbol
\enumerationisfinis
\end{definitio}

\begin{definitio}[Hamiltonian flow]
\label{definitio "Hamiltonian flow"}
Let us keep the conditions of the Definition \ref{definitio "Hamiltonian vector field or symplectic gradient"}, and go ahead with another definition.
\enumerationisinitium
\item Let 
\begin{equation}
	\frac{d}{dt}\varphi_t(x) = \vec{X}\bigl(\varphi_t(x)\bigr), \text{ with } x \in \mathcal{M} = \mathbb{R}^{2n},
\end{equation}
be the flow of a vector field $\vec{X}$. Let $\vec{X} \equival \vec{X}_\Hamiltonian$. The flow $\varphi^\Hamiltonian_t$, given by $\dot{x} = \vec{X}_\Hamiltonian(x)$ on $\mathcal{M}$, is said \emph{Hamiltonian flow} of $\Hamiltonian$, and it is defined by
\begin{equation}
\label{equation "Hamiltonian flow"}
	\frac{dx}{dt} = \dot{\varphi}^\Hamiltonian_t = \vec{X}_\Hamiltonian(x).
\end{equation}
For a symplectic manifold $(\mathcal{M}, \omega_\mathrm{s})$ of fixed $\dim(\mathcal{M}) = 2n$, the Eq. \eqref{equation "Hamiltonian flow"} reproduces the \emph{Hamilton–Jacobi equation(s)} \cite{Hamilton "On a General Method in Dynamics; by which the Study of the Motions of all free Systems of attracting or repelling Points is reduced to the Search and Differentiation of one central Relation or characteristic Function"} \cite{Hamilton "On the Application to Dynamics of a General Mathematical Method previously Applied to Optics"} \cite{Hamilton "Second Essay on a General Method in Dynamics"} \cite{Jacobi "Lettre sur quelques points d'analyse mathematique"} \cite{Jacobi "Vorlesungen uber Dynamik"}, a canonical expression of which is \eqref{equation "Hamilton's differential equations"}. 
\item Let $\Hamiltonian \in \mathscr{C}^\infty(\mathcal{M})$. 
\subenumerationisinitium
\item The Hamiltonian flow $\varphi^\Hamiltonian_t $ is the \emph{time evolution of $(x_\mathsf{q}, y_\mathsf{p}) \in \mathring{\mathcal{T}}^*\mathcal{M}$} that sets out a Hamiltonian vector field on the (unit) cotangent bundle or, more generally, on the phase space of $\mathcal{M}$ under the Eqq. \eqref{equation "Hamilton's differential equations"}.
\item The Hamiltonian flow $\varphi^\Hamiltonian_t$ is a \emph{specially geodesic flow} if it evolves on the (unit) tangent bundle, $\varphi^\Hamiltonian_t \colon \mathring{\mathcal{T}}\mathcal{M} \to \mathring{\mathcal{T}}\mathcal{M}$, the form of which is fully equivalent to Eq. \eqref{equation "Geodesic flow"}.
\subenumerationisfinis
\item Let 
\begin{equation}
	\Liederivative_{\vec{X}}\omega_\mathrm{s} = \iota_{\vec{X}}d\omega_\mathrm{s} + d\iota_{\vec{X}}\omega_\mathrm{s}
\end{equation} 
be the \emph{Lie derivative} of $\omega_\mathrm{s}$ along $\vec{X}$, according to the so-called \emph{Cartan's magic formula} \cite{Cartan "Lecons sur les invariants integraux"} and the explicit formula of Ślebodziński \cite{Slebodzinski "Republication of: On Hamilton's canonical equations"}. We say that $\vec{X}$ is \emph{symplectic} if the Lie derivative vanishes, $\Liederivative_{\vec{X}}\omega_\mathrm{s} = 0$, i.e. if $\omega_\mathrm{s}$ is \emph{flow invariant} with respect to the vector field $\vec{X}$. 

Let the vector field $\vec{X}$ be Hamiltonian, which means $\vec{X} \equival \vec{X}_\Hamiltonian$, so 
\begin{equation}	
	\Liederivative_{\vec{X}_\Hamiltonian}\omega_\mathrm{s} = 0
\end{equation}
(that is because a Hamiltonian vector field is symplectic). We can state the same reasoning writing
\begin{equation}
	\frac{d}{dt}\bigl(\varphi^\Hamiltonian_t\bigr)^*\omega_\mathrm{s} = \bigl(\varphi^\Hamiltonian_t\bigr)^*\Liederivative_{\vec{X}_\Hamiltonian}\omega_\mathrm{s} = 0.
\end{equation}
Conversely, if $\vec{X}$ is symplectic, and therefore is Hamiltonian in the sense of the Eq. \eqref{equation "Hamiltonian vector field"}, the \emph{flow of $\vec{X}$ preserves the symplectic form and the energy (or Hamiltonian) function}, so
\begin{equation}
	\Liederivative_{\vec{X}_\Hamiltonian}\Hamiltonian = 0,
\end{equation}
and
\begin{equation}
\label{equation "Hamiltonian flow preserving the symplectic form"}
	\bigl(\varphi^\Hamiltonian_t\bigr)^*\omega_\mathrm{s} = \omega_\mathrm{s}.
\end{equation}
To put it another way, the flow $\varphi^\Hamiltonian_t$ is \emph{volume preserving} (Theorem \ref{theorema "Hamiltonian flow preserving the symplectic form"}).
\item Letting $\varphi^\Hamiltonian_t \colon \Upsilon \to \mathcal{M}$ be a (local) flow of $\vec{X}$ on a neighborhood $\Upsilon_x \subset \mathcal{M}$ of $x \in \mathcal{M}$, we define the Lie derivative of $\omega_\mathrm{s}$ along $\vec{X}$ as  
\begin{align}
	\Liederivative_{\vec{X}}{\omega_\mathrm{s}}_{(x)} & = \lim_{t \to 0}\frac{\bigl(\varphi^\Hamiltonian_t\bigr)^*\Bigl((\omega_\mathrm{s})_{\varphi^\Hamiltonian_t(x)}\Bigr) - {\omega_\mathrm{s}}_{(x)}}{t} \\
	& = \lim_{t \to 0}\frac{\left(\bigl(\varphi^\Hamiltonian_t\bigr)^*\omega_\mathrm{s}\right)_x - {\omega_\mathrm{s}}_{(x)}}{t} = \frac{d}{dt}\bigg|_{t = 0}\left(\bigl(\varphi^\Hamiltonian_t\bigr)^*\omega_\mathrm{s}\right)_x,
\end{align}
by a \emph{pullback} of $(\omega_\mathrm{s})_{\varphi^\Hamiltonian_t(x)}$ through $\bigl(\varphi^\Hamiltonian_t\bigr)^*$. \definitiosymbol
\enumerationisfinis
\end{definitio}

We transcribe in terms of theorem and proof a key point of this definition.

\begin{theorema}
\label{theorema "Hamiltonian flow preserving the symplectic form"}
The flow of a Hamiltonian vector field preserves the symplectic form, as is the equality \eqref{equation "Hamiltonian flow preserving the symplectic form"}; that is to say: a symplectic form is conserved along $\varphi^\Hamiltonian_t$.
\end{theorema}

\begin{proof}
For fixed $t, s \in \mathbb{R}$, we know that   
\begin{equation}
	\frac{d}{dt}\bigl(\varphi^\Hamiltonian_t\bigr)^*\omega_\mathrm{s} = \frac{d}{ds}\bigl(\varphi^\Hamiltonian_{t + s}\bigr)^*\bigg|_{s = 0}\omega_\mathrm{s} =
	\frac{d}{ds}\left(\bigl(\varphi^\Hamiltonian_t\bigr)^*\bigl(\varphi^\Hamiltonian_s\bigr)^*\right)\bigg|_{s = 0}\omega_\mathrm{s}
	 = \bigl(\varphi^\Hamiltonian_t\bigr)^*\Liederivative_{\vec{X}_\Hamiltonian}\omega_\mathrm{s},
\end{equation}
and
\begin{equation}
	\Liederivative_{\vec{X}_\Hamiltonian}\omega_\mathrm{s} = \iota_{\vec{X}_\Hamiltonian}d\omega_\mathrm{s} + d\iota_{\vec{X}_\Hamiltonian}\omega_\mathrm{s} = 0.
\end{equation}
Then
\begin{equation}
	\left(\frac{d}{dt}\bigl(\varphi^\Hamiltonian_t\bigr)^*\omega_\mathrm{s} = 0\right) \viz \left(\bigl(\varphi^\Hamiltonian_t\bigr)^*\omega_\mathrm{s} = \bigl(\varphi^\Hamiltonian_t\bigr)^*\big|_{t = 0}\omega_\mathrm{s} = \omega_\mathrm{s}\right). 
\end{equation}
\end{proof}

After the Definitions \ref{definitio "Hamiltonian vector field or symplectic gradient"} and \ref{definitio "Hamiltonian flow"}, we are ready to exhibit without a hitch the result on the invariance of the Liouville measure under the Hamiltonian flow.

\begin{theorema}[Liouville invariance]
\label{theorema "Liouville invariance"}
Let $(\mathcal{M}, \omega_\mathrm{s})$ be a symplectic manifold, with the 2-form such as \eqref{align "Non-degenerate closed differential 2-form"}. We suppose $\mathcal{M}$ is a (smooth) Riemannian manifold. Let the volume measure $(x^1_\mathsf{q}, \mathellipsis, x^n_\mathsf{q}, y^1_\mathsf{p}, \mathellipsis, y^n_\mathsf{p})$ be a Liouville measure on $\mathcal{M} = \mathbb{R}^{2n}$. The Hamiltonian (specially geodesic) flow $\varphi^\Hamiltonian_t$ on the manifold of a Hamiltonian vector field $\vec{X}_\Hamiltonian$ preserves the symplectic volume form, or the phase volume, hence the Liouville measure on the phase space.
\end{theorema}

\begin{proof}
\begin{equation}
	\frac{d}{dt}\bigl(\varphi^\Hamiltonian_t\bigr)^*\omega_\mathrm{s} = \bigl(\varphi^\Hamiltonian_t\bigr)^*\Liederivative_{\vec{X}_\Hamiltonian}\omega_\mathrm{s} = \bigl(\varphi^\Hamiltonian_t\bigr)^*\left(d(\vec{X}_\Hamiltonian\lrcorner \thinspace \omega_\mathrm{s}) + (\vec{X}_\Hamiltonian\lrcorner \thinspace d\omega_\mathrm{s})\right) = 0.
\end{equation}
\end{proof}
	
\begin{scholium}
The flow $\varphi^\Hamiltonian_t$ is invariant under the time-reversal transformation $t \to \dot{t} = -t$, and the Hamiltonian system is the same in the reverse direction of time. \scholiumsymbol
\end{scholium}

We are adding another theorem a latere derived from the large basket of E. Noether \cite{Noether "Invariante Variationsprobleme"}.

\begin{theorema}[Noether—on symplectic transformations]
\label{theorema "Noether—on symplectic transformations"}
Via Theorem \ref{theorema "Liouville invariance"}, we declare that a Hamiltonian vector fields is of Liouville type, and we may as well assume that $\Hamiltonian \colon \mathcal{M} \to \mathbb{R}$ is a symplectic invariant function for a 1-parameter group of symplectic transformations (thanks to which the symplectic form is invariant) by a function $\rotatedxi$. Consequently, $\rotatedxi$ is also an integral of the Hamiltonian flow $\varphi^\Hamiltonian_t$.
\end{theorema}

\begin{proof}
Adopting the \emph{Poisson bracket} of $\rotatedxi, \Hamiltonian \in \mathscr{C}^\infty(\mathcal{M})$, that is an $\mathbb{R}$-bilinear operation, we have a combined function $\{\rotatedxi, \Hamiltonian\} \in \mathscr{C}^\infty(\mathcal{M})$ given by
\begin{equation}
	\{\rotatedxi, \Hamiltonian\} = \omega_\mathrm{s}(\vec{X}_\rotatedxi, \vec{X}_\Hamiltonian) = d\rotatedxi(\vec{X}_\Hamiltonian) = \vec{X}_\Hamiltonian(\rotatedxi).
\end{equation}
One could imagine that the function $\Hamiltonian$ is an integral with regard to the flow of $\rotatedxi$, so the function $\rotatedxi$ is an integral with regard to the flow of $\Hamiltonian$. Two steps are required.
\enumerationisinitium
\item The first is to prove the following proposition: $\rotatedxi$ is an integral of $\varphi^\Hamiltonian_t$ iff such a function has zero Poisson bracket with $\Hamiltonian$, i.e. $\{\rotatedxi, \Hamiltonian\} = 0$. One of the properties of the Poisson bracket is $\{\rotatedxi, \Hamiltonian\} = \Liederivative_{\vec{X}_\Hamiltonian}\rotatedxi$. Letting $x \in \mathcal{M}$ and $t, s \in \mathbb{R}$, from this we obtain
\begin{equation}
	\Liederivative_{\vec{X}_\Hamiltonian}\rotatedxi = \frac{d}{dt}\bigg|_{t = 0}\bigl(\varphi^\Hamiltonian_t\bigr)^*\rotatedxi = \{\rotatedxi, \Hamiltonian\} = 0,
\end{equation}
thereby
\begin{align}
	\frac{d}{dt}\rotatedxi\left(\varphi^\Hamiltonian_t(x)\right) & = \frac{d}{dt}\bigl(\varphi^\Hamiltonian_t\bigr)^*\rotatedxi(x) = \frac{d}{ds}\bigg|_{s = 0}\bigl(\varphi^\Hamiltonian_{t + s}\bigr)^*\rotatedxi(x)  \notag \\
	& = \frac{d}{ds}\left(\bigl(\varphi^\Hamiltonian_t\bigr)^*\bigl(\varphi^\Hamiltonian_s\bigr)^*\right)\bigg|_{s = 0}\rotatedxi(x) \\ \notag
	& = \bigl(\varphi^\Hamiltonian_t\bigr)^*\frac{d}{ds}\bigl(\varphi^\Hamiltonian_s\bigr)^*\bigg|_{s = 0}\rotatedxi(x) = \bigl(\varphi^\Hamiltonian_t\bigr)^*\Liederivative_{\vec{X}_\Hamiltonian}\rotatedxi = 0. 
\end{align}
\item The second step is revealed by another property of the Poisson bracket, the \emph{anti-symmetry}, or \emph{skew symmetry}: $\{\rotatedxi, \Hamiltonian\} = -\{\Hamiltonian, \rotatedxi\}$, and we are done.
\enumerationisfinis
\end{proof}

\begin{corollarium}[Poisson]
\label{corollarium "Poisson"}
By adopting the premises and the result of Theorem \ref{theorema "Noether—on symplectic transformations"}, that is 
\enumerationisinitium
\item $\Hamiltonian \colon \mathcal{M} \to \mathbb{R}$, 
\item $\vec{X}_\Hamiltonian\lrcorner \thinspace \omega_\mathrm{s} = d\Hamiltonian$
\item $\frac{dx}{dt} = \dot{\varphi}^\Hamiltonian_t = \vec{X}_\Hamiltonian$.
\enumerationisfinis
Let $\rotatedeta$ be a third smooth function on $\mathcal{M}$. Then the combined function $\{\rotatedxi, \rotatedeta\} \in \mathscr{C}^\infty(\mathcal{M})$ is an integral of $\varphi^\Hamiltonian_t$.
\end{corollarium}

\begin{proof}
By manipulating the \emph{Jacobi identity}, we see a further property of the Poisson bracket, 
\begin{equation}	
	\bigl\{\{\rotatedxi, \rotatedeta\}, \Hamiltonian\bigr\} + \bigl\{\{\rotatedeta, \Hamiltonian\}, \rotatedxi\bigr\} + \bigr\{\{\Hamiltonian, \rotatedxi\}, \rotatedeta\bigl\} = 0,
\end{equation} 
so we can write $\bigl\{\{\rotatedxi, \rotatedeta\}, \Hamiltonian\bigr\} = \bigl\{\rotatedxi, \{\rotatedeta, \Hamiltonian\}\bigr\} + \bigl\{\rotatedeta, \{\Hamiltonian, \rotatedxi\}\bigr\} = 0$, and the corollary is proven.
\end{proof}

The reader should be aware that the \emph{integrability of Hamiltonian vector fields} is still a \emph{ill-defined notion} (evidently, the integrable Hamiltonian systems are a special class of Hamiltonian systems), working principally from a physical perspective, as Hamiltonian systems are generally regarded as being of \emph{chaotic nature}.

\begin{margo}[Hamiltonian chaos through a geometrization of dynamics]
\label{margo "Hamiltonian chaos through a geometrization of dynamics"}
Let us define \emph{Hamiltonian chaos} as a deterministic Hamiltonian dynamics, which, however, is unstable and therefore unpredictability. Worth of mention is the \emph{geometric approach} to instability properties, that is, chaotic attributes, of Hamiltonian dynamics; see the work of M. Pettini et al. \cite{Pettini "Geometrical hints for a nonperturbative approach to Hamiltonian dynamics"} \cite{Casetti and Pettini "Analytic computation of the strong stochasticity threshold in Hamiltonian dynamics using Riemannian geometry"} \cite{Cerruti-Sola and Pettini "Geometric description of chaos in self-gravitating systems"} \cite{Casetti Clementi and Pettini "Riemannian theory of Hamiltonian chaos and Lyapunov exponents"} \cite{Pettini "Geometry and Topology in Hamiltonian Dynamics and Statistical Mechanics"}, see also \cite{Horwitz Zion Lewkowicz Schiffer and Levitan "Geometry of Hamiltonian Chaos"} \cite{Yahaloma Levitan Lewkowicz Horwitz "Lyapunov vs. geometrical stability analysis of the Kepler and the restricted three body problems"}. It identifies a Hamiltonian flow with a geodesic flow on a Riemannian manifold, with a suitable metric, and proves that some curvature properties, or rather the fluctuations of curvature along the geodesics, generate dynamical instability, i.e. chaos, in the geodesic flow (suppose e.g. there is a locally unstable geodesic flow; that entails a sensitive dependence on the initial conditions from neighboring geodesics, as they diverge exponentially).

Such a instability is investigated by means of the \emph{equation of geodesic deviation}, otherwise referred to as \emph{Levi-Civita equation} \cite{Levi-Civita "Sullo scostamento geodetico"} \cite{Levi-Civita "Sur l'ecart geodesique"}, or even \emph{Jacobi–Levi-Civita equation}, 
\begin{equation}
		\frac{\nabla}{ds}\frac{\nabla}{ds}\vec{J}^\mu + {\Riemann^\mu}_{\nu\xi\varrho}\frac{dx^\nu}{ds}\vec{J}^\xi\frac{dx^\varrho}{ds} = \frac{\nabla^2\vec{J}}{ds^2}(s) + \Riemann\left(\vec{J}(s), \dot{\gamma}_\mathrm{c}(s)\right)\dot{\gamma}_\mathrm{c}(s) = 0,
\end{equation}
where $\vec{J} \in \mathcal{T}_{\gamma_\mathrm{c}(s)}\mathcal{M}$, for each $s \in \mathbb{R}$, is a Jacobi field (see Margo \ref{margo "Jacobi field"}) acting as the vector field of geodesic deviation and serving to locally measure the distance between two nearby geodesics, $\frac{\nabla}{ds}$ is the covariant derivative along a geodesic, ${\Riemann^\mu}_{\nu\xi\varrho}$ is the Riemann curvature tensor, and 
\begin{equation}
	\dot{\gamma}_\mathrm{c}(s) = \frac{d}{ds}\gamma_\mathrm{c}(s). 
\end{equation}
This occurs chiefly when it comes to compact negatively curved manifolds, but there are also cases with positive curvature \cite{Knieper & Weiss "A surface with positive curvature and positive topological entropy"}. \margosymbol
\end{margo}

\subsection{Poincaré Recurrence Theorem}
\label{subsection "Poincaré Recurrence Theorem"}
	
\begingroup
\footnotesize
In the mechanistic hypothesis, all phenomena must be \emph{reversible} [\,\dots] A theorem, easy to prove, tells us that a bounded world, controlled only by the laws of mechanics, will always come back through a state very close to its initial state. On the contrary, according to accepted experimental laws [\,\dots] the universe tends toward a certain final state, from which it will never depart [as emblem of the irreversibility]. In this final state, which will be a kind of death, all bodies will be at rest at the same temperature [\,\dots] but [the universe] does not remain that way forever, for which the above-mentioned theorem is not violated; it solely stays there for an enormously long time, which is longer the more numerous are the molecules. This state will not be the definitive death of the universe, but a sort of slumber, from which it will awake after millions of millions of centuries.\footnote{
	One similar idea is also expressed in Kelvin \cite[§§ 13, 31 and postscript, pp. 559-560]{Kelvin "On Periodic Motion of a Finite Conservative System"}; read e.g. § 31, p. 559: «Our result proves that every path infinitely near to the orbit is unstable unless every root of the equation for $e$ has a real value between $1$ and $-1$. It does not prove that the motion \emph{is} stable when this condition is fulfilled. Stability or instability for this case cannot be tested without going to higher orders of approximation in the consideration of paths very nearly coincident with an orbit».
	} \\
\indent — \textsc{H. Poincaré} \cite[pp. 534, 536]{Poincare "Le Mecanisme et l'experience"}

\endgroup

\vspace{2mm}

It may be interesting to discuss a renowned application of Liouville's theorem, that is the Poincaré recurrence theorem (in the sense that the latter is a consequence of the former), and see how these two theorems are intertwined. The recurrence theorem, discussed by Poincaré in his groundbreaking paper on the \emph{three-body problem} \cite{Poincare "Sur le probleme des trois corps et les equations de la dynamique"}, and subsequently revisited and proved by C. Carathéodory \cite{Caratheodory "Uber den Wiederkehrsatz von Poincare"}, claims, broadly, that a dynamical systems, after a sufficiently long time (long but finite), will return to a state very close to its initial state. There is also a quantitative version of the theorem, and it is the work of M. Kac \cite{Kac "On the notion of recurrence in discrete stochastic processes"}.

On the whole, we can say that the Poincaré recurrence theorem is considered the first pivotal mathematical result, at least in the qualitative acceptation, of \emph{ergodic theory} (Section \ref{section "A Framework for Ergodicity"}).

\subsubsection{Step I}

\begin{theorema}[Poincaré—recurrence time]
\label{theorema "Poincaré—recurrence time"}
~\enumerationisinitium
\item 
\label{item "Primary version of Poincaré's theorem"}
Let $(\invertedbreve{\Omega}, \bbmu)$ be a $\sigma$-finite measure space and $\varphi_\bbmu \colon \invertedbreve{\Omega} \to \invertedbreve{\Omega}$ a measure preserving transformation, where $\invertedbreve{\Omega}$ is a non-empty set and $\bbmu$ is a finite measure invariant under $\varphi_\bbmu$. Let $\invertedbreve{E} \subset \invertedbreve{\Omega}$ be a measurable subset, with $\bbmu(\invertedbreve{E}) > 0$. Then $\bbmu$-almost any point $x \in \invertedbreve{E}$ returns (is recurrent) to $\invertedbreve{E}$ infinitely often; so, there exists an infinte set $\{n \in \mathbb{N} \mid \varphi_\bbmu^n(x \in \invertedbreve{E})\}$ (see \ref{item "Primary version of Poincaré's theorem: alternative exposition"} in Scholium \ref{scholium "Lindelöf space, Borel measure plus sigma-algebra"}).
\item 
\label{item "Topological version of Poincaré's theorem"}
Let $(\mathcal{X}, \distance)$ be a separable metric space, that means a set with a countable dense subset, and thereby a topological space with the Lindelöf property (see \ref{item "Lindelöf space"} in Scholium \ref{scholium "Lindelöf space, Borel measure plus sigma-algebra"}). Let $\bbmu$ denote a finite Borel measure on $\mathcal{X}$, based on the $\sigma$-algebra $\mathscr{B}_\sigma(\mathcal{X})$ of Borel sets (see \ref{item "Borel measure"} in Scholium \ref{scholium "Lindelöf space, Borel measure plus sigma-algebra"}). Then $\bbmu$-almost any point $x \in \mathcal{X}$ is recurrent, or rather there exist integers $0 < n_1 < n_2 < n_3 \cdots$ for which
\begin{equation}
	\lim_{\rotatedell \to \infty}\varphi_\bbmu^{n_\rotatedell}(x) = x.
\end{equation}
\enumerationisfinis  	
\end{theorema}

Before proceeding with the theorem, it is good to dwell on some concepts mentioned above.

\subsubsection[On the Lindelöf Space, Borel Measure plus $\sigma$-Algebra]{On the Lindelöf Space, Borel Measure plus $\mathbold{\sigma}$-Algebra}
\label{subsubsection "On the Lindelöf Space, Borel Measure plus sigma-Algebra"}

\begin{scholium}[Lindelöf space, Borel measure plus $\sigma$-algebra]
\label{scholium "Lindelöf space, Borel measure plus sigma-algebra"}
~\enumerationisinitium
\item 
\label{item "Lindelöf space"}
A topological space $\mathcal{X}$ is said to be a \emph{Lindelöf space} \cite{Lindelof "Sur quelques points de la theorie des ensembles}\footnote{
	E.L. Lindelöf's first writing in which a space named in his honour appears is \cite{Lindelof "Sur quelques points de la theorie des ensembles}, but the systematic study of Lindelöf-type space(s) begins later, with C. Kuratowski and W. Sierpiński \cite{Kuratowski et Sierpinski "Le theoreme de Borel-Lebesgue dans la theorie des ensembles abstraits"}, whilst in Ru. literature it appears as \textcyrillic{\emph{финально компактные пространство}} (finally compact space), see P.S. Aleksandrov \cite[p. 27, Ru. version]{Aleksandrov "Some results in the theory of topological spaces obtained within the last twenty-five years"}.
	} 
if it is possible to extract a countable subcover from every open cover of $\mathcal{X}$. Recall that
\subenumerationisinitium
\item a metric space $(\mathcal{X}, \distance)$ is a Lindelöf space if every open cover of $\mathcal{X}$ has a subcover with a countable number of members; 
\item a space with a countable basis is as such a Lindelöf space (it is therefore a second-countable space); 
\item a separable metric space has the property of being Lindelöf (as well as a Lindelöf metric space has the property of being separable);
\item a $\sigma$-compact space is a Lindelöf space.
\subenumerationisfinis 
\item
\label{item "Borel measure"}
A \emph{Borel measure} on $\mathcal{X}$, see e.g. \cite{Borel "Un theoreme sur les ensembles mesurables"}, is a measure in which the Borel sets are measurable. We can conceive of \emph{Borel sets} as being the members of the smallest collection containing open, or closed, sets formed through countable unions and countable intersections, or else, if the Borel sets are in a Hausdorff $\sigma$-compact space, as being the members of the smallest $\sigma$-ring including compact sets. The collection of all Borel sets on $\mathcal{X}$ is known as \emph{Borel $\sigma$-algebra} (sometimes \emph{Borel $\sigma$-field}), and it is denoted by $\mathscr{B}_\sigma(\mathcal{X})$, or simply by $\mathscr{B}_\sigma$, thus the $\sigma$-algebra $\mathscr{B}_\sigma(\mathcal{X})$ on $\mathcal{X}$ is the smallest $\sigma$-algebra ($\sigma$-field) that contains all the open, or closed, Borel sets and that is closed under the operations of countable union and countable intersection.
\item 
\label{item "Primary version of Poincaré's theorem: alternative exposition"}
The statement \ref{item "Topological version of Poincaré's theorem"} of Theorem \ref{theorema "Poincaré—recurrence time"} is equivalent to the first but is topologically structured; in fact, in the statement \ref{item "Primary version of Poincaré's theorem"}, it would be the same if a \emph{probability space} $(\invertedbreve{\Omega}, \mathscr{B}_\sigma, \bbmu)$, with a Borel $\sigma$-algebra on $\invertedbreve{\Omega}$, were used; we should just specify that there is a set $\invertedbreve{F} \subset \invertedbreve{E}$, with $\bbmu(\invertedbreve{F}) = \bbmu(\invertedbreve{E})$, and, for any $x \in \invertedbreve{F}$, one has a sequence $0 < n_1 < n_2 < n_3 \cdots$ for which $\varphi_\bbmu^{n_\rotatedell}(x \in \invertedbreve{E})$, for every value $\rotatedell \geqslant 1$. \scholiumsymbol
\enumerationisfinis
\end{scholium}

\subsubsection{Step II}

\begin{proof}
 This is the resolution for both versions of Theorem \ref{theorema "Poincaré—recurrence time"}.
\enumerationisinitium
\item Taking $k \in \mathbb{N}$, and imposing the set $\invertedbreve{P}^k \equival \{x \in \invertedbreve{E} \mid \varphi_\bbmu^n({x} \notin \invertedbreve{E})\}$, $n \geqslant k$, we define 
\begin{equation}
	\invertedbreve{P}^k = \invertedbreve{E} \backslash \bigcup_{i \geqslant k}\varphi_\bbmu^{-i}(\invertedbreve{E}),
\end{equation}
for any $i \geqslant k$, so that $\invertedbreve{P}^k$ is measurable and $\invertedbreve{P}^k \cap \varphi_\bbmu^{-i}(\invertedbreve{P}^k) = \varnothing$. Letting $i - \rotatedell \geqslant k$, it ensues that 
\begin{equation}
	\varphi_\bbmu^{-i}(\invertedbreve{P}^k) \cap \varphi_\bbmu^{-\rotatedell}(\invertedbreve{P}^k) = \varnothing, 
\end{equation}
and that there is a sequence $\{\varphi_\bbmu^{-ik}(\invertedbreve{P}^k)\}_{i \in \mathbb{N}_0}$ of disjoint sets. Since $\varphi_\bbmu$ preserves $\bbmu$, the measure $\bbmu$ of each set is the same. Finally, it appears that $\bbmu(\invertedbreve{P}^k) = 0$, given that $\bbmu(\invertedbreve{\Omega})$ is finite. 
\item Fix a countable basis $\mathcal{B}_n$ for the topology of $\mathcal{X}$, and this shows that $\mathcal{X}$ is second-countable and, for this reason, separable (is, in short, a Lindelöf space). Indicating by $\widetilde{\mathcal{X}}$ the full collection of infinitely recurrent points $x \in \mathcal{X}$ and by $\mathcal{Y}^n$ the set of not infinitely recurrent points $x \in \mathcal{B}_n$ to the countable basis, it happens that
\begin{equation}
	\mathcal{X} \backslash \widetilde{\mathcal{X}} = \bigcup_n\mathcal{Y}^n.
\end{equation}
The above-mentioned result indicates that $\bbmu(\mathcal{Y}^n) = 0$, from which derives $\bbmu(\mathcal{X} \backslash \widetilde{\mathcal{X}}) = 0$. 
\enumerationisfinis
\end{proof}

At this stage, let us take a glance at the intertwining of Liouville's \ref{theorema "Liouville invariance"} and Poincaré's \ref{theorema "Poincaré—recurrence time"} Theorems.

\begin{exemplum}[Liouvillian volume preserving system and stability à la Poisson in Poincareian method]
\label{exemplum "Liouvillian volume preserving system and stability à la Poisson in Poincareian method"}
Let $\Upsilon \subset \mathbb{R}^n$ be an open subset of an $n$-dimensional Euclidean space. Let $\bblambda$ be the \emph{Lebesgue measure} on $\Upsilon$. Let $\textgreek{\text{\ddigamma}} \in \mathscr{C}^\infty(\Upsilon)$ be a continuously differentiable map from $\Upsilon$ into $\mathbb{R}^n$, i.e. $\textgreek{\text{\ddigamma}} \colon \Upsilon \subset \mathbb{R}^n \to \mathbb{R}^n$, where $\textgreek{\text{\ddigamma}} = \textgreek{\text{\ddigamma}}_1\sigma_1 + \cdots + \textgreek{\text{\ddigamma}}_n\sigma_n$. Note that $\textgreek{\text{\ddigamma}}$ is also called \emph{vector function of position}, or more commonly \emph{vector field} on $\Upsilon$, since $\textgreek{\text{\ddigamma}}$ is a relation which associates one vector $\textgreek{\text{\ddigamma}}(x)$ with each point. We identify the \emph{divergence} (operator) of $\textgreek{\text{\ddigamma}}$ with the \emph{scalar field}, that is, the real function
 \begin{equation}
 	\divergence{\textgreek{\text{\ddigamma}}} = \nabla \cdot \textgreek{\text{\ddigamma}} = \frac{\partial\textgreek{\text{\ddigamma}}_1}{\partial{x_1}} + \cdots + \frac{\partial\textgreek{\text{\ddigamma}}_n}{\partial{x_n}} = \sum^n_{\nu = 1}\frac{\partial\textgreek{\text{\ddigamma}}_\nu}{\partial{x_\nu}},
\end{equation} 
indicating the outward \emph{flux density} of $\textgreek{\text{\ddigamma}}$. We know that
\enumerationisinitium
\item according to Liouville's Theorem \ref{theorema "Liouville invariance"}, a geodesic flow $\varphi_t$ within the Hamiltonian action-flow framework (meaning that the first flow is a special case of the second one) preserves the volume measure, or the phase volume, on the manifold, 
\item $\bblambda$ is flow invariant, it does not change under $\varphi_t$,
\enumerationisfinis
iff $\divergence{\textgreek{\text{\ddigamma}}} = 0$ at each point, or iff $\textgreek{\text{\ddigamma}}$ is divergence free. Denoting by $\invertedbreve{E}$ a compact (closed and bounded) set, for any value $\alpha \in \mathbb{R}$, it follows that 
\begin{equation}
	\bblambda\bigl(\varphi_{t + \alpha}(\invertedbreve{E})\bigr) = \int_{\varphi_{t + \alpha}(\invertedbreve{E})}dx = \int_{\varphi_t(\invertedbreve{E})} \bigg|\det\left(\frac{\partial\varphi_\alpha(x)}{\partial_x}\right)\bigg|dx.  
\end{equation}
Let $\mathscr{O}$ be the \emph{Bachmann–Landau} notation, or \emph{Landau symbol} (big-oh and little-oh) \cite[p. 61]{Landau "Handbuch der Lehre von der Verteilung der Primzahlen I"} \cite[p. 883]{Landau "Handbuch der Lehre von der Verteilung der Primzahlen II"}, and consider a \emph{Taylor series expansion} \cite{Taylor "Methodus Incrementorum Directa et Inversa"} at the point $x$ of a system $\dot{x} = \textgreek{\text{\ddigamma}}(x)$, in which $\dot{x}$ is the derivative of $\textgreek{\text{\ddigamma}}$ with respect to the single independent variable $x$. We have 
\begin{equation}
	\det\left(\frac{\partial\varphi_\alpha(x)}{\partial{x}}\right) = 1 + \divergence{\textgreek{\text{\ddigamma}}}(x)\alpha + \mathscr{O}(\alpha).
\end{equation}
Read $\mathscr{O}(\alpha)$ as \emph{oh of $\alpha$}, and it stands for \emph{order of magnitude}. Then
\begin{equation}
	\frac{d}{dt}\bblambda\bigl(\varphi_t(\invertedbreve{E})\bigr) = \lim_{\alpha \to 0}\int_{\varphi_t(\invertedbreve{E})}\frac{\divergence{\textgreek{\text{\ddigamma}}}(x)\alpha + \mathscr{O}(\alpha)}{\alpha}dx = \int_{\varphi_t(\invertedbreve{E})}\divergence{\textgreek{\text{\ddigamma}}}(x)dx.
\end{equation} 
Through the Poincaré recurrence Theorem \ref{theorema "Poincaré—recurrence time"}, for almost any $y \in \Upsilon$, the equation
\begin{equation}
	\liminf_{t \to \infty}d\bigl(\varphi_t(y), y\bigr) \leqslant \liminf_{n \to \infty}d\bigl(\varphi_n(y), y\bigr) = 0
\end{equation}
holds. It goes to show that in a dynamical system like this there is an \emph{orbital stability}, called \emph{Poisson stability} and unearthed by Poincaré \cite[chap. XXVI]{Poincare "Les Methodes nouvelles de la Mecanique celeste III"} in studying the celestial mechanics, so that almost all recurrent points are non-wandering, or \emph{Poisson-stable}. \exemplumsymbol
\end{exemplum}

\subsection{Continuous Flow on a Hausdorff Space Containing Recurrent Points}
\label{subsection "Continuous Flow on a Hausdorff Space Containing Recurrent Points"}

\begingroup
\footnotesize
A topological space is nothing other than a set of arbitrary elements (called “points” of the space) in which a concept of continuity is defined. Now this concept of continuity is based on the existence of relations, which may be defined as local or neighborhood relations—it is precisely these relations which are preserved in a continuous mapping of one figure onto another. Therefore, in more precise wording, a topological space is a set in which certain subsets are defined and are associated to the points of the space as their neighborhoods. Depending upon which axioms these neighborhoods satisfy, one distinguishes between different types of topological spaces. The most important among them are the so-called Hausdorff spaces. \\
\indent — \textsc{P.S. Aleksandrov} \cite[pp. 8-9]{Aleksandrov "Elementary Concepts of Topology"}, see also \cite{Aleksandrov "Zur Begrundung der $n$-dimensionalen mengentheoretischen Topologie"}

\endgroup

\vspace{2mm}

Poincaré's Theorem \ref{theorema "Poincaré—recurrence time"} can be efficiently applied to the highly generic notion of flow as \emph{continuous dynamical system} on spaces with a topology introduced by F. Hausdorff \cite[Kap. VII]{Hausdorff "Grundzuge der Mengenlehre"}.

\begin{margo}[Hausdorff space]
\label{margo "Hausdorff space"}
A topological space $\mathcal{X}$ is said \emph{Hausdorff space}, or \emph{$\mathscr{T}_2$ space},\footnote{
	The letter $\mathscr{T}$ (from the Ge. \emph{Trennungsaxiom}) is for \emph{separation axiom}, in the terminology of H. Tietze \cite[p. 291]{Tietze "Beitrage zur allgemeinen Topologie. I. Axiome fur verschiedene Fassungen des Umgebungsbegriffs"}, the subscript \emph{2} is just a classification for Hausdorffian spaces. 
	}
if there are two distinct points $x \neq y$ of $\mathcal{X}$ admitting disjoint neighborhoods, e.g. $E_\mu$ and $E_\nu$, such that $x \in E_\mu$, $y \in E_\nu$ and $E_\mu \cap E_\nu = \varnothing$. \margosymbol
\end{margo}

\begin{theorema}
Let us say that $\varphi_t$ is a generic flow representing a continuous map on a Hausdorff space $\mathcal{X}$. Let $\{\mathfrak{P}\}_\mathrm{set} \viz \{\mathfrak{P}\}_\mathrm{set}(\varphi_t)$ be the Poincareian set of recurrent $x\text{-points} \in \varphi_t$. Resorting to Poisson stability (Example \ref{exemplum "Liouvillian volume preserving system and stability à la Poisson in Poincareian method"}), one can see that $\{\mathfrak{P}\}_\mathrm{set}$ is flow invariant.
\end{theorema}

\begin{proof}
We have a Poincaré recurrent point $x \in \{\mathfrak{P}\}_\mathrm{set}$ of $\varphi_t$, a value $\alpha \in \mathbb{R}$, and a sequence $\{t_n\}_n \in \mathbb{R}$ with $\lim_{n \to \infty}t_n = \infty$. Then
\begin{equation}
	\varphi_\alpha(x) = \varphi_\alpha\left(\lim_{n \to \infty}\varphi_{t_n}(x)\right) = \lim_{n \to \infty}\varphi_\alpha\bigl(\varphi_{t_n}(x)\bigr) = \lim_{n \to \infty}\varphi_{t_n}\bigl(\varphi_\alpha(x)\bigr).
\end{equation}
We then infer that $\varphi_\alpha(x) \in \{\mathfrak{P}\}_\mathrm{set}$, so $\{\mathfrak{P}\}_\mathrm{set}$ is flow invariant on $\mathcal{X}$.
\end{proof}

\vspace{10mm}

\setcounter{secnumdepth}{0}  
\section{References and Bibliographic Details}
\setcounter{secnumdepth}{3}
\markright{References and Bibliographic Details}

\begingroup
\footnotesize
\noindent Section \ref{subsection "Geodesic Flow from Lobačevskijan Geometry"}

\begin{indent paragraph: 15pt}
As intro about the geodesic flow on surfaces of constant negative curvature, see P.[B.] Eberlein \cite{Eberlein "Geodesic Flows on Negatively Curved Manifolds I"}.
\end{indent paragraph: 15pt}

\noindent Section \ref{subsection "Projective Linear Transforms: Dynamics on the Modular Surface, and Horocycle Flow"} 

\begin{indent paragraph: 15pt}
The general readings for this Section are \cite[sec. 4.5]{Adler "Geodesic flows interval maps and symbolic dynamics"} \cite[pp. 57-60, 111-113]{Bekka Mayer "Ergodic Theory and Topological Dynamics of Group Actions on Homogeneous Spaces"} \cite[secc. 9.1.1, 9.3]{Bergeron "The Spectrum of Hyperbolic Surfaces"} \cite[sec. 4.4]{Coudene "Ergodic Theory and Dynamical Systems"} \cite{Dani "An introduction to flows on homogeneous spaces"} \cite[pp. 281-288]{Einsiedler Ward "Ergodic Theory with a view towards Number Theory"} \cite[pp. 267-270]{Einsiedler and Ward "Diophantine Problems and Homogeneous Dynamics"} \cite[sec. 2]{Fisher "Small-scale Structure via Flows"} \cite{Manning "Dynamics of geodesic and horocycle flows on surfaces of constant negative curvature"} \cite[pp. 424-428]{Marklof "Theta Sums Eisenstein Series and the Semiclassical Dynamics of a Precessing Spin"} \cite[pp. 147-148]{Mayer "Transfer Operators the Selberg Zeta Function and the Lewis-Zagier Theory of Period Functions"} \cite[sec. 10.3]{Nicholls "The Ergodic Theory of Discrete Groups"} \cite{Schapira "Dynamics of Geodesic and Horocyclic Flows"}. Look for references to $SL_2(\mathbb{R})$ in \cite{Dani "Flows on homogeneous spaces: a review"} \cite[sec. 10.1.2]{Einsiedler and Luethi "Kloosterman Sums Disjointness and Equidistribution"} \cite{Forni "Effective Equidistribution of Nilflows and Bounds on Weyl Sums"} \cite[sec. 6.1.1]{Kontorovich "Applications of Thin Orbits"} \cite[sec. 14.1.5]{Matheus "The Lagrange and Markov Spectra from the Dynamical Point of View"} \cite[pp. 170-172]{Spatzier "Harmonic Analysis in Rigidity Theory"} \cite[sec. 2]{Thouvenot "Some Properties and Applications of Joinings in Ergodic Theory"}. — On the horocycle flow, see e.g. \cite[chap. V]{Dal'Bo "Geodesic and Horocyclic Trajectories"} \cite{Forni "Limit theorems for horocycle flows"} \cite{Ratner "Rigidity of horocycle flows"} \cite[chap. 6]{Glasner "Ergodic Theory via Joinings"} \cite{Ratner "The rate of mixing for geodesic and horocycle flows"} \cite{Sarig "Unique Ergodicity for Infinite Measures"}; for a relation between horocycles (in the Beltrami–Poincaré disk model) and Apollonian gaskets in a geodesic dynamics context, see \cite{Oh "Dynamics on Geometrically Finite Hyperbolic Manifolds with Applications to Apollonian Circle Packings and Beyond"}. — For the horocycle orbits, see \cite{Dani and Smillie "Uniform distribution of horocycle orbits for Fuchsian groups"}.
\end{indent paragraph: 15pt}

\noindent Section \ref{subsection "Continuous Function on the Unit Tangent Bundle with Compact Support"}

\begin{indent paragraph: 15pt}
On the Fuchsian/discrete group, hyperbolic surfaces, and geodesic flow, cf. \cite[sec. 2]{Borthwick "Introduction to Spectral Theory on Hyperbolic Surfaces"} \cite[pp. 213-215]{Katok Hasselblatt "Introduction to the Modern Theory of Dynamical Systems"}, and especially S. Katok \cite{Katok "Closed geodesics periods and arithmetic of modular forms"} \cite{Katok "Fuchsian groups geodesic flows on surfaces of constant negative curvature and symbolic coding of geodesics"}. — On the continuous functions with compact support and related vector spaces, see e.g. \cite[pp. 54-55]{Bachman Narici Beckenstein "Fourier and Wavelet Analysis"} \cite[sec. 5.1]{Christensen "Functions Spaces and Expansions: Mathematical Tools in Physics and Engineering"} \cite[p. 178]{Conway "Functions of One Complex Variable II"} \cite[secc. 12.1-12.2]{Kubrusly "Essentials of Measure Theory"}; see also \cite[§ 3]{Schneider "Nonarchimedean Functional Analysis"}. — On the geodesic flows as measure preserving in this context, see \cite[pp. 160-162]{Gromoll Walschap "Metric Foliations and Curvature"}.
\end{indent paragraph: 15pt}

\noindent Section \ref{subsection "Leafy Stratifications for the Geodesic Flow by Tangent Vectors"}

\begin{indent paragraph: 15pt}
Example \ref{exemplum "Stable and unstable foliations, geodesic and horocycle flows"}: on the stable and unstable foliations, geodesic and horocycle flows, see \cite[pp. 12-13, 205-208]{Barreira Pesin "Introduction to Smooth Ergodic Theory"} \cite{Hubbard and Miller "Equidistribution of Horocyclic Flows on Complete Hyperbolic Surfaces of Finite Area"} \cite[sec. 7.3.2]{Parkkonen and Paulin "A Survey of Some Arithmetic Applications of Ergodic Theory in Negative Curvature"} \cite[pp. 29-46]{Pesin "Lectures on partial hyperbolicity and stable ergodicity"} \cite{Wilkinson "Conservative Partially Hyperbolic Dynamics"}; on the horocycle foliation, see \cite{Hirsch Pugh "Smoothness of horocycle foliations"} \cite{Bowen and Marcus "Unique ergodicity for horocycle foliations"} \cite{Ballmann Brin Burns "On the differentiability of horocycles and horocycle foliations"}, moreover, see \cite[pp. 266-269]{Kaimanovich "SAT Actions and Ergodic Properties of the Horosphere Foliation"} and \cite[p. 497]{Charitos and Papadoperakis "Parameters for generalized Teichmuller spaces"} \cite[pp. 664-666]{Fock and Goncharov "Dual Teichmuller and lamination spaces"} \cite[pp. 152-154, 186-187]{Papadopoulos and Theret "On Teichmuller's metric and Thurston's asymmetric metric on Teichmuller space"}.
\end{indent paragraph: 15pt}

\noindent Section \ref{subsection "Invariant and Geodesible Foliations"}

\begin{indent paragraph: 15pt}
On the fact that the induced Riemannian metric along the leaves is invariant, cf. \cite[pp. 39-40]{Barletta Dragomir Duggal "Foliations in Cauchy-Riemann Geometry"} \cite[p. 36]{Rovenskii "Foliations on Riemannian Manifolds and Submanifolds"} \cite[pp. 70-71]{Tondeur "Geometry of Foliations"}. — On the totally (and almost totally) geodesic foliation, see \cite[pp. 150-158]{Czarnecki Walczak "Extrinsic geometry of foliations"}. — A summary of the hyperbolic foliation is in \cite[sec. 18]{Hurder "Classifying foliations"}. — On the non-existence of totally geodesic foliations in (positively or negatively curved) pseudo-Riemannian manifolds, see \cite[p. 133]{Bejancu Farran "Foliations and Geometric Structures"}.
\end{indent paragraph: 15pt}

\noindent Section \ref{subsection "Diffeomorphism and Flow on Negatively Curved Surfaces"}

\begin{indent paragraph: 15pt}
On the Anosov flow, see \cite{Plante "Anosov Flows"} and \cite[pp. 8-11, 75-80]{Margulis "On Some Aspects of the Theory of Anosov Systems With a Survey by R. Sharp: Periodic Orbits of Hyperbolic Flows"}. —	Definition \ref{definitio "Anosov diffeomorphism"}: cf. \cite[pp. 113-119]{Pesin "General Theory of Smooth Hyperbolic Dynamical Systems"}. — Example \ref{exemplum "Hyperbolic toral automorphism"}: see \cite[pp. 70-71]{Franks "Anosov Diffeomorphisms"} \cite{Manning "There are No New Anosov Diffeomorphisms on Tori"}. — Definition \ref{definitio "Anosov flow"}: for a synopsis on the Anosov diffeomorphism/flow, see \cite[chap. III, sec. 2]{Mane "Ergodic Theory and Differentiable Dynamics"} \cite[sec. 7.1]{Oliva "Geometric Mechanics"} \cite[sec. 2.1]{Pesin "Lectures on partial hyperbolicity and stable ergodicity"}; on the pseudo-Anosov flows, see \cite{Calegari "Foliations and the Geometry of 3-Manifolds"}. — Scholium \ref{scholium "Tangent bundle in the Anosov system"}: see \cite[pp. 729, 732, 734-736, 743]{Plante "Anosov Flows"} \cite[pp. 54-55]{Walczak "Dynamics of Foliations Groups and Pseudogroups"}, and cf. \cite[pp. 8-9]{Barreira Pesin "Introduction to Smooth Ergodic Theory"} \cite[sec. 1.1]{Pinto Rand Ferreira Fine "Structures of Hyperbolic Diffeomorphisms"}.
\end{indent paragraph: 15pt}

\noindent Section \ref{subsection "Hölder Continuity of Subspaces in a Map with the Anosov Property"}

\begin{indent paragraph: 15pt}
On the Hölder continuity, see e.g. \cite[sec. 4.1]{Gilbarg Trudinger "Elliptic Partial Differential Equations of Second Order"} \cite[sec. 1.1]{Fiorenza "Holder and locally Holder Continuous Functions"}. — The proof of Brin is restated by L. Barreira and Ya.B. Pesin in several of their works, see e.g. \cite[pp. 48-51]{Barreira Pesin "Lyapunov Exponents and Smooth Ergodic Theory"}, alternatively, see \cite[pp. 92-93]{Ballmann "Lectures on Spaces of Nonpositive Curvature"}. 
\end{indent paragraph: 15pt}

\noindent Section \ref{subsection "Structural Stability: Andronov–Pontrjagin Criterion"}
	
\begin{indent paragraph: 15pt}
Example \ref{exemplum "Harmonic and van der Pol oscillator"}: on the dynamics of an oscillator in the presence of friction, see \cite[chap. I, § 4]{Andronov Vitt and Khaikin "Theory of Oscillators"}; about oscillators and chaos, see \cite{Guckenheimer "The Birth of Chaos"} \cite[secc. 5.3-5.6]{Lakshmanan Rajasekar "Nonlinear Dynamics: Integrability Chaos and Patterns"}. — Definition \ref{definitio "Structurally stable Anosov system"}: on the structural stability, see \cite[pp. 89-97, 124, 129-133]{Arnold "Geometrical Methods In The Theory Of Ordinary Differential Equations"} \cite[pp. 126-131]{Anosov "On the development of the theory of dynamical systems during the past quarter century"} \cite[sec. 1.3]{Guo Wu "Bifurcation Theory of Functional Differential Equations"} \cite[sec. 2.5]{Kuznetsov "Elements of Applied Bifurcation Theory"} \cite[pp. 15-16]{Pilyugin Sakai "Shadowing and Hyperbolicity"} \cite[pp. 170-171]{Ruelle "Elements of Differentiable Dynamics and Bifurcation Theory"}. — Definition \ref{definitio "structurally stable foliation"}: see \cite[pp. 21, 59-65]{Nikolaev "Foliations on Surfaces"}. — Scholium \ref{scholium "Morse–Smale system, or hyperbolic strange attractors"}: on the Peixoto's theorem, see \cite[pp. 60-65, 258-259]{Guckenheimer Holmes "Nonlinear Oscillations Dynamical Systems and Bifurcations of Vector Fields"}; for a classification of 2-dimensional basic sets, see \cite[pp. 180-185]{Grines Medvedev Pochinka "Dynamical Systems on 2- and 3-Manifolds"}; on the bifurcation theory, see e.g. \cite[chapp. 3, 6-7]{Guckenheimer Holmes "Nonlinear Oscillations Dynamical Systems and Bifurcations of Vector Fields"} \cite[chap. 4]{Perko "Differential Equations and Dynamical Systems"}; a study that specifically looks at the flows (including that of Anosov) in 3-dimensional spaces is \cite{Araujo Pacifico "Three-Dimensional Flows}.
\end{indent paragraph: 15pt}

\noindent Section \ref{section "Integrability and Recurrence of Flow: Models in Comparison"}

\begin{indent paragraph: 15pt}
On the (complete) integrable Hamiltonian systems (including integrable geodesic flows in the sense of Liouville), see e.g. \cite[chapp. 11-13]{Bolsinov and Fomenko "Integrable Hamiltonian Systems: Geometry Topology Classification"} \cite{Bolsinov Jovanovic "Integrable Geodesic Flows on Riemannian Manifolds: Construction and Obstructions"} \cite{Hitchin "Riemann surfaces and integrable systems"}. — On the appearance of chaos by introducing a Fuchsian group, cf. e.g. \cite[p. 526 ff.]{Chang and Mayer "An Extension of the Thermodynamic Formalism Approach to Selberg's Zeta Function for General Modular Groups"}. — Margo \ref{margo "Eulerian equations of motion for rigid bodies and incompressible fluids as geodesic flows"}: reports on the Euler–Arnold equation are in T. Tao \cite[sec. 5.4]{Tao "Compactness and Contradiction"}; on the incompressible flows, see e.g. \cite{Ma Wang "Geometric Theory of Incompressible Flows with Applications to Fluid Dynamics"} \cite{Bardos and Titi "Euler equations for incompressible ideal fluids"} \cite{Bardos "Equazioni di Eulero e di Navier-Stokes per fluidi ideali incomprimibili"} \cite{Ambrosio and Figalli "Lecture notes on variational models for incompressible Euler equations"} \cite{Brenier "Some Variational and Stochastic Methods for the Euler Equations of Incompressible Fluid Dynamics and Related Models"}.
\end{indent paragraph: 15pt}

\noindent Section \ref{subsection "Liouville Measure: Integral Invariant of Hamiltonian Dynamics"}

\begin{indent paragraph: 15pt}
Scholium \ref{scholium "On the correspondence between the geodesic and Hamiltonian flows"}: on the (Hamiltonian) cogeodesic flow, see \cite[pp. 27-28]{Geiges "An Introduction to Contact Topology"}; about the $n$-body problem, see Meyer's book \cite{Meyer "Periodic Solutions of the N-Body Problem"}.
\end{indent paragraph: 15pt}	

\noindent Section \ref{subsection "Symplectic Geometry; Liouville's, Noether's and Poisson's Theorems"}

\begin{indent paragraph: 15pt}
In support of Definitions \ref{definitio "Hamiltonian vector field or symplectic gradient"} and \ref{definitio "Hamiltonian flow"}, as well Theorems of \ref{theorema "Liouville invariance"} and \ref{theorema "Noether—on symplectic transformations"}, and of Corollary \ref{corollarium "Poisson"}, see \cite[pp. 41-42, 90-91]{Audin Cannas da Silva Lerman "Symplectic Geometry of Integrable Hamiltonian Systems"} \cite[pp. 167-172, 177-179, 443-449]{Feng Qin "Symplectic Geometric Algorithms for Hamiltonian Systems"} \cite[secc. 1.1.-1.2]{Gaeta Rodriguez "Lectures on Hyperhamiltonian Dynamics and Physical Applications"} \cite[pp. 9-23, 105-106, 143-150]{Hofer Zehnder "Symplectic Invariants and Hamiltonian Dynamics"} \cite[sec. 4.1]{Hu and Yang "Differentiable and Complex Dynamics of Several Variables"} \cite[chap. 5, sec. 5]{Katok Hasselblatt "Introduction to the Modern Theory of Dynamical Systems"} \cite[secc. 2.4-2.5]{Koszul Zou "Introduction to Symplectic Geometry"} \cite[chap. III, secc. 4-6, pp. 96-105]{Libermann and Marle "Symplectic Geometry and Analytical Mechanics"} \cite[secc. 4.4.2-4.4.5]{Viana Oliveira "Foundations of Ergodic Theory"} \cite[chap. V]{Zehnder "Lectures on Dynamical Systems: Hamiltonian Vector Fields and Symplectic Capacities"}, see also \cite{Trofimov Fomenko "Geometrie and Algebraic Mechanisms of the Integrability of Hamiltonian Systems on Homogeneous Spaces and Lie Algebras"}. Furthermore, the following is added: on the Hamilton's differential equations, cf. e.g. \cite[sec. 3.3.4, and pp. 358-359]{Lee Leok McClamroch "Global Formulations of Lagrangian and Hamiltonian Dynamics on Manifolds. A Geometric Approach to Modeling and Analysis"}; for a reconstruction of the Hamilton–Jacobi equation, see \cite{Nakane and Fraser "The Early History of Hamilton-Jacobi Dynamics 1834-1837"}; on the Lie derivative, see \cite{Trautman "Remarks on the History of the Notion of Lie Differentiation"}; about the Noether's theorem(s), see \cite{Kosmann-Schwarzbach "The Noether Theorems: Invariance and Conservation Laws in the Twentieth Century"}; on the chaotic motion, in comparison with the corresponding regular behavior, in Hamiltonian systems, see e.g. \cite{Varvoglis "Regular and Chaotic Motion in Hamiltonian Systems"} \cite{Zaslavsky "Hamiltonian Chaos and Fractional Dynamics"}; about the instability and chaotic action of geodesic flows, see \cite{Szydlowski "The Generalized Local Instability Criterion from the Geodesic Deviation Equation"}. — Margo \ref{margo "Hamiltonian chaos through a geometrization of dynamics"}: on the 3-body problem, cf. \cite{Safaai Saadat "On the Prediction of Chaos in the Restricted Three-Body Problem"}; review and insights on the Levi-Civita equation of geodesic deviation are in \cite{Rund "The equation of geodesic deviation of Levi-Civita: its generalizations and implications"}.	
\end{indent paragraph: 15pt}

\noindent Section \ref{subsection "Poincaré Recurrence Theorem"}

\begin{indent paragraph: 15pt}
A concise-careful presentation to Poincareian topological (chaotic) dynamics is in C. Bartocci \cite{Bartocci "Equazioni e orbite celesti: gli albori della dinamica topologica"}. — Theorem \ref{theorema "Poincaré—recurrence time"}: on Liouville's \ref{theorema "Liouville invariance"} and Poincaré's Theorems, cf. \cite[pp. 8-10]{Bekka Mayer "Ergodic Theory and Topological Dynamics of Group Actions on Homogeneous Spaces"} \cite[37-39]{Zhang "Integrability of Dynamical Systems: Algebra and Analysis"}; for recent developments on the Poincaré recurrence, see e.g. \cite{Barreira "Poincare recurrence: old and new"}. — Scholium \ref{scholium "Lindelöf space, Borel measure plus sigma-algebra"}: on the Lindelöf space, cf. e.g. \cite[pp. 144-145]{Bourbaki "Elements of Mathematics: General Topology 1"} \cite[pp. 61-64]{Heinonen Koskela Shanmugalingam Tyson "Sobolev Spaces on Metric Measure Spaces: An Approach Based on Upper Gradients"}; on the Borel $\sigma$-algebra, see e.g. \cite[sec. 1.2]{Bogachev "Measure Theory I"} and \cite[sec. 6.5]{Bogachev "Measure Theory II"} \cite[pp. 83-84]{Srivastava "A Course on Borel Sets"}.
\end{indent paragraph: 15pt}

\noindent Section \ref{subsection "Continuous Flow on a Hausdorff Space Containing Recurrent Points"}

\begin{indent paragraph: 15pt}
On the Theorems of Liouville \ref{theorema "Liouville invariance"} and Poincaré \ref{theorema "Poincaré—recurrence time"}, and Hausdorff space, cf. \cite[pp. 51-52, 159-161]{Alongi Nelson "Recurrence and Topology"}.
\end{indent paragraph: 15pt}

\endgroup

\chapter{Geometric and Topological Aspects of Complexity and Dynamics, Part II. Ergodicity and Entropy}
\chaptermark{Geometric and Topological Aspects of Complexity and Dynamics, Part II}{}
\label{chapter "Geometric and Topological Aspects of Complexity and Dynamics, Part II: Ergodicity and Entropy"}

\begingroup
\footnotesize
Every hypothesis must derive indubitable results from mechanically well-defined assumptions by mathematically correct methods. If the results agree with a large series of facts, \emph{we must be content, even if the true nature of facts is not revealed in every respect}. No one hypothesis has hitherto attained this last end, the Theory of Gases not excepted [\,\dots]. [I]n gases certain entities [\,\dots] [c]an it be seriously expected that they will behave exactly as aggregates of Newtonian centres of force, or as the rigid bodies of our Mechanics? And \emph{how awkward is the human mind in divining the nature of things, when forsaken by the analogy} [cf. Section \ref{section "The Role of Analogy"}] \emph{of what we see and touch directly?} \\
\indent — \textsc{L. Boltzmann} \cite[pp. 413-414, e.a.]{Boltzmann "On Certain Questions of the Theory of Gases}

\endgroup

\section{A Framework for Ergodicity}
\label{section "A Framework for Ergodicity"}

In this Section we explore the subject of ergodicity, initially with a premise regarding its theoretical origins, by drawing attention to the germinal input of Maxwell and Boltzmann. We have previously dealt with the Poincaré recurrence Theorem (Section \ref{subsection "Poincaré Recurrence Theorem"}), which is the qualitative foundation of ergodic theory. We shall take up here some more technical statements: the ergodic Theorems of Birkhoff  \ref{theorema "Birkhoff's ergodic Theorem"} and Anosov \ref{theorema "Anosov's Ergodic Theorem"}, and the Hopfian statistical process \ref{theorema "Hopfian statistical process"} about the ergodicity of the geodesic flow.

\subsection{Prior Knowledge: Maxwell–Boltzmann Probability Distribution, and Ergodic Hypothesis of Thermodynamics}
\label{subsection "Prior Knowledge: Maxwell–Boltzmann Probability Distribution, and Ergodic Hypothesis of Thermodynamics"}

\begingroup
\footnotesize
The only assumption which is necessary for the direct proof [of Boltzmann's \cite{Boltzmann "Studien uber das Gleichgewicht der lebendigen Kraft zwischen bewegten materiellen Punkten"} theorem about the solution of the problem of the equilibrium of kinetic energy among a finite number of material points, or the equilibrium of temperature in a liquid or solid system]\footnote{
	The problem of defining the mathematical condition of the equilibrium of energy as the dynamical representative of the physical condition of the equality of temperature e.g. for a system of gas molecules.
	}
is that the system, if left to itself in its actual state of motion, will, sooner or later, pass through every phase which is consistent with the equation of energy. \\
\indent — \textsc{J.C. Maxwell} \cite[p. 548]{Maxwell "On Boltzmann's Theorem on the average distribution of energy in a system of material points"}

\vspace{2mm}

As Maxwell \cite{Maxwell "On Boltzmann's Theorem on the average distribution of energy in a system of material points"} has demonstrated, if the distribution of systems is a completely stationary one, as long as the values of the slowly-varying coordinates are constant, the number of systems, for which the coordinates and momenta lie between the [above] limits, will always remain the same. (I have proposed the name of \emph{Ergoden} for such a totality of systems). \\
\indent — \textsc{L. Boltzmann} \cite[p. 208]{Boltzmann "Ueber die mechanischen Analogien des zweiten Hauptsatzes der Thermodynamik"}

\endgroup

\vspace{2mm}

\enumerationisinitium
\item It should first be said that the idea of ergodicity was born with L. Boltzmann \cite{Boltzmann "Ueber die mechanischen Analogien des zweiten Hauptsatzes der Thermodynamik"}, as regards the need to describe statistically the action of a transformation relating to a set of microstates of a thermodynamic system on a constant-energy surface. Microstates are  microscopic configurations (such as position and momentum, or energy) that the system can assume exhibiting the distinctive thermal fluctuations. The probability of finding the system in one of its microstates is described by a function that  furnishes the probabilities of occurrence of possible states with reference to a statistical ensemble of many microscopic configurations. This function is what characterizes the macroscopic properties (e.g. temperature, pressure, volume, density) of the system.
\subenumerationisinitium
\item The ergodicity is connected with the notion of \emph{probability distribution (uniformly distributed)} with random mechanical processes, and it takes as its springboard the revealing glimpses of J.C. Maxwell \cite{Maxwell "Illustrations of the Dynamical Theory of Gases. Part I", Maxwell "Illustrations of the Dynamical Theory of Gases. Part II"} \cite{Maxwell "On Boltzmann's Theorem on the average distribution of energy in a system of material points"} and Boltzmann \cite{Boltzmann "Studien uber das Gleichgewicht der lebendigen Kraft zwischen bewegten materiellen Punkten"} \cite{Boltzmann "Einige allgemeine Satze uber Warmegleichgewicht"} \cite{Boltzmann "Analytischer Beweis des zweiten Hauptsatzes der mechanischen Warmetheorie aus den Satzen uber das Gleichgewicht der lebendigen Kraft"} \cite{Boltzmann "Ueber die Eigenschaften monocyklischer und anderer damit verwandter Systeme"}; see also Kelvin \cite{Kelvin "On Some Test Cases for the Maxwell-Boltzmann Doctrine regarding Distribution of Energy"} \cite[postscript]{Kelvin "On Periodic Motion of a Finite Conservative System"}. According to the distribution of the system, the statistical \emph{ensemble average} of possible microstates and the respective \emph{time average} on the evolution of initial conditions are identical along the orbit of any point moving in the \emph{phase space} (the space in which to each point corresponds a possible state of the system), see Eq. \eqref{equation "Boltzmann–Birkhoff Ergodic Hypothesis"}.
\item 
\label{item "Ergodic hypothesis of thermodynamics, Boltzmann equation, and H-theorem"}
The supposition of Boltzmann, commonly called \emph{ergodic hypothesis} of thermodynamics, provides in origin that, for a thermal system, the orbit of each phase point equals the whole of the surface of constant-energy. This approach stems from the need to show that the evolution of a system proceeds towards the equilibrium state, starting from an initial non-equilibrium state, e.g. from a state that initially occupies a small region of the energy surface (volume). The \emph{Boltzmann (transport) equation} and the \emph{$\mathsf{H}$-theorem}\footnote{
	The letter originally chosen by Boltzmann is $\mathsf{E}$. The adoption of the letter $\mathsf{H}$ comes from S.H. Burbury \cite{Burbury "On some Problems in the Kinetic Theory of Gases"}, and then it is also accepted by Boltzmann \cite{Boltzmann "Vorlesungen uber Gastheorie I"}. E.g. in \cite[p. 59]{Boltzmann "Vorlesungen uber Gastheorie I"} Boltzmann writes explicitly: «[\,\dots] apart from a constant, $-\mathsf{H}$ represents the logarithm of the probability of the state of the gas analyzed [\textit{Logarithmus der Wahrscheinlichkeit des betreffenden Zustandes des Gases darstellt}]». The letter $\mathsf{H}$ can be interpreted as the La. H (\emph{aitch}) or even the Gr. \textgreek{Η, η} (\emph{eta}) \cite{Brush "Boltzmann's "Eta Theorem": Where's the Evidence?"}, as appears to be the case of J.W. Gibbs \cite[p. 309]{Gibbs "Graphical Methods In The Thermodynamics of Fluids"} \cite[e.g. chap. IV]{Gibbs "Elementary principles in statistical mechanics"}.
	} 
(see the quick reminder \ref{item "Boltzmann's H and H-theorem"} in Definition \ref{definitio "Kolmogorov–Sinai metric entropy"}), both presented in \cite{Boltzmann "Weitere Studien uber das Warmegleichgewicht unter Gasmolekulen"}, contribute to explain a similar evolutionary behavior, that if we adopt a probability distribution, the Maxwell–Boltzmann distribution is always obtained \emph{asymptotically}. 

The Boltzmann equation is a transport evolution equation for the probability density of the velocities of the molecules (with statistical behavior) in an ideal gas, simple fluid, etc. in a non-equilibrium state. The ergodic hypothesis serves precisely to mark out the transition, via transport equation, from non-equilibrium to equilibrium phenomena.
\item 
\label{item "Entropy in Prior Knowledge: Maxwell–Boltzmann Probability Distribution, and Ergodic Hypothesis of Thermodynamics"}
The natural logarithm of the number of microstates represents the well-known \emph{entropy}, that is a thermodynamic quantity. Boltzmann's \cite{Boltzmann "Bemerkungen uber einige Probleme der mechanischen Warmetheorie"} \cite{Boltzmann "Uber die Beziehung zwischen dem zweiten Hauptsatze der mechanischen Warmetheorie und der Wahrscheinlichkeitsrechnung resp. den Satzen uber das Warmegleichgewicht"} goal was to seek a relationship between the \emph{macroscopic nature (thermodynamics of irreversible processes)} and the \emph{microscopic nature (dynamics of reversible processes)},\footnote{
	\label{footnote "Boltzmann's grave formula"}
	The Boltzmann's formula carved onto his grave, $\mathsf{S} = k_\textsc{b}\log{W}$, is placed in perpetual memory of this relationship, where $\mathsf{S}$ is the entropy of a thermodynamic system, with macroscopic properties, and $W$ is the number of microscopic configurations, multiplied by the Boltzmann constant $k_\textsc{b}$. In spite of its name, it was not written by Boltzmann (in this form), but by M. Planck;\endnote{
	Below are excerpts from some works by M. Planck:

	\setlength\parindent{8pt}
	($\mathnormal{1}$) Planck \cite[p. 238]{Planck "Zur Theorie des Gesetzes der Energieverteilung im Normalspectrum"}: «Since the entropy of a resonator is determined by the way in which the energy is simultaneously distributed over many resonators, I assumed that this quantity could be evaluated through the introduction of probability considerations [\textit{Wahrscheinlichkeitsbetrachtungen}], the importance of which for the second law of thermodynamics was first discovered by Mr. Boltzmann \cite{Boltzmann "Uber die Beziehung zwischen dem zweiten Hauptsatze der mechanischen Warmetheorie und der Wahrscheinlichkeitsrechnung resp. den Satzen uber das Warmegleichgewicht"} in the electromagnetic radiation theory». 
	
	($\mathnormal{2}$) Planck \cite[p. 556]{Planck "Ueber das Gesetz der Energieverteilung im Normalspectrum"}: «We now set the entropy $\mathsf{S}_N$ of the system, within an arbitrary additive constant, proportional to the logarithm of the probability $W$ [\textit{Entropie des Systems \textnormal{[\,\dots]} proportional dem Logarithmus der Wahrscheinlichkeit}], so that the $N$ resonators all together have the energy $U_N$, therefore: $\mathsf{S}_N = k_{[\textsc{b}]}\log{W} + \text{const.}$». 
	
	($\mathnormal{3}$) Planck \cite{Planck "Vorlesungen uber die Theorie der Warmestrahlung"} = \cite{Planck "The Theory of Heat Radiation"}, §§ 119-120: «The logarithmic connection between entropy and probability was first stated by L. Boltzmann \cite[§ 6]{Boltzmann "Vorlesungen uber Gastheorie I"} in his kinetic theory of gases. Nevertheless our equation [$\mathsf{S} = k_{[\textsc{b}]}\log{W}$­] differs in its meaning from the corresponding one of Boltzmann in two essential points. Firstly, Boltzmann's equation lacks the factor $k_{[\textsc{b}]}$, which is due to the fact that Boltzmann always used gram-molecules, not the molecules themselves, in his calculations. Secondly, and this is of greater consequence, Boltzmann leaves an additive constant undetermined in the entropy $\mathsf{S}$ as is done in the whole of classical thermodynamics, and accordingly there is a constant factor of proportionality, which remains undetermined in the value of the probability $W$. In contrast with this we assign a definite absolute value to the entropy $\mathsf{S}$. This is a step of fundamental importance, [since it] leads necessarily to the “hypothesis of quanta” and moreover it also leads, as regards radiant heat, to a definite law of distribution of energy of black radiation [\,\dots]. We shall designate the quantity $W$ thus defined as the “thermodynamic probability”, in contrast to the “mathematical probability”, to which it is proportional but not equal. For, while the mathematical probability is a proper fraction, the thermodynamic probability is [\,\dots] always an integer». 
	
	($\mathnormal{4}$) Planck \cite[p. 412]{Planck "The Genesis and Present State of Development of the Quantum Theory"}: «This constant [$k_{[\textsc{b}]}$] is often referred to as Boltzmann's constant, although, to my knowledge, Boltzmann himself never introduced it—a peculiar state of affairs, which can be explained by the fact that Boltzmann, as appears from his occasional utterances, never gave thought to the possibility of carrying out an exact measurement of the constant». 
	
	($\mathnormal{5}$) Planck \cite[pp. 41-42]{Planck "Scientific Autobiography and Other Papers"}: about the thermal electromagnetic «radiation formula», «I began to devote myself to the task of in­vesting it with a true physical meaning. This quest automatically led me to study the interrela­tion of entropy and probability—in other words, to pursue the line of thought inaugurated by Boltzmann. Since the entropy $\mathsf{S}$ is an additive magnitude but the probability $W$ is a multiplicative one, I simply postulated that $\mathsf{S} = k_{[\textsc{b}]} \cdot \log{W}$­, where $k_{[\textsc{b}]}$ is a universal constant; and I investigated  whether the formula for $W$, which is obtained when $\mathsf{S}$ is replaced by its value corresponding to the above radiation law, could be interpreted as a measure of probability».
	} 
	it would thence not be wrong, if we called it \emph{Boltzmann–Planck formula}.
	} 
	through a probabilistic interpretation of thermodynamic, or law of increase of entropy, at the microscopic level.

The $\mathsf{H}$-theorem, for this purpose, is conceived for a gas composed of molecules in ceaseless and chaotic movements obeying the laws of Newtonian (classical) mechanics, and in Boltzmann's mind it implements such a relationship.
\subenumerationisfinis
\item Below are some remarks on the double nature of the ergodic hypothesis.
\subenumerationisinitium
\item Stricto sensu, the ergodic hypothesis is false, since it would be nearly impossible that the orbit of any phase point traverses all points of the surface. This is apparent e.g. from the Kolmogorov–Arnold–Moser (\textsc{kam}) theorem \cite{Kolmogorov "Preservation of Conditionally Periodic Movements with Small Change in the Hamilton Function"} \cite{Arnold "Proof of a theorem of A.N. Kolmogorov on the invariance of quasi-periodic motions under small perturbations of the Hamiltonian"} \cite{Moser "On invariant curves of area-preserving mappings of an annulus"} and the Fermi–Pasta–Ulam \& Tsingou (\textsc{fpu+t}) problem \cite{Fermi Pasta and Ulam "Studies of Nonlinear Problems"}. For instance, the \textsc{fpu+t} experiment with a 1-dimensional chain of non-linear oscillators or, equivalently, a vibrating string of $N$ particles reproducing the discretized structure of a flat crystal, exhibits an \emph{almost periodic} and \emph{non-ergodic} behavior. 

Another example of non-ergodicity is presented by \emph{spin glasses}. The \emph{Parisi solution} \cite{Parisi "The order parameter for spin glasses: a function on the interval 0-1"} \cite{Parisi "A sequence of approximated solutions to the S-K model for spin glasses"} \cite{Parisi "Order Parameter for Spin-Glasses"} of the Sherrington–Kirkpatrick model \cite{Sherrington and Kirkpatrick "Solvable Model of a Spin-Glass"} on a mean field approximation, shows that, in the low temperature phase, there is a \emph{ergodicity breaking}, forming an infinite number of pure equilibrium states with a non-trivial order parameter distribution, i.e. pure thermodynamic states with a hierarchical organization. The mathematical transcription of the physical conditions behind these states is the so-called \emph{replica symmetry breaking}, which is a spontaneous mechanism allied with the breakdown of ergodicity in the spin glass transition \cite{Mezard Parisi Sourlas Toulouse Virasoro "Replica symmetry breaking and the nature of the spin glass phase"} \cite[chap. III]{Mezard Parisi Virasoro "Spin Glass Theory and Beyond"}.
\item Lato sensu, namely in a \emph{weak form}, the ergodicity is a powerful statement, inasmuch as it allows to treat the probability of certain outcomes (occurring with asymptotic frequency) as a “measurable” property: A.I. Khinchin \cite[chap. III]{Khinchin "Mathematical Foundations of Statistical Mechanics"} proves that the (weak) ergodicity is a basically valid hypothesis for systems of many degrees of freedom.

Physically, in terms of practical calculus, it is more appropriate to talk about \emph{quasi-ergodic transformation(s)}, as suggested by P. and T. Ehrenfest \cite[p. 90]{P. and T. Ehrenfest "The Conceptual Foundations of the Statistical Approach in Mechanics"}, by enforcing the less restrictive condition under which there are microstates arbitrarily close to states that are compatible with the entire constant-energy surface. 

In this context, we can remember the contributions of G.D. Birkhoff \cite{Birkhoff "Proof of the Ergodic Theorem"} (see below) and J. von Neumann \cite{Neumann "Proof of the Quasi-Ergodic Hypothesis"}, providing the basis of a proof of correctness for a ergodic hypothesis, whose theorem can be summarised as follows: for almost all phase points along the corresponding orbits, there exists a 1-parameter group of measure preserving transformations (of a measure space) standing for the time evolution of the system, and the group is \emph{metrically transitive} \cite{Birkhoff and Smith "Structure Analysis of Surface Transformations"}, i.e. such that any measurable subset (of this space) has zero measure.
\subenumerationisfinis
\item The ergodic theory can be intended, ultimately, as a \emph{theory of the long-run probabilistic, or statistical, behavior of a dynamical system}, which is a \emph{way for analyzing and measuring} (in the widest sense of the term) \emph{chaotic phenomena}. The Poincaré recurrence Theorem \ref{theorema "Poincaré—recurrence time"} is illuminating in connection therewith. 
\item The couple Ehrenfest \cite[p. 89]{P. and T. Ehrenfest "The Conceptual Foundations of the Statistical Approach in Mechanics"} ascribe the origin of the word \emph{ergodic} to \textgreek{ἔργον-ὁδός}, “energy-path”, and so does N. Wiener \cite[p. 49]{Wiener "Cybernetics: or the Control and Communication in the Animal and the Machine"}, while G. Gallavotti, F. Bonetto and G. Gentile \cite[p. 2]{Gallavotti Bonetto Gentile "Aspects of Ergodic Qualitative and Statistical Theory of Motion"} to \textgreek{ἔργον-εἶδος}, “energy-that which is seen/measured”.
\enumerationisfinis

\subsection{Birkhoff's Ergodic Theorem}
\label{subsection "Birkhoff's Ergodic Theorem"}

\begin{theorema}
\label{theorema "Birkhoff's ergodic Theorem"}
Let $(\invertedbreve{\Omega}, \bbmu) \viz (\invertedbreve{\Omega}, \mathscr{B}_\sigma, \bbmu)$ be a $\sigma$-finite measure space and $\invertedbreve{E}$ a subset of $\invertedbreve{\Omega}$, where $\invertedbreve{\Omega}$ is a non-empty set equipped with a so-called $\sigma$-algebra (cf. Theorem \ref{theorema "Poincaré—recurrence time"} and Scholium \ref{scholium "Lindelöf space, Borel measure plus sigma-algebra"}), $\bbmu \colon \mathscr{B}_\sigma \to [0, \infty]$ is a finite measure on $\invertedbreve{\Omega}$, and $\mathscr{B}_\sigma$ is a $\sigma$-algebra of subsets of $\invertedbreve{\Omega}$. We can define a measure space to be a \emph{probability space} if $\bbmu(\invertedbreve{\Omega}) = 1$. Let $\varphi_\bbmu \colon \invertedbreve{\Omega} \to \invertedbreve{\Omega}$ be a measure preserving transformation of a measure space onto itself, and $\Lebesgue^1(\invertedbreve{\Omega}, \bbmu)$ \textnormal{\cite{Riesz "Untersuchungen uber Systeme integrierbarer Funktionen"}} the space of a measurable function $\tau_\bbmu \colon \invertedbreve{\Omega} \to \mathbb{R}$. To be noted that $\bbmu$ is $\varphi_\bbmu$-invariant because $\bbmu\bigl(\varphi_\bbmu^{-1}(\invertedbreve{E})\bigr) = \bbmu(\invertedbreve{E})$. We shall indicate by $\mathbbl{M}(\invertedbreve{\Omega})$ the set of such $\varphi_\bbmu$-invariant (probability) measures on $\invertedbreve{\Omega}$. Then the Birkhoff proposition \textnormal{\cite{Birkhoff "Proof of the Ergodic Theorem"}}\footnote{
	A further, classical, proof of this theorem, as well as that of Birkhoff, can also be found in N. Wiener \cite{Wiener "The ergodic theorem"}.
	} 
assumes that there exists a limit
\begin{align}
\label{equation "Boltzmann–Birkhoff Ergodic Hypothesis"}
	(\tau_\bbmu)_{\varphi_\bbmu}(x) & = \lim_{n \to \infty}\frac{1}{n}\sum^{n - 1}_{\nu = 0}\tau_\bbmu\bigl(\varphi_\bbmu^\nu(x)\bigr) \notag \\
	& = \int_{\invertedbreve{\Omega}}(\tau_\bbmu)_{\varphi_\bbmu}d\bbmu = \int_{\invertedbreve{\Omega}}{\tau_\bbmu}d\bbmu,
\end{align}
for each $\tau_\bbmu \in \Lebesgue^1(\invertedbreve{\Omega}, \bbmu)$ and $\bbmu$-almost any point $x \in \invertedbreve{\Omega}$, or rather, there exists a time average, for all measures $\bbmu \in \mathbbl{M}(\invertedbreve{\Omega})$, such that $\bbmu$ is ergodic, i.e. $\bbmu(\invertedbreve{E}) = 0$ or $\bbmu(\invertedbreve{\Omega} \backslash \invertedbreve{E}) = 0$. This is the Boltzmann–Birkhoff formula according to which the \emph{time average} equals the \emph{space average}. 
\end{theorema}

\begin{proof}
Let ${\zeta_\bbmu}_1 {\zeta_\bbmu}_2, {\zeta_\bbmu}_3, \mathellipsis \in \Lebesgue^1(\invertedbreve{\Omega}, \bbmu)$ denote a non-decreasing sequence $\{{\zeta_\bbmu}_n\}_{n \in \mathbb{N}}$ of $\bbmu$-integrable functions. Let us fix
\begin{equation}
	{\zeta_\bbmu}_n = \max_{1 \leqslant r \leqslant n}\left\{\sum^{r - 1}_{\nu = 0}\zeta_\bbmu \circ \varphi_\bbmu^\nu\right\},
\end{equation}
and
\begin{equation}
	\int_{\invertedbreve{\Omega}}\big|{\zeta_\bbmu}_n\big|d\bbmu \leqslant \sum^{n - 1}_{\nu = 0}\int_{\invertedbreve{\Omega}}\big|\zeta_\bbmu \circ \varphi_\bbmu^\nu\big|d\bbmu.
\end{equation}
If 
\begin{equation}
	{\zeta_\bbmu}_{n + 1} = \zeta_\bbmu + \max_{1 \leqslant r \leqslant n}\left\{0, \sum^r_{\nu = 1}\zeta_\bbmu \circ \varphi_\bbmu^\nu\right\} = \zeta_\bbmu + \max\bigl\{0, {\zeta_\bbmu}_n \circ \varphi_\bbmu\bigr\},
\end{equation}
one has
\begin{equation}
	\lim_{n \to \infty}{\zeta_\bbmu}_{n + 1}(x) = +\infty \text{ iff } \lim_{n \to \infty}{\zeta_\bbmu}_n\bigl(\varphi_\bbmu(x)\bigr) = +\infty,
\end{equation}
from which the $\varphi_\bbmu$-invariance of  
\begin{equation}
	\invertedbreve{E} = \left\{x \in \invertedbreve{\Omega} \mathrel{\Big|} \lim_{n \to \infty}{\zeta_\bbmu}_n(x) = +\infty\right\}
\end{equation}
is obtained, and ${\zeta_\bbmu}_{n + 1} - {\zeta_\bbmu}_n \circ \varphi_\bbmu = \zeta_\bbmu + \max\bigl\{0, {\zeta_\bbmu}_n \circ \varphi_\bbmu\bigr\} - {\zeta_\bbmu}_n \circ \varphi_\bbmu = \zeta_\bbmu - \min\bigl\{0, {\zeta_\bbmu}_n \circ \varphi_\bbmu\bigr\}$. What follows is ${\zeta_\bbmu}_{n + 1} - {\zeta_\bbmu}_n \circ \varphi_\bbmu \searrow \zeta_\bbmu$ on $\invertedbreve{E}$. The adoption of the \emph{Lebesgue's and Levi's dominated and monotone convergence theorems} (cf. Theorems \ref{theorema "Lebesgue's dominated convergence"} and \ref{theorema "Lebesgue–Levi monotone convergence"}) gives
\begin{align}
	& 0 \leqslant \int_{\invertedbreve{E}}\left({\zeta_\bbmu}_{n + 1} - {\zeta_\bbmu}_n\right)d\bbmu = \int_{\invertedbreve{E}}\left({\zeta_\bbmu}_{n + 1} - {\zeta_\bbmu}_n \circ \varphi_\bbmu\right)d\bbmu \xrightarrow[n \to \infty]{}\int_{\invertedbreve{E}}{\zeta_\bbmu}d\bbmu, \\
	& \int_{\invertedbreve{E}}{\zeta_\bbmu}d\bbmu \geqslant 0.
\end{align}
Let $\mathscr{U}_\sigma = \bigl\{\varphi_\bbmu^{-1}(\invertedbreve{E}) = \invertedbreve{E}\bigr\}$ be a $\sigma$-subalgebra of $\mathscr{B}_\sigma$ and $\tau_{\bbmu(\mathscr{U})} \in \Lebesgue^1(\invertedbreve{\Omega}, \bbmu)$ a $\mathscr{U}_\sigma$-measurable function such that
\begin{equation}
	\int_{\invertedbreve{E}}\tau_{\bbmu(\mathscr{U})}d\bbmu = \int_{\invertedbreve{E}}{\tau_\bbmu}d\bbmu.
\end{equation}
Taking the function $\zeta_\bbmu = \tau_\bbmu - \tau_{\bbmu(\mathscr{U})} - \epsilon$, with $\epsilon > 0$, we get  
\begin{equation}
	\int_{\invertedbreve{E}}{\zeta_\bbmu}d\bbmu = - \epsilon\bbmu(\invertedbreve{E}),
\end{equation}
consequently $\bbmu(\invertedbreve{E}) = 0$, and $\lim_{n \to \infty}{\zeta_\bbmu}_n(x) < +\infty$, at $\bbmu$-almost $x \in \invertedbreve{\Omega}$. Write
\begin{equation}
	\limsup_{n \to \infty}\frac{1}{n}\sum^{n - 1}_{\nu = 0}\zeta_\bbmu \circ \varphi_\bbmu^\nu \leqslant \limsup_{n \to \infty}\frac{{\zeta_\bbmu}_n}{n} \leqslant 0.
\end{equation}
We shall say that $\tau_{\bbmu(\mathscr{U})}$ is $\varphi_\bbmu$-invariant; this leads to 
\begin{equation}
	\limsup_{n \to \infty}\frac{1}{n}\sum^{n - 1}_{\nu = 0}\tau_\bbmu \circ \varphi_\bbmu^\nu \leqslant	\tau_{\bbmu(\mathscr{U})} + \epsilon \text{ and } \liminf_{n \to \infty}\frac{1}{n}\sum^{n - 1}_{\nu = 0}\tau_\bbmu \circ \varphi_\bbmu^\nu \geqslant \tau_{\bbmu(\mathscr{U})} - \epsilon,
\end{equation}
with $\tau_\bbmu$ replaced by $-\tau_\bbmu$ in the limit inferior of the sequence, $\bbmu$-almost everywhere on $\invertedbreve{\Omega}$. Then 
\begin{equation}
	\tau_{\bbmu(\mathscr{U})} - \epsilon \leqslant \liminf_{n \to \infty}\frac{1}{n}\sum^{n - 1}_{\nu = 0}\tau_\bbmu \circ \varphi_\bbmu^\nu \leqslant \limsup_{n \to \infty}\frac{1}{n}\sum^{n - 1}_{\nu = 0}\tau_\bbmu \circ \varphi_\bbmu^\nu \leqslant \tau_{\bbmu(\mathscr{U})} + \epsilon,
\end{equation}
and so
\begin{equation}
	\lim_{n \to \infty}\frac{1}{n}\sum^{n - 1}_{\nu = 0}\tau_\bbmu \circ \varphi_\bbmu^\nu = \tau_{\bbmu(\mathscr{U})}.
\end{equation}
The last equation tells us that $(\tau_\bbmu)_{\varphi_\bbmu}$ is the same as $\tau_{\bbmu(\mathscr{U})}$ $\bbmu$-almost everywhere on $\invertedbreve{\Omega}$. It is therefore possible to conclude that
\begin{equation}
	\int_{\invertedbreve{\Omega}}(\tau_\bbmu)_{\varphi_\bbmu}d\bbmu = \int_{\invertedbreve{\Omega}}\tau_{\bbmu(\mathscr{U})}d\bbmu = \int_{\invertedbreve{\Omega}}{\tau_\bbmu}d\bbmu, \text{ for all } \bbmu \in \mathbbl{M}(\invertedbreve{\Omega}).
\end{equation}
\end{proof}

\subsubsection{Addendum. Convergence Theorems: Lebesgue, Levi and Fatou's Lemma}
\label{subsubsection "Addendum. Convergence Theorems: Lebesgue, Levi and Fatou's Lemma"}

And now about the convergence theorems by H. Lebesgue \cite{Lebesgue "Sur la methode de M. Goursat pour la resolution de l'equation de Fredholm"} and B. Levi \cite{Levi B. "Sopra l'integrazione delle serie"}, as well as an inequality theorem by P. Fatou \cite{Fatou "Series trigonometriques et series de Taylor"}, commonly called \emph{Fatou's lemma}.

\begin{theorema}[Lebesgue's dominated convergence]
\label{theorema "Lebesgue's dominated convergence"}
Let $(\invertedbreve{\Omega}, \mathscr{B}_\sigma, \bbmu)$ be a measure space, and $\{{\psi_\bbmu}_n\} \colon \invertedbreve{\Omega} \to \mathbb{C}$ a sequence of complex-valued $\mathscr{B}_\sigma$-measurable functions on $\invertedbreve{\Omega}$ pointwise converging $\bbmu$-almost everywhere to $\psi_\bbmu$. We define ${\psi_\bbmu}(x) = \lim_{n \to \infty}{\psi_\bbmu}_n(x)$, for all points $x \in \invertedbreve{\Omega}$. If there is a non-negative $\bbmu$-integrable function $\zeta_\bbmu \in \Lebesgue^1(\invertedbreve{\Omega}, \bbmu) \colon \invertedbreve{\Omega} \to [0, \infty] $ such that $|{\psi_\bbmu}_n(x)| \leqslant \zeta_\bbmu(x)$, for any $n \in \mathbb{N}$, then $\psi_\bbmu \in \Lebesgue^1(\invertedbreve{\Omega}, \bbmu)$, i.e. the function $\psi_\bbmu$ is $\bbmu$-integrable, and
\begin{equation}
	\lim_{n \to \infty}\int_{\invertedbreve{\Omega}}{\psi_\bbmu}_n{d\bbmu} = \int_{\invertedbreve{\Omega}}\psi_\bbmu{d\bbmu}, 
\end{equation}
and also
\begin{equation}
	\lim_{n \to \infty}\int_{\invertedbreve{\Omega}}|{\psi_\bbmu}_n - \psi_\bbmu|d\bbmu = 0.
\end{equation}
\end{theorema}

\begin{proof}
Without loss of generality, assume that ${\psi_\bbmu}_n$ converges pointwise everywhere (and not $\bbmu$-almost everywhere), and that ${\psi_\bbmu}_n$ is real, separating the involved sequence of complex numbers into real and imaginary parts, so that $-\zeta_\bbmu \leqslant {\psi_\bbmu}_n \leqslant \zeta_\bbmu$. Thanks to Fatou's lemma (\ref{lemma "Fatou"}) we can write
\begin{equation}
	\int_{\invertedbreve{\Omega}}(\psi_\bbmu + \zeta_\bbmu)d\bbmu \leqslant \liminf_{n \to \infty}\int_{\invertedbreve{\Omega}}({\psi_\bbmu}_n + \zeta_\bbmu)d\bbmu.
\end{equation}
Since the integral $\int_{\invertedbreve{\Omega}}\zeta_\bbmu$ is finite, it is possible to subtract this quantity,
\begin{equation}
	\int_{\invertedbreve{\Omega}}\psi_\bbmu{d\bbmu} \leqslant \liminf_{n \to \infty}\int_{\invertedbreve{\Omega}}{\psi_\bbmu}_n{d\bbmu}.
\end{equation}
The same applies to $\zeta_\bbmu - {\psi_\bbmu}_n$,
\begin{equation}
	\int_{\invertedbreve{\Omega}}(\zeta_\bbmu - \psi_\bbmu)d\bbmu \leqslant \liminf_{n \to \infty}\int_{\invertedbreve{\Omega}}(\zeta_\bbmu - {\psi_\bbmu}_n)d\bbmu,
\end{equation}
from which, subtracting $\int_{\invertedbreve{\Omega}}\zeta_\bbmu$, we have
\begin{equation}
	\limsup_{n \to \infty}\int_{\invertedbreve{\Omega}}{\psi_\bbmu}_n{d\bbmu} \leqslant \int_{\invertedbreve{\Omega}}\psi_\bbmu{d\bbmu},
\end{equation}
and
\begin{equation}
	\limsup_{n \to \infty}\int_{\invertedbreve{\Omega}}|{\psi_\bbmu}_n - \psi_\bbmu|d\bbmu \leqslant 0.
\end{equation}
\end{proof}

\begin{theorema}[Lebesgue–Levi monotone convergence]
\label{theorema "Lebesgue–Levi monotone convergence"}
Let $(\invertedbreve{\Omega}, \mathscr{B}_\sigma, \bbmu)$ be a measure space, and $\{{\psi_\bbmu}_n\} \colon \invertedbreve{\Omega} \to [0, \infty]$ a sequence of $\mathscr{B}_\sigma$-measurable functions such that ${\psi_\bbmu}_n(x) \leqslant {\psi_\bbmu}_{n + 1}(x)$, for any $n \in \mathbb{N}$. We define $\psi_\bbmu(x) = \lim_{n \to \infty}{\psi_\bbmu}_n(x)$, for each $x \in \invertedbreve{\Omega}$. Suppose that $0 \leqslant {\psi_\bbmu}_1(x) \leqslant {\psi_\bbmu}_2(x) \leqslant \cdots \leqslant \infty$ (monotone non-decreasing sequence). Then 
\begin{equation}
	\lim_{n \to \infty}\int_{\invertedbreve{\Omega}}{\psi_\bbmu}_n{d\bbmu} = \int_{\invertedbreve{\Omega}}\psi_\bbmu{d\bbmu},
\end{equation}	
or
\begin{equation}
	\lim_{n \to \infty}\int_{\invertedbreve{\Omega}}{\psi_\bbmu}_n{d\bbmu} = \int_{\invertedbreve{\Omega}}\lim_{n \to \infty}{\psi_\bbmu}_n{d\bbmu}.
\end{equation}
\end{theorema}

\begin{proof}
The sequence ${\psi_\bbmu}_n$ is non-decreasing, from which follows that 	
\begin{equation}
	\lim_{n \to \infty}\int_{\invertedbreve{\Omega}}{\psi_\bbmu}_n{d\bbmu} \leqslant \int_{\invertedbreve{\Omega}}\psi_\bbmu{d\bbmu}.
\end{equation}
The reverse inequality is given by including a function $\zeta_\bbmu$ with a pointwise property,
\begin{equation}
	\int_{\invertedbreve{\Omega}}\zeta_\bbmu{d\bbmu} \leqslant \lim_{n \to \infty}\int_{\invertedbreve{\Omega}}{\psi_\bbmu}_n{d\bbmu}.
\end{equation}
Fixing 
\enumerationisinitium
\item $0 < c_\nu < \infty$ (in which $c_\nu$ are finite non-negative constants), 
\item some $\mathscr{B}_\sigma$-measurable subsets $[\invertedbreve{E}_1, \mathellipsis, \invertedbreve{E}_r] \subset \invertedbreve{\Omega}$, 
\item an indicator function $\mathbbl{1}_{\invertedbreve{E}} \colon \invertedbreve{\Omega} \to \{0, 1\}$, 
\enumerationisfinis
suppose that $\zeta_\bbmu$ is vertically-truncated, so we get
\begin{align}
	& \zeta_\bbmu = \sum^r_{\nu = 1}c_\nu \cdot \mathbbl{1}_{\invertedbreve{E}_\nu}, \\
	& \int_{\invertedbreve{\Omega}}\zeta_\bbmu{d\bbmu} = \sum^r_{\nu = 1}c_\nu\bbmu(\invertedbreve{E}_\nu),
\end{align}
and $\psi_\bbmu(x) = \sup_n{\psi_\bbmu}_n(x) > (1 - \varepsilon)c_\nu$, for a value $0 < \varepsilon < 1$. The upwards monotonicity allows us to determine $\lim_{n \to \infty}\bbmu(\invertedbreve{E}_{\nu, n}) = \bbmu(\invertedbreve{E}_\nu)$, where $\invertedbreve{E}_{\nu, n} = \bigl\{x \in \invertedbreve{E}_\nu \mid {\psi_\bbmu}_n(x) > (1 - \varepsilon)c_\nu\bigr\}$. Therefore
\begin{equation}
	\int_{\invertedbreve{\Omega}}{\psi_\bbmu}_n{d\bbmu} \geqslant (1 - \varepsilon)\sum^r_{\nu = 1}c_\nu\bbmu(\invertedbreve{E}_{\nu, n}),
\end{equation}
and
\begin{equation}
	\lim_{n \to \infty}\int_{\invertedbreve{\Omega}}{\psi_\bbmu}_n{d\bbmu} \geqslant (1 - \varepsilon)\sum^r_{\nu = 1}c_\nu\bbmu(\invertedbreve{E}_\nu),
\end{equation}
putting $n \to \infty$. The demonstration shall be provided through $\varepsilon \to 0$.
\end{proof}

\begin{lemma}[Fatou]
\label{lemma "Fatou"}	
Let $(\invertedbreve{\Omega}, \mathscr{B}_\sigma, \bbmu)$ be a measure space, and $\{\psi_\bbmu\}_n \colon \invertedbreve{\Omega} \to [0, \infty]$ a sequence of $\mathscr{B}_\sigma$-measurable functions. Then
\begin{equation}
	\int_{\invertedbreve{\Omega}}\liminf_{n \to \infty}{\psi_\bbmu}_n{d\bbmu} \leqslant \liminf_{n \to \infty}\int_{\invertedbreve{\Omega}}{\psi_\bbmu}_n{d\bbmu}.
\end{equation}
\end{lemma}

\begin{proof}
Setting ${\zeta_\bbmu}_n(x) = \inf_{k \geqslant n}{\psi_\bbmu}_k(x)$, one has $0 \leqslant {\zeta_\bbmu}_n \leqslant {\psi_\bbmu}_n$, ${\zeta_\bbmu}_n \leqslant {\zeta_\bbmu}_{n + 1}$, for which $\{\zeta_\bbmu\}_n$ is a monotone sequence of measurable functions. By Theorem \ref{theorema "Lebesgue–Levi monotone convergence"},
\begin{equation}
	\int_{\invertedbreve{\Omega}}\psi_\bbmu{d\bbmu} = \lim_{n \to \infty}\int_{\invertedbreve{\Omega}}{\zeta_\bbmu}_n{d\bbmu}. 
\end{equation}
Since
\begin{equation}
	\int_{\invertedbreve{\Omega}}{\zeta_\bbmu}_n{d\bbmu} \leqslant \int_{\invertedbreve{\Omega}}{\psi_\bbmu}_k{d\bbmu}, 
\end{equation}
we have 
\begin{equation}
	\int_{\invertedbreve{\Omega}}\psi_\bbmu{d\bbmu} = \lim_{n \to \infty}\int_{\invertedbreve{\Omega}}{\zeta_\bbmu}_n{d\bbmu} \leqslant \lim_{n \to \infty}\inf_{k \geqslant n}\int_{\invertedbreve{\Omega}}{\psi_\bbmu}_k{d\bbmu} = \liminf_{n \to \infty}\int_{\invertedbreve{\Omega}}{\psi_\bbmu}_n{d\bbmu}.
\end{equation}
\end{proof}

\subsection{Anosov's Ergodic Theorem}

We will adopt the Hopf–Anosov–Sinai modus operandi \cite{Hopf "Statistik der geodatischen Linien in Mannigfaltigkeiten negativer Krummung"} \cite[pp. 144-146]{Anosov and Sinai "Some smooth ergodic system"}, summarised by Ya.B. Pesin \cite[pp. 85-86]{Pesin "Lectures on partial hyperbolicity and stable ergodicity"} with a diffeomorphism of class $\mathscr{C}^2$. Other advantageous solutions are e.g. in Ya.G. Sinai \cite{Sinai "Markov partitions and $C$-diffeomorphisms"}, see R. Mañé \cite[pp. 180-189]{Mane "Ergodic Theory and Differentiable Dynamics"}, and in R. Bowen \cite[chap. 4]{Bowen "Equilibrium States and the Ergodic Theory of Anosov Diffeomorphisms"}.

\begin{theorema}
\label{theorema "Anosov's Ergodic Theorem"}
Let $\varphi \viz \varphi[\mathscr{A}] \colon \mathcal{M} \to \mathcal{M}$ be a $\mathscr{C}^2$ Anosov diffeomorphism (Definition \ref{definitio "Anosov diffeomorphism"}), and $\mathcal{M}$ a smooth compact connected Riemannian manifold. Let $\mathcal{W}^\mathrm{s}$ and $\mathcal{W}^\mathrm{u}$ be the stable and unstable foliations $\mathcal{F}$ of $\mathcal{M}$ with smooth leaves, then $\mathcal{W}^\mathrm{s}(x)$ and $\mathcal{W}^\mathrm{u}(x)$ the local leaves (of foliations) passing through $x \in \mathcal{M}$. Let $\mathscr{B}_\sigma(\mathcal{F})$ be a $\sigma$-algebra of sets invariant with reference to $\varphi$. Then $\varphi$ is ergodic, if $\varphi$ preserves the volume form.
\end{theorema}

\begin{proof}
Consider a continuous real-valued function $\tau \colon \mathcal{M} \to \mathbb{R}$, and write a trio of functions 
\begin{equation}
	\begin{drcases}
	\tau^+(x) = \lim_{n \to +\infty}\frac{1}{n}\sum^{n - 1}_{\nu = 0} \\
	\tau^-(x) = \lim_{n \to -\infty}\frac{1}{|n|}\sum^{n + 1}_{\nu = 0} \\
	\bar{\tau}(x) = \lim_{n \to +\infty}\frac{1}{2n +1}\sum^{n}_{\nu = -n} 
	\end{drcases}
	\tau\bigl(\varphi^\nu(x)\bigr).
\end{equation}	
The continuous trio, formed by $\tau^+(x)$, $\tau^-(x)$ and $\bar{\tau}(x)$, is 
\enumerationisinitium
\item \emph{identical} at almost any $x \in \mathcal{M}$ (this is due to Birkhoff's Theorem \ref{theorema "Birkhoff's ergodic Theorem"}), 
\item \emph{dense} in the space of $\bbmu$-functions on $\mathcal{M}$ concerning the $\sigma$-algebra $\mathscr{B}_\sigma(\mathcal{F}) \subset_\bbmu \mathscr{B}_\sigma(\mathcal{W}^\mathrm{s}) \cap \mathscr{B}_\sigma(\mathcal{W}^\mathrm{u})$. 
\enumerationisfinis

We shall say that a set is \emph{conull} if the set is the complement of a null set. Fixing a conull set $\mathcal{M}_0$ in $\mathcal{M}$, and taking a local leaf $\mathcal{W}^\mathrm{s}(x)$ of the stable foliation, i.e. a local stable manifold, for all $x \in \mathcal{M}_0$ and any point $z \in \mathcal{W}^\mathrm{s}(x)$, we observe that 
\begin{equation}
	\left|\tau\bigl(\varphi^{\nu \to \infty}(x)\bigr) - \tau\bigl(\varphi^{\nu \to \infty}(z)\bigr)\right| \to 0
\end{equation}
and then $\tau^+(x) = \tau^-(z)$. The Anosovian absolute continuity \cite[§ 17]{Anosov "Geodesic flows on closed Riemannian manifolds of negative curvature"} ensures that, for almost any $z \in \mathcal{W}^\mathrm{s}(x)$, the set membership $z \in \mathcal{M}_0$ holds, at almost any $x \in \mathcal{M}$. The condition that $z \notin \mathcal{M}_0$ gives a set of null measure, with a leaf of the unstable foliation: 
\begin{equation}
	\bigcup\subscript{\left\{z \in \mathcal{W}^\mathrm{s}(x), z \notin \mathcal{M}_0\right\}}\mathcal{W}^\mathrm{u}(z). 
\end{equation}
Letting $y \in \mathcal{M}$, for almost each $y$ close enough to $x$, it is clear that $z = \mathcal{W}^\mathrm{s}(x) \cap \mathcal{W}^\mathrm{u}(y)$ is a point of $\mathcal{M}_0$, and $\bar{\tau}(y) = \tau^-(y) = \tau^-(z) = \bar{\tau}(z) = \bar{\tau}(x)$. It follows that $\bar{\tau}$ appears to be locally constant almost everywhere on some neighborhood of $x$, and thence almost everywhere on $\mathcal{M}$, by means of locally path-connected space between any two points of the manifold, and we are done with the ergodicity of $\varphi \viz \varphi[\mathscr{A}]$.
\end{proof}

\subsection{Ergodicity of the Geodesic Flow: Hopfian Statistical Process}
\label{subsection "Ergodicity of the Geodesic Flow: Hopfian Statistical Process"}

Birkhoff's \ref{theorema "Birkhoff's ergodic Theorem"} and Anosov's \ref{theorema "Anosov's Ergodic Theorem"} Theorems give the cue to show that the geodesic flow 
\enumerationisinitium
\item on the tangent bundle of a negatively curved compact Riemannian manifold, or of a compact (and connected) surface of finite area and constant negative Gaussian curvature, 
\item or, more widely, for any hyperbolic $\mathscr{C}^2$ Anosov system, whether it be a flow or diffeomorphism,
\enumerationisfinis
is ergodic, and it is with respect to the Liouville measure (Theorem \ref{theorema "Liouville invariance"}).

To be accurate, the \emph{ergodicity of a geodesic flow with respect to the Liouville measure on (the unit tangent bundle of) surfaces of variable negative curvature} and \emph{on manifolds of constant negative curvature of arbitrary dimension}, was initially investigated and proved by G.A. Hedlund \cite{Hedlund "The dynamics of geodesic flows"} and, especially, E. Hopf \cite{Hopf "Statistik der geodatischen Linien in Mannigfaltigkeiten negativer Krummung"} \cite{Hopf "Statistik der Losungen geodatischer Probleme vom unstabilen Typus. II"}; about Hopf, see also \cite{Hopf "Ergodentheorie"} \cite{Hopf "Ergodic Theory and the Geodesic Flow on Surfaces of Constant Negative Curvature"}. Hopf's insight was extended by Anosov \cite{Anosov "Geodesic flows on closed Riemannian manifolds of negative curvature"} as it stands today.

\begin{theorema}[Hopfian statistical process]
\label{theorema "Hopfian statistical process"}
Let $\mathcal{S}_\mathbbl{\Gamma} \equival \mathbbl{\Gamma} \backslash \mathbb{U}^2_\mathbb{C}$ be a Riemann (also modular) surface, for some Fuchsian group $\mathbbl{\Gamma}$ \textnormal{\cite{Hopf "Fuchsian groups and ergodic theory"}} (Section \ref{subsection "Projective Linear Transforms: Dynamics on the Modular Surface, and Horocycle Flow"}), a non-elementary discrete subgroup of $PSL_2(\mathbb{R})$ or a conjugate of such a group in $PSL_2(\mathbb{C})$. Let $\mathring{\mathcal{T}}^1\mathbb{D}_\mathbb{C}$\footnote{
	Alternative notation: $\mathring{\mathcal{S}}^1\mathbb{D}_\mathbb{C}$.
	}
be the unit tangent bundle of the Beltrami–Poincaré unit disk (2-ball) $\mathbb{D}_\mathbb{C} \equival \mathbb{B}^2_\mathbb{C} = \{z \in \mathbb{C} \mid \|z\| < 1\}$, cf. Eq. \eqref{align "Unit disk in the complex plane"}. Assume further that the unit tangent bundle $\mathring{\mathcal{T}}^1\mathcal{S}_\mathbbl{\Gamma}$ of $\mathcal{S}_\mathbbl{\Gamma}$ can be identified with the quotient $\mathbbl{\Gamma} \backslash \mathring{\mathcal{T}}^1\mathbb{D}_\mathbb{C}$. The geodesic flow $\varphi_t$ on $\mathring{\mathcal{T}}^1\mathcal{S}_\mathbbl{\Gamma}$ is conservative and, consequently, ergodic, with respect to the Liouville-like measure $\bbmu$.
\end{theorema}

Note that the above identification is connect to the projection $\pi \colon \mathring{\mathcal{T}}^1\mathbb{D}_\mathbb{C} \to \mathring{\mathcal{T}}^1\mathcal{S}_\mathbbl{\Gamma}$, so 
\begin{equation}
	\pi \circ \varphi_t\big|_{\mathring{\mathcal{T}}^1\mathbb{D}_\mathbb{C}} = \varphi_t\big|_{\mathring{\mathcal{T}}^1\mathcal{S}_\mathbbl{\Gamma}}.
\end{equation}

\begin{proof}
We rely partly on the Patterson–Sullivan method \cite{Patterson "The limit set of a Fuchsian group"} \cite{Sullivan "The density at infinity of a discrete group of hyperbolic motions"} \cite{Sullivan "Discrete conformal groups and measurable dynamics"} \cite{Sullivan "Entropy Hausdorff measures old and new and limit sets of geometrically finite Kleinian groups"} to proceed with the demonstration.
~\enumerationisinitium
\item Let us introduce the \emph{limit set} $U_\mathbbl{\Gamma} \subset \partial\mathbb{U}^2_\mathbb{C}$, which is the set of limit points of all orbits in $\mathbb{U}^2_\mathbb{C}$. Take the case of $\mathbb{D}_\mathbb{C}$, and imagine an object formed by circles-horocycles tangent to the boundary circle. (Comprehensibly, we need to think about the correspondence of disk to the upper half-plane, within the Beltrami–Poincaré models: in these two solutions of 2-dimensional hyperbolic geometry, every horocycle corresponds to each other). Let $\widetilde{\mathcal{N}}_\mathbbl{\Gamma}$ be the set of a Fuchsian group $\mathbbl{\Gamma}$, called \emph{Nielsen region} \cite{Nielsen "Untersuchungen zur Topologie der geschlossenen zweiseitigen Flachen"} \cite{Nielsen "Uber Gruppen linearer Transformation"} \cite{Fenchel Nielsen "Discontinuous Groups of Isometries in the Hyperbolic Plane"}, corresponding to the \emph{convex hull} of $U_\mathbbl{\Gamma}$. The quotient 
\begin{equation}
	\mathcal{N}_\mathbbl{\Gamma} \equival \mathbbl{\Gamma} \backslash \widetilde{\mathcal{N}}_\mathbbl{\Gamma} 
\end{equation} 
is the \emph{convex core} of $\mathcal{S}_\mathbbl{\Gamma}$. Find a \emph{horocyclic region} $\mathcal{C}_\eta$ in the Beltrami–Poincaré disk, i.e. an open region bounded by a horocycle tangent to $\partial\mathbb{U}^2_\mathbb{C}$. Let 
\begin{equation}
	\widetilde{\mathcal{K}}_\mathbbl{\Gamma} \equival \widetilde{\mathcal{N}}_\mathbbl{\Gamma} - \{\mathcal{C}_\eta\}
\end{equation}
be the \emph{reduced Nielsen region}, and 
\begin{equation}
	\mathcal{K}_\mathbbl{\Gamma} \equival \mathbbl{\Gamma} \backslash \widetilde{\mathcal{K}}_\mathbbl{\Gamma} 
\end{equation}
the quotient region, known as \emph{compact core} of $\mathcal{S}_\mathbbl{\Gamma}$. Let us draw a geodesic on $\mathbb{D}_\mathbb{C}$ starting inside $\widetilde{\mathcal{K}}_\mathbbl{\Gamma}$. Since $\bbmu(\partial\mathbb{U}^2_\mathbb{C} - U_\mathbbl{\Gamma}) = 0$, almost any geodesic does not leave the Nielsen region $\widetilde{\mathcal{N}}_\mathbbl{\Gamma}$. Without dwelling too much on it, if the set of all geodesics has zero $\bbmu$-measure, associated with the fact that any geodesic ends in parabolic fixed points in $U_\mathbbl{\Gamma}$, then almost any geodesic on $\mathcal{S}_\mathbbl{\Gamma}$ will return to the compact core $\mathcal{K}_\mathbbl{\Gamma}$ after so many times. This proves that the geodesic flow on $\mathring{\mathcal{T}}^1\mathcal{S}_\mathbbl{\Gamma}$ is conservative with respect to $\bbmu$. 
\item Let $\lambda_\bbmu \in \Lebesgue^1(\mathring{\mathcal{T}}^1\mathcal{S}_\mathbbl{\Gamma}, d\bbmu)$ be a continuous, positive definite and $\bbmu$-integrable function. Let 
\begin{equation}
	\textcyrillic{\textit{ъ}} \colon \alpha, \dot{\alpha} \mapsto \textcyrillic{\textit{ъ}}(\alpha), \textcyrillic{\textit{ъ}}(\dot{\alpha})
\end{equation}
be a transformation acting on the space 
\begin{equation}
	\partial\mathbb{D}_\mathbb{C} \times \partial\mathbb{D}_\mathbb{C} \equival \{\alpha, \dot{\alpha} \in \partial\mathbb{D}_\mathbb{C} \times \partial\mathbb{D}_\mathbb{C} \mid \alpha \neq \dot{\alpha}\}. 
\end{equation}
If the geodesic flow is conservative, it follows that
\begin{equation}
\label{equation "Limit of integration in Hopfian statistical process"}
	\lim_{\textcyrillic{\textit{ъ}} \to \infty}\int^{\textcyrillic{\textit{ъ}}}_0\lambda_\bbmu\bigl(\varphi_t(x)\bigr)dt = +\infty,
\end{equation}
for almost every $x \in \mathring{\mathcal{T}}^1\mathcal{S}_\mathbbl{\Gamma}$. Let $\kappa_\bbmu \in \Lebesgue^1(\mathring{\mathcal{T}}^1\mathcal{S}_\mathbbl{\Gamma}, d\bbmu)$ be another $\bbmu$-integrable function. We set (via Birkhoff's \ref{theorema "Birkhoff's ergodic Theorem"} Theorem) the quotient function and the convergence
\begin{equation}
	\kappa_\bbmu\bigl(\textcyrillic{\textit{ъ}}, x\bigr) = \frac{\int^{\textcyrillic{\textit{ъ}}}_0\kappa_\bbmu\bigl(\varphi_t(x)\bigr)dt}{\int^{\textcyrillic{\textit{ъ}}}_0\lambda_\bbmu\bigl(\varphi_t(x)\bigr)dt}, \enspace \kappa_\bbmu(x) = \lim_{\textcyrillic{\textit{ъ}} \to \infty}\kappa_\bbmu\bigl(\textcyrillic{\textit{ъ}}, x\bigr),
\end{equation}
so
\begin{equation}
(\kappa_\bbmu)_\lambda(x) = \lim_{\textcyrillic{\textit{ъ}} \to \infty}\frac{\int^{\textcyrillic{\textit{ъ}}}_0\kappa_\bbmu\bigl(\varphi_t(x)\bigr)dt}{\int^{\textcyrillic{\textit{ъ}}}_0\lambda_\bbmu\bigl(\varphi_t(x)\bigr)dt}.
\end{equation}
The function $(\kappa_\bbmu)_\lambda$ is invariant under the geodesic flow $\varphi_t$, provided that $\lambda_\bbmu$ satisfies \eqref{equation "Limit of integration in Hopfian statistical process"}. The next step is to demonstrate that, in the conditions in which $\kappa_\bbmu$ is $\bbmu$-integrable, $(\kappa_\bbmu)_\lambda$ is constant almost everywhere, and as a result the geodesic flow is ergodic. Assuming that $\kappa_\bbmu$ is compactly supported continuous, the ergodicity of the geodesic flow is tied to the constancy of $(\kappa_\bbmu)_\lambda$ for $\kappa_\bbmu$, given that functions of $\kappa_\bbmu$ type are dense in $\Lebesgue^1(\mathring{\mathcal{T}}^1\mathcal{S}_\mathbbl{\Gamma}, d\bbmu)$. In other words, this implies that 
\[
	d\bigl(\varphi_t(x), \varphi_{t + s}(\dot{x})\bigr) \to 0, 
\]	
and it tends to zero as $t \to \infty$, for a value $s \in \mathbb{R}$. Let us treat the differences
\begin{align}
\label{align "Equality of differences in Hopfian statistical process"}
	& \frac{\int^{\textcyrillic{\textit{ъ}}}_0\kappa_\bbmu\bigl(\varphi_t(x)\bigr)dt}{\int^{\textcyrillic{\textit{ъ}}}_0\lambda_\bbmu\bigl(\varphi_t(x)\bigr)dt} - \frac{\int^{\textcyrillic{\textit{ъ}}}_0\kappa_\bbmu\bigl(\varphi_{t + s}(\dot{x})\bigr)dt}{\int^{\textcyrillic{\textit{ъ}}}_0\lambda_\bbmu\bigl(\varphi_{t + s}(\dot{x})\bigr)dt} = \frac{\int^{\textcyrillic{\textit{ъ}}}_0\left\{\frac{\kappa_\bbmu\bigl(\varphi_t(x)\bigr) - \kappa_\bbmu\bigl(\varphi_{t + s}(\dot{x})\bigr)}{\lambda_\bbmu\bigl(\varphi_t(x)\bigr)}\right\}\lambda_\bbmu\bigl(\varphi_t(x)\bigr)dt}{\int^{\textcyrillic{\textit{ъ}}}_0\lambda_\bbmu\bigl(\varphi_t(x)\bigr)dt} \notag \\
	& - \frac{\int^{\textcyrillic{\textit{ъ}}}_0\kappa_\bbmu\bigl(\varphi_{t + s}(\dot{x})\bigr)dt}{\int^{\textcyrillic{\textit{ъ}}}_0\lambda_\bbmu\bigl(\varphi_{t + s}(\dot{x})\bigr)dt} \cdot \frac{\int^{\textcyrillic{\textit{ъ}}}_0\left\{\frac{\lambda_\bbmu\bigl(\varphi_t(x)\bigr) - \lambda_\bbmu\bigl(\varphi_{t + s}(\dot{x})\bigr)}{\lambda_\bbmu\bigl(\varphi_t(x)\bigr)}\right\}\lambda_\bbmu\bigl(\varphi_t(x)\bigr)dt}{\int^{\textcyrillic{\textit{ъ}}}_0\lambda_\bbmu\bigl(\varphi_t(x)\bigr)dt}.
\end{align} 
Let us look at the the right-hand side of Eq. \eqref{align "Equality of differences in Hopfian statistical process"}: the first term and the last factor tend to 0 as $t \to \infty$, but the first factor is bounded. Thus we observe that 
\begin{equation}
	(\kappa_\bbmu)_\lambda(x) =	(\kappa_\bbmu)_\lambda(\dot{x}) = \lim_{\textcyrillic{\textit{ъ}} \to \infty}\frac{\int^{\textcyrillic{\textit{ъ}}}_0\kappa_\bbmu\bigl(\varphi_t(\dot{x})\bigr)dt}{\int^{\textcyrillic{\textit{ъ}}}_0\lambda_\bbmu\bigl(\varphi_t(\dot{x})\bigr)dt} = \lim_{\textcyrillic{\textit{ъ}} \to \infty}\frac{\int^{\textcyrillic{\textit{ъ}}}_0\kappa_\bbmu\bigl(\varphi_{t + s}(\dot{x})\bigr)dt}{\int^{\textcyrillic{\textit{ъ}}}_0\lambda_\bbmu\bigl(\varphi_{t + s}(\dot{x})\bigr)dt}.
\end{equation}
Briefly, after the action of $\mathbbl{\Gamma}$ on $\partial\mathbb{D}_\mathbb{C}$, it appears that $(\kappa_\bbmu)_\lambda$ is a $\mathbbl{\Gamma}$-invariant function on $\partial\mathbb{D}_\mathbb{C} \times \partial\mathbb{D}_\mathbb{C}$ which is constant almost everywhere on $\partial\mathbb{D}_\mathbb{C} \times \{\alpha^+\}$ and on $\{\alpha^-\} \times \partial\mathbb{D}_\mathbb{C}$, as well as on $\partial\mathbb{D}_\mathbb{C} \times \partial\mathbb{D}_\mathbb{C}$, thanks to \emph{Fubini's theorem} \cite{Fubini "Sugli integrali multipli"}. This proves that the geodesic flow on $\mathring{\mathcal{T}}^1\mathcal{S}_\mathbbl{\Gamma}$ is ergodic with respect to $\bbmu$. 
\enumerationisfinis
\end{proof}

Summarising: the geodesic flow on the unit tangent bundle of a surface (of finite area) and constant negative curvature represents a prototypal object of an ergodic flow. The Anosov flow is a generalization of this object, and it, too, is ergodic.

\section{Entropy within the Topological Thermodynamics: at the Frontier of Order and Chaos}
\label{section "Entropy within the Topological Thermodynamics: at the Frontier of Order and Chaos"}

\subsection{The Entropy-Energy Roots}
\label{subsection "The Entropy-Energy Roots"}

\begingroup
\footnotesize
I propose to call the quantity $\mathsf{S}$ the \emph{entropy} of the body, from the Greek word \textgreek{ἡ τροπή}, the transformation. I have deliberately formed the word \emph{entropy} to be as similar as possible to the word \emph{energy}, because the two quantities to be denoted by these words are so closely intertwined in their physical sense that a nominologic similarity of some sort seems appropriate. \\
\indent — \textsc{R. Clausius} \cite[p. 390]{Clausius "Ueber verschiedene fur die Anwendung bequeme Formen der Hauptgleichungen der mechanischen Warmetheorie"}

\endgroup

\vspace{2mm}

The birth of entropy tallies with the research put forward by L. \& S. Carnot \cite{Carnot L. "Principes fondamentaux de l'equilibre et du mouvement"} \cite{Carnot S. "Reflexions sur la puissance motrice du feu et sur les machines propres a developper cette puissance"}, but R. Clausius \cite{Clausius "Ueber verschiedene fur die Anwendung bequeme Formen der Hauptgleichungen der mechanischen Warmetheorie"} is the one to introduce this magnitude, starting with the name, and elaborating a mechanical theory of heat, in an attempt to better clarify the significance of the second law of thermodynamics.

Boltzmann was the first to establish a connection between entropy ($\mathsf{S}$) and microscopic states of a system ($W$) that are consistent with an observed macroscopic (thermal) phenomenon. He assumes the existence of a relationship between entropy and thermodynamic probability of a macroscopic state, by defining the thermodynamic probability as the number of ways in which this state can be realized with different microscopic configurations (see point \ref{item "Entropy in Prior Knowledge: Maxwell–Boltzmann Probability Distribution, and Ergodic Hypothesis of Thermodynamics"}, p. \pageref{item "Entropy in Prior Knowledge: Maxwell–Boltzmann Probability Distribution, and Ergodic Hypothesis of Thermodynamics"}). 

Subsequent studies  have shown that the entropy is a \emph{flexible} magnitude, with a wide application in the physical-mathematical disciplines; in general, it designates the \emph{measure of the degree of disorder,\footnote{
	\label{footnote "Relativity of the order/disorder concept"}
	The concept of order/disorder is \emph{relative}: it is subordinate  to some of our mental constructs. The atomic-molecular components of tea, when they are contained in a tea leaf, can be said to be less disordered, or more ordered (with, let we say, low entropy), compared to when they are released in hot water inside a teapot, appearing more disordered, or less ordered (with a higher entropy). But what is established to be a state of order/disorder is relative to an \emph{initial condition} (the atomic-molecular state of the leaf), as long as it is \emph{conventionally} taken as a \emph{postulate} (\textgreek{ὑπόθεσις}, cf. Section \ref{subsubsection "Scholium: Greco-Hellenistic Scientific Modus Operandi (the Origin of Hypotheses)"}), or a starting-principle. An ordered/disordered state subsequently \emph{derives} from this postulate. Cf. footnote \ref{footnote "order and chaos: separate concepts"}, p. \pageref{footnote "order and chaos: separate concepts"}, on the distinction between “disorder” and “chaos”.
	}
chaos, or uncertainty, and mixed-up-ness}—to use Gibbs' categorization \cite[p. 418]{Gibbs "Unpublished Fragments"}—\emph{in an isolated, or conservative, system}.

What we are interested in doing here is merely investigating the entropy in the following arguments: metric entropy (Kolmogorov–Sinai entropy), topological entropy and pressure (Ruelle–Walters free energy density), and related variational Theorems \ref{theorema "Variational principles for the topological entropy"} and \ref{theorema "Variational principles for the topological pressure"}. We will also take into consideration the entropy of the geodesic flow (Theorem of Pesin \& Freire–Mañé \ref{theorema "P+FM"}, with a look at the Lyapunov exponent).

\subsection{Kolmogorov–Sinai Metric Entropy (Quantity for a Measure Preserving Transformation)}
\label{subsection "Kolmogorov–Sinai Metric Entropy (Quantity for a Measure Preserving Transformation)"}

The \emph{topological entropy} corresponds to an \emph{invariant magnitude} in ergodic theory for continuous transformations defined
\enumerationisinitium
\item on a compact topological space, according to R.L. Adler, A.G. Konheim and M.H. McAndrew \cite{Adler Konheim and McAndrew "Topological entropy"},
\item on a (compact) metric space of finite dimension, according to R. Bowen \cite{Bowen "Entropy for Group Endomorphisms and Homogeneous Spaces"} and E.I. Dinaburg \cite{Dinaburg "On the relations among various entropy characteristics of dynamical systems"}, 
\item on any compact Hausdorff space, according to T.N.T. Goodman \cite{Goodman "Relating Topological Entropy and Measure Entropy"}, 
\item or on subsets of a compact space in a manner that resembles a Hausdorff fractal dimension, according to R. Bowen \cite{Bowen "Topological Entropy for Noncompact Sets"}.
\enumerationisfinis
The mathematical background of the topological entropy is offered by the \emph{Kolmogorov–Sinai metric entropy} \cite{Kolmogorov "A new metric invariant of transient dynamical systems and automorphisms in Lebesgue spaces"} \cite{Kolmogorov "On entropy per unit time as a metric invariant of automorphisms"} \cite{Sinai "On the notion of entropy of a dynamical system"}, also known as \emph{measure-theoretic entropy}.

We present some definitions of topological entropy, but first we need to say a quick word about the entropy of Kolmogorov and Sinai. We are starting with the assumption that the phase space of the dynamical system is a Lebesgue–Rohlin space. 

\begin{definitio}[Lebesgue–Rohlin space]
\label{definitio "Lebesgue–Rohlin space"}
Let $\mathcal{X}$ be a separable metric space, $\mathscr{B}_\sigma$ the $\sigma$-algebra of Borel sets of $\mathcal{X}$, and $\bbmu$ a Borel measure on $\mathcal{X}$. A probability space $(\mathcal{X}, \mathscr{B}_\sigma, \bbmu)$ is called a \emph{Lebesgue–Rohlin space} \cite{Rokhlin "On the fundamental ideas of measure theory"}, or \emph{Lebesgue (probability) space}, if it is a complete $\modulo 0$ (modulo zero) with respect to one of its countable bases $\mathcal{B}_n$, namely if there is a complete measurable space $(\widetilde{\mathcal{X}}, \widetilde{\mathscr{B}}_\sigma, \tilde{\bbmu})$ with respect to a basis $\widetilde{\mathcal{B}}_n$, a set $\mathcal{X}_0 \in \widetilde{\mathscr{B}}_\sigma$ of full $\tilde{\bbmu}$-measure, and $\mathcal{X} \xrightarrow{\pi} \mathcal{X}_0$ so that $
	\pi(\mathcal{B}_n) = \widetilde{\mathcal{B}}_n \cap \mathcal{X}_0, \enspace \bbmu \circ \pi^{-1} = \tilde{\bbmu}$. \definitiosymbol
\end{definitio}

\begin{definitio}[Kolmogorov–Sinai metric entropy]
\label{definitio "Kolmogorov–Sinai metric entropy"}
~\enumerationisinitium
\item 
\label{item "Boltzmann's H and H-theorem"}
We remind that the Boltzmann's $\mathsf{H}$ is a quantity with which to measure the extent to which the condition of a system is different from its corresponding equilibrium; the Boltzmann's $\mathsf{H}$-theorem is responsible for exhibiting that $\mathsf{H}$ tends to decrease with time to a minimum, and therefore the system is approaching the equilibrium condition (cf. \ref{item "Ergodic hypothesis of thermodynamics, Boltzmann equation, and H-theorem"} in Section \ref{subsection "Prior Knowledge: Maxwell–Boltzmann Probability Distribution, and Ergodic Hypothesis of Thermodynamics"}).
\item 
Denote by $\varphi_\bbmu \colon \mathcal{X} \to \mathcal{X}$ the measure preserving automorphisms of the Lebesgue–Rohlin space, or the transformation of $\mathcal{X}$ onto itself. The probability space (understood as in Definition \ref{definitio "Lebesgue–Rohlin space"}) and the measure preserving transformation generate a \emph{measure preserving dynamical system}, indicated by $(\mathcal{X}, \mathscr{B}_\sigma, \bbmu, \varphi_\bbmu)$. Take a finite or countable measurable partition $C^\mathcal{X} = \{C_1, \mathellipsis, C_r\}$ of $\mathcal{X}$. The theory of information, see the \emph{Shannon–McMillan theorem} \cite{Shannon "A Mathematical Theory of Communication"} \cite{McMillan "The Basic Theorems of Information Theory"}, in which the entropy is interpreted as a \emph{measure of uncertainty} about a system, tells that the entropy of $C^\mathcal{X}$ is fixed by 
\begin{equation}
	\mathsf{H}(C^\mathcal{X}) = - \sum^r_{\nu = 1}\bbmu(C^\mathcal{X}_\nu)\log{\bbmu}(C^\mathcal{X}_\nu).
\end{equation}
For each partition $C^\mathcal{X}$, there exist the limit
\begin{equation}
	\mathsf{h}_\bbmu(\varphi_\bbmu, C^\mathcal{X}) = \lim_{n \to \infty}\frac{1}{n}\mathsf{H}\left(\bigvee^{n - 1}_{\nu = 0}\varphi_\bbmu^{-\nu}(C^\mathcal{X})\right) = \lim_{n \to \infty}\mathsf{H}\left(C^\mathcal{X} \mathrel{\Bigg|} \bigvee^n_{\nu = 1}\varphi_\bbmu^{-\nu}(C^\mathcal{X})\right),
\end{equation}
through which an entropy of \emph{invariant metric} type appears. The Kolmogorov–Sinai metric entropy of $(\mathcal{X}, \mathscr{B}_\sigma, \bbmu, \varphi_\bbmu)$ is the upper bound of $\mathsf{h}_\bbmu$ over all $C^\mathcal{X}$,
\begin{equation}
	\mathsf{h}_\bbmu(\varphi_\bbmu) = \sup_{C^\mathcal{X}}\lim_{n \to \infty}\frac{1}{n}\mathsf{H}(C^\mathcal{X}) = \sup_{C^\mathcal{X}}\mathsf{h}_\bbmu(\varphi_\bbmu, C^\mathcal{X}),
\end{equation}   
and we can define it as the entropy of the automorphism $\varphi_\bbmu \colon \mathsf{h}_\bbmu = \sup_{C^\mathcal{X}}\mathsf{h}_\bbmu(C^\mathcal{X})$, i.e. the \emph{entropy of a measure preserving transformation}. \definitiosymbol
\enumerationisfinis
\end{definitio}

\subsection{Topological Entropy and Pressure, aka Ruelle–Walters Free Energy Density}
\label{subsection "Topological Entropy and Pressure, aka Ruelle–Walters Free Energy Density"}

\begingroup
\footnotesize
The purpose of this work is to introduce the notion of entropy as an invariant for continuous mappings [on a compact topological space]. \\
\indent — \textsc{R.L. Adler, A.G. Konheim} and \textsc{M.H. McAndrew} \cite[p. 309]{Adler Konheim and McAndrew "Topological entropy"}

\vspace{2mm}

Ergodic theory involves itself with the study of transformations of a measure space. Topological dynamics is involved with homeomorphisms of a topological space. The entropy of a measure preserving transformation is a gauge of its randomness. \\
\indent — \textsc{F. Hahn} and \textsc{Y. Katznelson} \cite[p. 335]{Hahn and Katznelson "On the Entropy of Uniquely Ergodic Transformations"}

\endgroup

\vspace{2mm}

\begin{definitio}[Topological entropy]
~\enumerationisinitium
\item Adler–Konheim–McAndrew solution. Suppose that the space $\mathcal{X}$ is compact and a topologic structure. Let $\varphi_\bbmu \colon \mathcal{X} \to \mathcal{X}$ be a continuous transformation, where $\bbmu$ is an $\varphi_\bbmu$-invariant Borel measure on $\mathcal{X}$. Let $\mathcal{U} = \{U_\nu\}$ be an open cover of $\mathcal{X}$ such that $\mathcal{X} = \bigcup_\nu{U_\nu}$. The entropy of $\varphi_\bbmu$ with respect to $\mathcal{U}$ is
\begin{equation}
	\mathsf{H}_\bbmu(\varphi_\bbmu, \mathcal{U}) = \lim_{n \to \infty}\frac{1}{n}\mathsf{H}_\bbmu\left(\mathcal{U} \vee \varphi_\bbmu^{-1}\mathcal{U} \vee \mathellipsis \vee \varphi_\bbmu^{-n+1}\mathcal{U}\right).	
\end{equation}
The topological entropy of $\varphi_\bbmu$ is the supremum of $\mathsf{H}(\varphi_\bbmu, \mathcal{U})$ over all open covers $\mathcal{U}$,
\begin{equation} 
	\mathsf{h}_\topological(\varphi_\bbmu) = \sup_\bbmu\mathsf{H}_\bbmu(\varphi_\bbmu, \mathcal{U}).
\end{equation} 
\item Bowen–Dinaburg solution. Let $\varphi_\bbmu \colon \mathcal{X} \to \mathcal{X}$ be a continuous transformation of a compact metric space $(\mathcal{X}, \distance)$, with a metric $\distance$ on $\mathcal{X}$, where $\bbmu$ is an $\varphi_\bbmu$-invariant Borel measure on $\mathcal{X}$, i.e. a probability measure on every Borel set $\invertedbreve{E} \subset \mathcal{X}$, with $\bbmu(\mathcal{X}) = 1$ and $\bbmu\bigl(\varphi_\bbmu^{-1}(\invertedbreve{E})\bigr) = \bbmu(\invertedbreve{E})$. For each number $n \in \mathbb{N}$, we impose on $\mathcal{X}$ a metric 
\begin{equation}
	\distance_n(x, y) = \max\left\{\distance\bigl(\varphi_\bbmu^\nu(x), \varphi_\bbmu^\nu(y)\bigr) \mid 0 \leqslant \nu \leqslant n - 1\right\}.
\end{equation}
Given a value $\varepsilon > 0$, a set $\invertedbreve{E}$ of $\mathcal{X}$ is said to be \emph{$(n, \varepsilon)$-separated} with respect to $\varphi_\bbmu$ if $\distance_n(x, y) > \varepsilon$, for each pair of distinct points $x, y \in \invertedbreve{E}$, $x \neq y$. Write as $N(\distance_n, \varepsilon)$ the maximum number of points of $(n, \varepsilon)$-separated sets $\invertedbreve{E} \subset \mathcal{X}$. Then the topological entropy of $\varphi_\bbmu$ is defined as
\begin{equation}	 
	\mathsf{h}_\topological(\varphi_\bbmu) \viz \mathsf{h}^\distance_\topological(\varphi_\bbmu) = \lim_{\varepsilon \to 0}\left(\limsup_{n \to \infty}\frac{1}{n}\log{N(\distance_n, \varepsilon)}\right).
\end{equation}
\item Summary solution. Let $\mathbbl{M}(\mathcal{X})$ be a family of Borel probability measures on $\mathcal{X}$. The topological entropy of $\varphi_\bbmu$ is the supremum of the Kolmogorov–Sinai metric entropy $\mathsf{h}_\bbmu$, 
\begin{equation}	
\label{equation "Topological entropy: summary solution"} 
	\mathsf{h}_\topological(\varphi_\bbmu) = \sup_\bbmu\mathsf{h}_\bbmu(\varphi_\bbmu),
\end{equation}
with $\bbmu \in \mathbbl{M}(\mathcal{X})$. The Eq. \eqref{equation "Topological entropy: summary solution"} is governed by a variational principle (see Theorem \ref{theorema "Variational principles for the topological entropy"}) under which $\mathsf{h}_\bbmu(\varphi_\bbmu) \leqslant \mathsf{h}_\topological(\varphi_\bbmu)$, as revealed by L.W. Goodwyn \cite{Goodwyn "Topological Entropy Bounds Measure-Theoretic Entropy"}.
\item $\mathsf{h}_\topological(\varphi_\bbmu)$ is said \emph{invariant} (more precisely, it consists of an invariant of topological conjugacy) in the sense that, given a transformation $\varphi_{\bbmu_{k = 1, 2}}$ of $\mathcal{X}$ into itself, and a homeomorphism 
\begin{equation}
	\vartheta \colon \mathcal{X}_1 \to \mathcal{X}_2, \text{ with } \vartheta \circ \varphi_{\bbmu_1} = \varphi_{\bbmu_2} \circ \vartheta, 
\end{equation}
then 
\begin{equation}
	\mathsf{h}_\topological(\varphi_{\bbmu_1}) = \mathsf{h}_\topological(\varphi_{\bbmu_2}). 
\end{equation}
In this way (since the homeomorphism is no more than a topological isomorphism), the topological $\mathsf{h}_\topological$-entropy can be described as an \emph{isomorphism invariant for a measure preserving transformation}. \definitiosymbol
\enumerationisfinis
\end{definitio}

\begin{margo}[Entropy and topology in Pettini's interpretation]
Also of note is the M. Pettini's \cite[pp. 248, 285-294]{Pettini "Geometry and Topology in Hamiltonian Dynamics and Statistical Mechanics"} solution, coming from a physical background, that combines entropy and topology as part of the involvement of topology at the origin of \emph{thermodynamic phase transitions}; it based on a Riemannian theory of Hamiltonian chaos, and  consists in studying \emph{curvature fluctuations} of an (enlarged) configuration space(-time) described by the Riemannian geometrization of (Hamiltonian) dynamics. In essence, it will be shown in an analytical way that a certain topological change in the configuration space(-time) is necessarily related to a phase transition phenomenon. \margosymbol
\end{margo}

We are introducing a further quantity about the thermodynamic formalism, the so-called \emph{topological pressure}, more correctly known also as \emph{free energy density}. The free energy, in this context, is the sum of the Kolmogorov–Sinai metric entropy and the integral of a continuous (Borel-measurable) function regarding the phase space probability distribution. It is a notion modelled on the grand canonical ensemble definition for lattice gas calculations \cite[pp. 942-943]{van Enter Fernandez Sokal "Regularity properties and pathologies of position-space renormalization-group transformations: Scope and limitations of Gibbsian theory"}, as a generalization of the topological entropy. The genesis of the topological pressure can be found in D. Ruelle \cite{Ruelle "Statistical mechanics on a compact set with $Z^nu$ action satisfying expansiveness and specification"} \cite{Ruelle "Statistical Mechanics on a Compact Set with $Z^p$ Action Satisfying Expansiveness and Specification"} and P. Walters \cite{Walters "A Variational Principle for the Pressure of Continuous Transformations"}, cf. \cite[6.20-22]{Ruelle "Thermodynamic Formalism: The Mathematical Structures of Equilibrium Statistical Mechanics"}.

\begin{definitio}[Topological pressure]
\label{definitio "Topological pressure"}
Letting $\varphi_\bbmu \colon \mathcal{X} \to \mathcal{X}$ be a continuous transformation of a compact metric space $(\mathcal{X}, \distance)$, the topological pressure of a continuous function 
\begin{equation}
	\underdot{\varpi} \colon \mathcal{X} \to \mathbb{R} 
\end{equation}
with respect to $\varphi_\bbmu$ is
\begin{equation}
	\mathsf{P}_\topological(\underdot{\varpi}) \viz \mathsf{P}_\topological(\varphi_\bbmu, \underdot{\varpi}) = \lim_{\varepsilon \to 0}\limsup_{n \to \infty}\frac{1}{n}\log\sup_{\invertedbreve{E} \subset \mathcal{X}}\sum_{x \in \invertedbreve{E}}\exp{\left\{\sum^{n - 1}_{\nu = 0}\underdot{\varpi}\bigl(\varphi_\bbmu^\nu(x)\bigr)\right\}},
\end{equation}
where $\invertedbreve{E} \subset \mathcal{X}$ is an $(n, \varepsilon)$-separated set in the way that the Bowen–Dinaburg solution shows. The expressions
\begin{equation}
	\mathsf{P}_\bbmu(\underdot{\varpi}) \viz \mathsf{P}_\bbmu(\varphi_\bbmu, \underdot{\varpi}) = \mathsf{h}_\bbmu(\varphi_\bbmu) + \int_\mathcal{X}\underdot{\varpi}{d\bbmu}. 
\end{equation}
defines the pressure of $\bbmu \in \mathbbl{M}(\mathcal{X})$. \definitiosymbol
\end{definitio}

\subsection{Application of the Calculus of Variations in the Topological Entropy and Pressure}
\label{subsection "Application of the Calculus of Variations in the Topological Entropy and Pressure"}

Next is the variational principle for these quantities.

\begin{theorema}[Variational principles for the topological entropy]
\label{theorema "Variational principles for the topological entropy"}
Let $\varphi_\bbmu \colon \mathcal{X} \to \mathcal{X}$ be a continuous transformation of a compact metric space $(\mathcal{X}, \distance)$, where $\bbmu$ is an $\varphi_\bbmu$-invariant Borel measure on $\mathcal{X}$. Then
\begin{equation}
	\mathsf{h}_\topological(\varphi_\bbmu) = \left\{\sup_\bbmu\mathsf{h}_\bbmu(\varphi_\bbmu) \mathrel{\Bigg|} \bbmu \in \mathbbl{M}(\mathcal{X})\right\}.
\end{equation}
\end{theorema}

\begin{proof}
Take two finite or countable measurable partitions $C^\mathcal{X} = \{C_1, \mathellipsis, C_r\}$ and $K^\mathcal{X} = \{K_0, \mathellipsis, K_r\}$ of $\mathcal{X}$, where 
\begin{equation}
	K_0 = \mathcal{X} \backslash \bigcup^r_{\nu = 1}K_\nu,
\end{equation}
according to the inclusion relation $K_\nu \subset C_\nu$, for $\nu = 1, \mathellipsis, r$, and a value $\delta > 0$ such that $\bbmu(C_\nu \backslash K_\nu) < \delta$. Then 
\begin{equation}
\label{equation "Metric and topological entropy + partitions"}	
	\mathsf{h}_\bbmu(\varphi_\bbmu^k, C^\mathcal{X}) \leqslant \mathsf{h}_\bbmu(\varphi_\bbmu^k, K^\mathcal{X}) + \mathsf{H}_\bbmu(C^\mathcal{X} \mid K^\mathcal{X}) < \mathsf{h}_\bbmu(\varphi_\bbmu^k, K^\mathcal{X}) + 1, 
\end{equation}
for any $k \in \mathbb{N}$. Let $\mathcal{U} = \{K_0 \cup K_1, \mathellipsis, K_0 \cup K_r\}$ and
\begin{equation}
	\mathcal{U}_{kn} = \left\{\bigcap^{n - 1}_{\nu = 0}\varphi_\bbmu^{-\nu{k}}U_\nu \mathrel{\bigg|} U_0, \mathellipsis, U_{n - 1} \in \mathcal{U}\right\}_{k, n \in \mathbb{N}}
\end{equation}
be open and finite covers of $\mathcal{X}$. Moreover, $K_0 \cup K_\nu \in \mathcal{U}$ is an union each element of which intersects at most the elements $K_0, K_\nu \in K^\mathcal{X}$, so $\#K^\mathcal{X}_{kn}\leqslant 2^n\#\mathcal{U}_{kn}$, where $\#$ denotes the cardinality. Since $\mathsf{H}_\bbmu(K^\mathcal{X}_{kn}) \leqslant \log\#K^\mathcal{X}_{kn}$, we see that $\mathsf{H}_\bbmu(K^\mathcal{X}_{kn}) \leqslant \log\#K^\mathcal{X}_{kn} \leqslant \log(2^n\#\mathcal{U}_{kn})$.

Let us say that $\varepsilon \viz \mathrm{Leb}(\mathcal{U})$ is the \emph{Lebesgue number} of $\mathcal{U}$, on the basis of which (for each open covering of a compact metric space) there exists a positive number $\varepsilon > 0$ such that any subset of $\mathcal{X}$ of diameter less than $\varepsilon$ is fully contained in one of the elements of $\mathcal{U}$. (This means, for example, that it is possible to cover $\mathcal{X}$ by a finite number of open balls of radius $\rho = \frac{\varepsilon}{3}$, in order to demonstrate the compactness in a sequentially compact metric space, and it appears that each of these balls lies in some element of $\mathcal{U}$, i.e. $\varepsilon$-$\mathbb{B}_\rho \subset U$). Thus $\varepsilon$ is also the Lebesgue number of $\mathcal{U}_{kn}$ with respect to the metric $\distance^{(*)} \viz \distance_n\varphi_\bbmu^k$. In addition, $\mathcal{U}_{kn}$ is a minimal covering, and every open set $U \in \mathcal{U}_{kn}$ contains at least a point $x_U$ not in any other element of $\mathcal{U}_{kn}$, therefore $\mathbb{B}_{\distance^{(*)}}(x_U, \varepsilon) \subset U$.

By Eq. \eqref{equation "Metric and topological entropy + partitions"} we have $\mathsf{h}_\bbmu(\varphi_\bbmu^k, C^\mathcal{X}) \leqslant \mathsf{h}_\topological(\varphi_\bbmu^k) + \log{2} + 1$. Since $\mathsf{h}_\bbmu(\varphi_\bbmu^k) = k\mathsf{h}_\bbmu(\varphi_\bbmu)$, we obtain 
\begin{equation}
	\mathsf{h}_\bbmu(\varphi_\bbmu) = \frac{1}{k}\mathsf{h}_\bbmu(\varphi_\bbmu^k) \leqslant \frac{1}{k}\left(\mathsf{h}_\topological(\varphi_\bbmu^k) + \log{2} + 1\right) = \mathsf{h}_\topological(\varphi_\bbmu) + \frac{1}{k}(\log{2} + 1).
\end{equation}
Imposing $k \to \infty$, it turns out that $\mathsf{h}_\bbmu(\varphi_\bbmu) \leqslant \mathsf{h}_\topological(\varphi_\bbmu)$.

Now, assume two probability measures
\begin{equation}
\label{equation "Probability measures"}
	\bbmu_{n \in \mathbb{N}} = \frac{1}{n}\sum^{n - 1}_{\nu = 0}\varphi_{\bbmu^\star}^\nu\bbsigma_n \text{ and } \bbsigma_n = \frac{1}{\#\invertedbreve{E}_n}\sum_{x \in \invertedbreve{E}_n}\delta_x.
\end{equation}
It is known that, for a compact metric space $\mathcal{X}$ and an $(n, \varepsilon)$-separated set $\invertedbreve{E}_n \subset \mathcal{X}$, there exists in the weak-$\star$ topology an accumulation point
\begin{equation}
	\bbmu = \lim_{n \to \infty} \bbmu_{r_n} \in \mathbbl{M}(\mathcal{X})
\end{equation}
of the sequence $\{r_n\}_{n \in \mathbb{N}}$ of measures $\bbmu_{r_n}$ with
\begin{equation}
	\lim_{n \to \infty}\frac{1}{r_n}\log\#\invertedbreve{E}_{r_n} = \limsup_{n \to \infty}\frac{1}{n}\log\#\invertedbreve{E}_n, 
\end{equation}
for which
\begin{equation}
	\limsup_{n \to \infty}\frac{1}{n}\log\#\invertedbreve{E}_n \leqslant \mathsf{h}_\bbmu(\varphi_\bbmu).
\end{equation}
The same goes for
\begin{equation}
	\limsup_{n \to \infty}\frac{1}{n}\log{N(\distance_n, \varepsilon)} \leqslant \mathsf{h}_\bbmu(\varphi_\bbmu) \text{ and } \limsup_{n \to \infty}\frac{1}{n}\log{N(\distance_n, \varepsilon)} \leqslant \sup_\bbmu\mathsf{h}_\bbmu(\varphi_\bbmu),
\end{equation}
so
\begin{equation}
	\limsup_{n \to \infty}\frac{1}{n}\log{N(\distance_n, \varepsilon)} \leqslant \sup_\bbmu\mathsf{h}_\bbmu(\varphi_\bbmu).
\end{equation}
The variational principle is demonstrated by imposing $\varepsilon \to 0$.
\end{proof}

\begin{theorema}[Variational principles for the topological pressure]
\label{theorema "Variational principles for the topological pressure"}
Let $\varphi_\bbmu \colon \mathcal{X} \to \mathcal{X}$ be a continuous transformation of a compact metric space $(\mathcal{X}, \distance)$, where $\bbmu$ is an $\varphi_\bbmu$-invariant Borel measure on $\mathcal{X}$. Let $\underdot{\varpi} \colon \mathcal{X} \to \mathbb{R}$ denote a continuous function, that is $\underdot{\varpi} \in \mathscr{C}^0(\mathcal{X})$. Then
\begin{align}
	\mathsf{P}_\topological(\underdot{\varpi}) & \viz \mathsf{P}_\topological(\varphi_\bbmu, \underdot{\varpi}) \notag \\ 
	& = \sup_\bbmu\left(\mathsf{h}_\bbmu(\varphi_\bbmu) + \int_\mathcal{X}\underdot{\varpi}{d\bbmu}\right) = \left\{\sup_\bbmu\mathsf{P}_\bbmu(\underdot{\varpi}) \mathrel{\Bigg|} \bbmu \in \mathbbl{M}(\mathcal{X})\right\}.
\end{align}
\end{theorema}

\begin{proof}
Same initial conditions of the previous Theorem \ref{theorema "Variational principles for the topological entropy"}. Take two finite or countable measurable partitions $C^\mathcal{X} = \{C_1, \mathellipsis, C_r\}$ and $K^\mathcal{X} = \{K_0, \mathellipsis, K_r\}$ of $\mathcal{X}$, where $K_0 = \mathcal{X} \backslash \bigcup^r_{\nu = 1}K_\nu$,
according to the inclusion relation $K_\nu \subset C_\nu$, for $\nu = 1, \mathellipsis, r$, and a value $\delta > 0$ such that $\bbmu(C_\nu \backslash K_\nu) < \delta$. For a choice of $\delta$ sufficiently small, we have $\mathsf{h}_\bbmu(\varphi_\bbmu, C^\mathcal{X}) < \mathsf{h}_\bbmu(\varphi_\bbmu, K^\mathcal{X}) + 1$.

For $n \in \mathbb{N}$, we select an $\bigl(n, \frac{\varepsilon}{2}\bigr)$-separated set $\invertedbreve{E} \subset \mathcal{X}$, with $\#\invertedbreve{E} = N\bigl(\distance_n, \frac{\varepsilon}{2}\bigr)$. Let $x_C$ such that $\underdot{\varpi}_n(x_C) = \sup\{\underdot{\varpi}_n(x) \mid x \in C\}$, where $\underdot{\varpi}_n = \sum^{n - 1}_{\iota = 0}\underdot{\varpi} \circ \varphi_\bbmu^\iota$, for $\iota = 0, \mathellipsis, n - 1$, and $y_C \in \invertedbreve{E}$ such that $\distance_n(x_C, y_C) < \varepsilon$. Since $|\underdot{\varpi}(x) - \underdot{\varpi}(y)| < 1$ whenever $d(x, y) < \varepsilon$, we observe that $\underdot{\varpi}_n(x_C) \leqslant \underdot{\varpi}_n(y_C) + n$, and $\#\{C \in K^\mathcal{X}_n \mid y_C = x\} \leqslant 2^n$, for each $x \in \invertedbreve{E}$. To follow, a lemma is required for the continuation of the proof.

\begin{lemma}
\label{lemma "Lemma for the variational principle on the topological pressure"}
Given some number $\alpha_\nu \geqslant 0$, with $\sum^r_{\nu = 1}\alpha_\nu = 1$, and $\beta_\nu \in \mathbb{R}$, for $\nu = 1, \mathellipsis, r$, the formula	
\begin{equation}
\label{equation "Formula in Lemma for the variational principle on the topological pressure"}
	\sum^r_{\nu = 1}\alpha_\nu(\beta_\nu - \log\alpha_\nu) \leqslant \log\sum^r_{\nu = 1}e^{\beta_\nu}
\end{equation}
holds with equality iff 
\begin{equation}
	\alpha_\nu = \frac{e^{\beta_\nu}}{\sum^r_{\nu = 1}e^{\beta_\nu}}, 
\end{equation}
where $e^{\beta_\nu} \viz \exp{\{\beta_\nu\}}$ is the exponential function.
\end{lemma}

\begin{proof}
Let 
\begin{equation}
	\gamma_\nu = \frac{e^{\beta_\nu}}{\sum^r_{\nu = 1}e^{\beta_\nu}} 
\end{equation}
and $x_\nu = \frac{\alpha_\nu}{\gamma_\nu}$. For a convex function $\upsilon \colon [0, 1] \to \mathbb{R}$, 
\begin{equation}
	\upsilon(x) =
	\begin{cases}
	x\log{x} \text{ if } 0 < x \leqslant 1, \\
	0 \text{ if} x = 0,
	\end{cases}
\end{equation}
it happens that
\begin{equation}
	0 = \upsilon\left(\sum^r_{\nu = 1}\gamma_\nu{x_\nu}\right) \leqslant \sum^r_{\nu = 1}\gamma_\nu\upsilon(x_\nu) = \sum^r_{\nu = 1}\alpha_\nu\left(\log\alpha_\nu + \log\sum^r_{\nu = 1}e^{\beta_\nu} - \beta_\nu\right)
\end{equation}
with equality iff $\alpha_\nu = \gamma_\nu$.
\end{proof}

The lemma raised enables us to move forward:
\begin{align}
	\mathsf{H}_\bbmu(K^\mathcal{X}_n) + \int_\mathcal{X}\underdot{\varpi}_n{d\bbmu} & \leqslant \sum_{C \in K^\mathcal{X}_n}\bbmu(C)\bigl(-\log\bbmu(C) + \underdot{\varpi}_n(x_C)\bigr) \notag \\
	& \leqslant \log\sum_{C \in K^\mathcal{X}_n}e^{\underdot{\varpi}_n(x_C)} \leqslant \log\sum_{C \in K^\mathcal{X}_n}e^{\underdot{\varpi}_n(y_C) + n} \notag \\ 
	& \hspace{33mm} \leqslant n + \log\left(2^n\sum_{x \in \invertedbreve{E}}e^{\underdot{\varpi}_n(x)}\right),
\end{align} 
and
\begin{align}
	\frac{1}{n}\mathsf{H}_\bbmu(K^\mathcal{X}_n) + \int_\mathcal{X}\underdot{\varpi}{d\bbmu} & = \frac{1}{n}\mathsf{H}_\bbmu(K^\mathcal{X}_n) + \frac{1}{n}\int_\mathcal{X}\underdot{\varpi}_n{d\bbmu} \notag \\
	& \leqslant 1 + \log{2} + \frac{1}{n}\log\left(\sup_{\invertedbreve{E}}\sum_{x \in \invertedbreve{E}}e^{\underdot{\varpi}_n(x)}\right), 
\end{align}
so
\begin{align}
	\mathsf{h}_\bbmu(\varphi_\bbmu, C^\mathcal{X}) + \int_\mathcal{X}\underdot{\varpi}{d\bbmu} & < \mathsf{h}_\bbmu(\varphi_\bbmu, K^\mathcal{X}) + 1 + \int_\mathcal{X}\underdot{\varpi}{d\bbmu} \notag \\ 
	& \leqslant 2 + \log{2} + \limsup_{n \to \infty}\frac{1}{n}\log\left(\sup_{\invertedbreve{E}}\sum_{x \in \invertedbreve{E}}e^{\underdot{\varpi}_n(x)}\right),
\end{align}
and, if $\varepsilon \to 0$,
\begin{equation}
	\mathsf{h}_\bbmu(\varphi_\bbmu) + \int_\mathcal{X}\underdot{\varpi}{d\bbmu} = \sup_{C^\mathcal{X}}\left(\mathsf{h}_\bbmu(\varphi_\bbmu, C^\mathcal{X}) + \int_\mathcal{X}\underdot{\varpi}{d\bbmu}\right) \leqslant 2 + \log{2} + \mathsf{P}_\topological(\underdot{\varpi}),
\end{equation}
where $	\mathsf{P}_\topological(\underdot{\varpi}) \viz \mathsf{P}_\topological(\varphi_\bbmu, \underdot{\varpi})$. Leveraging the equality $\mathsf{P}_\topological^k(\underdot{\varpi}_k) = k\mathsf{P}_\topological(\underdot{\varpi})$, for $k \in \mathbb{N}$, yields
\begin{align}
	\mathsf{h}_\bbmu(\varphi_\bbmu) + \int_\mathcal{X}\underdot{\varpi}{d\bbmu} = \frac{1}{k}\left(\mathsf{h}_\bbmu(\varphi_\bbmu^k) + \int_\mathcal{X}\underdot{\varpi}_k{d\bbmu}\right) & \leqslant \frac{1}{k}\bigl(2 + \log{2} + \mathsf{P}_\topological^k(\underdot{\varpi}_k)\bigr) \notag \\
	& = \frac{2 + \log{2}}{k} + \mathsf{P}_\topological(\underdot{\varpi}).
\end{align}
Letting $k \to \infty$, we arrive at
\begin{equation}
	\mathsf{h}_\bbmu(\varphi_\bbmu) + \int_\mathcal{X}\underdot{\varpi}{d\bbmu} \leqslant \mathsf{P}_\topological(\underdot{\varpi}),
\end{equation}
and
\begin{equation}
	\sup_\bbmu\left(\mathsf{h}_\bbmu(\varphi_\bbmu) + \int_\mathcal{X}\underdot{\varpi}{d\bbmu}\right) \leqslant \mathsf{P}_\topological(\underdot{\varpi}).
\end{equation}
We are only halfway along the road to the completion of the proof. At this stage, take
\enumerationisinitium
\item an $(n, \varepsilon)$-separated set $\invertedbreve{E}_n \subset \mathcal{X}$ of points in the metric $\distance_n$, with $\varepsilon > 0$, such that
\begin{equation}
\label{equation "Set $E$ of points for the variational principle on the topological pressure"}
	\log\sum_{x \in \invertedbreve{E}_n}e^{\underdot{\varpi}_n(x)} > \log\left(\sup_{\invertedbreve{E}}\sum_{x \in \invertedbreve{E}_n}e^{\underdot{\varpi}_n(x)}\right) - 1;
\end{equation}
\item two probability measures $\bbmu_n = \frac{1}{n}\sum^{n - 1}_{\nu = 0}\varphi_{\bbmu^\star}^\nu\bbsigma_n$, as in \eqref{equation "Probability measures"}, and $\bbsigma_n$, by imposing 
\begin{equation}
	\bbsigma_n = \frac{\sum_{x \in \invertedbreve{E}_n}e^{\underdot{\varpi}_n(x)}\delta_x}{\sum_{x \in \invertedbreve{E}_n}e^{\underdot{\varpi}_n(x)}},
\end{equation}
and a sequence $\{r_n\}_{n \in \mathbb{N}}$ of $\bbmu_{r_n}$ such that
\begin{equation}
\label{equation "Limit and limit superior with sequence for the variational principle on the topological pressure"}
	\lim_{n \to \infty}\frac{1}{r_n}\log\sum_{x \in \invertedbreve{E}_{r_n}}e^{\underdot{\varpi}_{r_n}(x)} = \limsup_{n \to \infty}\frac{1}{n}\log\sum_{x \in \invertedbreve{E}_n}e^{\underdot{\varpi}_n(x)}.
\end{equation}
\enumerationisfinis
Let be $W^\mathcal{X} = \{W_1, \mathellipsis, W_r\}$ a partition of $\mathcal{X}$ of diameter $W < \varepsilon$ and $\bbmu(\partial{W}) = 0$. Setting $\invertedbreve{E}_n = \{x_1, \mathellipsis, x_r\}, \alpha_\nu = \bbsigma_n\{x_\nu\}$, and $\beta_\nu = \underdot{\varpi}_n(x_\nu)$, for $\nu = 1, \mathellipsis, r$, we see that
\begin{equation}
	\alpha_\nu = \frac{e^{\underdot{\varpi}_n(x_\nu)}}{\sum_{x \in \invertedbreve{E}_n}e^{\underdot{\varpi}_n(x)}} = \frac{e^{\beta_\nu}}{\sum^r_{\nu = 1}e^{\beta_\nu}}.
\end{equation}
Now, it is quite clear that the formula \eqref{equation "Formula in Lemma for the variational principle on the topological pressure"} is an identity; thanks to Lemma \ref{lemma "Lemma for the variational principle on the topological pressure"},
\begin{align}
\label{align "Equation with partition $W$ for the variational principle on the topological pressure"}
	\mathsf{H}_{	\bbsigma_n}(W^\mathcal{X}_n) + n \int_\mathcal{X}\underdot{\varpi}{d\bbmu_n} & = \mathsf{H}_{\bbsigma_n}(W^\mathcal{X}_n) + \int_\mathcal{X}\underdot{\varpi}_n{d\bbsigma_n} \notag \\ 
	& = \sum_{x \in \invertedbreve{E}_n}\bbsigma_n\{x\}\bigl(-\log\bbsigma_n\{x\} + \underdot{\varpi}_n(x)\bigr) = \log\sum_{x \in \invertedbreve{E}_n}e^{\underdot{\varpi}_n(x)}.
\end{align}
Let $n = mk + u$, with $m \geqslant 0$ and $0 \leqslant u < k$, for $k, n \in \mathbb{N}$, so we can establish
\begin{equation}
	W^\mathcal{X}_n = W^\mathcal{X}_{mk + u} = \bigvee^{m - 1}_{\iota = 0}\varphi_\bbmu^{-\iota{k}}W^\mathcal{X}_k \vee \bigvee^{mk + u - 1}_{\iota = mk}\varphi_\bbmu^{-\iota}W^\mathcal{X},
\end{equation}
and, for $\nu = 0, \mathellipsis, k - 1$,
\begin{equation}
	C^\mathcal{X} = \bigvee^{m - 1}_{\iota = 0}\varphi_\bbmu^{-\iota{k} - \nu}W^\mathcal{X}_k \vee \bigvee^{mk + u - 1}_{\iota = mk}\varphi_\bbmu^{-\iota}W^\mathcal{X} \vee W^\mathcal{X}_\nu.
\end{equation}
Using Eq. \eqref{align "Equation with partition $W$ for the variational principle on the topological pressure"}, it appears that
\begin{align}
	\frac{k}{n}\log\sum_{x \in \invertedbreve{E}_n}e^{\underdot{\varpi}_n(x)} & = \frac{k}{n}\mathsf{H}_{\bbsigma_n}(W^\mathcal{X}_n) + k \int_\mathcal{X}\underdot{\varpi}{d\bbmu_n} \notag \\
	& = \frac{1}{n}\sum^{k - 1}_{\nu = 0}\mathsf{H}_{\bbsigma_n}(W^\mathcal{X}_n) + k \int_\mathcal{X}\underdot{\varpi}{d\bbmu_n} \notag \\
	& \leqslant \frac{1}{n}\sum^{k - 1}_{\nu = 0}\sum^{m - 1}_{\iota = 0}\mathsf{H}_{\bbsigma_n}(\varphi_\bbmu^{-\iota{k} - \nu}W^\mathcal{X}_k) + \frac{2k^2}{n}\log\#W^\mathcal{X} + k \int_\mathcal{X}\underdot{\varpi}{d\bbmu_n} \notag \\
	& \leqslant \mathsf{H}_{\bbmu_n}(W^\mathcal{X}_k) + \frac{2k^2}{n}\log\#W^\mathcal{X} + k \int_\mathcal{X}\underdot{\varpi}{d\bbmu_n},
\end{align}
and hence, under the Eq. \eqref{equation "Limit and limit superior with sequence for the variational principle on the topological pressure"},
\begin{align}
	\lim_{n \to \infty}\frac{1}{r_n}\log\sum_{x \in \invertedbreve{E}_{r_n}}e^{\underdot{\varpi}_{r_n}(x)} & \leqslant \lim_{n \to \infty}\left(\frac{1}{k}\mathsf{H}_{\bbmu_{r_n}}(W^\mathcal{X}_k) + \frac{2k}{r_n}\log\#W^\mathcal{X} + \int_\mathcal{X}\underdot{\varpi}{d\bbmu_{r_n}}\right) \notag \\
	& = \frac{1}{k}\mathsf{H}_\bbmu(W^\mathcal{X}_k) + \int_\mathcal{X}\underdot{\varpi}{d\bbmu}.
\end{align}
If $k \to \infty$
\begin{equation}
	\lim_{n \to \infty}\frac{1}{r_n}\log\sum_{x \in \invertedbreve{E}_{r_n}}e^{\underdot{\varpi}_{r_n}(x)} \leqslant \mathsf{h}_\bbmu(\varphi_\bbmu, W^\mathcal{X}) + \int_\mathcal{X}\underdot{\varpi}{d\bbmu} \leqslant \mathsf{h}_\bbmu(\varphi_\bbmu) + \int_\mathcal{X}\underdot{\varpi}{d\bbmu}.
\end{equation}
Therefore
\begin{equation}
	\limsup_{n \to \infty}\frac{1}{n}\log\sum_{x \in \invertedbreve{E}_n}e^{\underdot{\varpi}_n(x)} \leqslant \sup_{\bbsigma}\left(\mathsf{h}_\bbsigma(\varphi_\bbmu) + \int_\mathcal{X}\underdot{\varpi}{d\bbsigma}\right).
\end{equation}
From Eq. \eqref{equation "Set $E$ of points for the variational principle on the topological pressure"}, letting $\varepsilon \to 0$, we get
\begin{equation}
	\mathsf{P}_\topological(\underdot{\varpi}) \leqslant \sup_\bbsigma\left(\mathsf{h}_\bbsigma(\varphi_\bbmu) + \int_\mathcal{X}\underdot{\varpi}{d\bbsigma}\right) = \left\{\sup_\bbsigma\mathsf{P}_\bbsigma(\underdot{\varpi}) \mathrel{\Bigg|} \bbsigma \in \mathbbl{M}(\mathcal{X})\right\},
\end{equation}
and the proof is complete.
\end{proof}

\subsection{Entropy of the Geodesic Flow}

\subsubsection{Topological Entropy (via Mañé's Formula) and Geodesic Entropy}

We consider here the \emph{topological entropy of any geodesic flow} and the \emph{geodesic entropy of the Riemannian metric}, denoted by $\mathsf{h}_\topological(\varphi_t)$ and $	\mathsf{h}_\geodesic(g)$, respectively.
\enumerationisinitium
\item Let a closed connected $\mathscr{C}^\infty$ Riemannian manifold $\mathcal{M}$ and a geodesic flow on the unit tangent bundle $\varphi_t \colon \mathring{\mathcal{T}}^1\mathcal{M} \to \mathring{\mathcal{T}}^1\mathcal{M}$ be given; fix $\tau > 0$ and indicate with  $N_\tau(x, y)$ the number of geodesics parametrized by arc length between two points $x$ and $y$ in $\mathcal{M}$ with length $\leqslant \tau$. R. Mañé \cite{Mane "On the topological entropy of geodesic flows"} identifies the topological entropy as a \emph{Riemannian characterization}, which relates the exponential growth rate of $N_\tau(x, y)$, as a function of $\tau$, with the topological entropy $\mathsf{h}_\topological(\varphi_t)$. The function $N_\tau(x, y)$ is finite and locally constant on an open full measure subset in the product manifold $\mathcal{M} \times \mathcal{M}$. More specifically, the following statement, known as \emph{Mañé's formula}, is applicable,
\begin{align}
		\mathsf{h}_\topological(\varphi_t) & = \limsup_{\tau \to \infty}\frac{1}{\tau}\log\int_{\mathcal{M} \times \mathcal{M}}N_\tau(x, y)dxdy \notag \\
		& = \limsup_{\tau \to \infty}\frac{1}{\tau}\log\int_{\mathring{\mathcal{T}}^1\mathcal{M}}\expansion(d_\theta\varphi_t)d\theta \notag \\
		& = \limsup_{\tau \to \infty}\frac{1}{\tau}\log\int_{\mathring{\mathcal{T}}^1\mathcal{M}}\det\left(d_\theta\varphi_t\big|_{\mathring{\mathcal{V}}(\theta)}\right)d\theta \notag \\ 
		& = \limsup_{\tau \to \infty}\frac{1}{\tau}\log\int_\mathcal{M}\volume\bigl(\varphi_t\left(\mathcal{T}_x\mathcal{M}\right)\bigr)dx,
\end{align}
where $\mathring{\mathcal{V}}(\theta) = d_\theta\pi^{-1}(\{0\})$ represents the vertical fiber at $\theta = (x, v) \in \mathring{\mathcal{T}}^1\mathcal{M}$, with the projection map $\pi \colon \mathring{\mathcal{T}}^1\mathcal{M} \to \mathcal{M}$, and
\begin{equation}
	\volume\bigl(\varphi_t\left(\mathcal{T}_x\mathcal{M}\right)\bigr) = \int_{\mathcal{T}_x\mathcal{M}}\det\left(d_{(x, v)}\varphi_t\big|_{\mathring{\mathcal{V}}(x, v)}\right)dv,
\end{equation}
cf. \cite[pp. 82-92]{Paternain "Geodesic Flows"}. The referential background supporting the Mañé's formula consists partly in the theorem of Y. Yomdin \cite{Yomdin "Volume growth and entropy"} on the \emph{coincidence} of the (boundary of the) growth rate of volumes and the topological entropy (conceived as a growth rate under iteration), and partly in the Berger–Bott formula \cite{Berger and Bott "Sur les varietes a courbure strictement positive"} and Przytycki's inequality \cite{Przytycki "An Upper Estimation for Topological Entropy of Diffeomorphisms"} for the $\mathscr{C}^2$ flow on a closed Riemannian manifold.
\item Let a compact $\mathscr{C}^\infty$ Riemannian manifold $\mathcal{M}$, the Riemannian measure  $\bbmu$ (induced by the Riemannian structure), and $x, y \in \mathcal{M}$ be given. Let $\mathcal{N} \subset \mathcal{M}$ be a compact smooth submanifold, and $N_\tau(\mathcal{N}, y)$ the number of geodesics (i.e. geodesic arcs or segments) with length $\leqslant \tau$ that connect a point in $\mathcal{N}$ to $y$ and are orthogonal to $\mathcal{N}$. G.P. and M. Paternain \cite{Paternain "On the topology of manifolds with completely integrable geodesic flows"} \cite{Paternain Paternain "Topological entropy versus geodesic entropy"} define the \emph{geodesic entropy} as
\begin{equation}
	\mathsf{h}_\geodesic(g) = \limsup_{\tau \to \infty}\frac{1}{\tau}\log\int_\mathcal{M}N_\tau(\mathcal{N}, y)d\bbmu(y),
\end{equation}
or 
\begin{equation}
	\mathsf{h}_\geodesic(g) = \limsup_{\tau \to \infty}\frac{1}{\tau}\log\int_{\mathcal{M} \times \mathcal{M}}N_\tau(x, y)d\bbmu(x)d\bbmu(y),
\end{equation}
if $\mathcal{N}$ is the diagonal in $\mathcal{M} \times \mathcal{M}$. By means of Yomdin's theorem again, they show this inequality: $\mathsf{h}_\geodesic(g) \leqslant \mathsf{h}_\topological(\varphi_t)$. Mañé \cite{Mane "On the topological entropy of geodesic flows"} demonstrates instead that $\mathsf{h}_\geodesic(g) \geqslant \mathsf{h}_\topological(\varphi_t)$.

Remember that the notion of \emph{(exponential) growth rate $\lambda_\volume$ of volume} of a Riemannian manifold is developed by A. Manning \cite{Manning "Topological Entropy for Geodesic Flows"} in this way:
\begin{equation}
	\lambda_\volume(\mathcal{M}) = \lim_{\rho \to \infty}\frac{1}{\rho}\log\volume\bigl(\mathbb{B}_\rho(x)\bigr),
\end{equation}
where $\volume\bigl(\mathbb{B}_\rho(x)\bigr)$ is the volume of the ball $\mathbb{B}_\rho$ of radius $\rho$ and center $x$ in the universal covering $\widetilde{\mathcal{M}}$ of $\mathcal{M}$, such that $\rho^{-1}\log\volume\bigl(\mathbb{B}_\rho(x)\bigr)$ converges to a limit $\lambda_\volume \geqslant 0$ as $\rho \to \infty$ and $\lambda_\volume$ is independent of $x$. It was demonstrated that $\lambda_\volume(\mathcal{M}) \leqslant \mathsf{h}_\topological(\varphi_t)$, for a compact Riemannian manifold and a geodesic flow on $\mathring{\mathcal{T}}^1\mathcal{M}$, and $\lambda_\volume(\mathcal{M}) = \mathsf{h}_\topological(\varphi_t)$, if all sectional curvatures of $\mathcal{M}$ are $\leqslant 0$ (in this instance, the growth rate of volume on the universal covering is the \emph{volume entropy}). By putting $\rho = \tau$, and letting $\tilde{x}$ be a lift of a point $x \in \mathcal{M}$ to $\widetilde{\mathcal{M}}$, it can be noted that even 
\begin{equation}
	\int_\mathcal{M}N_\tau(x, y)dy \geqslant \volume\bigl(\mathbb{B}_\tau(\tilde{x})\bigr),
\end{equation}
and $\tau^{-1}\log\volume\bigl(\mathbb{B}_\tau(\tilde{x})\bigr)$ converges to $\lambda_\volume \geqslant 0$ and is independent of $\tilde{x}$.
\enumerationisfinis

\subsubsection{Positive vs. Zero Topological Entropy}

There are several examples showing that the topological entropy is positive, $\mathsf{h}_\topological(\varphi_t) > 0$; nevertheless, integrable Hamiltonian systems have null topological entropy, $\mathsf{h}_\topological(\varphi_t) = 0$; this includes verifying, on a case by case basis, that the integrability of the geodesic flow involves fading away of the topological entropy. But be careful: as it has been found out subsequently \cite[pp. 67-72]{Bolsinov Jovanovic "Integrable Geodesic Flows on Riemannian Manifolds: Construction and Obstructions"}, this does not mean that the integrability is  possible iff there is a zero topological entropy: the vanishing of the topological entropy is neither a necessary nor a sufficient condition for integrability.

\begin{exemplum}[Positive topological entropy in the case of geodesic flows on a 2-dimensional surface, and on a manifold which is not rationally elliptic]
About the positiveness of the topological entropy of geodesic flows, we mention the following works:
\enumerationisinitium
\item E.I. Dinaburg \cite{Dinaburg "On the relations among various entropy characteristics of dynamical systems"}. Assuming that the fundamental group $\pi_1(\mathcal{M})$ has exponential growth, for all smooth Riemannian metrics on $\mathcal{M}$ with a suitable topology, it is possible to prove that the entropy $\mathsf{h}_\topological$ of some geodesic flow on a 2-dimensional surface of genus $g > 1$ is positive.
\item G.P. Paternain \cite{Paternain "Entropy and completely integrable Hamiltonian systems"} \cite{Paternain "On the topology of manifolds with completely integrable geodesic flows"} \cite{Paternain "On the topology of manifolds with completely integrable geodesic flows II"}. Given a compact simply connected Riemannian manifold $\mathcal{M}$, if $\mathcal{M}$ is not rationally elliptic, then the entropy $\mathsf{h}_\topological$ of some geodesic flow is positive—a manifold is called \emph{rationally elliptic} if the total number of rational homotopy groups $\pi_z(\mathcal{M}) \otimes \mathbb{Q}$ is finite-dimensional, $z \in \mathbb{Z}$, or if there is a positive integer $z_0$ such that $\pi_z(\mathcal{M}) \otimes \mathbb{Q} = 0$, for any $z \geqslant z_0$. \exemplumsymbol
\enumerationisfinis
\end{exemplum}

\begin{exemplum}[Zero topological entropy in the case of completely integrable geodesic flows]
As regards the vanishing of the topological entropy $\mathsf{h}_\topological(\varphi_t)$, we describe briefly two results achieved again by Paternain.
\enumerationisinitium
\item For a compact Riemannian manifold whose geodesic flow is completely integrable with periodic integrals, the group $\pi_1(\mathcal{M})$ has sub-exponential growth. Suppose that $\pi_1(\mathcal{M})$ is finite, so $\mathcal{M}$ is rationally elliptic. It is possible to show that $\mathsf{h}_\topological$ of this flow is zero, cf. \cite{Paternain "On the topology of manifolds with completely integrable geodesic flows"}. 
\item Let $(\mathcal{M}, \omega_\mathrm{s})$ be a symplectic manifold (cf. \ref{item "Symplectic manifold"} in Definition \ref{definitio "Hamiltonian vector field or symplectic gradient"}), $\Hamiltonian$ a smooth Hamiltonian function, $\varphi^\Hamiltonian_t$ the flow of a Hamiltonian vector field $\vec{X}_\Hamiltonian$, cf. Eq. \eqref{equation "Hamiltonian flow"}, and $\vec{W}_\Hamiltonian \subset \vec{X}_\Hamiltonian$ a compact separable flow invariant subset. If $\Hamiltonian$ is completely integrable with non-degenerate first integrals, then $\mathsf{h}_\topological$ of $\varphi^\Hamiltonian_t\big|_{\vec{W}_\Hamiltonian}$ vanishes. More simply put, given a certain smooth compact Riemannian manifold whose geodesic flow is completely integrable with non-degenerate first integrals, the number $\mathsf{h}_\topological$ of this flow is zero, cf. \cite{Paternain "On the topology of manifolds with completely integrable geodesic flows II"}. \exemplumsymbol
\enumerationisfinis
\end{exemplum}

\begin{exemplum}[Positive topological entropy in the case of a $\mathscr{C}^\infty$ integrable geodesic flow]
This construction is due to A.V. Bolsinov and I.A. Taimanov \cite{Bolsinov Taimanov "Integrable geodesic flow with positive topological entropy"}; see also \cite{Butler "New Examples of Integrable Geodesic Flow"}. Take a 3-dimensional real-analytic Riemannian manifold diffeomorphic to the quotient of $\torus^2 = \mathbb{R}^2/\mathbb{Z}^2 \times \mathbb{R}^1$ concerning a free action of $\mathbb{Z}$, i.e. a homogeneous space that is a quotient of the 3-dimensional Lie group $\mathit{Sol}$, which is one of the eight homogeneous Thurston 3-geometries \cite{Thurston "The Geometry and Topology of Three-Manifolds"} \cite{Thurston "Hyperbolic geometry and 3-manifolds"} \cite{Thurston "Three dimensional manifolds Kleinian groups and hyperbolic geometry"} \cite{Thurston "Hyperbolic structures on 3-manifolds I: Deformation of acylindrical manifolds"} \cite{Thurston "Three-dimensional Geometry and Topology 1"} \cite{Thurston "Hyperbolic Structures on 3-manifolds II: Surface groups and 3-manifolds which fiber over the circle"} \cite{Thurston "Hyperbolic Structures on 3-manifolds III: Deformations of 3-manifolds with incompressible boundary"} (see Margo \ref{margo "The eight 3-geometries of Thurston"}); the group $\mathit{Sol}$ can be imagined as a split extension of $\mathbb{R}^2$ by $\mathbb{R}$, so the exact sequence is $0 \longrightarrow \mathbb{R}^2 \longrightarrow \mathit{Sol} \longrightarrow \mathbb{R} \longrightarrow 0$. Under these circumstances, the following holds:
\enumerationisinitium
\item the geodesic flow on the $\mathit{Sol}$-manifold is (Liouville) integrable by $\mathscr{C}^\infty$ smooth first integrals and not (Liouville) integrable by real-analytic first integrals;
\item the fundamental group of the $\mathit{Sol}$-manifold has an exponential growth;
\item the Kolmogorov–Sinai metric entropy of the geodesic flow vanish ($\mathsf{h}_\bbmu = 0$), especially for a Liouville-like measure;
\item the topological entropy of the geodesic flow is positive. \exemplumsymbol
\enumerationisfinis
\end{exemplum}

\subsubsection{Metric Entropy of the Geodesic Flow: on a Theorem of Pesin \& Freire–Mañé; Lyapunov Exponent}
\label{subsubsection "Metric Entropy of the Geodesic Flow: on a Theorem of Pesin and Freire–Mañé; Lyapunov Exponent"}

We talk again on the Kolmogorov–Sinai metric entropy (Definition \ref{definitio "Kolmogorov–Sinai metric entropy"}), but this time associated with the geodesic flow, and we treat a theorem summarizing its features, whose roots stretch back to the papers of Ya.B. Pesin \cite{Pesin "Equations for the entropy of a geodesic flow on a compact Riemannian manifold without conjugate points"} \cite{Pesin "Geodesic flows with hyperbolic behaviour of the trajectories and objects connected with them"} and Freire–Mañé \cite{Freire Mane "On the Entropy of the Geodesic Flow in Manifolds Without Conjugate Points"}; cf. the previous work of Eberlein \cite{Eberlein "When is a geodesic flow of Anosov type? I"}.

\begin{theorema}[\textsc{p+fm}]
\label{theorema "P+FM"}
Let $\varphi_t \colon \mathring{\mathcal{T}}^1\mathcal{M} \to \mathring{\mathcal{T}}^1\mathcal{M}$ be the geodesic flow on the unit tangent bundle of a compact $\mathscr{C}^4$ negatively curved surface $\kappa(v)$, and $\bbmu$ a Liouville-like measure. For each tangent vector $v \in \mathring{\mathcal{T}}^1\mathcal{M}$, the Kolmogorov–Sinai metric entropy of $\varphi_1$ is
\begin{equation}
\label{equation "Theorem P+FM"}
	\mathsf{h}_\bbmu(\varphi_1) = -\int_{\mathring{\mathcal{T}}^1\mathcal{M}}\kappa(v)d\bbmu(v).
\end{equation}
\end{theorema}

\begin{proof}
Let $v^\perp$ be the set of vectors $w \in \mathcal{M}$ orthogonal to $v$. For $w \in v^\perp$, take a vector $\vartheta_w \viz \vartheta(w)$. The set of the vectors  $w \in v^\perp$, $\vartheta_w \viz \vartheta(w)$ forms a 1-dimensional linear subspaces of the second (or double) tangent space $\mathcal{T}_v\mathring{\mathcal{T}}^1\mathcal{M}$, which is denoted by $\mathcal{E}^\pm(v)$. Comprehensibly, $\mathcal{T}_v\mathring{\mathcal{T}}^1\mathcal{M} = \mathcal{E}^+(v) \oplus \mathcal{E}^-(v)$. The subspaces $\mathcal{E}^\pm(v) \subset \mathcal{T}_v\mathring{\mathcal{T}}^1\mathcal{M}$ are invariant space under $d\varphi_t$, namely $d\varphi_t\mathcal{E}^+(v) = \bigl(\mathcal{E}^+\varphi_t(v)\bigr)$ and $d\varphi_t\mathcal{E}^-(v) = \bigl(\mathcal{E}^-\varphi_t(v)\bigr)$. Keep in mind that the vector $\vartheta_w \in \mathcal{E}^\pm(v)$ is such that $d\pi\vartheta_w = w$. Let $\kappa_x(v_1, v_2) \geqslant -\alpha^2$, for a value $\alpha > 0$ and all points $x \in \mathcal{M}$, so $\|{\kappa}\vartheta_w \leqslant \alpha\|d{\pi}\vartheta_w\|$. Fixing $t \geqslant 1$, we have therefore $\|d{\varphi_t}\vartheta_w\| \leqslant (1 + \alpha)\|d{\pi} \circ d{\varphi_t}\vartheta_w\|$, and the function
\begin{equation}
\label{equation "Lyapunov-like exponent"}
	\lambda_\textsc{l}^-(v, \vartheta_w) = \lim_{t \to \infty}\frac{1}{t}\log\|d{\varphi_t}\vartheta_w\| = \lim_{t \to \infty}\frac{1}{t}\log\|d\pi \circ d{\varphi_t}\vartheta_w\|,
\end{equation}
where $\lambda_\textsc{l}^-$ is a Lyapunov exponent (see Margo \ref{margo "On the Lyapunov exponent"}); note that each point in such a space is backward regular (which depends on whether the Lyapunov exponent $\lambda_\textsc{l}^-$ is backward regular and the filtration $\mathfrak{F}^-$ is coherent), as stated by the \emph{Lyapunov–Perron} method \cite{Lyapunov "The General Problem of the Stability Of Motion"} \cite{Perron "Die Ordnungszahlen linearer Differentialgleichungssysteme"} \cite{Perron "Die Stabilitatsfrage bei Differentialgleichungen"}.

Fix $s \neq 0$, and let 
\begin{equation}
	\chi_s(v) = \frac{\|d\pi \circ d{\varphi_s}\vartheta_w\|}{\|d{\pi}\vartheta_w\|}. 
\end{equation}
Thanks to Birkhoff's ergodic Theorem \ref{theorema "Birkhoff's ergodic Theorem"} and 
Eq. \eqref{equation "Lyapunov-like exponent"}, one gets 
\begin{align}
	\mathsf{h}_\bbmu(\varphi_s) & = -\int_{\mathring{\mathcal{T}}^1\mathcal{M}}\lambda_\textsc{l}^-(v, \vartheta_w)d\bbmu(v) \notag \\
	& = -\int_{\mathring{\mathcal{T}}^1\mathcal{M}}\lim_{n \to \infty}\frac{1}{sn}\log\|d\pi \circ d\varphi_{sn}\vartheta_w\|d\bbmu(v) \notag \\
	& = -\int_{\mathring{\mathcal{T}}^1\mathcal{M}}\lim_{n \to \infty}\frac{1}{sn}\sum^{n - 1}_{\nu = 1}\log\chi_s\bigl(\varphi_{sn}(v)\bigr)d\bbmu(v) \notag \\
	& = -\int_{\mathring{\mathcal{T}}^1\mathcal{M}}\log\chi_s(v)d\bbmu(v),
\end{align}
and, in view of the fact that $\mathsf{h}_\bbmu(\varphi_s) = |s|\mathsf{h}_\bbmu(\varphi_1)$,
\begin{equation}
	\mathsf{h}_\bbmu(\varphi_1) = \frac{1}{s}\mathsf{h}_\bbmu(\varphi_s) = \lim_{s \to 0}\frac{1}{s}\mathsf{h}_\bbmu(\varphi_s) = -\int_{\mathring{\mathcal{T}}^1\mathcal{M}}\lim_{s \to 0}\frac{1}{s}\log\chi_s(v)d\bbmu(v),
\end{equation}
for $s > 0$. We indicate by $\vec{J}_{\vartheta_w}$ the Jacobi field in regard to $\vartheta_w$ (see Margo \ref{margo "Jacobi field"}). It ensues that
\begin{align}
	|\vec{J}_{\vartheta_w}(s)|^2 & = |\vec{J}_{\vartheta_w}(0)|^2 + \int^s_0\frac{d}{dy}|\vec{J}_{\vartheta_w}(y)|^2dy \notag \\ 
	& = |\vec{J}_{\vartheta_w}(0)|^2 + 2 \int^s_0|\vec{J}_{\vartheta_w}(y)|\frac{d}{dy}|\vec{J}_{\vartheta_w}(y)|dy.
\end{align}
By setting
\begin{equation}
	\alpha(s) = \frac{1}{s}\int^s_0|\vec{J}_{\vartheta_w}(y)|\frac{d}{dy}|\vec{J}_{\vartheta_w}(y)|dy,
\end{equation}
we can write
\begin{equation}
	\alpha(0) = 2|\vec{J}_{\vartheta_w}(s)|\frac{d}{ds}|\vec{J}_{\vartheta_w}(s)| = 2\kappa(v),
\end{equation} 
for $s = 0$. By means of the last two equations, we get
\begin{equation}
	\|d\pi \circ d{\varphi_s}\vartheta_w\| = \sqrt{\|d\pi \circ d{\varphi_s}\vartheta_w\|^2} = \sqrt{|\vec{J}_{\vartheta_w}(s)|^2} = \sqrt{1 + s\alpha(s)},
\end{equation}
given that $|\vec{J}_{\vartheta_w}(0)|^2 = 1$. Then 
\begin{equation}
	\lim_{s \to 0}\frac{1}{s}\log\chi_s(v) = \kappa(v),
\end{equation}
from which we obtain \eqref{equation "Theorem P+FM"}, as required.
\end{proof}

\begin{margo}[On the Lyapunov exponent]
\label{margo "On the Lyapunov exponent"}
~\enumerationisinitium
\item Let us have a general definition. Given two vectors $v, w \in \mathbb{R}^n$, a function $\lambda_\textsc{l} \colon \mathbb{R}^n \to \mathbb{R} \cup \{-\infty\}$ is said to be a \emph{Lyapunov exponent} if
\subenumerationisinitium
\item $\lambda_\textsc{l}(\alpha{v}) = \lambda_\textsc{l}(v)$, for $\alpha \in \mathbb{R} \backslash \{0\}$, 
\item $\lambda_\textsc{l}(v + w) \leqslant \max\{\lambda_\textsc{l}(v), \lambda_\textsc{l}(w)\}$, 
\item $\lambda_\textsc{l}(0) = -\infty$. 
\subenumerationisfinis
One should replace $\mathbb{R}^n$ by an $n$-dimensional real vector space, and it means the same thing.
\item A function $\lambda_\textsc{l} \colon \mathbb{R}^n \to \mathbb{R} \cup \{-\infty\}$ is a Lyapunov exponent on $\mathbb{R}^n$ iff there are numbers $(\lambda_\textsc{l})_1 < \cdots < (\lambda_\textsc{l})_k$, for some integer $1 \leqslant k \leqslant n$, and a filtration $\mathfrak{F} = \{E_\nu \mid \nu = 0, \mathellipsis, k\}$, which is a collection of linear subspaces $E_\nu$ of $\mathbb{R}^n$, such that 
\subenumerationisinitium
\item $\lambda_\textsc{l}(v) \leqslant (\lambda_\textsc{l})_\nu$, for $v \in E_\nu$, 
\item $\lambda_\textsc{l}(v) = (\lambda_\textsc{l})_\nu$, for $v \in E_\nu \backslash E_{\nu - 1}$, 
\item $\lambda_\textsc{l}(0) = -\infty$. 
\subenumerationisfinis
In the following the \emph{proof} of this statement. Each vector $v \in E_\nu \backslash E_{\nu - 1}$ corresponds to $(\lambda_\textsc{l})_{\nu - 1} < \lambda_\textsc{l}(v) \leqslant (\lambda_\textsc{l})_\nu$. Since there are no values of $\lambda_\textsc{l}$ between $(\lambda_\textsc{l})_{\nu - 1}$ and $(\lambda_\textsc{l})_\nu$, then $\lambda_\textsc{l}(v) = (\lambda_\textsc{l})_\nu$. Because $\lambda_\textsc{l}$ attains the constant value $(\lambda_\textsc{l})_\nu$ on $E_\nu \backslash E_{\nu - 1}$, the linear subspace $E_\nu$ can be described as $E_\nu = \{v \in \mathbb{R}^n \mid \lambda_\textsc{l}(v) \leqslant (\lambda_\textsc{l})_\nu\}$. We see that if $v \in E_\nu \backslash E_{\nu - 1}$, subsequently $\alpha{v} \in E_\nu \backslash E_{\nu - 1}$ and $\lambda_\textsc{l}(\alpha{v}) = \lambda_\textsc{l}(v)$, for $v \in \mathbb{R}^n$ and $\alpha \in \mathbb{R} \backslash \{0\}$. Let the vectors $v_1, v_2 \in \mathbb{R}^n \backslash \{0\}$, and the function $\lambda_\textsc{l}(v_\xi) = (\lambda_\textsc{l})_{\nu_\xi}$, for $\xi = 1, 2$, be given. By putting $\nu_1 < \nu_2$, we have $v_1 + v_2 \in E_{\nu_1} \cup E_{\nu_2} = E_{\nu_2}$, and thus $\lambda_\textsc{l}(v_1 + v_2) \leqslant (\lambda_\textsc{l})_{\nu_2} = \max\{\lambda_\textsc{l}(v_1), \lambda_\textsc{l}(v_2)\}$, which proves that the function $\lambda_\textsc{l}$ is Lyapunov exponent, and also $\mathfrak{F} = (\mathfrak{F}_\lambda)_\textsc{l}$, assuming that $\lambda_\textsc{l}(v) = (\lambda_\textsc{l})_\nu$. \margosymbol
\enumerationisfinis
\end{margo}

\begin{margo}[Jacobi field]
\label{margo "Jacobi field"}
Incidentally, we remember that a \emph{Jacobi field} $\vec{J}_{\vartheta_w}$ is a vector field along a geodesic $\gamma_\mathrm{c}(t)$, with a parameter $t$, satisfying the second order equation
\begin{align}
	\frac{D^2\vec{J}_{\vartheta_w}}{dt^2} & \viz \nabla_{\dot{\gamma}_\mathrm{c}(t)}\nabla_{\dot{\gamma}_\mathrm{c}(t)}\vec{J}_{\vartheta_w} \notag \\
	& + \Riemann\left(\vec{J}_{\vartheta_w}(t), \dot{\gamma}_\mathrm{c}(t)\right)\dot{\gamma}_\mathrm{c}(t) = 0, 
\end{align}
where $D^2$ is the second covariant derivative along $\gamma_\mathrm{c}(t)$ with respect to the Levi-Civita connection (Section \ref{subsection "Levi-Civita Connection Theorem on a (pseudo-)Riemannian Manifold"}), $\dot{\gamma}_\mathrm{c}(t)$ is the velocity vector field of the geodesic, and $\Riemann$ is the Riemann curvature tensor. From the map $\vartheta_w \mapsto \vec{J}_{\vartheta_w}(t)$ it is possible to check that 
\begin{align}
	& \vec{J}_{\vartheta_w}(0) = d{\pi}\vartheta_w, \\
	& \frac{d}{dt}\vec{J}_{\vartheta_w}(0) = \kappa(\vartheta_w),
\end{align}
and
\begin{align}
	& \vec{J}_{\vartheta_w}(t) = d\pi \circ d\varphi_t\vartheta_w, \\
	& \frac{d}{dt}\vec{J}_{\vartheta_w}(t) = \kappa \circ d\varphi_t{\vartheta_w}, 
\end{align}
for $v \in \mathring{\mathcal{T}}^1\mathcal{M}$ and $\vartheta_w \in \mathcal{T}_v\mathring{\mathcal{T}}^1\mathcal{M}$. \margosymbol
\end{margo}

\vspace{10mm}

\setcounter{secnumdepth}{0}  
\section{References and Bibliographic Details}
\setcounter{secnumdepth}{3}
\markright{References and Bibliographic Details}

\begingroup
\footnotesize
\noindent Section \ref{subsection "Prior Knowledge: Maxwell–Boltzmann Probability Distribution, and Ergodic Hypothesis of Thermodynamics"}

\begin{indent paragraph: 15pt}
On the ergodic hypothesis, see e.g. \cite{Szasz "Boltzmann's Ergodic Hypothesis a Conjecture for Centuries?"}. — On the Boltzmann (transport) equation and surrounding areas, see e.g. G.E. Uhlenbeck \cite{Uhlenbeck "The Boltzmann Equation"}, C. Cercignani \cite[chapp. I-II]{Cercignani "The Boltzmann Equation and Its Applications"} \cite{Cercignani "The Boltzmann equation and fluid dynamics"} \cite{Cercignani "134 years of Boltzmann equation"} and C. Villani \cite{Villani "A Review of Mathematical Topics in Collisional Kinetic Theory"} \cite{Villani "Entropy production and convergence to equilibrium for the Boltzmann equation"} \cite{Villani "Entropy Production and Convergence to Equilibrium"}. —  On the $\mathsf{H}$-theorem, see e.g. \cite[chapp. VI (Boltzmann's $\mathsf{H}$-theorem), XII (The quantum mechanical $\mathsf{H}$-theorem)]{Tolman "The Principles of Statistical Mechanics"} \cite{Villani "$H$-theorem and beyond: Boltzmann's entropy in today's mathematics"}. — Readings on the entropy principle, micro- and micro-scopic nature: see e.g. \cite{Gallavotti "Entropy nonequilibrium chaos and infinitesimals"} \cite[chap. 1]{Gallavotti "Nonequilibrium and Irreversibility"} \cite{Lebowitz "From time-symmetric microscopic dynamics to time-asymmetric macroscopic behavior: an overview"}, see also \cite[chapp. 3-4]{Cercignani Illner Pulvirenti "The Mathematical Theory of Dilute Gases"}; an overview of the entropy and reversible vs. irreversible processes is in \cite{Villani "(Ir)reversibility and Entropy"}. — Fermi–Pasta–Ulam \& Tsingou problem: see \cite{Lichtenberg Livi Pettini and Ruffo "Dynamics of Oscillator Chains"} \cite{Cencini Cecconi Falcioni and Vulpiani "Role of Chaos for the Validity of Statistical Mechanics Laws: Diffusion and Conduction"} \cite{Benettin Carati Galgani and Giorgilli "The Fermi-Pasta-Ulam Problem and the Metastability Perspective"} \cite{Paleari and Penati "Numerical Methods and Results in the FPU Problem"}. — For a rigorous demonstration of the Parisi solution, see F. Guerra \cite{Guerra "Broken Replica Symmetry Bounds in the Mean Field Spin Glass Model"} and M. Talagrand \cite{Talagrand "The Parisi formula"}. — Birkhoff's and von Neumann's contributions to ergodic theorems/theory are summarized in \cite{Moore "Ergodic theorem ergodic theory and statistical mechanics}.	
\end{indent paragraph: 15pt}

\noindent Section \ref{subsection "Birkhoff's Ergodic Theorem"}

\begin{indent paragraph: 15pt}
Theorem \ref{theorema "Birkhoff's ergodic Theorem"}: cf. e.g. \cite[sec. 2.5]{Barreira "Ergodic Theory Hyperbolic Dynamics and Dimension Theory"} and \cite[pp. 83-84]{Glasner "Ergodic Theory via Joinings"}; a more thorough examination in \cite[pp. 89-95]{Mane "Ergodic Theory and Differentiable Dynamics"}.
\end{indent paragraph: 15pt}

\noindent Section \ref{subsubsection "Addendum. Convergence Theorems: Lebesgue, Levi and Fatou's Lemma"}

\begin{indent paragraph: 15pt}
On convergence theorems by Lebesgue, Levi and Fatou, refer to \cite[pp. 34-37]{Ambrosio Da Prato and Mennucci "Introduction to Measure Theory and Integration"} \cite[sec. 2.8]{Bogachev "Measure Theory I"} \cite[pp. 21-27]{Rudin "Real and Complex Analysis"} \cite[pp. 25-36]{Salamon D.A. "Measure and Integration"} \cite[§ 1.4]{Tao "An Introduction to Measure Theory"}.
\end{indent paragraph: 15pt}

\noindent Section \ref{subsection "Ergodicity of the Geodesic Flow: Hopfian Statistical Process"}

\begin{indent paragraph: 15pt}
On the ergodicity of geodesic flows, see \cite[app.]{Ballmann "Lectures on Spaces of Nonpositive Curvature"} \cite[sec. 14.2]{Borthwick "Spectral Theory of Infinite-Area Hyperbolic Surfaces"} \cite[pp. 302-306]{Gallavotti "Foundations of Fluid Mechanics"}; on the stable ergodicity, and references to the Anosov system, see e.g. \cite{Burns Pugh Wilkinson "Stable ergodicity and Anosov flows"} \cite[chap. 8]{Bonatti Diaz Viana "Dynamics Beyond Uniform Hyperbolicity: A Global Geometric and Probabilistic Perspective"}. — Theorem \ref{theorema "Hopfian statistical process"}: cf. \cite[pp. 285-296]{Nicholls "A Measure on the Limit Set of a Discrete Group"} \cite[pp. 334-335]{Borthwick "Spectral Theory of Infinite-Area Hyperbolic Surfaces"}.
\end{indent paragraph: 15pt}

\noindent Section \ref{section "Entropy within the Topological Thermodynamics: at the Frontier of Order and Chaos"}

\begin{indent paragraph: 15pt}
On entropy, thermodynamic probability, macroscopic state, and microscopic configurations, see e.g. \cite{Grandy "Entropy and the Time Evolution of Macroscopic Systems"}.
\end{indent paragraph: 15pt}

\noindent Section \ref{subsection "Kolmogorov–Sinai Metric Entropy (Quantity for a Measure Preserving Transformation)"}

\begin{indent paragraph: 15pt}
On the Kolmogorov–Sinai metric entropy (measure-theoretic entropy), see \cite{Gurevich "Entropy Theory of Dynamical Systems"}.
\end{indent paragraph: 15pt}

\noindent Section \ref{subsection "Topological Entropy and Pressure, aka Ruelle–Walters Free Energy Density"}

\begin{indent paragraph: 15pt}
Definition \ref{definitio "Topological pressure"}: cf. e.g. \cite[p. 162]{Jiang Qian Qian "Mathematical Theory of Nonequilibrium Steady States: On the Frontier of Probability and Dynamical Systems"}.
\end{indent paragraph: 15pt}

\noindent Section \ref{subsection "Application of the Calculus of Variations in the Topological Entropy and Pressure"}

\begin{indent paragraph: 15pt}
On the variational principle for the topological entropy and pressure, cf. e.g. \cite[secc. 4.5, 5.4]{Barreira "Ergodic Theory Hyperbolic Dynamics and Dimension Theory"} \cite[sec. 2.3]{Barreira "Thermodynamic Formalism and Applications to Dimension Theory"} \cite[pp. 181-182, 625-626]{Katok Hasselblatt "Introduction to the Modern Theory of Dynamical Systems"} \cite[chap. 9]{Neshveyev Stormer "Dynamical Entropy in Operator Algebras"} \cite[sec. 3.1.5]{Walczak "Dynamics of Foliations Groups and Pseudogroups"}.
\end{indent paragraph: 15pt}

\noindent Section \ref{subsubsection "Metric Entropy of the Geodesic Flow: on a Theorem of Pesin and Freire–Mañé; Lyapunov Exponent"}

\begin{indent paragraph: 15pt}
Theorem \ref{theorema "P+FM"}: cf. e.g. \cite[chap. 7, § 4, pp. 137-143]{Pesin "General Theory of Smooth Hyperbolic Dynamical Systems"} \cite[sec. 12.3]{Barreira Pesin "Nonuniform Hyperbolicity: Dynamics of Systems with Nonzero Lyapunov Exponents"}. 
\end{indent paragraph: 15pt}

\endgroup

\chapter{On the Chaos, Part I. Micro- and Macro-scales}
\label{chapter "On the Chaos, Part I. Micro- and Macro-scales"}

\begingroup
\footnotesize
Chinese tea is some kind of \emph{optimal function}, say \ZhTraditional{茶}-function, for the daily math-struggle for negative entropy:\footnote{
	\label{footnote "Negative entropy"}
	The concept of \emph{negative entropy}, indicating the amount of information (number of bits) that is ordered, was introduced by E. Schrödinger; it is sometimes expressed with a bad crasis, \emph{negentropy}, coined by L. Brillouin \cite[p. 1152]{Brillouin "The Negentropy Principle of Information"}, cf. \cite[p. 49]{Chaisson "Cosmic Evolution: The Rise of Complexity in Nature"}; alternatively, there is the word \emph{syntropy}, due to L. Fantappiè \cite[capp. II-III]{Fantappie "Principi di una teoria unitaria del mondo fisico e biologico"}.
	
	The syntropic phenomena of Fantappiè are the ones that converge to a state of complexity—«such as e.g. the formation of the eye and of many very complicated systems of living beings, the chlorophyll process, the ascent of the lymph in plants, the psychic phenomena of the human personality» \cite[pref.]{Fantappie "Principi di una teoria unitaria del mondo fisico e biologico"}—under variable stages of differentiation, up to a high degree of diversity. A syntropic system is, inherently, an ordered system, i.e. with low entropy, and richly complex. But beware:
	\enumerationisinitium
	\item All phenomena, including the syntropic ones, depend, from what we can tell, on an initial state of low entropy, which proceeds towards a state whose entropy is stringently increasing with time; so syntropy—a measure of the degree of order, or apparent absence of randomness, and presence of information—and entropy—a measure of the degree of disorder, or randomness, and lack of information (Section \ref{subsubsection "Information Flow of What? An Entropy Flux Question"})—are braided together: local phenomena producing order and complexity (think of biological evolution) coexist, without contradiction, with the entropy increase principle on a global scale, by the growing disorder, or randomness, of the universe. 
	\item The increasing amounts of order and complexity are not the only evolutionary directions of localized structures: we can witness a decrease of complexity, to wit, an increase of simplicity, in the course of evolution (one thinks of the manifestation of new circumstances more efficient, with a lower degree of complexity than the old ones, without losing an ordered state). 
	\enumerationisfinis
	
	On grounds of the definition of complexity, all this reasoning incorporates conceptual changes. We can say that the complexity of a system 
	
	· is a state in which several parts are connected, in an orderly fashion, to each other; or 
	
	· is the amount of information suitable to describe the regularities of that system, as proposed by M. Gell-Mann and S. Lloyd \cite{Gell-Mann Lloyd "Information Measures Effective Complexity and Total Information"} \cite{Gell-Mann Lloyd "Effective Complexity"}; after which we must agree on what is meant by “regularity” \cite{Gell-Mann "What is Complexity? Remarks on"} \cite{Gell-Mann "What Is Complexity?"}, as distinct from “randomness”, “fortuitousness”, etc., according to the context of study/application, cf. footnote \ref{footnote "Two examples of physico-mathematical faith or belief"} on p. \pageref{footnote "Two examples of physico-mathematical faith or belief"}.
	}
from chaos to order; as the Hatter says, \ZhTraditional{「因為老是吃茶」}· “It's always tea-time”.\footnote{
	L. Carroll \cite[VII. \ZhTraditional{瘋茶會} · \textit{A Mad Tea-Party}, p. \ZhTraditional{一百二}]{Carroll "Alice's Adventures in Wonderland"}.
	} \\
\indent — Personal application of the \textsc{Boltzmann–Schrödinger} law of life \cite[p. 40]{Boltzmann "Der zweiten Hauptsatz der mechanischen Warmetheorie"}\footnote{
	«The general struggle for existence of animate beings is [\,\dots] a struggle for [against] entropy [\textit{Der allgemeine Daseinskampf der Lebewesen ist \textnormal{[\,\dots]} ein Kampf um die Entropie}]».
	}
\cite[p. 71]{Schrodinger "What is Life? The Physical Aspect of the Living Cell"}\footnote{
	«[A] living organism continually increases its entropy [\,\dots] and thus tends to approach the dangerous state of maximum entropy, which is death. It can only keep aloof from it, i.e. alive, by continually drawing from its environment negative entropy».
	} via \textit{Camellia sinensis}

\endgroup

\section{Bohr's Tea Principle—Uncertainty and Entropy}
\label{section "Bohr's Tea Principle—Uncertainty and Entropy"}

\begingroup
\footnotesize 
In the rigorous formulation of the law of causality: “If we know the present precisely, we can calculate the future”, it is not the conclusion that is erroneous, but the premise. In principle, we \emph{cannot} know the present in all its parameters [\textit{Wir \emph{können} die Gegenwart in allen Bestimmungsstücken prinzipiell \emph{nicht} kennenlernen}]. Resultantly the whole of what is perceived is a selection from a totality of possibilities and a limitation of what is possible in the future [\textit{ist alles Wahrnehmen eine Auswahl aus einer Fülle von Möglichkeiten und eine Beschränkung des zukünftig Möglichen}]. \\
\indent — \textsc{W. Heisenberg} \cite[p. 197]{Heisenberg "Uber den anschaulichen Inhalt der quantentheoretischen Kinematik und Mechanik"}

\vspace{2mm}

Let us look at the properties of an individual hydrogen atom inside the teapot. Its temperature, if we can talk about it at all, is surely as high as that of the rest of the tea, [e.g.] 70 degrees, because it exchanges heat with all the other tea molecules. Its energy, however, must fluctuate and this precisely because it exchanges heat; hence we can only define a probability curve for its energy. If, conversely, we had measured the energy of the hydrogen atom and not the temperature of the tea, then we could not deduce the latter unequivocally from the former; once again we could only draw a probability curve — for the temperature [\,\dots]. [T]he lack of certainty as to the precise temperature or energy values, is relatively large in so small an object as a hydrogen atom, and hence significant. In a much larger object, for instance a small quantity of tea within the pot, it becomes considerably smaller and can therefore be neglected. \\
\indent — \textsc{N. Bohr} in the recollection of \textsc{W. Heisenberg} \cite[pp. 106-107]{Heisenberg "Physics and Beyond. Encounters and Conversations"}

\endgroup

\vspace{2mm}

Let us have a cup of tea. We want to pause on the fact that the correlation between macroscopic and microscopic phenomena in probabilistic terms, that we have found in the ergodic hypothesis and in the definitions of entropy (see Chapter \ref{chapter "Geometric and Topological Aspects of Complexity and Dynamics, Part II: Ergodicity and Entropy"}), is transferred and repurposed into quantum mechanics. Whether it is classical thermodynamics or quantum mechanics, we fail to fully understand the passage from micro- to macro-matter, and vice versa, and the so-called \emph{emergence} in nature (cf. Section \ref{section "How Far is it Possible to Analyze Nature? The Crux of the Mathematics of Emergence"}), especially in connection with complex dynamic systems, or the transition from discrete to continuous states, and vice versa, with the reduction of aggregate matter (macroscopic structures) at the level of subatomic constituents (elementary particles).

We can \emph{think} of temperature of tea within the pot as a measure of the average kinetic energy (or the average speed) of its atomic-molecular components; therefore the thermodynamic properties reflect the statistical and quantum-mechanical behavior in the underlying microstates. The entropy of the teapot, which is the production of energy but also of information associated with it, is a measure of the number of possible microscopic arrangements of the atomic-molecular components.

The mathematical knot that ties these two physical worlds, of micro- and macro-matter, is a collection of  probabilistic, or statistical, laws. As we mentioned previously (Chapter \ref{chapter "Geometric and Topological Aspects of Complexity and Dynamics, Part II: Ergodicity and Entropy"}), it was Boltzmann that forged an interconnection between the atomism of matter and the statistical theory. This knot is beautifully exemplified by Bohr's consideration, cited in epigraph, on the hydrogen atom in a teapot, and its energy and temperature, which recalls the \emph{Heisenberg's uncertainty principle} \cite{Heisenberg "Uber den anschaulichen Inhalt der quantentheoretischen Kinematik und Mechanik"}, but see also contributions by E.H. Kennard \cite{Kennard "Zur Quantenmechanik einfacher Bewegungstypen"} and H. Weyl \cite[pp. 77, 393-394]{Weyl "The Theory of Groups and Quantum Mechanics"}. Let us check this out with the language of mathematics. 

\subsection{Quantum Heisenberg–Weyl Inequality, and Schwartz Space}
\label{subsection "Quantum Heisenberg–Weyl Inequality, and Schwartz Space}

\begin{theorema}[\textgreek{α}. Heisenberg–Weyl Inequality]
~\enumerationisinitium
\item Let $\psi$ be a unit vector, identifying a state of a particle, in the Hilbert space $\mathfrak{H} = \Lebesgue^2(\mathbb{R})$, where $\Lebesgue^2(\mathbb{R})$ is the space of all square-integrable functions $\textcyrillic{\textit{я}} \colon \mathbb{R} \to \mathbb{C} \cup \{\infty\}$. 
\item Let $\hat{\mathrm{x}} \colon \mathfrak{D}(\hat{\mathrm{x}}) \to \Lebesgue^2(\mathbb{R})$ be a position operator, given by $(\hat{\mathrm{x}}\psi)(x) = x\psi(x)$, for any $\psi \in \mathfrak{D}(\hat{\mathrm{x}})$, and $\hat{\mathrm{p}} \colon \mathfrak{D}(\hat{\mathrm{p}}) \to \Lebesgue^2(\mathbb{R})$ a momentum operator on the real line (also known as position and momentum \emph{observables} of a particle), where $\mathfrak{D}(\cdot) = \{\psi \in \Lebesgue^2(\mathbb{R})\}$ is a linear subspace of $\mathfrak{H}$, i.e. a domain of $\hat{\mathrm{x}}$ and $\hat{\mathrm{p}}$. 
\item The $\mathfrak{D}$-domain of $[\hat{\mathrm{x}}, \hat{\mathrm{p}}]$ contains the \emph{Schwartz space} (Margo \ref{margo "Schwartz space"}), denoted by $\mathcal{S}_\mathfrak{c}(\mathbb{R})$, for which $\mathcal{S}_\mathfrak{c}(\mathbb{R}) \subset \mathfrak{D}([\hat{\mathrm{x}}, \hat{\mathrm{p}}])$ and $[\hat{\mathrm{x}}, \hat{\mathrm{p}}]\psi = i\hbar\psi$, with the reduced Planck's constant, for any $\psi \in \mathcal{S}_\mathfrak{c}(\mathbb{R})$. 
\enumerationisfinis

If $\hat{\mathrm{x}}$ and $\hat{\mathrm{p}}$ are symmetric, $\hat{\mathrm{x}}\hat{\mathrm{p}} = \hat{\mathrm{p}}\hat{\mathrm{x}}$, such that $[\hat{\mathrm{x}}, \hat{\mathrm{p}}] = \hat{\mathrm{x}}\hat{\mathrm{p}} - \hat{\mathrm{p}}\hat{\mathrm{x}} = i\hbar\idem$, and $\psi \in \mathfrak{D}(\hat{\mathrm{x}}\hat{\mathrm{p}}) \cap \mathfrak{D}(\hat{\mathrm{p}}\hat{\mathrm{x}}) \cap \mathfrak{D}(\hat{\mathrm{x}}^2) \cap \mathfrak{D}(\hat{\mathrm{p}}^2)$, then
\begin{equation}
\label{equation "Heisenberg uncertainty"}
	\mathrm{\Delta}^2_\psi(\hat{\mathrm{x}}) \mathrm{\Delta}^2_\psi(\hat{\mathrm{p}}) \geqslant \tfrac{1}{4}|\langle\psi, [\hat{\mathrm{x}}, \hat{\mathrm{p}}]\psi\rangle|^2.
\end{equation}
\end{theorema}

\begin{proof}
Setting $\langle(\hat{\mathrm{x}} + i\beta\hat{\mathrm{p}})\psi, (\hat{\mathrm{x}} + i\beta\hat{\mathrm{p}})\psi\rangle \geqslant 0$, for a value $\beta \in \mathbb{R}$, one has 
\begin{align}
	\langle(\hat{\mathrm{x}} + i\beta\hat{\mathrm{p}})\psi, (\hat{\mathrm{x}} + i\beta\hat{\mathrm{p}})\psi\rangle & = \beta^2\langle\psi, \hat{\mathrm{p}}^2\psi\rangle + i\beta\langle\psi, \hat{\mathrm{x}}\hat{\mathrm{p}}\psi\rangle - i\beta\langle\psi, \hat{\mathrm{p}}\hat{\mathrm{x}}\psi\rangle + \langle\psi, \hat{\mathrm{x}}^2\psi\rangle \notag \\
	& = \beta^2\langle\psi, \hat{\mathrm{p}}^2\psi\rangle + \beta(-\langle\psi, \psi\rangle) + \langle\psi, \hat{\mathrm{x}}^2\psi\rangle,
\end{align}
which holds when $\langle\psi, \psi\rangle^2 \leqslant 4\langle\psi, \hat{\mathrm{p}}^2\psi\rangle\langle\psi, \hat{\mathrm{x}}^2\psi\rangle$.
\end{proof}
From \eqref{equation "Heisenberg uncertainty"}, $\mathrm{\Delta}^2_\psi(\hat{\mathrm{x}})\mathrm{\Delta}^2_\psi(\hat{\mathrm{p}}) \geqslant \frac{\hbar^2}{4}$, and
\begin{equation}
\label{equation "Uncertainty relation"}
	\mathrm{\Delta}_\psi(\hat{\mathrm{x}})\mathrm{\Delta}_\psi(\hat{\mathrm{p}}) \geqslant \frac{\hbar}{2}, 
\end{equation}
in which $\mathrm{\Delta}_\psi(\hat{\mathrm{x}})$ and $\mathrm{\Delta}_\psi(\hat{\mathrm{p}})$ are the \emph{uncertainties} of $\hat{\mathrm{x}}$ and $\hat{\mathrm{p}}$ in the state $\psi$. The inequality \eqref{equation "Uncertainty relation"}, combining the \emph{standard deviations} $\mathrm{\Delta}_\psi$ of position and momentum, carries with it a break between small and large scales (e.g. of motion and distribution) in the matter.

\begin{theorema}[\textgreek{β}. Heisenberg–Weyl Inequality]
Let $\psi \in \mathfrak{D}(\hat{\mathrm{x}}) \cap \mathfrak{D}(\hat{\mathrm{p}})$ in $\mathfrak{H} = \Lebesgue^2(\mathbb{R})$, then $\mathrm{\Delta}_\psi(\hat{\mathrm{x}})\mathrm{\Delta}_\psi(\hat{\mathrm{p}}) \geqslant \frac{\hbar}{2}$.
\end{theorema}

\begin{proof}
We consider the expression $(\hat{\mathrm{p}}\psi)(x) = -i\hbar\lim_{u \to 0}\left(\frac{\psi(x + u) - \psi(x)}{u}\right)$, for any $\psi \in \mathfrak{D}(\hat{\mathrm{p}})$, so
\begin{align}
	\langle\hat{\mathrm{x}}\psi, \hat{\mathrm{p}}\psi\rangle & = \lim_{u \to 0}\left\langle\hat{\mathrm{x}}\psi, -i\hbar\left(\tfrac{\psi(x + u) - \psi(x)}{u}\right)\right\rangle \notag \\ 
	& = \lim_{u \to 0}\left\{\tfrac{1}{u}\langle{i}\hbar(y - u)\psi(y - u), \psi(y)\rangle + \tfrac{i\hbar}{u}\langle\hat{\mathrm{x}}\psi, \psi\rangle\right\} \notag \\
	& = \lim_{u \to 0}\left\{\left\langle{i}\hbar\left(\tfrac{\psi(x - u) - \psi(x)}{u}\right), \hat{\mathrm{x}}\psi(x)\right\rangle + i\hbar\langle\psi(x - u), \psi(x)\rangle\right\} \notag \\
	& = \langle\hat{\mathrm{p}}\psi, \hat{\mathrm{x}}\psi\rangle +i\hbar\langle\psi, \psi\rangle.
\end{align}
Based on the \emph{Cauchy–Schwarz inequality} \cite{Cauchy "Sur les Formules qui resultent de l'emploi du signe > ou < et sur les Moyennes entre plusieurs quantites"} \cite{Schwarz "Uber ein Flachen kleinsten Flacheninhalts betreffendes Problem der Variationsrechnung"}, for a value $\alpha, \beta \in \mathbb{R}$, one has 
\begin{align}
	\langle\psi, \psi\rangle & = \tfrac{1}{i\hbar}\{\langle(\hat{\mathrm{x}} - \alpha\idem)\psi, (\hat{\mathrm{p}} - \beta\idem)\psi\rangle - \langle(\hat{\mathrm{p}} - \beta\idem)\psi, (\hat{\mathrm{x}} - \alpha\idem)\psi\rangle\} \notag \\
	& = \tfrac{2}{\hbar}\Im\langle(\hat{\mathrm{x}} - \alpha\idem)\psi, (\hat{\mathrm{p}} - \beta\idem)\psi\rangle \leqslant \tfrac{2}{\hbar}\|(\hat{\mathrm{x}} - \alpha\idem)\psi\| \cdot \|(\hat{\mathrm{p}} - \beta\idem)\psi\|.
\end{align}
Putting $\alpha = \langle\psi, \hat{\mathrm{x}}\psi\rangle$ and $\beta = \langle\psi, \hat{\mathrm{p}}\psi\rangle$, there follows that $\|(\hat{\mathrm{x}} - \alpha\idem)\psi\|^2 = \mathrm{\Delta}^2_\psi(\hat{\mathrm{x}})$ and $\|(\hat{\mathrm{p}} - \beta\idem)\psi\|^2 = \mathrm{\Delta}^2_\psi(\hat{\mathrm{p}})$, ergo $1 \leqslant \frac{2}{\hbar}\mathrm{\Delta}_\psi(\hat{\mathrm{x}})\mathrm{\Delta}_\psi(\hat{\mathrm{p}})$. 
\end{proof}

\begin{margo}[Schwartz space]
\label{margo "Schwartz space"}
The Schwartz space $\mathcal{S}_\mathfrak{c}(\mathbb{R}^n)$ is named after A. Grothendieck \cite{Grothendieck "Sur les espaces (F) et (DF)"} in L. Schwartz's honour \cite{Dieudonne Schwartz "La dualite dans les espaces (F) et (LF)"} \cite{Schwartz "Produits tensoriels topologiques d'espaces vectoriels topologiques. Espaces vectoriels topologiques nucleaires"}. It is a locally convex topological vector space, or even a nuclear space, intended as a finite-dimensional vector space. Let $\nu = (\nu_1, \mathellipsis, \nu_n)$ and $\xi = (\xi_1, \mathellipsis, \xi_n)$ denote multi-indices consisting of $n$-tuples of non-negative integers, that is to say, of natural numbers 
\[
	\nu_{1, \mathellipsis, n}, \xi_{1, \mathellipsis, n} \in \mathbb{N}^n_0 = \mathbb{N}^n \cup \{0\}. 
\]
Technically, the Schwartz space is the space of complex-valued functions $\textcyrillic{\textit{я}} \in \mathcal{S}_\mathfrak{c}(\mathbb{R}^n)$ of rapid decrease on $\mathbb{R}^n$ if
\enumerationisinitium
\item $\textcyrillic{\textit{я}} \in \mathscr{C}^\infty(\mathbb{R}^n)$,
\item $\lim_{x \to \pm\infty}|x^\nu\partial^\xi\textcyrillic{\textit{я}}(x)|$ = 0, where $x^\nu = x^{\nu_1}_1 \cdots x^{\nu_n}_n$ and $\partial^\xi = \frac{\partial^{|\xi|}}{\partial{x}^{\xi_1}_1 \cdots \partial{x}^{\xi_n}_n}$ of order $|\xi| = \xi_1 + \cdots + \xi_n$, i.e.
\begin{equation}
	\lim_{x \to \pm\infty}\left|x^{\nu_1}_1 \cdots x^{\nu_n}_n\frac{\partial^{\xi_1}}{\partial{x}^{\xi_1}_1} \cdots \frac{\partial^{\xi_n}}{\partial{x}^{\xi_n}_n}\textcyrillic{\textit{я}}(x)\right| = 0, 
\end{equation}	
and
\begin{equation}
\label{equation "Schwartz space"}
	\mathcal{S}_\mathfrak{c}(\mathbb{R}^n) = \left\{\textcyrillic{\textit{я}} \in \mathscr{C}^\infty(\mathbb{R}^n) \mathrel{\bigg|} \sup_{x \in \mathbb{R}^n}\left|x^\nu\partial^\xi\textcyrillic{\textit{я}}(x)\right| < \infty\right\}.
\end{equation}
By \eqref{equation "Schwartz space"}, any element of $\mathcal{S}_\mathfrak{c}(\mathbb{R}^n)$ is square-integrable onto the $n$-dimensional real coordinate space, which means that the Schwartz space is a linear—but not topological—subspace of the Hilbert $\Lebesgue^2$-space, $\mathcal{S}_\mathfrak{c}(\mathbb{R}^n) \subset \Lebesgue^2(\mathbb{R}^n)$, and, in general, $\mathcal{S}_\mathfrak{c}(\mathbb{R}^n) \subset \Lebesgue^p(\mathbb{R}^n)$, for any $p \geqslant 1$. The Schwartz space includes the vector space of all smooth and compactly supported functions, the so-called \emph{bump functions}, denoted by $\mathscr{C}^\infty_0$, for which $\mathscr{C}^\infty_0(\mathbb{R}^n) \subset \mathcal{S}_\mathfrak{c}(\mathbb{R}^n) \subset \mathscr{C}^\infty(\mathbb{R}^n)$. \margosymbol
\enumerationisfinis
\end{margo}

\section{(In)deterministic Flow: Lorenz System in Comparison with Quantum Mechanics}
\label{section "(In)deterministic Flow: Lorenz System in Comparison with Quantum Mechanics"}

\begingroup
\footnotesize
[W]e shall assume [\,\dots] that the [idealized fluid] systems [in the small and large scales  of motion] with which we are dealing are deterministic [that is, the exact present state determines the exact state at any future time]. We shall acknowledge that the state of a system cannot be observed without error, but we shall assume [\,\dots] that there is no limit to how small the error may be made. We shall then produce evidence favoring the conclusion that the observable behaviour of certain deterministic [fluid] systems is \emph{indistinguishable} from that of indeterministic systems [\,\dots], in that they possess an intrinsic finite range of predictability which cannot be lengthened by reducing the error of observation to any value greater than zero [\,\dots]. 

It is appropriate to ask at this point whether real fluid systems possess a similar lack of predictability [\,\dots]. [T]he fact [is] that we do not know the governing equations for any real systems. We need not invoke Heisenberg's Principle of Uncertainty [cf. epigraph under Section \ref{section "Bohr's Tea Principle—Uncertainty and Entropy"}] to make such a statement, nor do we even need to recognize that fluids are collections of molecules rather than continua; there are processes of somewhat larger scale which are not completely understood. \\
\indent — \textsc{E.N. Lorenz} \cite[pp. 290, 306, 304, e.a.]{Lorenz "The predictability of a flow which possesses many scales of motion"}

\endgroup

\vspace{2mm}

The Lorenz system \cite{Lorenz "Deterministic Nonperiodic Flow"} \cite{Lorenz "The predictability of a flow which possesses many scales of motion"}, which originates from the weather forecasting, allows us to see that certain formally deterministic fluid systems possessing many scales of motion are not observationally distinguishable from indeterministic systems; we can visualize the motion as the flow of a fluid, or as a hydrodynamic flow, in the phase space. Because of small and negligible, or even undetectable, errors in observing the fluid system, the initial uncertainties in the smallest scales are amplified in the largest scales, so the predictability of the future of the system does end up containing some errors. 

Here the instability in meteorology (that is, the instability of the atmosphere) becomes, thanks to Lorenz's strange attractor (Section \ref{section "The Lorenz Flow: a Strange Attractor"}), the basal property of \emph{irregular} or \emph{chaotic} dynamical systems, and it is considered the root of the irregularity \cite{Lorenz "Irregularity: a fundamental property of the atmosphere"}. A distinctive feature of the irregularity is the absence of periodicity, such is the case e.g. of the instability of non-periodic flow in systems of deterministic equations.\footnote{
	The use of the word “chaos” in mathematics and physics appears in a paper written by T.-Y. Li and J.A. Yorke \cite{Li Yorke "Period Three Implies Chaos"} about a situation in which there is a non-periodic sequence in a Lorenz type fluid flow.
	}
Solutions of these equations are identified with the orbits in the phase space.  

It is also clear that there is a relationship between Lorenz type attractor and the simultaneous growth of both entropy and information, as it is shown in R. Shaw \cite{Shaw "Strange Attractors Chaotic Behavior and Information Flow"}, or that an information \emph{entropy of the attractor} is associated with the measure of its dimensionality, see Grassberger \& Procaccia \cite{Grassberger and Procaccia "Measuring the strangeness of strange attractors"}.

According to the study of chaos, given a certain dynamical system, we cannot achieve absolute precision in the measurement of the initial state of its evolution, for which there is (and remains) some approximation, or \emph{uncertainty} of the prediction about the future state of the system. This component, let us repeat, is closely related to the presence of one or more uncorrectable errors in the measurement of the initial data, because it is possible to measure all state variables, or magnitudes dependent on time, only with \emph{relative precision} (Section \ref{subsection "Sensitivity to Initial Conditions, and Weak Predictability"}). 

In quantum theory, the concept of \emph{indetermination}, as well as the concept of \emph{uncertainty}, or \emph{entropy}, have a connotation other than that emanating from dynamical systems investigated by  chaotic mathematical models. It is enough to mention the non-relativistic Schrödinger equation \cite{Schrodinger "Quantisierung als Eigenwertproblem (Erste Mitteilung)", Schrodinger "Quantisierung als Eigenwertproblem (Zweite Mitteilung)", Schrodinger "Quantisierung als Eigenwertproblem (Dritte Mitteilung)", Schrodinger "Quantisierung als Eigenwertproblem (Vierte Mitteilung)"}.

\subsection{Schrödinger Wave Equation in 1D and Solution via Fourier Transform}
\label{subsection "Schrödinger Wave Equation in 1D and Solution via Fourier Transform"}

\begingroup
\footnotesize
One could summarise [\,\dots] paradoxically: the movement of particles follows a probability law, but the probability itself evolves in accordance with the law of causality [under which the  punctual knowledge of a state at a certain instant determines the distribution of the state for all later times]. \\
\indent — \textsc{M. Born} \cite[p. 804]{Born "Quantenmechanik der Stossvorgange"}

\vspace{2mm}

One of the most satisfactory features of the present quantum theory is that the differential equations that express the causality of classical mechanics do not get lost, but are all retained in symbolic form, and indeterminacy appears only in the application of these equations to the results of observations. \\
\indent — \textsc{P.A.M. Dirac} \cite[p. 4]{Dirac "The Principles of Quantum Mechanics (1935)"}

\endgroup

\vspace{2mm}

Let us examine the most common situation. We show an exemplification.

\begin{exemplum}[Quantum Schrödinger particle]
\label{exemplum "Quantum Schrödinger particle"}
Setting the potential energy function to zero, we analyze the case of a particle of mass $m$ freely moving in a 1-dimensional space. Let $\psi(x)$ represent the state vector, or wave function, of the particle, and $\hat{\Hamiltonian}$ the Hamiltonian operator, on the assumption, known as \emph{de Broglie hypothesis} \cite{De Broglie "Recherches sur la theorie des Quanta"}, that there is a wave-like behavior of matter on the microscopic scale (here, wave packets directly represent the moving atomic corpuscles). The corresponding Schrödinger equation is
\begin{subequations}
\label{subequations "Schrödinger equation in 1D"}
\begin{align}
	i\hbar\frac{\partial\psi}{\partial{t}}(x, t) & = -\frac{\hbar^2}{2m}\frac{\partial^2}{\partial{x}^2}\psi(x, t), \\ \frac{\partial\psi}{\partial{t}} & = \frac{i\hbar}{2m}\frac{\partial^2\psi}{\partial{x}^2},
\end{align}
\end{subequations}
the second one is in simplified form, while the general form is
\begin{equation}
	i\hbar\frac{\partial\psi}{\partial{t}}(x, t) = \hat{\Hamiltonian}\psi(x, t), \text{ or } \frac{\partial\psi}{\partial{t}} = \frac{1}{i\hbar}\hat{\Hamiltonian}\psi. 
\end{equation}
The solution of \eqref{subequations "Schrödinger equation in 1D"} is the function 
\begin{equation}
\label{equation "Solutions of the Schrödinger equation in 1D"}
	\psi(x, t) = e^{ik\bigl(x - \frac{\omega(k)}{k}t\bigr)},
\end{equation}
where $k$ is the wave number, i.e. the spatial angular frequency of the wave, and $\omega(k) = \frac{\hbar{k}^2}{2m}$ is the angular frequency—in radians—per unit time. \exemplumsymbol
\end{exemplum}

We shall now give a propositional theorem.
 
\begin{propositio}[Solution of the Schrödinger equation through the Fourier transform method]
\label{propositio "Solution of the Schrödinger equation through the Fourier transform method"}
Let $\psi_0 \in \Lebesgue^2(\mathbb{R})$ be a Schwartz function, as an element of the Schwartz space (see Margo \ref{margo "Schwartz space"}), representing the initial (state) wave function, and $\widehat{\psi}_0$ the Fourier transform \textnormal{\cite{Fourier "Theorie analytique de la chaleur"}} of $\psi_0$ for which 
\begin{equation}
	\psi(x, t) = \frac{1}{\sqrt{2\pi}}\int^{+\infty}_{-\infty}\widehat{\psi}_0(k)e^{ik\bigl(x - \frac{\omega(k)}{k}t\bigr)}
\end{equation}
is fixed. Then $\psi(x, t)$ is the solution of \eqref{subequations "Schrödinger equation in 1D"} imposing $\psi_0$.
\end{propositio}

\begin{proof}
Knowing that $\widehat{\psi}_0$ is itself a Schwartz function, we know, too, that $\widehat{\psi}_0$ decays faster than $\frac{1}{k^4}$ as $k$ tends to plus-minus infinity. For an eigenfunction of the form 
\begin{equation}
	\psi_k(x) = e^{ikx}, \enspace k \to \pm\infty, 
\end{equation}
expressing a particle with momentum $\hbar{k}$, we integrate the derivative of $e^{ikx}$ to get an estimate like this: 
\begin{equation}
	\left|\frac{e^{ik(x + g)} - e^{ikx}}{g}\right| \leqslant |k|. 
\end{equation}
To take a derivative with respect to $x$  under the integral sign, we need to use the Lebesgue's dominated convergence theorem \cite{Lebesgue "Sur la methode de M. Goursat pour la resolution de l'equation de Fredholm"} (see Theorem \ref{theorema "Lebesgue's dominated convergence"}), so that the derivative can “pull down” a factor of $\pm{ik}$, and so on for the second derivative, owing to the rapid decay of the Fourier transform $\widehat{\psi}_0$ on the space of Schwartz functions. The process of differentiation under the integral sign is thus the method by which the Schrödinger equation \eqref{subequations "Schrödinger equation in 1D"} is solved by $\psi(x, t)$, given that \eqref{equation "Solutions of the Schrödinger equation in 1D"} satisfies this equation for any spatial frequency of the wave. 
\end{proof}

The time dependent Eq. \eqref{subequations "Schrödinger equation in 1D"} is a \emph{linear partial differential equation} of second order; it is of a deterministic type (for which the evolution of the wave function of a particle is deterministically described), and the unpredictability appearing in quantum mechanics, starting with the Copenhagen interpretation, is not chaotic, or at least, it is not chaotic under the above-mentioned Lorenzian acceptation (see Margo \ref{margo "Quantum chaos"}), but is connected to the nebulous notion of \emph{probability amplitude}, introduced by M. Born \cite{Born "Zur Quantenmechanik der Stossvorgange"} \cite{Born "Quantenmechanik der Stossvorgange"}.\endnote{
	The “probability”—of an event—is not something unconditionally divorced from a deterministic, or pseudo-deterministic, context. M. Born \cite[p. 84]{Born "Quantenmechanik der Stossvorgange"}: «The movement of particles follows a probability law, but the probability [\textit{Wahrscheinlichkeit}] itself evolves in accordance with the law of causality [\textit{Kausalgesetz}] [to wit], a [complete] knowledge of a state, at a certain instant, determines the distribution of the state for all later times».
	} 

In the view instead of deterministic chaos, the unpredictability of the evolution of a system is inherent in the fact that small initial differences, in the data collected, can turn into increasingly and dramatically large differences over time. Nonetheless, there are schemes, such as the one in the \emph{Ghirardi–Rimini–Weber theory} \cite{Ghirardi Rimini Weber "A model for a unified quantum description of macroscopic and microscopic systems"} \cite{Ghirardi Rimini Weber "Unified dynamics for microscopic and macroscopic systems"}, allowing the possibility to add stochastic and \emph{non-linear ramifications} in a Schrödinger dynamical framework; in that context, see J.S. Bell \cite{Bell "Are there quantum jumps?"} \cite{Bell "Against 'measurement'"}.

\begin{margo}[Quantum chaos]
\label{margo "Quantum chaos"}
Attempts at grafting a deterministic/classical chaos onto quantum mechanics are multiple. We report here T.N. Palmer \cite{Palmer "A local deterministic model of quantum spin measurement"} \cite{Palmer "A granular permutation-based representation of complex numbers and quaternions: elements of a possible realistic quantum theory"}, because he is in the wake of Lorenz's picture, and takes advantage of the Alexander–Yorke–You–Kan riddled basin \cite{Alexander Yorke You Kan "Riddled Basins"}. We remind also a quantum version of the classical Hamiltonian system involving the Anosov flow with chaotic behavior; this quantum dynamics, that is, a quantum manifestation of Hamiltonian chaos, is in keeping with the flow generated by \emph{Laplace–Beltrami operator} on a compact Riemannian manifold of negative curvature. \margosymbol
\end{margo}
 
\section{Non-perfect Fluid in the Teapot and Brownian Motion}
\label{section "Non-perfect Fluid in the Teapot and Brownian Motion"}

\begingroup
\footnotesize 
While examining the form of these particles [contained in the grains of pollen] immersed in water, I observed many of them very evidently in motion [\,\dots]. These motions were such as to satisfy me, after frequently repeated observation, that they arose neither from currents in the fluid, nor from its gradual evaporation, but belonged to the particle itself. \\
\indent — \textsc{R. Brown} \cite[pp. 466-467]{Brown "A Brief Account of Microscopical Observations"}

\vspace{2mm}

I think that the dancing motion of the extremely minute solid particles in a liquid, can be attributed to the different velocities that must be at the same temperature, both in these solid particles and in the molecules of the liquid that strike them on all sides [\,\dots]. And thereby the Brownian motion, thus declared, provides us with one of the most beautiful and direct experimental demonstrations of the fundamental principles of the mechanical theory of heat, manifesting the assiduous vibrational state necessarily present both in liquids and solids even when their temperature is not altered.\endnote{
	Original It. version: «Ebbene, io penso che il moto di danza delle particelle solide estremamente minute entro un liquido, possa attribuirsi alle differenti velocità che esser devono ad una medesima temperatura, sia in codeste particelle solide, sia nelle molecole del liquido che le urtano d'ogni banda [\,\dots]. E di tal modo il moto browniano, cosi dichiarato, ci fornisce una delle più belle e dirette dimostrazioni sperimentali dei fondamentali principii della teoria meccanica del calore, manifestando quell'assiduo stato vibratorio che esser deve e nei liquidi e nei solidi ancor quando non si muta in essi la temperatura».
	} \\
\indent — \textsc{G. Cantoni} \cite[pp. 163, 167]{Cantoni "Su alcune condizioni fisiche dell'affinita e sul moto browniano"}
	
\vspace{2mm}

[I]t will be shown that according to the molecular-kinetic theory of heat, bodies of microscopically-visible size suspended in a liquid will perform movements of such magnitude that they can be easily observed in a microscope, on account of the molecular motions of heat \cite[p. 549, transl. p. 1]{Einstein "Uber die von der molekularkinetischen Theorie der Warme geforderte Bewegung von in ruhenden Flussigkeiten suspendierten Teilchen"}. \\
\indent [T]he so-called Brownian motion is caused by the irregular thermal movements of the molecules of the liquid \cite[p. 371, transl. p. 19]{Einstein "Zur Theorie der Brownschen Bewegung"}. \\
\indent — \textsc{A. Einstein}
	
\endgroup

\subsection{From \emph{Camellia Sinensis} to \emph{Clarkia Pulchella}}
\label{subsection "From Camellia Sinensis to Clarkia Pulchella"}

We know that \emph{a drop of tea}, at the atomic-molecular level, \emph{is not a perfect (or ideal) fluid}, to wit, a fluid absolutely free from viscosity, in which there are no shear stresses (between the fluid particles). The equation of state of the perfect fluid is used in idealized models capable of describing the distribution of matter, as is the case with the Friedmann–Lemaître–Robertson–Walker (\textsc{flrw}) metric \cite{Friedman "Uber die Krummung des Raumes"} \cite{Friedman "Uber die Moglichkeit einer Welt mit konstanter negativer Krummung des Raumes"} \cite{Robertson "Kinematics and World-Structure"} 
\cite{Robertson "Kinematics and World-Structure. II"} \cite{Robertson "Kinematics and World-Structure. III"} \cite{Walker "On Milne's Theory of World-Structure"}, that is an exact and analytically solvable solution of Einstein's field equations, i.e. with the general relativity, whose geometric structure is a 4-dimensional continuum (the differentiable space-time manifold). 

A drop of tea split in two is still a drop of tea, the same holds true for the half of the latter, and so forth; but, when we get to the hydrogen and oxygen, we no longer have to deal with it as a infinitely (continuously, or in an arbitrary manner) divisible tea-substance, together with the additive of tannin (polyphenolic compounds) extracted from the leaves of \textit{Camellia sinensis} (\ZhSimplified{茶树}). What we are actually dealing with here is an aggregate of organic molecules and $N$ atoms. The image of a fluid as a \emph{homogeneous continuum}, \emph{spatially uniform}, falls into a macroscopic viewpoint (tea in the teapot); but, at a higher-resolution “peek”, it breaks into a \emph{granular structure}.  

We can take a cue from \emph{Brownian motion}. The observations through the microscope by R. Brown \cite{Brown "A Brief Account of Microscopical Observations"} of the random and incessant motion of particles contained in the grains of pollen of the plant \textit{Clarkia pulchella} suspended in water, are the simplest and most direct evidence that the macroscopic physico-chemical substance (the continuum fluid) consists of a structure of discrete molecules and elementary constituents. The irregular movements of microscopic grains of pollen arise from thermal molecular movement, so the Brownian motion is related to the molecular theory of heat. G. Cantoni was the first to give this explanation, and A. Einstein \cite{Einstein "Uber die von der molekularkinetischen Theorie der Warme geforderte Bewegung von in ruhenden Flussigkeiten suspendierten Teilchen"} \cite{Einstein "Zur Theorie der Brownschen Bewegung"} made a quantitative description of it.

On a micro-scale, the fluid is atomic-molecular, on a macro-scale, it is a continuum. The passage from one to another is a pressing problem in quantum mechanics (e.g. Schrödinger's cat \cite[p. 812]{Schrodinger "Die gegenwartige Situation in der Quantenmechanik"} and decoherence), but also in chaotic dynamical systems (e.g. weak predictability in Lorenz-like attractors).

The same trouble of passage from micro-scale to macro-scale is present in the Brownian motion, with a micro- and macro-physics of fluids. Brownian motion is considered a specific model of \emph{random walk}, this is because the diffusion of visible particles of organic origin, included in the grains of pollen, is due to their apparently random fluctuation in the liquid in which they are suspended. There are, however, some experiments \cite{Dettmann Cohen van Beijeren Grassberger Schreiber Gaspard Briggs Francis Sengers Gammon Dorfman Calabrese "Microscopic chaos from brownian motion?"} \cite{Briggs Sengers Francis Gaspard Gammon Dorfman Calabrese "Tracking a colloidal particle for the measurement of dynamic entropies"} \cite[chap. 18]{Mazo "Brownian Motion: Fluctuations Dynamics and Applications"} that, on a microscopic dynamics, seem to show a \emph{fractal nature of Brownian paths}, and a possible presence of deterministic chaos in Brownian-like motions, or a positive dynamic entropy congruous with microscopic chaos. These are outstanding issues in this regard.

\subsection{Fokker–Planck (Diffusion) Equation in Einstein's Theory of Brownian Motion}
\label{subsection "Fokker–Planck (Diffusion) Equation in Einstein's Theory of Brownian Motion"}

We now resume the discourse on the Einstein's theory \cite[§ 4]{Einstein "Uber die von der molekularkinetischen Theorie der Warme geforderte Bewegung von in ruhenden Flussigkeiten suspendierten Teilchen"} for Brownian particles. Denote by $\tau$ an interval of time, and by $\mathrm{\Delta}$ a positive or negative amount,\footnote{
	$\sqrt{\mathrm{\Delta}^2}$, to be exact \cite[p. 238]{Einstein "Elementare Theorie der Brownschen Bewegung"}, the square root of the mean value of $\mathrm{\Delta}$.
	} 
or better said, a certain displacement, i.e. a length of path, measuring the increase of each particle, intended as suspended \emph{pollen particle} (particle expelled from pollen grains in the liquid) along $x$-axis, that is in 1-dimensional space. (A \emph{solute molecule} is another terminology for a pollen particle, not, of course, to be confused with a \emph{solvent molecule}, like a water molecule, which is much smaller). The number of the particles with a displacement from $\mathrm{\Delta}$ to $\mathrm{\Delta} + d\mathrm{\Delta}$ occurring in $\tau$ can be expressed as 
\begin{equation}
	dn = n\mathscr{P}(\mathrm{\Delta}, \tau)d\mathrm{\Delta}, 
\end{equation}
where $\mathscr{P}(\mathrm{\Delta}, \tau)$ is the \emph{probability density function}, and
\begin{equation}
	\int^{+\infty}_{-\infty}\mathscr{P}(\mathrm{\Delta}, \tau)d\mathrm{\Delta} = 1.
\end{equation}
It is supposed that there is no external agent that exerts force on the Brownian system, for which there is a homogeneous distribution of mass. By adopting the symbol $\textit{\dh}$ for the solute density, the average number of particles per unit volume $\volume = \textit{\dh}(x, t)$ between $x$ and $x + dx$ at time $t + \tau$ can thus be defined as
\begin{equation}
	\textit{\dh}(x, t + \tau)dx = \int^{\mathrm{\Delta} = +\infty}_{\mathrm{\Delta} = -\infty}\Bigl\{\textit{\dh}(x - \mathrm{\Delta}, t)dx\Bigr\} \times \Bigl\{\mathscr{P}(\mathrm{\Delta}, \tau)d\mathrm{\Delta}\Bigr\},
\end{equation}
which corresponds to a \emph{Chapman–Kolmogorov equation}. In this equation, $\textit{\dh}(x - \mathrm{\Delta}, t)dx$ is the average number of particles in the interval $dx$ at $x - \mathrm{\Delta}$ at $t$, whilst $\mathscr{P}(\mathrm{\Delta}, \tau)d\mathrm{\Delta}$ represents the particles, which (having a displacement along $x$ from $\mathrm{\Delta}$ to $\mathrm{\Delta} + d\mathrm{\Delta}$) are in the interval $dx$ along $x$ at $t + \tau$. The density $\textit{\dh}(x - \mathrm{\Delta}, t)$ can be expanded under the integral sign into Taylor series, and one has
\begin{align}
	\textit{\dh}(x, t + \tau) & = \int^{+\infty}_{-\infty}\mathscr{P}(\mathrm{\Delta}, \tau)\textit{\dh}(x - \mathrm{\Delta}, t)d\mathrm{\Delta} \notag \\
	& = \int^{+\infty}_{-\infty}\mathscr{P}(\mathrm{\Delta}, \tau)\left(\textit{\dh}(x, t) + \sum^\infty_{\nu = 1}\frac{(-\mathrm{\Delta})^\nu}{\nu!}\frac{\partial^\nu\textit{\dh}(x, t)}{\partial{x^\nu}}\right)d\mathrm{\Delta} \notag \\
	& = \textit{\dh}(x, t)\int^{+\infty}_{-\infty}\mathscr{P}(\mathrm{\Delta}, \tau)d\mathrm{\Delta} \notag \\ 
	& + \sum^\infty_{\nu = 1}\frac{\partial{x^\nu}\textit{\dh}(x, t)}{\partial{x^\nu}}\left(1/\nu!\int^{+\infty}_{-\infty}(-\mathrm{\Delta})^\nu\mathscr{P}(\mathrm{\Delta}, \tau)d\mathrm{\Delta}\right),
\end{align}	
with $\nu \in \mathbb{Z}$. The integral 
\begin{equation*}
	\int^{+\infty}_{-\infty}\mathscr{P}(\mathrm{\Delta}, \tau) 
\end{equation*}
is equal to 1 on account of the properties of $\mathscr{P}$ in $\mathrm{\Delta}$, whereas the integral representation under the sign of summation vanishes, noticing that the function $\mathscr{P}(\mathrm{\Delta}, \tau)$ is an even function of $\mathrm{\Delta}$, i.e. 
\begin{equation}
	\mathscr{P}(\mathrm{\Delta}, \tau) = \mathscr{P}(-\mathrm{\Delta}, \tau). 
\end{equation}
It then proceeds by writing the time derivative of $\textit{\dh}(x, t)$ by means of the infinite series 
\begin{equation}
\label{equation "Einstein's convergent infinite series"}
	\frac{\partial\textit{\dh}(x, t)}{\partial{t}} = \sum^\infty_{\nu = 1}\left((1/\tau)\frac{1}{(2\nu)!}\int^{+\infty}_{-\infty}\mathrm{\Delta}^{2\nu}\mathscr{P}(\mathrm{\Delta}, \tau)d\mathrm{\Delta}\right)\frac{\partial^{2\nu}\textit{\dh}(x, t)}{\partial{x}^{2\nu}}.
\end{equation} 
The converge, from Einstein's perspective, is sufficiently rapid to ignore any terms other than the first. This truncation, as was noted, has no foundation, even if it looks reasonable; and yet it is in Einstein's derivation of the Brownian process. (This is a case in which the physical intuition, aimed at explaining an experimental fact, can be disengaged from the mathematical guide, without compromising the overall validity of the theory). By assumption, the coefficient of diffusion is defined as
\begin{equation}
	D_m = (1/2\tau)\int^{+\infty}_{-\infty}\mathrm{\Delta}^2\mathscr{P}(\mathrm{\Delta}, \tau)d\mathrm{\Delta} = (1/\tau)\int^{+\infty}_{-\infty}\frac{\mathrm{\Delta}^2}{2}\mathscr{P}(\mathrm{\Delta}, \tau)d\mathrm{\Delta}, 
\end{equation}
and from \eqref{equation "Einstein's convergent infinite series"} there is the 1-dimensional diffusion equation \cite{Fick "Ueber Diffusion"} (Fick's law),
\begin{equation}
	\frac{\partial\textit{\dh}(x, t)}{\partial{t}} = D_m\frac{\partial^2\textit{\dh}(x, t)}{\partial{x}^2},
\end{equation}
which corresponds to a \emph{Fokker–Planck equation} \cite{Fokker "Die mittlere Energie rotierender elektrischer Dipole im Strahlungsfeld"} \cite{Planck Uber einen Satz der statistischen Dynamik und seine Erweiterung in der Quantentheorie"}.

\section{Continuity and Discreteness—Differential Equations and Numerical Computing}
\label{section "Continuity and Discreteness—Differential Equations and Numerical Computing"}

\begingroup
\footnotesize
Properly speaking, there is no science that does not have its \emph{metaphysics}, if we understand by this word the general principles on which a science is based.\endnote{
	Original Fr. version: «A proprement parler, il n'y a point de science qui n'ait sa \emph{métaphysique}, si on entend par ce mot les principes généraux sur lesquels une science est appuyée».
	} \\
\indent — \textsc{J. le R. d'Alembert} \cite[p. 294]{d'Alembert "Oeuvres de d'Alembert"}

\vspace{2mm}

Science is what we understand well enough to explain to a computer. Art is everything else we do. \\
\indent — \textsc{D.E. Knuth} \cite[p. xi]{Knuth "Foreword"}

\vspace{2mm}

One does not believe that one has created a clear concept of what a \emph{continuum} is merely by employing this word or writing out a differential equation. On closer examination, the differential equation is only the expression for the fact that we need to think firstly of a finite number, and then this number must grow until every further [incremental] growth is no more relevant \cite[pp. 233-234]{Boltzmann "Ueber die Unentbehrlichkeit der Atomistik in der Naturwissenschaft"} = \cite[p. 144]{Boltzmann "Populare Schriften"}. \\
\indent  The concepts of differential and integral calculus, detached of any atomistic notions [\textit{atomistischen Vorstellung}], are of a truly metaphysical character, if, following a successful Mach's definition, by this we mean something that we have forgotten how we did arrive at our conceptions \cite[p. 792]{Boltzmann "Nochmals uber die Atomistik"} = \cite[p. 160]{Boltzmann "Populare Schriften"}. \\
\indent — \textsc{L. Boltzmann}

\vspace{2mm}

[F]rom the most remote ages to the present day, the idea of continuity has dominated the mathematical surveys and all their most interesting and productive applications. When the conditions of the problems have allowed it, the attempt has always been made to trace (sometimes even intuitively and unconsciously) the cases of discontinuity back to cases of continuity [\,\dots] induced by the power of infinitesimal methods [\,\dots]; at the same time, every infinitesimal question has been considered as a limiting case of questions concerning the discontinuous. \\
\indent — \textsc{V. Volterra} \cite[pp. 539-540]{Volterra "L'evoluzione delle idee fondamentali del calcolo infinitesimale"}

\endgroup

\vspace{2mm}

Data processing relating to one or more orbits e.g. in a Lorenz attractor with an executable \emph{computer program} is always a partial processing, since the collection of data items cannot be achieved through numbers with infinite decimal places, so it is truncated, and thereby approximated, at some level. Furthermore, perform a full computation means to do operations for manipulating data with the measurements at our disposal, that are never completely accurate, as they are sensitive to initial conditions. This implies that the behavior of a \emph{calculated} orbit will thus be different from that of the \emph{real} (or exact) orbit. This is why the mathematical physics is defined as a building models discipline, with the known laws which govern the behavior for example of a real hydrodynamical systems, or the set of idealizations of a hydrodynamical system, that is, idealized equations as the exact equations for a model of a real system, cf. \cite[p. 289]{Lorenz "The predictability of a flow which possesses many scales of motion"}.

Take some chaotic dynamical system; we say that it shall be expressed resorting to an integration of ordinary differential equations, or to partial differential equations requiring a continuous dependence of solutions on changes in the initial-value data, with given parameters, changes e.g. in the various coefficients, or in the boundary-value data. Once we have worked out the computability of these equations, when the analytical model is solved using finite length numerical data (that is, when the numerical value of the unknowns is calculated), the differential equations end up being discretized. The system of equations, from its continuous modeling, is, of course, brought back to the discrete, or “atomic”, origin of numbers.

\begin{margo}[Cellular automata]
\label{margo "Cellular automata"}
There are models, like \emph{cellular automata}, that are intrinsically discrete; it is possible to treat mathematically fluid flows as \emph{lattice gas}, see S. Wolfram \cite{Wolfram "Thermodynamics and Hydrodynamics of Cellular Automata"} \cite{Wolfram "Cellular Automaton Fluids: Basic Theory"}, allowing us to have a simulation, but still partial, of flows derived from the Navier–Stokes equations. The aggregate behavior of cellular automata can in fact be a tool for approximating continuum systems. Cellular automaton interpretations in quantum theory are in G. 't Hooft \cite{'t Hooft Isler and Kalitzin "Quantum field theoretic behavior of a deterministic cellular automaton"} \cite{G. 't Hooft "Classical cellular automata and quantum field theory"} \cite{'t Hooft "The Cellular Automaton Interpretation of Quantum Mechanics"}. \margosymbol
\end{margo}

\subsection{Hyperbolic Equation of a Vibrating String: d'Alembert's Formula for the 1-Dimensional Wave Phenomenon}
\label{subsection "Hyperbolic Equation of a Vibrating String: d'Alembert's Formula for the 1-Dimensional Wave Phenomenon"}

\begingroup
\footnotesize
\textgreek{Σημεῖόν}\footnote{
	See footnote \ref{footnote "Semeión vs. stigmé"}, p. \pageref{footnote "Semeión vs. stigmé"}.
	}
\textgreek{ἐστιν [\,\dots] ἢ πέρας ἀδιάστατον} · A point is [\,\dots] an extremity without extension. \\
\indent — \textsc{Heron (Diophantus?)} \cite[\textgreek{α´. ⟨Περὶ σημεῖόν⟩}, 11-12, p. 14]{Heron of Alexandria "Heronis definitiones"} = \cite[pp. 468-469]{Various authors "Greek Mathematical Works II: from Aristarchus to Pappus"}

\endgroup

\vspace{2mm}

To pick up the thread of the previous speech, there is a charming yet problematic coexistence of a \emph{continuous scheme}, relating to the macroscopic-like behavior of differential equations, and an \emph{atomic unit}, relating to the microscopic-like behavior of numbers or (as a geometric counterpart) points. As an example of this tension, we will choose an eminent result of \emph{hyperbolic partial differential equations}, the d'Alembert's equation for the oscillatory motion of a vibrating string, and the corresponding formula \cite{d'Alembert "Recherches sur la Courbe que forme une Corde tendue mise en vibration; Suite des Recherches sur la Courbe que forme une Corde tendue mise en vibration", d'Alembert "Addition au Memoire sur la Courbe que forme une Corde tendue mise en vibration}, that is a solution for the wave equation of dimension 1. Its application is wide: fluid dynamics, acoustics (sound waves), electromagnetic waves, such as light. This is the equation:
\begin{equation}
\label{equation "Wave equation"}
	\frac{\partial^2\upsilon}{\partial{t^2}} = \frac{\partial^2\upsilon}{\partial{s^2}},
\end{equation}
where $\upsilon = \upsilon(s, t)$ is the the unknown function, or the dependent variable, $s$ is the length parameter and $t$ the time, evidently using $s$ and $t$ as independent variables. Let $c$ be a constant concerning the speed of propagation of the wave. If we identify $s$ with the $x$-axis, and $\upsilon$ with the direction aligned with the $y$-axis, in order to have the height of the oscillating string, then the Eq. \eqref{equation "Wave equation"} can be written in the following equivalent forms, 
\begin{subequations}
\label{subequations "Wave equation + wave speed"}
\begin{align}
	& \frac{\partial^2\upsilon}{\partial{t^2}} = c^2\frac{\partial^2\upsilon}{\partial{x^2}}, \\
	& \frac{\partial^2\upsilon}{\partial{x^2}} = \frac{1}{c^2}\frac{\partial^2\upsilon}{\partial{t^2}}, \\
	& c^2\frac{\partial^2\upsilon}{\partial{x^2}} - \frac{\partial^2\upsilon}{\partial{t^2}} = \frac{\partial^2\upsilon}{\partial{x^2}} - \frac{1}{c^2}\frac{\partial^2\upsilon}{\partial{t^2}} = 0,
\end{align}
\end{subequations}
under which the function $\upsilon \equival y(x, ct)$ gives the vertical displacement of the string, or better, of the points of the string (see below), from a horizontal equilibrium at position $x$ and time $t$. Another way of writing the wave equation is with the d'Alembertian (d'Alembert operator), $\dAlembertian = \Laplacian - \frac{1}{c^2}\frac{\partial^2}{\partial{t^2}} = \frac{1}{c^2}\frac{\partial^2}{\partial{t^2}} - \nabla^2$,
\begin{equation}
\label{equation "d'Alembert's wave equation with d'Alembertian"}
	\dAlembertian\upsilon = \left(\Laplacian - \frac{1}{c^2}\frac{\partial^2}{\partial{t^2}}\right)\upsilon = 0, \text{ i.e. } \dAlembertian\upsilon = 0,
\end{equation}
where $\Laplacian = \nabla^2 = \nabla \cdot \nabla$ is the Laplace operator, or Laplacian. 

\begin{margo}
The Eq. \eqref{equation "d'Alembert's wave equation with d'Alembertian"} occurs extensively within the field of physics, see e.g. the formalism of Landau–Lifshitz \cite[§ 46]{Landau and Lifshitz "The Classical Theory of Fields: Course of Theoretical Physics II"}. A typical application is in Minkowski space-time, the well-known 4-dimensional flat space for the Einsteinian relativity, in which the d'Alembertian is consistent with 
\begin{equation}
\label{equation "d'Alembertian"}
	\dAlembertian = \eta^{\mu\nu}\partial_\mu\partial_\nu = \frac{1}{c^2}\frac{\partial^2}{\partial{t^2}} - \nabla^2 = \frac{1}{c^2}\frac{\partial^2}{\partial{t^2}} - \frac{\partial^2}{\partial{x^2}} - \frac{\partial^2}{\partial{y^2}} - \frac{\partial^2}{\partial{z^2}},
\end{equation} 
where $\eta^{\mu\nu}$ is the Minkowski metric. \margosymbol
\end{margo}
	
We say that the string is tied at both ends to some support, or that it is fixed at the end-points $x = 0$, and let $x = \length(\mathrm{s})$ be the length of the string. The boundary conditions are 
\begin{equation}
\label{equation "Boundary condition for the wave equation"}
	\upsilon(0, t) = \upsilon\bigl(\length(\mathrm{s}), t\bigr) = 0
	\begin{cases}
	\upsilon(0, t) = 0, \\
	\upsilon\bigl(\length(\mathrm{s}), t\bigr) = 0.
	\end{cases}
\end{equation}
Suppose that the initial speed of the string is zero, and let $\upsilon(x, 0) = \textcyrillic{\textit{я}}(x)$ at $t = 0$, for $0 \leqslant x \leqslant \length$. The solution of the wave equation, according to \eqref{equation "Boundary condition for the wave equation"}, is $\upsilon(x, t) = \frac{1}{2}\textcyrillic{\textit{я}}(x + ct) + \frac{1}{2}\textcyrillic{\textit{я}}(x - ct)$, where $\textcyrillic{\textit{я}}$ is an arbitrary function. Putting 
\begin{equation}
	\begin{cases}
	\upsilon(x, 0) = \textcyrillic{\textit{я}}(x), \\ 
	\frac{\partial\upsilon}{\partial{t}}(x, 0) = \textcyrillic{\textit{ю}}(x),	
	\end{cases}
\end{equation}
as initial conditions (\emph{Cauchy data}), with $-\infty < x \in \mathbb{R} < +\infty$ (string of infinite length), and $t \geqslant 0$, the d'Alembert's formula, representing the general solution to \eqref{subequations "Wave equation + wave speed"}, becomes
\begin{equation}
\label{equation "d'Alembert's formula"}
	\upsilon(x, t) = \frac{\textcyrillic{\textit{я}}(x + ct) + \textcyrillic{\textit{я}}(x - ct)}{2} + \frac{1}{2c}\int^{x + ct}_{x - ct}\textcyrillic{\textit{ю}}(z)dz.
\end{equation}

\begin{scholium}
~\enumerationisinitium
\item If $\textcyrillic{\textit{я}} \in \mathscr{C}^2(\mathbb{R})$ and $\textcyrillic{\textit{ю}} \in \mathscr{C}^1(\mathbb{R})$, then $\upsilon \in \mathscr{C}^2$ in $\mathbb{R} \times [0, \infty)$.
\item The Eq. \eqref{equation "d'Alembert's formula"} is also called \emph{solution of the Cauchy problem for the $1\mathrm{D}$ wave equation} (since the initial value conditions fall under the Cauchy boundary conditions). \scholiumsymbol
\enumerationisfinis 	
\end{scholium}

It is possible to find the solution by introducing a change of variables, $\gamma_{\mathrm{s}+} = x + ct$ and $\gamma_{\mathrm{s}-} = x - ct$. By the chain rule, one has
\begin{equation}	
	\begin{cases}
	\frac{\partial^2\upsilon}{\partial{x}^2} = \frac{\partial^2\upsilon}{\partial\gamma_{\mathrm{s}+}^2} + 2 \frac{\partial^2\upsilon}{\partial\gamma_{\mathrm{s}+}\partial\gamma_{\mathrm{s}-}} + \frac{\partial^2\upsilon}{\partial\gamma_{\mathrm{s}-}^2}, \\
	\frac{\partial^2\upsilon}{\partial{t}^2} = c^2\left(\frac{\partial^2\upsilon}{\partial\gamma_{\mathrm{s}+}^2} - 2 \frac{\partial^2\upsilon}{\partial\gamma_{\mathrm{s}+}\partial\gamma_{\mathrm{s}-}} + \frac{\partial^2\upsilon}{\partial\gamma_{\mathrm{s}-}^2}\right);
	\end{cases}
\end{equation}
consequently the wave equation is of the form
\begin{equation}
	\frac{\partial^2\upsilon}{\partial\gamma_{\mathrm{s}+}\partial\gamma_{\mathrm{s}-}} = 0,
\end{equation}
and its solution is
\begin{equation}
	\upsilon(\gamma_{\mathrm{s}+}, \gamma_{\mathrm{s}-}) = \textcyrillic{\textit{я}}(\gamma_{\mathrm{s}+}) + \textcyrillic{\textit{ю}}(\gamma_{\mathrm{s}-}) = \upsilon(x, t) = \textcyrillic{\textit{я}}(x + ct) + \textcyrillic{\textit{ю}}(x - ct).
\end{equation}

\subsection{Bi-punctuality}
\label{subsection "Bi-punctuality"}

\begingroup
\footnotesize
The mathematician considers only celestial bodies as fictitious, reducing them to simple material points, and subject exclusively to the action of their mutual gravitational attraction, which strictly obeys Newton's law. How will a similar system behave? Is it stable? The analyst faces a problem as difficult as it is interesting. And yet it is not the same problem which is present in the natural context. Real stars are not material points, and are also subject to forces other than Newtonian attraction. \\
\indent — \textsc{H. Poincaré} \cite[p. 540]{Poincare "Sur la stabilite du systeme solaire"}

\endgroup

\vspace{2mm}

From d'Alembert's proposal, the string \eqref{subequations "Wave equation + wave speed"} \eqref{equation "d'Alembert's wave equation with d'Alembertian"} is a continuous object (macroscopic-like behavior) but formed by a discrete sequence of bead-like corpuscles, usually called \emph{material points}, that have a certain mass; but each corpuscle is treated, for calculation purposes, as a point of the string (microscopic-like behavior), or a line devoid of width and thickness, an entity without dimension (0-dimensional). The d'Alembert's brainchild is to replace a finite number of corpuscles with an indefinitely increasing number for a fixed-length string, with a gradual mass-reduction of each of them that tends to 0 (punctiform state). Similarly, the distance of corpuscles from one another, i.e. the segment or interval-length $\mathrm{\Delta}x$ that separates two points, tends to 0. We have
\begin{equation}
	\frac{\partial^2\upsilon(x, ct)}{\partial{t}^2} = c^2 \left\{\frac{\upsilon(x, ct + \mathrm{\Delta}x) - 2\upsilon(x, ct) + \upsilon(x, ct - \mathrm{\Delta}x)}{\mathrm{\Delta}x^2}\right\},
\end{equation}
that leads to 
\begin{equation}
	\frac{\partial^2\upsilon(x, ct)}{\partial{t^2}} = c^2\frac{\partial^2\upsilon(x, ct)}{\partial{x^2}}, 
\end{equation}
namely the first form of \eqref{subequations "Wave equation + wave speed"}. In this way, a continuous line is generated; in other words, d'Alembert postulates that the string is a perfect infinitely (sub)divisible element. As a result, there is a conflicting overlapping of concepts, \emph{the concrete one of segment (corpuscle mass, or material point)} and \emph{the abstract one of geometric point (without extension) of the string}. The d'Alembert's formula is a \emph{hybrid specimen} of \emph{real} and \emph{idealized} vibrating string.

It is striking that a recourse to the material point, which is a geometric point-mass, can be found in the reasoning of Poincaré (see epigraph) regarding the celestial mechanics, on the stability or chaoticity of the Solar system. Later it will be shown by J. Laskar \cite{Laskar "A numerical experiment on the chaotic behaviour of the Solar System"} that the orbits of the planets are chaotic.

\begin{margo}
Studies on the existence of non-linear phenomena in vibrating strings,\footnote{
	Just by coincidence, one of the first examples of \emph{non-linearity} was discovered in the study of \emph{strings}, and dates back to Vincenzo Galilei, musician and musical theorist \cite{V. Galilei "Dialogo della Musica Antica Et Della Moderna"} \cite{V. Galilei "Fronimo Dialogo Sopra l'arte del bene intavolare"}, and father of Galileo. The frequency of the fundamental tone (which is the lowest frequency) of a vibrating string is 
	
	· directly proportional to the square root of the string tension (length and mass constant), which is a sign of non-linearity, and
	
	· inversely proportional to the square root of the mass per unit length of the string (tension and length constant), i.e. linear density.
	
	These laws go nowadays under the name of M. Mersenne \cite{Mersenne "Harmonie universelle contenant la theorie et la pratique de la musique"}.
	} 
including chaotic oscillations, are available in \cite{Tufillaro "Nonlinear and chaotic string vibrations"} \cite{Molteno Tufillaro "An experimental investigation into the dynamics of a string"} by N.B. Tufillaro and T.C. Molteno. \margosymbol
\end{margo}

\subsection{Geometro-physical Singularities: 1D Lines, 0D-like Elements, and the Point-electron, or any Particle as a Point-mass}
\label{subsection "Geometro-physical Singularities: 1D Lines, 0D-like Elements, and the Point-electron, or any Particle as a Point-mass"}

\begingroup
\footnotesize
Quoad continui aute[m] compositionem manifestum est ex præostensis ad ipsum ex indivisibilibus componendum nos minimè cogi, solum enim continua sequi indivisibilium proportionem, \emph{\&} è conversò.\footnote{
	«As regards the composition of the continuum, it is clear from the above that we are not obliged to think that it is composed of indivisibles: indeed, our only intention was to show that the continuum follows the proportional magnitudes of indivisibles, and vice versa».
	} \\
\indent — \textsc{B. Cavalieri} \cite[Liber septimus, p. 2]{Cavalieri "Geometria indivisibilibus continuorum Nova quadam ratione promota"}\endnote{
	In the 1653 edition this sentence is on p. 483.
	}

\vspace{2mm}

Suppono in limine (juxtâ Bonaventuræ Cavallerii \emph{Geometriam Indivisibilium}) Planum quodlibet quasi ex infinitis lineis parallelis conflari: Vel potiùs (quod ego mallem) ex infinitis P[ar]allelogrammis æquè altis; quorum quidem singulorum altitudo sit totius altitudinis $\frac{1}{\infty} [= 0]$, sive aliquota pars infinite parva.\footnote{
	«I suppose in advance (based on Bonaventura Cavalieri's geometry of indivisibles) that any plane is formed by an infinite number of parallel lines: Or rather (which I prefer) by an infinite number of parallelograms of equal altitude; the altitude of each of which can certainly be $\frac{1}{\infty} [= 0]$ of the whole altitude, or an infinitely small part».
	} \\
\indent — \textsc{J. Wallis} \cite[Pars Prima, Prop. I, p. 4]{Wallis "De Sectionibus Conicis"}

\endgroup

\vspace{2mm}

It should be observed \emph{en passant} that the above paradox, which arises from the definition of point, and it is debating the choice between the mathematical notion and the physical one, is at the root of the current understandings, with all the problematic consequences, of elementary particles within the Standard Model (the consideration of the electron as a geometric point-mass, so as a fundamental particle of null-valued dimension, the \emph{electron cloud} model, the question of zero-point energy, the renormalization procedures in \textsc{qft}, etc.). 

Mathematics, for its part (we can start, for convenience, from Cavalieri, and then Wallis), amphibolically conceives of space, take e.g. a plane, as an extent with geometric entities having an infinitesimal thickness or width (that is, small but non-zero quantities), and as an infinity of $1\mathrm{D}$ lines (Cavalieri's indivisibles), which, in their turn, are composed by $0\mathrm{D}$-like elements. Here \emph{geometric and arithmetic arguments are confused} with each other, without distinction. (Wallis' work is emblematic because in him the notions of infinity and zero are explicitly interlaced for the first time).\footnote{
	See \cite[Prop. CLXXXII, Scholium, p. 152]{Wallis "Arithmetica Infinitorum"}: «[T]he more terms there are supposed, the smaller becomes the difference of the base or the altitude [of the parallelogram], [so] if we proceed to infinity it vanishes: indeed $1/\infty$ (an infinitely small part) can be taken for nothing [i.e. zero] [\textit{ubi in infinitum proceditur evanescet, quippe $\frac{1}{\infty}$ (pars infinite parva) habenda erit pro nihilo}]»; see also \cite[Prop. CLXXXVIII, p. 169]{Wallis "Arithmetica Infinitorum"}: since $\frac{1}{\infty} = 0$ and $\frac{1}{0} = \infty$, «propterea [\,\dots] esset $\infty \times 0 = 1$».
	}

Physics, on the other side, describes nature by mathematics, albeit at different levels, i.e. by theoretico-physical models or by more purely mathematical approaches; and when the corpuscular properties of a particle are being described, in which unavoidably the idea of (physical) space is participating, it epitomizes the operative-creative spirit of this view. Otherwise expressed, physics re-elaborates and combines mathematical and physico-mathematical concepts such as \emph{zero} and \emph{infinite energy}, respectively, transferring them also to nature: infinite, in fact, and in a exquisitely mathematical sense, are the charge and mass of a 0-dimensional \emph{point-electron}, viz. of any particle contemplated as a point-mass (cf. Section \ref{subsection "Scholium: Point-charge/Point-mass of Electricity: Singularities (or Quasi-singularities) of Fields"}).

\subsection{Classical Ultraviolet Divergence: Electrodynamics of Charged Point-Particles, and Rowe's Renormalization}
\label{subsection "Classical Ultraviolet Divergence: Electrodynamics of Charged Point-Particles, and Rowe's Renormalization"}

\begingroup
\footnotesize
[C]he l punto $\mathrm{\pstroke}$ la sua indivisibilitade et imensurabile.\footnote{
	«Point because of its indivisibility is immeasurable». Note. The letter $\mathrm{\pstroke}$ (\emph{pee} with stroke through descender) is a medieval abbreviation of \emph{per}.
	} \\
\indent — \textsc{D. Alighieri} \cite[p. d ii-left, or unnumbered p. 52]{Alighieri "Convivio di Dante Alighieri fiorentino"}

\vspace{2mm}

Field-theoretic infinities—first encountered in Lorentz's computation of electron self-mass—have persisted in classical electrodynamics for seventy [years] and in quantum electrodynamics for some thirty-five years. These long years of frustration have left in the subject a curious affection for the infinities and a passionate belief that they are an inevitable part of nature. \\
\indent — \textsc{C.J. Isham, A. Salam, and J. Strathdee} \cite[Introduction, p. 2]{Isham Salam Strathdee "Infinity Suppression in Gravity-Modified Electrodynamics - II"}

\endgroup

\vspace{2mm}

There are techniques to circumvent, at least in some contexts, the problem related to the conception of \emph{punctiformity} of charged particles, such as electrons.\footnote{
	The \emph{first identification of the electron with a point} (point-particle, point-like particle, etc.) dates back to J. Frenkel \cite[p. 527]{Frenkel "Zur Elektrodynamik punktformiger Elektronen"}: «The inner equilibrium of an extended electron  becomes an insoluble riddle in electrodynamics. I hold this riddle [\,\dots] to be a wholly scholastic problem. It has emerged from an uncritical application to the elementary parts of matter (electrons) of a principle of division, which, if applied to composite systems (atoms, etc.), led precisely to these “smallest” parts. The electrons are not only physically but also geometrically indivisible [\textit{Die Elektronen sind nicht nur physikalisch, sondern auch, geometrisch unteilbar}]. They have no extension in space [\textit{Sie haben gar keine Ausdehnung im Raume}]. There are no inner forces between the elements of an electron; such elements do not exist. The electromagnetic interpretation of the mass is thereby eliminated, and with it all difficulties in the determination of exact equations of motion of an electron based on the (Lorentzian) principle disappear».
	} 
Here we can take a look at the issue of \emph{infinite energy of the electromagnetic field of a point-like charge}, with a E.G.P. Rowe's ad hoc solution. We will then give an overall appraisal on this solution separately, in Section \ref{subsection "Open/Unsolved Problem in Rowe's Solution"}.

\begin{exemplum}[Singularities of the energy-momentum tensor for the electromagnetic field, and Rowe's fix]
\label{exemplum "Singularities of the energy-momentum tensor for the electromagnetic field, and Rowe's fix"}
We start \emph{ex abrupto} from the problem. We adopt the perspective of K. Lechner and  P.A. Marchetti \cite{Lechner "Radiation reaction and 4-momentum conservation for point-like dyons"} \cite{Lechner Marchetti "Variational principle and energy-momentum tensor for relativistic electrodynamics of point charges"} \cite[chap. 16]{Lechner "Classical Electrodynamics: A Modern Perspective}.

Given a distance $\distance = |x - y(t)|$ from a point-like charge, the anti-symmetric \emph{electromagnetic tensor}, also called \emph{Maxwell tensor}, $F^{[\mu\nu]}$, i.e. $F^{\mu\nu} = - F^{\nu\mu}$, which represents the electromagnetic field, with asymptotic behavior \emph{diverges} in $y(t)$ as 
\begin{equation}
	^\mathrm{(div)}\F_{\mu\nu} \viz \F_{\mu\nu} \sim \frac{1}{\distance^2},
\end{equation}
so $\F_{\mu\nu}(y_\distance) = \infty$. We shall write the energy-momentum tensor of the charged point-particle for an electromagnetic field as
\begin{equation}
\label{equation "Electromagnetic energy-momentum tensor"}
	\Tau_\mathrm{em}^{\mu\nu} = \F^{\mu\xi}{\F_\xi}^\nu + \frac{1}{4}\eta^{\mu\nu}\F^{\xi\varrho}\F_{\xi\varrho},
\end{equation}
where $\eta^{\mu\nu}$ is the metric tensor of Minkowski space-time. The asymptotic $\Tau$-divergence will be 
\begin{equation}
	^\mathrm{(div)}\Tau_\mathrm{em}^{\mu\nu} \viz \Tau_\mathrm{em}^{\mu\nu} \sim \frac{1}{\distance^4}.
\end{equation}
Tensor \eqref{equation "Electromagnetic energy-momentum tensor"} is a \emph{non-integrable singularity}, i.e. it is not a $\mathcal{D}^*(\Upsilon)$-distribution element,\footnote{
	By $\mathcal{D}^*(\Upsilon)$ is denoted the linear \emph{space of distributions}; it is a locally convex topological vector space, that is, a continuous dual space of space $\mathcal{D}(\Upsilon)$ of test functions, for a  \emph{continuous linear functional} $\varphi_\mathcal{D} \colon \mathcal{D}(\Upsilon) \to \mathbb{F}$, with an open subset $\Upsilon \subset \mathbb{F}$.
	} 
for which there are not partial derivatives for the 4-divergence 
\begin{equation}
	\partial_\mu\Tau_\mathrm{em}^{\mu\nu} = \partial_\mu\F^{\mu\xi}{\F_\xi}^\nu + \F^{\mu\xi}\partial_\mu{\F_\xi}^\nu + \frac{1}{2}\F_{\xi\varrho}\partial^\nu\F^{\xi\varrho}; 
\end{equation}
while the electrostatic self-energy for an electrically charged point-particle e.g. in a static electromagnetic field is \emph{infinite}: 
\begin{equation}
\label{equation "Electrostatic self-energy for a static point-particle"}
	E_\mathrm{em} = \int\Tau_\mathrm{em}^{00}d^3x = \frac{1}{2}\left(\frac{e_\mathrm{c}}{4\pi}\right)^2\int\frac{1}{\distance^4}d^3x = \infty.
\end{equation}
One has actually a divergence of the integrals for the total 4-momentum vector $P_\mathrm{em}^\mu = \int\Tau_\mathrm{em}^{0\mu}d^3x$ of the electromagnetic field. In this context, the \emph{Lorentz–Dirac equation} \cite{Lorentz "The Theory of Electrons and Its Applications to the Phenomena of Light and Radiant Heat"} \cite{Dirac "Classical theory of radiating electrons"} 
\begin{equation}
	\frac{d\momentum^\mu}{ds} = \frac{e_\mathrm{c}^2}{6\pi}\left(\frac{da^\mu}{ds} + a^2u^\mu\right) + e_\mathrm{c}\F_\mathrm{ext}^{\mu\nu}(y)u_\nu 
\end{equation}
is also divergent, in which the 4-velocity $u^\mu = \frac{dy^\mu}{ds}$, and the derivatives of the 4-momentum $\momentum^\mu = mu^\mu$ and 4-acceleration $a^\mu = \frac{du^\mu}{ds}$ appear, whilst $m$ is the the particle's mass, and $e_\mathrm{c}\F_\mathrm{ext}^{\mu\nu}(y)u_\nu$ is an external 4-force. 

Indubitably, all these stumbling blocks, on account of the \emph{ultraviolet divergence} (\textsc{uv} since it occurs at infinitesimal distances), arise from the idealization of the particle intended as \emph{punctiformity} or 0-space.

The resolutive intuition comes from an article by Rowe \cite{Rowe "Structure of the energy tensor in the classical electrodynamics of point particles"}. The first step is to \emph{regularize} the various fields. Note that the electromagnetic tensor, in its entirety, is the sum of the Liénard–Wiechert field \cite{Lienard "Champ electrique et magnetique produit par une charge electrique concentree en un point et animee d'un mouvement quelconque"} \cite{Wiechert "Elektrodynamische Elementargesetze"} and of an external field (of an arbitrary nature): $\F^{\mu\nu} = \F_\textsc{lw}^{\mu\nu} + \F_\mathrm{ext}^{\mu\nu}$. We write the regularized Liénard–Wiechert field as 
\begin{equation}
	\F_{\textsc{lw}(\varepsilon)}^{\mu\nu} = \F_\textsc{lw}^{\mu\nu}\big|_{s(x) \to s_\varepsilon(x)}, 
\end{equation}
and the same can be done for the other fields, for any value $\varepsilon > 0$. Then 
\begin{equation}
	\F_\varepsilon^{\mu\nu} = \F_{\textsc{lw}(\varepsilon)}^{\mu\nu} + \F_{\mathrm{ext}(\varepsilon)}^{\mu\nu}, 
\end{equation}
under which the electromagnetic (Maxwell) tensor $\F_\varepsilon^{\mu\nu}$ is a \emph{regular distribution}, and its components are $\mathscr{C}^\infty(\mathbb{R}^4)$ bounded functions. This allows us to regularize, that is, to make \emph{finished}, the energy of the electromagnetic field, by replacing Eq. \eqref{equation "Electrostatic self-energy for a static point-particle"} with
\begin{equation}
	E_{\mathrm{em}(\varepsilon)} = \frac{1}{2}\int{E}_\varepsilon^2d^3x = \frac{1}{2}\left(\frac{e_\mathrm{c}}{4\pi}\right)^2\int\frac{\distance^2d^3x}{(\distance^2 + \varepsilon^2)^3},
\end{equation}
the divergence of which is $\frac{1}{\varepsilon}$, for $\varepsilon \to 0$, and the same we do with the energy-momentum tensor \eqref{equation "Electromagnetic energy-momentum tensor"} rewritten in terms of \emph{regular distribution},
\begin{equation}
	\Tau_{\mathrm{em}(\varepsilon)}^{\mu\nu} = \F_\varepsilon^{\mu\xi}{{\F_{(\varepsilon)}}_\xi}^\nu + \frac{1}{4}\eta^{\mu\nu}\F_\varepsilon^{\xi\varrho}{\F_{(\varepsilon)}}_{\xi\varrho},
\end{equation}
from which $\lim_{\varepsilon \to 0}\Tau_{\mathrm{em}(\varepsilon)}^{\mu\nu} = \Tau_\mathrm{em}^{\mu\nu}$ pointwise, if $x^\mu \neq y^\mu(s)$, with an arbitrary world line $y^\mu(s)$ of the point-particle; but if $x^\mu = y^\mu(s)$, this convergence limit does not apply.

Now, it is necessary to rely on the renormalization, subtracting the divergent part of $\Tau_\varepsilon^{\mu\nu}$, 
\begin{equation}
\label{equation "Renormalized electromagnetic energy-momentum tensor"}
	\Taustroke_\mathrm{em}^{\mu\nu} = \mathcal{D}^* - \lim_{\varepsilon \to 0}\left(\Tau_{\varepsilon}^{\mu\nu} - {}^\mathrm{div}\Tau_{\varepsilon}^{\mu\nu}\right), 
\end{equation}
denoting by $\Taustroke_\mathrm{em}^{\mu\nu}$ the \emph{renormalized} electromagnetic energy-momentum tensor, and by 
\begin{equation}
	^\mathrm{div}\Tau_\varepsilon^{\mu\nu} = \frac{1}{\varepsilon}\left(\frac{e_\mathrm{c}}{4\pi}\right)^2\int\textgreek{Ζ}^{\mu\nu}\delta^4(x - y)ds
\end{equation}
the symmetric and traceless \emph{divergent} part of $\Tau_\varepsilon^{\mu\nu}$, i.e. $^\mathrm{div}\Tau_\varepsilon^{[\mu\nu]}\eta_{\mu\nu} = 0$, including an arbitrary tensor $\textgreek{Ζ}^{\mu\nu} = c_\alpha{u}^\mu{u}^\nu + c_\beta\eta^{\mu\nu}$, where $c_\alpha$ and $c_\beta$ are constants, such that $c_\beta = -\frac{c_\alpha}{4}$, as long as $\textgreek{Ζ}^{\mu\nu}$ is symmetric and traceless. In \eqref{equation "Renormalized electromagnetic energy-momentum tensor"} $\Taustroke_\mathrm{em}^{\mu\nu}$ results to be a \emph{distribution} unequivocally. So we get
\begin{equation}
\label{equation "Renormalized electromagnetic energy-momentum tensor with constant"}
	\Taustroke_\mathrm{em}^{\mu\nu} = \mathcal{D}^* - \lim_{\varepsilon \to 0}\left(\Tau_\varepsilon^{\mu\nu} -\frac{c_\alpha}{\varepsilon}\left(\frac{e_\mathrm{c}}{4\pi}\right)^2\int\left(u^\mu{u}^\nu - \tfrac{1}{4}\eta^{\mu\nu}\right)\delta^4(x - y)ds\right). 
\end{equation}
The determination of $c_\alpha$ must ensure that $\Taustroke_\mathrm{em}^{\mu\nu}$ is conserved, and this happens iff $c_\alpha = \frac{\pi^2}{2}$.

Let us analyze the simplified case of a free point-particle, i.e. a particle not subject to external forces, so $\F_\mathrm{ext}^{\mu\nu} = 0$, but where there is only a radiation damping force $d\momentum^\mu/ds = e_\mathrm{c}^2/6\pi\left(\frac{da^\mu}{ds} + a^2u^\mu\right)$, and the active fields (regular distributions) are $\Tau_\varepsilon^{\mu\nu} = \F_{\textsc{lw}(\varepsilon)}^{\mu\nu}$. The individual components of $\Tau_\mathrm{em}^{\mu\nu}$ can then be reformulated in this way:
\begin{align}
	& \left\{\Tau_\mathrm{em}^{00} = \frac{1}{2}\left(\frac{e_\mathrm{c}}{4\pi}\right)^2\frac{1}{\bbrho^4}\right\} \refo \left\{\Tau_\varepsilon^{00} = \frac{1}{2}\left(\frac{e_\mathrm{c}}{4\pi}\right)^2\frac{\bbrho^2}{(\bbrho^2 + \varepsilon^2)^3}\right\}, \\
	& \left\{\Tau_\mathrm{em}^{0k} = 0\right\} \refo \left\{\Tau_\varepsilon^{0k} = 0\right\}, \\
	& \left\{\Tau_\mathrm{em}^{k\rotatedell} = \frac{1}{2}\left(\frac{e_\mathrm{c}}{4\pi}\right)^2\frac{1}{\bbrho^4}\left(\delta^{k\rotatedell} - 2\frac{x^kx^\rotatedell}{\bbrho^2}\right)\right\} \refo \left\{\Tau_\varepsilon^{k\rotatedell}\frac{1}{2}\left(\frac{e_\mathrm{c}}{4\pi}\right)^2\frac{\delta^{k\rotatedell}\bbrho^2 - 2x^kx^\rotatedell}{(\bbrho^2 + \varepsilon^2)^3}\right\}.
\end{align}
From these equations and \eqref{equation "Renormalized electromagnetic energy-momentum tensor with constant"}, we arrive at
\begin{align}	
	& 
	\label{align "Renormalized electromagnetic energy-momentum tensor (first form limit)"}
	\Taustroke_\mathrm{em}^{00} = \frac{1}{2}\left(\frac{e_\mathrm{c}}{4\pi}\right)^2\mathcal{D}^* - \lim_{\varepsilon \to 0}\left(\frac{\bbrho^2}{(\bbrho^2 + \varepsilon^2)^3} - \frac{3c_\alpha}{2\varepsilon}\delta^3(x)\right), \\
	& 
	\label{align "Renormalized electromagnetic energy-momentum tensor (second form limit)"}
	\Taustroke_\mathrm{em}^{0k} = 0, \\
	& 
	\label{align "Renormalized electromagnetic energy-momentum tensor (third form limit)"}
	\Taustroke_\mathrm{em}^{k\rotatedell} = \frac{1}{2}\left(\frac{e_\mathrm{c}}{4\pi}\right)^2\mathcal{D}^* - \lim_{\varepsilon \to 0}\left(\frac{\delta^{k\rotatedell}\bbrho^2 - 2x^kx^\rotatedell}{(\bbrho^2 + \varepsilon^2)^3} - \frac{c_\alpha}{2\varepsilon}\delta^{k\rotatedell}\delta^3(x)\right).
\end{align} 

The existence of $\Taustroke_\mathrm{em}^{00}$ in \eqref{align "Renormalized electromagnetic energy-momentum tensor (first form limit)"} is proven by showing that, for every test function $\varphi_\mathcal{D}(x) \in \mathcal{D} = \mathcal{D}(\mathbb{R}^3)$, there is a limit $\varepsilon \to 0$, by putting a tensor
\begin{subequations}
\label{subequations "Existence of the limit for energy density"}
\begin{align}
	\Taustroke_\mathrm{em}^{00}(\varphi) & = \frac{1}{2}\left(\frac{e_\mathrm{c}}{4\pi}\right)^2\lim_{\varepsilon \to 0}\left\{\int\frac{\bbrho^2\varphi(x)}{(\bbrho^2 + \varepsilon^2)^3}d^3x - \frac{3c_\alpha}{2\varepsilon}\varphi(0)\right\} \\
	& = \frac{1}{2}\left(\frac{e_\mathrm{c}}{4\pi}\right)^2\lim_{\varepsilon \to 0}\left\{\int\frac{\bbrho^2[\varphi(x) - \varphi(0)]}{(\bbrho^2 + \varepsilon^2)^3}d^3x + \frac{3}{2\varepsilon} \cdot \frac{\pi^2}{2} - c_\alpha \cdot \varphi(0)\right\},
\end{align}
\end{subequations}
in whose integral explicitation the Lebesgue's dominated convergence Theorem \ref{theorema "Lebesgue's dominated convergence"} is invoked, specifying the function $|\textcyrillic{\textit{я}}_n(x)| \leqslant \textcyrillic{\textit{ю}}(x) \in \Lebesgue^1(\mathbb{R}^3)$. In order to prove this, firstly we write a regularized distribution
\begin{equation}
	\textcyrillic{\textit{я}}_\varepsilon(x) = \frac{\bbrho^2[(\varphi(x) - \varphi(0) - x^k\partial_k\varphi(0)]}{(\bbrho^2 + \varepsilon^2)^3}\textgreek{Ζ}(1 - \bbrho) + \frac{\bbrho^2[(\varphi(x) - \varphi(0)]}{(\bbrho^2 + \varepsilon^2)^3}\textgreek{Ζ}(\bbrho - 1), 
\end{equation}
thanks to which the limit $\varepsilon \to 0$ is possible, for the integral 
\begin{equation}
	\int\frac{\bbrho^2[(\varphi(x) - \varphi(0)]}{(\bbrho^2 + \varepsilon^2)^3}d^3x = \int\textcyrillic{\textit{я}}_\varepsilon(x)d^3x.
\end{equation}
Setting $\bbrho \to 0$, then $\varphi(x) - \varphi(0) - x^k\partial_k\varphi(0)$ gives the value a zero as $\bbrho^2$, but $\textcyrillic{\textit{ю}}(x)$ is increasing as $\frac{1}{\bbrho^2}$. Setting $\bbrho \to \infty$, then $\textcyrillic{\textit{ю}}(x)$ gives the value a zero as $\frac{1}{\bbrho^4}$. We can thus take the limit under the integral sign to get
\begin{align}
	\lim_{\varepsilon \to 0}\int\frac{\bbrho^2[(\varphi(x) - \varphi(0)]}{(\bbrho^2 + \varepsilon^2)^3}d^3x = & \int_{\bbrho < 1}\frac{\varphi(x) - \varphi(0) - x^k\partial_k\varphi(0)}{\bbrho^4}d^3x \notag \\
	& + \int_{\bbrho > 1}\frac{\varphi(x) - \varphi(0)}{\bbrho^4}d^3x,
\end{align}
whence it follows that in \eqref{subequations "Existence of the limit for energy density"} there is a limit iff $c_\alpha = \frac{\pi^2}{2}$, and its renormalized final form is 
\begin{equation}
	\Taustroke_\mathrm{em}^{00}(\varphi) = \frac{1}{2}\left(\frac{e_\mathrm{c}}{4\pi}\right)^2\int\frac{\varphi(x) - \varphi(0)}{\bbrho^4}d^3x.
\end{equation}
The same procedure is valid to demonstrate the existence of a distributional limit in \eqref{align "Renormalized electromagnetic energy-momentum tensor (third form limit)"}, with the final form 
\begin{equation}
\label{equation "Existence of the renormalized electromagnetic energy-momentum tensor (third form limit)"}
	\Taustroke_\mathrm{em}^{k\rotatedell}(\varphi) = \frac{1}{2}\left(\frac{e_\mathrm{c}}{4\pi}\right)^2\int\frac{\varphi(x) - \varphi(0)}{\bbrho^4}\left(\delta^{k\rotatedell} - 2 \frac{x^kx^\rotatedell}{\bbrho^2}\right)d^3x.
\end{equation}

The conservation of the electromagnetic energy-momentum tensor is equivalent to  affirming that in \eqref{align "Renormalized electromagnetic energy-momentum tensor (first form limit)"} \eqref{align "Renormalized electromagnetic energy-momentum tensor (second form limit)"} \eqref{align "Renormalized electromagnetic energy-momentum tensor (third form limit)"} the \emph{continuity equation} $\partial_\mu\Taustroke_\mathrm{em}^{\mu\nu} = 0$ is satisfied. This occurs when the component is $\nu = 0$, in fact $\Taustroke_\mathrm{em}^{00}$ does not depend on the time and $\Taustroke_\mathrm{em}^{0k} = 0$. Different is the case of $\nu = \rotatedell$, for which it must apply $\partial_\mu\Taustroke_\mathrm{em}^{k\rotatedell} = 0$. Here are the steps to verify the validity of the continuity equation. From the divergence of Eq. \eqref{align "Renormalized electromagnetic energy-momentum tensor (third form limit)"}, with $c_\alpha = \pi^2/2$, one has
\begin{align}
	\partial\Taustroke_\mathrm{em}^{k\rotatedell} & = \frac{1}{2}\left(\frac{e_\mathrm{c}}{4\pi}\right)^2\mathcal{D}^* - \lim_{\varepsilon \to 0}\left\{\partial_k\left(\frac{\delta^{k\rotatedell}\bbrho^2 - 2x^kx^\rotatedell}{(\bbrho^2 + \varepsilon^2)^3} - \frac{\pi^2}{4\varepsilon}\partial_\rotatedell\delta^3(x)\right)\right\} \notag \\
		& = \frac{1}{2}\left(\frac{e_\mathrm{c}}{4\pi}\right)^2\partial_\rotatedell\left\{\mathcal{D}^* - \lim_{\varepsilon \to 0}\left(\frac{\varepsilon^2}{(\bbrho^2 + \varepsilon^2)^3} - \frac{\pi^2}{4\varepsilon}\delta^3(x)\right)\right\}. 
\end{align}
Such an equation presupposes that, for $\varepsilon \to 0$, is zero the limit of
\begin{equation}
	\int\frac{\varepsilon^2\varphi(x)}{(\bbrho^2 + \varepsilon^2)^3}d^3x - \frac{\pi^2}{4\varepsilon}\varphi(0) = \int\frac{\varepsilon^2[(\varphi(x) - \varphi(0)]}{(\bbrho^2 + \varepsilon^2)^3}d^3x - \frac{\pi^2}{4\varepsilon} =\int\gamma_\varepsilon(x)d^3x,
\end{equation}
where 
\begin{equation}
	\gamma_\varepsilon(x) = \frac{\varphi[\varepsilon(x)] - \varphi(0)}{\varepsilon(\bbrho^2 + 1)^3} 	
\end{equation}
is a sequence of functions. Once again is the dominated convergence theorem that is helpful, with which we can determine the pointwise limit 
\begin{equation}
	\lim_{\varepsilon \to 0}\gamma_\varepsilon(x)\frac{x^k\partial_k\varphi(0)}{(\bbrho^2 + 1)^3}. 	
\end{equation}
Accordingly, it is possible to take the limit under the integral sign to get
\begin{equation}
	\lim_{\varepsilon \to 0}\left\{\int\frac{\varepsilon^2\varphi(x)}{(\bbrho^2 + \varepsilon^2)^3}d^3x - \frac{\pi^2}{4\varepsilon}\varphi(0)\right\} = \int\lim_{\varepsilon \to 0}\gamma_\varepsilon(x)d^3x = \frac{x^k\partial_k\varphi(0)}{(\bbrho^2 + 1)^3}d^3x = 0.
\end{equation}
Ergo the continuity $\Taustroke$-equation is valid for $\Taustroke_\mathrm{em}^{k\rotatedell}$ in \eqref{equation "Existence of the renormalized electromagnetic energy-momentum tensor (third form limit)"}. 
\exemplumsymbol
\end{exemplum}

\subsection{Open/Unsolved Problem in Rowe's Solution}
\label{subsection "Open/Unsolved Problem in Rowe's Solution"}

\begingroup
\footnotesize
The \emph{continuum theories} make direct use of the ordinary concept of electric field strength, even for the fields in the interior of the electron. This field strength is however defined as the force acting on a test particle, and since there are no test particles smaller than an electron [\,\dots], the field strength at a [\emph{mathematical}] \emph{point} in the interior of such a particle would seem to be unobservable, by definition, and thus be \emph{fictitious and without physical meaning} [\textit{eine physikalisch inhaltslose Fiktion}] \cite[§ 67, p. 775]{Pauli "Relativitatstheorie"} = \cite[§ 67, p. 206, e.a.]{Pauli "Theory of Relativity"}.\footnote{
	Compare it to what P.W. Bridgman \cite[pp. 63, 149-150]{Bridgman "The Logic of Modern Physics"} writes: «The structure of our mathematics is such that we are almost forced, whether we want to or not, to talk about the inside of an electron, although physically we cannot assign any meaning to such statements [\,\dots]. [W]hen we get down to this scale of magnitude, our mathematics ought to be making statements about the relative behavior of discrete electrons, and not mention so much as by implication the density at points inside an electron. But this sort of thing we apparently cannot yet do; the proper mathematical language has not been developed». \cite[pp. 107-108]{Bridgman "The Logic of Modern Physics"}: «The question which interests in principle here is what meaning, if any, shall be attached to the mass of the elements of the electron. It is evident that we here go beyond any possible experience, at least for the present». \cite[pp. 145-146, e.a.]{Bridgman "The Logic of Modern Physics"}: «[T]he concept of the field at points inside the electron is an \emph{invention without physical reality}. Not only is the field concept meaningless at points inside the electron, but it is meaningless at points outside within a certain distance, because the exploring charge can never be made smaller than the electron itself, and so can never come closer than a certain distance».
	} \\
\indent We may be foolishly \emph{barking up the wrong tree if we pursue a theory of continuity within the electron} \cite[e.a.]{Pauli "Merkurperihelbewegung und Strahlenablenkung in Weyl's Gravitationstheorie"}. \\
\indent — \textsc{W. Pauli}

\endgroup

\vspace{2mm}

\enumerationisinitium
\item Rowe's strategy combines the renormalization of the electromagnetic energy-momentum tensor ($\Tau_\mathrm{em}^{\mu\nu} \refo \Taustroke_\mathrm{em}^{\mu\nu}$) with the interpretation of $\Taustroke_\mathrm{em}^{\mu\nu}$ not as a distributional limit, but rather as a sum of $n$-th derivative of a distribution. It is also explicit that he \emph{assumes the geometric model of punctiformity} of the charged particle, and therefore the problem of infinity, within this model, is not tackled at the root, that is, it is not resolved, but it is \emph{analytically fixed} to overcome the \textsc{uv} divergence.
\item The reference distributions, or linear functionals, in Example \ref{exemplum "Singularities of the energy-momentum tensor for the electromagnetic field, and Rowe's fix"} require the \emph{assumption of continuity}, which is a double-edged analytical weapon tied to the infinity  from which the problem originates. 
\item Rowe implicitly takes for granted that the \emph{physical space} (whatever that means) has the same characteristics of continuity and infinite divisibility as the \emph{space of geometry} (because this is where the singularity of point-like charges arises). Yet these characteristics are simultaneously denied in the elementary-particle description of matter, in favor of a discontinuous property.
\enumerationisfinis 

\subsection{Scholium: Point-charge/Point-mass of Electricity: Singularities (or Quasi-singularities) of Fields}
\label{subsection "Scholium: Point-charge/Point-mass of Electricity: Singularities (or Quasi-singularities) of Fields"}

\begingroup
\footnotesize
The question remains as to why Nature should have chosen [a] particular [spinning electron] model for the electron [with the pruriginous questions about the electron's internal structure] instead of being satisfied with the point-charge. \\
\indent — \textsc{P.A.M. Dirac} \cite[p. 610]{Dirac "The Quantum Theory of the Electron"}

\vspace{2mm}

The nuclear radius being of the same order of magnitude as the classical electron radius, it is very doubtful that the motion of the electron in the nucleus can still be considered as that of a point-charge, and therefore whether quantum mechanics can be applied. \\
\indent — \textsc{F. Rasetti} \cite[p. 149]{Rasetti "Il nucleo atomico"}

\endgroup

\vspace{2mm}

What we have explored and discussed above (Sections \ref{subsection "Geometro-physical Singularities: 1D Lines, 0D-like Elements, and the Point-electron, or any Particle as a Point-mass"}, \ref{subsection "Classical Ultraviolet Divergence: Electrodynamics of Charged Point-Particles, and Rowe's Renormalization"}, \ref{subsection "Open/Unsolved Problem in Rowe's Solution"}) encourages us to open a “Scholium” Section on the point-like electron (having zero radius). The two epigraphic bits, one from 1928 (Dirac), the other from 1936 (Rasetti), provide a historical insight, and do understand that the question has been open for a long time.

The following passage by P.W. Bridgman \cite[pp. 188-194, e.a.]{Bridgman "The Way Things Are"} is full of suggestions that make us reflect on the many pitfalls concealed in the concept of particle as a point-mass of electricity, charged point-mass, point-charge (of definite mass and charge), et similia. Its clear-headedness compels us to an entire reading.
 
\vspace{2mm}

\begingroup
\footnotesize
The idea of particle seems to imply a certain simplicity, but it [\,\dots] covers a growingly complicated experimental situation. The particle of Newtonian mechanics [\,\dots] was a mass point [whose motion] was governed by the equations of mechanics [\,\dots]; this of course was an idealization from experience. At first its value was mostly in treating the situations of astronomy, in which the dimensions of the planets or other heavenly bodies are so small compared with their distances apart that their motions can be calculated within the precision of measurement by treating their masses as all concentrated at the centers of gravity. The validity of the idea of mass-points was presently accepted for itself and projected toward the very small, where it found itself in congenial ground [\,\dots]. Here it proved of a value in attempts to explain the constitution and properties of matter in bulk, as, for example, in some of the speculations of Lord Kelvin, or the speculations of Newton himself in the \textit{Opticks}. Even this early mass-point developed complications, and almost from the beginning was invested with the property of impenetrability or infinite hardness in addition to the possession of mass and position [\,\dots]. But what is the experiment by which one could decide whether such mass particles are “really” impenetrable or not? [\,\dots]

	What about “identity”, or is it perhaps that impenetrability is merely another way of saying that the particle has identity? [\,\dots] Another question is “How many independent parameters may a particle have and still be a particle?” [\,\dots] How about velocity as another independent parameter? [\,\dots] When we admit velocity we at the same time admit other parameters, such as momentum and kinetic energy. These new parameters are indispensable in describing the behavior of our particle when brought into reaction with other particles or objects but they are not independent parameters because they may be computed in terms of velocity and mass. We recognize, nevertheless, that with the possession of these various parameters our particle is getting more complicated than the simple thing with which we started. We cannot mull over this situation without presently wanting to ask questions of “how” or “why” [\,\dots]. A Greek like Zeno would have been genuinely perplexed to find a satisfactory answer to the question “How is it that a thing characterized by position can also have velocity?” The modern physicist, on the other hand, does not regard this as a pressing or important question. He accepts \emph{as a brute fact} that ordinary bodies have velocities and, simultaneously, positions, and regards any difficulty of reconciling them as due to something in his thinking machinery, which \emph{he need not bother to straighten out for most of his purposes} [\,\dots]. His attempts at explanation do not prove very illuminating, however. Impenetrability might, for example, be explained in terms of the infinite forces brought into play when two particles are brought into close juxtaposition, but such infinite forces are themselves in need of explanation. If one attempts to explain them in ordinary terms, one is soon talking of infinite elastic constants of the material of the particle, which involves deformation of the particle and all such unwelcome logical consequences as the ultraviolet catastrophe which sparked the development of quantum theory [\,\dots].

	\emph{The association of the idea of particle with point is not necessary}—in fact we give numerical value to the diameter of an electron or proton. It is true that this diameter is somewhat nebulous; it may mark the order of magnitude of the distance from the associated mathematical point at which the forces begin to increase in a catastrophic surge, or it may mark the boundary of the region within which the charge must be concentrated in order to account for its mass, as in the Lorentz electron. \emph{The idea of mathematical point} thus \emph{does not appear to be essential—only that of a physically unanalyzable region}.

	How does mathematics handle particles? There seems to be a tacit ideal here which is not attained in practice. We would like to have a system of equations, some of the solutions of which have point singularities with unique properties which can be set into correspondence with the physical properties of the corresponding particle. That is, the existence of the singular point and the particle should be forced by the equation. But this is not what we have, as can be seen by looking at the simple electrostatic case for an electron [\,\dots]. \emph{We do not have equations, the singularities of which are forced by the equations themselves, but we have equations which respond by singularities in their solutions to other singularities which we impose from outside}. In other words, given only Laplace's equation [for the potential], there would be no way whatever of predicting the physical occurrence of electrons. So far as I can see, the same is true of Schrödinger's equation for wave mechanics [\,\dots].

	Physically, and for the physicist, we would appear to be in the presence of a particle when there is no experimental evidence demanding internal structure [\,\dots]. Mathematically it does not seem that the mathematics (the equations) of itself proclaims when we are dealing with a particle, but a “text” has to be added, stating that this or that property of the mathematics means a particle. The simplest example is a point singularity, which the text \emph{proclaims} to mean a particle. This is natural enough, but is it inevitable? Logically and humanly the concept of particle comes pretty close to containing a \emph{concealed contradiction}. Logically, “particle” is a verbal flag to indicate that we have come to the end [\,\dots].\endnote{
	Let us stop—literary interruption/integration—in the deceptively clapped-out prose of C.E. Gadda \cite[XI. \textit{L'atomo e l'infinito}, pp. 149-150, lines 63-123, e.a.]{Gadda "Meditazione milanese"}, which has the enjoyment of a fumesophical-like parody (cf. Section \ref{section "Interludio Giocoso. Against the Fumesophers, or the Tragicomic Smoke-sellers"}). I leave his text wilfully without En. translation, because, as A. Arbasino underlines \cite[see \textit{Genius Loci} (1977)]{Arbasino "L'Ingegnere in blu"}, the Gaddian language is an implacable \textit{foisonnement} of idiolects, so rich as to appear viscerally composite, intricately interwoven, and effusively Pantagruelian; a language that, somehow, is a reflected image of the complexity of the world, referred to as a “muddle” («garbuglio»), “ball of wool” («gomitolo», or «gnommero», in Romanesco), “filament of lumps” («filamento di grumi»), or “infinite tangle of relationships” («groviglio infinito di relazioni»), pertaining to the knotty «multiplicity of meanings of reality». Here is the jocular Gaddian excerpt: «[L]'umana conoscenza pone a sé medesima a dover spiegare relazioni sempre più ‘entrelacées’ integrandosi: e questa necessità, per quel che pertiene al sempre più piccolo, la spinse a dividere le molecole in atomi e gli atomi in joni [\,\dots]. Questa barocca idea del tagliare [\,\dots] è un'idea grossamente simbolistica che pur nel palese suo valore di simbolo ha ancora forza di prender la mano al guidatore, come un cattivo cavallo. Certe tendenze grosse, \emph{antropomorfiche} del pensiero greco sono come quei fiori spinosi a forma di pallottola che s'attaccano alle vesti lungo il sentiero, e uno non se ne libera più.

	\setlength\parindent{8pt}
	Ancor oggi il ‘témno’ [\textgreek{τέμνω}, “cut”] infierisce, e l'atomo e l'jone sono concepiti quantitativamente come frantumi o nuclei ultimi: e sta bene. Ma non si è voluto vedere il senso che la particola supposta attualmente infima (non dico atomo per ripicco) è soltanto ‘l'ultimo termine d'un sistema di giustificazioni attualmente noto’ e che questo termine è removibile, col deformarsi del sistema. Particola supposta infima o atomo è ciò al di là del quale non possiamo o non ci occorre attualmente di andare per la giustificazione della realtà, \emph{nel sistema che di essa ci siamo fatti}.

	Insisto su questo: l'espressione prende la mano al pensiero: e il barocco e il pleistocenico témno, se era degno di Anassimandro, non è degno di Lord Kelvin [Anaximander, or rather, Democritus, and Kelvin, are all under the same  chains of mental turmoil, and Gadda knows it well, actually: in science, problems change, but they do not disappear].
	
	Atomo è dunque un esempio caratteristico [\,\dots] di quei termini removibili che il sistema della conoscenza gradualmente rimuove o riscatta, decomponendoli. L'atomo può definirsi con una analogia infinitesimale. Come l'infinitesimo quantitativo non è un granulo o chicco piccolissimo (ché allora sarebbe pur sempre un finito) ma è definito in matematica come la quantità evanescente | ossia più piccola di ogni quantità finita, per quanto piccola; e cioè l'inconoscibile nella direzione del quantum, e caratteristica sua è quella di non essere misurabile su un metro o misura finita; così l'atomo è l'evanescente logico, cioè quel così piccolo logico che permane integro o vergine e non ulteriormente decomposto rispetto al sistema di relazioni costituente la realtà nota [\,\dots].

	Il critico: “Ma l'elettrone?”

	Rispondo: “Ci vuol pazienza con voi! Avrete ben capito che ho finito (per vivacità storica) a chiamar atomo quello che avevo con giusta testardaggine chiamato: ‘particola supposta attualmente infima’. Cioè ho chiamato atomo la molecola dei fisici, l'atomo dei fisici, l'elettrone dei fisici, in somma il limite semovente del sistema razionale totale semovente”».
		}

	“Particle” seems to be in some ways a necessity of thought. I have seen no evidence that any physicist, or any one else for that matter, is capable of thinking of an ostensible continuum, as in the equations of hydrodynamics for example, without inventing particles in the continuum to which to tie his thought. \emph{“Particle” for the physicist plays a role similar to that of “point” for the geometer}, and seems to be equally unavoidable.
	
\endgroup

\vspace{2mm}

The hydrodynamic case deserves a few more words \cite[p. 232]{Bridgman "How Much Rigor is Possible in Physics?"}:

\vspace{2mm}

\begingroup
\footnotesize
The equations of hydrodynamics, for instance, purportedly deal with continuous media, but the variables in the equations refer to the motion of “particles” of the fluid, which, whatever other properties they may have, at least have the property of identifiability». But it «would seem to violate the presumptive perfect homogeneity and continuity of the fluid. The two concepts are mutually contradictory and exclusive, but nevertheless our thinking seems to demand them.

\endgroup
 
\sectionmark{How Far is it Possible to \emph{Analyze} Nature? Crux of the Mathematics of Emergence}
\section{How Far is it Possible to \emph{Analyze} Nature? The Crux of the Mathematics of Emergence}
\sectionmark{How Far is it Possible to \emph{Analyze} Nature? Crux of the Mathematics of Emergence}
\label{section "How Far is it Possible to Analyze Nature? The Crux of the Mathematics of Emergence"} 

All previous Sections (of this Chapter) can be linked by a under a common thread, that of the relationship between the \emph{microscopic scale} and the \emph{macroscopic scale}, to which the distinction between macroscopic continuity and atomic-molecular discreteness is connected. 

What we call \emph{emergent phenomena} is our attempt to understand how matter—or even space—can change its properties as it passes from one scale to another. The problem is very enthralling. An excerpt by P.W. Bridgman \cite[pp. 220-221]{Bridgman "The Logic of Modern Physics"} gives an efficacious summing-up of it, with some biting questions:

\vspace{2mm}

\begingroup
\footnotesize
There is a certain thesis that is loosely related to the view that nature is finite downward [the microscopic world], namely, that an explanation of the universe is possible in which we start with small scale things, and explain large scale phenomena in terms of their small scale constituents, the thesis, in other words, that all the properties of the large are contained in the properties of the small and that the large may be constructed out of the small [\,\dots]. To maintain this thesis would demand that aggregates of things never acquire properties in virtue of their numbers which they do not already possess as individuals. Is this true? Consider, for example, the two-dimensional geometry on the surface of a sphere. This is non-Euclidean. Is the geometry of the individual elements of the surface of the sphere non-Euclidean, or do they acquire this property in changing scale? Is the kinetic energy of a number of electrons all moving together in such a way as to constitute an electric current the sum of the kinetic energies of the individual electrons, or is there an additional term? Is the mass of an electron the sum of the masses of its elements? \\
\indent A mathematical consideration is suggestive here. Those properties of a system which can be described in terms of linear differential equations have the property of additivity; the effect of a number of elements is the sum of the effects separately, and no new properties appear in the aggregate which were not present in the individual elements. But if there are combination terms (as in the electrical energy, which contains the square of the field), then the sum is more than (or different from) its parts, and new effects may appear in the aggregate.

\endgroup

\vspace{2mm}

This excerpt is befittingly entitled \textit{On the Possibility of Describing Nature Completely in Terms of Analysis}: “analysis” comes from the Gr. \textgreek{ἀναλύω}, “break down”, “unloose”, “resolve (into its elements)”. This is at the core of the mathematics of emergence: \emph{resolving} a system into its microscopic scale elements, and seeing how it takes the form that it presently has.

The emergence problem is \emph{immensely huge}. We can probe the secrets of molecular clusters \emph{regardless of} the underlying atomic nature—and this is what happens e.g. in condensed matter physics and, even more, in biology. But the questions that arise are wondrous. Here are a few examples.
	
	· Why are the water molecules not wet? 
	
	· Why do a lot of animals have perceptions but all the atoms of which they are composed are devoid of percipiency? 
	
	· What relationship exists between the emergent properties of a phenomenon, at a more macroscopic level, and its atomic ground? 
	
	· To what extent can an emergent order be examined as such (emergent)? 
	
	· Where is the “point” of separation that causes an independence between the two worlds, the micro- and macro-cosm? We can say it with a quantum paradox: where do all atoms “end” and where does a Schrödinger's cat (state) \cite[p. 812]{Schrodinger "Die gegenwartige Situation in der Quantenmechanik"} “begin”? Is this independence also an incompatibility, viz. an incommensurability/irreducibility?
	 
	Unavoidable reading on the emergence a propos of the «hierarchical structure of science» is the paper \cite{Anderson "More Is Different: Broken symmetry and the nature of the hierarchical structure of science"} by P.W. Anderson.

\vspace{10mm}

\setcounter{secnumdepth}{0}  
\section{References and Bibliographic Details}
\setcounter{secnumdepth}{3}
\markright{References and Bibliographic Details}

\begingroup
\footnotesize
\noindent Section \ref{subsection "Quantum Heisenberg–Weyl Inequality, and Schwartz Space}

\begin{indent paragraph: 15pt}
Margo \ref{margo "Schwartz space"}: see \cite{Terzioglu "On Schwartz spaces"}.
\end{indent paragraph: 15pt}

\noindent Section \ref{subsection "Schrödinger Wave Equation in 1D and Solution via Fourier Transform"}

\begin{indent paragraph: 15pt}
Example \ref{exemplum "Quantum Schrödinger particle"}: see e.g. \cite[chap. 4]{Hall "Quantum Theory for Mathematicians"}. — Proposition \ref{propositio "Solution of the Schrödinger equation through the Fourier transform method"}: on the Fourier transform, see e.g. \cite[chapp. 2, 4]{Steeb "Hilbert Spaces Wavelets Generalised Functions and Modern Quantum Mechanics"}. — Margo \ref{margo "Quantum chaos"}: for a framing of quantum chaos, see \cite{Degli Esposti Graffi Isola "Equidistribution of Periodic Orbits: an overview of classical vs quantum results"} \cite{Casati Chirikov (Eds.) "Quantum chaos between order and disorder: A selection of papers"} \cite{Hurt "Quantum Chaos and Mesoscopic Systems: Mathematical Methods in the Quantum Signatures of Chaos"} \cite[part. II]{de Almeida "Hamiltonian Systems: Chaos and Quantization"} \cite{Nakamura "Quantum versus Chaos. Questions Emerging from Mesoscopic Cosmos"} \cite{Nakamura Harayama "Quantum chaos and quantum dots"} \cite{Stockmann "Quantum Chaos: An Introduction"} \cite{Anantharaman Koch and Nonnenmacher "Entropy of Eigenfunctions"} \cite{Bolte Steiner (Eds.) "Hyperbolic Geometry and Applications in Quantum Chaos and Cosmology"} \cite{Nonnenmacher "Anatomy of Quantum Chaotic Eigenstates"} \cite{Haake Gnutzmann Kus "Quantum Signatures of Chaos"}.
\end{indent paragraph: 15pt}

\noindent Section \ref{subsection "From Camellia Sinensis to Clarkia Pulchella"}

\begin{indent paragraph: 15pt}
As regards the Brownian motion, a full account is in \cite{Morters and Peres "Brownian Motion"}, while a mathematical framing is in \cite[§ 6.3]{Stroock "Mathematics of Probability"}; about G. Cantoni, see \cite{Guareschi "Nota sulla storia del movimento browniano"} \cite[pp. 233-239]{Gallavotti "Statistical Mechanics: Short Treatise"}.
\end{indent paragraph: 15pt}

\noindent Section \ref{subsection "Fokker–Planck (Diffusion) Equation in Einstein's Theory of Brownian Motion"}

\begin{indent paragraph: 15pt}
A synopsis on Brownian motion, with reference to Einstein's theory, is in \cite[chapp. 4-5]{Nelson "Dynamical Theories of Brownian Motion"} \cite[sec. 2.3]{Lampo March Lewenstein "Quantum Brownian Motion Revisited: Extensions and Applications"} and especially \cite[chap. 3]{Gillespie and Seitaridou "Simple Brownian Diffusion: An Introduction to the Standard Theoretical Models"}.
\end{indent paragraph: 15pt}

\noindent Section \ref{section "Continuity and Discreteness—Differential Equations and Numerical Computing"}

\begin{indent paragraph: 15pt}
· A book distinctively devoted to problems of continuity and discreteness in dynamical systems with chaotic behavior is \cite{Blank "Discreteness and Continuity in Problems of Chaotic Dynamics"}. \\
· Margo \ref{margo "Cellular automata"}: with regard to the lattice gas and cellular automata, see \cite[chap. 3]{Wolf-Gladrow "Lattice-Gas Cellular Automata and Lattice Boltzmann Models: An Introduction"}, see also \cite[part II]{Succi "The Lattice Boltzmann Equation For Complex States of Flowing Matter"}.
\end{indent paragraph: 15pt}

\noindent Sections \ref{subsection "Hyperbolic Equation of a Vibrating String: d'Alembert's Formula for the 1-Dimensional Wave Phenomenon"} and \ref{subsection "Bi-punctuality"}

\begin{indent paragraph: 15pt}
· For an introduction to the eighteenth century passionate debate around mathematics \& mathematical physics of vibrating strings involving d'Alembert, L. Euler, Dan. Bernoulli, and J.-L. Lagrange, see C. Truesdell \cite[III, §§ 33-42]{Truesdell "The Rational Mechanics of Flexible or Elastic Bodies 1638-1788: Introduction To Leonhardi Euleri Opera Omnia Vol. X et XI Seriei Secundae"} and U. Bottazzini \cite[sec. 1.3]{Bottazzini "The Higher Calculus: A History of Real and Complex Analysis from Euler to Weierstrass"}. \\
· On the wave equation and d'Alembert's formula, see e.g. \cite[secc. 5.3, 6.8]{Buttazzo Giaquinta and Hildebrandt "One-dimensional Variational Problems: An Introduction"} \cite[sec. 7.1.2]{Selvadurai "Partial Differential Equations in Mechanics 1: Fundamentals Laplace's Equation Diffusion Equation Wave Equation"} \cite[sec. 2.4]{Evans "Partial Differential Equations"} \cite[secc. 5.4.1-3]{Salsa "Partial Differential Equations in Action. From Modelling to Theory"} \cite[sec. 1.2.2]{Giaquinta "Funzioni e numeri"}. \\ 
· In \cite[chap. 4.5]{Egorov Shubin "Foundations of the Classical Theory of Partial Differential Equations"} there is a summing-up of the formulæ to solve the wave equations on $1\mathrm{D}$, $2\mathrm{D}$, and $3\mathrm{D}$.
\end{indent paragraph: 15pt}

\noindent Section \ref{subsection "Geometro-physical Singularities: 1D Lines, 0D-like Elements, and the Point-electron, or any Particle as a Point-mass"}

\begin{indent paragraph: 15pt}
For those who love historical studies corroborated by a rigorous philology, on the figure of J. Wallis and his mathematical bequest, we invite you to read L. Maierù \cite[pp. 91-172]{Maieru "Fra Descartes e Newton: Isaac Barrow e John Wallis"} \cite{Maieru "John Wallis. Una vita per un progetto"}.
\end{indent paragraph: 15pt}

\endgroup

\chapter{On the Chaos, Part II. Non-linear Analysis}
\label{chapter "On the Chaos, Part II. Non-linear Analysis"}

\begingroup
\footnotesize
My purpose here is to make the case for a mathematical beauty [\,\dots] with the [\,\dots] intellectual pleasure of creating order from seeming chaos. \\
\indent — \textsc{R.P. Langlands} \cite[p. 43]{Langlands "Is there beauty in mathematical theories?"}

\endgroup

\section{The Lorenz Flow: a Strange Attractor} 
\label{section "The Lorenz Flow: a Strange Attractor"}

\subsection{Sensitivity to Initial Conditions, and Weak Predictability}
\label{subsection "Sensitivity to Initial Conditions, and Weak Predictability"}

\begingroup
\footnotesize
[T]he good [result] is obtained through many calculations by means of small differences [\textgreek{\textit{παρὰ μικρὸν διὰ πολλῶν ἀριθμῶν}}]. In the same way in this tekhne, because many calculations are needed, making a small change in the individual parts [\textgreek{\textit{μικρὰν ἐν τοῖς κατὰ μέρος παρέκβασιν}}] gives rise to a large error in the result [\textgreek{\textit{μέγα συγκεφαλαιοῦν ἐπὶ πέρας ἁμάρτημα}}].\endnote{
	Full original Gr. text: «\textgreek{τὸ γὰρ εὖ παρὰ μικρὸν διὰ πολλῶν ἀριθμῶν ἔφη γίνεσθαι. τὸν αὐτὸν δὴ τρόπον καὶ ἐπὶ ταύτης τῆς τέχνης συμβαίνει διὰ πολλῶν ἀριθμῶν συντελουμένων τῶν ἔργων μικρὰν ἐν τοῖς κατὰ μέρος παρέκβασιν ποιησαμένους μέγα συγκεφαλαιοῦν ἐπὶ πέρας ἁμάρτημα}».
	} \\
\indent — \textsc{Philo of Byzantium}, also known as \textsc{Philo Mechanicus} \cite[50, 7-13, p. 8]{Philo of Byzantium "Philons Belopoiika (viertes Buch der Mechanik)"}

\vspace{2mm}

When the state of things is such that an infinitely small variation of the present state will alter only by an infinitely small quantity the state at some future time, the condition of the system, whether at rest or in motion, is said to be stable; but when an infinitely small variation in the present state may bring about a finite difference in the state of the system in a finite time, the condition of the system is said to be unstable. It is manifest that the existence of unstable conditions renders impossible the prediction of future events, if our  knowledge of the present state is only approximate and not accurate [\,\dots]. There are certain classes of phenomena, as I have said, in which a small error in the data only introduces a small error in the result [\,\dots]. The course of events in these cases is stable [\,\dots]. There are other classes of phenomena which are more complicated, and in which cases of instability may occur, the number of such cases increasing, in an exceedingly rapid manner, as the number of variables increases. \\
\indent — \textsc{J.C. Maxwell} \cite[pp. 362, 364]{Maxwell "Does the progress of Physical Science tend to give any advantage to the opinion of Necessity (or Determinism) over that of the Contingency of Events and the Freedom of the Will?"}

\vspace{2mm}

Any change, however small, carried in the initial direction of a geodesic which remains at a finite distance is sufficient to produce any kind of variation in the final aspect of the [disturbed] curve \cite[pp. 70-71]{Hadamard "Les surfaces a courbures opposees et leurs lignes geodesiques"}. \\
\indent One of the important problems of Mechanics, namely that of the stability of the Solar System, can be characterized under the category of misplaced questions. If, indeed, we replace the study of stability of the Solar System with the analogous study of geodesics on [negatively curved] surfaces [\,\dots], we observe that any stable orbit can be transformed, by an infinitely small change in the initial data, into a completely unstable orbit [\,\dots]. Now, in astronomical problems, the initial data are never known except [within certain  error limits]. This error, no matter how small, can lead to a total and absolute perturbation \cite[p. 14]{Hadamard "Notice sur les travaux scientifiques"}. \\
\indent — \textsc{J. Hadamard}

\vspace{2mm}

[I]t may happen that small differences in the initial conditions [in the causes] generate very great ones in the final phenomena [in the effects]; a small error in the former will produce an enormous error in the latter. Prediction becomes impossible and we have the fortuitous phenomenon. \\
\indent — \textsc{H. Poincaré} \cite[I, chap. IV,\endnote{
	This chap. is a reproduction of the article \textit{Le hasard}, appeared on La Revue du Mois, Vol. III, 1907, pp. 257-276.
	} 
pp. 68-69]{Poincare "Science et methode"}

\vspace{2mm}

Small errors in the coarser structure of the weather pattern [\,\dots] tend to double in about three days [\,\dots]. Small errors in the finer structure—e.g., the positions of individual clouds—tend to grow much more rapidly, doubling in hours or less [\,\dots]. Errors in the finer structure, having attained appreciable size, tend to induce errors in the coarser structure. \\
\indent — \textsc{E.N. Lorenz} \cite[p. 3 in the original typescript]{Lorenz "Predictability: Does the Flap of a Butterfly's Wings in Brazil Set Off a Tornado in Texas?"}

\endgroup

\vspace{2mm}

\begin{figure}[h!]
\centering
\includegraphics[width = 0.75\textwidth]{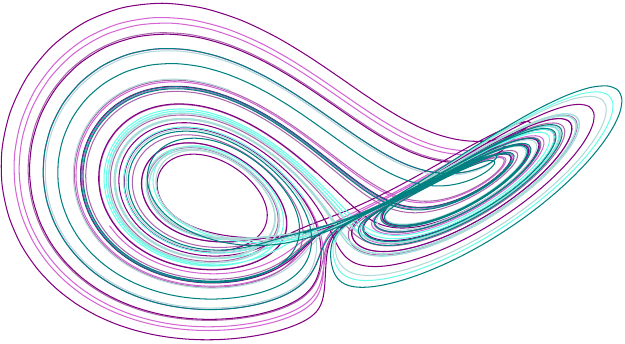}
\caption{Lorentz attractor via Euler method in 6 strokes, with \\
$\mathtt{local sigma [Pr] = 3}$, $\mathtt{local rho [Ra] = 26.5}$, \\
$\mathtt{local beta = 1}$, and \\
$\mathtt{return\{sigma*(y - x), - x*z + rho*x - y, x*y - beta*z\}}$. \\
The total number of orbits starting from various initial points is six: 
\textcolor{mallard}{•} 
\textcolor[HTML]{B2D8D8}{•}
\textcolor[HTML]{68FFE8}{•}
\textcolor{eggplant}{•}
\textcolor[HTML]{AC66AC}{•}
\textcolor[HTML]{D466D4}{•}}
\label{figure "Lorenz attractor"}
\end{figure}

We should like to say a couple of words about the Lorenz strange attractor. In its first appearance \cite[p. 137]{Lorenz "Deterministic Nonperiodic Flow"} it already has the standard double-winged shape, which coincidentally is a butterfly-like image (Fig. \ref{figure "Lorenz attractor"}); basically, it involves two spirals in a 3-dimensional phase space that map the condition of the dynamic system as it moves/evolves. The system under examination is an idealized hydrodynamic flow of \emph{deterministic ordinary non-linear differential equations}, focusing on its implementation in the thermal convection in atmospheric motion as a form of the convective fluid flow. The system's Lorenz equations are
\begin{equation}
\label{equation "Lorenz equations"}
	x, y, z \in \mathbb{R}^3
	\begin{cases}
	\dot{x} = \frac{dx}{dt} = - \mathrm{Pr}x + \mathrm{Pr}y, & \text{ with } \mathrm{Pr} = 10, \\
	\dot{y} = \frac{dy}{dt} = - xz + \mathrm{Ra}x - y, & \text{ with } \mathrm{Ra} = 28, \\
	\dot{z}	= \frac{dz}{dt} = xy - \beta z, & \text{ with } \beta = \frac{8}{3},
	\end{cases}
\end{equation}
indicating a time rate of change of three physical quantities, $x, y, z$, with three positive parameters: the Prandtl number $\mathrm{Pr}$, or the ratio of kinematic viscosity to thermal diffusivity, the Rayleigh number $\mathrm{Ra}$, or the ratio of buoyancy-driven flow to viscous and thermal dissipation, and a physical proportion $\beta$ of the attractor. 

The Eqq. \eqref{equation "Lorenz equations"} describe statistically the oscillatory behavior of a \emph{truncated (or finite, or else discrete) Fourier representation}; otherwise stated, these consist in a system of equations approximated in terms of Fourier modes and divided into a triad interaction, vertical, horizontal and vertical-horizontal, for the hydrodynamical simulation of atmospheric convection.

Lorenz shows that non-periodic solutions (if the solutions are bounded) are ordinarily \emph{unstable} in respect of small modifications, so that marginally different initial states can evolve into greatly different states; ergo a marginal uncertainty in the initial conditions can grow into a considerable uncertainty in the final state, for which the idea of predicting the future in the long-term becomes impossible. This is the so-called \emph{sensitive dependence upon initial conditions},\footnote{
	The mathematical unpredictability of certain physical phenomena, caused by the sensitivity of the initial conditions, is a limitation that can be represented from many different perspectives. One of the most common ones is that of the orbital representation (trajectories, paths, and so on). See e.g. L. Brillouin \cite[p. 125]{Brillouin "Scientific Uncertainty and Information"}: «It is impossible to study the properties of a single (mathematical) trajectory. The physicist knows only \emph{bundles of trajectories}, corresponding to slightly different initial conditions. É. Borel \cite[note II, pp. 94-101]{Borel "Introduction geometrique a quelques theories physiques}, for instance, computed that a displacement of 1 cm, on a mass of 1 gram, located somewhere in a not too distant star (say, Sirius) would change the gravitational field on the earth by a fraction $10^{-100}$. The present author went further and proved that any information obtained from an experiment must be paid for by a correspond­ing increase of entropy in the measuring device: infinite accuracy would cost an infinite amount of entropy increase and require infinite energy! This is absolutely unthinkable».
	} 
or the loss of stability through any small displacement, and it is what characterizes a \emph{system of deterministic chaos}. 

\subsection{Necessary and Sufficient Conditions for the Chaos, and More}
\label{subsection "Necessary and Sufficient Conditions for the Chaos, and More"}

\begingroup
\footnotesize
[A]t the macroscopic level, numerous phenomena present a certain type of instability, due to the fact that initial symmetry disappears. Thus a homogeneous disc allowed to fall freely through the air from a horizontal position will fall in a spiral. If one takes a cylindrical bath, full of water, and drains it through a central plug-hole, the liquid will drain with a rotary movement the sense of which is a priori unknown and unpredictable. In all cases of this type, minute variations in initial conditions may lead to very great variations in subsequent development [and nevertheless] it is quite possible to postulate that the phenomenon is determined, but this is properly speaking a \emph{metaphysical position}, impossible to verify experimentally.\footnote{
	Cf. P.W. Bridgman \cite[p. 211, e.a.]{Bridgman "The Logic of Modern Physics"}: «Determinism to the physicist is simply a way of stating certain implications of his \emph{conviction} of the connectivity of nature». This stance can act as a key to interpreting the purport of Laplace's demon \cite{Laplace "Essai philosophique sur les Probabilites"}, a \emph{célèbre} figura of scientific literature.\endnote{
	Let us assume that the final-core of nature is governed by determinism—to be sure, we repeat it again: it is a speculation, cf. footnote \ref{footnote "Two examples of physico-mathematical faith or belief"}, p. \pageref{footnote "Two examples of physico-mathematical faith or belief"}. Then \cite[pp. 3-4]{Laplace "Essai philosophique sur les Probabilites"}: «We ought to consider the present state of the universe as the effect of its anterior state, and as the cause of the one which is to follow. An intelligence which, for an instant, could comprehend all the forces by which nature is animated, and the respective situation of the beings who compose it, [and] if [this intelligence] were vast enough to submit these data to analysis, it would embrace in the same formula the movements of the greatest bodies of the universe and those of the lightest atom [\textit{embrasserait dans la même formule, les mouvemens des plus grands corps de l'univers et ceux du plus léger atome}]: nothing would be uncertain for it, and the future, as the past, would be present to its eyes [\textit{rien ne serait incertain pour elle, et l'avenir comme le passé, serait présent à ses yeux}]. The human mind [\textit{esprit}] offers [\,\dots] a faint sketch [\textit{faible esquisse}] of this [supreme] intelligence [\,\dots]. All [our] efforts in the search for truth tend to bring [our mind] closer to [such] a intelligence, but from which it will always remain infinitely distant [\textit{toujours infiniment éloigné}]».
	}
	} 
If we are only going to be happy with experimentally controllable properties, we shall be led to replace the unverifiable hypothesis of determinism by the empirically verifiable property of “structural stability” [see \cite{Andronov and Pontrjagin "Coarse systems"} in Section \ref{subsection "Structural Stability: Andronov–Pontrjagin Criterion"}]: a process $(P)$ is structurally stable if a small variation in initial conditions leads to a process $(P')$ isomorphic to $(P)$ (in the sense that a small transformation in space-time, an $\varepsilon$-homeomorphism in geometry, will bring the process $(P')$ back to the process $(P)$). \\
\indent — \textsc{R. Thom} \cite[p. 16]{Thom "A dynamic theory of morphogenesis"}
	
\endgroup

\vspace{2mm}

~\enumerationisinitium
\item The presence of the instability is a necessary but not sufficient condition for the chaos. A sufficient condition is provided by the persistence of instability (during the motion) in any length of time. 
\item The existence of a sensitivity to the initial states is a necessary but not sufficient condition for the occurrence of chaos. One of the differentiating traits of a chaotic system with \emph{strange} attractors is the generation of orbits that remain confined to a bounded region of the phase space, in addition to the above-mentioned sensitivity.
\item A Lorenz type attractor can be geometrically defined as a flow having \emph{local instability} paired with \emph{(global) non-persistent structural stability}, see. e.g. \cite{Guckenheimer Williams "Structural stability of Lorenz attractors"}. Weather pattern and climatological statistics are an example of this paradoxical combination.
\item There are many other examples of systems, in addition to that of Lorenz, appearing with irregular and non-periodic fluctuations, or else models with a quasi-cyclic motion or behaving quasi-periodically.  It is important to mention the dynamical structure of some biological systems built with \emph{Lotka–Volterra equations} \cite{Lotka "Analytical Note on Certain Rhythmic Relations in Organic Systems"} \cite{Lotka "Elements of Physical Biology"} \cite{Volterra "Fluctuations in the Abundance of a Species considered Mathematically"} \cite{Volterra "Variazioni e fluttuazioni del numero d'individui in specie animali conviventi"} \cite{Volterra "Una teoria matematica sulla lotta per l'esistenza"} \cite{Volterra D'Ancona "Les associations biologiques etudiees au point de vue mathematique"} for the predator-prey interaction.
\enumerationisfinis 

\section{Curves of Infinite Length in a Finite Volume: Three Examples from the Past}

\begingroup
\footnotesize 
One apparent contradiction requires further examination. It is difficult to reconcile the merging of two surfaces, one containing each spiral, with the inability of two trajectories to merge. \\ 
\indent — \textsc{E.N. Lorenz} \cite[p. 140]{Lorenz "Deterministic Nonperiodic Flow"}

\endgroup

\vspace{2mm}

The peculiarity of the Lorenz's set is that, in the double-winged geometry of an attractor, there is a merging of the surfaces, each with a spiral, while the orbits, which are the solutions of \eqref{equation "Lorenz equations"}, display an impossibility to merge into each other. That is the hallmark of its \emph{fractal} nature. It then follows that there is an apparent paradox, or self-contradiction. The orbits in the attractor are \emph{open curves of infinite length}, but they are \emph{enclosed in a limited volume}, i.e. \emph{restricted to a finite region in the space} of all possible system states; every curve evolves in a spiraling way over an infinite time horizon, but never intersects itself (never passes the same point twice) nor stays in the already occupied states. 

It is something which seems counter-intuitive, and yet this is not a novelty; it has notable precedents. Among all the examples available, we shall select two  of them: Torricelli's acute hyperbolic solid and Peano–Hilbert curve.

\subsection{Torricelli's Acute Hyperbolic Solid}

\begingroup
\footnotesize 
Incredibile videri potest, cum solidum hoc infinitam longitudinem habeat, nullam tamen ex illis superficiebus cylindricis quas nos consideramus, infinitam longitudinem habere; sed unamquamq; esse terminatam.\footnote{
	«Incredible though it may seem, this solid has an infinite length, and nevertheless none of the cylindrical surfaces we looked at has an infinite length but all of them are finite».
	} \\
\indent — \textsc{E. Torricelli} \cite[\textit{De Solido Hyperbolico Acuto}, Scholium, p. 116]{Torricelli "De Dimensione Parabolae"}

\vspace{2mm}

[Ho] goduto [\,\dots] de' saporitissimi frutti del suo [di Torricelli] ingegno, essendomi riuscito infinitamente ammirabile quel solido iperbolico infinitamente lungo, et uguale ad un corpo quanto a tutte e tre le dimensioni finito.\footnote{
	«[I] have enjoyed [\,\dots] the great savory fruits of your [of Torricelli] genius, because I have found endlessly admirable that infinitely long hyperbolic solid, which is equal to a finite body in all three dimensions».
	} \\
\indent — \textsc{B. Cavalieri} \cite[pp. 65-66, letter to \textsc{E. Torricelli}, 17 Dec. 1641]{Torricelli "Opere di Evangelista Torricelli III"}

\endgroup

\vspace{2mm}

The acute hyperbolic solid, later known as \emph{Torricelli's trumpet} was discovered by E. Torricelli \cite[\textit{De Solido Hyperbolico Acuto}, pp. 113-135]{Torricelli "De Dimensione Parabolae"}. It has infinite extension but finite (and measurable) volume. It is the cubature of the hyperboloid of revolution, a theorem under which there exists an equivalence between a certain acute solid of infinite length, generated by rotating one branch of the hyperbola around an asymptote (as around an axis), and a cylinder of finite height.

Building a Torricelli's trumpet is easy. Let the axis of revolution be the axis of $x$, for $x \geqslant 1$. Take a function 
\begin{equation}
	\textcyrillic{\textit{я}}(y) = \frac{1}{x}. 
\end{equation}
We call \emph{Torricelli's surface} the space that is created through rotation (to wit, cylinder plus rotated hyperbola), and we indicate by $\mathbbl{Tor}$ the resulting figure in $3\mathrm{D}$. So the $\mathbbl{Tor}$-space is the  Torricelli's trumpet we are looking for. It presents a \emph{finite volume},
\begin{align}
	\volume(\mathbbl{Tor}) & = \int^\infty_1\pi[\textcyrillic{\textit{я}}(y)]^2dx = \pi\int^\infty_1\frac{dx}{x^2} = \frac{\pi(-1)}{x}\Bigg|^\infty_1 = -\frac{\pi}{x}\Big|^\infty_1 \notag \\ 
	& = \pi \cdot 0 - (-1) = \pi(-1)^2 = \pi(1) = \pi,
\end{align}
and, simultaneously, \emph{infinite surface area},
\begin{align}
	A_\mathbbl{Tor} & = \int^\infty_1 2\pi \textcyrillic{\textit{я}}(y)\sqrt{1 + [\dot{\textcyrillic{\textit{я}}}(y)]^2}dx > 2\pi\int^\infty_1 \textcyrillic{\textit{я}}(y)dx \notag \\
	& = 2\pi\int^\infty_1\frac{dx}{x} = 2\pi \cdot \lim_{n \to \infty}\big|\ln(x)\big|^n_1 = 2\pi \cdot \ln(\infty - 0) = \infty.
\end{align}
$\mathbbl{Tor}$-space is not a fractal, but it likewise expresses the pseudo-paradox of the coexistence between finite and infinite in the same figure. We are not giving a graphic representation of it here because it closely resembles Beltrami's pseudosphere (Fig. \ref{figure "Beltrami's pseudosphere"}).

\subsection{Peano–Hilbert Curve}
\label{subsection "Peano–Hilbert Curve"}

\begingroup
\footnotesize
A continuous curve can fill a portion of space: this is one of the strangest facts in set theory, the discovery of which we owe to G. Peano. \\
\indent — \textsc{F. Hausdorff} \cite[p. 369]{Hausdorff "Grundzuge der Mengenlehre"}

\endgroup

\vspace{2mm}

\begin{figure}[h!]
\centering
\includegraphics[width = 0.65\textwidth]{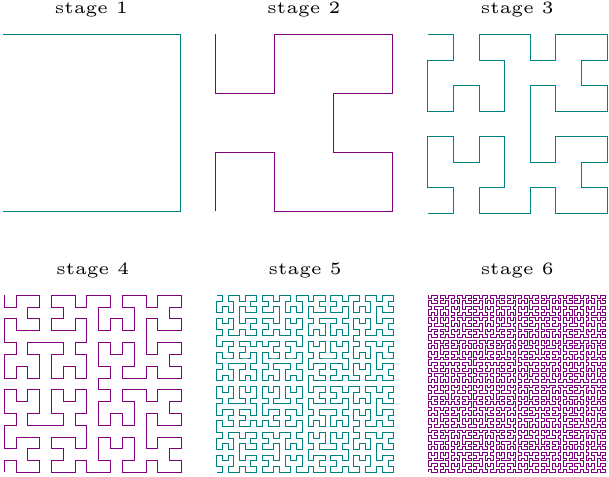}
\caption{The first six steps (or iterations) in the generation of a Hilbert space-filling curve into a $2\mathrm{D}$ area, whose Hausdorff dimension is $\mathrm{D_f} = \frac{\log[1/(1/4)]}{\log{2}} = 2$}
\label{figure "Hilbert curve"}
\end{figure}

\begin{figure}[h!]
\centering
\includegraphics[width = 0.75\textwidth]{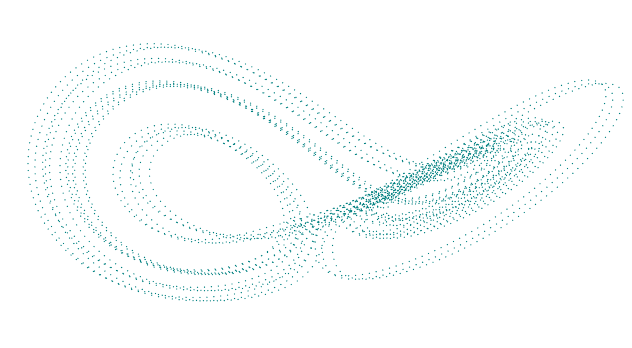}
\caption{\emph{Hazy attractor} (\textgreek{α}): a type of \emph{quasi-random} attractor of a stochastic-like Lorenz system. Here a \emph{nebula of multiple pseudo-orbits} (up to 18 strokes) sets the tone (cf. Fig. \ref{figure "Lorenz attractor"})}
\label{figure "Hazy attractor: quasi-random attractor of a stochastic-like Lorenz system I"}
\end{figure}

\begin{figure}[h!]
\centering
\includegraphics[width = 0.75\textwidth]{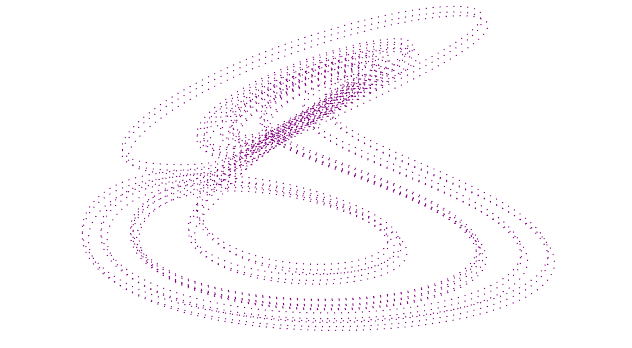}
\caption{\emph{Hazy attractor} (\textgreek{β}): an alternative version of the attractor in Fig. \ref{figure "Hazy attractor: quasi-random attractor of a stochastic-like Lorenz system I"}}
\label{figure "Hazy attractor: quasi-random attractor of a stochastic-like Lorenz system II"}
\end{figure}

Another example is the \emph{Peano curve} \cite{Peano "Sur une courbe qui remplit toute une aire plane"} \cite[Fasc. 1, Figg. $(c)$-$(d)$, p. 240]{Peano "Formulario Mathematico"}; it is a continuous curve but do not has a derivative in the 2-dimensional plane, which fills an entire flat area, e.g. the surface of a unit square, that is a \emph{curve of infinite length (at least potentially) within a finite area}, obtained by a space-filling construction. Similarly, it may get a Peano curve in $3\mathrm{D}$. The Hilbert curve \cite{Hilbert "Ueber die stetige Abbildung einer Linie auf ein Flachenstuck"} (Fig. \ref{figure "Hilbert curve"}) is a simplified variant of the Peano curve, with Hausdorff dimension 
\begin{equation}
	\mathrm{D_f} = \frac{\log\left(\frac{1}{f}\right)}{\log{2}} = 2, \enspace f = \tfrac{1}{4}.
\end{equation}

\subsection{Snowflake Curve of von Koch}
\label{subsubsection "Snowflake Curve of von Koch"}

A further quirk in the abstract world of curves is the \emph{Koch curve} \cite{von Koch "Sur une courbe continue sans tangente obtenue par une construction geometrique elementaire"} \cite[Figg. 1-2, p. 149]{Koch "Une methode geometrique elementaire pour l'etude de certaines questions de la theorie des courbes planes"}, aka \emph{Koch snowflake}, a continuous but non-differentiable closed curve in $\mathbb{R}^2$ with a \emph{finite area} bounded by an \emph{infinite perimeter}.

Take an equilateral triangle, say $\rotatedtriangle_1$. Remove the middle third of each side $\mathrm{L}$, and put three equilateral triangles without the bases in the gaps. One gets a 6-pointed polygon with 12 lines segments, the length of which is $\length = \frac{\mathrm{L}}{3}$, with a total length 
\begin{equation}
	\length_\mathrm{t} = 3 \cdot 4 \cdot \frac{\mathrm{L}}{3}. 
\end{equation}
Repeat the process for an unlimited period of time, by—endlessly—adding smaller and smaller triangles.

Let us draw some iterations of the Koch curve. To make things easier, let us imagine a line segment; divide it into three equal parts, and remove the middle part; now, put a triangle without the base in the gap (the unfilled space). The procedure shall be repeated from the start:
\begin{center}
\begin{tikzpicture}[scale = 2, decoration = Koch snowflake]
	\draw[thin, mallard]decorate{(0,0) -- (3,0)};
	\draw[thin, eggplant]decorate{decorate{(0,-1) -- (3,-1)}};
	\draw[thin, mallard]decorate{decorate{decorate{(0,-2) -- (3,-2)}}};
	\draw[thin, eggplant]decorate{decorate{decorate{decorate{(0,-3) -- (3,-3)}}}};
\end{tikzpicture}
\end{center}

It comes out a self-similar fractal, so the Koch curve can be written as 
\begin{equation}
	\Kochcurve = (\Kochcurve)_\infty = \lim_{n \to \infty}(\Kochcurve)_n. 	
\end{equation}
The number of sides, or lines segments, after $n$-th iterations is $N_n = 3 \cdot 4^n$, for which
\begin{equation}
	(\Kochcurve)_n = 3 \cdot 4^n \cdot \frac{\mathrm{L}}{3^n} = 3\mathrm{L}\left(\frac{4}{3}\right)^n \to \infty \text{ as } n \to \infty,	
\end{equation}
and $\length(\Kochcurve) = \infty$ (infinite length). The Hausdorff dimension of $\Kochcurve$ is
\begin{equation}
	\mathrm{D_f} = \lim_{\frac{\mathrm{L}}{3^n} \to 0}\frac{\log{N_n}}{\log\left(\frac{1}{\frac{\mathrm{L}}{3^n}}\right)} = \lim_{n \to \infty} \frac{\log 3 \cdot 4^n}{\log(3^n)} = \frac{\log{4}}{\log{3}} = \frac{2\log{2}}{\log{3}} =  1.26185950714291\cdots
\end{equation}  

And this is what the Koch snowflake looks like after a tot of iterations:
\begin{center}
\begin{tikzpicture}
	\draw[thin, eggplant, l-system = {rule set = {F -> F-F++F-F}, step = 2pt, angle = 60, axiom = F++F++F, order = 4}] lindenmayer system -- cycle;
	\end{tikzpicture}
\end{center}

The finite area of $\Kochcurve$ is equivalent to $\frac{8}{5}$ of the area of $\rotatedtriangle_1$.

\sectionmark{Strangeness—Fractal Dimension of the Orbit in the Phase Space Attractor}
\section{Strangeness—Fractal Dimension of the Orbit in the Phase Space Attractor: Grassberger–Procaccia Algorithm and Lyapunov–Kaplan–Yorke Dimension}
\sectionmark{Strangeness—Fractal Dimension of the Orbit in the Phase Space Attractor}
\label{section "Strangeness—Fractal Dimension of the Orbit in the Phase Space Attractor: Grassberger–Procaccia Algorithm and Lyapunov–Kaplan–Yorke Dimension"}

\begingroup
\footnotesize
A formula can be very simple and create a universe of bottomless complexity.\footnote{
	See Section \ref{subsection "Mandelbrot Set"}.
	} \\
\indent — \textsc{B. Mandelbrot}\endnote{
	From the short documentary by E. Morris, \textit{Big Brains. Small Films. Benoît Mandelbrot, The Father of Fractals}. The conversation dates back to September 2010.
	}

\vspace{2mm}

An attractor representing the flow of a viscous fluid is part of an infinite-dimensional space, but has itself only finite dimension [\,\dots]. According to the modes paradigm, a finite-dimensional space can describe only a finite number of modes. (Mathematically: a finite-dimensional space can contain only a finite-dimensional torus). Yet frequency analysis reveals a continuum of frequencies, which one would interpret as a continuum of modes. Is such a thing possible? \\
\indent — \textsc{D. Ruelle} \cite[pp. 64-65]{Ruelle "Chance and Chaos"}

\endgroup

\vspace{2mm}

An attractor is called \emph{strange} if it has a \emph{fractal dimension}, also referred to as \emph{Hausdorff} or \emph{Hausdorff–Besicovitch dimension} \cite{Hausdorff "Dimension und ausseres Mass"} \cite{Besicovitch "On linear sets of points of fractional dimension"}, that is, if its dimension is not integer, as defined by B. Mandelbrot \cite{Mandelbrot "Les objets fractals. Forme hasard et dimension"} \cite{Mandelbrot "Geometrie fractale de la turbulence. Dimension de Hausdorff dispersion et nature des singularites du mouvement des fluides"} \cite{Mandelbrot "The Fractal Geometry of Nature"}.\footnote{
	The Cantor set of points \cite{Cantor "Ueber unendliche lineare Punktmannichfaltigkeiten"}, or, as Mandelbrot \cite[p. 52]{Mandelbrot "Les objets fractals. Forme hasard et dimension"} likes to call it, the Cantor dust [\textit{poussière}] in its various versions, is an archetype of fractal objects.
	}\textsuperscript{,}\endnote{
	The fortune and applicative extensibility of fractals is well known, and is often invoked as a \emph{crocevia} of synthesis for criteria of unity between theories in (apparent) contrast, such as relativity and quantum mechanics, or between theories of macroscopic level and theories of microscopic level; see, in this direction, L. Nottale \cite{Nottale "Scale Relativity and Fractal Space-Time. A New Approach to Unifying Relativity and Quantum Mechanics"}, and Nottale \& Lehner \cite{Nottale and Lehner "Turbulence and Scale Relativity"}.
	} 
The term “strange” for the attractor—as it derives from a \emph{Hopf bifurcation} \cite[p. 270]{Poincare "Sur l'equilibre d'une masse fluide animee d'un mouvement de rotation"} of solutions of equations—is due to D. Ruelle and F. Takens \cite{Ruelle and Takens "On the Nature of Turbulence"} \cite{Ruelle and Takens "Note "On the Nature of Turbulence""}, in connection with the study on the hydrodynamic turbulence. The fractal dimension of the strange attractor can be proved on the so-called \emph{correlation dimension}, which generally is a \emph{measure for the strangeness of attractors}, and also serves to distinguish between deterministic chaos (to which the Lorenz attractor belongs) and random noise. The correlation dimension is thus intrinsically related to the fractal dimension, and it can be achieved from a time series of one or more variables, by introducing the Grassberger–Procaccia algorithm \cite{Grassberger and Procaccia "Characterization of Strange Attractors"} \cite{Grassberger and Procaccia "Measuring the strangeness of strange attractors"} \cite{Grassberger "Generalized dimensions of strange attractors"}. Let us look more closely at what it is. 

Consider the set $\{x_\mu = x(t + \mu\tau) \mid \mu = 1, \mathellipsis, N\}$ of $N$ points on the attractor, in which the (spatial) correlation between points of a time series is expressed, with an arbitrary time increment $\tau$ between successive measurements. This correlation is measurable through the \emph{correlation integral}, 
\begin{equation}
	 C(\distance) = \lim_{N \to \infty}\frac{1}{N^2}\sum^N_{\stackrel{\mu, \nu = 1}{\mu \neq \nu}}\Theta(\distance - \|x_\mu - x_\nu\|),
\end{equation}
where $\distance$ is a threshold distance, for which the distance $\|x_\mu - x_\nu\|$ between all pairs of points $(x_\mu, x_\nu)$ is less than $\distance$, $\Theta(\cdot)$ is the Heaviside step function, 
\begin{equation}
\label{equation "Heaviside step function"} 
	\Theta(x) =
	\left\{
	\!\begin{aligned}
	0 \text{ if } x < 0 \\
	\tfrac{1}{2} \text{ if } x = 0 \\
	1 \text{ if } x > 0
	\end{aligned}
	\right\} \text{for some value } x,
\end{equation}
and $\|\cdot\|$ is the norm on $x$. The Grassberger–Procaccia technique proves that $C(\distance)$ is like a power of $(\distance)$ for small $(\distance)$: $C(\distance) \propto \distance^\bbnu$. The exponent $\bbnu$ of the power law dependence of $C(\distance)$ is the correlation dimension:
\begin{equation}
	\bbnu = \lim_{\distance \to 0}\frac{\log{C}(\distance)}{\log\distance}.
\end{equation} 
The correlation dimension for the Lorenz attractor is $\bbnu = 2.05 \pm 0.01$, while its fractal dimension is $\mathrm{D_f} = 2.06 \pm 0.01$.

There are also other tools to approach the problem of determining the attractor dimensionality. The Lyapunov dimension, or Kaplan–Yorke dimension \cite{Kaplan and Yorke "Chaotic behavior of multidimensional difference equations"} \cite{Frederickson Kaplan Yorke and Yorke "The Liapunov Dimension of Strange Attractors"}, 
\begin{equation}
		\mathrm{D_L} = k + \frac{\sum_{\nu = 1}^k(\lambda_\textsc{l})_\nu}{|(\lambda_\textsc{l})_{k + 1}|},
\end{equation}
is one of them, where $k$ is the largest integer such that 
\begin{equation}
	\begin{cases}
	\sum_{\nu = 1}^k(\lambda_\textsc{l})_\nu \geqslant 0, \\ 
	\sum_{\nu = 1}^{k + 1}(\lambda_\textsc{l})_\nu < 0. 	
	\end{cases}
\end{equation}
It is thus a way to estimate the fractal dimension of an attractor. The Lorenz attractor exhibits \cite[p. 289]{Wolf Swift Swinney and Vastano "Determining Lyapunov Exponents from a Time Series"} a chaotic nature with these parameter values: $\mathrm{Pr} = 16.0$, $\mathrm{Ra} = 45.92$, $\beta = 4.0$; and its Lyapunov spectrum is: $(\lambda_\textsc{l})_1 = 2.16$, $(\lambda_\textsc{l})_2 = 0.00$, $(\lambda_\textsc{l})_3 = -32.4$; hence the Lyapunov dimension for the Lorenz attractor is $\mathrm{D_L} = 2.07$. See also \cite{Leonov Kuznetsov Korzhemanova Kusakin "Lyapunov dimension formula for the global attractor of the Lorenz system"}. Now, since, thanks to a Pesin's theorem \cite{Pesin "Characteristic Lyapunov exponents and smooth ergodic theory"}, the sum of positive Lyapunov exponents provides a valid estimate for the Kolmogorov–Sinai metric entropy (Definition \ref{definitio "Kolmogorov–Sinai metric entropy"}), the dimension $\mathrm{D_L}$ is an entropic-like measure, as an indicator of the degree of complexity in the attractor dynamics, i.e. of disorder (chaos) of the points on the Lorenz or other attractor; cf. J.-P. Eckmann and D. Ruelle \cite{Eckmann and Ruelle "Ergodic theory of chaos and strange attractors"}.

\subsection{Mandelbrot Set}
\label{subsection "Mandelbrot Set"}

Given a complex map, for a \emph{complex quadratic polynomial} $\varphi_c$, 
\begin{equation}
	\begin{cases}
	\label{cases "Complex quadratic polynomial"}
	\varphi_c \colon \mathbb{C} \to \mathbb{C}, \\
	\varphi_c(z) = z^2 + c,	
	\end{cases}
\end{equation}
i.e. a quadratic family $\varphi_c \colon z \mapsto z^2 + c$, the \emph{Mandelbrot set} is defined by 
\begin{equation}
	M_\mathbb{C} = \bigl\{c \in \mathbb{C} \mid \varphi_c^n(0) \sNeg[1mu]{\rightarrow} \infty \text{ as } n \to \infty\bigr\}.
\end{equation}

\section{Strangeness and Chaoticity}

The strangeness of an attractor depends on its geometry or shape, if it is fractal (non-integer dimension, orbits of infinite length within a finite volume of the phase space). The chaoticity of an attractor depends on its complexity (instability of the orbits, sensitivity on initial values of the mapping, presence of a positive Lyapunov exponent), see e.g. \cite{Grebogi Ott and Yorke "Chaos Strange Attractors and Fractal Basin Boundaries in Nonlinear Dynamics"}. Generally speaking, this appears to suggest that chaotic attractors are also strange and strange attractors are also chaotic; and it is true into a huge variety of species in the safari park of attractors, except in some cases. There are chaotic attractors that are not strange, as in \cite{Holden and Muhamad "A graphical zoo of strange and peculiar attractors"}, and strange attractors that are not chaotic, as in \cite{Grebogi Otta Pelik and Yorked "Strange attractors that are not chaotic"} \cite{Ditto Spano Savage Rauseo Heagy and E. Ott "Experimental Observation of a Strange Nonchaotic Attractor"}.

\section{Hyperbolicity, Singularity, and \textsc{srb} Measure in the Attracting Sets}

In this Section we want to take a glance at the geometry of Lorenz-like attractor, and its statistical behavior, as well as at the role of the \emph{physical} invariant measure, known as \emph{Sinai–Ruelle–Bowen} \emph{measure}, connected with it.

\subsection{Singular Hyperbolic Lorenz-like Attractor}
\label{subsection "Singular Hyperbolic Lorenz-like Attractor"}

The study of geometry of the Lorenz attractor, called \emph{geometric (model) Lorenz flow}, begins with V.S. Afraimovich, V.V. Bykov \& L.P. Shilnikov \cite{Afraimovich Bykov Shilnikov "On the origin and structure of the Lorenz attractor"}, and with J. Guckenheimer \& R.F Williams \cite{Guckenheimer Williams "Structural stability of Lorenz attractors"} \cite{Williams "The structure of Lorenz attractors"}. It is known that the \emph{Lorenz attractor is not hyperbolic}, since it contains a single equilibrium point of saddle type at the origin 
\begin{equation}
	\singularity = (0, 0, 0), 
\end{equation}
and regular orbits are formed near this point. This singularity prevents a hyperbolic structure. Nevertheless, the Lorenz(-like) attractor has a weak form of hyperbolicity, that takes the name of \emph{singular hyperbolicity}, in compliance with the discovery made by C.A. Morales, M.J. Pacífico and E.R. Pujals (\textsc{mpp}) \cite{Morales Pacifico Pujals "On $C^1$ robust singular transitive sets for three-dimensional flows"} \cite{Morales Pacifico Pujals "Singular hyperbolic systems"} \cite{Morales Pacifico Pujals "Robust transitive singular sets for 3-flows are partially hyperbolic attractors or repellers"}, but see before E.A. Sataev \cite{Sataev "Invariant measures for hyperbolic maps with singularities"}; cf. \cite{Araujo Galatolo and Pacifico "Statistical Properties of Lorenz-like Flows Recent Developments and Perspectives"}.

\begin{definitio}[Hyperbolic and partially hyperbolic invariant set]
\label{definitio "Hyperbolic and partially hyperbolic invariant set"}
\enumerationisinitium
Let $\mathcal{M}$ be a closed 3-manifold, and $\mathcal{X}^r(\mathcal{M})$ the space of differentiable vector fields $\vec{X}$ on $\mathcal{M}$ in the $\mathscr{C}^r$ topology, with $r \geqslant 1$. For $\vec{X} \in \mathcal{X}^r(\mathcal{M})$, the flow induced by $\vec{X}$ is denoted by $\varphi_t \colon \mathcal{M} \to \mathcal{M}$, with $t \in \mathbb{R}$.
\item A compact invariant set $\mathbbl{\Lambda} \subset \mathcal{M}$ is \emph{hyperbolic}, if there is a continuous ($\mathcal{T}\varphi_t$)-invariant splitting 
\begin{equation}
	\mathring{\mathcal{T}}_\mathbbl{\Lambda}\mathcal{M} \equival \mathring{\mathcal{E}}^\mathrm{s}_\mathbbl{\Lambda} \oplus \mathring{\mathcal{E}}^{\varphi_t}_\mathbbl{\Lambda} \oplus \mathring{\mathcal{E}}^\mathrm{u}_\mathbbl{\Lambda}, 
\end{equation}
for which the tangent bundle $\mathring{\mathcal{T}}_\mathbbl{\Lambda}\mathcal{M} $ decomposes in three $d\varphi_t$-invariant subbundles of dimension 1, where $\mathring{\mathcal{E}}^\mathrm{s}_\mathbbl{\Lambda}$ and $\mathring{\mathcal{E}}^\mathrm{u}_\mathbbl{\Lambda}$ are uniformly contracted and expanded, respectively, by the derivative $d\varphi_t$, with $t > 0$, and $\mathring{\mathcal{E}}^{\varphi_t}_\mathbbl{\Lambda}$ is the direction of the flow (cf. Scholium \ref{scholium "Tangent bundle in the Anosov system"}). 
\item 
\label{item "Partially hyperbolic invariant set"}
A compact invariant set $\mathbbl{\Lambda}$ of $\vec{X} \in \mathcal{X}^r(\mathcal{M})$ is \emph{partially hyperbolic} if there is a continuous and dominated splitting 
\begin{equation}
	\mathring{\mathcal{T}}_\mathbbl{\Lambda}\mathcal{M} \equival \mathring{\mathcal{E}}^\mathrm{s}_\mathbbl{\Lambda} \oplus \mathring{\mathcal{E}}^\mathrm{cu}_\mathbbl{\Lambda},
\end{equation}
and constants $c > 0$ and $0 < \lambda < 1$, where $\mathring{\mathcal{E}}^\mathrm{s}_\mathbbl{\Lambda}$ is the uniformly contracting 1-dimensional subbundle, and $\mathring{\mathcal{E}}^\mathrm{cu}_\mathbbl{\Lambda}$ is the the 2-dimensional subbundle, called \emph{central unstable direction} of $\mathring{\mathcal{T}}_\mathbbl{\Lambda}\mathcal{M}$, which is \emph{volume expanding} and includes the direction of the flow, such that the following conditions hold:
\subenumerationisinitium
\item a dominating state, 
\begin{equation}
	\|d_x\varphi_t \mid \mathring{\mathcal{E}}^\mathrm{s}_\mathbbl{\Lambda}(x)\|\cdot\|d_x\varphi_{-t} \mid \mathring{\mathcal{E}}^\mathrm{cu}_\mathbbl{\Lambda}(x)\| < c\lambda^t, 
\end{equation}
\item a contracting state, 
\begin{equation}
	\|d_x\varphi_t \mid \mathring{\mathcal{E}}^\mathrm{s}_\mathbbl{\Lambda}(x)\| \leqslant c\lambda^t,
\end{equation}
\subenumerationisfinis
for each $x \in \mathbbl{\Lambda}$. \definitiosymbol
\enumerationisfinis
\end{definitio}

So let us just skip ahead to the notion of singularity combined with the second definition above.

\begin{definitio}[Singular hyperbolic set]
Let $\setsingularities_{\vec{X}}(\mathbbl{\Lambda})$ be the set of singularities of $\vec{X}$ in $\mathbbl{\Lambda}$. A partially hyperbolic set (as in \ref{item "Partially hyperbolic invariant set"} of Definition \ref{definitio "Hyperbolic and partially hyperbolic invariant set"}) is \emph{singular hyperbolic} if any singularity $\singularity \in \setsingularities_{\vec{X}}(\mathbbl{\Lambda})$ is hyperbolic with volume expanding central direction. \definitiosymbol
\end{definitio}

The Morales–Pacífico–Pujals scheme provides the following result.

\begin{theorema}[\textsc{mpp}]
\label{theorema "MPP"}
~\enumerationisinitium
\item Any $\mathscr{C}^1$ robustly transitive set $\mathbbl{\Lambda}$ of $\vec{X} \in \mathcal{X}^1(\mathcal{M})$ for a 3-dimensional flow is a singular hyperbolic attractor for $\vec{X}$, or repeller for $-\vec{X}$, i.e. $\mathbbl{\Lambda}_\singularity$. 
\item The set $\mathbbl{\Lambda}_\singularity$ is Lorenz-like, for which any set of this type resembles a geometric Lorenz flow.
\item If there is no $\singularity \in \setsingularities_{\vec{X}}(\mathbbl{\Lambda})$, then $\mathbbl{\Lambda}$ is a hyperbolic set. 
\item 
\label{item "Uniformly hyperbolic set"}
The set $\mathbbl{\Lambda}$ is uniformly hyperbolic iff there is no attached $\singularity \in \setsingularities_{\vec{X}}(\mathbbl{\Lambda})$ either for $\vec{X}$ or $-\vec{X}$.
\enumerationisfinis
\end{theorema}

\begin{proof}
For the demonstration, we refer to \cite{Morales Pacifico Pujals "On $C^1$ robust singular transitive sets for three-dimensional flows"} \cite{Morales Pacifico Pujals "Singular hyperbolic systems"} \cite{Morales Pacifico Pujals "Robust transitive singular sets for 3-flows are partially hyperbolic attractors or repellers"}.
\end{proof}

It should be noted that the leaves of the invariant foliation associated with the attracting sets $\mathbbl{\Lambda}$ undergo a contraction by the flow. \emph{Robust} means that $\mathbbl{\Lambda}$ is not destroyed by arbitrarily small $\mathscr{C}^1$ perturbations of the original flow. 

\begin{definitio}[Lorenz-like singularity]
A singularity $\singularity \in \setsingularities_{\vec{X}}(\mathbbl{\Lambda})$ is Lorenz-like if its eigenvalues are real and satisfy $\lambda_2 < \lambda_3 < 0 < -\lambda_3 < \lambda_1$. \definitiosymbol
\end{definitio}

\subsection{Sinai–Ruelle–Bowen Measure: Uniformly Hyperbolic Attractor}
\label{subsection "Sinai–Ruelle–Bowen Measure: Uniformly Hyperbolic Attractor"}

One of the most widely used methods to comprehend and manage the probabilistic, or statistical, aspects of hyperbolic sets is the probability measure identified by Ya.G. Sinai, D. Ruelle and R. Bowen (\textsc{srb}) \cite{Sinai "Gibbs measures in ergodic theory"} \cite{Ruelle "A Measure Associated with Axiom-A Attractors"} \cite{Bowen and Ruelle "The Ergodic Theory of Axiom A Flows"}. Natural applications of the \textsc{srb} measure comprise the uniformly hyperbolic attractor. 
	
\begin{definitio}[\textsc{srb} measure for an attractor of hyperbolic diffeomorphism and flow]
~\enumerationisinitium
\item Let $\mathbbl{\Lambda} \subset \mathcal{M}$ be an attractor for a uniformly hyperbolic diffeomorphism $\varphi_\bbmu \colon \mathcal{M} \to \mathcal{M}$ of class $\mathscr{C}^2$ on a compact Riemannian manifold, where $\bbmu$ is an $\varphi_\bbmu$-invariant (Borel) probability measure on $\mathcal{M}$.
\subenumerationisinitium
\item Let $\mathcal{W}^\mathrm{s}$  be an open neighborhood of $\mathbbl{\Lambda}$, or more precisely, an open stable set containing a neighborhood of $\mathbbl{\Lambda}$. We can represent the attractor as a topological object,
\begin{equation}
\label{equation "Topological attractor"}
	\mathbbl{\Lambda} \viz \mathbbl{\Lambda}_\topological = \bigcap_{n \geqslant 0}\varphi_\bbmu(\mathcal{W}^\mathrm{s}).
\end{equation}
We say that $\mathcal{W}^\mathrm{s}$ represents the region of initial conditions in the phase space, better known as attractor's \emph{basin of attraction} with positive Lebesgue measure, denoted by 
\begin{equation}
	\mathcal{B}(\mathbbl{\Lambda}) = \{x \in \mathcal{M} \mid \text{set of } \bbmu\text{-points } x \subset \mathbbl{\Lambda}\},
\end{equation}
and that $\mathcal{B}(\mathbbl{\Lambda})$ is foliated with stable manifolds. 

If $\varphi_\bbmu$ is topologically transitive, and it has a positive Lyapunov exponent almost everywhere with respect to $\bbmu$, then the measure $\bbmu$ for an attractor \eqref{equation "Topological attractor"} is called a \emph{measure of Sinai–Ruelle–Bowen} which satisfies 
\begin{equation}
	\lim_{n \to \infty}\frac{1}{n}\sum^{n - 1}_{\nu = 0}\tau\bigl(\varphi_\bbmu^\nu(x)\bigr) = \int_{\mathbbl{\Lambda} \subset \mathcal{M}}\varphi_\bbmu{d\bbmu},
\end{equation}
for any continuous function $\tau \colon \mathcal{M} \to \mathbb{R}$ and almost all points $x \in \mathcal{B}(\mathbbl{\Lambda})$ of positive Lebesgue measure, $\bblambda\bigl(\mathcal{B}(\mathbbl{\Lambda})\bigr) > 0$.

\item We can also visualize the \textsc{srb} measure in another way, considering $\mathbbl{\Lambda}$ as a union of unstable manifolds $\mathcal{W}^\mathrm{u}$, or taking into account the leaves of the unstable foliations of the attractor. If $\bbmu$ is absolutely continuous (with respect to the Riemannian measure induced) on the unstable manifolds, i.e. if there is a uniform absolute continuity of unstable foliations of $\mathbbl{\Lambda}$, then the $\varphi_\bbmu$-invariant (Borel) probability measure $\bbmu$ is consistent with the measure of Sinai–Ruelle–Bowen, and $\varphi_\bbmu$ has an \emph{ergodic Sinai–Ruelle–Bowen measure} $\bbmu$ on $\mathbbl{\Lambda}$.
\subenumerationisfinis 
\item A twinned definition can be adopted for a $\mathscr{C}^2$ uniformly hyperbolic (Anosov) flow $\{\varphi_t\}_{t \in \mathbb{R}} \colon \mathbb{R} \to \mathbb{R}$ and an attractor 
\begin{equation}
	\mathbbl{\Lambda} \viz \mathbbl{\Lambda}_\topological = \bigcap_{t \geqslant 0}\varphi_t(\mathcal{W}^\mathrm{s}). 
\end{equation}
The Sinai–Ruelle–Bowen measure is such that
\begin{equation}
	\lim_{t \to \infty}\frac{1}{t}\int^t_0\tau\bigl(\varphi_t(x)\bigr)dt = \int_\mathbbl{\Lambda}\tau{d\bbmu},
\end{equation}
for $t \geqslant t_0, \varphi_\bbmu \colon \mathbbl{\Lambda} \to \mathbb{R}$. \definitiosymbol
\enumerationisfinis
\end{definitio}

\subsection{Sinai–Ruelle–Bowen Measure: Singular Hyperbolic Attractor}

The singular hyperbolic attractor ($\mathbbl{\Lambda}_\singularity$), of which the Lorenz-like attractor is the primary example, is a \emph{non-uniformly hyperbolic set}, because of the singularity (see point \ref{item "Uniformly hyperbolic set"} of Theorem \ref{theorema "MPP"}). We can even say that the hyperbolicity here is discontinuously uniform, in the sense that the origin $\singularity = (0, 0, 0)$ of $\mathbbl{\Lambda}_\singularity$ is the point where the uniformity breaks down. But this will not imply the absence of a Sinai–Ruelle–Bowen measure for it. 

We mention some contributions. Ya.B. Pesin \cite{Pesin "Dynamical systems with generalized hyperbolic attractors: hyperbolic ergodic and topological properties"} finds an analog of the \textsc{srb} measure for hyperbolic attractors and investigates on the ergodic properties of this measure. W. Tucker \cite{Tucker "The Lorenz attractor exists"} \cite{Tucker "A Rigorous ODE Solver and Smale's 14th Problem} explicitly demonstrates that Lorenz-like flows admit a unique finite \textsc{srb} measure. J.F. Alves, C. Bonatti and M. Viana \cite{Alves Bonatti Viana "SRB measures for partially hyperbolic systems whose central direction is mostly expanding"} build a \textsc{srb} measure for partially hyperbolic sets in which the tangent bundle splits into two invariant subbundles, and one of these is uniformly contracting, whereas the other is non-uniformly expanding. R.J. Metzger \cite{Metzger "Sinai-Ruelle-Bowen measures for contracting Lorenz maps and flows"} constructs a \textsc{srb} measure for contracting Lorenz-like flows with eigenvalues at the singularity satisfying $\lambda_1 + \lambda_3 < 0$. See also \cite{Luzzatto Melbourne Paccaut "The Lorenz Attractor is Mixing"}. 

\section{Margo. What is the Origin of the \emph{Complex Systems}?}

\begingroup
\footnotesize
Quantum mechanics has taught us to see in the exponential law of radioactive transformations an elementary law which cannot be reduced to a simpler causal mechanism. Of course the statistical laws known in classical mechanics and concerning \emph{complex systems}, retain their validity according to quantum mechanics. On the other hand, [quantum mechanics] modifies the rules for the determination of internal configurations in two different ways, depending on the nature of the physical systems, giving rise respectively to the statistical theories of Bose[–]Einstein, or Fermi. But the introduction into physics of a new kind of statistical law, or rather simply [a] probabilistic [law], which is hidden, instead of the supposed determinism, under the ordinary statistical laws[,] obliges us to reconsider the foundations of the analogy with the above-stated statistical social laws.\endnote{
	Original It. version: «La meccanica quantistica ci ha insegnato a vedere nella legge esponenziale delle trasformazioni radioattive una legge elementare non riducibile ad un più semplice meccanismo causale. Naturalmente anche le leggi statistiche note alla meccanica classica e riguardanti \emph{sistemi complessi}, conservano la loro validità secondo la meccanica quantistica. Questa modifica peraltro le regole per la determinazione delle configurazioni interne, e in due modi diversi, a seconda della natura dei sistemi fisici, dando luogo rispettivamente alle teorie statistiche di Bose-Einstein, o di Fermi. Ma l'introduzione nella fisica di un nuovo tipo di legge statistica, o meglio semplicemente probabilistica, che si nasconde, in luogo del supposto determinismo, sotto le leggi statistiche ordinarie obbliga a rivedere le basi dell'analogia che abbiamo stabilita più sopra con le leggi statistiche sociali». 
	} \\ 
\indent — \textsc{E. Majorana} \cite[p. 66]{Majorana "Il valore delle leggi statistiche nella fisica e nelle scienze sociali"} 

\endgroup

\vspace{2mm}

For all we know, Majorana \cite{Majorana "Il valore delle leggi statistiche nella fisica e nelle scienze sociali"} was the first to conceive and understand the so-called \emph{complex systems}, starting from the use of the expression itself.

\vspace{10mm}

\setcounter{secnumdepth}{0}  
\section{References and Bibliographic Details}
\setcounter{secnumdepth}{3}
\markright{References and Bibliographic Details}

\begingroup
\footnotesize
\noindent Section \ref{subsection "Sensitivity to Initial Conditions, and Weak Predictability"}

\begin{indent paragraph: 15pt}
About the (Lorenz) strange attractors, see e.g. \cite{Plykin Sataev Shlyachkov "Strange Attractors"}.
\end{indent paragraph: 15pt}

\noindent Section \ref{subsection "Necessary and Sufficient Conditions for the Chaos, and More"}

\begin{indent paragraph: 15pt}
The interplay between mathematics and dynamics of biological systems in the works of V. Volterra is analyzed in \cite{Israel "Volterra D'Ancona e la biologia matematica"} \cite{Manfredi Micheli "Ecologia Matematica e Matematica delle popolazioni"}.	
\end{indent paragraph: 15pt}

\noindent Section \ref{subsection "Peano–Hilbert Curve"}

\begin{indent paragraph: 15pt}
For the Peano curve, see e.g. \cite[chap. 3]{Sagan "Space-Filling Curves"} \cite[sec. 2.5, and pp. 358, 365]{Peitgen Jurgens Saupe "Chaos and Fractals: New Frontiers of Science Second Edition"} \cite[chap. 5]{Cannon "Two-Dimensional Spaces II: Topology as Fluid Geometry"}.	
\end{indent paragraph: 15pt}

\noindent Section \ref{section "Strangeness—Fractal Dimension of the Orbit in the Phase Space Attractor: Grassberger–Procaccia Algorithm and Lyapunov–Kaplan–Yorke Dimension"}

\begin{indent paragraph: 15pt}
About the Hausdorff dimension and measure, see e.g. \cite[chap. 4]{Mattila "Geometry of Sets and Measures in Euclidean Spaces. Fractals and rectifiability"} \cite{Mattila "Fourier Analysis and Hausdorff Dimension"}.	
\end{indent paragraph: 15pt}

\noindent Section \ref{subsection "Mandelbrot Set"}

\begin{indent paragraph: 15pt}
For the Mandelbrot set, see e.g. \cite[sec. 3.2]{Dang Kauffman Sandin "Hypercomplex Iterations: Distance Estimation and Higher Dimensional Fractals"} \cite{Lei "Local properties of the Mandelbrot set at parabolic points"} \cite{McMullen "The Mandelbrot set is universal"}.
\end{indent paragraph: 15pt}

\noindent Section \ref{subsection "Singular Hyperbolic Lorenz-like Attractor"}

\begin{indent paragraph: 15pt}
An overview on the singular hyperbolicity is in \cite[chapp. 3, 5, 6]{Araujo Pacifico "Three-Dimensional Flows}.
\end{indent paragraph: 15pt}
	
\noindent Section \ref{subsection "Sinai–Ruelle–Bowen Measure: Uniformly Hyperbolic Attractor"}

\begin{indent paragraph: 15pt}
The \textsc{srb} measure makes its appearance on \cite{Eckmann and Ruelle "Ergodic theory of chaos and strange attractors"}; see e.g. \cite[sec. 1.3]{Bonatti Diaz Viana "Dynamics Beyond Uniform Hyperbolicity: A Global Geometric and Probabilistic Perspective"} \cite[chap. 7]{Ghys "The Lorenz Attractor a Paradigm for Chaos"} \cite[pp. 282-283]{Guckenheimer Holmes "Nonlinear Oscillations Dynamical Systems and Bifurcations of Vector Fields"} \cite{Pesin "Sinai's Work on Markov Partitions and SRB Measures"}. 	
\end{indent paragraph: 15pt}

\endgroup

\chapter{Randomness and Stochastic Systems}
\label{chapter "Randomness and Stochastic Systems"}

\begingroup
\footnotesize
It may well be that the universe itself is completely deterministic (though this depends on what the “true” laws of physics are, and also to some extent on certain ontological assumptions about reality), in which case randomness is simply a mathematical concept, modeled using such abstract mathematical objects as probability spaces. Nevertheless, the concept of \emph{pseudorandomness}—objects which “behave” randomly in various statistical senses—still makes sense in a purely deterministic setting. A typical example are the digits of $\pi = 3.14159\cdots$; this is a deterministic sequence of digits, but is widely believed to behave pseudorandomly in various precise senses (e.g. each digit should asymptotically appear 10\% of the time) [this is the Borel-normality conjecture \cite{Borel "Les probabilites denombrables et leurs applications arithmetiques"} for $\pi$]. If a deterministic system exhibits a sufficient amount of pseudorandomness, then random mathematical models (e.g. statistical mechanics) can yield accurate predictions of reality, even if the underlying physics of that reality has no randomness in it. \\
\indent — \textsc{T. Tao}\endnote{
	T. Tao, \textit{Comment}, 9 Oct., 2007, in \textit{Simons Lecture I: Structure and randomness in Fourier analysis and number theory} = \cite[pp. 155-164]{Tao "Structure and Randomness: pages from year one of a mathematical blog"}, from \textit{What's new \textnormal{[weblog]}. Updates on my research and expository papers, discussion of open problems, and other maths-related topics, by T. Tao}.
	}

\endgroup

\section{Pullback and Random Attractors}

The lack of long-term predictability in the chaotic dynamics, such as the Lorenz system, is the property causing the \emph{apparent} randomness of chaotic orbits. For this reason, a Lorenz type attractor, that is a strange (fractal) and chaotic set, \emph{seems} to be governed by stochastic equations of motion, including a (white) noise variable; but actually it is a subset of the phase space of a system, the Lorenzian, which is still deterministic, although it is irregular, or non-periodic, and non-linear. In this system the growing divergence between of neighboring orbits is not (more or less) proportional to the number of variables, while in a stochastic process,\footnote{
	The adjective “stochastic” comes from the Gr. \textgreek{στοχαστικός}, and means “able to aiming at”, “skilful in guessing”.
	}
for some random phenomenon, this occurs. 

A stochastic system is a system of \emph{non-Lorenzian (in the classical sense)} type. However, when a stochastic dynamic is explicitly derived from a Lorenzian model, it is usually called \emph{stochastic Lorenz system}. We remember that, in a classical Lorenz system \cite{Lorenz "Deterministic Nonperiodic Flow"} \cite{Lorenz "The Predictability of Hydrodynamic Flow"} \cite{Lorenz "The problem of deducing the climate from the governing equations"} \cite{Lorenz "The predictability of a flow which possesses many scales of motion"} \cite{Lorenz "Irregularity: a fundamental property of the atmosphere"}, the random element is present at the moment of perturbation, since the initial conditions, or the states of the system, can be chosen randomly; in a stochastic Lorenz system, the random element is increasingly pervasive. Below we will concentrate on the concept of pullback attractor and then on that of random attractor, even if historically the second concept precedes the first.

\subsection{Pullback Attractor: Process and Skew Product Flow} 
\label{subsection "Pullback Attractor: Process and Skew Product Flow"}

For the genesis and development of the \emph{pullback attractor}, the reference papers are: D.N. Cheban, P.E. Kloeden, and B. Schmalfuß \cite{Cheban Kloeden and Schmalfuss "Pullback Attractors in Dissipative Nonautonomous Differential Equations Under Discretization"} \cite{Cheban Kloeden and Schmalfuss "The Relationship between Pullback Forward and Global Attractors of Nonautonomous Dynamical Systems"}, Kloeden \cite{Kloeden "Pullback Attractors in Nonautonomous Difference Equations"}, Kloeden, H. Keller, and Schmalfuß \cite{Kloeden Keller and Schmalfuss "Towards a Theory of Random Numerical Dynamics"}; see also \cite{Kloeden Potzsche and Rasmussen "Discrete-Time Nonautonomous Dynamical Systems"}.

That kind of attractors are part of non-autonomous systems. To begin with, we give some basic definitions concerning the autonomous systems. 

\subsubsection{Non-autonomous Dynamical Systems}

\begin{definitio}[Autonomous dynamical system: continuity and discreteness]
\label{definitio "Autonomous dynamical system: continuity and discreteness"}
Let $(\mathcal{X}, \distance)$ be a metric space, and $\mathbb{T}$ the time set. More specifically, the set $\mathbb{T} = \mathbb{R} = \mathbb{R}_- \cup \{0\} \cup \mathbb{R}_+$ is known as \emph{two-sided continuous time}, and the set $\mathbb{T} = \mathbb{Z} = \{0, \pm1, \pm2, \pm3, \mathellipsis\}$ as \emph{two-sided discrete time}. Denote by $\varphi \viz \varphi_\mathbb{T}$ the related dynamical system.
\enumerationisinitium
\item A \emph{dynamical system} on $\mathcal{X}$ is a continuous function $\varphi \colon \mathbb{T} \times \mathcal{X} \to \mathcal{X}$, with 
\subenumerationisinitium
\item an initial value condition, $\varphi(0, x_0) = x_0$, for $x_0 \in \mathcal{X}$,
\item a group property, $\varphi(s + t, x_0) = \varphi\bigl(s, \varphi(t, x_0)\bigr)$, for $s, t \in \mathbb{T}$ and $x_0 \in \mathcal{X}$.
\subenumerationisfinis 
We have a continuous dynamical system if $\mathbb{T} = \mathbb{R}$, and a discrete dynamical system if $\mathbb{T} = \mathbb{Z}$.
\item 
\label{item "Semi-dynamical system"}
A \emph{semi-dynamical system} on $\mathcal{X}$ maintains the same dynamical definition, except for the time set: $\varphi \colon \mathbb{T}_* \times \mathcal{X} \to \mathcal{X}$, where $\mathbb{T}_* = \{0\} \cup \mathbb{T}_+ = \{t \in \mathbb{T} \mid t \geqslant 0\}$. \definitiosymbol
\enumerationisfinis
\end{definitio}

A non-autonomous dynamical system (\textsc{nas}) can be defined with two formulations, one called the “process”, the other the “skew product flow”. Let us find them.

\begin{definitio}[Process and skew product flow for a \textsc{nas}]
\label{definitio "Process and skew product flow for a NAS"}
~\enumerationisinitium
\item 
\label{item "Process formulation"}
A \emph{process}, also known as a \emph{2-parameter semigroup}, on $\mathcal{X}$ is a continuous mapping $(t, t_0, x_0) \mapsto \varphi(t, t_0, x_0) \in \mathcal{X}$, for $t, t_0 \in \mathbb{T}$ and $x_0 \in \mathcal{X}$, $t \geqslant t_0$, with 
\subenumerationisinitium
\item an initial value condition, $\varphi(t_0, t_0, x_0) = x_0$, for $t_0 \in \mathbb{T}$ and $x_0 \in \mathcal{X}$,
\item a 2-parameter semigroup property, $\varphi(t_2, t_0, x_0) = \varphi\bigl(t_2, t_1, \varphi(t_1, t_0, x_0)\bigr)$, for $t_0 \leqslant t_1 \leqslant t_2$ and $x_0 \in \mathcal{X}$. 
\subenumerationisfinis
\item Let $\mathcal{Q}$ be a base space, and $\vartheta \viz \vartheta_\mathbb{T} = \{\vartheta_t\}_{t \in \mathbb{T}}$ a dynamical system on $\mathcal{Q}$, that is a group of homeomorphisms under composition on $\mathcal{Q}$. A \emph{skew product flow} is an autonomous semi-dynamical system $\pi \viz \pi_\mathbb{T}$, corresponding to a product mapping $\pi \colon \mathbb{T}_* \times \mathcal{Q} \times \mathcal{X} \to \mathcal{Q} \times \mathcal{X}$ on the extended phase space $(\mathcal{X}\mathcal{Q})_\bbmu = \mathcal{Q} \times \mathcal{X}$, where the mapping is given by $\pi\bigl(t, (q, x)\bigr) = \bigl(\vartheta_t(q), \varphi(t, q, x)\bigr)$, for $q, x \in \mathcal{Q} \times \mathcal{X}$. \definitiosymbol
\enumerationisfinis
\end{definitio}

The skew product flow formulation of a \textsc{nas} emerges from a \emph{driving mechanism} of $\vartheta$ (which means that with it the temporal change of the vector field of the system is generated) and a \emph{cocycle mapping} of $\varphi$.

\begin{definitio}[Non-autonomous dynamical system via $\vartheta$-driving and $\varphi$-cocycle]
Take two metric spaces, $(\mathcal{X}, \distance_\mathcal{X})$ and $(\mathcal{Q}, \distance_\mathcal{Q})$, where $\mathcal{X}$ is the phase space and $\mathcal{Q}$ is the base space, and two dynamical systems, $\vartheta \viz \vartheta_\mathbb{T}$ and $\varphi \viz \varphi_\mathbb{T}$, with $\mathbb{T} = \mathbb{R}$ or $\mathbb{T} = \mathbb{Z}$. Consider the autonomous dynamical system $\vartheta$ as a driving mechanism. A \emph{non-autonomous dynamical system} $(\vartheta, \varphi)$ is the cocycle mapping $\varphi \colon \mathbb{T}_* \times \mathcal{Q} \times \mathcal{X} \to \mathcal{X}$ on $\mathcal{X}$ driven by $\vartheta$ on $\mathcal{Q}$. In particular, $\vartheta$ and $\varphi$ are such that 
\enumerationisinitium
\item $\vartheta_0(q) = q$, for $q \in \mathcal{Q}$, and $\varphi(0, q, x) = x$, for $(q, x) \in \mathcal{Q} \times \mathcal{X}$,
\item $\vartheta_{s + t} = \vartheta_s\bigl(\vartheta_t(q)\bigr)$, for $s, t \in \mathbb{T}$, and $\varphi(t + s, q, x) = \varphi\bigl(t, \vartheta_s(q), \varphi(s, q, x)\bigr)$, for $s, t \in \mathbb{T}_*$, $(q, x) \in \mathcal{Q} \times \mathcal{X}$,
\item $(t, q) \mapsto \vartheta_t(q)$ and $(t, q, x) \mapsto \varphi(t, q, x)$ are continuous. \definitiosymbol
\enumerationisfinis
\end{definitio}

\begin{definitio}[Non-autonomous dynamical set]
\label{definitio "Non-autonomous dynamical set"}
~\enumerationisinitium
\item Let $\check{\mathbbl{A}} = \{A_t\}_{t \in \mathbb{T}}$ denote a family of subsets of $\mathcal{X}$, and let $\mathbbl{A} \subset (\mathbb{T}\mathcal{X})_\bbmu = \mathbb{T} \times \mathcal{X}$ be a subset induced by $\check{\mathbbl{A}}$, where $A_t = \{x \in \mathcal{X}\mid (t, x) \in \mathbbl{A}\}$ is the \emph{$t$-fiber} of $\mathbbl{A}$, and $(\mathbb{T}\mathcal{X})_\bbmu = \mathbb{T} \times \mathcal{X}$ is the extended phase space caused by a process $\varphi$ on $\mathcal{X}$. The subset $\mathbbl{A} = \{(t, x) \mid x \in A_t\}$ of $(\mathbb{T}\mathcal{X})_\bbmu$ is what we call a \emph{non-autonomous set of a process}. 
\subenumerationisinitium
\item If the $t$-fiber of $\mathbbl{A}$ is compact, then $\mathbbl{A}$ is \emph{compact}. 
\item If $\varphi(t, t_0, A_{t_0}) = A_t$, for $t \geqslant t_0$, then $\mathbbl{A}$ is \emph{invariant}.
\subenumerationisfinis 
\item A similar description applies to a skew product flow $\pi = (\vartheta, \varphi)$, in which the \emph{$q$-fiber} of $\mathbbl{A}$, i.e. $A_q = \{x \in \mathcal{X}\mid (q, x) \in \mathbbl{A}\}$, for $q \in \mathcal{Q}$, and the extended phase space $(\mathcal{X}\mathcal{Q})_\bbmu = \mathcal{Q} \times \mathcal{X}$ appear. Therefore the subset $\mathbbl{A} \subset (\mathcal{X}\mathcal{Q})_\bbmu = \mathcal{Q} \times \mathcal{X}$ is consistent with a \emph{non-autonomous set of a skew product flow}. 
\subenumerationisinitium
\item If the $q$-fiber of $\mathbbl{A}$ is compact, then $\mathbbl{A}$ is \emph{compact}. 
\item If $\varphi(t, q, A_q) = A_{\vartheta_t(q)}$, for $t \geqslant 0$, then $\mathbbl{A}$ is \emph{invariant}. \definitiosymbol
\subenumerationisfinis 
\enumerationisfinis
\end{definitio}

So we move on to integrate these definitions with that of pullback attractor.

\subsubsection{Theorems on the Pullback Attractor for a Process and Skew Product Flow}

\begin{definitio}[Pullback and forward attractors for a process]
Take a process $\varphi$ on a metric space $(\mathcal{X}, \distance)$ into account (see \ref{item "Process formulation"} in Definition \ref{definitio "Process and skew product flow for a NAS"}). Let $\mathbbl{\Lambda}$ be a non-empty, compact and invariant non-autonomous set (as in Definition \ref{definitio "Non-autonomous dynamical set"}).
\enumerationisinitium
\item First we have to set down the \emph{non-autonomous pullback or forward attractivity} of $\mathbbl{\Lambda}$.\subenumerationisinitium
\item We say that $\mathbbl{\Lambda}$ is \emph{pullback attracting} if
\begin{equation}
	\lim_{t_0 \to -\infty}\distance_\mathcal{X}\Bigl(\varphi(t, t_0, x_0), \Lambda_t\Bigr) = 0,
\end{equation}
for $x_0 \in \mathcal{X}$ and $t \in \mathbb{T}$, while $\mathbbl{\Lambda}$ is \emph{forward attracting} (under the Lyapunov stability criteria) if 
\begin{equation}
	\lim_{t \to +\infty}\distance_\mathcal{X}\Bigl(\varphi(t, t_0, x_0), \Lambda_t\Bigr) = 0,
\end{equation}
for $x_0 \in \mathcal{X}$ and $t_0 \in \mathbb{T}$, where $\distance = \dist_\mathcal{X}$.
\item We thus specify what a pullback or forward attractor is. We say that $\mathbbl{\Lambda}$ is a \emph{pullback/forward attractor} of a process $\varphi$ if $\mathbbl{\Lambda}$ pullback/forward attracts bounded subsets of $\mathcal{X}$.
\subenumerationisfinis
\item Let us try using the concept of family, so the previous point can be made less ambiguous. A family $\mathbbl{\Lambda} = \{\Lambda_t\}_{t \in \mathbb{T}}$ of non-empty, compact and $\varphi$-invariant subsets of $\mathcal{X}$ is called a \emph{pullback attractor} for $\varphi$ with respect to another family $\mathbbl{E} = \{E_t\}_{t \in \mathbb{T}}$ of bounded subsets of $\mathcal{X}$ if
\subenumerationisinitium
\item $\displaystyle{\lim_{t_0 \to -\infty}\distance_\mathcal{X}\Bigl(\varphi(t, t_0, E_{t_0}), \Lambda_t\Bigr) = 0}$, for $t \in \mathbb{T}$,
\item $\varphi(t, t_0, \Lambda_{t_0}) = \Lambda_t$, with $t \geqslant t_0$. 
\subenumerationisfinis 
The family $\mathbbl{\Lambda}$ is called a \emph{forward attractor} for $\varphi$  with respect to $\mathbbl{E} = \{E_t\}_{t \in \mathbb{T}}$ if 
\subenumerationisinitium
\item $\displaystyle{\lim_{t \to +\infty}\distance_\mathcal{X}\Bigl(\varphi(t, t_0, E_{t_0}), \Lambda_t\Bigr) = 0}$, for $t_0 \in \mathbb{T}$,
\item $\varphi(t, t_0, \Lambda_{t_0}) = \Lambda_t$, with $t \geqslant t_0$. \definitiosymbol
\subenumerationisfinis 
\enumerationisfinis
\end{definitio}

\begin{definitio}[Pullback and forward attractors for a skew product flow]
Let $\pi = (\vartheta, \varphi)$ be a skew product flow on the extended phase space $(\mathcal{X}\mathcal{Q})_\bbmu = \mathcal{Q} \times \mathcal{X}$, and $E$ a non-empty bounded subset of $\mathcal{X}$. Denote by $\mathbbl{\Lambda} = \{\Lambda_q \subset \mathcal{X}\}$ a non-empty, compact and invariant non-autonomous set, such that $\varphi(t, q, \Lambda_q) = \Lambda\bigl(\vartheta_t(q)\bigr)$, for $t \geqslant 0$ and $q \in \mathcal{Q}$. Then $\mathbbl{\Lambda}$ is a \emph{pullback attractor} of $(\vartheta, \varphi)$ if the convergence
\begin{equation}
	\lim_{t \to \infty}\distance_\mathcal{X}\Bigl(\varphi\bigl(t, \vartheta_{-t}(q), E\bigr), \Lambda_q\Bigr) = 0
\end{equation}  
shall apply to any $E \subset \mathcal{X}$. The same goes for the \emph{forward attractor} of $(\vartheta, \varphi)$, imposing
\begin{equation}
	\lim_{t \to \infty}\distance_\mathcal{X}\Bigl(\varphi(t, q, E), \Lambda_{\vartheta_t(q)}\Bigr) = 0.	
\end{equation} 
\definitiosymbol
\end{definitio}

Before proceeding to the existence theorems of attractors (we will put the pullback case), we need to establish the meaning of the absorbing set.

\begin{definitio}[On the absorbing set]
Let $F \subset \mathcal{X}$ a non-empty compact subset.
\enumerationisinitium
\item  We call $F$ an \emph{absorbing set} of a (semi-)dynamical system $\varphi$ on $\mathcal{X}$ if there is a time $\tau_E \viz \tau(E) \in \mathbb{T}$ (or $\mathbb{T}_+$, in the case in which the function is $\varphi \colon \mathbb{T}_* \times \mathcal{X} \to \mathcal{X}$, see \ref{item "Semi-dynamical system"} in Definition \ref{definitio "Autonomous dynamical system: continuity and discreteness"}), for any bounded subset $E \subset \mathcal{X}$, so that $\varphi(t, E) \subset F$, for $t \geqslant \tau$. Furthermore, $F$ is also an attracting set if $\varphi(t, F) \subseteqq F$, i.e. if $F$ is positively invariant.
\item  We call $F$ a \emph{pullback absorbing set of a process} $\varphi$ if there is a time $\tau_E(t) \viz \tau(t, E) > 0$, for any bounded subset $E \subset \mathcal{X}$ and $t \in \mathbb{T}$, so that $\varphi(t, t_0, E) \subset F$, for $t_0 \in \mathbb{T}$, with $t_0 \leqslant t - \tau$. 
\item We call $F$ a \emph{pullback absorbing set of a skew product flow} $(\vartheta, \varphi)$ if there is a time $\tau_E(q) \viz \tau(q, E) > 0$, for any bounded subset $E \subset \mathcal{X}$ and $q \in \mathcal{Q}$, so that $\varphi(t, \vartheta_{-t}(q), E) \subset F$, for $t \geqslant \tau$. \definitiosymbol
\enumerationisfinis
\end{definitio}

\begin{theorema}[Pullback attractor for a process]
Let $\varphi$ be a process on a metric space $(\mathcal{X}, \distance)$, and $F$ a compact set which is pullback absorbing for $\varphi$, so that $\varphi(t, t_0, F) \subset F$, for $t \geqslant t_0$. Then there is a pullback attractor $\mathbbl{\Lambda}$ with $F$-fibers uniquely defined by
\begin{equation}
	\Lambda_t = \bigcap_{\tau \geqslant 0}\overline{\bigcup_{t_0 \leqslant -\tau}\varphi(t, t_0, F)}, \text{ for } t, \tau \in \mathbb{T}.
\end{equation}
\end{theorema}

\begin{proof}
We refer to the proof of Theorem \ref{theorema "Pullback attractor for a skew product flow"}, because it works here, too.	
\end{proof}

\begin{theorema}[Pullback attractor for a skew product flow]
\label{theorema "Pullback attractor for a skew product flow"}
Let $(\vartheta, \varphi)$ be a skew product flow on a metric space $(\mathcal{X}, \distance)$, and $F$ a compact set which is pullback absorbing for $(\vartheta, \varphi)$, so that $\varphi(t, q, F) \subset F$, for $t \geqslant 0$ and $q \in \mathcal{Q}$. Then follow two results. 
\enumerationisinitium
\item There is a pullback attractor $\mathbbl{\Lambda}$ with $F$-fibers uniquely defined by
\begin{equation}
	\Lambda_q = \bigcap_{\tau \geqslant 0}\overline{\bigcup_{t \geqslant \tau}\varphi(t, \vartheta_{-t}(q), F)},
\end{equation}
for $t, \tau \in \mathbb{T}$ and $q \in \mathcal{Q}$.
\item Given a \emph{compact} metric space $(\mathcal{Q}, \distance_\mathcal{Q})$, the limit superior is 
\begin{equation}
\label{equation "Limit superior in pullback attractor for a skew product flow"}
	\lim_{t \to \infty}\sup_{q \in \mathcal{Q}}\distance\left(\varphi(t, q, E), \overline{\bigcup_{q \in \mathcal{Q}}\Lambda_q} \subset F\right) = 0
\end{equation} 
for any bounded subset $E \subset \mathcal{X}$. 
\enumerationisfinis
\end{theorema}

\begin{proof}
This demonstration is divided into several parts, and embraces the Kloeden–Rasmussen's way \cite[chap. 3, sec. 3]{Kloeden Rasmussen "Nonautonomous Dynamical Systems"}.
\enumerationisinitium
\item The first step to do is to prove that $\lim_{t \to \infty}\distance\bigl[\varphi\bigl(t, \vartheta_{-t}(q), F\bigr)\Lambda_q\bigr] = 0$. Suppose, by contradiction, we have $t_k \to \infty$ and $x_k \in \varphi\bigl(t_k, \vartheta_{-t_k}(q), F\bigr) \subset F$, whereby $\distance(x_k, \Lambda_q) > \varepsilon$, for $k \in \mathbb{N}$ and a certain value $\varepsilon$. Indicating by $\dot{k} \to \infty$ the subsequence, with $x_{\dot{k}} \to x_0$, we write $x_{\dot{k}} \in \bigcup_{t \geqslant \tau}\varphi\bigl(t, \vartheta_{-t}(q), F\bigr)$, for $\tau \geqslant 0$, with $t_{\dot{k}} \geqslant \tau$, from which
\begin{equation}
	x_0 \in \overline{\bigcup_{t \geqslant \tau}\varphi\bigl(t, \vartheta_{-t}(q), F\bigr)},
\end{equation} 
for $\tau \geqslant 0$, and thence $x_0 \in \Lambda_q$, which contradicts the starting assumption. 
\item By $\varphi(t, q, F) \subset F$ we get
\subenumerationisinitium
\item $G_\tau(q) = \bigcup_{s \geqslant \tau}\varphi\bigl(s, \vartheta_{-s}(q), F\bigr)$ in $F$, for $\tau \geqslant 0$,
\item $\Lambda_{\vartheta_{-t(q)}} = \bigcap_{\tau \geqslant 0}\overline{G_\tau\bigl(\vartheta_{-t}(q)\bigr)}$.
\subenumerationisfinis
Let $x^\tau \in \overline{G_\tau\bigl(\vartheta_{-t}(q)\bigr)} \subset F$, and $x = \varphi\bigl(t, \vartheta_{-t}(q), x^\tau\bigr)$, for $\tau \geqslant 0$. Suppose the point $x^\epsilon$ is a limit point of the set $\{x^\tau \mid \tau \geqslant 0\}$, and $x^\epsilon$ belongs to $\bigcap_{\tau \geqslant 0}\overline{G_\tau\bigl(\vartheta_{-t}(q)\bigr)}$. From the continuity condition of $\varphi\bigl(t, \vartheta_{-t}(q), \cdot\bigr)$ it is clear that $x = \varphi\bigl(t, \vartheta_{-t}(q), x^\epsilon\bigr)$ and $x \in \varphi\bigl[t, \vartheta_{-t}(q), \bigcap_{\tau \geqslant 0}\overline{G_\tau\bigl(\vartheta_{-t}(q)\bigr)}\bigr] = \varphi\bigl(t, \vartheta_{-t}(q), \Lambda_{\vartheta_{-t(q)}}\bigr)$. The triplet provided by
\subenumerationisinitium
\item $\displaystyle{\varphi\left(t, \vartheta_{-t}(q), \bigcap_{\tau \geqslant 0}\overline{G_\tau\bigl(\vartheta_{-t}(q)\bigr)}\right) = \bigcap_{\tau \geqslant 0}\varphi\left(t, \vartheta_{-t}(q), \overline{G_\tau\bigl(\vartheta_{-t}(q)\bigr)}\right)}$, 
\item $\overline{G_\tau\bigl(\vartheta_{-t}(q)\bigr)}$, with its compactness, 
\item and $\varphi\bigl(t, \vartheta_{-t}(q), \cdot\bigr)$, with its continuity,
\subenumerationisfinis
furnishes
\begin{align}
	\varphi(\cdot, \cdot, \cdot)_t & = \bigcap_{\tau \geqslant 0}\varphi\left(t, \vartheta_{-t}(q), \overline{G_\tau\bigl(\vartheta_{-t}(q)\bigr)}\right) \supset \bigcap_{\tau \geqslant 0}\overline{\varphi\Bigl(t, \vartheta_{-t}(q), G_\tau\bigl(\vartheta_{-t}(q)\bigr)\Bigr)} \notag \\
	& = \bigcap_{\tau \geqslant 0}\overline{\bigcup_{s \geqslant \tau}\varphi\bigl(t, \vartheta_{-t}(q)\bigr), \varphi\bigl(s, \vartheta_{-t-s}(q), F\bigr)} \notag \\
	& = \bigcap_{\tau \geqslant 0}\overline{\bigcup_{s \geqslant \tau}\varphi\bigl(t + s, \vartheta_{-t-s}(q), F\bigr)} = \bigcap_{\tau \geqslant t}\overline{\bigcup_{s \geqslant \tau}\varphi\bigl(s, \vartheta_{-s}(q), F\bigr)} \supset \Lambda_q,
\end{align}
where $\varphi(\cdot, \cdot, \cdot)_t = \varphi\bigl(t, \vartheta_{-t}(q), \Lambda_{\vartheta_{-t(q)}}\bigr)$, hence $\Lambda_q \subset \varphi\bigl(t, \vartheta_{-t}(q), \Lambda_{\vartheta_{-t(q)}}\bigr)$, for $t \geqslant 0$ and $q \in \mathcal{Q}$. Let $\Upsilon_{\Lambda}$ a neighborhood of $\Lambda_q$. By the cocycle property $x(s + t, q_0, x_0) = x\bigl(s, q(t, q_0), x(t, q_0, x_0)\bigr)$,  $s, t \geqslant 0$, and inserting $\vartheta_{-\tau}(q)$, we obtain 
\begin{align}
	\varphi(\cdot, \cdot, \cdot)_\tau & \subset \varphi\Bigl(\tau, \vartheta_{-\tau}(q), \varphi\bigl(t, \vartheta_{-\tau-t}(q), \Lambda_{\vartheta_{-\tau-t(q)}}\bigr)\Bigl) \notag \\
	& = \varphi\Bigl(t, \vartheta_{-t}(q), \varphi\bigl(\tau, \vartheta_{-\tau-t}(q), \Lambda_{\vartheta_{-\tau-t(q)}}\bigr)\Bigl) \notag \\
	& \subset \varphi\Bigl(t, \vartheta_{-t}(q), \varphi\bigl(\tau, \vartheta_{-\tau-t}(q), F\bigr)\Bigl) \subset \varphi\bigl(t, \vartheta_{-t}(q), F\bigr) \subset \Upsilon_{\Lambda},
\end{align}
where $\varphi(\cdot, \cdot, \cdot)_\tau = \varphi\bigl(\tau, \vartheta_{-\tau}(q), \Lambda_{\vartheta_{-\tau(q)}}\bigr)$. Finally, $\varphi\bigl(\tau, \vartheta_{-\tau}(q), \Lambda_{\vartheta_{-\tau(q)}}\bigr) \subset \Lambda_q$, for $\tau \geqslant 0$ and $q \in \mathcal{Q}$. To get the $\varphi$-invariance of the non-autonomous set $\mathbbl{\Lambda} = \{\Lambda_q\}$, it will be sufficient to put $\tau = t$. We note that $\Lambda_q \subset F$, so it is uniformly bounded, and we deal with a pullback attractor.
\item Given a positive value $\varepsilon > 0$, a sequence $t_n \to \infty$, $q^\epsilon_n \in \mathcal{Q}$, and $x_n \in F$, whereby 
\begin{equation}
	\distance\left(\varphi(t_n, q^\epsilon_n, x_n), \overline{\bigcup_{q \in \mathcal{Q}}\Lambda_q} \subset F\right) > \varepsilon. 
\end{equation} 
Letting $q_n = \vartheta_{t_n}(q^\epsilon_n)$, we outline a subsequence $q_{\dot{n}} \to q_0 \in \mathcal{Q}$, and a distance $\distance\bigl[\varphi\bigl(\tau, \vartheta_{-\tau}(q_0), F\bigr)\Lambda_{q_0}\bigl] < \frac{\varepsilon}{2}$. From the cocycle property it follows that $\varphi\bigl(t_n, \vartheta_{-t_n}(q_n), x_n\bigr) = \varphi\bigl[\tau, \vartheta_{-\tau}(q_n), \varphi\bigl(t_n - \tau, \vartheta_{-t_n}(q_n), x_n\bigr)\bigr]$, for $t_n > \tau$. Since $F$ is
\subenumerationisinitium
\item a positively invariant set, one has $\varphi\bigl(t_n - \tau, \vartheta_{-t_n}(q_n), x_n\bigr) \subset F$, 
\item a compact set, one has a subsequence $\ddot{n}$ of $\dot{n}$ on $\tau$, i.e. $s_{\ddot{n}} = \varphi\bigl(t_{\ddot{n}} - \tau, \vartheta_{-t_{\ddot{n}}}(q_{\ddot{n}}), x_{\ddot{n}}\bigr) \to s_0 \in F$.
\subenumerationisfinis  
Related to the continuity condition of $(\vartheta, \varphi)$ there is the requirement that $\|\varphi\bigl(\tau, \vartheta_{-\tau}(q_{\ddot{n}}), s_{\ddot{n}}\bigr) - \varphi\bigl(\tau, \vartheta_{-\tau}(q_0), s_0\bigr)\| < \frac{\varepsilon}{2}$, if $\ddot{n} > n(\varepsilon)$, from which 
\begin{align}
	\varepsilon > \distance\bigl[\varphi\bigl(t_{\ddot{n}}, \vartheta_{-t_{\ddot{n}}}(q), x_{\ddot{n}}\bigr), \Lambda_{q_0}\bigr] & = \distance\bigl[\varphi\bigl(t_{\ddot{n}}, q^\epsilon_{\ddot{n}}, x_{\ddot{n}}\bigr), \Lambda_{q_0}\bigr] \notag \\
	& \geqslant \distance\left[\varphi\bigl(t_{\ddot{n}}, q^\epsilon_{\ddot{n}}, x_{\ddot{n}}\bigr), \overline{\bigcup_{q \in \mathcal{Q}}\Lambda_q} \subset F\right].
\end{align} 
This is in contradiction with the previous inequality $\distance\bigl(\varphi(t_n, q^\epsilon_n, x_n), \overline{\bigcup_{q \in \mathcal{Q}}\Lambda_q})\bigr) > \varepsilon$, and thereby \eqref{equation "Limit superior in pullback attractor for a skew product flow"} is true.
\enumerationisfinis
\end{proof}

\subsection{Random Attractor} 
\label{subsection "Random Attractor"} 

We shall treat here the concept of attractor in a random dynamical system (\textsc{rds}),\footnote{
	 Variant labelling: stochastic dynamical system.
	} 
the so-called \emph{random attractor}. Before writing the  rigorous Definition (\ref{definitio "Random dynamical system"}), it is sufficient to report that a random dynamical system is a non-autonomous dynamical system (Section \ref{subsection "Pullback Attractor: Process and Skew Product Flow"}), which is thought as a skew product flow in relation to a probability space (see Margo \ref{margo "On the RDS and random attractor"}). Below are some published studies with a focus on it.
\enumerationisinitium
\item F. Ledrappier and L.-S. Young \cite{Ledrappier and Young "Entropy Formula for Random Transformations"} exhibit a strange attractor with a random nature and a random version of the Sinai–Ruelle–Bowen. What emerges is a stochastic framework related to the diffeomorphism considered. 
\item H. Morimoto \cite{Morimoto "Attractors of probability measures for semilinear stochastic evolution equations"} looks at the probability measures on the attractor and the semi-linear stochastic evolution equations with chaotic solutions.
\item H. Crauel and F. Flandoli \cite{Crauel Flandoli "Attractors for random dynamical systems"} establish a guide to detect an attractor for random dynamical systems with parabolic structure that supports a Markov invariant measure. It is demonstrated that a stochastic flow related to the reaction-diffusion equation with additive (white) noise, as well as a stochastic flow related to the Navier–Stokes equation with multiplicative (white) noise, both have an attractor of this kind, that is a compact random invariant set. Here is introduced the notion of $\invertedbreve{\Omega}$-limit set for \textsc{rds}, cf. Eq. \eqref{equation "Omega-limit set"}. 
\item In Crauel \cite{Crauel "Global Random Attractors are Uniquely Determined by Attracting Deterministic Compact Sets"} it will be shown that a random attractor coincides with the $\invertedbreve{\Omega}$-limit set of a compact deterministic (non-random) set with probability arbitrarily close to 1.
\item In Flandoli and B. Schmalfuß \cite{Flandoli and Schmalfuss "Random attractors for the 3D stochastic Navier-Stokes equation with multiplicative white noise"} the existence of a random attractor is proven for the stochastic $3\mathrm{D}$ Navier–Stokes equation. See also Schmalfuß \cite{Schmalfuss "The random attractor of the stochastic Lorenz system"} \cite{Schmalfuss "Measure attractors and random attractors for stochastic partial differential equations"}.
\item In Crauel, A. Debussche and Flandoli \cite{Crauel Debussche and Flandoli "Random Attractors"} the above applications are extended to systems with hyperbolic structure, such as the non-linear random wave equation.
\item In M.D. Chekroun, E. Simonnet and M. Ghil \cite{Chekroun Simonnet Ghil "Stochastic climate dynamics: Random attractors and time-dependent invariant measures"} the geometry of a (global) random attractor for \textsc{rds} under the influence of non-linear and stochastic parameters is taken into consideration, by resorting to models both with a stochastic forcing of the classical Lorenz model, and with a random version of the Sinai–Ruelle–Bowen measure.
\enumerationisfinis

Let us get down to the specifics.

\subsubsection{Random (Stochastic) Dynamical System}

\begin{definitio}[Random dynamical system]
\label{definitio "Random dynamical system"} 
Let $(\mathcal{X}, \distance)$ be a complete metric space, and $(\invertedbreve{\Omega}, \mathscr{B}_\sigma, \bbmu)$ a probability space, with a Borel $\sigma$-algebra on $\invertedbreve{\Omega}$, where $\invertedbreve{\Omega}$ is a non-empty set and $\bbmu$ is a  probability measure. Choose two dynamical systems, $\vartheta \viz \vartheta_\mathbb{T}$ and $\varphi \viz \varphi_\mathbb{T}$, with $\mathbb{T} = \mathbb{R}$ or $\mathbb{T} = \mathbb{Z}$. A \emph{random dynamical system} $(\vartheta, \varphi)$ on $\invertedbreve{\Omega} \times \mathcal{X}$ is a metric dynamical system $\vartheta$ on $\invertedbreve{\Omega}$, that is a measure preserving transformations $\{\vartheta_t\}_{t \in \mathbb{T}} \colon \invertedbreve{\Omega} \to \invertedbreve{\Omega}$, for which
\enumerationisinitium
\item $\vartheta_t \circ \vartheta_s = \vartheta_{t + s}$, $t, s \in \mathbb{T}$, and $\vartheta_0\invertedbreve{\omega} = \id_{\invertedbreve{\omega}}$, for $\invertedbreve{\omega} \in \invertedbreve{\Omega}$, 
\item $(t, \invertedbreve{\omega}) \mapsto \vartheta_t(\invertedbreve{\omega})$ is measurable and $\bbmu$-invariant, i.e. $\bbmu\bigl(\vartheta_t(\Lambda)\bigr) = \bbmu(\Lambda)$, for $\Lambda \in \mathscr{B}_\sigma$,
\item the cocycle mapping $\varphi \colon \mathbb{T}_* = \{0\} \cup \mathbb{T}_+ \times \invertedbreve{\Omega} \times \mathcal{X} \to \mathcal{X}$ entails that
\subenumerationisinitium
\item $\varphi(0, \invertedbreve{\omega}, x_0) = x_0$, for $\invertedbreve{\omega} \in \invertedbreve{\Omega}$ and $x_0 \in \mathcal{X}$,
\item $\varphi(s + t, \invertedbreve{\omega}, x_0) = \varphi\bigl(s, \vartheta_t(\invertedbreve{\omega}), \varphi(t, \invertedbreve{\omega}, x_0)\bigr)$, for $s, t \in \mathbb{T}_*$, $\invertedbreve{\omega} \in \invertedbreve{\Omega}$ and $x_0 \in \mathcal{X}$,
\item $\invertedbreve{\omega} \mapsto \varphi(t, \invertedbreve{\omega}, x_0)$ is $\mathscr{B}_\sigma$-measurable and $(t, x_0) \mapsto \varphi(t, \invertedbreve{\omega}, x_0)$ is a continuous map. \definitiosymbol
\subenumerationisfinis 
\enumerationisfinis
\end{definitio}

\begin{margo}
\label{margo "On the RDS and random attractor"}
Note that $(\vartheta, \varphi)$, which represents a random dynamical system here, is the same notation that was used for the skew product flow, since both systems have crossed properties; and in fact a \emph{random attractor} (an attractor of a \textsc{rds}) \emph{is a pullback attractor formulated as a random set}, by making use of the probability space. \margosymbol
\end{margo}

\begin{definitio}[Random set]
Let $(\mathcal{X}, \distance)$ be a complete and separable metric space, and $(\invertedbreve{\Omega}, \mathscr{B}_\sigma, \bbmu)$ a probability space. A \emph{random set} can be described either as 
\enumerationisinitium
\item a measurable subset $\mathbbl{E} \subset \invertedbreve{\Omega} \times \mathcal{X}$ with respect to the product $\sigma$-algebra on $\invertedbreve{\Omega} \times \mathcal{X}$,
\item or as a family $\mathbbl{E} = \{E_{\invertedbreve{\omega}}\}_{\invertedbreve{\omega} \in \invertedbreve{\Omega}}$ of subsets of $\mathcal{X}$, provided that the mapping $\invertedbreve{\omega} \mapsto \distance(x, E_{\invertedbreve{\omega}})$ is $\mathscr{B}_\sigma$-measurable, for $\invertedbreve{\omega} \in \invertedbreve{\Omega}$ and $x \in \mathcal{X}$. A random set $\mathbbl{E}$ is
\subenumerationisinitium
\item \emph{closed} if $E_{\invertedbreve{\omega}}$ acts as a closed $\invertedbreve{\omega}$-fiber, for $\invertedbreve{\omega} \in \invertedbreve{\Omega}$, 
\item \emph{compact} if $E_{\invertedbreve{\omega}}$ acts as a compact $\invertedbreve{\omega}$-fiber, for $\invertedbreve{\omega} \in \invertedbreve{\Omega}$,
\item \emph{tempered} if $E_{\invertedbreve{\omega}} \subset \{x \in \mathcal{X} \mid \distance(x, x_0) \leqslant f(\invertedbreve{\omega})\}$, for $x_0 \in \mathcal{X}$, putting a random variable $f(\invertedbreve{\omega})$. \definitiosymbol
\subenumerationisfinis
\enumerationisfinis
\end{definitio}

The set of all $\invertedbreve{\omega}$-limit points of $\mathbbl{E}$ is called \emph{$\invertedbreve{\omega}$-limit set} of $\mathbbl{E}$ and defined by 
\begin{equation}
\label{equation "Omega-limit set"}
	\invertedbreve{\Omega}_\mathbbl{E}(\invertedbreve{\omega}) = \bigcap_{\tau \geqslant 0}\overline{\bigcup_{t \geqslant \tau}\varphi\bigl(t, \vartheta_{-t}(\invertedbreve{\omega}), F_{\vartheta_{-t}(\invertedbreve{\omega})}\bigr)}.
\end{equation}

\subsubsection{On the Random Sets (a Definition and a Theorem)}

\begin{definitio}[Random attractor]
Let $U_{\mathbbl{E}} \in \mathcal{X}$ be a \emph{universe} of tempered random sets. A random compact set $\mathbbl{\Lambda} = \{\Lambda_{\invertedbreve{\omega}}\}_{\invertedbreve{\omega} \in \invertedbreve{\Omega}}$ from $U_{\mathbbl{E}}$ is a \emph{random attractor} of a random dynamical system $(\vartheta, \varphi)$ on $\invertedbreve{\Omega} \times \mathcal{X}$ in $\mathbbl{E}$ if 
\enumerationisinitium
\item $\varphi(t, \invertedbreve{\omega}, \mathbbl{\Lambda}_{\invertedbreve{\omega}}) = \mathbbl{\Lambda}_{\vartheta_t(\invertedbreve{\omega})}$, for $t \geqslant 0$ and $\invertedbreve{\omega} \in \invertedbreve{\Omega}$, so $\mathbbl{\Lambda}$ is strictly $\varphi$-invariant,
\item $\displaystyle{\lim_{t \to \infty}\distance\bigl[\varphi\bigl(t, \vartheta_{-t}(\invertedbreve{\omega}), E_{\vartheta_{-t}(\invertedbreve{\omega})}\bigr), \Lambda_{\invertedbreve{\omega}}\bigr] = 0}$, for $\invertedbreve{\omega} \in \invertedbreve{\Omega}$ and $\mathbbl{E} \in U_\mathbbl{E}$. \definitiosymbol
\enumerationisfinis
\end{definitio}

\begin{theorema}[On the random attractor]
\label{theorema "On the random attractor"}
Take 
\enumerationisinitium
\item a random dynamical system $(\vartheta, \varphi)$ on $\invertedbreve{\Omega} \times \mathcal{X}$, with a compact mapping $\mathcal{X} \xrightarrow{\varphi(t, \invertedbreve{\omega}, \cdot)} \mathcal{X}$, for $t > 0$ and $\invertedbreve{\omega} \in \invertedbreve{\Omega}$, 
\item a tempered random closed and bounded set $\mathbbl{F} = \{F_{\invertedbreve{\omega}}\}_{\invertedbreve{\omega} \in \invertedbreve{\Omega}}$, that coincides with a random type pullback absorbing set,
\item a value $\tau(\mathbbl{E}, \invertedbreve{\omega}) \geqslant 0$, where $\mathbbl{E}$ is a tempered random set, 
\enumerationisfinis 
for which $\varphi\bigl(t, \vartheta_{-t}(\invertedbreve{\omega}), E_{\vartheta_{-t}(\invertedbreve{\omega})}\bigr) \subset F_{\invertedbreve{\omega}}$ with respect to 
\begin{equation}
	\mathbbl{E} = \{E_{\invertedbreve{\omega}}\}_{\invertedbreve{\omega} \in \invertedbreve{\Omega}}, 
\end{equation}
for $t \geqslant \tau(\mathbbl{E}, \invertedbreve{\omega})$. Then there exists a random (pullback) attractor 
\begin{equation}
	\mathbbl{\Lambda} = \{\Lambda_{\invertedbreve{\omega}}\}_{\invertedbreve{\omega} \in \invertedbreve{\Omega}}, 
\end{equation}
for $(\vartheta, \varphi)$, determined by
\begin{equation}
	\Lambda_{\invertedbreve{\omega}} = \bigcap_{s > 0}\overline{\bigcup_{t \geqslant s}\varphi\bigl(t, \vartheta_{-t}(\invertedbreve{\omega}), F_{\vartheta_{-t}(\invertedbreve{\omega})}\bigr)}.
\end{equation}
\end{theorema}

\begin{proof}
Here again, the demonstration can be carried out by resuming the guidelines in the proof of Theorem \ref{theorema "Pullback attractor for a skew product flow"}.
\end{proof}

\section{Stochastic Systems: Itô and Stratonovich Calculi}
\label{section "Stochastic Systems: Itô and Stratonovich Calculi"}

\begingroup
\footnotesize
The moods of the waters of the river were always delightful to watch. To me, as a mathematician and a physicist, they had another meaning as well. How could one bring to a mathematical regularity the study of the mass of ever shifting ripples and waves, for was not the highest destiny of mathematics the discovery of order among disorder? [\,\dots] This problem of the waves was clearly one for averaging and statistics. \\
\indent — \textsc{N. Wiener} \cite[p. 33]{Wiener "I Am A Mathematician: The Later Life of a Prodigy"} 

\endgroup

\vspace{2mm}

A random dynamical system can be generated from a stochastic model; what this means is that a stochastic differential equation (\textsc{sde}), on how it developed in Itô \cite{Ito "Stochastic Integral"} \cite{Ito "On a Stochastic Integral Equation"} \cite{Ito "On a Formula Concerning Stochastic Differentials"} \cite{Ito "Stochastic Differential Equations in a Differentiable Manifold"} or Stratonovich \cite{Stratonovich "A New Representation for Stochastic Integrals and Equations"} calculus, can be transformed into a random differential equation; therefore, it is possible to build, in several cases, a random attractor from \textsc{sde}s. This rule applies also to the opposite operation, generating stochastic differential equations from a random dynamical system. (One of the substantial differences between the stochastic and random differential equations is that the first ones are treated with \emph{white noise}, while the others with \emph{real noise}).

Here one can understand that the words “stochastic” and “random” are very close, and are often easily exchanged. Let us give a few examples. 

\subsection{Navier–Stokes Equations in 2D with Noise, and Lorenz Equations with Wiener Process and Thermal Fluctuations}

\subsubsection{Example I. Stochastically Forced Navier–Stokes Equations with Additive and Multiplicative Noise}

Given a random set $\mathbbl{E}$ such that 
\[
	\invertedbreve{\omega} \mapsto E_{\invertedbreve{\omega}} 
\]
is compact, and another set $\mathbbl{L}$ on a complete and separable metric space $\mathcal{X}$, for which $\mathbbl{E}$ absorbs $\mathbbl{L}$, the set 
\begin{equation}
	\Lambda_{\invertedbreve{\omega}} = \overline{\bigcup_{\mathbbl{L} \subset \mathcal{X}}\invertedbreve{\Omega}_\mathbbl{L}(\invertedbreve{\omega})}
\end{equation}
is called \emph{stochastic attractor}, when its construction has to do with the stochastic flow associated e.g. with $2\mathrm{D}$ Navier–Stokes equations with additive \eqref{align "Equation with additive noise"} and multiplicative \eqref{align "Equation with multiplicative noise"} noise \cite{Crauel Flandoli "Attractors for random dynamical systems"},
\begin{align}
	\label{align "Equation with additive noise"}
	& dv = \bigl\{{A_\varphi}v + B_\varphi(v, v) + \textcyrillic{\textit{п}}\bigr\}dt + \sum^n_{\nu = 1}{(G_\varphi)_\nu}dw_\nu(t), \\
	\label{align "Equation with multiplicative noise"}
	& dv = \bigl\{{A_\varphi}v + B_\varphi(v, v) + \textcyrillic{\textit{п}}\bigr\}dt + \sum^n_{\nu = 1}{\beta_\nu}v \circ dw_\nu(t),
\end{align}
according to the Stratonovich interpretation, where 

· $A_\varphi \colon \mathfrak{D}(A_\varphi) \subset \mathfrak{H} \to \mathfrak{H}$ is a self-adjoint linear operator (a linear mapping that is equal to its own adjoint) in a real separable Hilbert space $\mathfrak{H}$, and $\mathfrak{D}(A_\varphi)$ is a linear subspace of $\mathfrak{H}$, that is the domain of $A_\varphi$,

· $B_\varphi \colon \vec{V} \times \vec{V} \to \mathbb{R}$ is a bilinear mapping, with 
\begin{equation}
	\vec{V} = \mathfrak{D}\left(-\tfrac{A_\varphi}{2}\right), \enspace \forall v \in \mathfrak{D}(A_\varphi),
\end{equation}

· $\textcyrillic{\textit{п}}$ is a polynomial function on $\mathfrak{H}$,

· $(G_\varphi)_\nu \colon \mathfrak{D} \to \mathbb{R}$, $1 \leqslant \nu \leqslant n$, $(g_\varphi)_1, \mathellipsis, (g_\varphi)_n \in \mathfrak{H}$,

· $\beta_1, \mathellipsis, \beta_n \in \mathbb{R}$.

\subsubsection{Example II. Stochastically Forced System's Lorenz Equations with Wiener Process and Time Dependent Thermal Fluctuations in Weather Forecasting}
\label{subsubsection "Example II. Stochastically Forced System's Lorenz Equations with Wiener Process and Time Dependent Thermal Fluctuations in Weather Forecasting"}

A further manner of seeing how a random attractor—or, more generally, the \textsc{rds} theory—is involved in the \textsc{sde}s lies in the stochastically forced system's Lorenz equations \eqref{equation "Lorenz equations"}. There follow two models of \emph{stochastic Lorenz system}.

\enumerationisinitium
\item That is one way \cite{Chekroun Simonnet Ghil "Stochastic climate dynamics: Random attractors and time-dependent invariant measures"}:
\begin{equation}
	\begin{cases}
	dx = (- \mathrm{Pr}x + \mathrm{Pr}y)dt + (\textgreek{\text{Θ}}x)d\mathsf{W}_t, \\
	dy = (- xz + \mathrm{Ra}x - y)dt + (\textgreek{\text{Θ}}y)d\mathsf{W}_t, \\
	dz = (xy - {\beta}z)dz + (\textgreek{\text{Θ}}z)d\mathsf{W}_t,
	\end{cases}
\end{equation}
in which three stochastic differential equations make their appearance, with a multiplicative noise, under the Itô criteria. The symbol $\textgreek{\text{Θ}} > 0$ designates the \emph{noise intensity}, and $\mathsf{W}_t$, with $t \geqslant 0$, the \emph{Wiener process} \cite{Paley Wiener and Zygmund "Notes on random functions"} \cite[chap. IX]{Paley and Wiener "Fourier Transforms in the Complex Domain"} \cite{Wiener "Nonlinear Problems in Random Theory"}, that is another form of Brownian motion, in terms of continuous time stochastic (or random) process.
\item Another result is available directly from Lorenz \cite{Lorenz "Irregularity: a fundamental property of the atmosphere"} \cite{Lorenz "Can chaos and intransitivity lead to interannual variability?}, by adding a time dependent process of the forcing terms in his weather prediction model,
\begin{equation}
	\begin{cases}
	\frac{dx}{dt} = - y^2 - z^2 - ax + a\delta_{\mathsf{T}_1}\bigl(\vartheta_t(\invertedbreve{\omega})\bigr), \\
	\frac{dy}{dt} = xy - bxz - y + \delta_{\mathsf{T}_2}\bigl(\vartheta_t(\invertedbreve{\omega})\bigr), \\
	\frac{dz}{dt} = bxy + xz - z,
	\end{cases}
\end{equation}
where 

· $x$ is a variable for the intensity or the strength of the westerly wind current and the equivalent poleward temperature gradient; $y$ and $z$ are the cosine and sine phases of a chain of superposed waves, i.e. eddies transporting heat poleward; $t$ is an independent variable representing time;

· $a$ is a coefficient indicating, if $a < 1$, the damping of the westerly wind current, which is less rapidly than the waves (eddies), and the coefficient $b$, if $b > 1$, indicates the translation of the wave pattern by the westerly wind current;

· $\delta_{\mathsf{T}_1}$ and $\delta_{\mathsf{T}_2}$ are random forcing terms, as for example the thermal fluctuations, and supposed to be constant; more precisely, $a\delta_{\mathsf{T}_1}$ is the symmetric thermal forcing and $\delta_{\mathsf{T}_2}$ the asymmetric one, and they vary in time under a dynamical system $\{\vartheta_t\}_{t \in \mathbb{T}}$, with $\mathbb{T} = \mathbb{R}$.
\enumerationisfinis

\subsection{Random Attractor and Stochastic Differential Equations Driven by Noise, plus Ornstein–Uhlenbeck Process} 

\subsubsection{Example I, with Additive Noise} 

The first example concerns the additive noise. We start from an \emph{Itô stochastic differential equation} with a Wiener process. Let $f_\mu \colon \mathbb{R} \to \mathbb{R}$ be a drift coefficient function, and write the \textsc{sde} in the paradigmatic form and in its integral version,
\begin{subequations}
\label{subequations "Itô stochastic differential and integral equations"}
\begin{align}
		& d\mathcal{X}_t = f_\mu(\mathcal{X}_t)dt + cd\mathsf{W}_t, \\
		& \mathcal{X}_t = \mathcal{X}_0 + \int^t_0f_\mu(\mathcal{X}_s)ds + cd\mathsf{W}_t,
\end{align}
\end{subequations}
with a constant $c > 0$. The solution paths in \eqref{subequations "Itô stochastic differential and integral equations"} are continuous but not differentiable, hence we introduce a difference $\mathcal{X}_t - \OrnsteinUhlenbeck_t = \mathcal{X}_0 - \OrnsteinUhlenbeck_0 + \int^t_0[f_\mu(\mathcal{X}_s) + \OrnsteinUhlenbeck_s]ds$, which has a pathwise differentiability, where 
\begin{equation}
\label{equation "Ornstein–Uhlenbeck process"}
	\OrnsteinUhlenbeck_t = ce^{-t}\int^t_{-\infty}e^sd\mathsf{W}_s
\end{equation}
is the \emph{Ornstein–Uhlenbeck process} \cite{Uhlenbeck and Ornstein "On the Theory of the Brownian Motion"}. The equation with the difference corresponds to $\frac{d}{dt}(\mathcal{X}_t - \OrnsteinUhlenbeck_t) = f_\mu(\mathcal{X}_t) + \OrnsteinUhlenbeck_t$. At this stage, denoting by $c_\textsc{l} > 0$ a constant, we use the \emph{Lipschitz condition}, so as to achieve 
\begin{align}
	& \frac{d}{dt}|\mathcal{X}_t - \OrnsteinUhlenbeck_t|^2 = 2\langle\mathcal{X}_t - \OrnsteinUhlenbeck_t, f_\mu(\mathcal{X}_t) - f_\mu(\OrnsteinUhlenbeck_t)\rangle + 2\langle\mathcal{X}_t - \OrnsteinUhlenbeck_t, f_\mu(\OrnsteinUhlenbeck_t) + \OrnsteinUhlenbeck_t\rangle \notag \\
	& \hspace{63pt} \leqslant - 2c_\textsc{l}|\mathcal{X}_t - \OrnsteinUhlenbeck_t|^2 + c_\textsc{l}\left|\mathcal{X}_t - \OrnsteinUhlenbeck_t|^2 + \frac{4}{c_\textsc{l}}\right|f_\mu(\OrnsteinUhlenbeck_t) + \OrnsteinUhlenbeck_t|^2, 
\end{align}
and
\begin{equation}
	|\mathcal{X}_t - \OrnsteinUhlenbeck_t|^2 \leqslant |\mathcal{X}_{t_0} - \OrnsteinUhlenbeck_{t_0}|^2e^{-c_\textsc{l}(t - t_0)} + \frac{4e^{-c_\textsc{l}t}}{c_\textsc{l}}\int^t_{t_0}e^{c_\textsc{l}s}|f_\mu(\OrnsteinUhlenbeck_s) + \OrnsteinUhlenbeck_s|^2ds.
\end{equation} 
Imposing $t_0 \to - \infty$ in the pullback convergence, one has
\begin{equation}
	|\mathcal{X}_t - \OrnsteinUhlenbeck_t|^2 \leqslant \rho^2_\mathcal{X}\bigl(\vartheta_t(\invertedbreve{\omega})\bigr) = 1 + \frac{4e^{-c_\textsc{l}t}}{c_\textsc{l}}\int^t_{-\infty}e^{c_\textsc{l}s}\big|f_\mu\bigl[\OrnsteinUhlenbeck_s\bigl(\vartheta_t(\invertedbreve{\omega})\bigr)\bigr] + \OrnsteinUhlenbeck_s\bigl(\vartheta_t(\invertedbreve{\omega})\bigr)\big|^2ds,
\end{equation} 
for any $t \geqslant \tau(\mathbbl{E}, \invertedbreve{\omega})$, with a tempered random set $\mathbbl{E}$ (cf. Theorem \ref{theorema "On the random attractor"}), where $\rho_\mathcal{X}$ is the radius of a family of compact balls $\mathbb{B}_{\invertedbreve{\omega}}$, the centre of which is $\OrnsteinUhlenbeck_0(\invertedbreve{\omega})$. It follows that 
\begin{equation}
	|\mathcal{X}_t(\invertedbreve{\omega}) - \OrnsteinUhlenbeck_t(\invertedbreve{\omega})| \leqslant \rho_\mathcal{X}\bigl(\vartheta_t(\invertedbreve{\omega})\bigr) 
\end{equation}
and 
\begin{equation}
	|\mathcal{X}_t(\invertedbreve{\omega}) \leqslant \OrnsteinUhlenbeck_t|(\invertedbreve{\omega}) + \rho_\mathcal{X}\bigl(\vartheta_t(\invertedbreve{\omega})\bigr). 
\end{equation}
The family of $\mathbb{B}_{\invertedbreve{\omega}}$ is therefore an absorbing family of sets, and this system contains a random (pullback) attractor $\mathbbl{\Lambda} = \{\Lambda_{\invertedbreve{\omega}}\}_{\invertedbreve{\omega} \in \invertedbreve{\Omega}}$.

\subsubsection{Example II, with Multiplicative Noise}

The second example concerns the multiplicative noise. It begins with a \emph{Stratonovich stochastic differential equation} accompanied by a Wiener process,
\begin{equation}
	d\mathcal{X}_t = (-\mathcal{X}_t + 1)dt + \mathcal{X}_t \circ d\mathsf{W}_t,
\end{equation}
which can be redrafted as a random ordinary differential equation, 
\begin{equation}
\label{equation "RDE from Stratonovich SDE"}
	\dot{x} = -x\bigl(1 + \OrnsteinUhlenbeck_t(\invertedbreve{\omega})\bigr) + e^{-\OrnsteinUhlenbeck_t(\invertedbreve{\omega})},
\end{equation}
with the Ornstein–Uhlenbeck process \eqref{equation "Ornstein–Uhlenbeck process"}. Since
\begin{align}
	\OrnsteinUhlenbeck_\tau(\invertedbreve{\omega}) - \OrnsteinUhlenbeck_0(\invertedbreve{\omega}) & = -\int^\tau_0\OrnsteinUhlenbeck_s(\invertedbreve{\omega})ds + \mathsf{W}_\tau(\invertedbreve{\omega}), \\
	\int^t_0\OrnsteinUhlenbeck_s(\invertedbreve{\omega})ds + \OrnsteinUhlenbeck_0 & = \int^t_0\OrnsteinUhlenbeck_s(\invertedbreve{\omega})ds + \int^0_{-\infty}\OrnsteinUhlenbeck_s(\invertedbreve{\omega})ds = \int^t_{-\infty}\OrnsteinUhlenbeck_s(\invertedbreve{\omega})ds,
\end{align}
the solution of \eqref{equation "RDE from Stratonovich SDE"} is 
\begin{align}
	x(t, \invertedbreve{\omega}) & = \Lbrack:\exp{\left\{-(t - t_0)\left(\frac{t - t_0 + 1}{t - t_0}\int^t_{t_0}\OrnsteinUhlenbeck_s(\invertedbreve{\omega})ds\right)\right\}}x_0:\Rbrack \notag \\ 
	& + \int^t_{t_0}\exp{\left\{-(t - \tau) - \int^t_\tau\OrnsteinUhlenbeck_s(\invertedbreve{\omega})ds - \OrnsteinUhlenbeck_\tau(\invertedbreve{\omega})\right\}}d\tau,
\end{align}
where all that is within the symbols $\Lbrack:$ and $:\Rbrack$ is to be repeated, similarly to the \emph{repeat sign} in musical notation. Then
\begin{align}
	x(t, \invertedbreve{\omega}) & = \Lbrack:\cdots:\Rbrack \notag \\
	& + \exp{\left\{-t - \int^t_0\OrnsteinUhlenbeck_s(\invertedbreve{\omega})ds\right\}}\int^t_{t_0}\exp{\left\{\tau - \int^0_\tau\OrnsteinUhlenbeck_s(\invertedbreve{\omega})ds - \OrnsteinUhlenbeck_\tau(\invertedbreve{\omega})\right\}}d\tau \notag \\
	& = \Lbrack:\cdots:\Rbrack \notag \\
	& + \exp{\left\{-t - \int^t_0\OrnsteinUhlenbeck_s(\invertedbreve{\omega})ds\right\}}\int^t_{t_0}\exp{\left\{\tau + \int^\tau_0\OrnsteinUhlenbeck_s(\invertedbreve{\omega})ds - \OrnsteinUhlenbeck_\tau(\invertedbreve{\omega})\right\}}d\tau \notag \\
	& = \Lbrack:\cdots:\Rbrack + \exp{\left\{-t - \int^t_0\OrnsteinUhlenbeck_s(\invertedbreve{\omega})ds - \OrnsteinUhlenbeck_0\right\}}\int^t_{t_0}\exp{\left\{\tau - \mathsf{W}_\tau(\invertedbreve{\omega})\right\}}d\tau \notag \\  
	& = \Lbrack:\cdots:\Rbrack + \exp{\left\{-t - \int^t_{-\infty}\OrnsteinUhlenbeck_s(\invertedbreve{\omega})ds\right\}}\int^t_{t_0}\exp{\left\{\tau - \mathsf{W}_\tau(\invertedbreve{\omega})\right\}}d\tau.
\end{align}
Imposing $t_0 \to - \infty$, one obtains the existence of pathwise pullback attracting stationary solution, with 
\[
	\exp{\left\{-t + \int^t_{-\infty}\OrnsteinUhlenbeck_s(\invertedbreve{\omega})ds\right\}}\int^t_{-\infty}\exp{\{\tau - \mathsf{W}_\tau(\invertedbreve{\omega})\}}d\tau.
\]

\section[Excursus: Araujo's Butterfly Flight Dynamics]{Excursus: Araujo's Butterfly\footnote{
	In homage to R. Araujo, Venezuelan architect and illustrator. His works include the graphic representation of a \emph{geometric net}, via reconstructive projections, behind the flutter of butterfly wings. The Araujo's illustrations are made by hand with classical drawing tools (compass, protractor, square and rule) and constitute the visual image of what we could call \emph{sympathique calculation} of patterns in Nature. See his collection of drawings \cite{Araujo "Golden Ratio Coloring Book"}.
	}
Flight Dynamics} 

\begingroup
\footnotesize
There are ways out, of course, but they require that your [theoretical] relation to reality be altered in one way or the other. Either you consider a mathematical problem analogous to the one you cannot handle, but easier, and forget about close contact with physical reality. Or you stick with physical reality but idealize it differently (often at the cost of forgetting about mathematical rigor or logical consistency). \\ 
\indent — \textsc{D. Ruelle} \cite[p. 124]{Ruelle "Chance and Chaos"}
	
\endgroup

\vspace{2mm}

\enumerationisinitium
\item The existence of some underlying schema, maybe sustained by a geometric support (e.g. self-similar, repetitive or fractal patterns), with which to single out one type of predictability (certainly in the short-term) on systems where there are chaotic solutions, is at the basis of the conception of the Lorenz attractor and its differential equations. These equations, when parameter values and initial conditions are set, lead to a specific and unique geometric shape that identifies the attractor itself. However, for reasons of greater adherence to reality, deterministic chaos models of dynamical systems may require, if necessary, a stochastic development, making creative use of stochastic and random differential equations. We should not forget that the question of chaotic systems comes from the difficulty of calculating, for certain phenomena, the real orbit (compared to the mathematical model) or the correct orbit, since there is no way of knowing  exactly the initial conditions and proceeding to the calculation with infinite precision.
\item The mathematical description of the flight of a butterfly, when it dancing and lands on a flower, regarding also the influence of external factors—e.g. a mild breeze can alter the flight path—is an almost impossible challenge. But this challenge arouses the dream of mathematics towards Nature, the dream of enclosing and encapsulating the physical reality in a geometric grid and/or in a group of mathematical formulæ. The distinction between (deterministic) chaos and randomness is, in some sense, summed up in this challenge, consisting in describing and determining the flight, or better, the different possible types of flight path, of a butterfly.
\enumerationisfinis
 
\vspace{10mm}

\setcounter{secnumdepth}{0}  
\section{References and Bibliographic Details}
\setcounter{secnumdepth}{3}
\markright{References and Bibliographic Details}

\begingroup
\footnotesize
\noindent Section \ref{subsection "Pullback Attractor: Process and Skew Product Flow"}

\begin{indent paragraph: 15pt}
Deepenings for non-autonomous systems, pullback attractors (attractors of processes and of skew product flows) are in \cite[chapp. 1-3]{Kloeden Rasmussen "Nonautonomous Dynamical Systems"} \cite[chapp. 1-2, 16]{Carvalho Langa Robinson "Attractors for infinite-dimensional non-autonomous dynamical systems"} \cite[chapp. 1-3]{Caraballo Han "Applied Nonautonomous and Random Dynamical Systems: Applied Dynamical Systems"}.
\end{indent paragraph: 15pt}

\noindent Section \ref{subsection "Random Attractor"} 

\begin{indent paragraph: 15pt}
On the random attractor, cf. \cite[pp. 483-484]{Arnold "Random Dynamical Systems"} \cite[chap. 14]{Kloeden Rasmussen "Nonautonomous Dynamical Systems"} \cite[sec. 1.7]{Carvalho Langa Robinson "Attractors for infinite-dimensional non-autonomous dynamical systems"} \cite{Crauel Kloeden "Nonautonomous and Random Attractors"} \cite[sec. 4.2]{Caraballo Han "Applied Nonautonomous and Random Dynamical Systems: Applied Dynamical Systems"}. See also \cite[chap. VII]{Liu Qian "Smooth Ergodic Theory of Random Dynamical Systems"}.
\end{indent paragraph: 15pt}
		
\noindent Section \ref{section "Stochastic Systems: Itô and Stratonovich Calculi"}

\begin{indent paragraph: 15pt}
About the generation of \textsc{rds'} from \textsc{sde}s and, vice versa, of \textsc{sde}s from \textsc{rds'}, see \cite[chapp. 2.3.4-2.3.7]{Arnold "Random Dynamical Systems"}.
\end{indent paragraph: 15pt}

\endgroup

\chapter{Galois' Legacy—Rules over the Calculations: the Pursuit of Generality} 
\label{chapter "Galois' Legacy—Rules over the Calculations: the Pursuit of Generality"}

\begingroup
\footnotesize
There exists, in fact, for these kinds of equations, a certain order of \emph{Metaphysical considerations} [\textit{considérations Métaphysiques}] which \emph{hover over all the calculations} [\textit{planent sur tous les calculs}], and often make them \emph{useless}. I will cite, for example, the equations which give the division of Elliptic functions and which the renowned Abel \cite{Abel "Recherches sur les fonctions elliptiques"} has solved [\,\dots]. All that creates the beauty and simultaneously the difficulty of this theory, is that one has ceaselessly to indicate the progress of the analyses and to predict the results without ever being able to carry them out [\textit{prévoir les résultats sans jamais pouvoir les effectuer}] \cite[p. 22, e.a.]{Galois "Discours preliminaire"}.\endnote{
	Original Fr. version: «Il existe, en effet, pour ces sortes d'équations, un certain ordre de considérations Métaphysiques qui planent sur tous les calculs, et qui souvent les rendent inutiles. Je citerai, par exemple, les équations qui donnent la division des fonctions Elliptiques et que le célèbre Abel a résolues [\,\dots]. Tout ce qui fait la beauté et à la fois la difficulté de cette théorie, c'est qu'on a sans cesse à indiquer l'[analyse] des calculs et à prévoir les résultats sans jamais pouvoir les effectuer».
	} \\
\indent There will be found here a \emph{general condition} [\textit{condition générale}] which satisfies \emph{every} equation that is solvable by radicals, and which reciprocally ensures their solvability \cite[p. 417, e.m.]{Galois "Memoire sur les conditions de resolubilite des equations par radicaux"}.\endnote{
	Original Fr. version: «On trouvera ici une \emph{condition} générale à laquelle \emph{satisfait toute équation soluble par radicaux}, et qui réciproquement assure leur résolubilité».
	} \\
\indent — \textsc{É. Galois} (1830-1831)

\vspace{2mm}

The criterion of research so splendidly asserted by Abel [and continued by Galois] “to put problems in the \emph{most general aspect} in order to discover their true nature”, designated the direction of Analysis which aims \emph{to break free the knowledge} [\textit{liberare la conoscenza}] of qualitative relations \emph{from the accidental complications of calculations} [\textit{dalle complicazioni accidentali dei calcoli}], that is precisely that direction of which the geometric theory of equations and algebraic functions is the maximum implementation.\endnote{
	Original It. version: «Il criterio di ricerca così splendidamente fatto valere da Abel “porre i problemi nell'aspetto più generale per scoprirne la vera natura”, designava l'indirizzo dell'Analisi che vuol liberare la conoscenza dei rapporti qualitativi dalle complicazioni accidentali dei calcoli, cioè appunto quell'indirizzo di cui è massima attuazione la teoria geometrica delle equazioni e delle funzioni algebriche». 
	} \\
\indent — \textsc{F. Enriques} \cite[p. x, e.a.]{Enriques "Lezioni sulla teoria geometrica delle equazioni e delle funzioni algebriche I"}

\endgroup

\vspace{2mm}

More than any other science, mathematics seeks \emph{generality}. This is because with a general solution it is possible to have \emph{control over the individual operations, avoiding the need for the brute force calculations}. Mathematics—especially pure mathematics, but also, albeit with greater restrictions, applied mathematics—shows a vocation, during its evolution, for the search for \emph{absolute and universal structures}, that in Galois go under the name of \textit{metaphysical considerations} (see epigraph). The case of \emph{Galoisian theory}, as a branch of \emph{abstract algebra}, is emblematic.

\section{Generality as a \emph{Métaphysique} Aspiration of Mathematics}

\begingroup
\footnotesize
It would seem that among all the natural sciences, it is only in mathematics that what I have called “the dream”, or “the waking dream” [\textit{rêve éveillé}], is struck by a prohibition [due to its propensity to rigor and precision]. Other sciences, including sciences that are reported to be “exact” like physics, the dream is at least tolerated, if not even encouraged [\,\dots] under names more “respectable” like: “speculations”, “hypotheses” (such as the famous “atomic hypothesis”, resulting from a dream[-speculation] of Democritus), “theories”, etc. The passage from the status of dream [\,\dots] to that of “scientific truth” takes place imperceptibly, by a consensus that widens gradually. [But this is not a prerogative of physics, because even in mathematics it is necessary to embark on a free reverie]. Which necessarily brings to mind the [mathematical] waking dream of Évariste Galois. \\
\indent — \textsc{A. Grothendieck} \cite[6.3. (7) \textit{L'héritage de Galois}, pp. 14-15 otm]{Grothendieck "Recoltes et Semailles. Reflexions et temoignage sur un passe de mathematicien"}

\endgroup

\vspace{2mm}

\subsection{Galoisian Algebra on Polynomial (Un)solvability I. Preliminary Overview}
\label{subsection "Galoisian Algebra on Polynomial (Un)solvability I. Preliminary Overview"}

\begingroup
\footnotesize
After Paolo Ruffini of Modena had demonstrated the impossibility of solving equations of degree higher than the fourth in general by radicals, Abel [was] the first [to] propose to determine the conditions to be verified, so that an equation of any degree was in particular solvable by radicals. The method he followed in these arduous researches is partially sketched out in a fragment of Memoir found among his papers [\,\dots]. Almost at the same time as Abel, Galois was meditating on the same problem, and 17 months after the death of [Abel], he presented a Memoir to the Academy of Sciences in Paris where he expounded a new and profound theory created by him to solve the problem taken from a point of view \emph{more general}. \\
\indent — \textsc{E. Betti} \cite[p. 49, e.a.]{Betti "Sulla risoluzione delle equazioni algebriche"}

\endgroup

\vspace{2mm}

\enumerationisinitium
\item É. Galois' innovation was the foundation of rules (\emph{nouvelles dénominations} and \emph{nouveaux caractères}), associated with new mathematical objects, which provides a criterion for determining the solvability of \emph{polynomial} (or \emph{algebraic}) \emph{equations} by radicals, with coefficients in a certain field. A polynomial e.g. with numerical coefficients is said to be \emph{solvable by radicals} if its roots can be expressed as radical functions of the coefficients. Let us step back, and look at how we got to the Galois theory. 
\item Solvability of polynomials by radicals includes equations of degree less than or equal to 4. 
\subenumerationisinitium
\item The «first solver» of the cubic equation was S. Dal Ferro,\endnote{
	A small marble commemorative plaque reports exactly this, Scipione Dal Ferro «primo solutore dell'equazione cubica. Lettore nello Studio dal 1496 al 1525». It can be found affixed to his paternal home in Bologna, in via S. Petronio Vecchio near the intersection of via Guerrazzi.
	}
 whose formula was made public by G. Cardano\footnote{
	It should be remembered that in Cardano the presence of square roots of a negative number degenerates into a «sophistico» case \cite[cap. XXXVII. \textit{De regula falsum ponendi}, fol. 66-ii]{Cardano "Artis magnae sive de regulis algebraicis"}. A decisive step forward, with the implication of complex numbers, will be taken by R. Bombelli \cite[p. 169]{Bombelli "L'Algebra parte maggiore dell'Arimetica divisa in tre libri di Rafael Bombelli da Bologna"}, by implementing operating rules for the manipulation of what we now call \emph{unit imaginary number} and \emph{imaginary roots}, which, for Cardano, had the status of \emph{quantitates sylvestres}, see footnote \ref{footnote "Cardano's quantitas silvestris"}, p. \pageref{footnote "Cardano's quantitas silvestris"}. Bombelli \cite[pp. 293-294]{Bombelli "L'Algebra parte maggiore dell'Arimetica divisa in tre libri di Rafael Bombelli da Bologna"} writes: «Et benche à molti parerà questa cosa stravagante, perche di questa opinione fui ancho [io] già un tempo parendomi più tosto fosse sofistica [\,\dots], nondimeno tanto cercai, che trovai la dimostratione».
	}
\cite[cap. I. \textit{De duabus æquationibus in singulis capitulis}, fol. 3; cap. XI. \textit{De cubo \& rebus æqualibus numero}, fol. 29-ii]{Cardano "Artis magnae sive de regulis algebraicis"},\endnote{
	Fol. 3: «Scipione Dal Ferro of Bologna has solved the case of the cubic equation [\textit{capitulum cubi}, “capitulum” is for a family of equations], excellent and admirable accomplishment; such an art surpasses all human subtlety [\,\dots]. In emulation of him, my friend Niccolò Tartaglia of Brescia, wanting not to be outdone, was able to solve the same case when he showed it in a challenge with his [Dal Ferro's] pupil, Antonio Maria Del Fiore and, pushed by my numerous entreaties, [Tartaglia] gave it to me»; fol. 29-ii: «Scipione Dal Ferro of Bologna almost thirty years ago discovered this rule and handed it on to Venetian Antonio Maria Del Fiore, whose challenge with Niccolò Tartaglia of Brescia was the occasion for Niccolò to discover it; and he [Tartaglia], in response to my entreaties, gave it to me, but kept the proof to himself. With the support of this assistance, I found the demonstration in multiple forms, which was very difficult».
	}
followed by the controversy with N. Tartaglia \cite[pp. 124-125]{Tartaglia "Quesiti et inventioni diverse de Nicolo Tartalea"}.\endnote{
	Within these pages is the Tartaglia's short composition in verse with a sibylline solution of the cubic: «Quando chel cubo con le cose apresso / Se aguaglia à qualche numero discreto / Trouan dui altri differenti in esso / Dapoi terrai questo per consueto / Ch'el lor produtto sempre sia eguale / Al terzo cubo delle cose neto / El residuo poi suo generale / Delli lor lati cubi ben sottratti / Varra la tua cosa principale».
	}
It is about an equation of this kind
\begin{equation}
	\Lbrack:x^3 + ax^2 + bx + c:\Rbrack = 0,
\end{equation}
involving a polynomial of degree 3 $\mathrm{p}(x) = \Lbrack:\cdots:\Rbrack$ (symbols $\Lbrack:$ and $:\Rbrack$ are for a repeat sign, see Glossary), with arbitrary coefficients $a, b, c$.
\item The quartic equation was solved by L. Ferrari \cite[cap. XV. \textit{De cubo \& quadratis æqualibus numero}, fol. 34-ii; fol. 72-ii]{Cardano "Artis magnae sive de regulis algebraicis"}, a pupil of Cardano. It is about an equation of this kind
\begin{equation}
\label{equation "Quartic equation"}
	\Lbrack:x^4 + ax^3 + bx^2 + cx + d:\Rbrack = 0,	
\end{equation}
implying a 4th degree polynomial $\mathrm{p}(x) = \Lbrack:\cdots:\Rbrack$, with arbitrary complex coefficients $a, b, c, d$. The solution of \eqref{equation "Quartic equation"} reduces to a solution of a resolvent (auxiliary) cubic equation.
\subenumerationisfinis
\item \emph{Ruffini–Abel theorem} \cite{Ruffini "Teoria generale delle Equazioni in cui si dimostra impossibile la soluzione algebraica delle equazioni generali di grado superiore al quarto"} \cite{Ruffini "Della insolubilita delle Equazioni algebraiche generali di grado superiore al quarto"} \cite{Abel "Memoire sur les equations algebriques ou l'on demontre l'impossibilite de la resolution de l'equation generale du cinquieme degre"} \cite{Abel "Demonstration de l'impossibilite de la resolution algebrique des equations generales qui passent le quatrieme degre"} establishes that \emph{there is no formula, intended as an algebraic solution, or a solution in radicals, for general polynomial equations of degree 5 or higher}, 
\begin{equation}
	x^5 + ax^4 + bx^3 + cx^2 + dx + e = 0.
\end{equation}
For instance, quintic equations 
\begin{align}
	& x^5 - x - 1 = 0, \\
	& x^5 + 20x + 16 = 0,	
\end{align}
are a terminus: they cannot be solved \emph{algebraically}, i.e. \emph{in radicals}. 
\item From here the Galois' analysis starts, and the question he answers is: is there a \emph{criterion}, or a \emph{procedure}, for determining whether a general fifth- or higher- degree polynomial equation can be solved by radicals? Galois writes \cite[p. 408]{Galois "Lettre a Auguste Chevalier"}: 

\vspace{2mm}

\begingroup
\footnotesize
In the theory of equations, I researched in which cases equations were solvable by radicals, this has given me the opportunity to deepen this theory, and to describe all possible transformations on an equation, even when it is not solvable by radicals. 

\endgroup

\vspace{2mm}

Galois, taking up and enlarging what J.L. Lagrange \cite{Lagrange "Reflexions sur la resolution algebrique des equations"} had already done, associates to each polynomial equation a given set of permutations of the roots; thereby he delineates the concept of \emph{permutation group}, later called \emph{Galois group}. Permutation is a way of ordering in succession certain elements of a particular set; and a permutation group is a group formed from the set of permutations of its elements, equipped with an algebraic structure with the symmetry property, which serves to define how the roots of a polynomial are connected to each other.

The bit where (see epigraph) Galois mentions a «general condition» that prevails over \textit{tous les calculs}, he has in mind a modus to manage each permutation symmetrically; transposed into an equational context, this is equivalent to seeing the number of polynomial permutations and \emph{substitutions} (the passage from one permutation to another),\footnote{
	See A.-L. Cauchy \cite{Cauchy "Memoire sur le Nombre des Valeurs qu'une Fonction peut acquerir lorsqu'on y permute de toutes les manieres possibles les quantites qu'elle renferme"}.
	} 
from which the im/possibility of solvability by radicals of a polynomial equation can be determined. 

Galois theory arrives, in its original statute, at the following upshot (see Theorem \ref{theorema "Galois"}): \emph{since the Galois group of a polynomial is isomorphic to a group of permutations, then a polynomial equation is solvable by radicals iff its Galois group, or group of permutations of the roots, is a solvable group}.
\enumerationisfinis

\subsection{Galoisian Algebra on Polynomial (Un)solvability II. Theorems for the Quintic, and the Icosahedral Equation}
\label{subsection "Galoisian Algebra on Polynomial (Un)solvability II. Theorems for the Quintic, and the Icosahedral Equation"} 
 
\begingroup
\footnotesize
As soon as we enter upon the task of studying the rotations [of a certain space as geometrical operations] [\,\dots], by which the configurations [\,\dots] are transformed into themselves, we are compelled to take into account the important and comprehensive theory which has been principally established by the pioneering works of Galois, and which we term the \emph{group-theory} [\textit{Gruppentheorie}]. Originally sprung from the theory of equations, and having a correspondent relation with the \emph{permutations} of any kind of elements, this theory includes [\,\dots] every question with which we are concerned in the case of a closed manifoldness of any kind of \emph{operations}. We say of any operations that they form a \emph{group}, if any two of the operations, compounded, again produce an operation included amongst those first given. In this sense we have at the outset the proposition: \emph{The rotations which bring one of the regular solids into coincidence with itself collectively form a group}. \\
\indent — \textsc{F. Klein} \cite[pp. 4-5]{Klein "Vorlesungen uber das Ikosaeder und die Auflosung der Gleichungen vom funften Grade"} = \cite[p. 5]{Klein "Lectures on the Ikosahedron and the Solution of Equations of the Fifth Degree"}

\vspace{2mm}

[The] farewell letter [of Galois] \cite{Galois "Lettre a Auguste Chevalier"} written to a friend on the eve of his death [\,\dots], if judged by the novelty and profundity of ideas it contains [about the genesis of a theory that amalgamates the field theory to the group theory], is perhaps the most substantial piece of writing in the whole literature of mankind. \\
\indent — \textsc{H. Weyl} \cite[p. 138]{Weyl "Symmetry"}

\endgroup

\vspace{2mm}

The first thing is to focus on the concept on which the whole question revolves. What is meant by \emph{solvability} referring to a group? 

\begin{definitio}
A group $G$ is said to be \emph{solvable} if it is non-trivial, and has \emph{a normal series with Abelian factors}, namely if there is a chain of subgroups\footnote{
	We can write the chain in other forms too, of course: $G = H_0 \supset H_1 \supset \cdots \supset H_n = \{\idem\}$, or $\{\idem\} = H_0 \subset H_1 \subset \cdots \subset H_n = G$, or even $\{\idem\} = H_n \subset H_{n - 1} \subset \cdots \subset H_ 1 \subset H_0 = G$. 
	}
\begin{equation}
\label{equation "Solvable group"}
	G = H_0 \supset H_1 \supset \cdots \supset H_{n - 1} \supset H_n = \{\idem\},
\end{equation}
where $\idem$ is the identity (or neutral) element of $G$, such that the subgroup $H_\rotatedell$ is normal in $H_{\rotatedell - 1}$, and the quotient group $H_{\rotatedell - 1}/H_\rotatedell$ is Abelian, for $\rotatedell = 1, \mathellipsis, n$. \definitiosymbol
\end{definitio}

Let us get right down to Galois's thought through a series of theorems.

\begin{theorema}[Solvability and non-solvability of the symmetric group]
\label{theorema "Solvability and non-solvability of the symmetric group"}
Let $\symmetric_n$ denote a symmetric group on a finite set of $n$-elements, consisting of the permutations performed on the $n$-elements. Groups $\symmetric_2$, $\symmetric_3$, and $\symmetric_4$ are solvable, while $\symmetric_n$ is non-solvable for $n \geqslant 5$. Id est: if $G = \symmetric_n$, $G$, or the symmetric group on $\{1, \mathellipsis, n\}$, is solvable for $n \leqslant 4$, whereas $\symmetric_1$ is trivial. 
\end{theorema}

\begin{proof}
~\enumerationisinitium
\item Given a field $F$, and a polynomial $\mathrm{p}(x) \in F[x]$ of degree $n \geqslant 1$, we denote by $\splittingfieldK$ a \emph{splitting field} of $\mathrm{p}(x)$ over $F$ of degree $n$, which is the \emph{minimal (smallest) field extension} of $F$ in which $\mathrm{p}(x)$ splits into linear factors.\footnote{
	A field $\splittingfieldK$ is said to be an \emph{extension} of the field $F$ if there is an immersion $F \hookrightarrow \splittingfieldK$, or an $\varphi$-isomorphism of $F$ in $\splittingfieldK$, that is, $\varphi \colon F \to \splittingfieldK$.
	} 
The field extension $F \subset \splittingfieldK$ is, on that account, a pair of fields, where $\splittingfieldK$ is an extension field of $F$ and $F$ is a subfield of $\splittingfieldK$. Fixing an ordering of the roots $\textcyrillic{\textit{к}}_1, \mathellipsis, \textcyrillic{\textit{к}}_n$ of $\mathrm{p}(x)$ in $\splittingfieldK$, wee see that

· $\splittingfieldK = F(\textcyrillic{\textit{к}}_1, \mathellipsis, \textcyrillic{\textit{к}}_n)$ contains the roots $\textcyrillic{\textit{к}}_\rotatedell \in \splittingfieldK$, for $\rotatedell = 1, \mathellipsis, n$, 

· $\splittingfieldK = F[x]/\{\mathrm{p}(x)\}$ is the splitting field of a separable polynomial $\mathrm{p}(x) \in F[x]$. 

Let $\varphi_{(F)}$ be a $F$-automorphism of $\splittingfieldK$, and $\rotatedpi^{\varphi_{(F)}} \in \symmetric_n$ a permutation belonging to the symmetric group. The group of all $F$-automorphisms of $\splittingfieldK$, written $\Galois_F(\splittingfieldK) \viz \Galois(\splittingfieldK/F)$, is called the \emph{Galois group} of $\splittingfieldK$ over $F$, or the \emph{Galois group of the extension}, for which $\splittingfieldK$ is a finite Galois extension of $F$. Then we set an injective homomorphism
\begin{equation}
	\textcyrillic{\textit{Г}} \colon \Galois_F(\splittingfieldK) \xrightarrow{\textcyrillic{\textit{Г}}\text{-homomorphism}} \symmetric_n,
\end{equation} 
so the Galois group of $\mathrm{p}(x)$ is \emph{isomorphic} to a subgroup $G_\mathrm{sub} \viz H$ of $\symmetric_n$, i.e.
\begin{equation}
	\Galois_F(\splittingfieldK) \cong G_\mathrm{sub} \viz H = \{\rotatedpi^{\varphi_{(F)}} \mid \varphi_{(F)} \in \Galois_F(\splittingfieldK)\} \subset \symmetric_n, 
\end{equation}
with $\varphi_{(F)} \mapsto \rotatedpi^\varphi$, such that $\varphi_{(F)}(\textcyrillic{\textit{к}}) = \textcyrillic{\textit{к}}$, for all roots $\textcyrillic{\textit{к}}$ of $\mathrm{p}(x)$, and $\varphi_{(F)}$ is the identity over $\splittingfieldK$. 
\item Finally, we have the advantage of Theorem \ref{theorema "Galois"}.
\enumerationisfinis
\end{proof}

\begin{scholium}
Theorem \ref{theorema "Solvability and non-solvability of the symmetric group"} implies, inter alia, that any polynomial over a field of characteristic zero 

· is solvable by radicals, for degree 2, 3, and 4, 

· is non-solvable by radicals, for degree 5 or higher (with arbitrary coefficients), as laid down by the \emph{Ruffini–Abel theorem} \cite{Ruffini "Teoria generale delle Equazioni in cui si dimostra impossibile la soluzione algebraica delle equazioni generali di grado superiore al quarto"} \cite{Ruffini "Della insolubilita delle Equazioni algebraiche generali di grado superiore al quarto"} \cite{Abel "Memoire sur les equations algebriques ou l'on demontre l'impossibilite de la resolution de l'equation generale du cinquieme degre"} \cite{Abel "Demonstration de l'impossibilite de la resolution algebrique des equations generales qui passent le quatrieme degre"}. \scholiumsymbol
\end{scholium}

\begin{scholium}[Polynomial with rational coefficients, and Galois group over $\mathbb{Q}$]
If $\mathrm{p}(x) \in \mathbb{Q}[x]$ is a polynomial with rational coefficients, and $G$ is a finite Abelian group over $\mathbb{Q}$, we will write a Galois extension in the form $\splittingfieldK/\mathbb{Q}$, with $\Galois_\mathbb{Q}(\splittingfieldK) \cong G$. It should be noted that $\Galois_\mathbb{Q}(\splittingfieldK)$ is isomorphic to $\symmetric_3$. Why? The splitting field $\splittingfieldK = \mathbb{Q}(\sqrt[3]2, \textcyrillic{\textit{к}})$ is generated over $\mathbb{Q}$ by three roots, $\sqrt[3]2$, $\sqrt[3]2\textcyrillic{\textit{к}}$, and $\sqrt[3]2\textcyrillic{\textit{к}}^2$, of the polynomial $\mathrm{p}(x) = x^3 - 2\in \mathbb{Q}[x]$, that is to say, $x^3 - 2$ has three roots in $\mathbb{Q}(\sqrt[3]2, \textcyrillic{\textit{к}})$ and only one root in $\mathbb{Q}(\sqrt[3]2)$, where $\textcyrillic{\textit{к}} = \exp(2\frac{\pi i}{3})$ is a non-trivial cube root of unity in $\mathbb{C}$, i.e. $\textcyrillic{\textit{к}}^3 = 1$ and $\textcyrillic{\textit{к}} \neq 1$, satisfying the equation $x^2 + x + 1$. Then $[\mathbb{Q}(\sqrt[3]2, \textcyrillic{\textit{к}}) \colon \mathbb{Q}] = 6$, i.e., we have $\xi$-six functions,\footnote{
	(1st) $\sqrt[3]2 \to \sqrt[3]2$, $\textcyrillic{\textit{к}} \to \textcyrillic{\textit{к}}$, 
	(2nd) $\sqrt[3]2 \to \sqrt[3]2$, $\textcyrillic{\textit{к}} \to \textcyrillic{\textit{к}}^2$, 
	(3rd) $\sqrt[3]2 \to \sqrt[3]2\textcyrillic{\textit{к}}$, $\textcyrillic{\textit{к}} \to \textcyrillic{\textit{к}}$, 
	(4th) $\sqrt[3]2 \to \sqrt[3]2\textcyrillic{\textit{к}}$, $\textcyrillic{\textit{к}} \to \textcyrillic{\textit{к}}^2$, 
	(5th) $\sqrt[3]2 \to \sqrt[3]2\textcyrillic{\textit{к}}^2$, $\textcyrillic{\textit{к}} \to \textcyrillic{\textit{к}}$, 
	(6th) $\sqrt[3]2 \to \sqrt[3]2\textcyrillic{\textit{к}}^2$, $\textcyrillic{\textit{к}} \to \textcyrillic{\textit{к}}^2$.
	}
from which it appears that the extension $\mathbb{Q}(\sqrt[3]2, \textcyrillic{\textit{к}})/\mathbb{Q}$ is Galois, being that 
\begin{equation}
	\Galois_\mathbb{Q}\left(\mathbb{Q}(\sqrt[3]2, \textcyrillic{\textit{к}})\right) \viz \Galois\left(\mathbb{Q}(\sqrt[3]2, \textcyrillic{\textit{к}})/\mathbb{Q}\right) = \left[\mathbb{Q}(\sqrt[3]2, \textcyrillic{\textit{к}}) \colon \mathbb{Q}\right].
\end{equation}
Furthermore, $\mathbb{Q}(\sqrt[3]2, \textcyrillic{\textit{к}})/\mathbb{Q}$ has Galois group isomorphic to the symmetric group of degree 3, 
\begin{equation}
	\Galois_\mathbb{Q}\left(\mathbb{Q}(\sqrt[3]2, \textcyrillic{\textit{к}})\right) = \Galois_\mathbb{Q}(\splittingfieldK) \cong \symmetric_3,
\end{equation}
which is glaring if we think about $\splittingfieldK$ as $\mathbb{Q}(\sqrt[3]2, \sqrt[3]2\textcyrillic{\textit{к}}, \sqrt[3]2\textcyrillic{\textit{к}}^2)$. Wanting to summarize with a \emph{lattice-kite diagram}, the symbolic representation is
\[
\begin{tikzcd}
	& \mathbb{Q}(\sqrt[3]2, \textcyrillic{\textit{к}}) \arrow[rrd, no head] \arrow[d, no head] & & \\
	\mathbb{Q}(\textcyrillic{\textit{к}}) \arrow[ru, no head] & \mathbb{Q}(\sqrt[3]2) \arrow[d, magenta, no head]
	& \mathbb{Q}(\sqrt[3]2\textcyrillic{\textit{к}}) \arrow[lu, no head] & \mathbb{Q}(\sqrt[3]2\textcyrillic{\textit{к}}^2) \\
	& \mathbb{Q} \arrow[ru, magenta, no head] \arrow[lu, no head] \arrow[lu, vernal-green, no head] \arrow[rru, magenta, no head] & &  
\end{tikzcd}
\]
where the quadratic field $\mathbb{Q}(\textcyrillic{\textit{к}})$ over $\mathbb{Q}$ is in \textcolor{vernal-green}{\texttt{green \#03E364}}, and the three cubic fields are in \textcolor{magenta}{\texttt{magenta \#E30382}}. \scholiumsymbol
\end{scholium}

\begin{theorema}[Galois]
\label{theorema "Galois"}
Let $F$ be a field of zero characteristic; then a polynomial $\mathrm{p}(x) \in F[x]$ is solvable by radicals iff its Galois group over $F$ is a solvable group.
\end{theorema}

\begin{proof}
Let $\splittingfieldL$ be an \emph{intermediate splitting field} of the extension $F \subset \splittingfieldK$, aka a subextension of $F \subset \splittingfieldK$, satisfying $F \subset \splittingfieldL \subset \splittingfieldK$. We consider the group $\Galois_F(\splittingfieldL)$. Let $\textcyrillic{\textit{к}}_\alpha$ denote a primitive $n$-th root of unity (a number whose $n$-th power is equal to 1). We must show that the extension $F \subset \splittingfieldL(\textcyrillic{\textit{к}}_\alpha)$ is radical. Now, $\splittingfieldL(\textcyrillic{\textit{к}}_\alpha)$ is a splitting field of $\mathrm{p}(x)$ over $F(\textcyrillic{\textit{к}}_\alpha)$, so the extension $F(\textcyrillic{\textit{к}}_\alpha) \subset \splittingfieldL(\textcyrillic{\textit{к}}_\alpha)$ is Galois, and the map
\begin{equation}
	\textcyrillic{\textit{Г}} \colon \Galois_{F(\textcyrillic{\textit{к}}_\alpha)}\bigl(\splittingfieldL(\textcyrillic{\textit{к}}_\alpha)\bigr) \xrightarrow{\textcyrillic{\textit{Г}}\text{-homomorphism}} \Galois_F(\splittingfieldL)
\end{equation}
is an injective homomorphism. Since $\Galois_{F(\textcyrillic{\textit{к}}_\alpha)}\bigl(\splittingfieldL(\textcyrillic{\textit{к}}_\alpha)\bigr)$ is isomorphic to a subgroup  of $\Galois_F(\splittingfieldL)$, it is solvable. There is therefore a chain of subgroups
\begin{equation}
	\Galois_{F(\textcyrillic{\textit{к}}_\alpha)}\bigl(\splittingfieldL(\textcyrillic{\textit{к}}_\alpha)\bigr) = G = H_0 \supset H_1 \supset \cdots \supset H_{n - 1} \supset H_n = \{\idem_\splittingfieldK\},
\end{equation}
such that the subgroup $H_\rotatedell = \Galois_{F_\rotatedell}\bigl(\splittingfieldL(\textcyrillic{\textit{к}}_\alpha)\bigr)$ is normal in $H_{\rotatedell - 1} = \Galois_{F_{\rotatedell - 1}}\bigl(\splittingfieldL(\textcyrillic{\textit{к}}_\alpha)\bigr)$, and the quotient group 
\begin{equation}
	H_{\rotatedell - 1}/H_\rotatedell = \Galois_{F_{\rotatedell - 1}}\bigl(\splittingfieldL(\textcyrillic{\textit{к}}_\alpha)\bigr)/\Galois_{F_\rotatedell}\bigl(\splittingfieldL(\textcyrillic{\textit{к}}_\alpha)\bigr) 
\end{equation}
is cyclic, for $\rotatedell = 1, \mathellipsis, n$. Note that a group is \emph{cyclic} if it has single element, called a \emph{generator}. Consequently, one has a chain of subfields 
\begin{equation}
	F_0 = F(\textcyrillic{\textit{к}}_\alpha) \subset F_1 \subset \cdots \subset F_{\rotatedell - 1} \subset F_\rotatedell \subset \cdots \subset F_n = \splittingfieldL(\textcyrillic{\textit{к}}_\alpha),
\end{equation}
under which the extension $F_{\rotatedell - 1} \subset F_\rotatedell$ is Galois, and 
\begin{equation}
	\Galois_{F_{\rotatedell - 1}}(F_\rotatedell) = H_{\rotatedell - 1}/H_\rotatedell 
\end{equation}
is cyclic. The field $F_{\rotatedell - 1}$ containing all the $n$-th roots, contains also all the $n_\rotatedell$-th roots of unity, whilst $F_\rotatedell$ is a pure $n_\rotatedell$-radical extension; ergo $F \subset \splittingfieldL(\textcyrillic{\textit{к}}_\alpha)$ is a radical extension.
\end{proof}

Let us move to another group, called \emph{alternating group} of degree $n$, and denoted by $\mathfrak{A}_n$. It is a normal subgroup of $\symmetric_n$, and corresponds to the set of all even permutations; it is generated by 3-cycles (cycles with three elements), and contains all the 3-cycles of $\symmetric_n$, whereas the latter, the symmetric group, is generated by 2-cycles (cycles with two elements), or transpositions.

\begin{theorema}
Symmetric and alternating groups, $\symmetric_n$ and $\mathfrak{A}_n$, are non-solvable for $n \geqslant 5$.
\end{theorema}

\begin{proof}
We put $G = \symmetric_n$, or even $G = \mathfrak{A}_n$, indifferently. Let us consider the chain of subgroups 
\begin{equation}
	G = H_0 \supset H_1 \supset \cdots \supset H_{n - 1} \supset H_n, 
\end{equation}
such that the subgroup $H_\rotatedell$ is normal in $H_{\rotatedell - 1}$, and the quotient group $H_{\rotatedell - 1}/H_\rotatedell$ is cyclic, for $\rotatedell \geqslant 1$. If each $H_\rotatedell$ on this chain contains all the 3-cycles of $\symmetric_n$, the chain is not finite, i.e. $G \neq \{\idem\}$, so $G = \symmetric_n$, or $G = \mathfrak{A}_n$, is non-solvable. If $G = \symmetric_n$, or $G = \mathfrak{A}_n$, were solvable, we should have the same result as the Eq. \eqref{equation "Solvable group"}.

Let us see the case where $H_\rotatedell$ contains all the 3-cycles. We can write $x = (\alpha, \beta, \gamma)$, $y = (\gamma, \delta, \varepsilon) \in H_{\rotatedell - 1}$, setting $\{\alpha, \beta, \gamma, \delta, \varepsilon\} = \{1, 2, 3, 4, 5\}$. From the canonical projection 
\begin{equation}
	\prj \colon H_{\rotatedell - 1}\to H_{\rotatedell - 1}/H_\rotatedell,
\end{equation}	 
one obtains $\prj(x)^{-1}\prj(y)^{-1}\prj(x)\prj(y) = 1$, from which
\begin{equation}
	(x^{-1}y^{-1}xy) = (\gamma, \beta, \alpha)(\varepsilon, \delta, \gamma)(\alpha, \beta, \gamma)(\gamma, \delta, \varepsilon) = (\gamma, \beta, \varepsilon) \in H_\rotatedell.
\end{equation} 
\end{proof}

The alternate group $\mathfrak{A}_5$ 

· is the \emph{smallest non-solvable group};

· is simple, more precisely, is the non-Abelian simple group of smallest order, which is equal to $60 = 2^2 \cdot 3 \cdot 5$;

· is characterized by these conjugacy classes, in addition to the identity: 5-cycle permutations et 24 elements, 3-cycle permutations et 20 elements, and 2-cycle permutations et 15 elements: $1 + 12 + 12 + 20 + 15 = 60 = |\mathfrak{A}_5|$ (note that: since $24 \nmid 60 = |\mathfrak{A}_5|$, in $\mathfrak{A}_5$ the 5-cycles cannot be all conjugate by way of a permutation, thus the 24 5-cycles split into two conjugacy classes of 5-cycles, each of which has 12 elements);

· is isomorphic to the isometry group of the \emph{icosahedron} (Scholium \ref{scholium "Icosahedron"}); 

· is isomorphic to the rotation group of the \emph{dodecahedron} (Scholium \ref{scholium "Dodecahedron"}), the group of even permutations of five elements; namely it is isomorphic to the group of even isometries of the dodecahedron.\footnote{
	The isometry group, in its entirety, is isomorphic to $\mathfrak{A}_5 \times \mathbb{Z}_2 \viz \mathbb{Z}/2\mathbb{Z}$;
	}
	
· is isomorphic to $PSL_2(\mathbb{F}_4)$, having order $4^3 - 4 = 60$, and $PSL_2(\mathbb{F}_5)$, having order $\frac{1}{2}(5^3 - 5) = 60$, which are two projective special linear groups of degree 2, hence $PSL_2(\mathbb{F}_4) \cong PSL_2(\mathbb{F}_5) \cong \mathfrak{A}_5$.	

\begin{figure}[h!]
\centering
	\begin{minipage}[b]{0.385\textwidth}
	\includegraphics[width = \textwidth]{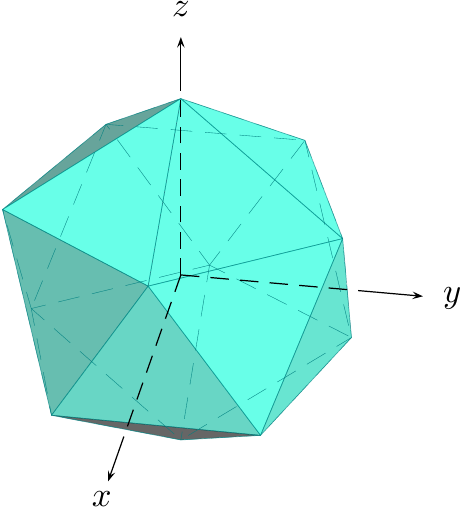}
	\caption{Icosahedron has 60 orientation preserving symmetries, and a full symmetry group of order $60 \times 2 = 120$}
	\label{figure "Icosahedron"} 
	\end{minipage}
	\hspace{30pt}
	\begin{minipage}[b]{0.410\textwidth}
	\includegraphics[width = \textwidth]{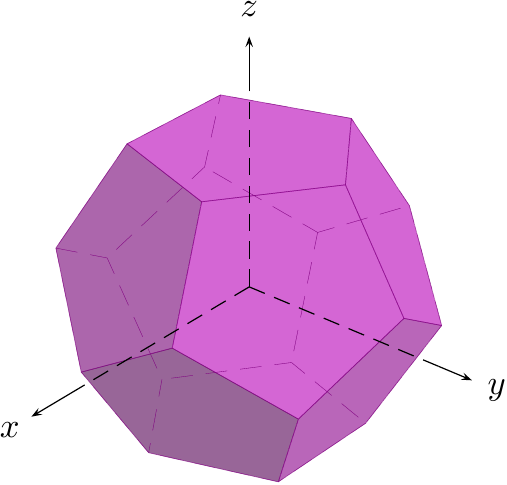}
	\caption{Dodecahedron has 100 space diagonals + 60 face diagonals}
	\label{figure "Dodecahedron"}
	\end{minipage}
\end{figure}

\begin{scholium}[Icosahedron]
\label{scholium "Icosahedron"}
We remember that the icosahedron has 20 triangular faces, 30 edges, and 12 vertices (Fig. \ref{figure "Icosahedron"}), or 60 orientation preserving symmetries, which correspond exactly to the 60 permutations of $\mathfrak{A}_5$. Here is what we have:

· the identity, or the unit element of the rotation group, 

· 24 5-cycles of $\mathfrak{A}_5$ corresponding to the rotations of the icosahedron whose axis passes through opposite vertices;

· 20 3-cycles of $\mathfrak{A}_5$ corresponding to the rotations of the icosahedron whose axis passes through midpoints of opposite faces,

· 15 double 2-cycles corresponding to the rotations of the icosahedron  whose axis passes through midpoints of opposite edges. 

Note. A classic reading on the icosahedral polyhedron and 5th degree equations is the work of F. Klein \cite{Klein "Vorlesungen uber das Ikosaeder und die Auflosung der Gleichungen vom funften Grade"} = \cite{Klein "Lectures on the Ikosahedron and the Solution of Equations of the Fifth Degree"}. \scholiumsymbol
\end{scholium}

\begin{scholium}[Dodecahedron]
\label{scholium "Dodecahedron"}
We remember that the dodecahedron has 12 pentagonal faces, 30 edges, and 20 vertices (Fig. \ref{figure "Dodecahedron"}), so its rotations are given by: 

· the identity, or the unit element of the rotation group, 

· 5-cycles of order 5 in $\mathfrak{A}_5$ corresponding to $\frac{5 \times 4 \times 3 \times 2}{5} = 24$ rotations by multiples of $\frac{2\pi}{5}$, with 6 axes that join the centers of opposite pentagonal faces, and 4 non-trivial rotations,

· 3-cycles of order 3 in $\mathfrak{A}_5$ corresponding to $\frac{5 \times 4 \times 3}{3} = 20$ rotations by multiples of $\frac{2\pi}{3}$, with 10 axes that join opposite vertices, and 2 non-trivial rotations,

· $\frac{(5 \times 4)}{2} \times \frac{3 \times 2}{2} = \frac{30}{2} = 15$ permutations in $\mathfrak{A}_5$ corresponding to the rotations by multiples of $\pi$, with 15 axes that join centres of opposite edges, and 1 non-trivial rotation. \scholiumsymbol
\end{scholium}

\begin{margo}[Small historical annotation on icosa- and dodeca-hedron]
Before Galois–Klein's works on abstract and gruppal algebra, the most important book, from the time of Academy in Athens, in which (together with other solids) the icosahedron and the dodecahedron appear, is perhaps the \textit{Divina proportione} of L. Pacioli, illustrated by Leonardo da Vinci: see tab. XXI: \textgreek{εἰκοσάεδρον ἐπίπεδον στερεόν} · Icosaedron Epipedon Stereon/Planum Solidum; and tab. XXVII: \textgreek{δ[ω]δεκάεδρον ἐπίπεδον στερεόν} · Dodecaedron Epipedon [S]tereon/Planum Solidum] in \cite{Pacioli Divina proportione}.\endnote{
	In the 1509 printed edition, both smooth breathing and acute accent are absent.
	} \margosymbol
\end{margo}

\section{Generality for Algebro-geometric Apparatuses}

\subsection{From Galois–Klein Icosahedron to Arkani-Hamed–Trnka Amplituhedron: a Link between Algebraic Geometry and \textsc{qft}} 

Acquiring a \textit{critère métaphysique} à la Galois (see above), a \emph{general method} thanks to which avoid long and tedious calculations, also interests theoretical physics. The objective is the same: it is a question of constructing a \emph{synthesis-structure}, or an \emph{abstract entity} which, so to speak, contains in itself and summarizes a whole series of mathematical operations needed to establish a certain result.

Among many examples, we choose that of the so-called  \emph{amplituhedron}, a polyhedron conceived by N. Arkani-Hamed and J. Trnka \cite{Arkani-Hamed and Trnka "The Amplituhedron"} \cite{Arkani-Hamed and Trnka "Into the amplituhedron"} \cite{Arkani-Hamed Bourjaily Cachazo Goncharov Postnikov and Trnka "Scattering Amplitudes and the Positive Grassmannian"}, which aims to provide a \emph{geometric visualization} of the perturbative calculation  of the S-matrix (scattering matrix). It   is a polytope, and it can be described as a \emph{positive Grassmannian}, namely a space equivalent, in such a context, to an \emph{algebraic variety}. The amplituhedron, more specifically, is an alternative to Feynman diagrams \cite{Feynman "Space-Time Approach to Quantum Electrodynamics"},\endnote{
	This is the first document in which Feynman diagrams appear.
	} 
and it is adopted for the computation of \emph{probability amplitudes}, that is, of all factors explaining the propagation of a particle for scattering events, as well as the various interactions between particles.

Each particle involved in the studied process, occupies a vertex of the amplituhedron (as many vertices must be drawn as there are particles scattering), while the momentum of a particle  shall be proportionate to the size of the affected face. The final value of the probability amplitude is instead equal (corresponds) to the volume of the amplituhedron, in accordance with symmetry rules.

\subsection{Twisted-type Cohomology for Multi-Loop Feynman Integrals} 

Something resembling the Arkani-Hamed–Trnka amplituhedron, but not as mold-breaking, occurs with the method of synthesis devised by P. Mastrolia and S. Mizerac \cite{Mastrolia and Mizera "Feynman integrals and intersection theory"} \cite{Frellesvig Gasparotto Laporta Mandal Mastrolia Mattiazzi and Mizera "Decomposition of Feynman integrals on the maximal cut by intersection numbers"} \cite{Frellesvig Gasparotto Mandal Mastrolia Mattiazzi and Mizera "Vector Space of Feynman Integrals and Multivariate Intersection Numbers"}, reinterpreting the inexhaustible calculations that derive from multi-loop Feynman integrals in an \emph{algebro-geometric structure}. It is about summarizing an almost-infinite computational operation of integrals in a \emph{shape-space scheme}, or producing a classification with certain topological invariants, on an algebraic basis, of possible values of path integrals within a \emph{twisted-type cohomology}, compare with the papers of M. Kita \& M. Yoshida \cite{Kita and Yoshida "Intersection Theory for Twisted Cycles"} \cite{Kita and Yoshida "Intersection Theory for Twisted Cycles II - Degenerate Arrangements"} \cite{Yoshida "Intersection Theory for Twisted Cycles III - Determinant Formulae"}. The heart of Mastrolia and Mizerac method is to interpret the Feynman loop integrals in terms of \emph{Aomoto–Gel'fand hypergeometric functions} \cite{Aomoto "Les equations aux differences lineaires et les integrales des fonctions multiformes"} \cite{Gel'fand "General theory of hypergeometric functions"}, in which they are grasped as pairings of twisted cycles \cite{Mimachi Yoshida "Intersection Numbers of Twisted Cycles and the Correlation Functions of the Conformal Field Theory"} \cite{Mimachi Yoshida "Intersection Numbers of Twisted Cycles Associated with the Selberg Integral and an Application to the Conformal Field Theory"}. As references to hypergeometric functions, see \cite{Aomoto Kita "Theory of Hypergeometric Functions"} \cite{Yoshida "Hypergeometric Functions My Love: Modular Interpretations of Configuration Spaces"}. 

\vspace{10mm}

\setcounter{secnumdepth}{0}  
\section{References and Bibliographic Details}
\setcounter{secnumdepth}{3}
\markright{References and Bibliographic Details}

\begingroup
\footnotesize
\noindent Sections \ref{subsection "Galoisian Algebra on Polynomial (Un)solvability I. Preliminary Overview"} and \ref{subsection "Galoisian Algebra on Polynomial (Un)solvability II. Theorems for the Quintic, and the Icosahedral Equation"}

\begin{indent paragraph: 15pt}
· For a picture of algebra from Cardano to Galois, between a historico-informative level and a technical one, see S. Maracchia \cite{Maracchia "Da Cardano a Galois. Momenti di storia dell'algebra"}. \\
· Preparatory readings to the Galois theory: besides the foundational lectures of E. Artin \cite{Artin "Galois Theory"}, see \cite{Morandi "Field and Galois Theory"} \cite{Serre "Topics in Galois theory"} \cite{Bergen "A Concrete Approach to Abstract Algebra: From the Integers to the Insolvability of the Quintic"} \cite{Cox "Galois Theory"} \cite{Khovanskii "Galois Theory Coverings and Riemann Surfaces"} \cite[chap. A-5]{Rotman "Advanced Modern Algebra 1"} \cite{Bosch "Algebra: From the Viewpoint of Galois Theory"}. \\
· On the historical level, it is worth mentioning the works of E. Betti \cite{Betti "Sulla risoluzione delle equazioni algebriche"} (the first paper of technical depth in the Galoisian area), A. Cayley \cite{Cayley "On the Theory of Groups as depending on the Symbolic Equation $theta^n = 1$"}, R. Dedekind \cite{Dedekind "Eine Vorlesung uber Algebra"}, C. Jordan \cite[I, § III, and III, chap. II, § II]{Jordan "Traite des substitutions et des equations algebriques"}, the above-mentioned F. Klein \cite{Klein "Vorlesungen uber das Ikosaeder und die Auflosung der Gleichungen vom funften Grade"} = \cite{Klein "Lectures on the Ikosahedron and the Solution of Equations of the Fifth Degree"}, and L. Bianchi \cite{Bianchi "Lezioni sulla teoria dei gruppi di sostituzioni e delle equazioni algebriche secondo Galois"}. \\
· The most important Grothendieck's works on Galois theory are \cite{Grothendieck "Esquisse d'un Programme"} and \cite{Grothendieck Raynaud "Revetements etales et groupe fondamental (SGA 1)"}. See (as a reference) \cite{Schneps and Lochak (Eds.) "Geometric Galois Actions 1. Around Grothendieck's Esquisse d'un Programme"}.
\end{indent paragraph: 15pt}

\endgroup

\chapter{Toroidal Fourier Analysis}
\label{chapter "Toroidal Fourier Analysis"}

\begingroup
\footnotesize
Et ignem regunt numeri.\footnote{
	«Numbers rule fire».
	} \\
\indent — \textsc{J.B.J. Fourier} \cite[p. 1]{Fourier "Theorie analytique de la chaleur"}, quote from \textsc{Plato}

\endgroup

\section{\emph{Exercices de (Fourierian) Style}}

From Fourier's metaphysics—his motto “Et ignem regunt numbers” (of Platonic derivation) is understandably a \emph{drool} (cf. Section \ref{subsubsection "Upside Down Fourier's Judgment"})—it is possible to derive many applications, the engineering result of which is amazing, see e.g. \cite{Prestini "The Evolution of Applied Harmonic Analysis: Models of the Real World"}; since among the loves of the author of this book there is music, for the presence of Fourier's frame in music theory and physics of musical instruments, see \cite[chap. 12.1]{Tonietti "And Yet It Is Heard: Musical Multilingual and Multicultural History of the Mathematical Sciences II} \cite{Amiot "Music Through Fourier Space Discrete: Fourier Transform in Music Theory"} \cite[chap. 2, Appendix C]{Bennett Jr. Morrison (Ed.) "The Science of Musical Sound 1: Stringed Instruments Pipe Organs and the Human Voice"}.

As purists, we will just give a geometric example among the many available; for convenience, we will choose one that has been meticulously explored by L. Grafakos \cite[chap. 3]{Grafakos "Classical Fourier Analysis"}; additionally it is a fine example.

\subsection{Summability of Fourier Series in Spherical Bochner–Riesz Means}
\label{subsection "Summability of Fourier Series in Spherical Bochner–Riesz Means"}

Let 
\begin{equation}
\label{equation "$n$-torus with [0, 1]"}
\torus^n = \underbrace{[0, 1] \times \cdots \times [0, 1]}_{n \text{ times}}
\end{equation}
be an $n$-torus corresponding to a cube $[0, 1]^n$, with $(x^1, \mathellipsis, 0, \mathellipsis, x^n)$ and $(x^1, \mathellipsis, 1, \mathellipsis, x^n)$, it being understood that $\torus^n$ can be defined as the set of all equivalence classes $\mathbb{R}^n/\mathbb{Z}^n$, or as a subset of the field of complex numbers, to wit, $\torus^n \subset \mathbb{C}^n$, with a function $\pi \colon \mathbb{R}^n \to \torus^n$, given by 
\begin{equation}
	\pi(x^1, \mathellipsis, x^n) = \left\{\left(e^{2\pi ix^1}, \mathellipsis, e^{2\pi ix^n}\right) \in \mathbb{C}^n \mathrel{\Big|} (x^1, \mathellipsis, x^n) \in [0, 1]^n\right\}.
\end{equation}

Compare Eq. \eqref{equation "$n$-torus with [0, 1]"} with Eq. \eqref{equation "$n$-torus"} in which 1-spheres, i.e. circles, $\mathbb{S}^1$ appear: $\torus^n \cong \mathbb{S}^1 \times \mathbb{S}^1 \cdots \times \mathbb{S}^1$, for $n$ times. We recall that a degenerate torus is but a double-covered spherical region.

Let
\begin{equation}
	\BochnerRiesz_\Rorder^\alpha\varphi(x) = \underset{|k| \leqslant \Rorder}{\sum_{k \in \mathbb{Z}^n}}\left(1 - \frac{|k|^2}{\Rorder^2}\right)^\alpha\widehat{\varphi}(k)e^{2\pi ik \cdot x}.
\end{equation}
be the \emph{Bochner–Riesz means} \cite{Bochner "Summation of Multiple Fourier Series by Spherical Means"} \cite{Riesz "Une methode de sommation equivalente a la methode des moyennes arithmetiques"} of index (or degree) $\alpha = \frac{n - 1}{2}$ and $\Rorder$-th order, where $\varphi$ is an integrable function on $\torus^n$, and $x \in \torus^n$. Note that $\Rorder$ designates the $\Rorder$-th spherical partial sum. 

\begin{theorema}
There is a $\varphi$ under which
\begin{equation}
	\limsup_{\Rorder \to \infty}\left|\BochnerRiesz_\Rorder^\alpha\varphi(x)\right| = \limsup_{\Rorder \to \infty}\left\{\underset{|k| \leqslant \Rorder}{\sum_{k \in \mathbb{Z}^n}}\left(1 - \frac{|k|^2}{\Rorder^2}\right)^\frac{n - 1}{2}\widehat{\varphi}(k)e^{2\pi ik \cdot x}\right\} = \infty,
\end{equation} 
for $n > 1$, and $x \in \torus^n$, such that $\varphi$ lies in arbitrarily small neighborhoods of the origin.
\end{theorema}

\begin{proof}
~\enumerationisinitium
\item[(\textgreek{α}) — \textbf{step I. Preparatory expressions}.]
~\subenumerationisinitium
\item Take a set $F = \{x \in \mathbb{R}^n\}$. We say that $F$ presents a full measure in $\mathbb{R}^n$. Let $x \in \mathbb{R}^n\backslash F$, $z \in \mathbb{Z}_+$, $k_1, \mathellipsis, k_z \in \mathbb{Z}_+$, $q_{k_1}, \mathellipsis, q_{k_z} \in \mathbb{Q}$ under wich $\sum^z_{\rotatedell = 1}q_{k_\rotatedell}|x - k_\rotatedell| = 0$. For this equation, we can identify a set $Q = \{k_1, \mathellipsis, k_z, q_{k_1}, \mathellipsis, q_{k_z}\}$ whose Lebesgue measure is 0, so
\begin{align}
	& \mathbb{R}^n\backslash F \subset \bigcup^\infty_{z = 1}\bigcup_{k_{1, \mathellipsis}}\bigcup_{q_{k_1, \mathellipsis}}Q, \\
	& k_{1, \mathellipsis} \viz k_1, \mathellipsis, k_z \in \mathbb{Z}^n, \notag \\
	& q_{k_1, \mathellipsis} \viz q_{k_1}, \mathellipsis, q_{k_z} \in \mathbb{Q}, \notag
\end{align}	
has Lebesgue measure 0.
\item Put $\textcyrillic{\textit{З}}_\Rorder^\alpha(x) = \sum_{|k| \leqslant \Rorder}\left(1 - \frac{|k|^2}{\Rorder^2}\right)^\alpha e^{2\pi ik \cdot x}$ as a kernel, $\alpha = \frac{n - 1}{2}$. Then
\begin{equation}
	\limsup_{\Rorder \to \infty}|\textcyrillic{\textit{З}}_\Rorder^\alpha(x)| = \infty,
\end{equation} 
for each $x \in F \cap \torus^{n \geqslant 2}$. Thereby we take a \emph{Bessel function} $\Bessel_{\frac{n}{2} + \alpha}$. If we choose $\Rorder \geqslant 1$, $x \notin \mathbb{Z}^n$, and $\alpha > 0$, then\footnote{
		\texttt{\textbackslash{Biggl}} $\textcolor{cynara-violet}{\lfloor}$ \emph{floor} and \texttt{\textbackslash{Biggr}} $\textcolor{cynara-violet}{\rceil}$ \emph{ceiling} notations are employed here improperly, viz. as simple splitters of an equation, when the latter is too long, and does not fit on a single line.
	}
\begin{align}
	& \Bessel_{\frac{n}{2} + \alpha}(2\pi\Rorder|x - k|) = \textcolor{cynara-violet}{\Biggl\lfloor} \notag \\
	& \frac{e^{2\pi i\Rorder|x - k|}e^{-i\frac{\pi}{2}(\frac{n}{2} + \alpha) - i\frac{\pi}{4}} + e^{-2\pi i\Rorder|x - k|}e^{i\frac{\pi}{2}(\frac{n}{2} + \alpha) + i\frac{\pi}{4}}}{\pi\sqrt{\Rorder|x - k|}} + \textgreek{\textit{ω}} \cdot \frac{1}{\left(\Rorder|x - k|\right)^\frac{3}{2}}\textcolor{cynara-violet}{\Biggr\rceil}, 
\end{align}
where $\textgreek{\textit{ω}}$ is a value related to the oscillation, since the Bessel function resembles an \emph{oscillating} (sine/cosine) \emph{function with a decay}. 

Thanks to the \emph{Poisson summation formula}, for any $x \in \torus^n\backslash\mathbb{Z}^n$, one has
\begin{equation}
	\textcyrillic{\textit{З}}_\Rorder^\alpha(x) = \frac{\textgreek{\text{Γ}}(\alpha + 1)}{\pi^\alpha}\Rorder^{\frac{n}{2} - \alpha}\sum_{k \in \mathbb{Z}^n}\frac{\Bessel_{\frac{n}{2} + \alpha}(2\pi\Rorder|x - k|)}{|x - k|^{\frac{n}{2} + \alpha}},
\end{equation}
where $\textgreek{\text{Γ}}$ is the \emph{gamma function}, with a convergence for $\alpha > \frac{n - 1}{2}$. Averaging over the $\Rorder$-th order, a factor of oscillation turns up, under which $\alpha = \frac{n - 1}{2}$ holds for the Bessel function. Setting $x \notin \mathbb{Z}^n$ and $\tau > 1$, we write
\begin{align}
	\frac{1}{\tau}\int^\tau_1\textcyrillic{\textit{З}}_\Rorder^\alpha(x)e^{2\pi i\beta\Rorder}d\Rorder & = \frac{\textgreek{\text{Γ}}(\alpha + 1)}{\pi^\alpha}\sum_{k \in \mathbb{Z}^n}\frac{e^{-i\frac{\pi}{2}(\frac{n}{2} + \alpha) - i\frac{\pi}{4}}}{|x - k|^{\frac{n + 1}{2} + \alpha}} \textcolor{cynara-violet}{\Biggl\lfloor} & \notag \\
	& \hspace{59pt} \frac{1}{\tau}\int^\tau_1e^{2\pi i\Rorder(\beta + |x - k|)}\left(\Rorder^{\frac{n - 1}{2} - \alpha}\right)d\Rorder\textcolor{cynara-violet}{\Biggr\rceil} \notag \\
	& + \frac{\textgreek{\text{Γ}}(\alpha + 1)}{\pi^\alpha}\sum_{k \in \mathbb{Z}^n}\frac{e^{i\frac{\pi}{2}(\frac{n}{2} + \alpha) + i\frac{\pi}{4}}}{|x - k|^{\frac{n + 1}{2} + \alpha}}\textcolor{cynara-violet}{\Biggl\lfloor} & \notag \\
	& \hspace{59pt} \frac{1}{\tau}\int^\tau_1e^{2\pi i\Rorder(\beta - |x - k|)}\left(\Rorder^{\frac{n - 1}{2} - \alpha}\right)d\Rorder\textcolor{cynara-violet}{\Biggr\rceil} \notag \\
	& + \frac{\textgreek{\text{Γ}}(\alpha + 1)}{\pi^\alpha}\sum_{k \in \mathbb{Z}^n}\textgreek{\textit{ω}} \cdot \frac{1}{|x - k|^{\frac{n + 3}{2} + \alpha}}\frac{1}{\tau}\int^\tau_1\left(\Rorder^{\frac{n + 3}{2} - \alpha}\right)d\Rorder.
\end{align}
For $\alpha \to \frac{n - 1}{2}$, the previous equation becomes 
\begin{align}
\label{align "Equation with oscillation (Bessel)"}
	\frac{1}{\tau}\int^\tau_1\textcyrillic{\textit{З}}_\Rorder^\frac{n - 1}{2}(x)e^{2\pi i\beta\Rorder}d\Rorder & = \frac{\textgreek{\text{Γ}}(\frac{n + 1}{2})}{\pi^\frac{n - 1}{2}}\sum_{k \in \mathbb{Z}^n}\frac{e^{-i\frac{\pi}{2}(\frac{2n - 1}{2}) - i\frac{\pi}{4}}}{|x - k|^n} \textcolor{cynara-violet}{\Biggl\lfloor} & \notag \\
	& \hspace{53pt} \frac{1}{\tau}\int^\tau_1e^{2\pi i\Rorder(\beta + |x - k|)}d\Rorder\textcolor{cynara-violet}{\Biggr\rceil} \notag \\
	& + \frac{\textgreek{\text{Γ}}(\frac{n + 1}{2})}{\pi^\frac{n - 1}{2}}\sum_{k \in \mathbb{Z}^n}\frac{e^{i\frac{\pi}{2}(\frac{2n - 1}{2}) + i\frac{\pi}{4}}}{|x - k|^n}\textcolor{cynara-violet}{\Biggl\lfloor} & \notag \\
	& \hspace{53pt} \frac{1}{\tau}\int^\tau_1e^{2\pi i\Rorder(\beta - |x - k|)}d\Rorder\textcolor{cynara-violet}{\Biggr\rceil} \notag \\
	& + \frac{\textgreek{\text{Γ}}(\frac{n + 1}{2})}{\pi^\frac{n - 1}{2}}\sum_{k \in \mathbb{Z}^n}\textgreek{\textit{ω}} \cdot \frac{1}{|x - k|^{n + 1}}\frac{1}{\tau}\int^\tau_1\frac{d\Rorder}{\Rorder}.
\end{align}
The first two sums show a convergence; this is due to the fact there exists a convergence of (Fourier) series for $\beta > \frac{n^2 + 1}{n + 1}$.

By that we have two options:

· if $\beta > 0$, and $\beta \neq |x - k_0|$, $k_0 \in \mathbb{Z}^n$, Eq. \eqref{align "Equation with oscillation (Bessel)"} is converging to 0 as $\tau \to \infty$, but

· if $\beta > 0$, and $\beta = \pm|x - k_0|$, Eq. \eqref{align "Equation with oscillation (Bessel)"} is converging to 
\begin{equation}
	\frac{\textgreek{\text{Γ}}(\frac{n + 1}{2})}{\pi^\frac{n - 1}{2}}\frac{e^{\pm i\{\frac{\pi}{2}(\frac{2n - 1}{2}) + i\frac{\pi}{4}\}}}{|x - k_0|^n} = \frac{\textgreek{\text{Γ}}(\frac{n + 1}{2})}{\pi^\frac{n - 1}{2}}\frac{e^{\pm i\frac{\pi(n)}{2}}}{|x - k_0|^n}.
\end{equation}
It follows that, for any point $x_0 \in F \cap \torus^n$,

· if $\beta = \beta_\rotatedell$, then $\lim_{\tau \to \infty}\frac{1}{\tau}\int^\tau_1\textcyrillic{\textit{З}}^\frac{n - 1}{2}_t(x_0)e^{2\pi i\beta t}dt = \frac{\textgreek{\text{Γ}}(\frac{n + 1}{2})}{\pi^\frac{n - 1}{2}}\frac{e^{i\frac{\pi(n)}{2}}}{\beta^n_\rotatedell}$,

· if $\beta = -\beta_\rotatedell$, then $\lim_{\tau \to \infty}\frac{1}{\tau}\int^\tau_1\textcyrillic{\textit{З}}^\frac{n - 1}{2}_t(x_0)e^{2\pi i\beta t}dt = \frac{\textgreek{\text{Γ}}(\frac{n + 1}{2})}{\pi^\frac{n - 1}{2}}\frac{e^{-i\frac{\pi(n)}{2}}}{\beta^n_\rotatedell}$,

· if $\beta \neq \pm\beta_\rotatedell$, then 0.

From there we can directly go to
\begin{align}
	& \lim_{\tau \to \infty}\frac{1}{\tau}\int^\tau_1\textcyrillic{\textit{З}}^\frac{n - 1}{2}_t(x_0)\prod^n_{\rotatedell = 1}\textcolor{cynara-violet}{\Biggl\lfloor} \notag \\ 
	& \frac{2 + e^{-i\frac{\pi(n)}{2}}e^{2\pi i(\beta_\rotatedell)_t} + e^{i\frac{\pi(n)}{2}}e^{-2\pi i(\beta_\rotatedell)_t}}{2}dt = \frac{\textgreek{\text{Γ}}(\frac{n + 1}{2})}{\pi^\frac{n - 1}{2}}\sum^n_{\rotatedell = 1}(\beta_\rotatedell)^{-n}\textcolor{cynara-violet}{\Biggr\rceil}.
\end{align}
Let us assume we have $\sup_{\Rorder \geqslant 1}\left|\textcyrillic{\textit{З}}_\Rorder^\frac{n - 1}{2}(x_0)\right| \leqslant Q_{x_0} < \infty$. Subsequently
\begin{align}
	\frac{\textgreek{\text{Γ}}(\frac{n + 1}{2})}{\pi^\frac{n - 1}{2}}\sum^n_{\rotatedell = 1}(\beta_\rotatedell)^{-n} & \leqslant Q_{x_0}\lim_{\tau \to \infty}\frac{1}{\tau}\int^\tau_1\prod^n_{\rotatedell = 1}\textcolor{cynara-violet}{\Biggl\lfloor} \notag \\ 
	& \frac{2 + e^{-i\frac{\pi(n)}{2}}e^{2\pi i(\beta_\rotatedell)_t} + e^{i\frac{\pi(n)}{2}}e^{-2\pi i(\beta_\rotatedell)_t}}{2}dt = Q_{x_0}\textcolor{cynara-violet}{\Biggr\rceil}, 
\end{align} 
which excludes $\sum^\infty_{\rotatedell = 1}(\beta_\rotatedell)^{-n} = \infty$. For that reason it must be 
\begin{equation}
	\sup_{\Rorder \geqslant 1}\left|\textcyrillic{\textit{З}}_\Rorder^\frac{n - 1}{2}(x_0)\right| = \infty,
\end{equation}
for each $x_0 \in F \cap \torus^n$.
\subenumerationisfinis
\item[(\textgreek{β}) — \textbf{step II. From Dirac delta to toroidal functions}.]  
Letting $\Diracdelta$ be the \emph{Dirac delta function} (Section \ref{subsection "Procrustean Bed: the Example of the Dirac Delta Function"}) for a point-mass at 0, we now rewrite the Bochner–Riesz means as 
\begin{equation}	
	\BochnerRiesz_\Rorder^\frac{n - 1}{2}\delta_0(x), 
\end{equation}
in which there is no convergence for almost $x \in \torus^n$. We need to find a sequence of integrable functions on the $n$-torus in place of the $\Diracdelta$-Dirac. 

We will start by using a radial function $\Theta_\rho \in \mathscr{C}^\infty$ on a Euclidean space $\mathbb{R}^n$, and putting 
\begin{equation}
	\theta_\epsilon(x) = \sum_{k \in \mathbb{Z}^n}\epsilon^{-n}\widehat{\Theta}_\rho\left(\frac{x + k}{\epsilon}\right) = \sum_{k \in \mathbb{Z}^n}\Theta(\epsilon \cdot k)e^{2\pi k \cdot x}, 
\end{equation}
so
\begin{equation}
\label{equation "Equation with radial function"}
	\sup_{x \in \torus^n}\sup_{\Rorder > 0}\left|\BochnerRiesz_\Rorder^\frac{n - 1}{2}\theta_\epsilon(x)\right| \leqslant \sum_{k \in \mathbb{Z}^n}\Theta(\epsilon \cdot k) \leqslant \sum_{k \in \mathbb{Z}^n}\frac{\dot{c}_n}{(1 + \epsilon|k|)^{n + 1}} \leqslant \left(\frac{c}{\epsilon}\right)^n,
\end{equation}
representing an estimate, for some constant $c_n$, where $\Theta(\epsilon \cdot k)$ denotes the $k$-th Fourier coefficient. Let $T_\rotatedell \subset \torus^n$ a $\rotatedell$-subset of the $n$-torus, under which 
\begin{equation}
	|T_\rotatedell| \geqslant \frac{\rotatedell - 1}{\rotatedell},
\end{equation}
and $0 < \Rorder_1 < \cdots < \Rorder_\rotatedell$. Hence let
\begin{equation}
\label{equation "equation with sup (intermediate step)"}
	\sup_{\Rorder \leqslant \Rorder_\rotatedell}\left|\BochnerRiesz_\Rorder^\frac{n - 1}{2}\left\{\sum^\infty_{m = 1}2^{-m}(\theta_{\epsilon_m} - \theta_{\Diracdelta_m})\right\}(x)\right| \geqslant \rotatedell, \text{with } \epsilon_\rotatedell \leqslant \Diracdelta_\rotatedell,
\end{equation}
for $x \in T_\rotatedell$. 

We adopt these values: $\Rorder_1 = 1$, $\epsilon_1 = \Diracdelta_1 = 1$, for $1 \leqslant \rotatedell \leqslant z - 1$ ($z > 1$). Given a distribution $\Diracdelta_z$, we proceed by defining the set
\begin{equation}
	Q_k = c_n \cdot 2^{-z}\delta^{-n}_z + c_n\sum^{z - 1}_{\rotatedell = 1}2^{-\rotatedell}(\epsilon_\rotatedell + \Diracdelta_\rotatedell)^{-n},
\end{equation}
which allows us to read Eq. \eqref{equation "Equation with radial function"} in these terms:
\begin{equation}
\label{equation "Equation with radial function (rewritten)"}
	\sup_{x \in \torus^n}\sup_{\Rorder > 0}\left|\BochnerRiesz_\Rorder^\frac{n - 1}{2}\left\{-2^{-z}\theta_{\Diracdelta_z} + \sum^{z - 1}_{\rotatedell = 1}2^{-\rotatedell}(\theta_{\epsilon_\rotatedell} - \theta_{\Diracdelta_\rotatedell})\right\}(x)\right| \leqslant Q_k.
\end{equation}
Applying Fatou's lemma \eqref{lemma "Fatou"}, we get
\begin{equation}
	\liminf_{n \to \infty}\left(\sup_{0 < \Rorder \leqslant n} \left|\BochnerRiesz_\Rorder^\frac{n - 1}{2}\Diracdelta_0(x)\right| > Q_z + z + 2\right) = 1, \enspace x \in \torus^n,
\end{equation}
where the $\Diracdelta_0$-Dirac is at the origin on the $n$-torus. If $\Rorder_z > \Rorder_{z - 1}$, it follows that 
\begin{equation}
	T_z = \left(\sup_{0 < \Rorder \leqslant \Rorder_z} \left|\BochnerRiesz_\Rorder^\frac{n - 1}{2}\Diracdelta_0(x)\right| > Q_z + z + 2\right), 
\end{equation}
whose measure is at least $\frac {z - 1}{z}$. Picking out $\epsilon_z \leqslant \Diracdelta_z$, we establish accordingly two expressions,
\begin{align}
	& \sup_{x \in \torus^n}\left|\BochnerRiesz_\Rorder^\frac{n - 1}{2}\Diracdelta_0(x) - \BochnerRiesz_\Rorder^\frac{n - 1}{2}\theta_{\epsilon_z}(x)\right| \textcolor{cynara-violet}{\Biggl\lfloor} \notag \\
	& \leqslant \sum_{|k| \leqslant \Rorder_z}\left(\frac{(\Rorder_z - |k|)(|k| + \Rorder_z)}{\Rorder_z^2}\right)^\frac{n - 1}{2}\left|1 - \widehat{\theta}_{\epsilon_z}(k)\right| \leqslant 1 \textcolor{cynara-violet}{\Biggl\rceil},
\end{align}
and
\begin{equation}
\label{equation "inf sup step"}
	\inf_{x \in T_z}\sup_{\Rorder \leqslant \Rorder_z}2^{-z}\left|\BochnerRiesz_\Rorder^\frac{n - 1}{2}\theta_{\epsilon_z}(x)\right| \geqslant Q_z + z + 1.
\end{equation}

Letting $C$ be a constant such that $|\Theta_\rho(x) - \Theta_\rho(y)| \leqslant C|x - y|$, $x, y \in \mathbb{R}^n$, and fixing $C \Diracdelta_z\bigl(\sum_{|k| \leqslant \Rorder_{z - 1}}|k| \leqslant 1 \bigr)$, we obtain 
\begin{align}
\label{align "sup sup step"}
	\sup_{x \in \torus^n}\sup_{\Rorder \leqslant \Rorder_{z - 1}}\left|\BochnerRiesz_\Rorder^\frac{n - 1}{2}\left(\theta_{\epsilon_z} - \theta_{\Diracdelta_z}\right)(x)\right| & \leqslant \sum_{|k| \leqslant \Rorder_{z - 1}}\left|\Theta_\rho(\epsilon_zk) - \Theta_\rho(\Diracdelta_zk)\right| \notag \\
	& \leqslant C(\Diracdelta_z - \epsilon_z)\sum_{|k| \leqslant \Rorder_{z - 1}}|k| \leqslant C\Diracdelta_z\sum_{|k| \leqslant \Rorder_{z - 1}}|k| \leqslant 1.
\end{align}

Eq. \eqref{equation "equation with sup (intermediate step)"} is now demonstrable for $\rotatedell = z$. First, we observe that
\begin{align}
	\BochnerRiesz_\Rorder^\frac{n - 1}{2}\left\{\sum^\infty_{m = 1}2^{-m}(\theta_{\epsilon_m} - \theta_{\Diracdelta_m})\right\}(x) & = \BochnerRiesz_\Rorder^\frac{n - 1}{2}\left\{-2^{-z}\theta_{\Diracdelta_z} + \sum^{z - 1}_{m = 1}2^{-m}(\theta_{\epsilon_m} - \theta_{\Diracdelta_m})\right\}(x) \notag \\
	& + \BochnerRiesz_\Rorder^\frac{n - 1}{2}\left(2^{-z}\theta_{\epsilon_z}\right)(x) \notag \\
	& + \BochnerRiesz_\Rorder^\frac{n - 1}{2}\left\{\sum^\infty_{m = z + 1}2^{-m}(\theta_{\epsilon_m} - \theta_{\Diracdelta_m})\right\}(x);
\end{align}
next, thanks to \eqref{equation "Equation with radial function (rewritten)"} \eqref{equation "inf sup step"} \eqref{align "sup sup step"}, we arrive at the supremum
\begin{equation}
	\sup_{\Rorder \leqslant \Rorder_{z - 1}}\left|\BochnerRiesz_\Rorder^\frac{n - 1}{2}\left\{\sum^\infty_{m = 1}2^{-m}(\theta_{\epsilon_m} - \theta_{\Diracdelta_m})\right\}(x)\right| \geqslant z, \enspace x \in T_z,
\end{equation}
which indicates the existence of \eqref{equation "equation with sup (intermediate step)"} under a selection of $(\rotatedell = z)$-values. After that, we determine the function
\begin{equation}
	\varphi = \sum^\infty_{m = 1}2^{-m}(\theta_{\epsilon_m} - \theta_{\Diracdelta_m}),
\end{equation}
where $\varphi$ maps into a Lebesgue space $\Lebesgue^{p = 1}(\torus^n)$, i.e. $\varphi$ is in $\Lebesgue^1(\torus^n)$. Finally, it is possible to demonstrate that
\begin{equation}
	\sup_{\Rorder > 0}\left|\BochnerRiesz_\Rorder^\frac{n - 1}{2}\varphi(x)\right| \infty,
\end{equation} 
Above we have established the value of $\epsilon_1$ at 1; but if we require that $\epsilon_1$ is arbitrarily small, $\varphi$ will have to lie in arbitrarily small neighborhoods of the origin and such that the main statement of the theorem is satisfied.
\enumerationisfinis
\end{proof}
 
\vspace{10mm}

\setcounter{secnumdepth}{0}  
\section{References and Bibliographic Details}
\setcounter{secnumdepth}{3}
\markright{References and Bibliographic Details}

\begingroup
\footnotesize
\noindent Section \ref{subsection "Summability of Fourier Series in Spherical Bochner–Riesz Means"}

\begin{indent paragraph: 15pt}
For an introduction to the Bochner–Riesz means, see \cite{Davis Chang "Lectures on Bochner-Riesz Means"} \cite{Lu Yan Bochner-Riesz "Means on Euclidean Spaces}.
\end{indent paragraph: 15pt}

\endgroup

\chapter{Variations on the Same Theme: Minima in the Calculus} 
\label{chapter "Variations on the Same Theme: Minima in the Calculus"} 

\begingroup
\footnotesize
A prejudice that must be eliminated considers mathematics interested only in quantitative aspects and not in qualitative aspects of things, it is believed that mathematics is the enemy of \emph{fantasy} and \emph{freedom}. Actually, mathematics, if well understood, broadens a person's grasp of imagination. For example, we would not have had all the development of modern physics if the \emph{mathematical imagination} had not arrived at the idea of infinite-dimensional spaces; likewise the idea of surface and manifold with various curvatures makes possible the imagination of a curved space-time.\endnote{
	Original It. version: «Un pregiudizio che va eliminato ritiene la matematica interessata solo agli aspetti quantitativi e non agli aspetti qualitativi delle cose, pensa che la matematica sia nemica della fantasia e della libertà. In realtà la matematica, se ben compresa, allarga le capacità di immaginazione di una persona. Per esempio non avremmo avuto tutto lo sviluppo della fisica moderna se l'immaginazione matematica non fosse arrivata all'idea di spazio a infinite dimensioni; parimenti l'idea di superficie e varietà con diverse curvature rende possibile l'immaginazione di uno spazio-tempo curvo».
	} \\
\indent — \textsc{E. De Giorgi} \cite[p. 723, e.a.]{De Giorgi "Lectio magistralis: Il valore sapienziale della matematica"}
	
\endgroup

\section{Prolegomenon: Geometrical Optics, Solid of Minimum Resistance, and Brachistochrone}
\sectionmark{Geometrical Optics, Solid of Minimum Resistance, and Brachistochrone}
\label{section "Prolegomenon: Geometrical Optics, Solid of Minimum Resistance, and Brachistochrone"}

In this chapter we will deal with the calculus of variations centered on the concept of \emph{minimum}; but first we will say a few words about some variational problems, which also include the concept of maximum.

By \emph{maximum} and \emph{minimum} problems we mean the research for optimality conditions or optimal approximation, both in the fields of the purely mathematical and physics sciences, or natural phenomena, always described by mathematics. These are just a few examples. 
\enumerationisinitium
\item The self-synchronization property of a certain number of mechanical metronomes on a left-right translating table: optimal automatic adjustment in oscillatory periodic motion with a ending in-phase synchrony (in left-right unison movement), starting from a non-synchronous condition.
\item The growth of a plant with a photoautotrophic capacity, i.e. the tendency to extract the maximum resources available in its terrestrial/aqua environment, evolving with an exclusive morphology. 
\item The hexagon-shaped cells in honeycombs: a bee's ability to pursue the least waste of space together with the greatest saving of wax.\footnote{
	The precise geometric comprehension of optimization in the bees' hexagonal wax cells, with maximum ratio for area/perimeter, is already in Pappus of Alexandria \cite[Liber quintus, pp. 73 recto and verso, 74 recto]{Pappus of Alexandria "Mathematicae Collectiones"}: \textit{tum maxime in apibus}, \textit{a tribus hexagonis, \& tribus hexagoni angulis}. Many centuries later, there is a description of J. Kepler \cite[Apum Alveoli, pp. 6-7]{Kepler "Strena Seu De Nive Sexangula"} on the \textit{ordine sexangulo} in honeycombs.
	}\textsuperscript{,}\footnote{
	In botanical research there are many experiments aimed at identifying geometric minimalities; from the past, see S. Hales \cite[Experiment XXXII, pp. 94-96]{Hales "Vegetable Staticks"}: an iron pot is filled with water and peas, and later covered with a leaden lid; after the necessary time, these peas end up assuming the shape of quasi-dodecahedra, as they dilated by imbibing the water.
	}
\item Investigation and experimentation on the most convenient construction of ship's hull structures: it is a minimization of water resistance; or on the best shape of an aircraft: aeroplane design to minimize the effects of drag forces (air resistance), and have an optimization of airfoils for a maximum value of lift coefficient.
\enumerationisfinis

\subsection{Fermat's Principle: Light Propagation}

\begingroup
\footnotesize
I do not claim nor have I ever claimed [\textit{n'ai jamais prétendu}] to be in Nature's secret confidence [\textit{être de la confidence secrète de la Nature}]. It has obscure and hidden ways [\textit{voies obscures et cachées}] which I have never set out to penetrate; I had only offered a little help in geometry on the subject of refraction [\,\dots]; I willingly abandon in your favor my pretend conquest in physics, and it is enough for me that you leave me in possession of my problem of geometry, entirely pure [\textit{tout pur}] and \emph{in abstracto}, whereby one can find the path of an object in motion which passes by two different media, and which seeks to complete its motion as soon as possible.\endnote{
	Original Fr. version: «[J]e ne prétends ni n'ai jamais prétendu être de la confidence secrète de la Nature. Elle a des voies obscures et cachées que je n'ai jamais entrepris de pénétrer; je lui avois seulement offert un petit secours de géométrie au sujet de la réfraction [\,\dots]; je vous abandonne de bon cœur ma prétendue conquête de physique, et il me suffit que vous me laissiez en possession de mon problème de géométrie tout pur et \emph{in abstracto}, par le moyen duquel on peut trouver la route d'un mobile qui passe par deux milieux différents et qui cherche d'achever son mouvement le plus tôt qu'il pourra».
	} \\
\indent — \textsc{P. de Fermat} \cite[lettre CXV to Clerselier, 21 mai 1662, p. 483]{de Fermat "Oeuvres de Fermat Tome Deuxieme (Correspondance)"}

\endgroup

\vspace{2mm}

The \emph{principle of least action} is closely linked with the \emph{principle of stationary action}, but beware, a stationary action does not always coincide with a minimum of a function. They are both rooted, back in time, in the earliest formulations of the \emph{calculus of variations}, at the end of the seventeenth century.

The calculus of variations is used to discover the maximum and minimum values of a functional, sometimes on the stimulation of responding to problems of description of certain natural phenomena. A historical example is the formulation of the laws of refraction, among which P. de Fermat's studies \cite{Fermat "Analyse pour les refractions"} \cite{Fermat "Synthese pour les refractions"} in geometrical optics stand out. It is assumed that a ray of light propagates from one point to another e.g. in homogeneous media, or in non-homogeneous media, undergoing a continuous variation of speed; after that, a mathematical elaboration is associated with a different behavior of the phenomenon. Fermat \cite[p. 125]{Fermat "Synthese pour les refractions"} writes:

\vspace{2mm}

\begingroup
\footnotesize
Our demonstration is based on this single postulate[,] that nature operates by the easiest and simplest means and routes [\,\dots] and not[,] as we keep saying[,] that nature always operates with the shortest lines [\textit{et non \textnormal{[\,\dots]} que la nature opère toujours par les lignes les plus courtes}].

\endgroup

\subsection{Newton's Problem of Fluid Resistance}

\begingroup
\footnotesize
\emph{Si Globus \& Cylindrus æqualibus diametris descripti, in Medio raro \& Elastico, secundum plagam axis Cylindri, æquali cum velocitate celerrime moveantur: erit resistentia Globi duplo minor quam resistentia Cylindri}.\footnote{
	\cite[Prop. XXXIV, Theor. XXVIII, pp. 117-119]{Newton "The Mathematical Principles of Natural Philosophy II"}: \emph{If in a rare medium, consisting of equal particles freely disposed at equal distances from each other, a globe and a cylinder described on equal diameters move with equal velocities, in the direction of the axis of the cylinder: the resistance of the globe will be but half so great as that of the cylinder}.
	} \\
\indent — \textsc{I. Newton} \cite[Liber II, Prop. XXXV, Theor. XXVIII, pp. 324-327]{Newton "Philosophiae Naturalis Principia Mathematica 1687"}

\endgroup

\vspace{2mm}

Optimization in fluid dynamics is a problem of Newtonian memory—in the above, reference was made to the naval architecture and hull construction. This is at the beginning of the variational calculus. Once reported in analytic language, with a system of orthogonal Cartesian axes, the problem posed by Newton is to determine a plane curve that joins two points, $A$ and $B$, and generates a solid of revolution about the $x$-axis (the $x$-axis coincides with the axis of rotation) in a homogeneous medium, encountering the \emph{smallest resistance}. 

The solution is to find a curve $y = y(x)$ which minimizes the integral
\begin{equation}
	F_{\textgreek{\text{μ}}} = \mathrm{v}_c\int^{x_b}_{x_a}\left(\frac{y\dot{y}^3}{1 + \dot{y}^2}\right)dx,
\end{equation}
letting $F_{\textgreek{\text{μ}}}$ the resistance of the rotating surface, where $\mathrm{v}_c$ is a constant depending on the speed of moving surface, while $x_a$ and $x_b$ are the abscissas of points $A$ and $B$ on the curve.

\subsection{Mechanics in the Shortest Time: the Cycloidal Curve}

\begingroup
\footnotesize
Problema novum ad cujus solutionem Mathematici invitantur—\emph{Datis in plano verticali duobus punctis $A$ \& $B$ assignare Mobili $M$, viam $AMB$, per quam gravitate sua descendens \& moveri incipiens a puncto $A$, brevissimo tempore perveniat ad alterum punctum $B$}.\footnote{
	«Given two points $A$ \& $B$ in a vertical plane[,] determine the path $AMB$ along which a moving [point-like mass, or particle] $M$, descending due to its gravity \& starting at $A$, reaches the other point $B$ in the shortest time».
	} \\
\indent — \textsc{Joh. Bernoulli} \cite[p. 269]{Bernoulli Joh. "Supplementum defectus Geometriae Cartesianae circa Inventionem Locorum"}

\endgroup

\vspace{2mm}

Another traditional problem of variational analysis is that of the brachistochrone curve, born from the intention to find the shortest path connecting two points, along which a point mass $M$ moves, or falls (assuming there is no friction): from a higher point, $A$, where the speed is zero, and all the $M$-energy is potential, to a lower one, $B$, the end-point, see epigraph of Joh. Bernoulli. The descent/travel time of $M$ is (with modern formalism)
\begin{equation}
	\mathscr{I}_t[\upsilon] = \frac{1}{\sqrt{2g}} \int^{x_2}_{x_1} \frac{\sqrt{1 + |\upsilon^{(\mathrm{i})}(x)|^2}}{\sqrt{y^1 - \upsilon(x)}}dx,
\end{equation}
where $g$ is the acceleration of gravity, $A = (x^1, y^1)$, $B = (x^2, y^2)$, letting $\upsilon \colon [x^1, x^2] \to \mathbb{R}$ be a function under which $x^1 < x^2$, $y^2 < y^1$, $\upsilon(x^1) = y^1$, $\upsilon(x^2) = y^2$, $\upsilon(x) < y^1$, for $x^1 < x \leqslant x^2$.  

The problem of the \emph{minimum descent/travel time}, for a point-like mass that slides down an inclined plane, under the graph of $\upsilon$, thus becomes,
\begin{equation}
	\min_{\upsilon \in U}\mathscr{I}_t[\upsilon], \enspace U = \Bigl\{\upsilon \in \mathscr{C}^1[x^1, x^2], \upsilon(x^1) = y^1, \upsilon(x^2) = y^2\Bigr\},
\end{equation} 
where $\min_{\upsilon \in U}\mathscr{I}_t[\upsilon]$ is the set of all curves joining $A$ with $B$. What we are looking for is the curve of least time in $\min_{\upsilon \in U}\mathscr{I}_t[\upsilon]$. The solution clarifies that the path $AMB$ is a \emph{cycloid}, so that a curve with the required \emph{temporal-minimality} satisfies an equation of the form 
\begin{equation}
	\frac{\partial}{\partial\upsilon^{(\mathrm{i})}}\left(\frac{\sqrt{1 + |\upsilon^{(\mathrm{i})}(x)|^2}}{\sqrt{y^1 - \upsilon(x)}}\right)\frac{d}{dx} - \frac{\partial}{\partial\upsilon}\left(\frac{\sqrt{1 + |\upsilon^{(\mathrm{i})}(x)|^2}}{\sqrt{y^1 - \upsilon(x)}}\right) = 0.
\end{equation}

This (cycloid) curve was named \emph{brachistochrone}, from the Gr. \textgreek{βράχιστος-χρόνος}, “shortest time”.
 
The \emph{problema} of Joh. Bernoulli was solved, among the giants, by I. Newton \cite{Newton "De Ratione Temporis quo grave labitur per rectam data duo puncta conjungentem"} \cite{Newton "Excerpta ex Transactionibus"}, G.W. Leibniz \cite{Leibniz "Communicatio suae pariter"}, the Bernoulli brothers, Johann \cite{Bernoulli Joh. "Curvatura radii in diaphanis non uniformibus"} \& Jacob \cite{Bernoulli Jac. "Solutio problematum fraternorum"}, and G.F.A. de L'Hôpital \cite{de L'Hopital "Solutio Problematis de Linea celerrimi Descensus"}.

But pay attention the two following matters.
\enumerationisinitium
\item The study of the space-path and the shortest traveling time of a moving point-like mass from the calm is nothing new; it is already in Galileo \cite[p. 150]{Galilei "Discorsi e dimostrazioni matematiche"}: «[\,\dots] naturalem motum gravium descendentium  continue accelerari [\,\dots], spatia à mobili descendente ex quiete».
\item The cycloid, before the puzzle of Bernoulli, was carefully analyzed by C. Huygens \cite[pp. 12-13, 155-156]{Huygens "Horologium oscillatorium"}, under the name of \emph{tautochrone} (or \emph{isochrone}) curve, i.e. \textgreek{ταὐτὸς-}, at the “same” (or \textgreek{ἰσο-}, in “equal”), \textgreek{-χρόνος}, “time”. The tautochrone is a curve along which a moving point-like mass reaches the lowest point on the curve always at the same time (assuming there is no friction), whatever the initial-point (or the starting position) is, for the gravitational acceleration. An example of tautochronism is the \emph{cycloidal pendulum clock}.\footnote{
	Joh. Bernoulli's comment \cite[p. 210]{Bernoulli Joh. "Curvatura radii in diaphanis non uniformibus"}, at the moment of the unexpected \emph{coincidence} between (his) brachistochrone and Huygens' tautochrone, is full of admiration: «Antequam finiam, non possum, quin iterum admirationem meam prodam, animo revolvens inexpectatam illam identitatem Tautochronæ \emph{Hugenianæ} nostræque \emph{Brachystochronæ}».
	}
\enumerationisfinis

\section{Theory of Minimal Surfaces}
\label{section "Theory of Minimal Surfaces"}

\subsection{Plateau's Problem: Soap Films and Bubbles}
\label{subsection "Plateau's Problem: Soap Films and Bubbles"}

\begingroup
\footnotesize
Another evident example of physical laws that translate principles of minimum or maximum is found in the laws of equilibrium of a weightless liquid, only subjected  to molecular forces. As Plateau has shown experimentally, by introducing olive oil in a mixture of water and alcohol, of equal specific weight, the weight of the oil is balanced by the buoyancy that, according to Archimedes' principle, it receives from the mixture in which it is immersed, and the oil behaves as if it were actually removed by the action of gravity [\,\dots]. Plateau proved, e.g., that a mass, in the previous and free conditions, assumes a spherical shape, that is, responding to the solution of the problem, in the Calculus of Variations, of a surface of minimal area bounding a given volume. And all the many experiences made by him [\,\dots] show that a minimum or maximum principle is always verified, the one of the minimum or maximum potential acting forces, corresponding, the minimum, at conditions of unstable equilibrium, the maximum, at conditions of stable equilibrium. And to this same conclusion also lead other Plateau's experiences on the equilibrium figures of very thin liquid films, obtained through the soap bubbles or by immersing iron wires in soap and water.\endnote{
	Original It. version: «Un altro esempio evidente di leggi fisiche che traducono principî di minimo o massimo si ha nelle leggi dell'equilibrio di un liquido senza peso, sottomesso soltanto alle forze molecolari. Come Plateau ha dimostrato sperimentalmente, introducendo dell'olio d'oliva in una miscela di acqua ed alcool, di ugual peso specifico, il peso dell'olio viene equilibrato dalla spinta che, secondo il principio di Archimede, esso riceve dalla miscela in cui è immerso, e l'olio si comporta come se fosse realmente sottratto all'azione della gravità [\,\dots]. Plateau provò, ad esempio, che una massa, nelle condizioni precedenti e libera, assume la forma sferica, rispondente cioè alla soluzione del problema, di Calcolo delle Variazioni, della superficie di area minima che racchiude un dato volume. E tutte le molteplici esperienze da lui fatte [\,\dots] mostrano che sempre risulta verificato un principio di minimo o massimo, quello del minimo o massimo del potenziale delle forze agenti, corrispondenti, il minimo, a condizioni di equilibrio instabile, il massimo, a condizioni di equilibrio stabile. Ed a questa medesima conclusione conducono anche altre esperienze, dello stesso Plateau, sulle figure di equilibrio delle lamine liquide sottilissime, ottenute mediante le bolle di sapone o immergendo in acqua e sapone dei supporti di fil di ferro».
	} \\
\indent — \textsc{L. Tonelli} \cite[pp. 9-10]{Tonelli "Fondamenti di Calcolo delle Variazioni I"}

\vspace{2mm}

[I]t became apparent that another of the classical extremum problems of analysis and geometry is intimately connected with Dirichlet's Principle [see Section \ref{subsubsection "An Example of Dirichlet's Assumption for the Minimal Surface Equation"}]. Since the early period of the calculus of variations, the problem of determining the surfaces of minimal area spanned in a given curve or subject to other boundary conditions has been attacked by many of the great mathematicians. Again physical experiments, such as those carried out by the Belgian physicist Plateau, lead immediately to the intuitive conviction that such problems can be solved. If a closed contour of wire is dipped into a soap solution, the liquid forms a film which, by virtue of the laws of surface tension, assumes as position of equilibrium the shape of a minimal surface spanned in the contour. \emph{But empirical evidence can never establish mathematical existence—nor can the mathematician's demand for existence proofs be dismissed by the physicist as useless rigor. (Only a mathematical existence proof can ensure that the mathematical description of a physical phenomenon is meaningful)}. \\
\indent — \textsc{R. Courant} \cite[p. 3, e.a.]{Courant "Dirichlet's Principle Conformal Mapping and Minimal Surfaces"}

\endgroup

\vspace{2mm}

Plateau's problem, already theoretically explored by Lagrange \cite{Lagrange "Essai d'une nouvelle methode pour determiner les maxima et les minima des formules integrales indefinies"}, has been meticulously treated by J.A.F. Plateau \cite{Plateau "Recherches experimentales et theorique sur les figures d'equilibre d'une masse liquide sans pesanteur"} \cite{Plateau "Statique experimentale et theorique des liquides soumis aux seules forces moleculaires I"} \cite{Plateau "Statique experimentale et theorique des liquides soumis aux seules forces moleculaires II"}, that we can generalize in the study of the behavior of \emph{soap films} (with boundary) and \emph{soap bubbles} (without boundary), and their explosion—suggestively called by Plateau «Théorie de l'explosion des bulles laminaires» \cite[§ 428]{Plateau "Statique experimentale et theorique des liquides soumis aux seules forces moleculaires II"}. This exercise was found to be an excellent method to visualize \emph{surfaces of minimum area}.

Prepare a soap-water blend, and a kit of (closed) wire frames with different shapes (e.g. ring, knot, helical line, Möbius strip, cube, etc.); immerse, one at a time, each wire in the soapy mixture. When it is pulled up, it is surprising to discover that soap films form in the portion bounded by the edge of the wire, each of which assumes a \emph{stable equilibrium position},\footnote{
	Of the many, one experimentally verifiable property is this: between component surfaces of the soap film, e.g. three minimal surfaces, the contact angle is $120^\circ$, whilst in a tetrahedral shape, the angle between lines of intersection (where the singular edges meet) is $109^\circ 28'16''$, an angle whose cosine is $-\frac{1}{3}$, known as \emph{Maraldi's angle}, see D.W. Thompson \cite[e.g. pp. 498, 549, 713]{Thompson "On Growth and Form"}.
	} 
and has a tendency to \emph{occupy minimum possible area}. Following Joh. \emph{Bernoulli's principle} of virtual displacements \cite{Bernoulli Joh. "Letter to Varignon 26 January 1717"}, the \emph{potential energy} in the \emph{surface tension} of the soap film is a \emph{minimum}.

\subsubsection{Quick Report on Solutions of Plateau's Problem}

\begingroup
\footnotesize
The Plateau Problem consists in showing that the greatest lower bound of the areas of surfaces with a given boundary is attained. This depends primarily on the \emph{meaning} we attach to the word \emph{surface}. \\
\indent — \textsc{E.R. Reifenberg} \cite[p. 1, e.a.]{Reifenberg "Solution of the Plateau problem for $m$-dimensional surfaces of varying topological type"}

\endgroup

\vspace{2mm}

\emph{Experimental} procedure à la Plateau (water, soap, and wire)—with a view to reproduce, simulate and determine, at least conceptually, a \emph{phenomenon of nature}—guide but do not verify the accuracy of the representation, as in mathematics, of the \emph{physical phenomenon observable} with soap films and bubbles. \emph{Plateau's problem} stems from the need to string together a good \emph{mathematical interpretation/understanding} of the generation of boundary minimal surfaces within this experimental guide.

Mathematical solutions to the Plateau problem were presented by various authors, with several techniques, see \cite{Harrison "Soap Film Solutions to Plateau's Problem"}. A roundup to follow.
\enumerationisinitium
\item R. Garnier \cite{Garnier "Le probleme de Plateau"} gives a demonstration for general polygonal boundary curves, or rather, for \emph{piecewise smooth Jordan curves} (which are piecewise smooth simple closed curves, i.e. loops), in Euclidean 3-space. 
\item T. Radó \cite{Rado "The problem of the least area and the problem of Plateau"} \cite{Rado "On Plateau's problem"}; his proof is in Euclidean 3-space. In \cite[p. 458]{Rado "On Plateau's problem"} he writes: 

\vspace{2mm}

\begingroup
\footnotesize
Plateau's problem implies that the minimal surface which we are seeking is \emph{of the type of the circle}. There remains therefore the more general problem of determining the minimal surfaces \emph{of all possible topological types} bounded by the given curve [\,\dots]. [T]here remains the problem of proving the existence of a surface, bounded by the given curve, with a minimum area; indeed, a minimal surface bounded by the given curve is not necessarily the solution of this problem of minimum area.

\endgroup

\vspace{2mm}

\item J. Douglas \cite{Douglas "The Mapping Theorem of Koebe and the Problem of Plateau"} \cite{Douglas "Solution of the problem of Plateau"} prepares another demonstration in Euclidean 3-space. His way is in the use of a functional of this form
\begin{align}
	\mathscr{I}_{\textsc{d}\mathrm{o}}[\upsilon] & = \frac{1}{4\pi}\int^{2\pi}_0\int^{2\pi}_0\frac{|\upsilon(\theta) - \upsilon(\phi)|^2}{4\sin^2\frac{1}{2}(\theta - \phi)}d\theta d\phi, \notag \\
	& = \frac{1}{4\pi}\int^{2\pi}_0\int^{2\pi}_0\frac{|\upsilon(e^{i\theta}) - \upsilon(e^{i\phi})|^2}{|e^{i\theta} - e^{i\phi}|^2}d\theta d\phi,
\end{align}
$e^{i\theta} = \upsilon(\cos{\theta}, \sin{\theta})$, videlicet, a function 
\begin{equation}
	\mathscr{I}_{\textsc{d}\mathrm{o}}[\upsilon] \colon \mathscr{C}^0_{2\pi}(\mathbb{R}, \mathbb{R}^3) \to \mathbb{R}, 
\end{equation}
where $\mathscr{I}_{\textsc{d}\mathrm{o}}$ is called \emph{Douglas functional}, $\upsilon \colon \mathbb{R} \to \mathbb{R}^3$ is some continuous function $2\pi$-periodic whose class is $\mathscr{C}^0_{2\pi}(\mathbb{R}, \mathbb{R}^3)$, such that $\upsilon(\theta + 2\pi) = \upsilon(\theta)$, for all $\theta \in \mathbb{R}$. 
\item The \emph{Radó–Douglas'} result is the first fairly general solution of Plateau's problem for many special boundary contours, that is, for many special layers of soapy liquid that span a wire contour, acting as a boundary. 

We here summarise it simply, employing a proposition of disks. \emph{Let $\Jordancurve \subset \mathbb{R}^3$ be a $\mathscr{C}^1$ piecewise smooth Jordan curves; there is a function $\upsilon \colon \mathbb{D} \subset \mathbb{R}^2 \to \mathbb{R}^3$ under which}
\subenumerationisinitium	
\item \emph{$\upsilon \colon \partial\mathbb{D} \to \Jordancurve$ is monotone, so the inverse image of a connected set (under a continuous function) is also connected,
\item $\upsilon \in \mathscr{C}^0(\overbar{\mathbb{D}}) \cap \Sobolev^{1, 2}(\mathbb{D})$, and $\upsilon \in \mathscr{C}^\infty(\mathbb{D})$,
\item the image of $\upsilon$ produces an area minimization among all disks with boundary contour $\Jordancurve$}.
\subenumerationisfinis
\item E.R. Reifenberg \cite{Reifenberg "Solution of the Plateau problem for $m$-dimensional surfaces of varying topological type"} \cite{Reifenberg "An Epiperimetric Inequality Related to the Analyticity of Minimal Surfaces"} \cite{Reifenberg "On the Analyticity of Minimal Surfaces"}  studies compact subsets with boundaries, associated with a \emph{minimization process of the Hausdorff measure in dimension 2}, from which both a \emph{generalization} of the minimal surface and a \emph{regularization} of it are attained. For subsets of higher dimension, the minimization process is applied directly on the \emph{$n$-dimensional Hausdorff measure}. The next step is to show that a minimal surface is \emph{analytic} at points where the surface density is (near) 1. 
\item C.B. Morrey \cite{Morrey "The problem of Plateau on a Riemannian manifold"} \cite{Morrey "The Higher-Dimensional Plateau Problem on a Riemannian Manifold"} extends Reifenberg's research inherent in a surface of minimal area bounded by a Jordan contour in Euclidean space, jumping to the condition of minimality in Riemannian spaces. 
\item By combining the above works, we have the \emph{Reifenberg–Morrey's} result. \emph{Given 
\subenumerationisinitium
\item[\textnormal{·}] a smooth, and metrically complete, Riemannian space $\mathcal{M}$, 
\item[\textnormal{·}] a compact measurable subset $\invertedbreve{N} \viz \mathcal{N} \subset \mathcal{M}$ of dimension $k - 1$, 
\item[\textnormal{·}] a compact Abelian group $G$, 
\subenumerationisfinis 
we assume that a measurable subset $\invertedbreve{Q} \hookrightarrow \mathcal{M}$ is embedded in the $\mathcal{M}$-space, with $G$ as a surface, and $\invertedbreve{N}$ as a boundary. Denoting by $\{\invertedbreve{Q}\}$ the class of all $G$-surfaces with boundary $\invertedbreve{N}$, then there exists a minimal surface $\invertedbreve{Q}_0$ in $\{\invertedbreve{Q}\}$ under which $\volume_k(\invertedbreve{Q}_0\backslash\invertedbreve{N})$. The surface $\invertedbreve{Q}_0\backslash\invertedbreve{N}$ is an open and smooth minimal space, i.e. it is differentiable infinitely many times, except for the set of points $\invertedbreve{\Sigma}$ whose $k$-dimensional measure is zero. Plus, if the Riemannian $\mathcal{M}$-space is analytic, the subspace $\invertedbreve{Q}_0\backslash(\invertedbreve{N} \cup \invertedbreve{\Sigma})$ is equally analytic.} 
\item E. De Giorgi \cite{De Giorgi "Complements to the $(n - 1)$-dimensional measure theory in a $n$-dimensional space"} \cite{De Giorgi "Area-minimizing oriented boundaries"} proves that certain \emph{minimal boundaries} are \emph{analytic}, more specifically, it is a case of \emph{locally analytic hypersurfaces},\footnote{
	A hypersurface is said to be a manifold or algebraic variety in $\mathbb{R}^n$ having one less dimension than that of the ambient (Euclidean) space, that is, $n - 1$.
	} 
with regard to \emph{area minimizing oriented boundaries}, through the theory of perimeters \& Caccioppoli (dimensionally oriented) sets \cite{Caccioppoli "Misura e integrazione sugli insiemi dimensionalmente orientati"} (see \cite{Ambrosio "La teoria dei perimetri di Caccioppoli-De Giorgi e i suoi piu recenti sviluppi} as a compendium); this is followed by the finding of \emph{regularity} of \emph{minimal surfaces}, or of \emph{local hypersurfaces of class $\mathscr{C}^\infty$}, supplying solutions to the problem of \emph{minimum $(n - 1)$-area in a Euclidean $n$-space} (see below, Section \ref{subsubsection "Locally Regular Hypersurface in the Caccioppoli–De Giorgi Theory of Finite Perimeters"}). In \cite{De Giorgi "Hypersurfaces of minimal measure in pluridimensional Euclidean spaces"} De Giorgi tackles the theory of minimal hypersurfaces embedded in Euclidean $n$-space, with $n > 3$, and restructures the Plateau's problem in Cartesian form.
\item F.J. Almgren \cite{Almgren "Plateau's Problem. An invitation to Varifold Geometry} re-contextualizes the problem of Plateau within the concept of \emph{varifold}, which is a differentiable manifold in geometric measure theory (mainly designed for the calculus of variations) \cite{Almgren "Questions and Answers about Area-Minimizing Surfaces and Geometric Measure Theory"}, cf. W.K. Allard \cite{Allard "On the first variation of a varifold: Boundary behavior"}. In \cite{Almgren Simon "Existence of Embedded Solutions of Plateau's Problem"} Almgren and L. Simon show that, \emph{given a minimal 2-surface $\surface^2$, and a uniformly convex open set $\Omega \subset \mathbb{R}^3$, if $\partial\Omega$ is a $\mathscr{C}^2$ surface with boundary, and $\Jordancurve$ is a rectifiable Jordan curve (a circle, by the way) of class $\mathscr{C}^3$ contained in $\partial\Omega$, so that $\partial\surface^2 = \Jordancurve \subset \partial\Omega$, then $\surface^2$ turns out to be embedded in $\Omega$, for which $\surface^2$ is diffeomorphic to a disk $\mathbb{D} \viz \mathbb{D}^2$, i.e., $\mathbb{B}^2$, and is area minimizing under a diffeomorphism $\varphi \colon \mathbb{D} \to \varphi(\mathbb{D})$, letting $\Jordancurve = \partial[\varphi(\mathbb{D})]$}.
\enumerationisfinis

\subsubsection{Locally Regular Hypersurface in the Caccioppoli–De Giorgi Theory of Finite Perimeters}
\label{subsubsection "Locally Regular Hypersurface in the Caccioppoli–De Giorgi Theory of Finite Perimeters"}

Here is a an example of hypersurface having regularity carried out in the manner of De Giorgi \cite{De Giorgi "Complements to the $(n - 1)$-dimensional measure theory in a $n$-dimensional space"} \cite{De Giorgi "Area-minimizing oriented boundaries"}, combined, in the end, with a H. Federer's elaboration \cite{Federer "The singular sets of area minimizing rectifiable currents with codimension one and of area minimizing flat chains modulo two with arbitrary codimension"}.

\begin{definitio}
A Caccioppoli set $\invertedbreve{C} \subset \mathbb{R}^n$ is a set of \emph{locally finite perimeter the boundary of which is measurable}. Note. A Caccioppoli set can coincide with a Borel set $\invertedbreve{E}$ (cf. point \ref{item "Borel measure"} in Section \ref{subsubsection "On the Lindelöf Space, Borel Measure plus sigma-Algebra"}); in the latter case, $\invertedbreve{C} = \invertedbreve{E}$ iff the perimeter of $\invertedbreve{E}$ is finite in any bounded open set $\Omega \subset \mathbb{R}^n$. \definitiosymbol
\end{definitio}

\begin{theorema}[Minimal boundary of Caccioppoli set, and regularity of its reduced boundary]
Take an open set $\Omega \subset \mathbb{R}^n$, and a Caccioppoli set $\invertedbreve{C} \subset \mathbb{R}^n$, with $n \geqslant 2$. Let $\partial^-\invertedbreve{C}$ denote the \emph{reduced boundary} of $\invertedbreve{C}$, i.e., the set of points $y$ satisfying
\subenumerationisinitium
\item an integral of the form
\begin{equation}
	\int_{\Omega_\rho(y)}|D\textgreek{\textit{χ}}(x, \invertedbreve{C})| > 0, 
\end{equation} 
where $\textgreek{\textit{χ}}(x, \invertedbreve{C})$ is the \emph{characteristic function} of $\invertedbreve{C}$, so that $\textgreek{\textit{χ}}$ will be 1 in $\invertedbreve{C}$ and 0 in $\mathbb{R}^n\backslash \invertedbreve{C}$, and $\rho$ is a positive number, 
\item a limit 
\begin{equation}
	\lim_{\rho \to 0}\frac{\int_{\Omega_\rho(y)}D\textgreek{\textit{χ}}(x, \invertedbreve{C})}{\int_{\Omega_\rho(y)}|D\textgreek{\textit{χ}}(x, \invertedbreve{C})|} = \hat{N}_{\invertedbreve{C}}(y),
\end{equation}
$\hat{N}_{\invertedbreve{C}}$ being the (unit) inner normal vector to $\invertedbreve{C}$,
\item and a condition $|\hat{N}_{\invertedbreve{C}}(y)| = 1$.
\subenumerationisfinis

If the boundary of $\invertedbreve{C}$ is minimal on $\Omega$, then $\partial^- \invertedbreve{C} \cap \Omega$ is a locally regular hypersurface, and it corresponds (Federer's elaboration) to an analytic $(n - 1)$-space. 
\end{theorema}

\begin{proof}
Given a point $y \in \partial^-\invertedbreve{C} \cap \Omega$, we denote by $\Pi(\invertedbreve{C})$ the perimeter of $\invertedbreve{C}$. It is possible to find some positive number $\rho$ that meets the following statement: consider a sequence of Caccioppoli sets $\{\invertedbreve{C}_\rotatedell\}$; when $\{\invertedbreve{C}_\rotatedell\}$ satisfies
\begin{subequations}
\begin{align}
	& \Pi(\invertedbreve{C}) = \lim_{\rotatedell \to \infty}\Pi(\invertedbreve{C}_\rotatedell), \\
	& \sum^\infty_{\rotatedell = 1}\int_{\mathbb{R}^n}|\textgreek{\textit{χ}}(x, \invertedbreve{C}) - \textgreek{\textit{χ}}(x, \invertedbreve{C}_\rotatedell)|dx < \infty,
\end{align}
\end{subequations}
these expressions hold 
\begin{subequations}
\begin{align}
	& \int_{\overbar{\mathbbl{B}}_\tau}|D\textgreek{\textit{χ}}(x, \invertedbreve{C})| = \lim_{\rotatedell \to \infty}\int_{\overbar{\mathbbl{B}}_\tau}|D\textgreek{\textit{χ}}(x, \invertedbreve{C}_\rotatedell)|, \\
	& \varphi(\invertedbreve{C}, \overbar{\mathbbl{B}}_\tau) = \lim_{\rotatedell \to \infty}\varphi(\invertedbreve{C}_\rotatedell, \overbar{\mathbbl{B}}_\tau),
\end{align}  
\end{subequations}
for almost every positive number $\tau$, where $\overbar{\mathbbl{B}}$ is a set of points, $\varphi$ is a function, with
\begin{equation}
	\lim_{\rotatedell \to \infty}\Pi(\invertedbreve{C}_\rotatedell \cap \overbar{\mathbbl{B}}_\tau) = \Pi(\invertedbreve{C} \cap \overbar{\mathbbl{B}}_\tau), \enspace \lim_{\rotatedell \to \infty}\Pi(\invertedbreve{C}_\rotatedell\backslash\overbar{\mathbbl{B}}_\tau) = \Pi(\invertedbreve{C}\backslash\overbar{\mathbbl{B}}_\tau). 
\end{equation}
\enumerationisinitium
\item We shall indicate by $\mathbbl{h}$ the \emph{Hausdorff measure}. If $\invertedbreve{C}$ is a quasi-regular domain in Euclidean space $\mathbb{R}^n$, ergo $\Pi(\invertedbreve{C})  = \mathbbl{h}^{n - 1}(\partial\invertedbreve{C})$, and $\mathbbl{h}^{n - 1}(\partial\invertedbreve{C}\backslash\partial^-\invertedbreve{C}) = 0$. 
\item Since we assume that $\partial\invertedbreve{C} \cap \Omega = \partial^-\invertedbreve{C} \cap \Omega$, and that the inner normal vector 
\begin{equation}
	\hat{N}_{\invertedbreve{C}}(x) = \frac{D\textgreek{\textit{χ}}(x, \invertedbreve{C})}{|D\textgreek{\textit{χ}}(x, \invertedbreve{C})|}
\end{equation} 
is continuous in $\partial\invertedbreve{C} \cap \Omega$, one understands that $\partial\invertedbreve{C} \cap \Omega$ is a locally regular hypersurface. 
\enumerationisfinis

The conclusion of all this is that the set $\partial^-\invertedbreve{C} \cap \Omega_{2^{-n}\rho}(y)$ is a locally regular hypersurface.
\end{proof}

\subsubsection{Min-max Conditions (Almgren–Pitts Theory) for Minimal Surfaces}

Closely related to the theory of minimal surfaces, and Plateau-like problems, there is the so-called \emph{min-max theory for minimal surfaces}, also known as \emph{Almgren–Pitts min-max theory}. It consists in calculating the existence of minimal surfaces with minimum points overlapping to maximum points (such as in the saddle points), so \emph{minimum and maximum coincide}. 
\enumerationisinitium
\item The min-max theory were initially analyzed by G.D. Birkhoff  \cite{Birkhoff "Dynamical Systems with Two Degrees of Freedom"} in 2-space—an equatorial curve on a sphere is interpretable as a (closed) geodesic on a minimal surface, which is, geometrically, a min-max condition; later, it was widely redefined and extended from 3- to 7-spaces by Almgren and J.T. Pitts \cite{Pitts "Existence and Regularity of Minimal Surfaces on Riemannian Manifolds"}. 
\item Generalizations of the min-max theory for minimal surfaces are in F.C. Marques and A. Neves, in collaboration with K. Irie \cite{Irie Marques and Neves "Density of minimal hypersurfaces for generic metrics"} and A. Song \cite{Marques Neves and Song "Equidistribution of minimal hypersurfaces for generic metrics"}: \emph{given a closed manifold $\mathcal{M}^{n + 1}$, putting $3 \leqslant (n + 1) \leqslant 7$, for almost every Riemannian metric $g$ of class $\mathscr{C}^\infty$, such that $\mathscr{C}^\infty$ is a Baire set \textnormal{\cite{Baire "Sur les fonctions de variables reelles"}} on the $(n + 1)$-manifold, it is proven that}

· \emph{the union of any closed, smooth, embedded minimal hypersurface has the property of density, with the implication of infinitely many minimal hypersurfaces—which solves a Yau's conjecture \textnormal{\cite{Yau "Problem Section"}}\footnote{
	\cite[problem № 88, p. 689-690]{Yau "Problem Section"}: \emph{Prove that any three-dimensional manifold must contain an infinite number of immersed minimal surfaces}. The complete proof of the Yau conjecture is the work of A. Song \cite{Song "Existence of infinitely many minimal hypersurfaces in closed manifolds"}, and it is true for all closed Riemannian manifolds of \emph{dimension at least 3 and at most 7}.
	} 
for generic metrics—under the min-max conditions;}

· \emph{a collection $\{\surface_k\}_{k \in \mathbb{N}}$ of closed, smooth, embedded, connected minimal hypersurfaces is uniform, or equidistributed, in the $(n + 1)$-manifold such that, for some $\textcyrillic{\textit{я}} \in \mathscr{C}^\infty$,}
\begin{equation}
	\lim_{r \to \infty}\frac{1}{\sum^r_{k = 1}\volume_g(\surface_k)}\sum^r_{k = 1}\int_{\surface_k}\textcyrillic{\textit{я}}d\surface_k = \frac{1}{\volume_g(\mathcal{M}^{n + 1})}\int_{\mathcal{M}^{n + 1}}\textcyrillic{\textit{я}}d\mathcal{M}^{n + 1}.
\end{equation}

Note. The volume considered of $\mathcal{M}^{n + 1}$ is a non-decreasing sequence of numbers $\{\textcyrillic{\textit{ч}}^\volume_k(\mathcal{M}^{n + 1})\}_{k \in \mathbb{N}}$, known as \emph{Weyl law for the volume spectrum} \cite{Weyl "Ueber die asymptotische Verteilung der Eigenwerte"},\footnote{ 
	Conjectured by M.L. Gromov \cite{Gromov "Dimension non-linear spectra and width"} \cite{Gromov "Isoperimetry of waists and concentration of maps"}, it is proved in \cite{Liokumovich Marques and Neves "Weyl law for the volume spectrum"}. 
	}
in keeping with a min-max procedure, under which
\begin{equation} 
	\lim_{k \to \infty}\textcyrillic{\textit{ч}}^\volume_k(\mathcal{M}^{n + 1})k^{-\frac{1}{n + 1}} = c_\textcyrillic{\textit{ч}}(n)\volume(\mathcal{M}^{n + 1})^{\frac{n}{n + 1}}, 
\end{equation}
setting $c_\textcyrillic{\textit{ч}}(n) > 0$ as a universal constant, where $c_\textcyrillic{\textit{ч}}(n) = 4\pi^2\text{volume of }\mathbb{B}_{(\rho = 1)}^{-\frac{2}{n + 1}}$, $\mathbb{B}_{(\rho = 1)}$ being the unit ball in $\mathbb{R}^{n + 1}$.
\enumerationisfinis

\subsubsection{Singularity and Radiolaria: a Math-Schema for a Biology}

\begingroup
\footnotesize
[T]here is a kind of microscopic sea life called Radiolaria which beautifully illustrates part of the main theorem of this paper. According to Thompson \cite{Thompson "On Growth and Form"}, these animals, when alive, are a small mass of protoplasm surrounded by a “froth” of cells [vacuoles, or alveoli]. As in soap films, the fluid in the interfaces of the froth accumulates most in the branchings, and the animal apparently acquires a skeletal structure by depositing a solid out of the fluid. When the animal dies, everything dissolves but the skeleton—in effect, the surface disappears, leaving just the singularities behind and provides a unique picture of singularity structure. \\
\indent — \textsc{J.E. Taylor} \cite[p. 493]{Taylor "The structure of singularities in soap-bubble-like and soap-film-like minimal surfaces"}

\endgroup

\vspace{2mm}

In a mathematical work, evocatively entitled \textit{The structure of singularities in soap-bubble-like and soap-film-like minimal surfaces}, J.E. Taylor \cite{Taylor "The structure of singularities in soap-bubble-like and soap-film-like minimal surfaces"} makes a classification of the local structure of singularities in a broad class of 2-dimensional minimal surfaces in $\mathbb{R}^3$, echoing the sorting of Almgren \cite{Almgren "Existence and regularity almost everywhere of solutions to elliptic variational problems with constraints"}. But the soapy singularity is not relegated to pure mathematics; its extension to biology is vast and suggestive. In support of his classification, Taylor recalls various shapes of \emph{Radiolaria}, already meticulously studied in the nineteenth century by E. Haeckel (zoologist),\footnote{
	A collection of Haeckel's marvelous drawings of Radiolaria is printed in his \textit{Report on the Radiolaria collected by H.M.S. Challenger during the years 1873-1876}, in \textit{Report on the Scientific Results of the Voyage of H.M.S. Challenger. Zoology—Vol. XVIII. Plates}, Order of Her Majesty's Government, London–Edinburgh–Dublin, 1887.
	} 
and in the twentieth century by D.W. Thompson \cite[pp. 426-452, 673-679, 694-740]{Thompson "On Growth and Form"} (biologist and mathematician).

The Radiolaria are marine protozoa, see Haeckel \cite[pp. 1-4]{Haeckel "Report on the Radiolaria collected by H.M.S. Challenger during the years 1873-1876"}, a class of the \emph{Protista} (unicellular organisms), divided by a porous membrane into an internal or intracapsular part, with nucleus, and an external or extracapsular part. The central capsule, the inner part of Radiolaria, is composed of

· a central nucleus, 
 
· an intracapsular or inner sarcode (endoplasm), or even a surrounding internal protoplasm,

· a capsule-membrane, that is, an enveloping porous membrane, 

· an internal or intracapsular skeleton,

· intracapsular vacuoles, or alveoli.

The \emph{extracapsulum}, the outer part of Radiolaria, is composed of

· a thick extracapsular jelly-veil (\emph{Calymma}), enveloping the whole central capsule,

· a maternal tissue of the external protoplasm, enveloping the capsule-membrane,

· pseudopodia, as needle-like protuberances, or filaments of protoplasm, radiating from the maternal tissue,

· extracapsular vacuoles, or alveoli.

The \emph{extracapsulum}, with its frothy vacuoles, or alveoli, exhibits an arrangement generating a reticular pattern, under a surface tension proportional to the area of the capsule-membrane. The radiolarian siliceous skeleton \cite[pp. lxviii-xcii]{Haeckel "Report on the Radiolaria collected by H.M.S. Challenger during the years 1873-1876"} frequently has spiny protrusions, while in some cases looks like a reproduction of regular polyhedra.

\subsection{Bernstein's Problem}

\begingroup
\footnotesize
If a minimal surface $S$ is represented by the equation $z = f(x, y)$[,] where $f$ admits continuous derivatives of the first two orders for any real value of $(x, y)$, [then] the surface $S$ reduces to a plan.\endnote{
	Original Fr. version: «Si une surface minima $S$ est représentée par l'équation $z = f(x, y)$ où $f$ admet des dérivées continues des deux premiers ordres pour toute valeur réelle de $(x, y)$, la surface $S$ se réduit à un plan».
	} \\
\indent — \textsc{S. Bernstein} \cite[p. 44]{Bernstein "Sur un theoreme de geometrie et son application aux equations aux derivees partielles du type elliptique"} 

\endgroup

\vspace{2mm}

In epigraph the Bernstein theorem \cite{Bernstein "Sur un theoreme de geometrie et son application aux equations aux derivees partielles du type elliptique"} is exposed in its original version, and it is solved in \emph{two dimensions}. We can reformulate it in another way (different words but same concept): 

\begin{theorema}[Bernstein]
\label{theorema "Bernstein"}
Given a $\mathscr{C}^2$ function $\upsilon(x, y) \colon \mathbb{R}^2 \to \mathbb{R}$ solving the minimal surface equation in the plane, namely on $\mathbb{R}^2$,
\begin{equation}
\label{equation "Minimal surface equation in the plane"}
	\frac{\partial}{\partial{x}}\left(\frac{\upsilon_x}{\sqrt{1 + \upsilon^2_x + \upsilon^2_y}}\right) + \frac{\partial}{\partial{y}}\left(\frac{\upsilon_y}{\sqrt{1 + \upsilon^2_x + \upsilon^2_y}}\right) = 0,
\end{equation}
it happens that $\upsilon(x, y) = c_{(1)}x + c_{(2)}y + c_{(3)}$ must be an affine linear function, for three constants $c_{(1)}, c_{(2)}, c_{(3)} \in \mathbb{R}$, and the graph of $\upsilon$ is a plane.
\end{theorema}

\begin{proof}
E. Hopf \cite{Hopf "On S. Bernstein's Theorem on Surfaces z(x y) of Nonpositive Curvature"} corrects a gap of topological kind in Bernstein's primary demonstration, and E.J. Mickle \cite{Mickle "A Remark on a Theorem of Serge Bernstein"} works out a simpler version of it. Here we follow the demonstrative line of J.C.C. Nitsche \cite{Nitsche Elementary Proof of Bernstein's Theorem on Minimal Surfaces"} because of its elementarity: it makes use of Liouville's Theorem for a simple holomorphic (hence analytic) function of one complex variable. 

We replace Eq. \eqref{equation "Minimal surface equation in the plane"} with
\begin{equation}
	(1 + \upsilon^2_y)\upsilon_{xx} - 2\upsilon_x\upsilon_y\upsilon_{xy} + (1 + \upsilon^2_x)\upsilon_{yy} = 0. 
\end{equation} 
We choose a convex polynomial $\mathrm{p}(x, y)$ satisfying $\mathrm{p}_{xx}\mathrm{p}_{yy} - \mathrm{p}^2_{xy} = 1$, with 
\begin{equation}
	\mathrm{p}_{xx} = \frac{1 + \upsilon^2_x}{\sqrt{1 + \upsilon^2_x + \upsilon^2_y}}, 
	\enspace 
	\mathrm{p}_{xy} = \frac{\upsilon_x\upsilon_y}{\sqrt{1 + \upsilon^2_x + \upsilon^2_y}}, 
	\enspace 
	\mathrm{p}_{yy} = \frac{1 + \upsilon^2_y}{\sqrt{1 + \upsilon^2_x + \upsilon^2_y}}.
\end{equation} 
After that, we set a map $(x, y) \mapsto (\zeta, \varphi)$ as a diffeomorphism from $\mathbb{R}^2$ onto itself, where $\zeta = x + \mathrm{p}_x(x, y)$ and $\varphi = y + \mathrm{p}_y(x, y)$. Putting $\varpi = \zeta + i\varphi$, we define a holomorphic function $\textgreek{\textit{ο}}$, under which $\textgreek{\textit{ο}}(\varpi) = x - \mathrm{p}_x(x, y) - i[y - \mathrm{p}_y(x, y)]$, so that in the equality 
\begin{equation}
	|\textgreek{\textit{ο}}^{(\mathrm{i})}(\varpi)|^2 = \frac{\mathrm{p}_{xx} + \mathrm{p}_{yy} - 2}{\mathrm{p}_{xx} + \mathrm{p}_{yy} + 2} < 1, 
\end{equation}
the function $\textgreek{\textit{ο}}^{(\mathrm{i})}$ is constant by \emph{Liouville's Theorem \ref{theorema "Liouville invariance"}}. But also the second derivatives
\begin{equation}
	\mathrm{p}_{xx} = \frac{|1 - \textgreek{\textit{ο}}^{(\mathrm{i})}|^2}{1 - |\textgreek{\textit{ο}}^{(\mathrm{i})}|^2}, \enspace \mathrm{p}_{yy} = \frac{|1 + \textgreek{\textit{ο}}^{(\mathrm{i})}|^2}{1 - |\textgreek{\textit{ο}}^{(\mathrm{i})}|^2}
\end{equation}
are constant, thereby $\mathrm{p}(x, y)$ is a \emph{quadratic polynomial}. 
\end{proof}

\subsubsection{Minimal Surfaces and Singular Minimal Cones}

Bernstein's problem leaps out with the following question. 

\begin{problema}[Bernstein's problem]
Besides a $\mathscr{C}^2(\mathbb{R}^2)$ Bernstein's solution, which is determined on the whole plane, are there other non-trivial solutions?
\end{problema}

Here is a rundown of some prominent solutions, or extensions of Bernstein's Theorem \ref{theorema "Bernstein"}.
\enumerationisinitium
\item An \emph{extension} of the theorem to $n$ dimensions, or to  hypersurfaces of any dimension, without using complex functions, is due to W.H. Fleming \cite[pp. 83-84]{Fleming "On the oriented Plateau problem"}, within the notion of \emph{integral currents} by Federer–Fleming \cite{Federer and Fleming "Normal and Integral Currents"}. Summarize the main achievements.
\subenumerationisinitium
\item The \emph{existence of minimal graphs}\footnote{
	A \emph{minimal graphs} is understood as a surface locally minimizing a specific perimeter.
	} 
in $\mathbb{R}^n$ involves the \emph{existence of singular minimal boundaries}, which we may call \emph{cones}, in $\mathbb{R}^n$ (\emph{Fleming conjecture}). This is equivalent to saying that Bernstein's theorem is valid in connection with the \emph{non-existence of singular minimal cones} of dimension $n$ in $\mathbb{R}^{n + 1}$. 
\item Fleming's final result, in a demonstrative key, is the non-existence of singular minimal cones in $\mathbb{R}^3$.
\subenumerationisfinis 
\item E. De Giorgi \cite{De Giorgi "Una estensione del teorema di Bernstein"} proves the Fleming conjecture in the case of hypersurfaces in Euclidean 4-space, with such an equation $x_4 = \upsilon(x_1, x_2, x_3)$, cf. D. Triscari \cite[p. 359]{Triscari "Sulle singolarita delle frontiere orientate di misura minima"}. If $\upsilon(x_1, x_2, x_3)$ is a continuous function on $\mathbb{R}^3$, with partial derivatives of any order, and if it verifies Euler–Lagrange equation
\begin{equation}
	\sum^3_{\rotatedell = 1}\frac{\partial}{\partial x_\rotatedell}
	\left\{
	\frac{\frac{\partial\upsilon}{\partial x_\rotatedell}}{\sqrt{1 + \sum^3_{m = 1}\left(\frac{\partial\upsilon}{\partial x_m}\right)^2}}
	\right\} = 0,
\end{equation} 
$\upsilon$ is a first degree polynomial. Below some of his greatest accomplishment.
\subenumerationisinitium
\item The existence of singular minimal cones is fixed (with an improvement of Fleming's thesis) in $\mathbb{R}^{n - 1}$.
\item The validity of Bernstein's theorem in dimension $n$ derives from the non-existence of singular minimal cones of dimension $(n - 1)$ in $\mathbb{R}^n$.
\item De Giorgi's final result: Bernstein's theorem holds in $\mathbb{R}^4$.
\subenumerationisfinis 
\item F.J. Almgren \cite{Almgren "Some interior regularity theorems for minimal surfaces and an extension of Bernstein's theorem"} proves the non-existence of singular minimal cones in $\mathbb{R}^4$. Let us put it in a more complete way: according to \emph{De Giorgi–Almgren result}, there are minimal graphs $\{\upsilon(x) \in \mathbb{R}^{n + 1} \mid x \in \mathbb{R}^n\}$, for $n \leqslant 4$, so the Bernstein's theorem is extended to $\mathbb{R}^5$.
\item R. Schoen, L. Simon and S.T. Yau \cite{Schoen Simon and Yau "Curvature estimates for minimal hypersurfaces"} give a proof of Bernstein's theorem in dimension less than or equal to 5, in virtue of a generalization of \emph{Heinz's estimate} \cite{Heinz "Uber die Losungen der Minimalflachengleichung"}, for $n \leqslant 5$.
\item J. Simons \cite{Simons "Minimal Varieties in Riemannian Manifolds"} proves the non-existence of minimal boundaries in dimension $n$ in $\mathbb{R}^{n + 1}$, for $n \leqslant 6$, for which Bernstein's theorem is still valid for functions of 7 independent variables, namely up to $\mathbb{R}^7$. In other words, there are \emph{regular} minimal hypersurfaces, or minimal graphs, in $\mathbb{R}^n$, with $n \leqslant 7$. But Simons finds an example of a 7-dimensional cone at least locally stable over $\mathbb{S}^3 \times \mathbb{S}^3 \subset \mathbb{S}^7 \subset \mathbb{R}^8$,
\begin{equation}
\label{equation "Simons' 7-cone"}
	\mathbb{C}\mathbbl{o}^7 = \left\{x \in \mathbb{R}^8 \mid x^2_1 + x^2_2 + x^2_3 + x^2_4 < x^2_5 + x^2_6 + x^2_7 + x^2_8\right\},
\end{equation}
that is, equivalently, a cone of codimension 1 in $\mathbb{R}^{2m}$, for $m \geqslant 4$.
\item E. Bombieri, De Giorgi, and E. Giusti \cite{Bombieri De Giorgi and Giusti "Minimal Cones and the Bernstein Problem"} prove that Simons' 7-cone is a \emph{minimal}, or rather, it is of locally minimal perimeter, and yet at the origin is singular, for which the existence of singular minimal boundaries in $\mathbb{R}^8$ is also proven. Moreover, they give non-linear solutions $\upsilon \colon \mathbb{R}^8 \to \mathbb{R}$. Let us look in some detail in the next Section.
\enumerationisfinis

\subsubsection{7-Conicity in Euclidean 8-Space, and Absence of Hyperplanes in Higher Dimensions: Bombieri–De Giorgi–Giusti Theorem}

To follow the major outcomes of Bombieri, De Giorgi, and Giusti \cite{Bombieri De Giorgi and Giusti "Minimal Cones and the Bernstein Problem"}.
\enumerationisinitium
\item Given a non-parametric minimal surface of codimension 1 having dimension $n$, one sees that Bernstein's theorem is valid for $n \leqslant 7$ (the minimal surface equation is affine). 
\item Confirmation of the existence of Simon's 7-cone \eqref{equation "Simons' 7-cone"} in $\mathbb{R}^8$, which \emph{falsifies Bernstein's theorem in 8-space}. 
\item Construction of non-trivial solutions of the minimal surface equation in 8 variables, but they are not affine. 
\item Construction of minimal graphs on $\mathbb{R}^8$, for $n \geqslant 8$, which are not hyperplanes. This places a limit on \emph{Bernstein's theorem}, which is \emph{no longer valid in 8 or higher-dimensional space}.
\enumerationisfinis

\begin{exemplum}[Simon's-like cones]
Given a sphere 
\begin{equation}
	\mathbb{S}^m_\rho = \{x^2_1 + \cdots + x^2_m = \rho^2\} \subset \mathbb{R}^m,
\end{equation}
where $\rho$ is the radius, and a set 
\begin{equation}
	\Omega = \{x^2_1 + \cdots + x^2_m \geqslant x^2_{m + 1} + \cdots + x^2_{2m}\} \subset \mathbb{R}^{2m}, 
\end{equation}
with an oriented boundary of least area. If $m \geqslant 4$, there is a cone 
\begin{equation}
	\mathbb{C}\mathbbl{o}^{2m}_\rho = \left\{x \in \mathbb{R}^{2m} \mid x^2_1 + \cdots + x^2_m = x^2_{m + 1} + \cdots + x^2_{2m} < \rho^2\right\}
\end{equation}
of codimension 1 in $\mathbb{R}^{2m}$ the boundary of which is $\mathbb{S}^m_\rho \times \mathbb{S}^m_\rho \subset \mathbb{S}^{2m}(\sqrt{2}\rho)$; the cone $\mathbb{C}\mathbbl{o}^{2m}_\rho$ is \emph{of least area}, and it has mean curvature zero at all points but \emph{at its vertex has a singular point}. \exemplasymbol
\end{exemplum}

\begin{theorema}[Bombieri–De Giorgi–Giusti]
There exist non-linear entire $\mathscr{C}^2(\mathbb{R}^n)$ solutions of the minimal surface equation
\begin{equation}
\label{equation "Minimal surface equation in dimension $n$"}
	\sum^n_{\rotatedell = 1}\frac{\partial}{\partial x_\rotatedell}
	\left(\frac{\frac{\partial\upsilon}{\partial x_\rotatedell}}{\sqrt{1 + |\nabla\upsilon|^2}}\right) = 0
\end{equation}
in $n$ independent variables, i.e. complete minimal graphs over $\mathbb{R}^n$,\footnote{
	$\upsilon(x_1, \mathellipsis  x_n) $ being a solution in $\mathbb{R}^n$.
	} 
which are not hyperplanes, for $n = 2m \geqslant 8$.
\end{theorema}

\begin{proof}[Proof (Sketch)]
Let us get right down to brass tacks. 
\enumerationisinitium
\item Relating to (the solvability of) the \emph{Dirichlet problem} \cite[§ 32, pp. 127-130]{Dirichlet "Vorlesungen uber die im umgekehrten verhaltniss des Quadrats der Entfernung wirkenden Krafte"} (see Section \ref{subsubsection "An Example of Dirichlet's Assumption for the Minimal Surface Equation"}) for an equation of the form 
\begin{equation}
	\sum^{2m}_{\rotatedell = 1}\frac{\partial}{\partial x_\rotatedell}
	\left(\frac{\frac{\partial\upsilon}{\partial x_\rotatedell}}{\sqrt{1 + |\nabla\upsilon|^2}}\right) = 0,
\end{equation}
with $\upsilon \in \mathscr{C}^2(\mathbb{B}_\mathrm{R})$, where $\mathbb{B}_\mathrm{R}$ is a ball ($\mathrm{R}$ indicates a radial distance), we exploit to the purpose a theorem by Bombieri–De Giorgi–Miranda \cite{Bombieri De Giorgi e Miranda "Una Maggiorazione A Priori Relativa Alle Ipersuperfici Minimali Non Parametriche} on a \emph{local estimate for the gradient} of the solutions of Eq. \eqref{equation "Minimal surface equation in dimension $n$"}. 
\item Take a sequence $\upsilon_1(x), \mathellipsis, \upsilon_k(x)$, $k = 1, \mathellipsis, n$. 
\item Let $\upsilon_1 = \max(\upsilon - \tau_c, 0)$, and $\upsilon_2 = \max(\upsilon, \tau_c)$, with $\tau_c$ as a real constant. Write an inequality 
\begin{equation}
	|\upsilon_1(x)| \leqslant |\upsilon_\mathrm{R}(x)| \leqslant |\upsilon_2(x)|,
\end{equation}
for $x \in \overbar{\mathbb{B}}_\mathrm{R}$, knowing that $\upsilon_\mathrm{R}(x) = 0$, for $x \in \mathbb{C}\mathbbl{o}^{2m} \cap \mathbb{B}_\mathrm{R}$.
\enumerationisfinis

From the three points above we obtain a new inequality
\begin{equation}
\label{equation "Bombieri De Giorgi Giusti inequality"}
	|\nabla\upsilon^k(x)| \leqslant c_{(1)}\exp{\left\{c_{(2)}\frac{1}{2j} \sup_{\mathbb{B}_{2j}}|\upsilon_2|\right\}} = c(j), 
\end{equation}
for all $k \geqslant j$, and $x \in \mathbb{B}_j$, where $c_{(1)}$ and $c_{(2)}$ are positive constants. We note that the right-hand side of \eqref{equation "Bombieri De Giorgi Giusti inequality"} is independent of $k$, from which we gain a subsequence $\{\upsilon_{k_z}(x)\}$, $z = 1, \mathellipsis, n$ of the sequence $\{\upsilon_k(x)\}$ with a uniform convergence in the closure of the unit ball $\overbar{\mathbb{B}}_{\rho = 1}$ to $\upsilon(x)$. It ensues that the function $\upsilon(x)$ appears analytic in $\mathbb{B}_{\rho = 1}$, which allows a verification of Eq. \eqref{equation "Minimal surface equation in dimension $n$"}.

In second place, by \emph{Ascoli–Arzelà theorem} \cite{Ascoli "Le curve limite di una varieta data di curve"} \cite{Arzela "Un'osservazione intorno alle Serie di funzioni"} \cite{Arzela "Sulle funzioni di linee"}, it is shown that the subsequence $\{\upsilon_{k_z}(x)\}$ is uniformly convergent in all compact set $\Omega \subset \mathbb{R}^{2m = 8}$. The function $\upsilon(x)$ is the limit of $\{\upsilon_{k_z}(x)\}$, and is analytic in $\mathbb{R}^{2m = 8}$, and here it provides a verification of the solution of the minimal surface Eq. \eqref{equation "Minimal surface equation in dimension $n$"}, as well as the inequality $|\upsilon_1(x)| \leqslant |\upsilon(x)| \leqslant |\upsilon_2(x)|$, namely $|\upsilon(x)| \geqslant |\upsilon(x)_1|$ in $\mathbb{R}^{2m = 8}$. From this fact we infer that 
\begin{equation}
	\limsup_{|x| \to \infty}|\upsilon_1(x)|/|x|^{2\reflectedepsilon} = 1, 
\end{equation}
for a positive constant $\reflectedepsilon > 1$. Finally, $\upsilon(x)$ is \emph{not a first degree polynomial}, and the graph of $\upsilon(x)$ is \emph{not a hyperplane}. It follows that Bernstein's Theorem \ref{theorema "Bernstein"} is not true in $\mathbb{R}^{2m = 8}$.
\end{proof}

\subsubsection{An Example of Dirichlet's Assumption for the Minimal Surface Equation}
\label{subsubsection "An Example of Dirichlet's Assumption for the Minimal Surface Equation"}

\begingroup
\footnotesize
[H]owever great may be our ignorance about how forces and states of matter vary into the infinitely small in space and time, we can surely assume that the functions to which Dirichlet's investigation did not extend, do not occur in nature [\textit{die Functionen, auf welche sich die Dirichlet'sche Untersuchung nicht erstreckt, in der Natur nicht vorkommen}]. \\
\indent — \textsc{B. Riemann} \cite[p. 100]{Riemann "Ueber die Darstellbarkeit einer Function durch eine trigonometrische Reihe"}

\vspace{2mm} 

Dirichlet's principle [is that inductive procedure] for which, from the existence of a lower limit values  of integration containing an indeterminate function, subject only to given boundary conditions of the integration field, one should get the existence of a limit function which satisfies the aforementioned conditions and which, substituted for the indeterminate function, ensures that the integral under consideration takes precisely the value of that lower limit. [T]he principle not only retains a particular suggestive force, but a very large value of deductive capacity [in the \emph{construction} of functions as solutions of equations, in which, under suitable hypotheses of continuity and derivability, the minimum conditions are reflected].\endnote{
	Extended original It. version: «È merito recente del signor Hilbert \cite{Hilbert "Uber das Dirichletsche Prinzip 1900"} di aver richiamata l'attenzione dei matematici sul procedimento induttivo che, sull'esempio del Riemann \cite{Riemann "Ueber die Darstellbarkeit einer Function durch eine trigonometrische Reihe"}, si usa ricordare col nome di \emph{principio} di Dirichlet; principio pel quale, dall'esistenza di un limite inferiore dei valori di un integrale contenente una funzione indeterminata, soggetta solo a date condizioni al contorno del campo di integrazione, dovrebbe concludersi l'esistenza di una funzione limite la quale soddisfaccia alle nominate condizioni e che, sostituita alla funzione indeterminata, faccia assumere all'integrale considerato precisamente il valore di quel limite inferiore. L'insufficienza del \emph{principio} fu rilevata con particolare evidenza dal Weierstrass \cite{Weierstrass "Uber das sogenannte Dirichlet'sche Princip"}, e dopo d'allora gli sforzi dei matematici rimpetto ai problemi che esso era destinato a risolvere si rivolsero a \emph{costruire} le funzioni richieste come soluzioni di equazioni, in cui, sotto convenienti ipotesi di continuità e di derivabilità, si traducevano le condizioni di minimo. Eppure il \emph{principio} non solo conserva una particolare forza suggestiva, ma un larghissimo valore di capacità deduttiva non si potrà disconoscergli per le dimostrazioni d'esistenza, ove appena si rifletta che in esso si assume come fondamento l'intuizione a priori dell'aggregato di tutte le funzioni».
	} \\
\indent — \textsc{B. Levi} \cite[p. 293]{Levi B. "Sul principio di Dirichlet. Memoria"}

\vspace{2mm}

Dirichle's minimum principle [\,\dots] is, together with the theory of integral equations, the most powerful tool, which today's analysis dispenses to establish the minimum theorems relating to the so-called boundary problems [\,\dots]. The deduction of  existence theorems [\,\dots] from the principle of minimum makes the calculation of variations much more harmonious and complete.\endnote{
	Extended original It. version \cite[pp. 121, 125]{Fubini "Sul principio di minimo di Dirichlet"}: «[Il] principio di minimo di Dirichlet [\,\dots] è, insieme alla teoria delle equazioni integrali, il più potente strumento, che l'analisi odierna fornisca per stabilire i teoremi di minimo relativi ai cosid[d]etti problemi al contorno [\,\dots]. Il dedurre poi i teoremi di esistenza [\,\dots] dal principio di minimo rende assai più armonico e completo il calcolo delle variazioni [\,\dots]. Per completare il nostro studio la parte essenziale è quella di dimostrare che la nostra funzione quasi-limite [cioè una funzione continua coincidente nei punti non eccezionali con la funzione limite di partenza, che è integrabile secondo Lebesgue (anziché secondo Riemann)] possiede derivate ed è armonica, e che essa sul contorno soddisfa alle condizioni imposte [\,\dots]. Per un tale studio i metodi finora applicati sono veramente deficienti, in quanto che ricorrono a vie indirette, girando, piuttosto che superando la difficoltà. Esse si basano in sostanza su certe proprietà \emph{integrali} delle funzioni armoniche, da cui scenda come conseguenza la proprietà \emph{differenziale} che la somma delle loro due derivate seconde non miste è uguale a zero. Tra le proprietà integrali che hanno servito a tale scopo noi ricorderemo p. es. la formola di Green, il teorema della media di Gauss o più generalmente la formola di Poisson per campi circolari [\,\dots]. Ma questi metodi, pure essendo sufficienti per i problemi di minimo che conducono a funzioni armoniche, e pure potendosi facilmente generalizzare a molte equazioni differenziali lineari, questi metodi, dico, sono insufficienti per casi più generali, p. es. per il problema di Plateau».
	} \\
\indent — \textsc{G. Fubini} \cite[p. 121]{Fubini "Sul principio di minimo di Dirichlet"}

\endgroup

\vspace{2mm}

Dirichlet problem concerns the need to find some function for a partial differential equation in the interior of a domain on whose boundary the function assumes certain (boundary) values. Cf. e.g. M. Miranda \cite{Miranda "Il problema di Dirichlet per l'equazione delle superfici minime"}.

\begin{problema}[A Dirichlet problem]
Given an open set $\Omega \subset \mathbb{R}^n$ of a Euclidean space, with $n \geqslant 2$, and given a real and continuous function $\textcyrillic{\textit{я}}$ defined on the boundary $\partial\Omega$ of $\Omega$, we have to find a real and continuous function $\upsilon(x)$ defined on $\Omega \cup \partial\Omega$, so that $\upsilon$ is harmonic in the interior of the domain, verifying
\begin{align}
	& \upsilon(x)|_{\partial\Omega} = \textcyrillic{\textit{я}}, \\
	& \upsilon(x) \in \mathscr{C}^2(\Omega), \enspace \divergence\frac{\nabla\upsilon(x)}{\sqrt{1 + |\nabla\upsilon(x)|^2}} = 0,
\end{align}
fo all $x \in \Omega$, where $\nabla$ is the gradient operator.
\end{problema}

\emph{Dirichlet problem} is somehow related to \emph{Dirichlet's principle},\endnote{
	When it was established, Dirichlet's principle was taken into consideration by B. Riemann. K. Weierstrass \cite{Weierstrass "Uber das sogenannte Dirichlet'sche Princip"} is among the first to blame the principle, asserting that its proof (in Riemann's original formulation) is mathematically unsatisfactory. At the end of the nineteenth century, and at the beginning of the following century, Dirichlet's principle, after a re-elaboration with \emph{direct methods}, is consolidated, and has a new life.

	\setlength\parindent{8pt}
	The renewed attention of scientific opinion on the principle is attracted by Hilbert, who mentions it in his list of 23 mathematical problems from  International Congress of Mathematicians of 1900 in Paris, see \cite[pp. 1-10]{Bartocci Betti Guerraggio Lucchetti (Eds.) "Mathematical Lives. Protagonists of the Twentieth Century From Hilbert to Wiles"}. This is the problem № XX \cite[p. 289]{Hilbert "Mathematische Probleme"} = \cite[pp. 103-104]{Hilbert "Sur les problemes futurs des Mathematiques"}. The first to launch the technique of direct methods, building upon the Dirichlet problem, is C. Arzelà \cite{Arzela "Il principio di Dirichlet"}; then there are D. Hilbert \cite{Hilbert "Uber das Dirichletsche Prinzip 1900"} \cite{Hilbert "Uber das Dirichletsche Prinzip 1904"}, J. Hadamard \cite{Hadamard "Sur le principe de Dirichlet"}, B. Levi \cite{Levi B. "Sul principio di Dirichlet. Memoria"} \cite{Levi B. "Sul principio di Dirichlet. Nota"}, G. Fubini \cite{Fubini "Sul principio di Dirichlet"} \cite{Fubini "Sul principio di minimo di Dirichlet"}, H. Lebesgue \cite{Lebesgue "Sur le probleme de Dirichlet"}. A story that mixes intuition and a search for rigor.
	} 
which is a criterion, but not the only one, to solve the Dirichlet problem. In the above condition, the objective is to \emph{minimize} an integral functional of the form
\begin{equation}
	\mathscr{I}_\textsc{d}[\upsilon] = \int_\Omega\Bigl(\sqrt{1 + |\nabla\upsilon(x)|^2}\Bigr)dx,
\end{equation}
known as \emph{energy functional}, or \emph{Dirichlet energy}, where $\nabla\upsilon \colon \Omega \to \mathbb{R}^n$ is the gradient vector field of $\upsilon$.

\begin{scholium}[Dirichlet's apparatus: out of nature]
Dirichlet's apparatus—problem \& principle—\emph{does not exist in nature}, of course, as instead it seems to transpire from the words of Riemann (see epigraph); it is only \emph{our} way of interpreting in the best way certain natural phenomena that we call \emph{phenomena on the boundary}. \scholiumsymbol
\end{scholium}

\section{De Giorgi's Theorem: Analytic Solutions in Variational Calculus}
\label{section "De Giorgi's Theorem: Analytic Solutions in Variational Calculus"}

Let us go and analyze a bit more closely, in this Section, a memorable question concerning the calculus of variations. For this purpose we choose a first-rate problem that has created the need to probe the analytic validity for solutions of regular variational problems. This is the Hilbert's 19th problem \cite[pp. 288-289]{Hilbert "Mathematische Probleme"} = \cite[pp. 101-103]{Hilbert "Sur les problemes futurs des Mathematiques"}. A resolution to this problem was provided by E. De Giorgi \cite{De Giorgi "Alcune applicazioni al Calcolo delle variazioni di una teoria della misura $K$-dimensionale"} \cite{De Giorgi "Sull'analiticita delle estremali degli integrali multipli"} \cite{De Giorgi "Sulla differenziabilita e l'analiticita delle estremali degli integrali multipli regolari"} \cite{De Giorgi "Un esempio di estremali discontinue per un problema variazionale di tipo ellittico"} and J.F. Nash \cite{Nash "Parabolic equations"} \cite{Nash "Continuity of Solutions of Parabolic and Elliptic Equations"} (see Scholium \ref{scholium "De Giorgi–Nash–Moser"}). We will limit ourselves to giving an account of the first of them.

\subsection{Hilbert's XIX Problem: Regularity for Elliptic Partial Differential Equations with Analytic Coefficients}
\label{subsection "Hilbert's XIX Problem: Regularity for Elliptic Partial Differential Equations with Analytic Coefficients"}

\begingroup
\footnotesize
\emph{Are the solutions of regular variational problems always necessarily analytic?} One of the conceptually most remarkable facts in the elements of the theory of analytic functions [is] that there are partial differential equations, the integrals of which are all of necessity \emph{analytic} functions of the independent variables, or, in short, there are [equations] only susceptible of analytic solutions. \\
\indent — \textsc{D. Hilbert} \cite[p. 288, e.a.]{Hilbert "Mathematische Probleme"} = \cite[p. 101]{Hilbert "Sur les problemes futurs des Mathematiques"}

\endgroup

\vspace{2mm}

As we read in epigraph, Hilbert's 19th problem asks if in the class of analytical solutions to elliptic partial differential equations—in the Euler–Lagrange primary typology, cf. Eqq. \eqref{subequations "Suitable form of Euler–Lagrange equations"} \eqref{subequations "Fundamental form of Euler–Lagrange equations"}, and the already mentioned works \cite{Eulero "Methodus inveniendi lineas curvas Maximi Minimive proprietate gaudentes sive solutio problematis isoperimetrici latissimo sensu accepti"} \cite{Lagrange "Essai d'une nouvelle methode pour determiner les maxima et les minima des formules integrales indefinies"}—there exists a \emph{universal character} that identifies \emph{analytically} any solution function relating to equations of this type.	

\subsection{On the Ground of Caccioppoli's Generality}

\begingroup
\footnotesize
For a complete proof of the existence and uniqueness theorems, it seems to me that it requires, first of all, a \emph{general method} of studying existential problems which, by making explicit the implications of the ancient methods of “extension”, analytical continuation and iterated procedure of successive approximations, allows \emph{to avoid} on a case by case the recourse to \emph{special artifices, laborious calculations} [\textit{evitare il ricorso caso per caso a speciali artifizî, calcoli laboriosi}], delicate demonstrations of convergence, real analytic superstructures masking often fundamentally simple facts; and then an in-depth treatment of elliptic partial differential equations, which would account in particular for the regularity properties of solutions depending on the analogous properties of the coefficients.\endnote{
	Original It. version: «Per una dimostrazione completa dei teoremi di esistenza e di unicità occorrevano, invero, mi sembra, innanzi tutto un metodo generale di studio dei problemi esistenziali che, esplicitando i sottintesi degli antichi metodi di “prolungamento”, prosecuzione analitica e procedimento iterato di approssimazioni successive, permettesse di evitare il ricorso caso per caso a speciali artifizî, calcoli laboriosi, dimostrazioni delicate di convergenza, vere soprastrutture analitiche mascheranti spesso fatti fondamentalmente semplici; e poi una trattazione approfondita delle equazioni ellittiche a derivate parziali, che rendesse conto in particolare delle proprietà di regolarità delle soluzioni in dipendenza dalle analoghe proprietà dei coefficienti».
	}\textsuperscript{,}\footnote{
	As it is easy to understand, we are not far from that need for \emph{generality} discussed in Chapter \ref{chapter "Galois' Legacy—Rules over the Calculations: the Pursuit of Generality"} on the Galoisian algebra and algebraic geometry in \textsc{qft}.
	} \\
\indent — \textsc{R. Caccioppoli} \cite[p. 1, e.a.]{Caccioppoli "Ovaloidi di metrica assegnata"}

\endgroup

\vspace{2mm}

A first way of dealing (defining and applying) an analytical character of the solutions of a class of variational problems is done, in De Giorgi \cite{De Giorgi "Definizione ed espressione analitica del perimetro di un insieme"} \cite{De Giorgi "Su una teoria generale della misura $(r - 1)$-dimensionale in uno spazio ad $r$ dimensioni"}, within the concept of \emph{perimeter of a set contained in an $n$-space}, on the ground of R. Caccioppoli's study \cite{Caccioppoli "Misura e integrazione sugli insiemi dimensionalmente orientati"} for oriented boundaries, with the involvement of domains and open sets.

In \cite{De Giorgi "Alcune applicazioni al Calcolo delle variazioni di una teoria della misura $K$-dimensionale"} De Giorgi presents a problem (suggested to him by G. Stampacchia), the resolution of which  exemplifies a condition for an analytic function in a subset of a Euclidean space.

\begin{exemplum}[Real analytic function in a subset of a Euclidean space]
Let $\upsilon(x) = \upsilon(x_1, \mathellipsis, x_n)$ be a function in $\Omega \subset \mathbb{R}^n$, such that $\upsilon(x)$ is absolutely continuous on almost all straight lines segments which are parallel to the coordinate axes, where the first order partial derivatives  are square summable in $\Omega$. Let $\varphi(y) = \varphi(y_1, \mathellipsis, y_n)$ be a real analytic function in $\mathbb{R}^n$, with $y \in \mathbb{R}^n$. Given a vector $w = (w_1, \mathellipsis, w_n)$, and two constants $c_{(1)} > 0$ and $c_{(2)} > 0$, one has the inequalities
\begin{equation}
	c_{(1)}|w|^2 \leqslant \sum^{1, n}_{\rotatedell, m} \frac{\partial^2\varphi}{\partial{y_\rotatedell}\partial{y_m}}w_\rotatedell w_m \leqslant c_{(2)}|w|^2.
\end{equation}

Assuming that $\breve{\upsilon}(x)$ is an extremal of the functional
\begin{equation}
\label{equation "Functional in De Giorgi's K-dimensional measure theory to the calculus of variations"}
	\mathscr{I}[\upsilon] = \int_{\Omega \subset \mathbb{R}^n}\varphi\left(\frac{\partial\upsilon}{\partial{x_1}}, \mathellipsis, \frac{\partial\upsilon}{\partial{x_n}}\right)dx_1, \mathellipsis, dx_n,
\end{equation}
(so $\mathscr{I}[\upsilon]$ is an extremum) one should find that $\breve{\upsilon}(x)$ is a \emph{real analytic} function in $\Omega$, or that there exists a real analytic function in $\Omega$ coincident with $\breve{\upsilon}(x)$ almost everywhere. \exemplasymbol
\end{exemplum}

\subsection{De Giorgi's Extremals with Hölder Continuous First Derivatives: Analyticity of Extremals of Regular Multiple Integrals}

In \cite{De Giorgi "Sull'analiticita delle estremali degli integrali multipli"} De Giorgi investigates some differential properties of an extremal of a multiple integral, again with a functional like the one in \eqref{equation "Functional in De Giorgi's K-dimensional measure theory to the calculus of variations"}. The function space (in which the extrema of multiple integrals are found) is treated with a direct methods of the calculus of variations. The result is that, within an  open bounded set $\Omega$, a certain function $\upsilon(x)$ is \emph{Hölder continuous} (cf. Definition \ref{definitio "Hölder condition}).

In the next paper \cite{De Giorgi "Sulla differenziabilita e l'analiticita delle estremali degli integrali multipli regolari"}, De Giorgi sets out the theorem enunciated in \cite{De Giorgi "Sull'analiticita delle estremali degli integrali multipli"}. By examining the extremals of some regular multiple integrals, he shows that the first order partial derivatives of square summable, whose existence is known a priori, are Hölder continuous; with this an infinitely differentiability and real analyticity of extremals is later proven. We will try to retrace the reasoning, in broad strokes, by dividing it into several steps.

\enumerationisinitium
\item Take a Euclidean $n$-space $\mathbb{R}^n$, plus an open subset $\Omega \subset \mathbb{R}^n$, and denote by $\mathscr{C}^{(2)}_\textsc{dg}(\Omega)$ the class of the functions $\textcyrillic{\textit{э}}(x)$ almost continuous in $\Omega$ such that
\subenumerationisinitium
\item $\textcyrillic{\textit{э}}(x)$ is absolutely continuous on almost all segments parallel to the coordinate axes contained in $\Omega$,
\item $\textcyrillic{\textit{э}}(x)$, and its first partial derivatives, are square summable functions in any closed and bounded set contained in $\Omega$.

For a number $\varepsilon > 0$, we indicate by $\mathscr{C}_\textsc{dg}(\Omega, \varepsilon)$ the class of functions $\textcyrillic{\textit{э}}(x)$ which,\footnote{
	The subscript $\textsc{dg}$ in $\mathscr{C}_\textsc{dg}$ is for \emph{De Giorgi} class, see Section \ref{subsubsection "De Giorgi Class via Sobolev Space"}.
	} 
in addition to conditions (i) and (ii), satisfy a third condition: 
\item given a point $y \in \Omega$, of which $\distance(y)$ is the distance from $\mathbb{R}\backslash\Omega$, and three numbers $m, \rho_1, \rho_2$, with $0 < \rho_1 < \rho_2 < \distance(y)$, we have
\begin{align}
	& \frac{\varepsilon}{(\rho_2 - \rho_1)^2} \int_{\textcyrillic{\textit{Ь}}_a(m) \cap \mathbb{B}(\rho_2, y)} [\textcyrillic{\textit{э}}(x) - m]^2 dx_1, \mathellipsis, dx_n \notag \\
	& \geqslant \int_{\textcyrillic{\textit{Ь}}_a(m) \cap \mathbb{B}(\rho_1, y)} |\nabla\textcyrillic{\textit{э}}|^2 dx_1, \mathellipsis, dx_n, \\
	& \frac{\varepsilon}{(\rho_2 - \rho_1)^2} \int_{\textcyrillic{\textit{Ь}}_b(m) \cap \mathbb{B}(\rho_2, y)} [\textcyrillic{\textit{э}}(x) - m]^2 dx_1, \mathellipsis, dx_n \notag \\
	& \geqslant \int_{\textcyrillic{\textit{Ь}}_b(m) \cap \mathbb{B}(\rho_1, y)} |\nabla\textcyrillic{\textit{э}}|^2 dx_1, \mathellipsis, dx_n,
\end{align}
where $\nabla\textcyrillic{\textit{э}} = \gradient\textcyrillic{\textit{э}}$, of course, $\textcyrillic{\textit{Ь}}_a(m)$ and $\textcyrillic{\textit{Ь}}_b(m)$ are the sets of points of $\Omega$ (i.e. $\textcyrillic{\textit{Ь}}_a, \textcyrillic{\textit{Ь}}_b \subset \Omega$) under which $\textcyrillic{\textit{э}}(x) > m$ and $\textcyrillic{\textit{э}}(x) < m$, respectively, and $\mathbb{B}$ indicates a ball with center $y$ and radius $\rho$.
\subenumerationisfinis
\item
\label{item "De Giorgi's Theorem I"}
After a series of lemmas, and also after a series of propositions aimed at understanding problems on elliptic differential equations, a theorem for a uniformly continuous and even Hölder function (in any compact subset of $\Omega$) is put forward: \emph{every function $\textcyrillic{\textit{э}}(x) \in \mathscr{C}_\textsc{dg}(\Omega, \varepsilon)$ is uniformly hölderian in any closed and bounded set contained in $\Omega$}.
\item We consider a continuous function $\varphi(p) = \varphi(p_1, \mathellipsis, p_n)$ in $\mathbb{R}^n$, with its first and second order partial derivatives, hence let
\begin{equation}
\label{equation "Continuous function and its first and second order partial derivatives"}
	\varphi_{\rotatedell, m}(p) = \frac{\partial^2\varphi}{\partial{p_\rotatedell}\partial{p_m}}, \enspace \varphi_\rotatedell(p) = \frac{\partial\varphi}{\partial{p_\rotatedell}}, 
\end{equation} 
for $\rotatedell, m = 1, \mathellipsis, n$. Two numbers $\alpha > 0$ and $\beta > 0$ are chosen, for each point $p \in \mathbb{R}^n$, and for each vector $w = (w_1, \mathellipsis, w_n)$, so that
\begin{equation}
\label{equation "Continuous function with sum and two positive numbers"}
	\alpha|w|^2 \leqslant \sum^{1, n}_{\rotatedell, m}\varphi_{\rotatedell, m}(p)w_\rotatedell w_m \leqslant \beta|w|^2.
\end{equation}
Given an open subset $\Omega \subset \mathbb{R}^n$, and a function $\breve{\upsilon}(x) \in \mathscr{C}^{(2)}_\textsc{dg}(\Omega)$, we will say that $\breve{\upsilon}(x)$ is extremal in $\Omega$ of the integral functional
\begin{equation}
	\mathscr{I}[\upsilon] = \int\varphi\left(\frac{\partial\upsilon}{\partial{x_1}}, \mathellipsis, \frac{\partial\upsilon}{\partial{x_n}}\right)dx_1, \mathellipsis, dx_n, 
\end{equation}
if, for any closed and bounded (i.e. compact) subset $\textcyrillic{\textit{Ь}}_c \subset \Omega$, and for any continuous function $\zeta(x)$ in $\Omega$, with its first derivatives, supposing $\zeta(x)$ is identically zero in $(\Omega\backslash\textcyrillic{\textit{Ь}}_c)$, we have
\begin{equation}
\label{equation "Condition equation"}
	\sum^n_{\rotatedell = 1} \int_{\Omega}\frac{\partial\zeta}{\partial{x_\rotatedell}}\varphi_\rotatedell\left(\frac{\partial\breve{\upsilon}}{\partial{x_1}}, \mathellipsis, \frac{\partial\breve{\upsilon}}{\partial{x_n}}\right)dx_1, \mathellipsis, dx_n = 0. 
\end{equation}
Which allows us to state the following.
\enumerationisfinis

\begin{theorema}[De Giorgi on the variational analyticity]
\label{theorema "De Giorgi on the variational analyticity"}
Every extremal in $\Omega$ of the integral functional $\mathscr{I}[\upsilon]$ has first order partial derivatives uniformly Hölder continuous in any closed and bounded set contained in $\Omega$. If $\varphi(p)$ is real analytic in $\mathbb{R}^n$, then every extremal is real analytic in $\Omega$.	
\end{theorema}

\begin{proof}
By Eqq. \eqref{equation "Continuous function and its first and second order partial derivatives"} and \eqref{equation "Continuous function with sum and two positive numbers"} one establishes that $|\varphi_\rotatedell(p)| \leqslant \beta|p| + c$, for $\rotatedell = 1, \mathellipsis, n$, and a constant $c > 0$, putting $|p|$ as the distance of $p$ from the origin of the coordinates. Since $\breve{\upsilon}(x) \in \mathscr{C}^{(2)}_\textsc{dg}(\Omega)$, all functions $\varphi_\rotatedell\bigl(\frac{\partial\breve{\upsilon}}{\partial{x_1}}, \mathellipsis, \frac{\partial\breve{\upsilon}}{\partial{x_n}}\bigr)$, no less than $\breve{\upsilon}(x)$, will have first order partial derivatives that are square summable in every closed and bounded set contained in $\Omega$.

Consequently, Eq. \eqref{equation "Condition equation"} can be verified if $\zeta(x) \in \mathscr{C}^{(2)}_\textsc{dg}(\Omega)$, and if $\zeta(x)$ vanishes identically outside a closed and bounded set contained in $\Omega$.

Now, let us consider a bounded open set $\textcyrillic{\textit{Ь}}_d \subset \Omega$ having positive distance from the boundary of $\Omega$, a number $\gamma > 0$ smaller than that distance, a positive integer $z$ not greater than $n$, and another positive integer $k$. So we write 
\begin{equation}
	\upsilon_k(x) = \breve{\upsilon}\left(x_1, \mathellipsis, x_z + \frac{\gamma}{k}, \mathellipsis, x_n\right), 
\end{equation}
where $\upsilon_k(x) \in \mathscr{C}^{(2)}_\textsc{dg}(\textcyrillic{\textit{Ь}}_d)$. For every $\textcyrillic{\textit{Ь}}_c \subset \textcyrillic{\textit{Ь}}_d$, and for every function $\chi(x) \in \mathscr{C}^{(2)}_\textsc{dg}(\textcyrillic{\textit{Ь}}_d)$ vanishing in $(\textcyrillic{\textit{Ь}}_d\backslash\textcyrillic{\textit{Ь}}_c)$, it is clear that
\begin{equation}
\label{equation "Vanishing equation w/ bounded open set"}
	\sum^n_{\rotatedell = 1} \int_{\textcyrillic{\textit{Ь}}_d}\varphi_\rotatedell\frac{\partial\chi}{\partial{x_\rotatedell}}\left(\frac{\partial\upsilon_k}{\partial{x_1}}, \mathellipsis, \frac{\partial\upsilon_k}{\partial{x_n}}\right)dx_1, \mathellipsis, dx_n = 0.
\end{equation}	
If $\breve{\textcyrillic{\textit{э}}}_k(x) = \upsilon_k(x) - \breve{\upsilon}(x)$, then $\breve{\textcyrillic{\textit{э}}}_k(x) \in \mathscr{C}^{(2)}_\textsc{dg}(\textcyrillic{\textit{Ь}}_d)$, for which
\begin{align}
	& \varphi_\rotatedell\left(\frac{\partial\breve{\upsilon}}{\partial{x_1}}, \mathellipsis, \frac{\partial\breve{\upsilon}}{\partial{x_n}}\right) - \varphi_\rotatedell\left(\frac{\partial\upsilon_k}{\partial{x_1}}, \mathellipsis, \frac{\partial\upsilon_k}{\partial{x_n}}\right) \notag \\
	& + \sum^{1, n}_{m}\Lbrack:\int^1_0\varphi_{\rotatedell, m}\left(\frac{\partial\breve{\upsilon}}{\partial{x_1}} + t \cdot \frac{\partial\breve{\textcyrillic{\textit{э}}}_k}{\partial{x_1}}, \mathellipsis, \frac{\partial\breve{\upsilon}}{\partial{x_1}} + t \cdot \frac{\partial\breve{\textcyrillic{\textit{э}}}_k}{\partial{x_n}}\right)dt:\Rbrack\frac{\partial\breve{\textcyrillic{\textit{э}}}_k}{\partial{x_m}} = 0.\footnotemark
\end{align}
\footnotetext{
	$\Lbrack:$ and $:\Rbrack$ are for a repeat sign, see Glossary.
	} 
From this Equation, by \eqref{equation "Continuous function with sum and two positive numbers"}, we get
\begin{equation}
	\alpha|w|^2 \leqslant \sum^{1, n}_{\rotatedell, m}\Lbrack:\cdots:\Rbrack w_\rotatedell w_m \leqslant \beta|w|^2,
\end{equation}
whilst by Eqq. \eqref{equation "Condition equation"} \eqref{equation "Vanishing equation w/ bounded open set"}, we get
\begin{equation}
	\sum^{1, n}_{\rotatedell, m}\int_{\textcyrillic{\textit{Ь}}_d}\Lbrack:\cdots:\Rbrack\frac{\partial\breve{\textcyrillic{\textit{э}}}_k}{\partial{x_m}}\frac{\partial\chi}{\partial{x_\rotatedell}}dx_1, \mathellipsis, dx_n = 0,
\end{equation}
and that goes for every function $\chi(x) \in \mathscr{C}^{(2)}_\textsc{dg}(\textcyrillic{\textit{Ь}}_d)$, which is identically zero outside a closed subset of $\textcyrillic{\textit{Ь}}_d$. Finally, $\breve{\textcyrillic{\textit{э}}}_k(x) \in \mathscr{C}_\textsc{dg}(\textcyrillic{\textit{Ь}}_d, \varepsilon)$, with $\varepsilon = \frac{\beta^2}{\alpha^2}$, by stressing that 
\begin{equation}
	\textcyrillic{\textit{э}}_k(x) = \frac{\breve{\textcyrillic{\textit{э}}}_k(x)k}{\gamma} \in \mathscr{C}_\textsc{dg}(\textcyrillic{\textit{Ь}}_d, \varepsilon).
\end{equation}
If the sequence $\textcyrillic{\textit{э}}_1(x), \mathellipsis,  \textcyrillic{\textit{э}}_k(x)$ converges to $\frac{\partial\breve{\upsilon}}{\partial{x}_z}$ in $\textcyrillic{\textit{Ь}}_d$, it happens that $\frac{\partial\breve{\upsilon}}{\partial{x}_z} \in \mathscr{C}_\textsc{dg}(\textcyrillic{\textit{Ь}}_d, \varepsilon)$. The conclusion is that all the first order prime derivatives of $\breve{\upsilon}(x)$ belong to $\mathscr{C}_\textsc{dg}(\Omega, \varepsilon)$, and they have a Hölderian continuity due to the theorem of point \ref{item "De Giorgi's Theorem I"}. If $\varphi(p)$ is real analytic, ergo $\breve{\upsilon}(x)$ appears to be real analytic, cf. G. Stampacchia \cite{Stampacchia "Sistemi di equazioni di tipo ellittico a derivate parziali del primo ordine e proprieta delle estremali degli integrali multipli"}, and C.B. Morrey \cite[p. 204]{Morrey "Second Order Elliptic Systems of Differential Equations"}. The answer to Hilbert's 19th question («Are the solutions of regular variational problems always necessarily analytic?», Section \ref{subsection "Hilbert's XIX Problem: Regularity for Elliptic Partial Differential Equations with Analytic Coefficients"}) is \emph{positive} within the conditions of the Theorem \ref{theorema "De Giorgi on the variational analyticity"}.
\end{proof}

\begin{scholium}[De Giorgi–Nash–Moser]
\label{scholium "De Giorgi–Nash–Moser"}
~\enumerationisinitium
\item[·] De Giorgi's result \cite{De Giorgi "Sulla differenziabilita e l'analiticita delle estremali degli integrali multipli regolari"} can be summarized by saying that, for it, there is a Hölderian continuity of \emph{weak} solutions of \emph{linear} elliptic differential equations.
\item[·] Besides De Giorgi, J.F. Nash \cite{Nash "Parabolic equations"} \cite{Nash "Continuity of Solutions of Parabolic and Elliptic Equations"} finds a second and independent solution to the Hilbert problem, working, in the first article, on proofs of theorems on a priori continuity of solutions of linear parabolic and elliptic partial differential equations, stimulated by existence problems for fluid and turbulent flows, and treating, in the second article, on \emph{non-linear} partial differential equations (again related to equations of flow for a viscous, compressible, \& heat conducting fluid, and turbulence phenomena). 
\item[·] J. Moser \cite{Moser "A New Proof of De Giorgi's Theorem Concerning the Regularity Problem for Elliptic Differential Equations"} writes, shortly after, an alternative proof to that of De Giorgi, thanks to the Sobolev embedding theorem (cf. Section \ref{subsubsection "Sobolev Embedding for a Null Trace Space, Orlicz Space, and General Sobolev Inequality"}).
\enumerationisfinis

With unified results and efforts (of the three authors), there is sometimes mention of the \emph{De Giorgi–Nash–Moser} (\textsc{dgnm}) \emph{theorem}. \scholiumsymbol
\end{scholium}

\subsubsection{De Giorgi Class via Sobolev Space}
\label{subsubsection "De Giorgi Class via Sobolev Space"}

The De Giorgi class 
\begin{align}
	\mathscr{C}_\textsc{dg}(\Omega, \varepsilon) & = \Bigl\{\left[\mathscr{C}_\textsc{dg}\right]^\pm_p(\Omega, \varepsilon) = \left[\mathscr{C}_\textsc{dg}\right]^+_p(\Omega, \varepsilon) \cap \left[\mathscr{C}_\textsc{dg}\right]^-_p(\Omega, \varepsilon)\Bigr\} \notag \\
	& = \mathscr{C}_\textsc{dg}^\pm = \mathscr{C}_\textsc{dg}^+ \cap \mathscr{C}_\textsc{dg}^-	
\end{align}
can be defined, via Sobolev space (see Section \ref{subsubsection "Sobolev Space $W^{k, p}(Omega)$"}), as the set (in a subset of a Euclidean space) of functions almost continuous $\textcyrillic{\textit{э}}(x) \in \Sobolev^{1, p}_\mathrm{loc}(\Omega)$,\footnote{
	Theorem \ref{theorema "De Giorgi on the variational analyticity"} also lays down that $\breve{\upsilon}(x) \in \mathscr{C}_\textsc{dg}(\Omega, \varepsilon)$.
	} for $p > 1$, which are \emph{locally Hölder continuous} in $\Omega$. Thus the two belonging-forms
\begin{equation}
\textcyrillic{\textit{э}}(x) \in	
	\begin{cases}
	\left[\mathscr{C}_\textsc{dg}\right]^\pm_p(\Omega, \varepsilon) \\
	\Sobolev^{1, p}_\mathrm{loc}(\Omega)
	\end{cases}
\end{equation}
are equivalent.

\subsection{De Giorgi's Exception: Discontinuous Extremals and Weak Solutions for Elliptic Systems}

In \cite{De Giorgi "Un esempio di estremali discontinue per un problema variazionale di tipo ellittico"} De Giorgi returns to the Hilbert problem (Section \ref{subsection "Hilbert's XIX Problem: Regularity for Elliptic Partial Differential Equations with Analytic Coefficients"}) elaborating an example of a \emph{weak} solution with \emph{discontinuous extremals} of a uniformly elliptic system. He replaces a real function $\textcyrillic{\textit{э}}(x)$\footnote{
	The extremals of
	$\int\left\{
	\left(\sum^n_{\rotatedell = 1}\mathbbl{b}_\rotatedell(x)\frac{\partial\textcyrillic{\textit{э}}}{\partial{x_\rotatedell}}\right)^2 + \sum^n_{\rotatedell = 1}\left(\frac{\partial\textcyrillic{\textit{э}}}{\partial{x_\rotatedell}}\right)^2
	\right\}dx$ have a Hölderian continuity.
	} 
with a \emph{vector function} $\textgreek{\text{\ddigamma}}(x) = \bigl(\textgreek{\text{\ddigamma}}_1(x), \mathellipsis, \textgreek{\text{\ddigamma}}_n(x)\bigr)$ having $n$ real components in $\mathbb{R}^n$, and uses the integral
\begin{equation}
\label{equation "Integral with discontinuous extremals"}
	\int\left\{
	\left(\sum^{1, n}_{\rotatedell, m}\mathbbl{b}_{\rotatedell m}(x)\frac{\partial\textgreek{\text{\ddigamma}}_\rotatedell}{\partial{x_m}}\right)^2 + \sum^{1, n}_{\rotatedell, m}\left(\frac{\partial\textgreek{\text{\ddigamma}}_\rotatedell}{\partial{x_m}}\right)^2
	\right\}dx,
\end{equation}
where $\mathbbl{b}(x)$ is a bounded measurable function. The extremals of the integral \eqref{equation "Integral with discontinuous extremals"} are discontinuous. The answer to Hilbert's 19th question is, within the conditions of this counterexample (and compared to the Theorem \ref{theorema "De Giorgi on the variational analyticity"}), \emph{negative}.

\vspace{5mm}

\vspace{10mm}

\setcounter{secnumdepth}{0}  
\section{References and Bibliographic Details}
\setcounter{secnumdepth}{3}
\markright{References and Bibliographic Details}

\begingroup
\footnotesize
\noindent Section \ref{section "Prolegomenon: Geometrical Optics, Solid of Minimum Resistance, and Brachistochrone"}

\begin{indent paragraph: 15pt}
· Observations on Newton's problem and on the brachistochrone curve, together with many other variational themes, date back, in Italy, to E. Pascal \cite[I, §§ 30-31]{Pascal E. "Calcolo delle variazioni e calcolo delle differenze finite"}; see also E.E. Levi \cite{Levi E. "Sulle condizioni sufficienti per il minimo nel calcolo delle variazioni"} \cite{Levi E. "Sui criterii sufficienti per il massimo e per il minimo nel Calcolo delle Variazioni"}, as exemplar papers on the maximum and minimum values in the calculus of variations. \\
· Is to be mentioned, historically, E.L. Lindelöf's work \cite{Lindelof "Lecons de calcul des variations"}. \\
· For an accurate survey on the genesis of the calculation of variations, see \cite{Freguglia Giaquinta "The Early Period of the Calculus of Variations"}, while for a rewarding treatment, see \cite{Giaquinta Hildebrandt "Calculus of Variations I: The Lagrangian Formalism"} \cite{Giaquinta Hildebrandt "Calculus of Variations II: The Hamiltonian Formalism"}. \\
· A varied and informative book but full of illustrations on the minimum principle is \cite{Hildebrandt Tromba "The Parsimonious Universe: Shape and Form in the Natural World"}.

\end{indent paragraph: 15pt}

\noindent Section \ref{section "Theory of Minimal Surfaces"}

\begin{indent paragraph: 15pt}
On the minimal surfaces, and variational calculus, with parts dedicated to Plateau's and Bernstein's problems, see \cite{Courant "Dirichlet's Principle Conformal Mapping and Minimal Surfaces"} \cite{Bombieri (Ed.) "Geometric Measure Theory and Minimal Surfaces"} \cite{Giusti "Minimal Surfaces and Functions of Bounded Variation"} \cite{Massari and Miranda "Minimal Surfaces of Codimension One"} \cite{Struwe "Plateau's Problem and the Calculus of Variations"} \cite{Nitsche "Lectures on Minimal Surfaces I: Introduction Fundamentals Geometry and Basic Boundary Value Problems"} \cite{Hildebrandt "Boundary Value Problems for Minimal Surfaces"} \cite{Simon "The Minimal Surface Equation"} \cite[chap. 4]{Giaquinta Modica Soucek "Cartesian Currents in the Calculus of Variations I. Cartesian Currents"} \cite[chap. 6]{Giaquinta Modica Soucek "Cartesian Currents in the Calculus of Variations II. Variational Integrals"} \cite{Osserman "A Survey of Minimal Surfaces"} \cite{Perez Ros "Properly embedded minimal surfaces with finite total curvature"} \cite{Fomenko Tuzhilin "Elements of the Geometry and Topology of Minimal Surfaces in Three-Dimensional Space"} \cite{Dierkes Hildebrandt Sauvigny "Minimal Surfaces"} \cite{Dierkes Hildebrandt Tromba "Global Analysis of Minimal Surfaces"} \cite{Dierkes Hildebrandt Tromba "Regularity of Minimal Surfaces"} \cite{Colding Minicozzi II "A Course in Minimal Surfaces"} \cite{Harrison and Pugh "Plateau's Problem"}.
\end{indent paragraph: 15pt}

\noindent Section \ref{subsection "Plateau's Problem: Soap Films and Bubbles"}

\begin{indent paragraph: 15pt}
For those interested in exploring the innumerable \textgreek{μῦθοι παράλληλοι} (parallel stories) of soap bubbles in mathematics, art, literature, music and architecture, can consult Emmer's works, e.g. \cite{Emmer "Bolle di sapone. Tra arte e matematica"}.
\end{indent paragraph: 15pt}

\noindent Section \ref{section "De Giorgi's Theorem: Analytic Solutions in Variational Calculus"}

\begin{indent paragraph: 15pt}
· On direct methods in the calculus of variations, see e.g. \cite{Giusti "Direct Methods in the Calculus of Variations"}; as classic texts, see L. Tonelli \cite{Tonelli "Sui massimi e minimi assoluti del Calcolo delle Variazioni"} \cite{Tonelli "Sur une methode directe du Calcul des Variations"} \cite{Tonelli "Fondamenti di Calcolo delle Variazioni I"} \cite{Tonelli "Fondamenti di Calcolo delle Variazioni II"}. \\
· Further reading on De Giorgi's theorem and surrounding issues, i.e. harmonic functions \& Hölder continuity, as well as on the regularity for elliptic equations: \cite{Ladyzhenskaya and Ural'tseva "Linear and Quasilinear Elliptic Equations"} \cite{Morrey "Multiple Integrals in the Calculus of Variations"} \cite{Giaquinta and Giusti "On the regularity of minima of variational integrals"} \cite{Eells and Fuglede "Harmonic Maps Between Riemannian Polyhedra"} \cite[chapp. 8, 9]{Giaquinta and Martinazzi "An Introduction to the Regularity Theory for Elliptic Systems Harmonic Maps and Minimal Graphs"}. \\
· About the De Giorgi classes, see \cite{Fan Zhao "A class of De Giorgi type and Holder continuity"} \cite{DiBenedetto and Gianazza "Some Properties of De Giorgi Classes"}.
\end{indent paragraph: 15pt}

\endgroup

\chapter{Outro—\emph{Parva Mathematica}: \emph{Libera Divagazione} \sfrac{1}{8}}
\label{chapter "Outro—Parva Mathematica: Libera Divagazione 1/8"}

\begingroup
\footnotesize
È stato detto [\,\dots] i Matematici esser come gli Amanti, i quali per poco che voi loro accordiate da principio, se ne sanno così bene approfittare, che insensibilmente là vi conducono, dove non avreste mai pensato.\footnote{
	\cite[p. 48]{Algarotti "Sir Isaac Newton's Philosophy Explain'd For the Use of Ladies. In Six Dialogues on Light and Colours I"} «The Mathematicians [\,\dots] are said to resemble Lovers. If what you grant them at first be ever so little, they know how to make so good an Advantage of it, as to lead you insensibly farther than you ever imagined».
	} \\
\indent — \textsc{F. Algarotti} \cite[p. 28]{Algarotti "Il newtonianismo per le dame ovvero Dialogo sopra la luce e i colori"}
	
\endgroup

\section{Excursion on Mathematical Objects}

\begingroup
\footnotesize
One should never forget that the functions, like all mathematical combinations of concepts [\textit{alle mathematischen Begriffszusammensetzungen}], are only our own creations [\textit{nur unsere eignen Geschöpfe}], and that where the definition with which one begins ceases to make sense, one should not really ask “What is it”? but “What is convenient to accept”? So that it can be consistent.\footnote{
	\label{footnote "Gauss' letter to Bessel 9 April 1830"}	
	But compare with Gauss' letter to Bessel dated 9 April 1830: \cite[p. 201]{Gauss' letter to Bessel 9 April 1830}: «According to my deepest belief, the theory of space [\textit{Raumlehre}] occupies a completely different position with regard to our knowledge a priori from that of the pure theory of magnitudes [\textit{Grössenlehre}, i.e. of numbers]; our knowledge of the former lacks entirely that complete conviction of its necessity (and therefore of its absolute truth) which belongs to the latter; we must humbly admit that if number is \emph{merely} a product of our mind [\textit{die Zahl bloss unseres Geistes Product ist}], space has a reality outside of our mind [\textit{der Raum auch ausser unserm Geiste eine Realität hat}], the laws of which we cannot fully prescribe a priori». 
	
	Please be advised: this is true, apparently, for \emph{physical 3-space}, and not for \emph{mathematical space}; think of the pluri-assorted abstract $n$-dimensional spaces, which reside in the mind, and are purely figment of our imagination.
	} \\
\indent — \textsc{C.F. Gauss} \cite[p. 363, Gauss' letter to F. Bessel, 21 November 1811]{Gauss' letter to Bessel 21 November 1811}

\vspace{2mm}

[L]a science [mathématique] est l'œuvre de l'esprit humain. \\
\indent — \textsc{É. Galois} \cite[p. 28]{Galois "Sciences Mathematiques: Discussions sur les progres de l'analyse pure"}

\endgroup

\subsection{Logical Creativity}

\begingroup
\footnotesize
[N]umbers are free creations of the human mind [\textit{Schöpfungen des menschlichen Geistes}]; they serve as a means of comprehending the difference of things more easily and sharply [\,\dots]. In this sense, which is well expressed in the paraphrase of a renowned aphorism \textgreek{ἀεὶ ὁ ἄνθρωπος ἀριθμητίζει}. \\
\indent — \textsc{R. Dedekind} \cite[pp. vii-viii, x]{Dedekind "Was sind und was sollen die Zahlen?"}
	
\endgroup

\vspace{2mm}

Dedekind \cite[pp. vii-viii]{Dedekind "Was sind und was sollen die Zahlen?"} talks about «arithmetic (algebra, analysis) as a part of logic», and a «purely logical process of building up the science of numbers [\textit{rein logischen Aufbau der Zahlen-Wissenschaft}]». His notion e.g. of natural number is based on logical notions, but this certainly does not indicate, with all due respect to Dedekind, that mathematics can be (entirely) founded on purely logical grounds; we reject logicism, especially the clumsy one (cf. Section \ref{subsection "Extra-logical Objects, and Gödelian Suggestions"}). 

It should be said nevertheless that, for Dedekind, “logic” means a “creative faculty” inherent in the mind, to wit, the art of reasoning, or the act of thinking. If we concede that there is an elementary logic, a natural logic, biologically congenital, which precedes a cultural logic, formally evolved, then certain basic arithmetic notions, according to Dedekind, can be anchored to the operations of this elementary logic. This appears admissible, albeit unsatisfactory. Logic, or thought, is not a faculty acting separately from arithmetic activity (\textgreek{ἀριθμέω}), as Dedekind seems to believe ingenuously. The clear-cut distinction of two levels—on the one hand, a science of reasoning, a logic as a foundational activity, and, on the other, an arithmetic activity—inescapably leads to a caricatural vision of the mathematical genesis of natural numbers. The \emph{shift from logic to number} is not a mechanism with such neatly distinguishable components.

\subsection{Axioms of Faith}
\label{subsection "Axioms of Faith"}

\begingroup
\footnotesize
The discovery of mathematical truths and the subsequent grasp of them by the understanding occurs in a much more “factual” [\textit{sachlicher}] way and much less “formal” [\,\dots]. It is merely a scientific \emph{faith} [\textit{wissenschaftlicher Glaube}] that, for example, all pertinent, general, true judgments about points, lines, and planes are derivable from the geometrical axioms. We are unable to grasp in genuine \emph{insight} that this is so or even to “prove” it in a logical way on the basis of the logical laws themselves. \\
\indent — \textsc{H. Weyl} \cite[§ 3, p. 11]{Weyl "Das Kontinuum: Kritische Untersuchungen uber die Grundlagen der Analysis"} = \cite[§ 3, p. 18]{Weyl "The Continuum: A Critical Examination of the Foundation of Analysis"} 
	
\endgroup

\vspace{2mm}

In contrast to logicism an intuitionist method is possible, and this was the approach taken e.g. by H. Weyl in a phase of his life. He too, like Dedekind, deals with the problem of the foundation of natural numbers, but he does it in a completely different way. 

We are convinced that the Weylian idea of a \emph{logical hole} within the \textit{Modell} of a rational science (concerning relationships and functions) logically organized as a \textit{deduktive Theorie}, the idea of what we might call \emph{axioms of faith},\endnote{
	\label{endnote "axiom"}
	Note the subtleness: the word \emph{axiom} comes from the Gr. \textgreek{ἄξιος}, “worthy” of “consideration”, “trust”, “belief”, or “faith”.
	} 
is a \emph{crucial punctum} in the understanding of the workings of mathematics, which is connected to the mathematico-primitive concepts encountered above (cf. Sections \ref{section "Point and Line as Primitive Ideas"} and \ref{section "Space-numeral Primitiveness: On the Continuum of Real Numbers"}), and then to some physical concepts (see point \ref{item "Physics is not mathematics (and vice versa), axioms of faith"}, p. \pageref{item "Physics is not mathematics (and vice versa), axioms of faith"}). H. Weyl \cite[§ 3, p. 12]{Weyl "Das Kontinuum: Kritische Untersuchungen uber die Grundlagen der Analysis"} = \cite[§ 3, p. 19]{Weyl "The Continuum: A Critical Examination of the Foundation of Analysis"} continues like this:

\vspace{2mm}

\begingroup
\footnotesize
The interpretation under consideration proves to be feasible only when one knows that the axioms are \emph{consistent} and \emph{complete}, in the sense that of two “antinomous” pertinent judgments [\,\dots] always one and only one is a logical consequence of the axioms. But we do not \emph{know} this (although we may believe it). And if this faith [\textit{Glaube}] is one day to be transformed into insight [\textit{Einsicht}], then, clearly, since logical inference consists of iterating certain elementary logical inferences, we will attain this insight only through our intuition [\textit{Anschauung}] of iteration, i.e., of the infinite repetition of a procedure. But from this intuition we also directly obtain the fundamental arithmetical insights into the natural numbers on the basis of which the whole Mathesis pura is logically constructed.

\endgroup

\vspace{2mm}

\subsubsection{A Task for Other Scientists: Proto-mathematics}
\label{subsubsection "A Task for Other Scientists: Proto-mathematics"}

How to define this insight/intuition, because Weyl never gives a completely satisfactory definition of it? 

This is not the place to analyze the transition from an intuitive nature of mathematics to its construction as a culture, or symbolic thought. We have also to admit honestly that this goes beyond the competence (and daily activity) of a mathematician, calling other disciplines into play, so we stop here. 

What is certain is that such a insight/intuition does not fall from a Platonic sky (Section \ref{subsection "Logomachy of Mathematicians, and Cock-and-Bull Stories"}), but, in parallel with its \emph{bio-evolutive} and \emph{physiological origin}, arising, shall we say, from some \emph{proto-mathematics}, it is the product of \emph{historical events}, and \emph{cultural residues}. 

As an Enriques already observed at the beginning of the twentieth century, it is expedient to investigate the \emph{physiological root of logic}; and equally, in geometry, we can identify a \emph{psycho-genetic ground} for abstract spaces, passing from visual, tactile and muscular origins of a physiological space to a geometric space, whence it follows that space in geometry is/turns out to be a sort of limit of physical space. Proto-mathematics surveys are of this type.\endnote{
	The works of G. Vallortigara, a neuroscientist, can give some hints on this. We point out, among many others, his popular book \cite{Vallortigara Panciera "Cervelli che contano"}, from which the following observation is taken on p. 18: «[O]ur symbolic knowledge of number rests on something more ancient and deeper, a pre-verbal and pre-symbolic, analogical and approximate representation that we share with other animal species and that is present in children before they can speak or have received any formal mathematical education. It would therefore seem that the “sense of number”, [a] non-verbal and non-symbolic [sense], is really a numerical competence, and not a perceptive-sensorial capacity of another nature».
	
	\setlength\parindent{8pt}
	S. Dehaene, a cognitive neuroscientist active in the field of mathematics, is also mentioned throughout the book (see Section \ref{subsubsection "Selection, Colors, and Understanding"}); he provides further food for thought.
	}

\subsection{Betwixt 0 and 1, and All the Rest}

\begingroup
\footnotesize
Intellectus igitur, non reperit, sed facit Numeros.\footnote{
	It may be curious to read, in this perspective, a part of the Stagiriticus tradition, in which the mental creation of numbers is a very common act. See e.g. J. Caramuel, which already thinks in a set key \cite[pp. xliii-xliv]{Caramuel "Meditatio Prooemialis: De Arithmeticarum numero e varietate}: «A guy was talking in his sleep, \& when the clock struck four, he said: “One, one, one, one. This clock is crazy: it has struck one four times over”. The guy had counted a hit four times, \& not four hits. He had in mind the one four times in place of four. The counting is one thing, the considering several things together is another. If I had four clocks in my Studio, \& if they all struck one, I will not say that they struck four but one four times. Four times one is not the same as four [that is, they are different]. And this difference is not a property of things, it is not independent of the operations of the mind: if anything, it depends on the mind of the person who is counting [\textit{Hæc differentia non est in rebus, à mentis operatione independens: ergo pendet à mente numerantis}]. Intellect, therefore, does not find but makes numbers [\textit{Intellectus igitur, non reperit, sed facit Numeros}]; it consider several things, as each distinct [\textit{ut discreta}] in itself, \& as intentionally united by thought [\textit{ut intentionaliter unita cogitando}]».
	} \\
\indent — \textsc{J. Caramuel} \cite[p. xliv]{Caramuel "Meditatio Prooemialis: De Arithmeticarum numero e varietate}

\vspace{2mm}

[T]he respective interpretations of the symbols 0 and 1 in the system of Logic are \emph{Nothing} and \emph{Universe}. \\
\indent — \textsc{G. Boole} \cite[p. 48]{Boole "An Investigation of the Laws of Thought"}

\vspace{2mm}

Nos cogita numero, ergo numero es. \\
\indent — \textsc{G. Peano} \cite[p. 365]{Peano "Super theorema de Cantor-Bernstein"} 
	
\vspace{2mm}

Nature does not count nor do integers occur in nature. Man made them all, integers and all the rest, Krone[c]ker to the contrary notwithstanding. \\
\indent — \textsc{P.W. Bridgman} \cite[p. 100]{Bridgman "The Way Things Are"}

\endgroup

\vspace{2mm}

\enumerationisinitium
\item Mathematics is mainly the science of numbers and measure(ment), that is, of space and spatial figures (geometry). 
\item The only numbers that may—mathematically—exist are 0 and 1, namely: 0 and 1 are, if you like, the only form (\textgreek{εἶδος}) of mathematics; but also for these two numbers it is necessary to talk about primary ideas, because even the base-2 numeral system,\footnote{
	The contours of binary calculus are expressly outlined by G.W. Leibniz \cite{Leibniz "Explication de l'Arithmetique Binaire"} (see also his earlier writing \textit{De organo sive arte magna cogitandi}, 1679), but its seeds were sown by the ancient Chinese culture of Fuxi (\ZhTraditional{伏羲}), here called “Fohy” \cite[pp. 87-88]{Leibniz "Explication de l'Arithmetique Binaire"}: «What is surprising in this calculus is that this Arithmetic by 0 \& 1 is found to contain the mystery of the lines of an ancient King \& Philosopher named \textit{Fohy}, who is believed to have lived more than four thousand years ago [\,\dots]. There are several Linear Figures attributed to him. All [of which] are found in this Arithmetic, but it is sufficient to show here the \emph{Figures of the Eight Cova}, as they are called, which are considered fundamental, \& to join to them the manifest explanation, provided that primarily a whole line ------ means unity or 1, and later that a broken line --- --- means zero or 0. The Chinese lost the meaning of the \textit{Cova} or Lineations of Fohy, perhaps more than a millennium; \& they have written commentaries on them in which they have sought I know not what hidden meanings. The true explanation now has to come from Europeans».
	} 
with 0s and 1s, is a product of \emph{human culture}, or a medley of primitive concepts, anyway connected to the evolution/experience, and fixed formally along it.
\item The remaining numbers, and all number types, are systems evolving from the branches of mathematics, in line with the dictates of powerful creation. 
\item Numbers are \emph{our} instrument (but let us safely call it \emph{filter}) to get to know the world. This is perhaps the primigenial meaning—or at least we like to perceive it that way—of the fragment attributed to Pythagorean Philolaus (\textit{Testimonia}, Part 2, D7 (B4), Stob. 1.21.7b):
\vspace{2mm}

\begingroup
\footnotesize

All the things that are known have a number, and we cannot think and know anything without it [\textgreek{\textit{καὶ πάντα γα μὰν τὰ γιγνωσκόμενα ἀριθμὸν ἔχοντι. οὐ γὰρ ὁτιῶν ⟨οἷόν⟩ τε οὐδὲν οὔτε νοηθῆμεν οὔτε γνωσθῆμεν ἄνευ τούτου}}]. 

\endgroup

\vspace{2mm}

Everything that is known is \emph{also} enumerated; this is because the very act of thought (\textgreek{νοεῖν}) has \textgreek{ἀριθμόν}: all things are known \emph{by} numbers, and without number (\textgreek{ἀριθμόν}) there is no knowledge (\textgreek{πάντα τὰ γιγνωσκόμενα}); cf. \cite[pp. 30-33]{Enriques De Santillana "Compendio di storia del pensiero scientifico dall'antichita fino ai tempi moderni"} \cite[pp. 25-26, 34-35]{Zellini "Gnomon. Una indagine sul numero"}. The number is ultimately our ability to \emph{distinguish} things, see \cite[chap. 6, and p. 64]{Klein "Greek Mathematical Thought and the Origin of Algebra"}.
\enumerationisfinis

\subsection{Logomachy of Mathematicians, and Cock-and-Bull Stories}
\label{subsection "Logomachy of Mathematicians, and Cock-and-Bull Stories"}

\begingroup
\footnotesize
There really is no such thing as [Mathematics]. There are only [mathematicians]. \\
\indent — Modified sentence of \textsc{E.H. Gombrich} \cite[p. 5]{Gombrich "The Story of Art"}\footnote{
	Gombrich is an art historian, so I replaced, in his phrase, the words “Art” and “artists” with “Mathematics” and “mathematicians”. The replacement of these words (and their interchangeability)—as it will be looked at later—is a painless operation.
	}

\endgroup

\vspace{2mm}

«Existence», for the objects of mathematics, shall be understood in a broad sense, and not ideally à la Hermite–É. Borel–Gödel–Thom–Manin (naïf realism), or à la Hardy–Connes–Penrose–Conway (descriptive realism), and nor à la Grothendieck (constructive realism), with the appropriate distinction among the various authors. 
\enumerationisinitium
\item 
\label{item "C. Hermite"}
C. Hermite \cite[lettre 410, p. 398]{Hermite "Correspondance d'Hermite et de Stieltjes II"}: 

\vspace{2mm}

\begingroup
\footnotesize
I believe that numbers and functions of analysis are not the arbitrary product of our mind [\textit{ne sont pas le produit arbitraire de notre esprit}]; I think that they exist outside of us [\textit{ils existent en dehors de nous}] with the same character of necessity as the things of objective reality [\textit{les choses de la réalité objective}], and that we encounter or discover them [\textit{nous les rencontrons ou les découvrons}], and that we study them, like physicists, chemists and zoologists [\textit{comme les physiciens, les chimistes et les zoologistes}].

\endgroup

\vspace{2mm}

\item É. Borel \cite[p. 33]{Frechet "La vie et l'oeuvre d'Emile Borel"}:

\vspace{2mm}

\begingroup
\footnotesize

My method [\,\dots] is a constant attention to study mathematical entities in themselves [\textit{êtres mathématiques en eux-mêmes}], as the biologist studies living beings [\textit{comme le biologiste étudie les êtres vivants}], to familiarize myself with them [\,\dots]. This working method led me to a realistic conception of mathematics [\textit{conception réaliste des mathématiques}] which distinguishes mathematical entities that can be effectively defined from those whose existence is purely hypothetical.\endnote{
	The following consideration from Borel via Fréchet \cite[pp. 25-26]{Frechet "La vie et l'oeuvre d'Emile Borel"} is less fanatic and historically more prudent, so it is more acceptable: «“I try to show that Mathematics is not a purely abstract game of the mind [\textit{un jeu purement abstrait de l'esprit}], but, on the contrary, is in close connection with concrete reality [\textit{étroite connexion avec la réalité concrète}]”.

	\setlength\parindent{8pt}
	It was the study of physical phenomena that suggested the notions of “continuity”, “derivative”, “integral”, “differential equation”, “vector” and “vector calculus”. And these notions, by a fair return, are part of the scientific baggage necessary for any physicist; through [these notions] he is capable of interpreting the results of his experiments [\,\dots]. If new physical phenomena suggest new mathematical models, mathematicians will have to devote themselves to the study of these new models and their generalizations with the legitimate hope that the new mathematical theories, thus constituted, will prove fruitful, by providing in turn to physicists useful forms of thought. In other words, the evolution of Physics must correspond to an evolution of Mathematics [\textit{à l'évolution de la Physique doit correspondre une évolution des Mathématiques}] which, without of course abandoning the study of classic and proven theories, have to develop taking into account the results of experience. 

	It is always in contact with Nature that mathematical Analysis has renewed itself [\textit{C'est toujours au contact de la Nature que l'Analyse mathématique s'est renouvelée}]; it is only thanks to this permanent contact that it [mathematical Analysis] was able to escape the danger of becoming a pure symbolism that turns circularly on itself [\textit{un pur symbolisme, tournant en rond sur lui-même}]».
	}

\endgroup

\vspace{2mm}

\item K. Gödel \cite[p. 128]{Godel "Russell's mathematical logic"}: 

\vspace{2mm}

\begingroup
\footnotesize
Classes [as collections of sets] and concepts [of formal logic as the properties and relations of things existing independently of our definitions and constructions] may, however, also be conceived as real objects.

\endgroup

\vspace{2mm}
 
\item R. Thom \cite[p. 100, it is an interview]{Thom "Predire n'est pas expliquer"}:  

\vspace{2mm}

\begingroup
\footnotesize
R. Thom: «Mathematical ideas are produced in our brains as long as we think of them. But since they exist [even] when we do not think about them, then they exist somewhere [\textit{quelque part}], and not only in our memory: they exist, I would say, elsewhere [\textit{également ailleurs}]». 
	
E. Noël: «Then do [mathematical ideas] already exist before they are discovered?». 
	
R. Thom: «Certainly! And they are realized [\textit{se réalisent}] [\,\dots] in this or that case, with this or that suitable material [\textit{matériau approprié}]. It is the old idea of participation in Plato['s  dialogues]».

\endgroup

\vspace{2mm}

\item Yu.I. Manin \cite[p. 4]{Manin "Mathematical Knowledge: Internal Social and Cultural Aspects"}: 

\vspace{2mm}

\begingroup
\footnotesize
[T]here is a noble vision of the great Castle of Mathematics, towering somewhere in the Platonic World of Ideas, which we humbly and devotedly discover (rather than invent). The greatest mathematicians manage to grasp outlines of the Grand Design, but even those to whom only a pattern on a small kitchen tile is revealed, can be blissfully happy. Alternatively, if one is inclined to use a semiotic metaphor, Mathematics is a proto-text whose existence is only postulated but which nevertheless underlies all corrupted and fragmentary copies we are bound to deal with.

\endgroup

\vspace{2mm}

\item G.H. Hardy: 
\label{item "G.H. Hardy"}

\vspace{2mm}

\begingroup
\footnotesize
\cite[p. 18]{Hardy "Mathematical Proof"} I have myself always thought of a mathematician as in the first instance an \emph{observer}, a man who gazes at a distant range of mountains and notes down his observations. His object is simply to distinguish clearly and notify to others as many different peaks as he can.

\cite[§ 24, p. 130]{Hardy "A Mathematician's Apology"} 317 is a prime, not because we think so, or because our minds are shaped in one way rather than another, but \emph{because it is so}, because mathematical reality is built that way.

\endgroup

\vspace{2mm}

\item A. Connes \cite[pp. 28, 40, 49]{Changeux Connes Matiere a pensee"}: 

\vspace{2mm}

\begingroup
\footnotesize
I think I am quite close to the realist point of view. For me, the sequence of prime numbers, for example, has a more stable reality than the material reality that surrounds us. We can compare the mathematician at work to an explorer discovering the world [\,\dots]. Let us compare mathematical reality [\textit{réalité mathématique}] to the material world around us. What proves the reality of this material world in addition to the perception that our brain has of it? Mainly, the coherence of our perceptions, and their permanence. More precisely, the coherence of touch and sight for one and the same individual. And the coherence between the [various] perceptions of several individuals. Mathematical reality is of the same nature [\,\dots]. I think mathematician develops a “sense”, irreducible to sight, hearing and touch, which allows him to perceive a constraining reality as it happens with physical reality but much more stable, because it is not localizable in space-time.

\endgroup

\vspace{2mm}

\item 
\label{item "R. Penrose"}
R. Penrose, whose position is recoverable in a few of his popular books, says this, summing up: the «existence» of mathematical objects is but the «objectivity of mathematical truth». For Penrose, “objectivity” means the presence of an «external standard» that is independent of our «individual opinions» and «particular culture»:

\vspace{2mm}

\begingroup
\footnotesize
\cite[p. 112]{Penrose "The Emperor's New Mind: Concerning Computers Minds and The Laws of Physics"} [O]ne can argue under the heading of ‘Platonism’ whether the objects of mathematical thought have any kind of actual ‘existence’ or whether it is just the concept of mathematical ‘truth’ which is absolute [\,\dots]. In my own mind, the absoluteness of mathematical truth and the Platonic existence of mathematical concepts are essentially the same thing. The ‘existence’ that must be attributed to the Mandelbrot set, for example, is a feature of its ‘absolute’ nature. Whether a point of the Argand plane does, or does not, belong to the Mandelbrot set is an absolute question, independent of which mathematician, or which computer, is examining it. It is the Mandelbrot set's ‘mathematician-independence’ that gives it its Platonic existence.\endnote{
	Penrose's belief is the one that will be replicated by Mandelbrot himself \cite[chap. 25, \textit{A Turning Point in Mathematics}]{Mandelbrot "The Fractalist: Memoir of a Scientific Maverick"}: «I don't feel I “invented” the Mandelbrot set: like all of mathematics, it has always been there, but a peculiar life orbit made me the right person at the right place at the right time to be the first to inspect this object, to begin to ask many questions about it, and to conjecture many answers. Though it had not been seen before, I had a very strong feeling that it existed but remained hidden because nobody had the insight to identify it».
	} 
Moreover, its finest details lie beyond what is accessible to us by use of computers. Those devices can yield only approximations to a structure that has a deeper and ‘computer-independent’ existence of its own.\footnote{
	Compare with R. Penrose \cite[p. 97]{Penrose Shimony Cartwright Hawking "The Large the Small and the Human Mind"}: «One of [my prejudices] is that the entire physical world can, in principle, be described in terms of mathematics. I am not saying that all of mathematics can be used to describe physics. What I am saying is that, if you choose the right bits of mathematics, these describe the physical world very accurately and so the physical world behaves according to mathematics. Thus, there is a small part of the Platonic world which encompasses our physical world».
	}

\cite[pp. 13-17]{Penrose "The Road to Reality: A Complete Guide to the Laws of the Universe} What I mean by this ‘existence’ is really just the objectivity of mathematical truth. Platonic existence, as I see it, refers to the existence of an objective external standard that is not dependent upon our individual opinions nor upon our particular culture. Such ‘existence’ could also refer to things other than mathematics, such as to morality or aesthetics [\,\dots]. Now, do we take the view that Fermat's assertion was always true, long before Fermat actually made it, or is its validity a purely cultural matter, dependent upon whatever might be the subjective standards of the community of human mathematicians? [\,\dots] The Mandelbrot set was certainly no invention of any human mind [\,\dots]. If it has meaning to assign an actual existence to the Mandelbrot set, then that existence is not within our minds, for no one can fully comprehend the set's endless variety and unlimited complication.

\endgroup

\vspace{2mm}

\item J.H. Conway \cite[p. 14]{Hargittai "John Conway: Mathematician of Symmetry and Everything Else"}:

\vspace{2mm}

\begingroup
\footnotesize
I'm perennially fascinated by mathematics, by how we can apprehend this amazing world that appears to be there, this mathematical world. How it comes about is not really physical anyway, it's not like these concrete buildings or the trees. No mathematician believes that the mathematical world is invented. We all [\textit{sic}] believe it's discovered. That implies a certain Platonism, implies a feeling that there is an ideal world. I don't really believe that.\endnote{
	It seems kooky that the man who wrote these words was that Conway \cite{Conway "On Numbers and Games"}, who built his fame pre-eminently as a wizard-inventor of numbers. See e.g. chapp. 0 (\textit{All Numbers Great and Small}) and 3 (\textit{The Structure of the General Surreal Number}) in \cite{Conway "On Numbers and Games"}, where the construction of the \emph{surreal numbers} is uncovered.
	}

\endgroup

\vspace{2mm}

\item A. Grothendieck \cite[2.5. \textit{Les héritiers et le bâtisseur}, p. 27 otm]{Grothendieck "Recoltes et Semailles. Reflexions et temoignage sur un passe de mathematicien"}:

\vspace{2mm}

\begingroup
\footnotesize
The structure of a [mathematical] object [\textit{chose}] is not in any manner something that we can “invent”. We can only patiently bring it to the daylight, humbly making it known, “\emph{discovering}” it.\footnote{
	\label{footnote "invention/creation and discovery"}
	The border between \emph{invention}/\emph{creation} and \emph{discovery}, in mathematics, is also blurred in the etymological suggestion; the La. \textit{invenīre} (from which \textit{inventĭo}, \textit{inventiones}) can be translated as “finding or discovering by investigating, or by construction”. For Grothendieck, inventiveness seems to coincide with a construction. A brief enquiry into the potpourri of words, with the interchange between “invention” and “discovery”, is in W.T. Gowers \cite{Gowers "Is mathematics discovered or invented?"}.
	} 
[Nevertheless] there is inventiveness in this work, and [\,\dots] we happen to perform as a blacksmith [\textit{forgeron}] or a tenacious builder [\textit{bâtisseur}] [\,\dots]. It is to \emph{express}, as faithfully as we can, these objects that we are at the core of discovery [\,\dots]. Thus we are lead to constantly “\emph{invent\textnormal{”} the language} capable of expressing, ever more finely, the intimate structure of the mathematical object, and to “construct”, with the help of this language, progressively and one step at a time, the “theories”  which are supposed to account for what has been apprehended and seen. There is a continual, uninterrupted back-and-forth movement between the \emph{apprehension} of objects and the \emph{expression} of what has been apprehended, by a language that is [repeatedly] refined and re-created.

\endgroup 
\enumerationisfinis
	
\vspace{2mm}

We believe that all the above-mentioned opinions—one of which even falls into a ludicrously divine pose—are to be put in the Big Book of \emph{cock-and-bull stories}, with the exception of Grothendieck, when he refers to an inventive step («constamment “\emph{inventer\textnormal{”} le langage} apte à exprimer de plus en plus finement la structure intime de la chose mathématique, et à “construire” à l'aide de ce langage [\,\dots] les “théories”»), and, partly, of Penrose (Section \ref{subsubsection "Many Truths I. Triangle with Two Line Segments"}). 

Questions about the concept of “existence” or “reality” in mathematics are a \emph{verbosa disputatio} \cite[between p. 16 and 17, under the Pleiadum Constellatio drawing]{Galilei "Sidereus Nuncius"} (a «wordy dispute», although fascinating), a logomachy (\textgreek{λογομαχία}), a vain and inconclusive dispute: depending on the meaning we give to this concept, the position taken can change. 

\subsubsection{The Road to Consensus}

The dispute can be closed specifying which terms representing what we call \emph{objectivity}, or what we think to be \emph{true} in mathematics, under a certain \emph{consensus}, in accordance with an \emph{organa­mento}\footnote{
	In this context, the quote from T. Landolfi \cite[p. 48]{Landolfi "Dialogo dei massimi sistemi"}, with the short story \textit{Dialogo dei massimi sistemi}, which talks about language, is an irresistibly pleasure: his irreverent sarcasm, dropped into a grotesque reconstruction, and enriched by a language full of pirouettes, is a stroke of genius.
	} 
(“organization”, “coordination”) that localizes and encodes—in time and space—our traces of a language of quantities, an \emph{imaginary language},\footnote{
	Imagination, in mathematics, is, neither more nor less, a \emph{representation} of an object through the medium of a symbolic \emph{strumentario}, from the La. \textit{imago}, which signifies “image”, “appearance”, “likeness”, “representation”, “imitation” (cf. \textgreek{μῖμος}), “thought”, “conception”.
	} 
thanks to which it is possible to identify certain attributes of \emph{shared reality}. And so A. Borel's \cite[pp. 13-14, e.a.]{A. Borel "Mathematics: Art and Science} attitude seems much more genuine to us:

\vspace{2mm}

\begingroup
\footnotesize
Many do [\,\dots] have a vague feeling that mathematics exists somewhere, even though, when they think about it, they cannot escape the conclusion that mathematics is exclusively a human creation. Such questions can be asked of many other concepts such as state, moral values, religion, etc. [\,\dots] we tend to posit existence on all those things which belong to a civilization or culture in that we \emph{share} them with other people and can exchange thoughts about them. Something becomes objective (as opposed to “subjective”) as soon as we are convinced that it exists in the minds of others in the same form as it does in ours, and that we can think about it and discuss it together. Because the language of mathematics is so precise, it is ideally suited to defining concepts for which such a \emph{consensus} exists. In my opinion, that is sufficient to provide us with a \emph{feeling} of an objective existence, of a reality of mathematics similar to that mentioned by Hardy and Hermite above, regardless of whether it has another origin, as Hardy and Hermite maintain.

\endgroup

\begin{margo}[And in theoretical physics?]
In theoretical physics one has the same thing. Here too there is the expectation of reaching a «consensus», an «agreement», or a «general mental attitude». Cf. W. Pauli \cite[p. 94]{Pauli "Phanomen und physikalische Realitat"}: 

\vspace{2mm}

\begingroup
\footnotesize
Man will have without exception the spontaneous experience of a reality [\textit{Erfahrung einer Wirklichkeit}] and will describe it in words that seem appropriate to him. However, he can recognize judgments on the being [\textit{Seinsurteile}] [of things] \textit{as conditioned} [\textit{alsbedingt}] by the efforts, hopes, desires, in short, by the general mental attitude [\textit{allgemeine seelische Einstellung}] of the individual or the group which make these statements. 

\endgroup

\vspace{2mm}

Read Pauli \cite[pp. 30-31, e.a.]{Pauli "Exclusion principle Lorentz group and reflexion of space-time and charge"} again, where he talks about how quantum mechanics has gradually taken its shape (cf. footnote \ref{footnote "Heisenberg's postcard to W. Pauli, 15 December 1924"}, p. \pageref{footnote "Heisenberg's postcard to W. Pauli, 15 December 1924"}): 

\vspace{2mm}

\begingroup
\footnotesize
After a brief period of spiritual and human confusion, caused by a provisional restriction to “Anschaulichkeit” [intuitive comprehension], a \emph{general agreement} was reached following the \emph{substitution of abstract mathematical symbols}, as for instance psi [$\psi$], \emph{for concrete pictures}. Especially the concrete picture of rotation has been replaced by mathematical characteristics of the representations of the group of rotations in three[-]dimensional [Euclidean] space. This group was soon amplified to the Lorentz group in the work of Dirac [see Section \ref{subsubsection "Lorentz Group plus Transformations"}].

I believe [\,\dots] that a rigorous mathematical formalism and epistemological analysis are both indispensable in physics [\,\dots]. While I try to use the former to connect all mentioned features of the theory with help of a richer “fulness” of plus and minus signs in an increasing “clarity”, the latter makes me aware that the final “truth” on the subject is still “dwelling in the abyss”. I refer here to Bohr's favourite verses of Schiller:\endnote{
	From the poem \textit{Spruch des Konfucius}.
	}
“Nur die Fülle führt zur Klarheit / Und im Abgrund wohnt die Wahrheit” [Only fullness leads to clarity / and truth lies in the abyss]. \margosymbol

\endgroup
\end{margo}

\subsubsection{Atiyah's No-counting Jellyfish}

In the magazine article by M.F. Atiyah \cite{Atiyah "Creation v Discovery: Conversations on Mind Matter and Mathematics"} there are several points of contact with my mathematical Weltanschauung (together with some phrases with which I disagree). The article is a direct response to Connes, to boot, in the above-mentioned conversation \cite{Changeux Connes Matiere a pensee"} with J.-P. Changeux, biologist and neurophysiologist. It is a bit long, but it deserves all the attention (e.a.):

\vspace{2mm}

\begingroup
\footnotesize
Does mathematics have an existence independent of our physical world? Do mathematicians discover theorems, rather than invent them? [\,\dots] Any mathematician must sympathise with Connes. We all feel that the integers, or circles, really exist in some abstract sense and the Platonist view is extremely seductive. But can we really defend it? Had the universe been one-dimensional or even discrete it is difficult to see how geometry could have evolved. It might seem that with the integers we are on firmer ground, and that counting is really a primordial notion.

But let us imagine that intelligence had resided, not in mankind, but in some vast solitary and isolated jellyfish, deep in the depths of the Pacific. It would have no experience of individual objects, only with the surrounding water. Motion, temperature and pressure would provide its basic sensory data. \emph{In such a pure continuum the discrete would not arise and there would be nothing to count}.

\endgroup

\vspace{2mm}

The jellyfish is the characterization of a symbolic being who lives in perfect continuity, immersed in a sort of ideal sea without discreteness; but we know—cf. point \ref{item "Continuity and discreteness"} in Section \ref{subsubsection "Euclidean Discretum, and Discrete Numerability"}—that (numerical) mathematics exists \emph{due to discreteness}. 

\vspace{2mm}

\begingroup
\footnotesize
Even more fundamentally, in a purely static universe without the notion of time, causality would disappear and with it that of logical implication and of mathematical proof [\,\dots]. It may be argued that such “gedanken universes” are not to be taken seriously. Our actual universe is a given datum and the inevitable background of all intelligent discussion. But this is tantamount to conceding that mathematics has evolved from the human experience. \emph{Man has created mathematics by idealising or abstracting elements of the physical world}.

For a Platonist like Connes mathematics lives in some ideal world. I find this a difficult notion to grasp and prefer to say, more pragmatically, that mathematics lives in the collective consciousness of mankind [\,\dots]: there are two essential components of mathematics. In the first place it deals with concepts and abstract processes which live in the conscious mind of the individual mathematician. Second, it must be communicable to other mathematicians. 

[\,\dots] Where does this point of view leave the dichotomy between creation and discovery in mathematics? By resisting the embrace of the Platonic world have we lost the possibility of making “discoveries”? Is every theorem man-made? Not at all [\,\dots]. [M]an creates the concepts of mathematics but he discovers the subsequent connections between them. The reason he can have it both ways is that mathematics is firmly rooted in the real world.

[\,\dots] In his dialogue with Connes, Changeux keeps hitting the Platonist rock. As a hard-headed experimental scientist Changeux wants to identify mathematics with what actually goes on in the brain. For him this is the only reality and the only place where mathematics exists. Connes disputes this extreme attitude and prefers to say that mathematical reality (which exists elsewhere) is reflected in the neurological processes of the brain. To confuse the two is like identifying a piece of literature or music with the ink and paper on which it is recorded. It is hard to disagree, but fascinating questions remain.

\endgroup

\vspace{2mm}

A poem or a song is not what is written on the paper, of course; but is what is written in our mind/heart. So we can rest assured: there is no a Platonic world in which the Poetry of a Zanzotto or the Music of the GY!BE hovers \& lives, also at the larval stage, and from which it descends; there is no a cosmos, in the Borgesian style, in which letters and notes floating in the air await to be captured by gifted writers and musicians. The same remark is replicable for mathematicians, who are not a privileged or «divine» race, as Dedekind mistakenly judges. Atiyah's article goes on like this:

\vspace{2mm}

\begingroup
\footnotesize
Man has been the ultimate winner of the evolutionary process and his brain has the structure needed to produce mathematics. Would a different neurological solution have led to a different kind of mathematics, or does mathematics depend only on the functional capacity of the brain, not on its biological mechanism?

If one views the brain in its evolutionary context then the mysterious success of mathematics in the physical sciences is at least partially explained. The brain evolved in order to deal with the physical world, so it should not be too surprising that it has developed a language, mathematics, that is \emph{well suited for the purpose}.

\endgroup

\vspace{2mm}

We will cover this subject-matter a bit more in Section \ref{subsubsection "Selection, Colors, and Understanding"}.

\subsubsection{Many Truths I. Triangle with Two Line Segments}
\label{subsubsection "Many Truths I. Triangle with Two Line Segments"}

\begingroup
\footnotesize
\textgreek{Σχήματα εὐθύγραμμά ἐστι τὰ ὑπὸ εὐθειῶν περιεχόμενα, τρίπλευρα μὲν τὰ ὑπὸ τριῶν}.\footnote{
	«Rectilinear figures are those contained by straight lines: trilateral figures being those contained by three [straight lines]».
	} \\
\indent — \textsc{Euclid} \cite[\textgreek{Ὅροι, ιθ´, Στοιχείων α´}, Book I, p. 6]{Euclidis "Elementa I"}

\vspace{2mm}

A triangle is a rectilinear figure included by three sides. \\
\indent — \textsc{O. Byrne} \cite[Book I, Euclid of Byrne, Def. XXI, p. xx]{Byrne "The First Six Books of The Elements of Euclid"} 

\endgroup

\vspace{2mm}

We ask a trivial question: is it possible to construct a polygon with three edges and three vertices having available two sides, or two line segments? No, that cannot be. The same occurs with numbers: if we say e.g. that a \emph{prime number} is a number $n \in \mathbb{N}_+$, greater than 1, that cannot be obtained by multiplying two smaller natural numbers, then 5 or 317 are prime numbers, whether we like it or not (cf. Hardy, point \ref{item "G.H. Hardy"} above).

By doing it this way, then, the \emph{triangle}, and the numbers \emph{5} or \emph{317}, exist in Penrose's sense, and Penrose is right to identify the “existence” of an object of mathematical thought with some «objectivity of mathematical truth» that «transcends» mere opinion. But this is not the \emph{heart} of the matter. Let us not be infatuated/misled with/by certain words, such as “existence”, “(transcendent) reality”, or “truth”.

It goes without saying that, in mathematics, a system of \emph{fundamental rules}—let us call them like that—is needed to go beyond individual or majority standpoints, or certain ideas which are «agreed by all», speaking of which, Penrose \cite[p. 13]{Penrose "The Road to Reality: A Complete Guide to the Laws of the Universe} stresses faultlessly a problematic circularity: «Do we mean ‘agreed by all’, for example, or ‘agreed by those who are in their right minds’, or ‘agreed by all those who have a Ph.D. in mathematics’ (not much use in Plato's day) and who have a right to venture an ‘authoritative’ opinion?». 

And yet, in mathematics, in its historical growth, there is a \emph{river of consensus}, carrying every thought, and imposing an evolution in the objectivity of the rules, so that the latter are pushed towards new horizons and perspectives, by enlarging them, or, on the contrary, by reducing them (Section \ref{section "Mathematical Evolutionism: What is a proof?"}). Not only that. There are rules, in a lot of mathematics, which are altogether arbitrary (Section \ref{section "Math-Inventiveness: Arbitrariness and Imaginative Endeavor"}). 

E. Frenkel, a specialist in the \emph{Langlands program}\footnote{
	The start of the Langlands program goes back to R. Langlands' letter \cite{Langlands "R. Langlands' letter to Prof. A. Weil"} to A. Weil, with the appearance of the ${^\Langlands}G$ group, or $\Langlands$-group of $G$, the Langlands dual group of a reductive algebraic group $G$, see \cite{Mueller "On the genesis of Robert P. Langlands' conjectures and his letter to Andre Weil"} \cite{Mueller "A Glimpse at the Genesis of the Langlands Program"} \cite[part IV, various authors, pp. 109-400]{Mueller Shahidi (Eds.) "The Genesis of the Langlands Program"}.
	} 
\cite{Frenkel "Lectures on the Langlands Program and Conformal Field Theory"} \cite[chapp. 1, 10]{Frenkel "Langlands Correspondence for Loop Groups"} \cite{Frenkel "Langlands Program Trace Formulas and their Geometrization"}, who reveals himself \cite[p. 234]{Frenkel "Love and Math: The Heart of Hidden Reality"} to be a Platonist à la Penrose, writes \cite[p. 8]{Frenkel "Mathematics Love and Tattoos"} e.g. that 

\vspace{2mm}

\begingroup
\footnotesize
In mathematics there is only one truth, and only one path to reach that truth. My mathematical work is perceived and interpreted in essentially the same way by everybody who reads it. Not so [\,\dots] in the arts in general. First, there isn't a single truth, and second, there are so many different paths to express the truth. And the viewer is always part of an artistic project: at the end of the day, it's all in the eye of the beholder. 

\endgroup

\vspace{2mm}

\emph{This is not correct}. Just take a look at the history (Section \ref{section "History and Mathematics"}). What does “one truth” mean? Could it mean that a triangle (every triangle) has three sides and three angles, and 5 is a prime, now and always? There again, we are facing a \textit{trivialis} necessity inherent in mathematics. But even in mathematics there are \emph{many truths}, just as there are many paths to reach them. Fortunately, Frenkel thinks (ibid.) that «Doing mathematics is a creative pursuit that requires passion, just like painting, music, and poetry» (cf. Section \ref{subsection "Fluvial- and Æolian-like Processes"}).

\subsubsection[Many Truths II. The Normality of $\pi$]{Many Truths II. The Normality of $\mathbold{\pi}$}

\begingroup
\footnotesize
This question raises difficulties for those who are too ready to identify truth and provability. If you look at actual mathematical practice, and in particular at how mathematical beliefs are formed, you find that \emph{mathematicians have opinions long before they have formal proofs}. \\
\indent — \textsc{W.T. Gowers} \cite[p. 194, e.a.]{Gowers "Does Mathematics Need a Philosophy?"} 

\endgroup

\vspace{2mm}

Such a \emph{multi-truth} also surfaces, in its own way, in number theory; see the case mentioned by W.T. Gowers, about the sequences of digits that occur with the right frequencies forming numbers called \emph{normal}. The normality of $\pi$ is, supposedly, an unsolved problem and, at the same time, an unproven theorem. The difficulty lies in finding and proving that the expansion of $\pi$ contains a million sevens in a row, for instance. But here is Gowers's comment \cite[ivi]{Gowers "Does Mathematics Need a Philosophy?"}: 

\vspace{2mm}

\begingroup
\footnotesize
So what, then, is the status of the reasonable-sounding heuristic argument that $\pi$ contains a million sevens in a row, an argument that convinces me and many others? This question raises difficulties for those who are too ready to identify truth and provability. If you look at actual mathematical practice, and in particular at how mathematical beliefs are formed, you find that mathematicians have opinions long before they have formal proofs. When I say that I think $\pi$ almost certainly has a million sevens somewhere in its decimal expansion, I am not saying that I think there is almost certainly a (feasibly short) \emph{proof} of this assertion—perhaps there is and perhaps there isn't. So it begins to look as though I am committed to some sort of Platonism.

\endgroup

\vspace{2mm}

Let us put it in other words: provability and truth take different roads; sometimes they are parallel, and other times they are crossed; but the road is never the same. This is how D.R. Hofstadter \cite[p. 19]{Hofstadter "Godel Escher Bach: an Eternal Golden Braid"} recaps  (referring to Gödel's revolution, cf. Section \ref{subsection "Extra-logical Objects, and Gödelian Suggestions"}): «Gödel showed that provability is a weaker notion than truth, no matter what axiomatic system is involved». But just be cautious: such a phrase, \emph{prima facie}, can instill the nefarious idea that truth is something monolithic; it is a wrong impression. The concept of mathematical truth is something \emph{prismatic}, \emph{multifaceted}.

\chapter{Outro—\emph{Parva Mathematica}: \emph{Libera Divagazione} \sfrac{2}{8}}
\label{chapter "Outro—Parva Mathematica: Libera Divagazione 2/8"}

\section{\emph{Naturæ Mæandri} and \emph{Filum Ariadneum}: a Botanical Comparison}
\label{section "Naturæ Mæandri and Filum Ariadneum: a Botanical Comparison"}

\begingroup
\footnotesize
Nomina si nescis, perit \& cognitio rerum \cite[VII, § 210, p. 158]{Linnaeus "Philosophia Botanica in qua explicantur Fundamenta Botanica cum Definitionibus Partium"}.\footnote{
	«If you do not know the names of things, the knowledge of them is lost too».
	} \\
\indent Nomina respondeant Methodo Systematicæ [\,\dots]. Scientia Naturæ innititur Cognitioni Naturalium Methodicæ \& Nomenclaturæ Systematicæ tamquam filo ariadneo, fecundum quod Naturæ mæandros unice tutoque permeare liceat \cite[pp. 7-8]{Linnaeus "Systema Naturae Per Regna Tria Naturae"}.\footnote{
	«Names respond to a systematic method. Science of nature rests on the method of knowledge [that we have] of it and on the systematic nomenclature, which is like an Ariadne's thread, only thanks to which it is possible to safely flow through the meanders of nature».
	} \\
\indent — \textsc{C. von Linné}

\vspace{2mm}

In their research, mathematicians study the facts of mathematics with a taxonomic zeal similar to a \emph{botanist} studying the properties of some \emph{rare plant}.\footnote{
	This passage is taken from \cite[\textit{The Double Life of Mathematics}, pp. 89-90, e.a.]{Rota "Indiscrete Thoughts"} that deserves to be fully read: «Are mathematical ideas invented or discovered? [\,\dots] [M]athematics has been leading a double life. In the first of its lives mathematics deals with \emph{facts}, like any other science. It is a fact that the altitudes of a triangle meet at a point; it is a fact that there are only seventeen kinds of symmetry in the plane; it is a fact that there are only five non-linear differential equations with fixed singularities; it is a fact that every finite group of odd order is solvable [\,\dots]. In its second life, mathematics deals with proofs. A mathematical theory begins with definitions and derives its results from clearly \emph{agreed-upon} rules of inference. Every fact of mathematics must be ensconced in an axiomatic theory and formally proved if it is to be \emph{accepted} as true. Axiomatic exposition is indispensable in mathematics because the facts of mathematics, unlike the facts of physics, are not amenable to experimental verification. The axiomatic method of mathematics is one of the great achievements of our culture. However, it is only a method. Whereas the facts of mathematics once discovered will never change, the method by which these facts are verified has changed many times in the past, and it would be foolhardy to expect that changes will not occur again at some future date». 
	
	Yeah, well, certainly, mathematics—as B. de Finetti \cite[p. 7]{de Finetti "Pirandello Maestro di Logica"} pointed out—must take care not to harbour «inveterate rationalistic illusions», considering itself as a \emph{complexus} of «absolute truths» that are extraneous to the \emph{historical relativism} of its evolution. But watch out: a mathematical \emph{fact}, in the sense of Rota, may very well be an \emph{artefact} (conceptual construct) of the mind whose process of discovery is just the role-playing in the process of creation, cf. Section \ref{subsection "Mathematics as a Technical Tool"}. Between a \emph{fact of mathematics} and a \emph{plant} it does open up a \emph{whole reality}.
	} \\
\indent — \textsc{G.-C. Rota} \cite[p. 89, e.a.]{Rota "Indiscrete Thoughts"} 

\endgroup

\vspace{2mm}

The \textit{naturæ mæandros} of Linnaeus bring to mind the \textit{oscuro laberinto} of Galilean nature (universe) \cite[p. 25]{Galilei "Il Saggiatore"}, in the celebrated piece in which instead of «names» there are «characters» belonging to the language of mathematics, that is, «triangles, circles, \& other Geometric figures» (see Section \ref{subsection "In the Bliss of Goddess Geometry"}).

In the natural sciences (from physics to botany), there are «things» (facts, events, organisms, and everything that presents itself to our eyes), but also «names» to identify them. Mathematics, because contains within itself a phytological-like propensity, does something similar to what Linnaeus' names do in botany: in the service of physics, identifing, fixing a nomenclature, and describing the labyrinths/meanders of nature.\footnote{
	But as soon as we skip from the Republic of pure mathematics to the mathematical physics, or to a mathematics with phenomenal velleities, the judgment changes promptly. In the 2nd book of his \textit{Exposition} (entitled \textit{Des mouvemens réels des corps célestes}), P.-S. de Laplace \cite[p. 94]{Laplace "Exposition du systeme du monde 1799"} writes: «If man were confined to collecting facts [\textit{recueillir des faits}], the sciences were but a sterile nomenclature [\textit{nomenclature stérile}], and he would never have known the great laws of nature».\endnote{
	Here is the Laplacian passage \cite[p. 94]{Laplace "Exposition du systeme du monde 1799"} in full: «Si l'homme s'étoit borné à recueillir des faits; les sciences ne seroient qu'une nomenclature stérile, et jamais il n'eût connu les grandes loix de la nature. C'est en comparant entr'eux les phénomènes, en cherchant à saisir leurs rapports; qu'il est parvenu à découvrir ces loix toujours empreintes dans leurs effets les plus variés». It is through the comparison between (variegated) phenomena, trying to capture their relationships, that, according to Laplace, we are led to the laws of nature.
	}
	}

\begin{margo}[Botanical nomenclature \& demon of order: a uniting thread for mathematics and morality]
\label{margo "Botanical nomenclature and demon of order: a uniting thread for mathematics and morality"}
To understand the closeness between botany and mathematics, or of its sense of order, and the familiarity with other disciplines, such as morality, we can recall an article by C. Magris \cite[p. 117, e.a.]{Magris "Utopia e disincanto"}:

\vspace{2mm}

\begingroup
\footnotesize
1735 is the year in which [Linnaeus] publishes the first edition of \textit{Systema Naturæ} \cite{Linnaeus "Systema Naturae Per Regna Tria Naturae"}, the great classification that will make him a ruler and a symbol of the natural sciences, a writer from whom Rousseau said, referring especially to his \textit{Philosophia Botanica} \cite{Linnaeus "Philosophia Botanica in qua explicantur Fundamenta Botanica cum Definitionibus Partium"}, to have drawn more profit than any book on morality. The great moralists, capable of thoroughly probing life and its anarchy, are pressed by the \emph{demon of order}, by the \emph{passion to catalog}, \emph{to define}; this \emph{passion for the totality is doomed to defeat}, because \emph{no system completely harnesses the unpredictable irregularity of existence}, but only \emph{the lucid and geometric love in the system allows us to truly understand the originality of life, its deviation from the law}.

It is the encyclopedia, with its \emph{rigorous alphabetical order} and its \emph{cadastre}, that evokes \emph{the chaotic and proliferating image of reality}; whoever flirts with disorder and shows off in confused poses, scattering the cards over the table to give himself a touch of ingenious recklessness, is a harmless and well-intentioned rhetorician, like one who exhibits his distraction or his own madcap youth and will hardly understand the demonicity of existence.

Rousseau was right to see in the great Swedish botanist a master of morality, viz. of conceptual procedures that \emph{educate \textnormal{[}our\textnormal{]} thought to penetrate the ambiguous and treacherous multiplicity of the world}.\endnote{
	Extended original It. version: «Nel 1735 Linneo, visitando un giardino ad Amburgo, annota sul suo taccuino l'epigrafe scritta sull'ingresso: “Non fare alcun male e non ne verrà alcuno a te, così come l'eco ti rimanda il tuo stesso grido nel bosco”. È l'anno in cui egli pubblica la prima edizione del \textit{Systema Naturæ}, la grande classificazione che farà di lui un sovrano e un simbolo delle scienze naturali, uno scrittore dal quale Rousseau diceva, riferendosi specialmente alla sua \textit{Philosophia Botanica}, di aver tratto più profitto che da qualsiasi libro di morale. I grandi moralisti, capaci di scandagliare a fondo la vita e la sua anarchia, sono incalzati dal demone dell'ordine, dalla passione di catalogare, definire; questa passione di totalità è votata alla sconfitta, perché nessun sistema imbriglia completamente l'imprevedibile irregolarità dell'esistenza, ma soltanto il lucido e geometrico amore del sistema permette di capire veramente l'originalità della vita, il suo scarto rispetto alla legge.
	
	\setlength\parindent{8pt}
	È l'enciclopedia, col suo rigoroso ordine alfabetico e col suo catasto, che evoca l'immagine caotica e proliferante della realtà; chi civetta col disordine e si atteggia in pose confuse, sparpagliando le carte sul suo tavolo per darsi un tocco di sregolatezza geniale, è un retore innocuo e benintenzionato, come chi esibisce la propria distrazione o la propria giovinezza scapestrata e difficilmente potrà comprendere la demonicità dell'esistenza.

	Rousseau aveva ragione di scorgere nel grande botanico svedese un maestro di morale ossia di procedure concettuali che educano il pensiero a penetrare l'ambigua e infida molteplicità del mondo».
	}

\endgroup
\end{margo}

It is as if physics («the chaotic and proliferating image of reality») emerged only from an encyclopedic assemblage—think about the ordering assemblage of mathematics. \margosymbol

\section{Rigorousness and Circularity}

\begingroup
\footnotesize
The theorems of mathematics motivate the definitions as much as the definitions motivate the theorems. A good definition is “justified” by the theorems that can be proved with it, just as the proof of the theorem is “justified” by appealing to a previously given definition. There is, thus, a \emph{hidden circularity} in formal mathematical exposition. The theorems are proved starting with definitions; but the definitions themselves are motivated by the theorems that we have previously decided ought to be correct. \\
\indent — \textsc{G.-C. Rota} \cite[p. 97, e.a.]{Rota "Indiscrete Thoughts"} 

\endgroup

\vspace{2mm}

Often one is led to confuse the axiomatic exposition of mathematics—which can erroneously translate into a lack of invention in the selection of adequate definitions—with the \emph{logicus strictus rigor}  (see Section \ref{subsection "Extra-logical Objects, and Gödelian Suggestions"}, and footnote \ref{footnote "Feynman, mathematics, language and logic"} on p. \pageref{footnote "Feynman, mathematics, language and logic"}); but on closer inspection, a \emph{circular} schema, dictated by the need to adjust definitions and theorems, appears in the genesis of mathematics, so that both go hand in hand. Far from being a lack of arbitrariness, the mathematical circularity is the affirmation of personal whim, within the limits, assuredly, of random choices that historically lead to this or that result.

\section{Grainy Music and Chance}
\label{section "Grainy Music and Chance"}

\begingroup
\footnotesize
I think one clearly discerns the internal grounds of the coincidence or parallelism [\,\dots] between the mathematical and musical \textgreek{ἔθος}. May not Music be described as the Mathematic of sense, Mathematic as Music of the reason? [\,\dots] Thus the musician \emph{feels} Mathematic, the mathematician \emph{thinks} Music,—Music the dream, Mathematic the working life—each to receive its consummation from the other. \\
\indent — \textsc{J.J. Sylvester} \cite[p. 613]{Sylvester "Algebraical Researches containing a disquisition on Newton's Rule for the Discovery of Imaginary Roots"}\endnote{
	An explicit reference to Sylvester is found in the English edition of J.A.E. Dieudonné's work \textit{Pour l'honneur de l'esprit humain. Les mathématiques aujourd'hui} (Hachette, Paris, 1987), transl. by H.G. and J.C. Dales as \textit{Mathematics – The Music of Reason} (Springer-Verlag, Berlin, Heidelberg, 1998\textsuperscript{2c.pr}), just to recollect the close and universal relationship between mathematics and music.
	}

\vspace{2mm}

[W]hen scientific and mathematical thought serve music, or any human creative activity, it should amalgamate dialectically with intuition. Man is one, indivisible, and total. He thinks with his belly and feels with his mind. \\
\indent — \textsc{I. Xenakis} \cite[p. 181]{Xenakis "Formalized Music: Thought and Mathematics in Composition"}

\endgroup

\vspace{2mm}

Scientific thought often proceeds ambiguously, and a background of \emph{ambiguity},\endnote{
	\label{endnote "Lolli, Ambiguità. Una viaggio fra letteratura e matematica"}
	An essay that tries to break into the \emph{unconfined empire of ambiguity} in mathematics, but also, parallelly, in literature, is \cite[see capp. VIII-XII, pp. 77-209]{Lolli "Ambiguita. Una viaggio fra letteratura e matematica"}.
	}
marked by a \emph{creative agency}, is, and remains, irremovable even with the completion of a theorem, or even a theory. Rarely an idea, related to a mathematical or mathematico-physical concepts, is born in a clear way, exactly as we see it today, but it is the result of multiple minds  which have succeeded one another, as e.g. G.V. Schiaparelli \cite[p. 11]{Schiaparelli "Scritti sulla storia della astronomia antica Parte prima - Scritti editi Tomo Secondo}\footnote{
	«If today, for us who are the last grandchildren of [\,\dots] illustrious masters, by making the most of their mistakes and their discoveries, and climbing to the top of the building erected by them, we were able to embrace a wider horizon with our eyes, it would be a dolt haughtiness to believe for this that we have a longer and sharper view compared to them. All our merit lies in having come into the world later [\textit{stolta superbia nostra sarebbe il credere per questo d'aver noi vista più lunga e più acuta di loro. Tutto il nostro merito sta nell'esser venuti al mondo più tardi}]».
	} 
reminds us; that is, it is the—uncontrolled—result of a cross-stratifying sediment, and of a reorganization by adventurous, if not fortuitous, actions (cf. Section \ref{subsection "Fluvial- and Æolian-like Processes"}). Certain solutions appear and then disappear, like a karst river, only to reappear in other places where they are needed. In brief, the advancement of scientific thought, along its shapes and nervaturæ, is a \emph{blind} concrescence, not teleologically determined, through deviations and variations that are Darwinianly \emph{accidental} or \emph{spontaneous} \cite[pp. 94, 213]{Darwin "On the Origin of Species by Means of Natural Selection Or The Preservation of Favoured Races in the Struggle for Life (1859)"} \cite[p. 205]{Darwin "On the Origin of Species by Means of Natural Selection Or The Preservation of Favoured Races in the Struggle for Life (1871)"}.

According to this analogy, theorems of mathematics and statements of physics (expressed in mathematical language) evolve a bit like a musical composition by I. Xenakis \cite[chapp. II-III, V]{Xenakis "Formalized Music: Thought and Mathematics in Composition"}\footnote{
	I. Markovian Stochastic Music—Theory, II. Markovian Stochastic Music—Applications, V. Free Stochastic Music by Computer.
	} 
\cite{Varga "Conversations with Iannis Xenakis"}: the evolutionary process is \emph{retrospectively predictable} in its entirety, but the individual events, which contribute to forming it, are \emph{wholly random}. Xenakis says \cite[pp. 73, 76, e.a.]{Varga "Conversations with Iannis Xenakis"}: 

\vspace{2mm}

\begingroup
\footnotesize
I felt that I could solve the \emph{slow change} in the large masses of sound events only with the help of \emph{probability} [theory] [\,\dots]. I, in trying to control mass events, naturally reached \emph{deter­minism and indeterminism}. In determinism the same cause always has the same effect. There's no deviation, no exception. The opposite of this is that the effect is always different, the chain never repeats itself. In this way we reach absolute chance—that is, indeterminism.\footnote{
	Xenakis later explains \cite[pp. 84-85, e.a.]{Varga "Conversations with Iannis Xenakis"} that: «In order for me to write \textit{Herma} \cite{Xenakis "Herma: Music symbolic pour piano"} I had to do logical operations with the sets. The basic operations are intersection, union and negation [\,\dots]. Natur­ally I had to place it all into time because so far I was only work­ing theoretically, with an outside-time structure (the logical functions) [\,\dots]. Time, then, is a means for us to unfold the outside­ time structure of the piece. How can we demonstrate the sounds of a set on the piano? By playing them one after the other. In what order? If we played them chromatically, upwards or downwards, we would be observing too strict a rule. If we want to be free the sounds should follow without any melodic law, independently of one another. So we have to play them at random. In other words, to demon­strate the elements of a set we have to use the stochastic method [\,\dots]. Let us say, as an example, that I have an $A$ set and a $B$ set. What can I do with them? I can combine them in a logical way. One such way, as I said, is their union—in other words I take all the notes of the two sets. Then I can take the sounds the two sets have in common. And finally I can take the sounds that the two sets don't have in common. There are, of course, other, more complex logical functions. In each case I get a new set. How can I demonstrate the elements of the sets? By playing them. But in order to remain neutral I have to play them at random. I emphasize: \emph{only the sounds that I play at random demonstrate the logical functions of the sets, nothing else}. The set can be amorphous as in \textit{Herma} or it may occur that its elements are connected in some way. In other words the sets may have an internal structure. That's what led me to group structures».
	}

\endgroup

\section{History and Mathematics}
\label{section "History and Mathematics"}

\begingroup
\footnotesize
[T]he nervous system is based on two types of communications: those which do not involve arithmetical formalisms, and those which do, i.e. communications of orders (logical ones) and communications of numbers (arithmetical ones). The former may be described as language proper, the latter as mathematics. It is only proper to realize that language is largely a historical accident. The basic human languages are traditionally transmitted to us in various forms, but their very multiplicity proves that there is nothing absolute and necessary about them. Just as languages like Greek or Sanskrit are historical facts and not absolute logical necessities, it is only reasonable to assume that \emph{logics and mathematics are} similarly \emph{historical, accidental forms of expression}.\footnote{
	It becomes thought-provoking to ask oneself if—since the language of modern humans (from homininian/australopith ancestors) has been guided by \emph{historical fortuity}—a different mathematical language, with a totally different linguistic evolution, can fatally lead to a different mathematical understanding of reality, videlicet, to a different physics. It is the debated \textit{controversia} sinking into the \emph{Sapir–Whorf hypothesis}. B. Lee Whorf \cite[pp. 212-213, e.a.]{Lee Whorf "Science and Linguistics"} points out that: «It was found that the background linguistic system (in other words, the grammar) of each language is not merely a reproducing instrument for voicing ideas but rather is itself the shaper of ideas, the program and guide for the individual's mental activity, for his analysis of impressions, for his synthesis of his mental stock in trade. Formulation of ideas is not an independent process, strictly rational in the old sense, but is part of a particular grammar, and differs, from slightly to greatly, between different grammars. \emph{We dissect nature} along lines laid down by \emph{our native languages}. The categories and types that we \emph{isolate} from the world of phenomena we do not find there because they stare every observer in the face; on the contrary, \emph{the world} is presented in a \emph{kaleidoscopic flux of impressions} which has to be \emph{organized by our minds}—and this means largely by the \emph{linguistic systems} in our minds. \emph{We cut nature up, organize it into concepts, and ascribe significances} as we do, largely because we are parties to an agreement to organize it in this way—an agreement that holds throughout our speech community and is codified in the patterns of our language».
	} \\
\indent — \textsc{J. von Neumann} \cite[pp. 81-82, e.a.]{Neumann "The Computer and the Brain"}

\vspace{2mm}

Mathematics is nothing if not a historical subject \emph{par excellence}. \\
\indent — \textsc{G.-C. Rota} \cite[p. 100]{Rota "Indiscrete Thoughts"} 

\endgroup

\vspace{2mm}

\enumerationisinitium
\item Mathematics, we said, grows without a method of control, \emph{grain by grain}, and varies by a relentless \emph{rearrangement}. This is because, as Rota says, mathematics is a historical subject-matter in an especially representative way.\footnote{
	\cite[p. 99]{Rota "Indiscrete Thoughts"}: «In short, no mathematician will ever dream of attacking a substantial mathematical problem without first becoming acquainted with the \emph{history} of the problem, be it the real history or an ideal history reconstructed by the gifted mathematician».
	} 
	It is enough to reveal the human-inventive aspect of mathematics, which has no intrinsic relationship with nature. \emph{Nature has no history}: it is just about events. History, by contrast, is \emph{narrative} activity on the part of man, that is, is a critical \emph{interpretation} (judgment) or \emph{representation} (investigation) of events, which takes shape as a «display of a [personal] inquiry» (\textgreek{ἱστορίης ἀπόδεξις}).\footnote{
	\label{footnote "Res gestæ and rerum gestarum narratio"}
	Herodotus \cite[A 1, p. 2]{Herodotus "Histories Vol. I Books I and II}. An unbiased distinction between history as \emph{res gestæ} and history as \emph{rerum gestarum narratio} tastes like the typical academic fluff. The two are united.
		}
\item The history of mathematics can be understood as a fortuitous/uncontrolled evolution of thought, which cannot be traced back to a univocal growth, or to the aseptic development. The history of mathematical thought (like any story) is conditioned by \emph{contingent stimuli and despondencies}, which have little to do with scientific rationality, psychological neutrality, or organic methodology (artificially reconstructed a posteriori).
\enumerationisfinis

\section{Mathematical Evolutionism: What is a \emph{proof}?}
\label{section "Mathematical Evolutionism: What is a proof?"}

\begingroup
\footnotesize
Le verità matematiche non sono come un continente che si va a mano a mano scoprendo. Le scoperte matematiche sono conseguenza di particolari \emph{creazioni della mente, varie e mutevoli} nella diuturna, affannosa indagine che tien dietro all'incessante evolversi ed affinarsi dell'intelligenza. È per questo che la Matematica, la più luminosa fra tutte le Scienze, è com'esse un \emph{perpetuo divenire}.\footnote{
	«Mathematical truths are not like a continent that is bit by bit being discovered. The mathematical discoveries are the consequence of particular \emph{creations of the mind, varied and changeable} in the daytime, frantic survey keeping up with the incessant evolution and refinement of intelligence. This is why Mathematics, the most luminous of all the Sciences, is like a \emph{perpetual becoming}».
	} \\
\indent — \textsc{M. Cipolla} \cite[p. 29, e.a.]{Cipolla "Sui fondamenti logici della Matematica secondo le recenti vedute di Hilbert"}  

\vspace{2mm}

We often hear that mathematics consists mainly in “proving theorems”. Is a writer's job mainly that of “writing sentences”? A mathematician's work is mostly a tangle of guesswork, analogy, wishful thinking and frustration, and proof, far from being the core of discovery, is more often than not a way of making sure that our minds are not playing tricks. \\
\indent — \textsc{G.-C. Rota} \cite[p. xviii]{Rota "Introduction to The Mathematical Experience"} 

\endgroup

\vspace{2mm}

The salient issue is that mathematics is \emph{not} something \emph{fixed}, but it evolves, from age to age and from culture to culture, so there is no external and universal archetype, with the exclusion of the fundamental rules (see Penrose, point \ref{item "R. Penrose"}, on p. \pageref{item "R. Penrose"}, and Section \ref{subsubsection "Many Truths I. Triangle with Two Line Segments"}). In the course of time, interests and questions mutate and adapt within the mathematical debate. The same concepts of \emph{definition} or \emph{proof} are historically conditioned,\footnote{
	We consider a bit of malarkey to say that the proof of a conjecture implies that the (conjectural) proposition has always been true—in the Platonic acceptation—even when the proof, in the past, was not yet in fermentation among the mathematicians' intuitions; one thinks of Fermat's last theorem, in its centuries-old history, proved by A. Wiles \cite{Wiles "Modular elliptic curves and Fermat's Last Theorem"}, with the support of R. Taylor \cite{Taylor and Wiles "Ring-theoretic properties of certain Hecke algebras"}.
	}
and suffer a (inter)subjective selection criterion,\footnote{
	G.-C. Rota \cite[pp. 189-190, e.a.]{Rota "The Phenomenology of Mathematical Proof"} does not beat about the bush: «G.H. Hardy wrote that every mathematical proof is a form of debunking [of the fakery that lies concealed underneath every logically correct proof]. We propose to change one word in Hardy's sentence, as follows: \emph{Every mathematical proof is a form of pretending}. Nowhere in the sciences does one find as wide a gap as that between the written version of a mathematical result and the discourse that is required in order to understand the same result. The axiomatic method of presentation of mathematics has reached in our time the zenith of fanaticism. A piece of mathematics, as it is written today, cannot be understood and appreciated without additional strenuous effort [\,\dots]. Do not get me wrong. I am not condemning the axiomatic method. There is at present no viable alternative to axiomatic presentation».
	}
or reveal some kind of social acceptance.\footnote{
	Cf. e.g. W.P. Thurston \cite[§ 4. \textit{What is a proof?}, pp. 168-169, e.a.]{Thurston "On proof and progress in mathematics"}: «Within any field, there are certain theorems and certain techniques that are generally known and generally accepted. When you write a paper, you refer to these without proof [\,\dots]. At first I was highly suspicious of this process. I would doubt whether a certain idea was really established [\,\dots]. [But then I realized that] mathematical knowledge and understanding were embedded in the minds and in the \emph{social fabric} of the community of people thinking about a particular topic. This knowledge was supported by written documents, but the written documents were not really primary».
	} 
There is no method of absolute validity, but many responsive practices extending throughout hypotheses (see Section \ref{subsubsection "Scholium: Greco-Hellenistic Scientific Modus Operandi (the Origin of Hypotheses)"}), presumptions, presuppositions, postulates, specifically chosen, or selected ad hoc. D. Hilbert's \cite[p. 85]{Hilbert "Die Grundlagen der Mathematik"} persuasion that «mathematics is a presuppositionless science [\textit{voraussetzungslose Wissenschaft}]» must be largely turned down. 

We do not have a mathematical truth that is crystallized, out of a historical epoch: what is valid, rigorous, or true is only a procedure that is shared by some members of the mathematical, or physico-mathematical, community, with personal judgments, linked to \emph{mental schemata}, and \emph{aesthetic sensitivities} (Section \ref{section "Beauty of Mathematics vs. Mathematics of Beauty"}), with {subjective evaluation}.\footnote{
	Nowadays, an uproarious example of how difficult it is to reach a harmony, a \emph{consensus}, on what \emph{proof} means in mathematics, is given by S. Mochizuki's \textit{Inter-universal Teichmüller theory} \cite{Mochizuki "Inter-universal Teichmuller Theory I: Construction of Hodge Theaters"} \cite{Mochizuki "Inter-universal Teichmuller Theory II: Hodge-Arakelov-Theoretic Evaluation"} \cite{Mochizuki "Inter-universal Teichmuller Theory III: Canonical Splittings of the Log-Theta-Lattice"} \cite{Mochizuki "Inter-universal Teichmuller Theory IV: Log-Volume Computations and Set-theoretic Foundations"} \cite{Mochizuki "On the Essential Logical Structure of Inter-universal Teichmuller Theory}, which, for Mochizuki, implies the proof of the \textit{abc conjecture}, or \emph{Oesterlé–Masser conjecture} \cite{Oesterle "Nouvelles approches du theoreme de Fermat"} \cite{Masser "Open problems"}, but not in the line of conviction coming from P. Scholze and J. Stix \cite{Scholze and Stix "Why abc is still a conjecture}.
	}

\section{The Role of Analogy}
\label{section "The Role of Analogy"}

\begingroup
\footnotesize
L'Analogia [commune à tutte le scienze mathematiche] è una \emph{conuenienza} (per dir cos[ì]) di alcune proportioni [\,\dots]. adunque se faremo comparatione di alcune cose fra loro, poniam caso, di due grandezze, esse saranno chiamate termini, \& il passaggio dall'una all'altra si dir[à] distanza. ma la comparatione è una \emph{conuenienza}, che gli antichi nominorono proportione. \& la comparatione, ò vero \emph{conuenienza} di questa proportione con un'altra proportione secondo una certa somiglianza chiamasi analogia.\footnote{
	«Analogy [common to all mathematical sciences] is a convenience (so to speak) of some proportions [\,\dots]. therefore if we make a comparison of some things among themselves, say, of two quantities, they will be called terms, \& the passage from one [quantity] to another will be called distance. but the comparison is a convenience, which the ancients defined proportion. \& the comparison, or the convenience of this proportion with another proportion according to a certain similarity, is called analogy».
	} \\
\indent — \textsc{F. Commandino} \cite[p. 61 verso, e.a.]{Commandino "De gli Elementi d'Euclide libri quindici"}

\vspace{2mm}

[\,\dots] Per accentuare certi movimenti e indicare le loro direzioni, s'impiegheranno i segni della matematica: $+ - \times : \, = \, > \, <$, e i segni musicali [\,\dots]. V'è in ciò una \emph{gradazione di analogie sempre più vaste}, vi sono dei rapporti sempre più profondi e solidi, quantunque lontanissimi. L'analogia non è altro che l'amore profondo che collega le cose distanti, apparentemente diverse ed ostili. Solo per mezzo di analogie vastissime uno stile orchestrale, ad un tempo policromo, polifonico e polimorfo, può abbracciare la vita della materia.\footnote{
	«[\,\dots] To accentuate certain movements and indicate their directions, mathematical [symbols] $+ - \times : \, = \, > \, <$ and musical [notation] will be employed [\,\dots]. There is in this a \emph{gradation of ever wider analogies}, there are ever deeper and more solid relationships, though very distant. The analogy is nothing more than the deep love that connects distant, apparently different and hostile things. Only by means of vast analogies an orchestral style, at once polychrome, polyphonic and polymorphic, can embrace the life of matter».
	} \\
\indent — \textsc{F.T. Marinetti} \cite[1st page]{Marinetti "Il manifesto tecnico della letteratura futurista"}

\endgroup

\vspace{2mm}

\enumerationisinitium
\item Let us face it: the mention of Marinetti is a literary amusement, a pleasant break; but what one uncovers in mathematics, and in all math-friendly physics techniques, is not that different. One of the guiding criterion, if not the main one, in mathematical science, since the dawn of history, is the \emph{analogy} (\textgreek{ἀριθμητική} or \textgreek{γεωμετρικὴ ἀναλογία}), see e.g. index graecitatis \cite[p. 71]{Pappus of Alexandria "Collectionis Vol. III Tomus II"} in the collection that contains some works of Pappus of Alexandria. 
It is the concept of \emph{relational equality} (\textgreek{ἰσότης λόγων}), of \emph{correspondence}, or \emph{proportionality}, and the notion of \emph{similarity relationship} of two elements. 

\item  In mathematical physics, see e.g. J.C. Maxwell \cite[p. 156, e.a.]{Maxwell "On Faraday's Lines of Force"}: 

\vspace{2mm}

\begingroup
\footnotesize

In order to obtain physical ideas without adopting [prejudicially] a physical theory we must make ourselves familiar with the existence of physical analogies. By a \emph{physical analogy} I mean that partial \emph{similarity} between the laws of one science and those of another which makes each of them illustrate the other. Thus all the mathematical sciences are founded on relations between physical laws and laws of numbers, so that the aim of exact science is to reduce the problems of nature to the determination of quantities by operations with numbers.

\endgroup

\vspace{2mm}

It should be noticed the jauntiness with which Maxwell passes from \emph{analogy} to \emph{partial similarity}, and leapfrogs to the (so-called) \emph{exact science}. 
\item Examples abound in theoretical physics, as a matter of fact. 
\subenumerationisinitium
\item One can read E. Fermi's words \cite[pp. 72, 75-77, e.a.]{Fermi "I fondamenti sperimentali della nuova meccanica atomica"}: 

\vspace{2mm}

\begingroup
\footnotesize
The most natural hypothesis that one is induced to make, when faced with something unknown, is that its behavior is \emph{analogous} to that of \emph{similar}, and already known, things.

[For example] Schr[ö]dinger [\,\dots] built [his] new [wave] mechanics based on a formal analogy between classical mechanics and geometrical optics [\,\dots]. A very close analogy can be established between geometrical optics and mechanics, by matching the path of a ray of light to the trajectory of a material point.\endnote{
	The same analogy is already in a 1926 paper \cite{Fermi ed Persico "Il principio delle adiabatiche e la nozione di forza viva nella nuova meccanica ondulatoria"} dedicated to the adiabatic principle and the \textit{vis viva} in the wave mechanics, co-written with E. Persico.
	} 
If the material point moves without forces acting on it, that is, if it crosses a region of space in which the potential is constant, its trajectory is a straight line: similarly a ray of light that crosses a region with a constant refractive index, follows a straight path. Therefore to a region in which the potential is constant corresponds, in the optical comparison, to a region in which the refractive index is constant. In a region of space where the potential is variable, a material point describes a curved trajectory; and similarly in a region of space where the refractive index varies from point to point, the rays of light follow curved paths (think e.g. of the phenomenon of the mirage). Given any distribution of potential, one can always \emph{imagine} that, in a region of space, the refractive index varies from point to point in such a way that the path traveled by light is identical to that traveled by a material point under the action of the potential [\,\dots]. Now Schr[ö]dinger observes that classical mechanics also fails when one tries to apply it to very small systems, such as atoms; in this case, in fact, the peculiar phenomena of quantum theory occur. He therefore tries to push beyond the analogy [\,\dots] between mechanics and optics and he comes to build the so-called wave mechanics, analogous to wave optics instead of geometrical optics [\,\dots]. The idea of considering mechanical facts as a manifestation of an undulatory phenomenon may perhaps seem very strange. The same idea, however, can also be suggested by another type of phenomena [\,\dots]. In fact the theory of light quanta, for the explanation of a certain group of phenomena, was induced to postulate a kind of corpuscular nature of light, or, at some level, to bring the phenomena of optics back to those of a swarm of corpuscles; this fact also certainly served to suggest that the analogy between mechanical and wave phenomena is deeper than it seems at first sight.

\endgroup

\vspace{2mm}

\item We can look back on a bit of \emph{Yang–Mills theory}. C.N. Yang and R. L. Mills \cite[p. 192, e.a.]{Yang and Mills "Conservation of Isotopic Spin and Isotopic Gauge Invariance"} 

\vspace{2mm}

\begingroup
\footnotesize
define \emph{isotopic gauge} as an arbitrary way of choosing the orientation of the isotopic spin axes at all space-time points, in \emph{analogy} with the electromagnetic gauge which represents an arbitrary way of choosing the complex phase factor of a charged field at all space-time points. We then propose that all physical processes (not involving the electromagnetic field) be invariant under an isotopic gauge transformation, $\psi \to \psi'$, $\psi' = S^{-1}\psi$, where $S$ represents a space-time dependent isotopic spin rotation [and $\psi$ is a 2-component wave function for a field with isotopic spin $\frac{1}{2}$].

\endgroup

\vspace{2mm}

It is a bet on creating a theory «[i]n analogy to the procedure of obtaining gauge invariant field strengths in the electromagnetic case», with the existence of a field, which has «the same relation to the isotopic spin that the electromagnetic field has to the electric charge». With this goal, they use $-\frac{1}{4}\F_{\mu\nu}\F^{\mu\nu}$ as a Lagrangian density, where 
\begin{equation}
	\F_{\mu\nu} = \partial_\mu A_\nu - \partial_\nu A_\mu + [A_\mu A_\nu] 
\end{equation}
is a Maxwell–Lorentz-type tensor, i.e. a strength tensor of the non-Abelian gauge field $A_\mu$, coming from the electromagnetic field $A^{(\gamma)}_\mu$. The analog of the Maxwell Lagrangian is
\begin{equation}
	\Lagrangian_\textsc{ym} = -\tfrac{1}{2}\trace\left(\F_{\mu\nu}\F^{\mu\nu}\right) = -\tfrac{1}{4}\F_{\mu\nu}^\alpha\F^{\mu\nu}_\alpha,
\end{equation}
in which $\F_{\mu\nu}\F^{\mu\nu}$ are invariant under global and local Lorentz transformations, and 
\begin{equation}
	\F_{\mu\nu}^\alpha = \partial_\mu A^\alpha_\nu - \partial_\nu A^\alpha_\mu + f_{\alpha\beta\gamma}A_\mu^\beta A_\nu^\gamma
\end{equation}
may be called \emph{Yang–Mills strength tensor}, having $f^{\alpha\beta\gamma}$ as structure constants of the gauge group (cf. point \ref{item "Gauge group structure constants"}, p. \pageref{item "Gauge group structure constants"}).
\item We can then revamp the \emph{Nambu–Jona-Lasinio program}. Y. Nambu and G. Jona-Lasinio \cite{Nambu and Jona-Lasinio "Dynamical Model of Elementary Particles Based on an Analogy with Superconductivity. I"} \cite{Nambu and Jona-Lasinio "Dynamical Model of Elementary Particles Based on an Analogy with Superconductivity. II"} base their model of strong interactions on an \emph{analogy} with the Bardeen–Cooper–Schrieffer–Bogoliubov theory of superconductivity. They consider \cite[p. 345, e.a.]{Nambu and Jona-Lasinio "Dynamical Model of Elementary Particles Based on an Analogy with Superconductivity. I"} a representation for a nucleon field $\psi$ of non-linear, point-like, chirally 4-fermion interaction allowing a $\gamma^5$-gauge group (cf. Section \ref{subsubsection "Dirac 4-Spinor Representation"}), so as to put

\vspace{2mm}

\begingroup
\footnotesize
an \emph{analogy} between the properties of Dirac particles and the quasi-particle excitations that appear in the theory of superconductivity.

\endgroup

\vspace{2mm} 

The Nambu–Jona-Lasinio-type Lagrangian density is something like this:

\begin{equation}
	\Lagrangian_\textsc{njl} = \bar{\psi}(i\slashed{\partial} - m)\psi + \couplingconstant_{\mathrm{d}}\Bigl((\bar{\psi}\psi)^2 + (\bar{\psi}i\gamma^5\vec{\sigmaPauli}\psi)^2\Bigr),
\end{equation}
where $\slashed{\partial}$ is the partial derivative in Feynman's slash notation, $m$ a nucleon mass, $\couplingconstant_{\mathrm{d}}$ a dimensionful coupling constant, and $\vec{\sigmaPauli} = (\sigmaPauli_1, \sigmaPauli_2, \sigmaPauli_3)$ are the Pauli matrices in isospin spaces.
\subenumerationisfinis
\item Pure mathematics, mathematical physics, and theoretical physics: three distinct disciplines that draw from the same “well” of analogy. As it is easy to comprise, many examples of knowledge proceed by analogy, jumping from one position to another.\footnote{
	E.g. analogies between Riemannian curvature and Cartan space (Section \ref{subsection "From g-Space Homogeneity to Blob-like Space"}), between spinor maps and other connections (Section \ref{subsection "Spinor Map (6-Dimensional Homomorphism): the Covering $SL_2(C)$ to $SO_{1, 3}^+(R)$"}), between magnetic force and gravitational force (Sections \ref{subsection "Vetturale of Energy Radiated in Gravitational Waves"} and \ref{subsubsection "Quadrupolarity and Transverse-Traceless Gauge"}), between (Regge) simplicial decompositions and ordinary polyhedra (Section \ref{subsection "Ex. 1. Regge Calculus: Simplicial Decompositions"}), between micro- and macro-scopic world (Section \ref{subsubsection "Sakharov's Elasticity of Space, and Liquid Space-Time"}), between equations of gravity and laws of thermo- and hydro-dynamics (Section \ref{subsubsection "Spatio-temporal/Gravitational Thermodynamics, and Entropic Gravity"}), between Ricci flow and heat-like equation (Section \ref{subsection "Evolution of Curvature: Stretching-Shrinking Processes"}). 
	}
But this is what gives «plaisir au chercheur», as A. Weil \cite[p. 408]{Weil "De la metaphysique aux mathematiques"} reminds us: 

\vspace{2mm}

\begingroup
\footnotesize
Nothing is more fruitful [\textit{fécond}], every mathematician knows it, than these obscure analogies [\textit{obscures analogies}], these troublesome [\textit{troubles reflets}] reflections of one theory into another, these furtive caresses, these inexplicable quarrels [\textit{brouilleries inexplicables}]; nothing also gives more pleasure to the researcher. 

\endgroup

\vspace{2mm}

\item The essence of mathematics, and—in different “measures”—of all math-based fields of study (such as mathematical physics, and theoretical physics), is not in logic (\textgreek{λογική}) but in the comprehension “upon” or “on” (\textgreek{ἀνά}) logic: here is the importance of analogy. The difference between us and a computer's logic, or any digital \textsc{ai}, lies completely in that \textgreek{ἀνά}. 

A computer, with a high level of performance, reaches hundreds of millions of billions of operations in a fraction of a second, whereas the human mind, after a little bit of repeated multiplication and division, normally encounters great difficulties. As S. Dehaene writes \cite[p. 119]{Dehaene "The Number Sense: How the Mind Creates Mathematics"}: 

\vspace{2mm}

\begingroup
\footnotesize
The Homo sapiens brain is to [logical-]formal calculation what the wing of the prehistoric bird Archaeopteryx was to flying: a clumsy organ, functional but far from optimal.
 
\endgroup

\vspace{2mm}

That is why the project of reproducing an \emph{analog mode} in \textsc{ai} is a captivate challenge to those who study the most complex mental processes via computer; see D.R. Hofstadter \cite{Hofstadter "Fluid Concepts and Creative Analogies"}.
\enumerationisfinis

\section{Math-Inventiveness: Arbitrariness and Imaginative Endeavor}
\label{section "Math-Inventiveness: Arbitrariness and Imaginative Endeavor"}

\subsection{Fluvial- and Æolian-like Processes}
\label{subsection "Fluvial- and Æolian-like Processes"}

\begingroup
\footnotesize
The \emph{imagination} acts no less in a geometer who creates than in a poet who invents. It is true that they operate differently on their object: the first shears it down and analyzes it, the second puts it together and embellishes it [\,\dots]. Of all the great men of antiquity, Archimedes is perhaps the one who most deserves to be placed beside Homer.\endnote{
	Original Fr. version: «L'imagination dans un géomètre qui crée, n'agit pas moins que dans un poète qui invente. Il est vrai qu'ils opèrent différemment sur leur objet: le premier le dépouille et l'analyse, le second le compose et l'embellit [\,\dots]. De tous les grands hommes de l'antiquité, Archimède est peut-être celui qui mérite le plus d'être placé à côté d'Homère».
	} \\
\indent — \textsc{J. le R. d'Alembert} \cite[p. 65, e.a.]{d'Alembert "Discours preliminaire de l'Encyclopedie"} 

\endgroup

\vspace{2mm}

It is like this. A mathematician proceeds in the same way as a poet, albeit in different ways.\footnote{
	Cf. W. Wordsworth \cite[Book II, 397-405, p. 31]{Wordsworth "The Prelude or Growth of a Poet's Mind"}: «The exercise [\,\dots] whose character [is] poetic as resembling more / Creative agency. I mean to speak / Of that interminable \emph{building rear'd} / \emph{By observation of affinities} / \emph{In objects} where no brotherhood exists / To common minds» (e.a.). The observation, by a poet, of affinities among objects of nature corresponds to the study, by a mathematician, of uncommon relations built through mathematical objects.
	}\textsuperscript{,}\endnote{
	Almost by osmosis, there are poetic styles full of jocose mathematics, see, for instance, the Oulipian group with its constraints of writing.
	}
\textit{Pace} G. Peano \cite[p. 67]{Peano "Osservazioni del Direttore sull'articolo precedente"}\footnote{	
	As Peano said, a mathematician «can make the hypotheses he likes best»; but once the hypotheses have been selected/chosen, «it is up to mathematics» to deduce the consequences rigorously; this is why, in lively controversy with C. Segre, Peano declared \cite[p. 67]{Peano "Osservazioni del Direttore sull'articolo precedente"} that «Whoever states consequences that are not contained in the premises, will be able to make poetry, but not mathematics [\textit{Chi enuncia delle conseguenze che non sono contenute nelle premesse, potrà fare della poesia, ma non della matematica}]». This sentence it is just a quip against Segre. Peano was under the chimera that the «perfected logic» which, according to him, is the telltale sign of mathematics, was sufficient to distinguish it from poetic literature. He was mistaken. Mind you: this does not mean that the celebrated and glorified «absolute rigor» of mathematics is removed or obscured. The “poetic” part of mathematics (cf. Intro, p. \pageref{subsection "Allegorical Figurations of Reality, Idealizations, and Techniques of Transcendence Playing: Physics Affairs"} \& p. \pageref{subsection "Through the Magnifier of Half-sleep"}, Section \ref{subsection "Figments of Imagination and Invention of Possible Worlds"}, \ref{subsection "The World is Not Mathematical"}, or p. \pageref{section "Siren's Song: No-return to Ithaca"}) is \emph{before} and \emph{after} the choice of hypotheses. It is a consubstantial quality.
	} 
and P.A.M. Dirac \cite[p. 21]{Oppenheimer "The Age of Science 1900-1950"},\footnote{
	J.R. Oppenheimer writes: «One evening more than 20 years ago Dirac, who was in Göttingen work­ing on his quantum theory of radiation, took me to task with characteristic gen­tleness. “I understand”, he said, “that you are writing poetry as well as work­ing at physics. I do not see how you can do both. In science one tries to say some­thing that no one knew before in a way that everyone can understand. Whereas in poetry\,\dots”» [one tries to say something that everybody knows already in words that nobody can understand].
	} 
there are not a few concealed ties and oblique intersections between the two disciplines.\footnote{
	The opinion of G. Leopardi \cite[3242, pp. 267-268]{Leopardi "Pensieri Vol. Quinto"} that poetry and mathematics are unconnected activities, with no intersecting elements, is an emeritus silliness: «Pure [\textit{sic}] or simple reason and mathematics never were and never will be able to discover anything poetic. Because all that is poetic [in nature] is felt rather than known and understood [\,\dots]. But pure reason and mathematics have no sensorium whatsoever. It is up to the imagination and sensitivity to discover or understand» the nature's spirit. That is not true. Mathematics \emph{is} imagination, is an act of the creative faculty. Reason (the «cold» [\textit{sic}] and «simple light of exact and geometric» rationality) and sensitivity are not distinctly separate as the fumesophers (Section \ref{section "Interludio Giocoso. Against the Fumesophers, or the Tragicomic Smoke-sellers"}) want. However, the motive which drives Leopardi \cite[586, p. 77]{Leopardi "Pensieri Vol. Secondo"} to this judgment is correct: «[T]he plan, the system, the machine of nature, is composed and organized in another way than that of reason and does not respond to mathematical exactness».
		
		In \cite[48, pp. 153-154]{Leopardi "Pensieri Vol. Primo"} Leopardi goes as far as to declare: «[T]here is nothing more enemy of nature than arid geometry, which takes away all the naturalness [with its “cold” and “bony” language] and the \textit{naïveté} [\,\dots] (where beauty is born), and the gracefulness». There is a bit of confusion here: geometry does not conflict with nature. If so, poetry would be too. And the contrast between the “dryness” of geometry (or of all of mathematics) and the “fertility” of poetry is an old cliché—mathematics also has a \emph{qualitative} side, joined with an imaginative aptitude. It would be more suitable to speak of \emph{inadequacy}, just in case: mathematics and poetry are both, in their own way, \emph{inadequate languages} to describe nature's quiddity, cf. Section \ref{subsection "Contextus II. Autobiographical Note"}.
	}
One of the common points between a statement in mathematics, e.g. a conjecture, or a theorem, and a composition in verse, as a poem, is that both of them start to exist as an eminently \emph{individual} act; they are the result of a completely partial and chance-generated view, and only then they become (if they become) a collective heritage, which is a mishmash of partial views and \emph{personal bias}. 

This notwithstanding, there are marked differences between literary composition and mathematics. A work of mathematics, as opposed to a poem or a novel, is not closed in on itself. In poetry, the \emph{form} is given \emph{once and for all}, whilst what changes is the interpretation, which is dispersed in the infinite underlying meanings. In science, things are slightly different. Scientific production is a mingling of \emph{fluvial- and æolian-like processes}, with sediments transport, erosions, and formations of new rock strata,\footnote{
	We could sum it up like this, with a Latin slogan: \textit{mĭttĕre, tŏllere, restruĕre}, which means \emph{putting, removing, restructuring}.
	} 
or something that looks like a \emph{Surtseyan activity};\footnote{
	Surtsey is a volcanic island, with an ever-evolving and chronic morphology, located in the \emph{Vestmannaeyjaklasanum}, off the south coast of Iceland.
	} 
it is about continuously re-writing processes, the \emph{form (formalism)} of which, together with the interpretation (the meaning), \emph{varies} grain by grain, and moulds itself from one generation, or even from one person, to another.

Mathematics is only one of the (many) possible expressions of \textgreek{λόγος}, and it is not outside of it; that is, mathematics is a \emph{human language} (see Section \ref{subsection "Math-Language and its Reasonably Effectiveness"}); and like any constructed language, \emph{incertitude} factors are involved: there is no strictly univocal correspondence between expression (symbolic notations) and content, in the awareness that the consensus on the meaning of symbolic notations is constantly evolving, and many of the notations undergo never-ending adjustments, under the urging of new interpretations. So, long story short, mathematics is an \emph{open opus}.

\subsection{Extra-logical Objects, and Gödelian Suggestions}
\label{subsection "Extra-logical Objects, and Gödelian Suggestions"}

\begingroup
\footnotesize
[T]he logical correctness of the reasonings that lead from axioms to theorems is not the only thing we had to worry about. Do the rules of perfect logic exhaust the whole of mathematics? [\textit{Les règles de la parfaite logique sont-elles toute la mathématique?}] 

[I]t is by logic that we demonstrate, [but] it is by intuition that we invent [\textit{c'est par la logique qu'on démontre, c'est par l'intuition qu'on invente}]. \\
\indent — \textsc{H. Poincaré} \cite[II, chap. III, pp. 158, and chap. II, p. 137]{Poincare "Science et methode"}\endnote{
	The first part of the passage cited is taken from a chap. entitled \textit{Les Mathématiques et la Logique}, whose source material is the article published in Rev. métaphys. et mor., Tome XIII, № 6, 1905, pp. 815-835. The second passage is taken from a chap. entitled \textit{Les définitions mathématiques et l'Enseignement}, whose source material dates back to a Poincaré's \emph{Conférences} du Musée Pédagogique (1904), under the title \textit{Les définitions générales en mathématiques}, and then published in \textit{L'Enseignement des sciences mathématiques et des sciences physiques}, par H. Poincaré, G. Lippmann, L. Poincaré, P. Langevin, É. Borel, F. Marotte, Imprimerie nationale, Paris, 1904, pp. 1-28; secondly it appeared in Enseign. Math., Tome VI, 1904, pp. 257-283. 
	}

\vspace{2mm}

Mathematics, as each of the other sciences, cannot be founded upon logic alone; rather, as a prerequisite for the use of logical inferences [\,\dots], something must already be given to us in the imagination [\textit{Vorstellung}]: certain extra-logical concrete objects [\textit{gewisse außer-logische konkrete Objekte}] that are present as immediate experience before any thought. \\
\indent — \textsc{D. Hilbert} \cite[p. 65]{Hilbert "Die Grundlagen der Mathematik"} 

\endgroup

\vspace{2mm}

\enumerationisinitium
\item Mathematics does not entirely end in the rigor of the rules of logic—take the renowned Gödel's paper \cite{Godel "Uber formal unentscheidbare Satze der Principia Mathematica und verwandter Systeme I"} = \cite{Godel "On Formally Undecidable Propositions of Principia Mathematica and Related Systems I"}, together with that of A.M. Turing \cite{Turing "On Computable Numbers with an Application to the Entscheidungsproblem"}, as a watershed (to be read in conjunction with the failure of logico-phagocytizing plans à la Whitehead–Russell): Gödel's \emph{incompleteness theorems} state,\footnote{
	Gödelian incompleteness theorems are divided into two:
	
	($\mathnormal{1}$) the first theorem says that: in any mathematical system MS containing (basic) arithmetic, there is a formula $F_\textsc{mt}$ such that, if MS is consistent, then neither $F_\textsc{mt}$ nor $\lnot{F}_\textsc{mt}$ is demonstrable in MS, whilst
	
	($\mathnormal{2}$) the second theorem says that: in any mathematical system MS containing (basic) arithmetic, if MS is consistent, it is not possible to prove the consistency of MS within MS.
	} 
among other things, that a succession of natural numbers, when treated axiomatically, in every system including arithmetic brings out formally undecidable (arithmetic) propositions.\endnote{
	\label{endnote "Gödel's Neumann and Goldstine"}
	One area in which Gödel's incompleteness theorems offer a vast service is the study of computational complexity. See on this J. von Neumann and H.H. Goldstine \cite{Neumann and Goldstine "Numerical Inverting of Matrices of High Order"} \cite{Neumann and Goldstine "Numerical Inverting of Matrices of High Order. II"} \cite{Neumann and Goldstine "On the Principles of Large Scale Computing Machines"}. The two authors investigate the problem of the inevitable transmission of errors in the arithmetical operations elaborated by an electronic computing machine, or “computer”. The appearance of truncated or rounded results, is imposed by the approximation of irrationals, with a deviation from a “true” numerical solution, as well as an increasing accumulation of inaccurate processes (a computer manages an immense amount of data better/faster than the human brain, but it performs operations by making use of numbers with finite decimal expansions). Other sources of error concerning current computers are:

		\setlength\parindent{8pt}
		· the translation of the mathematical continuum, of the field $\mathbb{R}$ of real numbers, into an arithmetico-computable discretuum (this is why any continuous analytic $\mathbb{R}$-value is finitized, and any irrational number is, at some “point”, truncated),
		
		· the replacement of non-linear partial differential equations with linear algebraic equations.
	}\textsuperscript{,}\endnote{
	In light of the endnote \ref{endnote "Gödel's Neumann and Goldstine"}, J.A.E. Dieudonné's comment \cite[pp. 27-28]{Dieudonne "Choix d'oeuvres mathematiques"}, although it dates back to 1981, on Gödel's results is perhaps too hasty, and it runs the risk of appearing superficial, even if the core of what Dieudonné says is partially true: «We can admire the acuteness and depth of the research that led to Gödel's, P. Cohen's [\,\dots] metamathematical theorems [\,\dots] but they did not have \emph{any} influence (neither positive nor negative) on the solution of the vast majority of the problems, which are of interest to mathematicians».
		}
Mathematics is \emph{broader} than logic (cf. Margo \ref{margo "Gödelian suggestions"} and footnote \ref{footnote "Feynman, mathematics, language and logic"} on p. \pageref{footnote "Feynman, mathematics, language and logic"}); it is something much more than a sum of tautological truths (\textgreek{ταὐτο-λογίαι}).\footnote{
	This brings us to the \emph{completeness theorem} in first-order logic; to put it with a catchword, here is the theorem: “All logical truths whatsoever are demonstrable”, when an equivalence between “logical truths” and “tautologies” is prescribed, and the two terms shall be considered to be synonymous: in this case, a “logical truth” is true solely as a result of its logical components. 
	
	(\textgreek{α}) For propositional logic, the completeness theorem was proved by E.L. Post \cite{Post "Introduction to a General Theory of Elementary Propositions"}, with reference to the \textit{Principia Mathematica} by N.A. Whitehead and B. Russell, so that all of Whitehead–Russell's theorems under logical rules of deduction, are \emph{tautologies} that can be verified through specific tables, called “truth-tables” \cite[pp. 166-168]{Post "Introduction to a General Theory of Elementary Propositions"}; in \cite[p. 164, e.a.]{Post "Introduction to a General Theory of Elementary Propositions"} one reads: «Our [\,\dots] theorem gives a uniform method for testing the truth of any proposition of the system; and by means of this theorem it becomes possible to exhibit certain general relations which exist between these propositions. These relations definitely show that the postulates of ‘Principia’ are capable of developing the \emph{complete system of the logic of propositions without} ever introducing \emph{results extraneous} to that system». The system of elementary propositions of \emph{Principia} is therefor consistent.
	
	(\textgreek{β}) For the predicate logic, the completeness theorem was proved by K. Gödel \cite{Godel "Uber die Vollstandigkeit des Logikkalkuls"}. 
	
	There is no such “completeness” condition—in (\textgreek{α})-(\textgreek{β}) Post–Gödel-style (deductive) systems—for mathematical truths.
	}
\item There is in fact a seeming contradiction of mathematics, icastically espoused by Poincaré \cite[pp. 9-10]{Poincare "La Science et l'Hypothese"}: if mathematics is not deductive, where does its \emph{logicus strictus rigor} come from? If, on the contrary, all the propositions which it enunciates may be derived from each other, deductively, through the rules of formal logic, how is it that mathematics is not reduced to an \textit{immense tautologie}, for which all the theorems represent indirect ways of affirming that $A = A$? The tautological semblance fades, or even dissipates, when we consider mathematics as a product of the imagination, a faculty of intuition.
\enumerationisfinis

\begin{margo}[Gödelian suggestions]
\label{margo "Gödelian suggestions"}
Earlier I accentuated the \emph{vastness} of mathematics, when it is dissected with the “clinical” tools of logic. We could, equally, talk about the \emph{inexhaustibility} of mathematics, with the auspicious addition—against all theories of everything—that something similar can be replicated in the physical realm. Believing in an ultimate theory is like being under the illusion that the genius of mathematics can be locked up forever in a lamp. See e.g. F.J. Dyson \cite[p. 449]{Dyson "Time without end: Physics and biology in an open universe"}: 

\vspace{2mm}

\begingroup
\footnotesize
Gödel proved [\,\dots] that the world of pure mathematics is inexhaustible; no finite set of axioms and rules of inference can ever encompass the whole of mathematics; given any finite set of axioms, we can find meaningful mathematical questions which the axioms leave unanswered. I hope that an analogous situation exists in the physical world [\,\dots]. [I]t means that the world of physics and astronomy is also inexhaustible; no matter how far we go into the future, there will always be new things happening, new information coming in, new worlds to explore, a constantly expanding domain of life, consciousness, and memory.

\endgroup

\vspace{2mm}

Another weird but cogent manner of illustrating the pillar of Gödel's theorematic block is to combine it with the insuperable hurdle that each one of us, as an individual, is not able to get away from himself. P.W. Bridgman \cite[pp. 6-7, e.a.]{Bridgman "The Way Things Are"} observes:

\vspace{2mm}

\begingroup
\footnotesize
The insight that we can never get away from ourselves is an insight which the human race through its long history has been deliberately, one is tempted to say wilfully, refusing to admit. But the ostensibly timeless absolutes are formulated and apprehended by us, and the vision which the mystic says is revealed by the direct intervention of God is still a vision apprehended by him. When we talk about getting away from ourselves it is we who are talking. All this is so obvious [\,\dots]. It is exceedingly suggestive to see in Gödel's theorem an application to our present problem, the problem of discovering the consequences of not being able to get away from ourselves. It is, of course, not a question of any formal and rigorous application of the theorem, but only of something qualitative and suggestive. The essence of the situation presented by Gödel's theorem seems to be that we are here concerned with a system dealing with itself—mathematics attempting to prove something about mathematics. [This theorem states that it is impossible to prove that a logical system, at least as complicated as arithmetic, contains no concealed contradictions by using only theorems which are derivable within the system]. It is tempting to generalize Gödel's theorem to read that whenever we have a system dealing with itself we may expect to encounter maladjustments and infelicities, if not downright paradox. The insight that we can never get away from ourselves obviously presents us with a situation of this sort. \emph{The brain that tries to understand is itself part of the world that it is trying to understand}. It seems that the situation cannot be dealt with satisfactorily in its entirety; the best, and well nigh all, we can do is to operate by successive approximations at different levels, isolating for treatment this or that group of phenomena.

\endgroup

\vspace{2mm}

All of this could qualify as a \emph{psychological} trouble, very far from the practice of physics; but it is the same difficulty that we saw in the Wheeler–DeWitt universe wave function, cf. footnote \ref{footnote "There is no wave function of the universe"} on p. \pageref{footnote "There is no wave function of the universe"}. \margosymbol
\end{margo}

\subsection{«Metaphysical Belief» and «Convenient Illusion»}

\begingroup
\footnotesize
Bourbaki sets off [\,\dots] from [\,\dots] an unprovable \emph{metaphysical belief} we willingly admit. It is that mathematics is fundamentally simple and that for each mathematical question there is, among all the possible ways of dealing with it, a best way, an optimal way. We can give examples where this is true and examples where we cannot say, because up to now we have not found the optimal method.\footnote{
	And then he goes on: «I cited, for example, group theory and analytical number theory [\,\dots]. In both one has a quantity of methods, each one more clever than the last [\,\dots] but we are sure that this is not the final way to deal with the question [\,\dots] but in the end, little by little, we manage to find one way which is better than the others. This is only a belief, I repeat, a metaphysical belief».
	}
 
[\,\dots] On foundations we believe in the reality of mathematics, but of course when philosophers [or fumesophers? (see Section \ref{section "Interludio Giocoso. Against the Fumesophers, or the Tragicomic Smoke-sellers"})] attack us with their paradoxes we rush to hide behind formalism and say: “Mathematics is just a combination of meaningless symbols” [\,\dots]. Finally we are left in peace [\,\dots] with the feeling each mathematician has that he is working with something real. This sensation is probably an \emph{illusion}, but is very convenient. \\
\indent — \textsc{J.A.[E.] Dieudonné} \cite[p. 145, e.a.]{Dieudonne "The Work of Nicholas Bourbaki"}

\endgroup

\vspace{2mm}

Directly linked to what we have seen above (Section \ref{subsection "Extra-logical Objects, and Gödelian Suggestions"}), one of the enticing dilemmas of the underlying structure(s)—\textgreek{λογική (τέχνη/τέχναι)}, and the axiomatic method—of mathematics, is this: is the logic of (used by) mathematics a \emph{logical formalism} (the logic itself) or a \emph{grammaticalism} whose formalism is but the syntax of (a) math-language? Namely, what is it that makes mathematics, or its language, so \emph{intelligible}? Or, what is the nature (and the burden) of mathematical formalism (in science)?

A mathematical logicians is more inclined towards the the first solution, whilst a (Bourbakist) mathematician (such as Dieudonné), with a rigor purposely created for the occasion, towards the second one.

What Dieudonné means by «metaphysical belief» is a way to plant, say, stakes (as reported by the miscellaneous prescriptions of the axiomatic method), and from which to start to do mathematics, without getting entangled in the foundations of mathematics—for this is the task, according to him, of the mathematical logician.

Concerning the «sensation» that mathematics is connected with reality («the feeling each mathematician has that he is working with something real»), whereas this is nothing more than a «convenient illusion», that is why a mathematician is said, not without mordancy, to be Platonic from Monday to Saturday, but not on Sunday, or when he/she takes a break from his/her research topics.

\subsection{Figments of Imagination—Mathematics as a «Fictional Activity» \& «Invention of Possible Worlds»}
\label{subsection "Figments of Imagination and Invention of Possible Worlds"}

\begingroup
\footnotesize
With regard to the conception of logical rigor [\,\dots] it is not a question for us of placing Science on a firmer basis, but simply of recognizing how fragile this basis is. We can make everything perfectly logical by keeping silent about our subjective judgments and replacing them with hypotheses, but these hypotheses have no value except as they derive from subjective judgments, and that is what we need to be talking about, as a crucial fact. 

We must \emph{invent the world} to frame our sensations, but we must never consider it as a rigid and static schema, as a definitive construction: it is only the provisional result of an effort aimed at synthesis. Our sensations, our fundamental concepts, starting with those of time and space, will never be the protagonists of a finished comedy. \\
\indent — \textsc{B. de Finetti} \cite[p. 124, e.a.]{de Finetti "L'invenzione della verita"}

\vspace{2mm}

Mathematical theories constitute fictional [imaginative] universes. \\
\indent — \textsc{C. Bartocci} \cite[p. xiii]{Bartocci "Introduzione" in Autori vari "Racconti matematici"}
	
\endgroup

\vspace{2mm}

Mathematics has its own \emph{self-consistency}, or logical core, which becomes explicit both through a logico-deductive modus and an intuitive ploy, inductively, with—let us say—experimental operations (cf. footnote \ref{footnote "Euler: quasi experimenta"} on p. \pageref{footnote "Euler: quasi experimenta"}), or step-by-step enumerations. And yet there is, in mathematics, a prevalent component of \emph{arbitrariness}, that is something more pregnant than Cantorian \textit{Freiheit} \cite[p. 564]{Cantor "Ueber unendliche lineare Punktmannichfaltigkeiten"}.\endnote{
	This is the line in question \cite[p. 564]{Cantor "Ueber unendliche lineare Punktmannichfaltigkeiten"}: «The \emph{essence} of \emph{mathematics} lies precisely in its \emph{freedom} (das \emph{Wesen} der \emph{Mathematik} liegt gerade in ihrer \emph{Freiheit})», and it dates back to 1883. We note in passing that Cantor also embraced another position (although without betraying the profound sense of freedom in mathematics), which is perhaps more Platonic; read the following La. phrases, in his papers:

· \cite[p. 62, Theses III, from 1869]{Cantor "De transformatione formarum ternariarum quadraticarum"}: «Numeros integros simili modo atque corpora coelestia totum quoddam legibus et relationibus compositum efficere», i.e., «Integers with their laws and relations constitute a totality in the same way as the celestial bodies».
 
· \cite[p. 481, from 1895]{Cantor "Beitrage zur Begrundung der transfiniten Mengenlehre (Erster Artikel)"}: «Neque enim leges intellectui aut rebus damus ad arbitrium nostrum, sed tanquam scribae fideles ab ipsius naturae voce latas et prolatas excipimus et describimus», i.e., «We do not make the laws of thought and things at [our] discretion, but, as faithful scribes, we receive and describe those [laws] established and transmitted by the voice of nature itself».
	} 
Cf. P. Valéry \cite[p. 42]{Valery "Cahiers 1894-1914 tome V"}: 

\vspace{2mm}

\begingroup
\footnotesize
Mathematics [\,\dots] teaches the determination against consequences, and the rigor of a route of [our] choice arbitrarily taken [\textit{rigueur de la route une fois choisie arbitrairement}]. It is thus the model of arbitrariness [\textit{le modèle de l'arbitraire}].

\endgroup

\vspace{2mm}

The well-known \emph{logicus strictus rigor Mathematicæ}, or rather, the rules for operating on the symbolic apparatus, come only after the inventive uplift, the \emph{human} activity of finding by the \emph{imagination} or \emph{ingenuity},\footnote{
	\label{footnote "Longinus, Novalis"}
	\textgreek{Λογγῖνος} (Longinus) \cite[XXXV, 3-4, p. 67]{Dionysius Longinus "De sublimitate libellus"}: «\textgreek{[Τ]ῇ θεωρίᾳ καὶ διανοίᾳ τῆς ἀνθρωπίνης ἐπιβολῆς οὐδ' ὁ σύμπας κόσμος ἀρκεῖ} ([T]he whole universe is not large enough [to contain] the impetus of human viewing and thought)». But, then again, as G.P.F. Freiherr von Hardenberg (Novalis) realizes \cite[p. 147]{von Hardenberg Novalis Schriften II"}, «Ohne Enthusiasmus [\textgreek{ἔνθεος}] keine Mathematik».
	} 
when particular rules are imposed, together with logical and convention procedures. Mathematical objects are, finally, a kind of \emph{figmenti dell'immaginazione}.\footnote{
	It is Bartocci again: Mathematics is a «fictional activity of our brain» \cite[p. xiv]{Bartocci "Introduzione" in Autori vari "Racconti matematici"} consisting mainly in the «invention of possible worlds» \cite[p. xii]{Bartocci "Introduzione" in Autori vari "Racconti matematici"}.
	
	It would be awesome to review the greatest \emph{fictional monsters} of mathematics (cf. footnote \ref{footnote "Cardano's quantitas silvestris"}, p. \pageref{footnote "Cardano's quantitas silvestris"}), but that would require a special book. Let us settle here for recalling a hitting example throughout the history of mathematical thought, that of Leibniz, when, in a letter to B. Des Bosses (11 March 1706) \cite[p. 32]{Leibniz and Des Bosses "The Leibniz-Des Bosses Correspondence"}, he defines the infinitesimal quantities as \emph{fictions of the mind}: «I do not believe [\textit{statuo}] in infinitely small magnitudes [\textit{magnitudines infinite parvas}] than infinitely large ones [\textit{infinite magnas}], that is, no more infinitesimals [\textit{infinitesimas}] than infinituples [\textit{infinituplas}]. For I hold both to be fictions of the mind [\textit{mentis fictionibus}] due to abbreviated ways of speaking, which are adapted to calculation, as imaginary roots in Algebra are too. In the meantime, I have demonstrated that these expressions are highly useful for abbreviating thinking, and thus for discovery [\textit{inventionem}]».
	}

So, when Hilbert \cite[p. 14]{Hilbert "Natur und mathematisches Erkennen"} says: 

\vspace{2mm}

\begingroup
\footnotesize
Mathematics is not like a game [\textit{Spiel}] the tasks of which are determined by arbitrarily [\textit{willkürlich}] devised rules, but a conceptual system with an inner necessity [\textit{innerer Notwendigkeit}] that can only be so and not in any other way, 

\endgroup

\vspace{2mm}

\noindent well, we consider this to be sort of sweetened Hermitianity (point \ref{item "C. Hermite"}, p. \pageref{item "C. Hermite"}), or the umpteenth recurrence of \textit{trivialis} necessity (Section \ref{subsubsection "Many Truths I. Triangle with Two Line Segments"}) pertaining to the foundations of mathematics.

\chapter{Outro—\emph{Parva Mathematica}: \emph{Libera Divagazione} \sfrac{3}{8}}
\label{chapter "Outro—Parva Mathematica: Libera Divagazione 3/8"}

\section{Mathematics in the Physical Sciences, and Nature of Reality I}
\label{section "Mathematics in the Physical Sciences, and Nature of Reality I"}

\begingroup
\footnotesize
[T]he word “reality”, understood as the name of everything that exists in nature independently from observers, should take into account the fact that each species only analyzes those external and internal stimuli that its sensors are able to capture. The reality of a bee differs, in this respect, from the one that is being debated by members of \emph{Homo sapiens sapiens}. Only the latter write books on what bees or bats do, but this is surely not a decisive reason to believe that [what we call] reality, mutably described by humans, is the “objective” reality, and that, as a consequence, human descriptions of what is there evolve chasing the aim of getting closer to the “truth”, i.e. to the labels that should be hung on real entities[,] which would be independent of the anatomies of living organisms. \\
\indent — \textsc{E. Bellone} \cite[pp. 121-122]{Bellone "L'origine delle teorie"}

\endgroup

\vspace{2mm}

Perhaps the best way to probe into the relationship between mathematics and physics is to declare in no uncertain terms that our understanding of the nature of reality is the result of our \emph{manner} of seeing and understanding; so it is not advisable to start from of the external world, outside of us, and unwarily to entrust to mathematics the arduous task of analyzing it “objectively”. It is more honest to keep the focus on mathematics, and its resilient definitions crashing inevitably onto the description of the world: we are aware that \emph{the nature of reality is irreducible to mathematics}, cf. \cite[p. 17]{Bellone "I modelli e la concezione del mondo nella fisica moderna: da Laplace a Bohr"}.

\subsection{Anthropoid Ways Ia. Regularity and Formal Relations}
\label{subsection "Anthropoid Ways Ia. Regularity and Formal Relations"}

\begingroup
\footnotesize
We do not need to know the intimate causes of the phenomena that Nature keeps hidden from us, but their relations [that is, mathematics]. \\
\indent — \textsc{G. Veronese} \cite[p. 23]{Veronese "Il vero nella matematica"}

\vspace{2mm}

Habit sometimes leads us to forget that what we study and classify are never the perceptions as such, the objects and their properties (that of being red, e.g.) but always exclusively the relations of our perceptions from each other. All our cognitions, although derived from experience, concerns exclusively \emph{their formal structure as a whole}. \\
\indent — \textsc{M. Ageno} \cite[§ 5, p. 35]{Ageno "La costruzione operativa della fisica"}

\endgroup

\vspace{2mm}

Similarly, we do not have access to an \emph{intrinsic} mathematical character of nature, because it does \emph{not} even exist. We must not confuse the \emph{regularity} (or uniformity) and \emph{repetitiveness of nature} with the \emph{ingenium geometricum} (i.e. with the regularity e.g. of a polyhedron, of a curve, or of a function), or with the \emph{mensura numeri} (correspondence between physical phenomenon and numerical sequence),\footnote{
	One of the most spectacular episodes of effectiveness of mathematics was U. Le Verrier's \cite{Le Verrier "Recherches sur les mouvements d'Uranus"} \cite{Le Verrier "Recherches sur les mouvements de la planete Herschel (dite Uranus)"} prediction of the existence of Neptune\endnote{
		C.T. Kowal \& S. Drake  \cite{Kowal and Drake "Galileo's observations of Neptune"} have historically reconstructed that the pre-discovery (first recorded observation) of Neptune belongs to Galileo; he  indicates Neptune by the letter $b$ in his notebook, on 27 December 1612, hour 15:46, mistaking it for a \emph{stella fissa}.
		} 
from unexplained anomalies in the motion of Uranus. He made use only of calculations and astronomical observations, according to Newton's laws of gravity. His was a triumph of numerical analysis. At a later date, with the same numerical method, Le Verrier succumbed to presumption of the existence of Vulcan, a new hypothetical planet, intending to give an account of some anomalies in the motion of Mercury; but its existence has never been confirmed. Vulcan does not exist. The anomalous rate of secular precession of Mercury's perihelion will become exhaustively intelligible, as it is known, within the relativistic paradigm \cite[§ 14, p. 804]{Einstein "Die Grundlage der allgemeinen Relativitatstheorie"} = \cite[§ 14, p. 145]{Einstein "The Foundation of the General Theory of Relativity"}, that is, with the \emph{creation of novel mathematics}, going beyond, where needed, Newton–Le Verrier's laws of motion \& numerical mathematics.
	}
that are anthropoid ways of studying (\textgreek{μαθήματα}) and describing the \emph{Naturæ regulæ}. On this, subtle considerations are in M. Ageno \cite[preface, secc. 1.1-2 + relative endnotes]{Ageno "Le origini della irreversibilita"}. 

Our knowledge relating to nature, even if it comes from an experiential ground, it regards not the intimate causes of the phenomena, but their \emph{relations}, which then, for us, are mathematical relations, namely \emph{formal entities} that are inherent in our intuitive and intellectual processes.\footnote{
	So we can nimbly return to pure mathematics; cf. e.g. E. Papperitz \cite[p. 36]{Papperitz "Ueber das System der rein mathematischen Wissenschaften"}: «The subject-matter of pure mathematics consists of the relations [\textit{Beziehungen}], which can be established between objects of thought [\textit{gedachten Elementen}], by considering them as contained in an ordered manifold; the law of order of this manifold must be dependent on our choice [\textit{Wahl}]».
	}

\begin{margo}
The etymology of \emph{number} insinuates an origin of the word in the practical and concrete life, something close to what M. Pasch \cite[Vorwort, p. iii]{Pasch "Vorlesungen uber neuere Geometrie"} calls \textit{empirischen Kern}, about the primeval stage of a geometric object that precedes the stage of \textit{künstlichen Begriffen}. It comes from the Gr. \textgreek{νέμω}, “dispense”, “distribute”, “assign” e.g. food, bliss, or anything else.\footnote{
	See Homer, \textit{Odyssey}, XIII, 50: «\textgreek{μέθυ νεῖμον}» («serve out wine»), XIV, 436: «\textgreek{τὰς δ᾽ ἄλλας νεῖμεν ἑκάστῳ}» («he distributed the rest [of the pieces of meat] to each»).
	} \margosymbol		
\end{margo}

\subsection{Anthropoid Ways Ib. Symmetry and Invariance in Physics: the Impact of Group Theory}
\label{subsection "Anthropoid Ways Ib. Symmetry and Invariance in Physics: the Impact of Group Theory"}

\begingroup
\footnotesize
Die Principien der Mechanik haben einen gruppentheoretischen Ursprung.\footnote{
	«The principles of mechanics have a group-theoretic origin». Note. Group structures are combined by Lie (ivi) with: \textit{Kinematik}, \textit{geodätischen Curven}, \textit{allgemeine Aequivalenzproblem in der Theorie der Differentialgleichungen}, \textit{Optik}, and \textit{mathematische Physik}.
	} \\
\indent — \textsc{M.S. Lie} \cite[p. vii]{Lie "Theorie der transformationsgruppen Dritter und Letzter Abschnitt"}

\endgroup

\vspace{2mm}

The principles of symmetry—the definition of which is prominently given with the principles of invariance, in keeping with a specific group of transformations (a principle of invariance, in turn, shall define a conservation law, for instance of a form, of a relation, of a physical quantity)—are \emph{our mode} of interpreting and summarizing the regularity of nature, or of what we call the “regularity of the physical world”, and, thereupon, the “regularity of the laws of nature”. But we must not confuse the symmetry groups with nature: a symmetry group is an \emph{algebro-geometric object} expressing a kind of regularity (if there is actually a regularity). 

Let do some inane examples. The gauge group of electromagnetic interactions (Sections \ref{subsection "Holonomy in Abelian Phase Factor: Gauge Group of Electromagnetic Interactions"}), the groupable synopsis \& the spinorial representation (Section \ref{section "Groupable Synopsis via Commutative Diagram"}), or the group of Lorentz transformations (Section \ref{subsubsection "Lorentz Group plus Transformations"}), are one thing, their physical occurrences are something else. The crystal lattices and the  crystallographic/Fedorov space groups, which dispense a representation of the symmetry of the crystals,\endnote{
	Cf. Intro, p. \pageref{subsection "Metaphorical Procedure: the Example of the Crystals"}.
	} 
are one thing (see e.g. epigraph under Section \ref{subsection "Discrete Gamma-Crystallographic Group, Killing–Hopf Theorem, and Isometric Action"}), a snowflake,\footnote{
	The first and paramount illustrated work on the snow crystals is the  reference book with selected photographs of W.A. Bentley \cite[chap. IV. \textit{Mysteries and Beauties of the Snow}, pp. 97-168]{Thompson "Water Wonders Every Child Should Know: Little Studies of Dew Frost Snow Ice and Rain Illustrated from photographs by W.A. Bentley"}.
	} 
a diamond, or the sodium chloride framework, with their manifestation of a uniform (ordered) distribution of elements, are something else; cf. H. Weyl \cite{Weyl "Symmetry"}.

When, in the treatises on physics, we find written, with a quasi-Biblical tone, that «The principles of symmetry determine the form of the laws of nature», that «The laws of nature [or rather, of physics] derive from symmetry (instead of the other way around)», that «It is an intrinsic symmetry, hierarchically distinct, that dictates the electromagnetic, weak, and strong interactions», or even that «The principles of symmetry, in fundamental physics, coincide with the particles themselves (as there is an intrinsic symmetry property of subatomic particles)», each of these expressions means that the operations—by and large Noetherian \cite{Noether "Invariante Variationsprobleme"}—of symmetry identify those properties both of individual configurations or states of a system and of the relations among them, that are consistent with a repetitive structure, say, a \emph{group structure}, in time and space, for which they remain unchanged. And yet, all this is an \emph{algebro-geometric} or \emph{algebro-topological principle}; a principle of invariance with respect to some transformation. 

In the lack of understanding of the above-mentioned distinction resides the \emph{danger} of believing that nature emanates—moreover “mysteriously” (Section \ref{subsubsection "Selection, Colors, and Understanding"}), almost “divinely”—from Mathematics; and we look like gullible.

\subsection{Anthropoid Ways Ic. Tangrams of Non-periodic Tiling: the Natural Quasi-crystals}

The history of quasi-crystals begins with the revelation, in mathematics, of \emph{non-periodic tilings}, so-called because its pattern does not repeat itself exactly, of which the best known is the \emph{Penrose's pentaplexity tiling} \cite{Penrose "Pentaplexity: A class of non-periodic tilings of the plane"}, with a 5-fold symmetry, and previously deemed prohibited.\footnote{
	A preparation of the non-periodic tiling \emph{in toto} coinciding with that of Penrose in \cite[p. 19 (1978)]{Penrose "Pentaplexity: A class of non-periodic tilings of the plane"}, made up of pentagons, thin rhombuses, half-stars ($\approx \frac{3}{5}$ of pentagrams, or 5-pointed stars), and pentagrams, is in J. Kepler \cite[third Fig. marked with “Aa”, in the second table of figures placed after p. 58]{Kepler "Harmonices Mundi Libri V"}.
	} 
The 5-plexity is built with a combination of \emph{kites} and \emph{darts}, and has the property of covering the entire plane: 

· a kite is a quadrilateral having $\angle_\mathrm{i}$ $72^\circ + 72^\circ + 72^\circ + 144^\circ$, which is composed of two Robinson triangles,\footnote{
	A paper by R.M. Robinson on tilings of the plane that is worth to mention is \cite{Robinson "Undecidability and nonperiodicity for tilings of the plane"}.
	}
each with $\angle_\mathrm{i}$ $36^\circ + 72^\circ + 72^\circ$, whilst

· a dart is a non-convex quadrilateral having $\angle_\mathrm{i}$ $36^\circ + 36^\circ + 72^\circ + 216^\circ$, which is composed of two Robinson triangles, each with $\angle_\mathrm{i}$ $36^\circ + 36^\circ + 108^\circ$.

Another Penrose tiling is the \emph{rhombus covering}, with two pieces: 

· a thin rhomb, or simply \texttt{rhomb}, having $\angle_\mathrm{i}$ $36^\circ + 36^\circ + 144^\circ + 144^\circ$, which is composed of two acute Robinson triangles, and 

· a thick rhomb, or simply \texttt{Rhomb}, having $\angle_\mathrm{i}$ $72^\circ + 72^\circ + 108^\circ + 108^\circ$, which is composed of two obtuse Robinson triangles.

\begin{figure}[h!]
\centering
\captionsetup{justification = centering}
	\includegraphics[width = 0.550\textwidth]{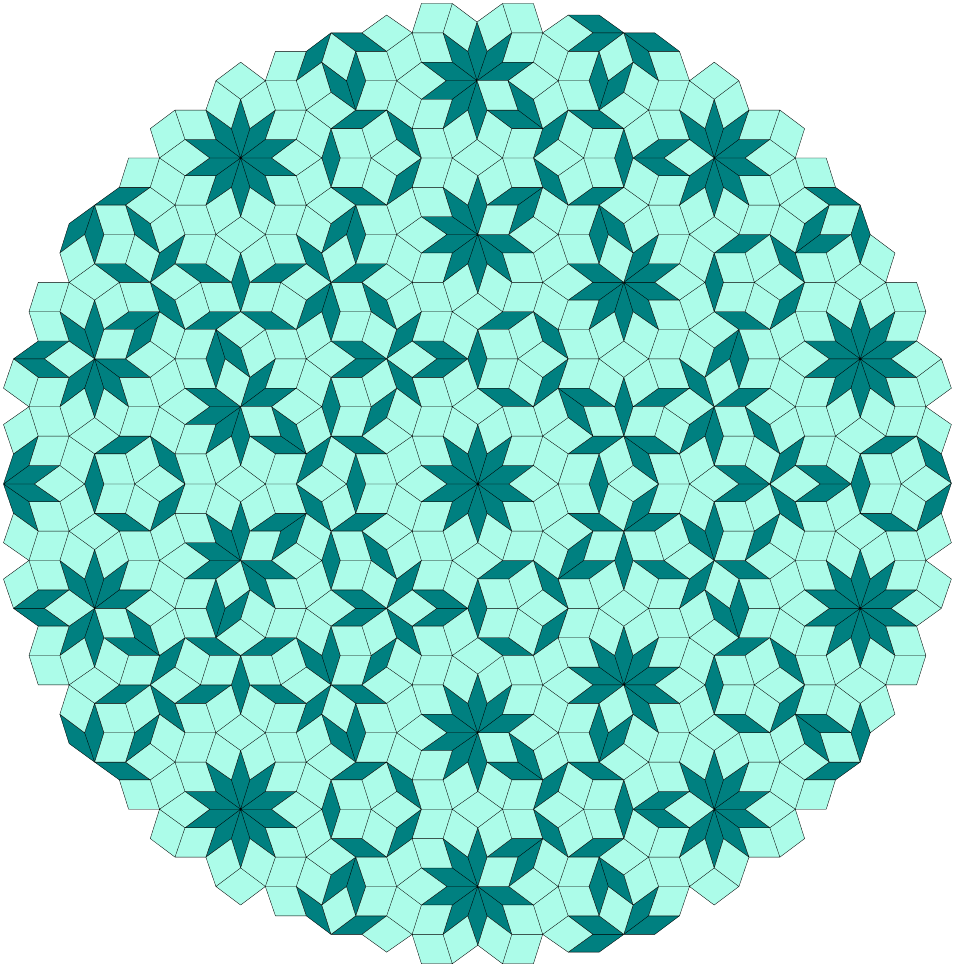}
	\caption{Arrangement № 1 of rhomboid tiling à la Penrose in 2-space, \\ with \texttt{346 rhombs} and \texttt{560 Rhombs}}
	\label{figure "Penrose Rhombus tiling 1"} 
\end{figure}

\begin{figure}[h!]
\centering
\captionsetup{justification = centering}
	\includegraphics[width = 0.550\textwidth]{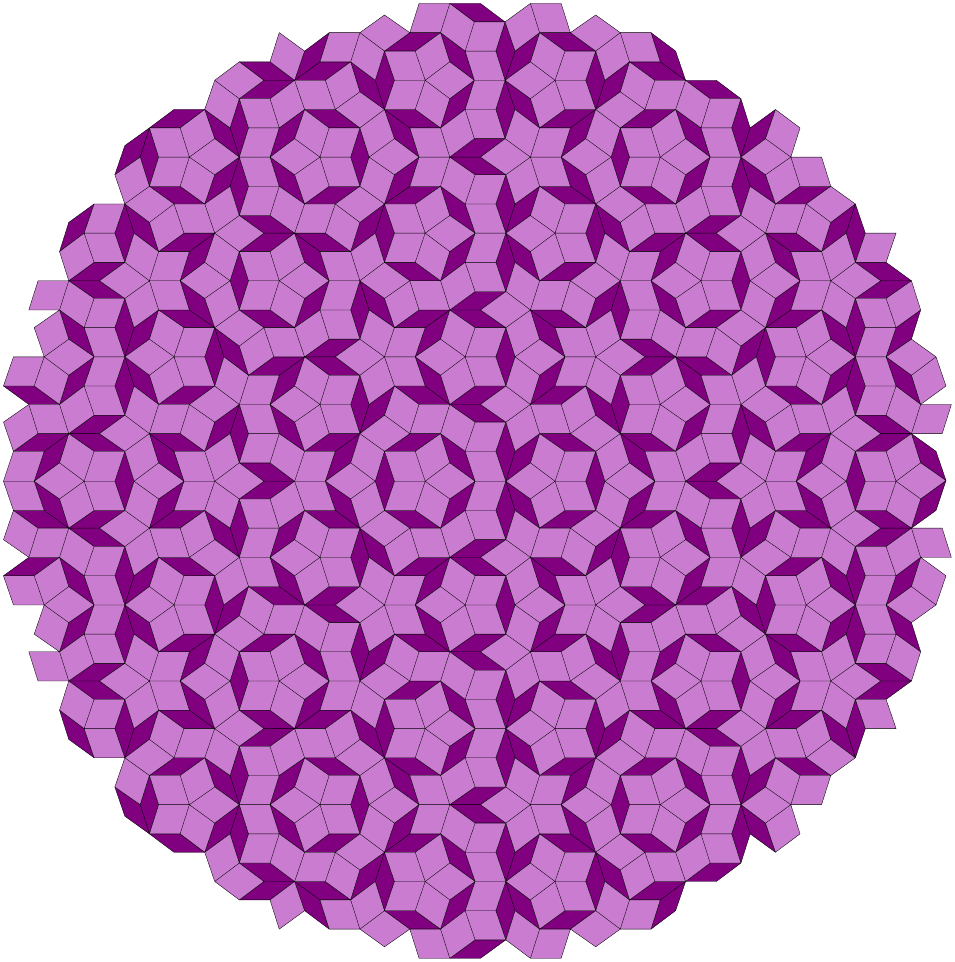}
	\caption{Arrangement № 2 of rhomboid tiling à la Penrose in 2-space, \\
	with \texttt{348 rhombs} and \texttt{560 Rhombs}}
	\label{figure "Penrose Rhombus tiling 2"} 
\end{figure}

\begin{figure}[h!]
\centering
\captionsetup{justification = centering}
	\includegraphics[width = 0.550\textwidth]{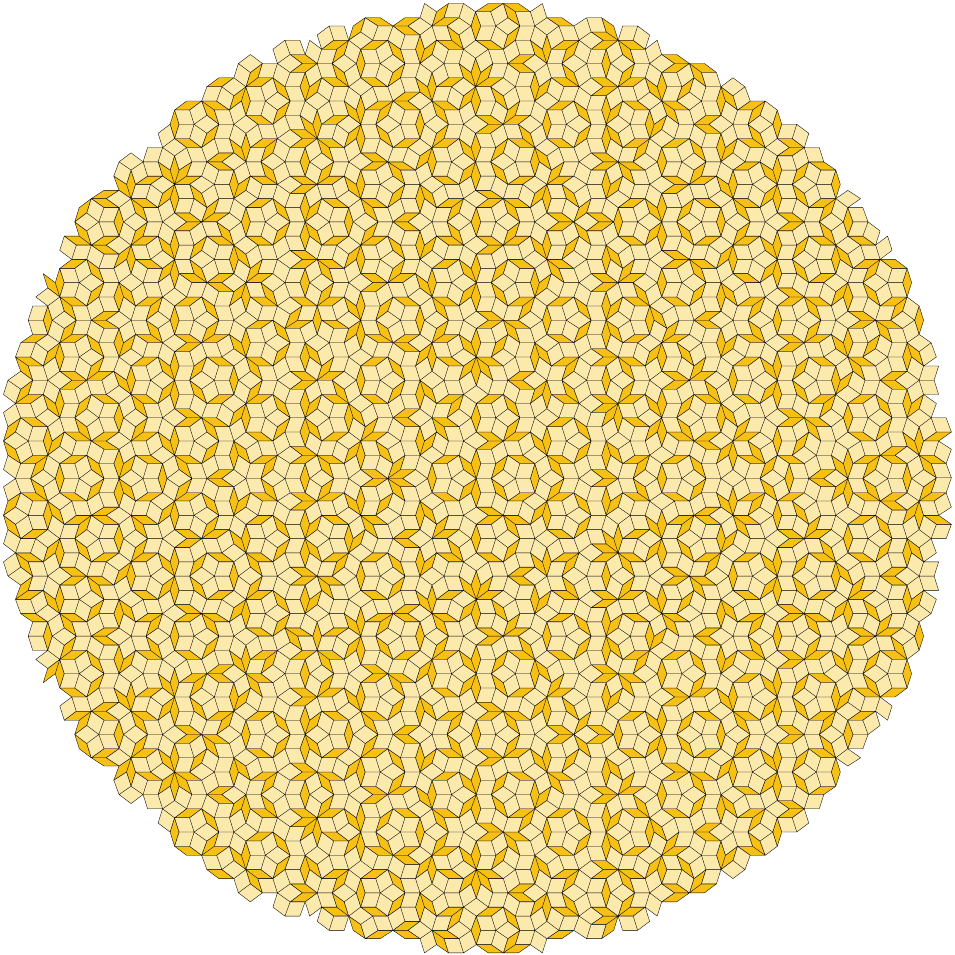}
	\caption{Arrangement № 3 of rhomboid tiling à la Penrose in 2-space, \\ 
	with \texttt{1458 rhombs} and \texttt{2356 Rhombs}}
	\label{figure "Penrose Rhombus tiling 3"} 
\end{figure}

In Figg. \ref{figure "Penrose Rhombus tiling 1"}, \ref{figure "Penrose Rhombus tiling 2"}, and \ref{figure "Penrose Rhombus tiling 3"}, there are three examples with three different combinations of pieces. 

The two pieces—that cause a non-periodic tile—are in a \emph{golden ratio} (cf. Section \ref{subsubsection "Phyllotaxis: From Helianthus Annuus to Muscari Comosum}) to each other, anyway. This is undoubtedly more evident in Fig. \ref{figure "Penrose Rhombus tiling 3"}, where more (star) pentagonal shapes emerge, to which, traditionally, one of the definitions of the golden ratio is traced back.

At a later time, physical structures were found in nature imitating the Penrose tilings, or other non-periodic tiles. These structures are called \emph{quasi-crystals}, and they are materials whose atoms are deterministically but not repetitively ordered, so they can be said to be \emph{non-periodic}, or better, \emph{quasi-periodic} (do not form a regular lattice), but, compared to (regular) periodic crystals, have forbidden symmetries, such as a \emph{5-symmetry in the plane} (\emph{5-fold symmetry in 2-space}), or an \emph{icosahedral symmetry in 3-space}.\footnote{
	A quick recap of the most important articles: 
	
	($\mathnormal{1}$) D. Shechtman, I. Blech, D. Gratias, and J.W. Cahn \cite{Shechtman Blech Gratias and J.W. Cahn "Metallic Phase with Long-Range Orientational Order and No Translational Symmetry"}: observation of a metallic solid, \ce{Al} with 10-14 at.\% \ce{Mn, Fe, or Cr} with long-range orientational order, and icosahedral point group symmetry, inconsistent with lattice translations; D. Shechtman and I.A. Blech \cite{Shechtman and Blech "The microstructure of rapidly solidified Al6Mn"}: study of a microstructure exhibiting icosahedral symmetry in electron diffraction patterns with 5-fold symmetry.
	
	($\mathnormal{2}$) D. Levine and P.J. Steinhardt \cite{Levine and Steinhardt "Quasicrystals: A New Class of Ordered Structures"}: classification of ideal 2- and 3-dimensional icosahedral quasi-crystals under their symmetry rotation, and comparison with an observed electron-diffraction pattern of an aluminum-manganese alloy.
	
	($\mathnormal{3}$) P. Gummelt \cite{Gummelt "Penrose Tilings as Coverings of Congruent Decagons"}: motivated by the arrangement of quasi-crystals, Gummelt (who is a mathematician) brings forth, geometrically, a tiling à la Penrose with overlaps for the regular decagon controlling non-periodic coverings of the Euclidean plane.
	
	($\mathnormal{4}$) P.J. Lu, K. Deffeyes, P.J. Steinhardt, and N. Yao \cite{Lu Deffeyes Steinhardt and Yao "Identifying and Indexing Icosahedral Quasicrystals from Powder Diffraction Patterns"}: presentation of a scheme to identify quasi-crystals contingent on powder diffraction patterns with a standardized indexing; see also \cite{Lu and Steinhardt "Decagonal and Quasi-Crystalline Tilings in Medieval Islamic Architecture"}.
	
	($\mathnormal{5}$) The search for natural quasi-periodic mineral starts with the study in 2009 of some grains of \emph{icosahedrite}, \ce{Al63Cu24Fe13}, the first specimen of natural quasi-crystal, having an \emph{icosahedral symmetry together with six distinct 5-fold symmetry axes}, found in the collection of the Museo di Storia Naturale of the Università degli Studi di Firenze (catalog number 46407/G); this alloy of aluminum, copper and iron appears as micrometer-sized grains set in a piece of khatyrkite, \ce{(Cu, Zn)Al2}, so-called because it comes from the Khatyrka river in the Koryak Mountains, along the northern half of Kamchatka Peninsula: see L. Bindi, P.J. Steinhardt, N. Yao, P.J. Lu \cite{Bindi Steinhardt Yao Lu "Natural Quasicrystals"} \cite{Bindi Steinhardt Yao Lu "Icosahedrite Al63Cu24Fe13 the first natural quasicrystal"}. 
	
	A spunky expedition (18 July-13 Aug 2011) to the Russian Far East \cite{Steinhardt and Bindi "Once upon a time in Kamchatka: the search for natural quasicrystals"} takes off, intended for finding other pieces of natural quasi-crystalline pattern. The exploration has brought the discovery of other icosahedral quasi-crystalline grains \cite{Bindi Eiler Guan Hollister MacPherson Steinhardt and N. Yao "Evidence for the extraterrestrial origin of a natural quasicrystal}, plus—associated with steinhardtite (\ce{Al38Ni32Fe30}), \ce{Fe}-poor steinhardtite (\ce{Al50Ni40Fe10}), \ce{Al}-bearing trevorite (\ce{NiFe2O4}), and \ce{Al}-bearing taenite (\ce{FeNi})—a natural quasi-crystal, whose composition is \ce{Al71Ni24Fe5}, with \emph{decagonal symmetry} comprising quasi-periodic atomic arrangements with \emph{10-fold symmetry} \cite{Bindi et al. "Natural quasicrystal with decagonal symmetry"}.
	}

The pseudo-conundrum is that the discovery of quasi-crystals is routinely accompanied by these kinds of questions: «How is it possible that mathematicians have anticipated the subsequent discoveries by physicists?», or «How can this be happening, that the explanation of the formation of natural quasi-crystals relies on the golden ratio, which was envisioned by Euclid for purely mathematical aims?». 

Similar questions are (rhetorically) tendentious and, most of all, wrong, because they mask the mathematical work with a shamanistic halo, portraying mathematicians as “oracles”, or “soothsayers”, capable of reading the hidden traces of nature before the factual confirmation by physicists, chemists, and geologists. All this is a good illustration of how man is \emph{misled by both periodic symmetry} (cf. Section \ref{subsection "Anthropoid Ways Ib. Symmetry and Invariance in Physics: the Impact of Group Theory"}) \emph{and non- or quasi-periodic symmetry}, the latter present in the quasi-crystals. Nature is so vast that many tangrams\footnote{
	The \emph{tangram} is a game of Chinese origin (\ZhSimplified{七巧板}), whose \emph{basic} version consists of a square-container containing 7 pieces of colored wood (5 triangles, 1 square, and 1 parallelogram), which can be combined in various ways to create different shapes, each time. 
	} 
in it have a chance to form and to evolve (and who knows how many of these puzzles are still unknown to us). And since man is within such a evolution, he is part of nature, there is no mathematical “epiphany” (see Section \ref{subsection "Math-Language and its Reasonably Effectiveness"}). 

\subsection{Anthropoid Ways II. Two Examples}

We should mention two debated topics: phyllotaxis and inverse-square laws.

\subsubsection{Phyllotaxis: From \emph{Helianthus Annuus} to \emph{Muscari Comosum}}
\label{subsubsection "Phyllotaxis: From Helianthus Annuus to Muscari Comosum}

\begingroup
\footnotesize
L'arrangement, le nombre, la force, \& les proportions des Folioles offrent bien des variétés \& des bizarreries, non seulement dans le même Individu, mais encore dans la même Feuille.\footnote{
	«Arrangement, number, strength, \& proportions of the Leaflets offer many varieties \& oddities, not only in the same Individual, but also in the same Leaf».
	} \\
\indent — \textsc{C. Bonnet} \cite[p. 193]{Bonnet "Recherches sur l'usage des feuilles dans les plantes"}

\endgroup

\vspace{2mm}

The word \emph{phyllotaxis} comes from the Gr. \textgreek{φυλλίς}, “leaf”, and \textgreek{τάξις}, “arrangement”, “order”. The third Mémoire of the Bonnet's book \cite[pp. 159-190]{Bonnet "Recherches sur l'usage des feuilles dans les plantes"}, entitled \textit{De l'Arrangement des Feuilles sur les Tiges, \& sur les Branches, \& de celui qu'on observe dans quelques autres parties des Plantes}, is entirely dedicated to explain this botanical manifestation. Some mathematically detailed studies on  phyllotaxis: \cite[chap. XIV]{Thompson "On Growth and Form"} \cite[pp. 72-73]{Weyl "Symmetry"} \cite[chap. 4]{Prusinkiewicz Lindenmayer "The Algorithmic Beauty of Plants"} \cite[sec. 5.9]{Peitgen Jurgens Saupe "Chaos and Fractals: New Frontiers of Science Second Edition"} \cite[sec. 1.10]{Stakhov Aranson "The "Golden" Non-Euclidean Geometry: Hilbert's Fourth Problem" "Golden" Dynamical Systems and the Fine-Structure Constant"}. 

The highly symmetric order characterizing the arrangement of leaves and flowers indicates that e.g. a sunflower (\textit{Helianthus annuus}), a  pineapple (\textit{Ananas comosus}), or a Romanesco broccoli (specimen of \textit{Brassica oleracea}) know \emph{geometry of phyllotactic patterns}, that they adopt the \emph{Fibonacci sequence} \cite[foglio 134 verso]{Fibonacci "De Abaco manuscript L.IV.20"} \cite[p. 404]{Fibonacci "Liber Abaci: A Translation into Modern English"},\footnote{
	The paragraph of interest to us is entitled \textit{Quot paria coniculorum in uno anno ex uno pario germinantur} («How many pairs of rabbits are created by one pair in one year»), where the Fibonacci sequence for the first time makes its appearance: «parium 1, primus 2, secundus 3, tercius 5, quartus 8, quintus 13, sextus 21, septimus 34, octavus 55, nonus 89, decimus 144, undecimus 233, duodecimus 377».
	} 
\begin{equation}
\label{equation "Fibonacci sequence"}
	\mathit{Fib}_0 = 0, \mathit{Fib}_1 = 1, \mathit{Fib}_n = \mathit{Fib}_{n - 1} + \mathit{Fib}_{n - 2}, n \geqslant 3,
\end{equation}
satisfying a \emph{recurrence relation}? Eq. \eqref{equation "Fibonacci sequence"} gives $0, 1, 1, 2, 3, 5, 8, 13, 21, 34, 55, \mathellipsis$, from which the identity 
\begin{equation}
	\mathit{Fib}_{n - 1}\mathit{Fib}_{n + 1} - \mathit{Fib}^2_n = (-1)^n.
\end{equation}

\textit{Helianthus}'s florets are magnificently sorted in a $\mathit{Fib}_n$-sequence, with \emph{34 clockwise and 55 counter-clockwise spirals}, or vice versa: reversing the counting pattern, one has 55 clockwise and 34 spirals counter-clockwise; furthermore, in the outermost part of the pseudanthium there are other \emph{21 counter-clockwise spirals}. 

\begin{figure}[h!]
\centering
\captionsetup{justification=centering}
	\includegraphics[width = 0.550\textwidth]{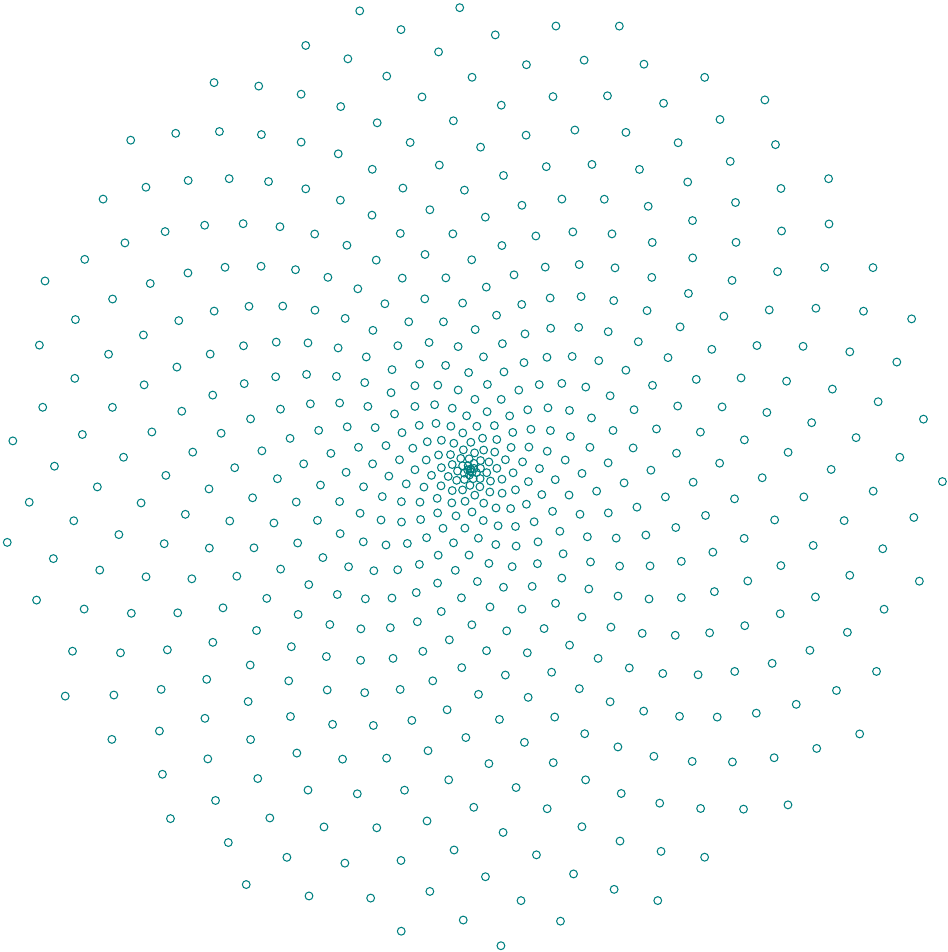}
	\caption{\textit{Helianthus}-like phyllotaxis: \\ spiral parastichy lines with \textcolor{mallard}{$\mathtt{Fib_n = 610}$ \texttt{circles}}, viz. florets, \\ via \texttt{"proportio aurea" (1 + sqrt(5))} command}
	\label{figure "phyllotaxis with 610 circles"} 
\end{figure}

\begin{figure}[h!]
\centering
\captionsetup{justification = centering}
	\includegraphics[width = 0.550\textwidth]{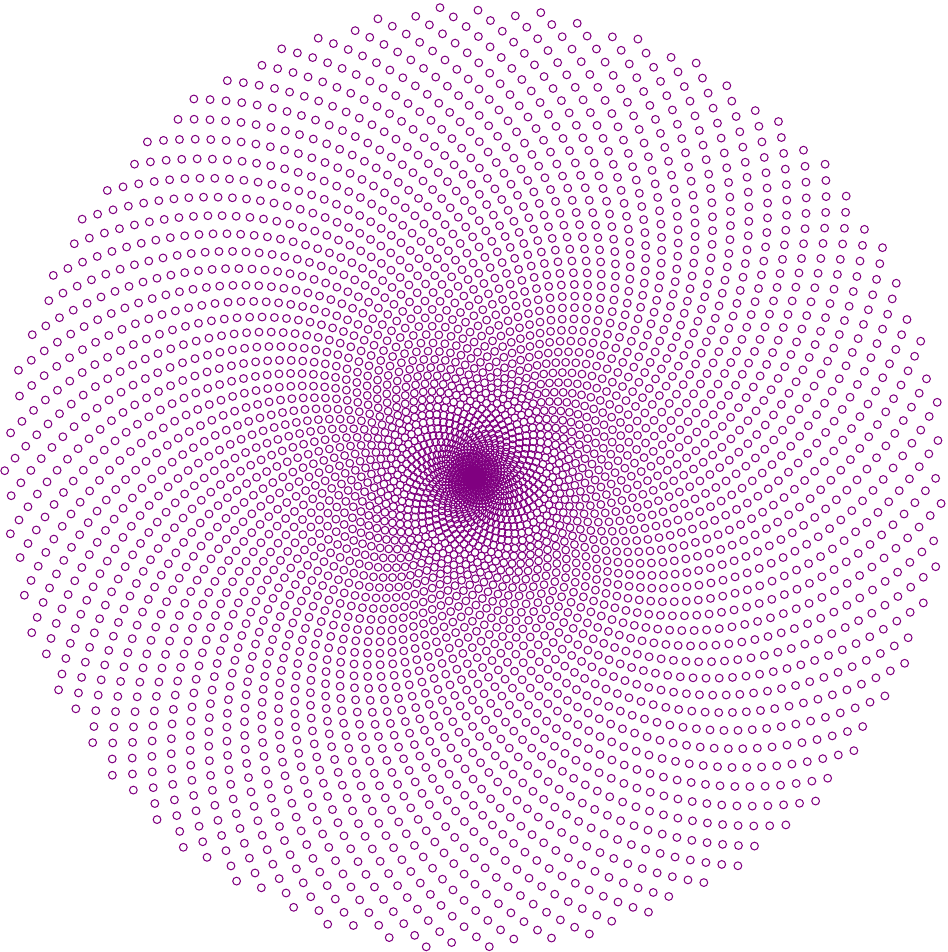}
	\caption{\textit{Helianthus}-like phyllotaxis: \\ spiral parastichy lines with \textcolor{eggplant}{$\mathtt{Fib_n = 4181}$ \texttt{circles}}, viz. florets, \\ via \texttt{"proportio aurea" (1 + sqrt(5))} command}
	\label{figure "phyllotaxis with 4181 circles"}
\end{figure}

By “florets” we mean “flower-heads” (\emph{capolini fiorali}), while the pseudanthium (aka capitulum) is the big flower head in which all mini flower-heads are grouped together to generate a floral configuration. Figg. \ref{figure "phyllotaxis with 610 circles"} and \ref{figure "phyllotaxis with 4181 circles"} are abstractions: it is assumed that the number of florets is equal to a Fibonacci number, which is an idealization, and florets are represented as button-like (circles), which is a further idealization. 

The spiral marking of florets (or leaves) are called \emph{parastichy} (from the Gr. \textgreek{παρά}, “beside”, “near”, and \textgreek{στίχος}, “row”, “file”, “line”), since florets (or leaves) are sorted in a spiraling pattern, one next to the other one. \emph{Parastichy numbers are always Fibonacci numbers}. An Helianthus, as we said earlier, has $(34, 55)$- and 21-parastichies.

Fibonaccian spirality, with different numbers, can be admire in many other plants, fruits, and vegetable organs. 

\textit{Cota tinctoria}'s flower-heads put together \emph{13 clockwise and 21 counter-clockwise spirals}, whilst other daisies of the family \textit{Asteraceae} present a $(21, 34)$-parastichy.

Strobilus in \textit{Pinaceae} conifers exhibits \emph{8 clockwise and 13 counter-clockwise spirals}, plus \emph{21 clockwise spirals} emerging farther from the central axis; which is equivalent to 8, 13- and 21-parastichies. 

Same parastichy numbers, 8, 13, and 21, are present in the \emph{Ananas comosus} along its helical tessellations. 

A $(8, 13)$-parastichy is found in the pointed shape of the Romanesco broccoli.

Other styles of orderly alignment are in bulbous plants: by way of example, the \emph{tassel hyacinth}, or \emph{lampascione} in It., its scientific name is \textit{Muscari comosum}, also known as \emph{Leopoldia comosa}, has violet-turning-blue flowers \emph{proportionally} exhibited on the stem, so as to form an astonishing multi-candelabrum.

All this symmetric order indicates that Nature follows an arrangement under the golden angle? Let us remember that, in a plant, a \emph{golden angle} is the angle between the \emph{radial directions} of two consecutive flowering stems or leaves; cf. L. and A. Bravais \cite[pp. 67, 74]{L. et A. Bravais "Essai sur la disposition des feuilles curviseriees"}: their calculated botanical golden angle may vary, in the diversity of species, between $\ang{137.06}$ and $\ang{137.47}$. 

As we know, since Kepler on, the Fibonacci numbers, and the Fibonacci sequence, are related with the golden number (\textit{proportio aurea})\footnote{
	L. Pacioli \cite[pars prima, p. 4 recto]{Pacioli Divina proportione} describes it thus: «questa [\,\dots] proportione non se po mai per numero intendibile asegnare ne per quantita alcuna rationale exprimere: ma sempre fia occulta e secreta e dali Mathematici chiamata irrationale».
	}
\begin{equation}
	\textgreek{\text{φ}} = \frac{1}{2}(1 + \sqrt{5}) = 1.6180339887498948\cdots
\end{equation}
It is possible \emph{approximate} in various manners. A very basic way is due to R. Simson's ancient intuition \cite{Simson "An Explication of an Obscure Passage in Albert Girard's Commentary upon Simon Stevin's Works"}, under which the successive convergents of a continued fraction for $\textgreek{\text{φ}}$ are numerically quasi-coincident with the $\mathit{Fib}_n$-sequence:
	\begin{equation}
		\lim_{n \to \infty}\left\{\frac{\mathit{Fib}_n}{\mathit{Fib}_{n - 1}}\right\} \aprx \textgreek{\text{φ}}, \text{ e.g. } \frac{102334155}{63245986} = 1.6180339887498947\cdots,	
	\end{equation}
where $\mathit{Fib}_{n = 40}$ (numerator) and $\mathit{Fib}_{n = 39}$ (denominator). For the same reason, from the Fibonacci sequence we can derive an approximation of the \emph{golden spiral}, called \emph{Fibonacci spiral}. The recursion relation is also satisfied by the golden ratio, so 
\begin{equation}
\textgreek{\text{φ}}^n = \mathit{Fib}_n\textgreek{\text{φ}} + \mathit{Fib}_{n - 1}.
\end{equation} 

Answer to previous questions: No, I do not think so at all (cf. Section \ref{subsubsection "How is it Possible?"}). What our eyes see as “mathematics” or “laws of symmetry”, for nature is simply an adaptive occupation of space, with different purposes depending largely on the surrounding environment, not to mention the principle, devised by P. Curie \cite{Curie "Sur la symetrie dans les phenomenes physiques symetrie d'un champ electrique et d'un champ magnetique},\footnote{
	\cite[pp. 393, 400]{Curie "Sur la symetrie dans les phenomenes physiques symetrie d'un champ electrique et d'un champ magnetique}: «I think it would be advantageous to introduce in the study of physical phenomena considerations on symmetry familiar to crystallographers. An isotropic body, for example, can be animated by a rectilinear or rotational movement; a liquid one can be the seat of vortex movements; a solid one can be compressed or twisted; it can be found in an electric or magnetic field; it can be crossed by an electric or heat current; it can be traversed by a ray of natural light or [a ray] polarized rectilinearly, circularly, elliptically, etc. [But] in any case, a certain characteristic asymmetry is necessary at each point of the body [\,\dots]. In other words, certain elements of symmetry can coexist with certain phenomena, but they are not necessary. What is necessary is that certain elements of symmetry do not exist. \emph{It is the asymmetry that creates the phenomenon} [\textit{C'est la dissymétrie qui crée le phénomène}]».
	} 
that \emph{asymmetry} is the kingpin of physics on which the origin of life rests.\footnote{
	With Curie, the \emph{regularity} of natural phenomena \emph{originates from a symmetry breaking}: a \emph{pattern} is not a “total” or “full” symmetry, which is something indistinguishable or unintelligible; it is a \emph{partial symmetry}. Stunning examples are the \emph{Chladni figures} \cite[Pl. 3-7]{Chladni "Traite d'acoustique} = \cite[App. C]{Chladni "Treatise on Acoustics"}.
	
	E.F.F. Chladni (1756-1827) is responsible for devising a method demonstrating that there are several modes of vibration of a mechanical surface, whether regular or irregular, covered with sand. His technique consists in observing the vibrations of glass (or metal) plates, having different forms (circles, ellipses, triangles, squares, rectangles, hexagons), produced by the friction of a violin bow along the edge of the plate, so that the grains of sand move away from the areas of greatest vibration, grouping together in \textit{lignes nodales}, and composing the above-named figures, which are \emph{patterns of vibration}: the nodal lines are points of virtually zero vibration amplitude. For the Chladni figures, see also M. Somerville \cite [Plate II-III]{Somerville "On the Connexion of the Physical Sciences"}.
	}\textsuperscript{,}\endnote{
	An early—if not the first mathematically accurate—example of \emph{symmetry breaking} was provided by C.G.J. Jacobi \cite{Jacobi "Ueber die Figur des Gleichgewich"}.
	}

But botanical phyllotaxis is not alone. Phyllotactic arrangements can also be proved \emph{analytically}, and even reproduced \emph{experimentally} in \emph{inorganic chemistry}. 
\enumerationisinitium
\item L.S. Levitov \cite{Levitov "Phyllotaxis of Flux Lattices in Layered Superconductors"} \cite{Levitov "Energetic Approach to Phyllotaxis"} shows that, studying the geometry of a \emph{flux lattice}, i.e. of a \emph{soft-lattice}, pinned by superconducting layers, we can witness—under variation of a magnetic field—an infinite sequence of continuous transitions of the lattice; these transactions are different manners of selection of shortest distances: that which is created, via analytic solution, is a \emph{hierarchical structure of quasi-bifurcations}, replicating in all respects the phyllotactic organization (e.g. with pairs of consecutive Fibonacci numbers). The phyllotaxis thus seems a general phenomenon, occurring in any dynamically accessible lattice subjected to strong deformations, where in addition a symmetry group $SL_2(\mathbb{Z}) \otimes \mathbb{Z}_2$ is involved \cite[§ 4]{Levitov "Energetic Approach to Phyllotaxis"}. An experimental demonstration of the Levitov's model is in \cite{Nisoli Gabor Lammert Maynard and Crespi "Static and Dynamical Phyllotaxis in a Magnetic Cactus"}.
\item S. Douady and Y. Couder \cite{Douady and Couder "Phyllotaxis as a Physical Self-Organized Growth Process"} bring us a laboratory evidence: if we slide some drops of ferrofluid at the center of a silicone oil-filled dish, they acquire a familiar disposition, and form \emph{leaf-like patterns}, e.g. distichous vs. spiral/alternate arrangements, see \cite[Fig. 2, p. 2099]{Douady and Couder "Phyllotaxis as a Physical Self-Organized Growth Process"}. The experiment consists of a horizontal dish placed in a vertical magnetic field created by two Helmholtz coils. The drops, resulting polarized, behave like small magnetic dipoles, generating a Fibonacci-type of series, with a convergence of the \emph{golden angle} e.g. from $\ang{180}$ to $\ang{137.47}$, very close to the golden section 
\begin{equation}
	\textgreek{\text{Φ}} = 2\pi(1 - \textgreek{\text{τ}}) \approx \ang{137.5}, \enspace \textgreek{\text{τ}} = \frac{1}{2}(-1 + \sqrt{5}) = 0.6180339887498948\cdots
\end{equation}
recognizing that $\textgreek{\text{τ}} = \frac{1}{\textgreek{\text{φ}}}$. In any case, the phyllotactic disposition appears to vary with the inserting speed of each drop. 

The inorganic pseudo-philotaxis emerges consequently from a \emph{self-organization} throughout iterative processes: it has \emph{emergent properties} very similar to those that permit the origination of fractal motifs in \emph{snowflakes}.
\enumerationisfinis

\subsubsection{Inverse-square Laws}
\label{subsubsection "Inverse-square Laws"}

\begingroup
\footnotesize
There is a substance, which I call the electric fluid, the particles of which repel each other and attract the particles of all other matter, with a force inversely as some less power of the distance than the cube: the particles of all other matter also, repel each other; and attract those of the electric fluid, with a force varying according to the same power of the distances. \\
\indent — \textsc{H. Cavendish} \cite[p. 585]{Cavendish "An attempt to explain some of the principal phaenomena of electricity by means of an elastic fluid"} 

\endgroup

\vspace{2mm}

The distance between charges (Coulomb's law) \cite[pp. 107-318]{Coulomb "Memoires de Coulomb"} and masses (Newton's law of gravitation) is $\frac{1}{r^2}$. Does this mean that 2, and precisely 2, is the (pre)chosen number by nature for electric and gravitational fields? It is clear that no. The number 2 is plainly due to the fact that these two laws, restricted to a \emph{space of three dimensions}, are formulated with a proportionality so that the intensity of radiation, or the force, varies inversely with the square of the distance from the source (cf. Question \ref{quaestio "Bridgman on the simplicity"}, p. \pageref{quaestio "Bridgman on the simplicity"}). There is no Pythagorean  mysticism in \emph{our} number 2, it is there because it is \emph{provided by equations of the Coulombian and Newtonian type} acting over the distance.

\subsection{Anthropoid Ways III. Laws of Nature}
\label{subsection "Anthropoid Ways III. Laws of Nature"}

\begingroup
\footnotesize
[N]ature certainly existed before man existed, but if nature existed before man, it is not the same as the natural sciences. For ex­ample, the concept of “the law of nature” \emph{cannot be com­pletely objective}, the word “law” being a \emph{purely human} principle. \\
\indent — \textsc{W. Heisenberg} \cite[p. 35, e.a.]{Heisenberg "Discussion of the Lecture of Werner Heisenberg"}

\endgroup

\vspace{2mm}

\enumerationisinitium
\item What we have defined above as an anthropoid way of describing the laws of nature does not detract from the \emph{physiological possibility} of some correspondence, or \emph{tuning}, between \emph{world} (external reality, environment around us) and \emph{mind}, the latter having co-evolutionarily adapted to the world of which it is a part, to register, more or less reliably, information on the external order/regularity, or on the (truest) nature of phenomena.
\item Reflect on the belief in a universe-program, now such a fashionable term—but the same can be said on the conviction of a geometric universe, when it shows geometrizable properties, or on the image of a universe as a thermodynamic machine.\footnote{
	\label{footnote "We see the nature of the world on the strength of the technology that we are capable of building"}
	Cf. e.g. \cite[p. 133]{Bridgman "The Way Things Are"}: «The picture of the nature of the world which we have evolved is heavily colored [scilicet: conditioned] by our experience with tools or instruments. We supplement and extend the evidence presented to us by our senses by the use of instruments, so that the very meaning of some of our most important concepts can be defined only in terms of operations with instruments».
	
	We see the nature of the world on the strength of the technology that we are capable of building, down the centuries. There is—in the history of science—the “universe-order” (\textgreek{κόσμος}), the “geometric universe”, the “clock-universe”, the “thermodynamic universe”, the “program-universe” (\textit{it from bit}), the “holographic universe”. All these “universes” reflect the changing fashions of the age.
	} 
Does the fact that the world is computable, and some of its information compressible, mean that the world is mathematical, or that reveals and intrinsic mathematical facet (since computability and compressibility are handled by mathematical logic)? No, because the “computability” of nature, or rather, a part of it, i.e. the chance of applying to a phenomenal area an algorithm that calculates a specific function, for each possible set of values of the independent variable, does not lie in the regularity of nature, but in \emph{our} way of reading some (computable) results in relation to that regularity.
\item Let us make two hypotheses. We say, first of all, that the laws of nature are what constrain their computable character, so that \emph{computability is a consequence of the laws of nature}; alternatively, we say that the laws of nature are constrained by some rules of computation, for which \emph{the laws of nature derive from computability}. In both cases, we arrive at the same evaluation: in the second case, and a fortiori in the first hypothesis, it is pretentious  and, at once, faulty to assert that mathematics is a peculiar part of the external world, or is about real objects.
\enumerationisfinis

\subsection{Math-Language and its Reasonably Effectiveness}
\label{subsection "Math-Language and its Reasonably Effectiveness"}

We will tackle below the pseudo-problem of the «unreasonable effectiveness» of mathematics in the physical sciences, under the well-known expression of E. Wigner \cite{Wigner "The Unreasonable Effectiveness of Mathematics in the Natural Sciences"}, and some kindred issues.

\subsubsection{The «Miracle» of Mathematics: the Wignerian Ignis Fatuus}

These are the—preposterous—words of Wigner \cite[pp. 7, 14]{Wigner "The Unreasonable Effectiveness of Mathematics in the Natural Sciences"}: 

\vspace{2mm}

\begingroup
\footnotesize
It is difficult to avoid the impression that a miracle confronts us here, quite comparable in its striking nature to the miracle that the human mind can string a thousand arguments together without getting itself into contradictions or to the two miracles of the existence of laws of nature and of the human mind's capacity to divine them [\,\dots]. The miracle of the appropriateness of the language of mathematics for the formulation of the laws of physics is a wonderful gift which we neither understand nor deserve.
 
\endgroup

\vspace{2mm}

Compare it to what Einstein had declared seven years earlier (letter to M. Solovine, 30 March 1952), see footnote \ref{footnote "Einstein's letter to M. Solovine, 30 March 1952"} on p. \pageref{footnote "Einstein's letter to M. Solovine, 30 March 1952"}.

\subsubsection{In the Bridgman's Furrow}

We share the P.W. Bridgman's vision \cite[pp. 60-62, e.a.]{Bridgman "The Logic of Modern Physics"} without reserve: 

\vspace{2mm}

\begingroup
\footnotesize
It is the merest truism [\,\dots] that mathematics is a \emph{human invention}. Furthermore, the mathematics in which the physicist is interested was developed for the explicit purpose of describing the behavior of the external world [nature], so that it is certainly no accident that there is a correspondence between mathematics and nature [\,\dots]. There is no longer any basis for the idealization of mathematics, and for the view that our imperfect knowledge of nature is responsible for failure to find in nature the precise relations of mathematics. It is the mathematics made by us which is imperfect and not our knowledge of nature. [From the operational point of view it is meaningless to attempt to separate “nature” from “knowledge of nature”].

\endgroup

\vspace{2mm}

\emph{Nature}, or better, an experimental side of it (its empirical content), \emph{is not mysteriously (pre)adapted to the concepts of mathematics}. The opposite is true: \emph{the concepts of mathematics are our creations adapted to better fit natural phenomena}. In the next Section, we will zoom in on this regard.

\subsubsection{Selection, Colors, and Understanding}
\label{subsubsection "Selection, Colors, and Understanding"}

\begingroup
\footnotesize
The evolution of mathematics is a fact. Science historians have recorded its slow rise, through trial and error, to greater efficiency. It may not be necessary, then, to postulate that the universe was designed to conform to mathematical laws. Isn't it rather our mathematical laws, and the organizing principles of our brain before them, that were selected according to how closely they fit the structure of the universe? The miracle of the effectiveness of mathematics, dear to Eugene Wigner, could then be accounted for by \emph{selective evolution}, just like the miracle of the adaptation of the eye to sight [\,\dots]. Numbers, like other mathematical objects, are \emph{mental constructions} whose roots are to be found in the adaptation of the human brain to the regularities of the universe [\,\dots]. The brain is not a [purely] logical, universal, and optimal machine. While evolution has endowed it with a special sensitivity to certain parameters useful to science, such as number, it has also made it particularly restive and inefficient in logic and in long series of calculations [see Chapter \ref{chapter "Galois' Legacy—Rules over the Calculations: the Pursuit of Generality"}]. It has biased it, finally, to project onto physical phenomena an \emph{anthropocentric framework} that causes all of us to see evidence for design where only evolution and randomness are at work. Is the universe really “written in mathematical language”, as Galileo contended? I am inclined to think instead that \emph{this is the only language with which we can try to read it}. \\
\indent — \textsc{S. Dehaene} \cite[pp. 232-233, e.a.]{Dehaene "The Number Sense: How the Mind Creates Mathematics"}

\endgroup

\vspace{2mm}

\enumerationisinitium
\item Mathematics is a \emph{language}, a specialized code of communication,\footnote{
We establish, once and for all, that by \emph{mathematical language} we do not mean, reductively, the counterpart, in mathematics, of the common system of signs suitable for communication in our daily life. Language, in mathematics, can also mean—if it is not better identifiable—a welter of «more or less clear images», a clutter of «psychological entities» freely combined \cite[p. 142]{Hadamard "The Psychology of Invention in the Mathematical Field"}.
	} 
in agreement with cognitive circuits,\footnote{
	We leave to others, each with its own expertise, the burden of analyzing the \emph{mode} in which this—(neuro)bio- and anthropo-logically—happens. If, however, the result of such an investigation is like the one in the book of G. Lakoff and R.E. Núñez (\textit{Where Mathematics Comes From: How the Embodied Mind Brings Mathematics into Being}, Basic Book, New York, 2000), then it is best to forget the whole thing. It is ludicrous to identify the conceptualization of mathematics—the manner in which mathematical objects are conceptualized (by mathematicians)—with the cognitive science of mathematics (under a cognitive linguist, and a psychologist), as if such a reduction process were fair. This kind of studies is suffering from self-referentiality: it is \emph{litterature} (Carroll's pun), rubbish. To paraphrase a memorable saying: the cognitive science of mathematics, set forth by Lakoff–Núñez, is as useful to mathematicians as ornithology is to birds. For a harsh criticism, see e.g. the free online papers by G. Lolli \cite{Lolli "La metafora in matematica"} \cite{Lolli "Da dove viene la matematica. Recensione a G. Lakoff e R.E. Nunez 'Where Mathematics comes from'"}.
	} 
which is why it is (more or less) \emph{effective}, like, to differing extents, \emph{any other} language;\footnote{
	\label{footnote "Feynman, mathematics, language and logic"}
	R.P. Feynman \cite[p. 40]{Feynman "The Character of Physical Law"} seems to disagree: «[M]athematics is \emph{not} just another language. Mathematics is a language plus reasoning; it is like a language plus logic. Mathematics is a tool for reasoning». But that is not the situation. Other non-mathematical languages are also supported by reasoning, so much so that logic is not mathematics, and mathematics is not a simple manifestation of logic, although mathematics is, in part, logical. If reasoning, in a broader sense, does not belong exclusively to mathematics, then mathematics does not have something more than other languages, but rather something different, just as each language adopts, in its own way, a type of organized reasoning, cf. Section \ref{subsection "Extra-logical Objects, and Gödelian Suggestions"}.
	}
for example, the \emph{cat's meow} too is effective, so s/he gets all the attention. This explains the so-called \emph{usefulness} of mathematics in physics,\footnote{
	It is an old story: cf. e.g. L. Belleri, \textit{In Geometrica problemata Simonis Stevinii}, in \cite[first page]{Stevin "Problematum geometricorum"}: «Vere igitur Diuam veteres dixêre Mathesin, / Cuius ab arte labor superas cognoscere sedes, / Terrarum, pelagi[que] vias, \& operta tenebris / Natura secreta dedit: coramque tuers», whose translation is: «The ancients quite rightly named the Mathesis ‘Divine’, which with the application of its technique allowed to learn the supreme places, the roads of the Earth, and the sea, and to unlock the secrets of Nature from darkness». Mathesis is for Maths, transliterated from the Gr. \textgreek{Mάθησις}, “the act of learning”, or “of knowledge”.
	}
which is not a «miracle», or something «mysterious», as Wigner \cite[pp. 2-3, 7, 14]{Wigner "The Unreasonable Effectiveness of Mathematics in the Natural Sciences"} seems to believe, but it is completely \emph{reasonable} (within the limits of reason).\endnote{
	See the article of F.E. Browder \cite{Browder "Does Pure Mathematics Have a Relation to the Sciences?"}, who deserves credit for filling seven and a half pages in three columns without saying anything relevant about the question that opens his article (\textit{Does pure mathematics have a relation to the sciences?}). He writes (p. 548): «Because of its origins and its nature, mathematics is not unreasonably effective [à la Wigner] in the physical sciences: it is simply (though surprisingly) effective». Right, and then? 
	}

In response to the Wignerian credo, M. Fabbrichesi \cite[p. 46]{Fabbrichesi "Pensare in formule. Newton Einstein e Heisenberg} makes an apt comparison between mathematics and music:

\vspace{2mm}

\begingroup
\footnotesize
But is this effectiveness of mathematics in grasping and explaining the world so surprising? Our brain was forged by natural selection and so it was for that piece of software, which our brain uses in deciphering the outside world, which we call mathematics. Why then should this [mathematics] not be perfectly suited to describing the same natural world that shaped it? The opposite would perhaps be even more surprising: how would a formal system have developed, a system that is understandable to us but with no relation to the environment in which we live? [\,\dots] A similar example [\,\dots] is the music. Its ability to reflect in giving voice to our deepest emotions seems sometimes supernatural and inexplicable [\,\dots]. Why does a series of notes and chords—air compression wave [propagating] in a concert hall—make us cry or comfort us, or both things at once? It must be because our emotional world finds its roots in the same ground where these sounds are processed and brought to our understanding. They evolved together, forged, maybe just randomly, by the same evolutionary pressure. The result is that a complex system of musical rules—in itself independent—shows a [precise] association and reflects with startling accuracy the same territory traced by our emotions. Mathematics looks like some kind of music of the mind and it does, for the territory of physical reality and of the objects that surround us, what music does for the territory of emotions.

\endgroup

\vspace{2mm}
	
G. Israel \cite[pp. 130-131]{Israel "La matematica e la realta. Capire il mondo con i numeri"}, to play against all the baloney and ignes fatui tied to the «miracle» of mathematics, remarks upon a simple issue: 

\vspace{2mm}

\begingroup
\footnotesize
The truth is that, for us, it is neither a miracle nor a mystery because any known mathematics from the seventeenth century onwards was created to study physical phenomena [\,\dots]. [T]he effectiveness of modern mathematics in physics is [\,\dots] a fact deriving from the appropriate way in which it arose from the context of the analysis of physical phenomena: its close bond with these phenomena provides a convincing and “reasonable” [\textit{razionale}] explanation of its success. [\,\dots] [P]hysics has abandoned the Aristotelian-style qualitative approach to adopt an essentially quantitative approach, and mathematics, in turn, has bowed to the description of physical phenomena, facing the problem of the representation of infinitesimal and infinite processes head-on. A new mathematics—that of the infinitesimal calculus or “calcolo sublime”—was created under the impulse of the ambition to quantitatively describe physical phenomena. But the subsequent axiomatic mathematics also had its roots firmly planted in earlier developments [\,\dots]: as Bourbaki has well-observed,\footnote{
	But cf. N. Bourbaki \cite[pp. 46-47]{Bourbaki "L'architecture des mathematiques La Mathematique ou les Mathematiques?}: «In the axiomatic conception, mathematics appears in short as a reservoir of abstract \emph{forms}—mathematical structures; and it is found—even though we do not really know why—that certain aspects of experimental reality are molded into some of these forms, as through a kind of pre-adaptation [\textit{certains aspects de la réalité expérimentale viennent se mouler en certaines de ces formes, comme par une sorte de préadaptation}]. It cannot be denied, of course, that most of these forms originally had a well-defined intuitive content; but precisely with a voluntary emptying of this content, we could give [these forms] all the effectiveness that they potentially carry, making them susceptible to receive new interpretations». Visibly, Bourbaki abandons himself to a foolery, too: all aspects of reality are not pre-adapted to the mathematical forms; the contrary one is worth: all mathematical forms are pre-adapted to the aspects of reality.
	} 
in the axiomatic method, a mathematician chooses axioms on which his theory is built, under the incitement of real problems.

\endgroup

\vspace{2mm}

One could ask why the world is, or rather, \emph{appears} mathematical, but this does not differ from asking why the flower of zucchini is, or rather, \emph{appears}, yellow-orange, whilst the \textit{Camellia sinensis'} flower becomes visible, to please our eyes, visible as a white surface (but who knows what colors the bumblebee sees\,\dots).\footnote{
 	What is the electromagnetic theory of a bumblebee (given that its perception of electromagnetic waves is different than ours)? Without doubt, we cannot envisage, or foresee, the “Maxwellian” interaction of a bumblebee. A Socratic Maxwell \cite[p. 245]{Maxwell "A Treatise on the Kinetic Theory of Gases"} suggests to us that: «[The] state of thoroughly conscious ignorance [\,\dots] is the prelude to every real advance in knowledge».
 	} 
 What we call a \emph{color} is a small portion, for a \emph{human observer}, of the electromagnetic spectrum. If the color sensation is not independent from our eyes,\footnote{
 	 See e.g. I. Newton \cite{Newton "Opticks 1704"}: Book I, Part II, p. 90: «For the rays to speak properly are not coloured. In them there is nothing else then a certain power and disposition to stir up a sensation of this or that Colour»; Book III, Qu[estion] 12, p. 134: «Do not the rays of Light in falling up on the bottom of the Eye excite vibrations in the \textit{Tunica retina}? Which vibrations, being propagated along the solid fibres of the optick Nerves into the Brain, cause the sense of seeing»; Book III, Qu[estion] 13, p. 135: «Do not several sort of rays make vibrations of several bignesses, which according to their bignesses excite sensations of several Colours, much after the manner that the vibrations of the Air, according to their several bignesses excite sensations of several sounds?».
 	}\textsuperscript{,}\footnote{
 	About the nature of colors and the relationship between colors and light, before the Newtonian systematization, F.M. Grimaldi's survey \cite[I-LX propp.]{Grimaldi "Physico-mathesis de lumine"} not infrequently shows subtleness and penetration, albeit with some inaccuracies, situated in its historical context; see e.g. the \textit{Sexaginta Propositiones} listed in the five-page \textit{Index Propositionum Primi Libri}.
 	} 
so likewise the disposition toward mathematics is not independent from our brain.
\item When we effectively describe certain aspects of nature of reality (external world) with mathematics, this does not indicate that nature is intrinsically mathematical. This must be reversed: \emph{all details of nature that we can mathematically describe are the only ones that we can understand}, by virtue of the \emph{fabrication} of mathematical tools, which are diversified, as the need arises, depending on the type of reality facing us, similar to how a screwdriver is different from a hammer. Substantially, we do not discover how nature is made; but, with mathematics, we figure out, and build, a way to interpret \emph{our} experience with it.

Nature of reality, the external world, for its part, has its own regularity (e.g. the constant number of spatial dimensions, at least in the macroscopic regime), along with random arrangements (see Chapters \ref{chapter "On the Chaos, Part I. Micro- and Macro-scales"} and \ref{chapter "Randomness and Stochastic Systems"}); and the regularity part is sufficient to grant a certain \emph{intelligibility}. This regularity is captured by mathematics, because it is the most \emph{convenient} language that we have developed (cf. Section \ref{subsection "Contextus I. Elements of Brachylogy—the Reverie of a Perfect Language, with a Margo on Music and Mathematics"}), starting from a biological substratum, and later we have enriched with layers of inventiveness.
\item A notation of symbols and indices bears no relation to nature's events more than a series of letters and characters, such as \emph{flower}, \textgreek{ἄνθος}, \textcyrillic{\textit{цветок}}, or \ZhTraditional{花}, has to do with an actual flower. Who would be doltish enough to confuse a word with the object designated by it?\footnote{
	The language, whether it is at the common stage or at the formal (symbolic) stage, is forged, is molded, by its use; because of this it changes, it is in unceasing flux. So the meaning of a word, or of a symbol, depends on \emph{how} and by \emph{whom} it is used.
	}
\item 
\label{item "A physics without symbolic apparatus etc."}
A physics without mathematics—with no mathematical creativity \& paraphernalia of math-symbols, I mean—is equivalent to nature (\textgreek{φυσική}, in the literal sense, “caused by nature”), which is a condition, for humans, devoid of representation: a nature that can be \emph{symbolically mute}, or rather, \emph{mathematically indescribable}, is, for us, \emph{physically incomprehensible} (cf. Section \ref{subsection "Contextus II. Autobiographical Note"}).

Mathematical symbols (the aforementioned Galilean «characters» in Section \ref{section "Naturæ Mæandri and Filum Ariadneum: a Botanical Comparison"}), having conventional roots and a formation in the mind of man, are part of \emph{culture} and not of nature—it makes no difference if their origin is a mélange between a \emph{cultural-like artifact} and a \emph{bio-makeup of our brain} (a biological complement): they are representative functions of reality (observable facts, external events, etc.), and share accordingly their own restrictions of \emph{infidelity} and \emph{distortion} in the various defining and descriptive stages, that pave the bridge between the combination of symbols (mathematical significance) and the content of reality (natural phenomena).
\item While studying the principles of physics, we see that mathematics 

· is \emph{not a (crude) translation} into formulas—into a rigorous language—of empirical laws deriving from observation, or a transposition into symbols of a trove of knowledge, and of statements about facts given by experimental physics; 

· is \emph{not a logical, moreover neutral, procedure} of deduction from phenomena that «takes nothing away and adds nothing», as a Poisson \cite[p. 5]{Poisson "Theorie mathematique de la chaleur"} believes,\endnote{
	S.D. Poisson \cite[p. 5]{Poisson "Theorie mathematique de la chaleur"}: «En lui donnant le titre de \textit{Théorie mathématique de la chaleur}, j'ai voulu indiquer qu'il s'agira de déduire, par un calcul rigoureux, toutes les conséquences d'une hypothèse générale sur la communication de la chaleur, fondée sur l'expérience et l'analogie. Ces conséquences seront alors une transformation de l'hypothèse même, à laquelle le calcul n'ôle et n'ajoute rien».
	}
to the assumptions previously gained from experience/analogy;

· is \emph{not a ready-made code}, to wit, a static and external code with respect to the practice of physical theory, but takes shape—from time to time—along its application; mathematics is more similar to an activity of reformulation, a \emph{living language} of reinterpretation of facts and phenomena.
\item 
\label{item "Selection of problems that are mathematically tractable"}
There is no «miracle» or «mystery», à la Wigner, concerning the effectiveness of mathematics in physics. Mathematics and physics fit together well, as they say, because mathematicians, and mathematical physicists, \emph{choose problems} that, under an evolutionary predisposition, they are capable of mathematically conceiving, and \emph{rule out} the rest.\footnote{
	Cf. G. Vallortigara, N. Panciera \cite[pp. 135-137]{Vallortigara Panciera "Cervelli che contano"}.
	}
Thence it is \emph{theatrically rhetorical} to ask questions like this, from R. Hersh \cite[p. 66]{Hersh "Inner Vision Outer Truth"}: «Is there some arcane psychological principle by which the most orig­inal and creative mathematicians find interesting or attractive just those directions in which Nature herself wants to go?». Nature for sure does not go where mathematicians go. It happens, much more simply, that mathematicians typically and mostly study—and apply to physics—those problems of nature that they can understand \emph{via mathematica}, so they labour under the \emph{illusion} that mathematics is a miracle. 

The affirmation \cite[p. 71]{Hersh "Inner Vision Outer Truth"} that a mathematical structure truly «captures» a «fundamental feature of nature» is equivalent to saying that a mathematical structure truly captures fundamental feature of the mathematical structure itself. But does that make sense?
\enumerationisfinis

\subsubsection{Seeing and Looking for: Elastic Lenses}
\label{subsubsection "Seeing and Looking for: Elastic Lenses"}

\begingroup
\footnotesize
We see what we look for [and we look for what we see]. No one is surprised if after putting on blue tinted glasses the world appears bluish [\,\dots]. [\emph{If}] \emph{the original phenomenon arises from the mathematical tools we use and not from the real world}, I am ready to strongly suggest that \emph{a lot of what we see comes from the glasses we put on}. \\
\indent — \textsc{R.W. Hamming} \cite[pp. 87-88, e.a.]{Hamming "The Unreasonable Effectiveness of Mathematics"} 

\endgroup

\vspace{2mm}

The comparison with the glasses, evoked by Hamming, is unhappy because it may suggest the misconception that our way of mathematically seeing the world is somehow static or rigid, as a fumesopher (Section \ref{section "Interludio Giocoso. Against the Fumesophers, or the Tragicomic Smoke-sellers"}) thought,\endnote{	
	I refer to Mr. K. (it is clear, is not it?). The a priori notion of space, in K.'s sense, went in crisis with the introduction of all non-Euclidean geometries (cf. Gauss in footnote \ref{footnote "Gauss' letter to Bessel 9 April 1830"} on p. \pageref{footnote "Gauss' letter to Bessel 9 April 1830"}). A demolishment of the a priori notion of time, in K.'s sense, is in G. Cantor \cite[p. 403]{Cantor "Fondements d'une theorie generale des ensembles}: «[I]n my opinion, the introduction of the notion of time or the idea of time should not serve to explain the much more primitive and more general notion of the continuum; time, in my opinion, is an idea that presupposes, in order to be clearly explained, the notion of the continuity, independent of that of time, and that, even with such a notion of continuity, [time] can neither be conceived objectively as a substance, nor subjectively as a necessary [and] a priori idea; this idea of time is only an auxiliary and relative idea, as serving to establish the relationship between the various movements occurring in nature and that we perceive. Thus there is nothing in nature resembling an objective or absolute time, and consequently one cannot take [the notion of] time as a measure of movement, but on the contrary one could consider [the notion of] movement as a measurement of time [\textit{le mouvement comme mesure du temps}]».
	} 
mistakenly believing that mathematics was a synthetic judgment a priori (see Section \ref{subsection "Contextus III. Helmholtz's Space-intuition, Poincaré's Inner 3-Dimensionality, and Grid Cells"}); mathematics is free or, if you prefer, \emph{elastic} as the mind, and \emph{co-evolves} with the “pliability” of the imagination. But such a comparison has the merit of being clear, and it has the goodness to illustrate immediately that physics passes through the \emph{introspective lenses} of (our) mathematics, anyhow. Luckily, Hamming (ivi) gives a series of examples that clarifie what he means by “glasses”:

\vspace{2mm}

\begingroup
\footnotesize
Not too long ago I was trying to put myself in Galileo's shoes, as it were, so that I might feel how he came to discover the law of falling bodies [\,\dots]. Since falling bodies do something, the only possible thing is that they all fall at the same speed—unless interfered with by other forces. There is nothing else they can do [\,\dots]. Galileo found his law not by experimenting but by simple, plain thinking, by scholastic reasoning.

I know that the textbooks often present the falling body law [and the inverse-square law] as an experimental observation; I am claiming that it is a logical law, a consequence of how we tend to think [\,\dots]. [I]f you believe in anything like the conservation of energy and think that we live in a three-dimensional Euclidean space, then how else could a symmetric central-force field fall off? Measurements of the exponent by doing experiments are to a great extent attempts to find out if we live in a Euclidean space, and not a test of the inverse square law at all [cf. Section \ref{subsubsection "Inverse-square Laws"}].

\endgroup

\vspace{2mm}

Another example follows, relating to the uncertainty principle:

\vspace{2mm}

\begingroup
\footnotesize
When you use the eigenfunctions [of the translation operator] you are naturally led to representing various functions, first as a countable number and then as a non-countable number of them—namely, the Fourier series and the Fourier integral. Well, it is a theorem in the theory of Fourier integrals that the variability of the function multiplied by the variability of its transform exceeds a fixed constant, in one notation $\frac{1}{2\pi}$. This says to me that in any linear, time invariant system you \emph{must} find an uncertainty principle. The size of Planck's constant is a matter of the detailed identification of the variables with integrals, but the inequality must occur.

\endgroup

\vspace{2mm}

Then the discussion drops on physical constants:

\vspace{2mm}

\begingroup
\footnotesize
As another example of \emph{what has often been thought to be a physical discovery but which turns out to have been put in there by ourselves}, I turn to the well-known fact that the distribution of physical constants is not uniform; rather the probability of a random physical constant having a leading digit of 1, 2, or 3 is approximately 60\%, and of course the leading digits of 5, 6, 7, 8, and 9 occur in total only about 40\% of the time. This distribution applies to many types of numbers, including the distribution of the coefficients of a power series having only one singularity on the circle of convergence. A close examination of this phenomenon shows that it is mainly an \emph{artifact} of the way we use numbers.

\endgroup

\vspace{2mm}

At a glance \cite[pp. 88-89]{Hamming "The Unreasonable Effectiveness of Mathematics"}, we can say that «we approach [mathematically] the situations with an intellectual apparatus so that we can only find what we do in many cases». 

Another explanation for why mathematics is effective is that, as we have already said above (point \ref{item "Selection of problems that are mathematically tractable"}, p. \pageref{item "Selection of problems that are mathematically tractable"}), «we select the kind of mathematics to use».\footnote{
	Which gives its historical status away. It is a platitude, but it is something that needs to be said: the postulates are chosen by us; and if these are not consistent with other parts of the mathematical architecture, it is appropriate to modify them \cite[pp. 86-87]{Hamming "The Unreasonable Effectiveness of Mathematics"}: «Mathematics has been made by man and therefore is apt to be altered rather continuously by him [\,\dots]. The idea that theorems follow from the postulates does not correspond to [the facts]. If the Pythagorean theorem were found to not follow from the postulates, we would again search for a way to alter the postulates until it was true».
	}
In conclusion, and inescapably, the effect of biological selection is underscored by Hamming:

\vspace{2mm}

\begingroup
\footnotesize
Just as there are odors that dogs can smell and we cannot, as well as sounds that dogs can hear and we cannot, so too there are wave lengths of light we cannot see and flavors we cannot taste. Why then, given our brains wired the way they are, does the remark, “Perhaps there are thoughts we cannot think”, surprise you? Evolution, so far, may possibly have blocked us from being able to think in some directions; there could be unthinkable thought.

\endgroup

\section{Three Scholia (pro Exhibendis Contextis Physicis et Mathematicis)}

Some issues dealt with in the previous two Sections (\ref{subsubsection "Selection, Colors, and Understanding"} and \ref{subsubsection "Seeing and Looking for: Elastic Lenses"}) may be complemented by further remarks. We will present three supporting Scholia below.

\subsection{Contextus I. Elements of Brachylogy—the Reverie of a Perfect Language, with a Margo on Music and Mathematics}
\label{subsection "Contextus I. Elements of Brachylogy—the Reverie of a Perfect Language, with a Margo on Music and Mathematics"}

One of the peculiarities of mathematical language is the brachylogy. P.-S. de Laplace, in his lectures (1795) \textit{Sur la numération et les opérations de l'arithmétique} \cite[p. 15]{Laplace "Premiere seance. Programme. Sur la numeration et les operations de l'arithmetique"}, tells us that  

\vspace{2mm}

\begingroup
\footnotesize
[T]he [\,\dots] most perfect language would be the one in which one could express the greatest number of ideas by the smallest number of words possible [\textit{le plus-grand nombre d'idées par le plus petit nombre de mots possible}]. Arithmetic is a particular language [which is somehow able to have this peculiarity].

\endgroup

\vspace{2mm}

L. Sinisgalli, poet, art critic but, as a mathematician and electronic engineer by training, in \cite[\textit{Calcolatrici}]{Sinisgalli "Furor Mathematicus"} annotates that:

\vspace{2mm}

\begingroup
\footnotesize
In each mathematical sign there is an indication of a movement, but of a movement shortened [\textit{movimento abbreviato}] to such an extent that it already contains in itself, so to speak, the result. The effort of mathematicians perhaps consisted in this: [the effort] to have built the most formidable system of abbreviations. Mathematicians have enclosed a concept and an operation in a sign.

\endgroup

\vspace{2mm}

E. Pound \cite[p. 28]{Pound "ABC of Reading"}, poet and critic, says that

\vspace{2mm}

\begingroup
\footnotesize
Great literature is simply language charged with meaning to the utmost possible degree.

\endgroup

\vspace{2mm}

Something of the kind can be repeated for the grammar and musical syntax (with its ingredients: melody, harmony, rhythm, and counterpoint). Herein one rediscovers the far-famed assertion by G.W. Leibniz \cite[IV, p. 437]{Leibniz "Epistola II ad C. Goldbachium"}: 

\vspace{2mm}

\begingroup
\footnotesize
Musica est exercitium arithmeticæ occultum nescientis se numerare animi.
 
\endgroup

\vspace{2mm}

The pith of the sentence is that \textit{exercitium occultum}: «Music is a secret exercise [in] arithmetic of the soul [mind], unaware that it is counting».

Of that there is no doubt: mathematics, poetry, and music, are, evenly, but each with its own modality, the three brachylogical languages having the greatest concentration of meaning, or of ideas, simultaneously with the smallest number of terms possible. It is the power of \textgreek{βραχυλογία}.
	
Before closing this Section, we add additional annotations about music.

\begin{margo}[Music and mathematics]
A piece of music is, say, a sound illustration of fractional expressions, as parts of arithmetic; more appropriately, musical fractions are “element” of the time in which a piece of music is played. Such an ancestral union—between music and mathematics—appears explicitly with the Pythagorean tuning; take the following frequency ratios for a major scale on D/Re, for instance: 
\begin{equation*}
	\text{non-tempered scale}
	\begin{cases}
	\text{G/Sol $\sfrac{4}{3}$, perfect 4th (P4), or diatessaron: $\frac{2}{3} \cdot 2$}, \\
	\text{D/Re $\sfrac{1}{1}$, unison, or prime (P1), with 0 size}, \\
	\text{A/La $\sfrac{3}{2}$, perfect 5th (P5), or diapente: $\frac{3}{2}$}, \\ 
	\text{E/Mi $\sfrac{9}{8}$, major 2nd (M2), or whole tone/step: $\left(\frac{3}{2}\right)^2 \cdot \frac{1}{2}$}, \\
	\text{B/Si $\sfrac{27}{16}$, major 6th (M6): $\left(\frac{3}{2}\right)^3 \cdot \frac{1}{2}$,} \\
	\text{F$\sharp$/Fa$\sharp$ $\sfrac{81}{64}$, major 3rd (M3): $\left(\frac{3}{2}\right)^4 \left(\frac{1}{2}\right)^2$}, \\
	\text{C$\sharp$/Do$\sharp$ $\sfrac{243}{128}$, major 7th (M7): $\left(\frac{3}{2}\right)^5 \left(\frac{1}{2}\right)^2$}, \\
	\text{G$\sharp$/Sol$\sharp$ $\sfrac{729}{512}$, augmented 4th (A4): $\left(\frac{3}{2}\right)^6 \left(\frac{1}{2}\right)^3$}.
	\end{cases}
\end{equation*}	 
	
The intervals of the minor scale, which we do not report here, also have their frequency ratios, of course—e.g. on the opposite side from A4, there is the 
\[
\text{diminished 5th (d5) A$\flat$/La$\flat$ $\sfrac{1024}{729}$, whose formula is $\left(\frac{2}{3}\right)^6 \cdot 2^4$}. 
\]

It is known that, in the Pythagorean tuning, the circle of fifths, after 12 tones/steps, does not close exactly with the starting note: it is an increasing and endless spiral; and it is a non-tempered scale, for that matter. 

In the reform of the construction of the musical scale, it should be remembered, for its importance, the Zarlino scale \cite{Zarlino "Le istitutioni harmoniche"}, to which the system of 5-limit tuning is connected, proposing an enlargement of the Pythagorean \textgreek{τετρακτύς}, where one finds a division of music in «Theorica, o Speculativa» \& in «Prattica». 

The fundamental revision is with the so-called \emph{temperament} of the musical scale, aimed at making adjustments, in “circumventing” the Pythagorean problem, through a systematic tuning instruments, by dividing the octave into 12 equal semitones, or half-steps, comprehensively elaborated by Zhu Zaiyu (\ZhTraditional{朱載堉})\footnote{
	See his monumental work: \ZhTraditional{朱載堉} (1536-1611), \ZhSimplified{乐律全书} [\textit{The Complete Book of Music}], \ZhSimplified{商务印书馆}, Shanghai, 1931.
	} 
and S. Stevin;\footnote{
	Stevin's manuscript is entitled \textit{Vande Spiegheling der Singconst} (circa 1605-1608), see \cite{Cohen "Simon Stevin's Equal Division of the Octave"}.
	}
this is the case of the equal-tempered scales, or 12-tone equal temperament,\footnote{ 
	Compare with the \textit{wohltemperierte Stimmung} (well temperament scheme), from J.S. Bach. 
	}
whose ratio is 
\[
	\sqrt[12]{2} = 1.0594630943592952\cdots
\]
	
One way or another, all intervals of intonations and changes in tonality are correlated with a skeletally mathematical, or quasi-mathematical, modulations; idem for dodecaphonic compositions, when every degree of the chromatic scale is used, see e.g. A. Schönberg \cite[XI, pp. 202-206]{Schoenberg "Theory of Harmony"}. I. Xenakis \cite[p. 170]{Xenakis "Formalized Music: Thought and Mathematics in Composition"}, a distinctly formal composer, combines algebra and \emph{granular synthesis} (a “cloud” of atomic sounds colliding with each other); getting to the nub of the matter, this is what he says:

\vspace{2mm}

\begingroup
\footnotesize
We have noted [\,\dots] three kinds of algebras:

1. The algebra of the components of a sonic event, with its vector language, independent of the procession of time, therefore an algebra \emph{outside-time}.

2. A \emph{temporal algebra}, which the sonic events create on the axis of metric time, and which is independent of the vector space.

3. An \emph{algebra in-time}, issuing from the correspondences and functional relations between the elements of the set of vectors $X$ and of the set of metric time, $T$, independent of the set of $X$.

\endgroup

\vspace{2mm}

J. Cross in \cite[\textit{Composing with numbers: sets, rows and magic squares}, p. 146]{Fauvel Flood and Wilson "Music and Mathematics: From Pythagoras to Fractals"} paints a portrait of the Xenakisian music: 

\vspace{2mm}

\begingroup
\footnotesize
Xenakis [\,\dots] [used] many mathematical models as well as computers to assist him in his pre-compositional calculations. He soon became interested in probability theory as a way of handling mass sound phenomena, and from this grew what he described as ‘large number’ or ‘stochastic’ music, where the operation of individual elements is unpredictable but the shape of the whole can be determined. For example, \textit{Pithoprakta}, the next work after \textit{Metastaseis}, drew (the composer claimed) on Maxwell–Boltzmann's kinetic theory of gases; \textit{Achorripsis} employed Poisson's law; and \textit{Duel} and \textit{Stratégie} used game theory—each work employed two conductors who ‘compete’ with one another. More recently, Xenakis developed what he called ‘symbolic music’ which drew on principles of symbolic logic. P. Griffiths has observed that “Xenakis's symbolic music has\,\dots the nature of a translation into sound of theorems of set theory”, first evident in \textit{Herma} for piano of 1960-1.
	
This may suggest that Xenakis's music is completely abstract and sterile. Not at all. His music, like the man, is all too human and he always asserted the primacy of music over mathematics—music, he believed, is never reducible to mathematics, even though they have many elements in common [\,\dots]. As another commentator has put it, “he gives us something only an artist can give—a dynamic picture of the universe informed by the science of today”.

Although Xenakis's use of a variety of mathematical models may have been undertaken in a more consistent and thoroughgoing manner than almost any other composer, it does not make his music any less exciting, challenging, creative [\,\dots]. Mathematics is a means to an end, not the end in itself. Composers today are as aware as have been thinkers of the past that music is inherently mathematical, but this does not mean to say that it \emph{is} mathematics. Composing with numbers is not an admission of compositional failure, a substitute for ‘inspiration’ or ‘musicality’, whatever those concepts may mean. Composers have composed with numbers as one way of generating new musical ideas, as a means of stimulating their creativity [\,\dots]. In Xenakis's words, this represents: “the effort to make ‘art’ while ‘geometrizing’”.
	
\endgroup

\vspace{2mm}

This comment is helpful in stating that music and mathematics, even in an \emph{osmotic} compound, or close to each other, are \emph{separate}. The Russian composer I.F. Stravinsky \cite[1. \textit{About Composing and Compositions}]{Craft and Stravinsky "Conversations with Igor Stravinsky"} clarifies, in his appraisal, that

\vspace{2mm}

\begingroup
\footnotesize
[Musical form] is far closer to mathematics than to literature—not perhaps to mathematics itself, but certainly to something like mathematical thinking and mathematical relationships [\,\dots]. I am not saying that composers think in equations or charts of numbers, nor are those things more able to symbolize music. But the way composers think—the way I think—is, it seems to me, not very different from mathematical thinking [\,\dots]. Musical form is mathematical because it is ideal, and form is always ideal [\,\dots]. But though it may be mathematical, the composer must not seek mathematical formulæ.

\endgroup

\vspace{2mm}

P. Schaeffer \cite[7.3]{Schaeffer "Traite des objets musicaux: essai interdisciplines"} = \cite[p. 97]{Schaeffer "Treatise on Musical Objects: An Essay Across Disciplines"}, avant-gardist composer (in electronic and experimental music) and musicologist, father of \textit{musique concrète}, does not fail to reiterate that

\vspace{2mm}

\begingroup
\footnotesize
[W]e should note that the analysis of music into abstract structures—that is, into terms that are meaningful to the intellect and not to perception—has tempted many a mind. Contemporary experiments show that it is possible to go a long way down that road, to the point of looking to mathematical functions or chance theories for the organizational rules of musical language. These attempts are scientific only \emph{a posteriori}: insofar as they are “experiments just to hear”. It is clear, however, that they are not of the first importance for us, since we wish to hear \emph{before} understanding and \emph{in order} to understand [\textit{nous voulons entendre \emph{avant} et \emph{afin} de comprendre}]. We take it as read that, even if the \emph{Art of the Fugue} can be completely reduced to a numbers game, the meaning of this game is in its manifestation in sound, because ultimately it is entirely based on criteria of \emph{musical perception}, which arithmetic may represent [\textit{traduit}] but certainly does not determine.

\endgroup

\vspace{2mm}

Then, he reports \cite[10.3]{Schaeffer "Traite des objets musicaux: essai interdisciplines"} = \cite[p. 134]{Schaeffer "Treatise on Musical Objects: An Essay Across Disciplines"} a passage from Helmholtz \cite[p. 58, e.a.]{Helmholtz "Die Lehre von den Tonempfindungen als physiologische Grundlage fur die Theorie der Musik"}:

\vspace{2mm}

\begingroup
\footnotesize
Fourier's theorem, shown here, demonstrates first that it is mathematically possible to consider a musical sound as a sum of simple sounds, in the sense we have given to these words, and, indeed, mathematicians have found it convenient to base their acoustical research on this way of analyzing vibrations. But it certainly does not follow from this that we are obliged to look at things in this way. Rather we should be asking: do these partial components of musical sound, demonstrated by mathematical theory and perceived by the ear, really exist in the mass of air outside the ear? Surely this method of analyzing vibrations, stipulated and made possible by Fourier's theorem, is simply a \emph{mathematical fiction} [\textit{bloss eine mathematische Fiction}], good for helping with calculations, but not necessarily having any real meaning [\textit{reellen Sinn}] for things themselves?

\endgroup

\vspace{2mm}

And here we return to Leibniz's quote («exercitium arithmeticæ occultum») given above. Neural circuits of the cerebral cortex process all molecular vibrations in air, turning a sequence of sound punctualities, which could be cacophonous, into an amalgam of \emph{non-informative sounds full of aesthetic content}, nay of meaning. In constructing coherent states, or “structured” sequences of sounds, music is blatantly a \emph{product of the brain}. Cf. footnote \ref{footnote "Listening to (a piece of) music..."} on p. \pageref{footnote "Listening to (a piece of) music..."}. 

In conclusion, mathematics, towards music, is a complex of organizational formulæ (based on a substratum of unawareness), which, as such, do not explain what music is, physiologically and emotionally \cite[36.13]{Schaeffer "Traite des objets musicaux: essai interdisciplines"} = \cite[p. 529]{Schaeffer "Treatise on Musical Objects: An Essay Across Disciplines"}: 

\vspace{2mm}

\begingroup
\footnotesize
[W]hy should Pythagoras's thinking not still be relevant today? The entire harmonic scale, which replicates the series of whole numbers, still presents the same enigma. What unacknowledged motives would turn away a whole physics- and mathematics-obsessed age from this fundamental thinking? The musical object, the most disembodied [\textit{désincarné}], the most abstract of all objects it is given to us to perceive, has, in fact, the virtue of being both the most mathematical and the most sensory [\textit{le plus mathématique et le plus sensible}] [\,\dots]. So the mystery of music and its dualism cannot, fortunately, be resolved [\,\dots]. Sound objects and musical structures, when they are authentic, have no informative mission [\textit{n'ont plus de mission de renseignement}]: they turn away from the descriptive world [\textit{s'écartent du monde descriptif}] with a sort of reticence in order to speak all the better about it to the senses, the heart and mind, to the whole being, ultimately about himself.

\endgroup

\vspace{2mm}

The \emph{fascination of the ambiguity} of musical structures comes from here. Now, we may effortlessly reminisce about the standpoint of an esteemed conductor and composer, L. Bernstein \cite[pp. 43, 39, e.a.]{Bernstein "The Unanswered Question: Six Talks at Harvard"}:

\vspace{2mm}

\begingroup
\footnotesize
[\,\dots] Ambiguity [in music is] the combination of [\,\dots] two contradictory forces, chromaticism and diatonicism, operating at the same time, that makes [a] passage so expressive [\,\dots]. But \emph{ambiguity} has always inhabited musical art (indeed, all the arts), because it is one of art's most potent \emph{aesthetic functions}. The more ambiguous, the more expressive, up to a certain point. \margosymbol

\endgroup
\end{margo}

\subsection{Contextus II. Autobiographical Note}
\label{subsection "Contextus II. Autobiographical Note"} 

\enumerationisinitium
\item 
\label{item "When I was twenty"}
When I was twenty—the age at which everything is taken to the extreme, the age of ardor, pulled along by fatuous exaggerations—I delighted in writing classical metrics and verses in prose. One of these compositions went like this: \textit{Per quasi flutti incanutiti candido spumeggia, \& ne' clivi pruinosi presto rameggia, il sol poeta che rapito all'affannoso scriver non s'attenta, ma lascia, cheto, gl'impeti ondosi reliqui a levarsi, e sonori a gonfiarsi, \& gli ontàni a stormir, verberati dall'aere ventoso} (the adjective \emph{reliqui} comes from the La. \textit{rĕlinquo}, “leave [behind]”, “let go”).
	
Something vaguely similar, if you will, is in G. Papini \cite[pp. 181-182]{Papini "Un uomo finito"} = \cite[pp. 200-202, partially modified translation]{Papini "The Failure"}—although he has quite another ambition in mind—when he asserts that 

\vspace{2mm}

\begingroup
\footnotesize
[\,\dots] it is not enough to write [the] names [of things] in books; it is not enough to classify them and to find their genealogy [\textit{non basta avere i nomi \textnormal{⟨}delle cose\textnormal{⟩} scritti nei libri; non basta averle classificate e genealogizzate}]; it is not enough to reduce them to general ideas and these generals to universal concepts, establishing the causative relations between the various groups of concepts [\textit{non basta averle ridotte a idee generali e le idee generali in concetti universali e aver formulati i rapporti di causa tra i diversi gruppi dei concetti}]. It is not enough to exhibit them in show-cases [\textit{vetrine}], each show-case labeled with the (inviolable?) law illustrated. To change reality, it is not enough to know its exteriors and through the categories of the reasoning intellect and the symbols of language [\textit{non basta conoscere \textnormal{⟨}la realtà, la natura delle cose\textnormal{⟩} dal di fuori e attraverso le forme dell'intelletto ragionante e i simboli del dizionario}].

\endgroup

\vspace{2mm}

So, let us sum it up thus: nature cannot be fully represented, because language, of any kind, is a faculty bringing into being differentiations, so that it prevents us from grasping, say, the unity of nature: a theme agreeingly loved by physicists,\footnote{
	Cf. e.g. D. Bohm \cite[chap. 1, e.m.]{Bohm "On Creativity"}, who mixes  together scientific creativity and discovery, in search of the archetypal “unity” in nature, the cosmic \textgreek{ἀρχέτυπον} (model, imago) of the “oneness”, “totality”, or “wholeness”, comparing the activity of a physicist to that of an artist, an architect, and a musician: «[What the scientist seeks is] a hitherto unknown lawfulness in the order of nature, which exhibits \emph{unity} in a broad range of phenomena. Thus, he wishes to find in the reality in which he lives a certain \emph{oneness} and \emph{totality}, or \emph{wholeness}, constituting a kind of \emph{harmony} that is felt to be beautiful. In this respect, the scientist is perhaps not basically different from the \emph{artist}, the \emph{architect}, the \emph{musical composer}, etc., who all want to \emph{create} this sort of thing in their work.

	To be sure, the scientist emphasizes the aspect of \emph{discovering} oneness and totality in nature. For this reason, the fact that his work can also be creative is often overlooked. But in order to discover oneness and totality, the scientist \emph{has to create} the new overall \emph{structures of ideas} which are needed to express the harmony and beauty that can be found in nature».
	} 
notably by the pursuers of the \textsc{gut}s (\emph{grand unified theories}) and of the elusory \emph{theory of everything}. 

\vspace{2mm}

\begingroup
\footnotesize
We must get inside [of the nature of reality]—Papini continues—[and] insert ourselves into it, become parts of it [\textit{Occorre entrarci dentro, inserirsi in essa, diventar parte di lei}], each of us an atom of its mass, a moment of its existence, a spark of its flame, a drop of its current [\textit{atomo della sua massa, momento della sua durata, scintilla della sua fiamma, gocciola della sua corrente}]. We must come in contact with all its aspects (even the most recondite, the most transitory, the least perceptible); blend ourselves with its fullness; abandon ourselves to its flow; lose ourselves in its immensity; become living realities in a living reality [\textit{Occorre entrare in contatto con tutti i suoi aspetti (anche i più nascosti, i più transitori, i meno visibili); fondersi nella sua pienezza; abbandonarsi al suo corso; perdersi nella sua immensità; farsi realtà viva nella viva realtà}]. We should not just stand in its presence like a thinking mechanism, a microscope reticle, a nomenclator and a [ratio] meter [\textit{Non già restare al suo cospetto come un meccanismo cerebrale, come una lente reticolata, come un nomenclatore e un misuratore}], rather we should dive into it headlong, penetrate into it and be penetrated by it; feel within our own selves the eternal multicolor, multisound, and multisavor of its flux, putting its pulse in rhythm with the pulsation of our blood, with our own heartbeat. Ensuring that it becomes wholly of us and that we all become [part] of it.\footnote{
	In the same year in which Papini published this “intellectual autobiography” (1913), the lyrical fragments of C. Rebora came out, in “Libreria della Voce” (the same Florentine book publisher of Papini). Similar Papinian feelings can also be discovered in Rebora's lyrics; which is readily explained: these are crosswise themes. See e.g. \cite[\textsc{xlvii}, pp. 89-90]{Rebora "Frammenti lirici"}: «Se come foglia in turbin si mulina / Volger potessi nella mia fatica, / Se come per ruscello all'acqua è moto / Fosse al mio ingegno eguale la sua china [\,\dots], Voce a un coro, stelo a un fiore / Trave a un palco, ghiaia al fango: / Esser qualcosa di adatto»; and \cite[\textsc{xxv}, p. 55]{Rebora "Frammenti lirici"}: «O realtà, essere in te vorrei: / Ma in un concreto e alterno / Svarïar perdo il senso / Del tuo vortice eterno. / Da te nascendo vano sfumo via».
	} 

Nobody aspires and tends to this mystical oneness [\textit{Nessuno aspira e tende a questa mistica immedesimazione}]. Not even the artists: they too, though they give expression to the particular, select, choose, eliminate, impoverish [\textit{scelgono, scartano, impoveriscono}]. There are sides, phases, flashes, of things which no one sees, which no one is trying to see [\textit{Vi son attimi e lati delle cose che nessuno vede, che nessuno cerca}] [\,\dots]. Philosophers could much better apply themselves to this patient excavation of the concrete particular than continue playing [\textit{gingillarsi}] with such kindergarten toys [\textit{giuochi froebeliani}] as a priori definitions and symmetrical systems [\,\dots]. 

If man, instead of detaching himself from reality, considering it merely as something to be measured and judged by him, were so to melt, so to dissolve himself, into the real as to feel its every atom and appearance kin of his kin, then his limited body would be absorbed into the immense body of the universe [\textit{Quando l'uomo, invece di separarsi dal reale, come qualcosa a sé che lo giudica e lo misura, si disfacesse nel reale in modo da sentir fratello ogni atomo e sorella ogni apparenza, allora il corpo limitato dell'uomo sparirebbe nel corpo smisurato dell'universo}]; his microcosm would become the very macrocosm [\textit{il microcosmo sarebbe effettualmente il macrocosmo}], and every part of the world would be as a part of him.

\endgroup

\vspace{2mm}

I am afraid that the same—impassable—limitations, which in literature provoke a poignant afflatus pushing towards a \emph{pancosmic mirage}, can be cogently laid bare, \textit{mutatis mutandis}, in physics: mathematical language is the key for a coherent description of reality; but, concurrently, mathematical language, inexorably, is the extent of our imperfect representation of the (ultimate) nature of reality, because \emph{a perfect representation is equivalent to the very nature of reality, that is, without mathematics}.
	
\textit{Hæc fabula docet} that \emph{literature, physics \& mathematics} are on equal footing: they \emph{are all flawed or inadequate representations}, albeit to varying degrees, of the things “out there”.
\item In the opposite direction, at least partly, goes the F. de Grisogono's crazy project (a sort of Faustian \textit{Anstoß}), a solitary and weirdo scientist, artfully described by his grandson, C. Magris \cite[\textit{Lagoons}, pp. 79-81]{Magris "Microcosms"}. This is the project of harnessing the power of chance in a gargantuan grid of pluri-knowledge, i.e. in a classification \& cataloguing system (just mention the Linnaean \textit{Cognitioni Naturalium Methodicæ \& Nomenclaturæ Systematicæ}, which serve as an Ariadne's thread, in order not to get lost in the \textit{Naturæ mæandri}, see Section \ref{section "Naturæ Mæandri and Filum Ariadneum: a Botanical Comparison"}), of putting a bridle on the random fury of events in the natural world, which are multiple and allegedly blind:

\vspace{2mm}

\begingroup
\footnotesize
Francesco de Grisogono [\,\dots] passing on to his grandson the nostalgia and the \emph{hybris} associated with the business of enclosing the world in a cage of signs and words [\textit{la \emph{hybris} di racchiudere il mondo in una gabbia di segni e di parole}].

[\,\dots] Above all else he worked at the fundamental dream of his life, the “conceptual calculus”, an \textit{ars combinatoria} based on rigorous mathematical foundations and capable of producing all the operations, the discoveries and the intuition of genius.

Francesco de Grisogono sought to free human creativity from the whims of chance and from the injustice of fate [\textit{liberare la creatività umana dai capricci del caso e dall'ingiustizia della sorte}] which, as he well knew, conditioned it and clipped its wings. To this task he brought a titanic impetus compounded with a genuine scientific rigour, a prophetic intuition, outdated impedimenta and a naïvety that was unavoidable in an isolated provincial. And if genius is inevitably subject to hazard, then conceptual calculus, with its machinery providing every possible operation, and imposing on them its inflexible logic, floats free of the randomness in which men, even geniuses, are ensnared.

The most interesting aspect of this Promethean design is the arrangement of the tables [\textit{stesura delle tabelle}] that the writer composes in his \textit{Seeds of a New Science},\footnote{
	F. de Grisogono, \textit{Germi di scienze nuove}, Voll. I-II, Guanda, Modena, 1944; Edizioni \textsc{lint}, Trieste, 1978\textsuperscript{2ed}.
	} 
to catalogue the infinite variety of the world [\textit{schedare l'infinita varietà del mondo}], in such a way as to organize the material for those combinations that will extract from reality all possible inventions and discoveries [\textit{in maniera da sistemare il materiale di quelle combinazioni che dovranno estrarre dalla realtà tutte le invenzioni e le scoperte possibili}]. It classifies types and subtypes of elements (unentwineable: bacillary, arched, contorted, “circuent”), the 36 determinations of a “ponderal” or the 21 determinations of an event, the locutions and the translocational operations, the “electriferous” instruments and the “sonifers”, the 17 parts of the “alteragifers”, the 143 modalities of an action, the 28 physiological phenomena and the same number of psychic phenomena, [the divers substances]—the friable, foliaceous, mucilaginous, foamy, mouth-puckering\,\dots. It suggests scientific research ranging from the brilliant to the hare-brained, enquiries into the influence of a vacuum on the variations in the electric resistance of selenium, through the effects of light or experiments to verify whether the given $\text{X}(2)^n$ contains properties that will arrest the decomposition of corpses.

Among those tables, those calculi and those mathematical signs, pigeon-holed and untouchable, sit the seduction and the prolixity of the world, the immensity of the celestial vaults and the chasms of the heart [\textit{Fra quelle tabelle, quei calcoli e quei segni matematici si affacciano, incasellate e inafferrabili, la seduzione e la prolissità del mondo, l'immensità degli spazi celesti e gli abissi del cuore}]. That all-encompassing \textit{hybris}, which toys with omnipotence, exposes the individual in his smallness and helplessness, lost as he is amid the infinite and even more so amid the enigmas of finite things, overwhelmed by love for life; all of this he tries to capture like a fisherman who seeks to capture the sea in his net [\textit{Quella \emph{hybris} totalizzante, che maneggia l'onnipotenza, mette a nudo l'indifesa piccolezza dell'individuo sperduto tra gli infiniti e ancor più fra le enigmatiche cose finite, il suo struggente amore per la vita, ch'egli cerca di afferrare come un pescatore che voglia catturare il mare con la sua rete}]. Only plain mathematics, with its signs as abstruse to the layman as hieroglyphics, can elicit the mysterious and terrible grace of living [\textit{Solo la nuda matematica, con i suoi segni astrusi per un profano come geroglifici, può far emergere la grazia misteriosa e terribile del vivere}]; here we have the glum, positivist, nineteenth-century honesty, with its rigour and its ingenuous faith in being able to eliminate metaphysics, which authenticates the sense of mystery—unstated and indeed doggedly banished like an error in a computation.

\endgroup

\vspace{2mm}

One has the dream of embracing life at every stage, as a whole, in its present, past, and future form. One more time. Papini and de Grisogono: different paths, which then intersect each other.
\enumerationisfinis

\subsection{Contextus III. Helmholtz's Space-intuition, Poincaré's Inner 3-Dimensionality, and Grid Cells}
\label{subsection "Contextus III. Helmholtz's Space-intuition, Poincaré's Inner 3-Dimensionality, and Grid Cells"}

Take the notion of space: it has a relationship with experience, as H. von Helmholtz already argued \cite[p. 31]{von Helmholtz "Ueber den Ursprung und die Bedeutung der geometrischen Axiome"}: space-intuition is 

\vspace{2mm}

\begingroup
\footnotesize
an intuition of the kind an artist has [\textit{eine solche Art der Anschauung}] of the objects to be portrayed; [this intuition is an] empirical knowledge gained by the accumulation and reinforcement of similar recurring impressions in our memory, and not a transcendental form given before any experience [\textit{keine transcendentale und vor aller Erfahrung gegebene Anschauungsform}]. 

\endgroup

\vspace{2mm}

In brief, our intuition of space is not a rigid cognitive faculty imposing a pre-established harmony between (a priori) form and reality. 

See also H. Poincaré \cite[II. chap. I,\endnote{
	This chap. is a reproduction, with minor changes, of the article \textit{La Relativité de l'Espace}, appeared on L'Année psychologique, Tome XIII, 1906, pp. 1-17.
	}
p. 121]{Poincare "Science et methode"}:

\vspace{2mm}

\begingroup
\footnotesize
[G]eometry [\,\dots] is a science born in connection with experience, we have created the space [\textit{nous avons créé l'espace}] that it studies, but adapting it to the world in which we live. We have chosen the most convenient [\textit{commode}] space, but it is experience that has guided our choice.

\endgroup

\vspace{2mm}

Poincaré argues \cite[pp. 117-119]{Poincare "Science et methode"} that the 3-dimensionality of space is «an internal property [\textit{propriété interne}] of human intelligence», although it is inextricably linked with an adaptive strategy; it is the «translation of a set of external facts»: «in nature there exist solid bodies». But the simple destruction of some associations of ideas, in his opinion, would be enough to acquire a different perceptual-cognitive picture, with the emergence of a space with two or four dimensions; such distorted physics would be «the same as ours», since it would be «the description of the same world in another language».
	
The experiential quality of space shall not preclude the existence of a \emph{innate} or \emph{neurobiological} space-knowledge starting from the chemico-physiological disposition; indeed, experience works precisely on a biological substrate. In this matter, see the works of E.I. \& M.-B. Moser \cite{Fyhn Molden Witter Moser Moser "Spatial Representation in the Entorhinal Cortex"} \cite{Hafting Fyhn Molden Moser and Moser "Microstructure of a spatial map in the entorhinal cortex"}, to whom we owe the discovery that the \emph{neuronal pattern} (in rats, mice, monkeys, humans, and other animals) is \emph{geometrically shaped}. In the entorhinal cortex there are special nerve cells, called \emph{grid cells}, which—along arbitrary/random trajectories—contribute to the determination and understanding of the location, distance, and direction in the surrounding space (there are also \emph{speed cells} for measuring the speed of movement); such a mapping takes place via local triangulations, which, in succession, generate pentagonal tessellations. More recently, several studies have been done on the $3\mathrm{D}$ grid cells, see \cite{Grieves Jedidi-Ayoub Mishchanchuk Liu Renaudineau Duvelle and Jeffery "Irregular distribution of grid cell firing fields in rats exploring a 3D volumetric space"} \cite{Ginosar Aljadeff Burak Sompolinsky Las and Ulanovsky "Locally ordered representation of 3D space in the entorhinal cortex"}.

\chapter{Outro—\emph{Parva Mathematica}: \emph{Libera Divagazione} \sfrac{4}{8}}
\label{chapter "Outro—Parva Mathematica: Libera Divagazione 4/8"}

\section{\emph{Interludio Giocoso}. Against the Fumesophers, or the Tragicomic Smoke-sellers}
\label{section "Interludio Giocoso. Against the Fumesophers, or the Tragicomic Smoke-sellers"}

\begingroup
\footnotesize
Et ie m'esbahis grandement d'un tas de fols Philosophes [\,\dots], \& mieux leur vaudroit s'aller frotter le cul au panicaut,\footnote{
	Panicaut is a perennial herbaceous plant of the genus \textit{Eryngium}, belonging to the Apiaceae family. There are various species; here are some of them: \textit{Eryngium maritimum}, \textit{E. campestre}, \textit{E. bourgatii}, \textit{E. spina-alba}. The leaves are predominantly \emph{tough}, the basal leaves are pinnately lobed and stiffly \emph{spiny}, with a broad stalk.
	} 
que de perdre ainsi le temps à disputer de ce dont ils ne sçavent l'origine.\footnote{
	«And I am greatly amazed at a rabble of foolish Philosophers [\,\dots], \& better were it for them to rub their ass against [the thorns of] \textit{Eryngium} than to waste away their time in disputing of that whereof they know not the origin».
	} \\
\indent — \textsc{F. Rabelais} \cite[Livre II, chap. XXXIII, p. 319]{Rabelais "Oeuvres 1558"}

\vspace{2mm}

If the hypoteposis of Geometric intuition, by prostergandering the prologomena of subconsciousness, were able to reintegrate its own subjectivism to the genesis of concominances, then the Ego would represent the self-phrasing of contemporary Mathematics, which would be nothing more than the exopolomaniacal transmification.\endnote{
	Original and not modified It. Petrolini's monologue: «Non fermarsi alla superficie, ascoltare bene quello che c'è dentro, quello che c'è sotto. È il mio motto: sempre più dentro, sempre più sotto\,\dots: “Se l'ipoteposi del sentimento personale, prostergando i prologomeni della subcoscienza, fosse capace di reintegrare il proprio subiettivismo alla genesi delle concominanze, allora io rappresenterei l'autofrasi della sintomatica contemporanea, che non sarebbe altro che la trasmificazione esopolomaniaca\,\dots” Che ve ne pare, eh?» (original audio recording; place and date of the show: unknown).
	} \\
\indent — Paraphrase of a monologue by \textsc{E. Petrolini} (1884-1936) interpreting \textsc{Gastone}

\endgroup

\vspace{2mm}

\emph{Fumesophy} is a \emph{smoky} and muddled thought activity, a hotchpotch of wooly ideas, a fluff-study (in It. one would say \textit{filosofume}, or \textit{fuffa}). A \emph{fumesopher} is someone practicing fumesophy. The majority of fumesophers, or smoke-sellers—paired with \emph{lifrelofres} of F. Rabelais \cite[Livre II, chap. II, p. 186]{Rabelais "Oeuvres 1558"}— strut around with their \emph{imbecillæ adsensiones}.\footnote{
	Cf. M.T. Cicero, \textit{Tusculanarum disputationum} \cite[liber IV, 7, 15, p. 171]{Cicero "Tusculanarum disputationum"}.
	}
And yet, borrowing a scathing banter from G. Prezzolini \cite[§ 29, p. 184]{Prezzolini "Codice della vita italiana"}, they are like politicians, or lawyers:

\vspace{2mm}

\begingroup
\footnotesize
To say nothing in many words has always been the first quality of [fumesophers]; and if they have merged [the ability] to say nothing with [the ability] to speak flowery, they have reached [the peaks of] perfection.\footnote{
	Prezzolini's phrase—in which I replace the term “politicians” with the term “fumesophers”—is this: «Il dire niente in molte parole è stata sempre la prima qualità dei [fumosofi]; che se hanno som­mato il dire niente al parlare fiorito, hanno rag­giunto la perfezione».
	}\textsuperscript{,}\footnote{
	Compare with P.W. Bridgman \cite[p. 30]{Bridgman "The Way Things Are"}: philosophy, like logic, is a verbal activity, but «logic is subject to a control that philosophy is not, the control of “truth”»; and so «it seems to me that [philosopher] sometimes tempted to treat verbal activity as a self-contained activity, worth pursuing for its own sake. It seems to me that he is inclined to hope that there must be some meaning in any grammatical combination of words, particularly when they deal with abstractions, and that he regards it as one of his problems to discover what this meaning may be». That is how the odd person (etymologically: \textit{persōna}, “theatrical mask”, “false face”) of the fumesopher springs up.
	}

\endgroup

\vspace{2mm}

In the vivid words of D. Giuliotti \& G. Papini \cite[p. 61]{Giuliotti e Papini "Dizionario dell'Omo Salvatico I (A-B)"}, they are a gang of merchant of «merda caramellata».

Shall we name a few? B. Croce,\footnote{
	A foretaste of chronic sequelae of Crocian disease is in \cite[cap. 3. (§ 1) \textit{Croce e la scienza: una eredità ingombrante}]{Israel "Chi sono i nemici della scienza?"}.
	}  
G. Gentile, M. Heidegger,\footnote{
	But, alas, there is not only this despicable fumesopher; but also his miserable epigones, with their lace knits. The burlesque continues via computer graphics: rummage through the internet, and find the drawing entitled \textit{Viola del pensiero debole} (2001) by T. Regge, sardonically dedicated to the notorious “weakened” Turinese.
	}
J. Lacan, K. Popper, T. Kuhn,\footnote{
	When Kuhn is a historian of science, what he writes may be fine: his books on the thermal electromagnetic radiation emitted by a black body and the ultraviolet catastrophe (which mark the birth of quantum mechanics)—\textit{Black-Body Theory and the Quantum Discontinuity 1894-1912} (1978), reprinted with a new Afterword, The Univ. of Chicago Press, Chicago and London, 1987—is a readable book; but when he is a theorist of science, the cogs of “garbage out” are operating. P.L. Galison \cite{Galison "Kuhn and the Quantum Controversy"} justifiably argues that the two Kuhns, one in the guise of a historian, the other in the guise of a theorist, are two sides of the same coin. But no historian makes history without walking on the tracks of some idea; the challenge is to resist the factiousness of fumesophy, without degenerating beyond the bounds of decency.
	} 
I. Lakatos, P. Feyerabend, G. Deleuze, P.-M. Foucault\footnote{
	See e.g. J.-M. Mandosio, \textit{Longévité d'une imposture: Michel Foucault, suivi de Foucaultphiles et foucaulâtres}, éditions de l'Encyclopédie des Nuisances, Paris, 2010\textsuperscript{2r.e}.
	}: a small gallery of freaks, each with his clout, his faults, and his «intellectual dishonesty», throughout the twentieth century. Against that background, see the book of A. Sokal \& J. Bricmont \cite{Sokal et Bricmont "Impostures intellectuelles"} = \cite{Sokal and Bricmont "Fashionable Nonsense: Postmodern Intellectuals' Abuse of Science"},\footnote{
 	The survey by Sokal \& Bricmont is brilliant, and the \emph{clownesque} parade of fumesophers is chiefly hilarious; but, once in a while, in the two authors' outlook on the methodology and history of physics a bit of crudeness, or even of viridity, transpires. We are far from the subtlety of a Bellone, or of an Acerbi, who, with M. Ugaglia, took care of the It. translation of this book \cite{Sokal et Bricmont "Impostures intellectuelles"} (Garzanti, Milano, 1999). It should, however, be added that their goal is to unmask the ludicrousness—tons of ideological and terminological guano—of clowns' fumesophy, and not to enter into historical and methodological (mis)interpretations of physics/mathematics.
 	} 
 and the pamphlet by E. Bellone \cite[parte II]{Bellone "La scienza negata. Il caso italiano"}; from the latter author, see also the essay \cite{Bellone "Il mondo di carta: ricerche sulla seconda rivoluzione scientifica"} on the complicated theory vs. experiment relationship, and on the misunderstandings of the history of physics and mathematical physics on the part of some fumesophical 20th century schools.

About the history of Italian thought, Enriques' controversy with the Crocian and Gentilian prattle is a milestone:\footnote{
	After 1918 the duo Croce–Gentile is loosened, and Gentile will be characterized by an increasing readiness for a close cooperation with the mathematical/scientific side, see \cite[pp. 66-104]{Guerraggio e Nastasi "Matematica cultura e potere nell'Italia postunitaria"}. Should not forget that his son, G. Gentile, Jr., was a talented physicist.
	} 
see \cite[pp. 47-65]{Guerraggio e Nastasi "Matematica cultura e potere nell'Italia postunitaria"} \cite[pp. 842-843]{Nastasi "Il contesto istituzionale"} \cite[p. 6, and cap. 3]{Bellone "La scienza negata. Il caso italiano"} \cite[pp. 27-28, 125-147]{Guerraggio Nastasi "Italian Mathematics Between the Two World Wars"} \cite[pp. 359-366]{Russo Santoni "Ingegni minuti. Una storia della scienza in Italia"}. Echoes of this quarrel are still to be found in F. Rasetti's memory \cite[pp. 307-308]{Goodstein "A Conversation with Franco Rasetti"}, who was one of the via Panisperna boys; his judgment (in an interview of 1982) is sharp:

\vspace{2mm}

\begingroup
\footnotesize
When I try to read something about the work of some [fumesophers], I have the impression that a [fumesophers] stands for this principle: that you have to discover the meaning of a word. That in a word there is something intrinsic, so to speak, which is different from the use of it—which of course is nonsense. As if words [\,\dots] have a sort of mystic content in themselves and you have to study what this meaning is. But words are only what they are used for, and there is nothing else in words [\,\dots]. We in the physics group in Rome had the deepest contempt for [fumesophy], and especially for Gentile. We had equal contempt for Gentile, who was a Fascist, and for Croce, who was an anti-Fascist, because we had a very poor opinion of [fumesophers] regardless of their political opinions.

\endgroup

\vspace{2mm}

The gallery of freaks can be fleshed out with ease. Just to give some examples, pungent observations on the fumesophy and bilge-thinking are in

· L. Boltzmann \cite[p. 385]{Boltzmann "Populare Schriften"}: the butt of his mockery is Schopenhauer, a «nonsense-spreader» (\textit{Unsinn schmierender}) with a «hollow verbiage» (\textit{hohlen Wortkram}), expressions already employed by the latter against another fumesopher, Hegel, see \cite[p. 378]{Boltzmann "Entgegnung auf einen von Prof. Ostwald uber das Gluck gehaltenen Vortrag"}; Boltzmann  \cite[642, pp. II 384-385]{Boltzmann "Leben und Briefe"} is even more abrasive when he writes to F. Brentano:

\vspace{2mm}

\begingroup
\footnotesize
Should not the irresistible desire to philosophize be compared to the vomiting caused by migraine, as [in each of these cases] there is something that is struggling to get out, even though there is nothing inside [\textit{dem Brechreiz bei Migräne zu vergleichen, der dort noch etwas herauswürgen will, wo gar nichts mehr drinn ist}]?

\endgroup

\vspace{2mm}

· R.P. Feynman \cite[p. 195]{Feynman "The Pleasure of Finding Things Out"}, who makes fun of Spinoza, because of laughable «pomposity», and for a «meaningless chewing around»; 

· M. Born \cite[p. 54]{Born "My Life: Recollections of a Nobel Laureate"}, who laconically remarks that fumesophers are like half-witted adventurers «blissfully unaware of the dangers» hidden in their argumentations.

Sardonic glosses coming from men of letters are not absent. See, for instance:

· C.E. Gadda \cite[XIV. \textit{Impossibile chiusura di un sistema}, p. 178, lines 10-25]{Gadda "Meditazione milanese"}:

\vspace{2mm}

\begingroup
\footnotesize
Every philosophical system, or every cognitive effort which integrates reality [\textit{ogni sforzo conoscitivo integratore della realtà}], has a malignant point or defective point, where the chickens come home to roost [\,\dots]. Only science does not seem to suffer from contradictions [here Gadda is childishly optimistic]: because it never constitutes a total system, but a plurality of positions (the various disciplines or sciences), each of which stands on its own (in the same fashion as Spanish fortresses in the Caribbean kingdom), while leaning, for that aspiration, on external positions: in the manner in which a new house | leans against the wall of the neighboring [house], already built. And each science sets its own terms, [which are] beautiful, neat, certain, finished, well combed, indisputable, without perplexity, without anguish [again, there is a bit too much Gaddian gullibility], without philosophical [fumesophical] cloud mass [\textit{nuvolaglie}].
 
\endgroup

\vspace{2mm}

· R. Musil \cite[p. 272]{Musil "The Man Without Qualities I"}: 

\vspace{2mm}

\begingroup
\footnotesize
Philosophers [viz. fumesophers] are violent criminals [\textit{Gewalttäter}] who have no armies to command, so they subject the world [to their tyranny] by locking it up in a system [\textit{in ein System sperren}] [of thought].\footnote{
	I must open a parenthesis. 
	The convoluted mechanisms of fumesophy try to enclose nature \& ephemeral life, within a system. This is where the complaint about Musil's fumesophical tyranny is razor-sharp.
	
	A system corresponds, in addition to an interconnected structure (\textgreek{σύστημα}), to a method of organization, which is certainly serviceable, but does not give access to the \textit{(datum) particulare}, or to the \textgreek{τόδε τι}, according to the Stagyritic vocabulary. This is because the objects of study, even where they are highly regular and symmetrically interacting, often exhibit random behaviors. The peculiarity of a phenomenon is something elusive, while the frantic and slippery multifariousness of life remains \emph{voiceless}. The only solution is \emph{silence} (cf. point \ref{item "When I was twenty"} in Section \ref{subsection "Contextus II. Autobiographical Note"}), since uniqueness is \emph{unspeakable} and the individual is \emph{ineffable} («individuum ineffabilis est»)—wise and lyrical ruminations are in C. Magris \cite[see \textit{Io sono indicibile} and \textit{Le monete della vita}]{Magris "Itaca e oltre"}. What characterizes the individual is not inside any \textgreek{σύστημα}. It is \emph{systemless}.
	
	A system, which is an instrument for knowledge, stands up to the vain hope of englobing things, (natural) phenomena and events within the length of a chain of elements, together with a conglomerate of laws, which enjoy some kind of generality. The englobation is usually expressed through a \emph{general predicate}, so that each systematology, which passes through language, does not grasp the wide variety and sophisticated multiplicity that is hidden behind the \textit{datum particulare}. There is a pleasantly sparkling German motto: \textit{Alle Sprache ist nur eine schlechte Übersetzung}, «All language is but a poor translation». Translation of what? Of the multiplicity of reality.

	Already in Aristotle this ambiguity is blatant. In his works, a quarrel is generated between language (\textgreek{λόγος}), with systematic ambitions, and scientific knowledge (\textgreek{ἐπιστήμη}), intentionally aimed at learning something concrete, tangible; it is easy to see that knowledge-\textgreek{ἐπιστήμη} soon slips into the inability to provide an exposition of the \textit{particulare}, of the \textgreek{τόδε τι}, which ends up as rarefied, stripped of its ground of individuality, and solidified in the abstraction of a general \emph{category} or \emph{species}.
	
	The substratum of everything, the ossified \textgreek{ὑποκείμενον}, oscillates in an equivocal way, from Aristotle onwards, between the experience of the singularity of the \textgreek{τόδε τι}, understood as its character of unrepeatability and uniqueness, and the form by means of the \textgreek{εἶδος} (“form”, “shape”, “class”, “kind”, “species”, for the classification of each individual), or of the \textgreek{σχῆμα} (“form”,“figure”).

	I take a description from an old Diary of mine, when I was a teenager: «[\,\dots] 19 May, 17:46, 1993. In the lower part of the longest branch, which points to the right, a \textit{Pyrrhocoris apterus} is moving quickly. The bright red of the firebug contrasts with the silvery-gray bark of the old twisted fig tree. From the window a whiff of camphor is perceived of passing with tangling clarity, etc.». The \textgreek{λόγος}, when it is a systematic language, shows its service in the classification of the fig tree, and facilitates its identification (it is a fig  and not a \textit{Citrus sinensis}); but, in addition to its symbolic value, the \textgreek{λόγος} describes, via \textgreek{εἶδος} or \textgreek{σχῆμα}, the life of a generic fig, rather than a specific tree (that fig, from 19 September, at 17:46, in 1993). 
	
	The form \& the system are \emph{inadequate} for the evaluation of life; they are \emph{not commensurate} with the size of the problems and the understanding of nature. In every form \& system the reality of the \textit{particulare}, that is, the concreteness of the \textgreek{τόδε τι}, disappears and is definitively lost.
	
	But compare with footnote \ref{footnote "Ideas, forms of vision, etc."} on p. \pageref{footnote "Ideas, forms of vision, etc."} in mathematical acceptation: there is a liberator \emph{revenge}.
	}

\endgroup

\vspace{2mm}

· E. Canetti \cite[p. 141, e.m.]{Canetti "Die Provinz des Menschen: Aufzeichnungen 1942-1972"}:

\vspace{2mm}

\begingroup
\footnotesize
What repels me most about the [fumesophers] is the process of evacuation [\textit{Entleerungsprozess}] of their thinking. The more frequently and skillfully they use their fundamental terms, the less of the world around them remains. They are like barbarians in a noble and vast palace full of gorgeous works. They [\,\dots] throw everything out the window, methodically and steadfastly: armchairs, pictures, plates, animals, children, until there is nothing left but empty rooms. Sometimes even the doors and windows are thrown away. [So] what remains is the bare house. They believe that these devastations have led to an improvement [\textit{Sie bilden sich ein, dass es um die Verwüstungen \emph{besser} steht}].

\endgroup

\vspace{2mm}

· C. Magris \cite[p. 122]{Magris "Utopia e disincanto"} tells us of a 

\vspace{2mm}

\begingroup
\footnotesize
Goethe [\,\dots] who mocks the philosophers locked in their rooms to rack their brains about “quibbles” [\textit{almanaccare gli “arzigogoli” del loro cervello}] without looking out of the window.

\endgroup

\vspace{2mm}

· G. Ceronetti \cite[p. 106]{Ceronetti "La pazienza dell'arrostito. Giornale e ricordi (1983-1987)"}: 

\vspace{2mm}

\begingroup
\footnotesize
What can [fumesophers] understand\,\dots Two colored clothes drying in an alley fluttering in the wind are enough to give an idea \emph{of the inadequacy}, the powerlessness to tighten up, of their doctrines.\footnote{
	Ceronetti's terse quip deserves to be reported in the original language: «Cosa mai possono capire i filosofi\,\dots Bastano due panni colorati che asciugano in un vicolo agitati dal vento a dare un'idea \emph{dell'inadeguatezza}, dell'impotenza a stringere, delle loro dottrine».
	}

\endgroup

\vspace{2mm}

Same suspicion is already in \cite[p. 101]{Ceronetti "Il silenzio del corpo. Materiali per studio di medicina"}:

\vspace{2mm}

\begingroup
\footnotesize
The most civilized of philosophers wields the club of the pure barbarian when he sentences [\,\dots]: Spinoza geometries the infernal [\textit{geometrifica l'infernale}],

\endgroup

\vspace{2mm}

and reiterated with more clarity in \cite[§ 1]{Ceronetti "Per le strade della Vergine"}:

\vspace{2mm}

\begingroup
\footnotesize
Inexhaustible walk [immersed] in the beauty of the Chiossone Museum [\,\dots]. Always a divine emotion, in front of the vision of the streets and houses of Edo at night [\,\dots]. For centuries, in a similar figurative samsaric tangle, one has to probe into the life that passes in a dream, the ineffable repetition of the immutable motif, [all] trades, lamps, [and] women\,\dots How dim is [\textit{Com'è fioca}] Spinoza's \textit{Ethics} in comparison [\textit{al paragone}], which wants to encircle and capture the whole of life [\textit{che vuole accerchiare e acchiappare l'intera vita}]\,\dots\footnote{
	Einsteins was an admirer of Spinoza for his pantheistic vision, but also for his \textit{more geometrico} deterministic \textgreek{σύστημα}. But Ceronetti is right. This time a man of letters beats a scientist on how the world—or even the physical aspect of nature—goes round.
	} 

\endgroup

\vspace{2mm}

Compare with \cite[§ 79]{Ceronetti "L'occhio del barbagianni"}: 

\vspace{2mm}

\begingroup
\footnotesize
The challenge of science to [fumesophy] is this: “Be my servant if you want to survive”. To remain free and not have to humiliate itself, [fumesophy] retreats into the shadows and waits for the pre-Socratic thinkers to come back as its future.

\endgroup

\vspace{2mm}

Now, let us get back to what mathematicians and physicists say. 

In sum, as R. Courant and H. Robbins \cite[intro, last page]{Courant and Robbins "What is Mathematics?"} reassert, the answer to the question \emph{What is mathematics?} is not to be sought in the doctrine of fumesophers but in an «active experience in mathematics itself». Mutatis mutandis, the same goes for physics; A. Einstein \cite[p. 349]{Einstein "Physics and Reality"} expresses it well: 

\vspace{2mm}

\begingroup
\footnotesize
[T]he physicist cannot simply surrender to the [fumesopher] the critical contemplation of the theoretical foundations; for, he himself knows best, and feels more surely where the shoe pinches.

\endgroup

\vspace{2mm}

See also Einstein \cite[p. 1]{Einstein "The Meaning of Relativity"}, who here pushes on the empirical relevance: 

\vspace{2mm}

\begingroup
\footnotesize
I am convinced that the [fumesophers] have had a harmful effect upon the progress of scientific thinking in removing certain fundamental concepts from the domain of empiricism, where they are under our control, to the intangible heights of the a priori. For even if it should appear that the universe of ideas cannot be deduced from experience by logical means, but is, in a sense, a creation of the human mind, without which no science is possible, nevertheless this universe of ideas is just as little independent of the nature of our experiences as clothes are of the form of the human body. This is particularly true of our concepts of time and space, which physicists have been obliged by the facts to bring down from the Olympus of the a priori in order to adjust them and put them in a serviceable condition.\footnote{
	Which tallies with a classical source, the EPR paper \cite[p. 777]{Einstein Podolsky and Rosen "Can Quantum-Mechanical Description of Physical Reality Be Considered Complete?"}: «The elements of the physical reality cannot be determined by \emph{a priori} philosophical considerations, but must be found by an appeal to results of experiments and measurements. A comprehensive definition of reality is, however, unnecessary for our purpose».
	}
	
\endgroup

\vspace{2mm}

Miserably, when philosophy goes into the hands of fumesophers, it happens that it does not lead to any good reflection/solution, and instead of acting as a guide in favor of physics (which is what every good natural philosophy should do), it turns out to be a blathering crowd on scientific discoveries; so that Dirac \cite{Dirac "Oral History Interview  Session II"} is right to give this verdict:

\vspace{2mm}

\begingroup
\footnotesize
I feel that philosophy will never lead to important discoveries. It's just a way of talking about discoveries which have already been made. 

\endgroup

\vspace{2mm}

The bitter critic is that, as noted by J.A. Wheeler \cite[p. 44]{Misner Thorne and Zurek "John Wheeler relativity and quantum information"}, 

\vspace{2mm}

\begingroup
\footnotesize
Philosophy is too important to be left to the philosophers [scilicet: fumesophers].

\endgroup

\vspace{2mm}

In that vein, S. Weinberg \cite[chap. VII. \textit{Against philosophy}]{Weinberg "Dreams of a Final Theory"} points out that every physicist carries around with him a «working philosophy», and that physics continues to be troubled by epistemological «biases», personal «preconceptions», «social» influences, or even «metaphysical presuppositions», which, taken together, sometimes guide and other times hinder the path of science. But all this is one of the characteristics of man, and does not go with fumesophy \& its load of codswallop. It is an open secret that «observation can never be freed of theory»; those who do science, know well «how theory-laden are all experimental data»: it is part of the game out there.

\chapter{Outro—\emph{Parva Mathematica}: \emph{Libera Divagazione} \sfrac{5}{8}}
\label{chapter "Outro—Parva Mathematica: Libera Divagazione 5/8"}

\section{Mathematics in the Physical Sciences, and Nature of Reality II}
\label{section "Mathematics in the Physical Sciences, and Nature of Reality II"}

\begingroup
\footnotesize
E di Serva divenni io già Padrona.\footnote{
	«And from Servant I have become a Mistress». 
	} \\
\indent — Libretto by \textsc{G.A. Federico}, music by \textsc{G.B. Pergolesi} \cite[penultimate p. not numbered]{Pergolesi and Federico "La serva padrona"}

\endgroup

\vspace{2mm}

We next consider the \emph{vexata quæstio} on the relationship between mathematics and physics: which is mistress and which is the servant? E. Bellone \cite[pp. 35, 61]{Bellone "Il mondo di carta: ricerche sulla seconda rivoluzione scientifica"} writes:

\vspace{2mm}

\begingroup
\footnotesize
Is mathematics the servant or the mistress of physics and experimental investigation? If physics knows the world thanks to the painstaking and repeated observation of facts, and if mathematics comes after observation [or follows the discovery made inductively] and is reduced to rules for writing out the laws—already known—in rigorous form, then are we not forced to admit that mathematics is a mere tool of [physical] thought [working in the study of things]? And, on the other hand, is it not a justifiable claim that no facts of experience [experiment] can be devised without recourse to already mathematized theories [so that mathematics is also able to give shape to the facts and plays not only an instrumental role]?

\endgroup

\vspace{2mm}

In the next Sections, we will try to make things clearer, yet, if we are in the midst of the fray, it will seem difficult at first to escape the tangle that A.G. Bierce was able to glimpse at the bottom of a marriage.\footnote{
	A.G. Bierce \cite[p. 213]{Bierce "The Devil's Dictionary"}: «\textsc{Marriage}. The state or condition of a community consisting of a master, a mistress and two slaves, making in all, two».
	} 
To paraphrase him, we could say that we are facing a condition consisting of two mistress and two servants, making in all, two.

\subsection{Factiveness of the Formal Structure}

\begingroup
\footnotesize
The problem is that facts per se do not constitute science: this last one derives its reality, its life, from an \emph{activity of the spirit} [i.e. from the imagination]. The facts are interpreted, connected, rethought in an entirely new unity. What we currently call “facts” are usually very complicated \emph{mental elaborates} in which the sensory and experience data are combined in an exceedingly complex way with largely \emph{arbitrary} mental elements. And the description of the facts is joined by increasingly restrictive needs of logical chaining and operational significance, that is, of the effective possibility of acting in a determinate manner.\endnote{
	Original It. version: «Il problema è che i fatti da soli non costituiscono scienza: questa trae la sua realtà, la sua vita, da una attività dello spirito. I fatti vengono interpretati, collegati, ripensati in una unità affatto nuova. Quelli che noi chiamiamo correntemente “fatti” sono di solito elaborati mentali complicatissimi in cui il dato sensoriale e dell'esperienza è combinato in modo estremamente complesso con elementi mentali in gran parte arbitrari. E alla descrizione dei fatti si aggiungono esigenze sempre più vincolanti di concatenazione logica e di significatività operativa, cioè di effettiva possibilità di agire in un determinato modo».
	} \\
\indent — \textsc{M. Ageno} \cite[§ 2.7, p. 20, e.a.]{Ageno "La costruzione operativa della fisica"}

\endgroup

\vspace{2mm}

Ageno's observation is enlightened. We will take it as a cue for our reflection.

\enumerationisinitium
\item To supplement what has been said above, it turns out that we do not record facts, we \emph{create} them; we do not collect data, we interpret them in a recombination of pieces that is our knowledge. The world is a \emph{suppellettile} of the biology of the mind.\footnote{
	A study on the «creation» referring to an external world of facts, and «reconstruction» of reality by the brain together with the sensory aggregate, is deepened by E. Bellone in several of his popular works \cite{Bellone "I corpi e le cose. Un modello naturalistico della conoscenza"}  \cite{Bellone "Molte nature. Saggio sull'evoluzione culturale"} \cite{Bellone "Qualcosa la fuori. Come il cervello crea la realta"}, with special attention to physical theories.
	}

Here is an illustrative excerpt by A. Einstein \cite[pp. 350-351]{Einstein "Physics and Reality"}: 

\vspace{2mm}

\begingroup
\footnotesize
I believe that the first step in the setting of a “real external world” [\textit{realen Aussenwelt}] is the formation [\textit{Bildung}] of the concept of bodily objects [\,\dots]. Out of the multitude of our sense experiences we take, mentally and arbitrarily, certain repeatedly occurring complexes of sense impression [\,\dots], and we attribute to them a meaning [\textit{Begriff}]—the meaning of the bodily object. Considered logically this concept is not identical with the totality of sense impressions referred to; but it is an arbitrary creation of the human (or animal) mind [\textit{freie Schöpfung des menschlichen (oder tierischen) Geistes}] [\,\dots].

The second step is to be found in the fact that, in our thinking (which determines our expectation), we attribute to this concept of the bodily object a significance [\textit{Bedeutung}], which is to a high degree independent of the sense impression which originally gives rise to it. This is what we mean when we attribute to the bodily object “a real existence”. The justification of such a setting rests exclusively on that fact that, by means of such concepts and mental relations between them, we are able to orient ourselves in the labyrinth of sense impressions. These notions and relations, although free statements of our thoughts, appear to us as stronger and more unalterable than the individual sense experience itself [\,\dots]. On the other hand, these concepts and relations, and indeed the setting of real objects and, generally speaking, the existence of “the real world” [\textit{realen Welt}], have justification only in so far as they are connected with sense impressions between which they form a mental connection.

\endgroup

\vspace{2mm}

\item Interestingly, in the nineteenth century a mechanical approach was still prevalent for the investigation of the phenomena of nature. For example, in H. Hertz, at the beginning of the Introduction to his \textit{Prinzipien} \cite{Hertz "Die Prinzipien der Mechanik in neuen Zusammenhange dargestellt"}, the thought-nature relationship (or observer-phenomenon relationship), is presented as a complex of organs, assuming an interlocking between the parties; he speaks of «conformity», that is, of congruence, as if they were gears: 

\vspace{2mm}

\begingroup
\footnotesize
\cite[pp. 1-2]{Hertz "Die Prinzipien der Mechanik in neuen Zusammenhange dargestellt"} = \cite[pp. 1-2]{Hertz "The Principles of Mechanics Presented in a New Form"} We form for ourselves images or symbols of external objects; and the form which we give them is such that the necessary [scilicet: logical] consequents [\textit{denknotwendigen Folgen}] of the images in thought are always the images of the necessary [\textit{sic}] consequents in nature of the things pictured. In order that this requirement may be satisfied, there must be a certain conformity [\textit{Übereinstimmungen}] between nature and our thought [\textit{Geiste}]. Experience teaches us that the requirement can be satisfied [\,\dots]. The images which we here speak of are our conceptions of things. With the things themselves they are in conformity in one important respect, namely, in satisfying the above-mentioned requirement [\,\dots]. As a matter of fact, we do not know, nor have we any means of knowing, whether our conceptions of things are in conformity with them in any other than this one fundamental respect. The images which we may form of things are not determined without ambiguity [\textit{Eindeutig\,\dots noch nicht bestimmt}] by the requirement that the [logical] consequents of the images must be the images of the consequents [in nature]. Various images of the same objects are possible, and these images may differ in various respects.\footnote{
	\label{footnote "Hertz's Preface"}
	Which is congruous with this observation in his Preface \cite[p. xxii]{Hertz "Die Prinzipien der Mechanik in neuen Zusammenhange dargestellt"} = \cite[p. xxi]{Hertz "The Principles of Mechanics Presented in a New Form"}: «All physicists agree that the problem of physics consists in tracing the phenomena of nature back to the simple laws of mechanics. But there is not the same agreement as to what these simple laws are [\,\dots]. [W]e have here \emph{no certainty} as to what is simple and permissible, and what is not» (e.a.).
	}

\endgroup

\vspace{2mm}

Except that the process of elaboration of the nature of reality, which belongs ultimately to the mathematical system, is not rigidly mechanical; it has, if anything, an elastic-evolutionary organicity, dynamically dispersed in fragmented ever-changing streams of spiritual creativity.
\item To be honest, a thought-nature relationship does not have a guarantee of a \textgreek{λογικός} type connection. From our direct experience of the world to some \emph{formal definition} (that is, to some system with axioms and statements) there is no \emph{logical route}. The passage from the \emph{facts} of experience to any \emph{theory}, it is a connection-procedure, as Einstein notes \cite[letter to M. Solovine, 7 May 1952]{Einstein "Letter to Maurice Solovine 7 May 1952"},\footnote{
	His «epistemological» schema \cite[p. 166]{Krull "Albert Einstein in seinen erkenntnistheoretischen Ausserungen"} = \cite[p. 272]{Einstein "Letter to Maurice Solovine 7 May 1952"} is as follows. 
	\enumerationisinitium
	\item[(\textgreek{α})] Firstly, there are our «direct experiences [\textit{Erlebnisse}]».
	\item[(\textgreek{β})] Then come the «axioms [\textit{Axiome}]» from which «we draw our conclusions [\textit{Folgerungen}]». «Psychologically» the axioms rest on the experiences. But «there is no logical route [\textit{es gibt keinen logischen Weg}]» leading from the experiences to the axioms, but only an «intuitive (psychological) connection [\textit{einen intuitiven (psychologischen) Zusammenhang}]».
	\item[(\textgreek{γ})] «From the axioms, by a logical route [\textit{auf logischem Wege}], are deduced individual assertions [\textit{Einzel-Aussagen\,\dots abgeleitet}]» that can lay claim to exactness.
	\item[(\textgreek{δ})] The assertions are connected to the experiences «(verification through experience [\textit{Prüfung an der Erfahrung}])». At a closer look, «this procedure also belongs to the extra-logical (intuitive) sphere [\textit{extra-logischen (intuitiven) Sphäre}]», because the relation between the concepts occurring in any assertion and the experiences «are not of a logical nature [\textit{nicht logischer Natur sind}]». But this relation between the assertions and the experiences is «(pragmatically) much less uncertain» than the relation between the axioms and the experiences.
	\enumerationisfinis
	}\textsuperscript{,}\endnote{
	Compare the above-mentioned letter with A. Einstein \cite[p. 684]{Einstein "Reply to Criticisms"}: «The scientist [\,\dots] must appear to the systematic epistemologist as a type of unscrupulous opportunist: he appears as \emph{realist} insofar as he seeks to describe a world independent of the acts of perception; as \emph{idealist} insofar as he looks upon the concepts and theories as the free inventions of the human spirit (not logically derivable from what is empirically given); as \emph{positivist} insofar as he considers his concepts and theories justified \emph{only} to the extent to which they furnish a logical representation of relations among sensory experiences».
	} 
that is inherent to an \emph{extra-logical (intuitive, or psychological) range}. And \emph{it is in this extra-logical womb where mathematics is conceived}.
\enumerationisfinis

\subsection{Bohrism in a Right Perspective}
\label{subsection "Bohrism in a Right Perspective"}

\begingroup
\footnotesize
There is no quantum [or spatio-temporal] world. There is only an abstract quantum [or spatio-temporal] physical description. It is wrong to think that the task of physics is to find out how nature is. Physics concerns what we can say about nature [by means of some language, including first of all the math-language]. \\
\indent — \textsc{N. Bohr} \cite[p. 12]{A. Petersen "The Philosophy of Niels Bohr"}

\endgroup

\vspace{2mm}

\enumerationisinitium 
\item As far as I am concerned, the phrase of Bohr should not be taken in an idealistic motive as he does elsewhere, see e.g. \cite[p. 485]{Bohr "Wirkungsquantum und Naturbeschreibung"} in which Bohr says that a «strict separation [\textit{Trennung}] between object and subject» cannot be maintained.\endnote{
	The Ge. version \cite[pp. 484-485]{Bohr "Wirkungsquantum und Naturbeschreibung"} is: «[D]er Unmöglichkeit einer strengen Trennung von Phänomen und Beobachtungsmittel [\,\dots]; keine strenge Trennung zwischen Objekt und Subjekt aufrecht zu erhalten ist, da ja auch der letztere Begriff dem Gedankeninhalt angehört». His credo is: we are dealing with «the impossibility of a strict separation of phenomenon and instrument for observation [\,\dots]; no strict separation between object and subject can be maintained», since everything «belongs to the thought content».
	} 
The focus is this: a physical theory that wants to connect some clues about the structure of nature, can \emph{only} do by means of a \emph{math-language}, so what one has, at the end of the fair, is but an \emph{image of reality}, an image constructed by a math-language, for which what is called \emph{reality}, in physics, is a “manifestation”, “appearance”, “occurrence”, or “emergence” of something, a \emph{phenomenon} (\textgreek{φαινόμενον}), precisely.

Needless to say, the Moon exists even if a mouse (or any other being) does not observe it; see H. Everett III \cite[p. 116]{Everett III "The Theory Of The Universal Wave Function"}: «[Einstein] could not believe that a mouse could bring about drastic changes in the universe simply by looking at it», and  A. Pais \cite[pp. 5-6]{Pais "'Subtle is the Lord...' The Science and the Life of Albert Einstein"}.
\item 
\label{item "Bertuglia and Vaio passage"}
Compare it with the exposition of C.S. Bertuglia and F. Vaio \cite[p. 240, e.a.]{Bertuglia Vaio "Nonlinearity Chaos and Complexity: The Dynamics of Natural and Social Systems"}: 

\vspace{2mm}

\begingroup
\footnotesize
To describe, interpret and predict phenomena of the environment (or of the world, the universe or whatever we wish to call it) in which we are immersed, we generally use models, where the meaning of the term “model”, in a very general context, is a \emph{mental representation} that, in a certain sense, replaces the environment (the world, the universe). In other words, since we have no means of going beyond the restricted window that our senses provide of reality, the latter is too complicated and substantially inaccessible to complete knowledge. All that we can do is to limit ourselves to constructing a representation of reality based on information gained from experience. In this vision, therefore, \emph{our knowledge is always relative to a perspective} and is \emph{conditioned by a point of view}, \emph{a product of the human mind}, and \emph{not something inherent to the order of things}. Science, therefore, does not study the physical world \emph{per se}, but rather \emph{our way of depicting} some regularities selected from our experience, that can be observed in certain conditions and from a particular perspective. The models that we form of the world are \emph{not authentic copies} of the latter [\,\dots]; in any event, their correspondence with a “real world”, that can only be known by means of representations, can \emph{never} be verified. The best we can do is to assess our model-representations of the world on a pragmatic level, judging whether they guide our understanding in a useful way towards the objective of an effective description of the observed phenomenology [cf. Section \ref{subsubsection "Math-Model in the Perseus Mythology: No Aberration-free Mirror of Reality"}].

\endgroup
\enumerationisfinis
	
\subsection{Physics is (Not) Mathematics}
\label{subsection "Physics is (Not) Mathematics"}

\begingroup
\footnotesize
\emph{Physics is mathematical}, not because we know so much about the physical world, but because we know so little; it is \emph{only} its mathematical properties that we can discover. For the rest, our knowledge is negative. In places where there are no eyes or ears or brains there are no colours or sounds, but there are events having certain characteristics which lead them to cause colours and sounds in places where there are eyes and ears and brains. We cannot find out what the world looks like from a place where there is nobody [\,\dots]; the attempt is as hopeless as trying to jump on one's own shadow. \\
\indent — \textsc{B. Russell} \cite[pp. 163-164, e.a.]{Russell "The Nature of Our Knowledge of Physics"}

\endgroup

\vspace{2mm}

Russell's position, bluntly declared in the above passage, goes well with ours, and it blends harmoniously with the thesis presented here. Let us consider, in the following Sections, some related arguments.

\subsubsection{Galilean Heritage}

\begingroup
\footnotesize
[I] nomi, e gl'attributi si deuono accomodare all'essenza delle cose, e non l'essenza à i nomi; perch[é] prima furon le cose, e poi i nomi [\,\dots]; ricordandoci, che la Natura sorda, \& inesorabile à nostri preghi.\footnote{
	«Names and attributes must be accommodated to the essence of things, and not the essence to the names, since things come first, and names afterwards [\,\dots]; reminding us that the Nature [is] deaf and inexorable to our prayers».
	} \\
\indent — \textsc{G. Galilei} \cite[pp. 12, 131]{Galilei "Istoria e dimostrazioni intorno alle macchie solari e loro accidenti comprese in tre lettere"}

\endgroup

\vspace{2mm}

The above-mentioned Galilean excerpt seems to contradict the Linnean passage \cite{Linnaeus "Systema Naturae Per Regna Tria Naturae"} reported in Section \ref{section "Naturæ Mæandri and Filum Ariadneum: a Botanical Comparison"}. But this apparent reversal is part of the physicist's cultural heritage,\footnote{
	A modern example of a mindset in the Galilean mold is offered by E. Segrè \cite[p. 419]{Segre "Personaggi e scoperte della fisica contemporanea. Da Rutherford ai quark"}: «It is the subtle play [\textit{gioco sottile}] between theory and experiment that brings forward physics [\,\dots]. We proceed by giving one knock on the hoop and another on the barrel. Physics wants to describe nature and predict phenomena. It is impossible to do this starting with a priori theories; we would stop after a few steps and every error would move us further away from the goal. On the other hand, using only experiments, after a few experiences we would find ourselves immersed in a wood [\textit{selva}] of disconnected facts with no hope for escape. It is the combination of experiment and theory, mediated by [our] mathematics, that permits the marvelous progress [of physics]».
	}
and certainly it is not a problem for those who believe in some kind of equality between nature and mathematics. Physicist is usually reluctant to subordinate «things» to «names» (which catalog and define «things»): he, in fact, does the opposite. And where does mathematics fit in all this? Let us try to see better this relationship, thanks to a series of stories that treat  these issues.

\subsubsection{Facts, Experiences, and Formulæ}

\begingroup
\footnotesize
The contradiction does not exist in reality [\textit{La contradiction n'est pas dans la réalité}], which is always in agreement with itself [\textit{toujours d'accord avec elle-même}]; it is lies in the theories [adopted] to express this reality.\endnote{
	Duhem's sentence has been a little modified, but the meaning of what he says has not been betrayed. The original sentence is this: «Que de discussions scientifiques où chacun des deux tenants prétend écraser son adversaire sous le témoignage irrécusable des faits! On s'oppose l'un à l'autre des observations contradictoires. La contradiction n'est pas dans la réalité, toujours d'accord avec elle-même; elle est entre les théories par lesquelles chacun des deux champions exprime cette réalité». A thought like that, even if in a completely different context, echoes in a more recent statement by Feynman, about the Space Shuttle Challenger disaster (1986): «[R]eality must take precedence over public relations, for nature cannot be fooled» (\textit{Personal Observations on Reliability of Shuttle}, in \textit{Report of the Presidential Commission on the Space Shuttle Challenger Accident, Vol. 2, Appendix F}).
	} \\
\indent — \textsc{P. Duhem} \cite[p. 261]{Duhem "La theorie physique. Son objet et sa structure"}
	
\endgroup

\vspace{2mm}

\enumerationisinitium
\item We often hear it said that «The development of a physics theory needs experimental data for guidance», or «Physics is not mathematics, and mathematics is not physics»—a quote from Feynman \cite[p. 55]{Feynman "The Character of Physical Law"}. Physics, without the «necessary foothold in facts», as M. Born says \cite[p. 90]{Born "Natural Philosophy of Cause and Chance"}, does not take a single step forward.\footnote{
	The Anglo-Saxon physics tradition, in the nineteenth century, is the modern cradle of this conception of science; by way of example, see:
	\enumerationisinitium	
	\item J.F.W. Herschel \cite[§§ 66-67, pp. 75-76]{Herschel "A Preliminary Discourse on the Study of Natural Philosophy"}: «Into abstract science [mathematics] [\,\dots] the notion of cause does not enter. The truths [in mathematics] it is conversant with are \emph{necessary} ones, and exist independent of cause. There may be no such real \emph{thing} as a right-lined triangle marked out in space; but the moment we conceive one in our minds, we cannot refuse to admit the sum of its three angles to be equal to two right angles [\,\dots]. To assert the contrary, would not be to rebel against a power, but to deny our own words. But in natural science \emph{cause} and \emph{effect} are the ultimate relations we contemplate; and \emph{laws}, whether imposed or maintained, which, for aught we can perceive, might have been other than they are. This distinction is very important. A clever man, shut up alone and allowed unlimited time, might reason out for himself all the truths of mathematics, by proceeding from those simple notions of space and number of which he cannot divest himself without ceasing to think. But he could never tell, by any effort of reasoning, what would become of a lump of sugar if immersed in water, or what impression would be produced on his eye by mixing the colours yellow and blue. We have thus pointed out to us, as the great, and indeed only ultimate source of our knowledge of nature and its laws, \textsc{experience}».
	\item P.G. Tait \cite[pp. 6, 25]{Tait "Lectures on Some Recent Advances in Physical Science"}: «There is nothing physical to be learned \emph{a priori} [\,\dots]. We have to face the question, where to draw the line between that which is physical and that which is utterly beyond physics [i.e. that which is mathematical abstraction]. And, again, our answer is—Experience alone can tell us; for experience is our only possible guide».
	\enumerationisfinis	

	The sin of this conception is its virtue: let us call it “guilelessness”. 
	} 
Claims of this kind are true; but a clarification must be made. 
\subenumerationisinitium
\item 
\label{item "Physics is not mathematics (and vice versa), axioms of faith"}
Physics is not mathematics (and vice versa), if by physics we mean a heuristic procedure that «pays but little attention to the precise [mathematical] reasoning from fixed axioms» \cite[p. 54]{Feynman "The Character of Physical Law"}, or that does not rigidly accept the hypothetical-deductive method. Feynman, unlike an Einstein \cite{Einstein "Letter to Maurice Solovine 7 May 1952"},\endnote{
	A. Einstein in W. Heisenberg \cite[p. 63]{Heisenberg "Physics and Beyond. Encounters and Conversations"}: «[I]t may be heuristically useful to keep in mind what one has actually observed. But on principle, it is quite wrong to try founding a theory on observable magnitudes alone. In reality the very opposite happens. It is the theory which decides what we can observe».
	
	\setlength\parindent{8pt}
	We are at the polar opposite of a phenomenological physics of Fermi. He did not have theoretical preset schemes, but convenient reasonings, or flexible models (guidelines, not strict rules), which are able, from time to time, to adapt plastically, and pragmatically, to the experiments (observable magnitudes); see V. Barone \cite[pp. xxxiv-xxxix]{Barone "Il sollievo della semplicita. Enrico Fermi fisico e divulgatore"}. The art of “shaping”—\textgreek{πλαστική (τέχνη)}—is ancient.	
	} 
or a Dirac \cite{Dirac "La bellezza come metodo. Saggi e riflessioni su fisica e matematica"}, does not give too much credit to an axiomatic physics, under the rules of Euclidean mathematics,\endnote{
	\textit{Nihil sub sole novum}. As J.F.W. Herschel \cite[§ 108, p. 100]{Herschel "A Preliminary Discourse on the Study of Natural Philosophy"} already wrote: the «radical error» of the Greek thought it was «to imagine that the same method which proved so eminently successful in mathematical, would be equally so in physical, enquiries, and that, by setting out from a few simple and almost self-evident notions, or \emph{axioms}, every thing could be reasoned out». Actually, this is only partly true, because the variegated richness of Hellenistic science (with its peaks of excellence, in the pre-Imperial age) gets forgotten: reading e.g. Euclid—all Euclid, not only the “geometric” one—exclusively with the lens of the axiomatic method, disconnected from the observed phenomena, is nothing short of reductive.
	} 
but prefers an empirical method, under a mathematics called «Babylonian», working from the ground, inductively,\footnote{
	\label{footnote "Euler: quasi experimenta"}
	Inductive techniques in mathematics are common; see e.g. \textit{Summarium}, pp. 19-20, to L. Euler \cite{Euler "Specimen de usu observationum in Mathesi pura"}: «It does not seem a little paradox to ascribe a great weight to observations [\textit{multum observationibus tribui}] even in that part of mathematics which is usually called pure, since people think that observations are restricted to external objects impressing our senses. As numbers in themselves must refer to the pure intellect alone, we can hardly understand the value of observations and quasi-experiments [\textit{quasi experimenta}] in the investigation of the nature of numbers. Yet, in fact [\,\dots], the properties of the numbers known to us have been mostly discovered by observation, long before their truth has been confirmed by strict proofs [\textit{rigidis demostrationibus}] [\,\dots]. Such knowledge which is supported only by observations, in case a proof is lacking, must be carefully distinguished from the truth, and gained only by induction. There is no lack of examples in which induction alone led to error [\,\dots]. Nevertheless, we can use this inductive method as an opportunity to investigate more accurately a [certain] property [of numbers], and to ascertain its truth or falsity». This excerpt is presented as a manifesto by G. Pólya \cite[p. 3]{Polya "Mathematics and Plausible Reasoning I: Induction and Analogy in Mathematics"}.
	} 
or phenomenologically, with particular cases;\endnote{
	This distinction within the community of theoretical physicists is already patent in mathematical physics. See e.g. V. Volterra \cite[pp. 912-913]{Volterra "Henri Poincare"}: «There are two kinds of mathematical physics [\,\dots]. In most cases the people who are greatly interested in one despise somewhat the other. The first kind [of mathematical physics] consists in a difficult and subtle analysis connected with physical questions. Its scope is to solve in a complete and exact manner the problems which it presents to us. It endeavors also to demonstrate by rigorous methods statements which are fundamental from mathematical and logical points of view [\,\dots]. The other kind of mathematical physics has a less analytical character, but forms a subject inseparable from any consideration of phenomena. We could expect no progress in their study without the aid which this brings them». 
	} 
but also inductive physics is full of \emph{axioms of faith or belief},\footnote{
	\label{footnote "Two examples of physico-mathematical faith or belief"}
	This sentence of Planck \cite[p. 214]{Planck "Where Is Science Going? The Universe In The Light of Modern Physics"} is happily blatant on this behalf: «Anybody who has been seriously engaged in scientific work of any kind realizes that over the entrance to the gates of the temple of science are written the words: \emph{Ye must have faith}», cf. Intro, p. \pageref{subsection "Allegorical Figurations of Reality, Idealizations, and Techniques of Transcendence Playing: Physics Affairs"}. See also R. Penrose \cite[chap. 2]{Penrose "Fashion Faith and Fantasy in the New Physics of the Universe"}. To follow are two examples of physico-mathematical faith or belief.
	\enumerationisinitium
	\item Any form of invariance, such as conservation laws (e.g. of energy, of linear/angular momentum, etc.), and gauge theories, or, more generally, laws of symmetry relating to the conservation laws.
	\item The awkward concept of chance, or causality. The distinction e.g. between random and pseudo-random strings encompasses a previous knowledge of what is (defined as) random. To this end, the calculation of probability is not helpful, because the latter is—presumed—founded on the concept of randomness (chance, causality), whereby a \emph{circular viciousness} is created. In spite of that, it is possible to appeal to efficacious definitions (escamotage), such as the ones used in information theory (cf. C. Shannon in Section \ref{subsubsection "Information Flow of What? An Entropy Flux Question"}), from which we derive this: if the information content (self-information) of a string has a length equal to that of the string, then the string is random; otherwise, the string is non-random, which is reflected in the practice of compressing (abbreviating) its information (e.g. M. Gell-Mann \cite[p. 16]{Gell-Mann "What is Complexity? Remarks on"} asserts that: «A bit string that is incompressible has no [\,\dots] regularities and is defined as “random”»)—we can call the second case \emph{squeezing or compressibility theorem}. Discernibly, this is not an absolute method to define what is random and what is pseudo-random, or non-random; it is only an \emph{anthropomorphic} operational parameter. We do not get to know the probability distributions, which serve to set the concept of probability, together with the notions of random variables, and stochastic processes; \emph{we do not know what the randomness is in nature}, because we have no idea what the rules of nature are \emph{at the bottom}.\endnote{
		We are still stuck where E. Fermi was, in 1929 \cite[p. 41, Barone's Ed.]{Fermi "Problemi attuali della fisica"}: «In the physics of macroscopic phenomena, all future states of a system subtracted from any external perturbation are univocally determined by the knowledge of the initial state of the system itself (determinism). Instead, there are various reasons that seem to indicate that such a deterministic principle is not valid in the microscopic world of atoms. The apparent macroscopic determinism would be solely the result of the fact that, in [our] macroscopic observations, only averages of numerous phenomena[,] occurring in the atoms constituting the different bodies[,] are observed. No one will escape the enormous importance of this question and the profound change that would take place in our views on natural phenomena, the day we were actually forced to realize that physical determinism is only a property of average, no longer valid when one operates over a single atom. It is precisely from a profound study of elementary phenomena that the answer to this exciting problem can be obtained; that is, it will be possible to see whether the apparent indeterminism of elementary phenomena derives from the fact that the observation of some cause has been omitted, or whether it represents one of the fundamental laws of the atomic world».
		}
	\enumerationisfinis
	
	And this is why any theory on the deterministic or indeterministic character of nature is limited to the theory itself, and not to nature, see e.g. G. 't Hooft's papers \cite{'t Hooft "A mathematical theory for deterministic quantum mechanics"} \cite{'t Hooft "Entangled quantum states in a local deterministic theory"}, where he sketches a determinist theory for quantum mechanics.
	}
of principles that are considered to be (more or less) self-evident,\footnote{
	Cf. H. Poincaré \cite[p. 24]{Poincare "La Science et l'Hypothese"}: «Induction applied to the physical sciences is always uncertain, because it rests on the belief [\textit{croyance}] in a general order of the Universe [\textit{ordre général de l'Univers}], an order outside of us [\textit{ordre qui est en dehors de nous}]. Mathematical induction, namely demonstration by recurrence, on the contrary, imposes itself necessarily because it is only the affirmation of a property of the mind itself [\textit{propriété de l'esprit lui-même}]». 
		} 
or with a deep-seated plausibility, of \emph{non-experimental hypotheses},\footnote{
	See G. Israel \cite[pp. 112-113]{Israel "La visione matematica della realta. Introduzione ai temi e alla storia della modellistica matematica"} \cite[pp. 98-99]{Israel "Meccanicismo. Trionfi e miserie della visione meccanica del mondo"} in outspoken polemic with V.I. Arnold \cite[chap. I]{Arnold "Mathematical Methods of Classical Mechanics"} on some «metaphysical hypotheses»—homogeneity and isotropy properties of space-time, principle of determinism (translated, wholly, into the existence and uniqueness theorems)—masquerading as «experimental facts».
	}
of \emph{prejudices},\footnote{
	Feynman himself \cite[pp. 199-200]{Feynman "The Pleasure of Finding Things Out"} states in an interview (1979) in \emph{Omni maga­zine}: «Forget what you hear about science without prejudice. Here, in an interview, talking about the Big Bang, I have no prejudices—but when I'm working, I have a lot of them. — \textit{Omni}: Prejudices in favor of \dots what? Symmetry, simplic­ity \dots? — Feynman: In favor of my mood of the day. One day I'll be convinced there's a certain type of symmetry that everybody believes in, the next day I'll try to figure out the conse­quences if it's not, and everybody's crazy but me».
	}
or even of \emph{swindles}.\footnote{
	\label{footnote "Heisenberg's postcard to W. Pauli, 15 December 1924"}
	It is not uncommon for a theory, when it is in its infancy, to be brought up with \emph{physico-mathematical subterfuges} and \emph{swindles}. The foundation of quantum mechanics is an impressive example. Read this postcard by W. Heisenberg to W. Pauli, sent from København on 15 December 1924 \cite[(76) pp. 192-193, e.m.]{Pauli "Wissenschaftlicher Briefwechsel I: 1919-1929"}: «Dear Pauli! Today I have read your new paper \cite{Pauli Uber den Zusammenhang des Abschlusses der Elektronengruppen im Atom mit der Komplexstruktur der Spektren"} [it contains the Pauli exclusion principle] and sure enough I am the person who most of all  brightens up about it, not only because you have pushed the \emph{swindle} to a previously unexpected height [\textit{weil Sie den Schwindel auf eine bisher ungeahnte}], by breaking all antecedent records, for which you used to insult me ([e.g.] for the introduction of single electrons with 4 degrees of freedom), but especially I am jubilant over the fact that you too (et tu, Brute!) are returned with a bowed head to the land of formalism philistines; but do not be sad, you will be welcomed with open arms. And if you think that you have written something against the previous sorts of swindle, that is of course a misunderstanding; swindle $\times$ swindle gives nothing correct and thereby two swindles can never contradict each other. So my congratulations».
	}\textsuperscript{,}\endnote{
	\label{endnote "Bacon's idola"}
	Let us pause a moment, and widen our vision. Prejudices—biases, illusions, misapprehensions, false impressions, chimeras, dogmas, linguistic swindles, etc.—are acutely categorized by F. Bacon as \emph{idols} \cite[Lib. I, XXXIX, p. 213]{Bacon "Novum organum"}. Mathematics \& physics, and all the sciences fall into this idolatrous chasm. Four types of idols are outlined in the Baconian classification.

	\setlength\parindent{8pt}
	· \textit{Idola tribus} \cite[Lib. I, XLI, pp. 214-216]{Bacon "Novum organum"}: «\emph{The idols of the tribe} are founded in human nature itself and in the very tribe or race of mankind. The assertion that the human senses are the measure of things is false [\textit{Falso enim asseritur sensum humanum esse mensuram rerum}]; on the contrary, all perceptions, both of sense and the mind, are on a human scale, and not on a scale of the universe [\textit{omnes perceptiones, tam sensus quam mentis, sunt ex analogia hominis, non ex analogia universi}]. The human understanding resembles an uneven mirror [\textit{intellectus humanus instar speculi inaequalis}] receiving rays [irregularly] from things and merging its own nature with the nature of things, which [resultantly] distorts and corrupts it [\textit{distorquet et inficit}]».

	· \textit{Idola specus} \cite[Lib. I, XLII, pp. 216-217]{Bacon "Novum organum"}: «The \emph{idols of the cave} are the misconceptions of the individual man. Every single one of us (in addition to the aberrations of human nature in general) has a kind of individual cave or cavern [\textit{specum sive cavernam}] which fragments and corrupts the light of nature [\textit{quae lumen naturae frangit et corrumpit}]: either from the unique and particular (pre)disposition of each one; or from his education and the company with others; or from his reading of books, and the authority of those whom he cultivates and admires; or from the different impressions encountered on the mind, [be the mind] preoccupied and prejudiced, or calm and distant, and so forth: so that the human spirit [\textit{spiritus humanus}] (under the dispositions of individual men) is a variable thing, and quite irregular, and almost haphazard [\textit{res varia, et omnino perturbata, et quasi fortuita}]».

	· \textit{Idola fori} \cite[Lib. I, XLIII, p. 217]{Bacon "Novum organum"}: «There are also illusions that are being born from [interpersonal] intercourse and human association, which we call \emph{idols of the market}, because of the trade and the community between men. Mankind, in fact, associates through conversation; words are chosen with the purpose of being understood by the common people. And thus a bad and inept application of words obstructs the intellect in admirable ways [\textit{mala et inepta verborum impositio miris modis intellectum obsidet}]. Neither the definitions nor the explanations, with which learned men are wont to defend and somehow protect themselves, are able to restore [the situation]. Plainly words do violence to the understanding, and perturb everything [\textit{verba plane vim faciunt intellectui, et omnia turbant}]; and lead men into stupid and innumerable disputes and spaciousnesses [\textit{inanes et innumeras controversias et commenta}]».

	· \textit{Idola theatri} \cite[Lib. I, XLIV, p. 218]{Bacon "Novum organum"}: «To finish off, there are idols, which have crept into men's minds from the various dogmas of philosophies [\textit{ex diversis dogmatibus philosophiarum}], and even from the perverted rules of demonstration [\textit{ex perversis legibus demonstrationum}]; these we call \emph{idols of the theatre}; for all the philosophies received or invented [\textit{receptae aut inventae}] are, for us, so many fables [\textit{tot fabulas}] produced and performed, which have created fictitious and scenic worlds [\textit{mundos effecerunt fictitios et scenicos}] [\,\dots]. And we are referring not only to the general [systems of] philosophy, but also to the numerous principles and axioms of the sciences [\textit{sed etiam de principiis et axiomatibus compluribus scientiarum}], which exhibit a vigorous growth from tradition, belief and negligence [\textit{quae ex traditione et fide et neglectu invaluerunt}]».
	}  
Besides, all first principles are selected not on a celestial decree but, humanly, in relation to what is thought/deemed fit, as the very word \textgreek{ἀξίωμα} (der. from \textgreek{ἄξιος}, cf. endnote \ref{endnote "axiom"}) suggests.
\item In terms of science, or knowledge, physics is still (a form of) mathematics, which from within turns its gaze outside itself, toward the so-called laws of nature. Physics \emph{of nature}, to wit, nature itself, is not mathematics, which is obvious; but \emph{our} physics, what we know as “physics”, \emph{is} mathematics, in the sense that, through mathematics, we construct physical theories, and interpret and reproduce experimental results.
\item At the root of it all, as L. Boltzmann \cite[p. 222]{Boltzmann "Uber die Entwicklung der Methoden der theoretischen Physik in neuerer Zeit"} reminds us (mentioning Goethe), lies the fact that «only half of our experience is ever experience»,\endnote{
	The exact phrase of Goethe is: «die Erfahrung nur die Hälfte der Erfahrung ist», in \cite[p. 262]{Goethe "sammtliche Werke in vierzig Banden Dritter Band"}.
	}
since \emph{any experimental observation is full of theory}; which was steadily emphasized by P. Duhem \cite[II, chapp. IV-V]{Duhem "La theorie physique. Son objet et sa structure"}:\footnote{
	P. Duhem \cite[pp. 245-246]{Duhem "La theorie physique. Son objet et sa structure"}: «Between an abstract symbol and a concrete fact there can be a correspondence, not complete parity; the abstract symbol cannot be the adequate representation of the concrete fact, the concrete fact cannot be the exact realization of the abstract symbol [\,\dots]. This disparity between the \emph{practical fact}, really observed, and the \emph{theoretical fact}, or the symbolic and abstract formula stated by the [mathematical] physicist [leads to this]: \emph{an infinity of distinct practical facts can correspond to the same theoretical fact} [\,\dots], and: \emph{an infinity of logically incompatible theoretical facts can correspond to the same practical fact}».
	} 
no fact of experience (or \emph{sensata esperienza} \cite[p. 24]{Galilei "Dialogo sopra i due Massimi Sistemi del Mondo Tolemaico e Copernicano"}, we might say), no experiment, is naked, and speaks for itself, but—so that it can have a physical significance—it is always charged with an \textit{interprétation théorique}.\footnote{
	The physiological bottom of this condition is articulated in H. von Helmholtz \cite[p. 168]{von Helmholtz "Die Tatsachen in der Wahrnehmung"}: «Even the most elementary representations [\textit{elementaren Vorstellungen}] contain a mental element [\textit{in sich ein Denken}] and occur according to the laws of thought. Everything that is added, in intuition [\textit{Anschauung}], to the raw materials of sensation can be regarded as a mental process».
	} 
The skein of \textit{réalité concrète} becomes a part of the study of physics when it is covered with \emph{abstrait et symbolique} representations, which are but mathematical relations (under varying degrees of abstraction).
\subenumerationisfinis
\item 
\label{item "Yang on gauge fields and fiber bundles"}
We quote one passage of C.N. Yang \cite[p. 97]{Yang "Magnetic Monopoles Gauge Fields and Fiber Bundles"}:

\vspace{2mm}

\begingroup
\footnotesize
Reflecting on how the concepts basic to gauge fields were formulated by physicists, we see that at every step, the development was tied to the problem of the conceptual description of the physical world. Firstly, Maxwell equations [that give life to an elementary example of Abelian gauge field] originated with the four fundamental experimental laws of electricity and magnetism and with Faraday's introduction of the concepts of field and flux. Maxwell's equations and the principles of quantum mechanics led to the idea of gauge invariance. Attempts to generalize this idea, motivated by physical concepts of phases, symmetry, and conservation laws, led to the theory of non-Abelian gauge fields [which are inherently non-linear]. That non-Abelian gauge fields are conceptually identical to ideas in the beautiful theory of fiber bundles, developed by mathematicians \emph{without reference to the physical world}, was a great marvel to me.

\endgroup

\vspace{2mm}
 
Identity between gauge fields and connections on fiber bundles should not be a source of marvel. What physicists know—and so what they call \emph{physics}—does not go beyond what they know with mathematics: the two are in all fused together. The «conceptual description of the physical world» is not anything other than the description of the physical world \emph{by} mathematical concepts/formulæ.
\enumerationisfinis

\subsubsection{Bell's Bafflement}

\begingroup
\footnotesize
It is when arbitrary mathematical possibilities are given equal status in this way that it becomes obscure to me that any physical interpretation has either emerged from, or been imposed on, the mathematics. \\
\indent — \textsc{J.S. Bell} \cite[pp. 137-138]{Bell "Quantum mechanics for cosmologists"}
	
\endgroup

\vspace{2mm}

Bell's bafflement is not rare for those who make theoretical physics, both in micro (quantum mechanics) and in macro (cosmology): is there a clear dividing line between imposition of mathematics (see Sections and \ref{section "Mathematics by Mathematicians vs. Mathematics by Physicists"} and \ref{section "Beauty of Mathematics vs. Mathematics of Beauty"}) on reality and reality more or less accurately disclosed by mathematics, sharply separating them?

\subsubsection{Upside Down Fourier's Judgment}
\label{subsubsection "Upside Down Fourier's Judgment"}

\begingroup
\footnotesize
In-depth study of nature is the most fertile source of mathematical discoveries [\,\dots]. Mathematical analysis has therefore necessary relations with sensible phenomena; its object is not created by human intelligence [\textit{son objet n'est point créé par l'intelligence de l'homme}], it is a pre-existing element of the universal order [\textit{élément préexistant de l'ordre universel}], and has nothing of contingent and fortuitous; it is imprinted in all nature [\textit{il est empreint dans toute la nature}] \cite[pp. xiij, 17]{Fourier "Theorie analytique de la chaleur"}\endnote{
	Original Fr. version: «L'étude approfondie de la nature est la source la plus féconde des découvertes mathématiques [\,\dots]. L'analyse mathématique a donc des rapports nécessaires avec les phénomènes sensibles; son objet n'est point créé par l'intelligence de l'homme, il est un élément préexistant de l'ordre universel, et n'a rien de contingent et de fortuit; il est empreint dans toute la nature». It is telling that the motto of this work \cite[p. 1]{Fourier "Theorie analytique de la chaleur"} is \textit{Et ignem regunt numeri}.
	} \\
\indent The differential equation expresses a relationship between functions of one or several variables, and fluxions of different orders taken in accordance with some of these variables. It is recognised that these relationships do not belong exclusively to the abstract science of calculus [\textit{ces relations n'appartiennent pas seulement à la science abstraite du calcul}]: they exist in the properties of curves and surfaces, in the motions of solids and fluids, in the distribution of heat, and in many other natural phenomena. The most general laws of the physical world are expressed by differential equations \cite[pp. 20-21]{Fourier "Analyse des equations determinees"}.\endnote{
	Original Fr. version: «L'équation différentielle est celle qui exprime une relation entre les fonctions d'une ou de plusieurs variables, et les fluxions de divers ordres prises par rapport à certaines de ces variables. On a reconnu que ces relations n'appartiennent pas seulement à la science abstraite du calcul: elles existent dans les propriétés des courbes et des surfaces, dans les mouvements des solides et des fluides, dans la distribution de la chaleur, et dans la plupart des phénomènes naturels. Les lois les plus générales du monde physique sont exprimées par des équations différentielles».
	} \\
\indent — \textsc{J.B.J. Fourier}

\endgroup

\vspace{2mm}

Fourier's judgment, as mentioned in epigraph, is exemplary, cf. \cite[pp. 145-147]{Bellone "Caos e armonia. Storia della fisica moderna e contemporanea"}. It is diametrically opposite to what we claim. Mathematics, for us,

· is a human creation,\footnote{
	Before hearing the cries of the Boeotians: it is a \textgreek{δόξα} (judgement), not a \textgreek{θεώρεμα} (theorem), so there is no a testing ground.
	}

· does not pre-exist to the universal order, because the mathematical order is our ideation,

· is not found in nature, making a proper distinction between \textgreek{τέχνη} (artifice, invention, technique) and \textgreek{φύσις} (nature of things, world, universe, or even outward form),

· has a contingent and fortuitous character, in the sense that it is historically (scilicet: randomly) determined.

Fourier's credo \cite[p. xiv]{Fourier "Theorie analytique de la chaleur"} is that «mathematical analysis is as extensive as nature itself [\textit{l'analyse mathématique est aussi étendue que la nature elle-même}]». But the opposite is true: \emph{nature}—at least for us—\emph{is as extensive as mathematical analysis}, viz. within the extension of a model.\footnote{
	By \emph{model} I mean here an “image” of reality, a “representation” of ideas (a knowledge) of any phenomenon.
	}
	
It may seem paradoxical, but from a judgment like that of Fourier comes out the obsessive and highfalutin idea that nature “speaks” in the human language of (our) numbers, although Nature mysteriously (!) precedes our math-language; except that the overall nature of the universe had the kindness to wait billions of years until the appearance of \textit{Homo mathematicus}.\footnote{
	This is reminiscent of a stinging thrust of J.S. Bell \cite[p. 117]{Bell "Quantum mechanics for cosmologists"}: «Was the world wave function waiting to jump for thousands of millions of years until a single-celled living creature appeared? Or did it have to wait a little longer for some more highly qualified measurer — with a Ph.D.?». Let us push the provocation even further: the measurement of a quantum wave function, when it comes to mathematics, depends, for all intents and purposes, on the observational/reasoning equipment of \textit{Homo physico-mathematicus}.
	} 

This whimsical idea is also deeply ingrained in many contemporary theoretical and mathematical physicists; see e.g. F.J. Dyson \cite[p. 213]{Dyson "Birds and Frogs"}: 

\vspace{2mm}

\begingroup
\footnotesize
One of the most profound jokes of nature is the square root of minus one that the physicist Erwin Schrödinger put into his wave equation when he invented wave mechanics in 1926 [\,\dots]. Starting from wave optics as a model, he wrote down a differential equation for a mechanical particle, but the equation made no sense. The equation looked like the equation of conduction of heat in a continuous medium [\,\dots]. Schrödinger's idea seemed to be going nowhere. But then came the surprise. Schrödinger put the square root of minus one into the equation, and suddenly it made sense. Suddenly it became a wave equation instead of a heat conduction equation [\,\dots]. 

All through the nineteenth century, mathematicians from Abel to Riemann and Weierstrass had been creating a magnificent theory of functions of complex variables [\,\dots]. But they always thought of complex numbers as an artificial construction, invented by human mathematicians as a useful and elegant abstraction from real life. It never entered their heads that this artificial number system that they had invented was in fact the ground on which atoms move. They never imagined that nature had got there first. 

\endgroup

\vspace{2mm}
	
Of course, we do not share such a \emph{pompously anthropocentric} opinion. May the reader permit us to make a witty remark: what does nature know about the square root of $-1$? What does an atom know of complex numbers, or complex analysis? The illiterate atom should be educated on elliptic functions (Legendre, Gauss, Abel, Jacobi), Cauchy's analysis, geometric and analytic function theories, by Riemann and Weierstrass, respectively, and much more.

\chapter{Outro—\emph{Parva Mathematica}: \emph{Libera Divagazione} \sfrac{6}{8}}
\label{chapter "Outro—Parva Mathematica: Libera Divagazione 6/8"}

\section{Mathematics in the Physical Sciences, and Nature of Reality III}
\label{section "Mathematics in the Physical Sciences, and Nature of Reality III"}

\begingroup
\footnotesize
\textgreek{Ἡ περὶ τῶν μαθηματῶν εἴδων τέχνη} · The technique of mathematical forms.\footnote{
	Cf. footnote \ref{footnote "Ideas, forms of vision, etc."} on p. \pageref{footnote "Ideas, forms of vision, etc."}. 
	} \\
\indent — Phrase created ad hoc 

\endgroup

\subsection{Mathematics as a Technical Tool}
\label{subsection "Mathematics as a Technical Tool"}

\begingroup
\footnotesize
[I]l est indispensable de distinguer entre la mathématique en tant qu'instrument [\,\dots], et l'étude de la nature, qui est une fin pour laquelle est forgé cet instrument. Le miracle de la science, c'est qu'on puisse édifier une mathématique abstraite, capable de s'appliquer ensuite avec efficacité aux lois de la nature. C'est guidé par les phénomènes naturels que le mathématicien, en fin de compte, choisit les axiomes qui donneront naissance à une théorie efficace.\footnote{
	«[I]t is essential to distinguish the mathematics  as a tool [\textit{instrument}] [\,\dots] from the study of nature, which is a purpose for which this tool is forged. The miracle of science consists of the possibility of building [\textit{édifier}] an abstract mathematics, which can then be effectively applied to the laws of nature. It is thanks to the guidance of natural phenomena that the mathematician, after all, chooses the axioms which could give rise to a powerful theory».
	} \\
\indent — \textsc{H. Cartan} \cite[p. 11]{Cartan H. "Sur le fondement logique des mathematiques"}

\vspace{2mm}

Mathematics is not the daughter of nature, but of art [\textgreek{τέχνη}].\footnote{
	\textgreek{Τέχνη}, which means “art”, is to be understood as a “set of rules”, “system/method of doing” something.
	} \\
\indent — \textsc{E. Giusti} \cite[p. 26]{Giusti "Ipotesi sulla natura degli oggetti matematici"}\footnote{
	More extensively Giusti notes \cite[pp. 26-27]{Giusti "Ipotesi sulla natura degli oggetti matematici"}: «Mathematical objects come not from the abstraction of real objects [\,\dots] but from a process of \emph{objectification of procedures}. They do not derive from an external reality, independent of man, of which they would represent the essence purged of material impurities, but formalize human activity». It is always «a process of abstraction [\,\dots] in a few invariable bits of the infinite variety of operations actually carried out; but the abstraction occurs not starting from the data of reality, but from the operations of techniques», that is, of technology as a knowledge of techniques and collection of procedures. «The same mechanism could include numbers, not abstractions from objects that do not exist [\,\dots] but objectifications of the activity of counting». 
	
	From here, he continues \cite[p. 75, e.a.]{Giusti "Ipotesi sulla natura degli oggetti matematici"}, «it also emerges in what sense one can speak of “discovery”, and not of “creation” or “invention” of mathematical objects [\,\dots]. They are \emph{first invented} as demonstrative procedures, and then \emph{later discovered} as objects mathematicians». Let us put it another way: discovery, in mathematics, is a re-elaboration of inventions in procedures of measuring and counting, which are typical of human activity, always under the influence of historical circumstances, and occasional vicissitudes.
	}

\endgroup

\vspace{2mm}

\subsubsection{How is it Possible?}
\label{subsubsection "How is it Possible?"}

\begingroup
\footnotesize
[I]t is tempting, if the only tool you have is a hammer, to treat everything as if it were a nail. \\
\indent — \textsc{A.H. Maslow} \cite[p. 15]{Maslow "The Psychology of Science: A Reconnaissance"}\endnote{
	It is a popular proverb, aka \emph{law of the instrument}, which can be a real device/tool or a state of mind. Before Maslow, it was formulated by other authors; but its purport is about the same. It reads as follows: “If the only tool you have (in your hand) is a hammer, everything will seem like a nail”.
	}

\endgroup

\vspace{2mm}

Mathematics explains little, and laboriously, about biology, except in the case of biological systems with a certain regularity, see the aforementioned (Section \ref{subsection "Necessary and Sufficient Conditions for the Chaos, and More"}) Lotka–Volterra equations \cite{Lotka "Analytical Note on Certain Rhythmic Relations in Organic Systems"} \cite{Lotka "Elements of Physical Biology"} \cite{Volterra "Fluctuations in the Abundance of a Species considered Mathematically"} \cite{Volterra "Variazioni e fluttuazioni del numero d'individui in specie animali conviventi"} \cite{Volterra "Una teoria matematica sulla lotta per l'esistenza"} \cite{Volterra D'Ancona "Les associations biologiques etudiees au point de vue mathematique"}. For this purpose, we can recall, by way of example, the daring essay by the astronomer G.V. Schiaparelli \cite{Schiaparelli "Studio Comparativo tra le Forme Organiche Naturali e le Forme Geometriche Pure"}, which ventures a comparison between «natural organic forms» and «pure geometric forms», between biological individuals and geometric (algebraic) curves,\footnote{
	Schiaparelli's reductionism in biology appears paroxysmal today, as it seems inadequate to reconstruct the full complexity of the primitive situation; even so, in addition to the honesty with which he presents, hat in hand, his «scientific hypothesis», he does no more, in principle, than some current biomedical modeling does; and there is no need to mention what happens in physics (and in mathematical physics), where reductionism is prevailing. Either way, it must not be forgotten that reductionism, even in the face of emergent phenomena, still maintains a profitable \emph{heuristic} value. Read his presentation dedicated to T. Vignoli (director of the Museo Civico di Storia Naturale in Milano), who gave Schiaparelli the incentive to write this comparative \textit{opusculo}. In it he speaks \cite[pp. 269-270]{Schiaparelli "Studio Comparativo tra le Forme Organiche Naturali e le Forme Geometriche Pure"} to us of a «persuasion that living matter could originally be ordered only in [\,\dots] four forms; just as mineral substances cannot crystallize in more than seven systems of polyhedral figures», and «the cause of such a division is to be sought in the necessary relationships of living matter with definite geometric forms of structure». «I had come to conjecture relationships between organic structures and that Geometry, which all informs the Cosmos, both in the large and in the small [\textit{congetturare relazioni fra le strutture organiche e quella Geometria, che tutto informa il Cosmo, così nel grande come nel piccolo}]. Considering the systematic ordering that reigns everywhere in the field of living beings, and the manifest correlations and connections that reveal themselves in every part, I was led to assimilate the set of organic forms to a system of pure geometric forms [\textit{assimilare l'insieme delle forme organiche ad un sistema di forme pure geometriche}] [A geometric form is said to be \emph{pure} when all its points derive from the same law, id est, from the same construction method, see ivi p. 273]. [J]ust as in a system of geometric forms the infinite variety of these [forms] derives from the variation of the parameters (or discriminatory elements) of one and the same fundamental form, so the organic types of nature (or at least of a kingdom of it) can all derive from the variations of a certain number of discriminatory elements according to a single formula or law; so that all common characters are due to the formula, all special and individual characters to the diversity of the above-named elements».
	} 
see \cite{Freguglia "Considerazioni sul modello di Giovanni V. Schiaparelli per una interpretazione geometrica delle concezioni darwiniane"}.

There is certainly a mathematics of biology, which studies living organisms, and builds models of organic structures; a notable example are the researches on the \emph{entrancing} formation flight of large and cohesive flocks of starlings (\textit{sturni vulgares}), conducted by G. Parisi's team \cite{Ballerini Cabibbo Candelier Cavagna Cisbani Giardina Lecomte Orlandi Parisi Procaccini Viale and Zdravkovic "Interaction ruling animal collective behavior depends on topological rather than metric distance: Evidence from a field study"} \cite{Ballerini Cabibbo Candelier Cavagna Cisbani Giardina Orlandi Parisi Procaccini Viale Zdravkovic "Empirical investigation of starling flocks: a benchmark study in collective animal behaviour"} \cite{Cavagna Cimarelli Giardina Parisi Santagati Stefanini and Viale "Scale-free correlations in starling flocks"} \cite{Procaccini Orlandi Cavagna Giardina Zoratto Santucci Chiarotti Hemelrijk Alleva Parisi Carere "Propagating waves in starling Sturnus vulgaris flocks under predation"}: it is, for me, a breathtaking dance. But \emph{a model is not an aberration-free mirror of reality} (see below, Section \ref{subsubsection "Math-Model in the Perseus Mythology: No Aberration-free Mirror of Reality"}), i.e., in the case of biological systems, of some organisms (there is no one-to-one correspondence between a behavior of a living organism and its bio-mathematical model). R.[M.] May \cite[p. 216, e.a.]{May "The Best Possible Time to be Alive. The Logistic Map"} candidly affirms that

\vspace{2mm}

\begingroup
\footnotesize
The models of biological communities [\,\dots] are \emph{caricatures of reality}. Just as a good caricature catches the essential truth behind the thing it is trying to depict but is forgivably vague about the unimportant details, so the most we can expect of the equations of population biology is that they capture the key points of the situation they are describing. So, for biologists studying animal populations, their equations are \emph{cartoons of reality}, \emph{not the perfect mirror images} sought by physicists. 
	
\endgroup

\vspace{2mm}
	
A model is not a representation deduced from reality. This is because, in general, mathematics does not explain reality (see below, Section \ref{subsection "Physico-mathematical Reality"}), but only what, in the knowledge, goes through technical tools (straightedge and compass, telescope, microscope, steam engine, computer, etc.). Let us not forget that mathematical knowledge itself is a technical tool: it is a \emph{technology}, i.e. an \emph{art of reasoning}.\footnote{ 
	If we wanted to adopt a noun to define the “mathematical object”, probably the choice would be \textgreek{χρῆμα} (cf. the vb. \textgreek{χράομαι}), which means a “thing” that one “needs”, something to “use”, or a “tool” for a wide range of applications.
	}

We know that algebra and geometry are a coherent apparatus of arbitrary symbols and notions, and that, by means of a phenomenological criterion (observation, induction, deduction, and empirical test, known as \textit{cimento}),\footnote{
	This is a precept whose source is Galilean, then inherited from the Newtonian tradition; it has a large following, see e.g. A.-M. Ampère \cite[p. 176]{Memoire sur la theorie mathematique des phenomenes electrodynamiques uniquement deduite de l'experience"}, according to which the recipe of mathematical physics consists of inducing general laws from phenomena and, subsequently, deducing mathematical formulæ: «Observe the facts first, while varying their circumstances as much as possible, accompany this first task with precise measurements to deduce general laws, based solely upon experience, and deduce from these laws, independently of any hypothesis on the nature of the forces that produce the phenomena, the mathematical value of these forces, namely, the formula that represents them [\textit{déduire de ces lois \textnormal{⟨}fondées sur l'expérience\textnormal{⟩} la valeur mathématique de ces forces, c'est-à-dire la formule qui les représente}], such is the course followed by Newton». By distinguishing coarsely physics from mathematics, the Newton–Ampère approach
	
	($\mathnormal{1}$) identifies the \emph{fact} with a \emph{phenomenon}; but a phenomenon is \emph{not} a fact, as the Greeks knew, it is only the result of the interaction between the observer and the observed;
	
	($\mathnormal{2}$) identifies the \emph{hypothesis} with 
	
	(i) a \emph{formulation} that in itself is \emph{empirically not testable as a fact}, and that must be \emph{rejected aprioristically}, see the famous Newton's assertion «Hypotheses non fingo» \cite[Scholium generale, p. 484]{Newton "Philosophiae Naturalis Principia Mathematica 1713"}, 
	
	(ii) or a \emph{statement without factual truth}, or even awaiting confirmation. \\
	Conversely, the hypothesis is, originally, in Greco-Hellenistic science, 
	
	($\mathnormal{1}$) a \emph{postulate}, i.e. a \emph{premise of a hypothetical-deductive system} (see Section \ref{subsubsection "Scholium: Greco-Hellenistic Scientific Modus Operandi (the Origin of Hypotheses)"}),
	
	($\mathnormal{2}$) a \emph{principle}, or a statement that is \emph{taken to be true/self-evident}, or is \emph{chosen heuristically as a starting point}, wherefore a simple supposition  serving as a foundation, literally (\textgreek{ὑπόθεσις}, from \textgreek{ὑποτίθημι}), whose beginning point lies in the perceptive sphere, for every possible theoretical argument, from which a \emph{deductive chain} follows. (A hypothesis is valid if it is possible to derive the observed phenomena from it). For more on this topic, see L. Russo \cite[cap. XIV]{Russo "Perche la cultura classica. La risposta di un non classicista"}.
	} 
or of a construction of adequate and plausible models of our  sense perception (what we see, hear, etc.),\footnote{
	See e.g. J. von Neumann \cite[p. 492]{Neumann "Method in the Physical Sciences"}: «[T]he sciences do not try to explain, they hardly even try to interpret, they mainly make models. By a model is meant a mathematical construct which, with the addition of certain verbal interpretations, describes observed phenomena. The justification of such a mathematical construct is solely and precisely that it is expected to work—that is, correctly to describe phenomena from a reasonably wide area. Furthermore, it must satisfy certain esthetic criteria—that is, in relation to how much it describes, it must be rather simple. I think it is worth while insisting on these vague terms—for instance, on the use of the word \emph{rather}. One cannot tell exactly how “simple” simple is» (cf. Hertz in footnote \ref{footnote "Hertz's Preface"}, p. \pageref{footnote "Hertz's Preface"}, for a same pronouncement). This is because “simple” is not an absolute concept but a relative one, and such vagueness is reflected in the construction of a model.
	} 
they can be applied to real objects, so as to have some coherent results on the way nature (the real world) operates. 

Why it happens? How is it possible? This applicability is not a mystery; this is possible because abstract numbers and postulates involving the properties of space (intuitively, or within a more formal axiomatic system) are \emph{our technical tools}, which, in addition to being used for the solution of practical problems, also provide at the same time the theoretical framework for understanding \emph{that solution}. So, au fond, mathematics is effective because it is about technology, the development of which is based on mathematical theories, being themselves a technology (or, better yet, the first technology). A Euclid, or even a Mandelbrot, are of little use in phyllotaxis (see Section \ref{subsubsection "Phyllotaxis: From Helianthus Annuus to Muscari Comosum}) because a plant is not a creation—a technology—of ours, but of nature, and it is much more complex, or much less elementary, than a photon, or an electron.

\subsubsection{Degree of Objectivity of Mathematics}
\label{subsubsection "Degree of Objectivity of Mathematics"}

There is a way to summarize the above, by saying that the \emph{degree of objectivity of mathematics is much lower than that of botany}: no flower is born from the head of a botanist, like Athena was born from the head of Jupiter, whilst this turns up for a proof of a theorem, because mathematics is our figment, whereas a flower is not.\footnote{
	Compare with P.W. Bridgman \cite[p. 78]{Bridgman "The Way Things Are"}: «[T]he matter of proof bulks so large for me in the enterprise of mathematics, and proof is so completely a personal matter, which cannot be communicated, that I would put mathematics on a level of lower “objectivity” than physics or chemistry. Mathematics is peculiarly and exclusively a human enterprise [\,\dots]. Logic is in much the same situation as mathematics [\,\dots]. The biological sciences may perhaps be put on the next level beyond classical physics, chemistry, and so forth».
	}

The same applies to everything else: it is much easier to solve the Poincaré conjecture (Section \ref{subsection "Poincaré Conjecture"}) than to predict the behavior of a butterfly—in front of me, hic et nunc, on a blossoming apricot branch—in the next five minutes.

As J.T. Schwartz \cite[pp. 21-22, e.a.]{Schwartz "The Pernicious Influence of Mathematics on Science"} (who happens to be a mathematician) picks out:

\vspace{2mm} 

\begingroup
\footnotesize
Mathematics [one of whose distinguishing features is its single-mindedness] is able to deal successfully only with the simplest of situations, more precisely, with a complex situation only to the extent that rare good fortune makes this complex situation hinge upon a few dominant simple factors. Beyond the well-traversed path, mathematics loses its bearings in a \emph{jungle} of unnamed special functions and impenetrable combinatorial particularities. Thus, the mathematical technique can only reach far if it starts from a point close to the simple essentials of a problem which has simple essentials. That form of wisdom which is the opposite of single-mindedness, the ability to keep many threads in hand, to draw for an argument from many disparate sources, is quite foreign to mathematics. 

\endgroup

\vspace{2mm} 

The old theme of the “obscure wood” is back, and of the “right path”,\footnote{
	If this Schwartz's argumentation is taken to the extreme, we can go even further, and assert that there is no real method (from the Gr. \textgreek{μέθοδος}, composed of \textgreek{μετα-}, incorporating the idea of “pursuing”, of “seeking”, and \textgreek{ὁδός}, “path”) in mathematics.
	}
which can be swiftly lost.

\subsubsection{Math-Model in the Perseus Mythology: No Aberration-free Mirror of Reality}
\label{subsubsection "Math-Model in the Perseus Mythology: No Aberration-free Mirror of Reality"}

Let us indulge in the Perseus mythology, at the suggestion of G. Lolli \cite[pp. 68-70]{Lolli "Discorso sulla matematica. Una rilettura delle Lezioni americane di Italo Calvino"}, on the stimulus of a re-reading of I. Calvino's \textit{Six Memos for the Next Millennium} \cite[pp. 4-7]{Calvino "Six Memos for the Next Millennium"}. This is a genuinely inspiring \textgreek{μῦθος} (“myth”, “story”, “fiction”). We cannot face reality, or be there vis-à-vis with the world; similarly to Medusa, the hideous and monstrous Gorgon with the \textit{anguiferumque caput}, the reality of physical entities/quantities, the world out there, is a mutable complex that petrifies, is chaos (\textgreek{χάος}, “infinite abyss”). Perseus does not turn his gaze on the face of Medusa but only on her image reflected in the bronze shield. Mathematics does the same: it works through an \emph{indirect vision}. The reflecting shield is a metaphor for the \emph{(mathematical) representation} of the external reality. Mathematics is an \emph{indirect thought} of the world. This is why it is said to be a \emph{model} (cf. point \ref{item "Bertuglia and Vaio passage"} in Section \ref{subsection "Bohrism in a Right Perspective"}). To put it jokingly, there are no magic formulæ, in the mathematical field, to break the spell of petrification.
	
In so doing, we may mention a name of Antiquity, on the back of the Perseusian mythology: Geminus of Rhodes \cite[cap. I, § 23, p. 12]{Geminus of Rhodes "Introduction to the Phenomena"}, in his \textgreek{Εἰσαγωγὴ}, makes a clear distinction between \emph{physical reality} of celestial motions, which belongs to the investigations of physicists, and \emph{mathematical model/artifice}, which belongs to astronomers-mathematicians.
	
On the contrary, whoever embraces a Platonic-Galilean scheme, under which Nature is written in (a) mathematical language (cf. Section \ref{subsection "In the Bliss of Goddess Geometry"}), this distinction is bogus; see Galileo \cite[p. 49]{Galilei "Diversi fragmenti attenenti al trattato delle cose che stanno su l'acqua"}: 

\vspace{2mm}

\begingroup
\footnotesize
Here I expect a terrible rebuff from some of the adversaries; and I already seem to hear intoning in my ears that it is one thing to deal with things physically but quite another mathematically, and that geometers must remain among their windmills, and not join together with the philosophical subject-matters [of nature], whose truths are different from mathematical truths; almost as if the truth could be more than one; almost as if geometry in our days prejudices the acquisition of true philosophy [of nature], as if it were impossible to be [simultaneously] a geometer and a philosopher [of nature, that is, a physicist], so that by a necessary consequence it is inferred that who knows geometry cannot know physics, nor can he discuss and deal with physical subject-matters physically.\endnote{
	Original It. version: «Qua io m'aspetto un rabbuffo terribile da qualcuno de gli avversarii; e già parmi di sentire intonar negli orecchi che altro è il trattar le cose fisicamente ed altro matematicamente, e che i geometri doveriano restar tra le lor girandole, e non affratellarsi con le materie filosofiche [della natura], le cui verità sono diverse dalle verità matematiche; quasi che il vero possa esser più di uno; quasi che la geometria a i nostri tempi progiudichi all'aqquisto della vera filosofia [della natura], quasi che sia impossibile esser geometra e filosofo [della natura, ossia fisico], sì che per necessaria conseguenz[a] si inferisca che chi sa geometria non possa saper fisica, nè possa discorrere e trattar delle materie fisiche fisicamente».
	} 

\endgroup

\subsubsection{What Myth Does Mathematics Tell Us? The Example of the Stefan Problem (Ice-Water Phase-boundary)}
\label{subsubsection "What Myth Does Mathematics Tell Us? The Example of the Stefan Problem (Ice-Water Phase-boundary)"}

\begingroup
\footnotesize
There are more things in heaven and earth [\,\dots] / Than are dreamt of in your [mathematics].\footnote{
	In the original Hamletian tragedy, the word “philosophy” appears instead of “mathematics”.
	}\textsuperscript{,}\footnote{
	A petty curiosity. G. Toraldo di Francia in \cite[p. 7]{Toraldo di Francia "Un universo troppo semplice. La visione storica e la visione scientifica del mondo"} overturns Hamlet's assertion, and he writes: «There are fewer things in heaven and earth than are dreamt of in your philosophy». I cannot but reject Toraldo di Francia's position; but I can excuse him, cf. footnote \ref{footnote "Longinus, Novalis"}, p. \pageref{footnote "Longinus, Novalis"}. The mind of a physicist, who copes mostly with—highly schematized—elementary laws, is much simpler than that of a mathematician or, a fortiori, of a plant \& animal biologist.

	Let me be clear: Toraldo di Francia's cogitations make a lot of sense, and many of his arguments are worthy of support, such as the survey on the “possible worlds” of the scientific imagination, on the “virtualities” of the human fancy, which lay upon the “objective” factuality of the event.\endnote{
	Personal note. His description \cite[pp. 48-56]{Toraldo di Francia "Un universo troppo semplice. La visione storica e la visione scientifica del mondo"} of the imagination, always ready to \emph{run wild}, while the gaze is turned out of the window during a train journey, corresponds to my experience. The contemplation of the landscape—when I travel alone by train—is my second, copious, fount of inspiration; the first one is the Morpheusian trip in the state of half-sleep (cf. Intro, p. \pageref{subsection "Through the Magnifier of Half-sleep"}).
	}
	What I cannot accept is the association of the mechanisms of nature with a «simple childish game». Nature is not within our imagination, or in any mechanism \textit{ad formam hominis}. If, in the biological field, the “instruction book” of life contains simple instructions \cite[pp. 111-122]{Toraldo di Francia "Un universo troppo semplice. La visione storica e la visione scientifica del mondo"}, or if there is an imposition of the same solutions for the same problems, does this mean that the universe itself is «strangely simple»? Nope. And yet he is well-aware of all this; read what he writes here \cite[p. 57]{Toraldo di Francia "L'indagine del mondo fisico"}: «It is not easy to say what \textit{simpler} means [\,\dots]. The reference to the simplicity of the \textit{calculation} is a gross anthropomorphism; nature makes no calculations» (cf. ivi, p. 62), or here \cite[p. 533, where the original Hamletian assertion appears]{Toraldo di Francia "L'indagine del mondo fisico"}: «The universe is something much deeper to study and understand than a simple set of particles moving according to certain laws».
	} \\
\indent — \textsc{W. Shakespeare}'s amended sentence \cite[\textit{Hamlet}, Act I, Sc. V, p. 504]{Shakespeare "Shakespeare's Comedies Histories Tragedies and Poems"} 

\endgroup

\vspace{2mm}

\enumerationisinitium
\item The key thing that must be understood is that when mathematics attempts to investigate natural phenomena, in the physical world, it does not explain the nature of them, but \emph{explains to itself}—with an artificial language, invented ad hoc—how it is possible to arrive at a \emph{coherent description} of this or that real phenomenon.
\item Take the \emph{Stefan problem} \cite{Stefan "Ueber die Theorie der Eisbildung insbesondere uber die Eisbildung im Polarmeere}, which is a \emph{free boundary problem} between the phases transitions of a substance having a phase change: ice melting to water is the most common case, where the ice is immersed in the water contained in a beaker. Thence one has a function of space and time (unknown evolutionary datum), coinciding with the \emph{temperature distribution} of the water, and the \emph{moving boundary}, i.e. the ice-water interface (other unknown datum): two spatial regions occupy two different physical states of matter of water, liquid and solid. 
 
Let us proceed with some illustrations.

The 2-phase Stefan problem, for a linear parabolic equation, is written as
\begin{subequations}
\begin{align}
	& (\upsilon_1)_{xx} + \reflectedepsilon^2_1(x, t)(\upsilon_1)_x + \reflectedepsilon^3_1(x, t)\upsilon_1 - \reflectedepsilon^1_1(x, t)(\upsilon_1)_t = 0, \enspace -\infty < x < y(t), \\
	& (\upsilon_2)_{xx} + \reflectedepsilon^2_2(x, t)(\upsilon_2)_x + \reflectedepsilon^3_2(x, t)\upsilon_2 - \reflectedepsilon^1_2(x, t)(\upsilon_2)_t = 0, \enspace y(t) < x < +\infty, \\
	& \upsilon_1(x, 0) = \varphi_1(x), \enspace -\infty < x < 0, \\
	& \upsilon_2(x, 0) = \varphi_2(x), \enspace 0 < x < +\infty, \\
	& \upsilon_1\bigl(y(t), t\bigr) = \upsilon_2\bigl(y(t), t\bigr) = 0,
\end{align}
\end{subequations}
with $\reflectedepsilon$-constants, for $t > 0$. The expression $\dot{y}(t)$ is equal to 
\begin{equation}
	(\upsilon_1)_x\bigl(y(t), t\bigr) - (\upsilon_2)_x\bigl(y(t), t\bigr),
\end{equation}
and
\begin{equation}
	y(0) = 0.\footnotemark
\end{equation}
\footnotetext{
	If we place $\upsilon(x, t)$ and $y(t)$, by setting $0 < x < y(t)$, for $t > 0$, the 2-phase Stefan problem becomes \cite[sup. III]{Rubinstein "The Stefan Problem"}
	\begin{subequations}
	\begin{align}
		& \frac{\partial^2\upsilon}{\partial x^2} = \frac{\partial\upsilon}{\partial t}, \\
		& \upsilon|_{x = 0} = -1, \enspace \upsilon|_{t = 0} = 4\bigl(x - y(0)\bigr), \enspace \upsilon|_{x = y(t)} = 0,
	\end{align}
	\end{subequations}
	where $\upsilon = \frac{\partial}{\partial x}\upsilon\bigl(y(t), t\bigr)$ is the temperature, and the value of $\upsilon$ at the boundary $x = 0$ is equal to $-1$, as the expression $\dot{y}(t)$ is equal to $\upsilon(t)$.
	}

The study of the regularity of the free boundary problems, and thus the formation of \emph{singularities}, videlicet, of \emph{icy cuspidate points}, in a block of ice along its progressive melting at 0 \textcelsius{},\footnote{
	Attention: the freezing point depends on the pressure level; at atmospheric pressure, water freezes at 0 \textcelsius{}. But under 20~000 atmospheres, water freezes at 75 \textcelsius{}.
	} 
are conventional part of research in the Stefan problem; see L.A. Caffarelli \cite{Caffarelli "The regularity of free boundaries in higher dimensions"} \cite{Caffarelli "Some Aspects of the One-Phase Stefan Problem"}, and C. \& A. Friedman \cite{Caffarelli and Friedman "Continuity of the Temperature in the Stefan Problem"}. Icy singularities are interlinked through the so-called \emph{parabolic obstacle problem}. 

Let 
\begin{equation}
	\upsilon[\mathsf{T}] = \upsilon[\mathsf{T}](x, t) 
\end{equation}
be a function of the temperature distribution of the water in which a block of ice is immersed, at some point $x \in \Omega \subset \mathbb{R}^n$ and at time $t \in \mathbb{R}_+$. If $\upsilon[\mathsf{T}] \geqslant 0$ in $\mathbb{R}_+ \times \Omega$, then 
\begin{align*}
	& \{\upsilon[\mathsf{T}] = 0\} \text{ is the ice region}, \\
	& \{\upsilon[\mathsf{T}] > 0\} \text{ is the water region}.
\end{align*}

Given

· an initial condition $\upsilon[\mathsf{T}](x, 0) \geqslant 0$, at an initial time $t = 0$, 

· a boundary condition $\upsilon[\mathsf{T}] = \upsilon[\mathsf{T}](x, t) \geqslant 0$, for $x \in \partial\Omega$ and $t \geqslant 0$, \\
the temperature evolution of the water follows the heat equation (cf. Chapter \ref{chapter "The Ricci Flow, or the Hamilton–Perelman Metric Evolution Machinery"})
\begin{subequations}
\begin{align}
	& \partial_t\upsilon = \Laplacian\upsilon, \\
	& \partial_t\upsilon - \Laplacian\upsilon = 0,
\end{align}
\end{subequations}
in $\{\upsilon[\mathsf{T}] > 0\}$. The moving boundary, or the ice-water interface, fulfills the condition
\begin{equation}
	\dot{x}(t) = -\nabla\upsilon\bigl(x(t), t\bigr),
\end{equation}
for any $x(t) \in \partial\{\upsilon(t) > 0\}$, $\partial\{\upsilon(t) > 0\}$ being the free boundary, and $\nabla\upsilon$ the gradient of $\upsilon(t)$. Putting 
\begin{equation}
	\textgreek{\text{υ}}(x, t) = \int^t_0\upsilon(x, y)dy
\end{equation}
as a \textit{Baiocchi–Duvaut transformation} \cite{Baiocchi "Sur un probleme a frontiere libre traduisant le filtrage de liquides a travers des milieux poreux"} \cite{Duvaut "Resolution d'un probleme de Stefan (Fusion d'un bloc de glace a zero degre)"} \cite{Baiocchi "Free Boundary Problmes in the Theory of Fluid Flow Through Porous Media"}, and denoting by $\textgreek{\textit{χ}}$ a characteristic function, the parabolic obstacle problem—which is \emph{locally equivalent} to the Stefan problem—is
\begin{subequations}
\begin{align}
	& \partial_t\textgreek{\text{υ}} = \Laplacian\textgreek{\text{υ}} - \textgreek{\textit{χ}}_{\{\textgreek{\text{υ}} > 0\}}, \\
	& \textgreek{\text{υ}} \geqslant 0, \\
	& \partial_t\textgreek{\text{υ}} \geqslant 0,
\end{align}
\end{subequations}
admitting a function 
\begin{equation}
	\textgreek{\text{υ}} \colon \mathbb{R}_+ \times \Omega \to \mathbb{R}. 
\end{equation}

As Caffarelli demonstrated in the above-mentioned papers, 
\begin{align*}
	\begin{cases}
	\mathscr{C}^{0, 1} \text{ in time}, \\
	\mathscr{C}^{1, 1} \text{ in space}, 
	\end{cases}
\end{align*}
are solutions to the parabolic obstacle, thereby showing that there is a regularity, at least in this $\Lebesgue^\infty_\mathrm{loc}$-scenario, of the free boundary for the Stefan problem.

Subsequently, the existence of \emph{complete regularities}, conforming to the appearance and the sudden \emph{disappearance} of icy singularities, was proved by A. Figalli, X. Ros-Oton, and J. Serra \cite{Figalli Ros-Oton Serra "The singular set in the Stefan problem"}, by digging out the Almgren-type \emph{branch sets} \cite{Almgren "Almgren's Big Regularity Paper"} for singularities in area minimizing surfaces, with these results:

· the parabolic Hausdorff dimension of a singular set is $\mathrm{D_f} \atmo n - 1$,

· there exists a $\mathscr{C}^\infty$-type regularity for a $\textgreek{\text{υ}}$-like function (under the Baiocchi–Duvaut transformation), or an expansion of $\mathscr{C}^\infty$ at all singular points, anent a set of parabolic Hausdorff dimension $\mathrm{D_f} \atmo n - 2$,

· the free boundary $\partial\{\textgreek{\text{υ}}(t) > 0\}$, the region of separation between ice and water, in $\mathbb{R}^3$ is smooth for almost every time $t$, where a set of singular times has Hausdorff dimension $\mathrm{D_f} \atmo \frac{1}{2}$.
\item Let us go back to the considerations at the opening of this Section. The mathematical demonstration does not touch physical reality, but concerns only the \emph{internal fidelity} of the equations (logical consistency) as compared to what we see with our eyes, and process with our mind, in the world outside us. That is to say: what mathematics proves is not the physical melting behavior of the ice into water. Maybe the opposite is true, in the sense that mathematics is limited to demonstrating, internally, that the behavior of ice can be \emph{consistent} within a set of equations, whilst the ice behavior has no special relationship to these equations, any more than it does to a fictional or poetic description of the physical world.

I find it \emph{staggering} that a bunch of mathematicians could seriously believe that some string of abstract symbols actually have to do with the chemico-physical behavior of a piece of ice floating in the water. If there is an actuality, it is all in our head, in the mathematical \textgreek{μῦθος} (see Section \ref{subsubsection "Math-Model in the Perseus Mythology: No Aberration-free Mirror of Reality"}) that we tell ourselves, not in the ice-water phase transition.
\enumerationisfinis

\subsubsection[Gromov Conjecture on the Flexibility and Rigidity of $\mathscr{C}^{1, \alpha = \frac{1}{2}}$]{Gromov Conjecture on $\protect\pseudobold{\mathscr{C}^{1, \alpha = \frac{1}{2}}}$ (Flexibility and Rigidity)}

\begingroup
\footnotesize
The terrain of isometric embeddings and the fields surrounding this terrain are vast and craggy with valleys separated by ridges of unreachable mountains; people cultivating their personal gardens in these “valleys” only vaguely aware of what happens away from their domains and the authors of general accounts on isometric embeddings have a limited acquaintance with the original papers. \\
\indent — \textsc{M.[L.] Gromov} \cite[p. 173]{Gromov "Geometric algebraic and analytic descendants of Nash isometric embedding theorems} 

\endgroup

\vspace{2mm}

The above problem (Section \ref{subsubsection "What Myth Does Mathematics Tell Us? The Example of the Stefan Problem (Ice-Water Phase-boundary)"}) does not arise in the territory of pure mathematics, at least as long as it remains pure, because there is no need for a comparison with reality, which slaps any theory. I choose one example among many: Gromov conjecture on $\mathscr{C}^{1, \frac{1}{2}}$ (criticality of the exponent $\frac{1}{2}$) \cite{Gromov "Geometric algebraic and analytic descendants of Nash isometric embedding theorems}.

\vspace{2mm}

\begin{coniectura}[Gromov conjecture]
\label{coniectura "Gromov conjecture"}
We are in the field of Nashian isometric embeddings (cf. Section \ref{section "On Nash's Embeddings: (Curved) Spaces in Euclidean Spaces"}). Suppose we add curves to a sphere, and assume that it is possible to add an infinite number of twists—Nash embeddings twisted  \textnormal{\cite{Nash "The Imbedding Problem for Riemannian Manifolds"} \cite{Nash "Analyticity of the solutions of implicit function problems with analytic data"}}, to be fair—to these curves. To what extent is it possible to crumple a sphere down to an $n$-ball without creasing or tearing it? Or: what is the $\mathscr{C}$-solution, as a numerical limit, between flatness/smoothness and twistedness/tortuosity, under which a sphere can be crumpled, without losing the preservation of its lengths?

The Gromov threshold, according to a procedure called \emph{convex integration}, is estimated to be $\mathscr{C}^{1, \alpha}$, for $\alpha = \frac{1}{2}$.
\end{coniectura}

The conjecture \ref{coniectura "Gromov conjecture"} was corroborated by C. De Lellis and D. Inauen \cite{De Lellis Inauen "C1 alpha Isometric Embeddings of Polar Caps"}, who note that, for $\alpha > \frac{1}{2}$, the Levi-Civita connection (Section \ref{subsection "Levi-Civita Connection Theorem on a (pseudo-)Riemannian Manifold"}) of all isometric immersions is induced by the Euclidean connection, whilst, for any $\alpha < \frac{1}{2}$, a standard 2-sphere does not retain this property. Shortly thereafter, the conjecture was generalized by W. Cao and D. Inauen \cite{Cao Inauen "Rigidity and Flexibility of Isometric Extensions"}. The Hölder space (cf. Section \ref{subsection "Hölder Continuity of Subspaces in a Map with the Anosov Property"}) $\mathscr{C}^{1, \alpha = \frac{1}{2}}$ is actually the critical value under which a topological space in $\mathbb{R}^N$ begin to crease; to wit, we are talking about the quantity between \emph{flexibility} and \emph{rigidity} of $\mathscr{C}^{1,\frac{1}{2}}$ isometric extensions.

Here the decisive intuition consists in making use of the physico-mathematical theory of turbulence, or the mathematical fluid dynamics, to verify where the flow in the manifold becomes turbulent; in such a case, the topological space is led to deform. This is related to Onsager's conjecture \cite{Onsager "Statistical Hydrodynamics"} (see Margo \ref{margo "Onsager's conjecture"}), proved by P. Isett \cite{Isett "A proof of Onsager's conjecture"} and T. Buckmaster, De Lellis, L. Székelyhidi Jr., \& V. Vicol \cite{Buckmaster De Lellis  Szekelyhidi Jr. and V. Vicol "Onsager's Conjecture for Admissible Weak Solutions"}.

Read, in parallel, the almost lyrical language in Gromov's epigraph, to underline that, even when one works in the abstract, the references to the \emph{physicality of the images} (“vast and craggy terrain”, “valleys” separated by “ridges” of “mountains”) are inevitable.

\begin{margo}[Onsager's conjecture]
\label{margo "Onsager's conjecture"}
This conjecture \cite{Onsager "Statistical Hydrodynamics"} is about the statistical hydrodynamics: the threshold for the validity of the energy conservation of a weak solutions of incompressible Euler equations,
\begin{align}
\begin{cases}
	\partial_t\upsilon + \upsilon \cdot \nabla\upsilon + \nabla\mathsf{P} = 0, \\
	\divergence\upsilon = 0, 
\end{cases}
\end{align}
in the periodic setting 
\[
	\torus^3 = \mathbb{R}^3 \backslash \mathbb{Z}^3,
\]
corresponds to the exponent $\alpha = \frac{1}{3}$, where $\upsilon$ is a vector field symbolizing the velocity of the fluid, viz. is a Hölder-continuous weak solution (to the Euler equations in which both viscosity and compressibility in the fluid tend to zero), and $\mathsf{P}$ is the pressure. \margosymbol
\end{margo}

\subsubsection{The Unabating Tension: Mathematics vs. Nature}

In Section \ref{subsubsection "How is it Possible?"} we had the opportunity to  invoke the so-called law of the instrument (“If the only tool you have—in your hand—is a hammer, everything will seem like a nail”), and to talk about mathematics as a \textgreek{τέχνημα}. In Section \ref{subsubsection "Degree of Objectivity of Mathematics"} we stressed the simplification techniques typical of the mathematical language. In Section \ref{subsubsection "Math-Model in the Perseus Mythology: No Aberration-free Mirror of Reality"} we compared the technique of mathematics to a distorting mirror. In Section \ref{subsubsection "What Myth Does Mathematics Tell Us? The Example of the Stefan Problem (Ice-Water Phase-boundary)"} we saw the mythos behind all this at work. There are two paragraphs from E. Giusti \cite[pp. 36-37, e.a.]{Giusti "Euclides reformatus. La teoria delle proporzioni nella scuola galileiana"} that deserve to be read in full, as they give a synthesis of each of these distinctive traits:

\vspace{2mm}

\begingroup
\footnotesize
The choice of a mathematical language [\,\dots] is not without consequences in the description of the world, and in the very image forming in the scientist's mind, with the concepts used: to the teeming multiplicity of real bodies that move, weigh, and balance each other, the mathematical physicist \emph{replaces} a \emph{crystallized universe} of invariable figures, quantities, motions, in which the infinite variety of things is replaced by a \emph{systematic game} of \emph{simple relations}. And if it is true that the physicist chooses one or another mathematical theory in relation to their adherence to reality that should be studied [\,\dots], it is no less true that, once a decision has been made, [a mathematical theory] substantially influences [\textit{condiziona}] the understanding of the phenomena [\,\dots]. [W]hen [a] theory [\,\dots] has been chosen for the description of physical phenomena, the only possible relationships between natural bodies are those that the theory foresees between abstract quantities, and it is on these [quantities] that our images of nature must be modeled, with all possible resulting \emph{distortions}. The aphorism “translator, traitor” does not apply only in literature.

If physics [our description of nature] is in a certain sense crystallized by the underlying mathematical structure (but, for the avoidance of doubt, we repeat that this is the only way to get out of the [sense of] “wonder” [\textgreek{θαυμασιότης}] and move towards the understanding of natural phenomena), and it becomes, so to speak, a “model” of the mathematical theory adopted, however, it does not remain inert, and it does not fail to submit its needs, especially when the simplifications introduced prevent an adequate understanding of the phenomena being considered. At this point we observe a symmetrical \emph{action of the physical world on the language} that describes it. If mathematics had claimed to \emph{freeze} the universe in a \emph{simple game} of \emph{abstract concepts}, by \emph{excluding} everything that was not attributable to relationships between [such concepts], there are natheless some questions, which indispensably are calling for an answer, and which cannot be circumvented only because they are not formulated in the \emph{chosen language}. The same mathematical concepts on which the basic theory was founded are thus brought to a \emph{ceaseless tension}, in an attempt to \emph{force} the interpretative framework, and to assume a broader one, in which to pose and possibly solve the problems whose enunciation in the previous [framework] was prevented. And when a more general theory is not available, the [physicist] becomes a mathematician, and engages with the rigidity of his method.

\endgroup

\subsection{Physico-mathematical Reality}
\label{subsection "Physico-mathematical Reality"}

\begingroup
\footnotesize
\textgreek{Ἐζητεῖτο δὲ καὶ παρὰ τοῖς γεωμέτραις, τίνα ἄν τις τρόπον τὸ δοθὲν στερεὸν διαμένον ἐν τῷ αὐτῷ σχήματι διπλασιάσειεν}.\footnote{
	«In what manner one might double a given solid, [the solid] keeping the same shape, it became a subject of investigation among geometers».
	} \\
\indent — \textsc{Eutocius} of Ascalon \cite[p. 104, 6-8]{Eutocius of Ascalon "Commentarium de sphaera et cylindro"}
 
\vspace{2mm}

Eadem mutata resurgo.\footnote{
	«I rise the same, though changed», with reference to the logarithmic spiral (\textit{spira mirabilis}); except that the engraver has carved an Archimedean spiral \cite[see e.g. \textgreek{ιβ´, ιϛ´, ιζ´, κα´, κε´}]{Archimedes "De lineis spiralibus"}.
	} \\	
\indent — \textsc{Jac. Bernoulli}'s tomb inscription in the Minster of Basel

\endgroup

\vspace{2mm}

If mathematics is the study of ideal constructions, together with the imagination of geometric shapes, and the understanding of numerical relations, then its study coincides with the pursuit of invariance, to wit, of identity (equality), and of a permanent character along the transformations of such forms and relations. The same type of pursuit is, in consequence, poured into the study of natural phenomena (when it expresses its results through mathematics).

A mathematician, for his part, has a lot of fun provoking the physics community. See e.g. G.H. Hardy's \cite[§ 24, pp. 128-129]{Hardy "A Mathematician's Apology"} annotation: 

\vspace{2mm}

\begingroup
\footnotesize
[I]t is the physicist who deals with the subject-matter usually described as “real”; but a very little reflection is enough to show that the physicist's reality, whatever it may be, has few or none of the attributes which common sense ascribes instinctively to reality. A chair may be a collection of whirling electrons [but this definition does not conform] at all closely to the suggestions of common sense. [No physicist has] ever given any convincing account of what “physical reality” is, or of how the physicist passes, from the confused mass of fact or sensation with which he starts, to the construction of the objects which he calls “real”. Thus we cannot be said to know what the subject-matter of physics is.\endnote{
	In light of this, the Tagore–Einstein conversation \cite[pp. 42-43]{Tagore and Einstein "Nature of Reality"} on the nature of reality is riveting, especially in the context of the quantum measurement problem:

	\setlength\parindent{8pt}
	A. Einstein: «There are two different conceptions about the nature of the Universe: ($\mathnormal{1}$) The world as a unity dependent on humanity, ($\mathnormal{2}$) the world as a reality independent of the human factor». 

	R. Tagore: «This world is a human world—the scientific view of it is also that of the scientific man. Therefore, the world apart from us does not exist; it is a relative world, depending for its reality upon our consciousness».

	A. Einstein: «If there is a \emph{reality} independent of man, there is also a truth relative to this reality; and in the same way the negation of the first engenders a negation of the existence of the latter [\,\dots]. [I]n our everyday life we feel compelled to ascribe a reality independent of man to the objects we use. We do this to connect the experiences of our senses in a reasonable way. For instance, if nobody is in this house, yet that table remains where it is [\,\dots]. If nobody would be in the house the table would exist all the same».

	R. Tagore: «Science has proved that the table as a solid object is an appearance and therefore that which the human mind perceives as a table would not exist if that mind were naught. At the same time it must be admitted that the fact that the ultimate physical reality is nothing but a multitude of separate revolving centres of electric force, also belongs to the human mind [\,\dots]. There is the reality of paper, infinitely different from the reality of literature. For the kind of mind possessed by the moth that eats that paper, literature is absolutely non-existent, yet for Man's mind, literature has a greater value of truth than the paper itself. In a similar manner, if there be some truth which has no sensuous or rational relation to human mind, it will ever remain as nothing so long as we remain human beings».

	A. Einstein: «Then I am more religious than you are!».
	
	The same stance is presented by Einstein in \cite[p. 274, e.a.]{Einstein "Ideas and Opinions"}: «If it is true that the axiomatic basis of theoretical physics cannot be extracted from experience but must be \emph{freely invented}, can we ever hope to find the right way? Nay, more, has this right way any existence outside our illusions? [\,\dots]. I answer without hesitation that there is, in my opinion, a right way, and that we are capable of finding it. Our experience hitherto justifies us in believing that nature is the realization of the simplest conceivable mathematical ideas. I am convinced that we can discover by means of purely mathematical constructions the concepts and the laws connecting them with each other, which furnish the key to the understanding of natural phenomena [\,\dots]. Experience remains, of course, the sole criterion of the physical utility of a mathematical construction. But the \emph{creative} principle resides in mathematics. In a certain sense, therefore, I hold it true that pure thought can grasp reality». 
	
	We agree with the viewpoint that mathematics is a creative principle, but we reject the persuasion that mathematics, or pure thought, can grasp reality, if this means “reality \emph{in its entirety}”, wherefore we reject, as a sort of \emph{idolatry} (cf. endnote \ref{endnote "Bacon's idola"}), the conviction that nature is the «realization» of (the simplest conceivable) mathematical ideas. No, it is just the opposite: (the simplest conceivable) mathematical ideas “realize”—in the sense that they \emph{represent}, or \emph{reproduce}, with all the limitations that this involves—the idea of nature, or rather, of small pieces of nature.
	} 

\endgroup

\vspace{2mm}

This is not new, and it should not be surprising; this has distant roots, in the famous Galilean action of defalcating the impediments of matter \cite[it is required that the scientist-geometer «difalchi gli impedimenti della materia», p. 202]{Galilei "Dialogo sopra i due Massimi Sistemi del Mondo Tolemaico e Copernicano"},\endnote{
	\label{endnote "Difalcare gli impedimenti della materia"}
	The passage \cite[pp. 198-203]{Galilei "Dialogo sopra i due Massimi Sistemi del Mondo Tolemaico e Copernicano"} containing the before-mentioned phrase is of capital importance; for this, we have an obligation, blended with pleasure (in the opinion of I. Calvino, Galileo is the greatest Italian prose writer), to recopy it in a large part, by attempting to make a translation, under the requirement to remain, as closely as possible, faithful to the original text.

	\setlength\parindent{8pt}
	Simp.: «[\,\dots] perchè finalmente queste sottigliezze mattematiche Sign. Salv.[iati] son vere in astratto, ma applicate alla materia sensibile, e fisica, non rispondono; perchè dimostrerranno ben'i mattematici con i lor principij, per esempio, che \textit{Sphęra tangit planum in puncto} [\,\dots]; ma come si viene alla materia, le cose vanno per un'altro verso; e così voglio dire di quest'angoli del contatto, e di queste proporzioni; che tutte poi vanno a monte, quando si viene alle cose materiali, e sensibili».
	
	Simp.: «[\,\dots] because finally these mathematical subtleties Sig. Salv.[iati] are true in the abstract, but applied to sensible, and physical, matter, they do not [cor]respond; because mathematicians will well demonstrate with their principles, for example, that \textit{Sphęra tangit planum in puncto} [\,\dots]; but as soon as one comes to the matter, things go in another way; and so I may say of these angles of contact, and of these proportions; that everything goes awry, when you have to deal with material, and sensible things».
	
	Salv.: «[\,\dots] Hor per mostrarvi quanto sia grande l'error di coloro, che dicono, che una sfera v.g. di brŏzo non tocca un piano v.g. d'acciaio in un punto; Ditemi qual concetto voi vi formeresti di uno, che dicesse, e costantemente asseverasse, che la sfera non fusse veramente sfera?».
	
	Salv.: «[\,\dots] Now to show you how great is the error of those, who say that v.g. a bronze sphere does not touch v.g. a steel plane at a [single] point; Tell me what [idea] you would have of one, who should say, and constantly assert, that the sphere is not truly a sphere?».

	Simp.: «Lo stimerei per privo di discorso affatto».
	
	Simp.: «I would esteem him completely devoid of [reason]».

	Salv.: «In questo stato è colui, che dice, che la sfera materiale non tocca un piano, pur materiale, in un punto, perchè il dir questo è l'istesso, che dire, che la sfera non è sfera. E che ciò sia vero, ditemi in quello, che voi costituite l'essenza della sfera, cioè, che cosa è quella, che fà differir la sfera da tutti gli altri corpi solidi».
	
	Salv.: «This is the state of the one who says, that the material sphere does not touch a plane, even if it is material, at a [single] point, for to say this is the same, as to say, that the sphere is not a sphere. And [if] this is true, tell me what [in your opinion] constitutes the essence of the sphere, that is, what is it that [sphere], what it is that makes the sphere differ from all other solid bodies».
	
	Simp.: «Questa dimostrazione cŏclude delle sfere in astratto, e non delle materiali [\,\dots]. Le sfere materiali son soggette a molti accidenti, a i quali non soggiacciono le immateriali; E perchè non può esser, che posandosi una sfera di metallo sopra un piano, il proprio peso non calchi in modo, che il piano ceda qualche poco, ò vero, che l'istessa sfera nel contatto si ammacchi? In oltre, quel piano difficilmente potrà esser perfetto, quando non per altro, almeno per esser la materia porosa; e forse non sarà men difficile il trovare una sfera così perfetta, che abbia tutte le linee dal centro alla superficie egualissime per l'appunto».
	
	Simp.: «This demonstration holds for spheres in the abstract, and not for materials [\,\dots]. The material spheres are subject to many accidents, [whilst] the immaterial ones are not subject; And why should it not be, that by placing a metal sphere on a plane, its own weight should not cause a dip in the plane, or that the sphere itself should not bruise in the contact? In addition, that plane can hardly be perfect, if for nothing else, yet at least [due to the fact] that matter is porous; and perhaps it will be no less difficult to find such a perfect sphere, which has all the lines from the center to the surface exactly equal».

	Salv.: «Oh tutte queste cose ve le concedo io facilmente, ma elle sono assai fuor di proposito; perchè mentre voi volete mostrarmi, che una sfera materiale non tocca un piano materiale in un punto, voi vi servite d'una sfera, che non è sfera, e d'un piano, che non è piano, poichè, per vostro detto, ò queste cose non si trovano al mondo, ò se si trovano si guastano nell'applicarsi a far l'effetto [\,\dots]. Adunque tuttavolta che in concreto voi applicate una sfera materiale a un piano materiale, voi applicate una sfera nŏ perfetta a un piano non perfetto; e questi dite, che non si toccano in un punto. Ma io vi dico, che anco in astratto una sfera immateriale, che non sia sfera perfetta può toccare un piano immateriale, che non sia piano perfetto, non in un punto, ma con parte della sua superficie, talchè sin quì quello, che accade in concreto, accade nell'istesso modo in astratto [\,\dots]. Sì come a voler, che i calcoli tornino sopra i Zuccheri, le Sete, e le Lane, bisogna, che il computista faccia le sue tare di casse, invoglie, \& altre bagaglie: Così, quando il filosofo Geometra vuol riconoscere in concreto gli effetti dimostrati in astratto, bisogna che, difalchi gli impedimenti della materia [\,\dots]. Però, quando voi haveste una sfera, \& un piano perfetti, benchè materiali, nŏ habbiate dubbio, che si toccherebbero in un punto. E se questo era, \& è impossibile ad haversi, molto fuor di proposito fu il dire, che \textit{Sphęra ænea non tangit in puncto}».
	
	Salv.: «Oh all these things I grant you easily, but they are far beyond [our] purpose; because while you want to show me that, a material sphere does not touch a material plane at a [single] point, you are using a sphere, that is not a sphere, and a plane, that is not plane, whereas, according to you, these things are not found in the world, or if they are found, they fail in applying themselves to have the effect [\,\dots]. Therefore whenever you concretely apply a material sphere to a material plane, you apply a non-perfect sphere to a non-perfect plane; and these you say, do not touch [each other] in one [sole] point. But I tell you, that even in the abstract an immaterial sphere, that should not be a perfect sphere may touch an immaterial plane, that is not a perfect plane, not in a [single] point, but with [a] part of its surface, so that what happens in the concrete, it happens in the the same way in the abstract [\,\dots]. Just like the reckoner has to calculate an amount of tare on the chests, casings, \& other baggage, to make the numbers work for Sugars, Silks, and Wools: So, when the philosopher of Geometry wants to recognize in concrete the effects demonstrated in the abstract, he must defalcate the impediments of matter [\,\dots]. Hence, if you had a perfect sphere and plane, even though they were material, you need not doubt, that they would touch in one point. And if [such a condition] was, \& is impossible to have, it was much besides the purpose to say, that \textit{Sphęra ænea non tangit in puncto}».
	} 
of removing accessory data («ostacoli accidentarÿ») \cite[p. 139]{Galilei "Dialogo sopra i due Massimi Sistemi del Mondo Tolemaico e Copernicano"}, that is, the «defects» of real objects, with the intent to consider only certain «perfect» (ideal) qualities, under constant characters (conservation laws) and invariance principles. The real object is \emph{replaced} with its mathematical abstraction. E. Torricelli \cite[lettera a M. Ricci, 10 Febbraio 1646, p. 20]{Torricelli "Lettere fin qui inedite precedute dalla vita di lui scritta da G. Ghinassi} abridges in a few lines, and in a fetching manner, the master's thought, the one on the opportunity to “remove” (\textit{difalcare}) the obstacles:

\vspace{2mm}

\begingroup
\footnotesize
I care very little [\textit{a me importa pochissimo}] if the principles of the doctrine \textit{de motu} are true or false. For if they are not true, let us pretend that that they are true [\textit{fingasi che sian veri}] according to what we have supposed, and then take all the other speculations derived from these principles, not [only] with a mixed practice but also with a geometric practice. I pretend or suppose [\textit{Io fingo o suppongo}] that some body or point moves downwards and upwards with a known proportion and horizontally with equal motion. When this is the case, I say that everything Galileo said, and what I [said] in addition [to his words], will come as a consequence. And if the lead, iron and stone balls do not observe that supposed proportion, with their damage, we will say that we are not talking about them [\textit{diremo che non parliamo di esse}].\endnote{
	Original It. version: «Che i principii della dottrina \textit{de motu} siano veri o falsi a me importa pochissimo. Poichè se non son veri, fingasi che sian veri conforme abbiamo supposto, e poi prendansi tutte le altre specolazioni derivate da essi principii, non come così miste, ma pure geometriche. Io fingo o suppongo che qualche corpo o punto si muova all'ingiù et all'insù con la nota proporzione et orizzontalmente con moto equabile. Quando questo sia io dico che seguirà tutto quello che ha detto il Galileo et io ancora. Se poi le palle di piombo, di ferro, di pietra non osservano quella supposta proporzione, suo danno, noi diremo che non parliamo di esse».
	}	

\endgroup

\vspace{2mm}

Besides, the Torricellian letter has the merit of alluding to the recurring but highly variable motifs, proper to the class of figments (which we have already encountered in Section \ref{subsection "Figments of Imagination and Invention of Possible Worlds"}), in the—relatively—free production of hypotheses.

\begin{scholium}[Before Galileo, there were the Greeks]
\label{scholium "Concept of model in ancient Greek science"}
But beware: Galileo was not the first to introduce the technique of “removing”, or “subtracting”, which is the wellspring of mathematical modeling; the concept of \emph{model} is already \emph{fully} elaborated in many different nuances and ductilely adopted in ancient Greek science, and this is because mathematics is the privileged reference to which to reduce observable facts (phenomena), see F. Acerbi \cite{Acerbi "Concetto ed uso dei modelli nella scienza greca antica"}.

And there is more. It is mathematics, consciously for scientists of the Greco-Hellenistic culture, that defines the structure of a model, that determines what can be formalized and expressed in a model, and what instead is destined to end up in the \emph{limbo} of the negligible/non-essential aspects. Let us put it differently: what is not mathematizable—what we are unable to insert into the model—\emph{moves away} from the sphere of knowability. \scholiumsymbol
\end{scholium}

With the deliberate use of models in the theories of \emph{classical physics}, the modeling construct acquires, or reacquires, a new, full, awareness; e.g. L. Boltzmann \cite[p. 324]{Boltzmann "Uber die Prinzipien der Mechanik I"} writes:

\vspace{2mm}

\begingroup
\footnotesize
Our ideas of things are never identical with their essence [\textit{Wesen}]. They are mere images [\textit{bloße Bilder}], or rather, signs [\textit{Zeichen}] for them, which necessarily represent what has been designated one-sidedly, so much so that  they cannot choice but imitate certain kinds of connections in them, whereby the essence remains completely unaffected.

\endgroup

\vspace{2mm}

The following two passages are worthy of being reported on this topic, namely on the \emph{disappearance} of reality that occurs in physics, in favor of a formal reality \emph{via} mathematics. The first is by H. Weyl, taken from the 3rd edition of his \textit{Raum-Zeit-Materie} \cite[§ 35, pp. 262-263]{Weyl "Raum-Zeit-Materie: Vorlesungen Uber Allgemeine Relativitatstheorie 1919"}:

\vspace{2mm}

\begingroup
\footnotesize
The more physics develops, the more it becomes clear that the relations between the phenomena of reality that each of us lives and those objective entities [\textit{objektiven Wesenheiten}] operating in physics through mathematical symbols are not as simple as it appears in an ingenuous conception, and that fundamentally nothing of the content of reality directly experienced goes in the physical world [\,\dots]. In the end all physical reality appears as a mere form [\textit{ganze physikalische Realität doch als eine bloße Form}]; it is not geometry that has become physics, but physics has become geometry [\,\dots]; the entire physical world has become a form [\textit{die gesamte physische Welt ist zur Form geworden}] the content of which grows from completely different areas than those of the physical world. Physics has no further significance for reality than formal logic has for the realm of truth. What formal logic teaches is surely based on the essence of truth, and no truth violates its laws. But whether a statement is true or not, it teaches absolutely nothing about it, it leaves entirely indefinite the content of truth [\,\dots]. I think that the description of physics is very similar [to that of logic], that is, it corresponds to a formal construction of reality [\textit{formale Verfassung der Wirklichkeit}]. Its laws are never actually violated, just as there are no truths inconsistent with logic [\,\dots]; the \textit{Grund} of reality is not grasped by them.\footnote{
	\label{footnote "Weyl Raum-Zeit-Materie, 4th edition"}
	In the 4th edition this bit was removed. And the book stops at a more triumphalist mood of mathematical physics (anyway, such a mood is already present in the 3rd edition). In the 4th edition \cite[p. 284]{Weyl "Raum-Zeit-Materie: Vorlesungen Uber Allgemeine Relativitatstheorie 1921"} = \cite[pp. 311-312]{Weyl "Space-Time-Matter"} we read: «Whoever looks back over the ground that has been traversed, leading from the Euclidean metrical structure to the mobile metrical field which depends on matter, and which includes the field phenomena of gravitation and electromagnetism; whoever endeavours to get a complete survey of what could be represented only successively and fitted into an articulate manifold, must be overwhelmed by a feeling of freedom [\textit{Gefühl Freiheit}] won [\,\dots]. He must feel imbued with the conviction that reason is not only a human, a too human, makeshift in the struggle for existence, but that, in spite of all disappointments and errors, the \textgreek{λογική} structure that permeates the world [\textit{Weltvernunft}] has increased, and that the consciousness of each one of us is the centre at which the One Light and Life of Truth comprehends itself in Phenomena. Our ears have caught a few of the fundamental chords from that harmony of the spheres of which Pythagoras and Kepler once dreamed [\textit{Ein paar Grundakkorde jener Harmonie der Sphären sind in unser Ohr gefallen, von der Pythagoras und Kepler träumten}]».

	A comment is needed without delay: if man is the center of truth, then there is no Truth with a capital T—and this is but the Pythagorean dream, later transfused into the Galilean–Keplerian project (cf. Section \ref{subsection "In the Bliss of Goddess Geometry"}). Au contraire, man is the place where the light of Truth (whatever is meant by this heavy word) is shattered, reduced to powder, and \emph{dissolves} into our mind.
	}

\endgroup

\vspace{2mm}

The other passage is by A.S. Eddington \cite[pp. 198-199, e.a.]{Eddington "Space Time and Gravitation. An Outline of the General Relativity Theory"}: 

\vspace{2mm}

\begingroup
\footnotesize
Mind filters out matter from the meaningless jumble of qualities, as the prism filters out the colours of the rainbow from the chaotic pulsations of white light. Mind exalts the permanent and ignores the transitory; and it appears from the mathematical study of relations that the only way in which mind can achieve her object is by picking out one particular quality as the permanent substance of the perceptual world, partitioning a perceptual time and space for it to be permanent in, and, as a necessary consequence of this Hobson's choice, the laws of gravitation and mechanics and geometry have to be obeyed. Is it too much to say that mind's search for permanence has \emph{created} the world of physics? So that the world we perceive around us could scarcely have been other than it is? 

[\,\dots] Are there then no genuine laws in the external world? Laws inherent in the substratum of events, which break through into the phenomena otherwise regulated by the \emph{despotism} of the mind?

\endgroup

\vspace{2mm}

In Weyl and Eddington, for the historical course of scientific thought, there is no trace of Galileo's naturalism, and their mathematics is free from the heavy burden of analyzing reality through a naive correspondence between natural objects and mathematical objects (which, in Galileo, are abstraction of natural ones); but, nevertheless, they maintain the idea of a description and understanding of the world through a network of abstract entities (formularies, schemes, models), with a formal construction, i.e. through a creation of the mind in search of identities (equalities) and of permanent shapes (structures) of the world.

\subsection{Dissolution of the Objective World: the Clamorous Incident of the Wave Function. An Authentic Story of \emph{Aesopian Fables} and \emph{Theater of the Absurd}}
\label{subsection "Dissolution of the Objective World: the Clamorous Incident of the Wave Function. An Authentic Story of Aesopian Fables and Theater of the Absurd"}

\begingroup
\footnotesize
[T]he Schrödinger wave-function bears to (the unknowable) physical reality the same relationship that a weather forecast bears to the weather. \\
\indent — \textsc{J.A. Wheeler} in the recollection of \textsc{G. Preparata} \cite[p. 2]{Preparata "An Introduction to a Realistic Quantum Physics"}

\vspace{2mm}

C'est ainsi que nous voyons le monde: nous le voyons à l'extérieur de nous-mêmes, et cependant nous n'en avons qu'une représentation en nous.\footnote{
	«Which is how we see the world: we see it as being outside ourselves, and yet we have only a representation of it within us».
} \\
\indent — \textsc{R. Magritte} \cite[p. 184]{Magritte "Catalogue Raisonne. II: Oil Paintings and Objects 1931-1948"}

\endgroup
 
\vspace{2mm}

\enumerationisinitium
\item Our musings in the above Section \ref{subsection "Physico-mathematical Reality"}, together with the previous Sections, propelled us into a burning issue: the \emph{dissolution of reality}, or a part of reality, in the mental trickles of the observer. Wheeler's salacious joke, in epigraph, makes us think back to the experience—hidden in the work of mathematicians and mathematical physicists—of finitude in the \emph{approximation}, \emph{disappearance} and \emph{reinvention} of the objective world. The state of the atmosphere is one thing, the prediction of the conditions of the atmosphere is quite another. Lorenz knew something about it (cf. Sections \ref{section "(In)deterministic Flow: Lorenz System in Comparison with Quantum Mechanics"}, \ref{section "The Lorenz Flow: a Strange Attractor"}, and \ref{subsubsection "Example II. Stochastically Forced System's Lorenz Equations with Wiener Process and Time Dependent Thermal Fluctuations in Weather Forecasting"}). It is a good and ironic way to close this Chapter.
\item Physico-mathematical production is not a “replication” of the objective (physical) world, out there; it is a process of (re)construction and elaboration of it, which ordinarily introduces “alterations”, like a distorting mirror (cf. Section \ref{subsubsection "Math-Model in the Perseus Mythology: No Aberration-free Mirror of Reality"}); as written by F. Bacon, \textit{intellectus humanus instar speculi inaequalis} («the human understanding resembles an uneven mirror»), see endnote \ref{endnote "Bacon's idola"}. This refers, albeit with the necessary differences, to both the micro- and macro-representation.
\item Scientific activity is tied to the condition of \emph{finitude} of the scientist or, expanding the field further, to the \textit{Condition humaine}, to quote R. Magritte \cite[p. 184]{Magritte "Catalogue Raisonne. II: Oil Paintings and Objects 1931-1948"}, see also \cite[pp. 65-66]{Magritte "Life Line"}:

\vspace{2mm}

\begingroup
\footnotesize
[An object, a “piece” of reality] for the observer [\textit{spectateur}], exists, simultaneously, in his mind [\textit{par la pensée}], as inside the room in the painting, and outside in the real landscape. 

\endgroup

\vspace{2mm}

Which also applies when we admire a “realistic” paintings, or even “hyperrealistic” images.

\item It is desirable to clarify that Preparata reports that expression from Wheeler with a critical spirit, because he has a “realistic” slant in quantum physics. Au contraire, I espouse the Wheelerian attitude, descending from Bohr's school of thought (cf. Section \ref{subsection "Bohrism in a Right Perspective"}). Realism is a noble impulse, but it can sometimes err through ingenuousness; see, however, below, point \ref{item "Realistic description"}. Let us repeat a lesson from L. Boltzmann \cite[p. 179]{Boltzmann "Uber die Frage nach der objektiven Existenz der Vorgange in der unbelebten Natur"}, entitled \textit{Über die Frage nach der objektiven Existenz der Vorgänge in der unbelebten Natur} (On the question of the objective existence of processes in inanimate nature):

\vspace{2mm}

\begingroup
\footnotesize
Our target will not be to establish the truth or falsity of one or the other world picture [\,\dots] in order to represent the objective world picture [\textit{Darstellung des objektiven Weltbildes}], but we will wonder whether either [of these picture] is usefulness [\textit{Zweckmäßigkeit}] for this or that purpose [\,\dots].\footnote{
	Pay heed: “truth” or “falsity” are not criteria or rules required by modern physics; what counts are “usefulness” and “fruitfulness”. Cf. e.g. O.M. Corbino \cite[p. 26]{Corbino "I fondamenti sperimentali delle nuove teorie fisiche"}: «It is very fortunate that physicists have become accustomed to lose interest for the definitive or provisional quality of their theoretical constructions, persuaded that these [constructions] do not cease to function as powerful instruments of progress [\,\dots]. They no longer ask themselves [\,\dots] whether the theories are true or not; they only require that these [theories] be fruitful, allowing for a certain economy of thought in the coordination [of the external world] of facts».
	} 
We give the most easily comprehensible rules for constructing this world picture without bothering how we subjectively [\textit{subjektiv}] arrived at these rules: its justification lies solely in the correspondence [\textit{Übereinstimmung}] between the world picture and the facts. 
 
\endgroup

\vspace{2mm}

It is the old \textit{adæquatio}, which recurs with a modern twist, between “world”, out there, and “picture”, or “representation”, of it. In spite of this, I am agree with Preparata \cite[pp. 63-64]{Preparata "An Introduction to a Realistic Quantum Physics"} \cite[pp. 199-200]{Preparata "Dai quark ai cristalli. Breve storia di un lungo viaggio dentro la materia"} \emph{toto cœlo} upon the following facts.
\subenumerationisinitium
\item Quantum mechanics is not a «complete» and «self-consistent» theory of reality—one only need look at the Bohrian «invention» of the wave-particle complementarity, aimed at fulfilling two irreconcilable aspects of physics, with a double appearance, the undulatory one and the punctiform one: particle behaves “intermittently” as a particle and as a wave (the behavior of a particle is subordinate to the kind of measurements we perform upon it).

Here there is the \emph{mistake} of believing that phenomena—be they at the micro- or macro-scale(s)—that are described by the same mathematics must, \emph{perforce}, belong to the same physical nature. We mulishly insist on calling “undulatory” and “corpuscular” the behavior of an electron, just because some sort of mathematical equation, initially created/adopted \emph{for} macroscopic studies, is sufficiently satisfactory for the description of microscopical observations. This happens because our mathematics was formed, along the phylogenetic and ontogenetic path, on the macroscopic world; and the laws of macrophysics, voluntarily or otherwise, constitute our point of departure towards microscopic depths. The wave-particle paradox is \emph{not in nature}, but \emph{in ourselves}.\footnote{
	Dirac \cite[p. 49]{Dirac "The origin of quantum field theory"} gives us gems, where he pontificates: «Instead of working with a picture of the photons as particles, one can use instead the components of the electromagnetic field. One thus gets a complete harmonizing [\textit{sic}] of the wave and corpuscular theories of light. One can treat light as composed of electromagnetic waves, each wave to be treated like an oscillator; alternatively, one can treat light as composed of photons, the photons being bosons and each photon state corresponding to one of the oscillators of the electromagnetic field. One then has the reconciliation [\textit{sic}] of the wave and corpuscular theories of light. They are just two mathematical descriptions of the same physical reality». The kink is squarely this: it is “mathematics” and not “physical reality”, or “nature”. These Diracian words are so diplomatic, that one ends up recognizing their \emph{math-hooey}.
	} 
Actually, the wave-particle paradox is one of the many-sided marks of the \textit{ambiguus}\footnote{
	The La. adjective \textit{ambiguus} is for “having ‘double’, ‘equivocal’, ‘shifting’, ‘interchangeable’ meaning”.
	} 
extent of human creativity, which pervades, with a variable profit, any production, in the scientific and artistic domain (cf. Sections \ref{section "Grainy Music and Chance"} and \ref{subsection "Contextus I. Elements of Brachylogy—the Reverie of a Perfect Language, with a Margo on Music and Mathematics"}).

What is the \textit{fabulæ moralis}? Once again, the story\footnote{
	It is assumed that there is no difference between “story” (fiction) and “history” (historic truth), cf. footnote \ref{footnote "Res gestæ and rerum gestarum narratio"} on p. \pageref{footnote "Res gestæ and rerum gestarum narratio"}. 
	} 
shows that (\textgreek{ὁ μῦθος δηλοῖ ὅτι}) we are faced with one of the picturesque examples where mathematics is misrepresented, i.e., is wrongly understood as the executive “code” of nature, or as an indistinct and pervasive “medulla” of the totality of natural phenomena. Note. With the fabulous/mythical \textgreek{μῦθος} \cite[p. xviii]{Aesop "Fabulae Aesopiae"}, the reference to Sections \ref{subsubsection "Math-Model in the Perseus Mythology: No Aberration-free Mirror of Reality"} and \ref{subsubsection "What Myth Does Mathematics Tell Us? The Example of the Stefan Problem (Ice-Water Phase-boundary)"} is plain; in this regard, my slogan could be: \emph{Aesopian fables also exist in (mathematical and physical) sciences}.

Not surprisingly one has, as a loophole, the construct of \emph{quantum field}, that seeks to rectify the quantum complementarity: one imagines that it is a continuous quantity (distributed everywhere in space) but also granularized in its particle version, so that “continuous” and “point-like” are two apparently contradictory aspects of the same reality, as the Bohr motto reads, in his coat of arms, featuring a taijitu (\ZhTraditional{太極圖}): \textit{Contraria sunt complementa}. I am sorry to say, but, in a mathematical key, the whole thing is not so simply resolved (calculated), quite the contrary.

The secret is to cross from one aspect to the other, more or less surreptitiously, by depicting a framework in which the particle is a punctual “condensation”, an energetic “knot” of the field, and the field is a continuous “flow” of the point-charge. Weyl's \cite[p. 171, e.a.]{Weyl "Philosophy of Mathematics and Natural Science} verbal portrayal is cogent:

\vspace{2mm}

\begingroup
\footnotesize
According to the [field theory of matter] a material particle such as an \emph{electron} is merely a \emph{small domain} of the electrical field within which the field strength assumes enormously high values, indicating that a comparatively \emph{huge field energy} is \emph{concentrated in a very small space}. Such an \emph{energy knot}, which by no means is clearly delineated against the remaining field, propagates through empty space like a water wave across the surface of a lake [\,\dots]. According to this view, there exists but one kind of natural law, namely, field laws of the same transparent nature as Maxwell had established for the electromagnetic field. The obscure problem of laws of interaction between matter and field does not arise. This conception of the world can hardly be described as dynamical any more, since the field is neither generated by nor acting upon an agent separate from the field, but following its own laws is in a quiet \emph{continuous flow}.

\endgroup

\vspace{2mm}
\item In the theoretical subsoil of quantum mechanics

\vspace{2mm}

\begingroup
\footnotesize
 the objective world [whatever the word “objective” may mean] dissolves into a more manageable subjective world, to which the [wave function] $\psi$ belongs \cite[p. 199]{Preparata "Dai quark ai cristalli. Breve storia di un lungo viaggio dentro la materia"}. 
 
\endgroup

\vspace{2mm}
The «subjectivism» permeating the Copenhagen interpretation of quantum mechanics,\footnote{
	Anyone who wants to learn more about the “Copenhagen interpretation”, at its birth, with the lectures of M. Born and N. Bohr from the Volta Conference, held at Lake Como, Pavia and Roma, 11-27 September 1927, can consult S. Boffi \cite{Boffi "Il postulato dei quanti e il significato della funzione d'onda"}. For a historical reconstruction, some quarrels and heartbreaking misapprehensions that accompanied the inception of quantum mechanics are narrated in Lindley's book \cite{Lindley "Uncertainty: Einstein Heisenberg Bohr and the Struggle for the Soul of Science"}.
	}
in the footsteps of Bohr and Heisenberg, can actually lead, in the paroxysm, to the

\vspace{2mm}

\begingroup
\footnotesize
disaster of a skepticism in which reality dissolves [into our mind], science becomes a social game, and every sort of sect and church can claim its piece of “truth”, even [in the] scientific [literature] \cite[p. 200]{Preparata "Dai quark ai cristalli. Breve storia di un lungo viaggio dentro la materia"}.
 
\endgroup

\vspace{2mm}

I will not go into details with respect to some delicate impasses, because it is easy to get caught up in the minutiæ; e.g. the sectarianism that blazes up in the debate about the “collapse” of the wave function often falls into ridiculous blunders. 

Not only that. As a consequence, one of the performances at the Theater of the Absurd can go on stage, with the identification of contradictions and paradoxes, or at least of seemingly contradictory and paradoxical concepts. This situation is historically well-documented in the evolution of quantum mechanics. See, for example, W. Heisenberg \cite[p. 42]{Heisenberg "Physics and Philosophy"}:
\vspace{2mm}

\begingroup
\footnotesize
[\,\dots] [A]n intensive study of all questions concerning the interpretation of quantum theory in Copenhagen finally led to a complete [\,\dots]. I remember discussions with Bohr which went through many hours till very late at night and ended almost in despair; and when at the end of the discussion I went alone for a walk in the neighboring park I repeated to myself again and again the question: Can nature possibly be as absurd as it seemed to us in these atomic experiments?

\endgroup

\vspace{2mm}

The speech of R. Feynman \cite[p. 10]{Feynman "QED. The Strange Theory of Light and Matter"} is even clearer: 
	
\vspace{2mm}

\begingroup
\footnotesize
The theory of quantum electrodynamics describes Nature as absurd from the point of view of common sense. And it agrees fully with experiment. So I hope you can accept Nature as She is—absurd [\,\dots]. Please don't turn yourself off because you can't believe Nature is so strange. 

\endgroup

\vspace{2mm}
	
In this, science appears to be stuck in the catchline of the \emph{credo quia absurdum}.\footnote{
	The phrase is attributed to Tertullian, but it is a misquotation. He never wrote it; he wrote, however, something similar (in his \textit{De Carne Christi}): «prorsus credibile [est], quia ineptum est» (it is immediately credible, because it is inconvenient), and «certum est, quia impossibile [est]» (it is certain, because it is impossible».
	
	I discover that L. Russo \cite[pp. 52-53]{Russo "Appunti per una storia dei concetti di "matematica" e "fisica""} holds a viewpoint close to mine.
	}
But how do we know if Nature is truly absurd? Well, we do not know that (yet), because there are no diriment experiments (\textit{experimenta crucis}), but a progression of experimental procedures along that direction, starting from T. Young's \cite{Young "The Bakerian Lecture. Experiments and calculations relative to physical optics"} double-slit experiment (the publication of which goes back to 1804). And then “absurd” compared to what? To common sense? Can we take the “common sense” (\textgreek{τῆς κοινῆς αἰσθήσεως πάθος}, in the ancient culture) as a universal yardstick?

It may be that the core of Nature is absurd, or that reality is paradoxical; but if it is, such an absurdity, or paradoxicality, is seen in relation to us, to our experience of it. What we define as “absurd”, “logically contradictory”, or “paradoxical”, in reference to certain aspects of reality, especially at the subatomic scales, is directly dependent on our common sense, logic, or intellect, and finally on our language; so this is about human affairs. Saying that “absurdities”, “contradictions”, or “paradoxes”, belong to the «intrinsic structure» of subatomic nature is an act of \emph{conceit}, of comical haughtiness. Assertions like this are aprioristic judgments that emanate from believing that our way of thinking are realities of nature. It is not correct to turn our “shape” of thought into a property of all nature. 

The pristine wisdom of Laozi \cite[\textsc{lxxi}, p. 159]{Laotze "The Book of The Simple Way"}, on a humble ignorance, seems lost: «To know one's ignorance is the best part of knowledge», namely, \ZhSimplified{“知之为知之, 不知为不知, 是知也”}. That reminds me of an excerpt from a letter by L. Euler \cite[lettre XXVIII, 15 Juillet 1760, p. 108]{Eulero "Lettres a une princesse d'Allemagne I"}:

\vspace{2mm}

\begingroup
\footnotesize
[À] entendre parler les savans, on s'imagine qu'ils poss[è]dent les plus profonds myst[è]res [de la nature], quoiqu'ils n'en sachent pas plus que le païsan, \& peut-être encore moins. V.[otre] A.[ltesse] reconnaîtra aisément, que ces apparentes subtilités ne sont que des chicanes.\footnote{
	There is a gleaming eighteenth-century It. transl. of these Eulerian words \cite[p. 145]{Eulero "Lettere scritte ad una principessa d'Alemagna"}: «[A] sentir parlare i dotti su di quello punto voi credereste che abbiano penetrato ne' misterj più ascosi della Natura, e pure essi non ne sanno più de' contadini, e forse anche meno. Ma V. A. può facilmente conoscere che tai sottigliezze apparenti non sono altro che sofismi».
	}
	
\endgroup

\vspace{2mm}	
\item Additionally, Preparata's disapproval \cite[pp. 5, 19-20, 39]{Preparata "An Introduction to a Realistic Quantum Physics"} towards the “quantum particle” as a Newtonian point-mass, and as a «truly metaphysical object», is correct, and I am of the same opinion (cf. Section \ref{subsection "Geometro-physical Singularities: 1D Lines, 0D-like Elements, and the Point-electron, or any Particle as a Point-mass"} and \ref{subsection "Scholium: Point-charge/Point-mass of Electricity: Singularities (or Quasi-singularities) of Fields"}).
\item
\label{item "Realistic description"}
A “realistic” description is the ambition of physics, surely.\footnote{
	But it does not have to be that for everybody. In certain physicists, viscerally forged by a sort of imprinting of mathematizing the theory of nature (“Think \textgreek{καλῶς}” could be their dictum), such as Dirac, there is a tasty paroxysm: he, not being interested in the objective reality of the external world, once declared that «the question of whether the wave [functions] $\psi$ were real or [fictitious]» was not a cause for concern for him, as he considered it a «metaphysical problem», cf. Boffi \cite[pp. 13-14]{Boffi "Le forme di Dirac"}.
	}
It must attempt to reveal, in its own way, the laws of nature, right? Against this background, J.S. Bell's reflection \cite[pp. 687-688]{Bell "Subject and Object"} is commendable: 

\vspace{2mm}

\begingroup
\footnotesize
It would be foolish to expect that the next basic development in theoretical physics will yield an accurate and final theory. But it is interesting to speculate on the possibility that a future theory will not be \emph{intrinsically} ambiguous and approximate. Such a theory could not be fundamentally about ‘measurements’, for that would again imply incompleteness of the system and unanalyzed interventions from outside. Rather it should again become possible to say of a system not that such and such may be \emph{observed} to be so but that such and such \emph{be} so. The theory would not be about ‘\emph{observ}ables’ but about ‘\emph{be}ables’. These beables need not of course resemble those of, say, classical electron theory; but at least they should, on the macroscopic level, yield an image of the everyday classical world, for [as Bohr says] ‘it is decisive to recognize that, however far the phenomena transcend the scope of classical physical explanation, the account of all evidence must be expressed in classical terms’.

\endgroup

\vspace{2mm}
 
The universe—echoing the words of F. Bacon—is not to be narrowed down to the limits of our understanding; but rather the understanding must be stretched, enlarged,\footnote{
	The widening of our knowledge is a critical topic. The snag is that this widening is accompanied by a \textit{reductio ad ordinem}; for example, N. Bohr \cite[p. 1]{Bohr "Introductory Survey"} writes that: «The task of science is both to extend the range of our experience and to reduce it to order». Turning to the notion of order, it is a recurring obsession in mathematics and in physics (cf. Chapters \ref{chapter "On the Chaos, Part I. Micro- and Macro-scales"}, \ref{chapter "On the Chaos, Part II. Non-linear Analysis"}, \ref{chapter "Randomness and Stochastic Systems"}).
	} 
to take in the picture, in the image, of the universe, as it is discovered. 

\subenumerationisfinis
\enumerationisfinis

\chapter{Outro—\emph{Parva Mathematica}: \emph{Libera Divagazione} \sfrac{7}{8}}
\label{chapter "Outro—Parva Mathematica: Libera Divagazione 7/8"}

\section{Mathematics by Mathematicians vs. Mathematics by Physicists}
\label{section "Mathematics by Mathematicians vs. Mathematics by Physicists"}

\begingroup
\footnotesize
Usually I do not trust physicists until I find my own proof or, at least, an explanation of their results. For this reason, a big part of theoretical physics remains outside my understanding [\,\dots]. [T]he worlds of mathematicians and physicists are quite different and there is a boundary which separates them. This boundary is very individual, and everybody chooses it for himself. \\
\indent — \textsc{Ya.G. Sinai} \cite[p. 565]{Sinai "Mathematicians and Physicists = Cats and Dogs?"}\endnote{
	From the same author, see also \cite{Sinai "How Mathematicians and Physicists Found Each Other in the Theory of Dynamical Systems and in Statistical Mechanics"}.
	}

\endgroup

\vspace{2mm}

In the following Sections we will examine the relationship between mathematics and physics, or rather, between mathematics made by mathematicians and mathematics done by physicists, and we will see that, depending on a personal taste, they may be far apart from and close to each other, or melted into one.

\subsection{Proof vs. Empirical Datum}
\label{subsection "Proof vs. Empirical Datum"}

\begingroup
\footnotesize
In mathematics there is an empty canvas before you which can be filled without reference to external reality [\,\dots]. The only value of mathematics lies in its internal structure, [whilst] physics is basically an empirical science [\,\dots]. \\
\indent Soon after coming to Princeton I became aware that my work on the Lorentz group was based on somewhat shaky arguments. I had naively manipulated unbounded operators without paying any attention to their domains of definition. I once complained to Dirac about the fact that my proofs were not rigorous and he replied, “I am not interested in proofs but only in what nature does”. This remark confirmed my growing conviction that I did not have the mysterious sixth sense which one needs in order to succeed in physics and I soon decided to move over to mathematics. \\
\indent — \textsc{Harish-Chandra} in the recollection of \textsc{R.P. Langlands} \cite[pp. 202, 205]{Langlands "Harish-Chandra: 11 October 1923-16 October 1983"}

\endgroup

\vspace{2mm}

Mathematics for physicists is not the same as that of mathematicians (or even of mathematicians interested in theoretical physics). Physicists sometimes do a sophisticated-type mathematics; but more often they do an uncouth-type mathematics suffering from a lack of rigor. Hence the not rare misunderstanding between mathematicians and physicists. 

A physicist does not aim at the accuracy of theorems and their proof; it is a priority, for him, to mathematically encapsulate the experimental knowledge of a phenomenon of nature, or to sew certain laws on the body of nature, by adopting and adapting some theorems at his convenience. Mathematical rigor, for a physicist, thus becomes secondary; which, for a (pure) mathematician, is hardly acceptable. 

Be careful though: someone, from this speech, might jump to the wrong conclusion. Rigor is a consubstantial feature of mathematics, but, at the same time, is also an \emph{ancillary} trait of it. Mathematics is more than its rigor, as we saw in Sections \ref{section "Mathematics in the Physical Sciences, and Nature of Reality I"}, \ref{section "Mathematics in the Physical Sciences, and Nature of Reality II"} and \ref{section "Mathematics in the Physical Sciences, and Nature of Reality III"}. 

\subsection{Procrustean Bed: the Example of the Dirac Delta Function}
\label{subsection "Procrustean Bed: the Example of the Dirac Delta Function"}

\begingroup
\footnotesize
To get a picture of $\Diracdelta(x)$, take a function of the real variable $x$ which vanishes everywhere except inside a small domain, of length $\epsilon$ say, surrounding the origin $x = 0$, and which is so large inside this domain that its integral over this domain is unity [\,\dots]. Then in the limit $\epsilon \to 0$ this function will go over into $\Diracdelta(x)$. \\
\indent — \textsc{P.A.M. Dirac} \cite[p. 58]{Dirac "The Principles of Quantum Mechanics (1958)"}

\vspace{2mm}

It is been more than 50 years that the engineer Heaviside \cite{Heaviside "On Operators in Physical Mathematics. Part I"} \cite{Heaviside "On Operators in Physical Mathematics. Part II"} has introduced his rules of symbolic calculus, in a daring report where a mathematical calculus [in many cases] not at all justified was used for the solution of physical problems [\,\dots]. Engineers are using it in a systematic way, everyone with his own personal conception, with a more or less tranquil conscience; it has become a technique “which is not rigorous but rather successful”. Since the famous function $\Diracdelta(x)$ it was introduced by Dirac \cite{Dirac "The Physical Interpretation of the Quantum Dynamics"}, which is zero everywhere except at $x = 0$ and infinite at $x = 0$ so that $\int^{+\infty}_{-\infty}\Diracdelta(x)dx = +1$, the formulas of symbolic calculus have become even more unacceptable under the rigor of mathematicians. Writing that the Heaviside [step] function $Y(x)$ equal to 0 for $x < 0 $ and to 1 for $x \geqslant 0$[,] has as [its] derivative the Dirac function $\Diracdelta(x)$\footnote{
	The derivative of the Heaviside step function—cf. Eq. \eqref{equation "Heaviside step function"}—is the Dirac delta function. 
	}
whose definition is mathematically contradictory [\textit{la définition même est mathématiquement contradictoire}], and talking about derivatives $\Diracdelta '(x), \Diracdelta''(x), \mathellipsis$ of this function devoid of real existence [\textit{dénuée d'existence réelle}], it is to exceed the prescribed limits. \\
\indent — \textsc{L. Schwartz} \cite[p. 3]{Schwartz "Theorie des distributions"}

\endgroup

\vspace{2mm}

An example of mathematics at the service of physics is the \emph{Dirac $\Diracdelta$ function} \cite[p. 625]{Dirac "The Physical Interpretation of the Quantum Dynamics"} \cite[§ 15]{Dirac "The Principles of Quantum Mechanics (1958)"}. For a well-behaved continuous function $\varphi(x)$ of $x$, there is a function $\Diracdelta(x)$ such that
\begin{equation}
	\int^{+\infty}_{-\infty}\Diracdelta(x - x_0)\varphi(x)dx = \varphi(x_0),
\end{equation}
which transforms $\varphi(x)$ into $\varphi(x_0)$. Delta function is defined by
\begin{equation}
\begin{cases}
	\Diracdelta(x) = 0, \text{ when } x \neq 0, \\
	\int^{+\infty}_{-\infty}\Diracdelta(x)dx = 1,
\end{cases}
\end{equation}
and thereby it is a quantity equal to zero everywhere except at a single point, $x = 0$, and in that zero-point it is infinitely large, whilst the integral over the real line is 1. However, if $\Diracdelta(x)$ is null everywhere except for $x = 0$, the integral of $\Diracdelta(x)$ cannot be other than 0; and yet, for Dirac, it has integral 1 (with a finite value). This is a brutal \emph{contradiction} to which Schwartz (in epigraph) refers. Stricto sensu, $\Diracdelta(x)$ is not a function but a functional; a corrective way is to interpret $\Diracdelta(x)$ as a limit of a sequence of functions that produces a sequence of numbers, e.g.
\begin{equation}
	\Diracdelta_\rotatedell(x) =  
	\begin{cases}
	0, \text{ for } x < -\frac{1}{2\rotatedell}, \\
	\rotatedell, \text{ for } -\frac{1}{2\rotatedell} < x < \frac{1}{2\rotatedell}, \\
	0, \text{ for } x > \frac{1}{2\rotatedell}.
	\end{cases}
\end{equation}

\subsection{Straitjacket for Feynman Path Integrals}
\label{subsection "Straitjacket for Feynman Path Integrals"}

\begingroup
\footnotesize
[T]o form the differential equations, one would need to know not only the one [the solution of which] is realized in nature [\textit{réalisée dans la nature}], but all those that are [infinitely] possible [\textit{toutes celles qui sont possibles}]. \\
\indent — \textsc{H. Poincaré} \cite[p. 44]{Poincare "Dernieres Pensees"}

\vspace{2mm}

In quantum mechanics the probability of an event which can happen in several different ways is the absolute square of a sum of complex contributions, one from each alternative way [\,\dots] within a region of space[-]time \cite[p. 367]{Feynman "Space-Time Approach to Non-Relativistic Quantum Mechanics"} \\
\indent The game I play is a very interesting one. It's imagination, in a tight straitjacket, which is this: that it has to agree with the known laws of physics \cite[p. 98]{Sykes (Ed.) "No Ordinary Genius: The Illustrated Richard Feynman"}. \\
\indent — \textsc{R.P. Feynman}
	
\vspace{2mm}

The Feynman path integral is the mathematicians' \textit{pons asinorum}.\footnote{
	It is the \emph{isosceles triangle theorem} in Euclid \cite[Proposition \textgreek{ε´, Στοιχείων α´}, Book I, p. 20]{Euclidis "Elementa I"}: «In isosceles triangles the angles at the base are equal to one another, and if the equal sides are prolonged [under the base] then the angles under the base will be equal to one another (\textgreek{Τῶν ἰσοσκελῶν τριγώνων αἱ πρὸς τῇ βάσει γωνίαι ἴσαι ἀλλήλαις εἰσίν, καὶ προσεκβληθεισῶν τῶν ἴσων εὐθειῶν αἱ ὑπὸ τὴν βάσιν γωνίαι ἴσαι ἀλλήλαις ἔσονται})».	
	} 
Attempts to put it on a sound footing have generated more mathematics than any subject in physics since the hydrogen atom. To no avail. The mystery remains, and it will stay with us for a long time.\endnote{
	There are also interpretations of the Feynman path integral in terms of reality, see e.g. R.D. Sorkin \cite{Sorkin "Quantum Measure Theory and its Interpretation"}. As far as we know, we are on a slippery slope here, on the fringe of a mix-up with mathematics and reality.
	} \\
\indent The Feynman integral, one of the most useful ideas in physics, stands as a challenge to mathematicians. While formally similar to Brownian motion [see Section \ref{section "Non-perfect Fluid in the Teapot and Brownian Motion"} and Margo \ref{margo "Pioneering work for the path integral formulation"}], and while admitting some of the same manipulations as the ones that were made rigorous long ago for Brownian motion, it has withstood all attempts at rigor. Behind the Feynman integral there lurks an even more enticing (and even less rigorous) concept: that of an amplitude which is meant to be the quantum-mechanical analog of probability (one gets probabilities by taking the absolute values of amplitudes and squaring them: hence the slogan “quantum mechanics is the imaginary square root of probability theory”). \\
\indent — \textsc{G.-C. Rota} \cite[p. 229]{Rota "Indiscrete Thoughts"}

\endgroup

\vspace{2mm}

The task of a physicist, generally, is to \emph{reproduce} certain experimental results in a regular (invariant) way, or to find a \emph{coincidence} between measurement data and theoretical predictions.\footnote{
	Cf. e.g. S.W. Hawking in \cite[chap. 7, p. 121]{Hawking and Penrose "The Nature of Space and Time"}: «I don't demand that a theory correspond to reality because I don't know what it is. Reality is not a quality you can test with litmus paper. All I'm concerned with is that the theory should predict the results of measurements».
	
	Hawking's mental line is the fastest road (and also it is got the least amount of dust); but there are other roads, with more tortuous curves. The debate between Hawking and Penrose is engrossing (ibid.): «These lectures have shown very clearly the difference between Roger [Penrose] and me. He's a Platonist and I'm a positivist. He's worried that Schrödinger's cat \cite[p. 812]{Schrodinger "Die gegenwartige Situation in der Quantenmechanik"} is in a quantum state, where it is half alive and half dead. He feels that can't correspond to reality [for Penrose the problem of \emph{ontology} is imperative in quantum mechanics]. But that doesn't bother me».
	} 
Physicists, even when they possess dissimilar mental predispositions, go in this same direction, like J. Schwinger \cite{Schwinger "Quantum Electrodynamics. I. A Covariant Formulation"} \cite{Schwinger "Quantum Electrodynamics. II. Vacuum Polarization and Self-Energy"} \cite{Schwinger "Quantum Electrodynamics. III. The Electromagnetic Properties of the Electron-Radiative Corrections to Scattering"}, always in search of mathematical rigor, and R.P. Feynman \cite{Feynman "Space-Time Approach to Non-Relativistic Quantum Mechanics"} \cite{Feynman and Hibbs "Quantum Mechanics and Path Integrals"}, little inclined to mathematical systematization, but attracted to the intuitive aspect of theories.

Feynman's original definition of the path integral, in its early days, before a better rigorization and refinement (still underway) for work of F.J. Dyson \cite{Dyson "The Radiation Theories of Tomonaga Schwinger and Feynman"}, C. Morette \cite{Morette "On the Definition and Approximation of Feynman's Path Integrals"} \cite{DeWitt-Morette Maheshwari Nelson "Path integration in non-relativistic quantum mechanics"} \cite{Dewitt-Morette "A Reasonable Method for Computing Path Integrals on Curved Spaces"}, S. Albeverio \& R. Høegh-Krohn, et al. \cite{Albeverio and Hoegh-Krohn "Oscillatory Integrals and the Method of Stationary Phase in Infinitely Many Dimensions with Applications to the Classical Limit of Quantum Mechanics I"} \cite{Albeverio and Hoegh-Krohn "Feynman Path Integrals and the Corresponding Method of Stationary Phase"} \cite{Albeverio Blanchard and Hoegh-Krohn "Feynman Path Integrals and the Trace Formula for the Schrodinger Operators"} \cite{Albeverio Guatteri and Mazzucchi "Phase space Feynman path integrals"} \cite{Albeverio Hoegh-Krohn Mazzucchi "Mathematical Theory of Feynman Path Integrals: An Introduction"} \cite{Albeverio Cangiotti Mazzucchi "A Rigorous Mathematical Construction of Feynman Path Integrals for the Schrodinger Equation with Magnetic Field"}, was unanimously considered to be a \emph{physically coherent} technique (or even extraordinarily explanatory) but \emph{mathematically inconsistent}. The path integral approach is an example of how the physical view, drawing from intuitive images of common experience, initially \emph{forces} and \emph{forges} some mathematical kit to explain particle phenomena.

\begin{exemplum}[Probability amplitude functional as sum over histories]
\label{exemplum "Probability amplitude functional as sum over histories"}
Two postulates summarize the Feynman's formulation \cite[sec. 3]{Feynman "Space-Time Approach to Non-Relativistic Quantum Mechanics"}.
\enumerationisinitium
\item Postulate I. \emph{The probability that a particle make a path in a region of space-time is given by the absolute square of a sum of complex contributions}. 

Suppose we have a particle which can have several values at $x$. Let $x_j$ be the result of the measurement of the coordinate $x$ at time $\tau_j$. The probability of a path, for successive values, $x_1, x_2, x_3, \mathellipsis$, at successive times, $\tau_1, \tau_2, \tau_3, \mathellipsis$, with a limit $\epsilon \to 0$ (where $\epsilon$ is the interval separating all subsequent position measurements), is a function of $x_1, \mathellipsis, x_j$, and let $\mathscr{P}(x_1, \mathellipsis, x_j, x_{j + 1}, \mathellipsis, x_\rotatedell)$ be this probability. Let $\alpha$ and $\beta$ be two results of two distinct measurements. Hence the probability that a path lies in a region $\Bbbk \subset \mathbb{R}^4_{1, 3}$ of space-time, say between $\alpha_j$ and $\beta_j$, for $x_j$, and between $\alpha_{j + 1}$ and $\beta_{j + 1}$, for $x_{j + 1}$, etc., is
\begin{align}
	& \int^{\beta_1, \mathellipsis, \beta_j}_{\alpha_1, \mathellipsis, \alpha_j} \int^{\beta_{j + 1}, \mathellipsis, \beta_\rotatedell}_{\alpha_{j + 1}, \mathellipsis, \beta_\rotatedell} \mathscr{P}\Lbrack:\cdots:\Rbrack = \int_{\Bbbk \subset \mathbb{R}^4_{1, 3}} \mathscr{P}\Lbrack:\cdots:\Rbrack, \notag \\
	& \text{setting } \mathscr{P}\Lbrack:(x_1, \mathellipsis, x_j, x_{j + 1}, \mathellipsis, x_\rotatedell)dx_1, \mathellipsis, dx_jdx_{j + 1}, \mathellipsis, dx_\rotatedell:\Rbrack,
\end{align} 
where the symbols $\Lbrack:$ and $:\Rbrack$ are for a repeat sign, see Glossary. For a measurement that allows to avoid possible uncertainties in the system (an ideal measurement), the path integral becomes
\begin{equation}
\label{equation "Path integral for ideal measurement"}
	\textcyrillic{\textit{ч}}_\mathbb{C}(\Bbbk) = \lim_{\epsilon \to 0} \int_{\Bbbk \subset \mathbb{R}^4_{1, 3}} \textcyrillic{\textit{Ч}}_\mathbb{C}(x_1, \mathellipsis, x_j, x_{j + 1}, \mathellipsis, x_\rotatedell)dx_1, \mathellipsis, dx_jdx_{j + 1}, \mathellipsis, dx_\rotatedell,
\end{equation}
where $\textcyrillic{\textit{ч}}_\mathbb{C}$ and $\textcyrillic{\textit{Ч}}_\mathbb{C}$ are complex numbers; specifically, $\textcyrillic{\textit{ч}}_\mathbb{C}$ designates a probability amplitude for $\Bbbk \subset \mathbb{R}^4_{1, 3}$, and $\textcyrillic{\textit{Ч}}_\mathbb{C}$ is the \emph{probability amplitude functional} of $x_j$ defining the entire path $x_\tau$. Here

· a path is determined by the positions $x_j$, or rather by a succession of points $x_j$, through which it passes at successive times $\tau_j$, i.e. at a sequence of equally spaced times 
\begin{equation}
	\tau_j = \tau_{j - 1} + \epsilon \to 0,
\end{equation}
· the probability that a particle is found to be in $\Bbbk \subset \mathbb{R}^4_{1, 3}$ with a measurement plus integral \eqref{equation "Path integral for ideal measurement"} is the square of $|\textcyrillic{\textit{ч}}_\mathbb{C}|^2$. 

\item Postulate II. \emph{Each path in the region contributes in equal way in magnitude, and the phase of any contribution is the Lagrangian action expressed in units of $\hbar$}.

The contribution $\textcyrillic{\textit{Ч}}_\mathbb{C}(x_\tau)$ from a certain path is proportional to $\exp{(\frac{i}{\hbar})}\mathscr{S}(x_\tau)$, in which the action
\begin{equation}
	\mathscr{S}(x_\tau) = \int\Lagrangian(\dot{x}_\tau, x_\tau)d\tau
\end{equation}
is the time integral of the Lagrangian $\Lagrangian(\dot{x}, x)$ along the path, cf. Dirac \cite{Dirac "The Lagrangian in Quantum Mechanics"}. A path here is determined by at all points, not just at $x_j$ as above, so it is necessary to assume that the function of $x_\tau$ is the path of a \emph{classical particle}, with $\Lagrangian$, between $\tau_j$ and $\tau_{j + 1}$, starting from $x_j$ at $\tau_j$ and reaching $x_{j + 1}$ at $\tau_{j + 1}$. Being that, under the \emph{classical path} the action is a \emph{minimum}, we can put 
\begin{equation}
	\mathscr{S} = \sum_j\left(\mathscr{S}(x_{j + 1}, x_j) = \min\int^{\tau_{j + 1}}_{\tau_j}\Lagrangian(\dot{x}_\tau, x_\tau)d\tau(x_{j + 1}, x_j)\right).
\end{equation}
Since the sum, in this equation, is infinite, even if $\epsilon$ is finite, we must limit the operation to a finite time interval, albeit arbitrarily long. Lastly, what can be inferred from the two postulates is
\begin{equation}
	\textcyrillic{\textit{ч}}_\mathbb{C}(\Bbbk) = \lim_{\epsilon \to 0} \int_{\Bbbk \subset \mathbb{R}^4_{1, 3}} \exp{\left\{\frac{i}{\hbar}\sum_j\mathscr{S}(x_{j + 1}, x_j)\right\}} \frac{dx_{j + 1}}{c_\textit{\dh}}\left(\frac{dx_j}{c_\textit{\dh}}\right),
\end{equation}
for a value 
\begin{equation}
	c_\textit{\dh} = (2\pi\hbar i/\momentum_0)^\frac{n}{2}\textit{\dh}^{-\frac{1}{2}}, 	
\end{equation}
where $\momentum_0$ is a constant in the number of particles $\frac{d\momentum}{\momentum_0}$ that have a given component of momentum in $d\momentum$, and $n$ degrees of freedom in spatio-temporal coordinates $x_{j + 1}$ at $\tau_{j + 1}$. \exemplumsymbol
\enumerationisfinis
\end{exemplum}

\begin{exemplum}[Hamiltonian path integral—phase space formulation]
\label{exemplum "Hamiltonian path integral—phase space formulation"}
A generalization of the Feynman's formulation is possible writing the amplitude as a product of integrals by recourse to phase spaces, 
\begin{align}
	& \langle x_{\rotatedell + 1}|\hat{U}(\tau_{\rotatedell + 1}, \tau_0)|x_0\rangle = \Lbrack:\prod^{\rotatedell}_{j = 1}\left(\int^{+\infty}_{-\infty}dx_j\right)\prod^{\rotatedell + 1}_{j = 1}\left(\int^{+\infty}_{-\infty}\frac{d\momentum_j}{2\pi\hbar}\right):\Rbrack \notag \\
	& \hspace{91pt} \times \exp{\left\{\frac{i}{\hbar}\sum^{\rotatedell + 1}_{j = 1}\Bigl(\momentum_j(x_j - x_{j -1}) - \epsilon\Hamiltonian(\momentum_j, x_j, \tau_j)\Bigr)\right\}},
\end{align}
with 
\begin{equation}
	\epsilon = \tau_j - \tau_{j - 1} = \frac{\tau_{\rotatedell + 1} - \tau_0}{\rotatedell + 1} > 0,
\end{equation}
where $\hat{U}(\tau_{\rotatedell + 1}, \tau_0)$ is the time evolution operator (propagator), $\momentum_j$ is a momentum variable, $\epsilon$ is an infinitesimal time interval, $\Hamiltonian$ is the \emph{classical Hamiltonian}, and $\textcyrillic{\textit{Ч}}$ is a functional integral (that is, an integration over all paths), for $j = 1, \mathellipsis, \rotatedell$. The limiting process will be
\begin{equation}
	\lim_{\rotatedell \to \infty}\Lbrack:\cdots:\Rbrack = \int^{x(\tau_{\rotatedell + 1}) = x_{\rotatedell + 1}}_{{x(\tau_0)} = x_0}\textcyrillic{\textit{Ч}}^{(\mathrm{i})}x\int\frac{\textcyrillic{\textit{Ч}}\momentum}{2\pi\hbar}.
\end{equation}
By combining both equations, the amplitude becomes
\begin{equation}
	\langle x_{\rotatedell + 1}|\hat{U}(\tau_{\rotatedell + 1}, \tau_0)|x_0\rangle = \int^{x(\tau_{\rotatedell + 1}) = x_{\rotatedell + 1}}_{{x(\tau_0)} = x_0}\textcyrillic{\textit{Ч}}^{(\mathrm{i})}x\int\frac{\textcyrillic{\textit{Ч}}\momentum}{2\pi\hbar}\exp{\bigl\{i\mathscr{S}(\momentum, x)/\hbar\bigr\}},
\end{equation}
and that is all. \exemplumsymbol
\end{exemplum}

\begin{margo}
\label{margo "Pioneering work for the path integral formulation"}
We are all quite clear that the Feynman's formulation is but a functional integration, the mathematical construction of which was inaugurated by P.J. Daniell \cite{Daniell "Integrals in An Infinite Number of Dimensions"}, and by N. Wiener \cite{Wiener "The Average of an Analytic Functional"} \cite{Wiener "The Average of an Analytic Functional and the Brownian Movement"} \cite{Wiener "Differential-Space"} \cite{Wiener "The Average value of a Functional"} \cite{Wiener "Generalized harmonic analysis"} for the  treatment of the \emph{Brownian motion} within the the analytic functional. \margosymbol
\end{margo}

\subsection{Ends and Means for Metric Functions}
\label{subsection "Ends and Means for Metric Functions"}

\begingroup
\footnotesize
The point to remember is that when we \emph{prove} a result without under­standing it—when it drops unforeseen out of a maze of mathematical formulae—we have no ground for hoping that it will apply except when the mathematical premises are rigorously fulfilled—that is to say, never [\,\dots]. But when we obtain by mathematical analysis an \emph{understanding} of a result—when we discern which of the conditions are essentially contributing to it and which are relatively unimportant—we have obtained knowledge adapted to the fluid premises of a natural physical problem [\,\dots]. [W]hereas for the mathematician insight is one of the tools and proof the finished product, for the physicist proof is one of the tools and insight the finished product. \\
\indent — \textsc{A.S. Eddington} \cite[p. 103]{Eddington "The Internal Constitution of the Stars"} 
	
\endgroup

\vspace{2mm}

Ends and means. For a physicist, nature (\textgreek{φύσις}) is the end of his own research, but it does not necessarily have to be for a mathematician. For a physicist, mathematics is a means, a tool.\footnote{
	Fermi's mindset is an Occamian paragon of \emph{parsimony}. He was a theoretical physicist, and yet he liked to \emph{shun} every \emph{redundancy}—or \emph{superfluous refinement}—of \emph{abstract mathematics}. Here is a direct evidence from Telegdi \cite[p. 97]{Telegdi "Fermi at Chicago"}: «Although endowed with remarkable analytic powers, Fermi often affected an aversion to abstract mathematics. [An] anecdot[e] may serve to illustrate his attitude: Once a notice appeared on the bulletin board announcing a course on the fundamentals of quantum mechanics. This notice read, “Students should be familiar with the mathematics of Hilbert spaces and Banach spaces.” Fermi commented, “Unfortunately I cannot learn about the fundamentals of quantum mechanics; I know about Hilbert spaces but not about Banach spaces”».
	} 
For a mathematician, this—working—tool is primarily used as an end in itself, when (pure) mathematics creates its own abstract structures; thence, end and means become the same thing. 

Set out below (Sections \ref{subsubsection "Schwarzschild, Gödel and Kerr Metrics"} and \ref{subsubsection "Dirac's Prediction of Anti-matter from Klein–Gordon Equation"}) are some minute considerations that broaden the above-stated topic a little. 

\subsubsection{Schwarzschild, Gödel and Kerr Metrics}
\label{subsubsection "Schwarzschild, Gödel and Kerr Metrics"}

\begingroup
\footnotesize
When all thermonuclear sources of energy are exhausted a sufficiently heavy star will collapse [\,\dots], this contraction will continue indefinitely [\,\dots]. The star thus tends to close itself off from any communication with a distant observer; only its gravitational field persists. \\
\indent — \textsc{J.R. Oppenheimer} and \textsc{H.[S.] Snyder} \cite[pp. 455-456]{Oppenheimer and Snyder "On Continued Gravitational Contraction"}\endnote{
	Historical note. This astrophysical investigation by Oppenheimer and Snyder is preceded by two other papers, on the nature of stellar neutron cores (each co-written with two different collaborators, R. Serber and G.M. Volkoff), which are inspired by a terse article on the origin of stellar energy \cite{Landau "Origin of Stellar Energy"}, by L. Landau.
	}

\vspace{2mm}

[A]fter a certain critical condition has been fulfilled, deviations from spherically symmetry cannot prevent space-time singularities from arising. \\
\indent — \textsc{R. Penrose} \cite[p. 58]{Penrose "Gravitational Collapse and Space-Time Singularities"}

\vspace{2mm}

If gravitational collapse is [\,\dots] inescapable in a star [\,\dots] [w]hat was once the core of a star is no longer visible. The [collapsing] core like the Cheshire cat fades from view. One leaves behind only its grin, the other, only its gravitational attraction [\,\dots]. [L]ight and particles incident from outside emerge and go down the black hole only to add to its mass and increase its gravitational attraction. \\
\indent — \textsc{J.A. Wheeler} \cite[pp. 8-9]{Wheeler "Our Universe: The Known and the Unknown"}

\endgroup

\vspace{2mm}

There are also cases, contrary to what Eddington (in epigraph) says, in which mathematics leads to apparently incomprehensible results, whose physical sensibleness emerges only with the passing of the years and the advance of technology: a suggestive example is the \emph{Schwarzschild metric} \cite{Schwarzschild "Uber das Gravitationsfeld eines Massenpunktes nach der Einsteinschen Theorie"} \cite{Schwarzschild "Uber das Gravitationsfeld einer Kugel aus inkompressibler Flussigkeit nach der Einsteinschen Theorie"}, 
\begin{equation}
\label{equation "Schwarzschild metric"}
	g \viz g_\textsc{s} = \left\{ds^2 = -\left(1 - \frac{2mG_\textsc{n}}{\rho}\right)dt^2 + \frac{d\rho^2}{1 - \frac{2mG_\textsc{n}}{\rho}} + \rho^2\left(d\theta^2 + \sin^2\theta d\phi^2\right)\right\},
\end{equation}
specifying that $\rho$ is the radial coordinate of a spherical surface around a body of mass $m$, $\theta$ denotes the polar angle (colatitude), and $\phi$ is the longitude on the spherical surface. The Eq. \eqref{equation "Schwarzschild metric"} provides examples of \emph{black holes} \cite{Abbott et al. LIGO Virgo "Observation of Gravitational Waves from a Binary Black Hole Merger"} \cite{Abbott et al. LIGO Virgo "Directly comparing GW150914 with numerical solutions of Einstein's equations for binary black hole coalescence"} \cite{The Event Horizon Telescope Collaboration "First M87 Event Horizon Telescope Results. I. The Shadow of the Supermassive Black Hole"}, and the occurrence of \emph{singularities} at $\rho = 2mG_\textsc{n}$. 

Conversely, there are models where the mathematical propositions are rigorously and fully satisfied but the physical sense is absent, and this is the case of the \emph{Gödel metric} \cite{Godel "An Example of a New Type of Cosmological Solutions of Einstein's Field Equations of Gravitation"},
\begin{equation}
\label{equation "Gödel metric"}
	g \viz g_\textsc{g} = \left\{ds^2 = \omega_\mathrm{v}^2\left(dx^2 + \frac{1}{2}e^{2x}dy^2 + dz^2 - (e^xdy + dt)^2\right)\right\},
\end{equation}
with its \emph{closed time-like curves} generating a rotating universe, where $\omega_\mathrm{v} > 0$ is a constant indicating the magnitude of the vorticity, and $(t, x, y, z)$ are the cylindrical coordinates, see \cite[sec. 5.7]{Hawking Ellis "The large scale structure of space-time"}. 

In the midst of these two geometries, there is e.g. the \emph{Kerr metric} \cite{Kerr "Gravitational Field of a Spinning Mass as an Example of Algebraically Special Metrics"}, 
\begin{equation}
\label{equation "Kerr metric"}
	g \viz g_{\textsc{k}\mathrm{e}} = 
	\left\{\!\begin{aligned}
    ds^2 = & -\left(1 - \frac{2m\rho}{\rho^2 + \alpha^2\cos^2\theta}\right)dt^2 \\
    & - 4m\rho\alpha\frac{\sin^2\theta}{\rho^2 + \alpha^2\cos^2\theta}dtd\phi \\
	& + \frac{\rho^2 + \alpha^2\cos^2\theta}{\rho^2 - 2m\rho + \alpha^2}d\rho^2 + (\rho^2 + \alpha^2\cos^2\theta)d\theta^2 \\
	& + \left(\rho^2 + \alpha^2 + \frac{2m\rho \alpha^2\sin^2\theta}{\rho^2 + \alpha^2\cos^2\theta}\right)\sin^2\theta d\phi^2
    \end{aligned}\right\},
\end{equation}
expressed in \emph{Boyer–Lindquist coordinates} \cite{Boyer and Lindquist "Maximal Analytic Extension of the Kerr Metric"}, for a parameter $\alpha = \frac{J}{m}$, where $J$ is the angular momentum. The Eq. \eqref{equation "Kerr metric"} is above all an algebraic solution, and it does not guarantee a realistic process for the external field of a rotating collapsing star \cite[p. 260]{Thorne "Relativistic Stars Black Holes and Gravitational Waves (Including an in-Depth Review of the Theory of Rotating Relativistic Stars)"}, but shows, as B. Carter \cite{Carter "Complete Analytic Extension of the Symmetry Axis of Kerr's Solution of Einstein's Equations"} \cite{Carter "Global Structure of the Kerr Family of Gravitational Fields"} had guessed, an asymptotic value for the formation of a rotating black hole: for $\alpha > m$, the Kerr metric is unrealistic, for $\alpha = m$, it represents an extremal. 

The Schwarzschild, Gödel and Kerr metrics \eqref{equation "Schwarzschild metric"} \eqref{equation "Gödel metric"} \eqref{equation "Kerr metric"} are three exact solutions of the Einstein field equations, so they are \emph{mathematically coequal}, but they have \emph{different physical implications}.

\subsubsection{Dirac's Prediction of Anti-matter from Klein–Gordon Equation}
\label{subsubsection "Dirac's Prediction of Anti-matter from Klein–Gordon Equation"}

\begingroup
\footnotesize
[T]he mathematician plays a game in which he himself invents the rules while the physicist plays a game in which the rules are provided by Nature.\footnote{
	We do not agree, in any way (for the reasons already exposed), with the continuation of Dirac's sentence: «but as time goes on it becomes increasingly evident that the rules which the mathematician finds interesting are the same as those which Nature has chosen».
	} \\
\indent — \textsc{P.A.M. Dirac} \cite[p. 124]{Dirac "The Relation between Mathematics and Physics"}

\endgroup

\vspace{2mm}

The Dirac model of the anti-matter is another resounding case that starts from a mathematically provable but physically incomprehensible result. And it is a further evidence of how complicated the attempt is to «interpret the mathematics physically» \cite[p. 243]{Dirac "The Quantum Theory of the Emission and Absorption of Radiation"}, that is, to «interpret the new [pure] mathematical features in terms of physical entities» \cite[p. 60]{Dirac "Quantised Singularities in the Electromagnetic Field"}, combined with the attempt to interpret nature mathematically, that is, to «formulate the experimental data in mathematical terms» \cite[ivi]{Dirac "Quantised Singularities in the Electromagnetic Field"}. For Dirac, the departure reasoning is the \emph{Klein–Gordon} equation \cite{Klein "Elektrodynamik und Wellenmechanik vom Standpunkt des Korrespondenzprinzips"} \cite{Gordon "Der Comptoneffekt nach der Schrodingerschen Theorie"}, 
\begin{subequations}
\label{subequations "Klein–Gordon equation"}
\begin{align}
	& \left(\frac{1}{c^2}\frac{\partial^2}{\partial{t^2}} - \nabla^2  + \frac{m^2c^2}{\hbar^2}\right)\psi = 0, \\
	& \dAlembertian + m^2\psi = 0,
\end{align} 
\end{subequations}
where $\nabla^2$ is the Laplacian, and 
\begin{equation}
	\dAlembertian = \frac{1}{c^2}\frac{\partial^2}{\partial{t^2}} - \nabla^2 
\end{equation}
is the d'Alembertian, which constitutes the basement for his Eq. \eqref{equation "Dirac equation"}. The Eq. \eqref{subequations "Klein–Gordon equation"} refers «equally well» to an electron with charge $-e$ and as to one with charge $[+]e$ \cite[p. 612]{Dirac "The Quantum Theory of the Electron"}, but the second type appears to have no physical meaning (it having never been observed in practice). 

Dirac's solution, as is well-known, comes from the proposal of the “holes” or “vacant (unoccupied) states” in the negative energy distribution \cite{Dirac "A Theory of Electrons and Protons", Dirac "On the Annihilation of Electrons and Protons"}, which then gave rise, at the suggestion of J.R. Oppenheimer \cite{Oppenheimer "On the Theory of Electrons and Protons"}, to the conjecture about the existence of an anti-matter particle, with the same mass but opposite charge to an electron, called «anti-electron» \cite[p. 61]{Dirac "Quantised Singularities in the Electromagnetic Field"}.\footnote{
	The discovery of the anti-electron, renamed «positron», follows with the experiments of C.D. Anderson \cite{Anderson "The Apparent Existence of Easily Deflectable Positives"} \cite{Anderson "The Positive Electron"}, and P.M.S. Blackett \& G.P.S. Occhialini \cite{Blackett and Occhialini "Some Photographs of the Tracks of Penetrating Radiation"}. See the summary \cite[§ 73]{Dirac "The Principles of Quantum Mechanics (1958)"}.
	}

\subsection{What is Out There? Inside the Head of a Mathematician}
\label{subsection "What is Out There? Inside the Head of a Mathematician"}

\subsubsection{Imagery-generated Spaces}

\begingroup
\footnotesize
In nature, we actually know only the motion [\textcyrillic{\textit{В природе мы познаем собственно только движение}}], without which sensory impressions are impossible. As a result, all other concepts, e.g. those relating to geometry [\textcyrillic{\textit{понятия\,\dots Геометрические}}], are artificially produced by our mind [\textcyrillic{\textit{произведены нашим умом искусственно}}], once they are drawn from the properties of motion; that would explain why the space in itself, [taken] separately, does not exist [\textcyrillic{\textit{пространство, само собой, отдельно, для нас не существует}}]. After which there can be no contradiction in our mind, if we assume that some forces in nature follow a geometry, others [follow] their own particular geometry [\,\dots]. \\
\indent In nature there are neither straight nor curved lines, there are neither planes nor curved surfaces [\textcyrillic{\textit{в природе нет ни прямых, ни кривых линий, нет плоскостей и кривых поверхностей}}]: we find only bodies [\textcyrillic{\textit{тела}}] in it, so that everything else, created by our imagination, exists in theory alone [\textcyrillic{\textit{все прочее, созданное нашим воображением, существует в одной теории}}]. \\
\indent — \textsc{N.I. Lobačevskij} \cite[pp. 64, 69]{Lobacevskij "Novyye nachala geometrii"}

\vspace{2mm}

If to the intuitive space, that is, to the order of our sensitivity [\textit{ordine della nostra sensibilità}], corresponds an order of external things [\textit{ordine delle cose esterne}], i.e. a real space for which the postulates express real properties, it is an issue closely linked with the problem of knowledge, which transcends the field of Geometry [\,\dots]. It is indifferent [\,\dots] that the issue is resolved in a skeptical (or idealistic) sense [\,\dots] or that instead, by attributing to space an objective reality, as Helmholtz \cite{von Helmholtz "Ueber den Ursprung und die Bedeutung der geometrischen Axiome"} [says], the value of postulates considered as physical truths is discussed. From these matters nothing can arise that destroys the \emph{logical} value of Geometry, and that changes our concept of intuitive space; therefore not only the \emph{possibility} but also the \emph{mathematical importance} of Geometry itself are independent of such issues.\footnote{
	Cf. the letter of Enriques to Castelnuovo \cite[№ 292, 12 maggio 1896, pp. 266-267]{Enriques "Riposte armonie. Lettere di Federigo Enriques a Guido Castelnuovo"}: «[there are very] few [works on space] that deserve some consideration; the more ones mope around the eternal question whether or not space is given a priori, and do not understand that space can be understood in two ways, “subjective space” [intuitive space, from our sensitivity] and “objective” [or real space]».
	} \\
\indent — \textsc{F. Enriques} \cite[pp. 3-4]{Enriques "Conferenze di Geometria. Fondamenti di una Geometria iperspaziale"}

\endgroup

\vspace{2mm}

\enumerationisinitium
\item From a narrowly mathematical perspective, the issue of the consideration of phenomenal reality does not exist, or it does not exist as a problem of ordering of experimental data. That is why a mathematician—see Enriques in epigraph—can allow himself to think about the space, for example, without heeding the implications of the physical world; he enjoys the opportunity of conceiving the space as a mental network of connections without reference to an external ordered collection of matter,\footnote{
	\label{footnote "Einstein's letter to M. Solovine, 30 March 1952"}
	The physicist, unlike the mathematician, feels the urge of giving prominence to the existence of a (self-)ordered reality. Iconic is the opinion of Einstein, which is much more unbalanced than that of Enriques. Einstein \cite[letter to M. Solovine, 30 March 1952]{Einstein "Letter to Maurice Solovine 30 March 1952"} writes: «You find it strange that I consider the comprehensibility of the world (to the extent that we are authorized to speak of such) as a miracle [\textit{Wunder}] or as an eternal mystery [\textit{ewiges Geheimnis}]. Well, a priori one should expect a chaotic world that is in no way comprehensible by thought. One could (yes \emph{one should}) expect the world to be subjected to law only to the extent that we order it through our intelligence. Ordering of this kind would be like the alphabetical ordering of the words of a language. By contrast, the kind of order created, for instance by Newton's theory of gravitation, is wholly different. Even if the axioms of the theory are proposed by man, the success of such a project presupposes a high degree of ordering of the objective world, which one has absolutely no right to expect a priori. So here lies the “miracle” which is being constantly reinforced as our knowledge expands».

	As we have seen above (Sections \ref{subsection "Anthropoid Ways Ia. Regularity and Formal Relations"}, \ref{subsection "Anthropoid Ways Ib. Symmetry and Invariance in Physics: the Impact of Group Theory"} and \ref{subsection "Anthropoid Ways III. Laws of Nature"}), there is a regularity of nature, or \textgreek{ἁρμονία} of the physical world. And this is conceivably what Einstein is talking about in that letter. But the way in which Einstein expresses himself is unacceptable to me. Frankly, I find it difficult to approve such a “mysterious” and “miraculous” degree of ordering of the «objective world», which is independent from experience:

	· to be right, there is no \emph{a priori} in nature, but only in the minds of men: “a priori/a posteriori” are human terms, to wit, fumesophy;

	· and then, why should the “mystery” of an ordered world be more astounding than that of a chaotic system? Why should a chaotic world be any less “miraculous”? Based on what we know, the physical world is \emph{stratiform} or \emph{stratum-shaped} (emergence, levels of organization, complex systems, etc.); 
	
	· due to this, in nature there is no axiology that puts an ordered system in the first place, as the pinnacle of evolution of the universe, and a chaotic one on the sidelines. Seemingly, order and chaos go hand in hand.
	}  
from corpuscles to celestial bodies.\footnote{
	Compare with V. Benci and P. Freguglia \cite[cap. 1]{Benci e Freguglia "Modelli e realta. Una riflessione sulle nozioni di spazio e tempo"}.
	}
\item Alternatively, a mathematician—see Lobačevskij in epigraph—can specify the existence of bodies in nature and their motion (our experience with the surrounding world), but this does not affect the imaginative construction in his mind about spatial relations, which are all artificial, or about representation in space, and interpretation of space, because the space in itself does not exist.
\enumerationisfinis

\subsubsection{No One, One, and One Hundred Thousand Spaces}

\begingroup
\footnotesize
If mathematicians, instead of talking about “space of $n$ dimensions”, had said, e.g., “continuum of $n$ coordinates”, nobody would have dealt with their novelty. But in this way, however, all the help that hyperspaces can give in the representation of those facts—even on the mechanical or physical side—would have been lacking, [and such facts], depending on multiple elements of variability, are not easily schematized in ordinary space [\,\dots]. Hyperspaces provide an appropriate and suggestive language that brings together apparently disparate facts, arousing analogies and hence fruitful inductions.\endnote{
	Original It. version: «[S]e i matematici invece di parlare di “spazio ad $n$ dimensioni”, avessero detto, per es., “continuo ad $n$ coordinate”, nessuno si sarebbe occupato delle loro novità. Ma in tal modo però sarebbe mancato tutto l'aiuto che possono dare gl'iperspazi nella rappresentazione di quei fatti — anche meccanici o fisici — i quali, dipendendo da molteplici elementi di variabilità, non sono facilmente schematizzabili nello spazio ordinario [\,\dots]. [G]l'iperspazî forniscono un linguaggio opportuno e suggestivo che ravvicina fatti apparentemente disparati, suscitando analogie e quindi induzioni feconde».
	} \\
\indent — \textsc{F. Severi} \cite[pp. 9-10]{Severi "Ipotesi e realta nelle scienze geometriche"}
 
\endgroup

\vspace{2mm}

Another difference that disconnects a mathematical space from a fiscal one is the freedom of the dimensional number, which dates back, historically, to the explicitly abstract works of J. Plücker \cite{Plucker "Theoremes generaux concernant les equations d'un degre quelconque entre un nombre quelconque d'inconnues"} and, mainly, of A. Cayley \cite{Cayley "Chapters in the Analytical Geometry of $(n)$ Dimensions"} and H. Grassmann \cite{Grassmann "Die lineale Ausdehnungslehre ein neuer Zweig der Mathematik"} \cite{Grassmann "Anhang III (1877) Kurze Uebersicht uber das Wesen der Ausdehnungslehre"};\footnote{
	H. Grassmann \cite[pp. 277-278]{Grassmann "Anhang III (1877) Kurze Uebersicht uber das Wesen der Ausdehnungslehre"}: «My theory of extension [\textit{Ausdehnungslehre}] sets up the abstract foundation of the theory of space (geometry), i.e. it is a purely mathematical science, detached from all spatial intuition [\,\dots]. The theory of space, since it is somehow connected to something given in nature, that is, to the [three-dimensional] space, is not a branch of pure mathematics, but an application of it to nature [\,\dots]. The propositions of the theory of extension are not mere translations of geometric propositions into an abstract language, but have a much more general meaning; for while the theory of [ordinary] space remains bound to the three dimensions of space, the abstract science [\textit{abstrakte Wissenschaft}] is free of these limits».
	}
wherefore many of the unified  theoretical framework of physics in a multi-dimensional geometry, all descendants of the Kaluza–Klein theory \cite{Kaluza Zum Unitatsproblem in der Physik} \cite{Klein "Quantentheorie und funfdimensionale Relativitatstheorie"}, see e.g. E. Witten \cite{Witten "Search for a Realistic Kaluza-Klein Theory"}, are cataloged, for now (on the edge of our current understanding), by a host of physicists as mathematical (scilicet: non-physical) theories, namely—to borrow from Veltman \cite[p. 308]{Veltman "Facts and Mysteries in Elementary Particle Physics"}—as \emph{figments of the theoretical mind}. Trenchant criticisms of string theory are in R. Penrose \cite[secc. 1.9-1.16]{Penrose "Fashion Faith and Fantasy in the New Physics of the Universe"}, cf. Margo \ref{margo "Margarita credulitatis"}.

\begin{margo}[Margarita credulitatis]
\label{margo "Margarita credulitatis"}
The following judgment of the mathematician R. Thomas—email message in the early 2000s reported by R. Penrose \cite[p. 90, see also pp. 92-93]{Penrose "Fashion Faith and Fantasy in the New Physics of the Universe"}—is a \emph{pearl of credulousness}. Thomas talks about the \emph{mirror symmetry} \cite{Greene Plesser "Duality in Calabi-Yau moduli space"} \cite{Candelas Lynker Schimmrigk "Calabi-Yau manifolds in weighted P4"} that permits to find an equivalence between two geometrically different Calabi–Yau manifolds, by counting the number of rational curves on such topological spaces (cf. Chapter \ref{chapter "Calabi–Yau Theorem: a Non-linear Complex Equation of Monge–Ampère Type on Compact Kähler Manifolds"}). It is a familiar story for insiders, so I shall refrain from revealing all the hidden details of this affair; additionally, the story is already narrated by B. Greene, in his best-seller,\footnote{
	B. Greene, \textit{The Elegant Universe: Superstrings, Hidden Dimensions, and the Quest for the Ultimate Theory}, Vintage Books, New York, 2000, pp. 259-262.
	} 
and by S.-T. Yau \& S. Nadis.\footnote{
	S.-T. Yau and S. Nadis, \textit{The Shape of Inner Space: String Theory and the Geometry of the Universe's Hidden Dimensions}, Basic Books, New York, 2010, pp. 163-173. 
	}
R. Thomas writes: 

\vspace{2mm}

\begingroup
\footnotesize
I can't emphasise enough how deep some of these dualities are, they constantly surprise us with new predictions. They show up structure never thought possible. Mathematicians confidently predicted several times that these things weren't possible, but [physicists] like [P.] Candelas, [X.C. De La Ossa, P.S. Green, L. Parkes] \cite{Candelas De La Ossa Green Parkes "A pair of Calabi-Yau manifolds as an exactly soluble superconformal theory"} have shown this to be wrong. Every prediction made, suitably interpreted mathematically [when compared with the result of mathematicians G. Ellingsrud and S.A. Strømme, who employed more rigorous techniques, but not related to the mirror symmetry technique], has turned out to be correct. And \emph{not} for any \emph{conceptual} maths reason so far—we have no idea why they're true, we just compute both sides independently and indeed find the same structures, symmetries and answers on both sides. To a mathematician these things cannot be coincidence, they must come from a higher reason [\textit{sic}]. And that reason is the \emph{assumption that this big mathematical theory \textnormal{[he is referring to the string theory]} describes nature}\,\dots.

\endgroup

\vspace{2mm}

I beg your pardon? How can people be so guileless? What does the mathematics of the moduli spaces for a Calabi–Yau manifold and its mirror have to do with nature? It is an extraordinary \emph{mathematical tool}, which made it possible, in the previously mentioned case, to count the exact number of degree 3 curves ($k = 3$) on $\mathbb{CP}^4[5]$, a hypersurface of degree 5 and dimension 3, with a real 6-dimensionality, viz. a quintic Calabi–Yau 3-fold in a projective 4-space $\mathbb{P}^4$. This number, incidentally, corresponds to 317~206~375 \cite[p. 60, Table 4]{Candelas De La Ossa Green Parkes "A pair of Calabi-Yau manifolds as an exactly soluble superconformal theory"}. Well, \emph{nature has nothing to do with this}. 

To learn more about the quintic 3-fold, see D.A. Cox, S. Katz \cite[chap. 2]{Cox Katz "Mirror Symmetry and Algebraic Geometry"}. Some mathematical demonstrations of mirror symmetry—under the physical suggestion revealed by Candelas et al., and the subsequent its reformulation (in conjecturable terms of homological algebra of mirror symmetry) by M. Kontsevich \cite{Kontsevich "Homological Algebra of Mirror Symmetry"}—are due to A. Givental \cite{Givental "A Mirror Theorem for Toric Complete Intersections"} and B.H. Lian, K. Liu \& S.-T. Yau \cite{Lian Liu and Yau "Mirror principle I"} \cite{Lian Liu and Yau "Mirror principle II"} \cite{Lian Liu and Yau "Mirror principle III"} \cite{Lian Liu and Yau "Mirror principle IV"}. \margosymbol
\end{margo}

\subsection{Intersections and Grafts}
\label{subsection "Intersections and Grafts"}

The above contrast (Sections \ref{subsection "Proof vs. Empirical Datum"}, \ref{subsection "Procrustean Bed: the Example of the Dirac Delta Function"}, \ref{subsection "Straitjacket for Feynman Path Integrals"}, \ref{subsection "Ends and Means for Metric Functions"}, and \ref{subsection "What is Out There? Inside the Head of a Mathematician"}) between mathematics of mathematicians and mathematics for physicists, is easily overcome in special cases. There are higher intellects, with their transversal work, at the top of the three disciplines, more or less separated from each other: mathematics (in its pure aspiration), mathematical physics, and theoretical physics. 

\begin{scholium}[Disciplinary clarification]
Let us try to get some clarification.
	
· \emph{Mathematics}, including its \emph{pure} domain, and \emph{physics} in the Greco-Hellenistic period belong together, they are not distinct. 

· \emph{Mathematical physics}, as opposed to \emph{experimental physics}, was born in France, between 1800 and 1820.
	
· The so-called \emph{pure} mathematics, as a discipline in its own right, appears only in the nineteenth century (beginning with Abel, Jacobi, and Galois); in the previous centuries, the distinction between \emph{pure} mathematics and \emph{applied} mathematics existed but it was occasional and not disciplinary. 
	
· \emph{Theoretical physics}, as separate from mathematical physics, was born in the German-speaking areas, between 1870 and 1890. Estimating the date of birth of a discipline is never a clean operation. Be that as it may, this distinction is already clear in E. Mach \cite{Mach "Die Mechanik in ihrer Entwicklung: historisch-kritisch dargestellt"} \cite{Mach "The Science of Mechanics. A Critical and Historical Exposition of Its Principles"} (1883-1889).

These dates are only indicative, and can not be taken too rigidly. A variegated reconstruction of the nascent relationship between pure mathematics and physical researches in the nineteenth century is in \cite[cap. I, pp. 13-48]{Lolli "Le ragioni fisiche e le dimostrazioni matematiche"}. \scholiumsymbol
\end{scholium}

Here some names.

· The Euclidean corpus. Alongside the \textit{Elements}, it should be remembered his works on optics and catoptrics \cite{Euclid "Prospettiva di Euclide"} \cite{Euclid "Euclidis Optica Opticorum recensio Theonis Catoptrica cum Scholiis antiquis"}.

· Archimedes of Syracuse, in whose work mathematics and physics are all one (see Sections \ref{subsubsection "Two Exemplary Evidences"} and \ref{subsubsection "Scholium: Greco-Hellenistic Scientific Modus Operandi (the Origin of Hypotheses)"}). The same can be said, relatively to the unity of mathematics \& physics, of Apollonius of Perga \cite{Apollonius of Perga "Francisci Maurolyci Messanensis Emendatio"} \cite{Apollonius of Perga "Apollonii Pergaei quae graece exstant cum commentariis antiquis I"}.

· Galileo, «nato per le Mattematiche» \cite[p. lxv]{Viviani "Racconto Istorico della vita del Sig. Galileo Galilei"} (but actually he is not a mathematician in a very real sense), and E. Torricelli, with his mathematical Cavalieri's principle (method of indivisibles), have developed many applications in the natural sciences; and, indeed, this is the origination of the \emph{geometrization of physical quantities}, the description of a \emph{physical reality} as/through a \emph{geometrical concept} \cite[cap. 2.2]{Giusti "Euclides reformatus. La teoria delle proporzioni nella scuola galileiana"}.

· I. Newton, whose productions have ranged from infinitesimal calculus\endnote{
	The attribution of the invention of infinitesimal calculus is an exciting topic among historians of mathematics: in addition to the works of I. Barrow and the Leibniz–Newton controversy, if there is anybody to mention, that is P. de Fermat: he pushed right up to the gates of differential/integral calculus; see e.g. \cite{Cajori "Who Was the First Inventor of the Calculus?"} \cite{Giusti "Il calcolo infinitesimale tra Leibniz e Newton"} \cite[pp. 30-52]{Giusti "Piccola storia del calcolo infinitesimale dall'antichita al Novecento"} \cite{Giusti "Dalla Geometrie al calcolo: il problema delle tangenti e le origini del calcolo infinitesimale"}. Yet there is a but: the presence of highly refined \emph{infinitesimal methods} is undoubtedly already in Archimedes, see in particular \cite{Archimedes "De lineis spiralibus"} \cite{Archimedes "Quadratura parabolae"}. The truth is that the concept of infinity is not rejected by Greek culture, as some historiography would like us to believe; e.g. the infinite sequence of additions is strongly present in Archimedean reasoning. He uses both the actual infinity (in his \textit{Method} \cite{Archimedes "The Method of mechanical theorems of Archimedes to Eratosthenes"} \cite{Archimedes "The Method of Archimedes Recently Discovered by Heiberg"}), to meet heuristic needs, and the \emph{potential infinity} (in the rest of his works), to meet the requirements of rigor, in accordance with the modern analysis. Integration theory and differential geometry, we can say, were born by generalizing Archimedes' techniques \cite{Archimedes "De lineis spiralibus"}, cf. \cite[pp. 55, 387-388]{Russo "The Forgotten Revolution: How Science Was Born in 300 BC and Why it Had to Be Reborn"} \cite[cap. 8]{Russo "Archimede. Un grande scienziato antico"}.
	} 
to celestial mechanics.

· The fab Bernoullis the pure inclination of which acted as a bank for natural sciences.

· L. Euler, who touched all branches of (pure) mathematics available at the time, going even into engineering field-work.

· J. le Rond d'Alembert, J.-L. Lagrange, P.-S. de Laplace, J.-.B.J. Fourier, J.C.F. Gauss, W.R. Hamilton, B. Riemann, E. Beltrami, F. Klein, G. Ricci Curbastro, H. Poincaré, D. Hilbert, V. Volterra, É. Cartan, F. Enriques, T. Levi-Civita, H. Weyl, A.N. Kolmogorov, J. von Neumann, and V.I. Arnold, all trained as mathematicians but repeatedly stimulated by physical problems; e.g. Levi-Civita is primarily a mathematician, and also a mathematical physicist (since he mentioned a lively interest in physics, with typical themes of subsequent theoretical physics).

· N.H. Abel, although his name is rooted in the unsolvability of the general quintic by radicals; there is no shortage of contributions to physical theories.

· É. Galois, whose \emph{abstract algebra} represents the conceptual beginning for the group theory, which plays a vital role in physics (e.g. classical and quantum mechanics, theory of relativity, high energy physics, gauge theory) coming from his magic pithos. The other two prominent mathematicians in group theory are F. Klein and the ubiquitous M.S. Lie.
 
· J.C. Maxwell, L. Boltzmann, both of them, albeit in different manners, with a strong propensity to mathematize physics (the latter was accused by Kelvin and P.G. Tait of not doing physics but only mathematics, or, say, an abstract theory with no relation to experimental observations).
	
· G. Cantor, with his hierarchy of infinites, leads us to reflect on which order type of mathematical infinity is more suitable for representing the laws of nature: continuous (real) infinity, or discontinuous?

· E. Majorana, a theoretical physicist but with a natural vocation to mathematical calculation.

· Dau's school, viz. L.D. Landau \& collaborators, who in their encyclopedic \textit{Course of Theoretical Physics} exhibit an indissoluble amalgam between physics and mathematics.\footnote{
	Here is what M.I. Kaganov \cite[p. 351]{Kaganov "Encyclopaedia of theoretical physics (Ru.)"} = \cite[p. 292]{Kaganov "Encyclopaedia of theoretical physics"} says about this mammoth work: «[T]he \textit{Course} is by no means just a handbook of mathematical methods. The whole exposition is based on physical concepts, either general ones (such as conservation laws) or model ones (ideal gas, collisionless plasma, a strictly periodic crystal, etc.). However, in reading the \textit{Course} there arises (or is enhanced) the understanding of the fact that there is no theoretical physics, and that there cannot be any, without a rigorous mathematical apparatus [\textcyrillic{\textit{теоретической физики нет и не может быть без строгого математического аппарата}}]. Estimates, suggestive arguments are needed specifically as suggestive arguments, utilizing which one can construct a rigorous theory whose result must necessarily be a formula (or a curve) relating physical quantities. If in the initial formulations the authors have indulged in a certain deliberate haste (in any event: essentially, the fundamental equations of any physical theory cannot be derived, they are a mathematical distillation of our experience [\textcyrillic{\textit{основные уравнения любой физической теории не могут быть выведены, они — математическая концентрация нашего опыта}}]), then later in going on to the development of the theory the authors are rigorous and very punctilious, although nowhere (over the whole \textit{Course}) do they indulge in purely mathematical ‘epsilontics’ taking the point of view (with perfect justification) that the aim of theoretical physics is not to prove existence theorems for solutions, but to obtain the solutions directly».
	}

· S. Ramanujan, a pure mathematician, whose intuitions have had an impact on physics. It is the case of \emph{mock modular forms}, dating back to the Ramanujan's papers \cite{Ramanujan "Mock theta-functions"} \cite{Andrews Berndt "Ramanujan's Lost Notebook I"}, reinterpreted to compute \cite{Dabholkar Murthy and Zagier "Quantum Black Holes Wall Crossing and Mock Modular Forms"} the black hole entropy \cite{Bekenstein "Black Holes and the Second Law"} \cite{Bekenstein "Black Holes and Entropy"} \cite{Hawking "Gravitational Radiation from Colliding Black Holes"} \cite{Hawking "Black hole explosions?"}. 

· R. Caccioppoli and E. De Giorgi: it is known that the calculus of variations switches easily from analysis to applications.

· R.P. Feynman: there are those who make a reckless use of mathematics, as a \emph{non}-mathematician: Feynman \cite[\textit{A Different Box of Tools}]{Feynman "Surely You're Joking Mr. Feynman!"} is one of them. One thinks of the path integral formulation and, in the main, the monomials of non-commutative algebra, reworked by G.-C. Rota in combinatorics previously associated with the invariant theory, see e.g. \cite{Rota "Combinatorial Theory and Invariant Theory"}, and then subsumed under the letterplace algebra and superalgebra, see e.g. \cite{Grosshans Rota Stein "Invariant Theory and Superalgebras"}.

· R. Thom and B. Mandelbrot: the first mathematician (a topologist), starting from the Bourbakist structuralism, lays the foundation for the catastrophe theory \cite{Thom "Structural Stability and Morphogenesis: An Outline of a General Theory of Models"} \cite{Thom "Mathematical Models of Morphogenesis"}, devoted an interpretation of natural phenomena, especially in biology; Mandelbrot, as it is well known, with its fractal geometry \cite{Mandelbrot "Les objets fractals. Forme hasard et dimension"} \cite{Mandelbrot "The Fractal Geometry of Nature"} \cite{Mandelbrot "Fractals and Chaos: The Mandelbrot Set and Beyond"} \cite{Mandelbrot "The Fractalist: Memoir of a Scientific Maverick"}, gave birth to a steady dialogue in the most disparate areas of application.

· A. Grothendieck: even in his work, nestled in the deepest abstractions, it is possible to detect some solutions that are adaptable in natural sciences. This is even more true for other Bourbakist, e.g. just think about the (direct or indirect) contributions from L. Schwartz and A. Connes to \textsc{qft}. 

· J.F. Nash, Jr., although his contributions mostly concern differential geometry, and partial differential equations, in some of his pages the focus is paid to game theory.

· M.F. Atiyah: the Atiyah–Singer index theorem (Section \ref{subsubsection "Margo. Atiyah–Singer Index Theorem"}), and his papers \cite{Atiyah "Collected Works Vol. 5 Gauge Theories"} on the gauge field theory (with particular attention to the Yang–Mills theory, see e.g. \cite{Atiyah Hitchin and Singer "Self-duality in four-dimensional Riemannian geometry"} \cite{Atiyah "Geometry of Yang-Mills Fields"}), show various facets of the link between, on the one hand, topology, geometry and analysis, and, on the other, its theoretical correspondent in physics.\footnote{
	As an example of versatility, the Atiyah–Singer theorem has the quality of being valid along multiple demonstrations: e.g. J.-M. Bismut \cite{Bismut "The Atiyah-Singer index theorem for families of Dirac operators: Two heat equation proofs"} gave a couple of fine proofs of the index theorem via the heat equation, for a family of Dirac operators; see also \cite{Bismut "The Atiyah-Singer Theorems: A Probabilistic Approach. I. The Index Theorem"} \cite{Bismut "The Atiyah-Singer Theorems: A Probabilistic Approach. II. The Lefschetz Fixed Point Formulas"}, where the Atiyah–Singer theorem—with Lefschetz fixed point formulæ—is proved for classical elliptic complexes by exploiting probabilistic methods.
	}

· R. Penrose's studies shuffle between mathematics and theoretical physics; 

· S.-T. Yau's several researches in pure mathematics are applied in some pieces of fundamental physics; suffice it to mention E. Witten's works  on Calabi–Yau manifolds, as well as Yau \& collaborators' studies \cite{Schoen and Yau "Positivity of the Total Mass of a General Space-Time"} \cite{Wang Yau "Quasilocal Mass in General Relativity"} \cite{Chen Wang Yau "Conserved Quantities in General Relativity: From the Quasi-Local Level to Spatial Infinity"} \cite{Chen Wang Wang and Yau "Supertranslation invariance of angular momentum"} on the quasilocal mass and angular momentum in general relativity.

· G. Perelman, whose articles in geometry and topology intermingle with matters of physics.

\subsubsection{Margo. Atiyah–Singer Index Theorem}
\label{subsubsection "Margo. Atiyah–Singer Index Theorem"}

The index theorem \cite{Atiyah and Singer "The index of elliptic operators on compact manifolds"} \cite{Atiyah and Singer "The index of elliptic operators: I"} \cite{Atiyah and Segal "The index of elliptic operators: II"} \cite{Atiyah and Singer "The index of elliptic operators: III"} \cite{Atiyah and Singer "The index of elliptic operators: IV"} \cite{Atiyah and Singer "The index of elliptic operators: V"} blooms from the need to find the \emph{Riemannian version of the Dirac equation in a context of algebraic geometry}, cf. \cite[p. 1]{Atiyah "Papers on Index Theory 56-93a (1963-84)"}. It makes use of a Dirac-type operator of a spin manifold, and verifies that this operator coincides with an \emph{elliptic differential operator} of first order on spinor fields, or rather, between spaces of spinorial field equations; so the index of the Dirac-type operator, aka \emph{Dirac–Atiyah–Singer operator} \cite[sec. 5]{Atiyah and Singer "The index of elliptic operators: III"},
\begin{equation}
	D_\textsc{das} \colon \Gamma_{\sezione}\left(\mathring{\mathcal{P}}_\textit{ß}^+\right) \to \Gamma_{\sezione}\left(\mathring{\mathcal{P}}_\textit{ß}^-\right)
\end{equation}	
is but a \emph{topological invariant}, letting $\mathcal{M}$ be a spin manifold with a spinor bundle (cf. Section \ref{subsection "Spinor Map (6-Dimensional Homomorphism): the Covering $SL_2(C)$ to $SO_{1, 3}^+(R)$"})
\begin{equation}
\mathring{\mathcal{P}}_\textit{ß} = \mathring{\mathcal{P}}_\Spin  \times_{\Spin(n)}\mathfrak{U}^n_{(\mathbb{C})}, \enspace \pi \colon \mathring{\mathcal{P}}_\Spin \to \mathcal{M},
\end{equation}
where 
\[
	\mathring{\mathcal{P}}_\Spin \overset{\pi}{\longrightarrow} \mathcal{M}
\]
denotes a $\Spin(n)$-principal bundle, and
\begin{equation}
	\mathfrak{U}^n_{(\mathbb{C})} \viz \mathbb{C}^{2^m} = \underbrace{\mathbb{C}^2 \otimes \mathellipsis \otimes \mathbb{C}^2}_{m \text{ times}}, \enspace n = 2m, 2m + 1,
\end{equation}
is the vector space of $n$-spinors. Otherwise expressed, the spinor bundle $\mathring{\mathcal{P}}_\textit{ß}$ is a complex vector bundle $\varsigma \colon \mathring{\mathcal{P}}_\textit{ß} \to \mathcal{M}$. Note that a spinor field is a section $\sezione\bigl(\mathring{\mathcal{P}}_\textit{ß}^\pm\bigr)$ of $\mathring{\mathcal{P}}_\textit{ß}$. 

This permits us to answer one of the questions that turned up crucial to the birth of the index theorem. Recalling that the number 
\begin{equation}
	\hat{A}(\mathcal{M}) = \hat{\mathtt{c}}(\mathcal{M})[\mathcal{M}],	
\end{equation}
called the \emph{Hirzebruch $\hat{A}$-genus}, with a characteristic class $\hat{\mathtt{c}}$, is the $\hat{A}$-genus of $\mathcal{M}$ \cite[p. 428]{Atiyah and Singer "The index of elliptic operators on compact manifolds"} \cite[p. 571]{Atiyah and Singer "The index of elliptic operators: III"}, one must pause to consider why the $\hat{A}$-genus of a spin manifolds is an \emph{integer} rather than a rational number (as is usually the case), in accordance with the proof in A. Borel and F. Hirzebruch \cite{A. Borel and Hirzebruch "Characteristic Classes and Homogeneous Spaces I"} \cite[chap. VII, § 25]{A. Borel and Hirzebruch "Characteristic Classes and Homogeneous Spaces II"} \cite{A. Borel and Hirzebruch "Characteristic Classes and Homogeneous Spaces III"}. Well, the index theorem gives a response on the \emph{integrality} of the $\hat{A}$-genus of certain homogeneous spaces: the Hirzebruch $\hat{A}$-genus of a spin manifold is identical to the index of its Dirac-type (Dirac–Atiyah–Singer) operator,
\begin{equation}
	\indice(D_\textsc{das}) = \left\{\hat{A}(\mathcal{M}) = \hat{\mathtt{c}}(\mathcal{M})[\mathcal{M}]\right\}.
\end{equation}

\subsubsection{Two Exemplary Evidences}
\label{subsubsection "Two Exemplary Evidences"}

\begingroup
\footnotesize
I felt it opportune to write out [\,\dots] the peculiarity of a certain method [\,\dots] to start an investigation by mechanics of certain problems in mathematics [\textgreek{\textit{ἐν τοῖς μαθήμασι θεωρεῖν διὰ τῶν μηχανικῶν}}]. I am convinced that this method is likewise helpful for the demonstration of the theorems themselves [\textgreek{\textit{εἰς τὴν ἀπόδειξιν αὐτῶν τῶν θεωρημάτων}}]. Some things first became [clear] to me by mechanics, although they had later to be proved geometrically [\textgreek{\textit{ὕστερον γεωμετρικῶς ἀπεδείχθη διὰ τὸ χωρὶς ἀποδείξεως}}] due to the fact that an investigation with this method does not equate to actual proof; but it is easier to provide a demonstration when some knowledge of the things sought has been acquired with this method rather than to seek it with no preliminary knowledge [\textgreek{\textit{μηδενὸς ἐγνωσμένου ζητεῖν}}]. \\
\indent — \textsc{Archimedes of Syracuse} \cite[letter to Eratosthenes (\textgreek{\textit{Ἀρχιμήδους Περὶ τῶν μηχανικῶν θεωρημάτων πρὸς Ἐρατοσθένη ἔφοδος}}), p. 386]{Archimedes "The Method of mechanical theorems of Archimedes to Eratosthenes"}

\endgroup

\vspace{2mm}

In epigraph reference is made to a description of the famed mechanical method (\textgreek{ἔφοδος}) of Archimedes for geometric theorems \cite{Archimedes "The Method of mechanical theorems of Archimedes to Eratosthenes"} \cite{Archimedes "The Method of Archimedes Recently Discovered by Heiberg"}: it is an exceptional testimony about the harmonious combination of mathematics and physics in the ancient Greek science. About the \textit{Method}, see \cite{Acerbi "Introduzione"}.

One of the traits of the Archimedean genius is the fusion of geometry, physics \& mathematics, and mechanics: three in one. To use G. Montanari's \cite[p. 81]{Montanari "Prostasi Fisicomatematica"} phrase: his proofs turn out to be true, geometrically, physico-mathematically, mechanically («[le sue] Dimostrazioni esser vere, \emph{Geometricamente, Fisicomatematicamente, Mecanicamente}»).

Something similar to what Archimedes tells Eratosthenes can be found in B. Riemann's activity. There is an introduction, written R. Courant and H. Robbins \cite[pp. 385-386]{Courant and Robbins "What is Mathematics?"}, about the experimental solutions of minimum problems (soap film experiments \& Plateau's problem, see Section \ref{subsection "Plateau's Problem: Soap Films and Bubbles"}), for the calculus of variations, where a persuasive narration is given, and it seems to proceed on this track:
\vspace{2mm}

\begingroup
\footnotesize
It is usually very difficult, and sometimes impossible, to solve variational problems explicitly in terms of formulas or geometrical constructions involving known simple elements [\,\dots]. In many cases, when [\,\dots] an existence proof turns out to be more or less difficult, it is stimulating to realize the mathematical conditions of the problem by corresponding physical devices, or rather, to consider the mathematical problem as an interpretation of a physical phenomenon. The existence of the physical phenomenon then represents the solution of the mathematical problem. Of course, this is only a plausibility consideration and not a mathematical proof, since the question still remains whether the mathematical interpretation of the physical event is adequate in a strict sense, or whether it gives only an inadequate image of physical reality. Sometimes such experiments, even if performed only in the imagination, are convincing even to mathematicians. In the nineteenth century many of the fundamental theorems of function theory were discovered by Riemann by thinking of simple experiments concerning the flow of electricity in thin metallic sheets. 

\endgroup

\vspace{2mm}

Under a comparative reading, this excerpt and the Archimedes' letter are not that far away.

\subsubsection{Scholium: Greco-Hellenistic Scientific Modus Operandi (the Origin of Hypotheses)}
\label{subsubsection "Scholium: Greco-Hellenistic Scientific Modus Operandi (the Origin of Hypotheses)"}

\begingroup
\footnotesize
\textgreek{[Σ]τέλεχος· αὕτη γὰρ οἷον ὑπόθεσις καὶ φύσις δένδρων} · [T]runk [down to the roots]: is but the foundation [\textit{hypothesis}]\footnote{
	Literally “a placing (\emph{thesis}) under (\emph{hypo}-)”, i.e. “base”, “ground”.
	}
	and the very nature of trees. \\
\indent — \textsc{Theophrastus} \cite[IV, \textsc{xiii}, 4, p. 129]{"Theophrasti Eresii Opera quae supersunt omnia I: Historia plantarum"} = \cite[pp. 388-389]{Theophrastus "Enquiry into Plants"}\footnote{
	The reference edition is \textgreek{Θεοφράστου}, \textgreek{\textit{Περί Φυτών Ιστορίας το βιβλίον Α' έως και το Ι', Περί Φυτών Αιτιών το βιβλίον Α' έως και το Ζ'}}, published by \textgreek{Ά.τ. Μανουτίου} [A.P. Manutius], \textgreek{Εν Ενετίαις} [in Venezia], 1495.
	}

\endgroup

\vspace{2mm}

The unity of the Greco-Hellenistic scientific method consists in developing a mathematical theory based on \emph{hypotheses} (\textgreek{ὑπόθεσις}), i.e. «foundation», «(theoretical) basis», or rather, «postulate», «principle».\footnote{
	Cf. L. Russo \cite[chap. 6.2]{Russo "The Forgotten Revolution: How Science Was Born in 300 BC and Why it Had to Be Reborn"}.
	} 
A mathematical theory of this kind aims to \emph{save the phenomena} (\textgreek{σῴζειν τὰ φαινόμενα}), which constitute the heuristic initiation of the modus operandi of classical antiquity. And when a mathematical theory of the phenomena of the exterior world is correctly calibrated on the observable phenomena, the phenomena are subsequently deducible from the hypotheses (postulates). Note that
\enumerationisinitium 
\item «hypothesis» means “being (placing) under” (\textgreek{ὑποτίθημι}); that is why Theophrastus calls the trunk, also including the roots, of trees \emph{hypotheses} (\textgreek{ὑπόθεσις [\,\dots] δένδρων});
\item «phenomenon» (\textgreek{φαινόμενον}) does not indicate a “fact” (an objective occurrence), but an “appearance” (an observed occurrence, with interaction between observer and object).
\enumerationisfinis

\chapter{Outro—\emph{Parva Mathematica}: \emph{Libera Divagazione} \sfrac{8}{8}}
\label{chapter "Outro—Parva Mathematica: Libera Divagazione 8/8"}

\section{Beauty of Mathematics vs. Mathematics of Beauty}
\label{section "Beauty of Mathematics vs. Mathematics of Beauty"}

\subsection{A Roundup of Physicists' Accounts}

The interlacing of beauty and mathematics in physics is long-standing. As to mention some names of the recent past: 
\enumerationisinitium
\item A. Einstein, printed obituary for E. Noether (\textit{The New York Times}, May 4, 1935): 

\vspace{2mm}

\begingroup
\footnotesize
Pure mathematics is, in its way, the poetry of logical ideas. One seeks the most general ideas of operation which will bring together in simple, logical and unified form the largest possible circle of formal relationships. In this effort toward logical beauty spiritual formulas are discovered necessary for the deeper penetration into the laws of nature.

\endgroup

\vspace{2mm}

\item P.A.M. Dirac:

\vspace{2mm}

\begingroup
\footnotesize
\subenumerationisinitium
\item \cite[pp. 122, 124]{Dirac "The Relation between Mathematics and Physics"} The physicist, in his study of natural phenomena, has two methods of making progress: [\textgreek{α}] the method of experiment and observation, and [\textgreek{β}] the method of mathematical reasoning. The former is just the collection of selected data; the latter enables one to infer results about experiments that have not been performed. There is no logical reason why the second method should be possible at all, but one has found in practice that it does work and meets with remarkable success. This must be ascribed to some \emph{mathematical quality in Nature} [\,\dots].\footnote{
	\label{footnote "Parolame-ciarpame"}
	From where I stand, the expression “mathematical quality in Nature” is \textit{parolame-ciarpame}.
	} 

One might describe the mathematical quality in Nature by saying that the universe is so constituted that mathematics is a useful tool in its description. However, recent advances in physical science show that this statement of the case is too trivial. The connection between mathematics and the description of the universe goes far deeper than this.\footnoteref{footnote "Parolame-ciarpame"} 

[\,\dots] The research worker, in his efforts to express the fundamental laws of Nature in mathematical form, should strive mainly for mathematical beauty [\,\dots]. [T]he mathematician plays a game in which he himself invents the rules while the physicist plays a game in which the rules are provided by Nature, but as time goes on it becomes increasingly evident that the rules which the mathematician finds interesting are the same as those which Nature has chosen. It is difficult to predict what the result of all this will be. Possibly, the two subjects will ultimately unify.\footnote{
	The tautological viciousness present in this Diracian reasoning is noticeable: physics can only see nature with the \textgreek{τέχνη} of mathematics (see Sections \ref{subsection "Mathematics as a Technical Tool"} and \ref{subsubsection "How is it Possible?"}).
	} 
\item Physical law should have mathematical beauty.\footnote{
	 \textcyrillic{Автографы мелом на стенах комнаты 4-59} · Phrase written on the chalkboard at Moscow State University, 3 October 1956.
	}
\item \cite[pp. 652-653]{Dirac "The relativistic electron wave equation"} Most physicists [\,\dots] say that all that a physicist needs is to have some theory giving results in agreement with observation. I say, that is not all that a physicist needs. A physicist needs that his equations should be mathematically sound [\,\dots]. The only feature of the new theory which one can be sure of is that it must be based on sound and beautiful mathematics.
\subenumerationisfinis

\endgroup

\vspace{2mm}

\item C.N. Yang \cite[pp. 237-238]{Yang "The Law of Parity Conservation and Other Symmetry Laws of Physics"}:\footnote{
	A treatment on beauty and physics in the work of Yang is in \cite{Shi "Beauty and Physics: 13 Important Contributions of Chen Ning Yang"}.
	} 

\vspace{2mm}

\begingroup
\footnotesize
Nature seems to take advantage of the simple mathematical representations of the symmetry laws. When one pauses to consider the elegance and the beautiful perfection of the mathematical reasoning involved and contrast it with the complex and far-reaching physical consequences, a deep sense of respect for the power of the symmetry laws never fails to develop.

\endgroup

\vspace{2mm}

\item W. Heisenberg \cite[p. 183]{Heisenberg "The Meaning of Beauty in the Exact Sciences"}: 

\vspace{2mm}

\begingroup
\footnotesize
[I]n exact science, no less than in the arts, [beauty] is the most important source of illumination and clarity.

\endgroup

\vspace{2mm}

\item W.E. Baylis and G. Jones \cite[p. 129]{Baylis and Jones "The Pauli-Algebra Approach to Special Relativity"}: 

\vspace{2mm}

\begingroup
\footnotesize
Beauty in physics is less subjective than in art, because physicists agree on some objective criteria. A beautiful physical theory 

· displays and exploits the inherent symmetry,
	
· allows many results to be derived from a few assumptions and minimal labor, and
	
· avoids unnecessary mathematical baggage and doesn't predict phenomena which are not realized physically.

\endgroup

\vspace{2mm}

\item For further discussion on the theme of beauty in science, see S. Chandrasekhar \cite{Chandrasekhar "The Perception of Beauty and the Pursuit of Science"} \cite{Chandrasekhar "Beauty and the quest for beauty in science"}.\endnote{
	For those who are interested, other pages devoted to capture some “sympathies” that tie together (i) sense of beauty, (ii) ability to produce verbal compositions in verse (poetry) and (iii) ability to produce formal equivalences (mathematics), are in \cite[pp. ix-xvi]{Farmelo (Ed.) "It Must Be Beautiful: Great Equations of Modern Science"}.
	}
\enumerationisfinis	

\subsection[In the Bliss of Goddess Geometry: Space-Universe in the Galilean–Keplerian Shadow, and the Poincaré–Weber–Seifert Dodecahedral Space Topology]{In the Bliss of Goddess Geometry: Space-Universe in the Galilean–Keplerian Shadow,\footnote{
	The “Galilean–Keplerian” wording is a (loutish) simplification. Galileo and Kepler had considerably distant mentalities, starting with the aesthetic judgments in science. On the differences between these two protagonists of modernity, see M. Bucciantini \cite{Bucciantini "Galileo e Keplero. Filosofia cosmologia e teologia nell'Eta della Controriforma"}, who is a historian.
	} 
and the Poincaré–Weber–Seifert Dodecahedral Space Topology}
\label{subsection "In the Bliss of Goddess Geometry"}

\begingroup
\footnotesize
The exploration of topological spaces is the exploration of the last great frontier—the human mind. These spaces have been \emph{created by humans} for the purpose of understanding the world in which we live. But ultimately they lead to an understanding of our mind, for it can only be understood in terms of its creations. \emph{Topological space\textnormal{[}s\textnormal{]}}, then, are at once a \emph{form of art} and a \emph{form of science}, and as such they \emph{reflect our deepest intellect}. \\
\indent — \textsc{J.S. Carter} \cite[p. x, e.a.]{Carter "How Surfaces Intersect in Space: An Introduction to Topology"}

\endgroup

\vspace{2mm}
	
The above examples may suffice. Be forewarned: the \emph{beauty of mathematics} is one thing, the \emph{mathematics of beauty} is another. In the second case, the chances—for applications (as in physics)—of goof-ups are high. The first striking example is offered by Kepler's belief \cite[tabula III, between p. 24 and 25]{Kepler "Prodromus dissertationum"}, where a Solar System is presented under a geometric schematization of the five Platonic solids (octahedron, icosahedron, dodecahedron, tetrahedron, and cube).\footnote{
	The polyhedra, whose faces all have congruent regular polygons, known as the Platonic solids, are only 5 as a consequence of Proposition 21 of Book XI of Euclid's \textit{Elements} \cite[\textgreek{κα´, Στοιχείων ια´}, Book XI, p. 54]{Euclidis "Elementa IV"}: «Any solid angle is contained by plane angles whose sum is less than four right angles (\textgreek{Ἅπασα στερεὰ γωνία ὑπὸ ἐλασσόνων [ἢ] τεσσάρων ὀρθῶν γωνιῶν ἐπιπέδων περιέχεται})»; the sum of the interior angles at each vertex is less than \ang{360}, otherwise at \ang{360} the resulting shape is flat: 
	 
	· tetrahedron has 3 triangles at each vertex, for $3 \times \ang{60} = \ang{180}$,
	
	· octahedron has 4 triangles at each vertex, for $4 \times \ang{60} = \ang{240}$, 
	
	· cube has 3 squares at each vertex, for $3 \times \ang{90} = \ang{270}$,
	
	· icosahedron has 5 triangles at each vertex, for $5 \times \ang{60} = \ang{300}$,
	
	· dodecahedron has 3 pentagons at each vertex, for $3 \times \ang{108} = \ang{324}$.
	
	Let us now imagine a polyhedron composed of $n$ hexagonal faces; for a regular 6-gon, each interior angle has a measure of \ang{120}. If we say that in this polyhedron 3 6-gons at each vertex meet, it comes out that $3 \times \ang{120} = \ang{360}$.
	} 
The distance of the six then known planets (Mercury, Venus, Earth, Mars, Jupiter, and Saturn) reflects the distance obtained by inscribing and circumscribing each of the regular convex polyhedrons by celestial spheres, according to a nesting \textgreek{σχῆμα}, placing one solid inside another.\footnote{
	At the center of the cosmos the Sun dominates; then there is Mercury with its circumsolar rotation. The celestial sphere containing the orbit of Mercury is inscribed in an octahedron, to which the orbit of Venus is circumscribed. It continues with the icosahedron, for the Earth, the dodecahedron, for Mars, the tetrahedron, for Jupiter. The orbit of Jupiter is inscribed in a cube, to which the celestial sphere corresponding to the orbit of Saturn is circumscribed.
	} 
It is a pleasant and symmetric scheme; nevertheless, as Kepler himself later realized, it is not a good transposition of experimentally detected quantities.

How far have we gone, nowadays, from the Keplerian theory of beauty \cite[pp. 194, 222, 241]{Kepler "Harmonices Mundi Libri V"}, under which the \emph{Geometrica pulchritudo} coincides with the \emph{pulchritudo Mundi}, and serves as the \emph{Archetypo Mundi}? Or from the Galilean \emph{Geometry of nature}? I am referring to the \emph{celeberrimo} piece \cite[p. 25]{Galilei "Il Saggiatore"}:

\vspace{2mm}

\begingroup
\footnotesize
Philosophy [of nature] is written in this grand book, which stands continually open before our eyes (I say the universe), but it cannot be understood if one does not first learn to understand the language, and know the characters, in which it is written. It is written in the mathematical language, and the characters are triangles, circles, \& other Geometric figures [\textit{Egli è scritto in lingua matematica, e i caratteri son triangoli, cerchi, \& altre figure Geometriche}], without which [without these means] it is impossible to humanly understand a word of it [\textit{senza i quali mezi è impossibile à intenderne umanamente parola}]; without these it is a vain wandering through a dark labyrinth [\textit{senza questi è un'aggirarsi vanamente per un'oscuro laberinto}]».\endnote{
		Original It. version: «La Filosofia [della natura] è scritta in questo grandissimo libro, che continuamente ci stà aperto innanzi à gli occhi (io dico l'universo), ma non si può intendere se prima non s'impara à intender la lingua, e conoscer i caratteri, ne' quali è scritto. Egli è scritto in lingua matematica, e i caratteri son triangoli, cerchi, \& altre figure Geometriche, senza i quali mezi è impossibile à intenderne umanamente parola; senza questi è un'aggirarsi vanamente per un'oscuro laberinto».
		}

\endgroup

\vspace{2mm}

Galileo is one of the great minds I admire (his genius spans the centuries); and yet this depiction is a \emph{load of bunkum}. Math-language is a product of mankind, and nature knows nothing of “triangles”, “circles”, or, in algebraic terms, of equations. How can we have the \emph{arrogance}—a risible revisitation of the mythological \textgreek{ὕβρις}—of putting a language (of geometric-, algebraic-, or analytic-type) of our own invention into the mouth of nature?\footnote{
	And yet Galileo (in another work), under the guise of Sagredo is far from any \textgreek{ὕβρις}. In fact, elsewhere \cite[p. 94]{Galilei "Dialogo sopra i due Massimi Sistemi del Mondo Tolemaico e Copernicano"}, we find lucidly written that: «It always seems to me extreme temerity on the part of some when they want to make human capacity the measure of what nature can do [\textit{Estrema temerità mi è parsa sempre quella di coloro, che voglion far la capacità umana misura di quanto possa, e sappia operar la natura}]».
	}

We are still there; we are still stuck in this Platonic–Galilean–Keplerian ideology.\endnote{
	Cf. G. Toraldo di Francia \cite[pp. 16-20]{Toraldo di Francia "L'indagine del mondo fisico"}, who seems to have a different opinion; but he does not address the issue (p. 20), and cautiously does not comment on this.
	}
It is all part of it. Terribly evincive is this verdict of Heisenberg \cite[p. 32]{Heisenberg "Natural Law and the Structure of Matter"}: 

\vspace{2mm}

\begingroup
\footnotesize
I think that modern physics has definitely decided in favour of Plato. These smallest units of matter are not in fact physical objects in the ordinary sense; they are forms, ideas which can be expressed unambiguously only in mathematical language.

\endgroup

\vspace{2mm}

Take a recent example. In the first half of the twenty-first century, J.-P. Luminet \& collaborators \cite{Luminet Weeks Riazuelo Lehoucq Uzan "Dodecahedral space topology as an explanation for weak wide-angle temperature correlations in the cosmic microwave background"} worked on a model of a finite universe, built via the \emph{Poincaré dodecahedron space}, primarily known as \emph{Poincaré homology 3-sphere} \cite{Poincare "Second complement a l'Analysis Situs}.
	
We can quickly remember that 
	
· a 3-sphere $\mathbb{S}^3$ is a set of nested 2-spheres, that is, a 3-dimensional surface of a 4-dimensional ball,
	
· a homology 3-sphere is a closed oriented 3-manifold whose homology—to be more precise, it is a $\mathbb{Z}$-homology—is the same as that of $\mathbb{S}^3 \cong SU_2(\mathbb{C})$, 
\begin{equation}
	\mathbb{S}^3_\homologygroup
		\left\{\homologygroup_1(\mathcal{X}^3, \mathbb{Z}) = 0\right\}, 
\end{equation}
where $\homologygroup$ is the homology group, and $\mathcal{X}^3$ is a topological 3-space. Let $\mathfrak{Ic}$ be the group of isometries leaving invariant a regular icosahedron, or a dodecahedron, in $\mathbb{R}^3$. The fundamental group of the Poincaré homology sphere is the binary icosahedral group $\widetilde{\mathfrak{Ic}}$ having order 120. Note that $\widetilde{\mathfrak{Ic}}$ is obtained by lifting $\mathfrak{Ic}$ to $SU_2(\mathbb{C})$. The covering morphisms
\[
	SU_2(\mathbb{C}) \to SO_3(\mathbb{R})
\] 
(see Section \ref{subsubsection "The Covering Morphisms $SU_2(C)$ to $SO_3(R)$"}) is directly linked to the map 
\begin{equation}
	\mathbb{S}^3 \to \mathbb{RP}^3 \cong SO_3(\mathbb{R}). 
\end{equation}
And so the Poincaré homology 3-sphere is isomorphic to $\mathbb{S}^3/\widetilde{\mathfrak{Ic}}$, 
\begin{equation}
	\mathbb{S}^3_\homologygroup \cong \mathbb{S}^3/\widetilde{\mathfrak{Ic}}.
\end{equation}
	
The rigorousness of a dodecahedral structure in the homological Poincaré sphere occurs later, with C. Weber and H. Seifert \cite{Weber und Seifert "Die beiden Dodekaederraume"}. The Poincaré homology sphere, as a \emph{spherical dodecahedral space} (optionally, we can call it \emph{Poincaré–Weber–Seifert space}), is a dodecahedral block of space having opposite faces \emph{glued} together. What does it look like? Imagine having a set of 120 spherical dodecahedra, each of which is a dodecahedral tile of the surface of $\mathbb{S}^3$. All spherical dodecahedra fit together, since their edge angles are \ang{120}. (Bear in mind that the spherical dodecahedral space and $\mathbb{S}^3$ maintain the same homology groups). 
	
Living in a positively curved space-universe of this kind is enough, for a Keplerian spirit, to rejoice in the bliss of Goddess Geometry.\footnote{
		Luminet's proposal, beyond its physical validity, has kindled a bit of curiosity for topology in astronomy/cosmology. (To notice the role of adventurer assigned to Riemann in \cite[p. 5, arXiv vr.]{Luminet Weeks Riazuelo Lehoucq Uzan "Dodecahedral space topology as an explanation for weak wide-angle temperature correlations in the cosmic microwave background"}: «In 1854 Georg [B.] Riemann cut the Gordian knot by proposing the hypersphere as a model of a finite universe with no troublesome boundary»). Cf. Luminet \cite[p. 81]{Luminet "Geometry and Topology in Relativistic Cosmology"}: «General relativity does not allow one to specify the topology of space, leaving the possibility that space is multiply rather than simply connected»; and \cite[p. 15]{Luminet "Cosmic topology: Twenty Years After"}: «What is the shape of the Universe? Is it finite or infinite? Is space multi-connected to create ghost images of faraway cosmic sources? After a “dark age” period, the field of cosmic topology has now become one of the major concerns in astronomy and cosmology, not only from theorists but also from observational astronomers».
		}

\subsection{von Neumann's Warning}

Aesthetic inclination is connate to science; but geometry is not \emph{of} nature; it is \emph{of} man, or other geometrizing beings.

In this regard, we can read J. von Neumann's remark \cite[pp. 8-9, e.a.]{von Neumann "The Mathematician. "The Works of the Mind""}:

\vspace{2mm}

\begingroup
\footnotesize
The mathematician has a wide variety of fields to which he may turn, and he enjoys a very considerable freedom in what he does with them [\,\dots]: I think that it is correct to say that his criteria of selection, and also those of success, are mainly \emph{aesthetical} [\,\dots]. [T]he aesthetical character is even more prominent [\,\dots] in the case of theoretical physics. One expects a mathematical theorem or a mathematical theory not only to describe and to classify in a simple and elegant way numerous and a priori disparate special cases. One also expects “elegance” in its “architectural”, structural makeup [\,\dots]. Also, if the deductions are lengthy or complicated, there should be some simple general principle involved, which “explains” the complications and detours, reduces the apparent arbitrariness to a few simple guiding motivations, etc. These criteria are clearly those of any \emph{creative art}, and the existence of some underlying empirical, worldly motif in the background—often in a very remote background—overgrown by “aestheticizing” developments and followed into a multitude of labyrinthine variants—all this is much more akin to the atmosphere of \emph{art pure} and simple than to that of the empirical sciences. 

\endgroup

\vspace{2mm}

But a little further down (ibid.), he warns against possible involutions that skulk behind such a mentality:

\vspace{2mm}

\begingroup
\footnotesize
I think that it is a relatively good approximation to truth [\,\dots] that \emph{mathematical ideas originate in empirics}, although the genealogy is sometimes long and obscure. But, once they are so conceived, the subject begins to live a peculiar life of its own and is better compared to a creative one, governed by almost entirely aesthetical motivations, than to anything else and, in particular, to an empirical science [\,\dots]. As a mathematical discipline travels far from its empirical source, or still more, if it is a second and third generation only indirectly inspired by ideas coming from “reality”, it is beset with very grave dangers. It becomes more and more purely aestheticizing, more and more purely \emph{l'art pour l'art}. This need not be bad, if the field is surrounded by correlated subjects, which still have closer empirical connections [\,\dots]. But there is a grave danger that the subject will develop along the line of least resistance, that the stream, so far from its source, will separate into a multitude of insignificant branches.

\endgroup
	
\subsection{Math-art: \emph{Prigioni} of the Mind: a Suggestion by Weil}
\label{subsection "Math-art: Prigioni of the Mind: a Suggestion by Weil"}

\begingroup
\footnotesize
Non ha l'ottimo artista alcun concetto / Ch'un marmo solo in se non circoscriva / Col suo soverchio,\footnote{
	From the La. \textit{supercŭlus}, and it means what is “too much”, “excess”, “superfluous”, “surplus”.
	} 
e solo a quello arriva / La man che obbedisce all'intelletto.\footnote{
	«The excellent artists never has a concept / That a single marble does not contain in itself / With its excess, and to it attains only / [if] Hand obeys the intellect [thought]». Here the etymological  hint holds: from the La. \textit{intellectus}, composed by \textit{intŭs}, “within”, “inside” (or \textit{intĕr}, “in the midst”), and \textit{lĕgĕre}, “to read”, “to gather”.
	} \\
\indent — \textsc{Michelangelo} \cite[p. 1]{Buonarroti "Rime di Michelangelo Buonarroti"} 

\vspace{2mm}

Mathematics [\,\dots] is nothing other than an art; a kind of sculpture [\textit{une espèce de sculpture}] in an extremely hard and resistant material (like certain porphyries used sometimes, I believe, by sculptors). Michelangelo expressed, in the first quatrain of an admirable sonnet, this idea (which I imagine to be more or less Platonic) that the block of marble contains, at the exit of the quarry, the hand-sculpted work, and that the artist's work consists in removing what is too much [\,\dots]. The mathematician is so subject to the grain, and cross-grain,\endnote{
	We use to say “grain” (\textit{fil}) and “cross-grain” (\textit{contrefil}) for the woodworking.
	} 
to any curvature and even accidents of the subject-matter he is working on, that it gives his work a kind of objectivity. But the work being done [\,\dots] is a work of art [\textit{œuvre d'art}], thereby inexplicable (in it alone lies its own explanation). \\
\indent — \textsc{A. Weil} \cite[p. 255]{Weil "Extrait d'une lettre du 29 fevrier 1940"}

\endgroup

\vspace{2mm}

The danger of aestheticization to which von Neumann refers (see above) is more present in the physical realm, but in pure mathematics this is not necessarily a peril. Indeed, this is the beauty of the freedom of pure thought. Weil's association between mathematics and sculpture is redolent: just as the artist's task is to remove the excess circumscribing an artwork, ideally already enclosed in the marble, for liberating—beneath the chisel—the \emph{shape} from the prison of rock, so the mission of the mathematician-sculptor is to bring out from his head the many \emph{forms} of mathematics, for liberating abstract objects thanks to the imaginative and, at the same time, conceptualizing capability of thought.\footnote{
	There is no contrast between \emph{logical rigor} and \emph{mathematical imagination}: the creative power of the mathematician is accomplished under an \emph{intuitive impulse}, and this is evident e.g. in the conception of new postulates; and yet this act of creation—let us say \emph{fantasy}—is not disconnected from an \emph{abstractive force}, which provides rigor to the flash of intuition.
	} 
Ça va sans dire, our interpretation of the mathematician-sculptor is not Platonico-realistic at all (cf. Section \ref{subsection "Logomachy of Mathematicians, and Cock-and-Bull Stories"}).

The combination of mathematics-art is taken up by E. Bombieri \cite[pp. xi-xii]{Bombieri "Introduzione" in A. Weil "Teoria dei numeri. Storia e matematica da Hammurabi a Legendre""} in his preface to the It. edition of Weil's book on number theory \cite{Weil "Number Theory. An approach through history from Hammurapi to Legendre"}: mathematics, like art, is beyond explanation, because its foundation and justification reside in itself.

\subsection{Inexplicability and Ideality of Forms: in the Homeland of the Dream}
\label{subsection "Inexplicability and Ideality of Forms: in the Homeland of the Dream"}

\begingroup
\footnotesize
[A]s for anything else, so for a mathematical theory—beauty can be perceived, but not explained. \\
\indent — \textsc{A. Cayley} \cite[p. 449]{Cayley "Presidential Address to the British Association September 1883"}

\endgroup

\vspace{2mm}

If mathematics is \textit{inexplicable}, since «elle seule est à elle-même son explication», this sort of self-foundation makes it compatible with an intuition of an artistic kind, and, most importantly, turns it into a lively thought of seeing, through its technique, an \emph{ideal} projection—what we call a \emph{dream}—of the world.\footnote{
	\label{footnote "Ideas, forms of vision, etc."}
	We know that, ideas are \emph{forms of vision}, ideal forms (idealizations) of concrete objects, from the Gr. \textgreek{ἰδέαι}, which comes, in turn, from \textgreek{ἰδεῖν}, inf. of \textgreek{ὁράω}, “to see”, “to look”. And mathematics consists of \textgreek{ἰδέαι}, and nothing else. (The mind, for its part, is somehow directed towards ideal perfection). From this one can also facilely infer that a form theory (\textgreek{θεωρία τῶν ἰδεῶν}), in the sense just illustrated, in which mathematical ideas reign, is but an old wording for the current set theory.
	} 
See e.g. what G.-C. Rota \cite[p. 13, originally p. 182]{Rota "The Phenomenology of Mathematical Beauty"} writes on this matter:

\vspace{2mm}

\begingroup
\footnotesize
Talk of mathematical beauty is a cop-out to [\,\dots] keep our description of mathematics as close as possible to the description of a mechanism. This cop-out is one step in a cherished activity of mathematicians, that of building a perfect world immune to the messiness of the ordinary world, a world where what we think \emph{should} be true turns out to \emph{be} true, a world that is free from the disappointments, ambiguities,\footnote{
	But compare with the Lolli's essay \cite{Lolli "Ambiguita. Una viaggio fra letteratura e matematica"}, already mentioned in endnote \ref{endnote "Lolli, Ambiguità. Una viaggio fra letteratura e matematica"}. 
	} 
and failures of that other world in which we live. 

\endgroup

\vspace{2mm}

He thus concludes his paper. We could not agree more. Mathematician-artist is graced with an imaginative foresight.
	
\subsection{The World is Not Mathematical}
\label{subsection "The World is Not Mathematical"}

\begingroup
\footnotesize
[A] mathematician who is not somewhat of a poet, will never be a perfect mathematician.\endnote{
	Original Ge. version: «[E]in Mathematiker, der nicht etwas Poet ist, wird nimmer ein vollkommener Mathematiker sein».
	} \\
\indent — \textsc{K. Weierstrass} \cite[letter to S. Kovalevskaya, 27 August 1883, p. 149]{Weierstrass "Letter dated 27 August 1883 to S. Kovalevskaya"} quoted by \textsc{M.G. Mittag-Leffler}
 
\vspace{2mm}

\textsc{le matematiche come arte} · [T]he object of mathematics—immanent order in Nature—is revealed to the mind through a process of abstraction; that is why [the various kinds of] mathematics are not only science, representation of that object, but also art, that is, expression of the person who constructs them, according to its intimate laws. The profound sense of order, proportion and measure is expressed precisely in it, [and] it will make a cosmos from the chaos of phenomena.\endnote{
	Original It. version: «[L]'oggetto delle matematiche — ordine immanente nella Natura — si discopre alla mente traverso un processo d'astrazione; appunto per ciò le matematiche non sono soltanto scienza, rappresentazione di quell'oggetto, sì anche arte, cioè espressione del soggetto che le costruisce, secondo le sue intime leggi. Si esprime proprio in essa il senso profondo dell'ordine, della proporzione e della misura, che farà un cosmo del caos dei fenomeni».
	} \\ 
\indent — \textsc{F. Enriques} \cite[§ 40, p. 155]{Enriques "Le matematiche nella storia e nella cultura"}

\endgroup

\vspace{2mm}

The world is not mathematical; but mathematics is, say, worldly, viz. of this world (\textgreek{τοῦ κόσμου τούτου}). Mathematics, like art, is an act of poetic imagination, an art of doing (\textgreek{ποιέω}), of producing thoughts, ideas (\textgreek{ἰδέαι}), about the world. And every idea is part of the art of dreaming; and, back at the dim beginnings of Myth, this art is musical (\textgreek{μουσική}), as it falls under the protection of the Muse.

\begin{margo}
The art of the Muses (\textgreek{Μοῦσαι}), in an all-encompassing sense: every art \& science. Some etymologists derive the name \textgreek{Μοῦσαι} from \textgreek{Μόνσαι}, whose root is to be found in \textgreek{μεν-} (or \textgreek{μεν, μαν}), consider e.g. the word \textit{mens} (mind); if so, the Muses are “those who meditate”, “those who create through their imagination” (but cf. p. \pageref{section "Siren's Song: No-return to Ithaca"}, where the sirens come after, with their enchanting and often \textgreek{δαιμονικός} \textit{chant}). \margosymbol
\end{margo}

\vspace{10mm}

\setcounter{secnumdepth}{0}  
\section{References and Bibliographic Details from \sfrac{1}{8} to \sfrac{8}{8}}
\setcounter{secnumdepth}{3}
\markright{References and Bibliographic Details}

\begingroup
\footnotesize

\noindent Sections \ref{section "Mathematical Evolutionism: What is a proof?"} and \ref{subsection "Fluvial- and Æolian-like Processes"}
\begin{indent paragraph: 15pt}
For an accessible account of the \emph{magmatic movement} about the notion of \emph{proof} in mathematics and, inevitably, of the magma-movement of mathematics itself, see \cite{Lolli "Matematica in movimento. Come cambiano le dimostrazioni"}.
\end{indent paragraph: 15pt}

\noindent Section \ref{subsection "Extra-logical Objects, and Gödelian Suggestions"}

\begin{indent paragraph: 15pt}
· Decidability/undecidability in Gödel and Turing, incompleteness, logic \& foundations of mathematics, are discussed in \cite[capp. IX-X, pp. 259-314]{Lolli "Le ragioni fisiche e le dimostrazioni matematiche"}. \\
· On the bickering between mathematics and logic interspersed with the theme of infinity, see \cite{Lolli "Il formalismo e l'infinito"}.
\end{indent paragraph: 15pt}

\noindent Sections \ref{chapter "Outro—Parva Mathematica: Libera Divagazione 3/8"}, \ref{chapter "Outro—Parva Mathematica: Libera Divagazione 5/8"}, \ref{chapter "Outro—Parva Mathematica: Libera Divagazione 6/8"}

\begin{indent paragraph: 15pt}
Further stimuli on the relationship between mathematics and natural sciences are in \cite{Boncinelli Bottazzini "La serva padrona. Fascino e potere della matematica"}.	
\end{indent paragraph: 15pt}

\noindent Section \ref{subsubsection "Inverse-square Laws"}

\begin{indent paragraph: 15pt}
A report on the inverse-square law and the Coulomb's law is in \cite[pp. 113-120]{Bellone "Caos e armonia. Storia della fisica moderna e contemporanea"}.
\end{indent paragraph: 15pt}

\noindent Section \ref{subsubsection "Selection, Colors, and Understanding"}

\begin{indent paragraph: 15pt}
Another book edited by S. Dehaene, which investigates, according to mental processes, the roots of mathematical thought, is this \cite{Dehaene Brannon (Eds.) "Space Time and Number in the Brain: Searching for the Foundations of Mathematical Thought"}.
\end{indent paragraph: 15pt}

\noindent Section \ref{subsection "Contextus I. Elements of Brachylogy—the Reverie of a Perfect Language, with a Margo on Music and Mathematics"}

\begin{indent paragraph: 15pt}
A book that probes the relationship between music and mathematics is \cite{Fauvel Flood and Wilson "Music and Mathematics: From Pythagoras to Fractals"}.	
\end{indent paragraph: 15pt}

\noindent Sections \ref{subsubsection "How is it Possible?"}

\begin{indent paragraph: 15pt}
A book in which there is an intersection of mathematics (geometry and topology) and biology plus various branches of natural sciences is \cite{Boi (Ed.) "Geometries of Nature Living Systems and Human Cognition: New Interactions of Mathematics with Natural Sciences and Humanities"}, edited by L. Boi.	
\end{indent paragraph: 15pt}

\noindent Section \ref{subsection "Straitjacket for Feynman Path Integrals"}

\begin{indent paragraph: 15pt}
On the Feynman path integrals, here are some insights: \\
· in the field of mathematics (vast but systematizing approach) \cite{Albeverio Hoegh-Krohn Mazzucchi "Mathematical Theory of Feynman Path Integrals: An Introduction"}, \\
· in functional analysis \cite{Gill Zachary "Functional Analysis and the Feynman Operator Calculus"}, \\
· in quantum mechanics \cite{Roepstorff "Path Integral Approach to Quantum Physics: An Introduction"} \cite{Zinn-Justin "Path Integrals in Quantum Mechanics"}, \\
· in quantum statistical mechanics \cite{Albeverio Kondratiev Kozitsky Rockner "The Statistical Mechanics of Quantum Lattice Systems: A Path Integral Approach"}, \\
· in probability theory, stochastic and physical random process \cite{Chaichian and Demichev "Path Integrals in Physic I. Stochastic Process and Quantum Mechanics"} \cite{Chaichian and Demichev "Path Integrals in Physic II. Quantum Field Theory Statistical Physics and other Modern Applications"} \\
· in curved space-time \cite{Bastianelli and van Nieuwenhuizen "Path Integrals and Anomalies in Curved Space"}, \\
· in geometry and topology \cite{Szabo "Equivariant Cohomology and Localization of Path Integrals"} \cite{Grosche "Path Integrals Hyperbolic Spaces and Selberg Trace Formulae"}, \\
· wide-ranging reading \cite{Grosche Steiner "Handbook of Feynman Path Integrals"} \cite{Johnson Lapidus "The Feynman Integral and Feynman's Operational Calculus"} \cite{Kleinert "Path Integrals in Quantum Mechanics Statistics Polymer Physics and Financial Markets"}.
\end{indent paragraph: 15pt}

\noindent Section \ref{subsection "Intersections and Grafts"}

\begin{indent paragraph: 15pt}
· Here are some monographic works on 
Euclid \cite{Acerbi "Introduzione" a Euclide Tutte le opere"},
Archimedes \cite[pp. xv-clxxxvi]{Heath "The Works of Archimedes"} \cite{Dijksterhuis "Archimedes"},
post-Apollonian theory (i.e. rediscovery of the theory of conic sections in the seventeenth century) \cite{Maieru "Le sezioni coniche nel Seicento"},
Galileo \& Torricelli \cite{Giusti "Euclides reformatus. La teoria delle proporzioni nella scuola galileiana"},
Euler \cite{Bradley Sandifer (Eds.) "Leonhard Euler: Life Work and Legacy"},
Gauss \cite{Dunnington "Carl Friedrich Gauss: Titan of Science"},
Riemann \cite{Laugwitz "Bernhard Riemann 1826-1866: Turning Points in the Conception of Mathematics"} \cite[cap. 7]{Bartocci "Una piramide di problemi. Storie di geometria da Gauss a Hilbert"},
Galois \cite{Toti Rigatelli "Evariste Galois 1811-1832"},
Maxwell \cite{Campbell and Garnett "The Life of James Clerk Maxwell with Selections from His Correspondence and Occasional Writings"},
Beltrami \cite{Tazzioli "Beltrami e i matematici "Relativisti". La meccanica in spazi curvi nella seconda meta dell'Ottocento"},
Boltzmann \cite{Cercignani "Ludwig Boltzmann: The Man Who Trusted Atoms"},
Ricci Curbastro \cite{Tonolo "Commemorazione di Gregorio Ricci-Curbastro nel primo centenario della nascita"},
Poincaré \cite{Verhulst "Henri Poincare: Impatient Genius"},
Volterra \cite{Guerraggio Paoloni "Vito Volterra"},
Levi-Civita \cite{Nastasi Tazzioli "Toward a scientific and personal biography of Tullio Levi-Civita (1873-1941)"} \cite{Caparrini Tazzioli "Alle origini della Fisica Teorica. La corrispondenza tra Augusto Righi e Tullio Levi-Civita (1901-1920)"},
Kolmogorov \cite[pp. 117-121]{Bartocci Betti Guerraggio Lucchetti (Eds.) "Mathematical Lives. Protagonists of the Twentieth Century From Hilbert to Wiles"},
von Neumann \cite{Ulam "John von Neumann 1903-1957"},
Majorana \cite{Amaldi "La vita e l'opera di E. Majorana (1906-1938)"} \cite{Guerra Robotti "Ettore Majorana: Aspects of his Scientific and Academic Activity"} \cite{Recami "Il caso Majorana"} \cite{Recami "Ettore Majorana: his work and his life"} \cite{Esposito "Ettore Majorana: Unveiled Genius and Endless Mysteries"},
Landau \cite{Khalatnikov (Ed.) "Landau: The Physicist and the Man. Recollections of L.D. Landau"},
Caccioppoli \cite[cap. 4.2]{Guerraggio "L'Analisi"} \cite{Sbordone "Renato Caccioppoli nel centenario della nascita"} \cite[pp. 97-107]{Bartocci Betti Guerraggio Lucchetti (Eds.) "Mathematical Lives. Protagonists of the Twentieth Century From Hilbert to Wiles"},
Schwartz \cite{Schwartz "Un Mathematicien aux prises avec le siecle"} = \cite{Schwartz "A Mathematician Grappling with His Century"},
Thom (applications of catastrophe theory) \cite{Woodcock and Davis "Catastrophe theory"}, 
Mandelbrot (and much more) \cite{Peitgen Jurgens Saupe "Chaos and Fractals: New Frontiers of Science Second Edition"},
De Giorgi \cite{Ambrosio Dal Maso Forti Miranda Spagnolo "Ennio De Giorgi"} \cite[pp. 147-155]{Bartocci Betti Guerraggio Lucchetti (Eds.) "Mathematical Lives. Protagonists of the Twentieth Century From Hilbert to Wiles"},
Grothendieck \cite[pp. 171-180]{Bartocci Betti Guerraggio Lucchetti (Eds.) "Mathematical Lives. Protagonists of the Twentieth Century From Hilbert to Wiles"},
Nash \cite{Nash "The Essential John Nash"} \cite[pp. 137-146]{Bartocci Betti Guerraggio Lucchetti (Eds.) "Mathematical Lives. Protagonists of the Twentieth Century From Hilbert to Wiles"},
Atiyah \cite[pp. 197-208]{Bartocci Betti Guerraggio Lucchetti (Eds.) "Mathematical Lives. Protagonists of the Twentieth Century From Hilbert to Wiles"}, 
Arnold \cite[pp. 209-211]{Bartocci Betti Guerraggio Lucchetti (Eds.) "Mathematical Lives. Protagonists of the Twentieth Century From Hilbert to Wiles"}.
\end{indent paragraph: 15pt}

\noindent Section \ref{subsubsection "Margo. Atiyah–Singer Index Theorem"}

\begin{indent paragraph: 15pt}
About the Dirac operator, Dirac–Atiyah–Singer operator (on spin manifolds), and Index theorem(s), see \cite[chapp. 3-6]{Berline Getzler Vergne "Heat Kernels and Dirac Operators"} \cite[sec. 6]{Booss-Bavnbek Wojciechowski "Elliptic Boundary Problems for Dirac Operators"} \cite[chapp. 1-3]{Esposito "Dirac Operators and Spectral Geometry"} \cite[chap. 3]{Friedrich "Dirac Operators in Riemannian Geometry"} \cite{Freed "The Atiyah-Singer Index Theorem"}.
\end{indent paragraph: 15pt}

\noindent Section \ref{subsection "In the Bliss of Goddess Geometry"}

\begin{indent paragraph: 15pt}
For the homology 3-sphere(s), see \cite[chap. 1]{Saveliev "Invariants for Homology 3-Spheres"}.
\end{indent paragraph: 15pt}

\endgroup

\backmatter
\chapter[\textbf{Ulterius Elementum in Cauda}]{Ulterius Elementum in Cauda}
\markboth{Ulterius Elementum in Cauda}{Ulterius Elementum in Cauda}

\phantomsection
\addcontentsline{toc}{section}{Siren's Song: No-return to Ithaca}

\thispagestyle{empty}

\begin{center}
\textsc{siren's song: \\ no-return to ithaca}
\end{center} 
\label{section "Siren's Song: No-return to Ithaca"}

Hereunder are transcribed—after adjustment of some Greek words in the classical spelling—two Greek elegiac couplets that, together with the drawing of a two-tailed siren,\footnote{
	Initially, in Greek mythology, the sirens were typified as birds, or bird-like beings, with a woman's head; the iconographic mutation into marine creatures, as women-fish, is later.
	}
appear on the title page of several Venetian sixteenth-century editions, printed by the publisher G. Varisco:\footnote{
	See e.g. \textit{Isocratis Orationes tres, cum interpretatione latina, ad verbum addita, ad discentium utilitatem. Ad Demonicum, de moribus adolescentium. Ad Nicoclem Cypri reg\~e, de princip\~u institutione. Nicocles, de principum \& subditorum officio}, Venetiis apud Ioannem Variscum, \& socios, \textsc{mdlxvii} (1567).
	}

\begin{figure}[h!]
\centering
\includegraphics{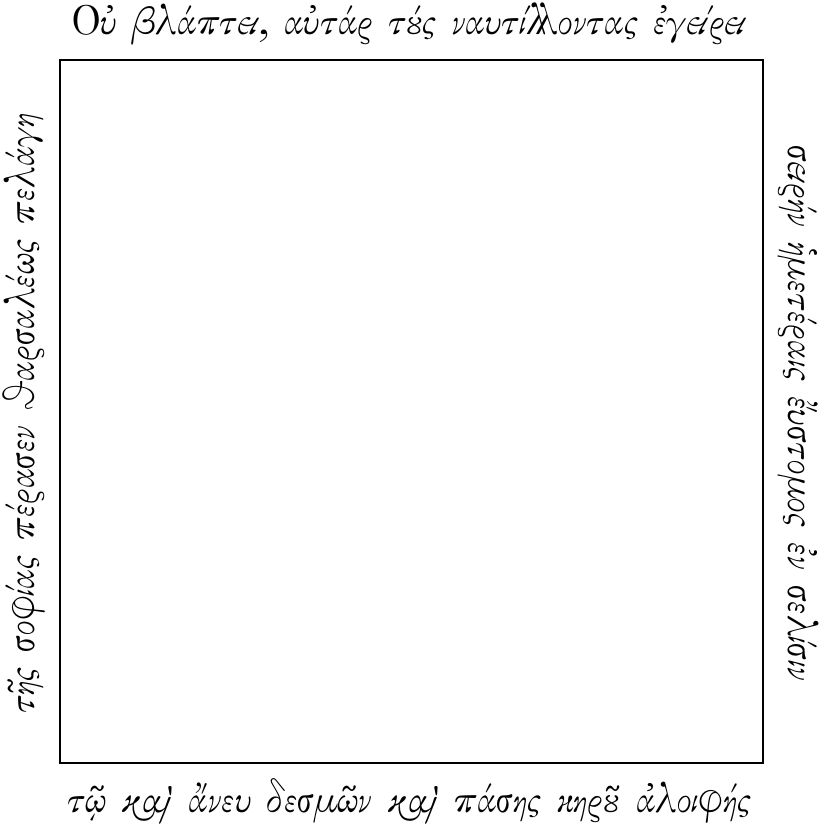}
\end{figure}

It should be read in a clockwise direction, starting at 12 o'clock;\footnote{
	\textgreek{«Οὐ βλάπτει, αὐτάρ τούς ναυτίλλοντας ἐγείρει /
	σειρήν ἡμετέραις ἔυστομος ἐν σελίσιν /
	τῷ καί ἄνευ δεσμῶν καὶ πάσης κηροῦ ἀλοιφής / 
	τῆς σοφίας πέρασεν θαρσαλέως πελάγη»}.
	}
of which the translation is: «A siren [\textgreek{\textit{σειρήν}}] with a beautiful voice [\textgreek{\textit{ἔυστομος}}]\footnote{
	\textgreek{Ἔυστομος} literally means “with a beautiful mouth”.
	}
in our pages [\textgreek{\textit{ἡμετέραις ἐν σελίσιν}}] does not harm [\textgreek{\textit{oὐ βλάπτει}}], on the contrary [\textgreek{\textit{αὐτάρ}}] she incites sailors [\textgreek{\textit{τούς ναυτίλλοντας ἐγείρει}}], for the one who [\textgreek{\textit{τῷ}}], without ropes [\textgreek{\textit{καὶ ἄνευ δεσμῶν}}] and [without] all the wax ointment [\textgreek{\textit{καὶ πάσης κηροῦ ἀλοιφής}}], has bravely [\textgreek{\textit{θαρσαλέως}}] crossed [\textgreek{\textit{πέρασεν}}] the seas of wisdom [\textgreek{\textit{τῆς σοφίας πελάγη}}]».

In other books of the Variscan editions, there is also a Latin version, which differs considerably from the Greek one:

\vspace{-1.3mm}

\begin{center}
\footnotesize
Non nocet hec nautis: verum excitat, \& regit ipsos, \\
	Siren, quæ in nostris conspicitur tabulis. \\
	Non vinclis opus est Ithacis, non unguine ceræ: \\ 
	virtutis per eam (nam potes) arripe iter.
\end{center}

\vspace{-1.3mm}

Odysseus' encounter with the sirens is here reinterpreted in an alternative way, reversing the Homeric tale.\footnote{
	These are the Homeric verses of the Odyssey \cite[XII, 39-54, pp. 435-437]{Homer "The Odyssey I"}: «[The] queenly Circe spoke to me [Odysseus] and said: “[\,\dots] To the Sirens first shalt thou come, who beguile all men whosoever comes to them. Whoso in ignorance draws near to them and hears the Sirens' voice, he nevermore returns [\,\dots], the Sirens beguile him with their clear-toned song [\,\dots]. But do thou row past them, and anoint the ears of thy comrades with sweet wax, which thou hast kneaded, lest any of the rest may hear. But if thou thyself hast a will to listen, let them bind thee in the swift ship hand and foot upright in the step of the mast, and let the ropes be made fast at the ends to the mast itself, that with delight thou mayest listen to the voice of the two Sirens. And if thou shalt implore and bid thy comrades to loose thee, then let them bind thee with yet more bonds”».
	
	For lovers of the It. language, there is the tantalizing translation (1822) by I. Pindemonte \cite[libro duodecimo, pp. 333-334]{Omero "Odissea Vol. I"}: «Le Sirene [\,\dots] / Mandano un canto dalle argute labbra, / Che alletta il passeggier [\,\dots]. Tu veloce oltrepassa, e con moll[ì]ta / Cera de' tuoi così l'orecchio tura, / Che non vi possa penetrar la voce. / Odila tu, se vuoi; sol che diritto / Te della nave all'albero i compagni / Leghino, e i piedi str[ì]nganti, e le mani: / Perchè il diletto di sentir la voce / Delle Sirene tu non perda».
	} 
The two poetic compositions (in the two ancient languages), commanding my admiration, give me the opportunity to reiterate that, when I am talking about mathematics as a \textgreek{τέχνη} or \textgreek{τέχνημα} (Section \ref{subsection "Mathematics as a Technical Tool"}), actually, I consider mathematics as \emph{culture}, to wit, as a sublime\footnote{	
	Literally: “lofty”, “borne aloft”, “uplifted”.
	} 
art of knowledge (such is the cognitive value of mathematics), or, if one prefers, as a tempestuous odyssey, an adventurous challenge of the mind \& a perilous rapture of the heart; and not, inanely, as a blind-cocooned technical know-how (cf. Intro, p. \pageref{subsection "Blind Specialism: Cocoon Syndrome"}), in an aberrant mercantile key, similar to what occurs in many Institutions.\footnote{
	As a general principle, when science is incorporated singly as a mere practical value, pure idiocy  is now firmly at the helm of the socio-economic degeneracy, and the extravagant dictates of bibliometrics (cf. footnote \ref{footnote "Ravings of bibliometrics"}, p. \pageref{footnote "Ravings of bibliometrics"}) accrete as teratological outgrowths.
	}
	
The image of mathematics, as an art that designedly chooses the magic of \emph{perdition} in the beauty of the song (\emph{canto dalle argute labbra}), and breaks the bonds of the chained will, is the best way to remember the union of the “two cultures”—I am referring to C.P. Snow's essay \cite{Snow "The Two Cultures: A Second Look}.\footnote{
	Cf. footnote \ref{footnote "C.P. Snow"} on p. \pageref{footnote "C.P. Snow"}.
	}

This is to lay stress on the fact that the \textgreek{τέχνη} of mathematics is on the \emph{border} of the «incomprehensible» (\textit{unerfaßlich}) and «unutterable» (\textit{unaussprechbar}), and aims, such as poetry,\footnote{
	It is even possible paroxysmally to say that mathematics is an epiphany, an emanation, of poetry; compare e.g. with P.B. Shelley \cite[p. 38, vv. 16-21]{Shelley "A Defense of Poetry"}, who speaks generically of science: «Poetry is [\,\dots] at once the centre and circumference of knowledge; it is that which comprehends all science, and that to which all science must be referred. It is at the same time the root and blossom of all other systems of thought».
	} 
at a word/concept, at a \textgreek{λόγος}, which is \emph{beyond that border}, i.e. \textit{jenseits der Sprache}, «beyond language», as H. Broch subtly wrote \cite[p. 533]{Broch "Der Tod des Vergil"}, for his dying Virgil.\footnote{
	Broch \cite[ibid.]{Broch "Der Tod des Vergil"}: «[T]he word [floating] beyond the expressible [\textit{Ausdrückbarem}] and the inexpressible [\textit{Nicht-Aus-drückbarem}] [\,\dots] into the flooding sound [\,\dots], a floating sea, sea-heavy, sea-light».
	}

\phantomsection
\addcontentsline{toc}{section}{A (Mathematical) Odyssey of Infinite Possibilities}

\begin{center}
\textsc{a (mathematical) odyssey of infinite possibilities}
\end{center} 

This wandering Odysseus seems to be able to reverse Horace's monitus \cite[Lib. I, XI, 27, p. 267]{Horatii Flacci "Epistularum"}: \textit{Caelum non animum mutant qui trans mare currunt} («Those who cross the sea change the sky [over their heads] but not their souls [minds]»); his traveling challenge overseas is not the way to reach a goal, but a means to reach oneself. The voyage of mathematics—the roamings of the mind—does approximately the same thing. In the perilous (mathematical) journey there is an unremitting change of mind. 

Now, we can lustily reconnects us to the Intro on p. \pageref{subsection "Through the Magnifier of Half-sleep"}: doing mathematics is like embarking on an \emph{Odyssean journey} into the \emph{kaleidoscopic and proteiform recesses of the unconscious}.

There is a paragraph by C. Magris \cite[pref., § 5, pp. 10-11, transl. slightly modified]{Magris "Journeying"}, which sums up well what is contained in the concept of traveling, and of our wandering Odysseus. Let us read it in full; it is worth the effort:

\vspace{2mm}

\begingroup
\footnotesize
Utopia and disillusionment. Many things are lost, when traveling: certainties, values, feelings, expectations fallen by the wayside—the road is harsh, but it is also a good teacher. Other things, other values and feelings are encountered, gathered,  picked up along the way. Like traveling, writing also involves disassembling, rearranging, putting back together; one moves through reality as in a theater of prose, shifting the curtains, opening new passages, getting lost in blind alleys and blocked by fake doors painted on the wall.

Reality, so often impenetrable, suddenly gives way, crumbles; the traveler, Cees Noteboom says, feels “the draft blowing through the cracks in the structure of causality”. What is real proves to be probabilistic, indeterministic, subject to sudden quantum collapses which cause some of its elements to vanish, swallowed up, sucked into spatio-temporal vortices, whirlpools of the mortality of all things, but also of the unforeseeable emergence of new life.

Traveling is a Musilian experience, committed to the sense of possibility rather than to the reality principle. One discovers, as in an archaeological dig, different strata of reality, concrete possibilities which were not materially realized but which existed and survived as shards forgotten in the rush of time, in still unlock opening, in still fluctuating states. Traveling entails coming to terms with reality but also with its alternatives, its gaps; with History and with a different history or other histories precluded and deterred by it, but not entirely stamped out.

Since the \textit{Odyssey}, travel and literature have appeared closely related;\footnote{
	We are closing off this Chapter with a warning, and so we can annularly connect us to what we denounced in the Intro, pp. \pageref{subsection "Metaphorical Procedure: the Example of the Crystals"} and \pageref{subsection "Blind Specialism: Cocoon Syndrome"}. The task of the intellectual (\textgreek{πολυμαθής}, as we used to say once), unique figure who clasps scientific and classical disciplines, is to make people understand that science and literature are like a train, with many (and different) wagons, traveling on the same rail. C. Magris \cite[pp. 25-26, e.a.]{Magris "Utopia e disincanto"} points out that: «Writers such as Musil, Joyce, Proust, Kafka, Svevo, Mann, Broch, Faulkner and others have asked [the realm of] fiction [a] knowledge of the world that precisely the enormous development of the sciences did not allow to entrust to the latter, because they—with their \emph{extreme specialization} that made each [field of study] \emph{inaccessible} to lovers of all other [fields] and even more to the average man—had shattered any \emph{sense of unity of the world} [\,\dots]. Today, literature has a new challenge, which arises from the gap with respect to science [one should think of quantum mechanics] and from the gap between scientific knowledge and any possibility [of this knowledge] of becoming part of the common cultural heritage [\,\dots]. The universe does not necessarily have to be organized under the laws corresponding to the structures of the mind and human perception; transforming the increasingly abstract knowledge of an indeterministic nature [one should think of present-day mathematical physics] into a \emph{poetic metaphor} is the arduous cultural challenge that is posed to literature today».
	} 
an analogous exploration, deconstruction, and recognition of the world and of the I. Writing continues the relocation, it packs and unpacks, rearranges things, shifts plenums and voids, discovers—invents? finds?—elements that escaped the list, and even the perception of what is real, almost as though putting them under a magnifying glass.\endnote{
	Original It. version \cite[pref., § 5]{Magris "L'infinito viaggiare"}: «Utopia e disincanto. Molte cose cadono, quando si viaggia; certezze, valori, sentimenti, aspettative che si perdono per strada — la strada è una dura, ma anche buona maestra. Altre cose, altri valori e sentimenti si trovano, s'incontrano, si raccattano per via. Come viaggiare, pure scrivere significa smontare, riassestare, ricombinare; si viaggia nella realtà come in un teatro di prosa, spostando le quinte, aprendo nuovi passaggi, perdendosi in vicoli ciechi e bloccandosi davanti a false porte disegnate sul muro. 

	\setlength\parindent{8pt}
	La realtà, così spesso impenetrabile, d'improvviso cede, si sfalda; il viaggiatore, dice Cees Noteboom, sente “gli spifferi dalle fessure dell'edificio causale”. Il reale si rivela probabilistico, indeterministico, soggetto a improvvisi collassi quantici che fanno sparire alcuni suoi elementi, inghiottiti, risucchiati in vortici dello spazio-tempo, mulinelli della mortalità di tutte le cose, ma anche dell'imprevedibile emergere di nuova vita.

	Viaggiare è un'esperienza musiliana, affidata al senso delle possibilità piuttosto che al principio di realtà. Si scoprono, come in uno scavo archeologico, altri strati del reale, le possibilità concrete che non si sono materialmente realizzate ma esistevano e sopravvivono in brandelli dimenticati dalla corsa del tempo, in varchi ancora aperti, in stati ancora fluttuanti. Viaggiare significa fare i conti con la realtà ma anche con le sue alternative, con i suoi vuoti; con la Storia e con un'altra storia o con altre storie da essa impedite e rimosse, ma non del tutto cancellate.

	Fin dall'\textit{Odissea}, viaggio e letteratura appaiono strettamente legati; un'analoga esplorazione, decostruzione e ricognizione del mondo e dell'io. La scrittura continua il trasloco, impacca e disfa, aggiusta, sposta i vuoti e i pieni, scopre — inventa? trova? — elementi sfuggiti all'inventario e perfino alla percezione del reale, quasi li ponesse sotto una lente d'ingrandimento».
	}

\endgroup

\vspace{2mm}

Which is flawlessly suited to the writing of mathematics; this last word comes from the Gr. noun \textgreek{μάθημα}, “that which is learned”, and so “that which has never a conclusion”, being an infinite process, scattered into countless possibilities—concisely, mathematics as \emph{a science, viz. an art, of possibilities rather than of reality}.

\phantomsection
\addcontentsline{toc}{section}{Fantasy and Invention: the Art of the Possible (On some verses of Galileo)}

\begin{center}
\textsc{fantasy and invention: the art of the possible \\ (on some verses of galileo)}
\end{center} 

Galileo, in his spare time, was also a poet; in \cite[\textit{Contro il portar la Toga},\footnote{
	\textit{Against the donning of the Gown} (1590), where the academic gown, or rather, all those academics who wear it, are despised in no uncertain terms \cite[298-301, p. 223]{Galileo "Poesie e frammenti"}: «[Quegli accademici] ch'han quelle veste delicate, / Se tu gli tasti, o son pieni di vento, / O di belletti o d'acque profumate, / O son fiascacci da pisciarvi drento». Not unexpectedly, it follows that the academic cosmetics, here symbolized by the gown, for Galileo (but also for me), is the enemy of fantasy and invention.
	} 
11-13, pp. 213-214]{Galileo "Poesie e frammenti"} he writes:
	
\vspace{-1.3mm}
	
\begin{center}
\footnotesize
A chi vuol una cosa ritrovare, \\
Bisogna adoperar la fantasia, \\
E giocar d'invenzione, e 'ndovinare.
\end{center}
	
\vspace{-1.3mm}

In translation: «Whoever wants to find something / must use [his] fantasy, / play with invention, and he has to guess». They are prodigious verses that sum up the spirit of many previous Chapters, in which the role of fantasy and invention, in mathematical and physical research, is stoutly exalted. Indeed: the art of the possible,\footnote{
	It is not accurate, therefore, to define the fantasy as the faculty of imagining impossible things.	
		} 
on which mathematics—and any mathematical structure in physics—feeds, passes through the doors of fantasy and invention.
\theendnotes
\setcounter{footnote}{0} 

\cleardoublepage
\phantomsection
\addcontentsline{toc}{chapter}{\bibname}



\newpage\null\thispagestyle{empty}\newpage

\thispagestyle{empty}

\vspace*{\fill}
\begingroup
\begin{center}
\textsc{finis}
\end{center}
\endgroup
\vspace*{\fill}

\end{document}